\PassOptionsToPackage{labelfont=bf,labelsep=period}{caption}
\PassOptionsToPackage{colorlinks=true,linkcolor=SaddleBrown,urlcolor=SeaGreen,citecolor=Crimson,hyperfigures,hyperfootnotes}{hyperref}
\documentclass[10pt,a4paper,english,twoside]{llncs}

\usepackage{etoolbox}
\usepackage{imakeidx}
\usepackage{enumerate}
\usepackage[vlined]{algorithm2e}
\usepackage[utf8]{inputenc}
\usepackage[T1]{fontenc}
\usepackage{multirow}
\usepackage{url}
\usepackage{amsmath,amssymb}
\usepackage{epsfig}
\usepackage[usenames,svgnames]{xcolor}
\usepackage{listings}
\usepackage{acronym}
\usepackage{wrapfig}
\usepackage{graphicx}
\usepackage{rotating}
\usepackage{colortbl}
\usepackage{pifont}
\usepackage{flafter}
\usepackage{tabularx}
\usepackage{nth}
\usepackage{fourier}
\usepackage{marvosym}
\usepackage{lettrine}
\usepackage{xkeyval}
\usepackage{makecell}
\usepackage{bookmark}
\usepackage{subfig}
\usepackage{fancyhdr}
\usepackage[strict]{changepage}
\usepackage{eso-pic}
\usepackage{lastpage}
\usepackage{trfsigns}
\usepackage[nomain,acronym,shortcuts]{glossaries} 
\usepackage[all]{hypcap}
\usepackage{cleveref}
\usepackage{caption}
\usepackage{longtable}

\noist

\definecolor{notparticipating}{RGB}{225,225,225}

\definecolor{nofailure}{RGB}{0,0,0}
\definecolor{txtnofailure}{RGB}{255,255,255}

\definecolor{bestscore}{RGB}{90,90,90}
\definecolor{txtbestscore}{RGB}{255,255,255}

\definecolor{ednf}{RGB}{255,255,255}
\definecolor{txtednf}{RGB}{0,0,0}

\definecolor{movf}{RGB}{255,255,255}
\definecolor{txtmovf}{RGB}{0,0,0}

\definecolor{sovf}{RGB}{255,255,255}
\definecolor{txtsovf}{RGB}{100,100,100}

\input{macros-al.tex}

\date{}
\title{
  Report on the Model Checking Contest
\\
  at Petri Nets 2013
}
\author{
     F. Kordon\inst{1}
\and A. Linard\inst{2}
\and
    \\
     M. Beccuti\inst{3}
\and D. Buchs\inst{4}
\and \L. Fronc\inst{5}
\and L.M. Hillah\inst{8}
\and F. Hulin-Hubard\inst{2}
\and F. Legond-Aubry\inst{8}
\and \\
     N. Lohmann\inst{6}
\and A. Marechal\inst{4}
\and E. Paviot-Adet\inst{9}
\and F. Pommereau\inst{4}
\and C. Rodr\'\i guez\inst{2}
\and C. Rohr\inst{7}
\and \\
     Y. Thierry-Mieg\inst{1}
\and H. Wimmel\inst{6}
\and K. Wolf\inst{6}
}

\institute{
     LIP6, CNRS UMR 7606, Université P. \& M. Curie~--~Paris~6 \\
     4, place Jussieu, F-75252 Paris Cedex 05, France\\
     \email{Fabrice.Kordon@lip6.fr},
     \email{Yann.Thierry-Mieg@lip6.fr}
\and
    LSV, Inria/\'Ecole Normale Sup\'erieure de Cachan,\\
    61, avenue du Président Wilson, Cachan, France \\
    \email{Alban.Linard@lsv.ens-cachan.fr},
	\email{fhh@lsv.ens-cachan.fr},
	\email{cesar.rodriguez@lsv.ens-cachan.fr}
\and
     Dipartimienti di Informatica, Univ. di Torino \\
     Corso Svizzera, 185, 10149 Torino, Italy\\
     \email{beccuti@di.unito.it}
\and
     Centre Universitaire d'Informatique, Université de Genève \\
     7, route de Drize, CH-1227 Carouge, Switzerland \\
     \email{Didier.Buchs@unige.ch} \\
\and
     IBISC, Université d'Évry Val d'Essonne \\
     22 Boulevard de France, 91037 Évry Cedex France\\
     \email{lfronc@ibisc.univ-evry.fr}, \email{franck.pommereau@ibisc.univ-evry.fr} \\
\and
     Universität Rostock, 
     18051 Rostock, Germany \\
     \email{niels.lohmann@uni-rostock.de},
     \email{Wimmel@Informatik.Uni-Oldenburg.de}, \email{karsten.wolf@uni-rostock.de}
\and
     Brandenburg University of Technology at Cottbus \\
     Postbox 10 13 44, 03013 Cottbus, Germany \\
     \email{rohrch@tu-cottbus.de}
\and
    LIP6, CNRS UMR 7606 and Universit\'e Paris Ouest Nanterre La D\'efense\\
	200, avenue de la R\'epublique, F-92001 Nanterre CEDEX, FRANCE\\
	\email{Lom-Messan.Hillah@lip6.fr}, \email{Fabrice.Legond-Aubry@lip6.fr}
\and
    LIP6, CNRS UMR 7606 and Universit\'e Ren\' Descartes\\
    143 Avenue de Versailles, 75016, Paris, France\\
	\email{Emmanuel.Paviot-Adet@lip6.fr}
}

\crefname{algocf}{Algorithm}{Alg.}
\DontPrintSemicolon

\newcommand{\BK}[0]{\textsf{\emph{B}ench\emph{K}it}}

\lstdefinelanguage{command}%
{ basicstyle=\normalsize\ttfamily
, flexiblecolumns=true
}%

\lstdefinelanguage{title}%
{ basicstyle=\normalsize\ttfamily
, flexiblecolumns=true
}%
\lstdefinelanguage{body}%
{ basicstyle=\footnotesize\ttfamily
, flexiblecolumns=true
}%

\makeindex[options=-s mcc, intoc, columns=2]
\makeglossaries

\newacronym[user1=MCC]{MCC}{%
  Model Checking Contest%
}{%
  Model Checking Contest @ Petri nets%
}
\newacronym[user1=MCC!2011]{MCC2011}{%
  MCC'2011%
}{%
  Model Checking Contest 2011%
}
\newacronym[user1=MCC!2012]{MCC2012}{%
  MCC'2012%
}{%
  Model Checking Contest 2012%
}
\newacronym[user1=MCC!2013]{MCC2013}{%
  MCC'2013%
}{%
  Model Checking Contest 2013%
}

\newacronym{SUMo}{SUMo}{International Workshop on Scalable and Usable Model Checking for Petri nets and other models of concurrency}
\newacronym{DD}{DD}{decision diagram}
\newacronym{IDD}{IDD}{Interval Decision Diagram}
\newacronym{SigmaDD}{\ensuremath{\Sigma}DD}{\ensuremath{\Sigma}~Decision Diagram}
\newacronym{ZDD}{ZDD}{Zero-Suppressed Decision Diagram}
\newacronym{ZMDD}{ZMDD}{Zero-Suppressed Multi-Terminal Decision Diagram}
\newacronym{BDD}{BDD}{Binary Decision Diagram}
\newacronym{DDD}{DDD}{Data Decision Diagram}
\newacronym{SDD}{SDD}{Set Decision Diagram}
\newacronym{PN}{Petri net}{Petri net}
\newacronym{GSPN}{GSPN}{Generalized Stochastic Petri Nets}
\newacronym{PT}{P/T}{Place/Transition net}
\newacronym{SN}{SN}{Symmetric net}
\newacronym{SNB}{SNB}{Symmetric nets with Bags}
\newacronym{CN}{Colored}{Colored net}
\newacronym{APN}{APN}{Algebraic Petri net}
\newacronym{HLPN}{HLPN}{High Level Petri net}
\newacronym{ITS}{ITS}{Instantiable Transition Systems}
\newacronym{AADT}{AADT}{Abstract Algebraic Data Type}
\newacronym{CTL}{CTL}{Computation Tree Logic}
\newacronym{LTL}{LTL}{Linear Temporal Logic}
\newacronym[user1=PNML]{PNML}{PNML}{Petri Net Markup Language}
\newacronym[user1=Model! Definition!CSRepetitions]{CSRepetitions}{%
  CSRepetitions%
}{%
  Client/Server with repetitions%
}

\newacronym[user1=Model!Colored!CSRepetitions]{CSRepetitions-COL}{%
  CSRepetitions (colored)%
}{%
  Client/Server with repetitions (colored)%
}

\newacronym[user1=Model!P/T!CSRepetitions]{CSRepetitions-PT}{%
  CSRepetitions (P/T)%
}{%
  Client/Server with repetitions (P/T)%
}

\newacronym[user1=Model! Definition!Echo]{Echo}{%
  Echo%
}{%
  Echo algorithm%
}

\newacronym[user1=Model! Definition!Echo!P/T]{Echo-PT}{%
  Echo (P/T)%
}{%
  Echo algorithm (P/T)%
}

\newacronym[user1=Model! Definition!Eratosthenes]{Eratosthenes}{%
  Eratosthenes%
}{%
  Eratosthenes' sieve%
}

\newacronym[user1=Model! Definition!FMS]{FMS}{%
  FMS%
}{%
  Flexible Manufacturing System%
}

\newacronym[user1=Model! Definition!GlobalRessAlloc]{GlobalRessAlloc}{%
  GlobalRessAlloc%
}{%
  Global Allocation Resource Management%
}

\newacronym[user1=Model! Definition!Kanban]{Kanban}{%
  Kanban%
}{%
  Kanban System%
}

\newacronym[user1=Model! Definition!LamportFastMutEx]{LamportFastMutEx}{%
  LamportFastMutEx%
}{%
  Lamport's fast mutual exclusion algorithm%
}

\newacronym[user1=Model! Definition!MAPK]{MAPK}{%
  MAPK%
}{%
  Mitogen-Activated Protein Kinase Kascade%
}

\newacronym[user1=Model! Definition!NeoElection]{NeoElection}{%
  NeoElection%
}{%
  Neo Election Protocol%
}

\newacronym[user1=Model! Definition!Peterson]{Peterson}{%
  Peterson%
}{%
  Peterson's mutual exclusion algorithm%
}

\newacronym[user1=Model! Definition!PhilosophersDyn]{PhilosophersDyn}{%
  PhilosophersDyn%
}{%
  Dynamic Dining Philosophers%
}

\newacronym[user1=Model! Definition!Philosophers]{Philosophers}{%
  Philosophers%
}{%
  Dining Philosophers%
}

\newacronym[user1=Model! Definition!Planning]{Planning}{%
  Planning%
}{%
  AI Planning%
}

\newacronym[user1=Model! Definition!Railroad]{Railroad}{%
  Railroad%
}{%
  Railroad%
}
\newacronym[user1=Model! Definition!Ring]{Ring}{%
  Ring%
}{%
  Three-Module Ring%
}

\newacronym[user1=Model! Definition!RwMutex]{RwMutex}{%
  RwMutex%
}{%
  Reader/Writer Mutual Exclusion%
}

\newacronym[user1=Model! Definition!SharedMemory]{SharedMemory}{%
  SharedMemory%
}{%
  Shared Memory%
}

\newacronym[user1=Model! Definition!SimpleLoadBal]{SimpleLoadBal}{%
  SimpleLoadBal%
}{%
  Simple Load Balancing System%
}

\newacronym[user1=Model! Definition!TokenRing]{TokenRing}{%
  TokenRing%
}{%
  Token Ring%
}

\newacronym[user1=Model! Definition!Dekker]{Dekker}{%
  Dekker%
}{%
  Dekker's algorithm%
}

\newacronym[user1=Model! Definition!DotAndBoxes]{DotAndBoxes}{%
  DotAndBoxes%
}{%
  DotAndBoxes%
}

\newacronym[user1=Model! Definition!DrinkVendingMachine]{DrinkVendingMachine}{%
  DrinkVendingMachine%
}{%
  Drinks Vending Machine%
}

\newacronym[user1=Model (surprise)! Definition!HouseConstruction]{HouseConstruction}{%
  HouseConstruction%
}{%
  House Construction%
}

\newacronym[user1=Model (surprise)! Definition!IBMB2S565S3960]{IBMB2S565S3960}{%
  IBMB2S565S3960%
}{%
  IBMB2S565S3960%
}

\newacronym[user1=Model! Definition!PermAdmissibility]{PermAdmissibility}{%
  PermAdmissibility%
}{%
  permutation admissibility%
}

\newacronym[user1=Model (surprise)! Definition!QuasiCertifProtocol]{QuasiCertifProtocol}{%
  QuasiCertifProtocol%
}{%
  quasi certification protocol%
}

\newacronym[user1=Model! Definition!RessAllocation]{RessAllocation}{%
  RessAllocation%
}{%
  Resource Allocation%
}

\newacronym[user1=Model (surprise)! Definition!Vasy2003]{Vasy2003}{%
  Vasy2003%
}{%
  Vasy2003%
}
\newacronym[user1=Model!P/T!Eratosthenes]{Eratosthenes-PT}{%
  Eratosthenes (P/T)%
}{%
  Eratosthenes' sieve (P/T)%
}

\newacronym[user1=Model!P/T!FMS]{FMS-PT}{%
  FMS (P/T)%
}{%
  Flexible Manufacturing System (P/T)%
}

\newacronym[user1=Model!Colored!GlobalRessAlloc]{GlobalRessAlloc-COL}{%
  GlobalRessAlloc (colored)%
}{%
  Global Allocation Resource Management (colored)%
}

\newacronym[user1=Model!P/T!GlobalRessAlloc]{GlobalRessAlloc-PT}{%
  GlobalRessAlloc (P/T)%
}{%
  Global Allocation Resource Management (P/T)%
}

\newacronym[user1=Model!P/T!Kanban]{Kanban-PT}{%
  Kanban (P/T)%
}{%
  Kanban System (P/T)%
}

\newacronym[user1=Model!P/T!LamportFastMutEx]{LamportFastMutEx-PT}{%
  LamportFastMutEx (P/T)%
}{%
  Lamport's fast mutual exclusion algorithm (P/T)%
}

\newacronym[user1=Model!Colored!LamportFastMutEx]{LamportFastMutEx-COL}{%
  LamportFastMutEx (colored)%
}{%
  Lamport's fast mutual exclusion algorithm (colored)%
}

\newacronym[user1=Model!P/T!MAPK]{MAPK-PT}{%
  MAPK (P/T)%
}{%
  Mitogen-Activated Protein Kinase Kascade (P/T)%
}

\newacronym[user1=Model!Colored!NeoElection]{NeoElection-COL}{%
  NeoElection (colored)%
}{%
  Neo Election Protocol (colored)%
}

\newacronym[user1=Model!P/T!NeoElection]{NeoElection-PT}{%
  NeoElection (P/T)%
}{%
  Neo Election Protocol (P/T)%
}

\newacronym[user1=Model!Colored!Peterson]{Peterson-COL}{%
  Peterson (colored)%
}{%
  Peterson's mutual exclusion algorithm (colored)%
}

\newacronym[user1=Model!P/T!Peterson]{Peterson-PT}{%
  Peterson (P/T)%
}{%
  Peterson's mutual exclusion algorithm (P/T)%
}

\newacronym[user1=Model!Colored!PhilosophersDyn]{PhilosophersDyn-COL}{%
  PhilosophersDyn (colored)%
}{%
  Dynamic Dining Philosophers (colored)%
}

\newacronym[user1=Model!P/T!PhilosophersDyn]{PhilosophersDyn-PT}{%
  PhilosophersDyn (P/T)%
}{%
  Dynamic Dining Philosophers (P/T)%
}

\newacronym[user1=Model!Colored!Philosophers]{Philosophers-COL}{%
  Philosophers (colored)%
}{%
  Dining Philosophers (colored)%
}

\newacronym[user1=Model!P/T!Philosophers]{Philosophers-PT}{%
  Philosophers (P/T)%
}{%
  Dining Philosophers (P/T)%
}

\newacronym[user1=Model!P/T!Planning]{Planning-PT}{%
  Planning (P/T)%
}{%
  AI Planning (P/T)%
}

\newacronym[user1=Model!P/T!Railroad]{Railroad-PT}{%
  Railroad (P/T)%
}{%
  Railroad (P/T)%
}
\newacronym[user1=Model!P/T!Ring]{Ring-PT}{%
  Ring (P/T)%
}{%
  Three-Module Ring (P/T)%
}

\newacronym[user1=Model!P/T!RwMutex]{RwMutex-PT}{%
  RwMutex (P/T)%
}{%
  Reader/Writer Mutual Exclusion (P/T)%
}

\newacronym[user1=Model!P/T!Dekker]{Dekker-PT}{%
  Dekker (P/T)%
}{%
  Dekker's algorithm (P/T)%
}

\newacronym[user1=Model (surprise)!P/T!HouseConstruction]{HouseConstruction-PT}{%
  HouseConstruction (P/T)%
}{%
  House Construction (P/T)%
}

\newacronym[user1=Model (surprise)!P/T!IBMB2S565S3960]{IBMB2S565S3960-PT}{%
  IBMB2S565S3960 (P/T)%
}{%
  IBMB2S565S3960 (P/T)%
}

\newacronym[user1=Model!P/T!RessAllocation]{RessAllocation-PT}{%
  RessAllocation (P/T)%
}{%
  Resource Allocation (P/T)%
}

\newacronym[user1=Model (surprise)!P/T!Vasy2003]{Vasy2003-PT}{%
  Vasy2003 (P/T)%
}{%
  Vasy2003 (P/T)%
}

\newacronym[user1=Model!P/T!SharedMemory]{SharedMemory-PT}{%
  SharedMemory (P/T)%
}{%
  Shared Memory (P/T)%
}

\newacronym[user1=Model!P/T!SimpleLoadBal]{SimpleLoadBal-PT}{%
  SimpleLoadBal (P/T)%
}{%
  Simple Load Balancing System (P/T)%
}

\newacronym[user1=Model!P/T!TokenRing]{TokenRing-PT}{%
  TokenRing (P/T)%
}{%
  Token Ring (P/T)%
}

\newacronym[user1=Model!P/T!DotAndBoxes]{DotAndBoxes-PT}{%
  DotAndBoxes (P/T)%
}{%
  DotAndBoxes (P/T)%
}

\newacronym[user1=Model!P/T!DrinkVendingMachine]{DrinkVendingMachine-PT}{%
  DrinkVendingMachine (P/T)%
}{%
  Drinks Vending Machine (P/T)%
}

\newacronym[user1=Model!P/T!PermAdmissibility]{PermAdmissibility-PT}{%
  PermAdmissibility (P/T)%
}{%
  permutation admissibility (P/T)%
}

\newacronym[user1=Model (surprise)!P/T!QuasiCertifProtocol]{QuasiCertifProtocol-PT}{%
  QuasiCertifProtocol (P/T)%
}{%
  quasi certification protocol (P/T)%
}

\newacronym[user1=Model!Colored!SharedMemory]{SharedMemory-COL}{%
  SharedMemory (colored)%
}{%
  Shared Memory (colored)%
}

\newacronym[user1=Model!Colored!SimpleLoadBal]{SimpleLoadBal-COL}{%
  SimpleLoadBal (colored)%
}{%
  Simple Load Balancing System (colored)%
}

\newacronym[user1=Model!Colored!TokenRing]{TokenRing-COL}{%
  TokenRing (colored)%
}{%
  Token Ring (colored)%
}

\newacronym[user1=Model!Colored!DotAndBoxes]{DotAndBoxes-COL}{%
  DotAndBoxes (colored)%
}{%
  DotAndBoxes (colored)%
}

\newacronym[user1=Model!Colored!DrinkVendingMachine]{DrinkVendingMachine-COL}{%
  DrinkVendingMachine (colored)%
}{%
  Drinks Vending Machine (colored)%
}

\newacronym[user1=Model!Colored!PermAdmissibility]{PermAdmissibility-COL}{%
  PermAdmissibility (colored)%
}{%
  permutation admissibility (colored)%
}

\newacronym[user1=Model (surprise)!Colored!QuasiCertifProtocol]{QuasiCertifProtocol-COL}{%
  QuasiCertifProtocol (colored)%
}{%
  quasi certification protocol (colored)%
}

\newacronym[user1=Tool!ACTIVITY-LOCAL]{ACTIVITY-LOCAL}{%
  \lstinline!ACTIVITY-LOCAL!%
}{%
  \lstinline!ACTIVITY-LOCAL!%
}

\newacronym[user1=Tool!AlPiNA]{alpina}{%
  \lstinline!AlPiNA!%
}{%
  \lstinline!AlPiNA!%
}

\newacronym[user1=Tool!Crocodile]{crocodile}{%
  \lstinline!Crocodile!%
}{%
  \lstinline!Crocodile!%
}

\newacronym[user1=Tool!Cunf]{cunf}{%
  \lstinline!Cunf!%
}{%
  \lstinline!Cunf!%
}

\newacronym[user1=Tool!GreatSPN]{greatspn}{%
  \lstinline!GreatSPN!%
}{%
  \lstinline!GreatSPN!%
}

\newacronym[user1=Tool!Helena]{helena}{%
  \lstinline!Helena!%
}{%
  \lstinline!Helena!%
}

\newacronym[user1=Tool!ITS-Tools]{its-tools}{%
  \lstinline!ITS-Tools!%
}{%
  \lstinline!ITS-Tools!%
}

\newacronym[user1=Tool!LoLA!(pessimistic)]{lola-pessimistic}{%
  \lstinline!LoLa pess!%
}{%
  \lstinline!LoLa pessimistic!%
}

\newacronym[user1=Tool!LoLA!(optimistic)]{lola-optimistic}{%
  \lstinline!LoLa opt!%
}{%
  \lstinline!LoLa optimistic!%
}

\newacronym[user1=Tool!LoLA!optimistic incomplete]{lola-optimistic-incomplete}{%
  \lstinline!LoLa opt inc!%
}{%
  \lstinline!LoLa optimistic incomplete!%
}

\newacronym[user1=Tool!LoLA]{lola}{%
  \lstinline!LoLA!%
}{%
  \lstinline!LoLA!%
}

\newacronym[user1=Tool!Marcie]{marcie}{%
  \lstinline!Marcie!%
}{%
  \lstinline!Marcie!%
}

\newacronym[user1=Tool!Neco]{neco}{%
  \lstinline!Neco!%
}{%
  \lstinline!Neco!%
}

\newacronym[user1=Tool!PNXDD]{pnxdd}{%
  \lstinline!PNXDD!%
}{%
  \lstinline!PNXDD!%
}

\newacronym[user1=Tool!PeTe]{pete}{%
  \lstinline!PeTe!%
}{%
  \lstinline!PeTe!%
}

\newacronym[user1=Tool!Sara]{sara}{%
  \lstinline!Sara!%
}{%
  \lstinline!Sara!%
}

\newacronym[user1=Tool!SMART]{smart}{%
  \lstinline!SMART!%
}{%
  \lstinline!SMART!%
}

\newacronym[user1=Tool!YASPA]{yaspa}{%
  \lstinline!YASPA!%
}{%
  \lstinline!YASPA!%
}

\newacronym[user1=Tool!UPPAAL]{uppaal}{%
  \lstinline!UPPAAL!%
}{%
  \lstinline!UPPAAL!%
}

\renewcommand\subsubsection{\@startsection{subsubsection}{3}{\z@}%
 {1em}%
 {.5em}%
 {\reset@font\normalsize\bfseries}}

\makeatletter
\renewcommand\section{\@startsection{section}{1}{\z@}%
 {1em}%
 {.5em}%
 {\reset@font\large\bfseries}}
\renewcommand\subsection{\@startsection{subsection}{2}{\z@}%
 {1em}%
 {.5em}%
 {\reset@font\normalsize\bfseries}}
\renewcommand\subsubsection{\@startsection{subsubsection}{3}{\z@}%
  {1em}%
  {-1em}%
  {\normalfont\normalsize\bfseries}}
\makeatother

\makeatletter
\apptocmd{\@gls@link}{\index{\glsentryuseri{#2}}}{}{}
\makeatother

\begin{document}
\pagenumbering{roman}
\thispagestyle{empty}
\input Starburst.fd
\newcommand\CoverPicture{
\put(-5,-100){
\parbox[b][\paperheight]{\paperwidth}{%
\vfill
\centering
\includegraphics[width=23.5cm,keepaspectratio]{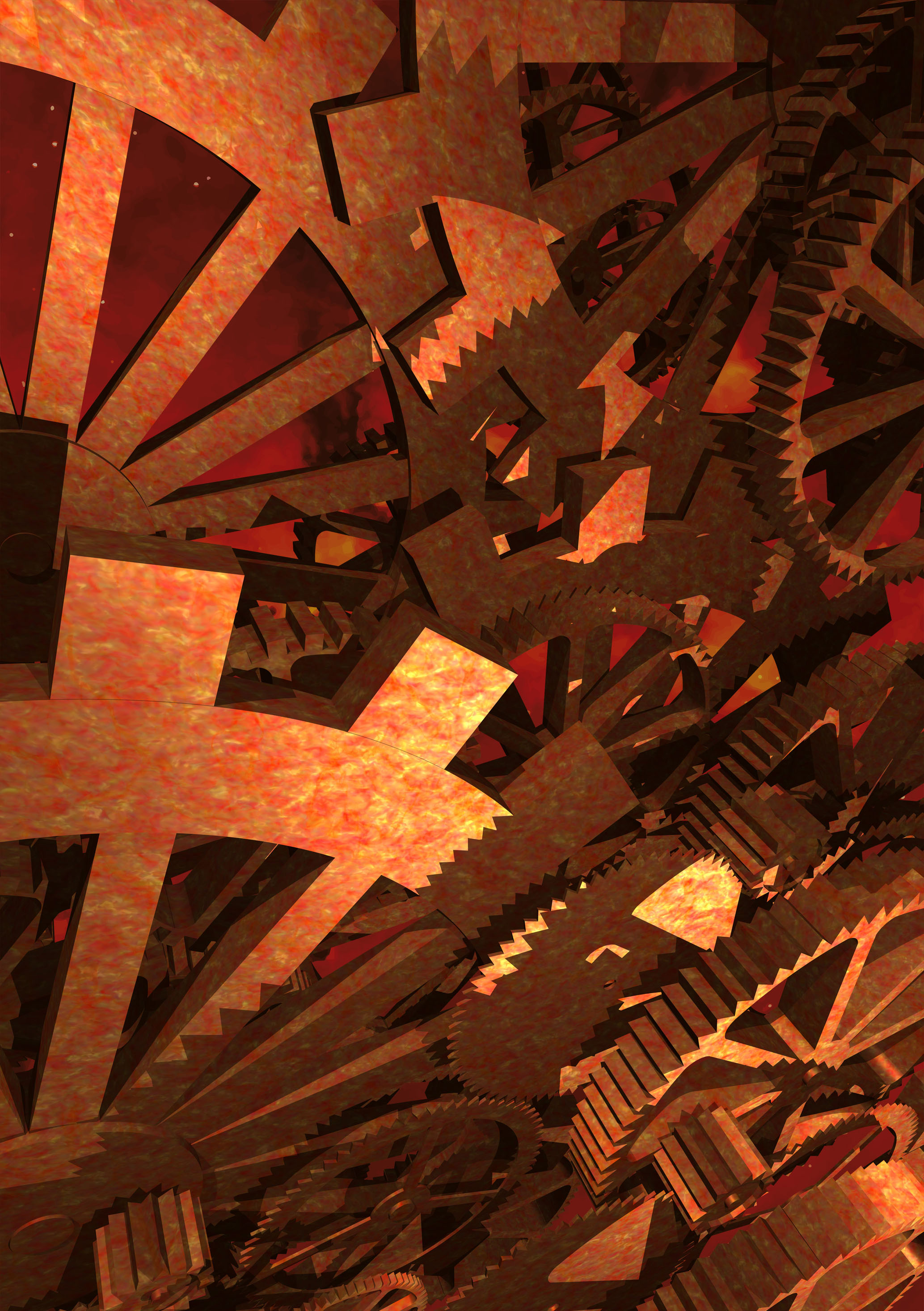}%
\vfill
}}}
\AddToShipoutPicture{\CoverPicture}
\begin{center}
\color{white}
{\bfseries\usefont{U}{Starburst}{xl}{n}
\mbox{}
\vfill
\Large
\textsf{* * *}
\vfill 
{\fontsize{10mm}{10mm}\selectfont M\fontsize{7mm}{7mm}\selectfont odel
\fontsize{10mm}{10mm}\selectfont C\fontsize{7mm}{7mm}\selectfont hecking
\fontsize{10mm}{10mm}\selectfont C\fontsize{7mm}{7mm}\selectfont ontest}
\\\huge
\textsf{@} Petri Nets
\vskip 0.5cm
Report on the 2013 edition
\vskip 1.0cm}
\huge
September 2013
\vskip 2.0cm
\begin{minipage}{.9\textwidth}
\Large\centering
     F. Kordon,
     A. Linard,
     \par
     M. Beccuti,
     D. Buchs,
     \L. Fronc,
     L.M. Hillah,
     \par
     F. Hulin-Hubard,
     F. Legond-Aubry,
     N. Lohmann,
     A. Marechal,
     \par
     E. Paviot-Adet,
     F Pommereau,
     C. Rodr\'\i guez,
     C. Rohr,
     \par
     Y. Thierry-Mieg,
     H. Wimmel,
     K. Wolf
\end{minipage}
\vfill
\vskip 0.4cm
\textsf{\Large* * *}
\vfill
\vskip 9.0cm
\mbox{}
\end{center}
\clearpage
\renewcommand\CoverPicture{}
\thispagestyle{empty}
\cleardoublepage

\setcounter{tocdepth}{2}
\tableofcontents
\cleardoublepage

\cfoot{\thepage}
\pagenumbering{arabic}
\maketitle


\begin{abstract}

This document presents the results of the Model Checking Contest held at Petri
Nets 2013 in Milano. This contest aimed at a fair and experimental evaluation
of the performances of model checking techniques applied to Petri nets. This
is the third edition after two successful editions in 2011~\cite{mcc2011} and
2012~\cite{mcc2012}.

The participating tools were compared on several examinations (state space
generation and evaluation of several types of formul{\ae} -- reachability,
LTL, CTL for various classes of atomic propositions) run on a set of common
models (Place/Transition and Symmetric Petri nets).

After a short overview of the contest, this paper provides the raw results
from the contest, model per model and examination per examination. An HTML
version of this report is also provided~\cite{mcc2013web}.

\end{abstract}


\part{Organization of the Model~Checking~Contest}
\label{part:one}
\section{Introduction}
\label{sec:intro}

When verifying by model checking a system with formal methods, such as
\aclp{PN}, one may have several questions such as:

\begin{quotation}
``When creating the model of a system,
should we use structural analysis or an explicit model checker to debug the model?''

``When verifying the final model of a highly concurrent system,
should we use a symmetry-based or a partial order reduction-based model checker?''

``When updating a model with large variable domains,
should we use a decision diagram-based or an abstraction-based model checker?''
\end{quotation}

Results that help to answer these questions are spread among numerous papers
in numerous conferences. The choice of the models and tools used in benchmarks
is rarely sufficient to answer these questions. Benchmark results are
available a long time after their publication, even if the computer
architecture has changed a lot. Moreover, as they are executed over several
platforms and composed of different models, conclusions are not easy.

The objective of the \acl{MCC} is to compare the efficiency of verification
techniques according to the characteristics of the models. To do so, the
\acs{MCC} compares tools on several classes of models, often with scaling
capabilities, \emph{e.g.}, values that set up the ``size'' of the associated
state space.

Through a benchmark, our goal is to identify the techniques that can tackle a
given type of problem identified in a ``typical model'', for a given class of
problem (\emph{e.g.}, state space generation, evaluation of reachability or
temporal formula{\ae}, etc.).

After Newcastle and Hamburg, the third edition of the \acl{MCC} took place
within the context of the Petri Nets 2013 conference, in Milano, Italy. The
original submission procedure was published early mid-February 2013 and
submissions gathered by early May 2013. After some tuning of the execution
environment, the evaluation procedure was operated on a cluster early June.
Results were presented during on June \nth{25}, 2013.

\bigskip The goal of this document is to report the raw data provided by this
third edition of the Model Checking Contest. It reflects the vision of the
\acs{MCC2013} organizers, as it was first presented in Milano. All tool
developers are listed in~\Cref{sec:conclusion}.

Please note that a web version of this this report (with hyperlinks) is also
available at \url{http://mcc.lip6.fr}~\cite{mcc2013web}.

\subsubsection{Structure of this report} The report for the \acs{MCC2013} is
divided in three volumes:

\begin{itemize}

\item the main document (the on you read now) that contains all the main
data gathered during this event,

\item two annexes that only report memory and CPU consumption of tool
executions (these should mostly interest tool developers).

Annex 1 concerns state space generation and reachability examinations (1378
pages) while annex 2 deals with CTL and LTL examinations (1732 pages).

\end{itemize}

The main document is structured in five parts. The first one deals with
factual information about models (section~\ref{sec:models}), involved tools
(section~\ref{sec:tools}), the methodology (section~\ref{sec:methodo}) and a a
short conclusion.

Other parts are almost completely generated automatically from the outputs
gathered during the model checking contest. They deal with the state space
examination (part~\ref{part:two}), Reachability analysis examinations
(part~\ref{part:three}), CTL analysis examinations (part~\ref{part:four}) and
LTL analysis examinations (part~\ref{part:five}).

\newpage
\input{02-models.tex}
\newpage
\section{Participating Tools}
\label{sec:tools}

Twelve tools were submitted at \acs{MCC2013} (including the variants of one
tool). We list them here by alphabetical order. For each submitted tool, the
disk image used to operate them is available from the MCC web page (section
Participating tools)~\cite{mcc2013web}.

\subsection{\acl{alpina} (Univ. Geneva, Switzerland)}

AlPiNA~\cite{buchs:petrinets:2010,buchs:tacas:2010,2011-hostettler-0} stands
for Algebraic Petri Nets Analyzer and is a model checker for Algebraic Petri
Nets created by the SMV Group at the University of Geneva. It is 100\% written
in Java and it is available under the terms of the GNU general public license.
Our goal is to provide a user friendly suite of tools for checking models
based of the Algebraic Petri Net formalism. AlPiNA provides a user-friendly
user interface that was built with the latest metamodeling techniques on the
eclipse platform.

Usually, the number of states of concurrent systems grows exponentially in
relation to the size of the system. This is called the State Space Explosion.
Symbolic Model Checking (SMC) and particularly SMC based on Decision Diagrams
is a proven technique to handle the State Space Explosion for simple
formalisms such as P/T Petri nets.

Algebraic Petri Nets (APN : Petri Nets + Abstract Algebraic Data Types) are a
powerful formalism to model concurrent systems. The State Space Explosion is
even worse in the case of the APNs than in the P/T nets, mainly because their
high expressive power allows end users to model more complex systems. To
tackle this problem, AlPiNA uses evolutions of the well known Binary Decision
Diagrams (BDDs), such as Data Decision Diagrams, Set Decision Diagrams and
Sigma-DDs. It also includes some optimizations specific to the APN formalism,
such as algebraic clustering and partial algebraic unfolding, to reduce the
memory footprint. With these optimizations, AlPiNA provides a good balance
between user-friendliness, modeling expressivity and computational
performances.

AlPiNA official web page is \url{http://alpina.unige.ch}.

\subsection{\acl{cunf} (\'Ecole Normale Sup\'erieure de Cachan, France)}

Cunf is a set of programs for carrying out unfolding-based verification of
Petri nets extended with read arcs, also known as contextual nets, or c-nets.
The package specifically contains the following tools:

\begin{itemize}

\item Cunf: constructs the unfolding of a c-net,

\item Cna: performs reachability and deadlock analysis using unfoldings
constructed by Cunf,

\item Scripts such as pep2dot or grml2pep to do format conversion between
various Petri net formats, unfolding formats, etc.

\end{itemize}

The unfolding of a c-net is another well-defined c-net of acyclic structure
that fully represents the reachable markings of the first. Because unfolding
represent behavior by partial orders rather than by interleaving, for highly
concurrent c-nets, unfolding are often much (exponentially) smaller, which
makes for natural interest in them for the verification of concurrent systems.

Cunf requires that the input c-net is 1-safe (i.e., no reachable marking puts
more than one token on every place), and for the time being the tool will
blindly assume this. It implements the c-net unfolding procedure proposed by
Baldan et al. in~\cite{cunf:ref1}, the algorithms and data structures actually
implemented have been partially described in~\cite{cunf:ref2}.

Cna (Contextual Net Analyzer), checks for place coverability or
deadlock-freedom of a c-net by examining its unfolding. The tool reduces these
problems to the satisfiability of a propositional formula that it generates
out of the unfolding, and uses Minisat as a back-end to solve the formula.

You may download the tool's manual from the tool's webpage, where you will
find detailed instructions for installation. The tool is integrated in the
Cosyverif environment, whose graphical editor you may want to use to analyze
nets constructed by hand. Cunf also comes with Python libraries for producing
c-nets programmatically, see Sec. 7 of the manual.

Cunf official web page is \url{http://code.google.com/p/cunf}. Cunf is also
distributed within the CosyVerif environment (\url{http://cosyverif.org}).

\subsection{\acl{greatspn} (Univ. Torino, Italy)}

GreatSPN \cite{greatspn:ref1,greatspn:ref2} is a suite of tools for the design
and analysis (qualitative and quantitative) of Generalized Stochastic Petri
Nets and Stochastic Well-formed nets. First released by the University of
Torino in the late 1980's, GreatSPN has been a widely used tool in the
research community since it provides a breadth of solvers for computing net
structural properties, the set of Reachable States (RS), the Reachability
Graph (RG) with and without symmetry exploitation, and performance evaluation
indices using both simulation and numerical solution for steady state and
transient measures.

Over the years, GreatSPN functionality has been extended, also thanks to the
collaboration with University of Paris 6 and the Universit\`a del Piemonte
Orientale, by improving and enhancing its solution algorithms, and by
providing new solution methods for new formalisms and property languages
defined over the years.

The last enhancements include:

\begin{itemize}

\item Model checking. A Computational Tree Logic (CTL) model checker for Petri
nets with priorities and a CSL-TA stochastic model checker for SPN. The CTL
model checker implementation is based on the Meedly library from University of
Iowa.

\item Optimization problem analyzer. Integration of the Markov Decision
Well-formed Net formalism and associated solution algorithms, which allow the
study of optimization problems based on Discrete Time Markov Decision Process.

\item Fluidification analysis. The addition of the PN2ODE module, which allows
to automatically derive from an SPN model a corresponding set of ODEs (in
Matlab format), whose solution provides the expected values of the performance
indices, as a function of time.

\item Dynamic SRG and Extended SRG. The algorithms for the construction of the
Symbolic RG have been extended to include Dynamic SRG and Extended SRG
construction, two non trivial extensions of the SRG construction which can
provide a reduction of the state space size in case of partially symmetrical
SWN models.

GreatSPN official web page is \url{http://www.di.unito.it/~greatspn}.

\end{itemize}

\subsection{\acl{its-tools} (Univ. Pierre \& Marie Curie, France)}

ITS-tools is a suite of model-checking tools, developed in the team MoVe at
LIP6. Written in C++, it is available under the terms of the GNU General
Public License.

It features state-space generation, reachability analysis, LTL and CTL
checking. ITS-tools accept a wide range of inputs: (Time) Petri Nets, ETF
(produced by the tool LTSmin), DVE (input format to the tool DiVinE, used in
the BEEM database), and a dedicated C-like format known as GAL. The input
models can also be given as compiled object files. This allows for large
possibilities of interaction with other tools.

Models, even in different formats, can also be easily composed, through the
formalism of Instantiable Transition Systems (ITS)~\cite{TPHK}. This ease the
modeling process. ITS-tools also features a graphical interface, as an Eclipse
plug-in, to further help the modeler, especially with compositions.

As for the back-end, ITS-tools rely on decision diagrams~\cite{FORTE2005} to
handle efficiently the combinatorial explosion of the state space. The
decision diagrams manipulation is performed by the libDDD library, that
features several mechanisms for the efficient manipulation of decision
diagrams~\cite{FI2009,its:cav2013}.

ITS-Tool official web page is \url{http://move.lip6.fr/software/DDD/itstools.php}.

\subsection{\acl{lola} (Univ. Rostock, Germany)}

LoLA~\cite{Wolf_2007_atpn} provides explicit state space verification for
place/transition nets. It supports various simple properties. For the contest,
mainly the reachability verification features are used.

LoLA offers several techniques for alleviating state explosion, including
various stubborn set methods, symmetries (which it can determine fully
automated), the sweep-line method (where it computes its own progress
measure), bloom filters, and linear algebraic compressions. To our best
knowledge, LoLA is the only tool worldwide that provides this large number of
explicit state space techniques in this high degree of automaton, and with
these possibilities for joint application.

In the MCC, we neither use symmetries nor the sweep-line method. For these
methods, performance is too volatile for the black box scenario implemented in
the MCC.

NOTE: associated to he main version, three variants (described below) were
provided.

LoLA official web page is \url{http://www.service-technology.org/lola}.

\subsubsection{Variant: \acl{lola-optimistic}} It uses a goal oriented stubborn set
method and linear algebraic state compression. Goal oriented stubborn sets
perform best on reachability queries that are ultimately satisfied in the net
under investigation. A heuristics takes care that a satisfying state is
reached with high probability already in very early stages of state space
exploration. This way, only a tiny portion of the state space is actually
explored. If the satisfying states are missed, however, or no satisfying state
is reachable, a significantly larger state space is produced than the one
produced by lola-pessimistic. Witness paths tend to be very small.

\subsubsection{Variant: \acl{lola-optimistic-incomplete}} In addition to
lola-optimistic, we use a bloom filter for internal representation of states.
That is, only the hash value of a state is marked in several hash tables, each
belonging to an independent hash function. The state itself is not stored at
all. This way, we can handle a larger number of states within a given amount
of memory. In the rare case of a hash collision, the colliding state is not
explored, so parts of the state space may be missed and false negative results
are possible

The user can specify the number of hash tables to be used and thus control the
probability of hash collisions.

\subsubsection{Variant: \acl{lola-pessimistic}}

This variant computes stubborn sets using a standard deletion algorithm.
Deletion algorithms are much slower than goal-oriented stubborn sets
(quadratic instead of linear) but yield better reduction. This better
reduction pays off when the whole state space needs to be explored (i.e. there
are no reachable satisfying states). If reachable states exist, this variant
is outperformed by the optimistic variant since it has no goal-orienting
heuristics and tends to miss reachable states in early phases of state space
exploration.

Witness paths are often much longer than in the optimistic variant.

\subsection{\acl{marcie} (Univ. Cottbus, Germany)}

MARCIE~\cite{marcie2013} is a tool for the analysis of Generalized Stochastic
Petri Nets, supporting qualitative and quantitative analyses including model
checking facilities. Particular features are symbolic state space analysis
including efficient saturation-based state space generation, evaluation of
standard Petri net properties as well as CTL model checking.

Most of MARCIE's features are realized on top of an Interval Decision Diagram
(IDD) implementation~\cite{tovchigrechko2008}. IDDs are used to efficiently
encode interval logic functions representing marking sets of bounded Petri
nets. This allows to efficiently support qualitative state space based
analysis techniques~\cite{schwarick2010}. Further, MARCIE applies heuristics
for the computation of static variable orders to achieve small DD
representations.

For quantitative analysis MARCIE implements a multi-threaded on-the-fly
computation of the underlying CTMC~\cite{srh2011}. It is thus less sensitive
to the number of distinct rate values than approaches based on, e.g.,
Multi-Terminal Decision Diagrams. Further it offers symbolic CSRL model
checking and permits to compute reward expectations. Additionally MARCIE
provides simulative and explicit approximative numerical analysis techniques.

MARCIE official web page is
\url{http://www-dssz.informatik.tu-cottbus.de/DSSZ/Software/Marcie}.

\subsection{\acl{neco} (Univ. Evry-Val-d'Essone, France)}

Neco is a suite of Unix tools to compile high-level Petri nets into libraries
for explicit model-checking. These libraries can be used to build state
spaces. It is a command-line tool suite available under the GNU Lesser GPL.

Neco compiler is based on SNAKES~\cite{SNAKES} toolkit and handles high-level
Petri nets annotated with arbitrary Python objects. This allows for big degree
of expressivity. Extracting information form models, Neco can identify object
types and produce optimized Python or C++ exploration code. The later is done
using the Cython language. Moreover, if a part of the model cannot be compiled
efficiently a Python fallback is used to handle this part of the model.

The compiler also performs model based optimizations using place
bounds~\cite{FP11} and control flow places for more efficient firing
functions. However, these optimizations are closely related to a modeling
language we use which allows them to be assumed by construction. Because the
models from the contest were not provided with such properties, these
optimizations could not be used.

The tool suite provides tools to compile high-level Petri nets and build state
spaces but this year we also provide a LTL model-checker: Neco-spot. It builds
upon Neco and upon Spot~\cite{spot2004,neco2013}, a C++ library of
model-checking algorithms.

Neco official web page is \url{https://code.google.com/p/neco-net-compiler}.

\subsection{\acl{pnxdd} (Univ. Pierre \& Marie Curie, France)}

PNXDD is CTL model-checker based on Set Decision Diagrams (SDD)~\cite{TPHK}
for PT-nets, a variant of the decision diagrams (DD) family with hierarchy.
Symmetric Petri Nets are handled via an optimized unfolding~\cite{p2006linar}
(removing places structurally detected as always empty and the associated
transitions).

Hierarchy paradigm, used together with DDs offers greater sharing
possibilities compared to traditional DDs. The ordering of variables in the
diagram, a crucial parameter to obtain good performances in DDs, becomes a new
challenge since portions of the model offering similar comportments must be
statically identified to obtain a good hierarchical order. PNXDD relies on
heuristics that are described in~\cite{2011-hong-0}.

PNXDD official web page is \url{http://cosyverif.org} (it is integrated in the
CosyVerif Verification Environment).

\subsection{\acl{sara} (Univ. Rostock, Germany)}

Sara~\cite{sara11} answers reachability queries using the Petri net state
equation. From this equation and inequations derived from the query, a linear
programming problem is generated and solved using a standard package. If the
system does not have solutions, we conclude that there are no reachable states
satisfying the query. Other wise, we obtain a firing count vector which
describes candidate firing sequences.

We check whether there is an executable firing sequence for the given vector.
If so, we have a reachable satisfying state and a witness path. If not, we add
inequalities that are not satisfied by the spurious solution. We result in one
or more new linear programming problems which enable less serious solutions
but still cover all feasible solutions. We proceed recursively with the new
problems.

Sara has excellent performance if the state equation as such rules out
reachability, or if an early solution reveals reachability. It may be used for
unbounded Petri nets. since it does not try to represent or to explore the
state space.

In worst case, Sara will not terminate (otherwise, our approach would
contradict the known EXPSPACE hardness of the general reachability problem).

SARA official web page is \url{http://www.service-technology.org/sara}.

\subsection{Summary of the Techniques Reported by Participating Tools}

During the \acs{MCC2013}, tools could report the use of identified techniques.
We summarize in table~\ref{tab:tool:techniques}. Identified techniques were:

\begin{itemize}

\item Abstractions: the tool exploits the use of abstractions (on the fly state elimination),

\item Decision Diagrams: the tool uses any kind of decision diagrams,

\item Explicit: the tool does explicit model checking,

\item Net Unfolding: the tool uses MacMillan unfolding,

\item Parallel Processing\footnote{In fact, only one core was allocated to each Virtual Machine so no parallelism could be enabled in practice (but no tool reported this feature).}: the tool uses multithreading,

\item Structural Reduction: the tool uses structural reductions (Berthelot, Haddad, etc.),

\item SAT/SMT: the tool uses a constraint solver,

\item State Compression: the tool uses some compression technique (other than decision diagrams),

\item Stubborn Sets: the tool uses partial order technique,

\item Symmetries: the tool exploits symmetries of the system,

\item Topological: the tool uses structural informations on the Petri net itself (e.g. siphons, traps, S-invariants or T-invariants, etc.) to optimize model checking,

\item Unfolding to P/T net: the tool transforms colored nets into their equivalent P/T,

\end{itemize}

\begin{table}
	\centering
	\begin{tabular}{|c|c|}
		\hline
		\textbf{Tool Name} & \textbf{Reported Technique} \\
		\hline
		AlPiNA & Decision Diagrams \\
		\hline
		\multirow{2}{*}{Cunf} & Net Unfolding \\
							  & SAT/SMT \\
		\hline
		greatSPN & Decision Diagrams \\
		\hline
		\multirow{2}{*}{ITS-Tools} & Decision Diagrams \\
							  & Structural Reductions \\
		\hline
		\multirow{3}{*}{LoLA (all variants)} & Explicit model checking \\
		  & State compression \\
		  & Stubborn sets \\
		\hline
		Marcie & Decision Diagrams \\
		\hline
		Neco & Explicit model checking \\
		\hline
		\multirow{2}{*}{PNXDD} & Decision Diagrams \\
							  & Topological \\
		\hline
		\multirow{3}{*}{Sara} & SAT/SMT \\
		  & Stubborn sets \\
		  & Topological \\
		\hline
	\end{tabular}
	\caption{Summary of the techniques reported to be used by tools for the \acs{MCC2013}\label{tab:tool:techniques}}
\end{table}

No tool did report a non listed technique (this was possible).

\newpage
\section{Evaluation Methodology}
\label{sec:methodo}

Roughly, the evaluation methodology was the same as for \acs{MCC2011} and
\acs{MCC2012} (it is presented in~\cite{mcc2011}). The main differences are
the following:

\begin{enumerate}

\item we created more categories for the formula evaluation examinations, to
enable a more precise support by tools;

\item since the virtual machine-based monitoring experimented in 2012 was a
success, a dedicated environment to operate tests (multi-purpose and thus
usable in other contexts) was implemented: {\BK}~\cite{benchkit:2013}. So, the
MCC execution environment is now composed of {\BK} and numerous post-analysis
scripts that gather and integrate data from the outputs delivered by tools.

\end{enumerate}

\subsection{The Examinations}

The \acs{MCC2011} reported numerous problems with formula classification
(reachability, CTL, LTL). One of them was the absence of classification for
atomic propositions in formula to be verified. Thus, many tools had troubles
to support a significative set of formulas.

For the \acs{MCC2013}, we thus also classified the type of atomic propositions
they involved. This lead to the list of examination reported in
table~\ref{tab:methodo:exam}

\begin{table}
\centering
{\footnotesize\begin{tabular}{|p{5.5cm}|p{8.4cm}|}
	\hline
	\multicolumn{1}{|c|}{\textbf{Value}} & \multicolumn{1}{c|}{\textbf{Signification}} \\
	\hline
	\texttt{StateSpace} & we ask for state space generation only\\
	\texttt{CTLCardinalityComparison} & we evaluate CTL properties dealing with checking cardinality of marking only \\
	\texttt{CTLFireability} & we evaluate CTL properties dealing with transition fireability only \\
	\texttt{CTLMarkingComparison} & we evaluate CTL properties dealing with marking comparison only \\
	\texttt{CTLPlaceComparison} & we evaluate CTL properties dealing with the comparison of places marking only \\
	\texttt{CTLMix} & we evaluate CTL properties dealing with all the previous type of atomic proposition \\
	\texttt{LTLCardinalityComparison} & we evaluate LTL properties dealing with checking cardinality of marking only \\
	\texttt{LTLFireability} & we evaluate LTL properties dealing with transition fireability only \\
	\texttt{LTLMarkingComparison} & we evaluate LTL properties dealing with marking comparison only \\
	\texttt{LTLPlaceComparison} & we evaluate LTL properties dealing with the comparison of places marking only \\
	\texttt{LTLMix} & we evaluate LTL properties dealing with all the previous type of atomic proposition \\
	\texttt{ReachabilityDeadlock} & we evaluate reachability properties dealing with transition deadlocks only \\
	\texttt{ReachabilityCardinalityComparison} & we evaluate reachability properties dealing with checking cardinality of marking only \\
	\texttt{ReachabilityFireability} & we evaluate reachability properties dealing with transition fireability only \\
	\texttt{ReachabilityMarkingComparison} & we evaluate reachability properties dealing with marking comparison only \\
	\texttt{ReachabilityPlaceComparison} & we evaluate reachability properties dealing with the comparison of places marking only \\
	\texttt{ReachabilityMix} & we evaluate reachability properties dealing with all the previous type of atomic proposition \\
	\hline
\end{tabular}}
\caption{List of the examinations proposed at the \acs{MCC2013}\label{tab:methodo:exam}}
\end{table}

\subsection{Execution Scheme}

The main execution loop is very simple. It is presented
in~\Cref{algo:mainscript}. For each model/instance and for each examination,
we perform a test and extract data from the {\BK} monitor (CPU and memory
consumption, if time confinement was reached or not) and from the tool
(stdout, result had to be formatted in dedicated lines).

\begin{algorithm}
	\KwIn{$M$, a set of scalable models to be processed}
	\ForEach{$m \in M$ and $v \in ScalingParameters(M)$}{
		launch the virtual machine for $m/v$\;
		report information to the database\;
		halt the virtual machine
		}
	\caption{Actions performed for each tool by the invocation script
	\label{algo:mainscript}}
\end{algorithm}

The main difficulty, handled by {\BK} was to dispatch the executions over the
set of involved machines. To keep consistency of the executions and make these
comparable, all the examinations related to a given model $m$ ware operated on
the same host.

\subsection{Involved Machines}

Three computers were made available to operate the submitted tools by various
institutions: cluster1\footnote{For cluster1, only 18 of the 23 available
nodes where allocated due to parallel experimentations.}, ebro and quadhexa-2.
Their characteristics are reported in table~\ref{tab:methodo:machines}.

Memory confinement was 4Gbyte of memory and 45mn of CPU for all examinations.

\begin{table}
	\centering
	\begin{tabular}{|c|c|c|}
		\hline
		\textbf{cluster1 (Univ P. \& M. Curie)} &
		\textbf{ebro (Univ. Rostock)} &
		\textbf{quadhexa-2 (Univ. Nanterre)} \\
		\hline
		\includegraphics[height=5cm]{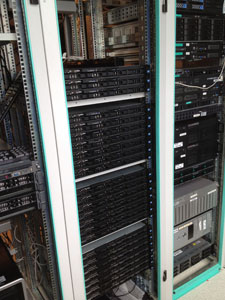} &
		\includegraphics[height=5cm,keepaspectratio]{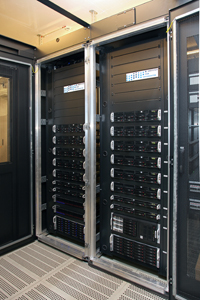} &
		\includegraphics[height=5cm,keepaspectratio]{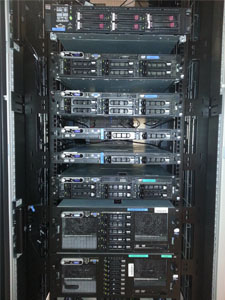}\\
		\hline
		\multicolumn{3}{|c|}{\textbf{Characteristics of the CPU}}\\
		total of 46 CPU &
		total of 64 CPU &
		total of 24 CPU \\
		23 $\times$ Intel Xeon E5645 &
		4 $\times$ AMD Opteron 6200 Series &
		4 $\times$ Intel Xeon X7460 \\
		2.4 GHz, 6-Core,&
		2.7 GHz, 16-Core,&
		2.66 GHz, 6-Core,\\
		6$\times$ 1536KB/12288KB L2/L3 &
		16$\times$ 1024KB/16MB L2/L3 &
		3 $\times$ 3MB/16NB L2/L3 \\
		\hline
		\multicolumn{3}{|c|}{\textbf{Memory}}\\
		23 $\times$ 8GB (2$\times$4GB) DDR3 / PC1333 &
		512GB (32$\times$ 16GB) DDR3 / PC1600 &
		128BG (8$\times$ 16BG) DDR3 / PC1333 \\
		\hline
		\multicolumn{3}{|c|}{\textbf{Disks}}\\
		23 $\times$ 500GB SATA 7200 +&
		2 $\times$ 1TB SAS2-Server-RAID +  &
		4 $\times$ 400GB RAID 1 (mirror)\\
		1TB SATA 7200 &
		2 $\times$ 128GB SSD Samsung 830  &
		Seagate SAS Cheetah \\
		 &
		SERIES SATA III MLC &
		\\
		\hline
		\multicolumn{3}{|c|}{\textbf{Linux Kernel}}\\
		\texttt{2.6.38.8-server-10.mga} &
		\texttt{2.6.32-358.11.1.el6.x86\_64} &
		\texttt{3.8.1-server-1.mga3} \\
		\hline
	\end{tabular}
\caption{Characteristics of the machines used for the \acs{MCC2013}\label{tab:methodo:machines}}
\end{table}

The organizers thank the Universities of
Paris-Ouest-Nanterre\footnote{\url{http://www.u-paris10.fr}},
Rostock\footnote{\url{http://www.uni-rostock.de}} and Pierre \& Marie
Curie\footnote{\url{http://www.upmc.fr}} for letting us using their computers.

\subsection{The Executions}

\Cref{tab:methodo:executions} provides a summary concerning the executions
over the proposed benchmark, for both the ``Known'' and the ``Surprise''
models. Please note that the number of execution does not includes a large
number of preliminary tests in cooperation with the tool developers?

\begin{table}
	\centering
	\begin{tabular}{|c|c|}
		\hline
		\textbf{``Known'' Models} &
		\textbf{``Surprise'' Models} \\
		\hline
		\multicolumn{2}{|c|}{\textbf{Total Number of tool executions}}\\
		49\,380 &
		4\,913 \\
		\hline
		\multicolumn{2}{|c|}{\textbf{Execution per Machine}}\\
		cluster1: 24\,937 &
		ebro: 1\,640 \\
		quadhexa-2: 24\,443 &
		quadhexa-2: 3\,273 \\
		\hline
		\multicolumn{2}{|c|}{\textbf{Total CPU Time required}}\\
		80 days, 18 hours, 17 minutes, 11 seconds &
		3 days, 11 hours, 45 minutes, 12 seconds \\
		\hline
		\multicolumn{2}{|c|}{\textbf{Size of collected raw data (CSV, outputs, etc. excluding charts)
		}}\\
		1.77GBytes	 &
		122.3MBytes \\
		\hline
		\multicolumn{2}{|c|}{\textbf{Produced Performance Charts (for models)
		}}\\
		1\,182 &
		177 \\
		\hline
		\multicolumn{2}{|c|}{\textbf{Produced Execution Charts (for relevant executions)}}\\
		13\,763 &
		1\,541 \\
		\hline
	\end{tabular}
\caption{Characteristics of the machines used for the \acs{MCC2013}\label{tab:methodo:executions}}
\end{table}

\subsection{Know Issues}

This section reports the open issues identified during the discussion
following the presentation of results on June 25th, 2013 in Milano. We
differentiate organizational matters from technical issues.

\subsubsection{Organizational Matters} are listed below:

\begin{itemize}

\item[Io1] Live Event: this event, lately announced, had to be canceled, which
is a pity. Its objective is to provide feedback on tools from a "usability"
point of view (look and feel, quality of documentation and tutorials, etc). We
will announce it earlier for the next edition in 2014.

\item[Io2] People: so far, active people in the model checking contest are to
few...

\item[Io3] Global schedule: the submission deadline should be pushed earlier
to allow more time for analysis of the tools.

\item[Io4] "known" and "surprise" models: A strong suggestion is, after the
call for model, to decide that all new models will be "surprise" and thus not
submitted after the call for models. Models of previous years will then be the
only "known" models. This should ease the management of the MCC and help to
relax the agenda and have the call for tools submission out earlier.

\item[Io5] Rules: some people suggest to clean up the rules in order to make
clear what is possible and what is not allowed. In particular, all precomputed
aspects should be carefully investigated

\item[Io6] Trophies: we all agreed on the fact that the formulas used this
year are mainly temporary. This formula should be discussed and should
introduce more aspects on the results like, time and memory consumption, how
correct the outputs are (only for the state space examination this year),
support of P/T and/or colored models, etc.

\end{itemize}

\subsubsection{Technical issues} are listed below:

\begin{itemize}

\item[It1] Generation of formulas: this is an issue for the second year and
it was not yet solved. The problem is to select a large number of formulas for
which we can state their result: satisfied or unsatisfied. Considering the
number of models and instances of these models (a total of 255 in 2013), these
mush be generated automatically

It seems that SPOT~\cite{spot2004} could provide a basis of solution for LTL
formulas. In particular, it offers a mechanism to generate random formulas and
to select them according to numerous criteria (\emph{e.g.} size of the related
B\"uchi automata)~\cite{spot2013}.

During the discussion, several other possible solutions. One is the definition
of formula patterns that could be repeated and arranged randomly and combined
together with atomic propositions.

Another possibility is to propose manually sort of "parameterized" formulas
that are scaled up for each instances, however, if this requires less
formulas, there are yet numerous formulas to provide.

The participants agree on the fact that purely random formulas are not good
but it is necessary to generate formulas automatically in "a good way". We can
insert existing formulas when they are available (this was done for the
surprise model "Vasy2003"). Another important point is that the output values
of formula should be known in advance so that: 1) their veracity could be
checked, and 2) there can be the same amount of satisfied and unsatisfied
formulas to be processes (maybe separately).

There is however a real problem due to issue Io2; manpower is quite low and
must be extended to let time for these tasks.

\item[It2] Grammar for formula: It could be made less ambiguous (e.g. fully
braketed expressions). Some people reported difficulties of interpretation and
then translation. The idea is to have a small task force that will bring out
proposals, especially for the atomic propositions. Somebody suggested to use
PNML identifiers of objects instead of their labels (but this may cause
problems with the equivalences between colored nets and P/T ones).

\item[It3] BenchKit \textcolor{red}{(resolved at this stage\footnote{A new
release of {\BK} is available at \url{http://benchkit.cosyverif.org}.})}: if
this benchmarking tool appear to be operational (it was successfully used to
operate the 54293 executions required this year), its usage remains difficult
for the non-developers. A new version should appear, making its "individual
use" easier, thus allowing the community to reuse outputs from this contest
and later ones.

\item[It4] High-level colored nets: a solution should be proposed to have
high-level colored nets (the problem in 2012 and 2013 was how to produce their
PNML representation). This would allow some tools using such models to be
"more on their playground" than with lower level Petri nets.

\end{itemize}

\newpage
\section{Conclusion}
\label{sec:conclusion}

This documen reported our experience with the \acl{MCC2013} (third edition).

From the tool developers' point of view, such an event allows to compare tools
on a common benchmark that could become a public repository. Also, some
mechanisms established for the contest, such as a language to elaborate the
formula to be verified could become, over the years, a common way to provide
formul{\ae} to the various tools developed by the community.

If the results for the state space generation are clear and can be
interpreted, we still faced, as last year, some troubles with formul{\ae}. The
main problem is the quite complex execution chain and, more particularly, the
translation from the provided formula language to the one of the tool. Since
formul{\ae} were generated automatically (and distinct for each instance of a
given model) it was impossible to predict their result and, most often, no
consensus was found between the participating tools.

For that reason, we only consider the fact that at least one value could be
completed by a given tool for these examinations (this is mentioned in the
corresponding sections).

Let us note that both the benchmarks and the tool submissions are available on
the \acs{MCC2013}'s web site (\url{http://mcc.lip6.fr/2013}). Experiments can
thus be reproduced thanks to the {\BK} confined run
environment~\cite{benchkit:2013} available at
\url{http://benchkit.cosyverif.org}.

\subsubsection{Acknowledgements} The \acl{MCC} organizers would like to thank
the following people for the help they provided in setting up this event:
Fabrice Legon-Aubry and Harro Wimmel (multi-core machine management), and
Lom Hillah (definition of properties).

\smallskip\noindent The \acs{MCC} organizers would also like to thank the
following institutions that borrowed a powerful multi-core or cluster machine
for the numerous execution required for the \acs{MCC2013}: Universit\'e Pierre
\& Marie Curie, Universit\'e Paris Ouest Nanterre and Universit\"at Rostock.

\smallskip\noindent The \acs{MCC} organizers would finally like to thank the tool
developers who made possible such a contest. They are:

\begin{itemize}
\item \acs{alpina}: Steve Hostettler, Alexis Marechal, and Edmundo Lopez;
\item \acs{cunf}: C\'esar Rodr\'iguez;
\item \acs{greatspn}: Elvio Amparore and Marco Beccuti;
\item \acs{its-tools}: Yann Thierry-Mieg, Maximilien Colange et. al.;
\item \acs{lola} (all variants): Niels Lohmann and Karsten Wolf;
\item \acs{marcie}: Alexey Tovchigrechko, Martin Schwarick, and Christian Rohr;
\item \acs{neco}: Lukasz Fronc;
\item \acs{pnxdd}: Silien Hong and Emmanuel Paviot-Adet;
\item \acs{sara}: Harro Wimmel and Karsten Wolf.
\end{itemize}

\part{State Space Generation}
\label{part:two}
\newpage

\section{The StateSpace Examination}
\label{sec:exam:StateSpace}
\index{Results!StateSpace}

This examination deals with state space generation only.
We first show a summary on the handling of models by the participating tools.
Then, we present the computed outputs and the associated scores for this
examination prior to a summary of relevant executions.

\subsection{Handling of Models by Tools}
\index{Performances!StateSpace}

\subsubsection{\acs{CSRepetitions-COL}}
The charts below respectively show how tools compete with this ``Known'' model (memory and CPU).

\index{Performances!StateSpace!CSRepetitions (Colored)}
\begin{center}
   \includegraphics[width=7.2cm]{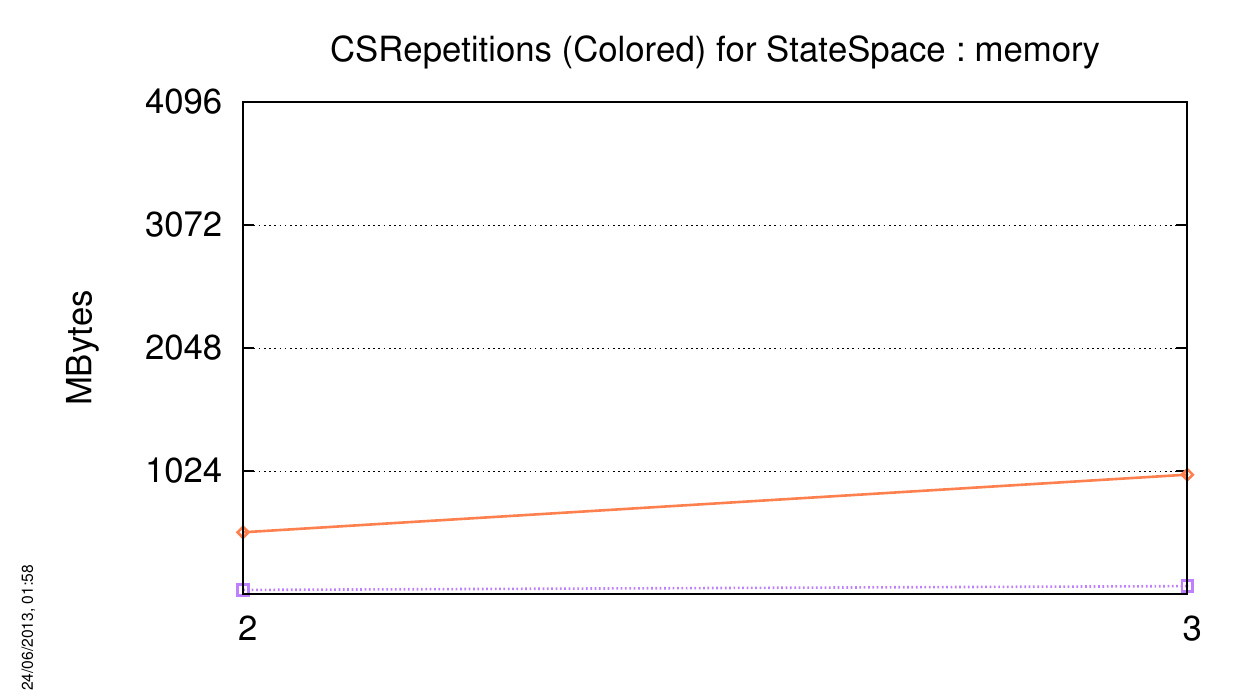}
   \includegraphics[width=7.2cm]{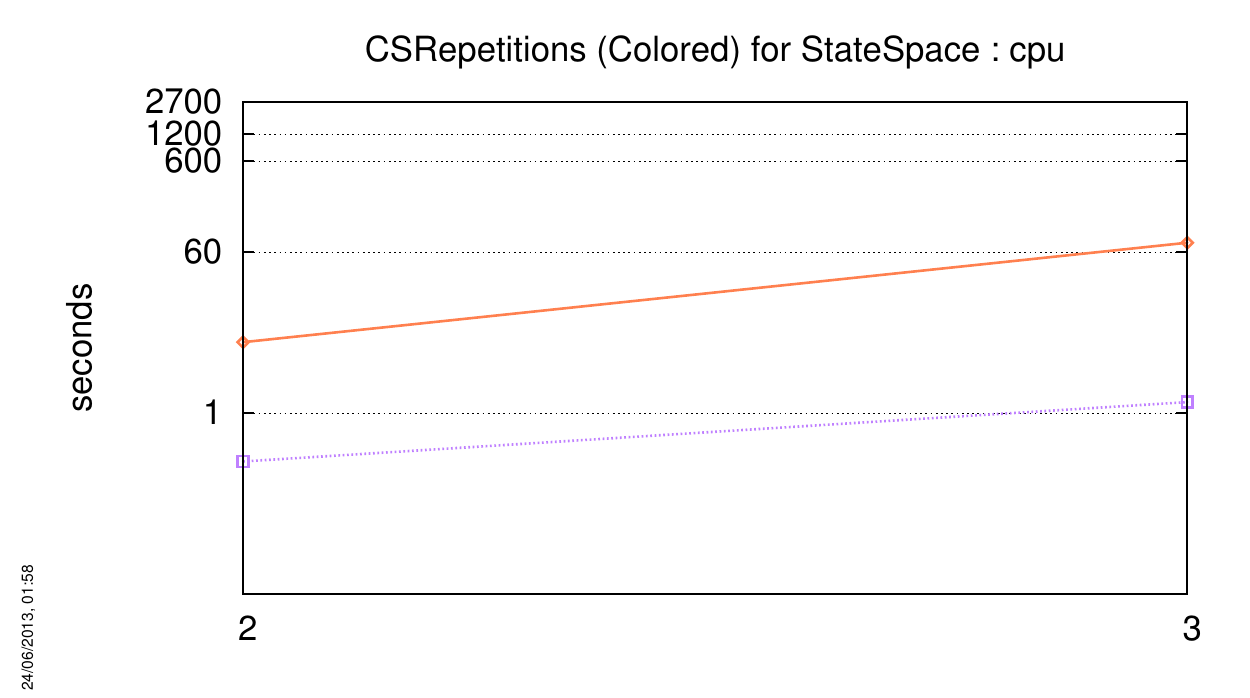}

   \includegraphics[height=1cm]{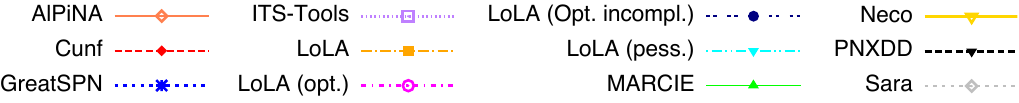}
\end{center}

\subsubsection{\acs{CSRepetitions-PT}}
The charts below respectively show how tools compete with this ``Known'' model (memory and CPU).

\index{Performances!StateSpace!CSRepetitions (P/T)}
\begin{center}
   \includegraphics[width=7.2cm]{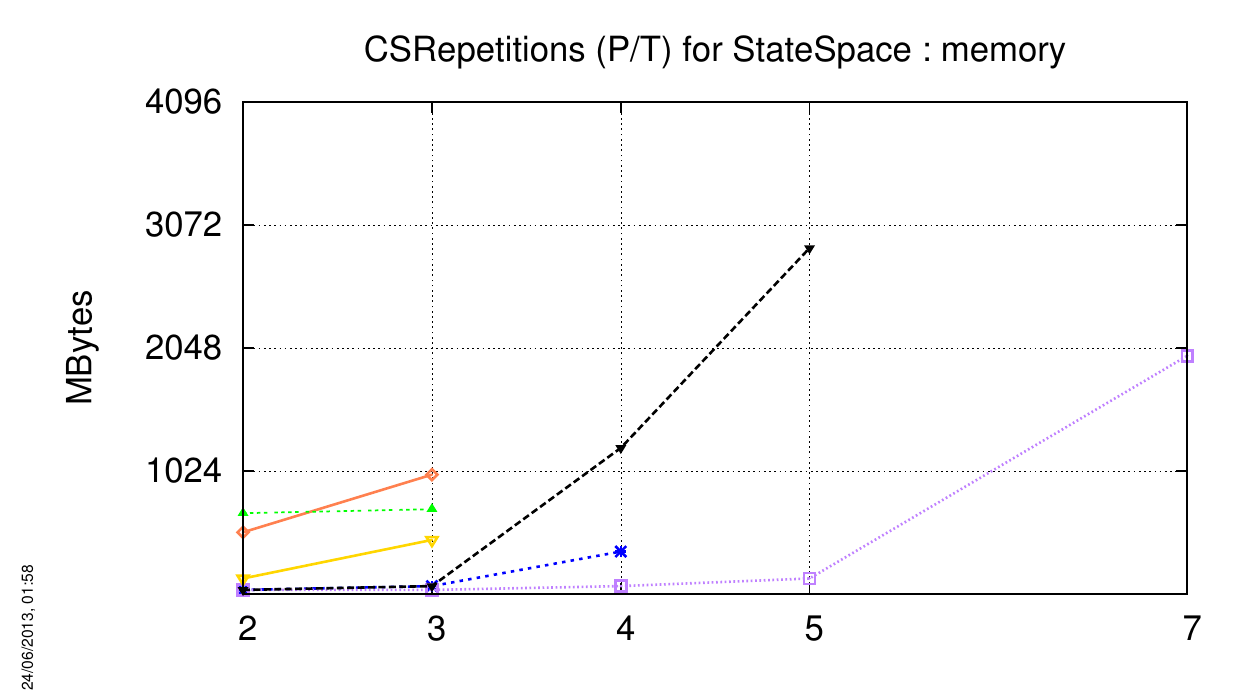}
   \includegraphics[width=7.2cm]{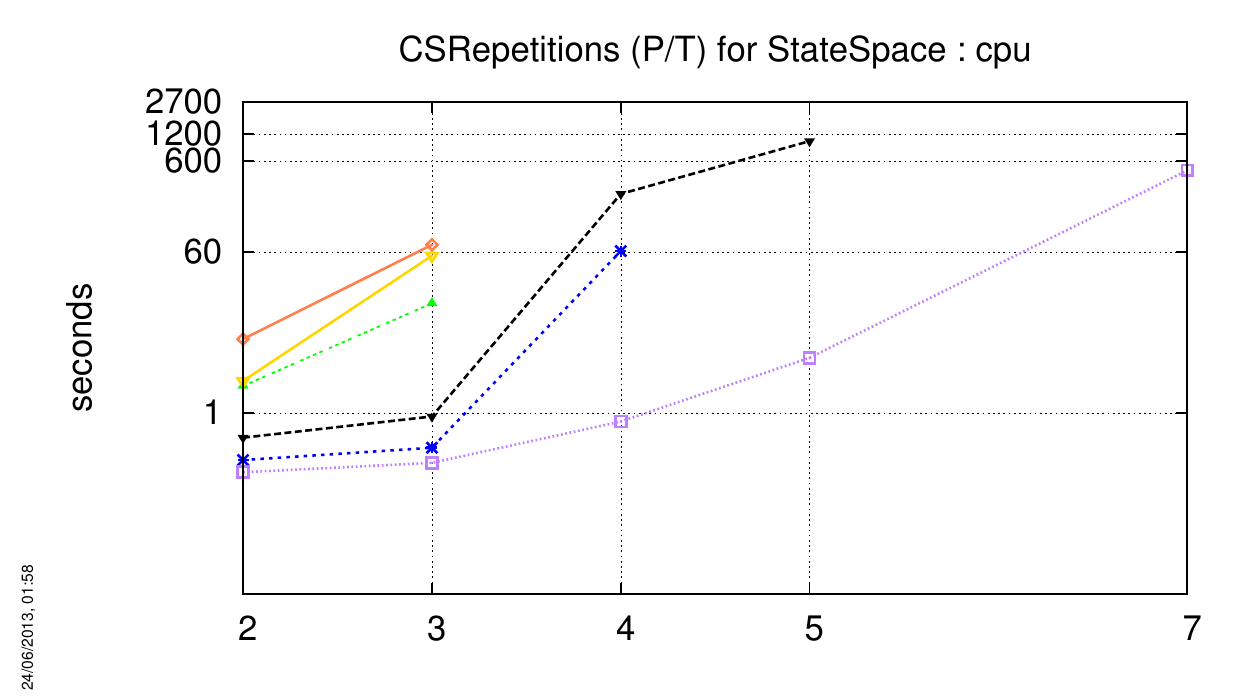}

   \includegraphics[height=1cm]{figures/tools-legend.pdf}
\end{center}

\subsubsection{\acs{Dekker-PT}}
The charts below respectively show how tools compete with this ``Known'' model (memory and CPU).

\index{Performances!StateSpace!Dekker (P/T)}
\begin{center}
   \includegraphics[width=7.2cm]{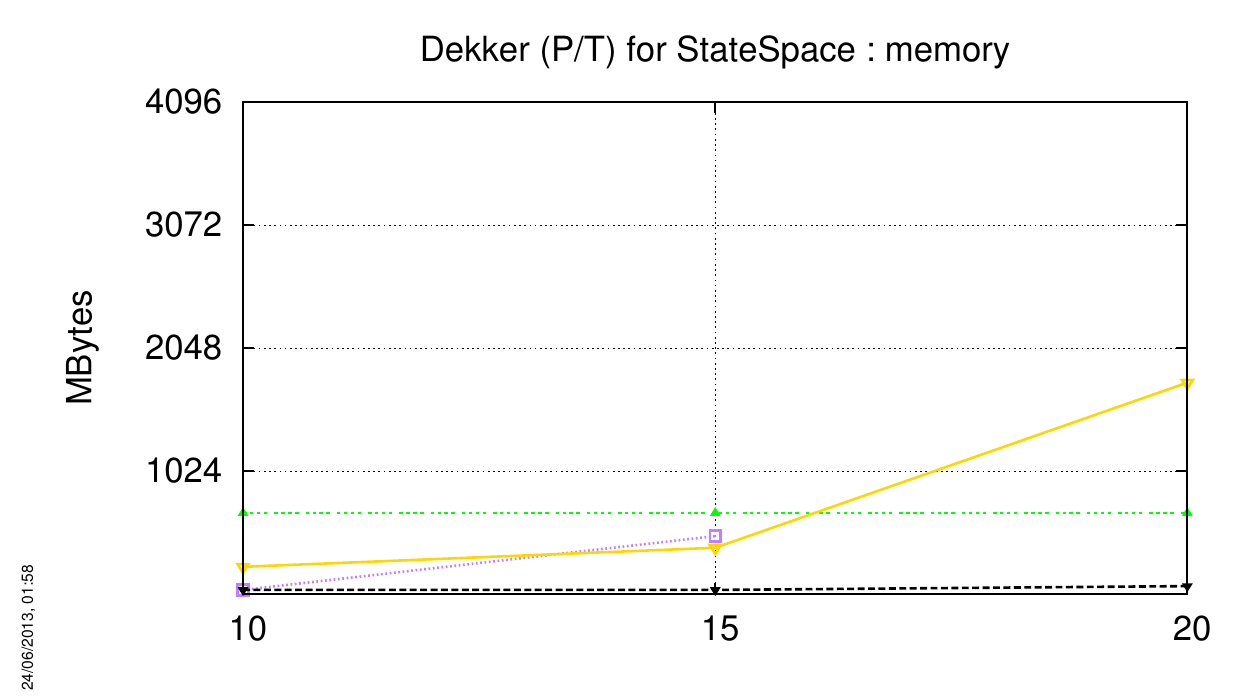}
   \includegraphics[width=7.2cm]{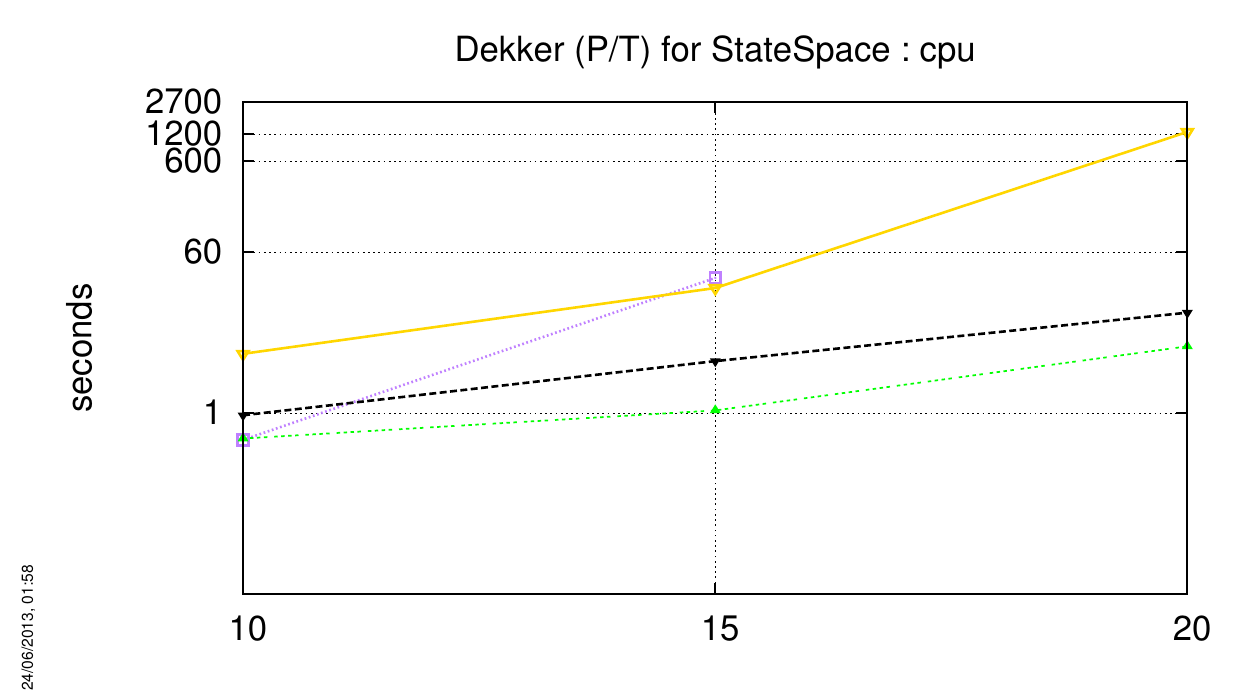}

   \includegraphics[height=1cm]{figures/tools-legend.pdf}
\end{center}

\subsubsection{\acs{DotAndBoxes-COL}}
The charts below respectively show how tools compete with this ``Known'' model (memory and CPU).

\index{Performances!StateSpace!DotAndBoxes (Colored)}
\begin{center}
   \includegraphics[width=7.2cm]{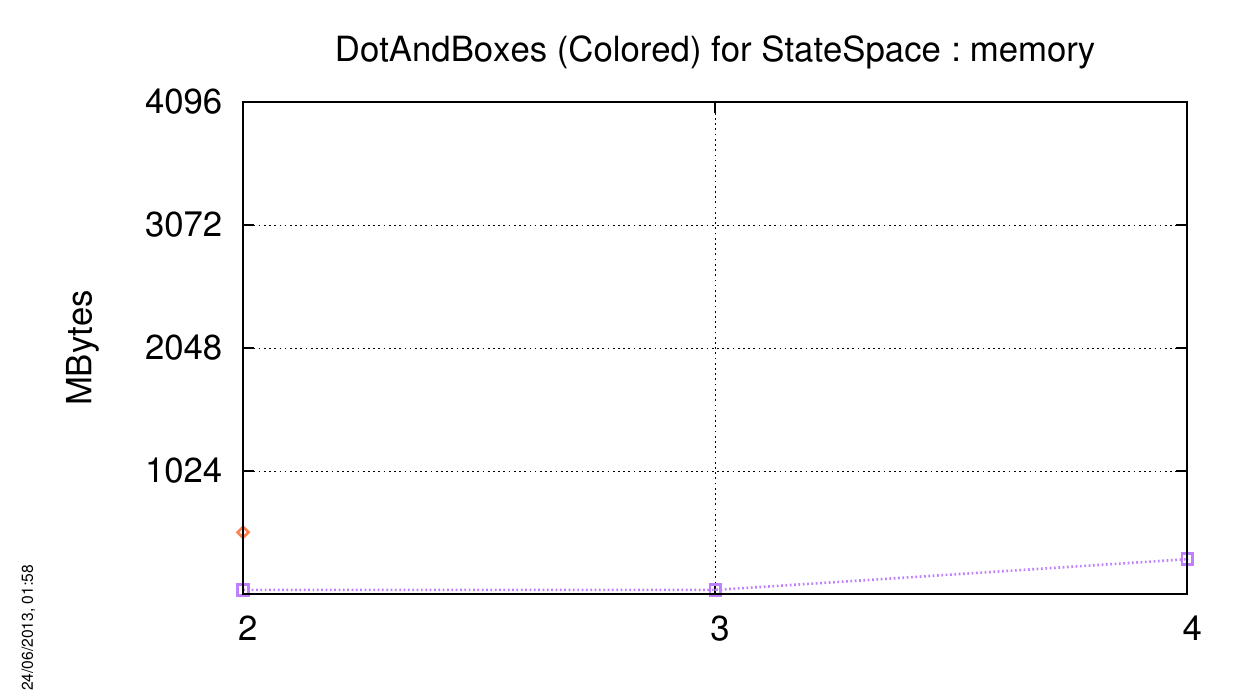}
   \includegraphics[width=7.2cm]{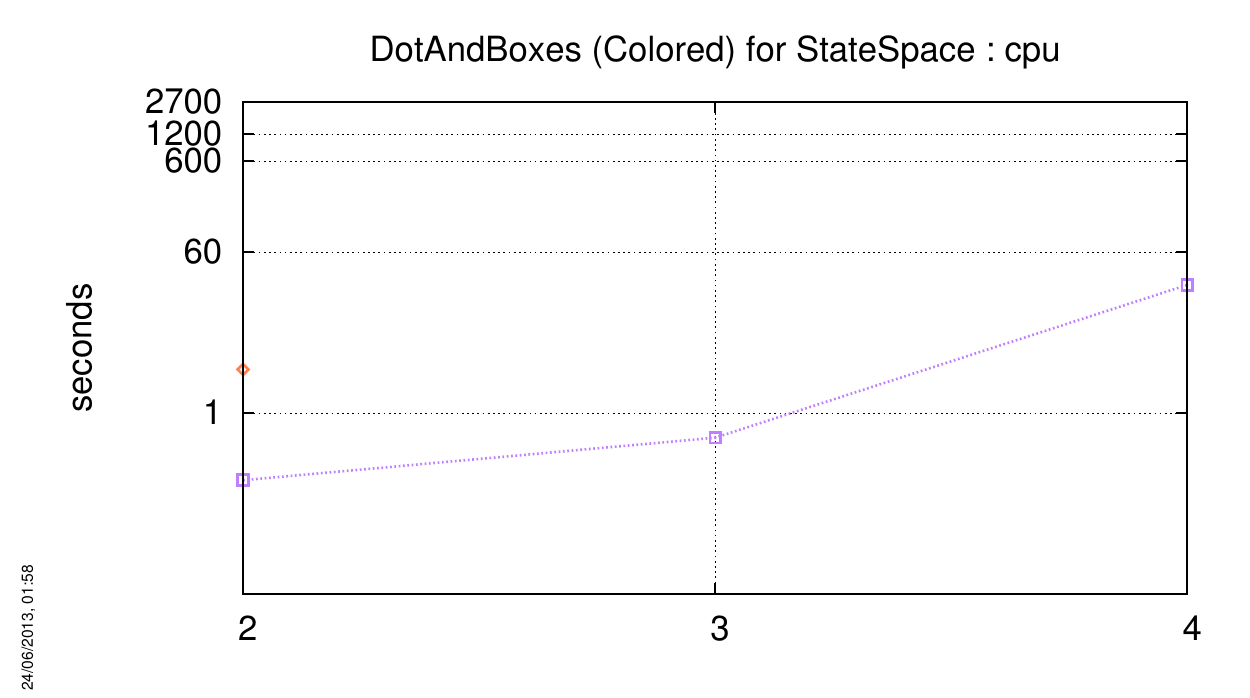}

   \includegraphics[height=1cm]{figures/tools-legend.pdf}
\end{center}

\subsubsection{\acs{DrinkVendingMachine-COL}}
The charts below respectively show how tools compete with this ``Known'' model (memory and CPU).

\index{Performances!StateSpace!DrinkVendingMachine (Colored)}
\begin{center}
   \includegraphics[width=7.2cm]{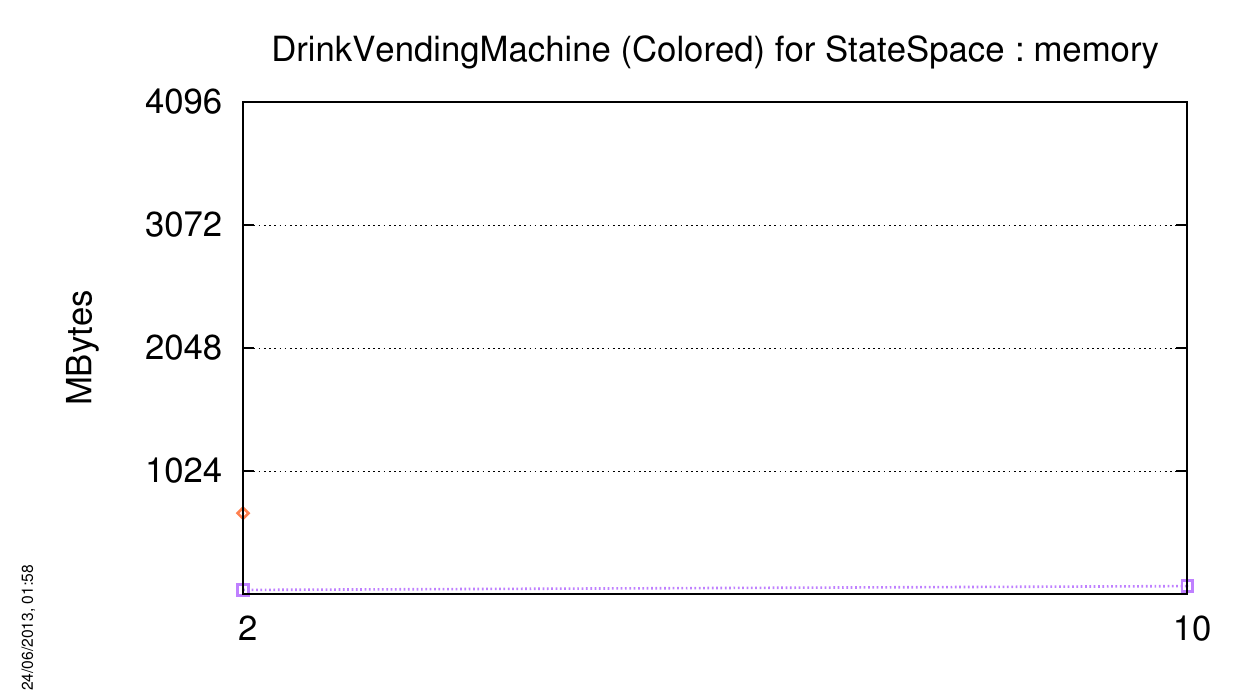}
   \includegraphics[width=7.2cm]{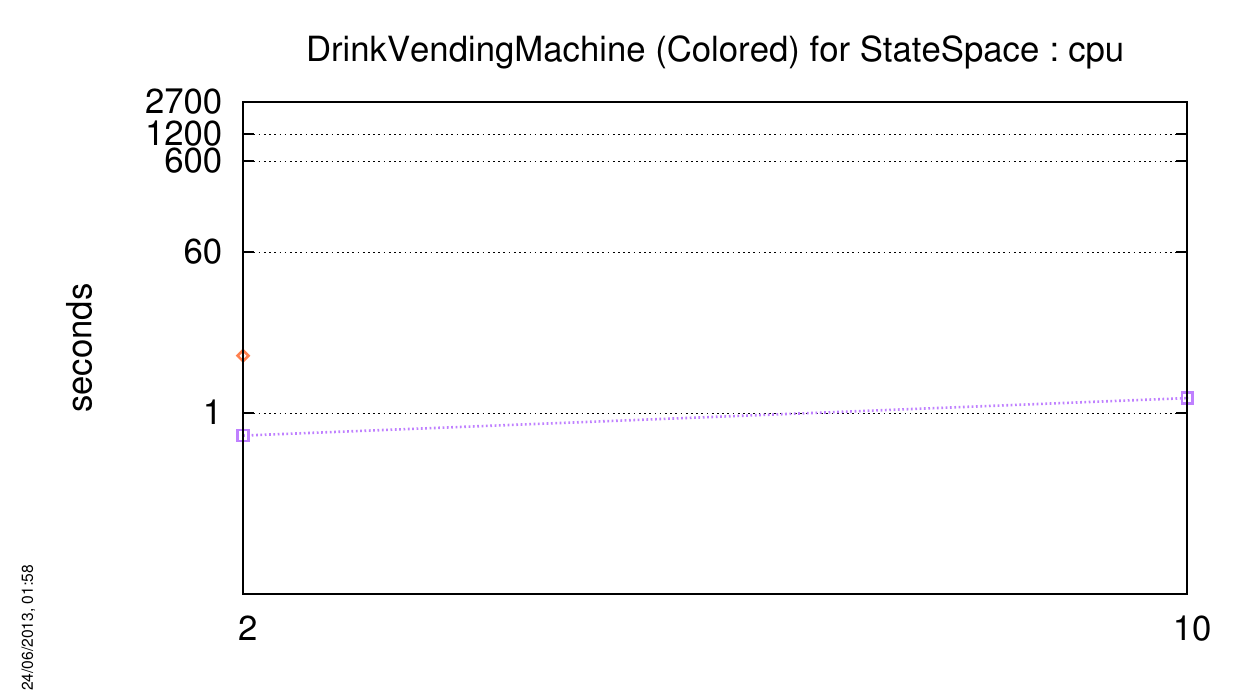}

   \includegraphics[height=1cm]{figures/tools-legend.pdf}
\end{center}

\subsubsection{\acs{DrinkVendingMachine-PT}}
The charts below respectively show how tools compete with this ``Known'' model (memory and CPU).

\index{Performances!StateSpace!DrinkVendingMachine (P/T)}
\begin{center}
   \includegraphics[width=7.2cm]{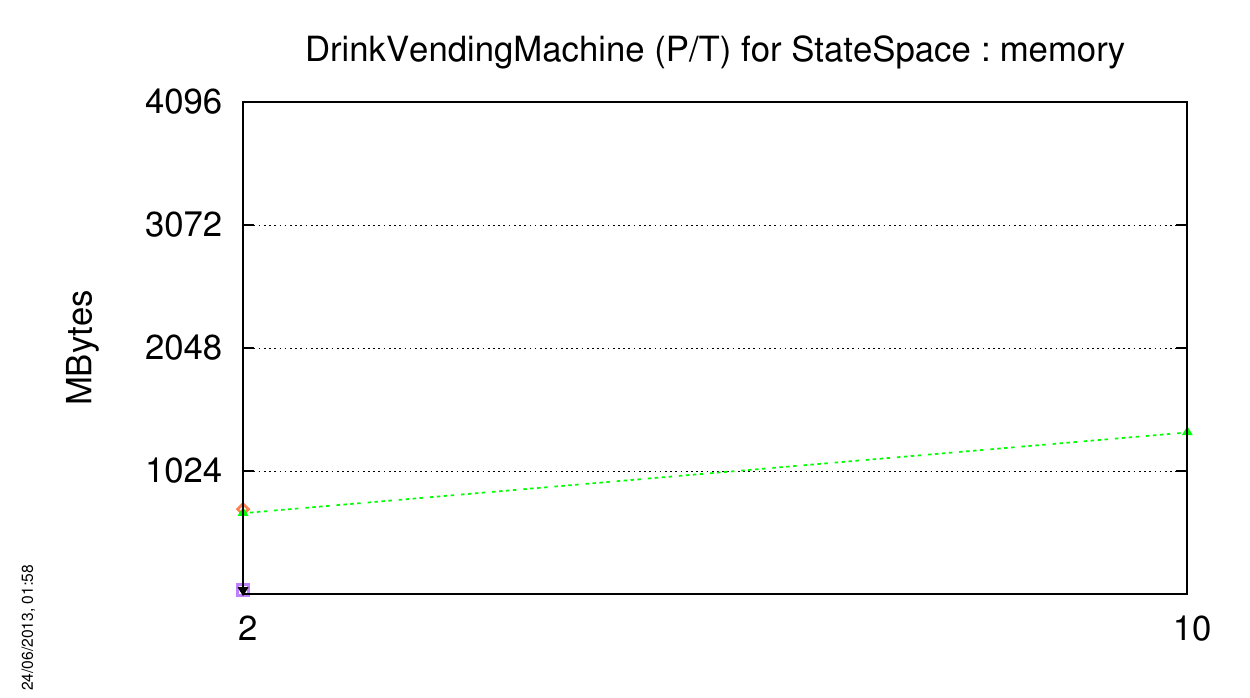}
   \includegraphics[width=7.2cm]{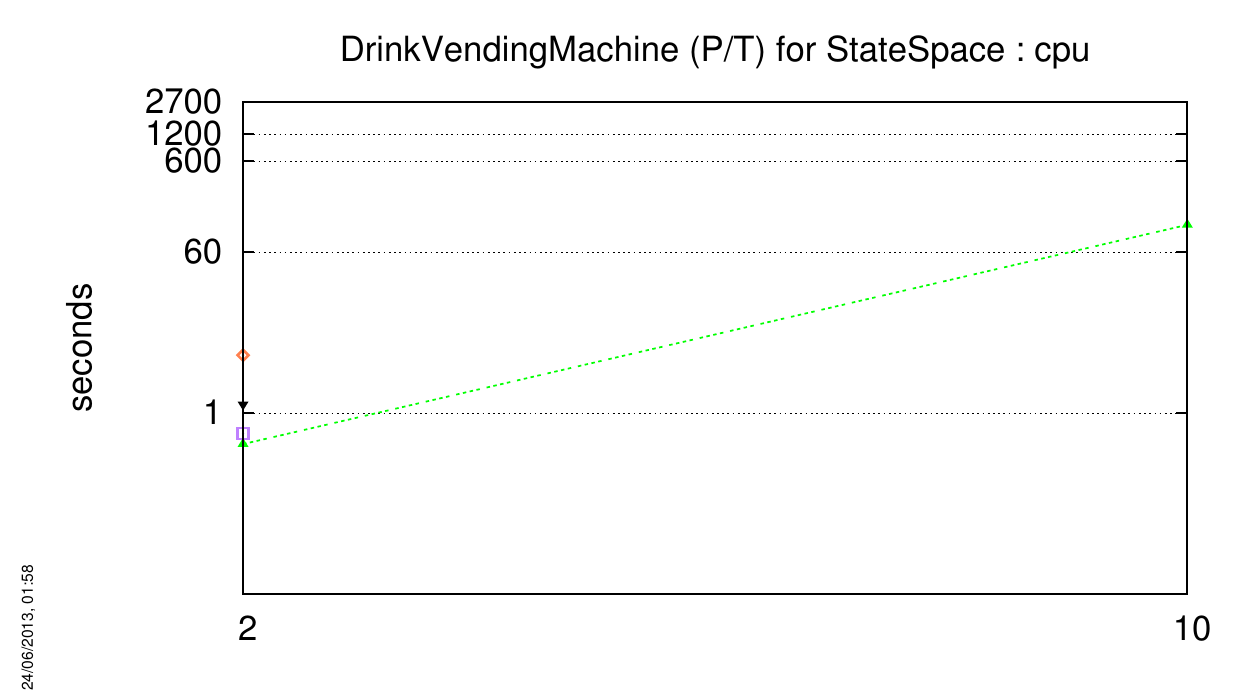}

   \includegraphics[height=1cm]{figures/tools-legend.pdf}
\end{center}

\subsubsection{\acs{Echo-PT}}
No instance of this model could be computed for the \textbf{StateSpace} examination.

\subsubsection{\acs{Eratosthenes-PT}}
The charts below respectively show how tools compete with this ``Known'' model (memory and CPU).

\index{Performances!StateSpace!Eratosthenes (P/T)}
\begin{center}
   \includegraphics[width=7.2cm]{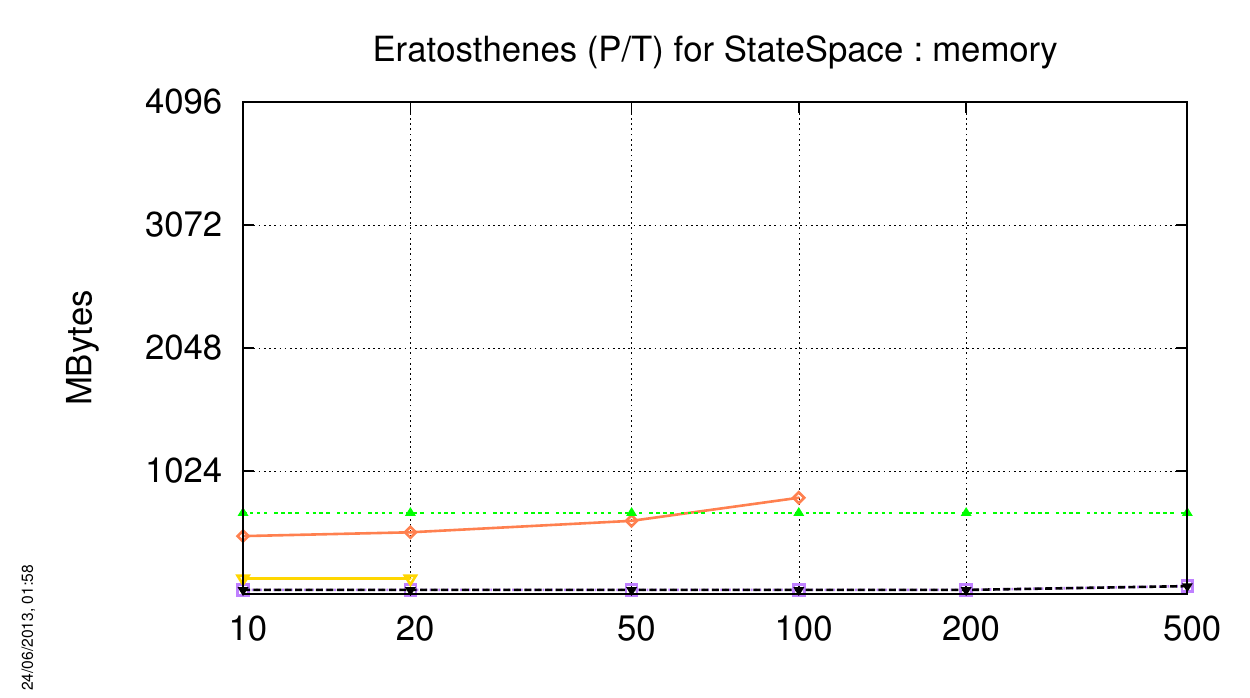}
   \includegraphics[width=7.2cm]{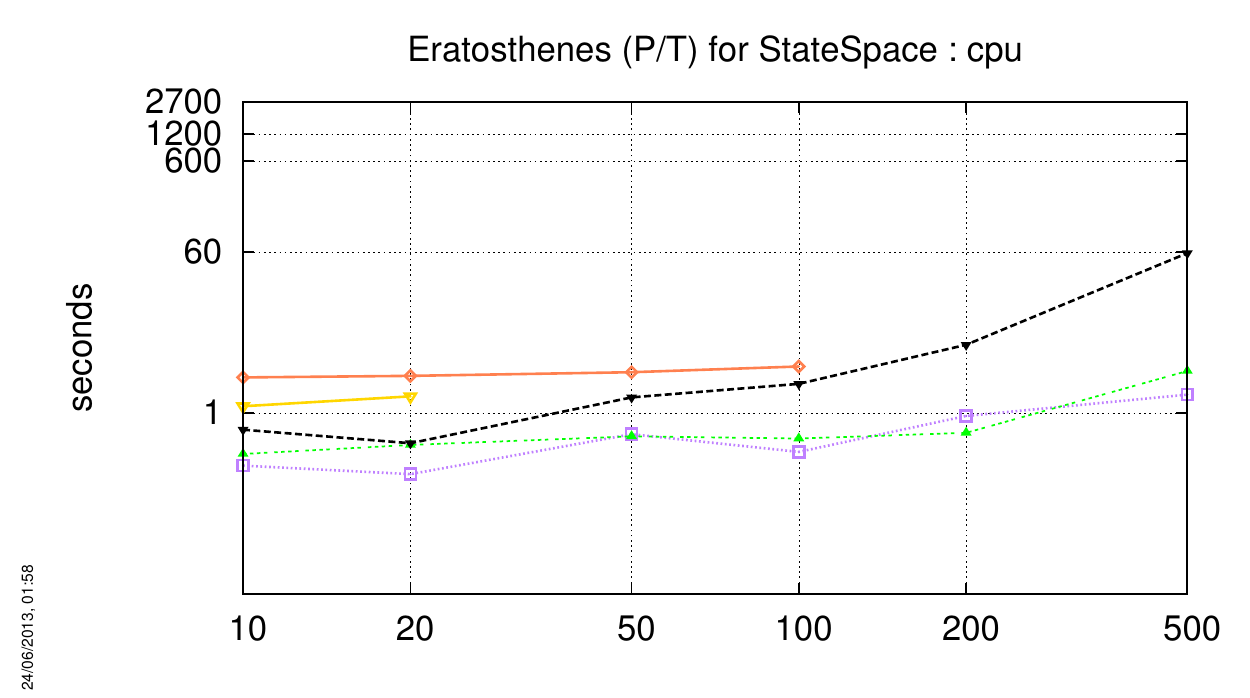}

   \includegraphics[height=1cm]{figures/tools-legend.pdf}
\end{center}

\subsubsection{\acs{FMS-PT}}
The charts below respectively show how tools compete with this ``Known'' model (memory and CPU).

\index{Performances!StateSpace!FMS (P/T)}
\begin{center}
   \includegraphics[width=7.2cm]{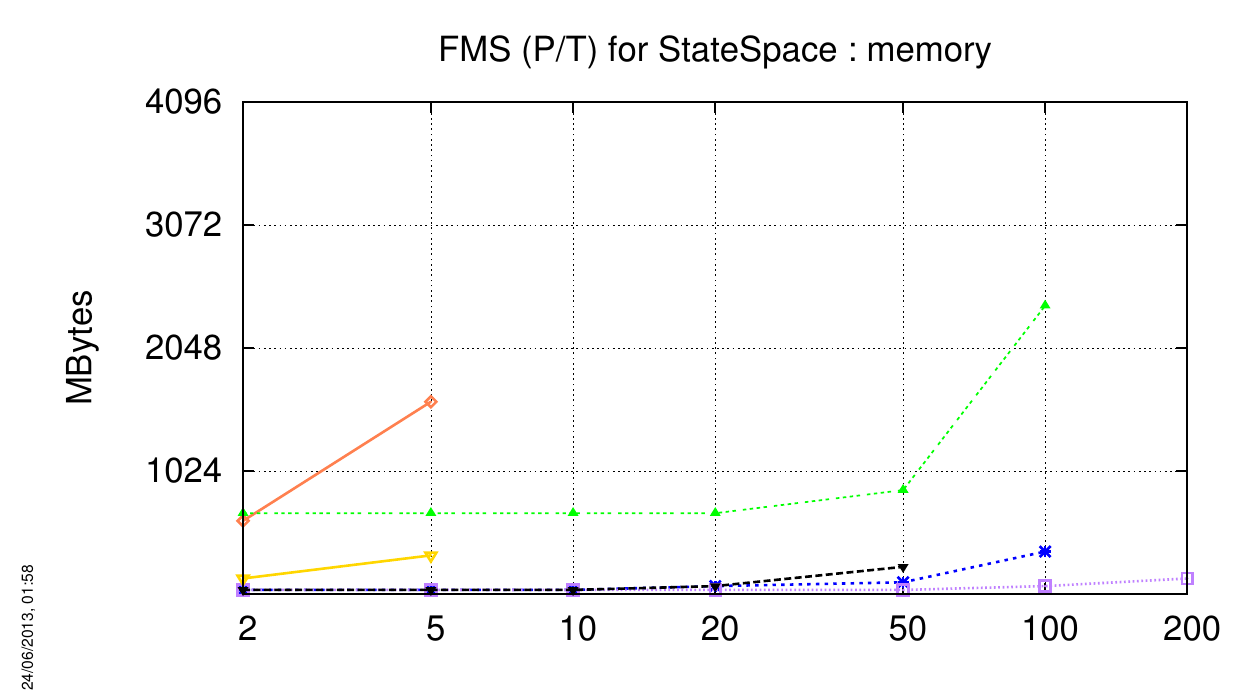}
   \includegraphics[width=7.2cm]{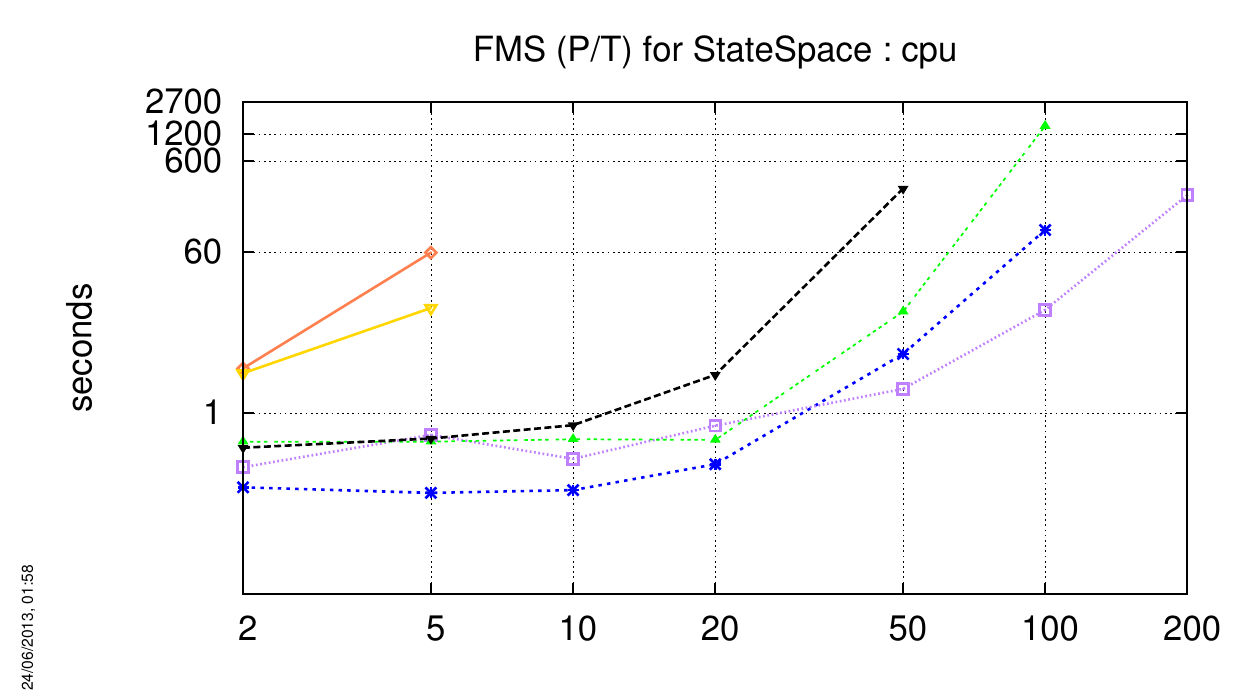}

   \includegraphics[height=1cm]{figures/tools-legend.pdf}
\end{center}

\subsubsection{\acs{GlobalRessAlloc-COL}}
The charts below respectively show how tools compete with this ``Known'' model (memory and CPU).

\index{Performances!StateSpace!GlobalRessAlloc (Colored)}
\begin{center}
   \includegraphics[width=7.2cm]{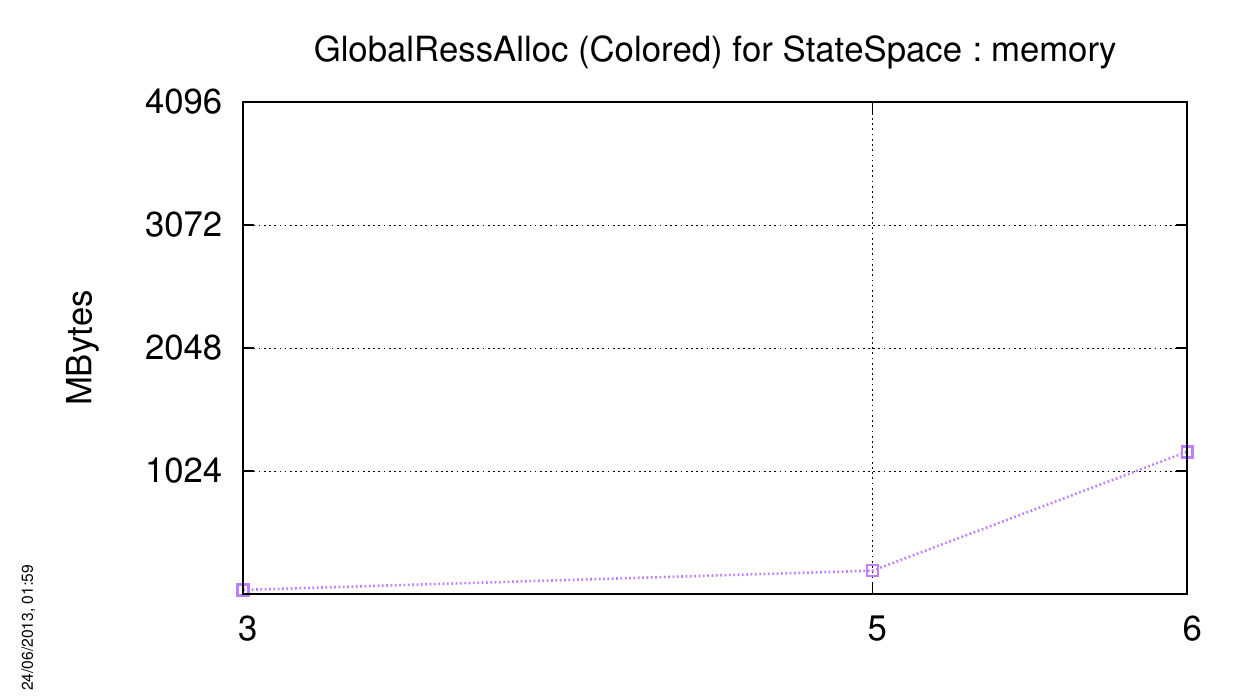}
   \includegraphics[width=7.2cm]{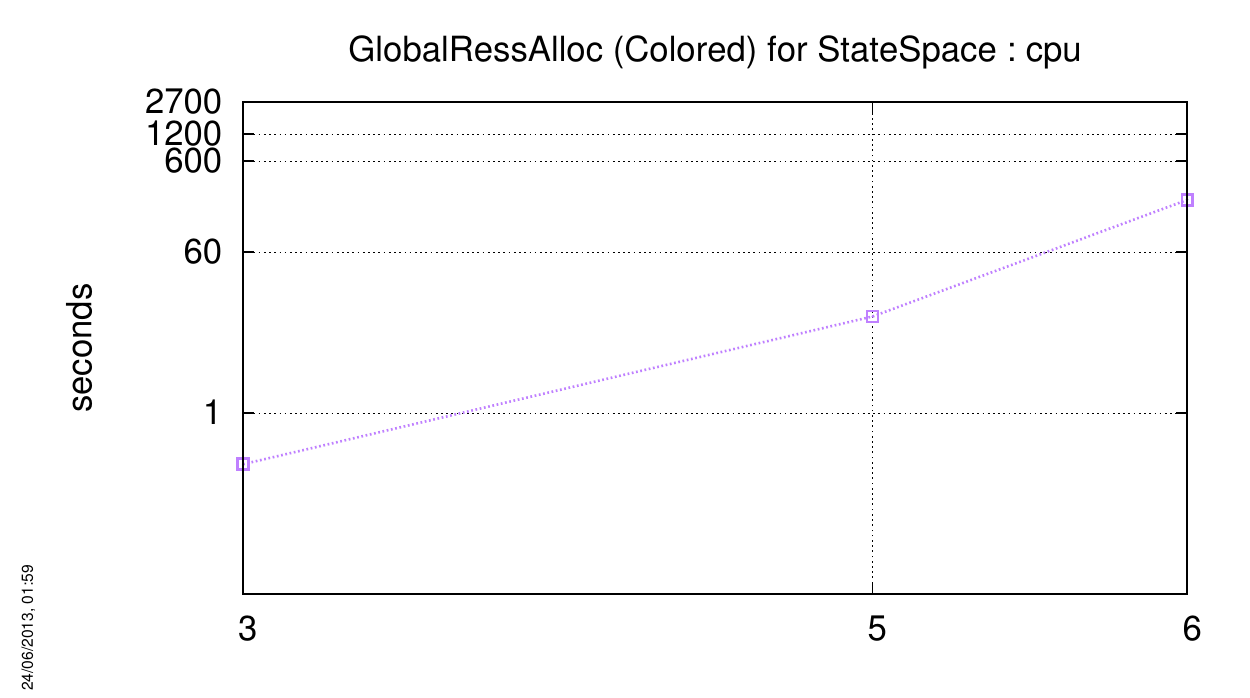}

   \includegraphics[height=1cm]{figures/tools-legend.pdf}
\end{center}

\subsubsection{\acs{GlobalRessAlloc-PT}}
The charts below respectively show how tools compete with this ``Known'' model (memory and CPU).

\index{Performances!StateSpace!GlobalRessAlloc (P/T)}
\begin{center}
   \includegraphics[width=7.2cm]{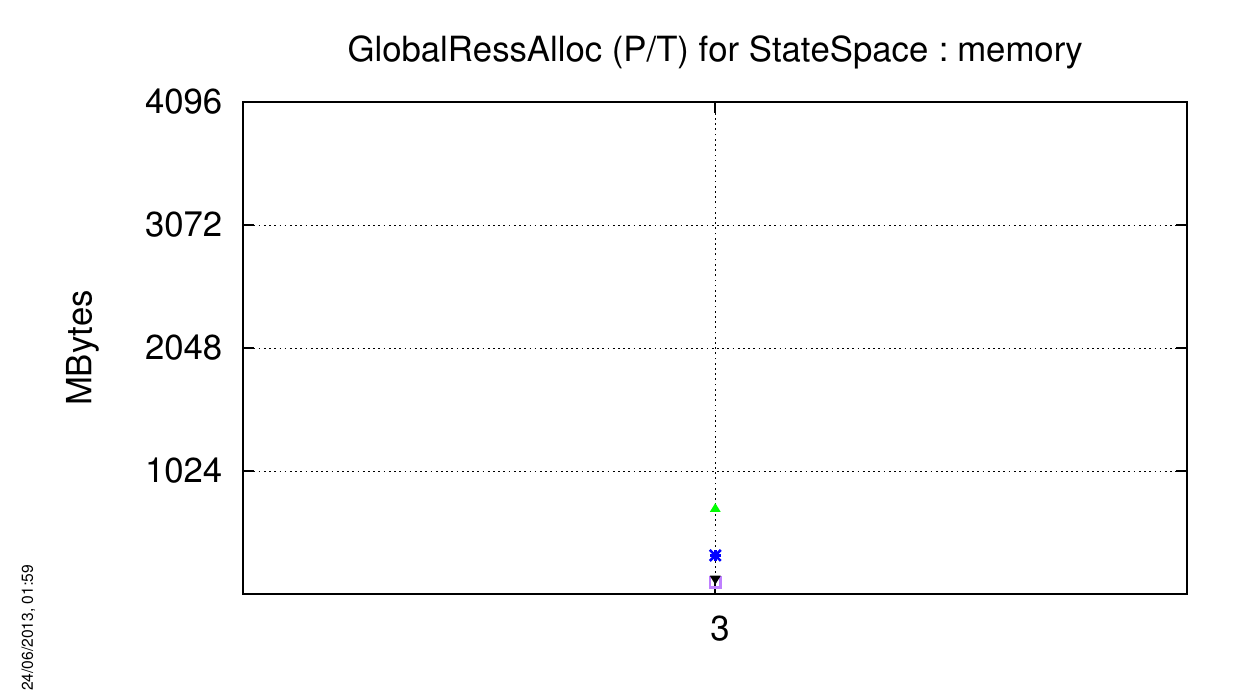}
   \includegraphics[width=7.2cm]{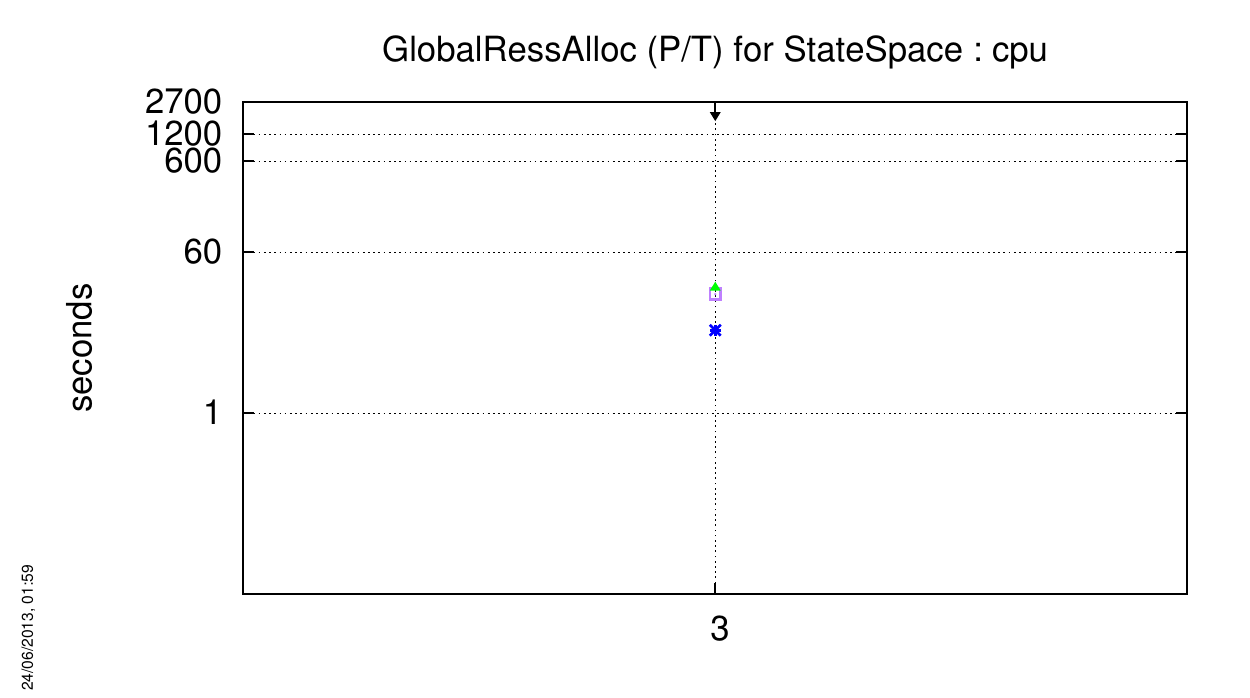}

   \includegraphics[height=1cm]{figures/tools-legend.pdf}
\end{center}

\subsubsection{\acs{Kanban-PT}}
The charts below respectively show how tools compete with this ``Known'' model (memory and CPU).

\index{Performances!StateSpace!Kanban (P/T)}
\begin{center}
   \includegraphics[width=7.2cm]{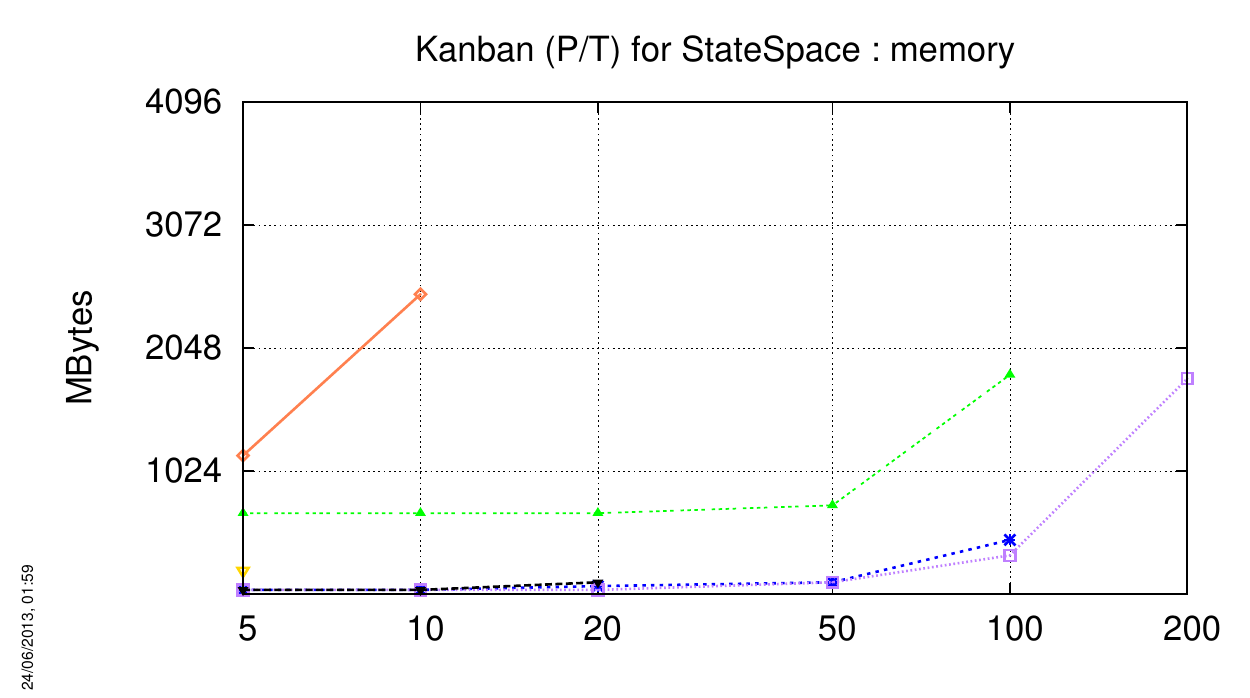}
   \includegraphics[width=7.2cm]{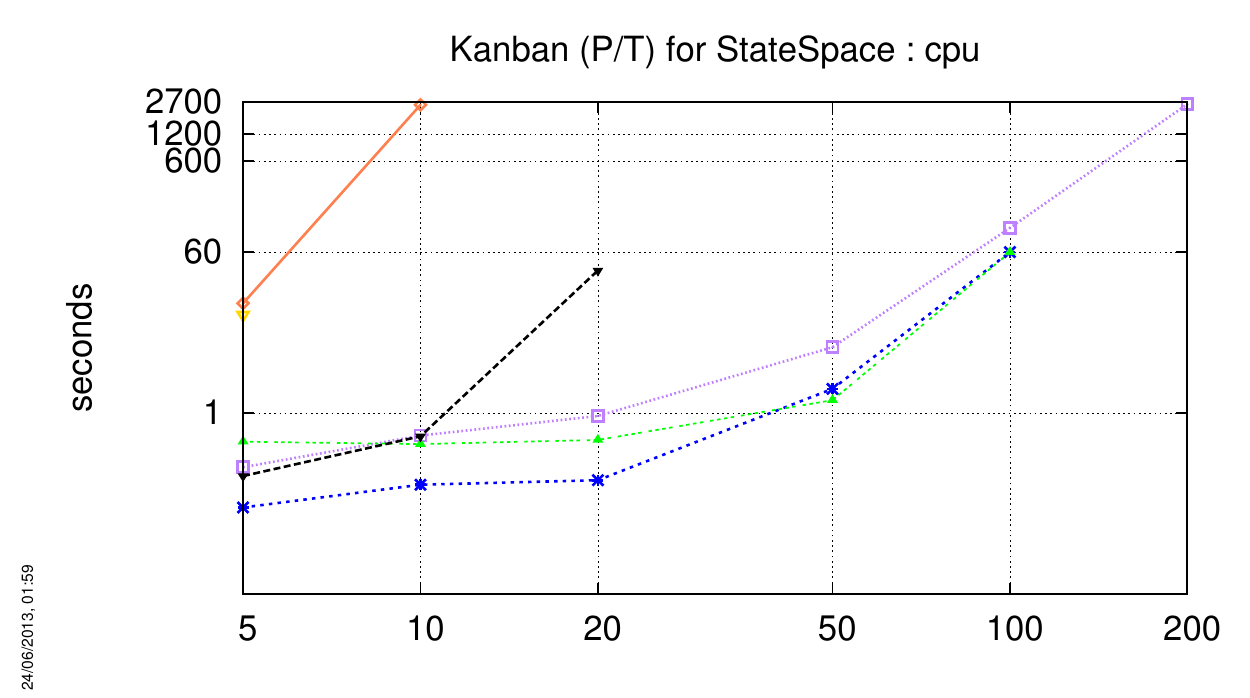}

   \includegraphics[height=1cm]{figures/tools-legend.pdf}
\end{center}

\subsubsection{\acs{LamportFastMutEx-COL}}
The charts below respectively show how tools compete with this ``Known'' model (memory and CPU).

\index{Performances!StateSpace!LamportFastMutEx (Colored)}
\begin{center}
   \includegraphics[width=7.2cm]{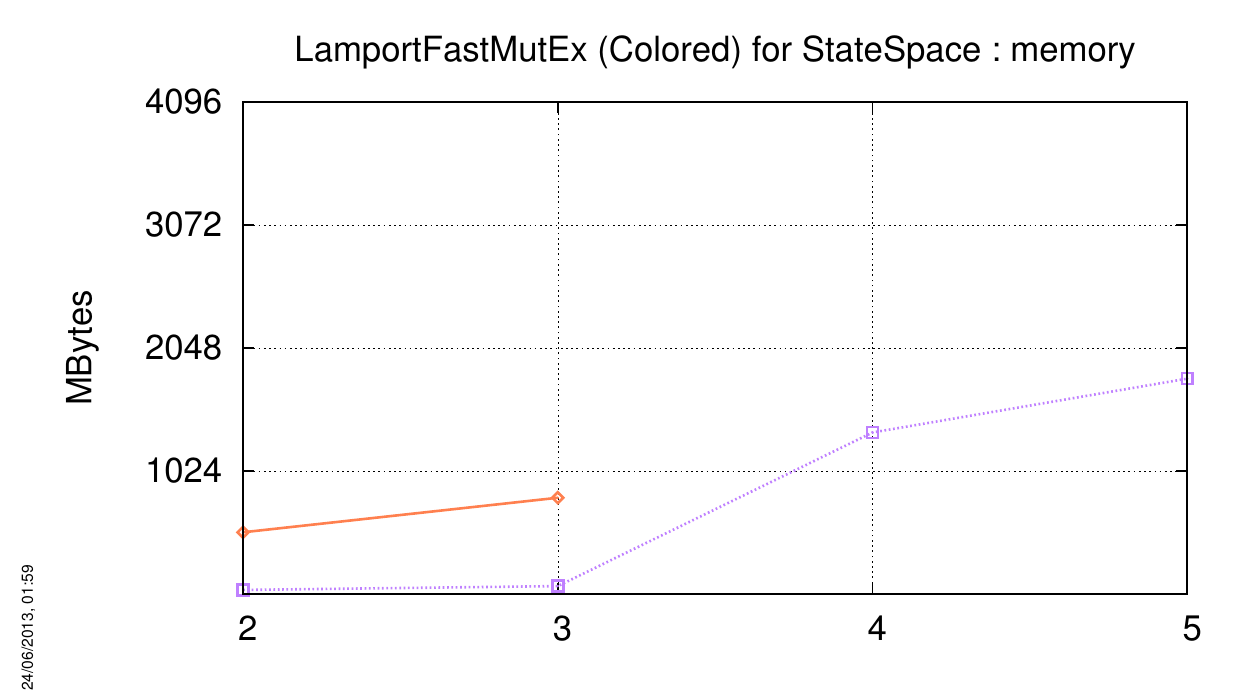}
   \includegraphics[width=7.2cm]{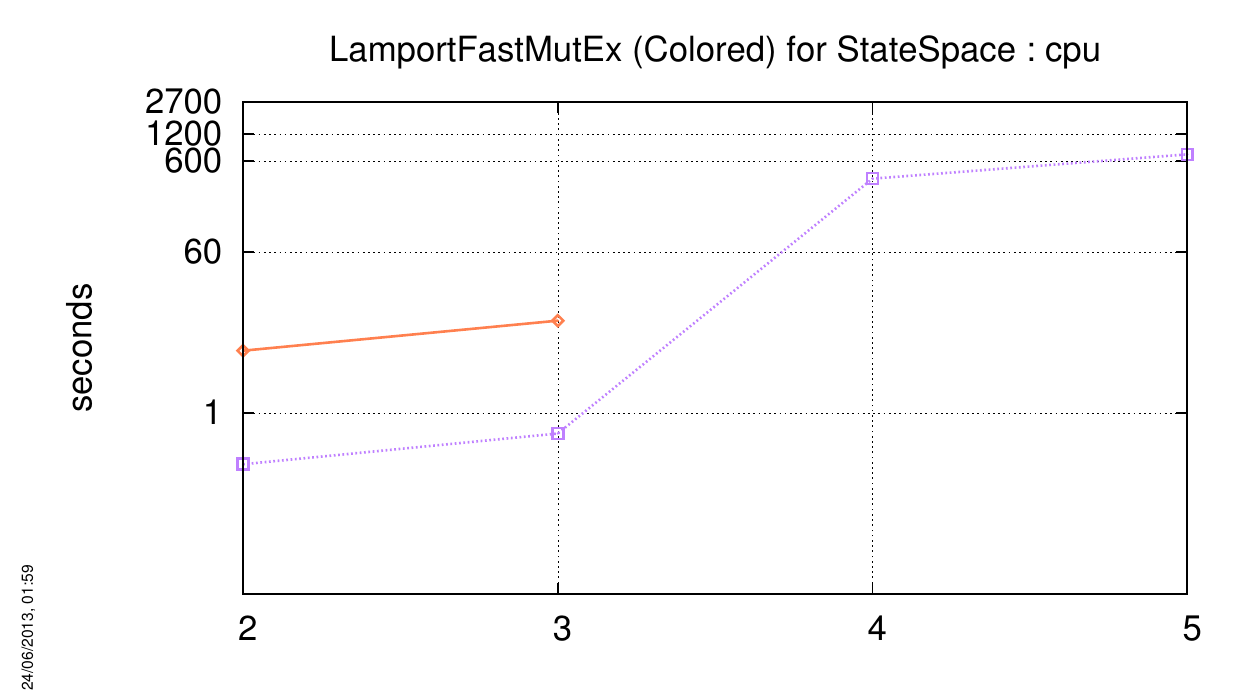}

   \includegraphics[height=1cm]{figures/tools-legend.pdf}
\end{center}

\subsubsection{\acs{LamportFastMutEx-PT}}
The charts below respectively show how tools compete with this ``Known'' model (memory and CPU).

\index{Performances!StateSpace!LamportFastMutEx (P/T)}
\begin{center}
   \includegraphics[width=7.2cm]{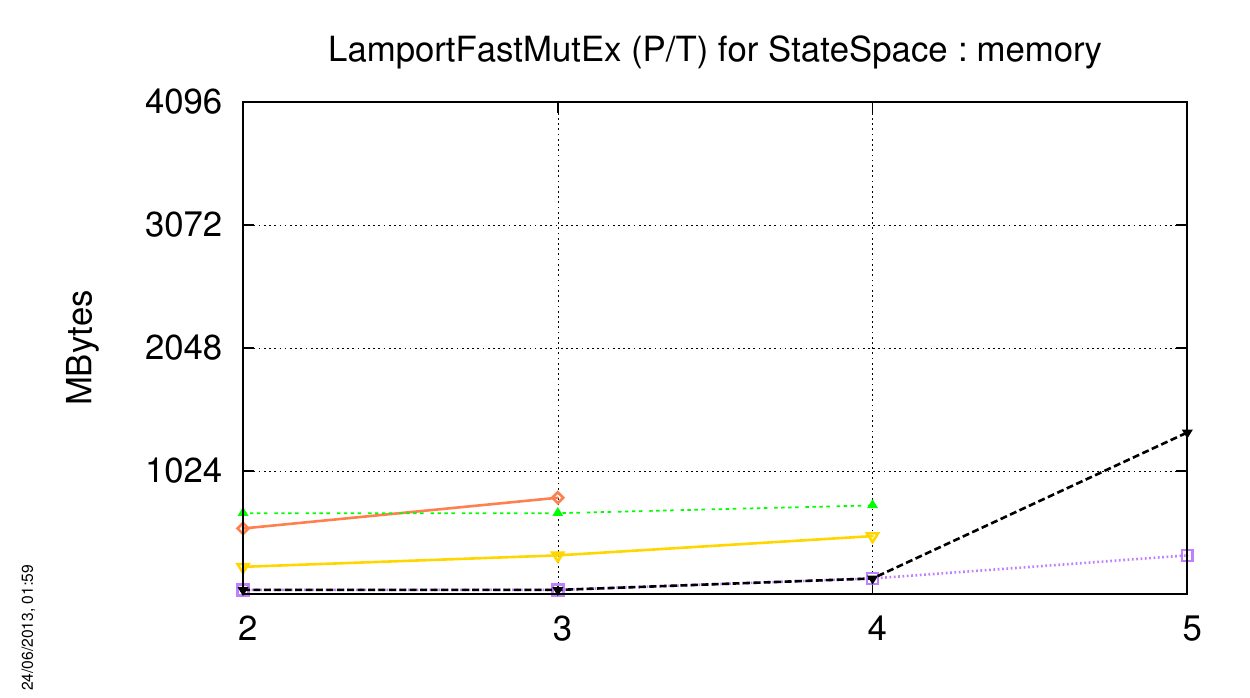}
   \includegraphics[width=7.2cm]{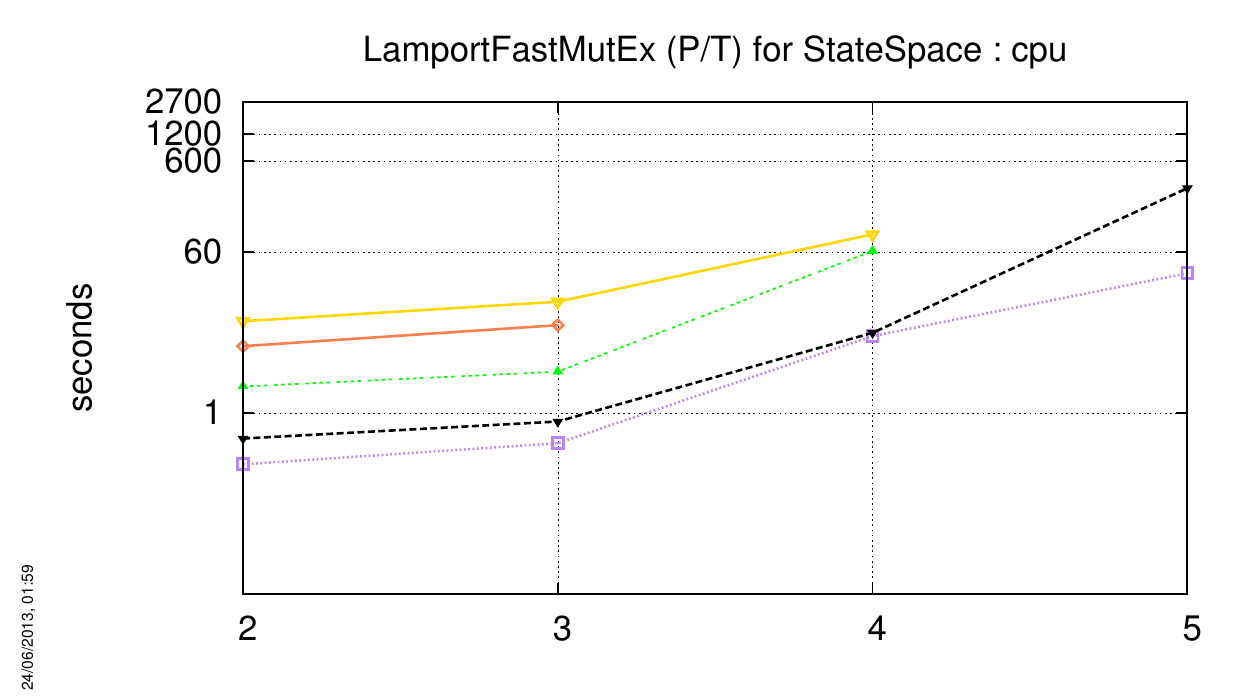}

   \includegraphics[height=1cm]{figures/tools-legend.pdf}
\end{center}

\subsubsection{\acs{MAPK-PT}}
The charts below respectively show how tools compete with this ``Known'' model (memory and CPU).

\index{Performances!StateSpace!MAPK (P/T)}
\begin{center}
   \includegraphics[width=7.2cm]{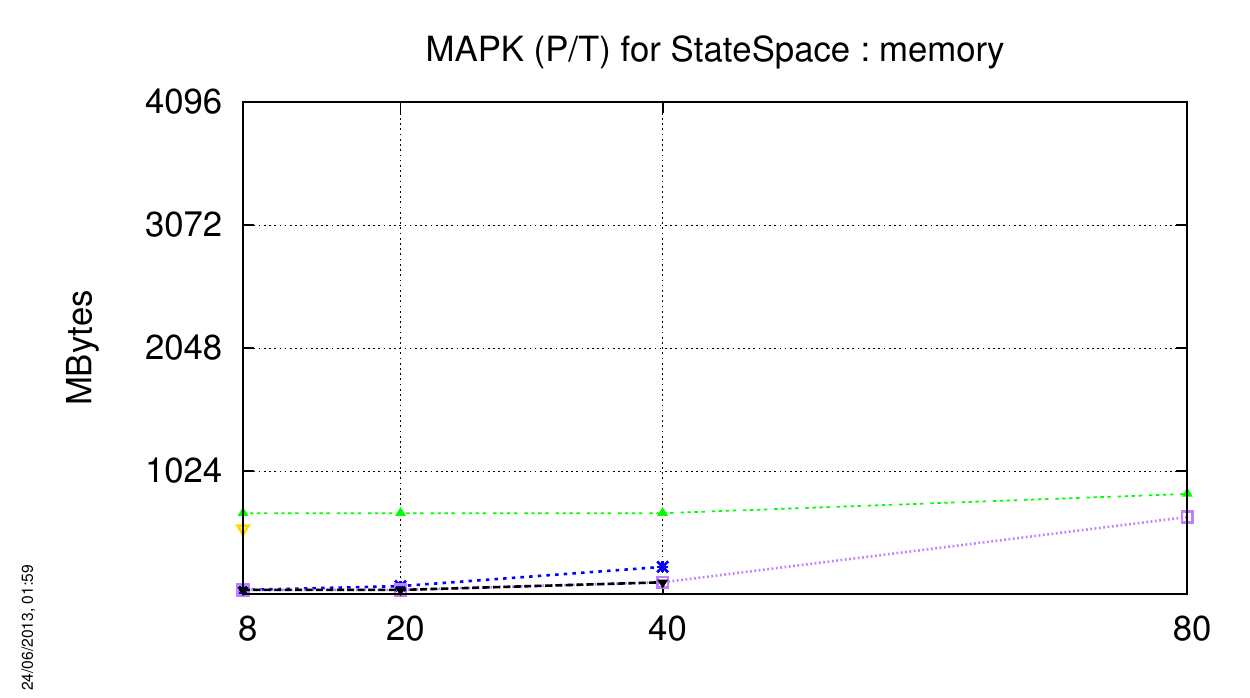}
   \includegraphics[width=7.2cm]{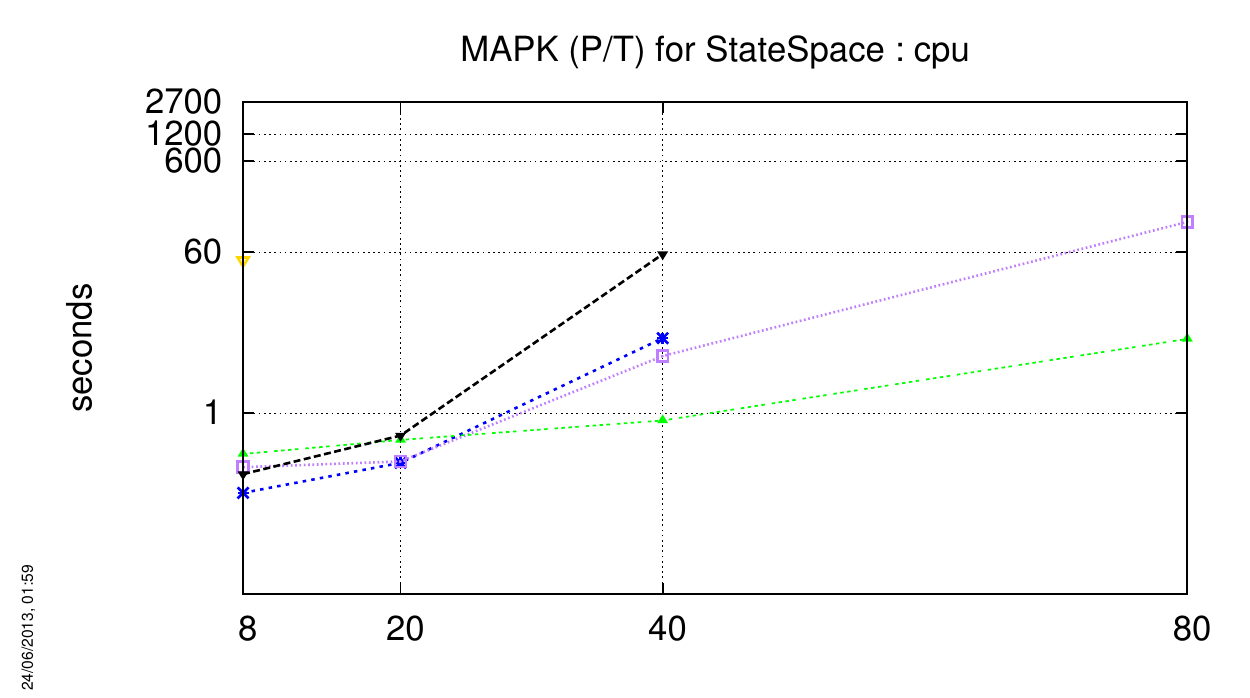}

   \includegraphics[height=1cm]{figures/tools-legend.pdf}
\end{center}

\subsubsection{\acs{NeoElection-COL}}
The charts below respectively show how tools compete with this ``Known'' model (memory and CPU).

\index{Performances!StateSpace!NeoElection (Colored)}
\begin{center}
   \includegraphics[width=7.2cm]{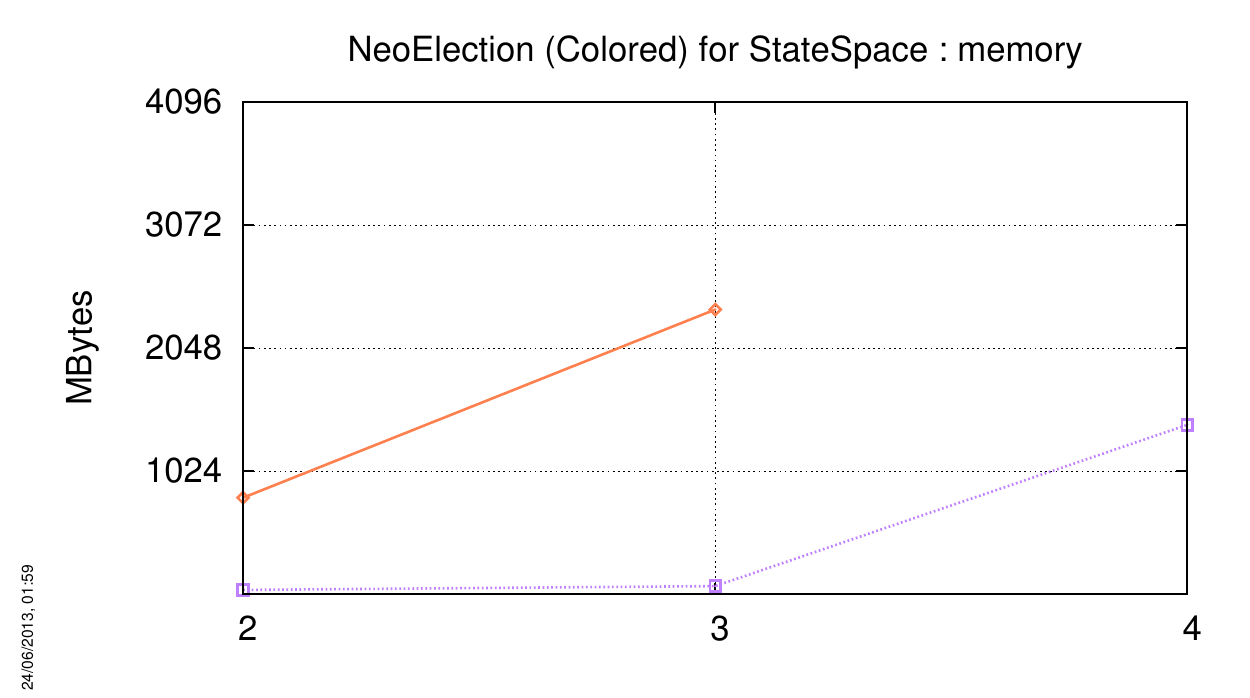}
   \includegraphics[width=7.2cm]{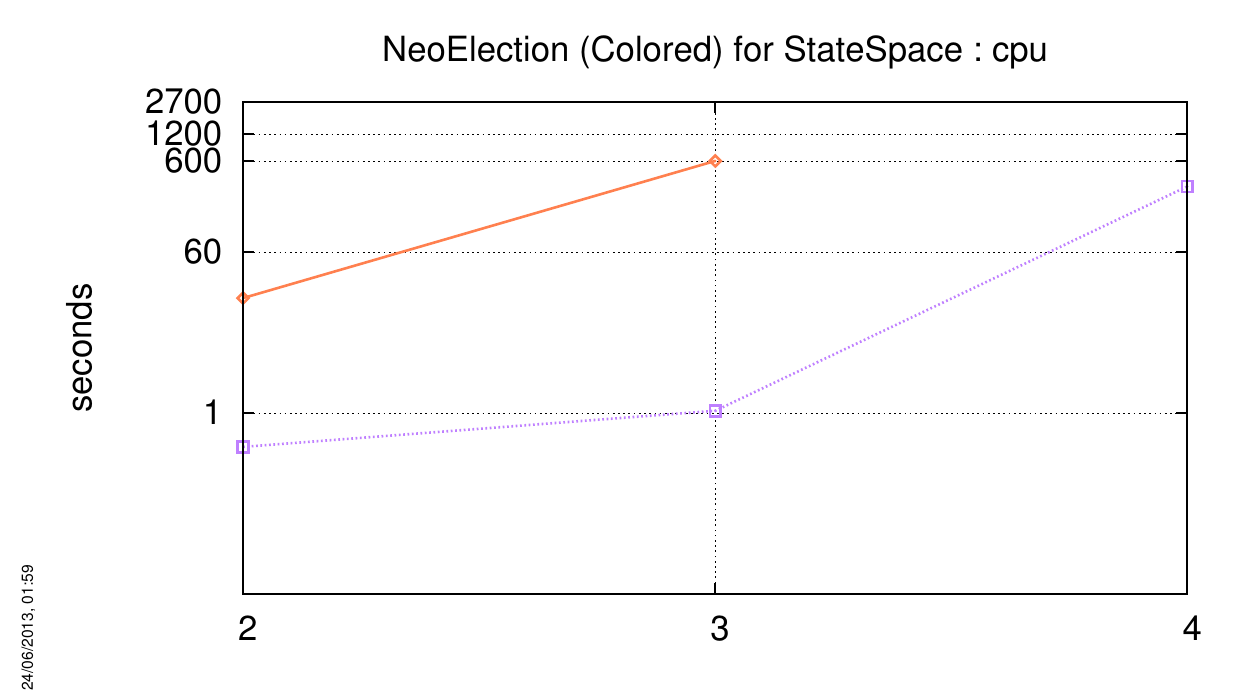}

   \includegraphics[height=1cm]{figures/tools-legend.pdf}
\end{center}

\subsubsection{\acs{NeoElection-PT}}
The charts below respectively show how tools compete with this ``Known'' model (memory and CPU).

\index{Performances!StateSpace!NeoElection (P/T)}
\begin{center}
   \includegraphics[width=7.2cm]{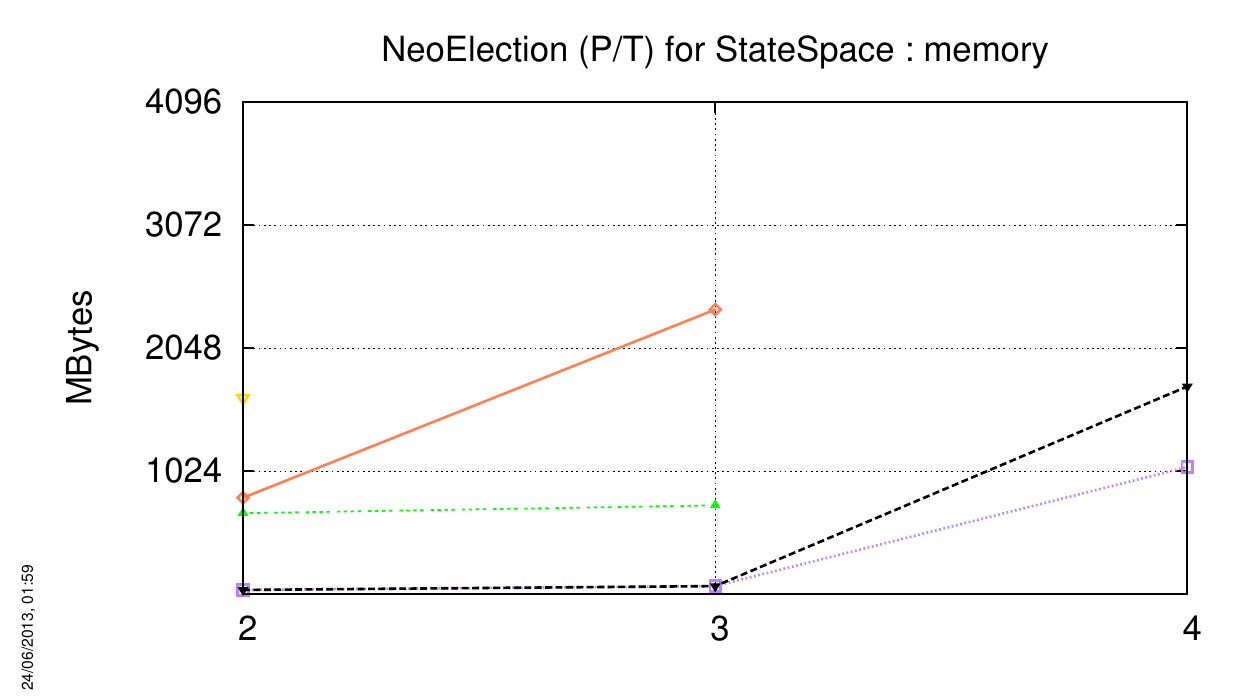}
   \includegraphics[width=7.2cm]{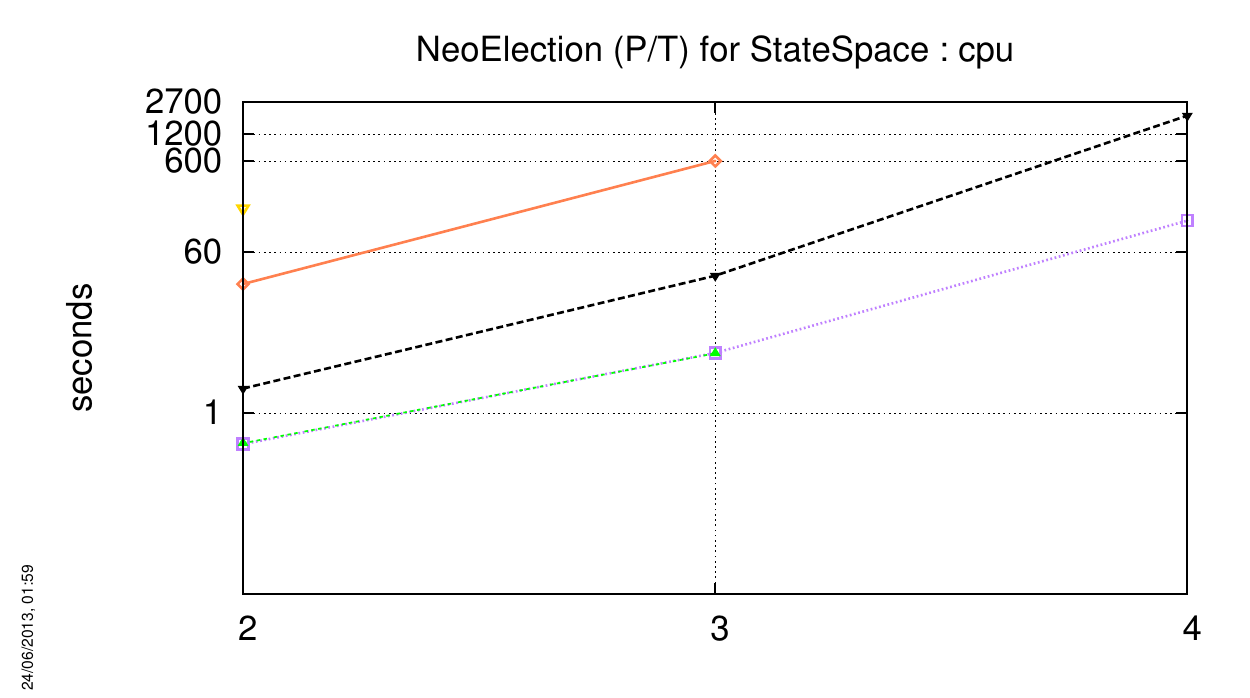}

   \includegraphics[height=1cm]{figures/tools-legend.pdf}
\end{center}

\subsubsection{\acs{PermAdmissibility-COL}}
The charts below respectively show how tools compete with this ``Known'' model (memory and CPU).

\index{Performances!StateSpace!PermAdmissibility (Colored)}
\begin{center}
   \includegraphics[width=7.2cm]{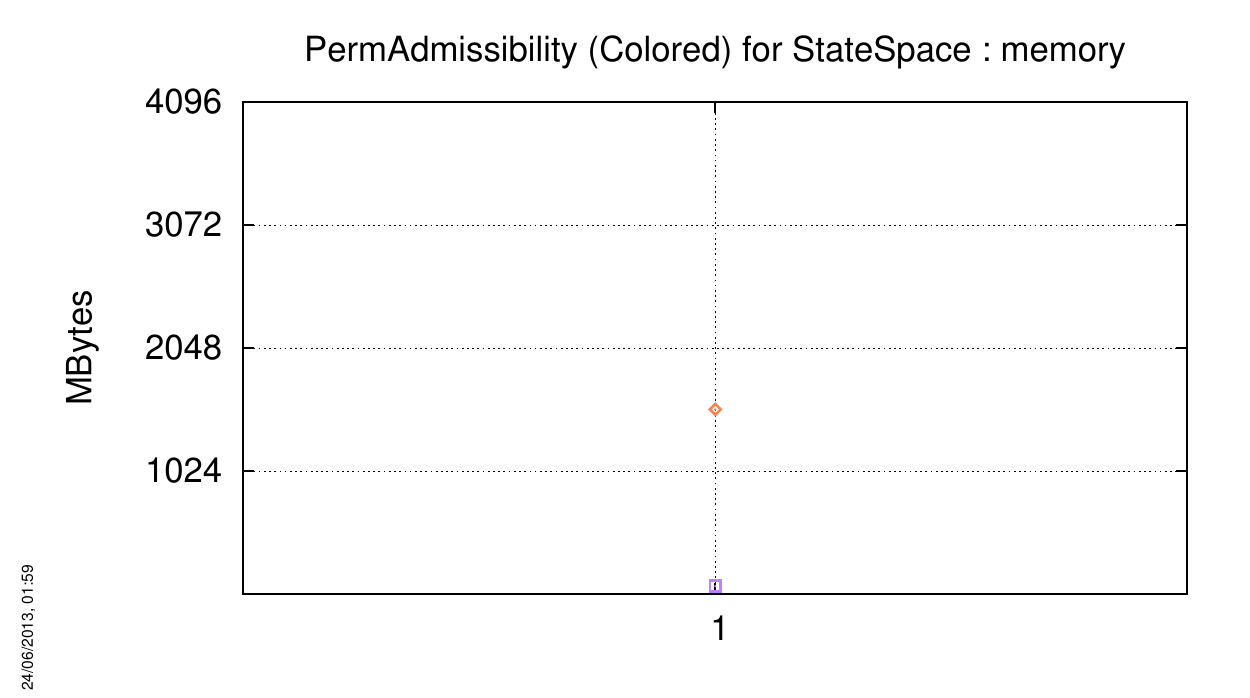}
   \includegraphics[width=7.2cm]{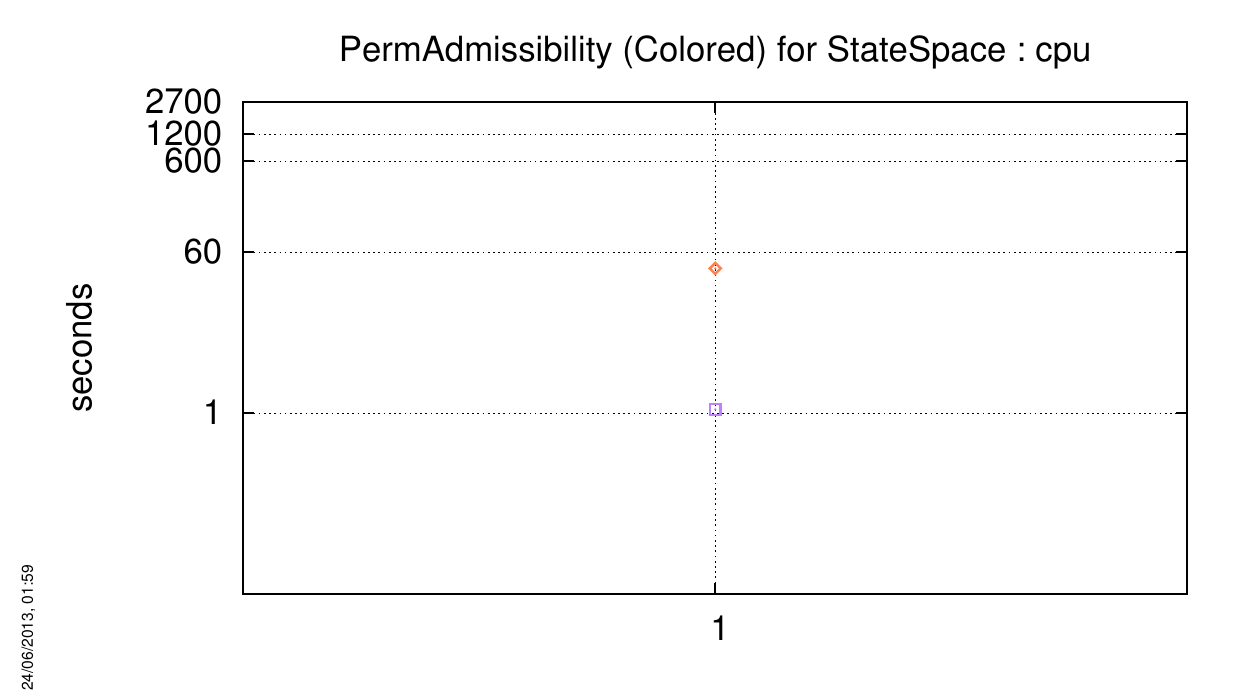}

   \includegraphics[height=1cm]{figures/tools-legend.pdf}
\end{center}

\subsubsection{\acs{PermAdmissibility-PT}}
The charts below respectively show how tools compete with this ``Known'' model (memory and CPU).

\index{Performances!StateSpace!PermAdmissibility (P/T)}
\begin{center}
   \includegraphics[width=7.2cm]{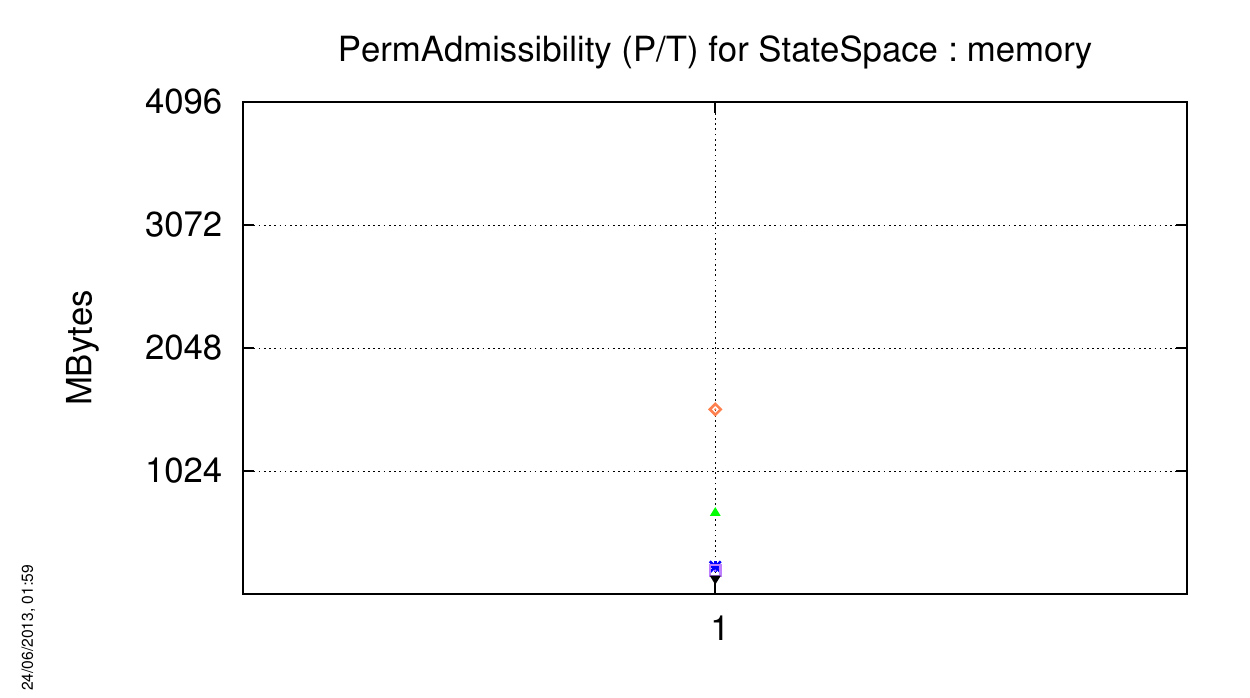}
   \includegraphics[width=7.2cm]{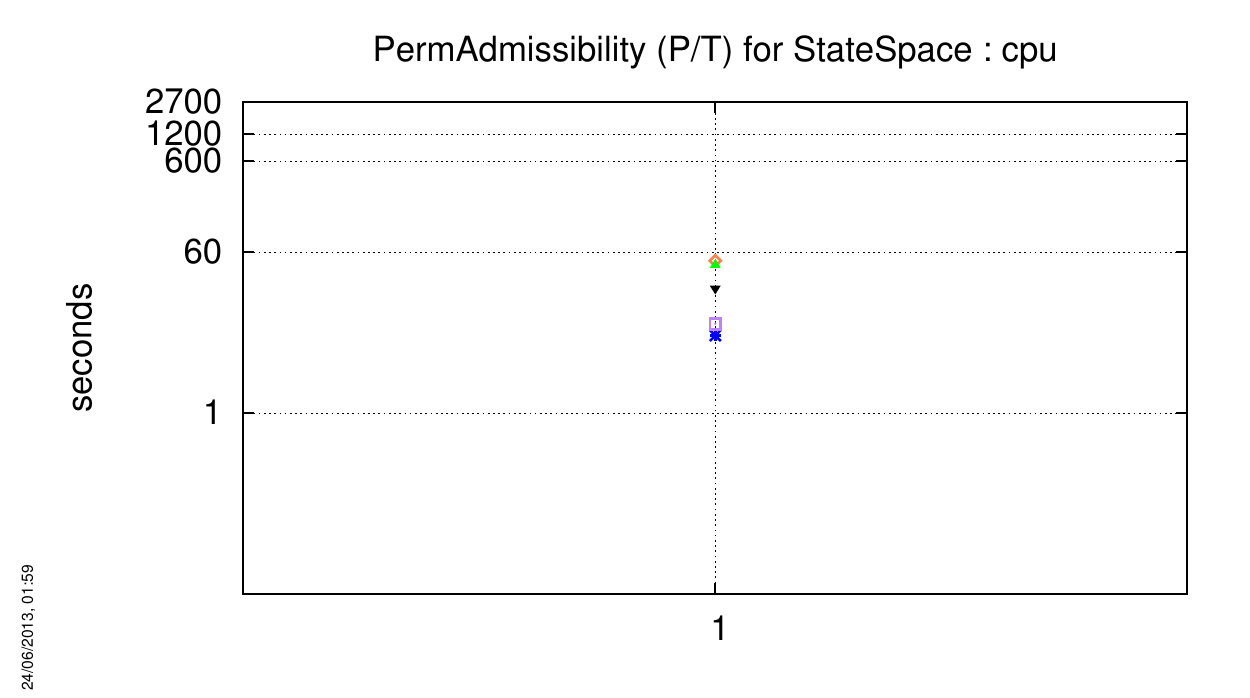}

   \includegraphics[height=1cm]{figures/tools-legend.pdf}
\end{center}

\subsubsection{\acs{Peterson-COL}}
The charts below respectively show how tools compete with this ``Known'' model (memory and CPU).

\index{Performances!StateSpace!Peterson (Colored)}
\begin{center}
   \includegraphics[width=7.2cm]{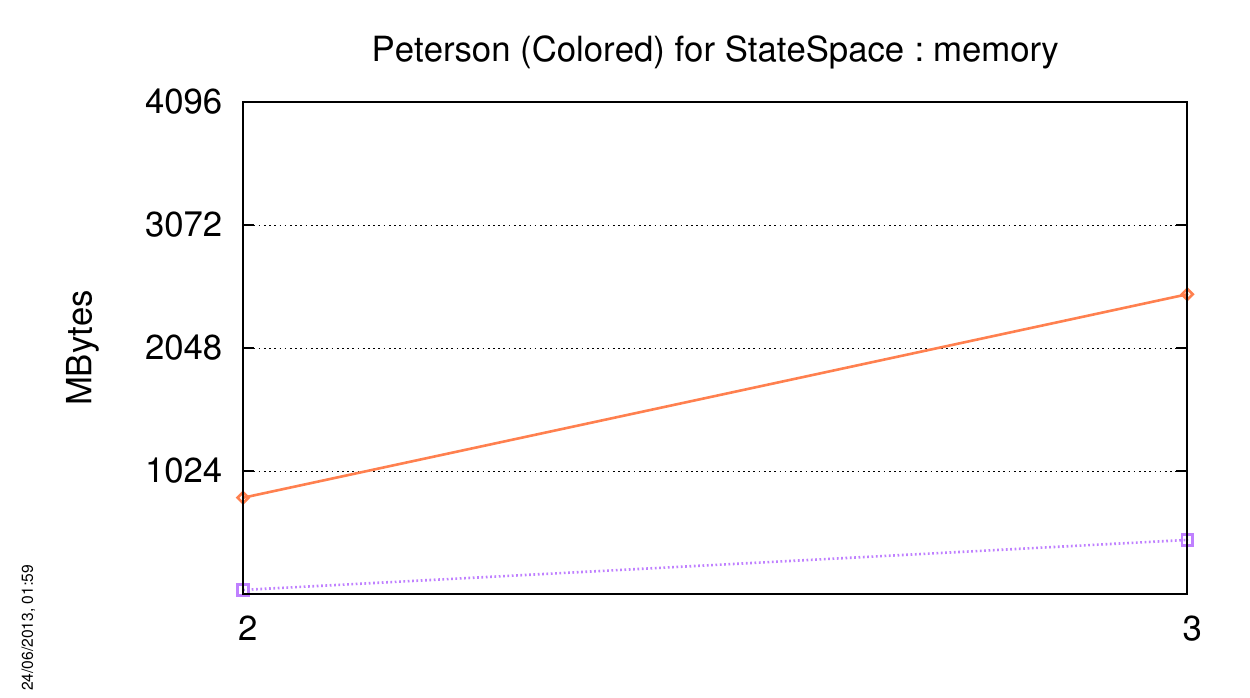}
   \includegraphics[width=7.2cm]{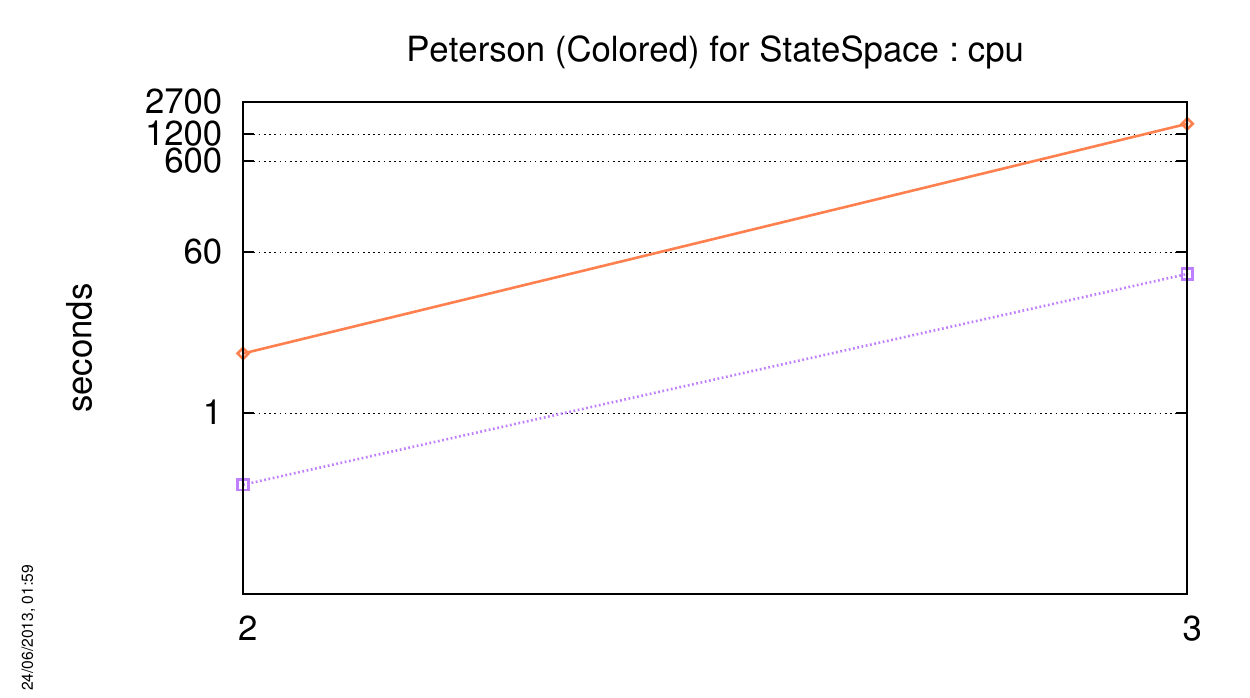}

   \includegraphics[height=1cm]{figures/tools-legend.pdf}
\end{center}

\subsubsection{\acs{Peterson-PT}}
The charts below respectively show how tools compete with this ``Known'' model (memory and CPU).

\index{Performances!StateSpace!Peterson (P/T)}
\begin{center}
   \includegraphics[width=7.2cm]{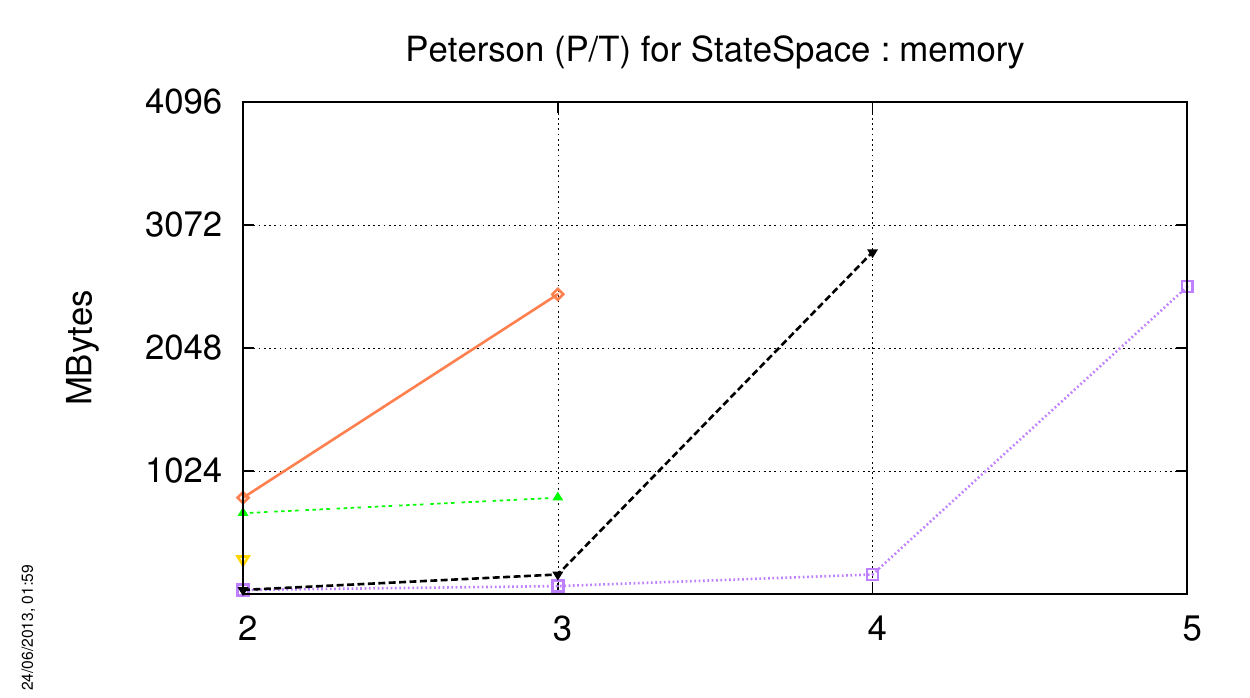}
   \includegraphics[width=7.2cm]{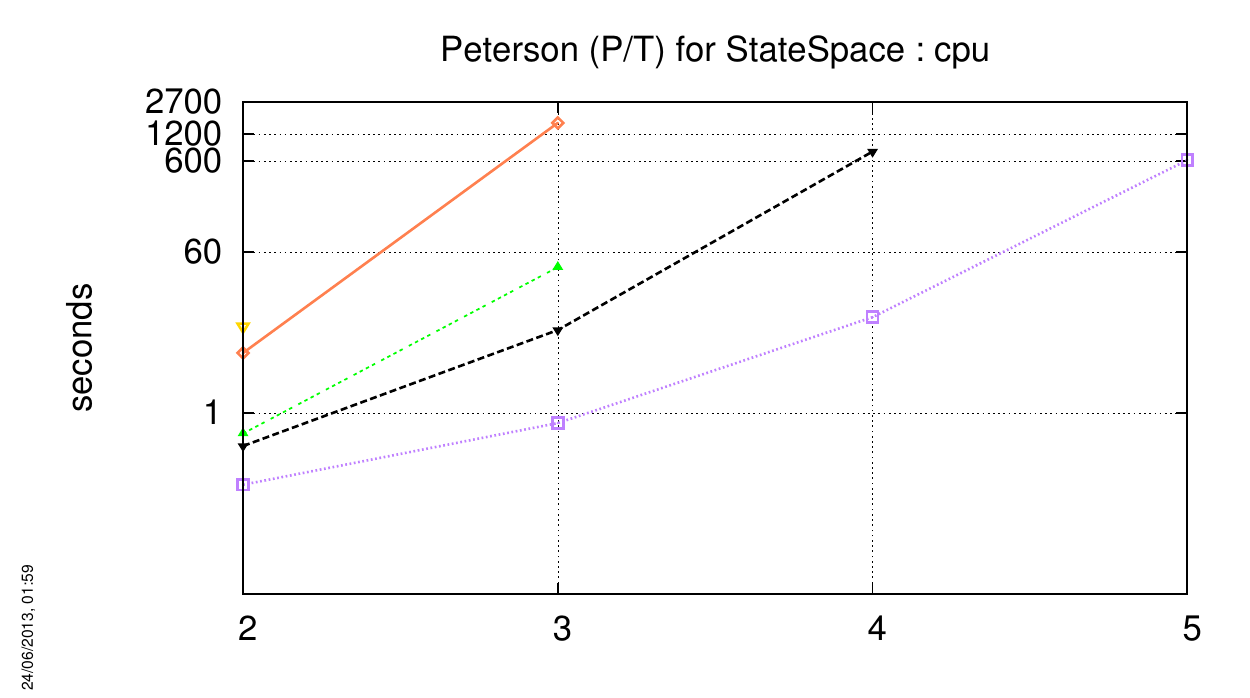}

   \includegraphics[height=1cm]{figures/tools-legend.pdf}
\end{center}

\subsubsection{\acs{Philosophers-COL}}
The charts below respectively show how tools compete with this ``Known'' model (memory and CPU).

\index{Performances!StateSpace!Philosophers (Colored)}
\begin{center}
   \includegraphics[width=7.2cm]{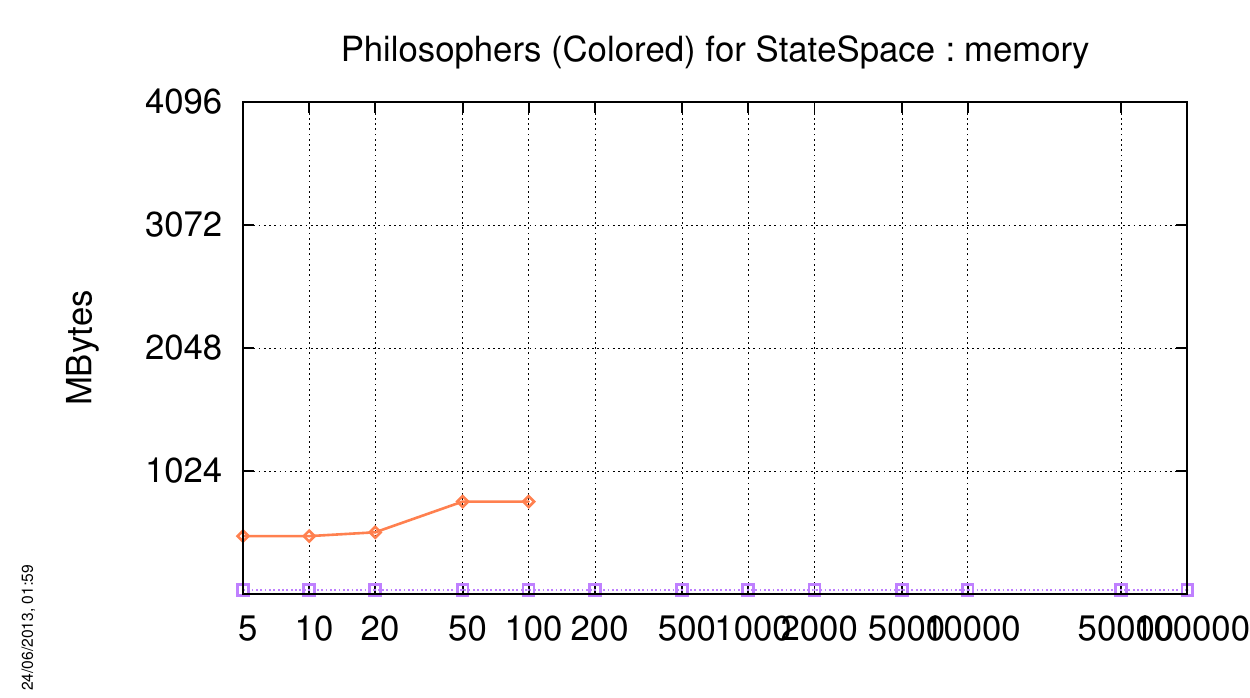}
   \includegraphics[width=7.2cm]{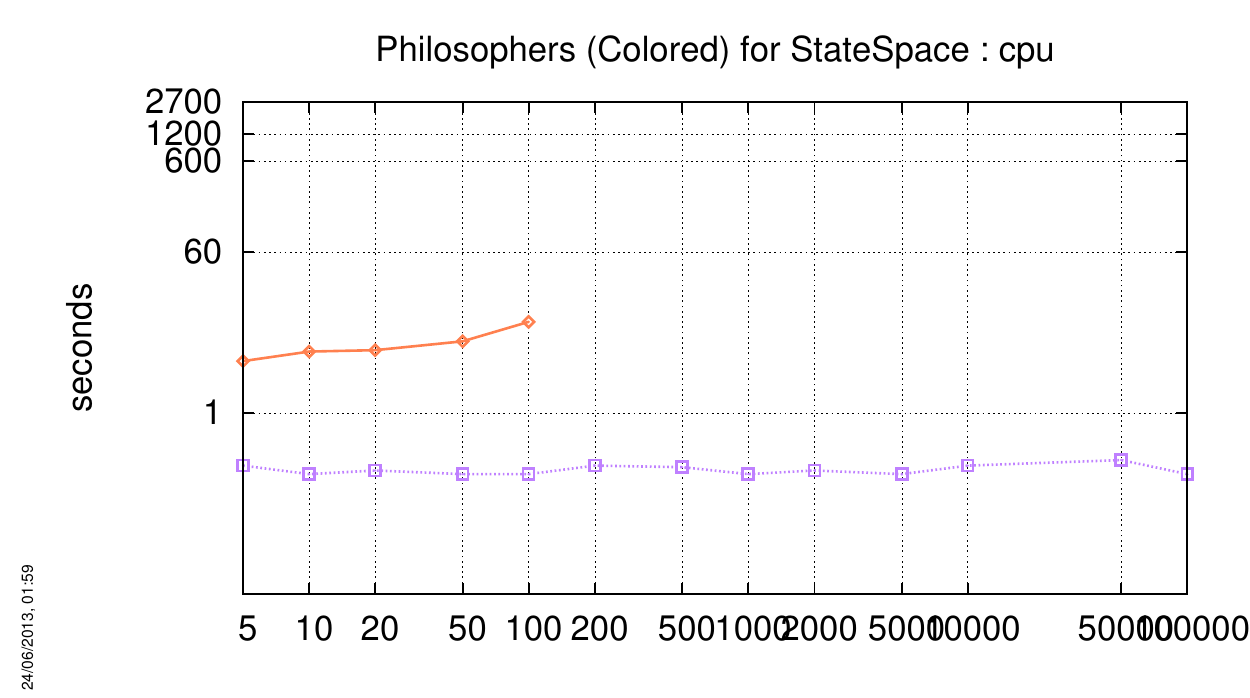}

   \includegraphics[height=1cm]{figures/tools-legend.pdf}
\end{center}

\subsubsection{\acs{Philosophers-PT}}
The charts below respectively show how tools compete with this ``Known'' model (memory and CPU).

\index{Performances!StateSpace!Philosophers (P/T)}
\begin{center}
   \includegraphics[width=7.2cm]{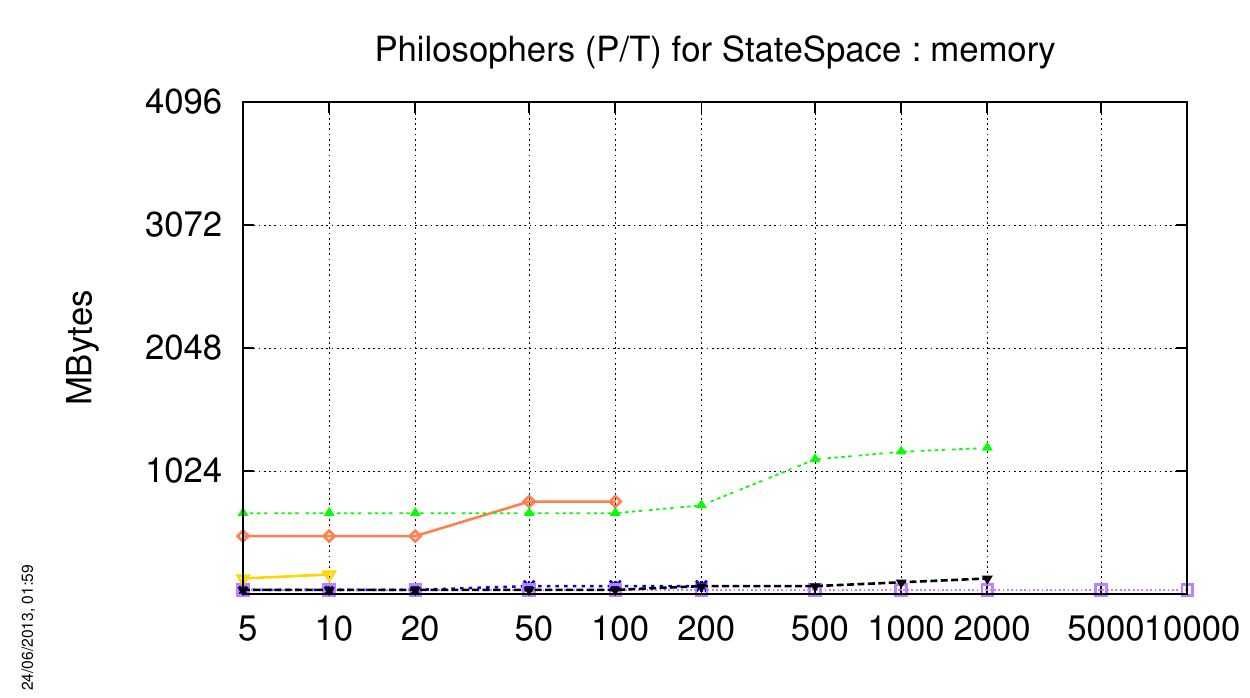}
   \includegraphics[width=7.2cm]{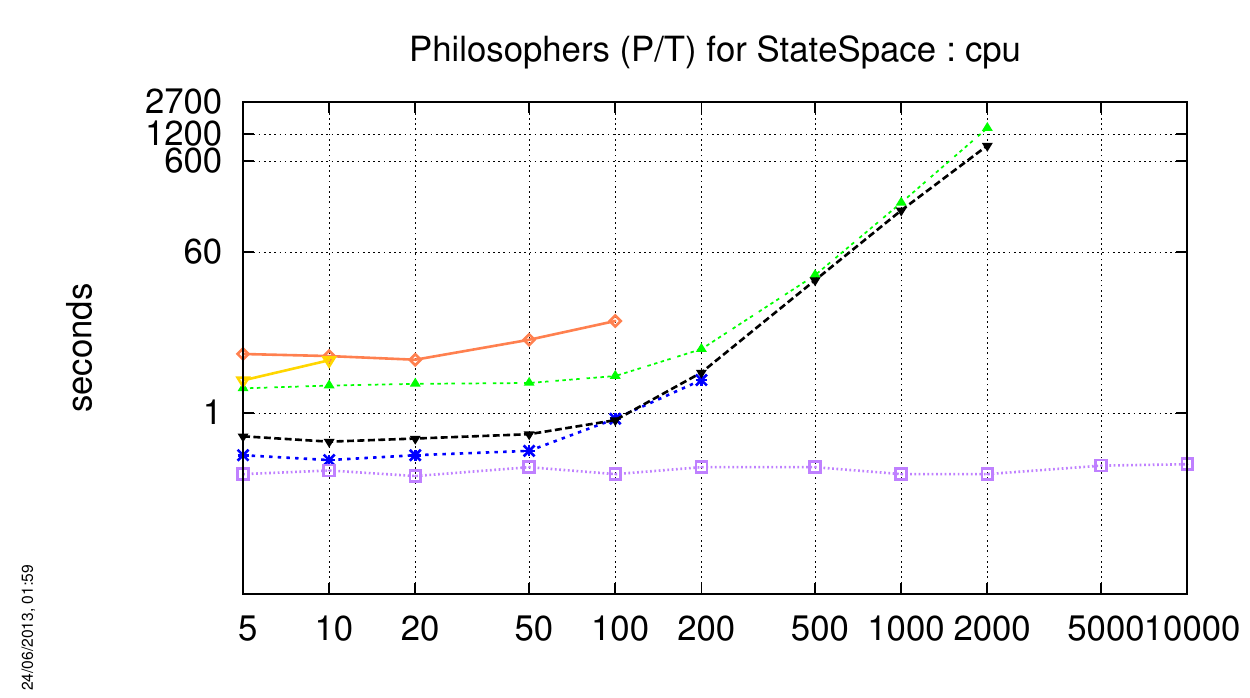}

   \includegraphics[height=1cm]{figures/tools-legend.pdf}
\end{center}

\subsubsection{\acs{PhilosophersDyn-COL}}
The charts below respectively show how tools compete with this ``Known'' model (memory and CPU).

\index{Performances!StateSpace!PhilosophersDyn (Colored)}
\begin{center}
   \includegraphics[width=7.2cm]{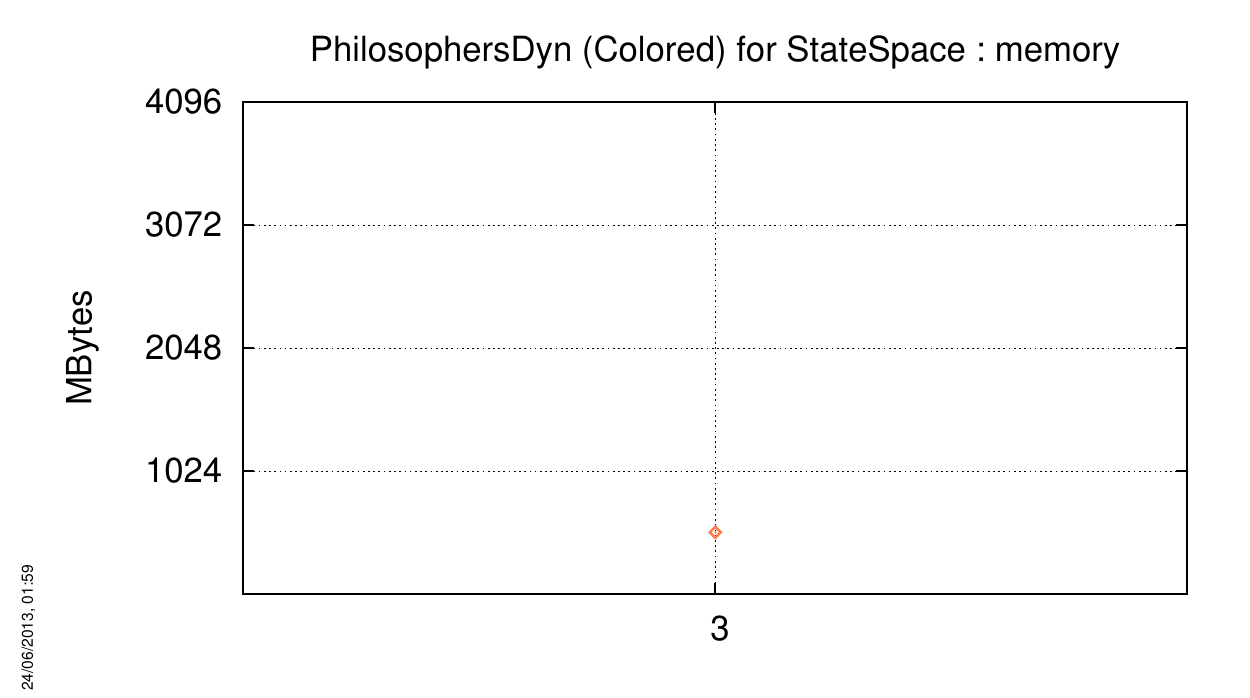}
   \includegraphics[width=7.2cm]{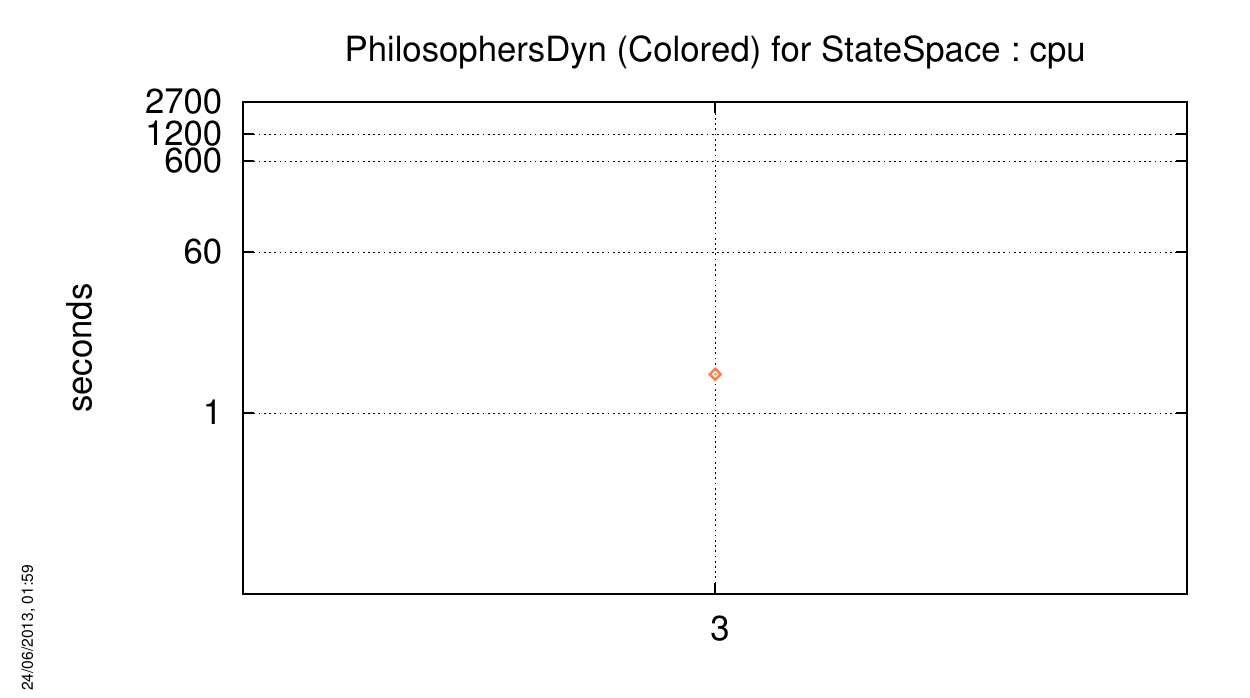}

   \includegraphics[height=1cm]{figures/tools-legend.pdf}
\end{center}

\subsubsection{\acs{PhilosophersDyn-PT}}
The charts below respectively show how tools compete with this ``Known'' model (memory and CPU).

\index{Performances!StateSpace!PhilosophersDyn (P/T)}
\begin{center}
   \includegraphics[width=7.2cm]{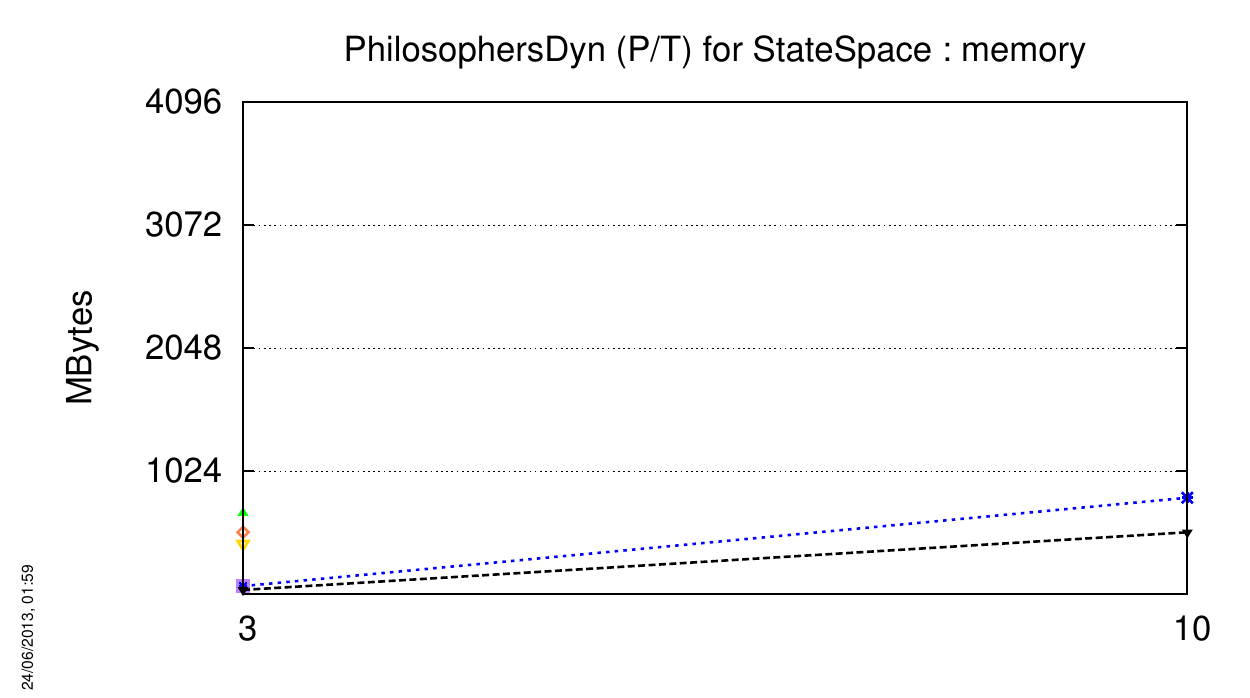}
   \includegraphics[width=7.2cm]{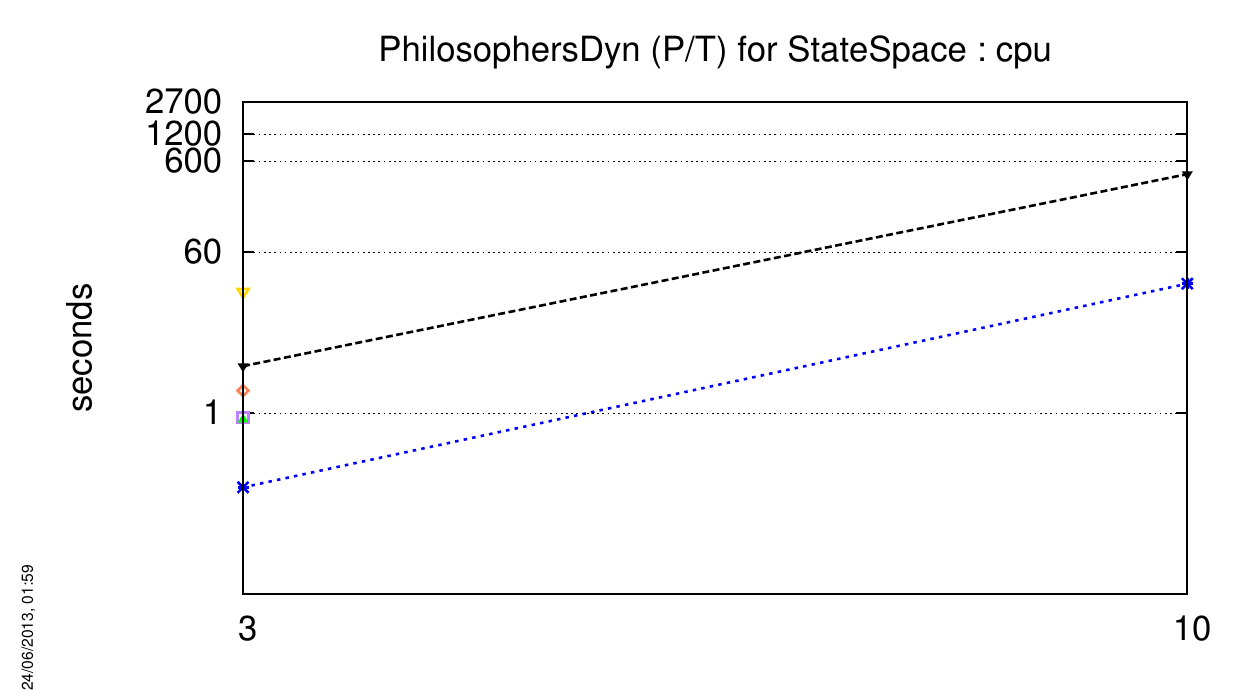}

   \includegraphics[height=1cm]{figures/tools-legend.pdf}
\end{center}

\subsubsection{\acs{Planning-PT}}
No instance of this model could be computed for the \textbf{StateSpace} examination.

\subsubsection{\acs{Railroad-PT}}
The charts below respectively show how tools compete with this ``Known'' model (memory and CPU).

\index{Performances!StateSpace!Railroad (P/T)}
\begin{center}
   \includegraphics[width=7.2cm]{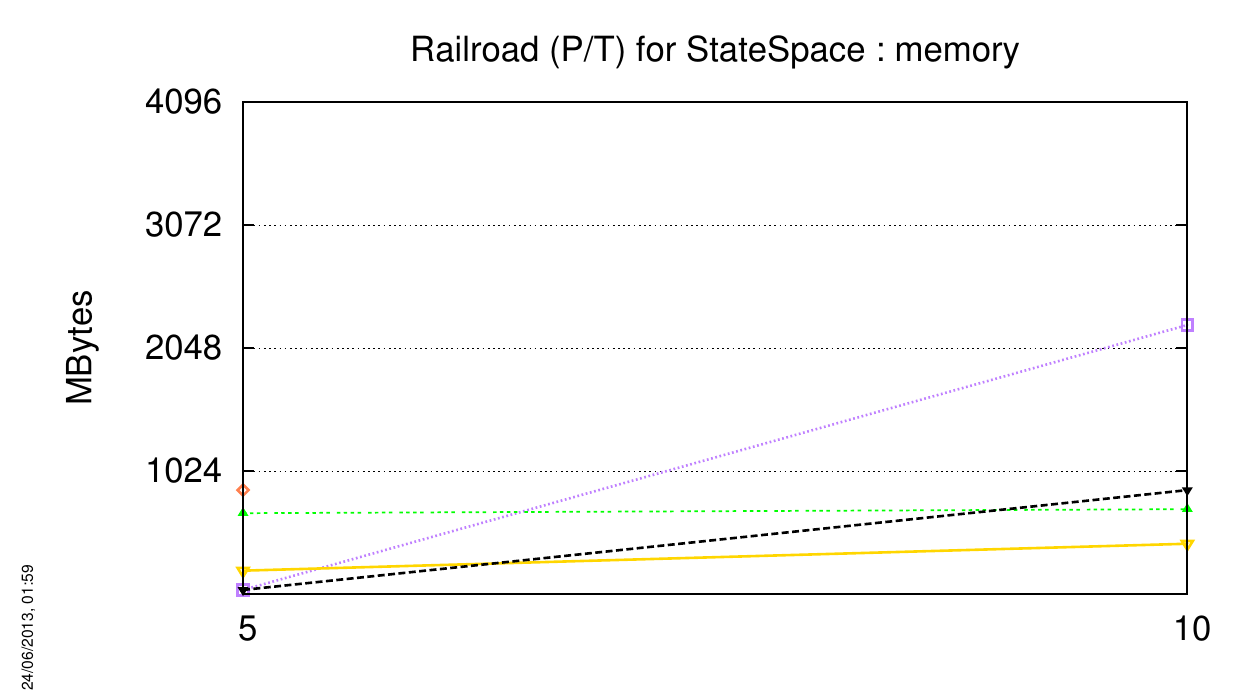}
   \includegraphics[width=7.2cm]{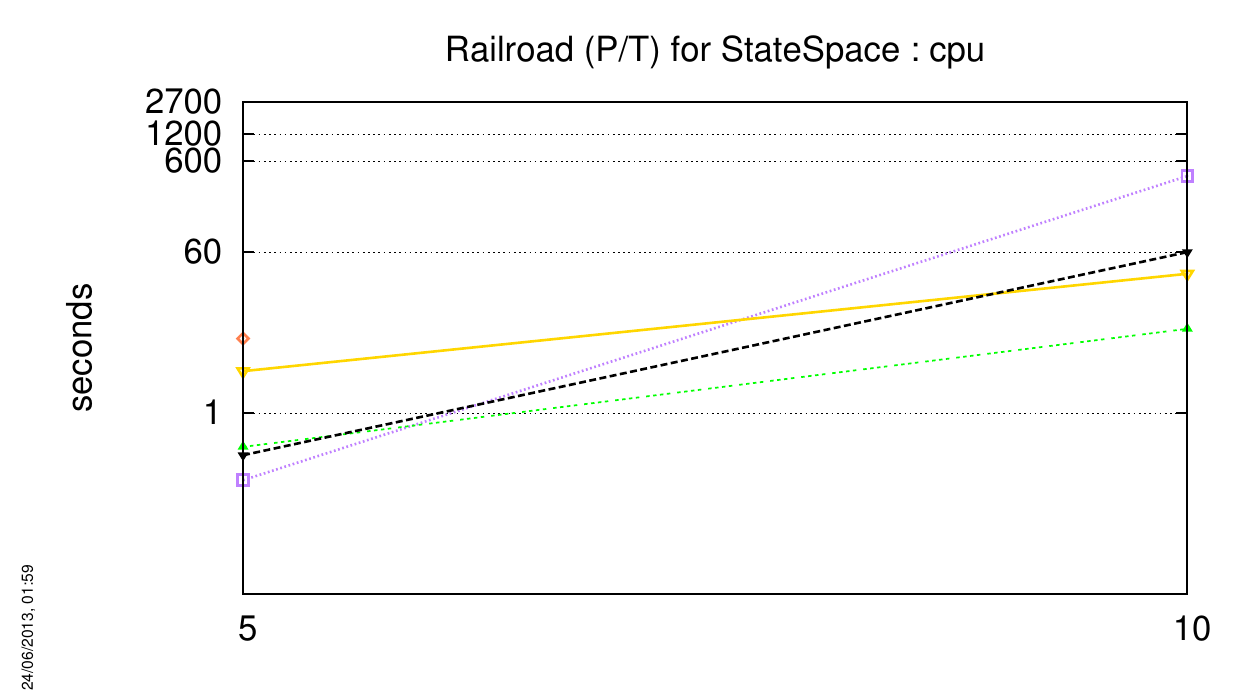}

   \includegraphics[height=1cm]{figures/tools-legend.pdf}
\end{center}

\subsubsection{\acs{RessAllocation-PT}}
The charts below respectively show how tools compete with this ``Known'' model (memory and CPU).

\index{Performances!StateSpace!RessAllocation (P/T)}
\begin{center}
   \includegraphics[width=7.2cm]{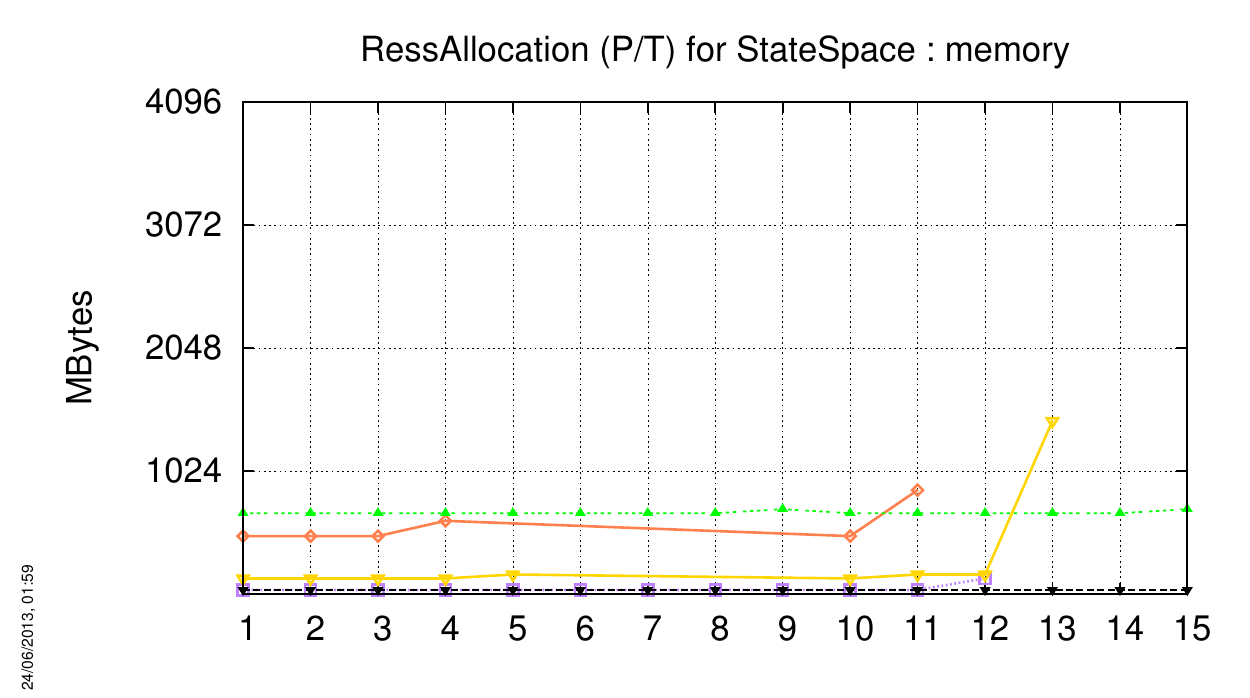}
   \includegraphics[width=7.2cm]{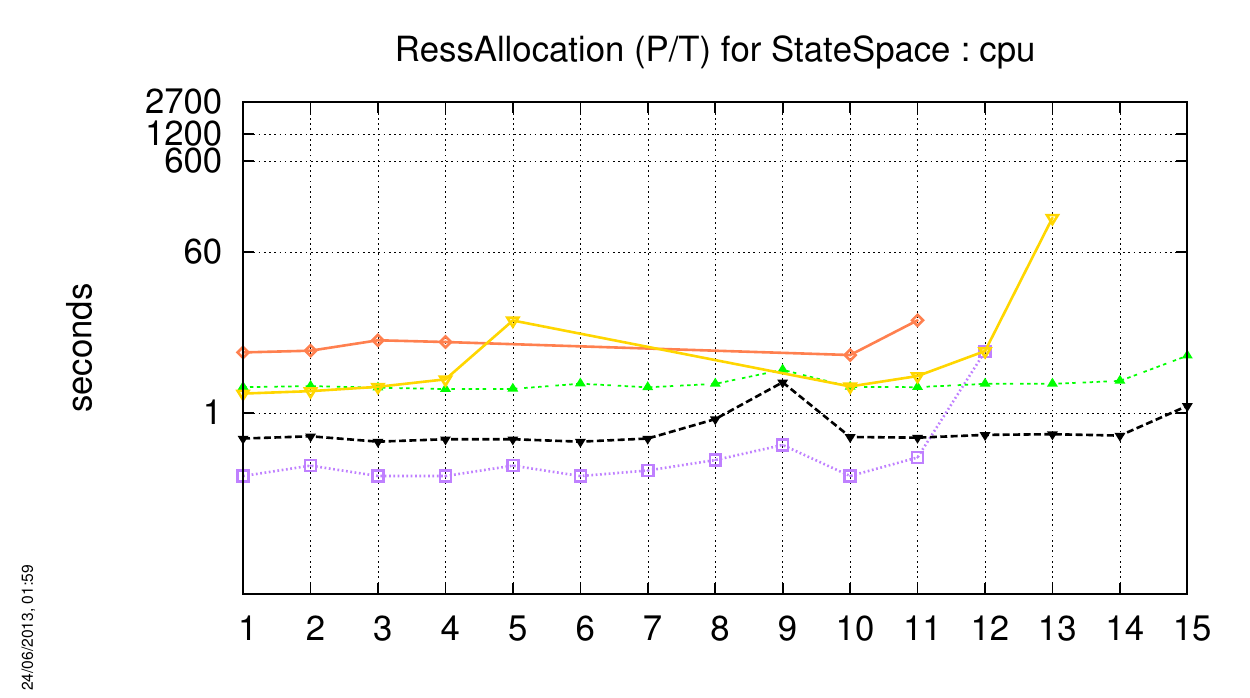}

   \includegraphics[height=1cm]{figures/tools-legend.pdf}
\end{center}

\subsubsection{\acs{Ring-PT}}
The charts below respectively show how tools compete with this ``Known'' model (memory and CPU).

\index{Performances!StateSpace!Ring (P/T)}
\begin{center}
   \includegraphics[width=7.2cm]{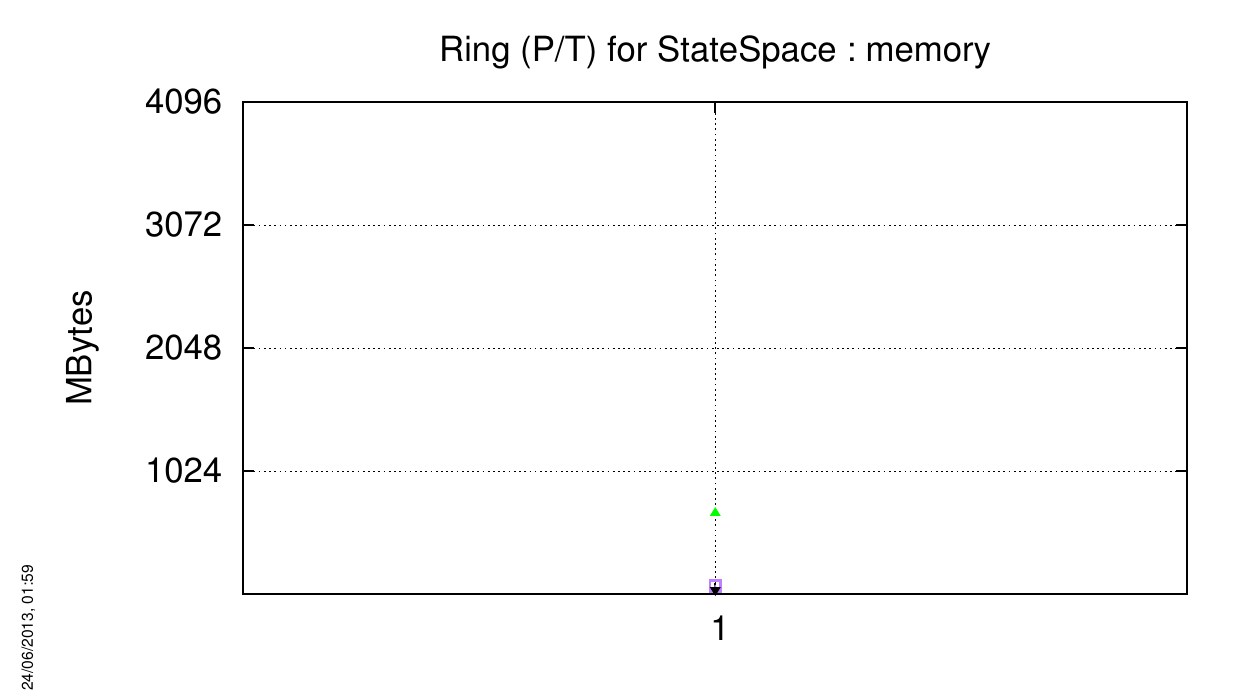}
   \includegraphics[width=7.2cm]{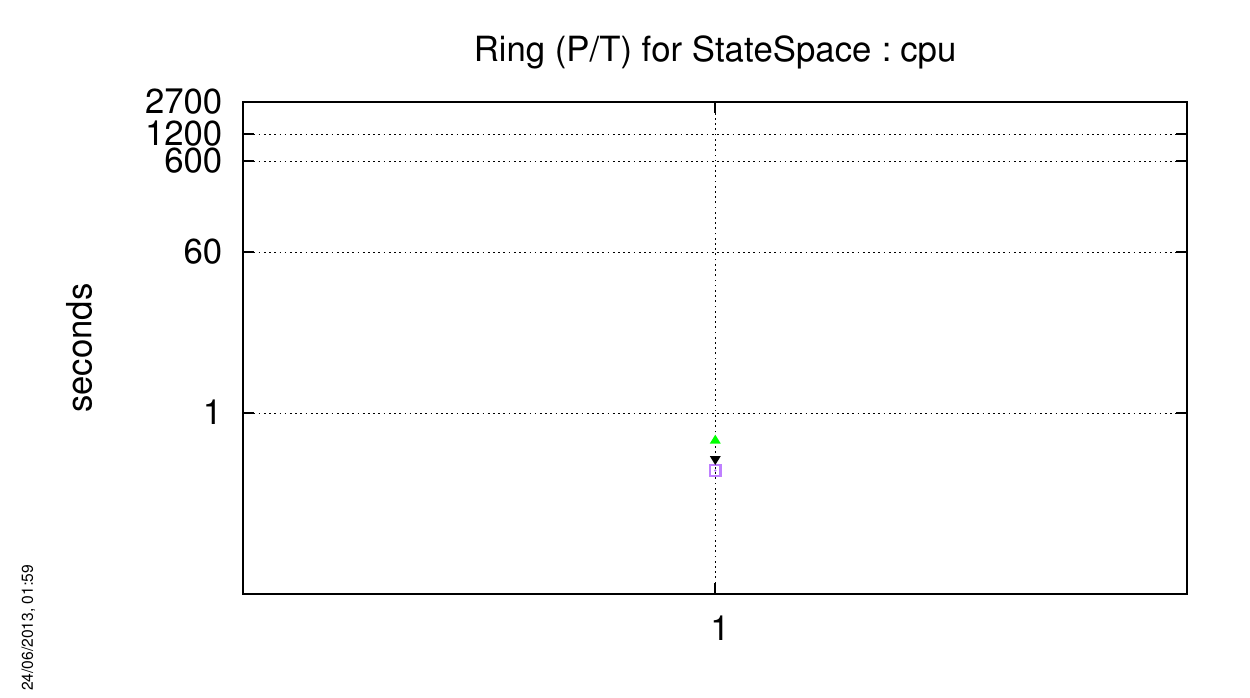}

   \includegraphics[height=1cm]{figures/tools-legend.pdf}
\end{center}

\subsubsection{\acs{RwMutex-PT}}
The charts below respectively show how tools compete with this ``Known'' model (memory and CPU).

\index{Performances!StateSpace!RwMutex (P/T)}
\begin{center}
   \includegraphics[width=7.2cm]{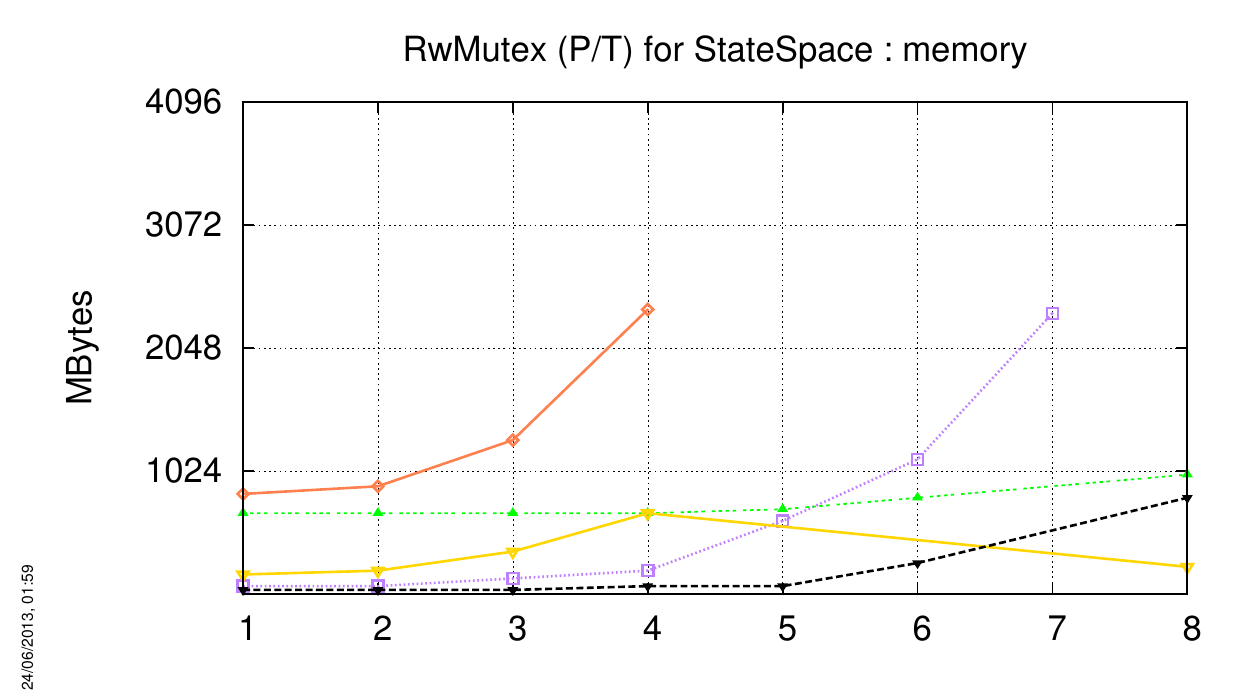}
   \includegraphics[width=7.2cm]{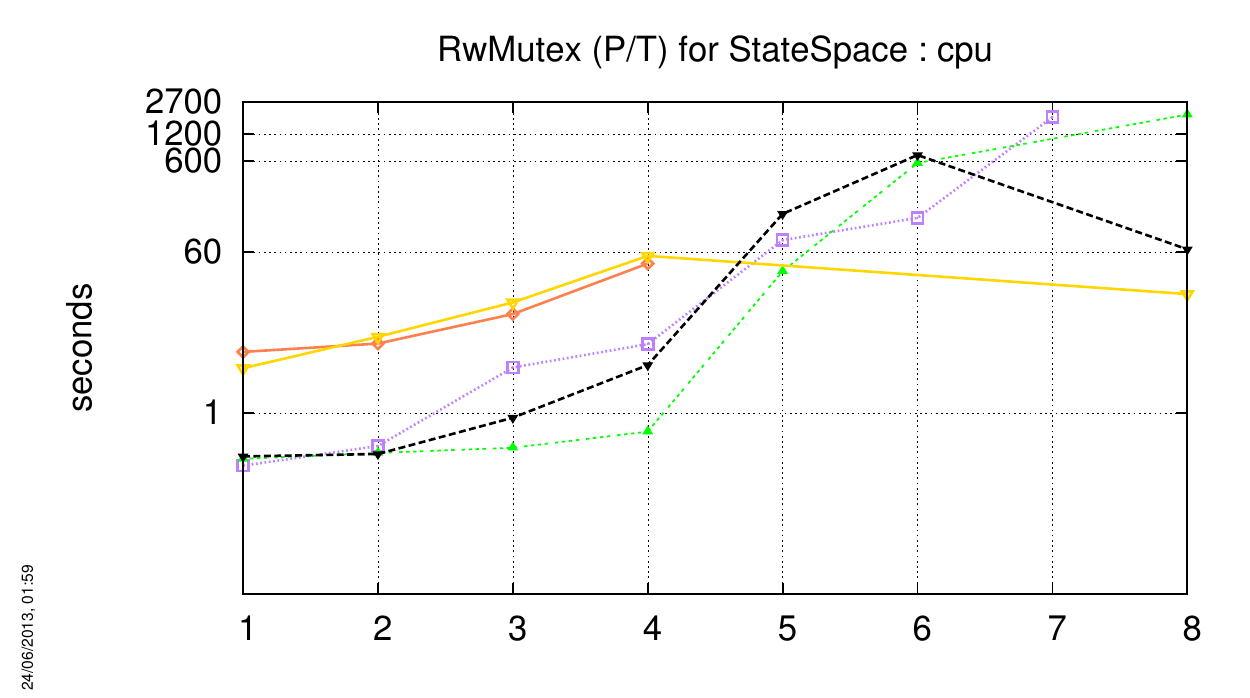}

   \includegraphics[height=1cm]{figures/tools-legend.pdf}
\end{center}

\subsubsection{\acs{SharedMemory-COL}}
The charts below respectively show how tools compete with this ``Known'' model (memory and CPU).

\index{Performances!StateSpace!SharedMemory (Colored)}
\begin{center}
   \includegraphics[width=7.2cm]{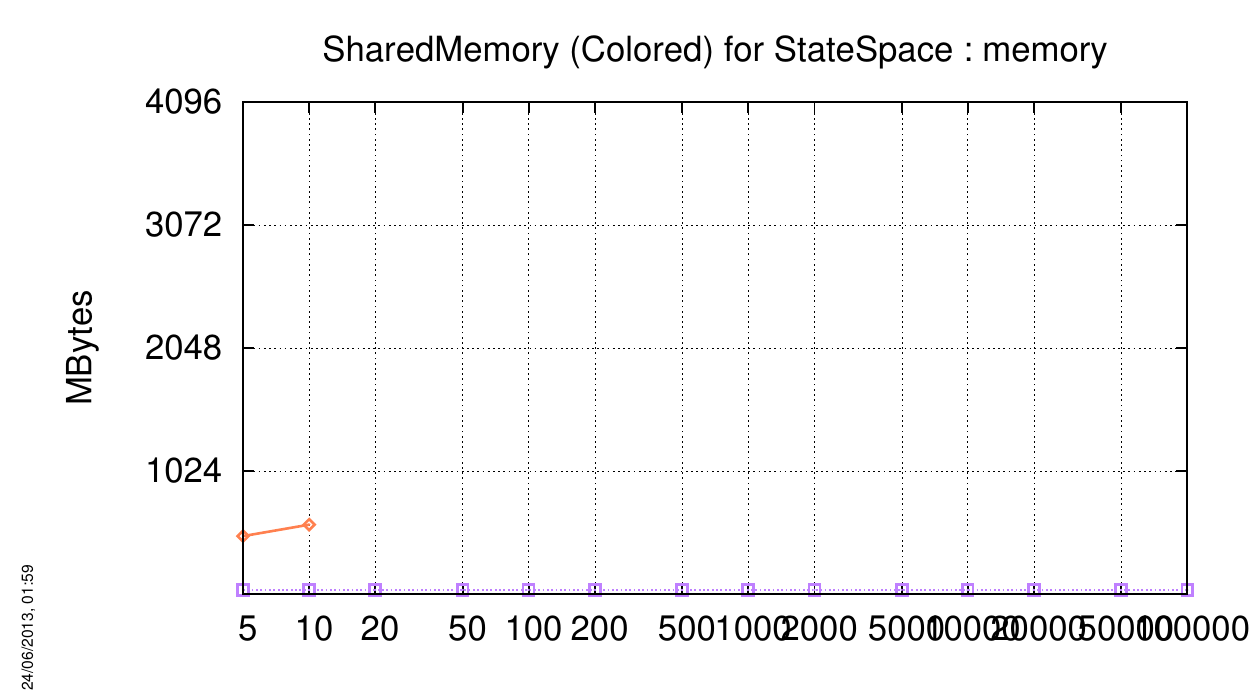}
   \includegraphics[width=7.2cm]{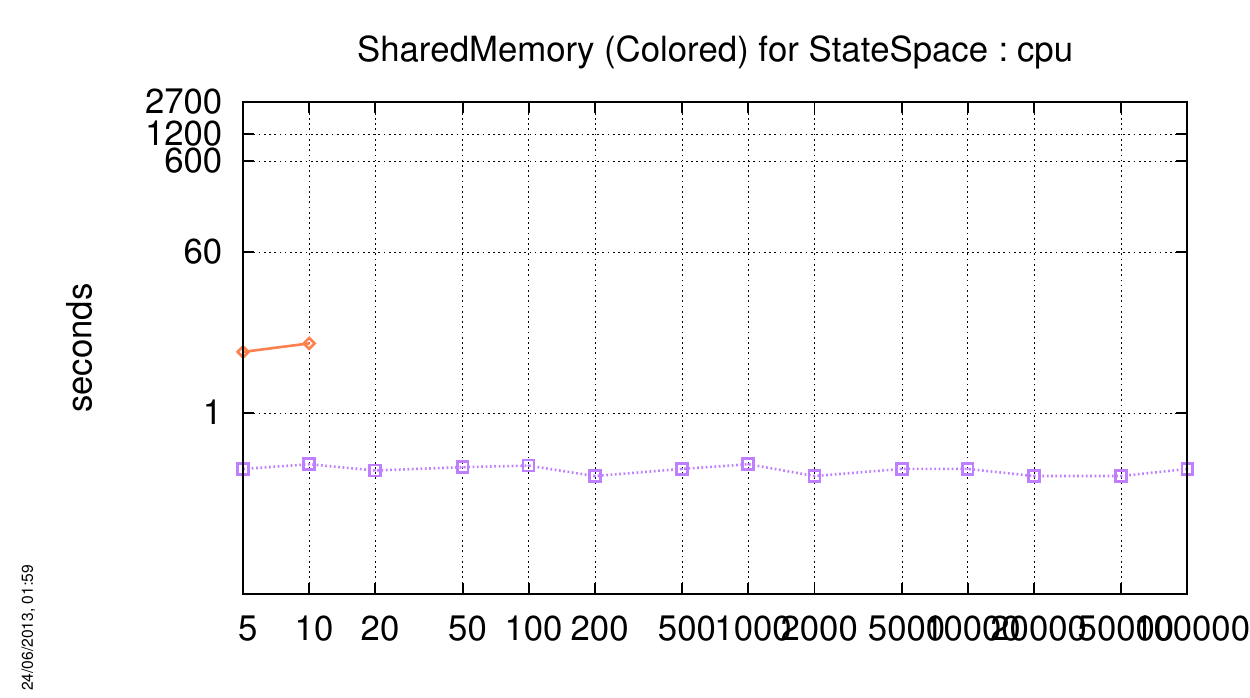}

   \includegraphics[height=1cm]{figures/tools-legend.pdf}
\end{center}

\subsubsection{\acs{SharedMemory-PT}}
The charts below respectively show how tools compete with this ``Known'' model (memory and CPU).

\index{Performances!StateSpace!SharedMemory (P/T)}
\begin{center}
   \includegraphics[width=7.2cm]{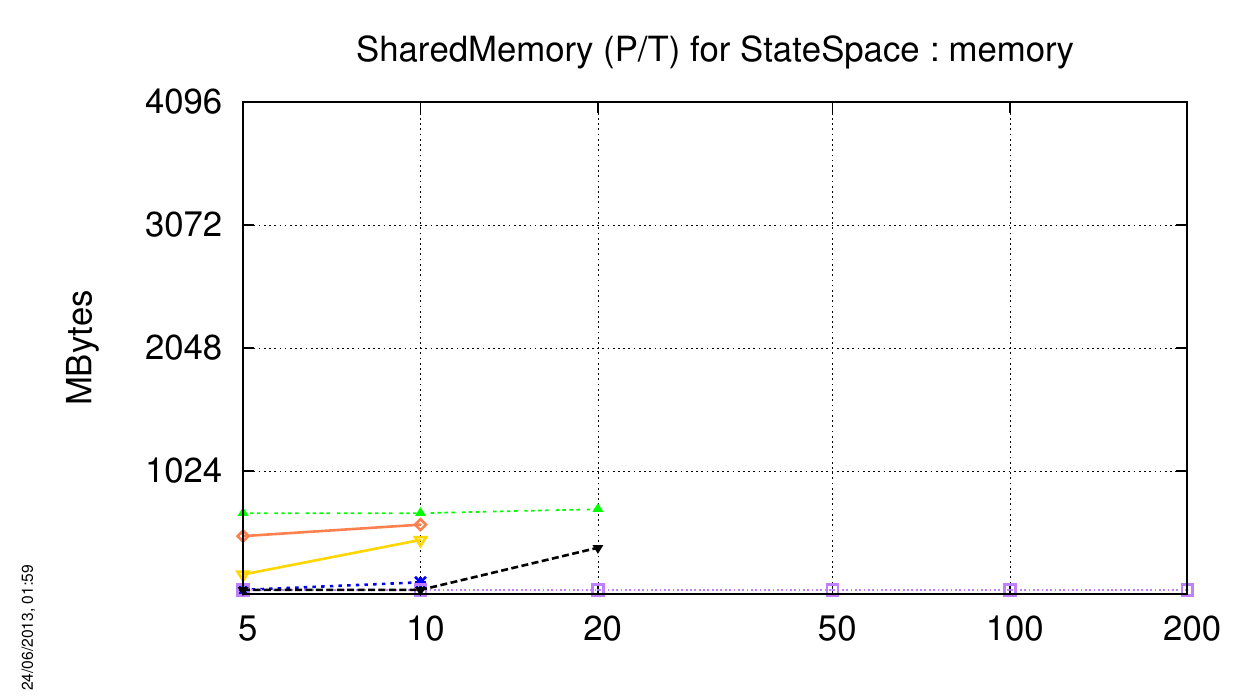}
   \includegraphics[width=7.2cm]{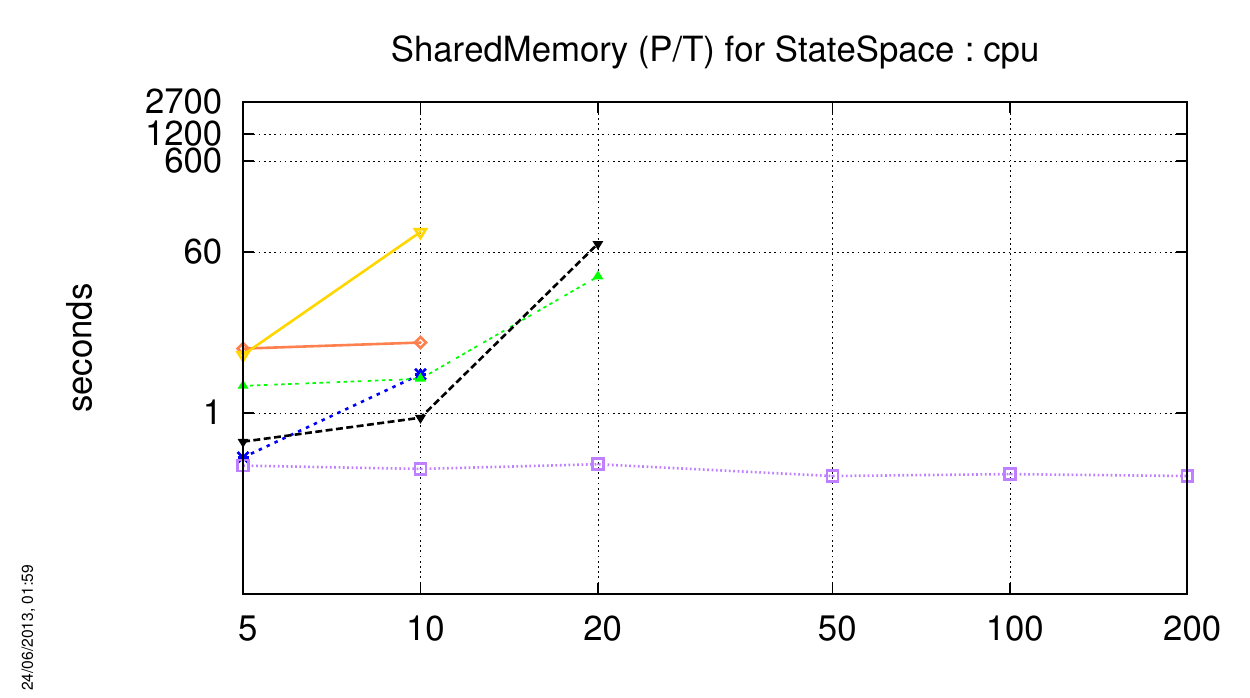}

   \includegraphics[height=1cm]{figures/tools-legend.pdf}
\end{center}

\subsubsection{\acs{SimpleLoadBal-COL}}
The charts below respectively show how tools compete with this ``Known'' model (memory and CPU).

\index{Performances!StateSpace!SimpleLoadBal (Colored)}
\begin{center}
   \includegraphics[width=7.2cm]{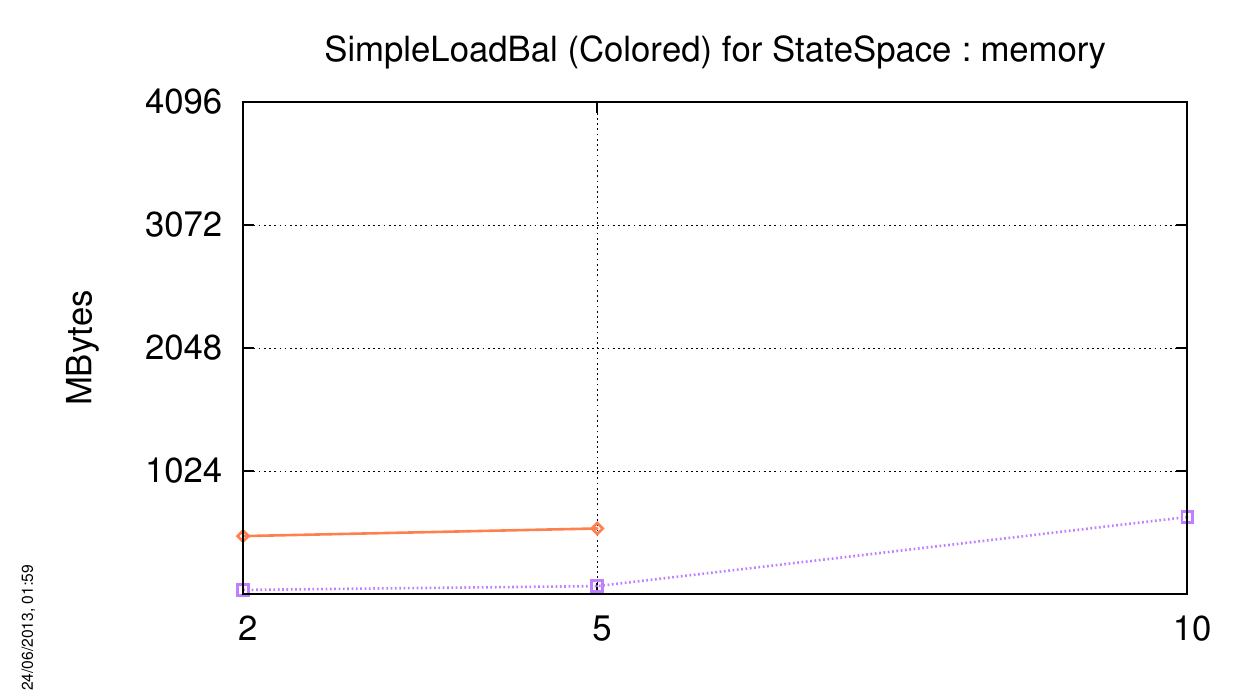}
   \includegraphics[width=7.2cm]{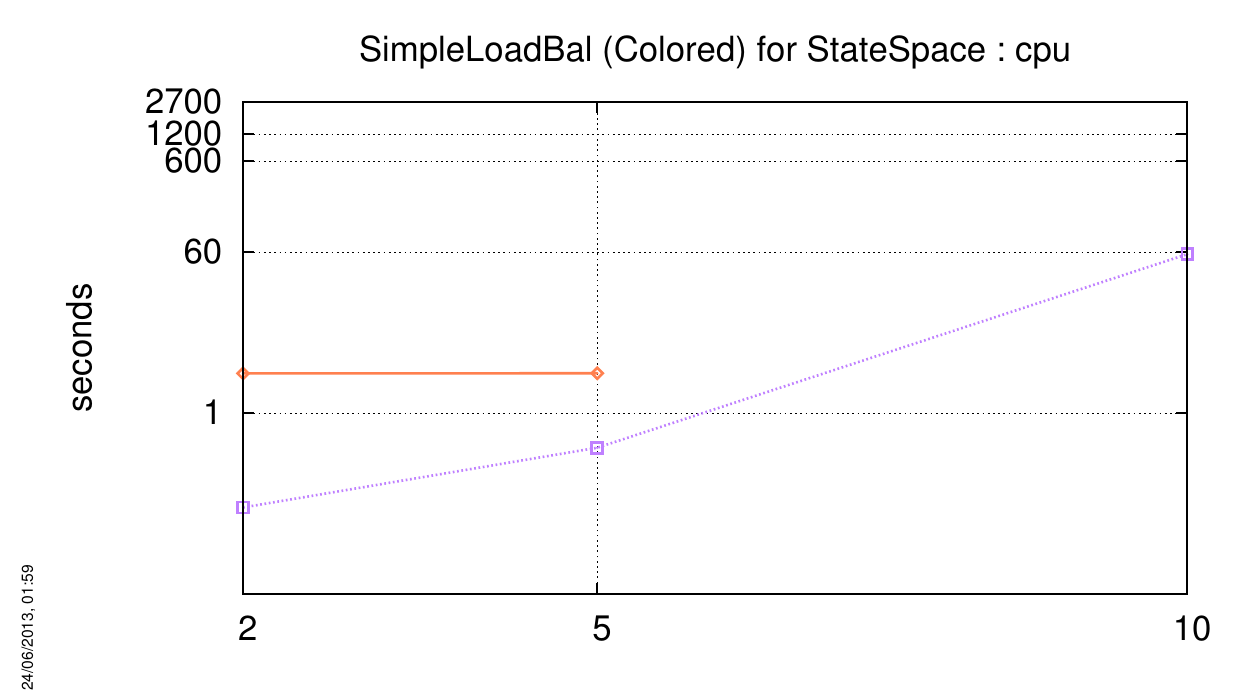}

   \includegraphics[height=1cm]{figures/tools-legend.pdf}
\end{center}

\subsubsection{\acs{SimpleLoadBal-PT}}
The charts below respectively show how tools compete with this ``Known'' model (memory and CPU).

\index{Performances!StateSpace!SimpleLoadBal (P/T)}
\begin{center}
   \includegraphics[width=7.2cm]{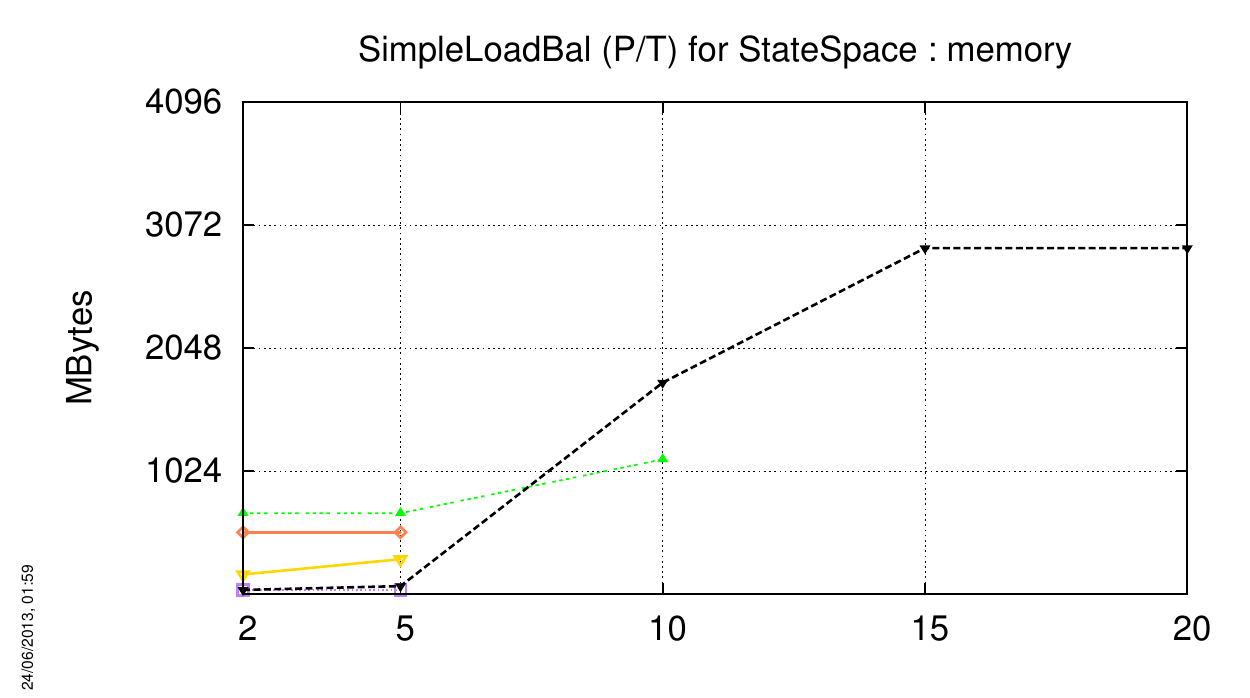}
   \includegraphics[width=7.2cm]{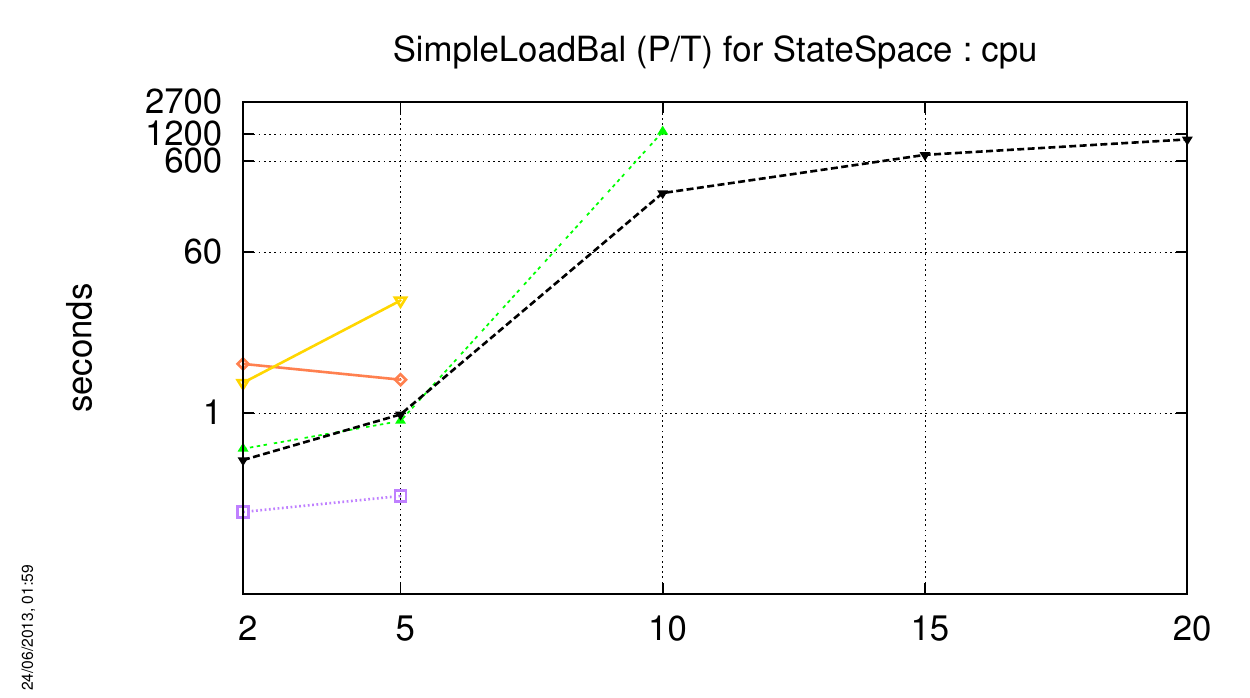}

   \includegraphics[height=1cm]{figures/tools-legend.pdf}
\end{center}

\subsubsection{\acs{TokenRing-COL}}
The charts below respectively show how tools compete with this ``Known'' model (memory and CPU).

\index{Performances!StateSpace!TokenRing (Colored)}
\begin{center}
   \includegraphics[width=7.2cm]{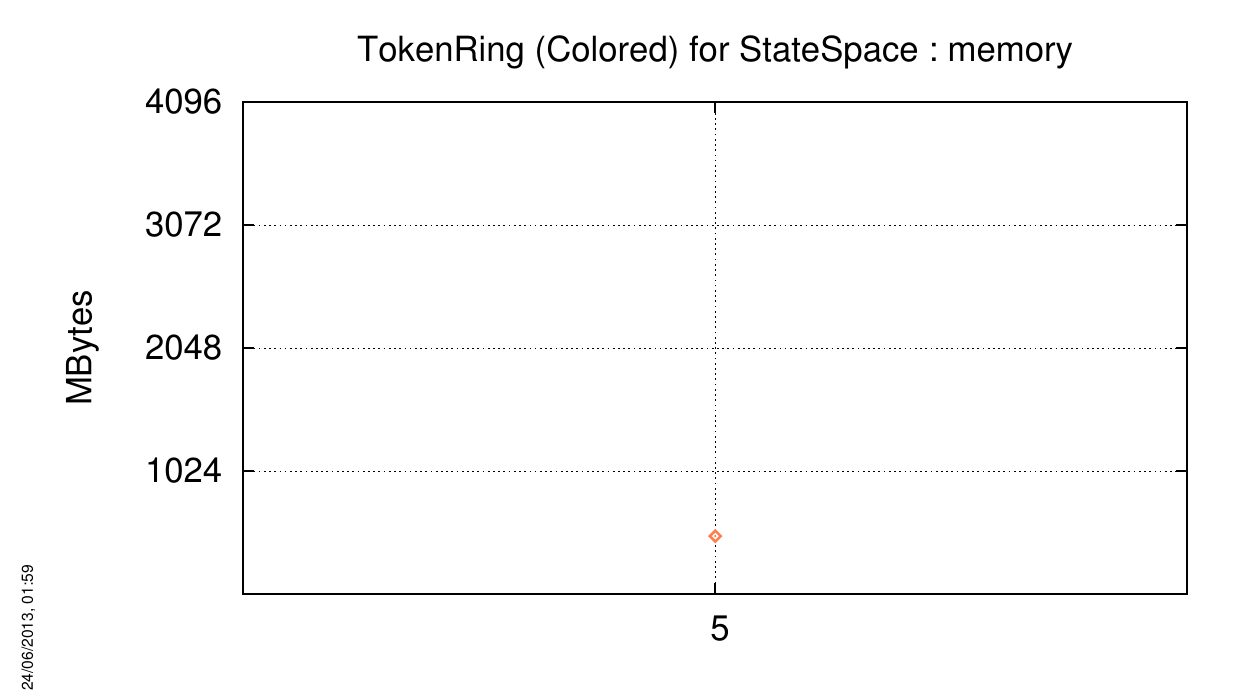}
   \includegraphics[width=7.2cm]{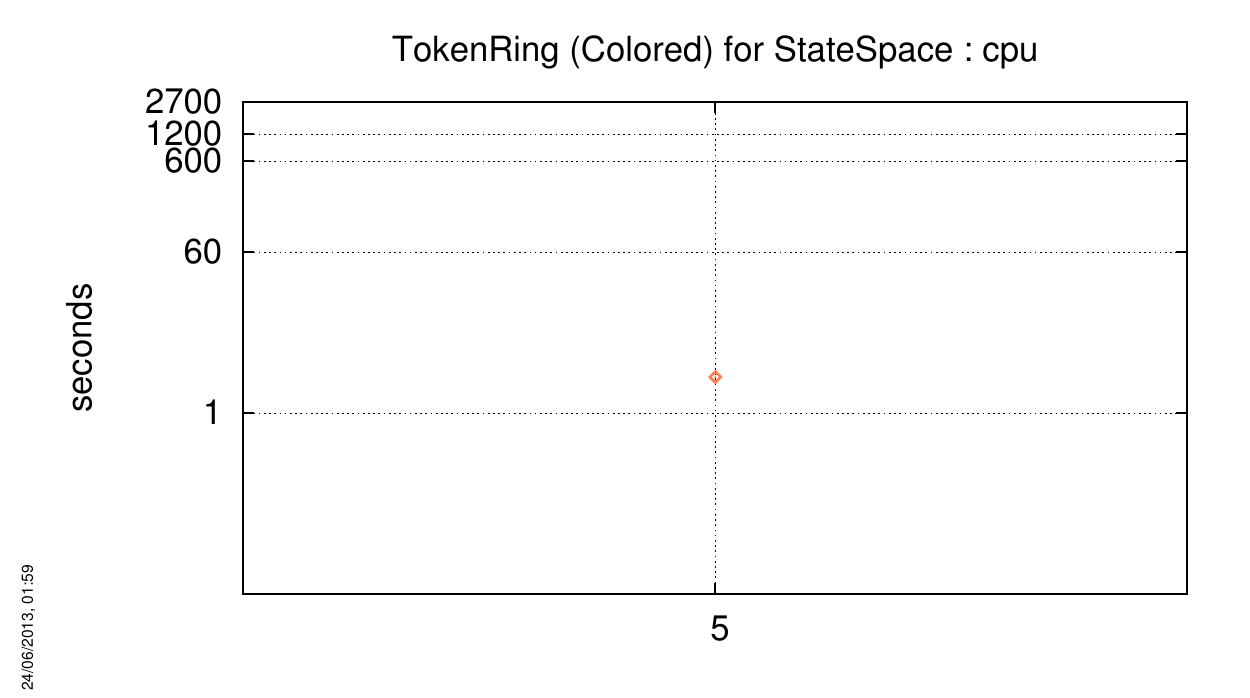}

   \includegraphics[height=1cm]{figures/tools-legend.pdf}
\end{center}

\subsubsection{\acs{TokenRing-PT}}
The charts below respectively show how tools compete with this ``Known'' model (memory and CPU).

\index{Performances!StateSpace!TokenRing (P/T)}
\begin{center}
   \includegraphics[width=7.2cm]{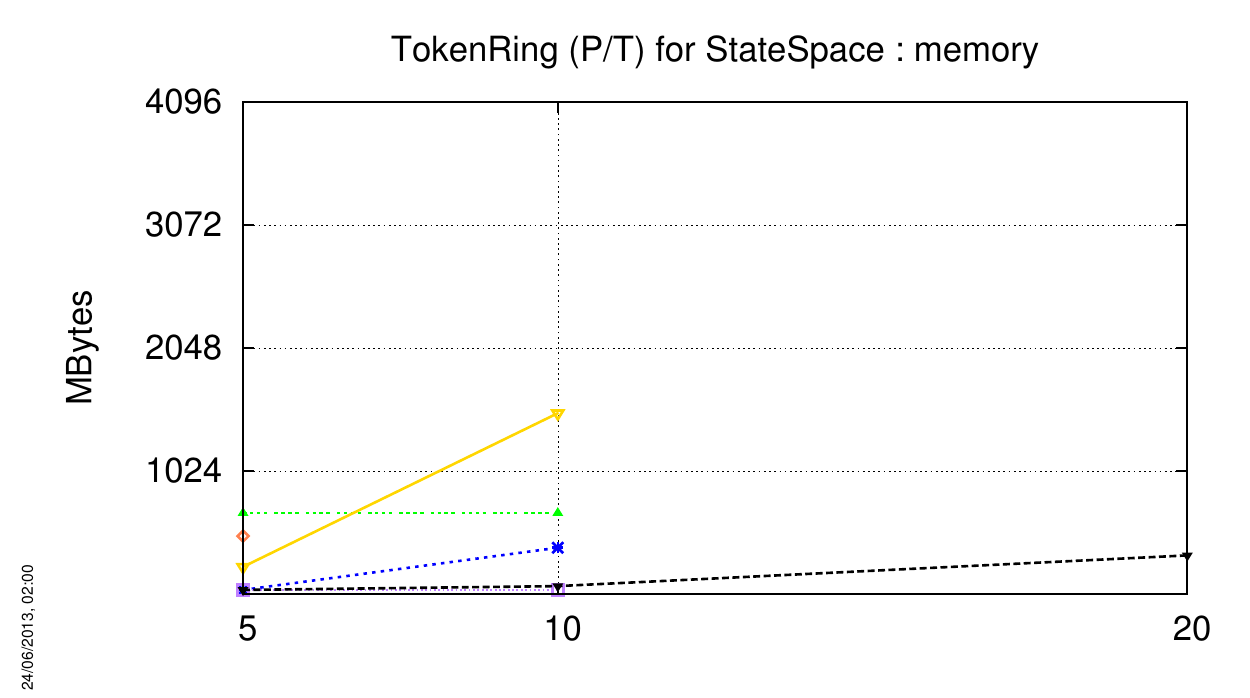}
   \includegraphics[width=7.2cm]{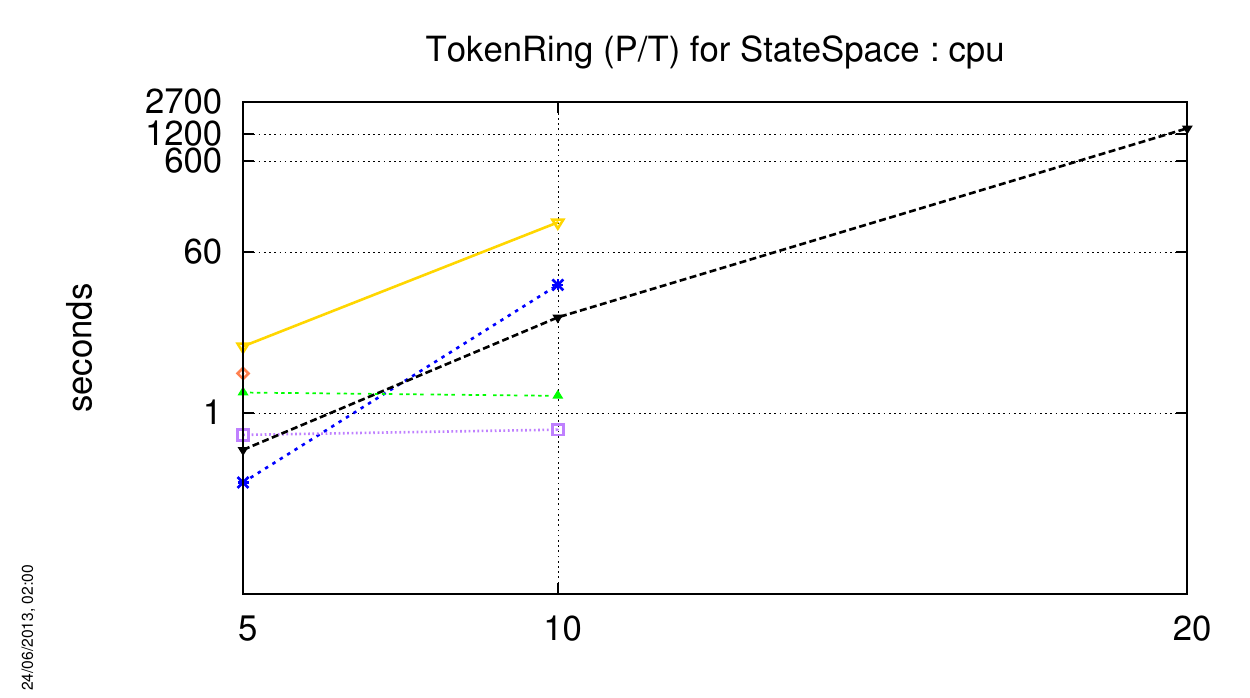}

   \includegraphics[height=1cm]{figures/tools-legend.pdf}
\end{center}

\subsubsection{\acs{HouseConstruction-PT}}
The charts below respectively show how tools compete with this ``Suprise'' model (memory and CPU).

\index{Performances!StateSpace!HouseConstruction (P/T)}
\begin{center}
   \includegraphics[width=7.2cm]{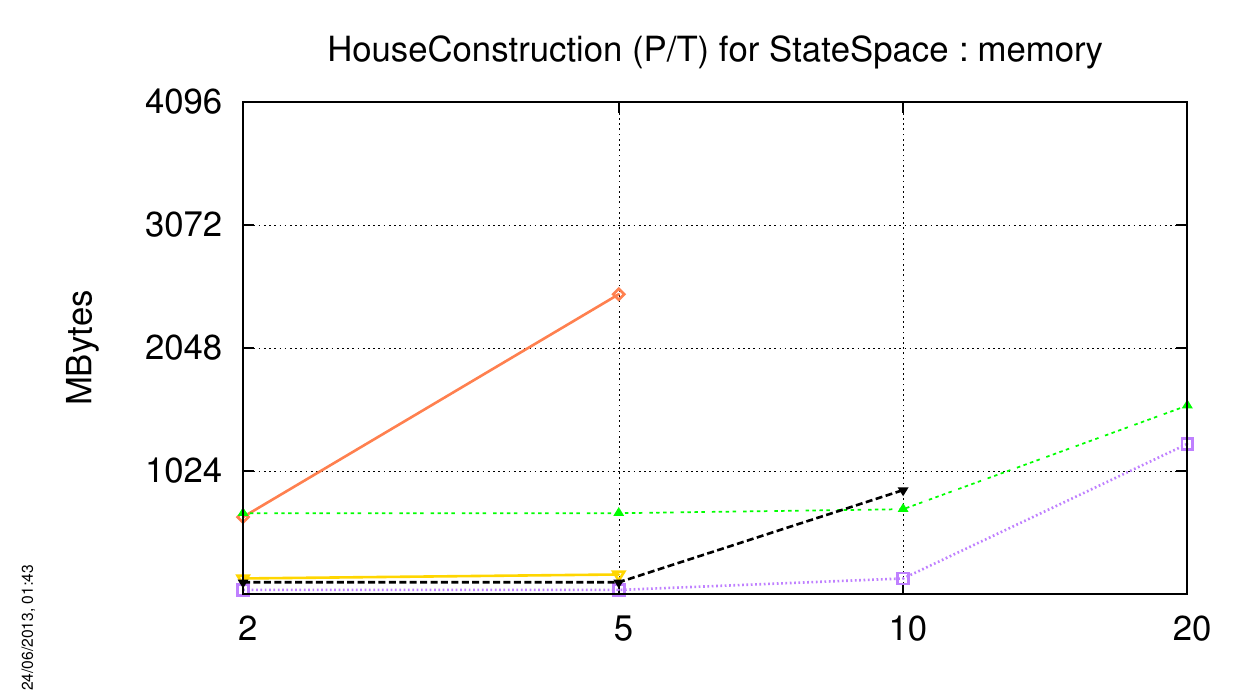}
   \includegraphics[width=7.2cm]{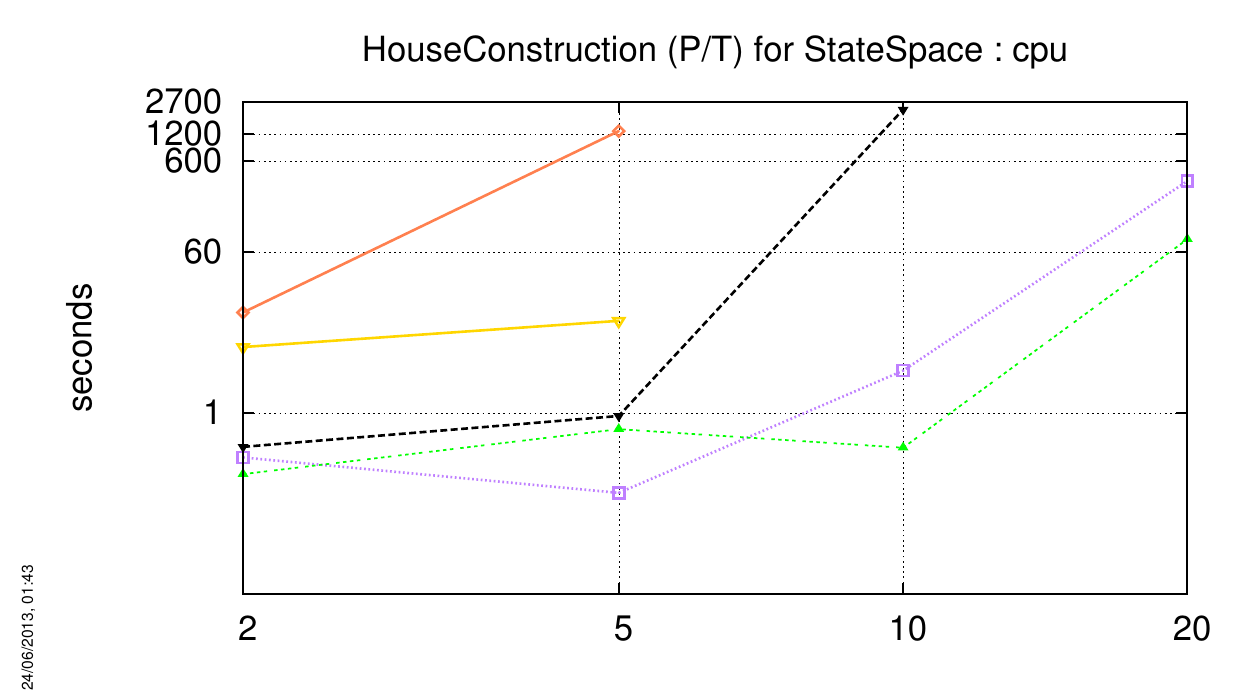}

   \includegraphics[height=1cm]{figures/tools-legend.pdf}
\end{center}

\subsubsection{\acs{IBMB2S565S3960-PT}}
The charts below respectively show how tools compete with this ``Suprise'' model (memory and CPU).

\index{Performances!StateSpace!IBMB2S565S3960 (P/T)}
\begin{center}
   \includegraphics[width=7.2cm]{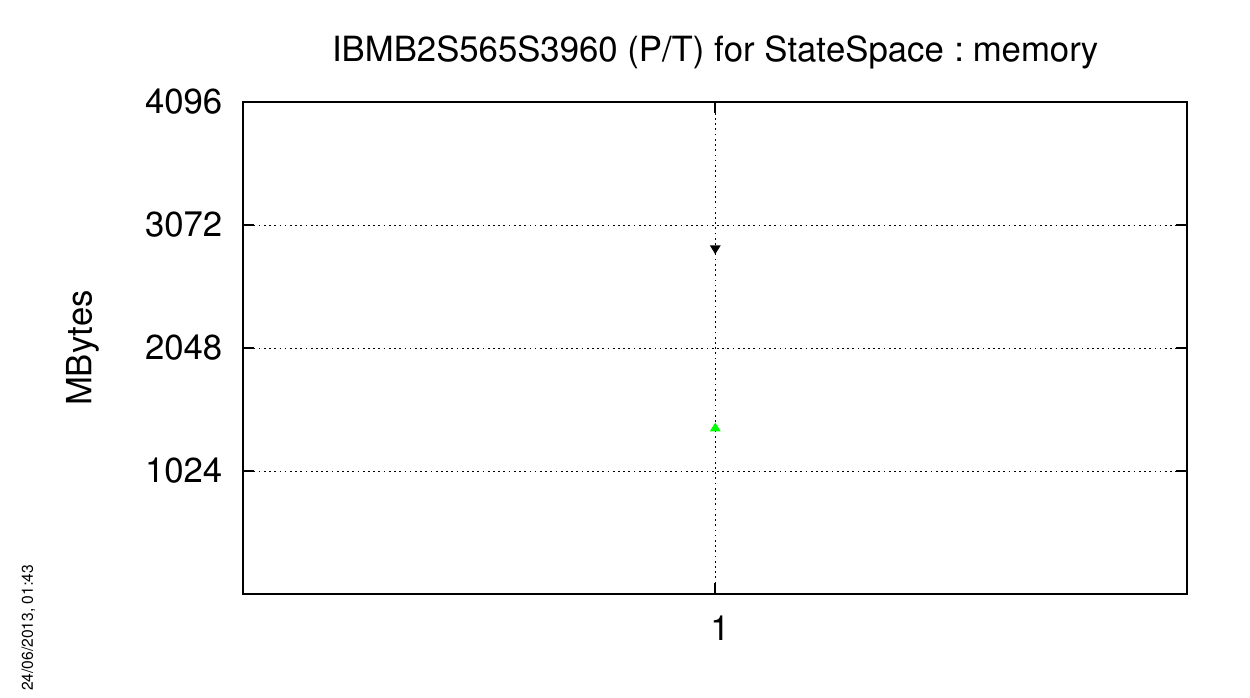}
   \includegraphics[width=7.2cm]{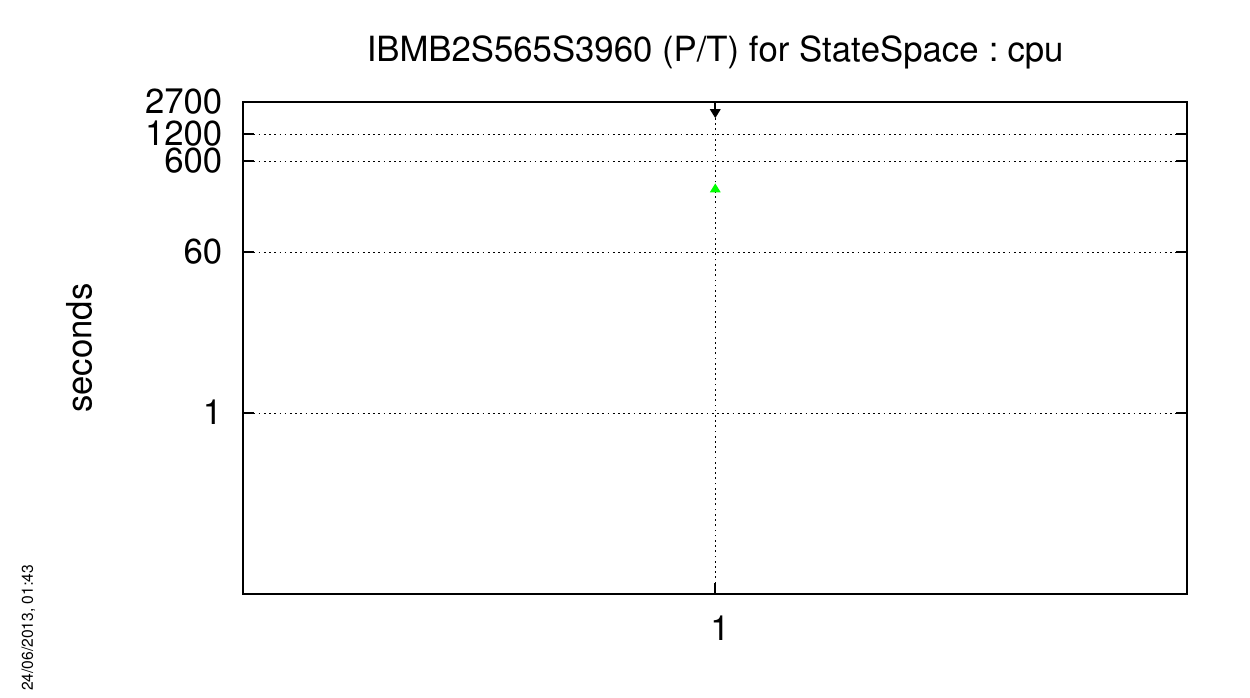}

   \includegraphics[height=1cm]{figures/tools-legend.pdf}
\end{center}

\subsubsection{\acs{QuasiCertifProtocol-COL}}
No instance of this model could be computed for the \textbf{StateSpace} examination.

\subsubsection{\acs{QuasiCertifProtocol-PT}}
The charts below respectively show how tools compete with this ``Suprise'' model (memory and CPU).

\index{Performances!StateSpace!QuasiCertifProtocol (P/T)}
\begin{center}
   \includegraphics[width=7.2cm]{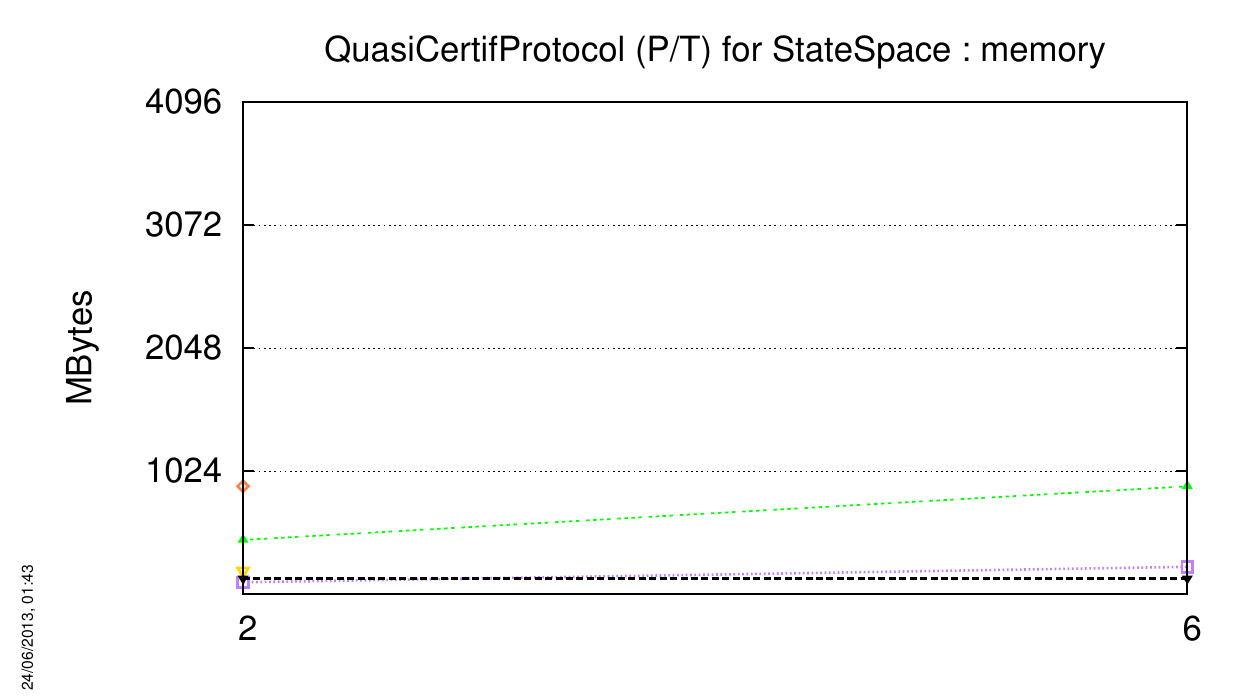}
   \includegraphics[width=7.2cm]{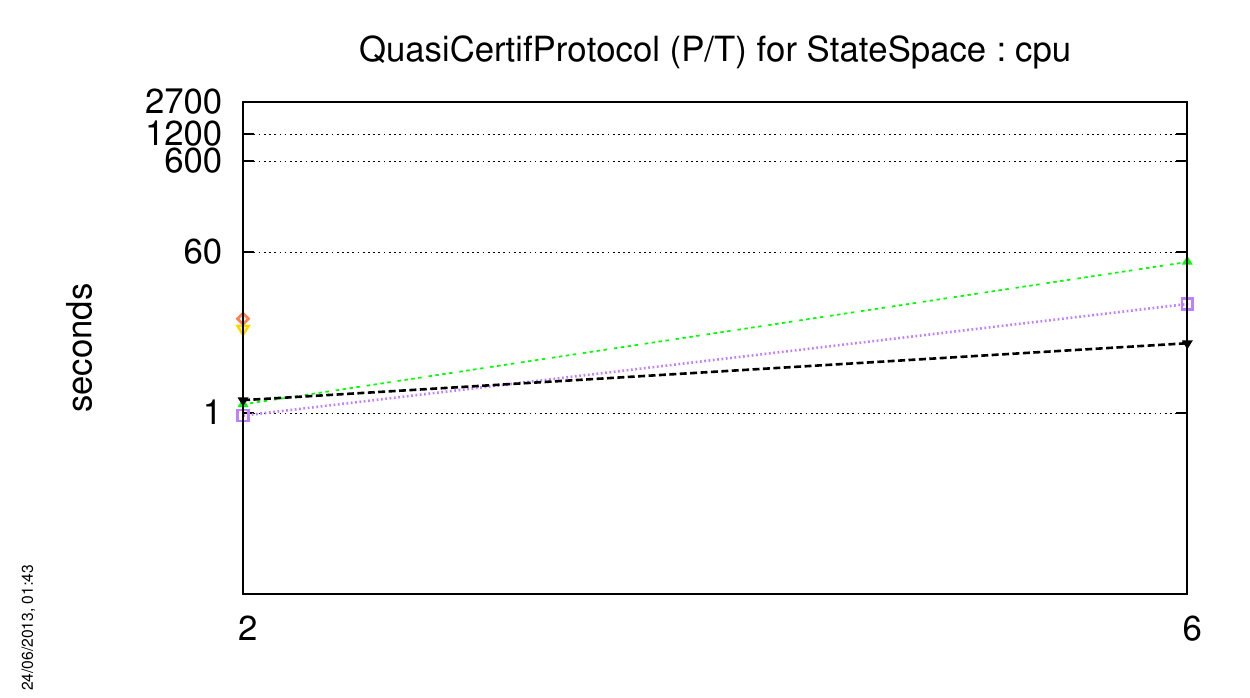}

   \includegraphics[height=1cm]{figures/tools-legend.pdf}
\end{center}

\subsubsection{\acs{Vasy2003-PT}}
The charts below respectively show how tools compete with this ``Suprise'' model (memory and CPU).

\index{Performances!StateSpace!Vasy2003 (P/T)}
\begin{center}
   \includegraphics[width=7.2cm]{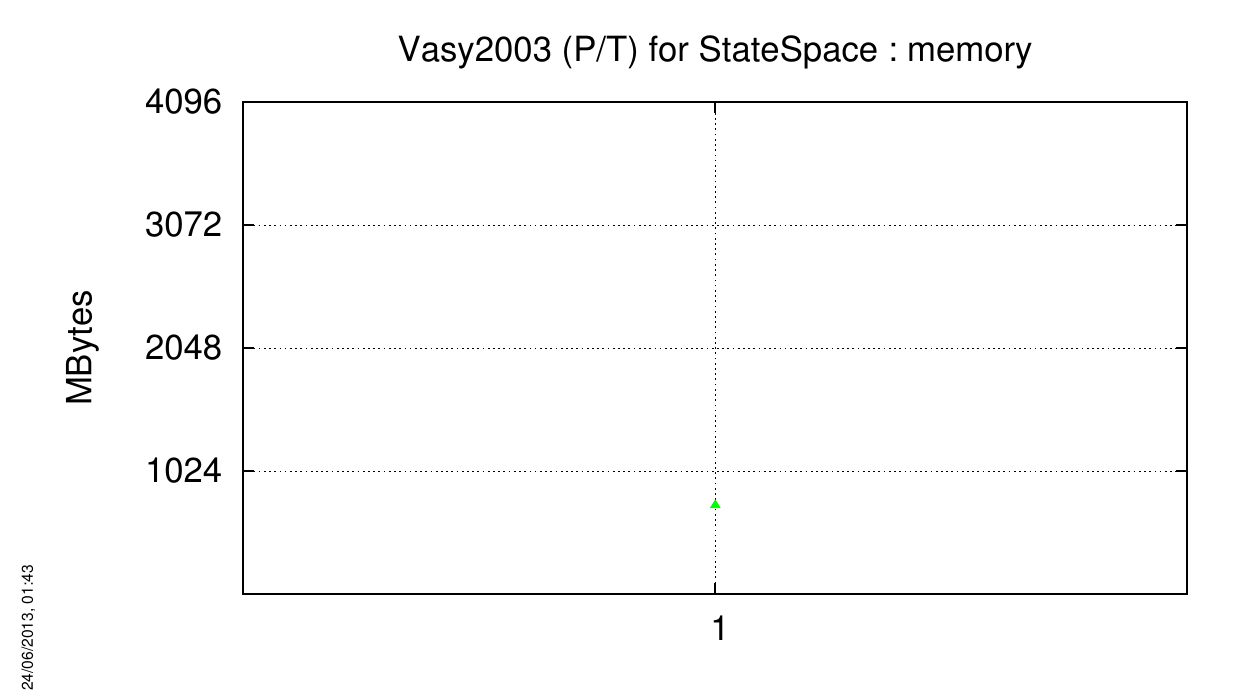}
   \includegraphics[width=7.2cm]{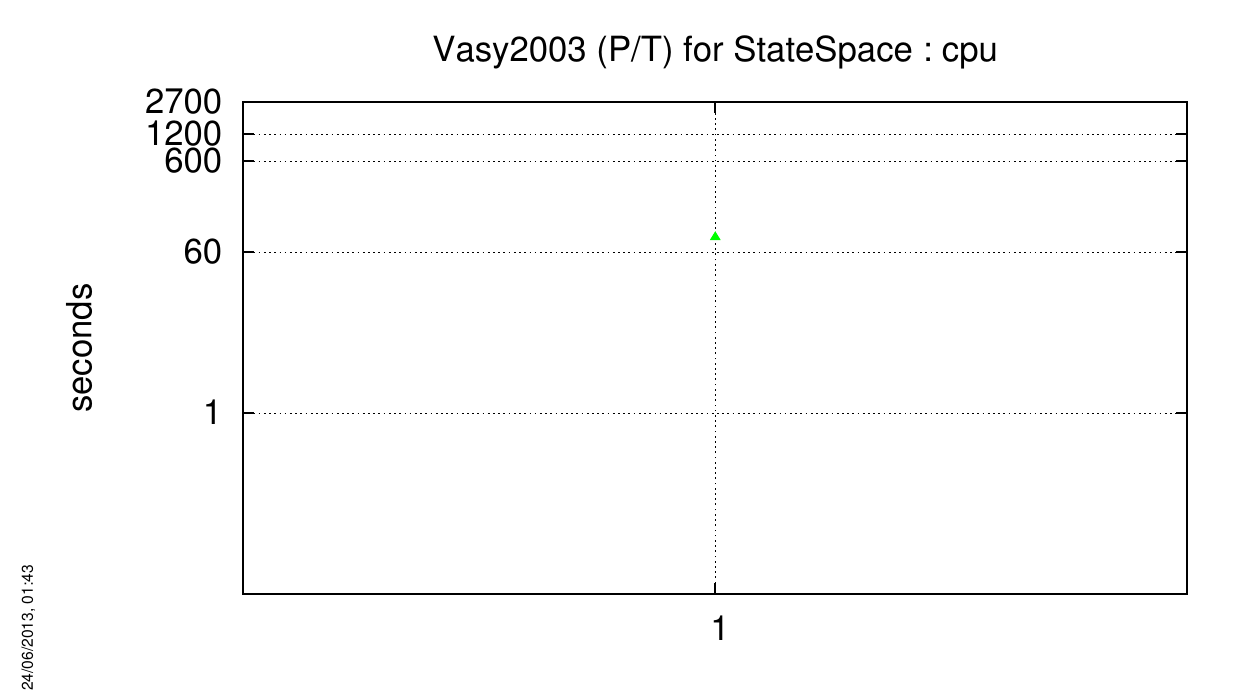}

   \includegraphics[height=1cm]{figures/tools-legend.pdf}
\end{center}

\subsection{Outputs for the StateSpace Examination}
\index{Outputs!StateSpace}

Please find enclosed the brute results for this examination (``Known'' and ``Surprise'' models).
We display only the score of tools that provide a results for at least one instance of one model.
The legend for the values is provided below:
\begin{itemize}
   \item\textbf{nc}: the tool does not compete this examination for this model/instance,
   \item\textbf{cc}: the tool cannot compute this examination for this model/instance,
   \item\textbf{to}: the tool cannot compute this examination for this model/instance within the maximum allowed time,
   \item\textbf{mp}: the tool encountered a memory problem (stack overflow or memory full),
   \item\textbf{nf}: there is no formula available for this type of examination (typically, this concerns P/T nets where
       comparing marking cardinality has no signification when there is no equivalent colored net).
\end{itemize}

Please note that, for some models/instances, we could not reformat the number of the state space (apparently over $10^{239}$ states) and then provide ``$\infty$ (ovf)'' as an answer.

\subsubsection{``Known'' Models}

Results are summarized in the table below.
{\footnotesize
\begin{longtable}{|cccccccc|}
   \hline
      \multicolumn{8}{|c|}{\bf \acs{CSRepetitions-COL}} \\
      \textbf{Instances}\index{Outputs!StateSpace!CSRepetitions (colored)} &
      \acs{alpina} &
      \acs{greatspn} &
      \acs{its-tools} &
      \acs{lola-pessimistic} &
      \acs{marcie} &
      \acs{neco} &
      \acs{pnxdd} \\
      02 &
      \textcolor{red}{\tt \scriptsize 7424} &
      nc &
      \textcolor{red}{\tt \scriptsize 7424} &
      nc &
      nc &
      nc &
      nc \\
      03 &
      \textcolor{red}{\tt \scriptsize $1.341\times 10^{08}$} &
      nc &
      \textcolor{red}{\tt \scriptsize $1.341\times 10^{08}$} &
      nc &
      nc &
      nc &
      nc \\
      04 &
      \textcolor{red}{\tt \scriptsize to} &
      nc &
      \textcolor{red}{\tt \scriptsize mp} &
      nc &
      nc &
      nc &
      nc \\
      05 &
      \textcolor{red}{\tt \scriptsize to} &
      nc &
      \textcolor{red}{\tt \scriptsize mp} &
      nc &
      nc &
      nc &
      nc \\
      07 &
      \textcolor{red}{\tt \scriptsize cc} &
      nc &
      \textcolor{red}{\tt \scriptsize mp} &
      nc &
      nc &
      nc &
      nc \\
      10 &
      \textcolor{red}{\tt \scriptsize cc} &
      nc &
      \textcolor{red}{\tt \scriptsize mp} &
      nc &
      nc &
      nc &
      nc \\
\hline
\end{longtable}
}

\subsubsection{``Surprise'' Models}

Results are summarized in the table below.
{\footnotesize
\begin{longtable}{|cccccc|}
   \hline
      \multicolumn{6}{|c|}{\bf \acs{HouseConstruction-PT}} \\
      instances\index{Outputs!StateSpace!HouseConstruction (P/T)} &
      \acs{alpina} &
      \acs{its-tools} &
      \acs{marcie} &
      \acs{neco} &
      \acs{pnxdd} \\
      002 &
      \textcolor{red}{\tt \scriptsize 1501} &
      \textcolor{red}{\tt \scriptsize 1501} &
      \textcolor{red}{\tt \scriptsize 1501} &
      \textcolor{red}{\tt \scriptsize 1501} &
      \textcolor{red}{\tt \scriptsize 1501} \\
      005 &
      \textcolor{red}{\tt \scriptsize 1187984} &
      \textcolor{red}{\tt \scriptsize 1187984} &
      \textcolor{red}{\tt \scriptsize 1187984} &
      \textcolor{red}{\tt \scriptsize 1187984} &
      \textcolor{red}{\tt \scriptsize 1187980} \\
      010 &
      \textcolor{red}{\tt \scriptsize to} &
      \textcolor{red}{\tt \scriptsize $1.664\times 10^{09}$} &
      \textcolor{red}{\tt \scriptsize $1.664\times 10^{09}$} &
      \textcolor{red}{\tt \scriptsize cc} &
      \textcolor{red}{\tt \scriptsize $1.664\times 10^{09}$} \\
      020 &
      \textcolor{red}{\tt \scriptsize to} &
      \textcolor{red}{\tt \scriptsize $1.367\times 10^{13}$} &
      \textcolor{red}{\tt \scriptsize $1.367\times 10^{13}$} &
      \textcolor{red}{\tt \scriptsize cc} &
      \textcolor{red}{\tt \scriptsize to} \\
      050 &
      \textcolor{red}{\tt \scriptsize to} &
      \textcolor{red}{\tt \scriptsize mp} &
      \textcolor{red}{\tt \scriptsize cc} &
      \textcolor{red}{\tt \scriptsize cc} &
      \textcolor{red}{\tt \scriptsize to} \\
      100 &
      \textcolor{red}{\tt \scriptsize to} &
      \textcolor{red}{\tt \scriptsize mp} &
      \textcolor{red}{\tt \scriptsize cc} &
      \textcolor{red}{\tt \scriptsize cc} &
      \textcolor{red}{\tt \scriptsize to} \\
\hline
\end{longtable}
}

\subsection{Score for the StateSpace Examination}
\index{Scores!StateSpace}

Please find enclosed the scores for this examination (``Known'' and ``Surprise'' models).
We display only the score of tools that provide a results for at least one instance of one model.
The total is first listed in the table below followed by a detail, for each proposed model.
Meaning of the line labels are:
\begin{itemize}
\item\textbf{1st instance}: the tool gets a bonus for having processed the first instance of this model (+1 point),
\item\textbf{instances}: the tool gets 1 point per instances treated 
(for that, we assume that at least one formula has been successfully computed),
\item\textbf{max reached}: the tool could process all the instances for the model (+2 points),
\item\textbf{best}: the tool is among the ones that processed a maximum of instances within the time and memory confinement (+2 points).
\end{itemize}

\subsubsection{``Known'' Models}

Scores are summarized in the table below.
For each model in this category, a tool may collect up to 411 points.

{\footnotesize
\begin{longtable}{|cccccccc|}
   \hline
      \multicolumn{8}{|c|}{\bf Total score of the tools for this examination} \\
        &
      \acs{alpina} &
      \acs{greatspn} &
      \acs{its-tools} &
      \acs{lola-pessimistic} &
      \acs{marcie} &
      \acs{neco} &
      \acs{pnxdd} \\
      Total Score &
      {\bf 103} &
      {\bf 47} &
      {\bf 234} &
      {\bf 0} &
      {\bf 129} &
      {\bf 64} &
      {\bf 139} \\
   \hline
      \multicolumn{8}{|c|}{\bf CSRepetitions (Colored)} \\
      \ \index{Scores!StateSpace!CSRepetitions (colored)} &
      \acs{alpina} &
      \acs{greatspn} &
      \acs{its-tools} &
      \acs{lola-pessimistic} &
      \acs{marcie} &
      \acs{neco} &
      \acs{pnxdd} \\
      1st instance &
      \textcolor{red}{1} &
      \textcolor{red}{0} &
      \textcolor{red}{1} &
      \textcolor{red}{0} &
      \textcolor{red}{0} &
      \textcolor{red}{0} &
      \textcolor{red}{0} \\
      instances &
      \textcolor{red}{2} &
      \textcolor{red}{0} &
      \textcolor{red}{2} &
      \textcolor{red}{0} &
      \textcolor{red}{0} &
      \textcolor{red}{0} &
      \textcolor{red}{0} \\
      max reached &
      \textcolor{red}{0} &
      \textcolor{red}{0} &
      \textcolor{red}{0} &
      \textcolor{red}{0} &
      \textcolor{red}{0} &
      \textcolor{red}{0} &
      \textcolor{red}{0} \\
\hline
\end{longtable}
}

\subsubsection{``Surprise'' Models}

Scores are summarized in the table below.
For each model in this category, a tool may collect up to 49 points.

{\footnotesize
\begin{longtable}{|cccccc|}
   \hline
      \multicolumn{6}{|c|}{\bf Total score of the tools for this examination} \\
        &
      \acs{alpina} &
      \acs{its-tools} &
      \acs{marcie} &
      \acs{neco} &
      \acs{pnxdd} \\
      Total Score &
      {\bf 5} &
      {\bf 12} &
      {\bf 24} &
      {\bf 5} &
      {\bf 15} \\
   \hline
      \multicolumn{6}{|c|}{\bf HouseConstruction (P/T)} \\
      \ \index{Scores!StateSpace!HouseConstruction (P/T)} &
      \acs{alpina} &
      \acs{its-tools} &
      \acs{marcie} &
      \acs{neco} &
      \acs{pnxdd} \\
      1st instance &
      \textcolor{red}{1} &
      \textcolor{red}{1} &
      \textcolor{red}{1} &
      \textcolor{red}{1} &
      \textcolor{red}{1} \\
      instances &
      \textcolor{red}{2} &
      \textcolor{red}{4} &
      \textcolor{red}{4} &
      \textcolor{red}{2} &
      \textcolor{red}{3} \\
      max reached &
      \textcolor{red}{0} &
      \textcolor{red}{0} &
      \textcolor{red}{0} &
      \textcolor{red}{0} &
      \textcolor{red}{0} \\
\hline
\end{longtable}
}

\subsection{Trophies for this Examination}
\index{Trophies!StateSpace}

Trophies are divided in three categories: ``Known'' models,
``Surprise'' models, and the global trophies (formula is then
$score_{global} = score_{known} + 2 \times score_{surprise}$).

\subsubsection{For ``Known'' Models} \ \\

\begin{tabular}{c|c|c}
      1 & 2 & 3 \\
   \includegraphics[width=2cm]{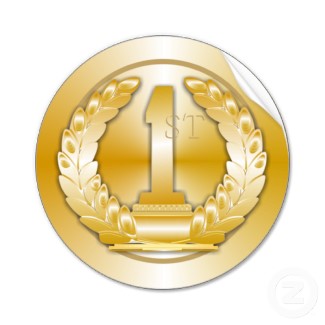} &
   \includegraphics[width=2cm]{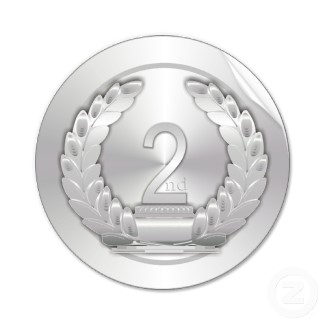} &
   \includegraphics[width=2cm]{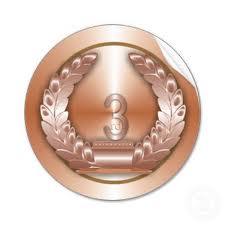} \\
   \acs{its-tools} &
   \acs{pnxdd} &
   \acs{marcie} \\
   234 points &
   139 points &
   129 points \\
\end{tabular}

\subsubsection{For ``Surprise'' Models}\  \\

\begin{tabular}{c|c|c}
      1 & 2 & 3 \\
   \includegraphics[width=2cm]{figures/gold.jpg} &
   \includegraphics[width=2cm]{figures/silver.jpg} &
   \includegraphics[width=2cm]{figures/bronse.jpg} \\
   \acs{marcie} &
   \acs{pnxdd} &
   \acs{its-tools} \\
   24 points &
   15 points &
   12 points \\
\end{tabular}

\subsubsection{Global} \ \\

\begin{tabular}{c|c|c}
      1 & 2 & 3 \\
   \includegraphics[width=2cm]{figures/gold.jpg} &
   \includegraphics[width=2cm]{figures/silver.jpg} &
   \includegraphics[width=2cm]{figures/bronse.jpg} \\
   \acs{its-tools} &
   \acs{marcie} &
   \acs{pnxdd} \\
   258 points &
   177 points &
   169 points \\
\end{tabular}

\part{Reachability Analysis}
\label{part:three}
\newpage

\section{The ReachabilityCardinalityComparison Examination}
\label{sec:exam:ReachabilityCardinalityComparison}
\index{Results!ReachabilityCardinalityComparison}

This examination deals with reachability properties dealing with checking cardinality of marking only.
We first show a summary on the handling of models by the participating tools.
Then, we present the computed outputs and the associated scores for this
examination prior to a summary of relevant executions.

\subsection{Handling of Models by Tools}
\index{Performances!ReachabilityCardinalityComparison}

\subsubsection{\acs{CSRepetitions-COL}}
The charts below respectively show how tools compete with this ``Known'' model (memory and CPU).

\index{Performances!ReachabilityCardinalityComparison!CSRepetitions (Colored)}
\begin{center}
   \includegraphics[width=7.2cm]{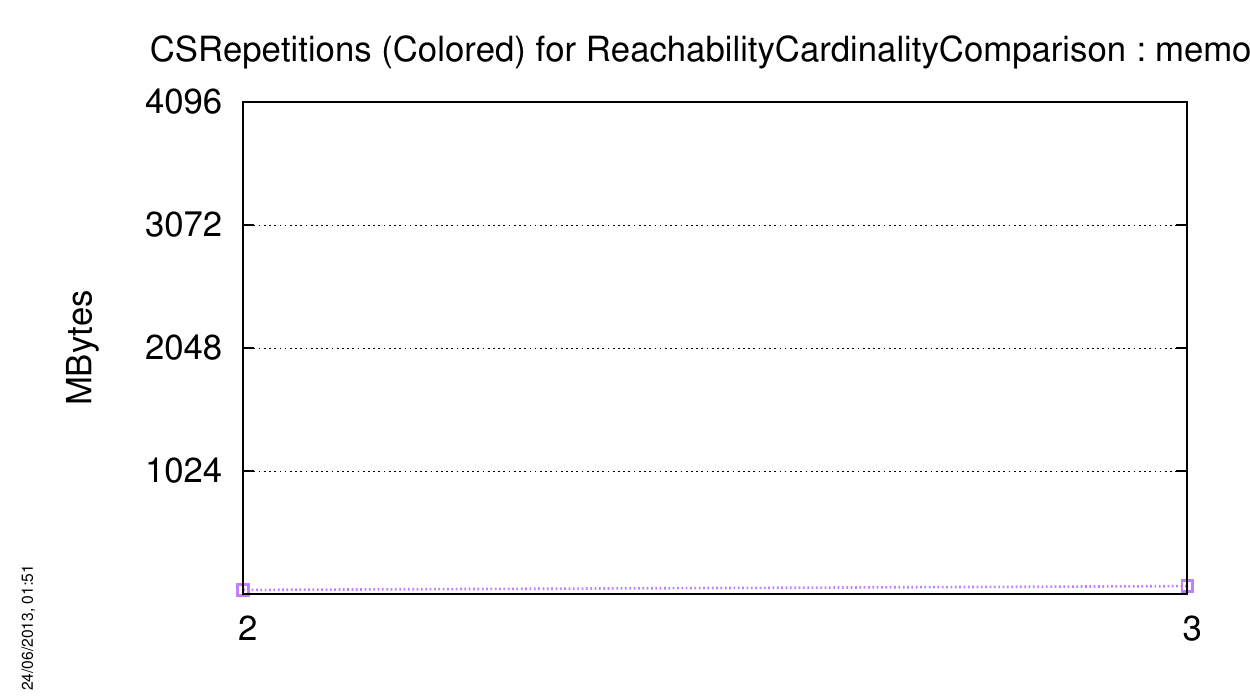}
   \includegraphics[width=7.2cm]{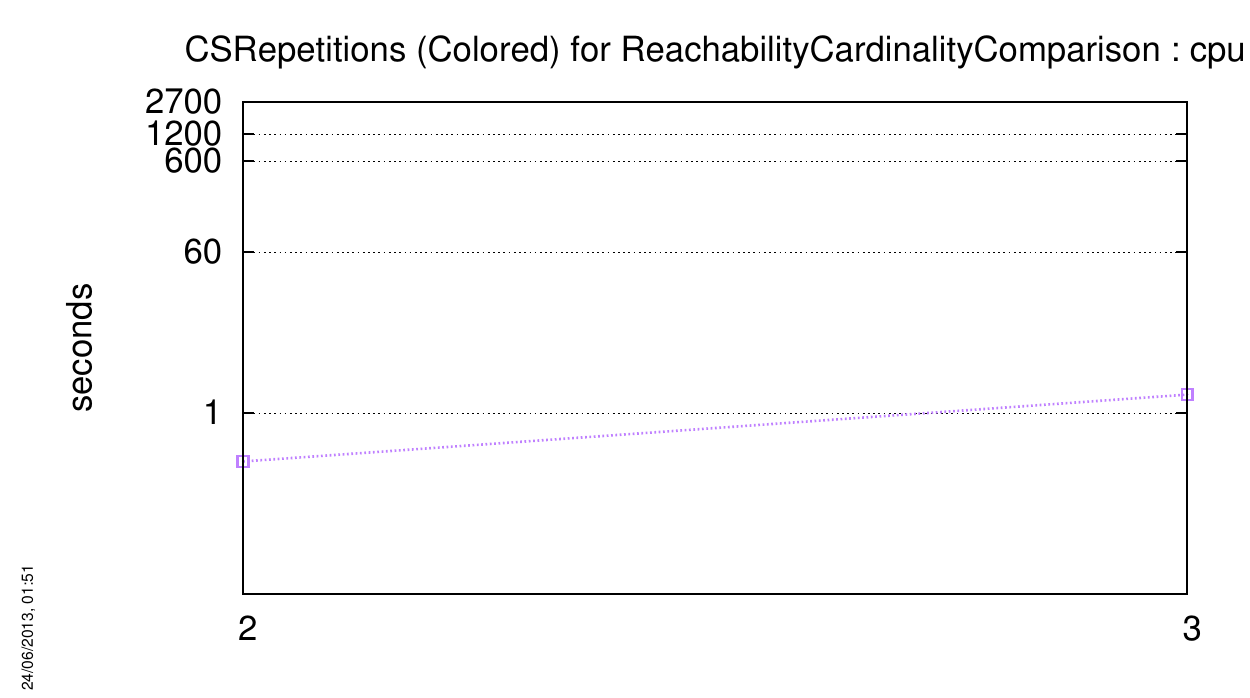}

   \includegraphics[height=1cm]{figures/tools-legend.pdf}
\end{center}

\subsubsection{\acs{CSRepetitions-PT}}
The charts below respectively show how tools compete with this ``Known'' model (memory and CPU).

\index{Performances!ReachabilityCardinalityComparison!CSRepetitions (P/T)}
\begin{center}
   \includegraphics[width=7.2cm]{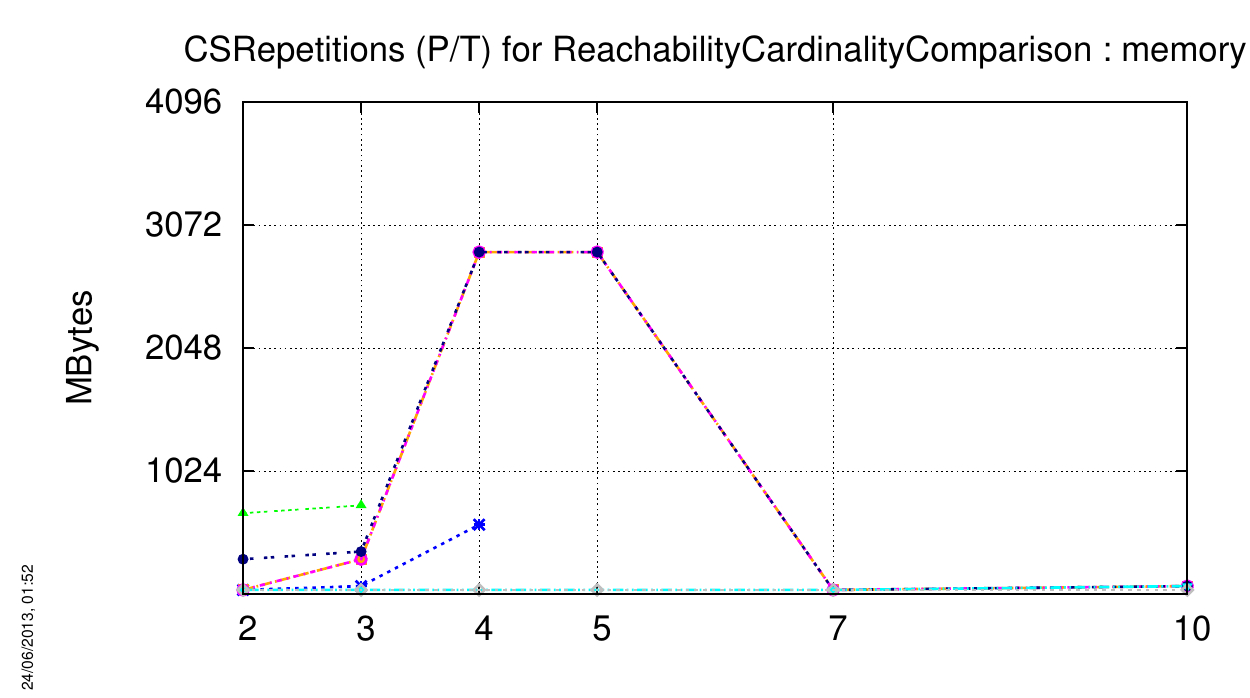}
   \includegraphics[width=7.2cm]{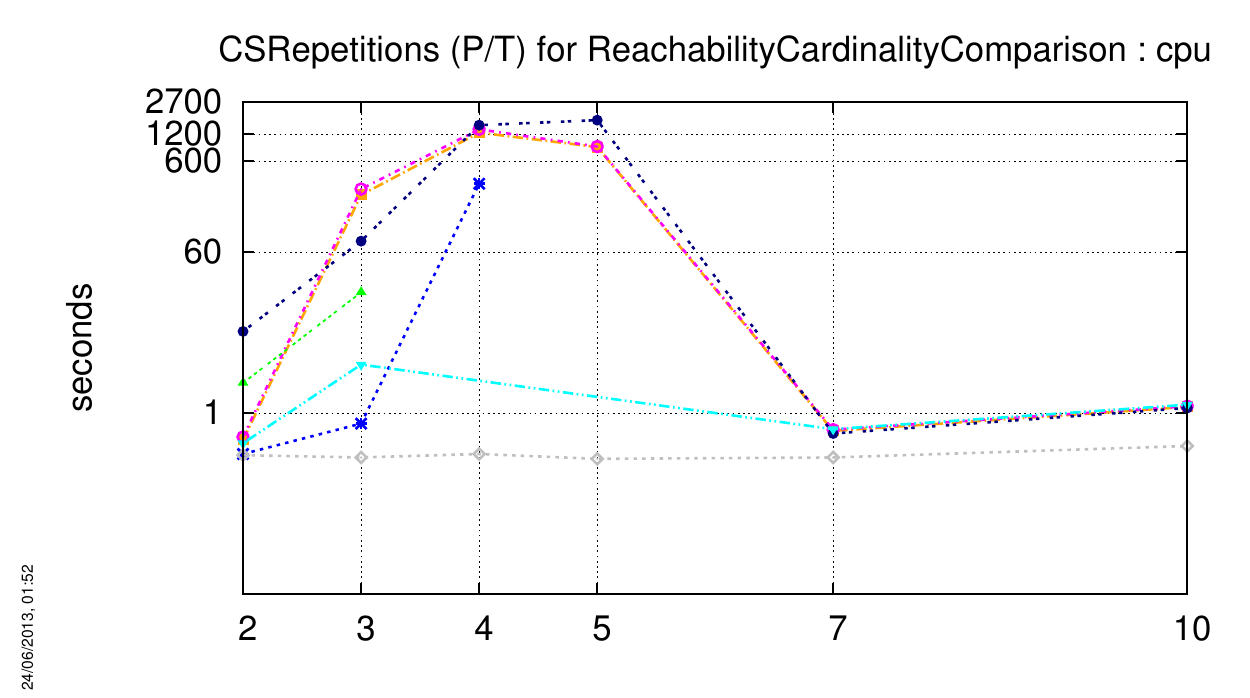}

   \includegraphics[height=1cm]{figures/tools-legend.pdf}
\end{center}

\subsubsection{\acs{Dekker-PT}}
The charts below respectively show how tools compete with this ``Known'' model (memory and CPU).

\index{Performances!ReachabilityCardinalityComparison!Dekker (P/T)}
\begin{center}
   \includegraphics[width=7.2cm]{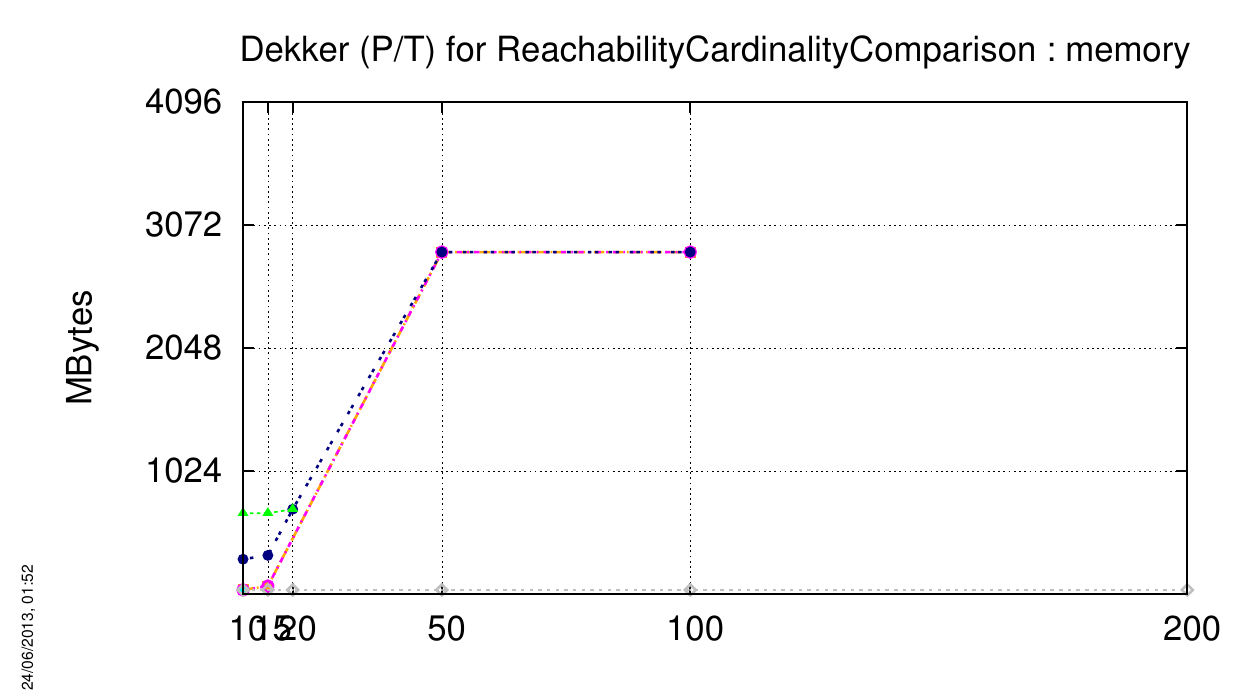}
   \includegraphics[width=7.2cm]{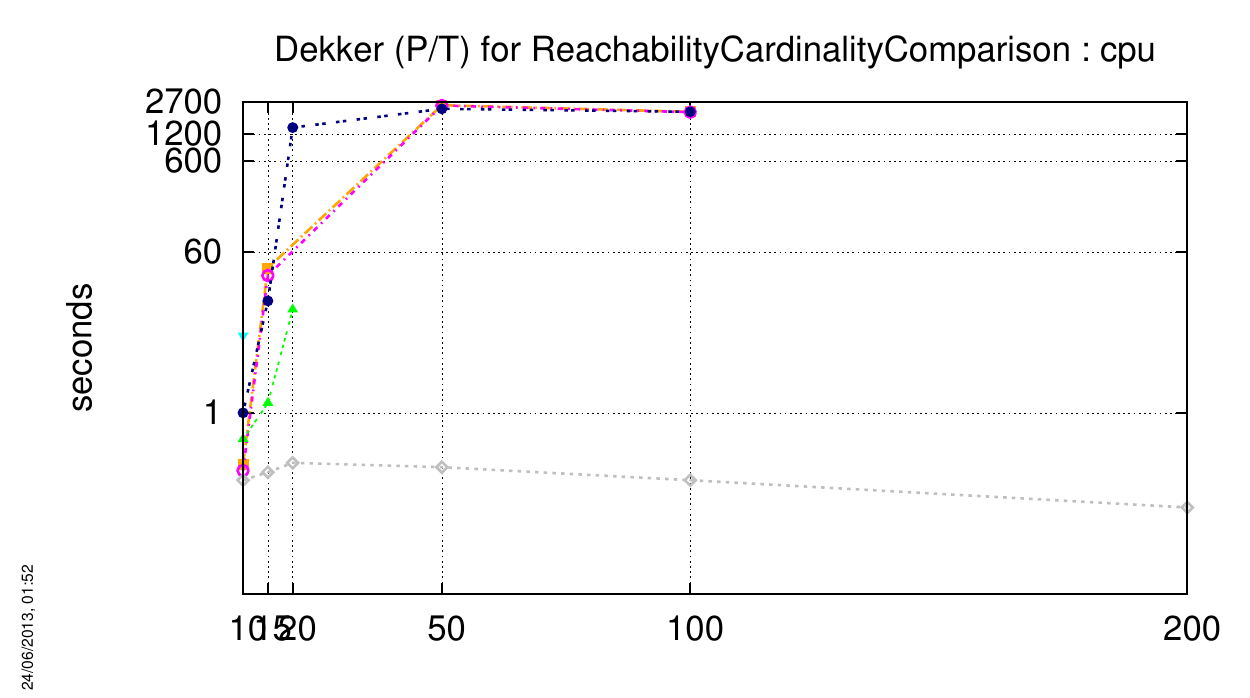}

   \includegraphics[height=1cm]{figures/tools-legend.pdf}
\end{center}

\subsubsection{\acs{DotAndBoxes-COL}}
The charts below respectively show how tools compete with this ``Known'' model (memory and CPU).

\index{Performances!ReachabilityCardinalityComparison!DotAndBoxes (Colored)}
\begin{center}
   \includegraphics[width=7.2cm]{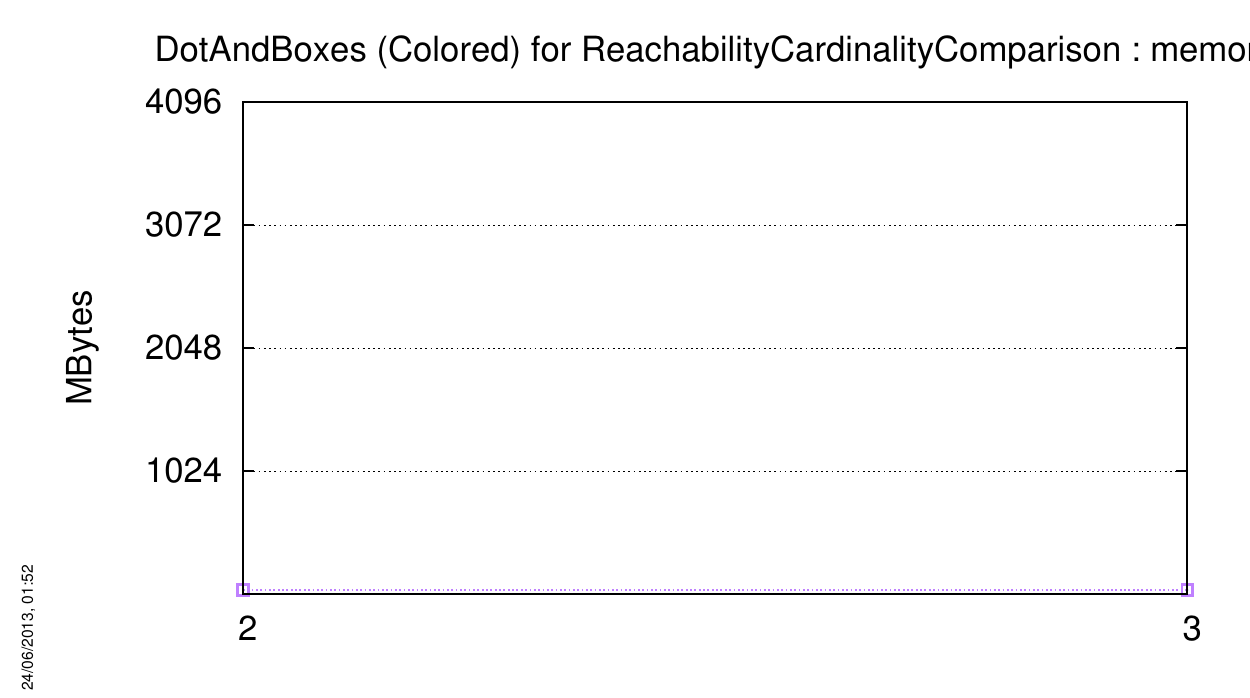}
   \includegraphics[width=7.2cm]{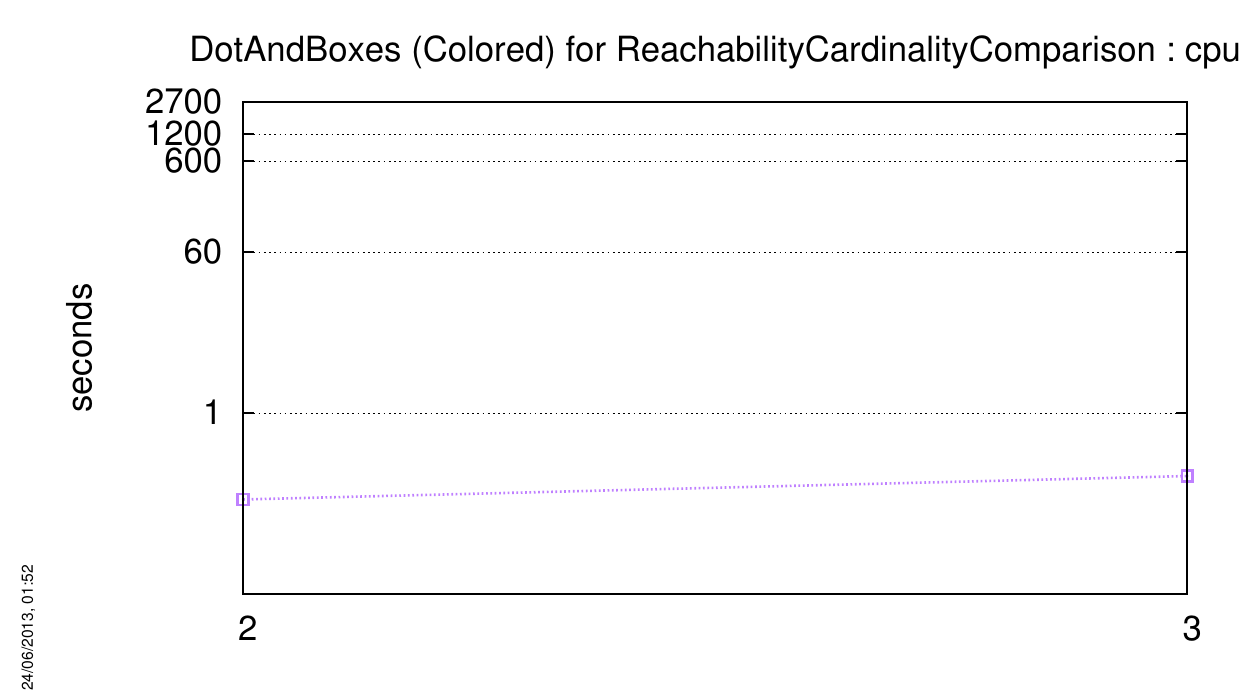}

   \includegraphics[height=1cm]{figures/tools-legend.pdf}
\end{center}

\subsubsection{\acs{DrinkVendingMachine-COL}}
The charts below respectively show how tools compete with this ``Known'' model (memory and CPU).

\index{Performances!ReachabilityCardinalityComparison!DrinkVendingMachine (Colored)}
\begin{center}
   \includegraphics[width=7.2cm]{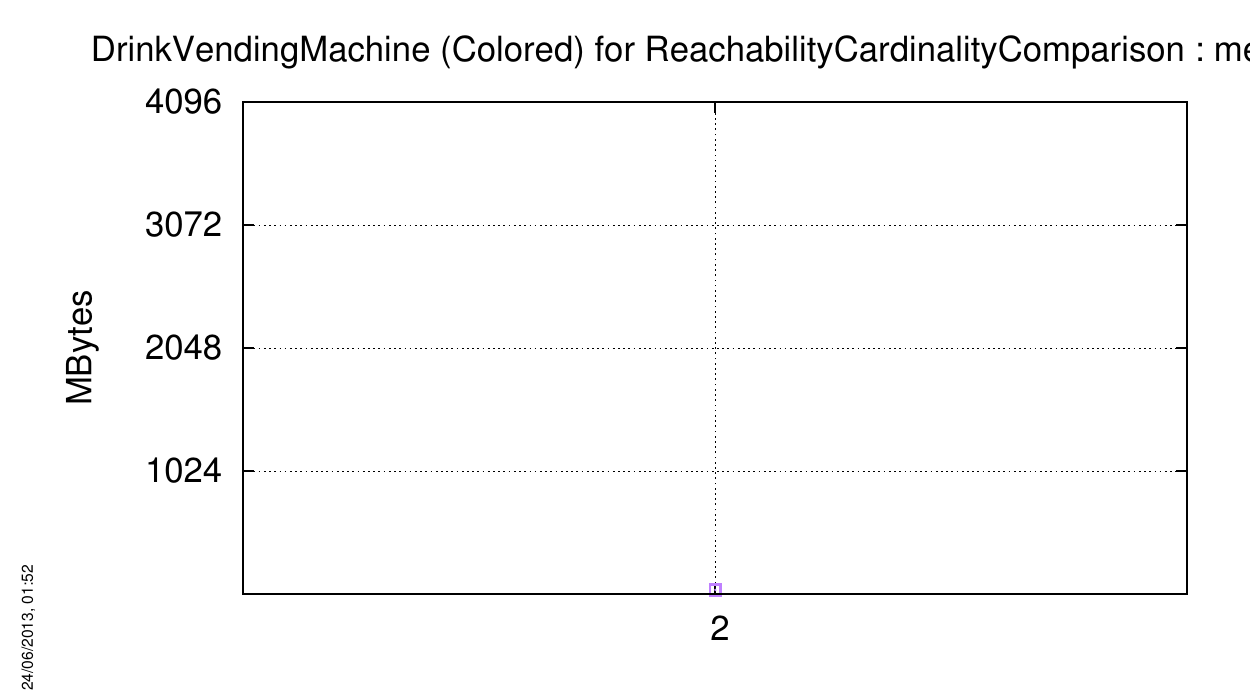}
   \includegraphics[width=7.2cm]{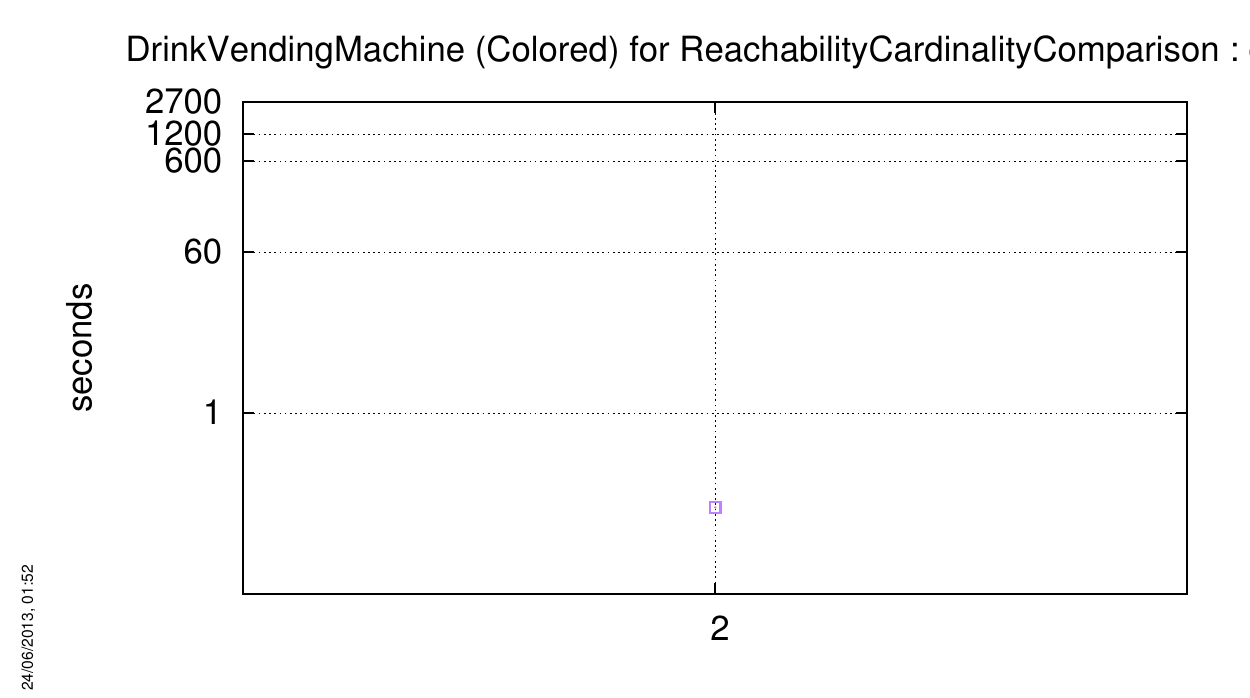}

   \includegraphics[height=1cm]{figures/tools-legend.pdf}
\end{center}

\subsubsection{\acs{DrinkVendingMachine-PT}}
The charts below respectively show how tools compete with this ``Known'' model (memory and CPU).

\index{Performances!ReachabilityCardinalityComparison!DrinkVendingMachine (P/T)}
\begin{center}
   \includegraphics[width=7.2cm]{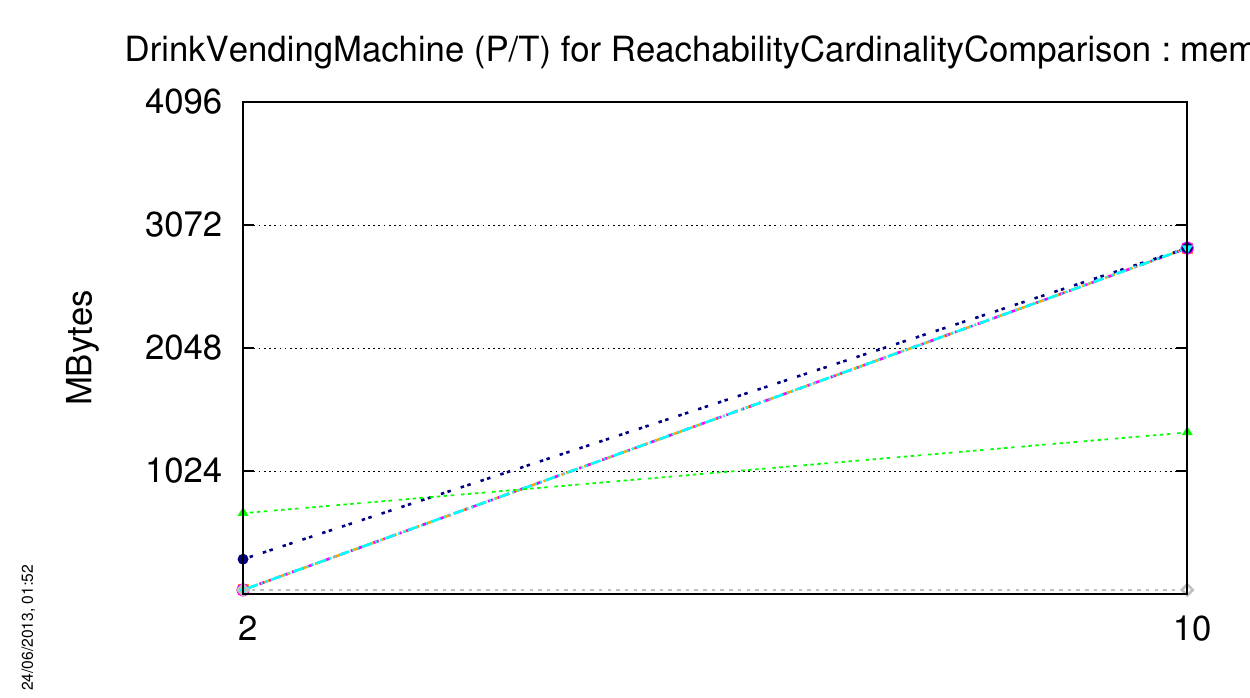}
   \includegraphics[width=7.2cm]{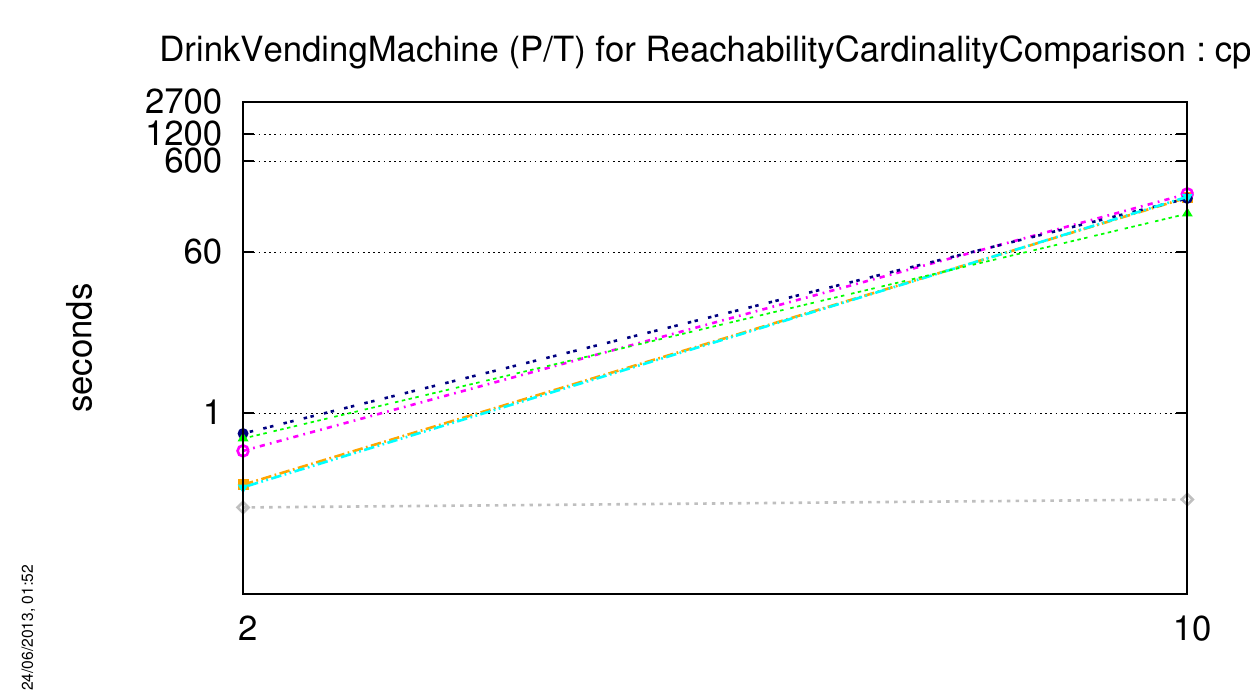}

   \includegraphics[height=1cm]{figures/tools-legend.pdf}
\end{center}

\subsubsection{\acs{Echo-PT}}
The charts below respectively show how tools compete with this ``Known'' model (memory and CPU).

\index{Performances!ReachabilityCardinalityComparison!Echo (P/T)}
\begin{center}
   \includegraphics[width=7.2cm]{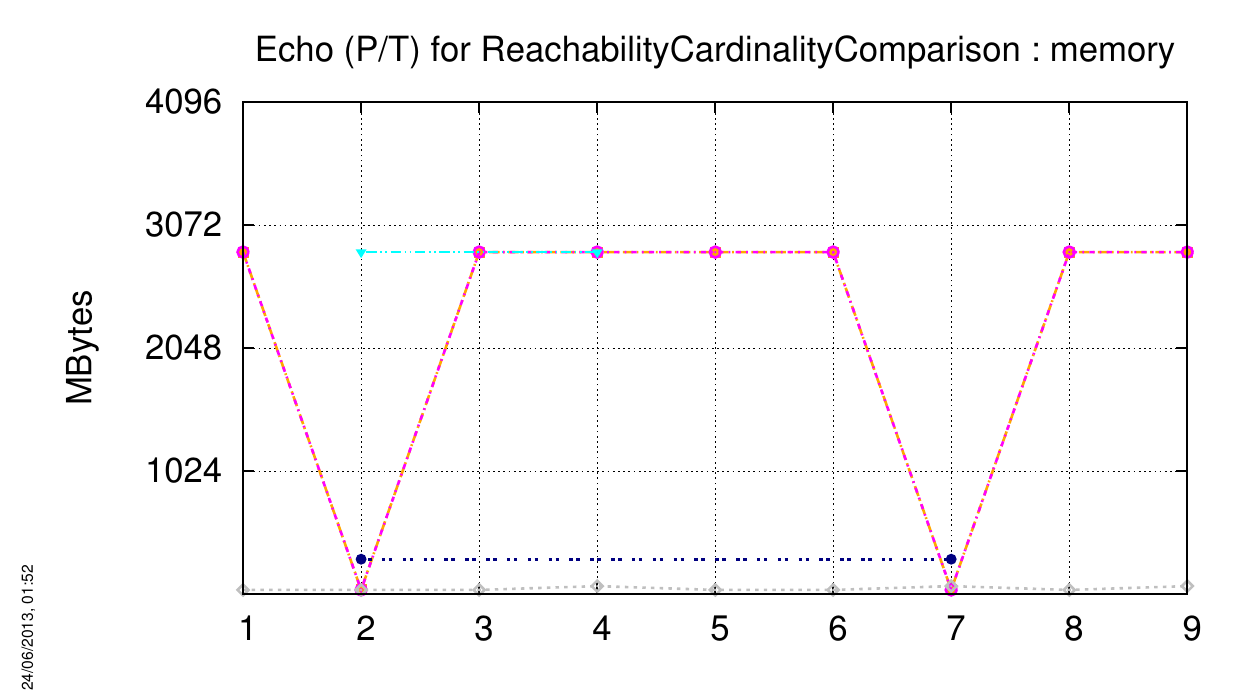}
   \includegraphics[width=7.2cm]{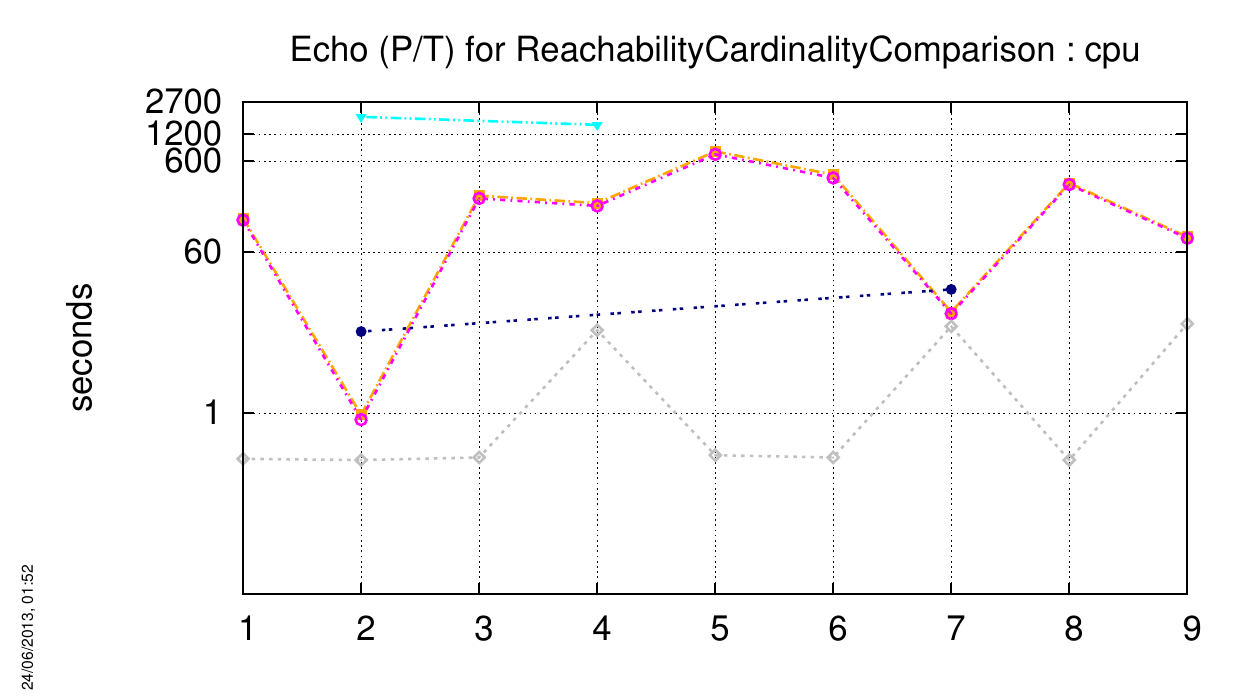}

   \includegraphics[height=1cm]{figures/tools-legend.pdf}
\end{center}

\subsubsection{\acs{Eratosthenes-PT}}
The charts below respectively show how tools compete with this ``Known'' model (memory and CPU).

\index{Performances!ReachabilityCardinalityComparison!Eratosthenes (P/T)}
\begin{center}
   \includegraphics[width=7.2cm]{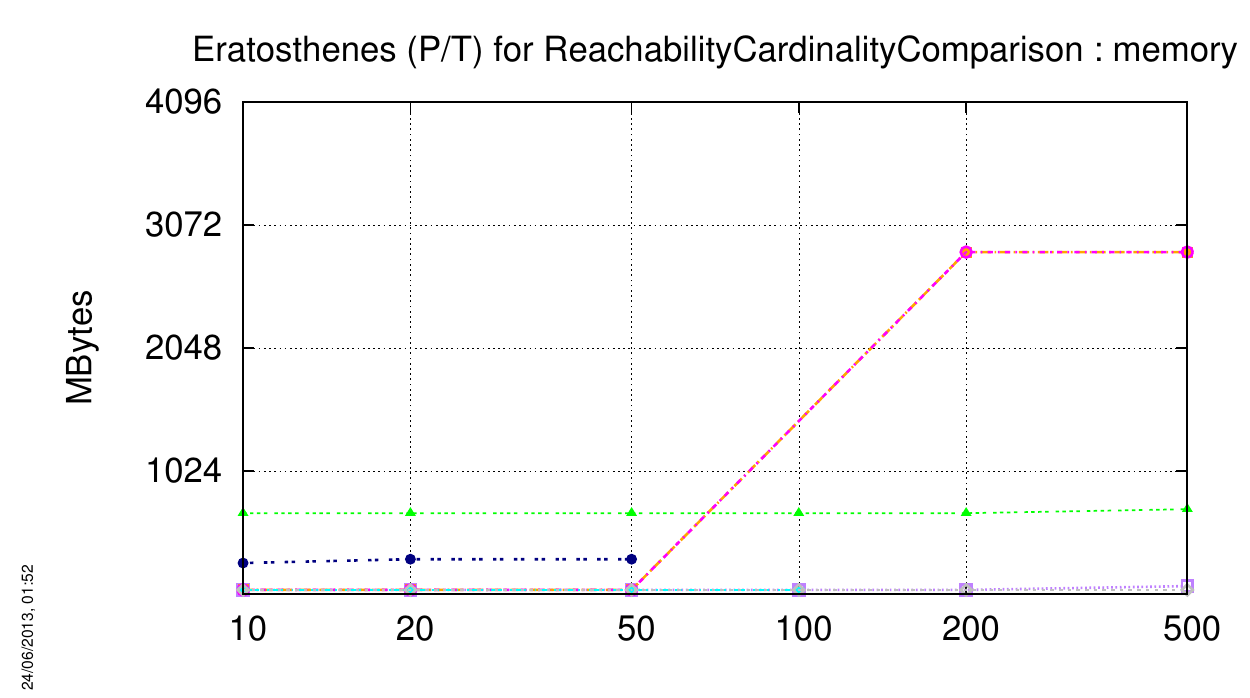}
   \includegraphics[width=7.2cm]{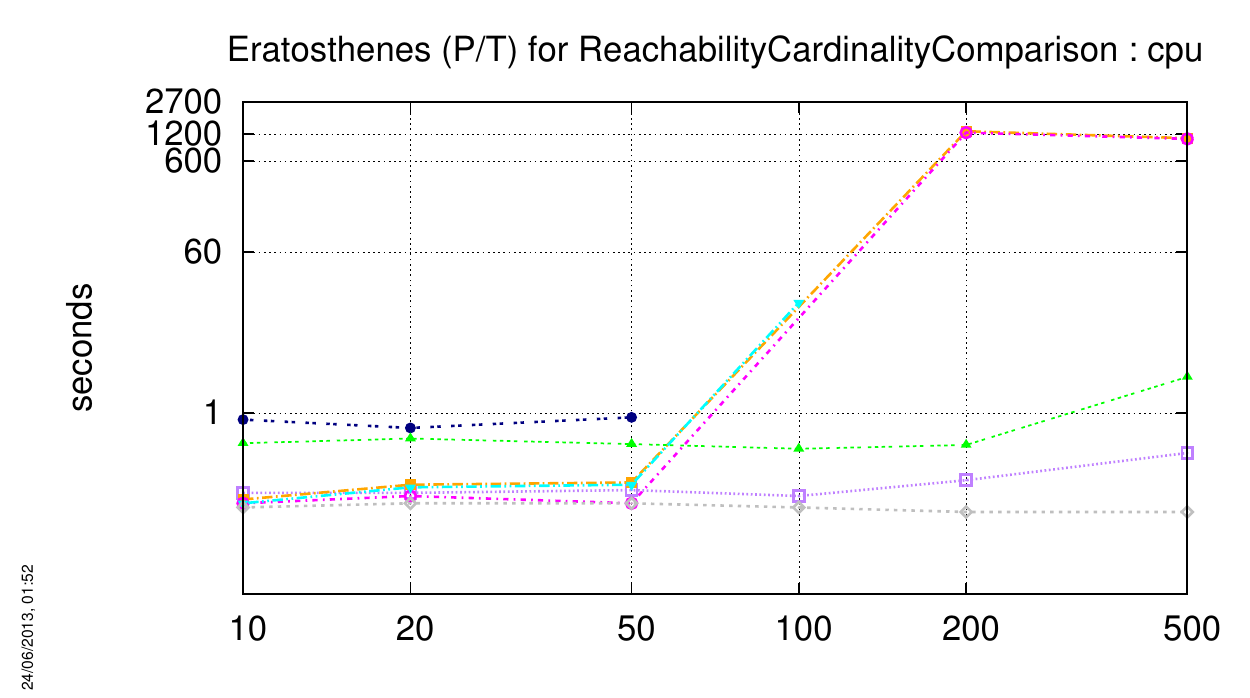}

   \includegraphics[height=1cm]{figures/tools-legend.pdf}
\end{center}

\subsubsection{\acs{FMS-PT}}
The charts below respectively show how tools compete with this ``Known'' model (memory and CPU).

\index{Performances!ReachabilityCardinalityComparison!FMS (P/T)}
\begin{center}
   \includegraphics[width=7.2cm]{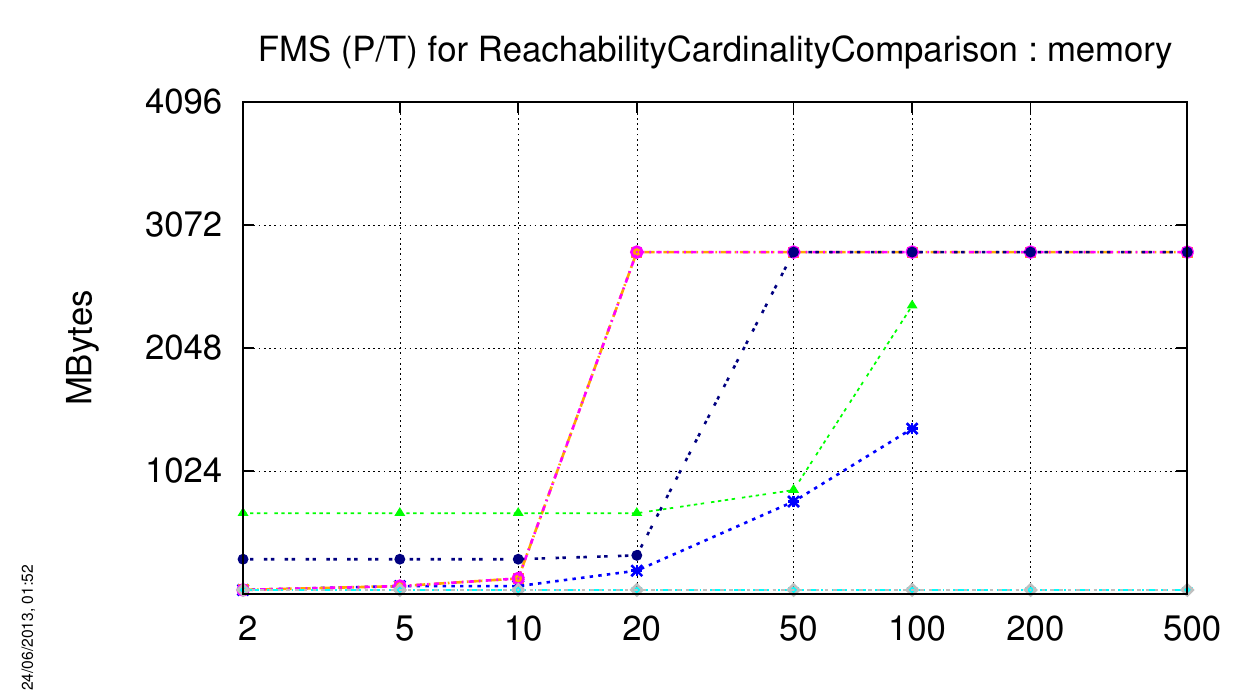}
   \includegraphics[width=7.2cm]{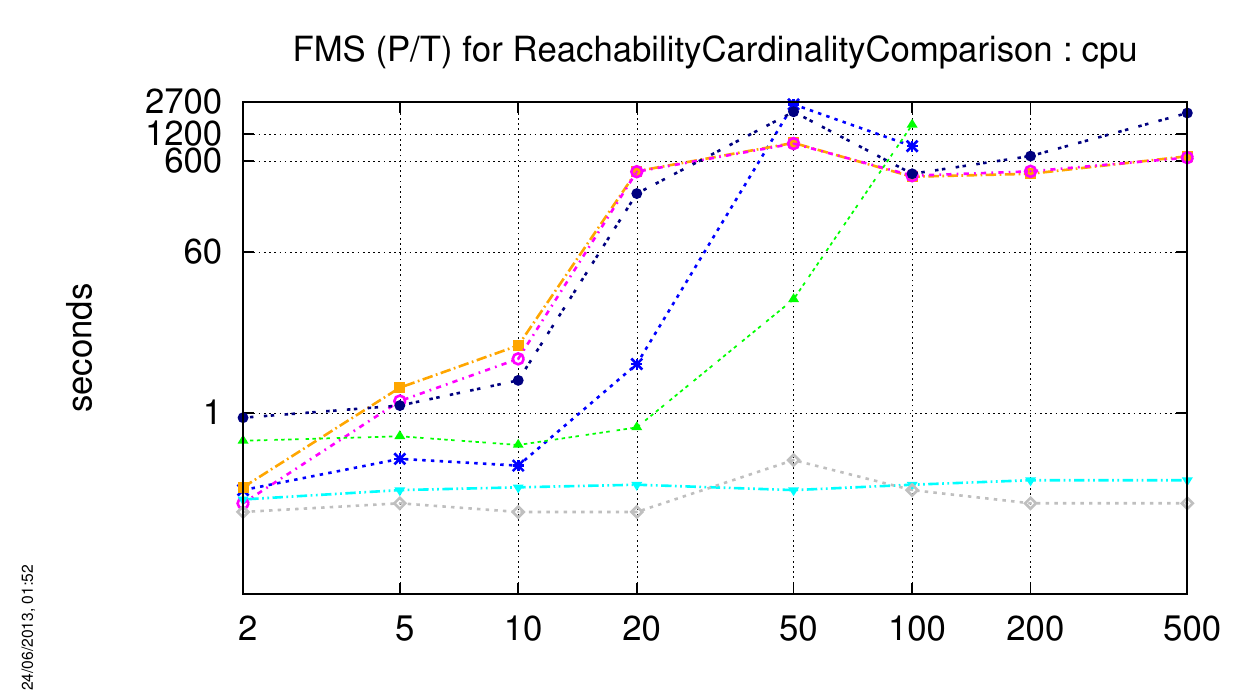}

   \includegraphics[height=1cm]{figures/tools-legend.pdf}
\end{center}

\subsubsection{\acs{GlobalRessAlloc-COL}}
The charts below respectively show how tools compete with this ``Known'' model (memory and CPU).

\index{Performances!ReachabilityCardinalityComparison!GlobalRessAlloc (Colored)}
\begin{center}
   \includegraphics[width=7.2cm]{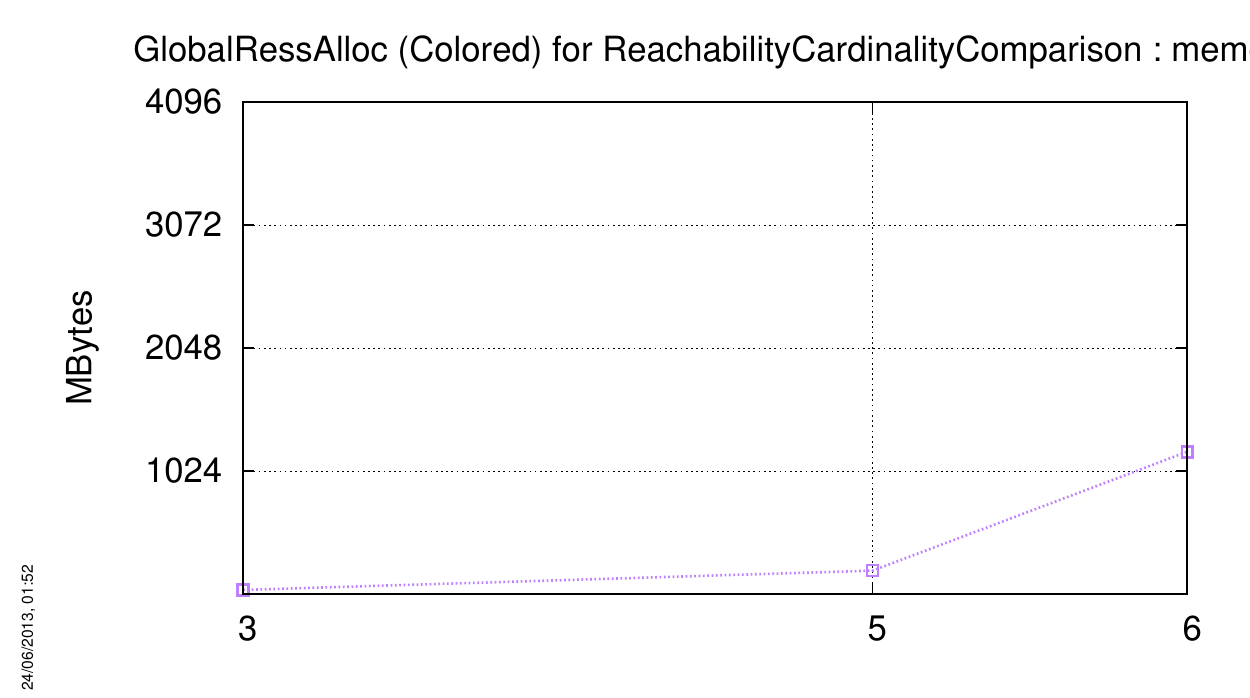}
   \includegraphics[width=7.2cm]{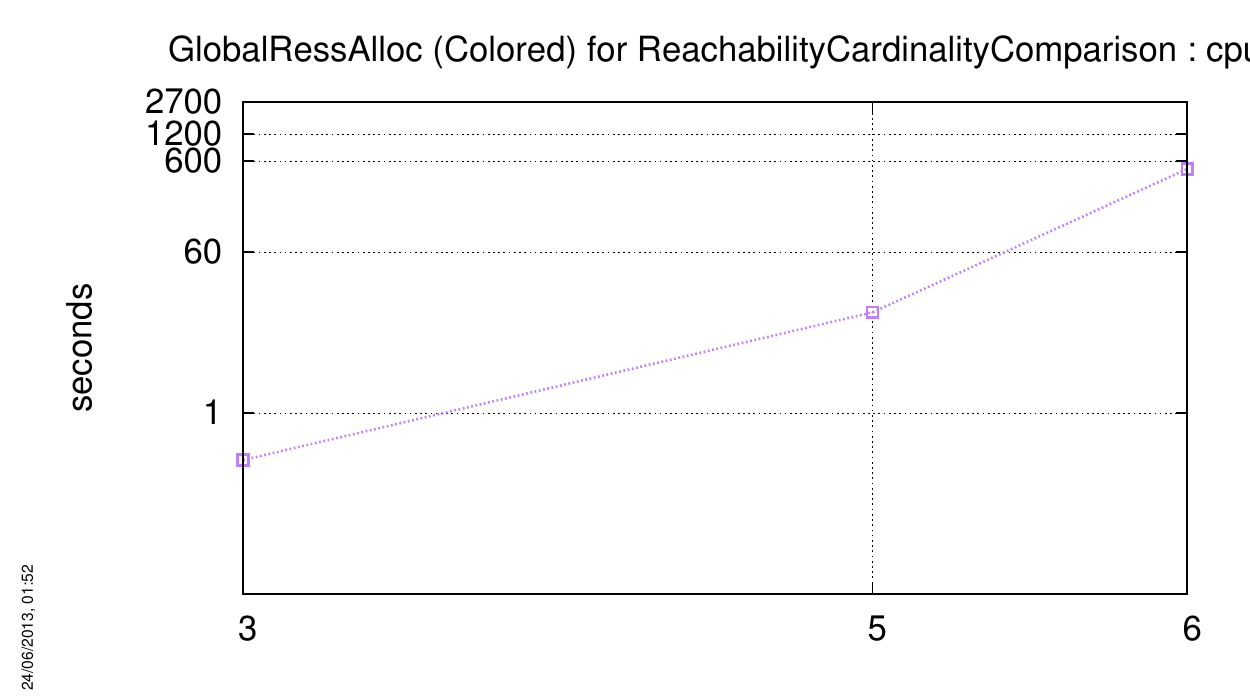}

   \includegraphics[height=1cm]{figures/tools-legend.pdf}
\end{center}

\subsubsection{\acs{GlobalRessAlloc-PT}}
The charts below respectively show how tools compete with this ``Known'' model (memory and CPU).

\index{Performances!ReachabilityCardinalityComparison!GlobalRessAlloc (P/T)}
\begin{center}
   \includegraphics[width=7.2cm]{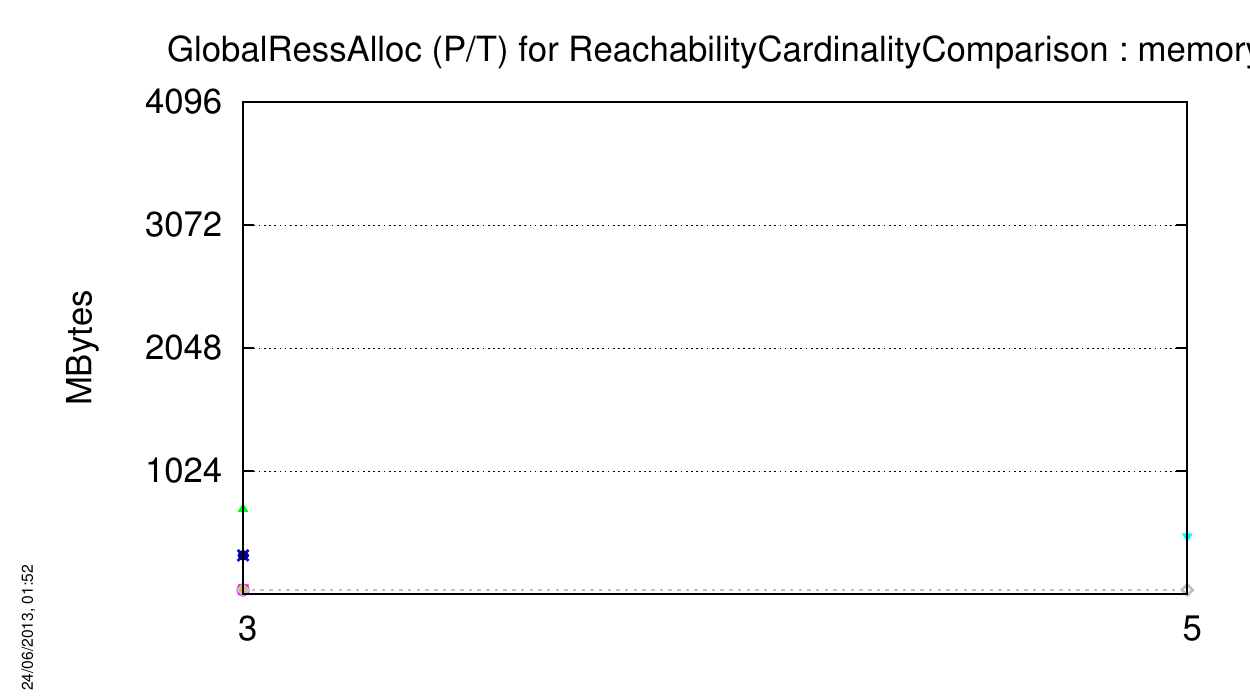}
   \includegraphics[width=7.2cm]{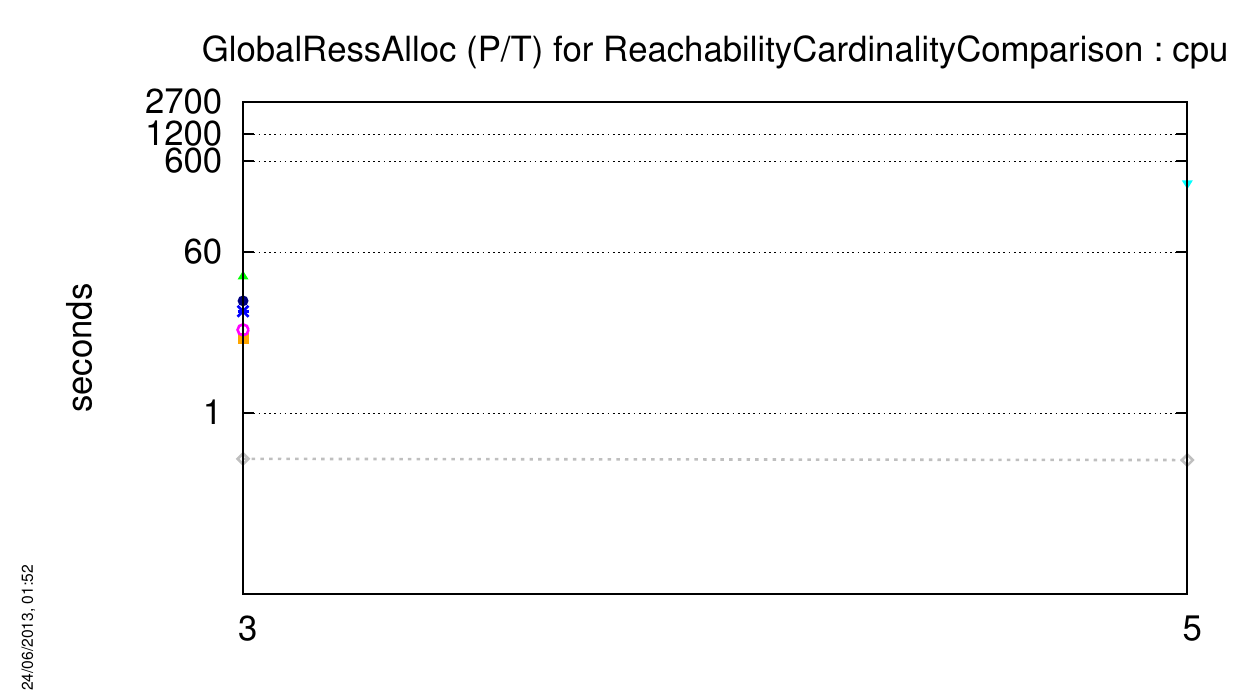}

   \includegraphics[height=1cm]{figures/tools-legend.pdf}
\end{center}

\subsubsection{\acs{Kanban-PT}}
The charts below respectively show how tools compete with this ``Known'' model (memory and CPU).

\index{Performances!ReachabilityCardinalityComparison!Kanban (P/T)}
\begin{center}
   \includegraphics[width=7.2cm]{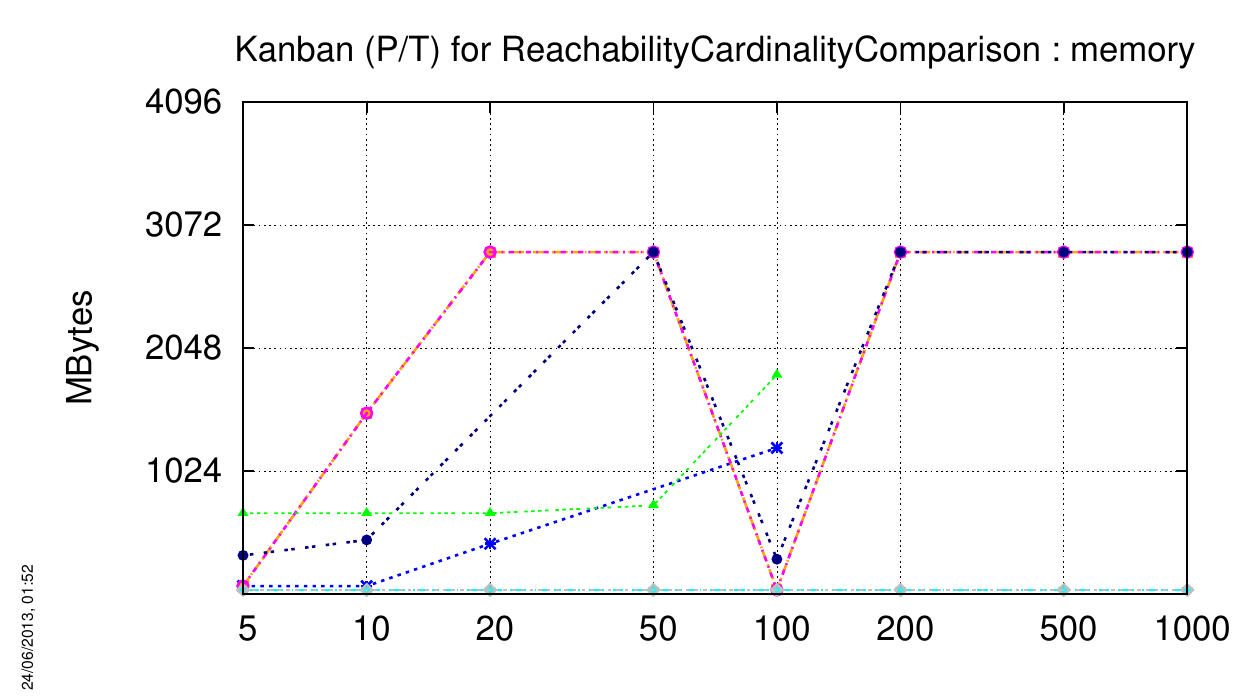}
   \includegraphics[width=7.2cm]{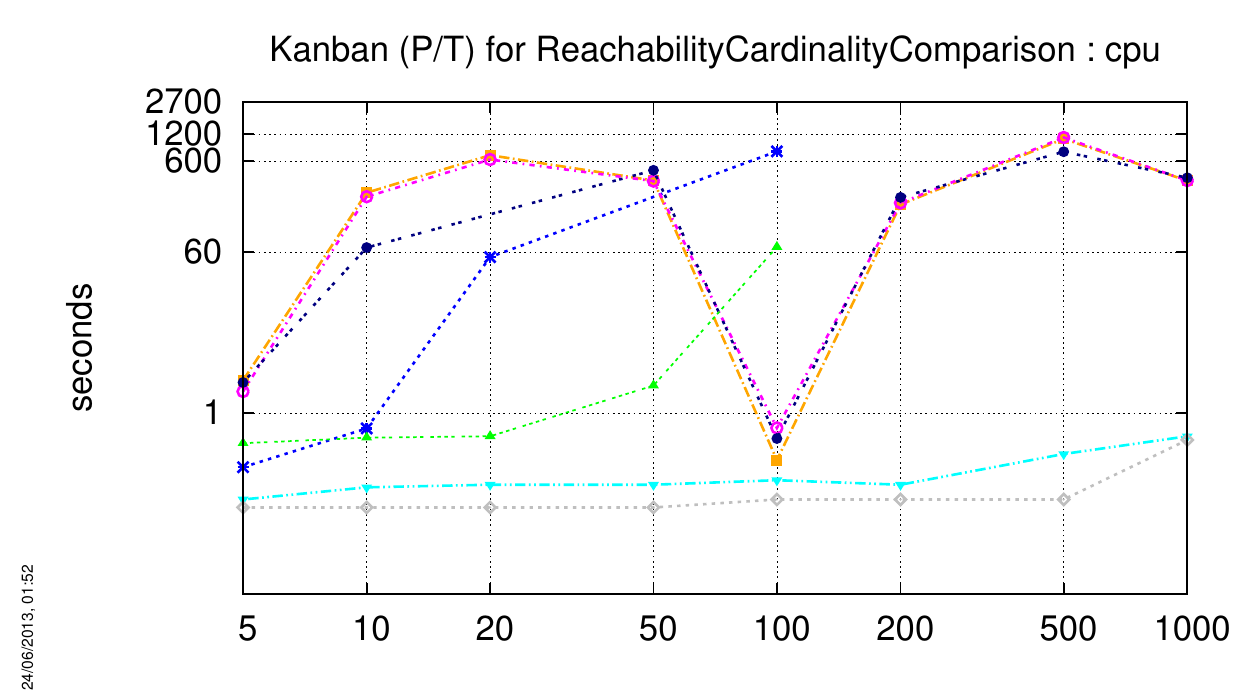}

   \includegraphics[height=1cm]{figures/tools-legend.pdf}
\end{center}

\subsubsection{\acs{LamportFastMutEx-COL}}
The charts below respectively show how tools compete with this ``Known'' model (memory and CPU).

\index{Performances!ReachabilityCardinalityComparison!LamportFastMutEx (Colored)}
\begin{center}
   \includegraphics[width=7.2cm]{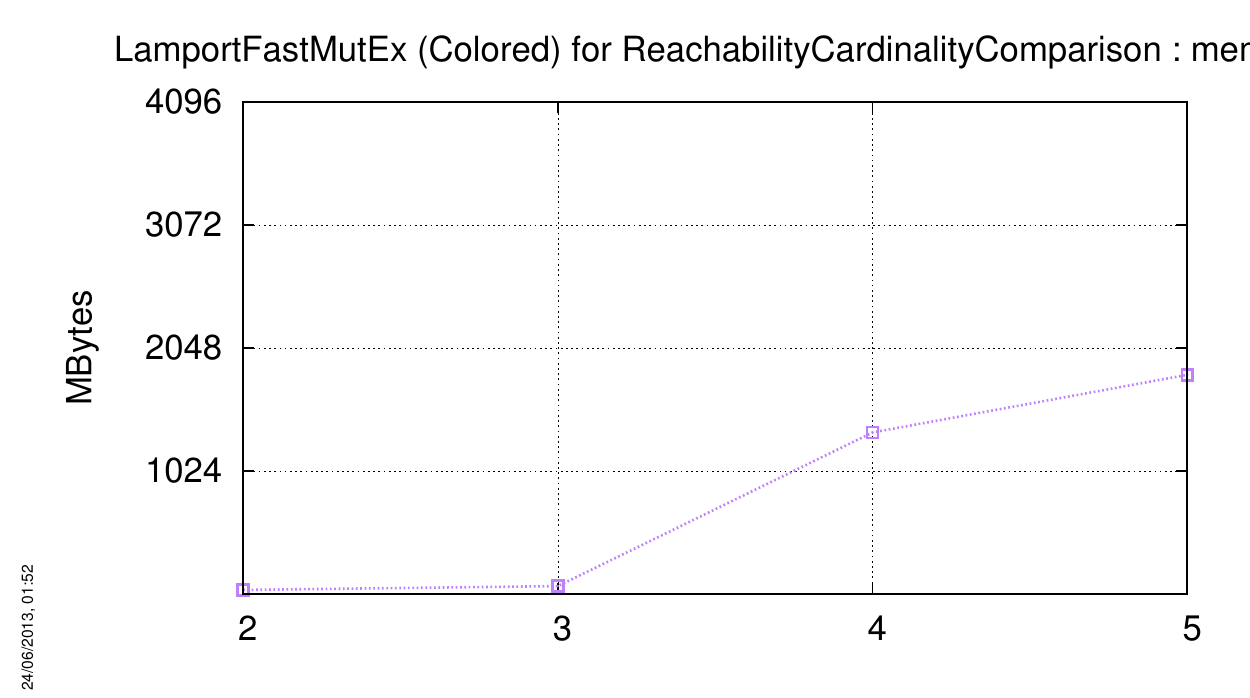}
   \includegraphics[width=7.2cm]{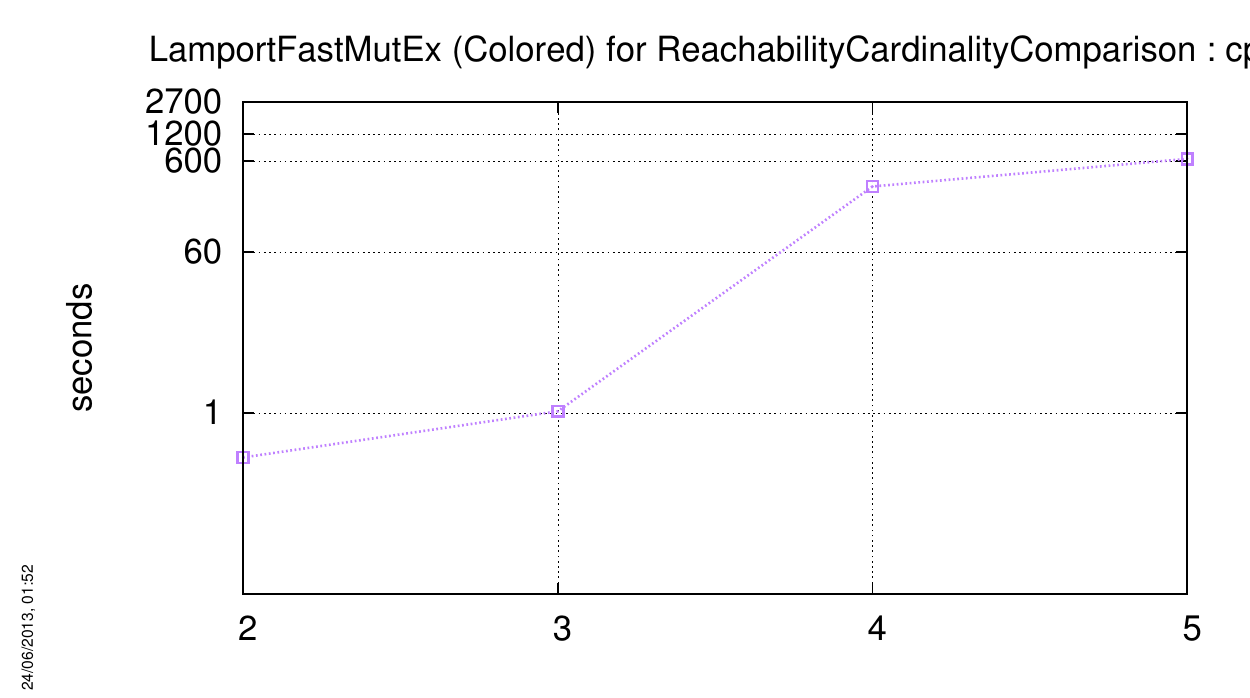}

   \includegraphics[height=1cm]{figures/tools-legend.pdf}
\end{center}

\subsubsection{\acs{LamportFastMutEx-PT}}
The charts below respectively show how tools compete with this ``Known'' model (memory and CPU).

\index{Performances!ReachabilityCardinalityComparison!LamportFastMutEx (P/T)}
\begin{center}
   \includegraphics[width=7.2cm]{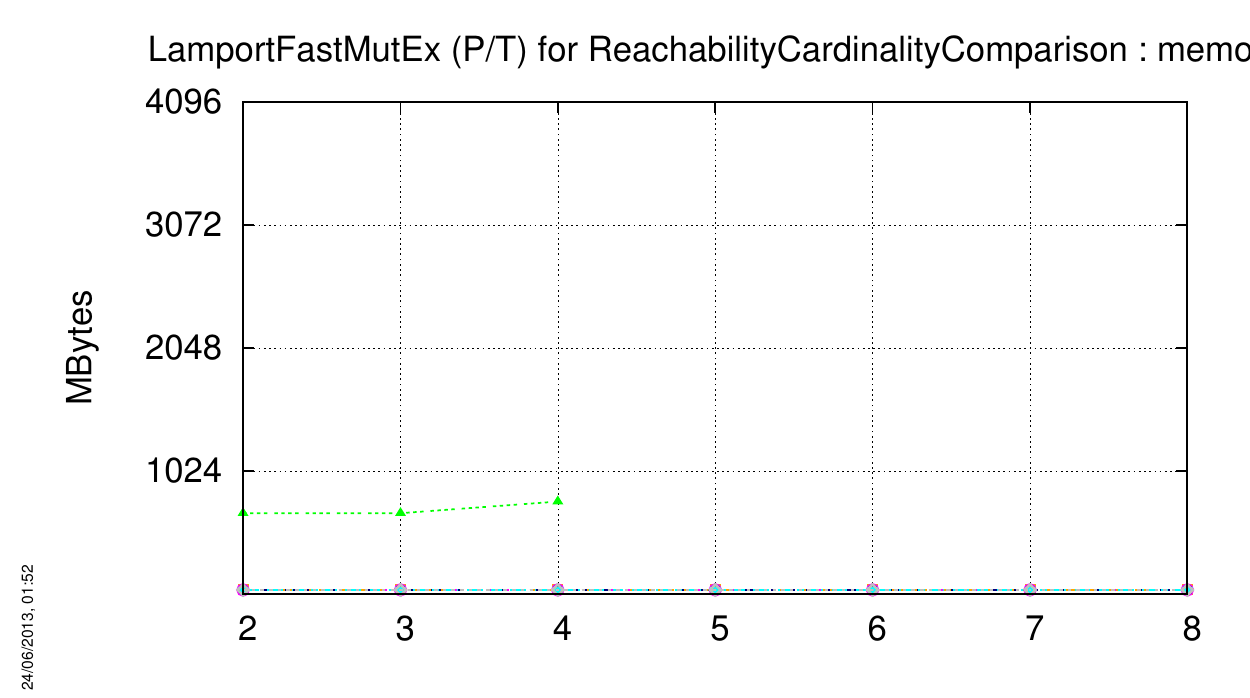}
   \includegraphics[width=7.2cm]{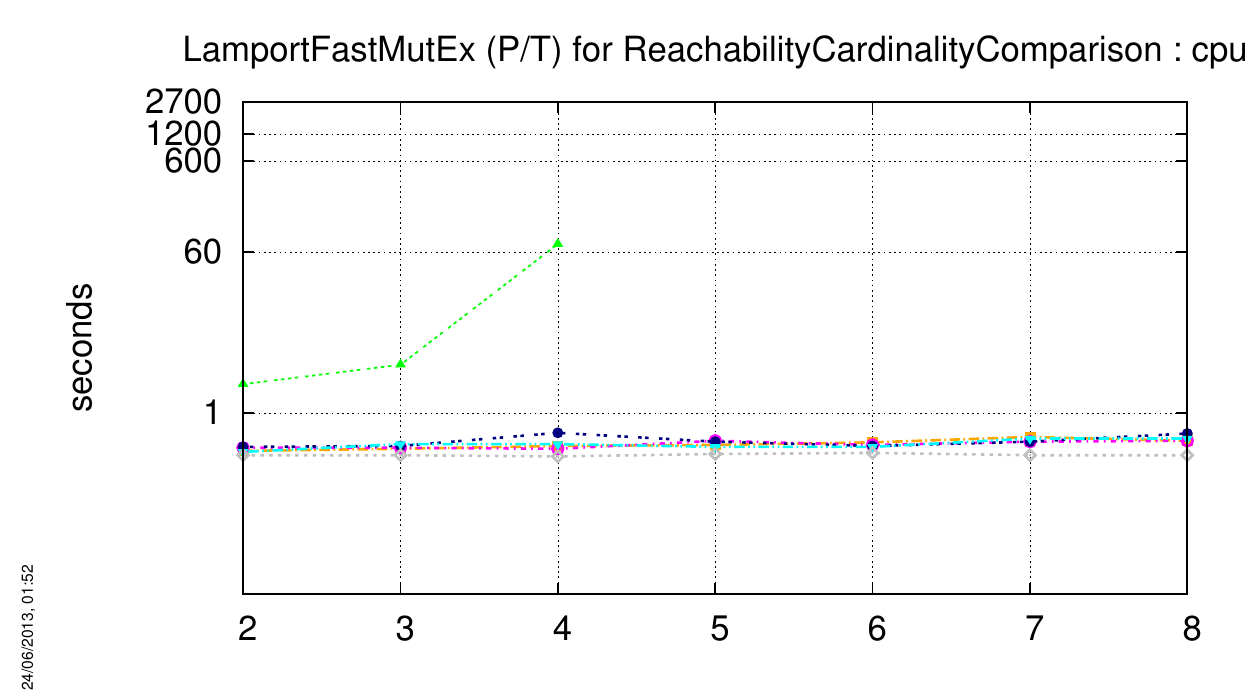}

   \includegraphics[height=1cm]{figures/tools-legend.pdf}
\end{center}

\subsubsection{\acs{MAPK-PT}}
The charts below respectively show how tools compete with this ``Known'' model (memory and CPU).

\index{Performances!ReachabilityCardinalityComparison!MAPK (P/T)}
\begin{center}
   \includegraphics[width=7.2cm]{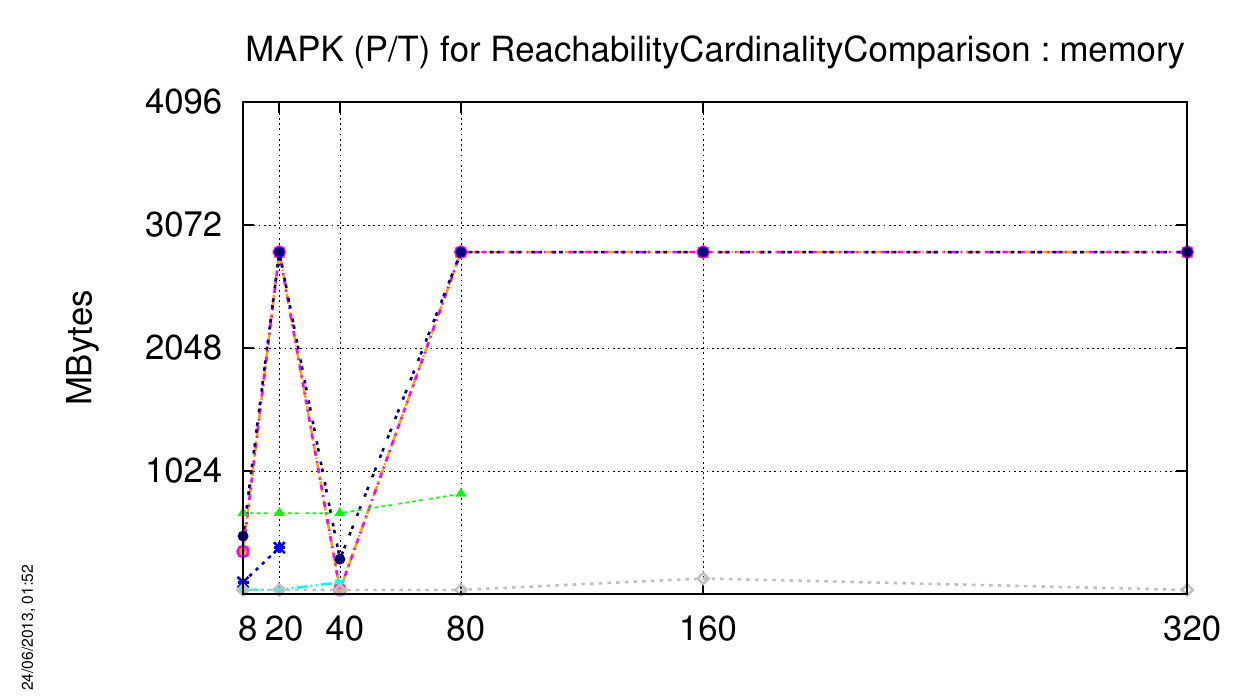}
   \includegraphics[width=7.2cm]{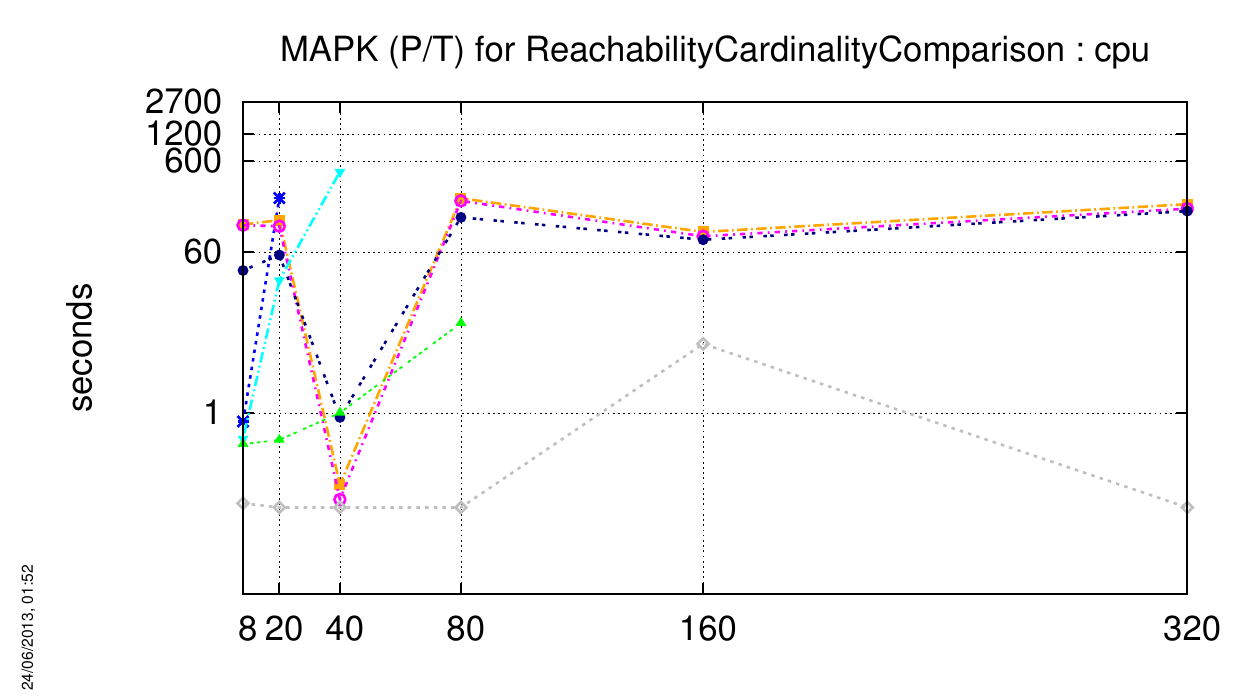}

   \includegraphics[height=1cm]{figures/tools-legend.pdf}
\end{center}

\subsubsection{\acs{NeoElection-COL}}
The charts below respectively show how tools compete with this ``Known'' model (memory and CPU).

\index{Performances!ReachabilityCardinalityComparison!NeoElection (Colored)}
\begin{center}
   \includegraphics[width=7.2cm]{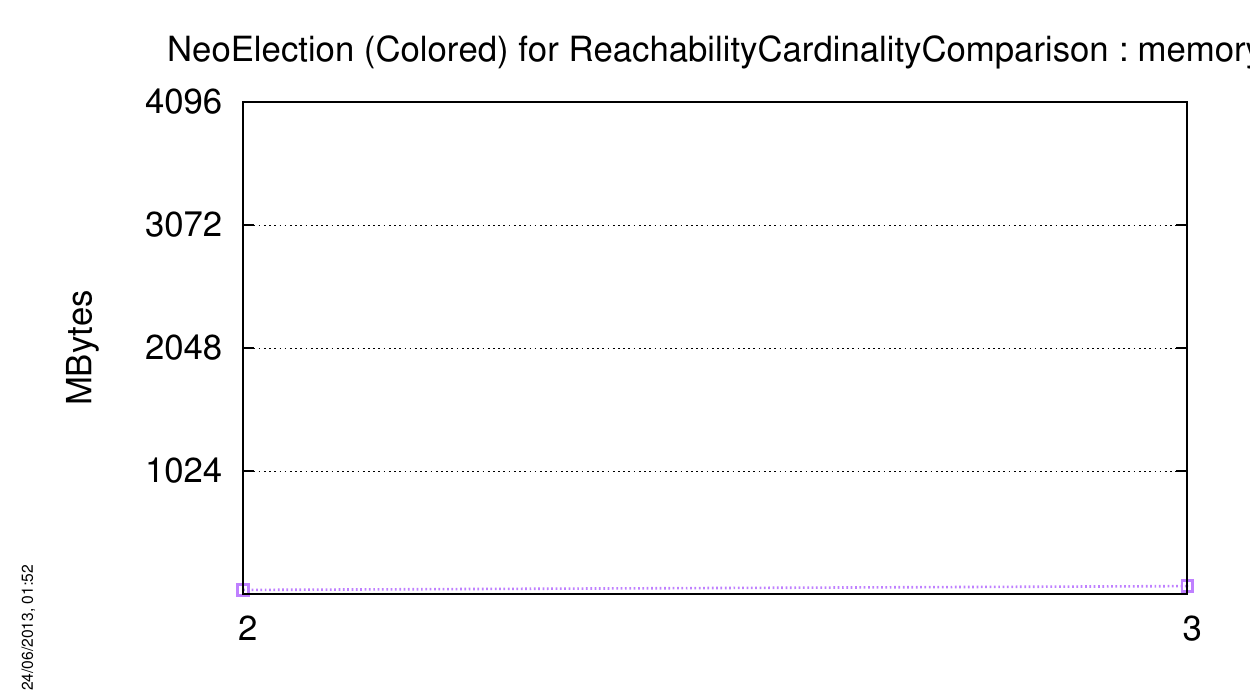}
   \includegraphics[width=7.2cm]{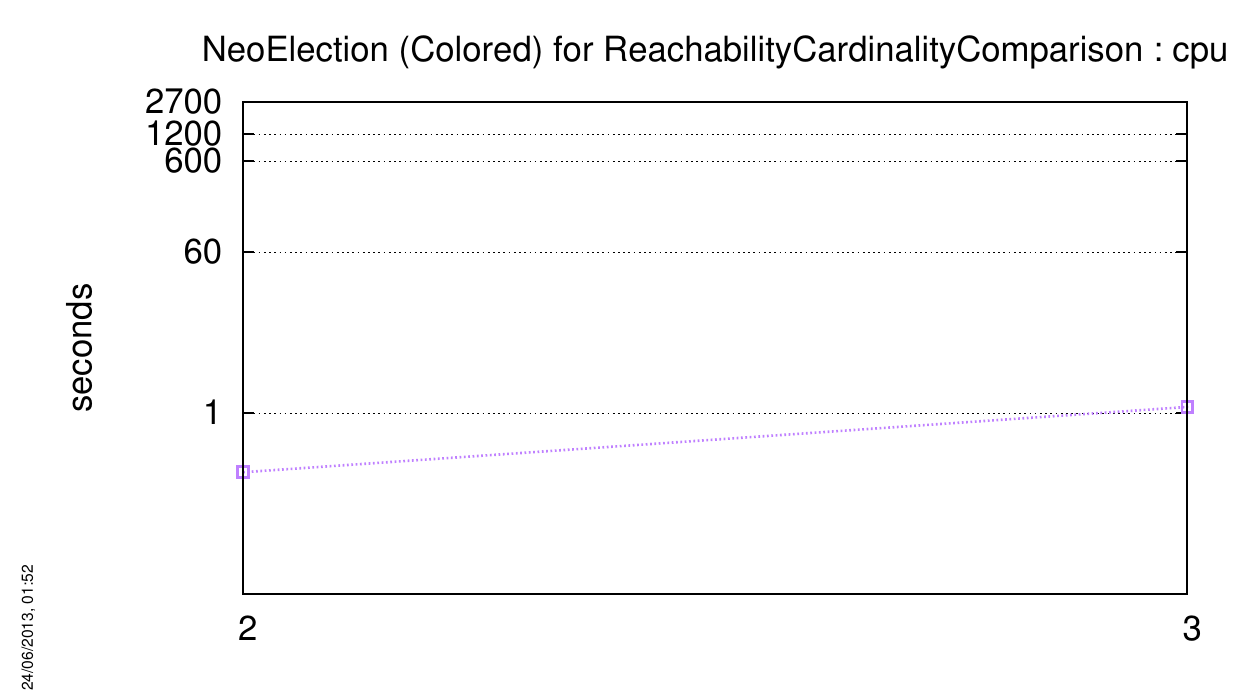}

   \includegraphics[height=1cm]{figures/tools-legend.pdf}
\end{center}

\subsubsection{\acs{NeoElection-PT}}
The charts below respectively show how tools compete with this ``Known'' model (memory and CPU).

\index{Performances!ReachabilityCardinalityComparison!NeoElection (P/T)}
\begin{center}
   \includegraphics[width=7.2cm]{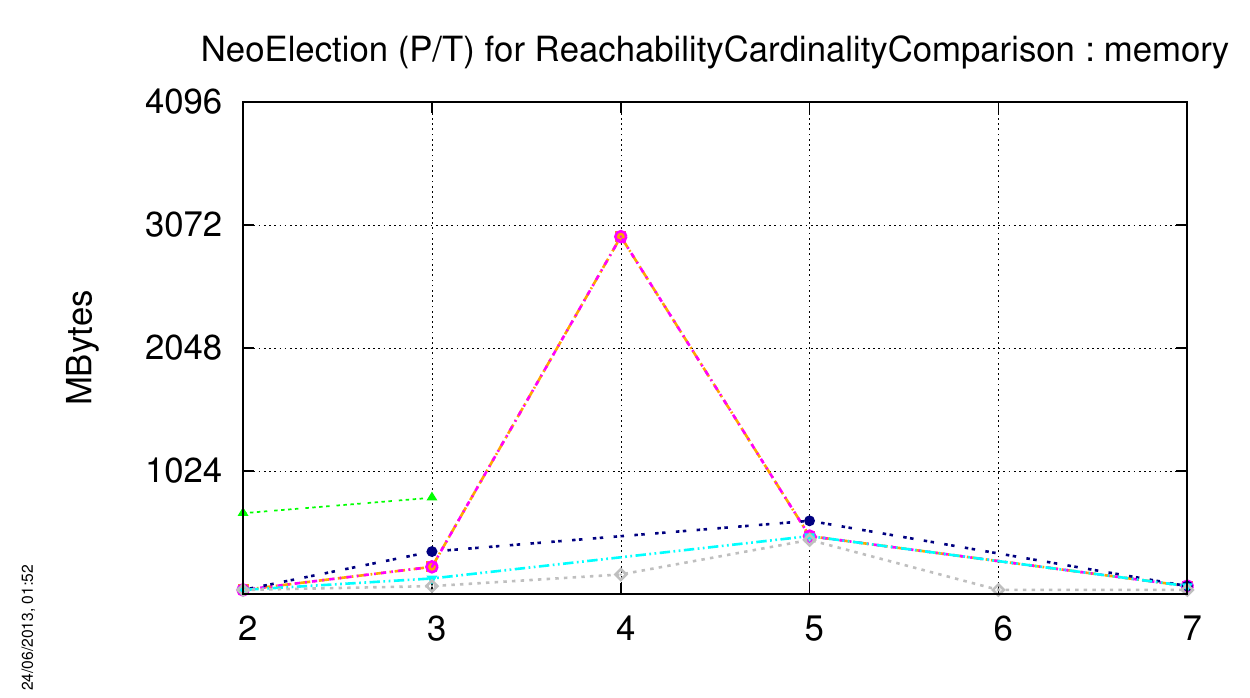}
   \includegraphics[width=7.2cm]{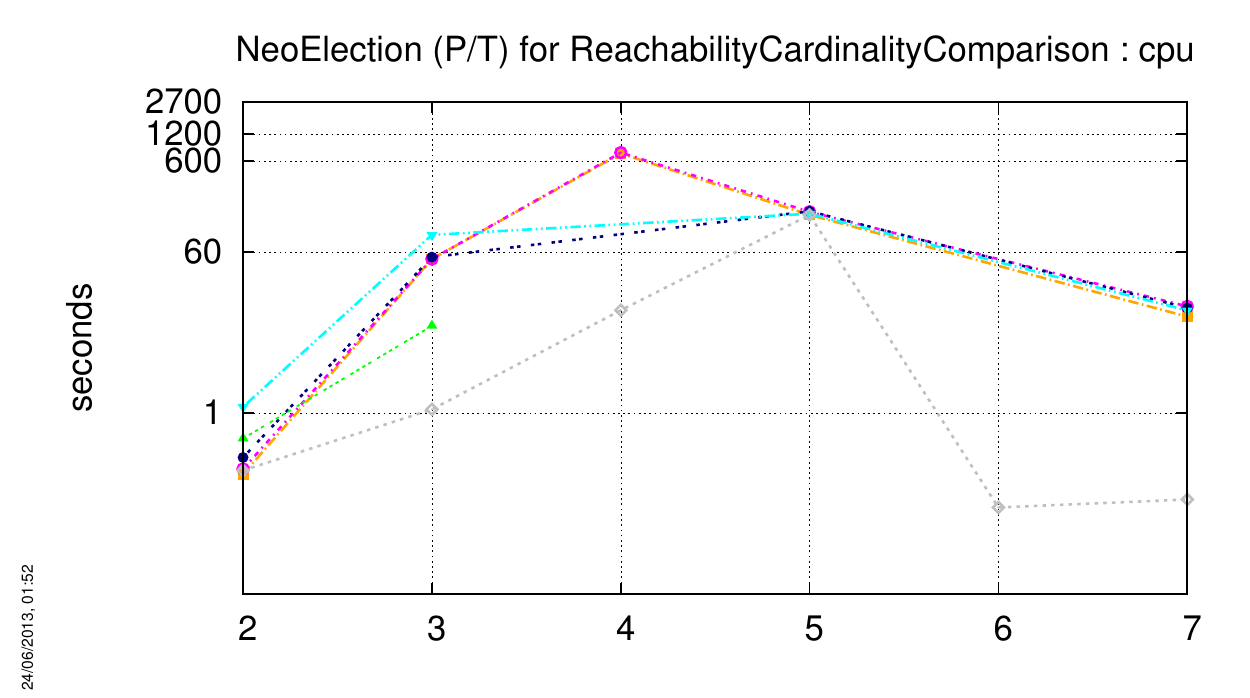}

   \includegraphics[height=1cm]{figures/tools-legend.pdf}
\end{center}

\subsubsection{\acs{PermAdmissibility-COL}}
The charts below respectively show how tools compete with this ``Known'' model (memory and CPU).

\index{Performances!ReachabilityCardinalityComparison!PermAdmissibility (Colored)}
\begin{center}
   \includegraphics[width=7.2cm]{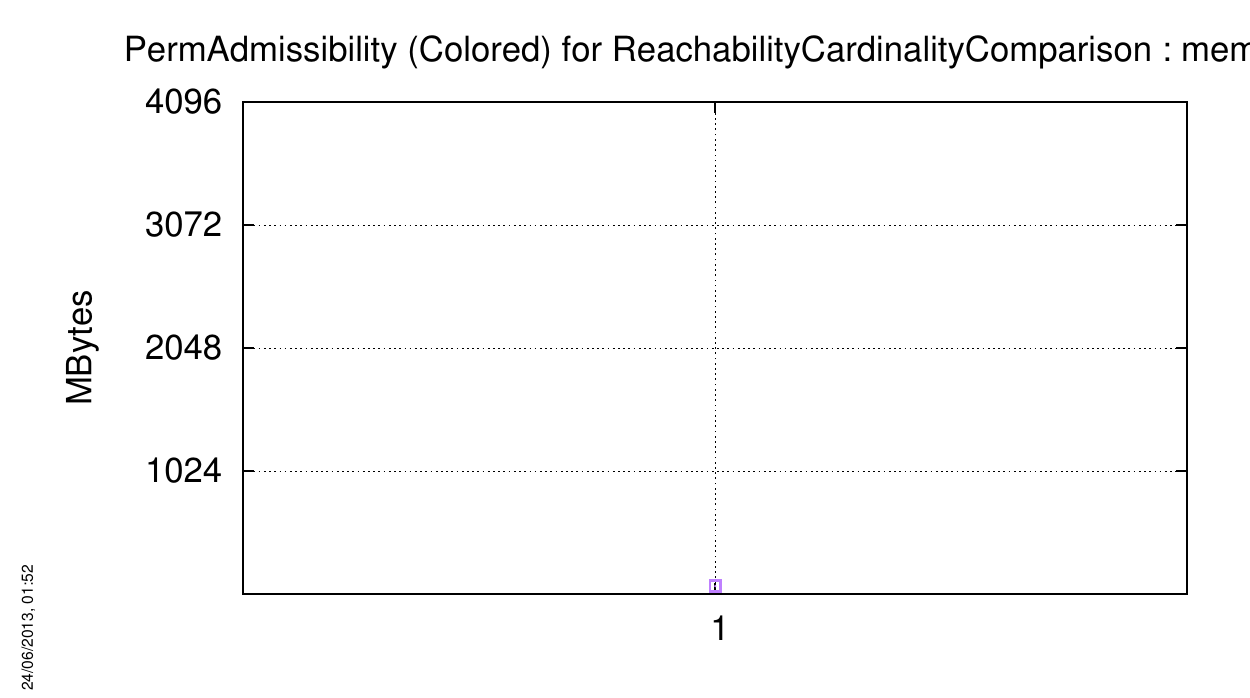}
   \includegraphics[width=7.2cm]{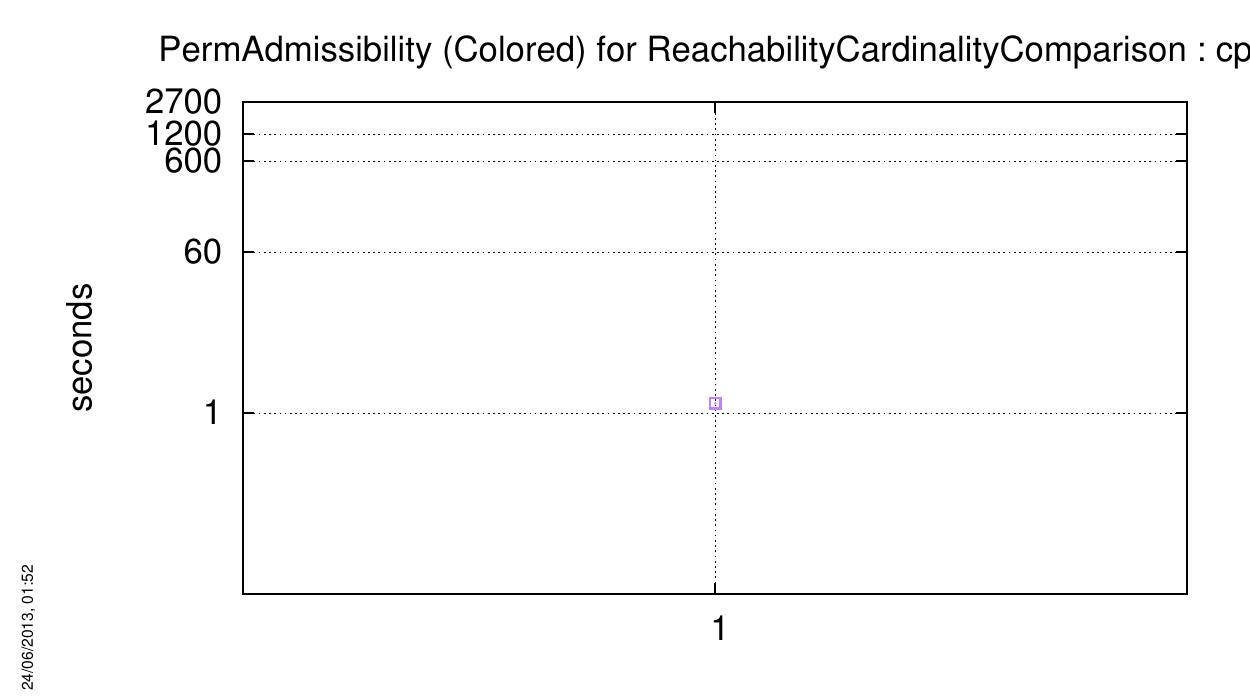}

   \includegraphics[height=1cm]{figures/tools-legend.pdf}
\end{center}

\subsubsection{\acs{PermAdmissibility-PT}}
The charts below respectively show how tools compete with this ``Known'' model (memory and CPU).

\index{Performances!ReachabilityCardinalityComparison!PermAdmissibility (P/T)}
\begin{center}
   \includegraphics[width=7.2cm]{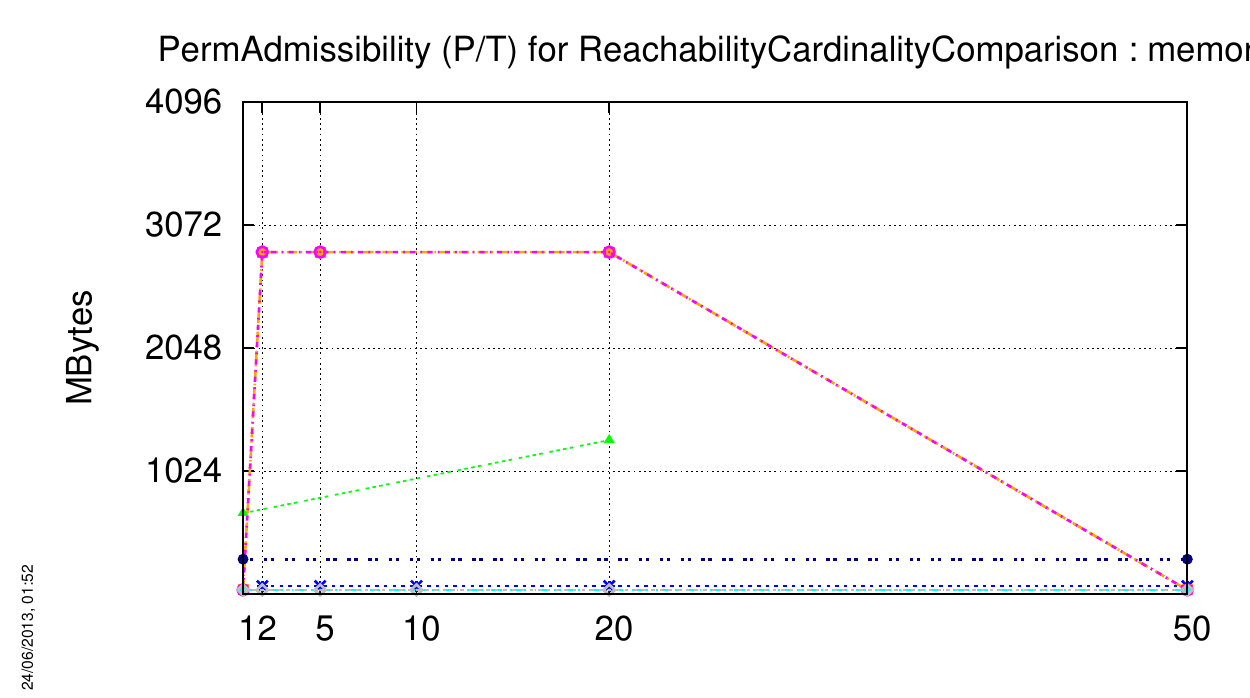}
   \includegraphics[width=7.2cm]{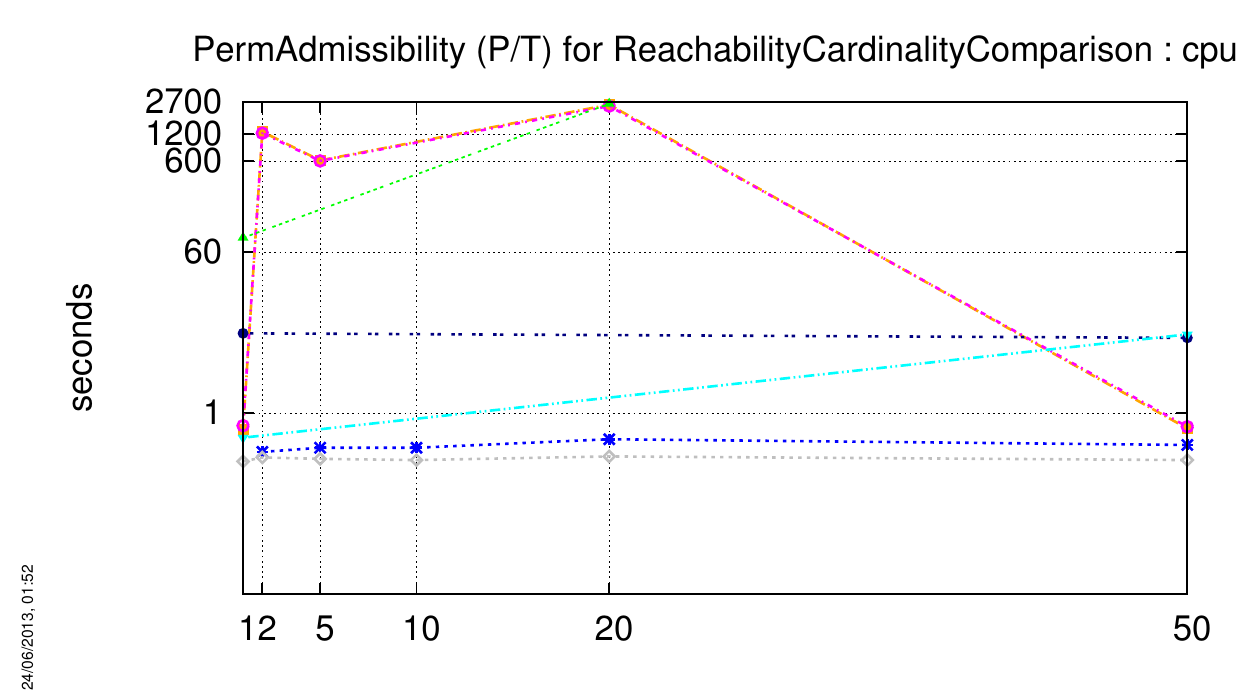}

   \includegraphics[height=1cm]{figures/tools-legend.pdf}
\end{center}

\subsubsection{\acs{Peterson-COL}}
The charts below respectively show how tools compete with this ``Known'' model (memory and CPU).

\index{Performances!ReachabilityCardinalityComparison!Peterson (Colored)}
\begin{center}
   \includegraphics[width=7.2cm]{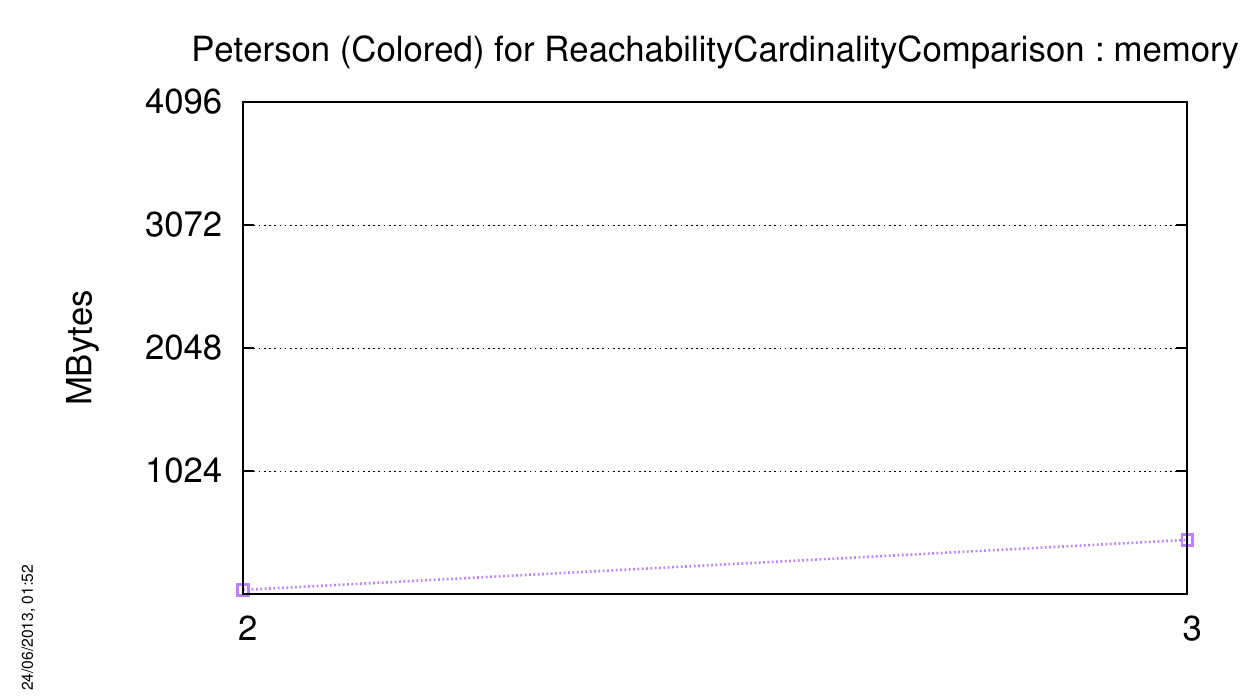}
   \includegraphics[width=7.2cm]{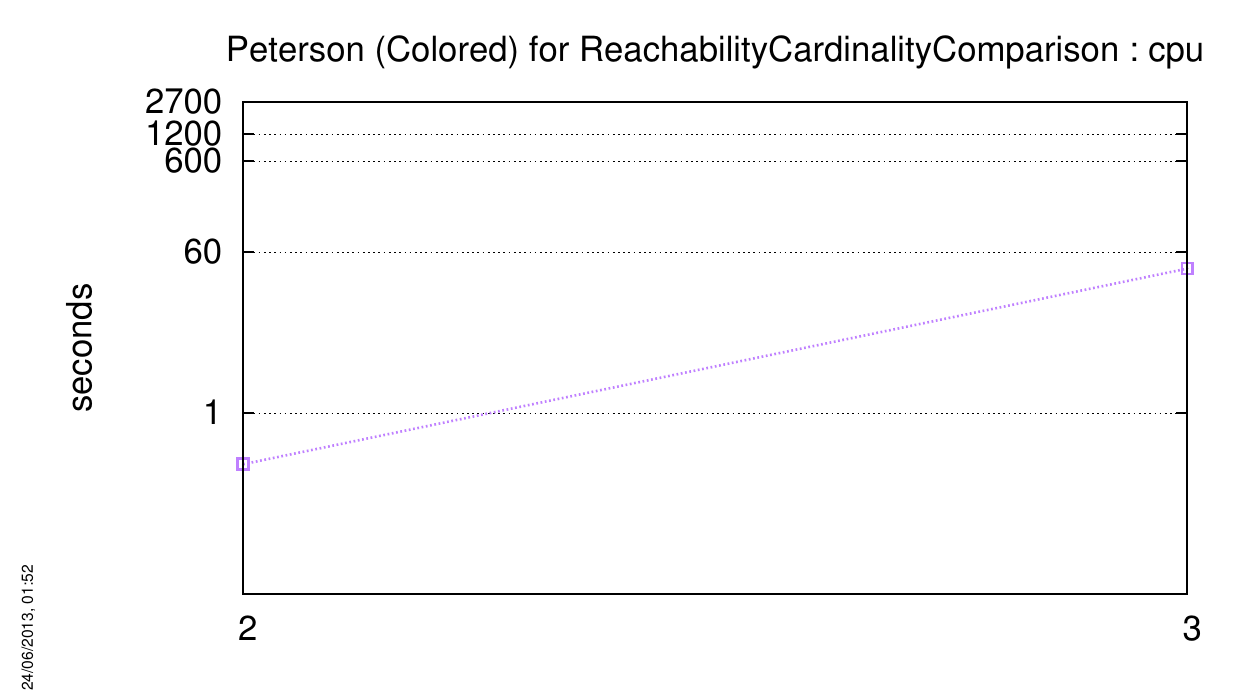}

   \includegraphics[height=1cm]{figures/tools-legend.pdf}
\end{center}

\subsubsection{\acs{Peterson-PT}}
The charts below respectively show how tools compete with this ``Known'' model (memory and CPU).

\index{Performances!ReachabilityCardinalityComparison!Peterson (P/T)}
\begin{center}
   \includegraphics[width=7.2cm]{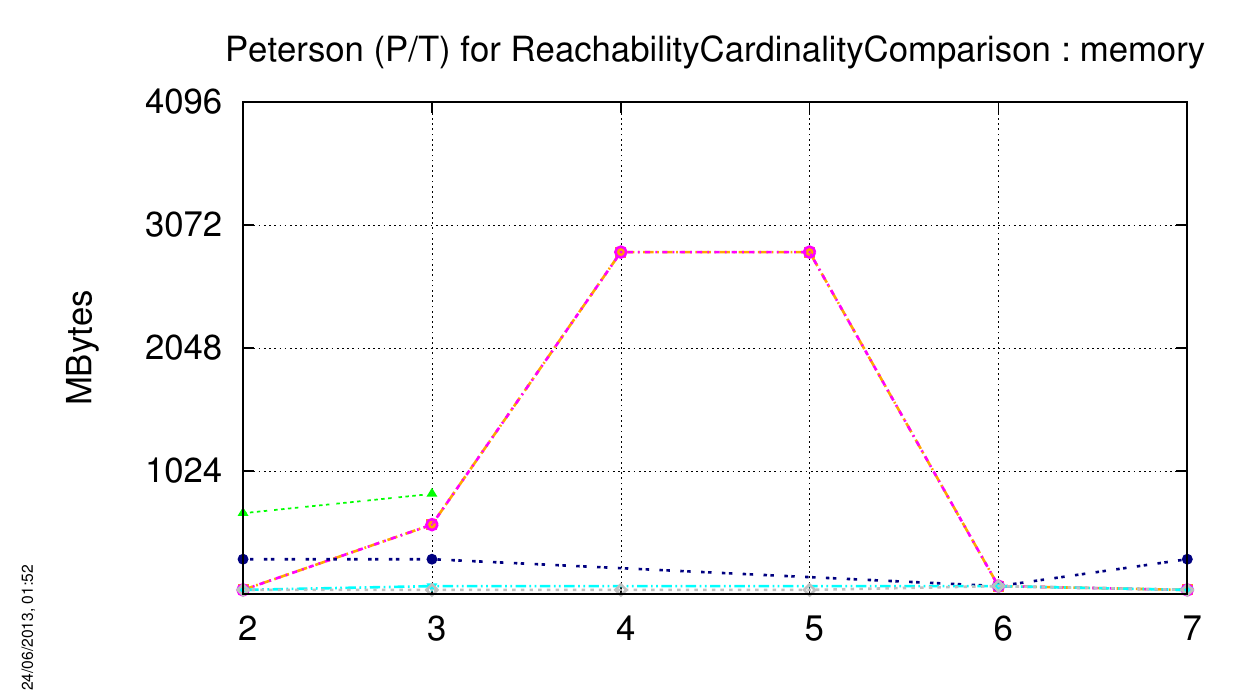}
   \includegraphics[width=7.2cm]{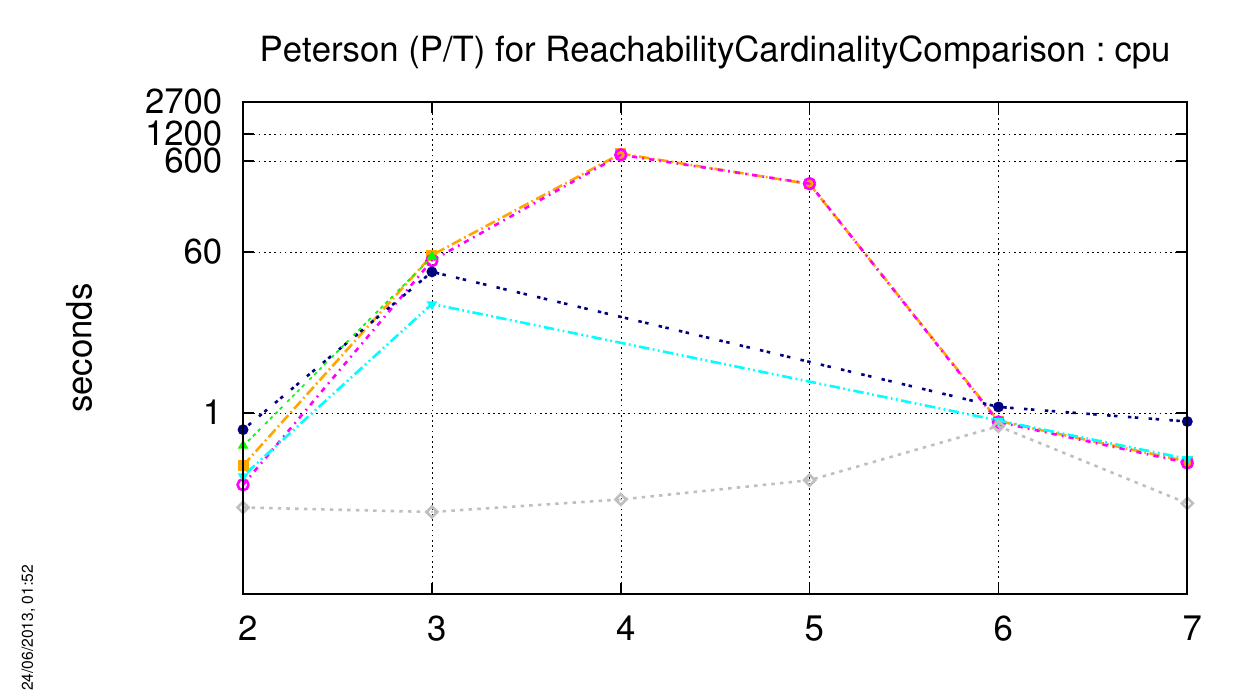}

   \includegraphics[height=1cm]{figures/tools-legend.pdf}
\end{center}

\subsubsection{\acs{Philosophers-COL}}
No instance of this model could be computed for the \textbf{ReachabilityCardinalityComparison} examination.

\subsubsection{\acs{Philosophers-PT}}
The charts below respectively show how tools compete with this ``Known'' model (memory and CPU).

\index{Performances!ReachabilityCardinalityComparison!Philosophers (P/T)}
\begin{center}
   \includegraphics[width=7.2cm]{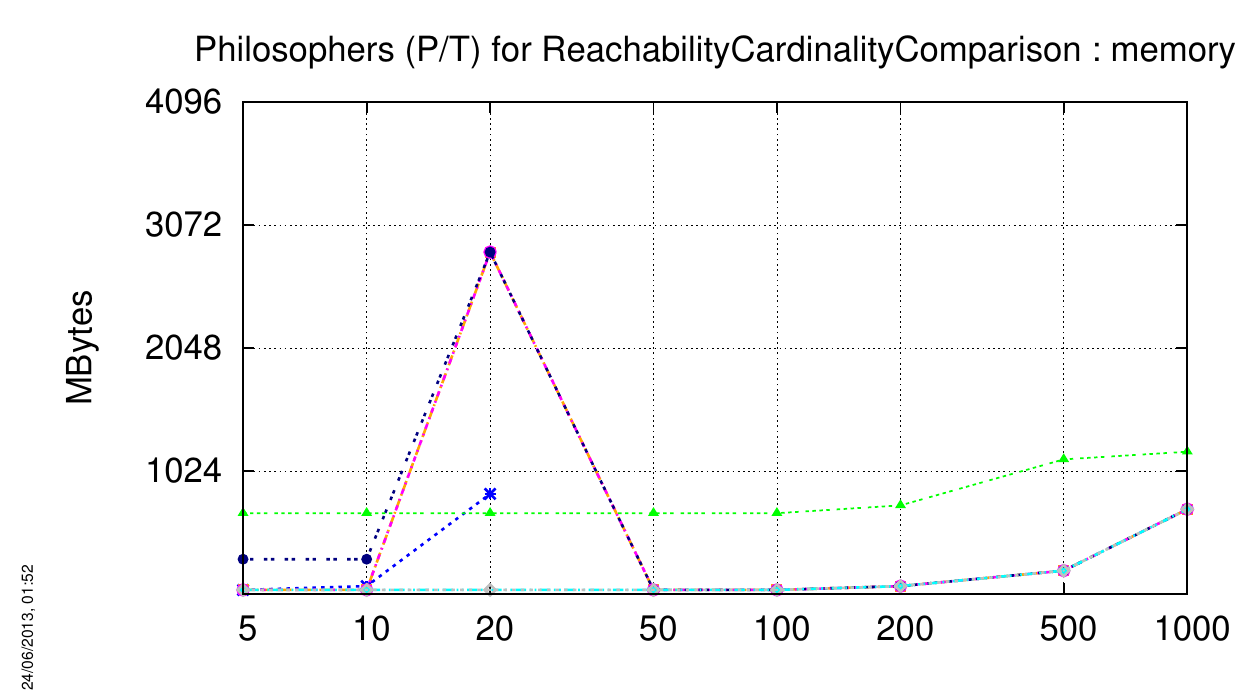}
   \includegraphics[width=7.2cm]{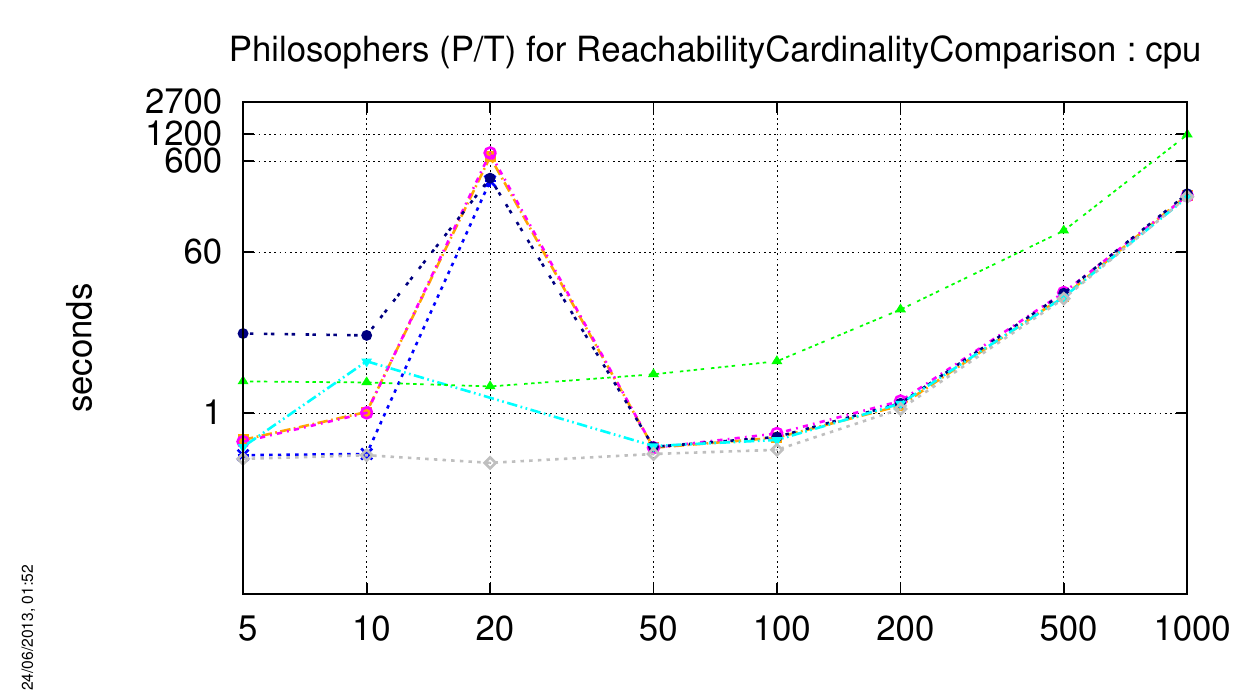}

   \includegraphics[height=1cm]{figures/tools-legend.pdf}
\end{center}

\subsubsection{\acs{PhilosophersDyn-COL}}
No instance of this model could be computed for the \textbf{ReachabilityCardinalityComparison} examination.

\subsubsection{\acs{PhilosophersDyn-PT}}
The charts below respectively show how tools compete with this ``Known'' model (memory and CPU).

\index{Performances!ReachabilityCardinalityComparison!PhilosophersDyn (P/T)}
\begin{center}
   \includegraphics[width=7.2cm]{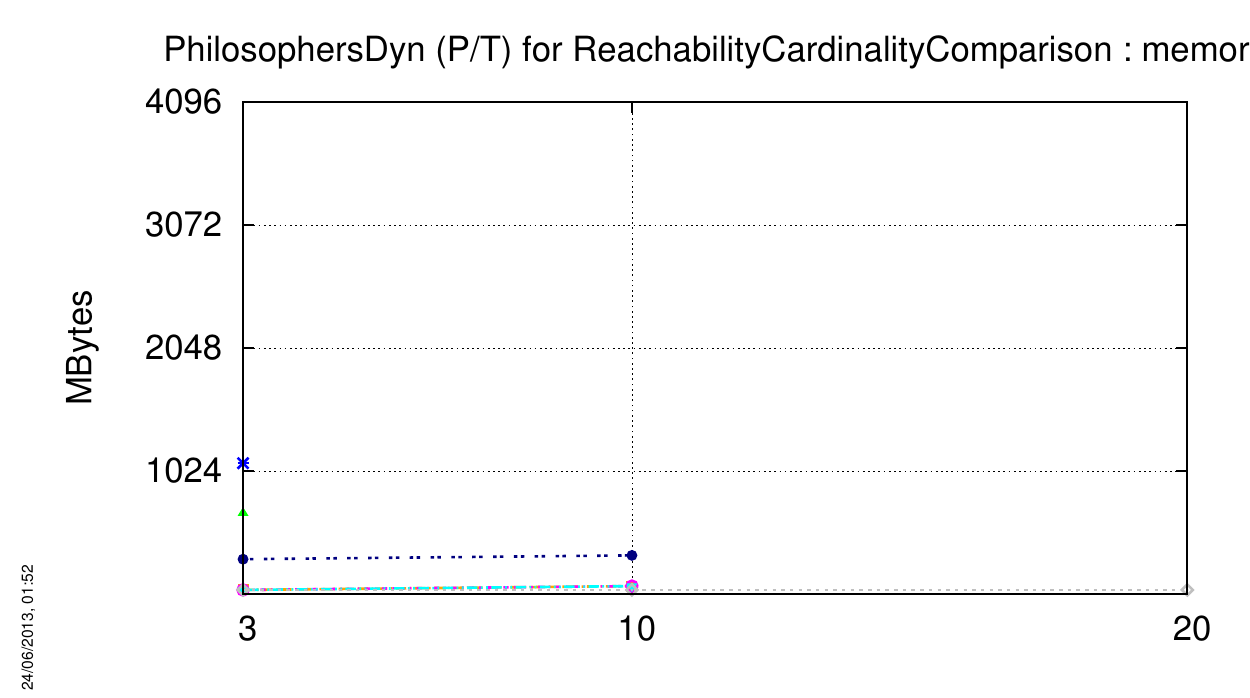}
   \includegraphics[width=7.2cm]{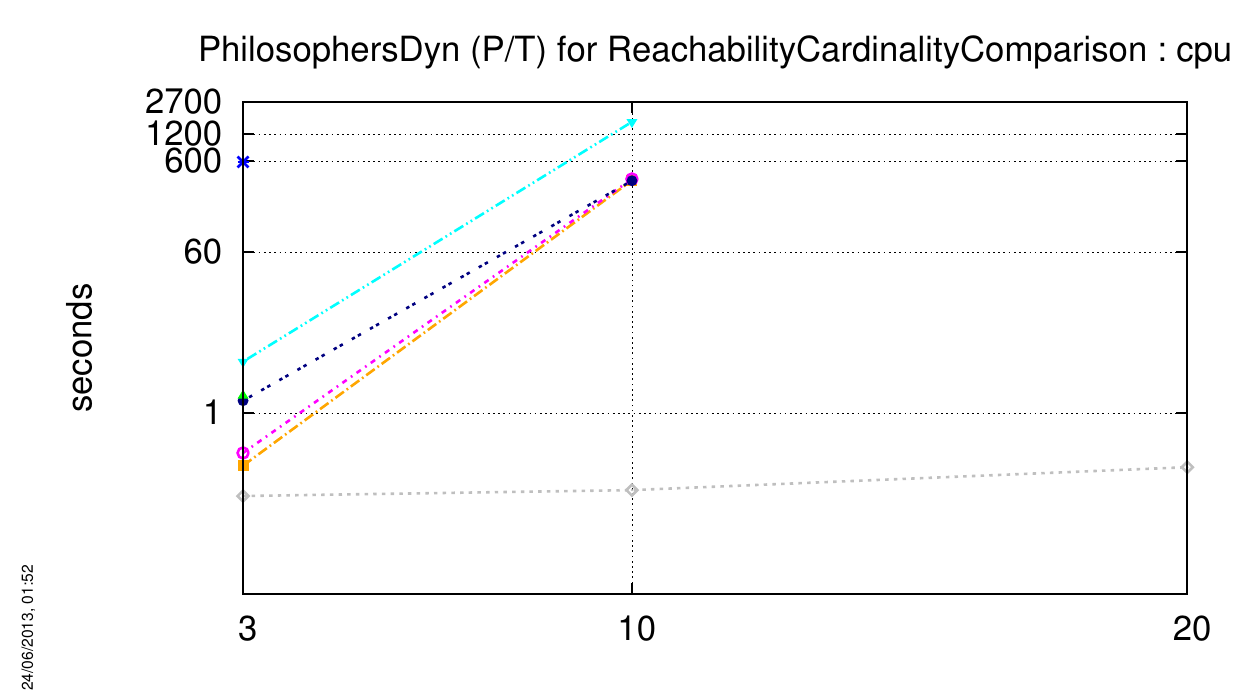}

   \includegraphics[height=1cm]{figures/tools-legend.pdf}
\end{center}

\subsubsection{\acs{Planning-PT}}
No instance of this model could be computed for the \textbf{ReachabilityCardinalityComparison} examination.

\subsubsection{\acs{Railroad-PT}}
The charts below respectively show how tools compete with this ``Known'' model (memory and CPU).

\index{Performances!ReachabilityCardinalityComparison!Railroad (P/T)}
\begin{center}
   \includegraphics[width=7.2cm]{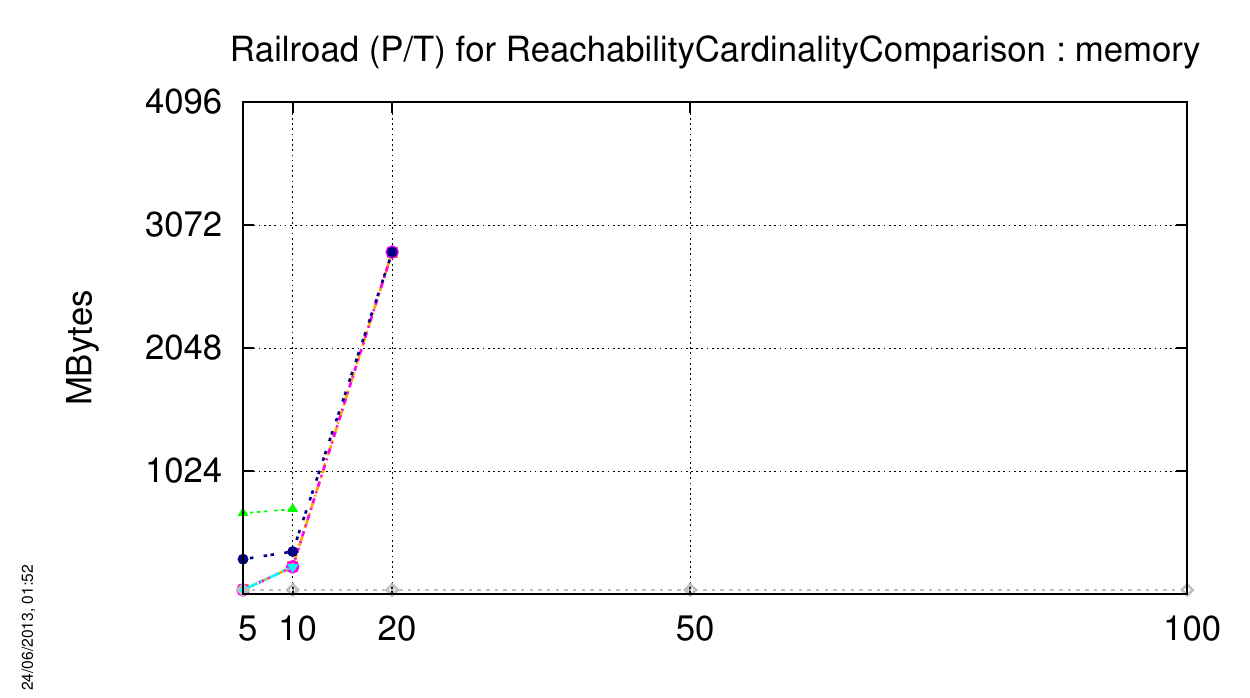}
   \includegraphics[width=7.2cm]{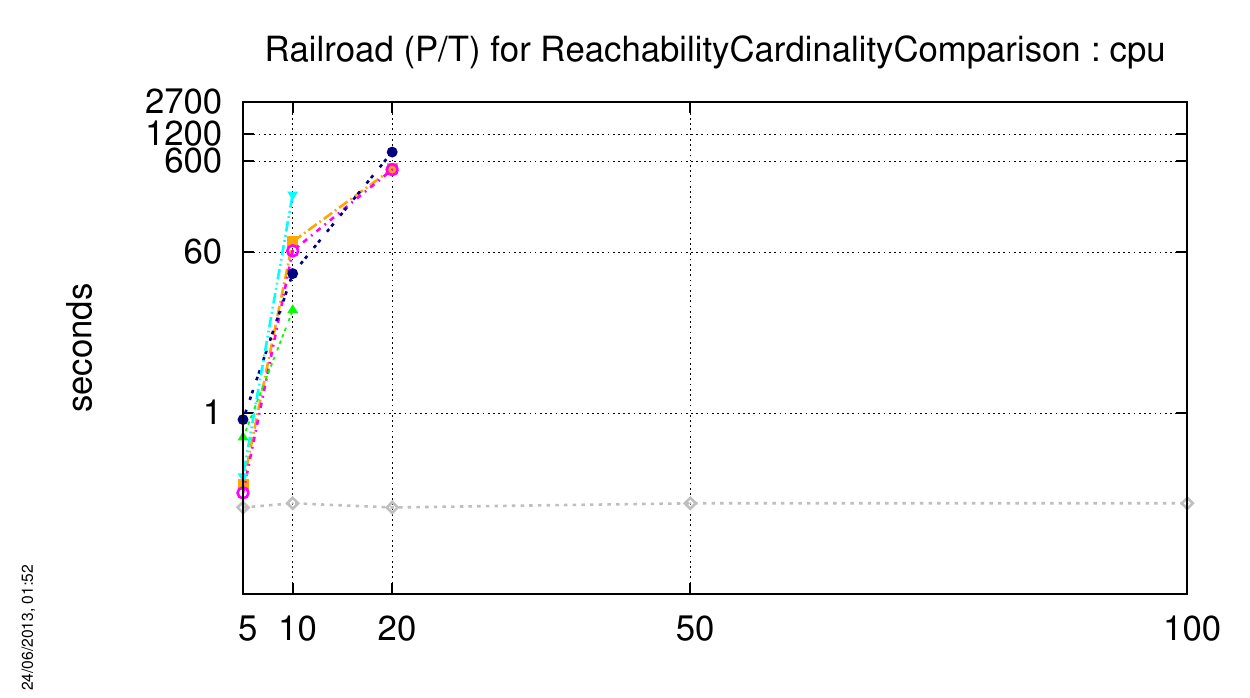}

   \includegraphics[height=1cm]{figures/tools-legend.pdf}
\end{center}

\subsubsection{\acs{RessAllocation-PT}}
The charts below respectively show how tools compete with this ``Known'' model (memory and CPU).

\index{Performances!ReachabilityCardinalityComparison!RessAllocation (P/T)}
\begin{center}
   \includegraphics[width=7.2cm]{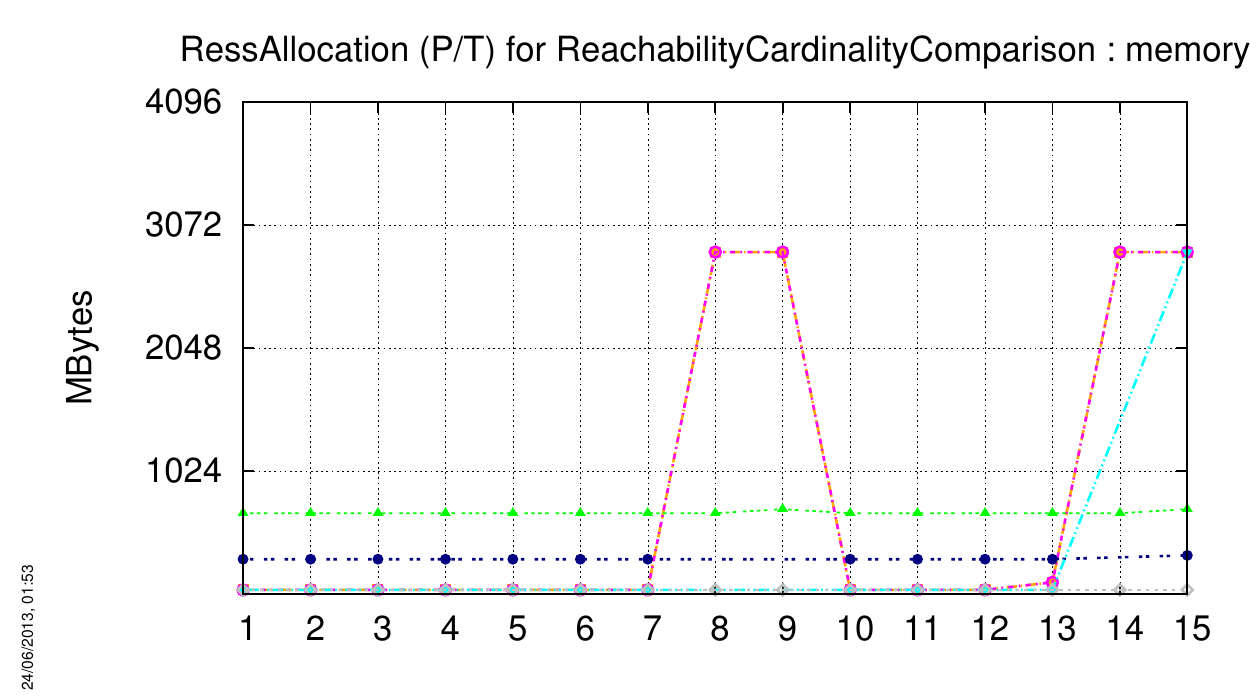}
   \includegraphics[width=7.2cm]{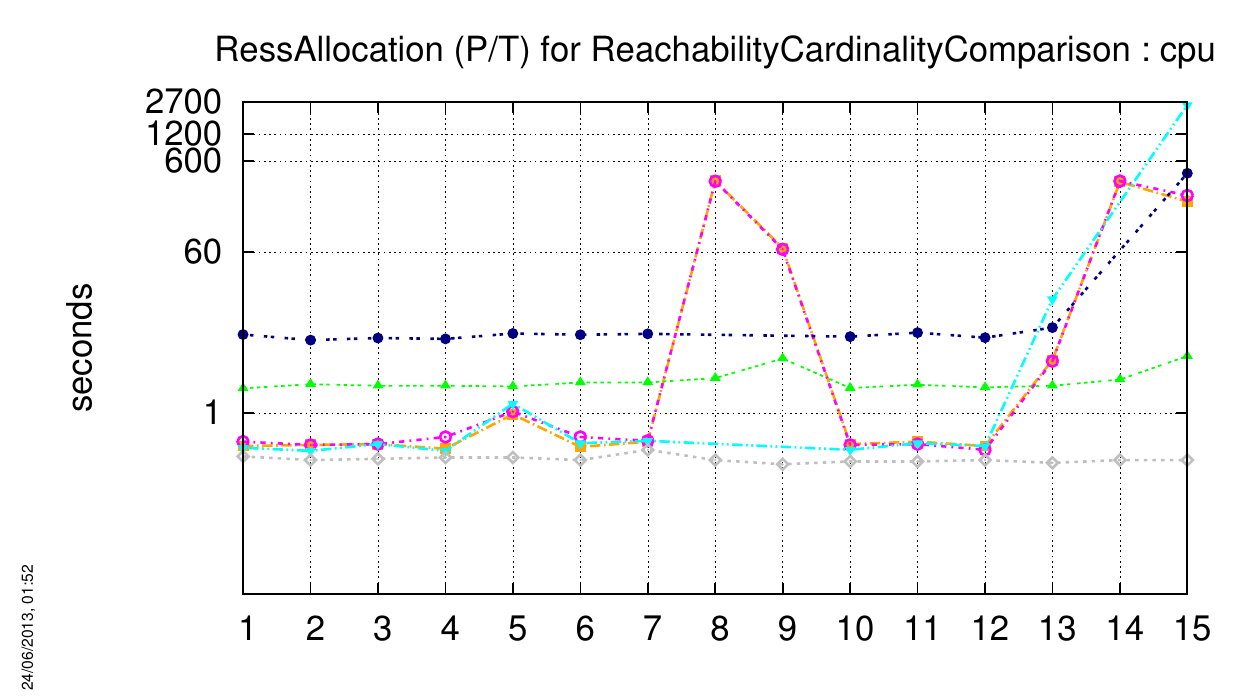}

   \includegraphics[height=1cm]{figures/tools-legend.pdf}
\end{center}

\subsubsection{\acs{Ring-PT}}
The charts below respectively show how tools compete with this ``Known'' model (memory and CPU).

\index{Performances!ReachabilityCardinalityComparison!Ring (P/T)}
\begin{center}
   \includegraphics[width=7.2cm]{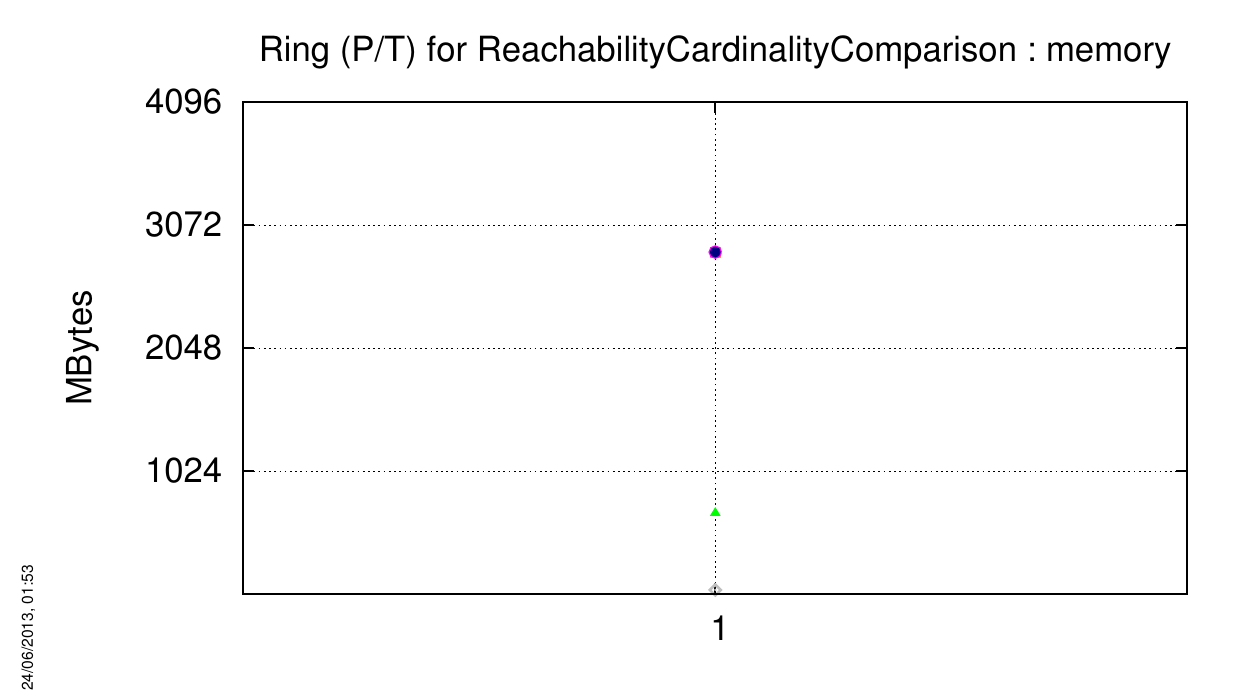}
   \includegraphics[width=7.2cm]{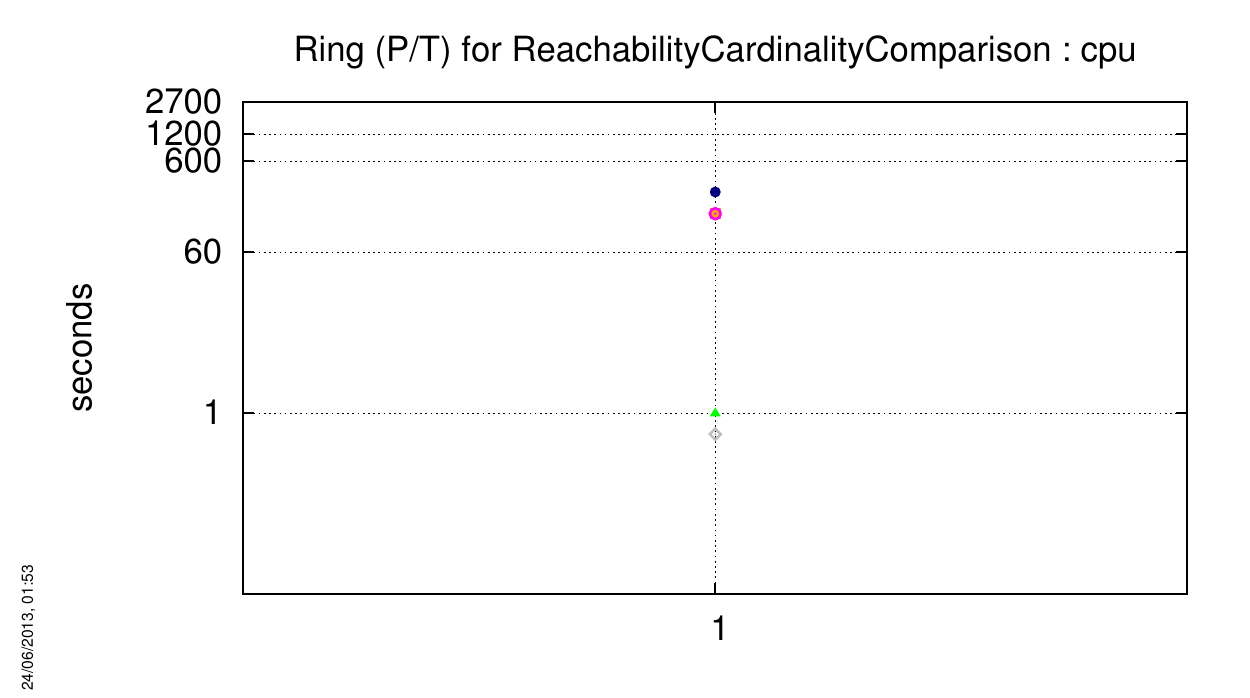}

   \includegraphics[height=1cm]{figures/tools-legend.pdf}
\end{center}

\subsubsection{\acs{RwMutex-PT}}
The charts below respectively show how tools compete with this ``Known'' model (memory and CPU).

\index{Performances!ReachabilityCardinalityComparison!RwMutex (P/T)}
\begin{center}
   \includegraphics[width=7.2cm]{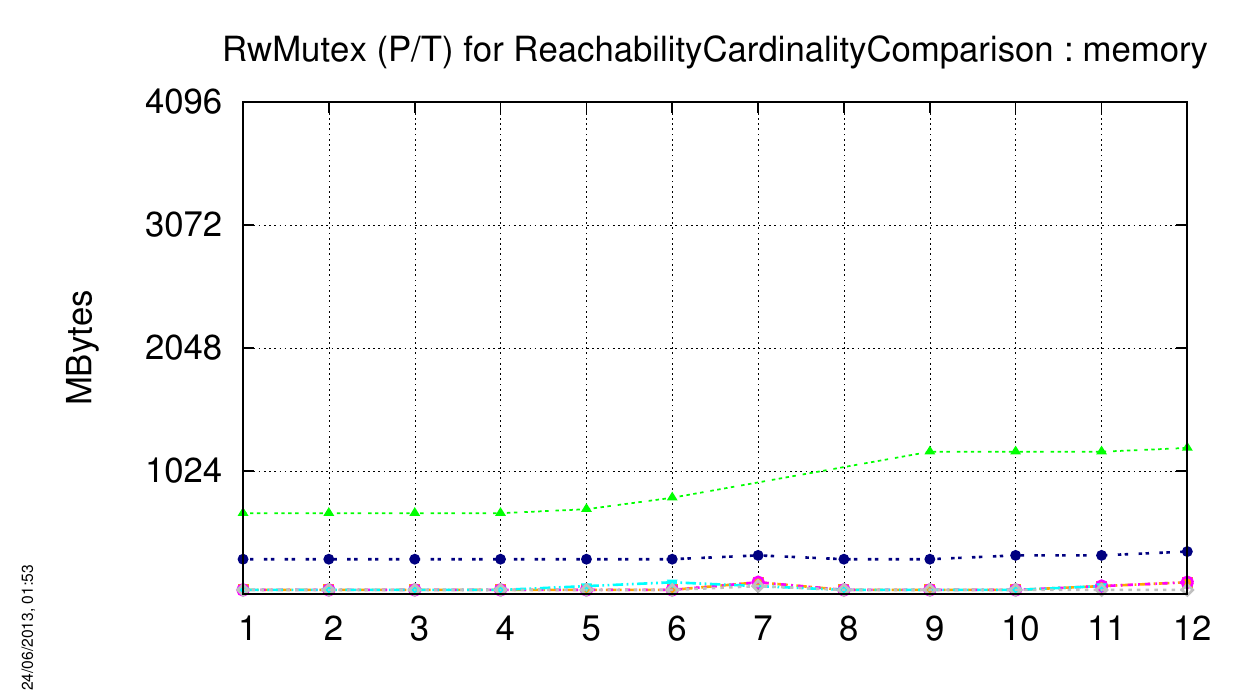}
   \includegraphics[width=7.2cm]{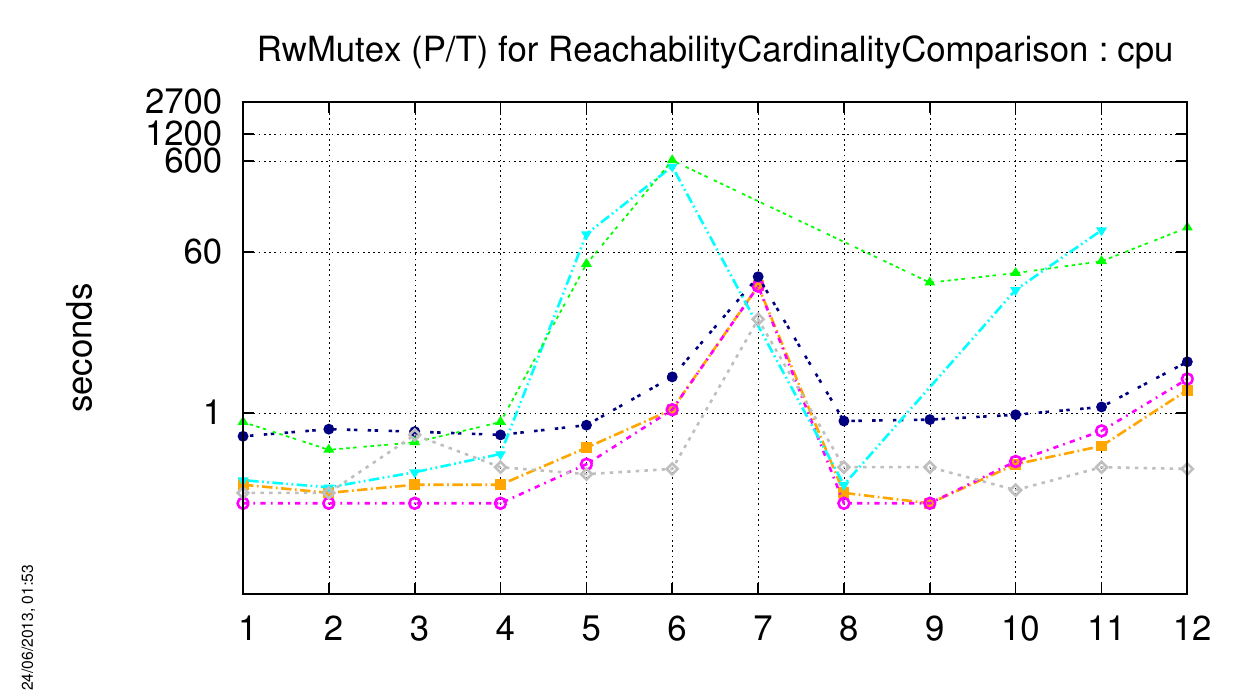}

   \includegraphics[height=1cm]{figures/tools-legend.pdf}
\end{center}

\subsubsection{\acs{SharedMemory-COL}}
No instance of this model could be computed for the \textbf{ReachabilityCardinalityComparison} examination.

\subsubsection{\acs{SharedMemory-PT}}
The charts below respectively show how tools compete with this ``Known'' model (memory and CPU).

\index{Performances!ReachabilityCardinalityComparison!SharedMemory (P/T)}
\begin{center}
   \includegraphics[width=7.2cm]{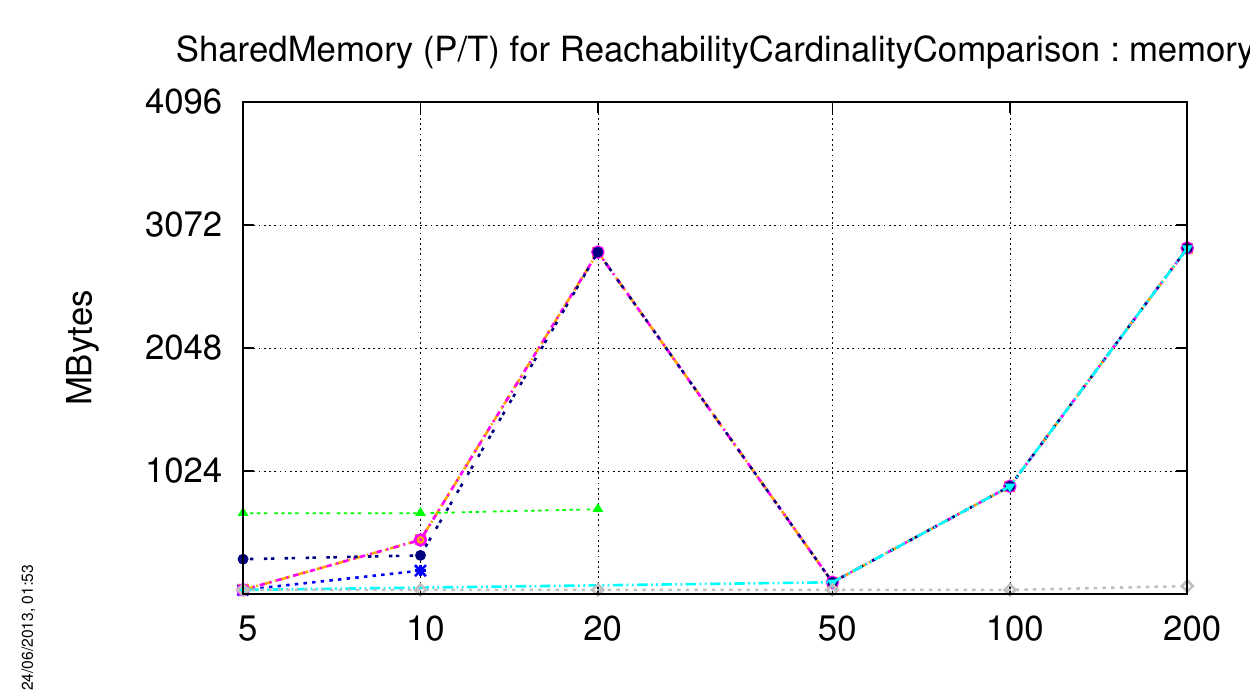}
   \includegraphics[width=7.2cm]{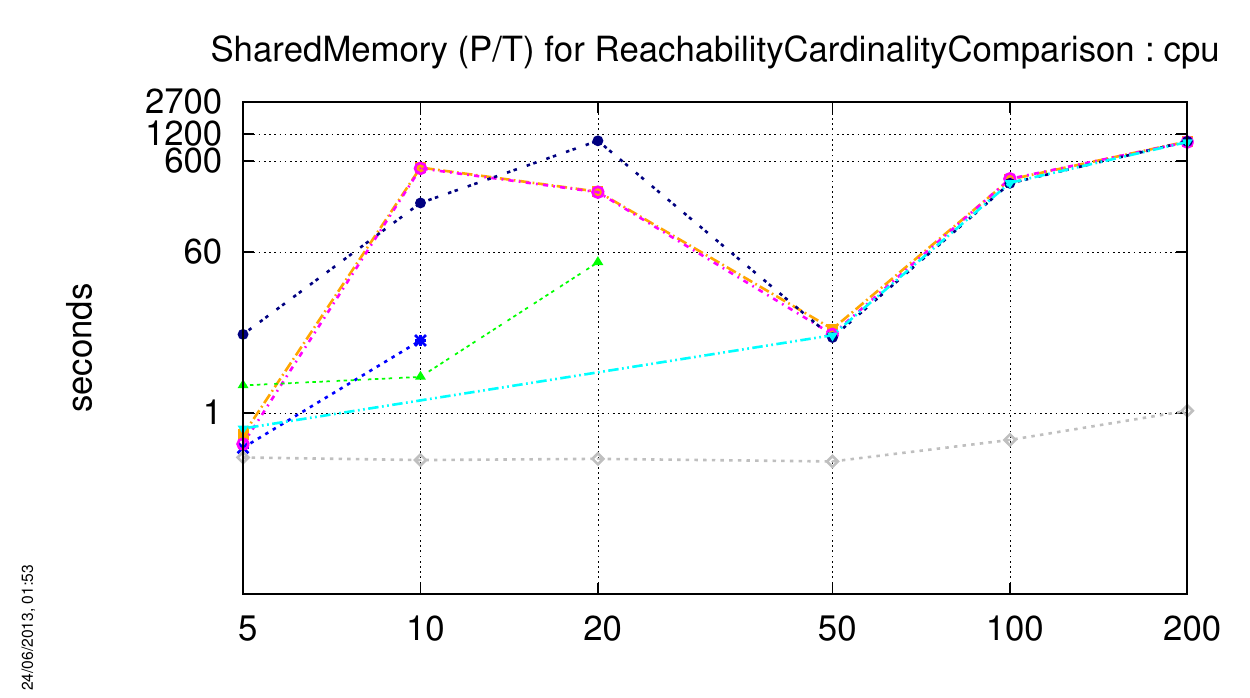}

   \includegraphics[height=1cm]{figures/tools-legend.pdf}
\end{center}

\subsubsection{\acs{SimpleLoadBal-COL}}
The charts below respectively show how tools compete with this ``Known'' model (memory and CPU).

\index{Performances!ReachabilityCardinalityComparison!SimpleLoadBal (Colored)}
\begin{center}
   \includegraphics[width=7.2cm]{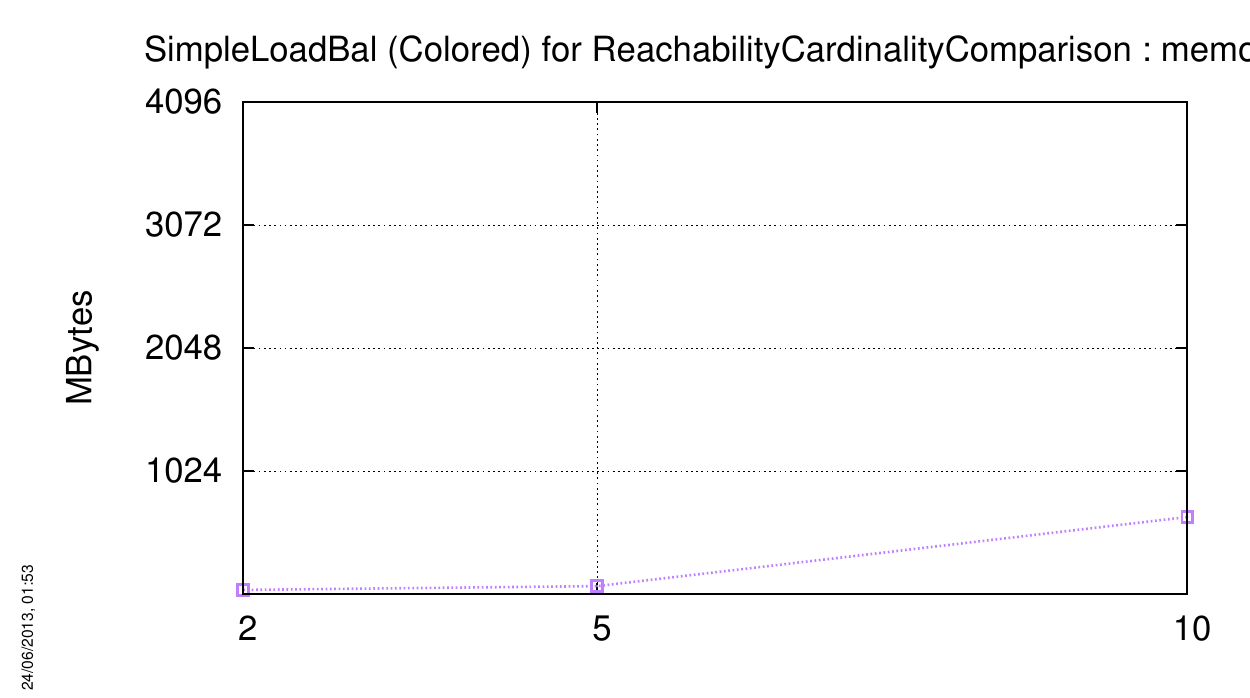}
   \includegraphics[width=7.2cm]{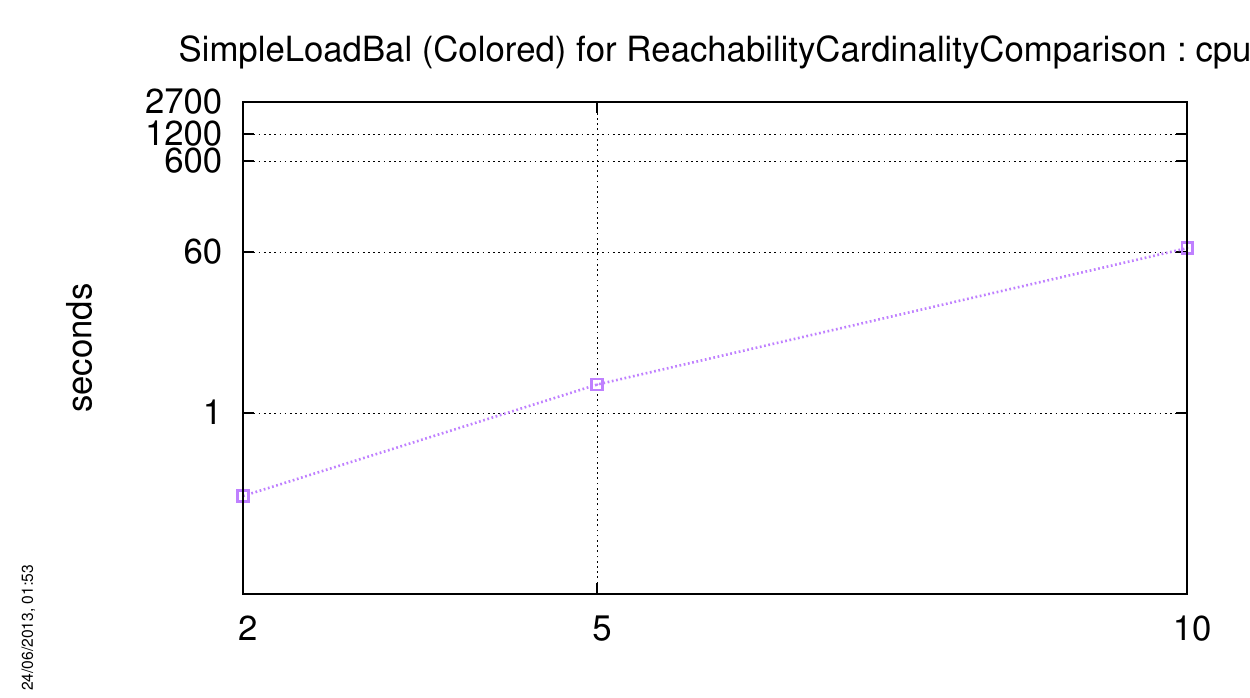}

   \includegraphics[height=1cm]{figures/tools-legend.pdf}
\end{center}

\subsubsection{\acs{SimpleLoadBal-PT}}
The charts below respectively show how tools compete with this ``Known'' model (memory and CPU).

\index{Performances!ReachabilityCardinalityComparison!SimpleLoadBal (P/T)}
\begin{center}
   \includegraphics[width=7.2cm]{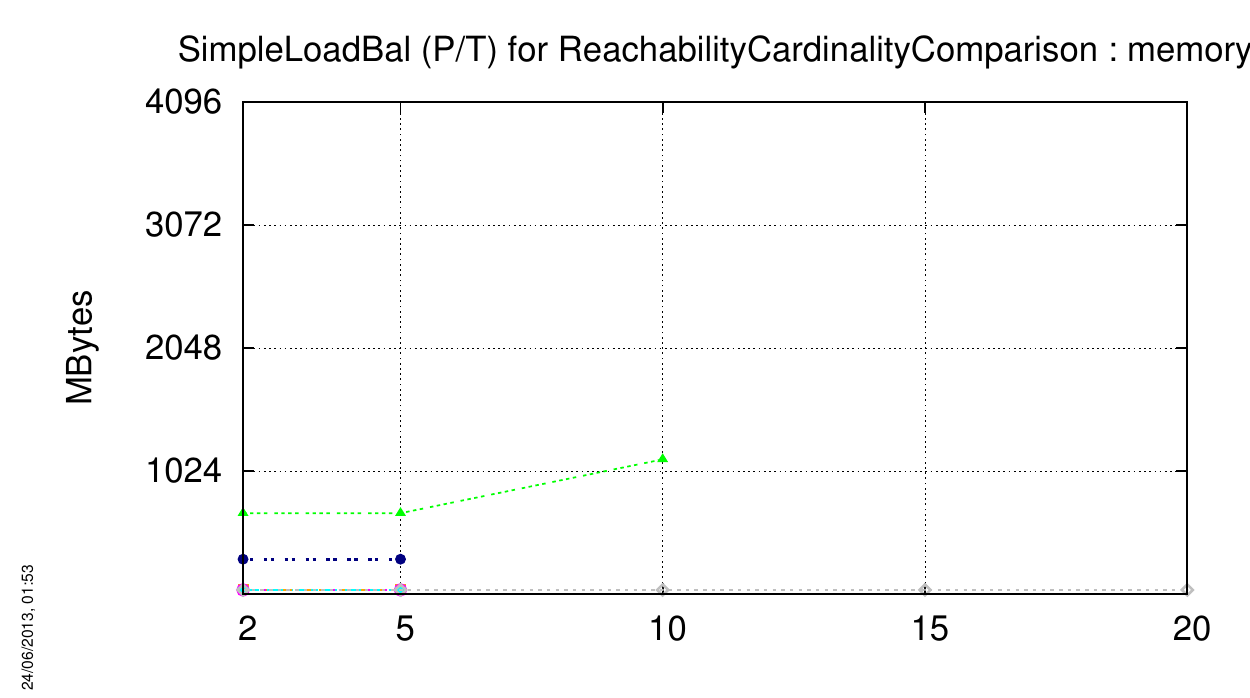}
   \includegraphics[width=7.2cm]{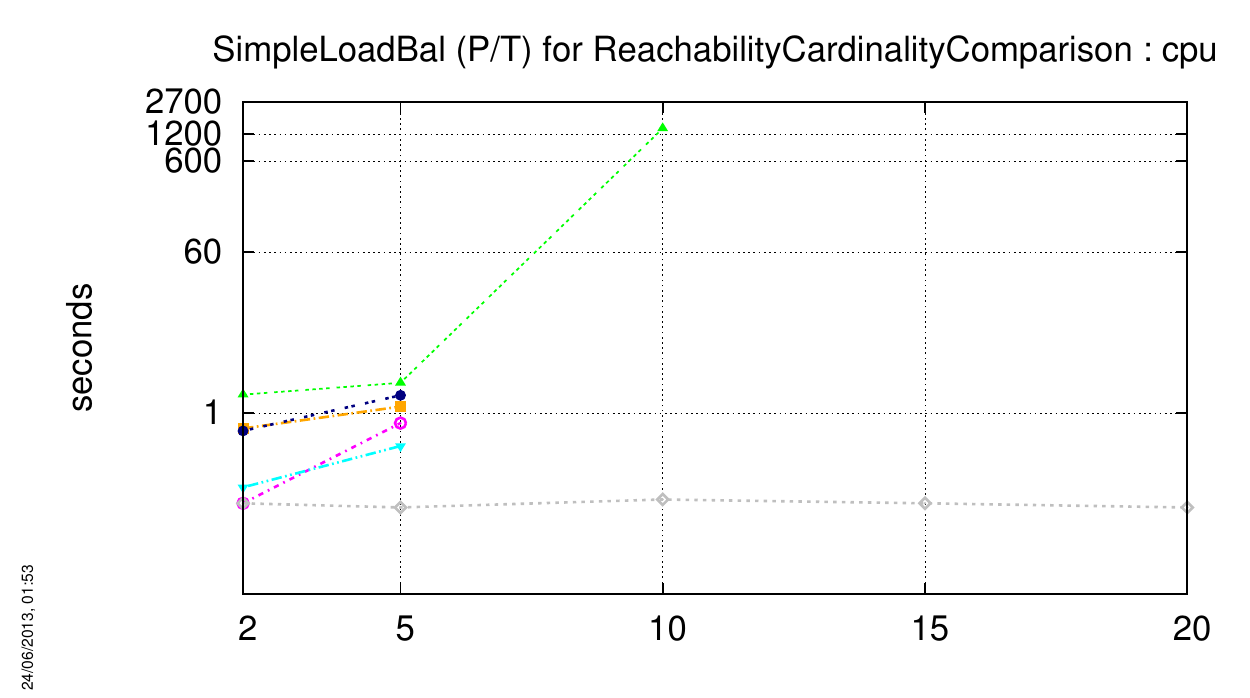}

   \includegraphics[height=1cm]{figures/tools-legend.pdf}
\end{center}

\subsubsection{\acs{TokenRing-COL}}
No instance of this model could be computed for the \textbf{ReachabilityCardinalityComparison} examination.

\subsubsection{\acs{TokenRing-PT}}
No instance of this model could be computed for the \textbf{ReachabilityCardinalityComparison} examination.

\subsubsection{\acs{HouseConstruction-PT}}
The charts below respectively show how tools compete with this ``Suprise'' model (memory and CPU).

\index{Performances!ReachabilityCardinalityComparison!HouseConstruction (P/T)}
\begin{center}
   \includegraphics[width=7.2cm]{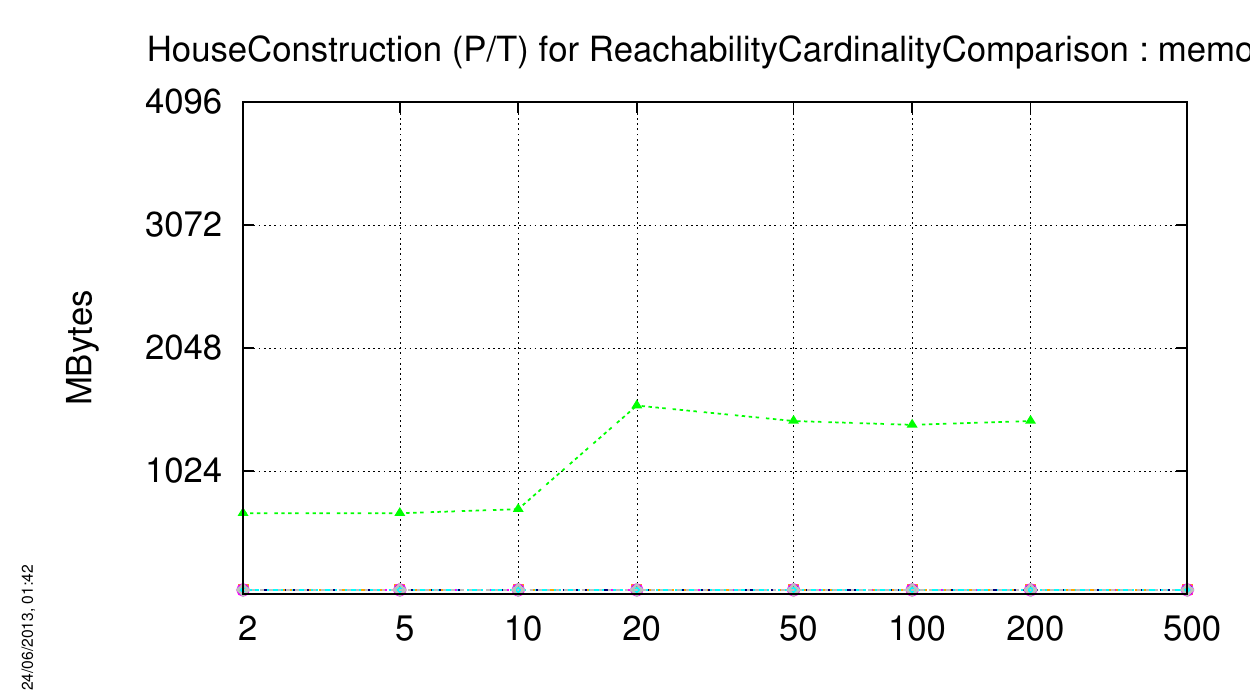}
   \includegraphics[width=7.2cm]{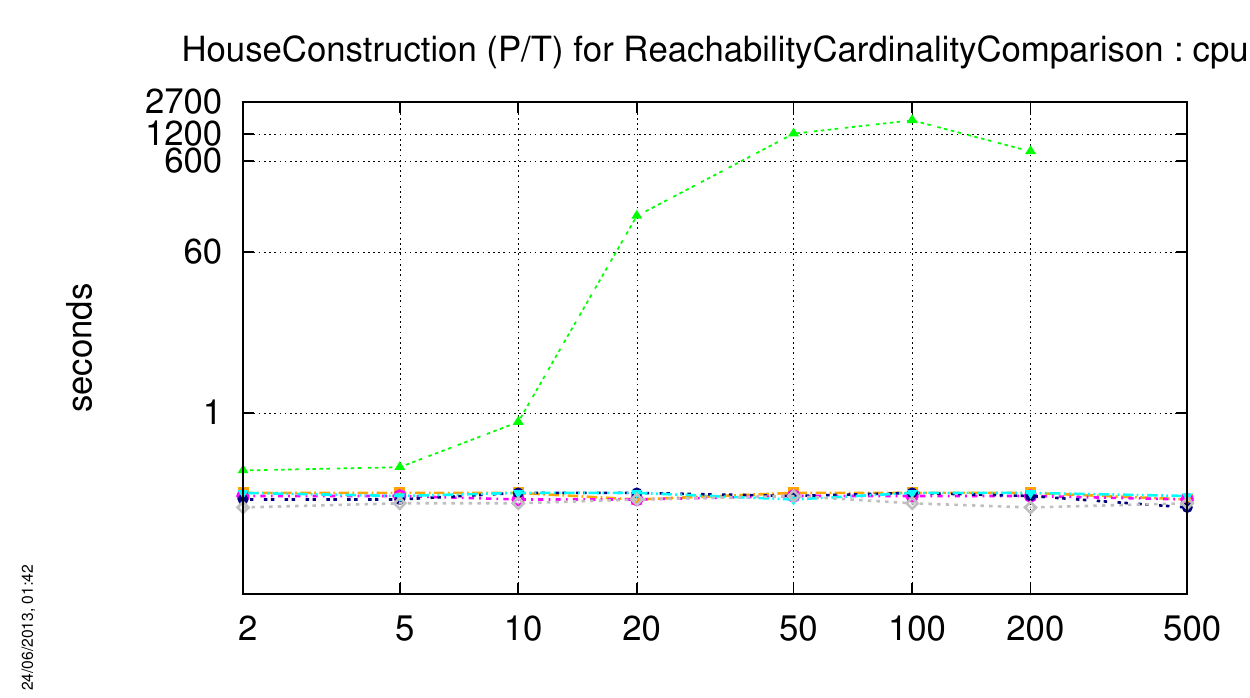}

   \includegraphics[height=1cm]{figures/tools-legend.pdf}
\end{center}

\subsubsection{\acs{IBMB2S565S3960-PT}}
The charts below respectively show how tools compete with this ``Suprise'' model (memory and CPU).

\index{Performances!ReachabilityCardinalityComparison!IBMB2S565S3960 (P/T)}
\begin{center}
   \includegraphics[width=7.2cm]{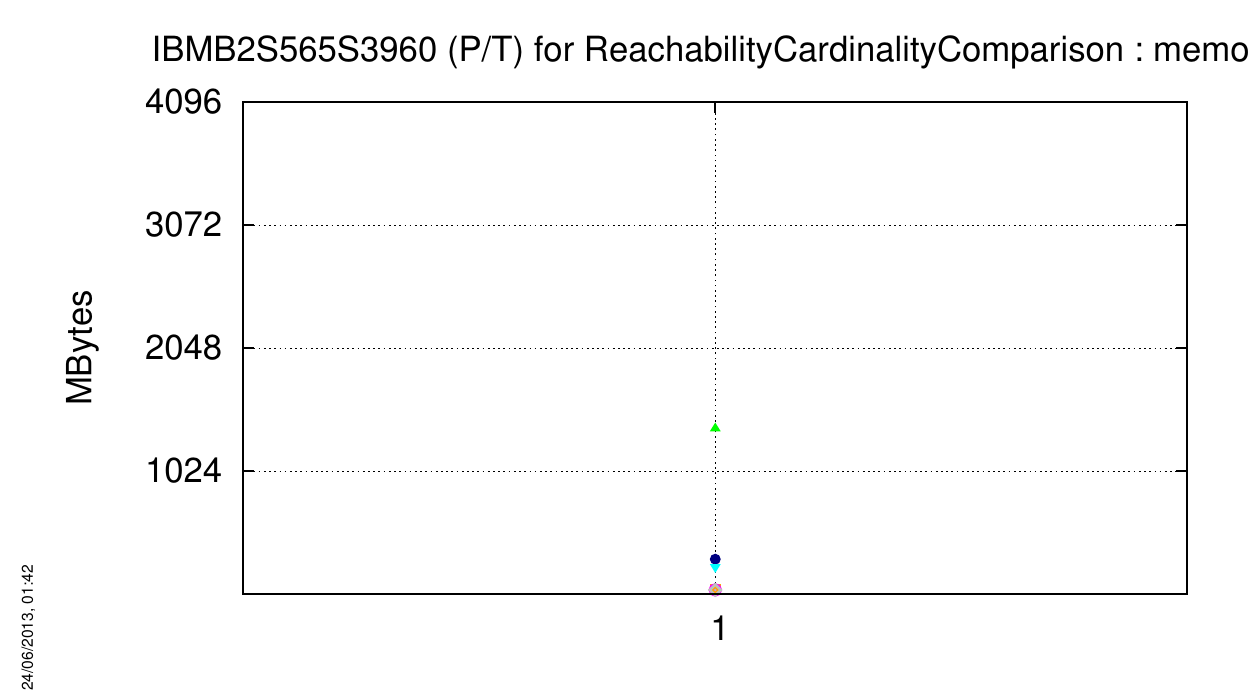}
   \includegraphics[width=7.2cm]{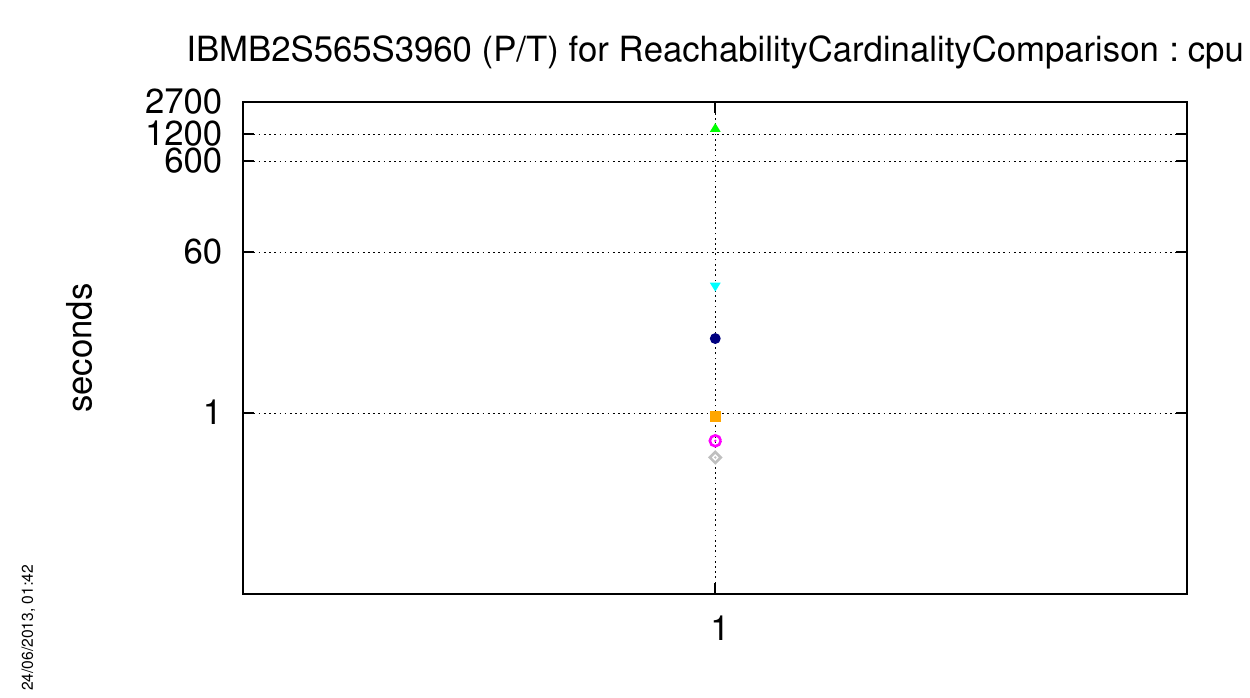}

   \includegraphics[height=1cm]{figures/tools-legend.pdf}
\end{center}

\subsubsection{\acs{QuasiCertifProtocol-COL}}
No instance of this model could be computed for the \textbf{ReachabilityCardinalityComparison} examination.

\subsubsection{\acs{QuasiCertifProtocol-PT}}
The charts below respectively show how tools compete with this ``Suprise'' model (memory and CPU).

\index{Performances!ReachabilityCardinalityComparison!QuasiCertifProtocol (P/T)}
\begin{center}
   \includegraphics[width=7.2cm]{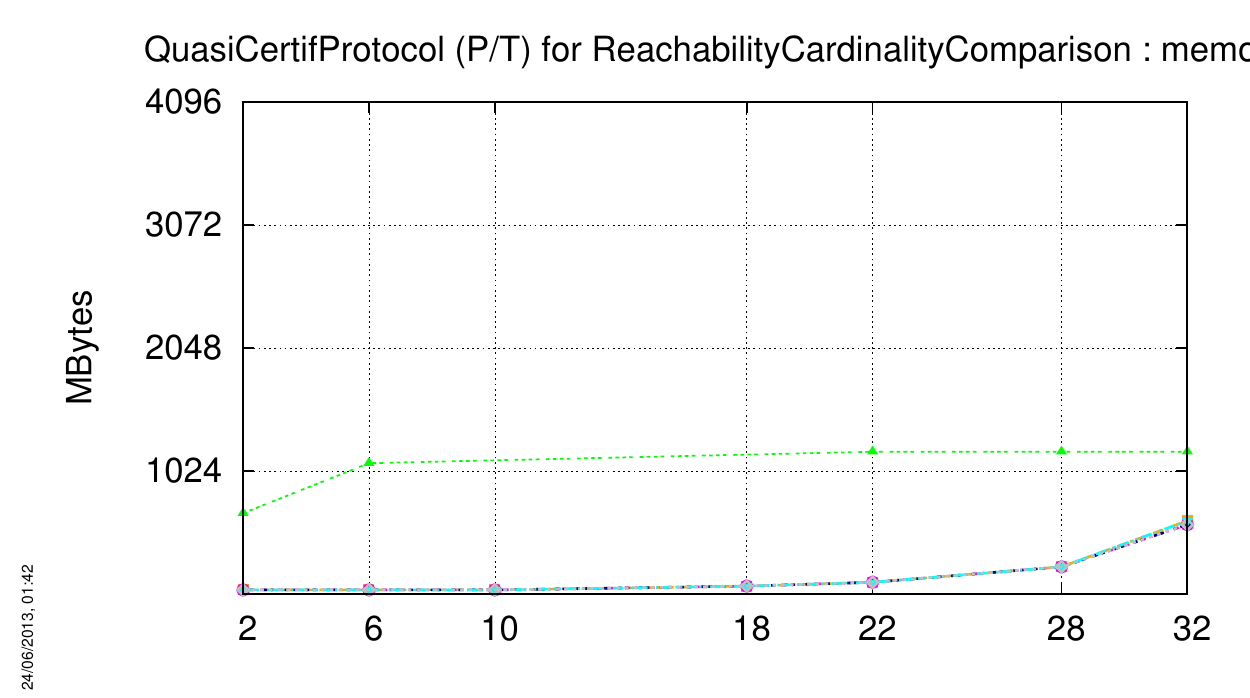}
   \includegraphics[width=7.2cm]{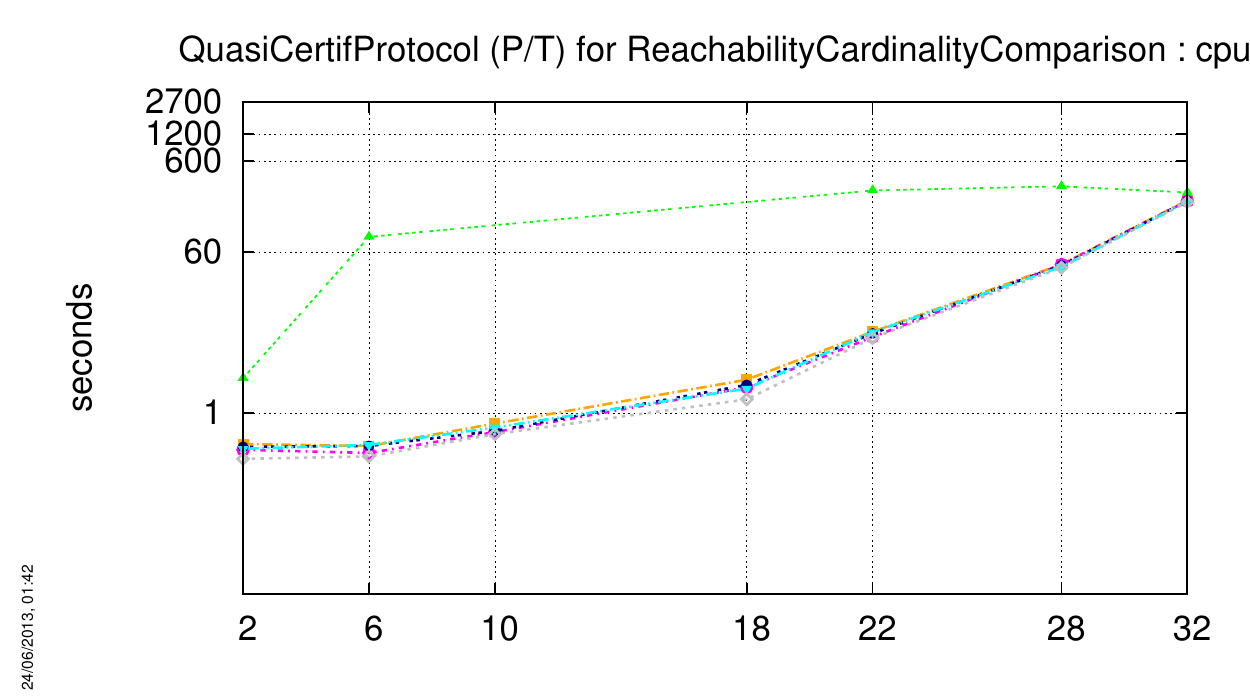}

   \includegraphics[height=1cm]{figures/tools-legend.pdf}
\end{center}

\subsubsection{\acs{Vasy2003-PT}}
The charts below respectively show how tools compete with this ``Suprise'' model (memory and CPU).

\index{Performances!ReachabilityCardinalityComparison!Vasy2003 (P/T)}
\begin{center}
   \includegraphics[width=7.2cm]{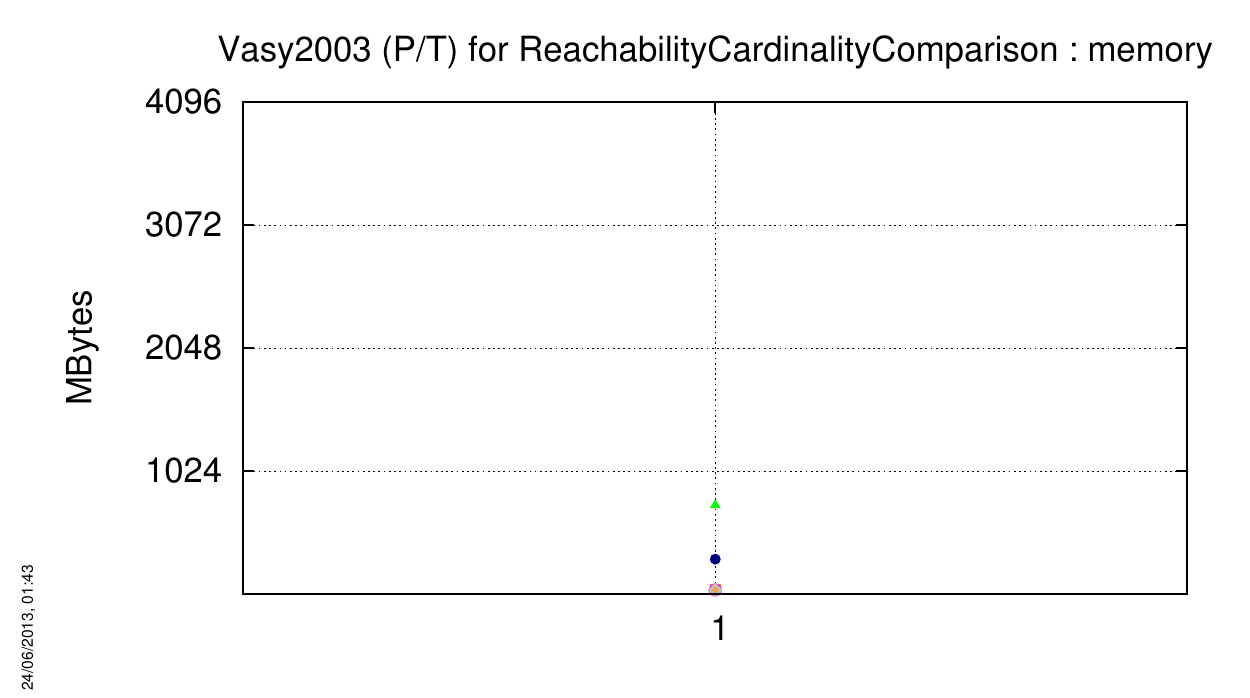}
   \includegraphics[width=7.2cm]{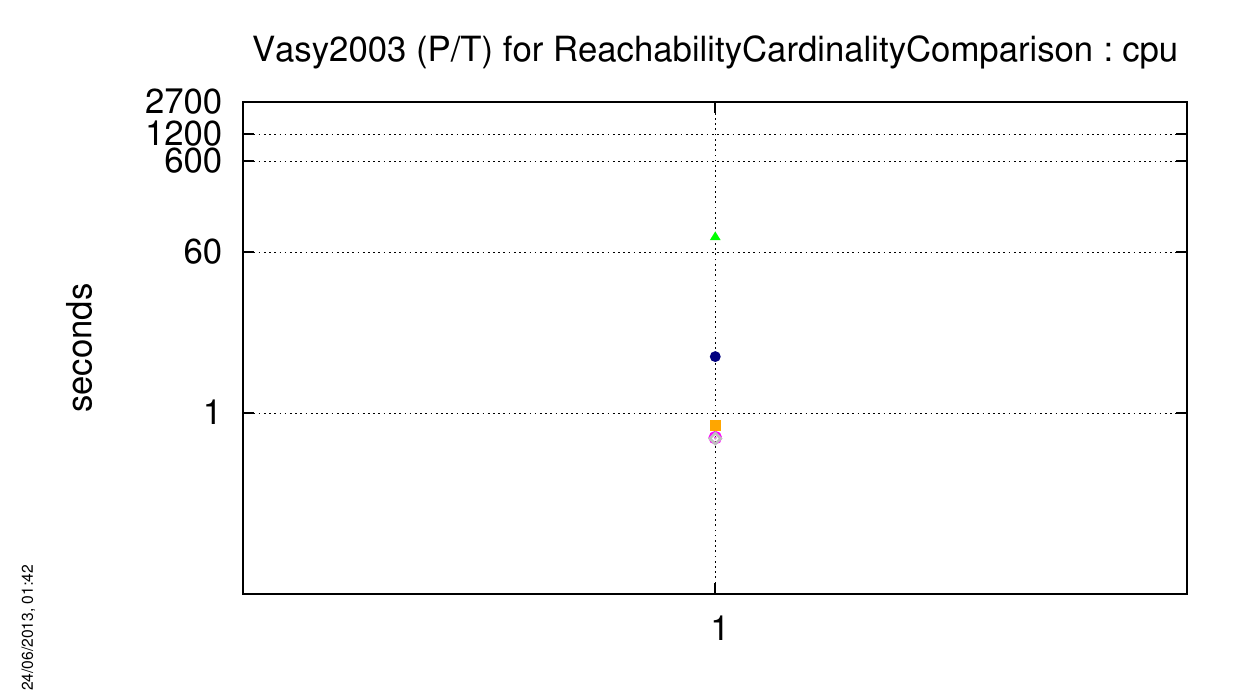}

   \includegraphics[height=1cm]{figures/tools-legend.pdf}
\end{center}

\subsection{Outputs for the ReachabilityCardinalityComparison Examination}
\index{Outputs!ReachabilityCardinalityComparison}

Please find enclosed the brute results for this examination (``Known'' and ``Surprise'' models).
We display only the score of tools that provide a results for at least one instance of one model.
The legend for the values is provided below:
\begin{itemize}
   \item\textbf{nc}: the tool does not compete this examination for this model/instance,
   \item\textbf{cc}: the tool cannot compute this examination for this model/instance,
   \item\textbf{to}: the tool cannot compute this examination for this model/instance within the maximum allowed time,
   \item\textbf{mp}: the tool encountered a memory problem (stack overflow or memory full),
   \item\textbf{nf}: there is no formula available for this type of examination (typically, this concerns P/T nets where
       comparing marking cardinality has no signification when there is no equivalent colored net).
\end{itemize}

\textbf{Note on the display of results for formulas:} each formula is considered as a flag (F if false, T if true, - or ?
when the value cannot be determined). These values are concatenated in the order they appear (we assume it is the order of formulas as they were provided).

\subsubsection{``Known'' Models}

Results are summarized in the table below.
{\footnotesize
\begin{longtable}{|ccccccccc|}
   \hline
      \multicolumn{9}{|c|}{\bf \acs{CSRepetitions-COL}} \\
      \textbf{Instances}\index{Outputs!ReachabilityCardinalityComparison!CSRepetitions (colored)} &
      \acs{greatspn} &
      \acs{its-tools} &
      \acs{lola} &
      \acs{lola-optimistic} &
      \acs{lola-optimistic-incomplete} &
      \acs{lola-pessimistic} &
      \acs{marcie} &
      \acs{sara} \\
      02 &
      nc &
      \textcolor{red}{\tt \scriptsize FFFFFFTF} &
      nc &
      nc &
      nc &
      nc &
      nc &
      nc \\
      03 &
      nc &
      \textcolor{red}{\tt \scriptsize FTFTFFFF} &
      nc &
      nc &
      nc &
      nc &
      nc &
      nc \\
      04 &
      nc &
      \textcolor{red}{\tt \scriptsize mp} &
      nc &
      nc &
      nc &
      nc &
      nc &
      nc \\
      05 &
      nc &
      \textcolor{red}{\tt \scriptsize mp} &
      nc &
      nc &
      nc &
      nc &
      nc &
      nc \\
      07 &
      nc &
      \textcolor{red}{\tt \scriptsize mp} &
      nc &
      nc &
      nc &
      nc &
      nc &
      nc \\
      10 &
      nc &
      \textcolor{red}{\tt \scriptsize mp} &
      nc &
      nc &
      nc &
      nc &
      nc &
      nc \\
\hline
\end{longtable}
}

\subsubsection{``Surprise'' Models}

Results are summarized in the table below.
{\footnotesize
\begin{longtable}{|ccccccc|}
   \hline
      \multicolumn{7}{|c|}{\bf \acs{HouseConstruction-PT}} \\
      instances\index{Outputs!ReachabilityCardinalityComparison!HouseConstruction (P/T)} &
      \acs{lola} &
      \acs{lola-optimistic} &
      \acs{lola-optimistic-incomplete} &
      \acs{lola-pessimistic} &
      \acs{marcie} &
      \acs{sara} \\
      002 &
      \textcolor{red}{\tt \scriptsize cc} &
      \textcolor{red}{\tt \scriptsize cc} &
      \textcolor{red}{\tt \scriptsize cc} &
      \textcolor{red}{\tt \scriptsize cc} &
      \textcolor{red}{\tt \scriptsize FFFTFFFF} &
      \textcolor{red}{\tt \scriptsize cc} \\
      005 &
      \textcolor{red}{\tt \scriptsize cc} &
      \textcolor{red}{\tt \scriptsize cc} &
      \textcolor{red}{\tt \scriptsize cc} &
      \textcolor{red}{\tt \scriptsize cc} &
      \textcolor{red}{\tt \scriptsize FFFTFFFF} &
      \textcolor{red}{\tt \scriptsize cc} \\
      010 &
      \textcolor{red}{\tt \scriptsize cc} &
      \textcolor{red}{\tt \scriptsize cc} &
      \textcolor{red}{\tt \scriptsize cc} &
      \textcolor{red}{\tt \scriptsize cc} &
      \textcolor{red}{\tt \scriptsize FFFFTFFF} &
      \textcolor{red}{\tt \scriptsize cc} \\
      020 &
      \textcolor{red}{\tt \scriptsize cc} &
      \textcolor{red}{\tt \scriptsize cc} &
      \textcolor{red}{\tt \scriptsize cc} &
      \textcolor{red}{\tt \scriptsize cc} &
      \textcolor{red}{\tt \scriptsize FFFTTFFF} &
      \textcolor{red}{\tt \scriptsize cc} \\
      050 &
      \textcolor{red}{\tt \scriptsize cc} &
      \textcolor{red}{\tt \scriptsize cc} &
      \textcolor{red}{\tt \scriptsize cc} &
      \textcolor{red}{\tt \scriptsize cc} &
      \textcolor{red}{\tt \scriptsize cc} &
      \textcolor{red}{\tt \scriptsize cc} \\
\hline
\end{longtable}
}

\subsection{Score for the ReachabilityCardinalityComparison Examination}
\index{Scores!ReachabilityCardinalityComparison}

Please find enclosed the scores for this examination (``Known'' and ``Surprise'' models).
We display only the score of tools that provide a results for at least one instance of one model.
The total is first listed in the table below followed by a detail, for each proposed model.
Meaning of the line labels are:
\begin{itemize}
\item\textbf{1st instance}: the tool gets a bonus for having processed the first instance of this model (+1 point),
\item\textbf{instances}: the tool gets 1 point per instances treated 
(for that, we assume that at least one formula has been successfully computed),
\item\textbf{max reached}: the tool could process all the instances for the model (+2 points),
\item\textbf{best}: the tool is among the ones that processed a maximum of instances within the time and memory confinement (+2 points).
\end{itemize}

\subsubsection{``Known'' Models}

Scores are summarized in the table below.
For each model in this category, a tool may collect up to 411 points.

{\footnotesize
\begin{longtable}{|ccccccccc|}
   \hline
      \multicolumn{9}{|c|}{\bf Total score of the tools for this examination} \\
        &
      \acs{greatspn} &
      \acs{its-tools} &
      \acs{lola} &
      \acs{lola-optimistic} &
      \acs{lola-optimistic-incomplete} &
      \acs{lola-pessimistic} &
      \acs{marcie} &
      \acs{sara} \\
      Total Score &
      {\bf 42} &
      {\bf 58} &
      {\bf 174} &
      {\bf 174} &
      {\bf 148} &
      {\bf 106} &
      {\bf 122} &
      {\bf 34} \\
   \hline
      \multicolumn{9}{|c|}{\bf CSRepetitions (Colored)} \\
      \ \index{Scores!ReachabilityCardinalityComparison!CSRepetitions (colored)} &
      \acs{greatspn} &
      \acs{its-tools} &
      \acs{lola} &
      \acs{lola-optimistic} &
      \acs{lola-optimistic-incomplete} &
      \acs{lola-pessimistic} &
      \acs{marcie} &
      \acs{sara} \\
      1st instance &
      \textcolor{red}{0} &
      \textcolor{red}{1} &
      \textcolor{red}{0} &
      \textcolor{red}{0} &
      \textcolor{red}{0} &
      \textcolor{red}{0} &
      \textcolor{red}{0} &
      \textcolor{red}{0} \\
      instances &
      \textcolor{red}{0} &
      \textcolor{red}{2} &
      \textcolor{red}{0} &
      \textcolor{red}{0} &
      \textcolor{red}{0} &
      \textcolor{red}{0} &
      \textcolor{red}{0} &
      \textcolor{red}{0} \\
      max reached &
      \textcolor{red}{0} &
      \textcolor{red}{0} &
      \textcolor{red}{0} &
      \textcolor{red}{0} &
      \textcolor{red}{0} &
      \textcolor{red}{0} &
      \textcolor{red}{0} &
      \textcolor{red}{0} \\
\hline
\end{longtable}
}

\subsubsection{``Surprise'' Models}

Scores are summarized in the table below.
For each model in this category, a tool may collect up to 49 points.

{\footnotesize
\begin{longtable}{|ccccccc|}
   \hline
      \multicolumn{7}{|c|}{\bf Total score of the tools for this examination} \\
        &
      \acs{lola} &
      \acs{lola-optimistic} &
      \acs{lola-optimistic-incomplete} &
      \acs{lola-pessimistic} &
      \acs{marcie} &
      \acs{sara} \\
      Total Score &
      {\bf 12} &
      {\bf 12} &
      {\bf 12} &
      {\bf 6} &
      {\bf 24} &
      {\bf 6} \\
   \hline
      \multicolumn{7}{|c|}{\bf HouseConstruction (P/T)} \\
      \ \index{Scores!ReachabilityCardinalityComparison!HouseConstruction (P/T)} &
      \acs{lola} &
      \acs{lola-optimistic} &
      \acs{lola-optimistic-incomplete} &
      \acs{lola-pessimistic} &
      \acs{marcie} &
      \acs{sara} \\
      1st instance &
      \textcolor{red}{0} &
      \textcolor{red}{0} &
      \textcolor{red}{0} &
      \textcolor{red}{0} &
      \textcolor{red}{1} &
      \textcolor{red}{0} \\
      instances &
      \textcolor{red}{0} &
      \textcolor{red}{0} &
      \textcolor{red}{0} &
      \textcolor{red}{0} &
      \textcolor{red}{4} &
      \textcolor{red}{0} \\
      max reached &
      \textcolor{red}{0} &
      \textcolor{red}{0} &
      \textcolor{red}{0} &
      \textcolor{red}{0} &
      \textcolor{red}{0} &
      \textcolor{red}{0} \\
\hline
\end{longtable}
}

\subsection{Trophies for this Examination}
\index{Trophies!ReachabilityCardinalityComparison}

Trophies are divided in three categories: ``Known'' models,
``Surprise'' models, and the global trophies (formula is then
$score_{global} = score_{known} + 2 \times score_{surprise}$).

\subsubsection{For ``Known'' Models} \ \\

\begin{tabular}{c|c|c}
      1 & 1 & 3 \\
   \includegraphics[width=2cm]{figures/gold.jpg} &
   \includegraphics[width=2cm]{figures/gold.jpg} &
   \includegraphics[width=2cm]{figures/bronse.jpg} \\
   \acs{lola} &
   \acs{lola-optimistic} &
   \acs{lola-optimistic-incomplete} \\
   174 points &
   174 points &
   148 points \\
\end{tabular}

\subsubsection{For ``Surprise'' Models}\  \\

\begin{tabular}{c|c|c|c}
      1 & 2 & 2 & 2 \\
   \includegraphics[width=2cm]{figures/gold.jpg} &
   \includegraphics[width=2cm]{figures/silver.jpg} &
   \includegraphics[width=2cm]{figures/silver.jpg} &
   \includegraphics[width=2cm]{figures/silver.jpg} \\
   \acs{marcie} &
   \acs{lola} &
   \acs{lola-optimistic} &
   \acs{lola-optimistic-incomplete} \\
   24 points &
   12 points &
   12 points &
   12 points \\
\end{tabular}

\subsubsection{Global} \ \\

\begin{tabular}{c|c|c}
      1 & 1 & 3 \\
   \includegraphics[width=2cm]{figures/gold.jpg} &
   \includegraphics[width=2cm]{figures/gold.jpg} &
   \includegraphics[width=2cm]{figures/bronse.jpg} \\
   \acs{lola} &
   \acs{lola-optimistic} &
   \acs{lola-optimistic-incomplete} \\
   198 points &
   198 points &
   172 points \\
\end{tabular}

\newpage

\section{The ReachabilityDeadlock Examination}
\label{sec:exam:ReachabilityDeadlock}
\index{Results!ReachabilityDeadlock}

This examination deals with reachability properties dealing with transition deadlocks only.
We first show a summary on the handling of models by the participating tools.
Then, we present the computed outputs and the associated scores for this
examination prior to a summary of relevant executions.

\subsection{Handling of Models by Tools}
\index{Performances!ReachabilityDeadlock}

\subsubsection{\acs{CSRepetitions-COL}}
The charts below respectively show how tools compete with this ``Known'' model (memory and CPU).

\index{Performances!ReachabilityDeadlock!CSRepetitions (Colored)}
\begin{center}
   \includegraphics[width=7.2cm]{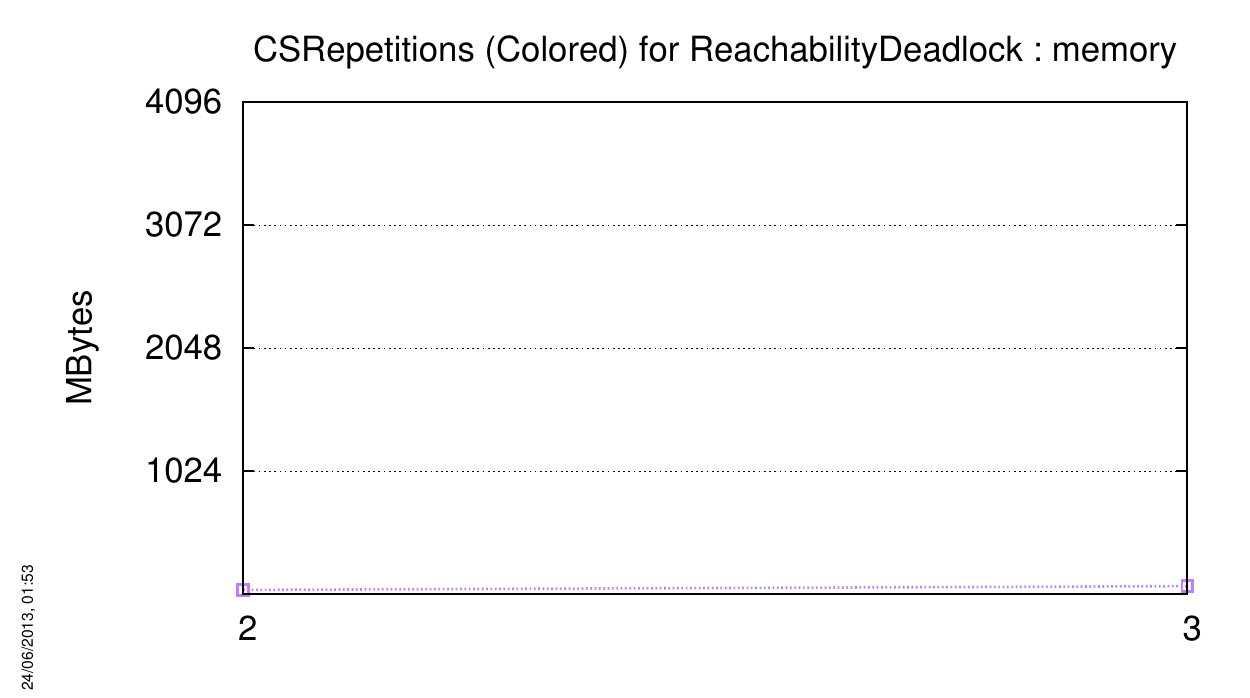}
   \includegraphics[width=7.2cm]{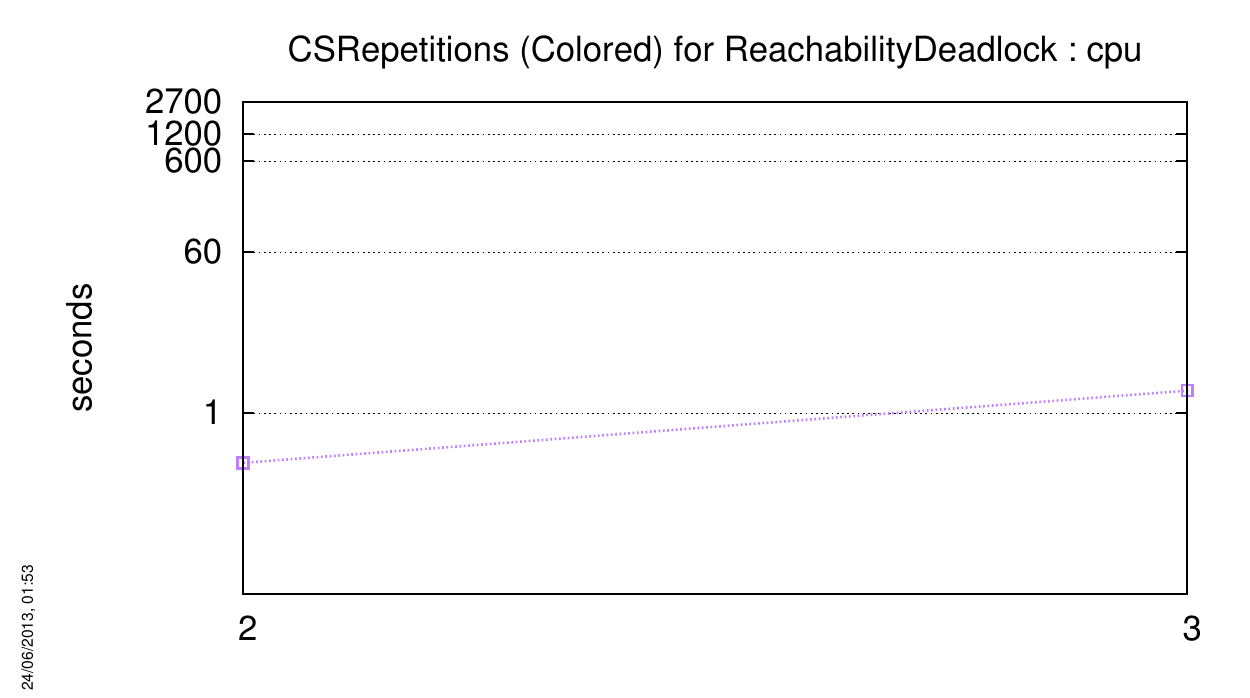}

   \includegraphics[height=1cm]{figures/tools-legend.pdf}
\end{center}

\subsubsection{\acs{CSRepetitions-PT}}
The charts below respectively show how tools compete with this ``Known'' model (memory and CPU).

\index{Performances!ReachabilityDeadlock!CSRepetitions (P/T)}
\begin{center}
   \includegraphics[width=7.2cm]{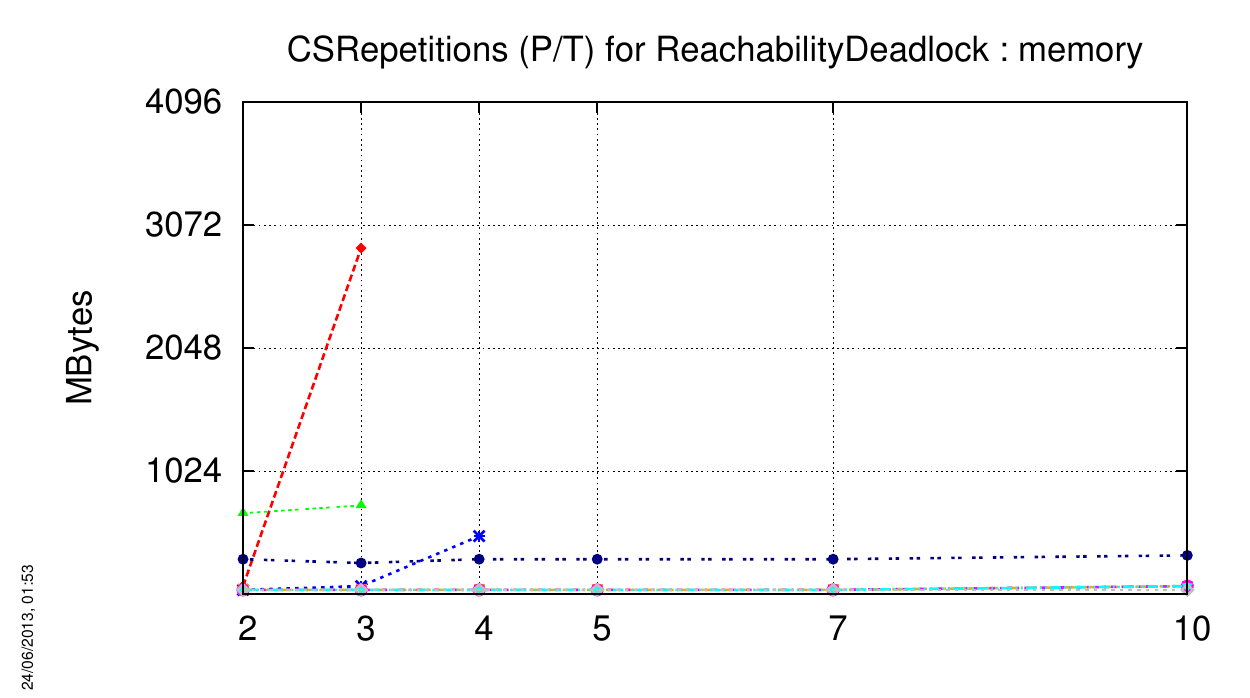}
   \includegraphics[width=7.2cm]{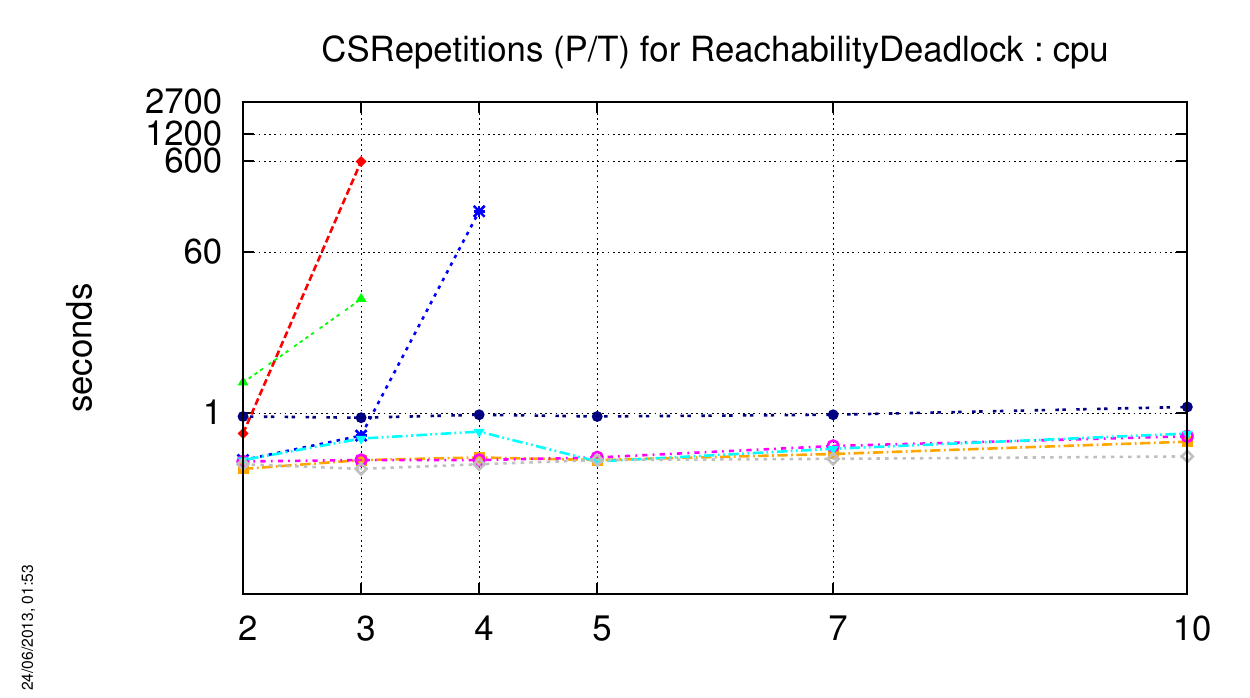}

   \includegraphics[height=1cm]{figures/tools-legend.pdf}
\end{center}

\subsubsection{\acs{Dekker-PT}}
The charts below respectively show how tools compete with this ``Known'' model (memory and CPU).

\index{Performances!ReachabilityDeadlock!Dekker (P/T)}
\begin{center}
   \includegraphics[width=7.2cm]{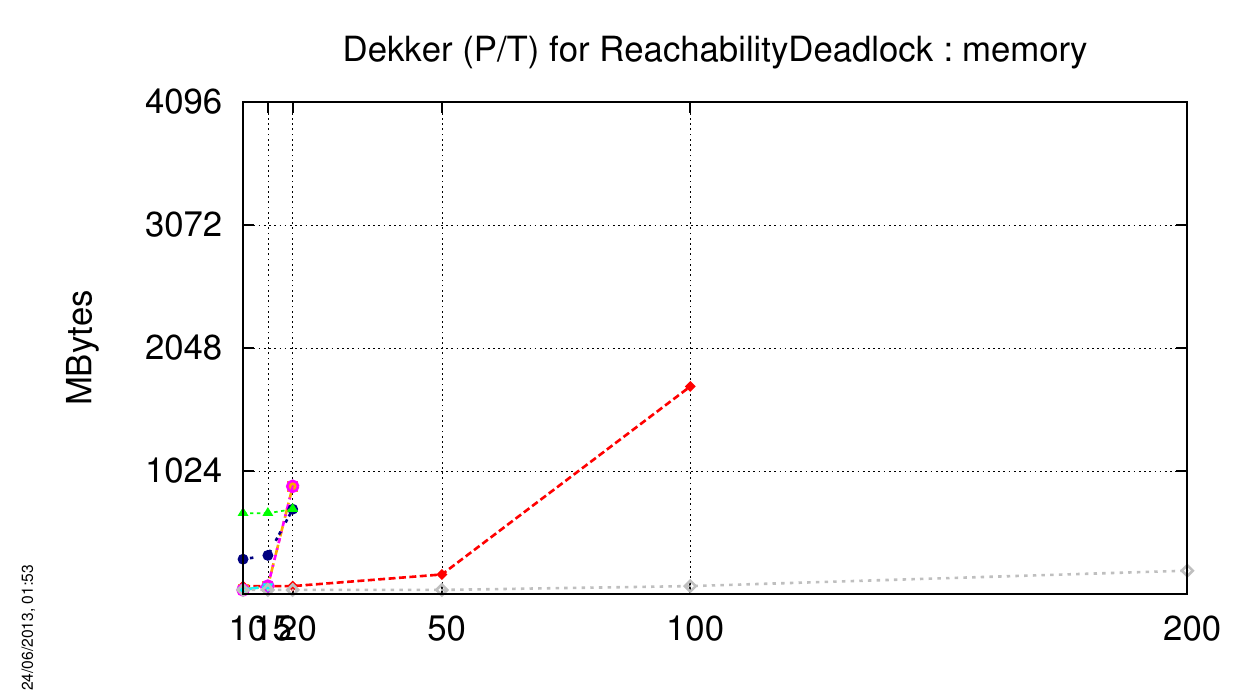}
   \includegraphics[width=7.2cm]{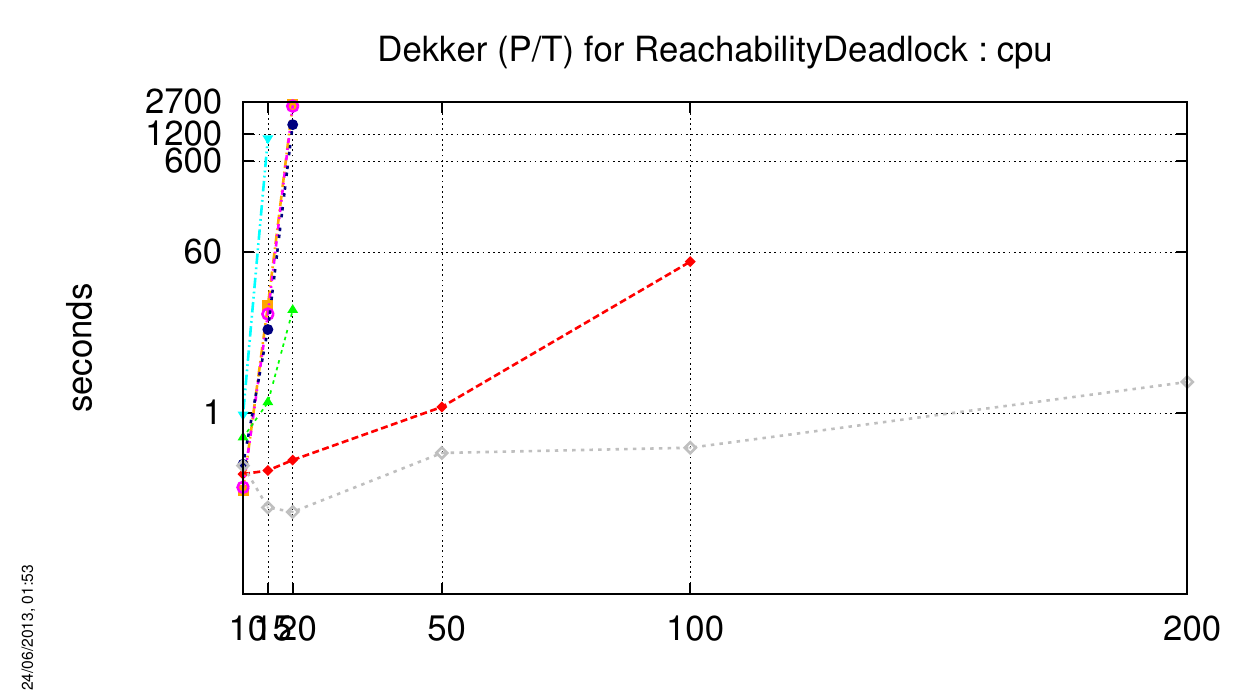}

   \includegraphics[height=1cm]{figures/tools-legend.pdf}
\end{center}

\subsubsection{\acs{DotAndBoxes-COL}}
The charts below respectively show how tools compete with this ``Known'' model (memory and CPU).

\index{Performances!ReachabilityDeadlock!DotAndBoxes (Colored)}
\begin{center}
   \includegraphics[width=7.2cm]{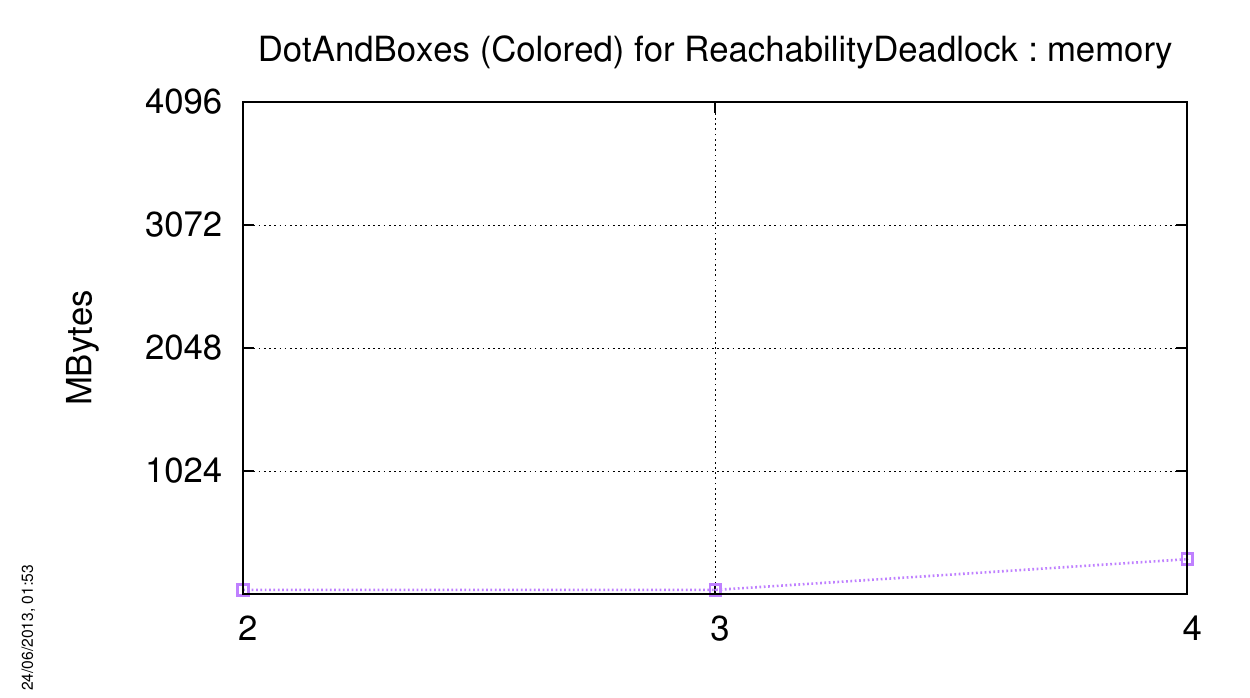}
   \includegraphics[width=7.2cm]{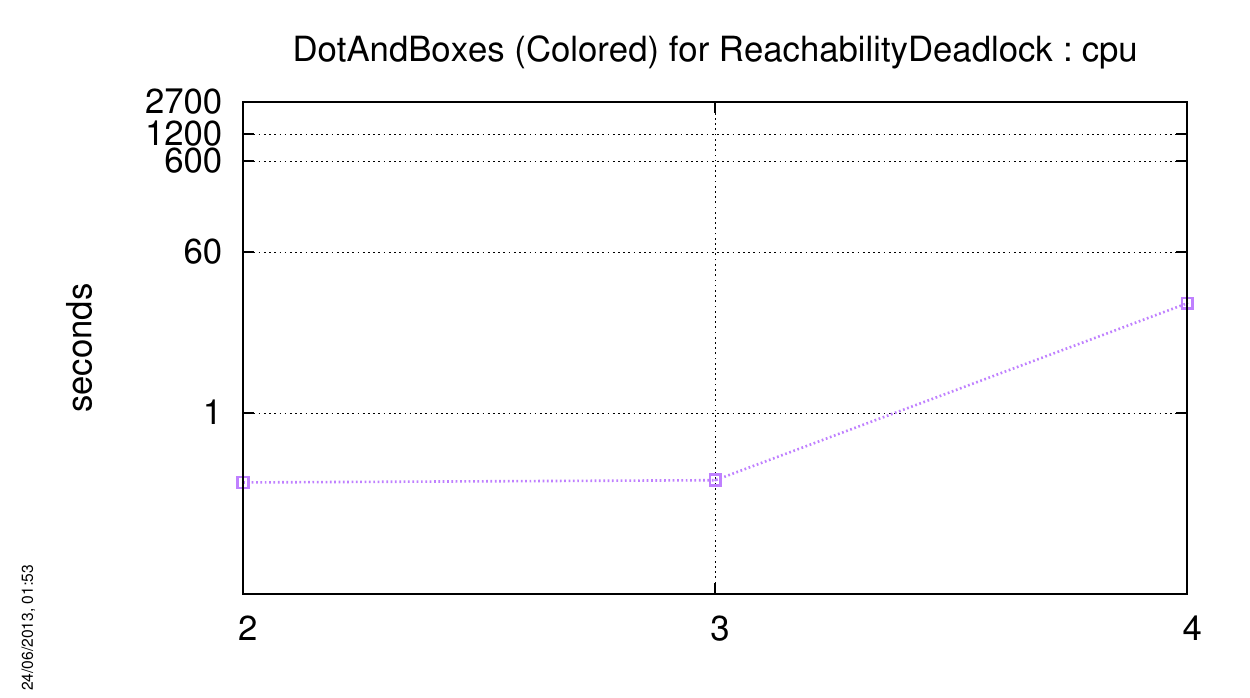}

   \includegraphics[height=1cm]{figures/tools-legend.pdf}
\end{center}

\subsubsection{\acs{DrinkVendingMachine-COL}}
The charts below respectively show how tools compete with this ``Known'' model (memory and CPU).

\index{Performances!ReachabilityDeadlock!DrinkVendingMachine (Colored)}
\begin{center}
   \includegraphics[width=7.2cm]{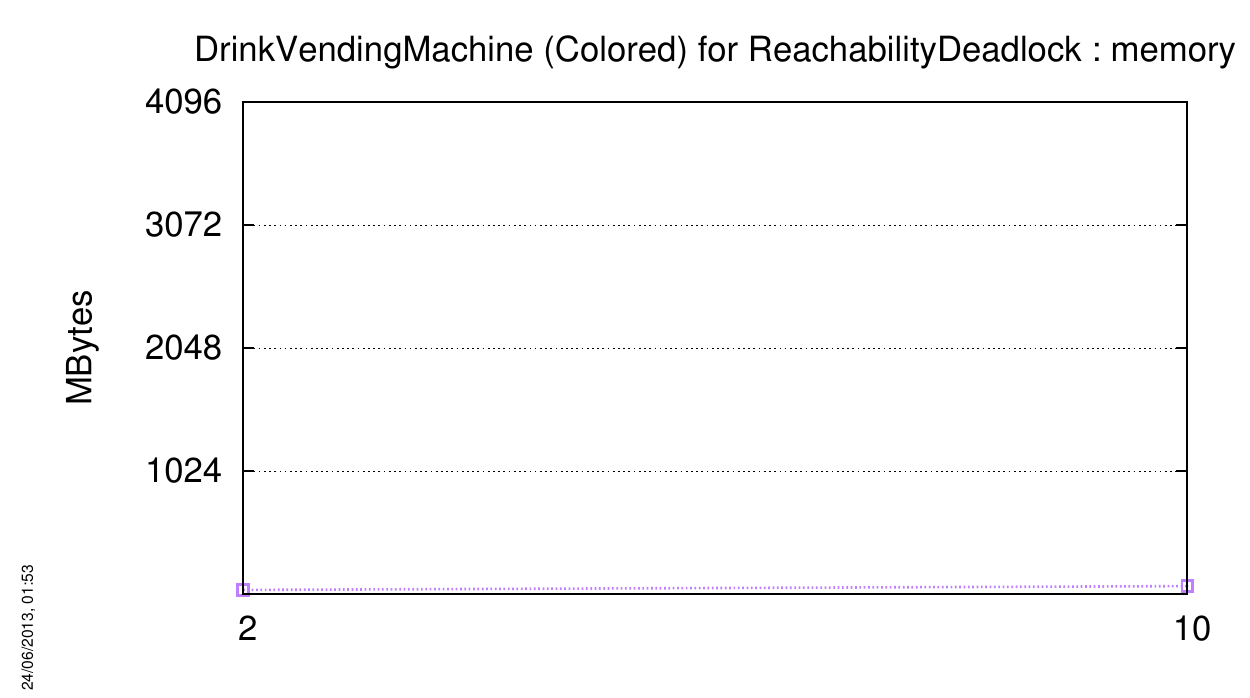}
   \includegraphics[width=7.2cm]{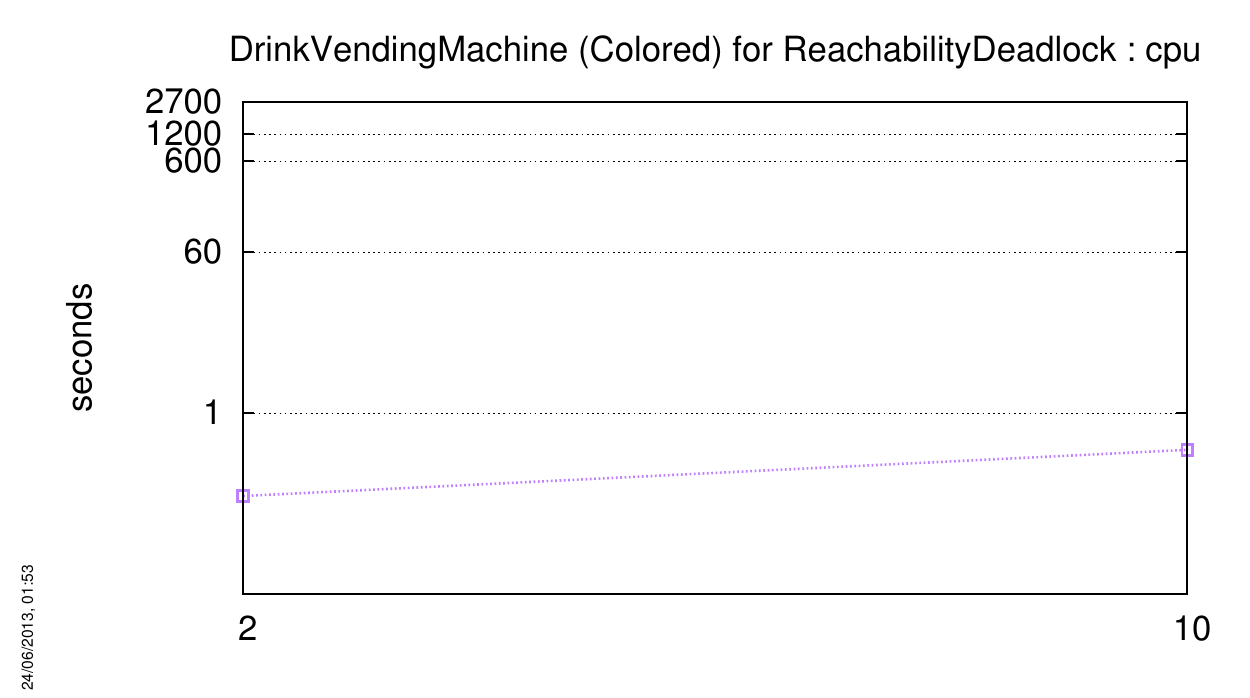}

   \includegraphics[height=1cm]{figures/tools-legend.pdf}
\end{center}

\subsubsection{\acs{DrinkVendingMachine-PT}}
The charts below respectively show how tools compete with this ``Known'' model (memory and CPU).

\index{Performances!ReachabilityDeadlock!DrinkVendingMachine (P/T)}
\begin{center}
   \includegraphics[width=7.2cm]{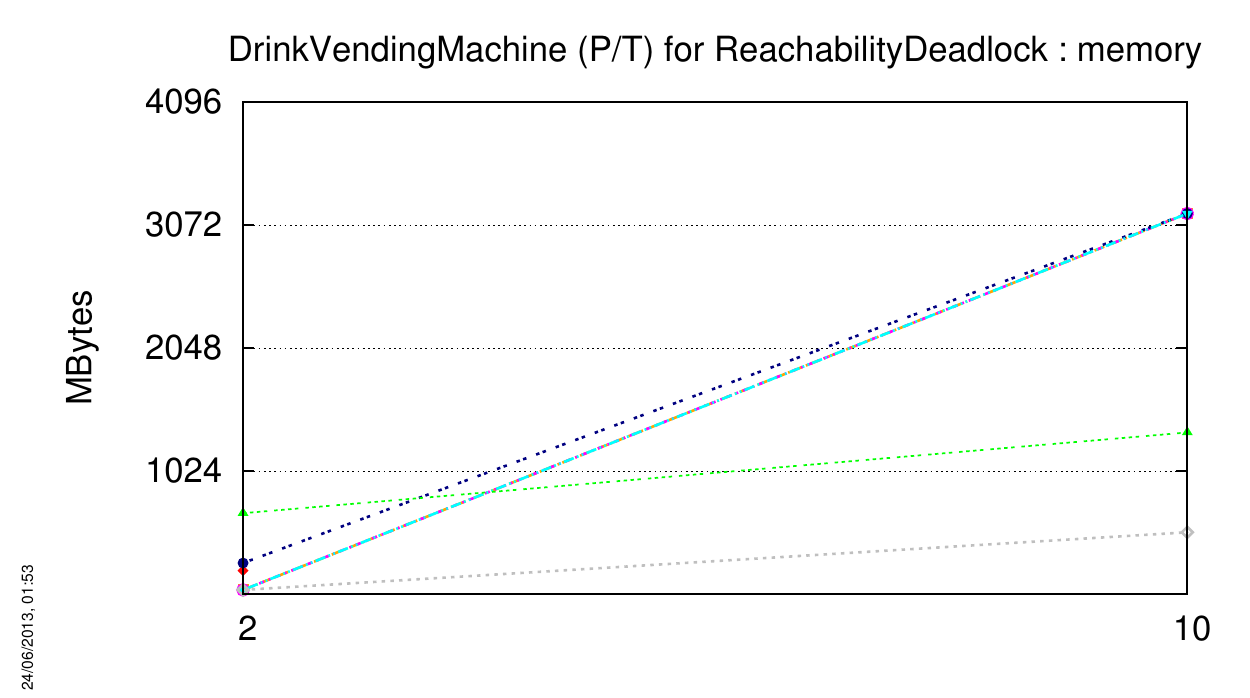}
   \includegraphics[width=7.2cm]{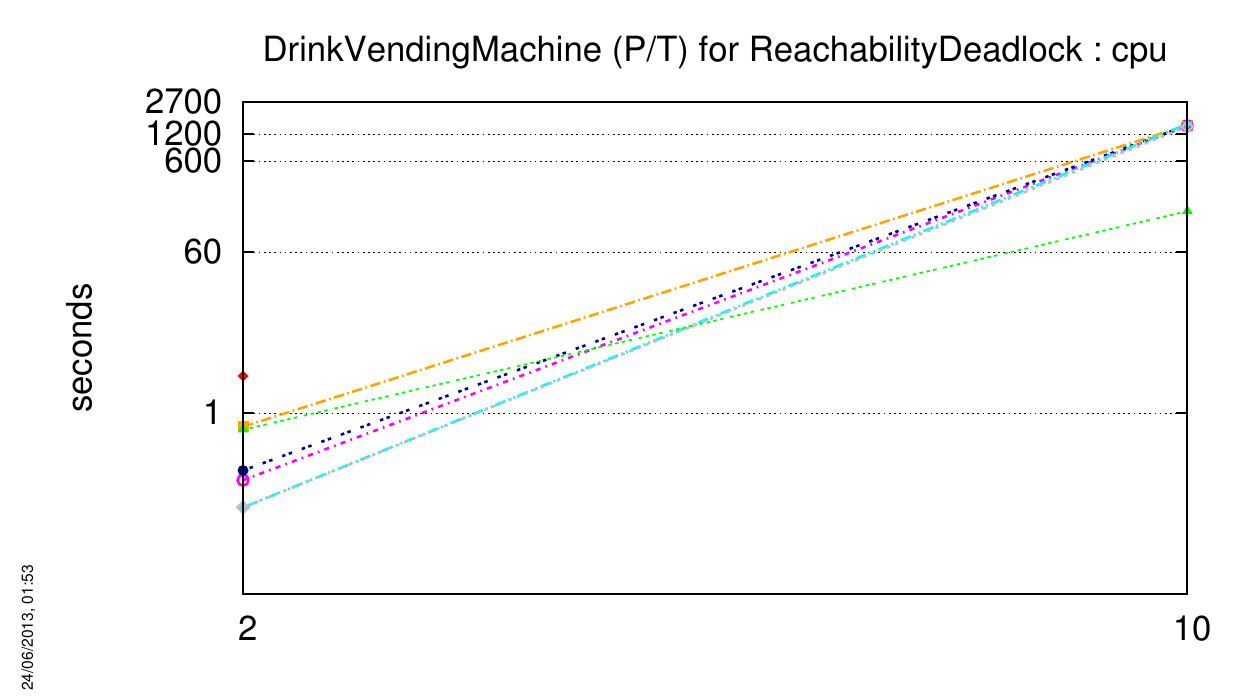}

   \includegraphics[height=1cm]{figures/tools-legend.pdf}
\end{center}

\subsubsection{\acs{Echo-PT}}
The charts below respectively show how tools compete with this ``Known'' model (memory and CPU).

\index{Performances!ReachabilityDeadlock!Echo (P/T)}
\begin{center}
   \includegraphics[width=7.2cm]{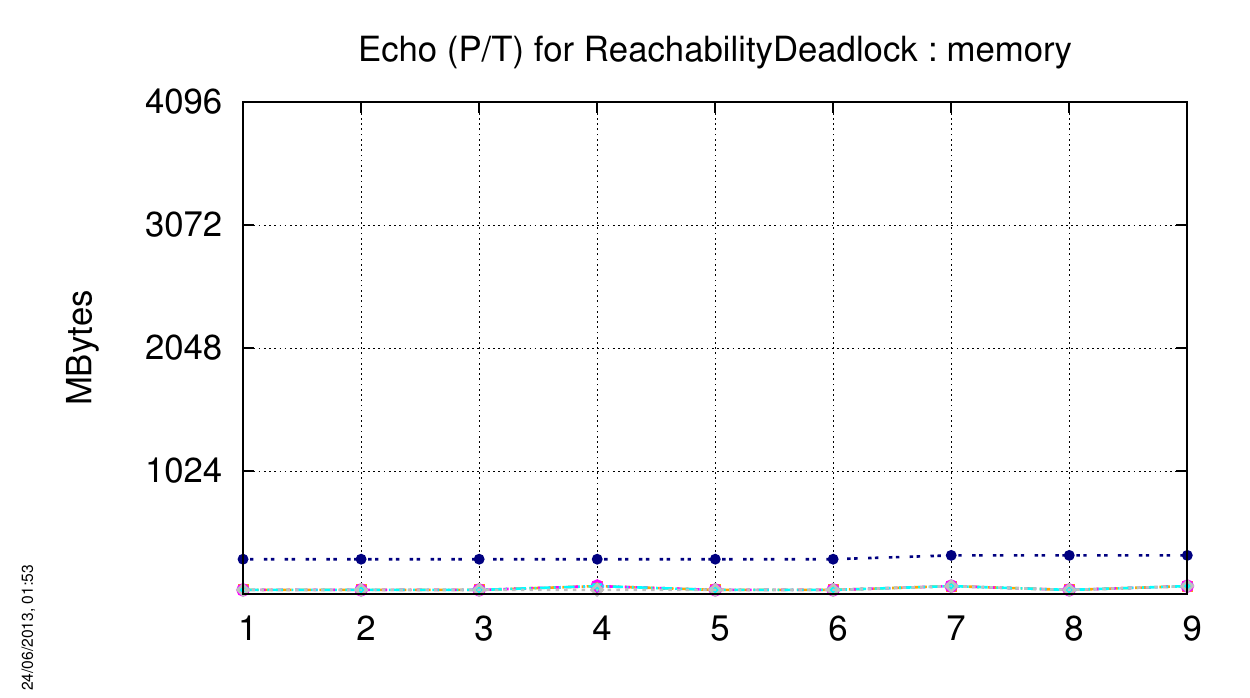}
   \includegraphics[width=7.2cm]{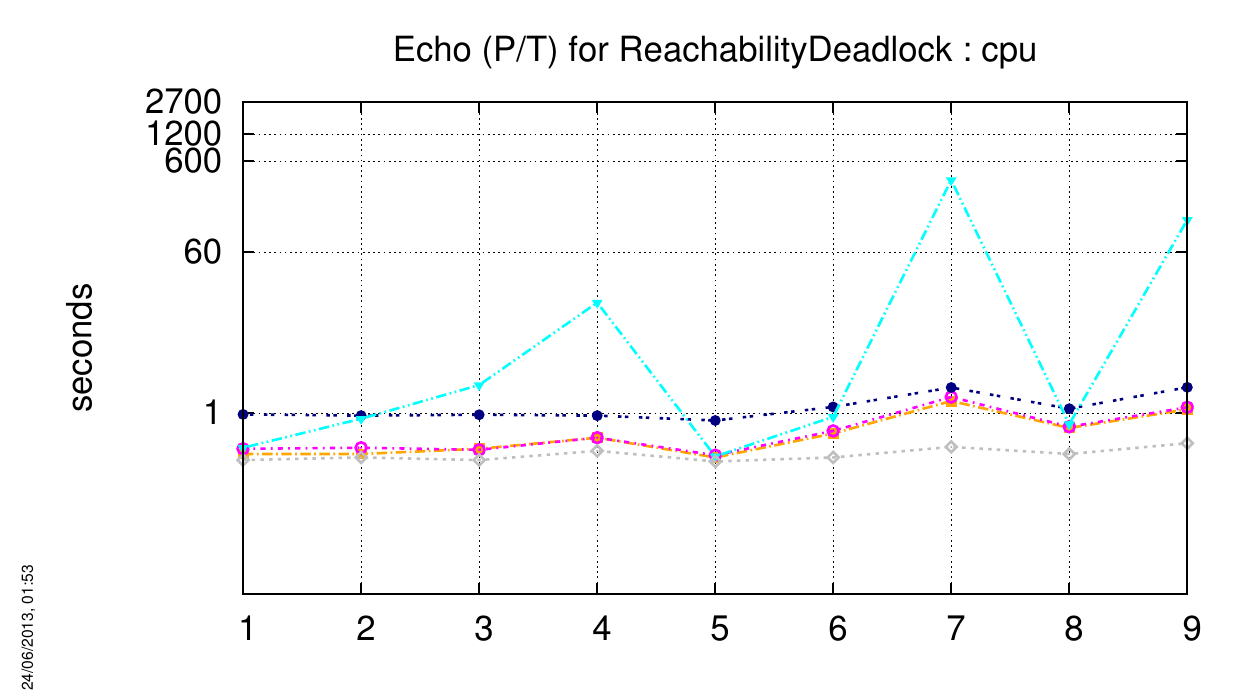}

   \includegraphics[height=1cm]{figures/tools-legend.pdf}
\end{center}

\subsubsection{\acs{Eratosthenes-PT}}
The charts below respectively show how tools compete with this ``Known'' model (memory and CPU).

\index{Performances!ReachabilityDeadlock!Eratosthenes (P/T)}
\begin{center}
   \includegraphics[width=7.2cm]{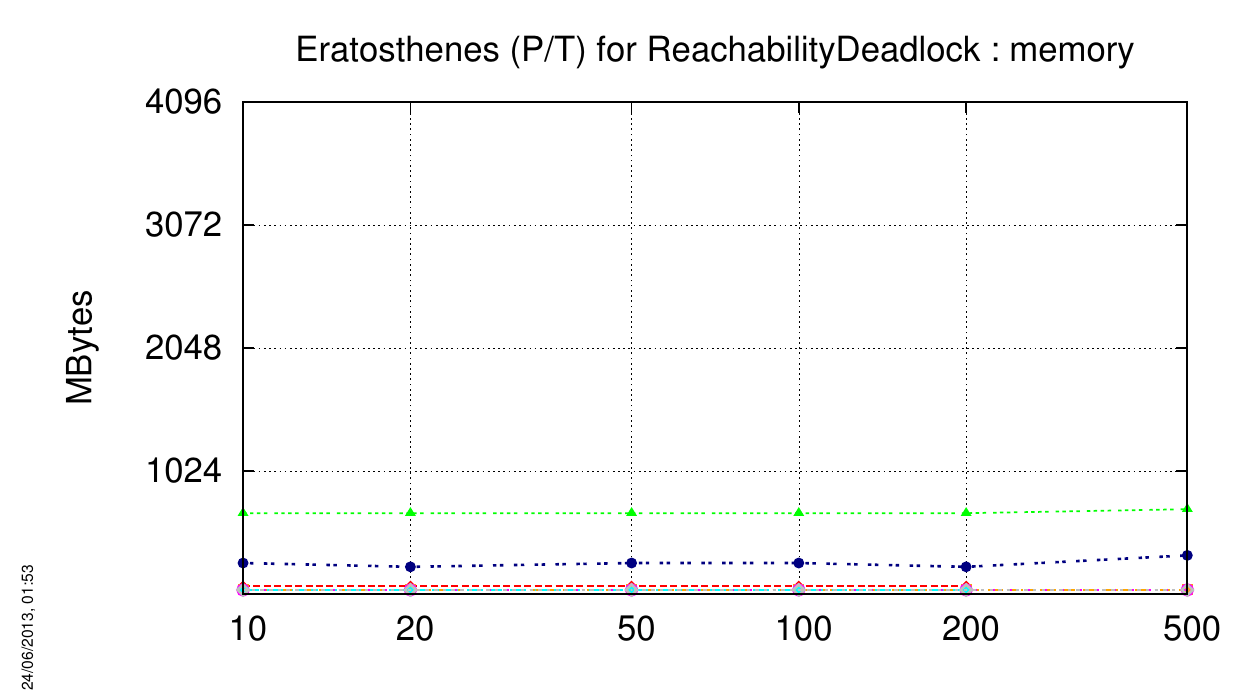}
   \includegraphics[width=7.2cm]{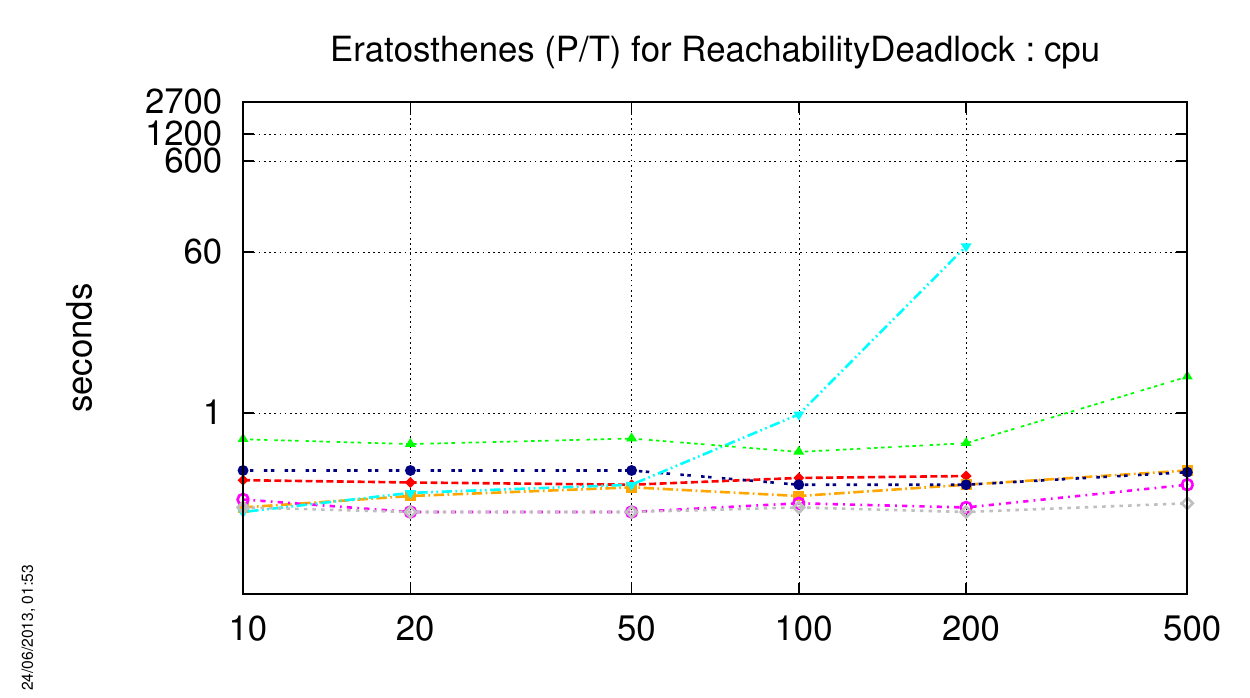}

   \includegraphics[height=1cm]{figures/tools-legend.pdf}
\end{center}

\subsubsection{\acs{FMS-PT}}
The charts below respectively show how tools compete with this ``Known'' model (memory and CPU).

\index{Performances!ReachabilityDeadlock!FMS (P/T)}
\begin{center}
   \includegraphics[width=7.2cm]{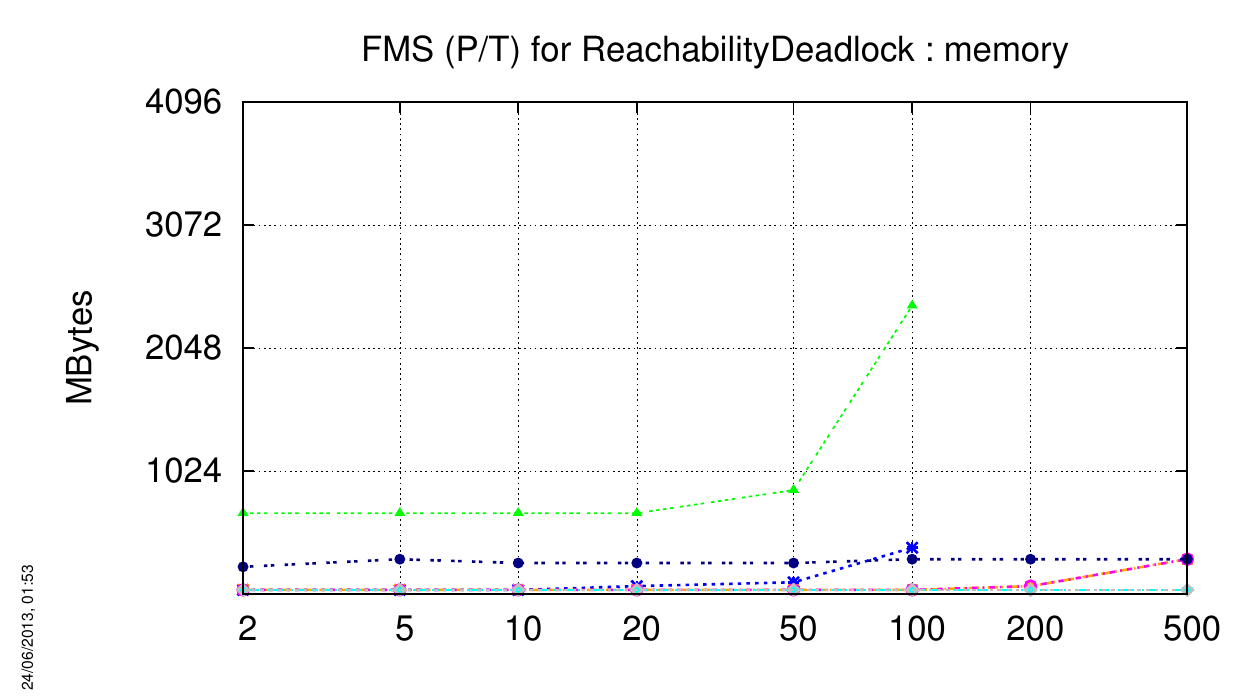}
   \includegraphics[width=7.2cm]{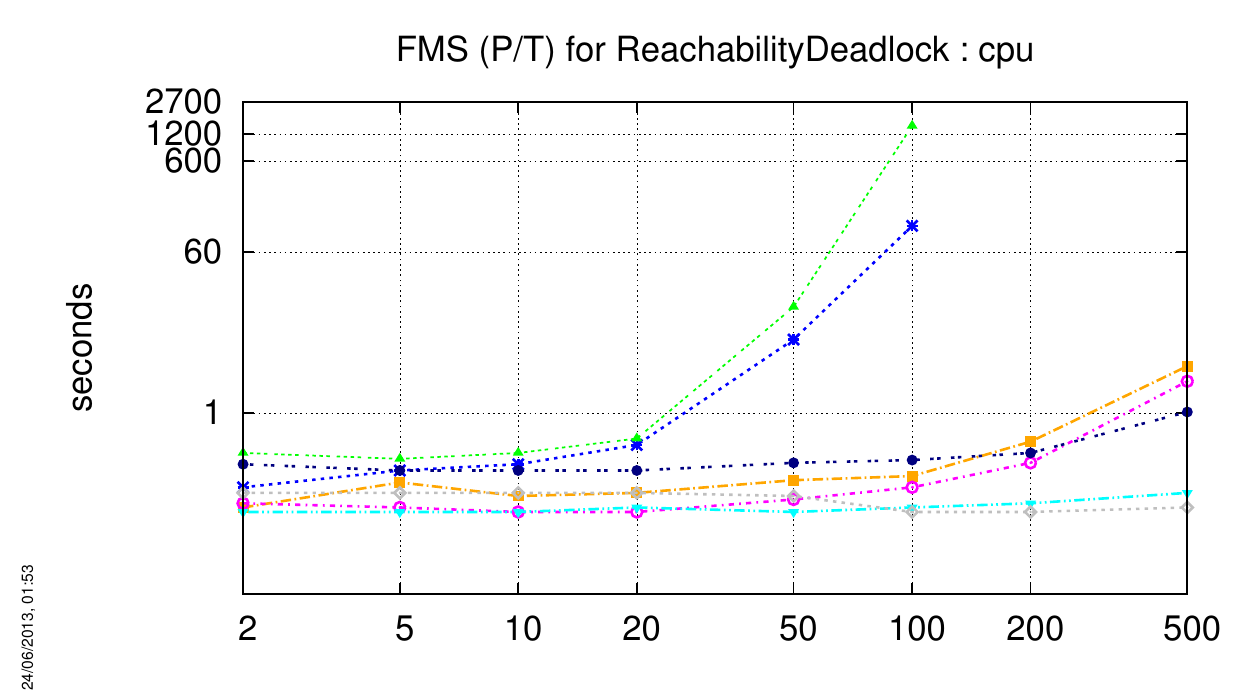}

   \includegraphics[height=1cm]{figures/tools-legend.pdf}
\end{center}

\subsubsection{\acs{GlobalRessAlloc-COL}}
The charts below respectively show how tools compete with this ``Known'' model (memory and CPU).

\index{Performances!ReachabilityDeadlock!GlobalRessAlloc (Colored)}
\begin{center}
   \includegraphics[width=7.2cm]{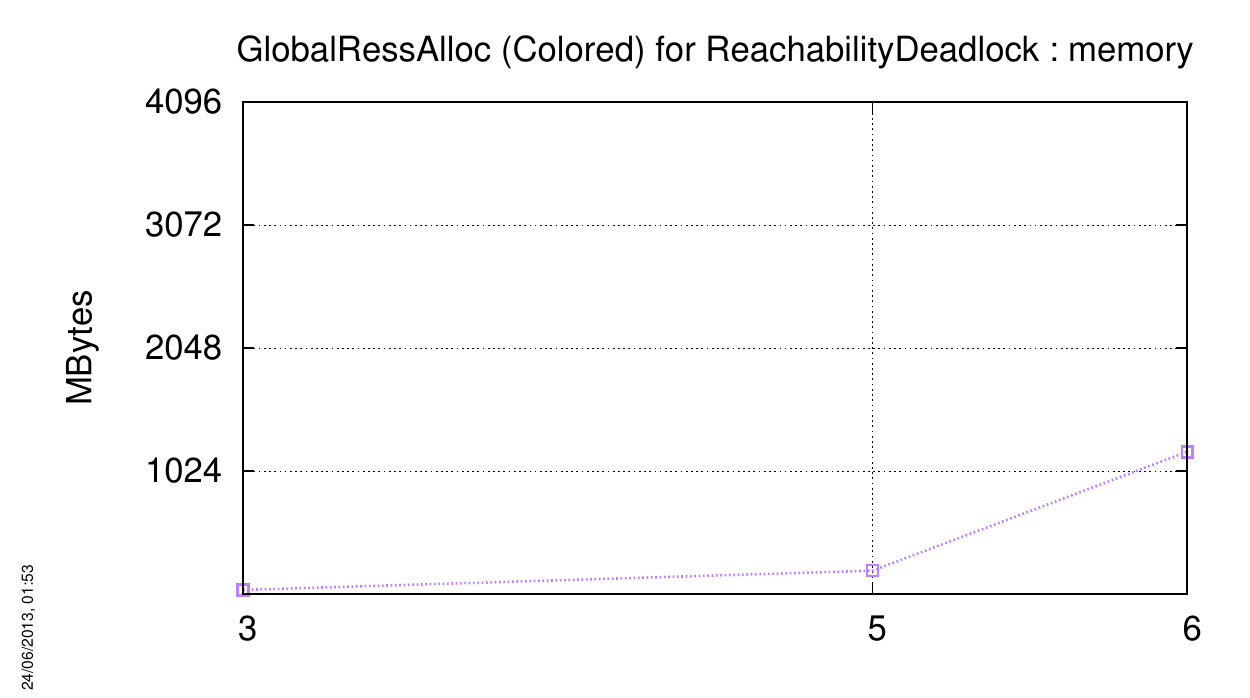}
   \includegraphics[width=7.2cm]{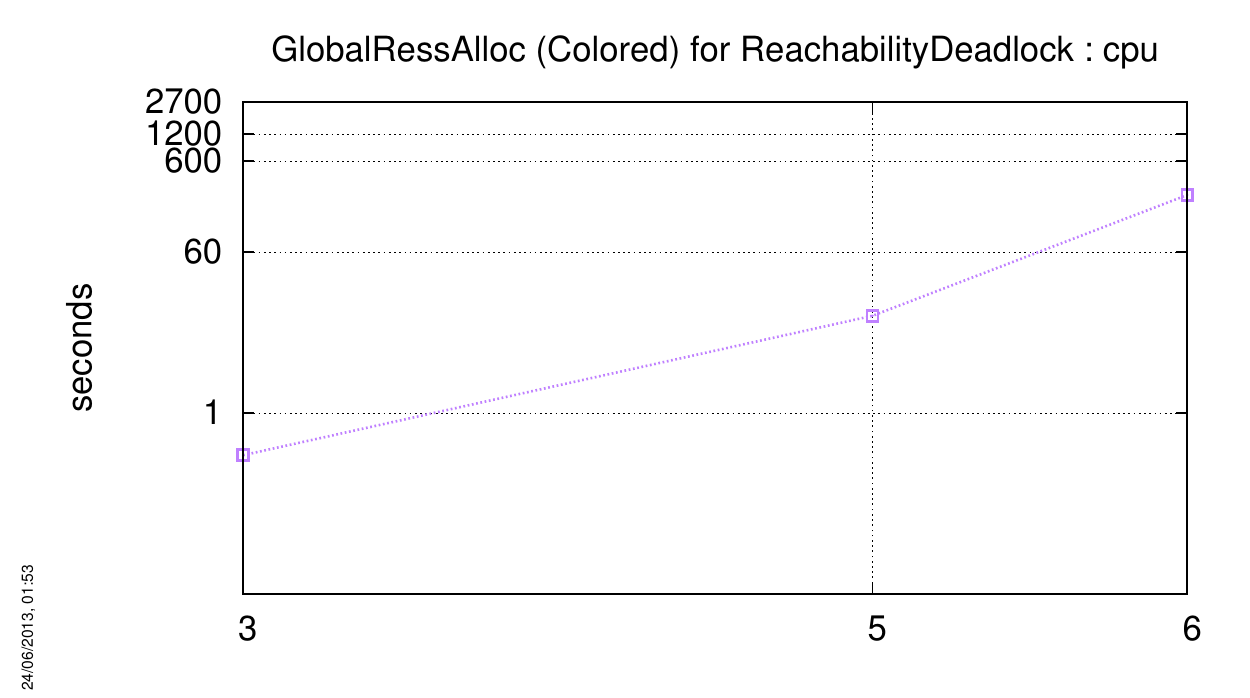}

   \includegraphics[height=1cm]{figures/tools-legend.pdf}
\end{center}

\subsubsection{\acs{GlobalRessAlloc-PT}}
The charts below respectively show how tools compete with this ``Known'' model (memory and CPU).

\index{Performances!ReachabilityDeadlock!GlobalRessAlloc (P/T)}
\begin{center}
   \includegraphics[width=7.2cm]{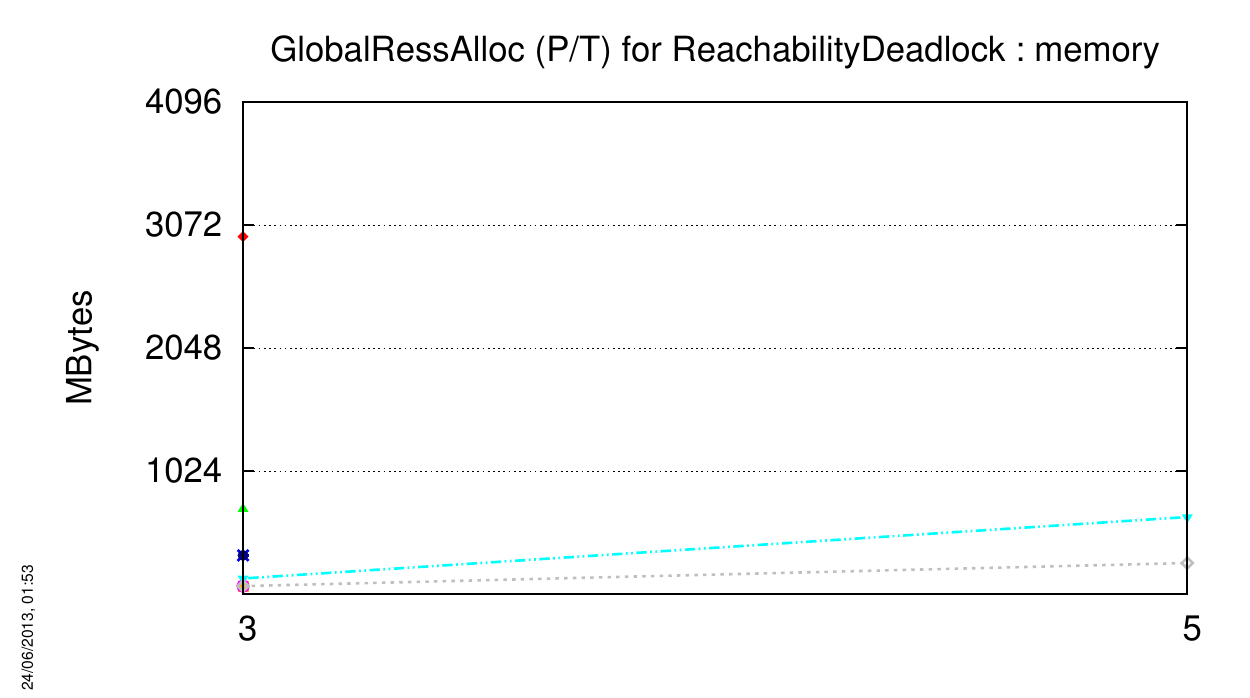}
   \includegraphics[width=7.2cm]{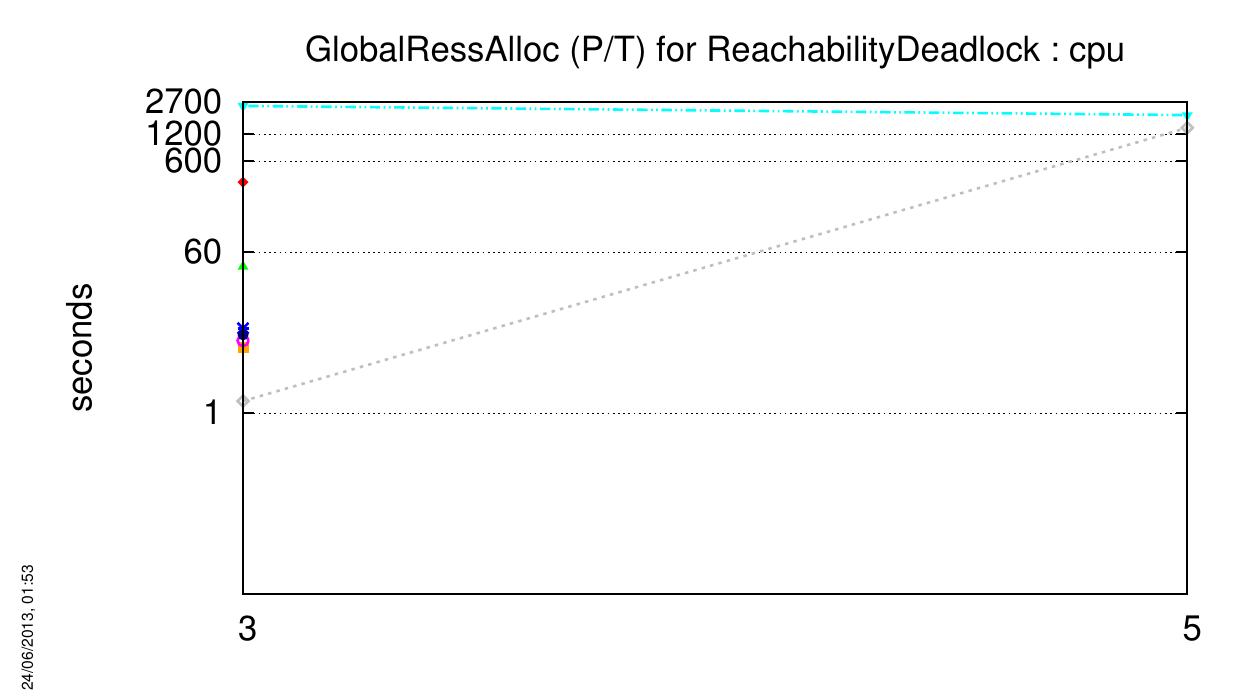}

   \includegraphics[height=1cm]{figures/tools-legend.pdf}
\end{center}

\subsubsection{\acs{Kanban-PT}}
The charts below respectively show how tools compete with this ``Known'' model (memory and CPU).

\index{Performances!ReachabilityDeadlock!Kanban (P/T)}
\begin{center}
   \includegraphics[width=7.2cm]{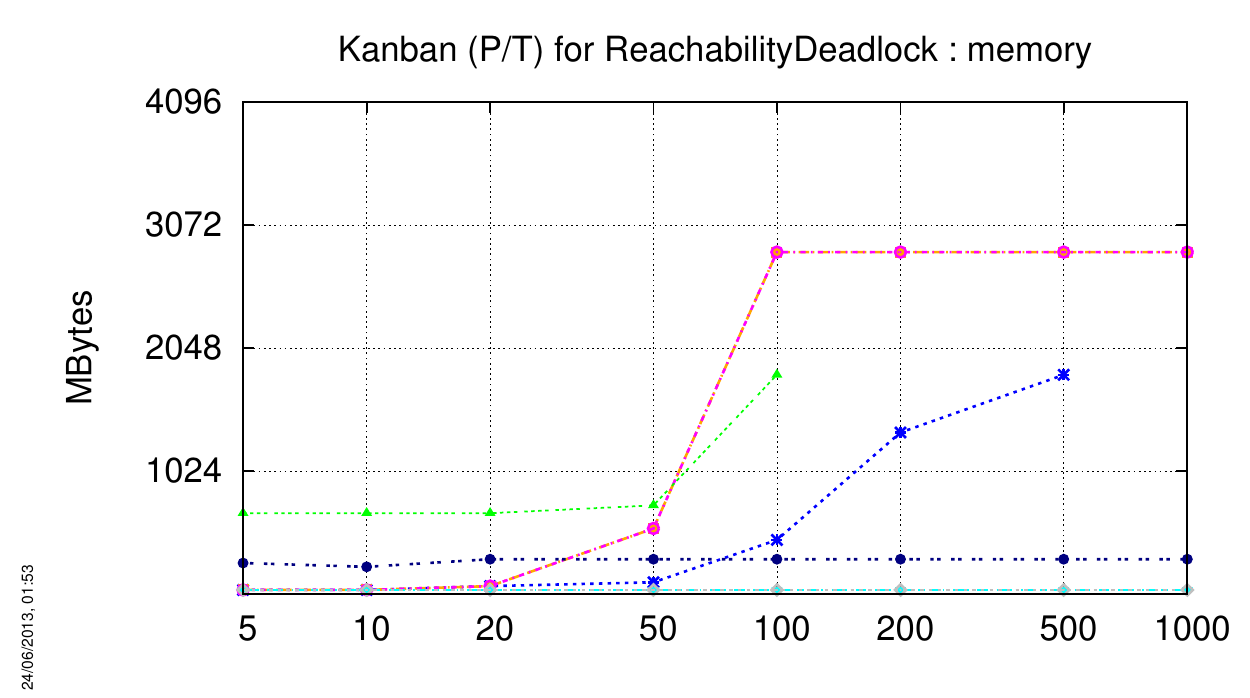}
   \includegraphics[width=7.2cm]{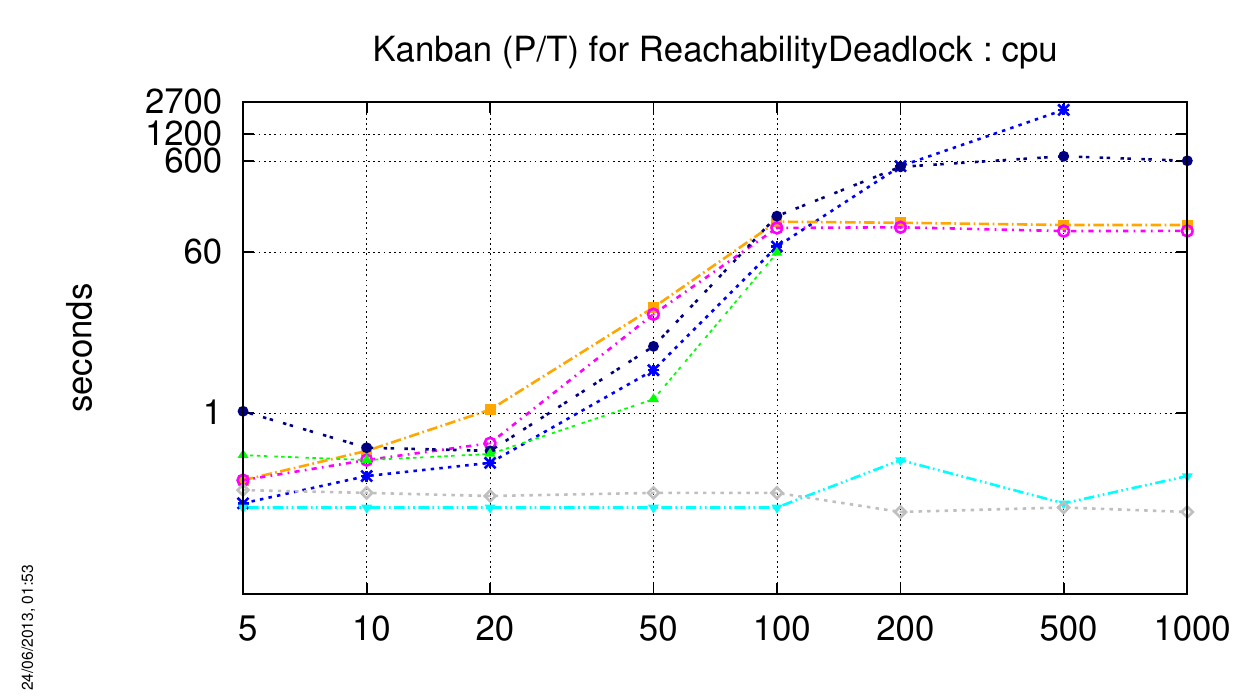}

   \includegraphics[height=1cm]{figures/tools-legend.pdf}
\end{center}

\subsubsection{\acs{LamportFastMutEx-COL}}
The charts below respectively show how tools compete with this ``Known'' model (memory and CPU).

\index{Performances!ReachabilityDeadlock!LamportFastMutEx (Colored)}
\begin{center}
   \includegraphics[width=7.2cm]{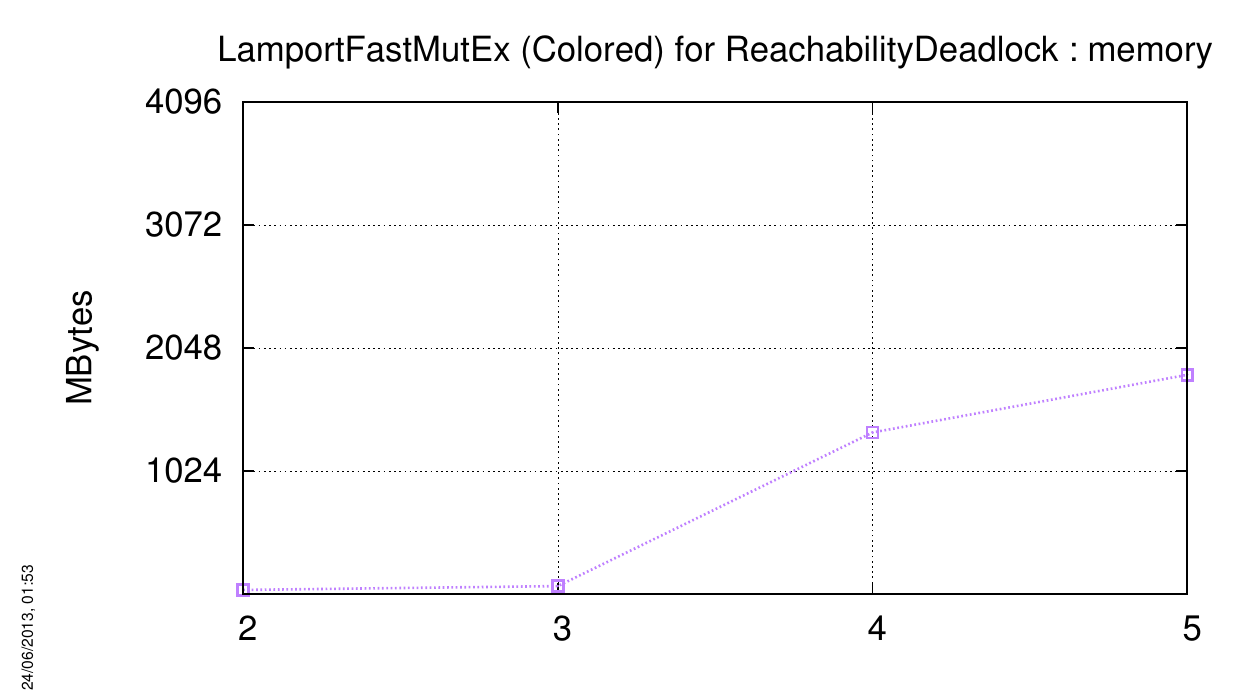}
   \includegraphics[width=7.2cm]{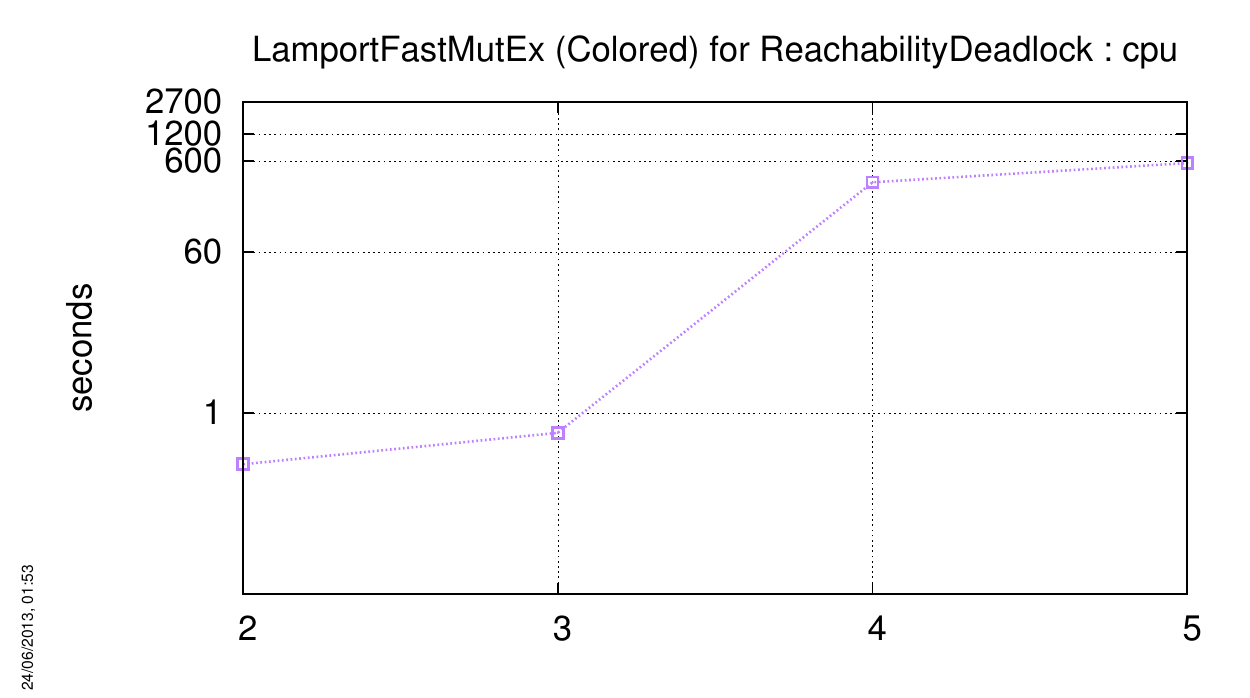}

   \includegraphics[height=1cm]{figures/tools-legend.pdf}
\end{center}

\subsubsection{\acs{LamportFastMutEx-PT}}
The charts below respectively show how tools compete with this ``Known'' model (memory and CPU).

\index{Performances!ReachabilityDeadlock!LamportFastMutEx (P/T)}
\begin{center}
   \includegraphics[width=7.2cm]{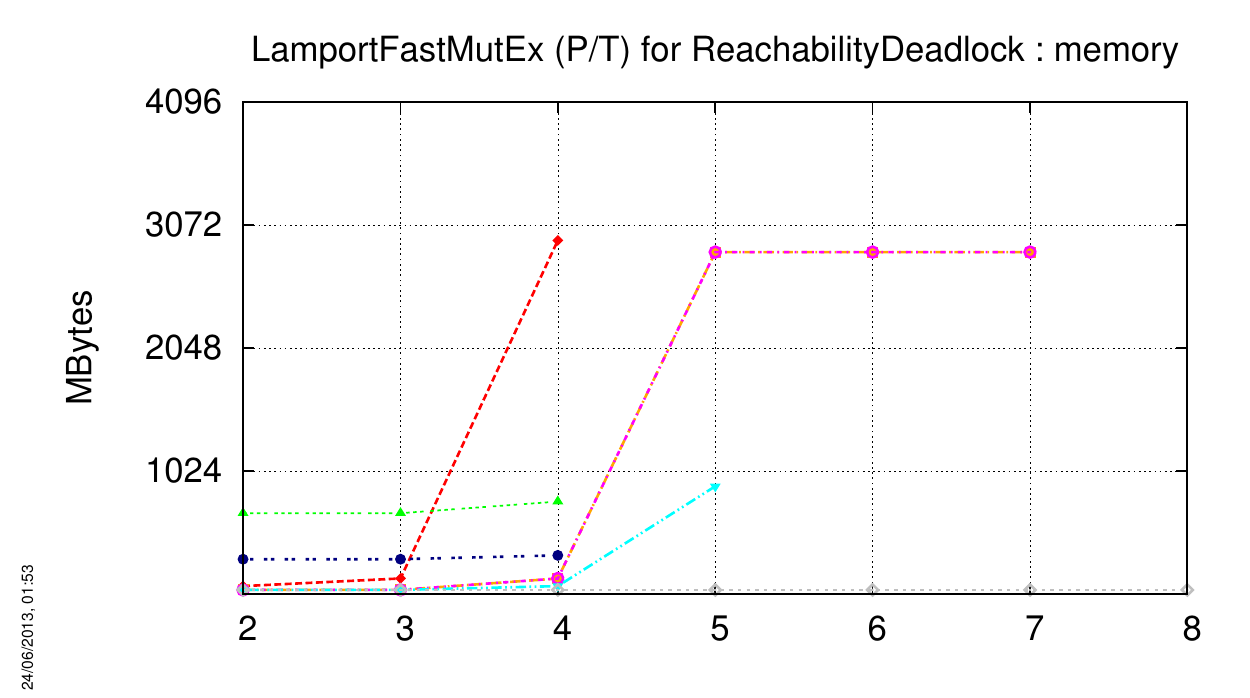}
   \includegraphics[width=7.2cm]{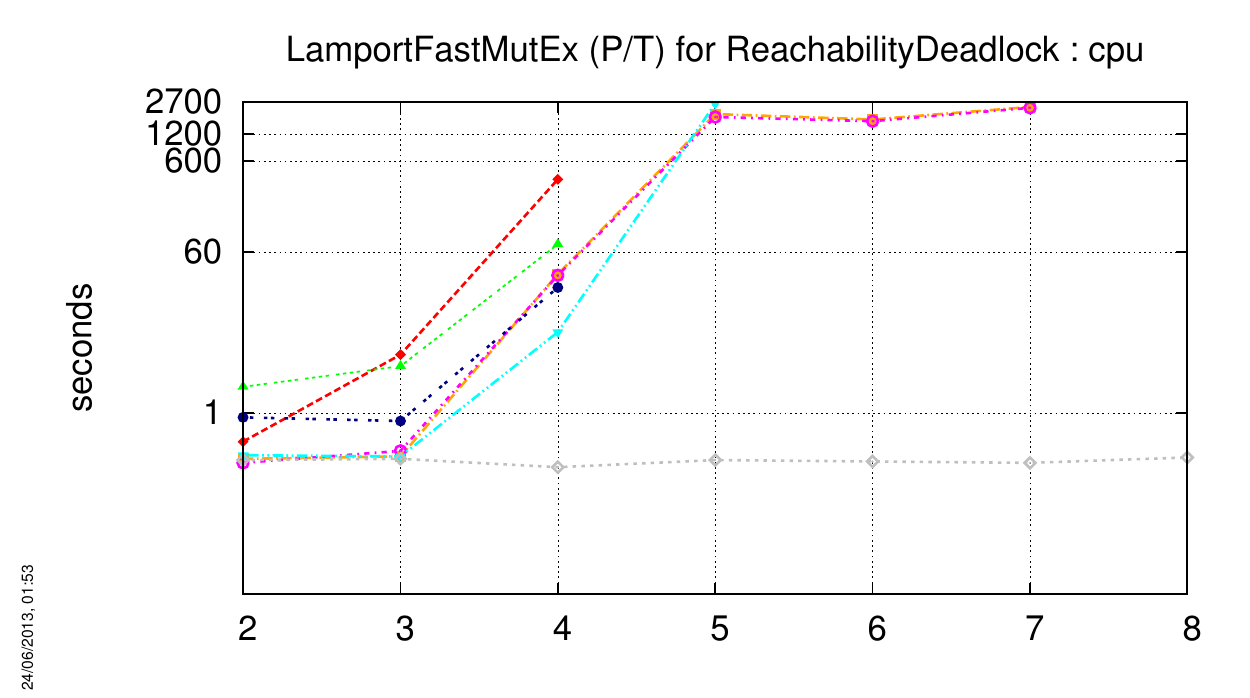}

   \includegraphics[height=1cm]{figures/tools-legend.pdf}
\end{center}

\subsubsection{\acs{MAPK-PT}}
The charts below respectively show how tools compete with this ``Known'' model (memory and CPU).

\index{Performances!ReachabilityDeadlock!MAPK (P/T)}
\begin{center}
   \includegraphics[width=7.2cm]{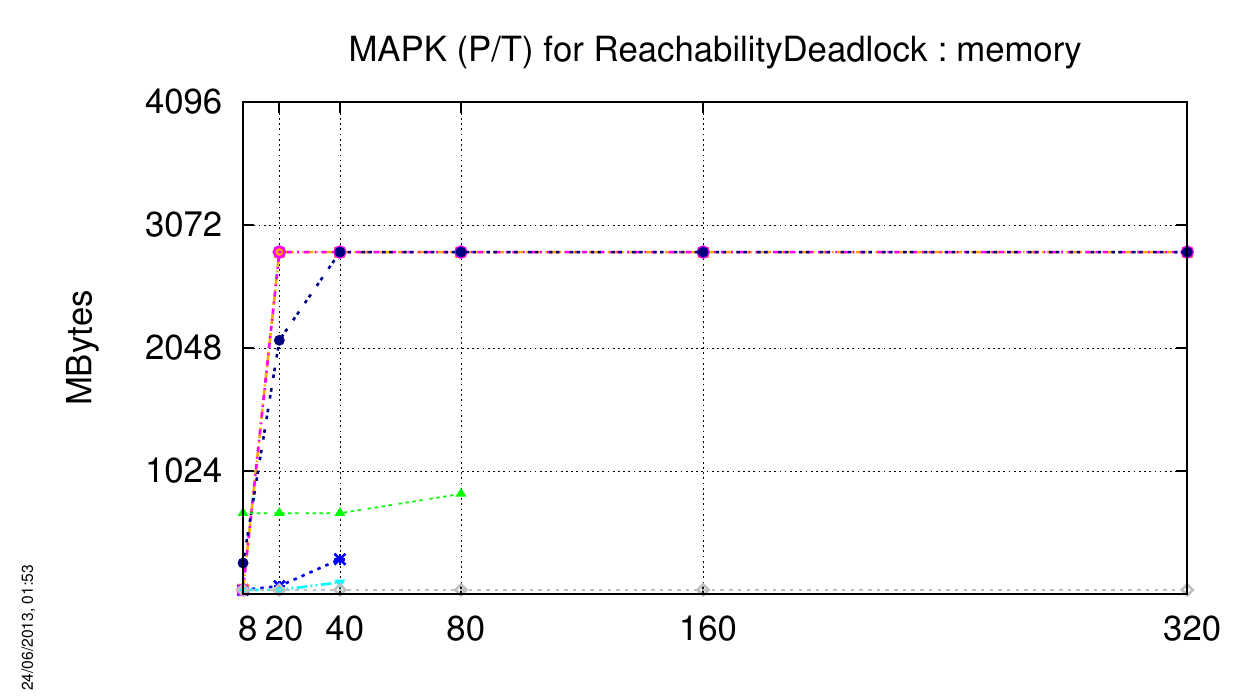}
   \includegraphics[width=7.2cm]{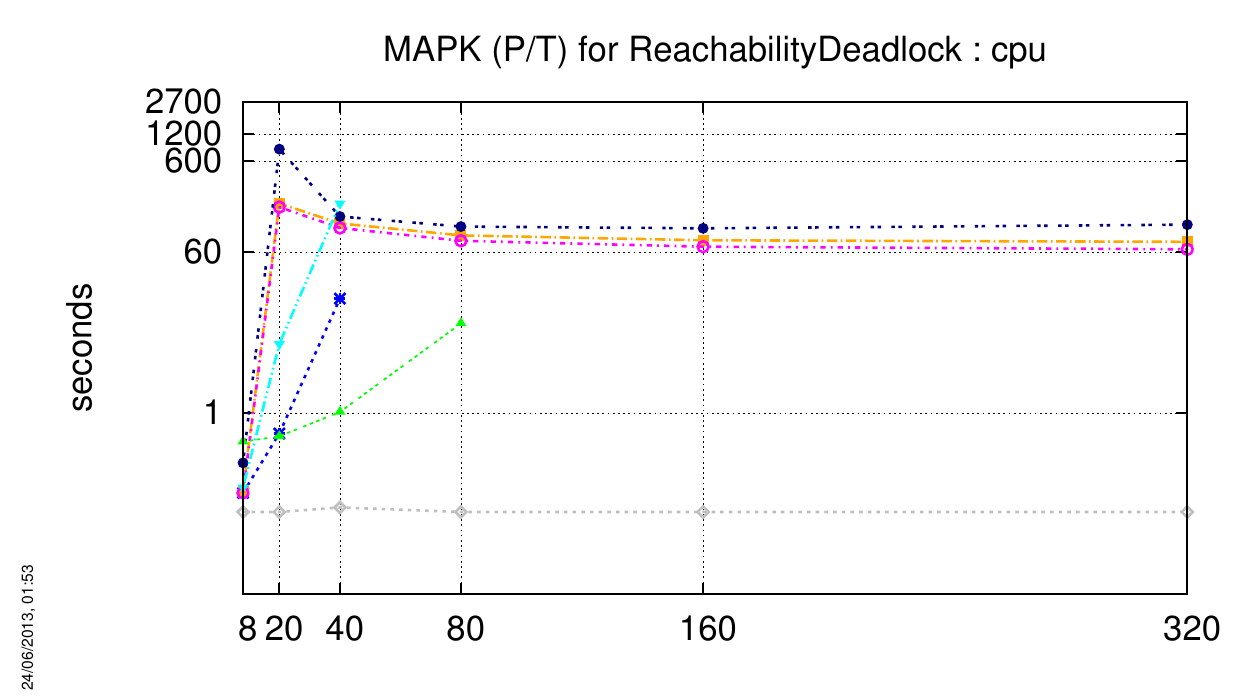}

   \includegraphics[height=1cm]{figures/tools-legend.pdf}
\end{center}

\subsubsection{\acs{NeoElection-COL}}
The charts below respectively show how tools compete with this ``Known'' model (memory and CPU).

\index{Performances!ReachabilityDeadlock!NeoElection (Colored)}
\begin{center}
   \includegraphics[width=7.2cm]{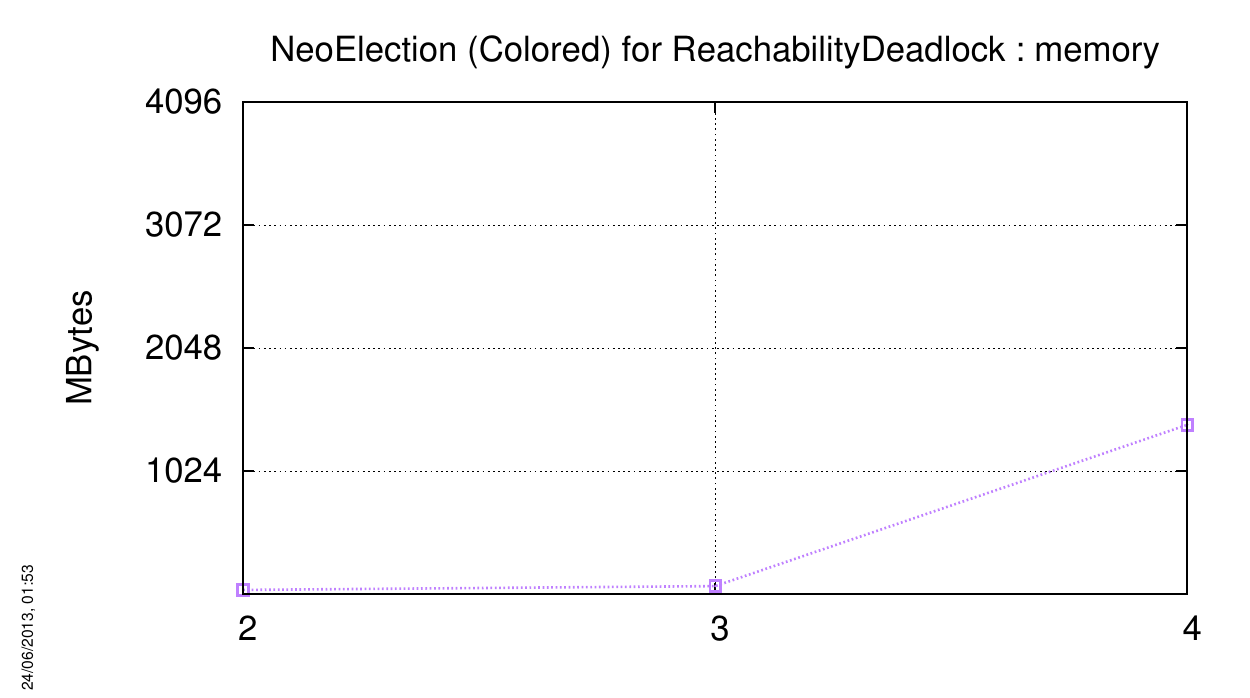}
   \includegraphics[width=7.2cm]{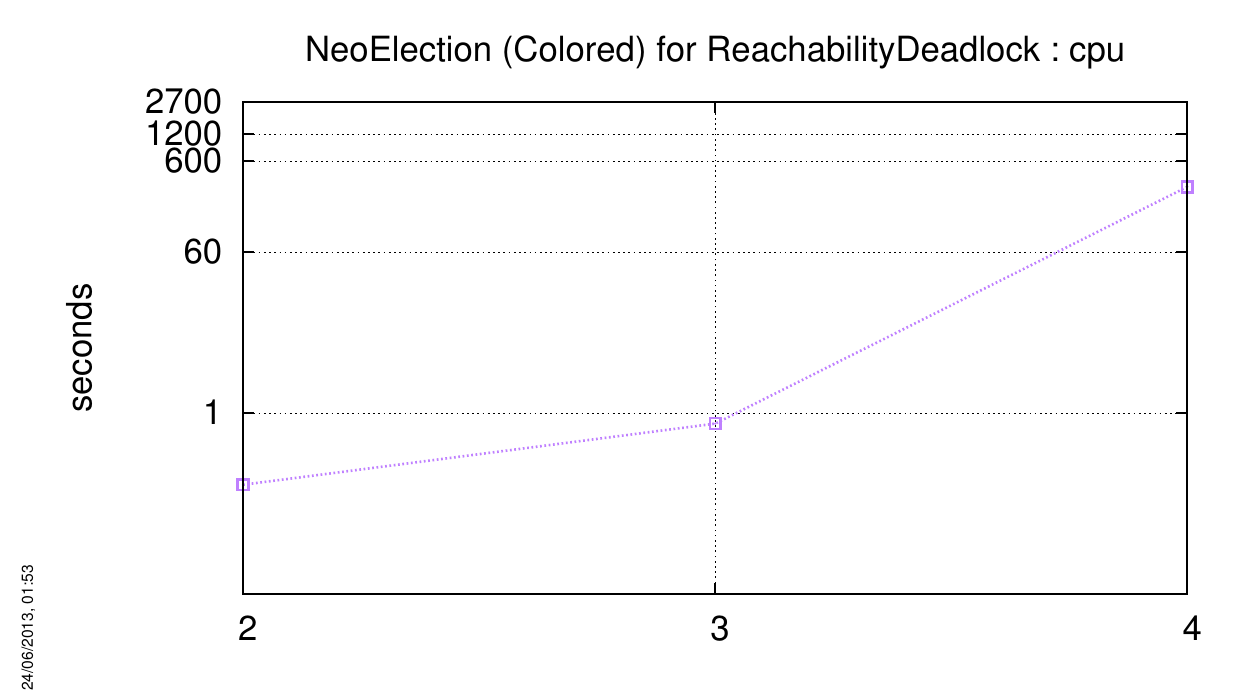}

   \includegraphics[height=1cm]{figures/tools-legend.pdf}
\end{center}

\subsubsection{\acs{NeoElection-PT}}
The charts below respectively show how tools compete with this ``Known'' model (memory and CPU).

\index{Performances!ReachabilityDeadlock!NeoElection (P/T)}
\begin{center}
   \includegraphics[width=7.2cm]{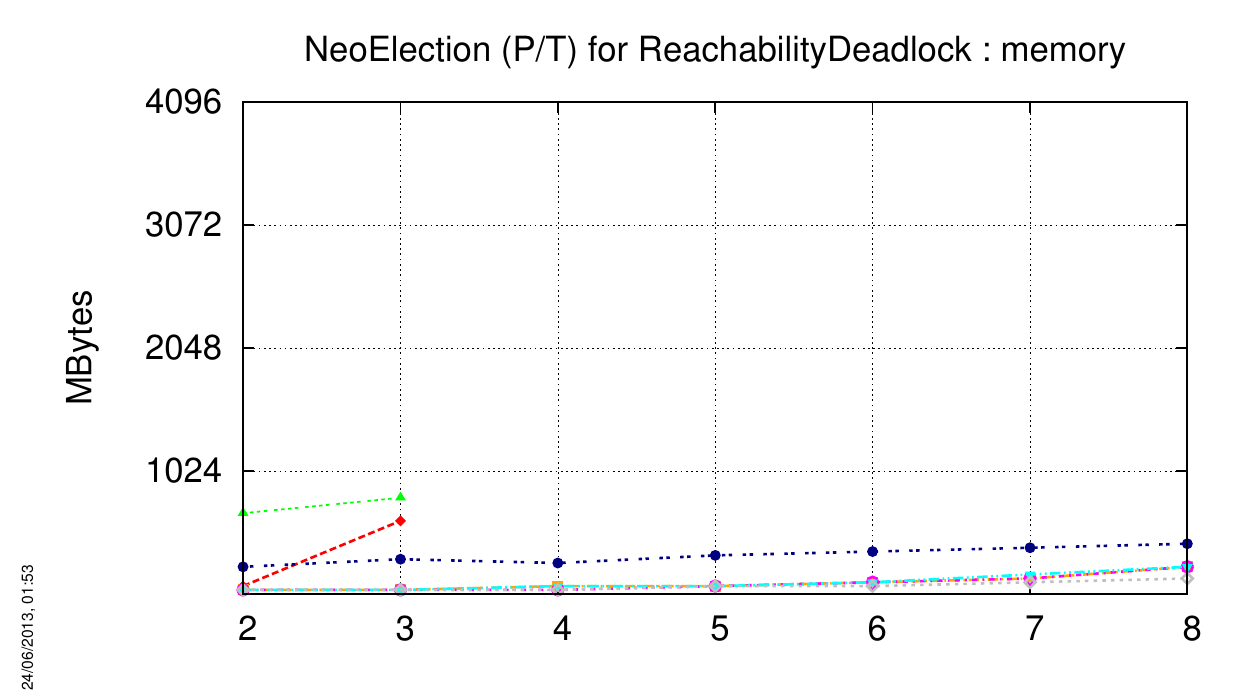}
   \includegraphics[width=7.2cm]{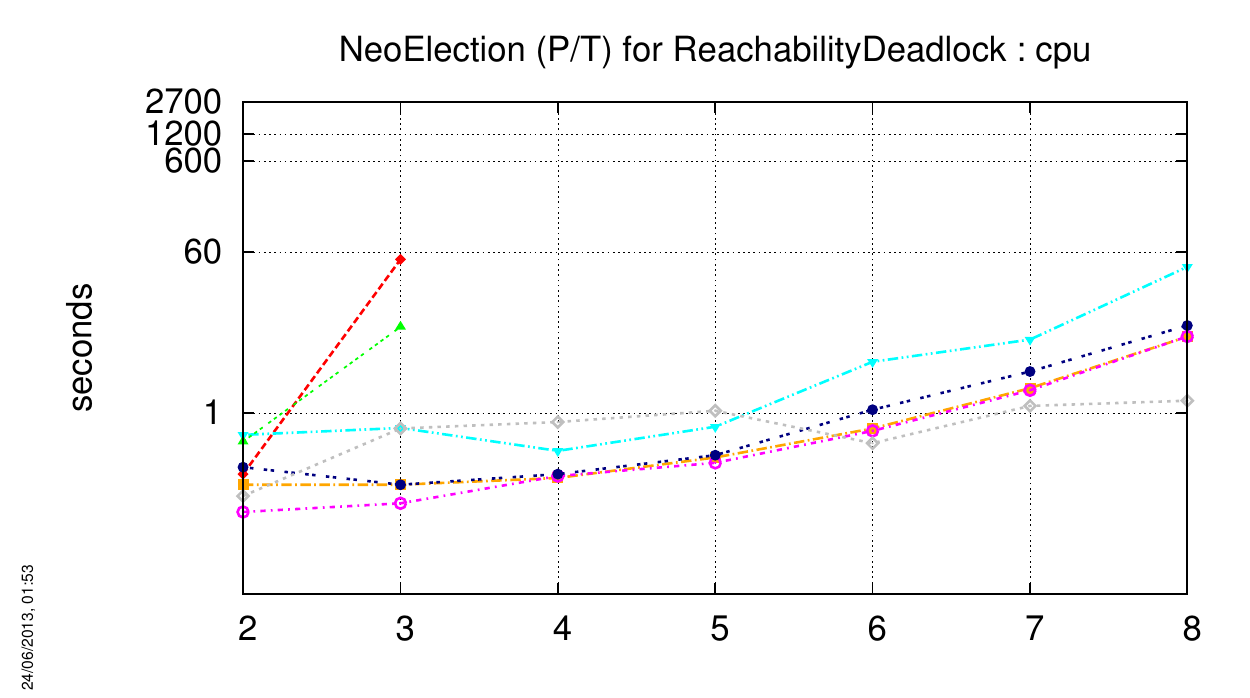}

   \includegraphics[height=1cm]{figures/tools-legend.pdf}
\end{center}

\subsubsection{\acs{PermAdmissibility-COL}}
The charts below respectively show how tools compete with this ``Known'' model (memory and CPU).

\index{Performances!ReachabilityDeadlock!PermAdmissibility (Colored)}
\begin{center}
   \includegraphics[width=7.2cm]{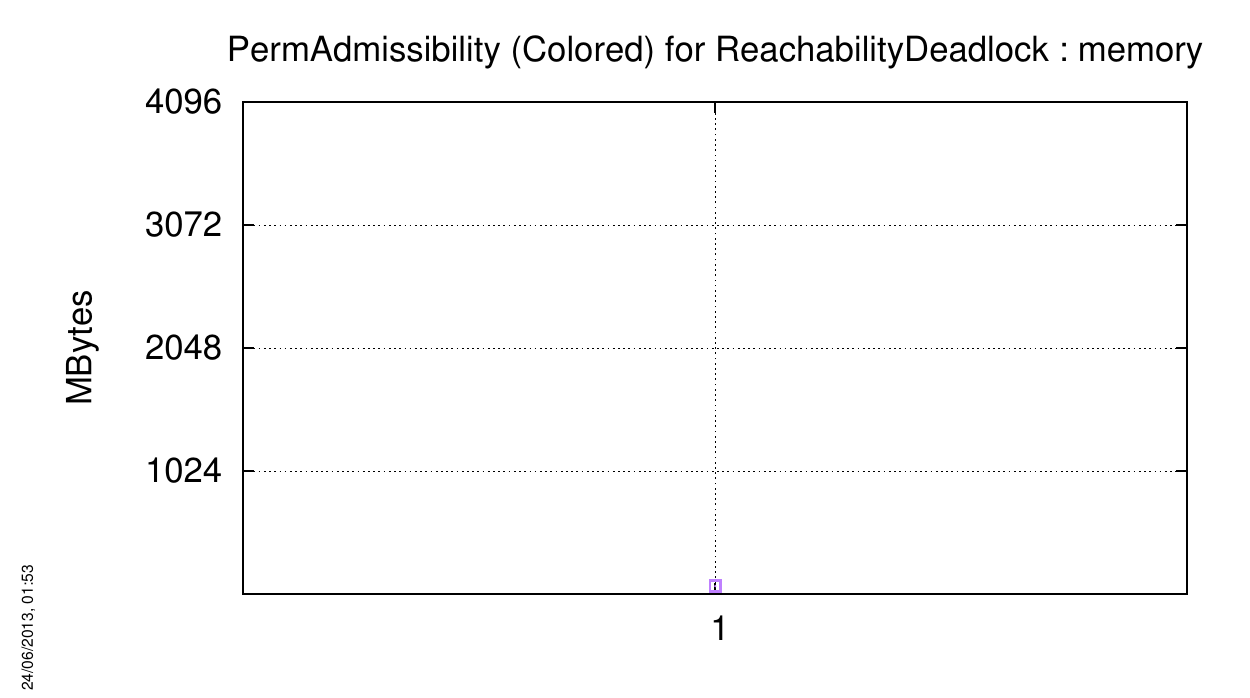}
   \includegraphics[width=7.2cm]{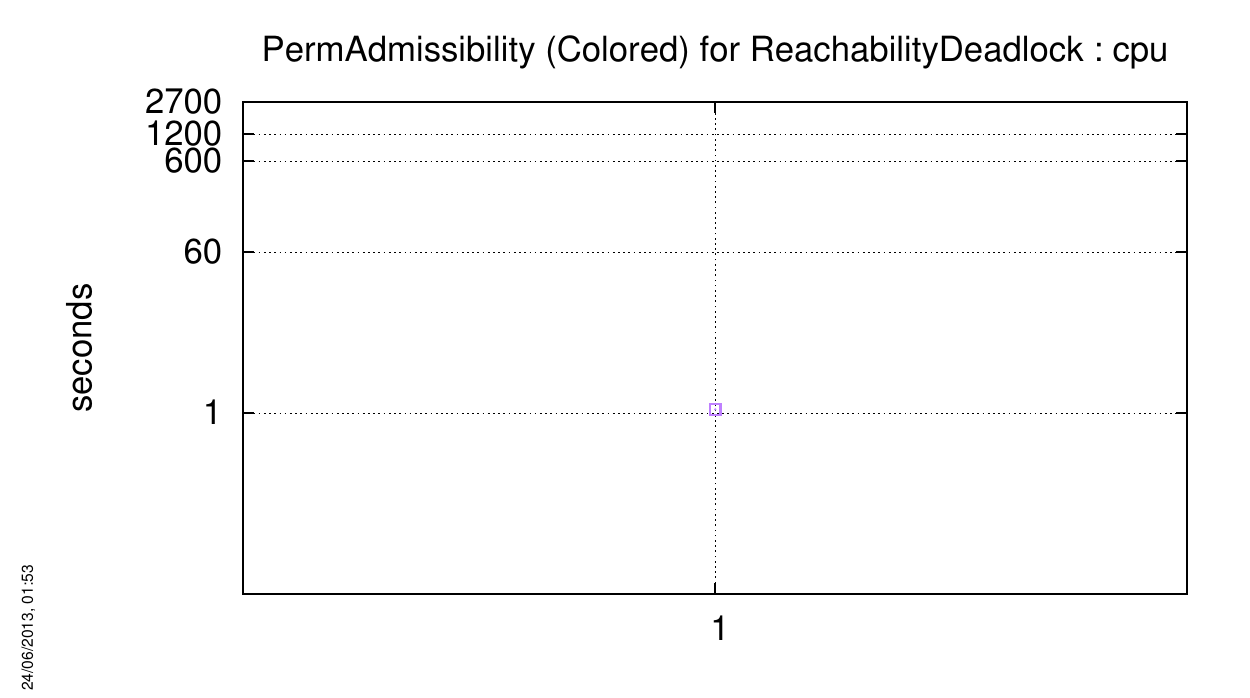}

   \includegraphics[height=1cm]{figures/tools-legend.pdf}
\end{center}

\subsubsection{\acs{PermAdmissibility-PT}}
The charts below respectively show how tools compete with this ``Known'' model (memory and CPU).

\index{Performances!ReachabilityDeadlock!PermAdmissibility (P/T)}
\begin{center}
   \includegraphics[width=7.2cm]{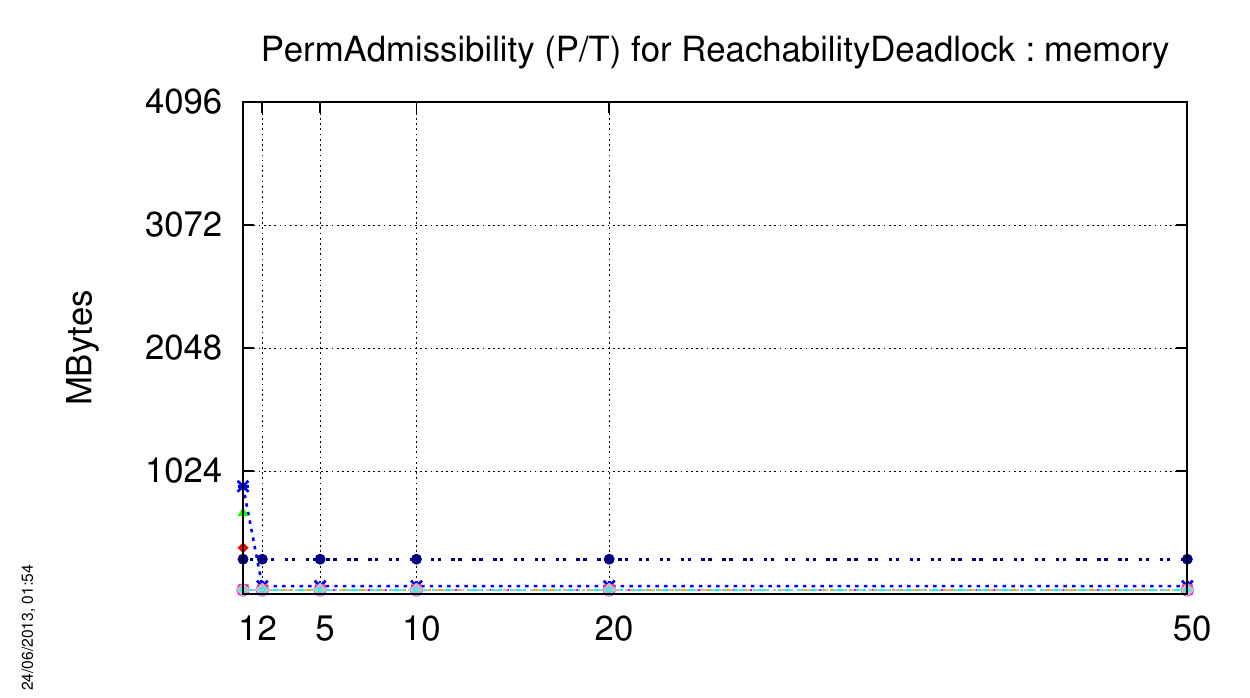}
   \includegraphics[width=7.2cm]{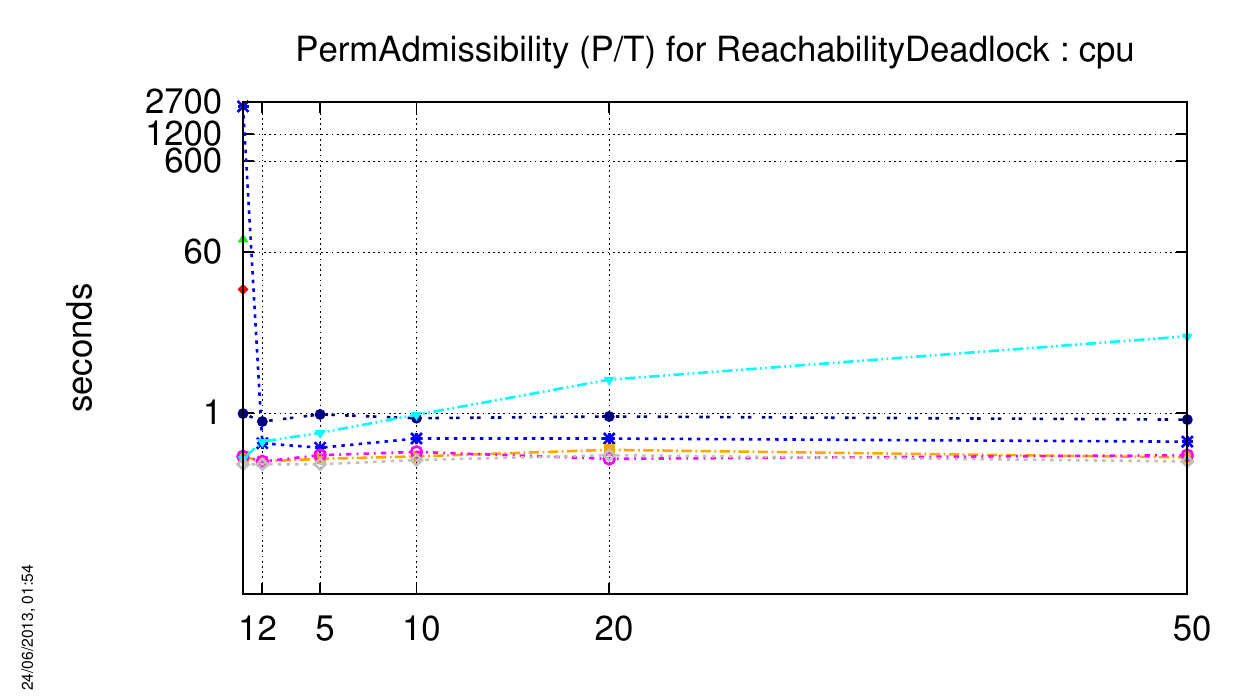}

   \includegraphics[height=1cm]{figures/tools-legend.pdf}
\end{center}

\subsubsection{\acs{Peterson-COL}}
The charts below respectively show how tools compete with this ``Known'' model (memory and CPU).

\index{Performances!ReachabilityDeadlock!Peterson (Colored)}
\begin{center}
   \includegraphics[width=7.2cm]{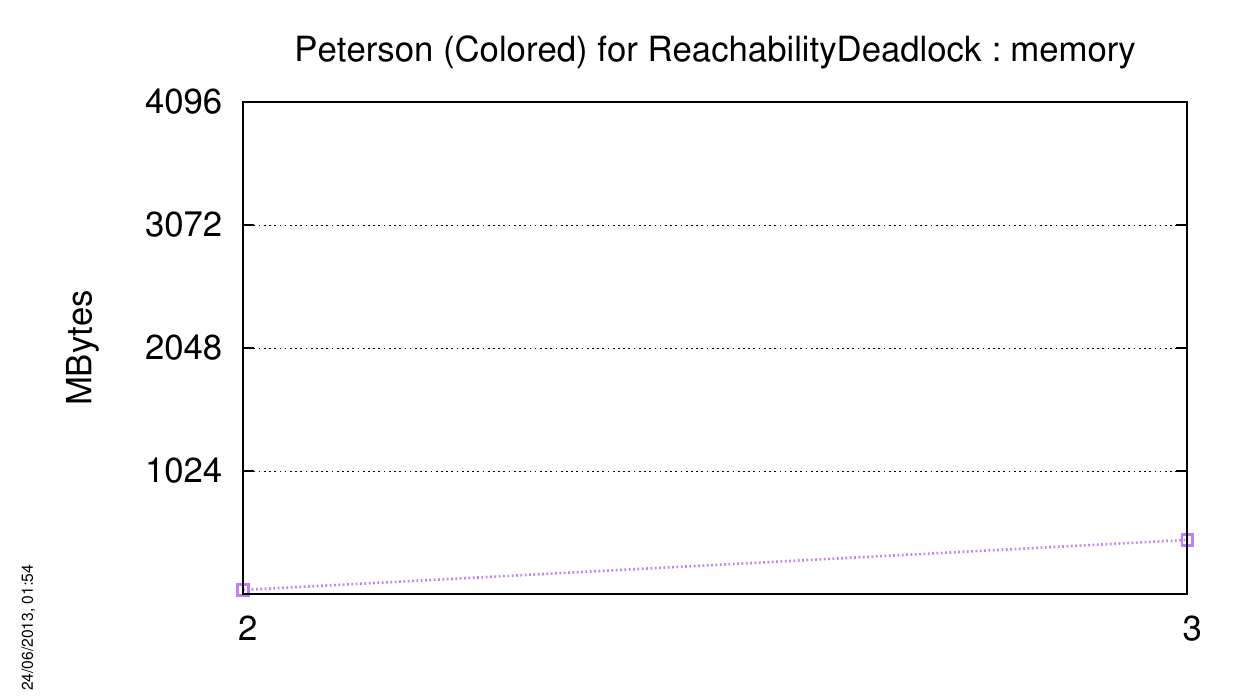}
   \includegraphics[width=7.2cm]{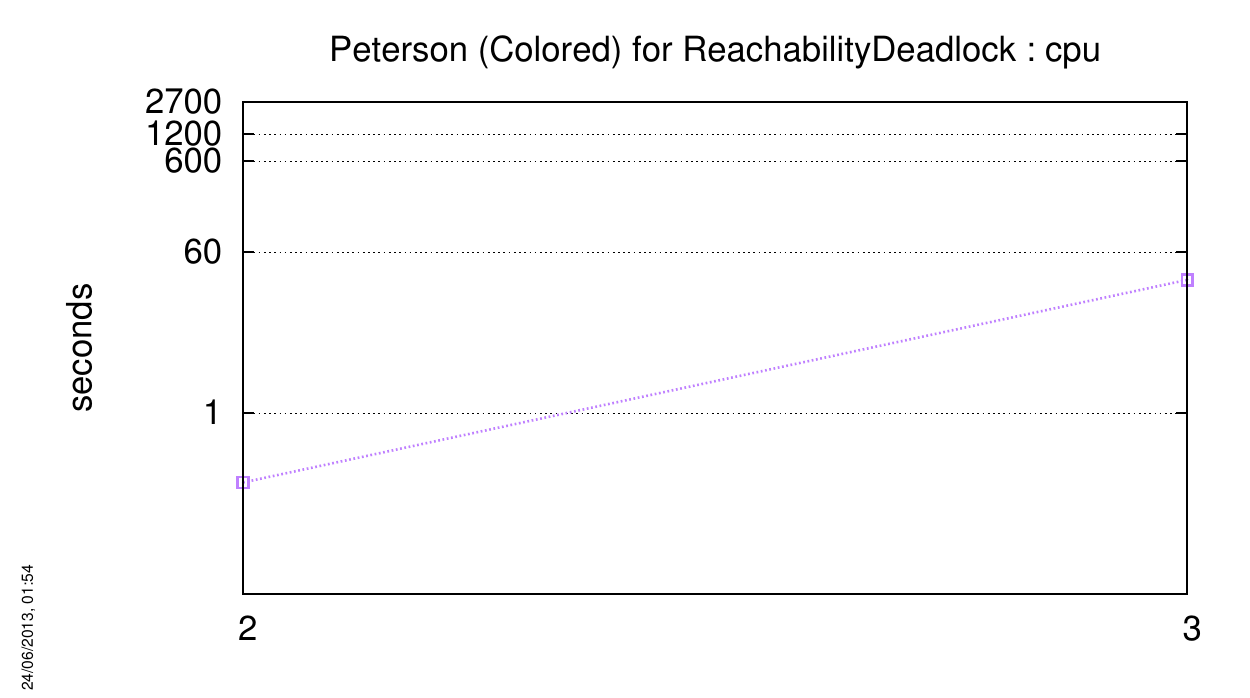}

   \includegraphics[height=1cm]{figures/tools-legend.pdf}
\end{center}

\subsubsection{\acs{Peterson-PT}}
The charts below respectively show how tools compete with this ``Known'' model (memory and CPU).

\index{Performances!ReachabilityDeadlock!Peterson (P/T)}
\begin{center}
   \includegraphics[width=7.2cm]{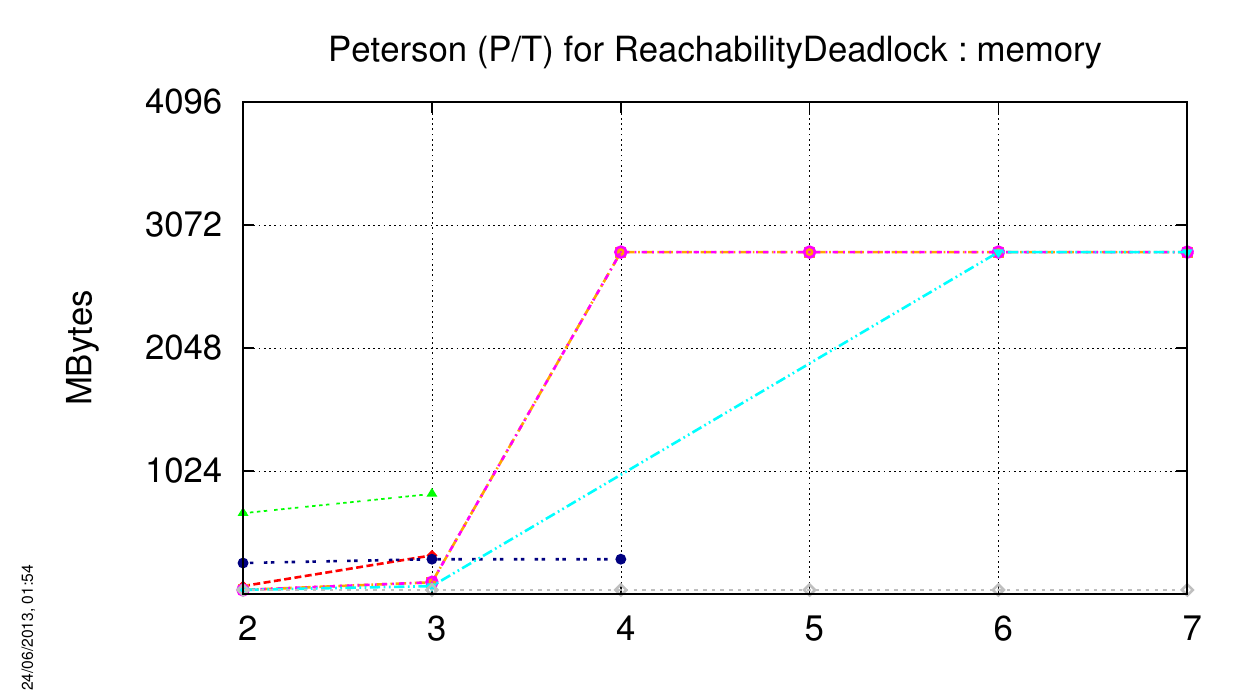}
   \includegraphics[width=7.2cm]{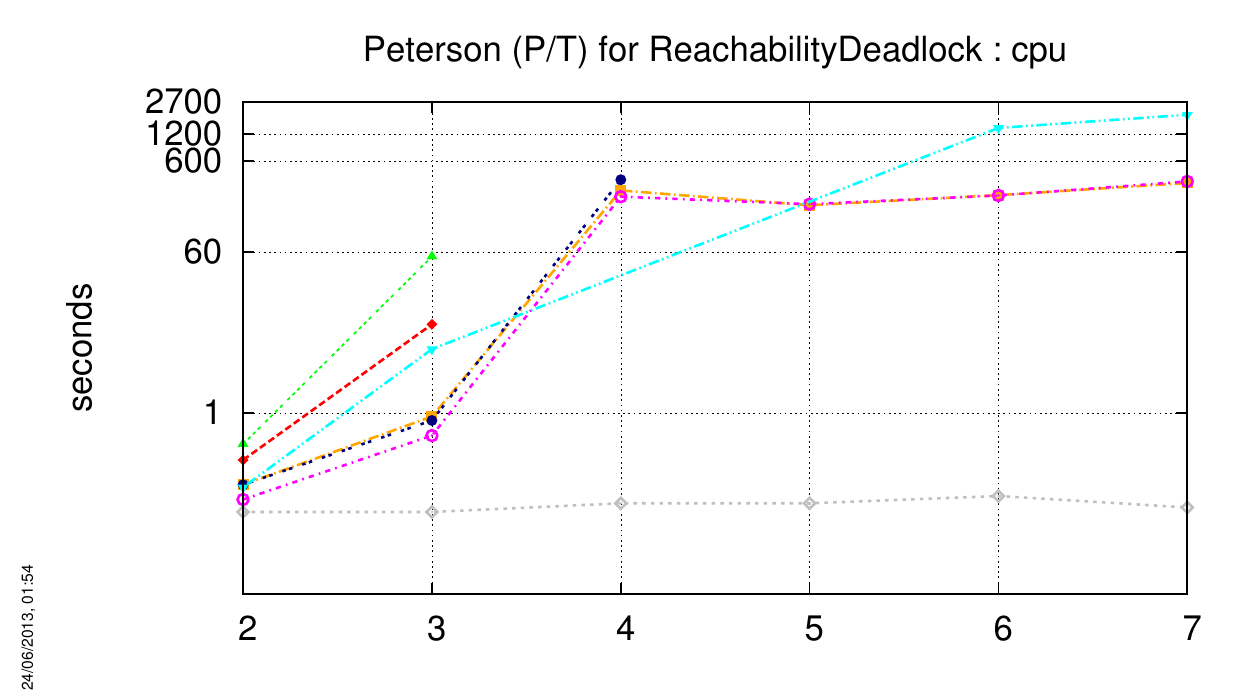}

   \includegraphics[height=1cm]{figures/tools-legend.pdf}
\end{center}

\subsubsection{\acs{Philosophers-COL}}
No instance of this model could be computed for the \textbf{ReachabilityDeadlock} examination.

\subsubsection{\acs{Philosophers-PT}}
The charts below respectively show how tools compete with this ``Known'' model (memory and CPU).

\index{Performances!ReachabilityDeadlock!Philosophers (P/T)}
\begin{center}
   \includegraphics[width=7.2cm]{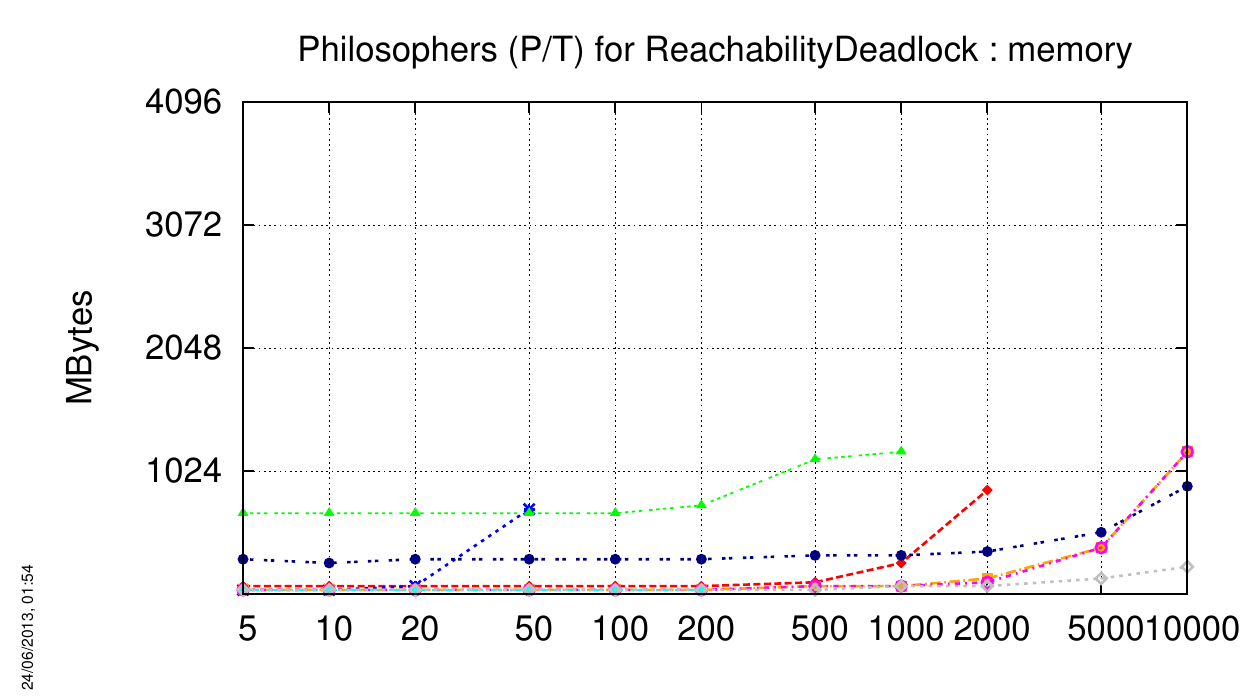}
   \includegraphics[width=7.2cm]{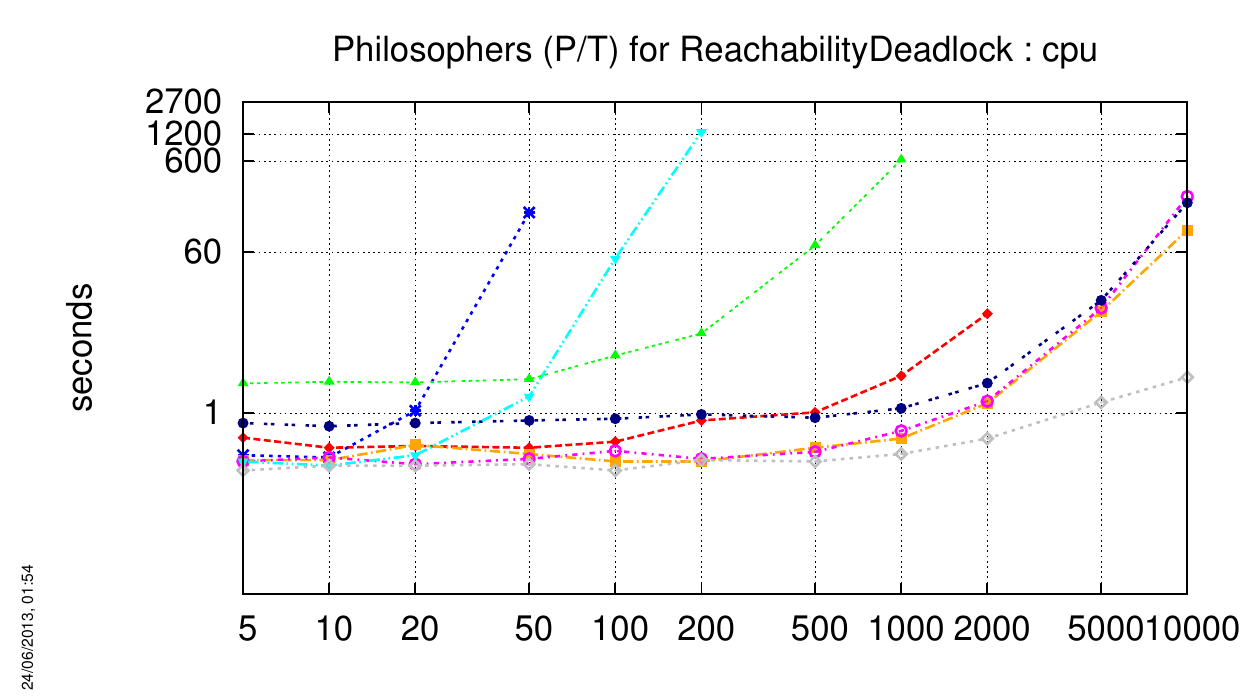}

   \includegraphics[height=1cm]{figures/tools-legend.pdf}
\end{center}

\subsubsection{\acs{PhilosophersDyn-COL}}
No instance of this model could be computed for the \textbf{ReachabilityDeadlock} examination.

\subsubsection{\acs{PhilosophersDyn-PT}}
The charts below respectively show how tools compete with this ``Known'' model (memory and CPU).

\index{Performances!ReachabilityDeadlock!PhilosophersDyn (P/T)}
\begin{center}
   \includegraphics[width=7.2cm]{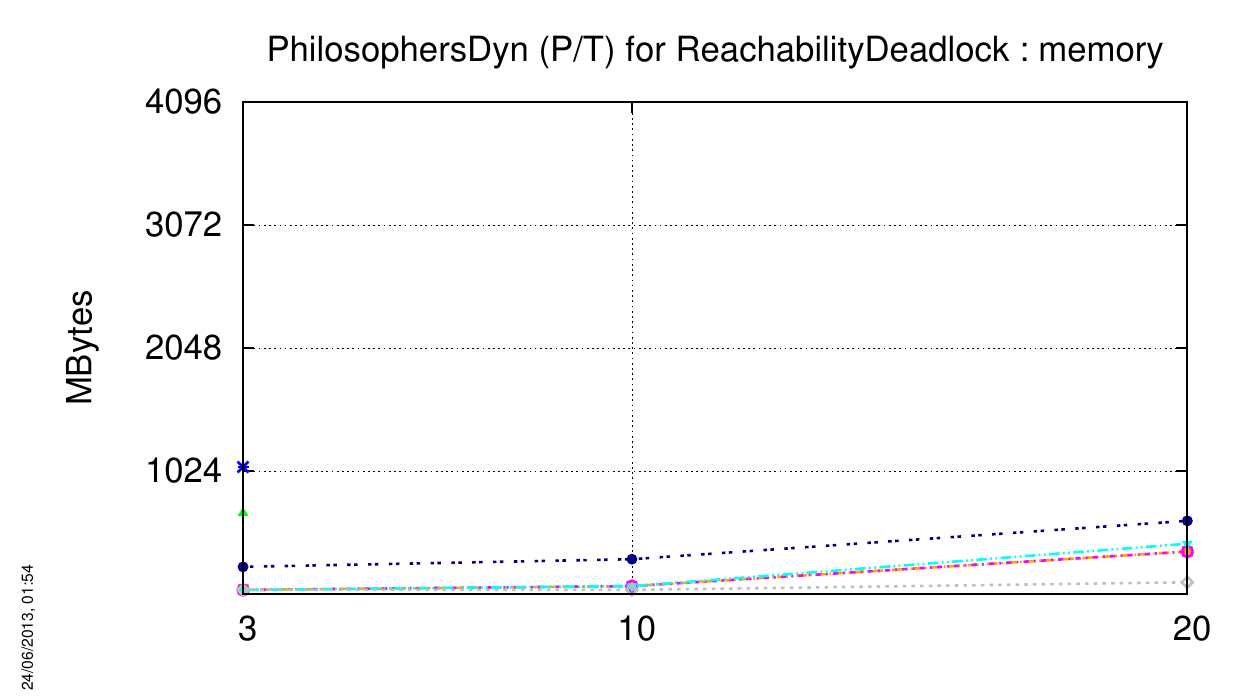}
   \includegraphics[width=7.2cm]{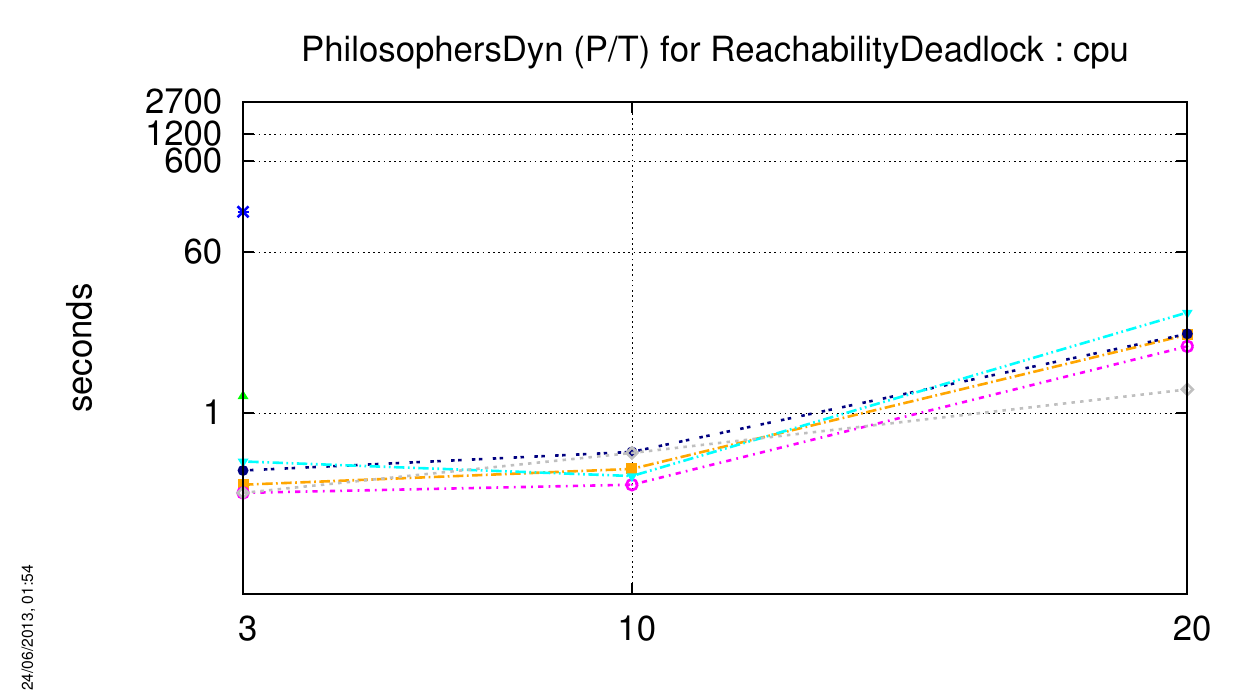}

   \includegraphics[height=1cm]{figures/tools-legend.pdf}
\end{center}

\subsubsection{\acs{Planning-PT}}
No instance of this model could be computed for the \textbf{ReachabilityDeadlock} examination.

\subsubsection{\acs{Railroad-PT}}
The charts below respectively show how tools compete with this ``Known'' model (memory and CPU).

\index{Performances!ReachabilityDeadlock!Railroad (P/T)}
\begin{center}
   \includegraphics[width=7.2cm]{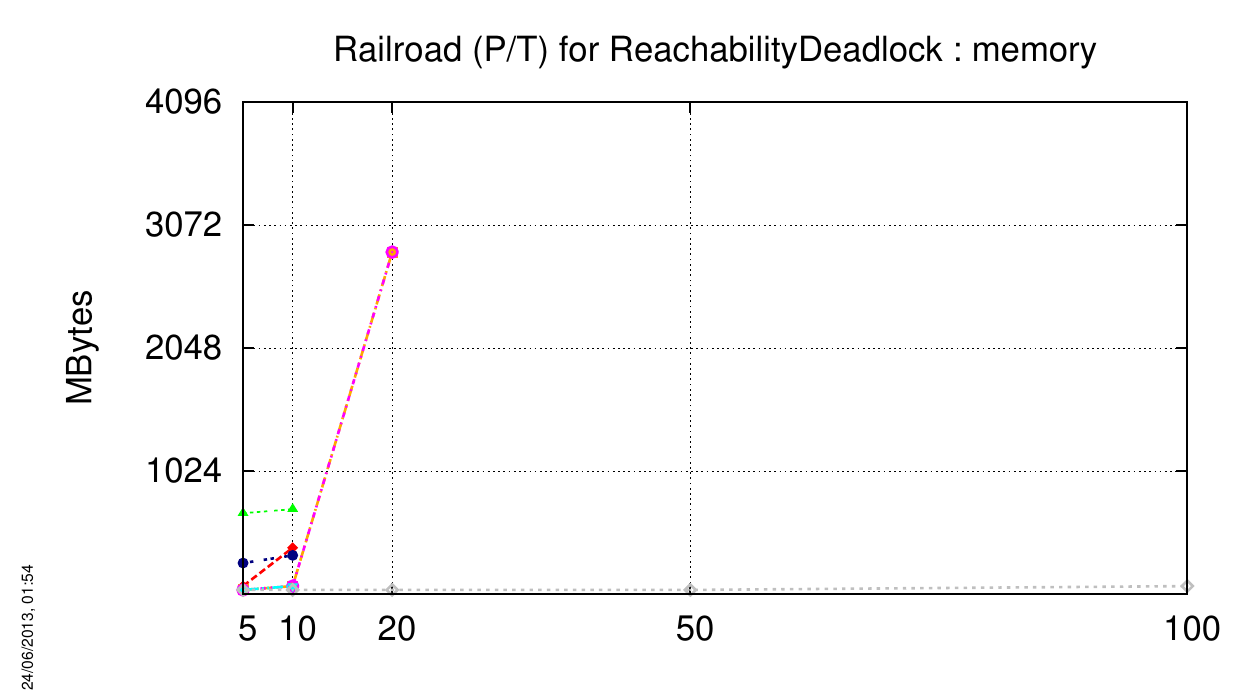}
   \includegraphics[width=7.2cm]{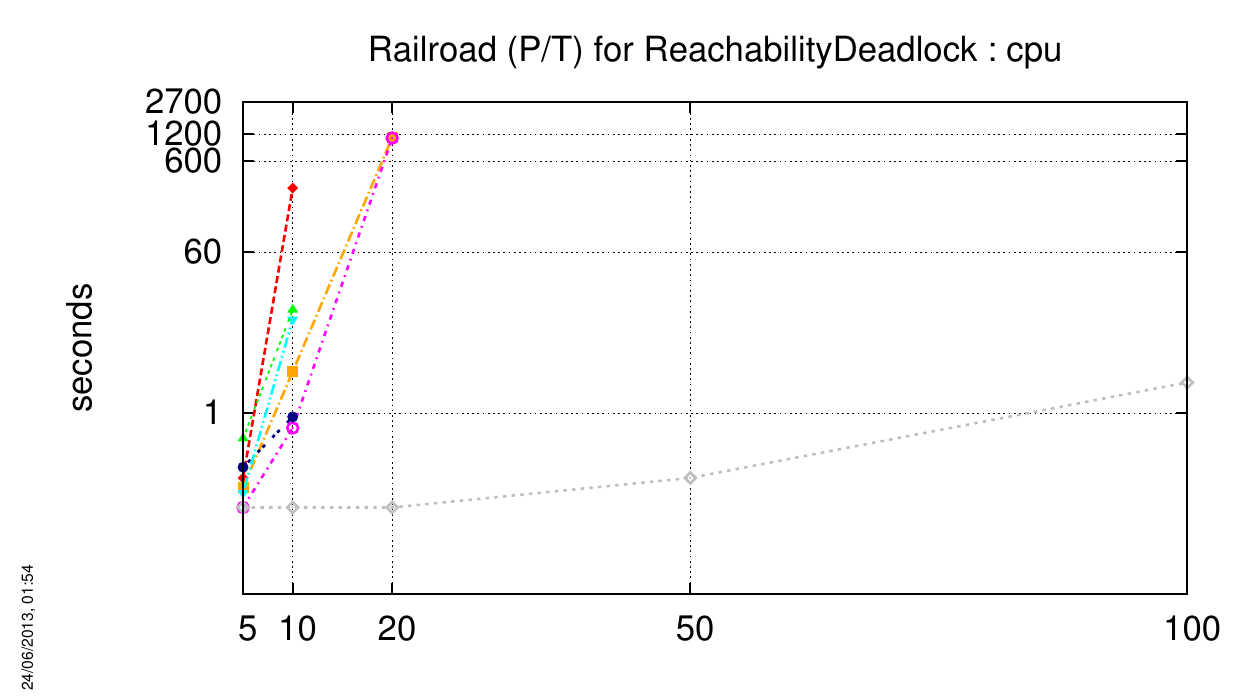}

   \includegraphics[height=1cm]{figures/tools-legend.pdf}
\end{center}

\subsubsection{\acs{RessAllocation-PT}}
The charts below respectively show how tools compete with this ``Known'' model (memory and CPU).

\index{Performances!ReachabilityDeadlock!RessAllocation (P/T)}
\begin{center}
   \includegraphics[width=7.2cm]{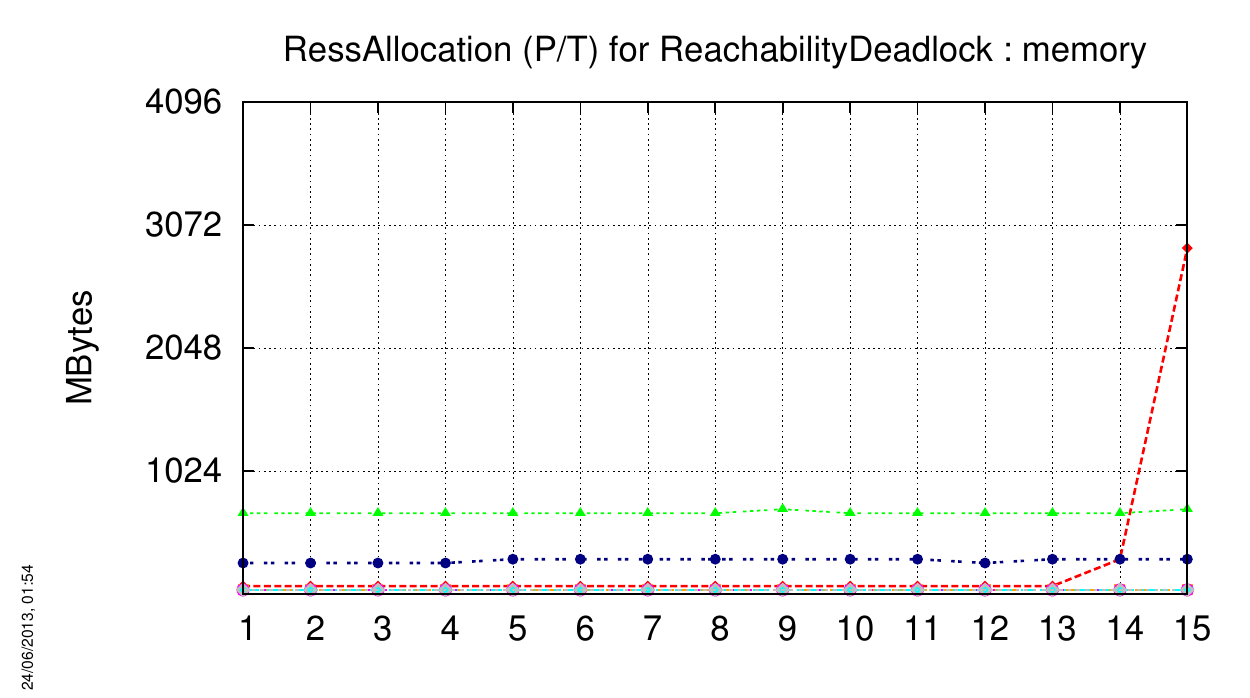}
   \includegraphics[width=7.2cm]{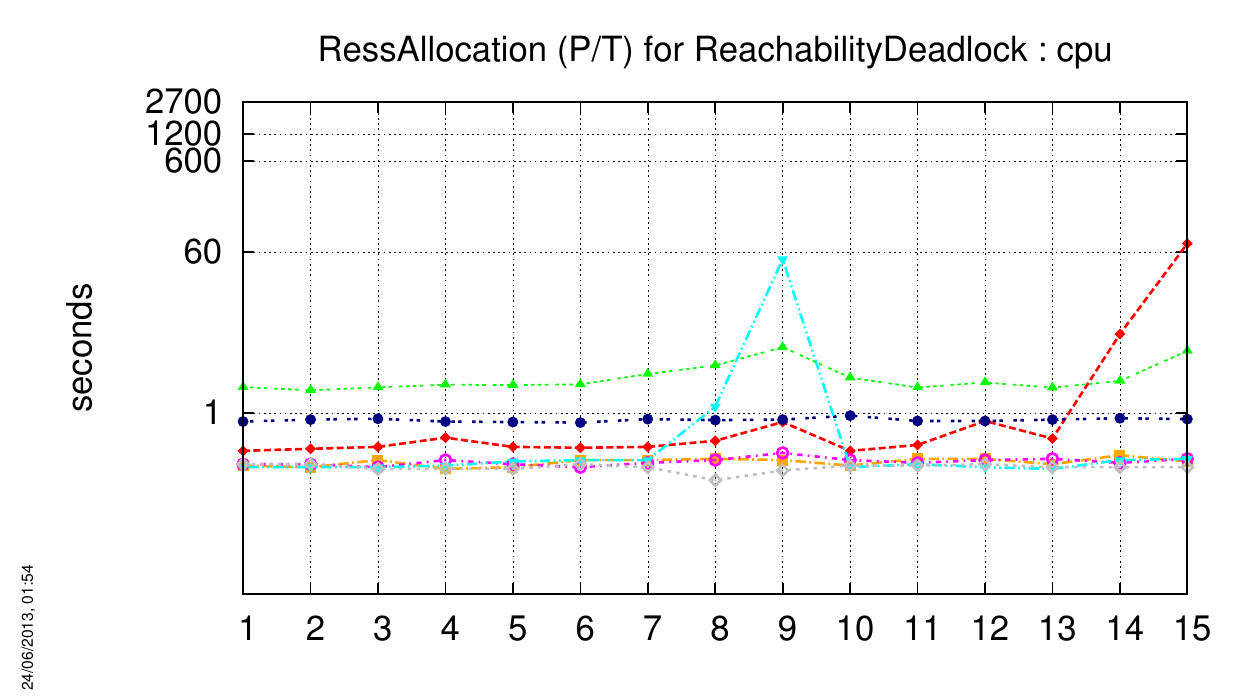}

   \includegraphics[height=1cm]{figures/tools-legend.pdf}
\end{center}

\subsubsection{\acs{Ring-PT}}
The charts below respectively show how tools compete with this ``Known'' model (memory and CPU).

\index{Performances!ReachabilityDeadlock!Ring (P/T)}
\begin{center}
   \includegraphics[width=7.2cm]{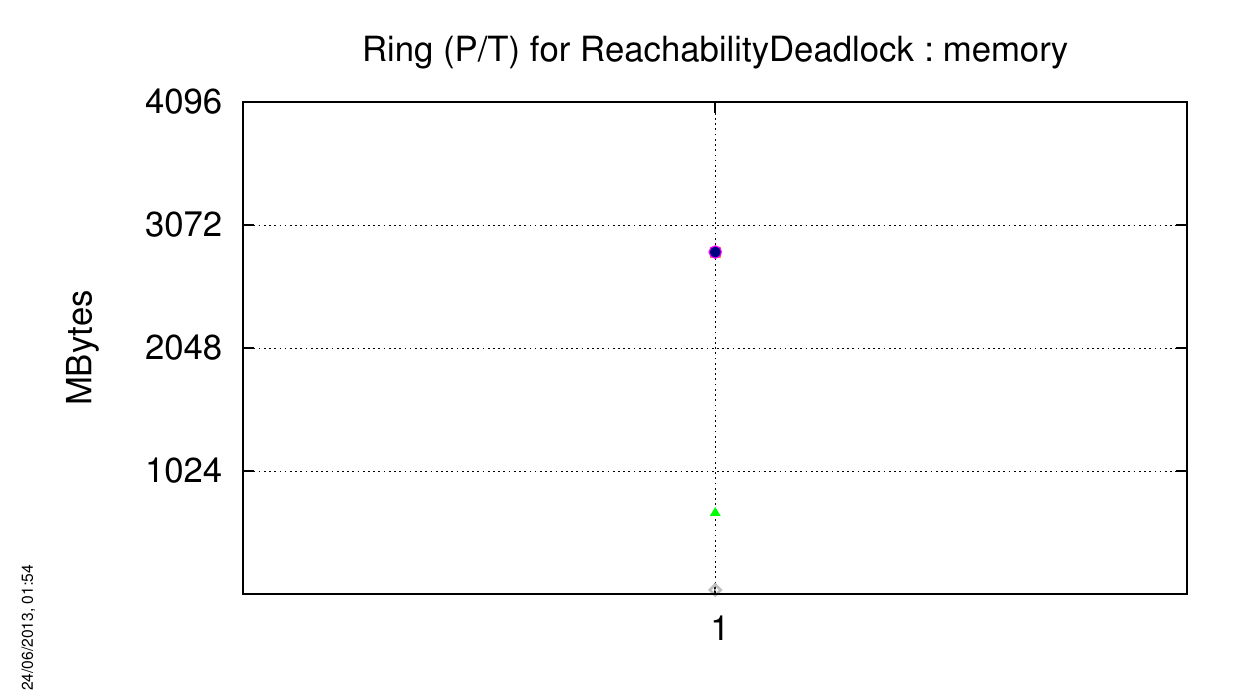}
   \includegraphics[width=7.2cm]{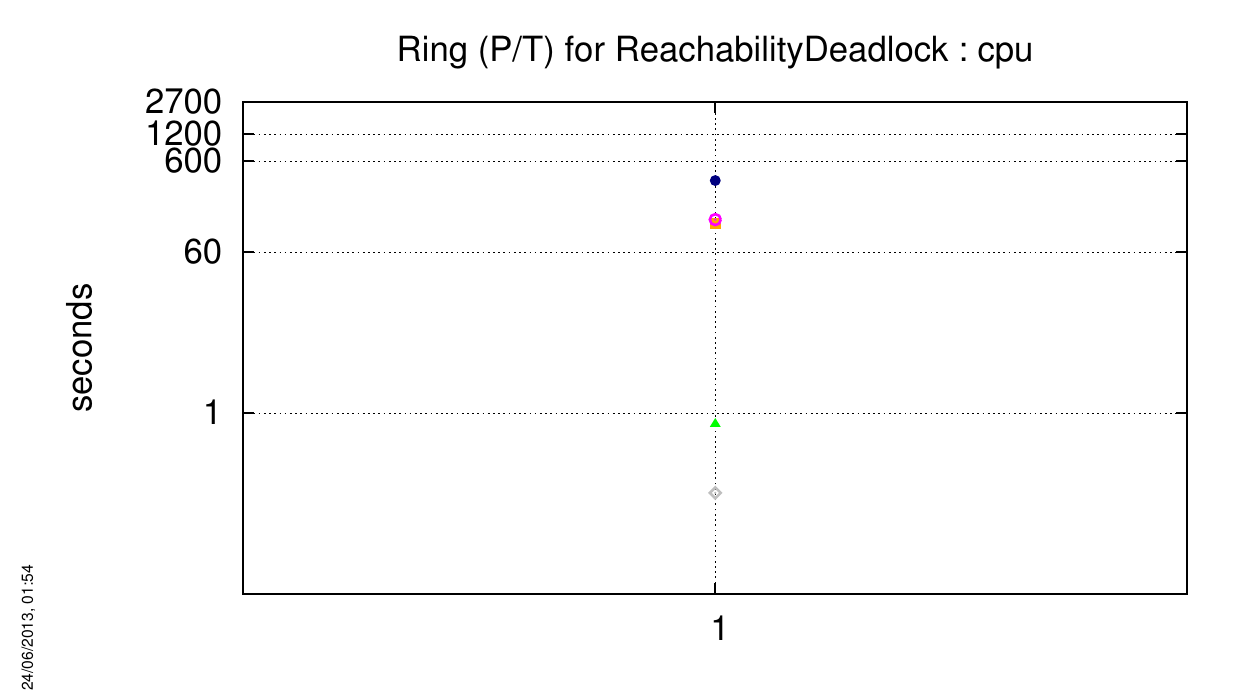}

   \includegraphics[height=1cm]{figures/tools-legend.pdf}
\end{center}

\subsubsection{\acs{RwMutex-PT}}
The charts below respectively show how tools compete with this ``Known'' model (memory and CPU).

\index{Performances!ReachabilityDeadlock!RwMutex (P/T)}
\begin{center}
   \includegraphics[width=7.2cm]{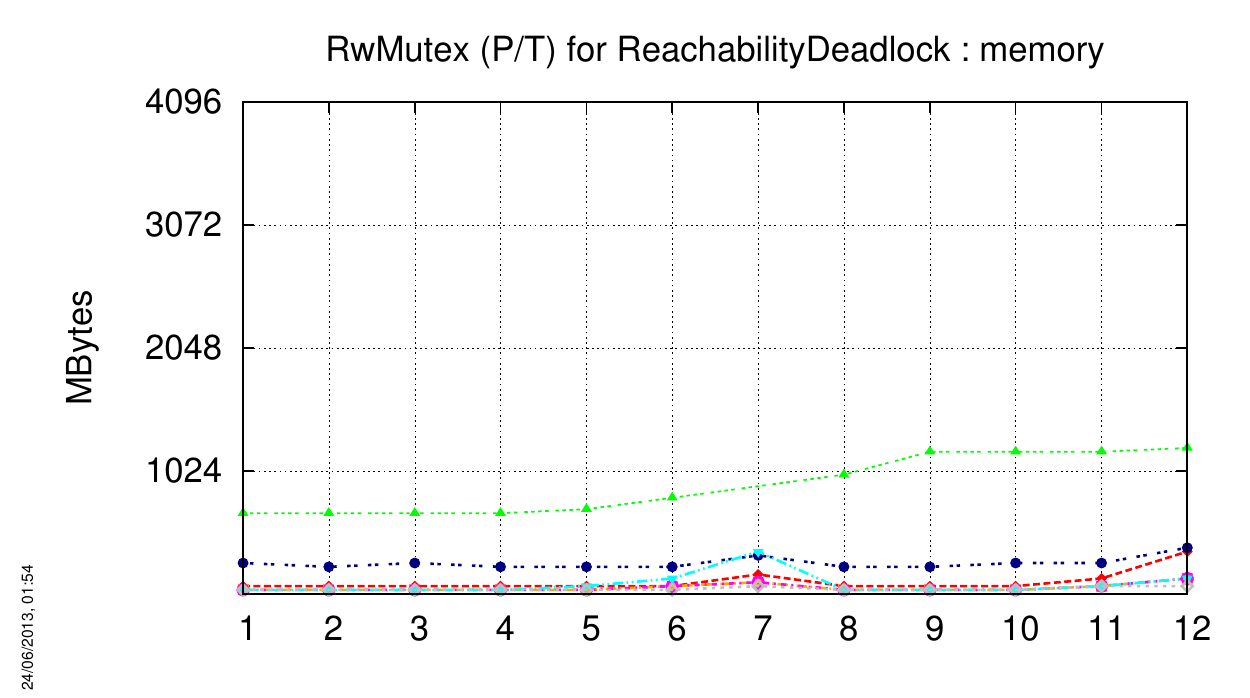}
   \includegraphics[width=7.2cm]{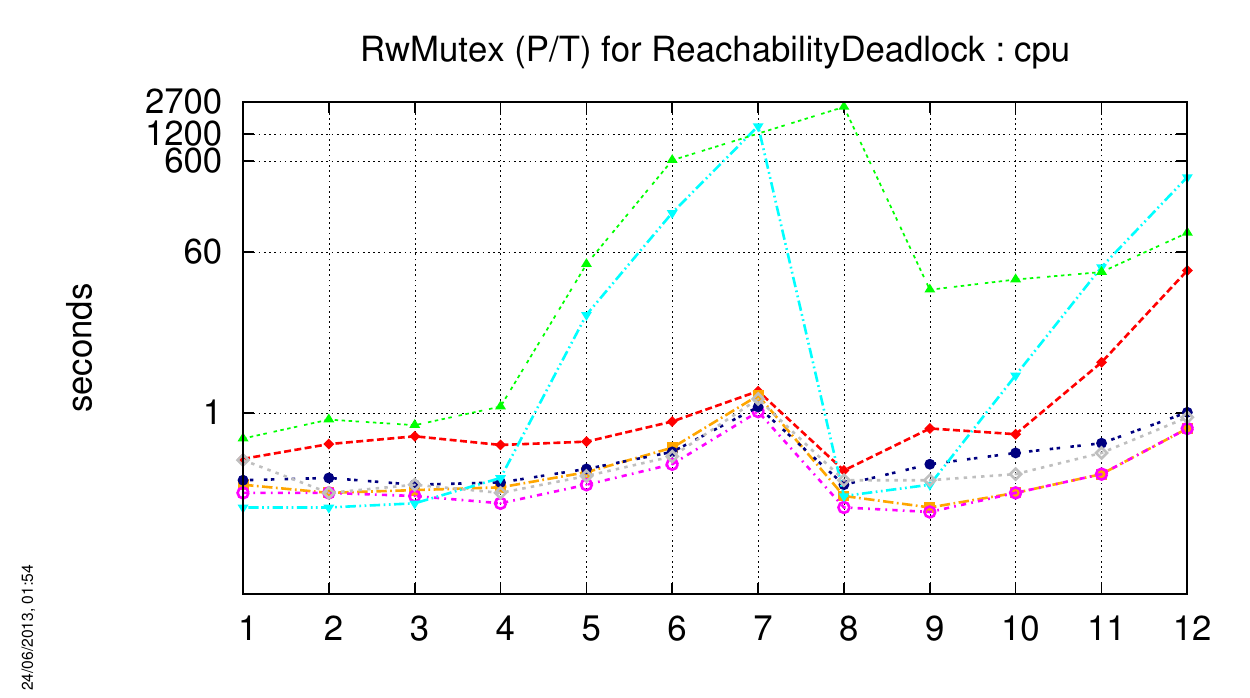}

   \includegraphics[height=1cm]{figures/tools-legend.pdf}
\end{center}

\subsubsection{\acs{SharedMemory-COL}}
No instance of this model could be computed for the \textbf{ReachabilityDeadlock} examination.

\subsubsection{\acs{SharedMemory-PT}}
The charts below respectively show how tools compete with this ``Known'' model (memory and CPU).

\index{Performances!ReachabilityDeadlock!SharedMemory (P/T)}
\begin{center}
   \includegraphics[width=7.2cm]{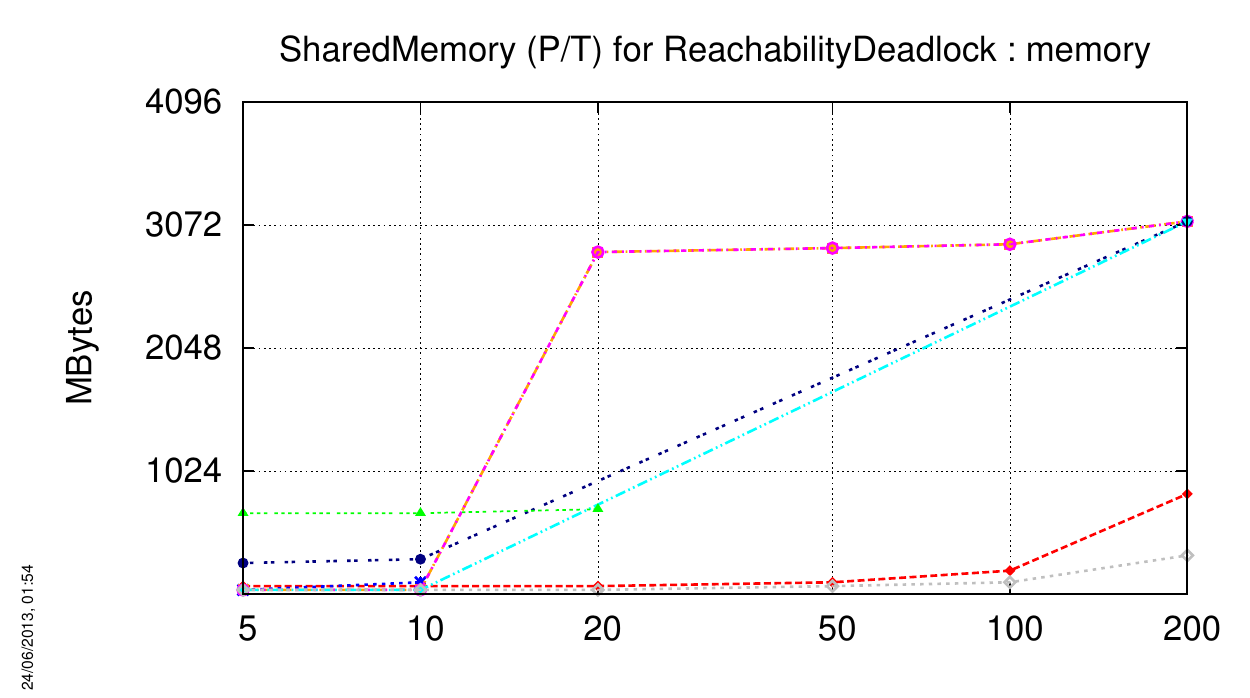}
   \includegraphics[width=7.2cm]{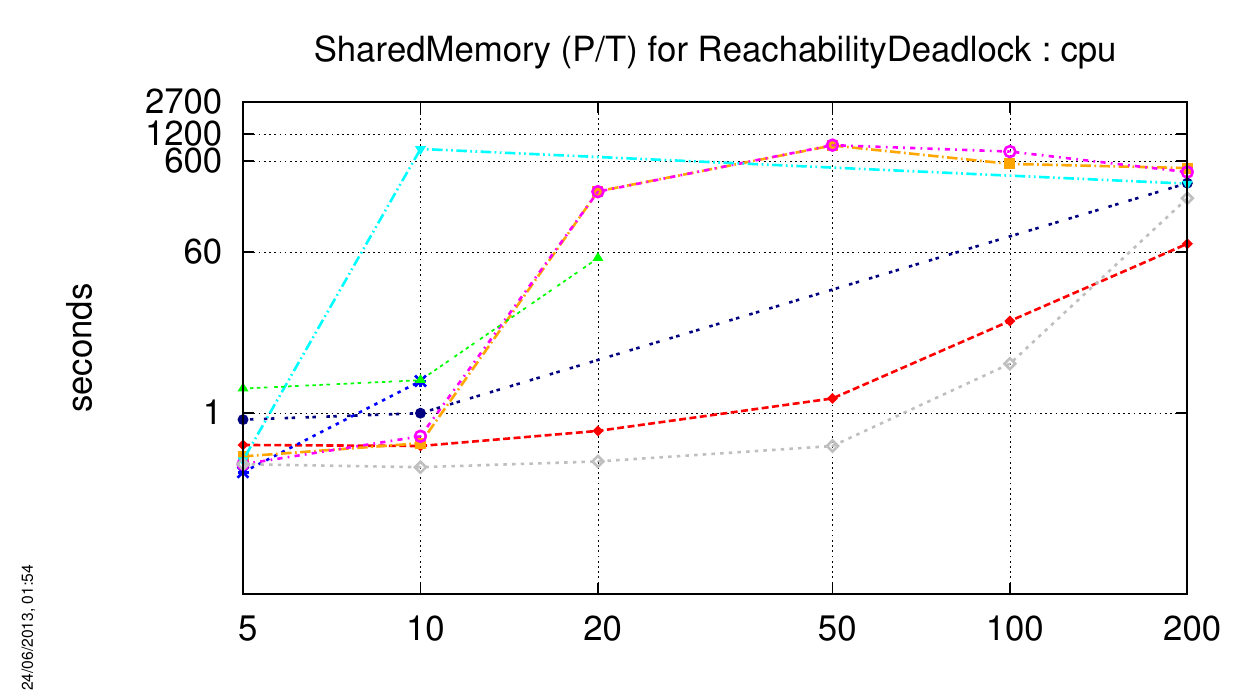}

   \includegraphics[height=1cm]{figures/tools-legend.pdf}
\end{center}

\subsubsection{\acs{SimpleLoadBal-COL}}
The charts below respectively show how tools compete with this ``Known'' model (memory and CPU).

\index{Performances!ReachabilityDeadlock!SimpleLoadBal (Colored)}
\begin{center}
   \includegraphics[width=7.2cm]{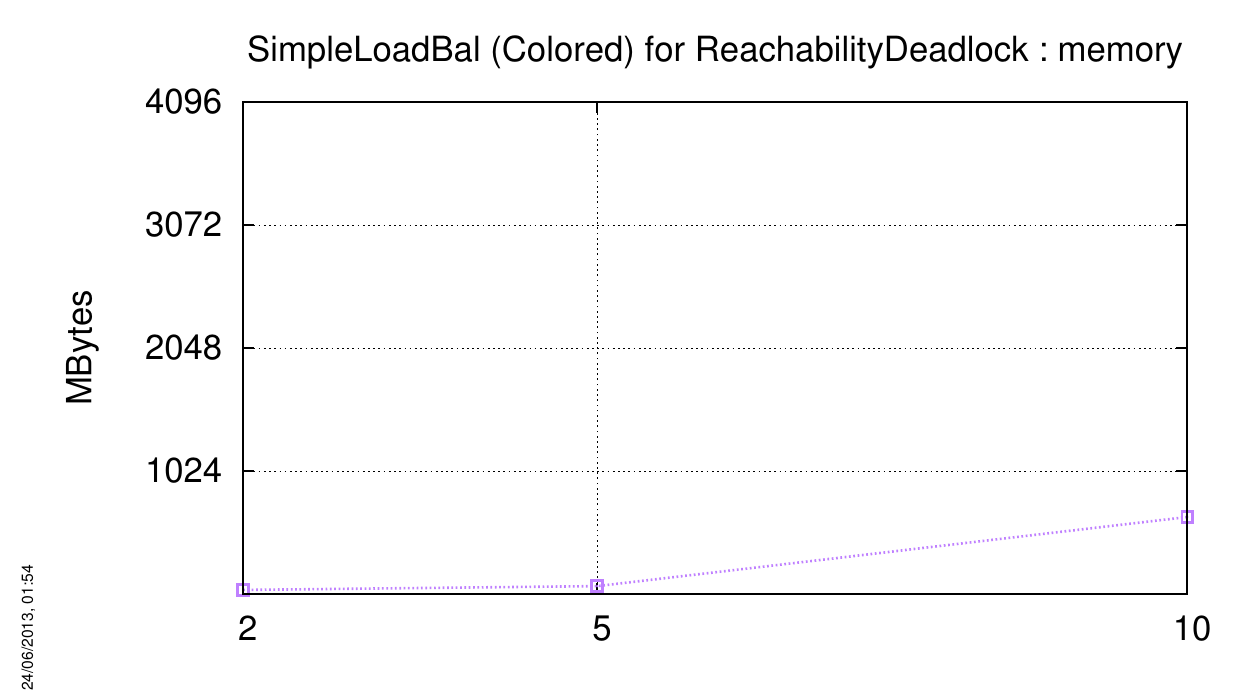}
   \includegraphics[width=7.2cm]{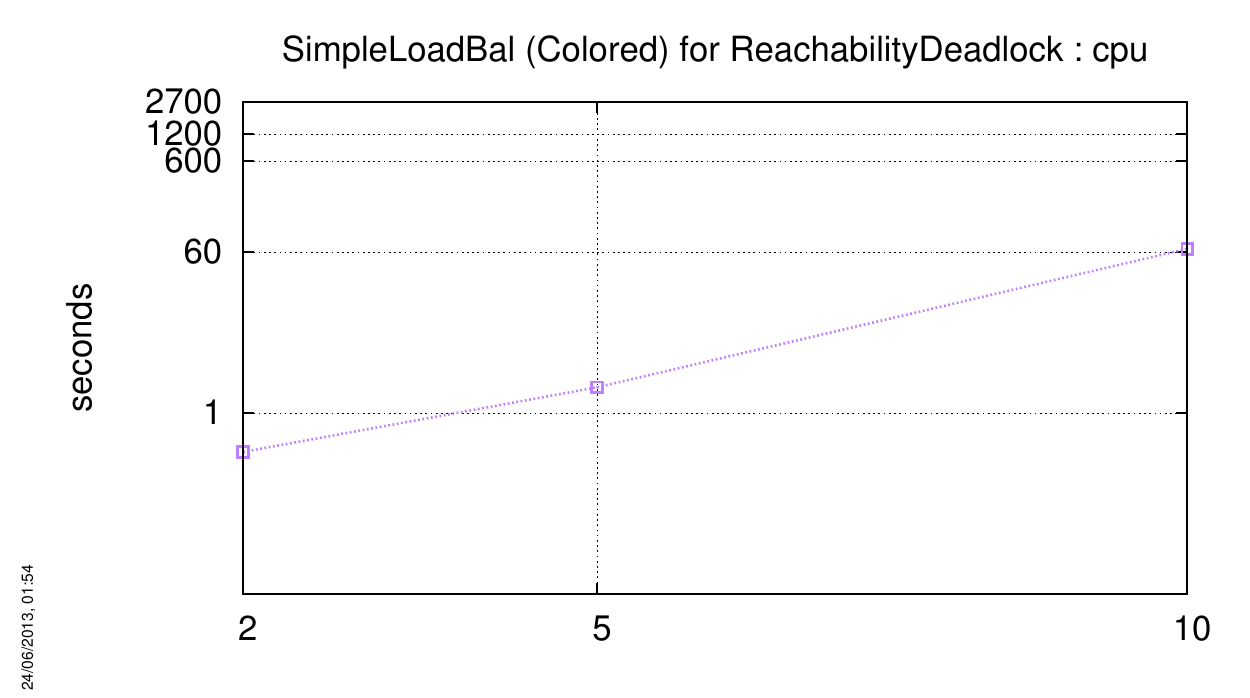}

   \includegraphics[height=1cm]{figures/tools-legend.pdf}
\end{center}

\subsubsection{\acs{SimpleLoadBal-PT}}
The charts below respectively show how tools compete with this ``Known'' model (memory and CPU).

\index{Performances!ReachabilityDeadlock!SimpleLoadBal (P/T)}
\begin{center}
   \includegraphics[width=7.2cm]{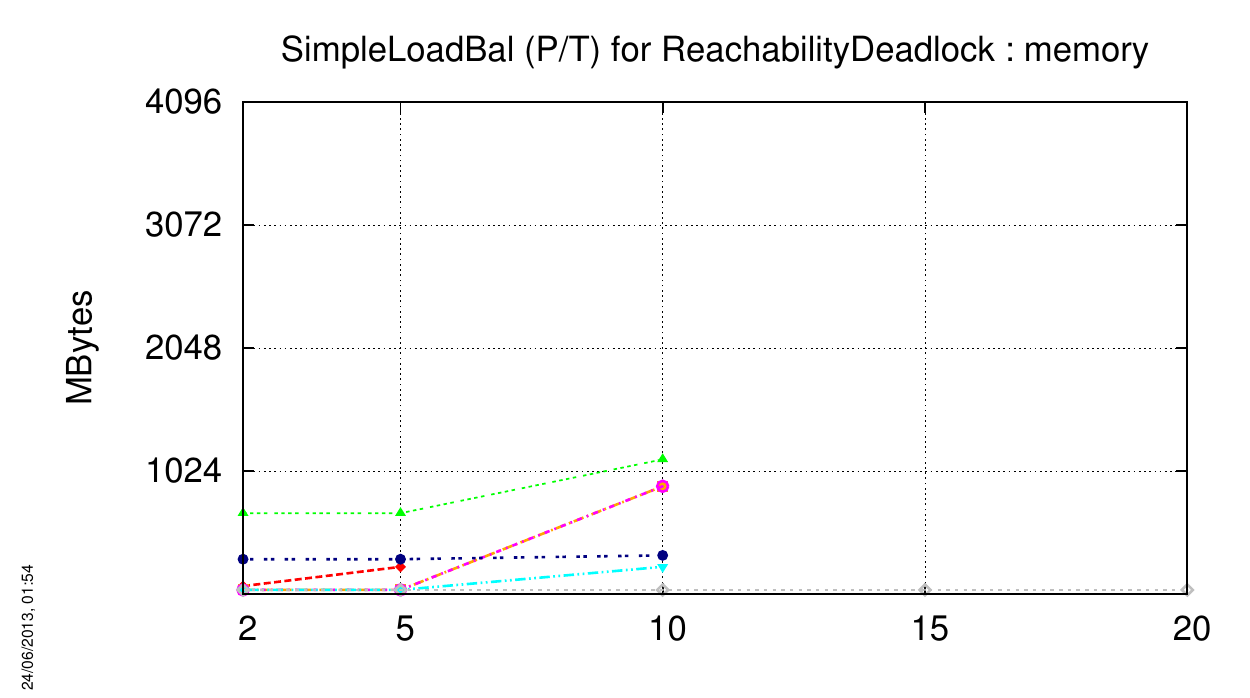}
   \includegraphics[width=7.2cm]{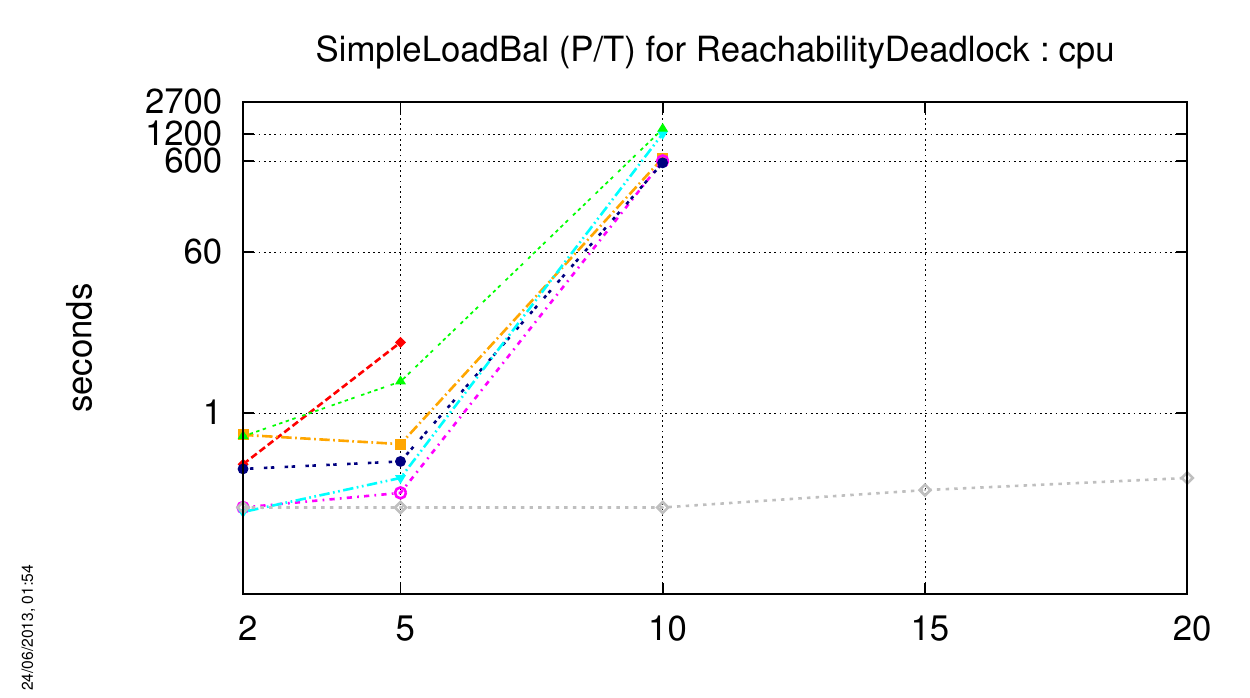}

   \includegraphics[height=1cm]{figures/tools-legend.pdf}
\end{center}

\subsubsection{\acs{TokenRing-COL}}
No instance of this model could be computed for the \textbf{ReachabilityDeadlock} examination.

\subsubsection{\acs{TokenRing-PT}}
The charts below respectively show how tools compete with this ``Known'' model (memory and CPU).

\index{Performances!ReachabilityDeadlock!TokenRing (P/T)}
\begin{center}
   \includegraphics[width=7.2cm]{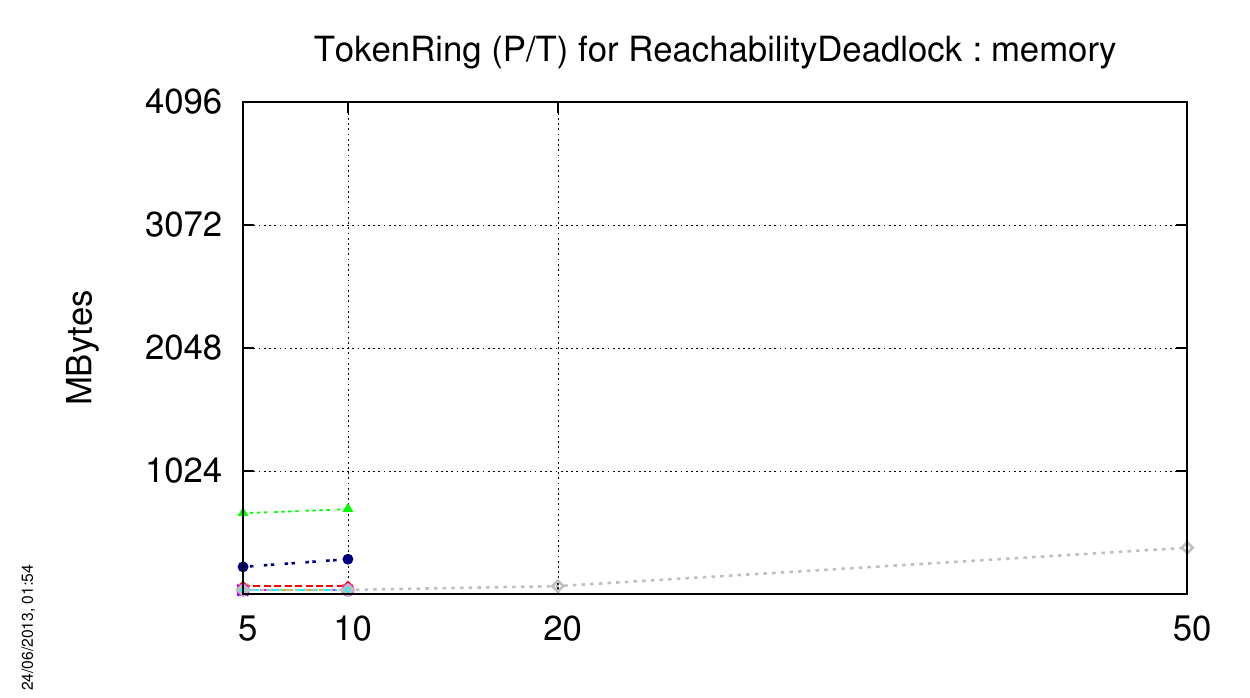}
   \includegraphics[width=7.2cm]{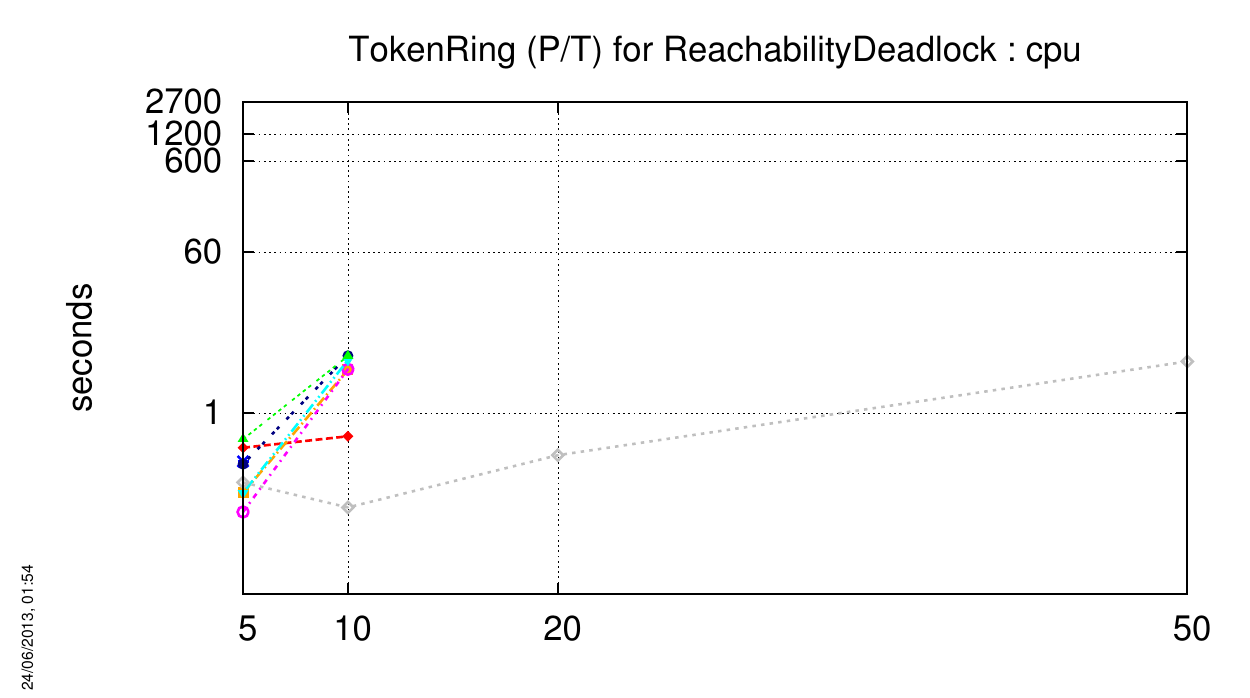}

   \includegraphics[height=1cm]{figures/tools-legend.pdf}
\end{center}

\subsubsection{\acs{HouseConstruction-PT}}
The charts below respectively show how tools compete with this ``Suprise'' model (memory and CPU).

\index{Performances!ReachabilityDeadlock!HouseConstruction (P/T)}
\begin{center}
   \includegraphics[width=7.2cm]{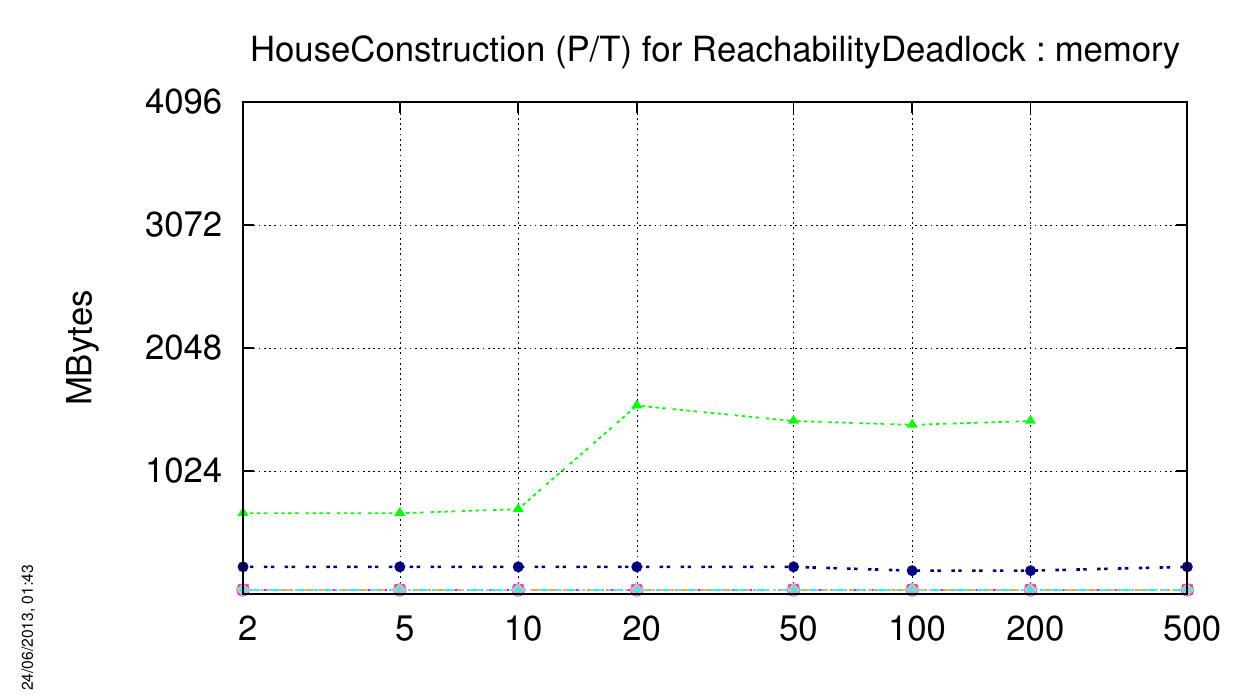}
   \includegraphics[width=7.2cm]{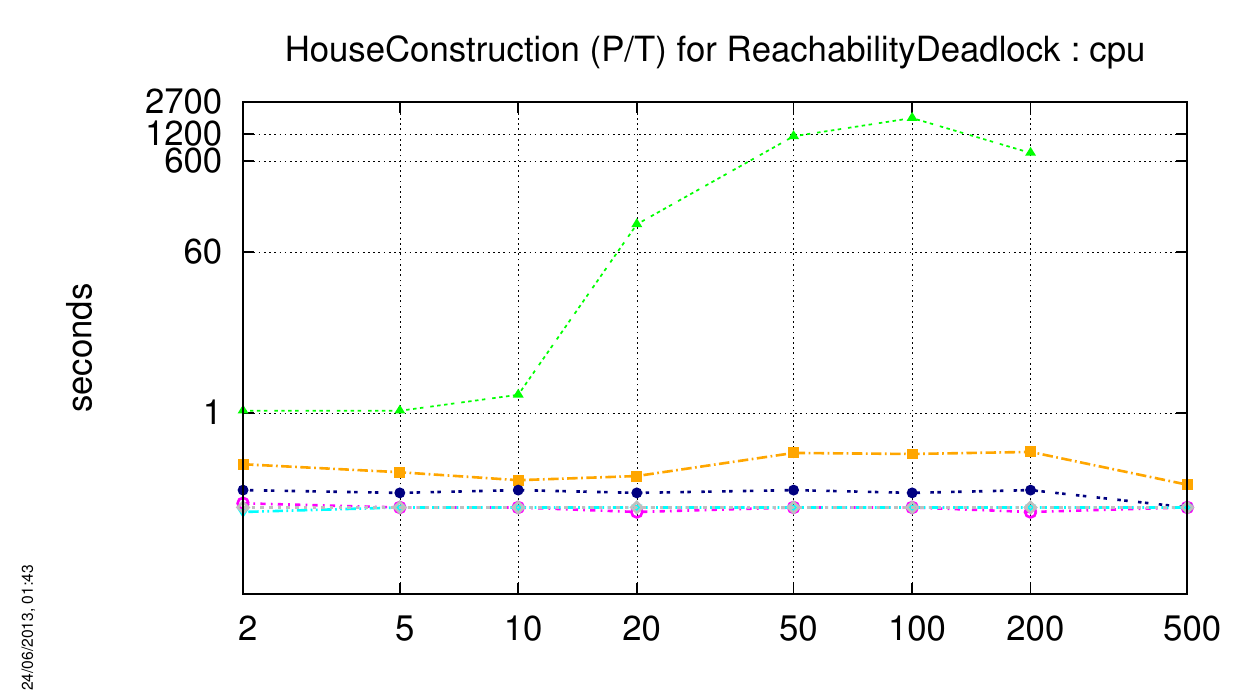}

   \includegraphics[height=1cm]{figures/tools-legend.pdf}
\end{center}

\subsubsection{\acs{IBMB2S565S3960-PT}}
The charts below respectively show how tools compete with this ``Suprise'' model (memory and CPU).

\index{Performances!ReachabilityDeadlock!IBMB2S565S3960 (P/T)}
\begin{center}
   \includegraphics[width=7.2cm]{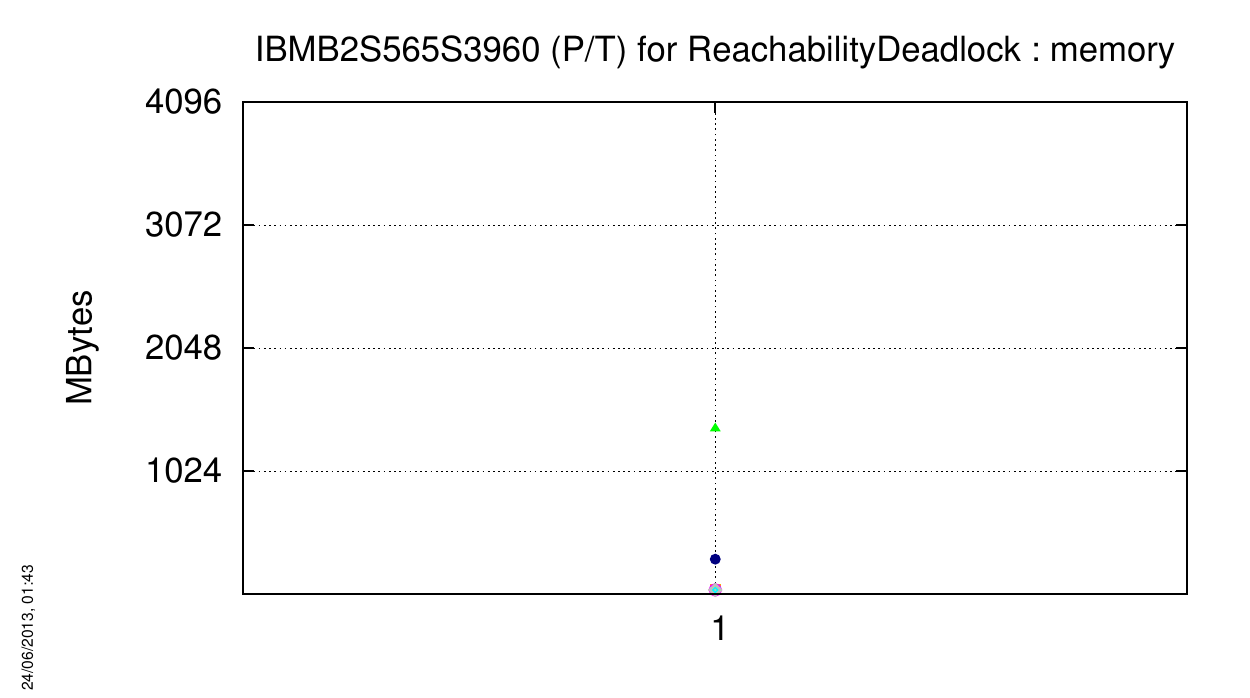}
   \includegraphics[width=7.2cm]{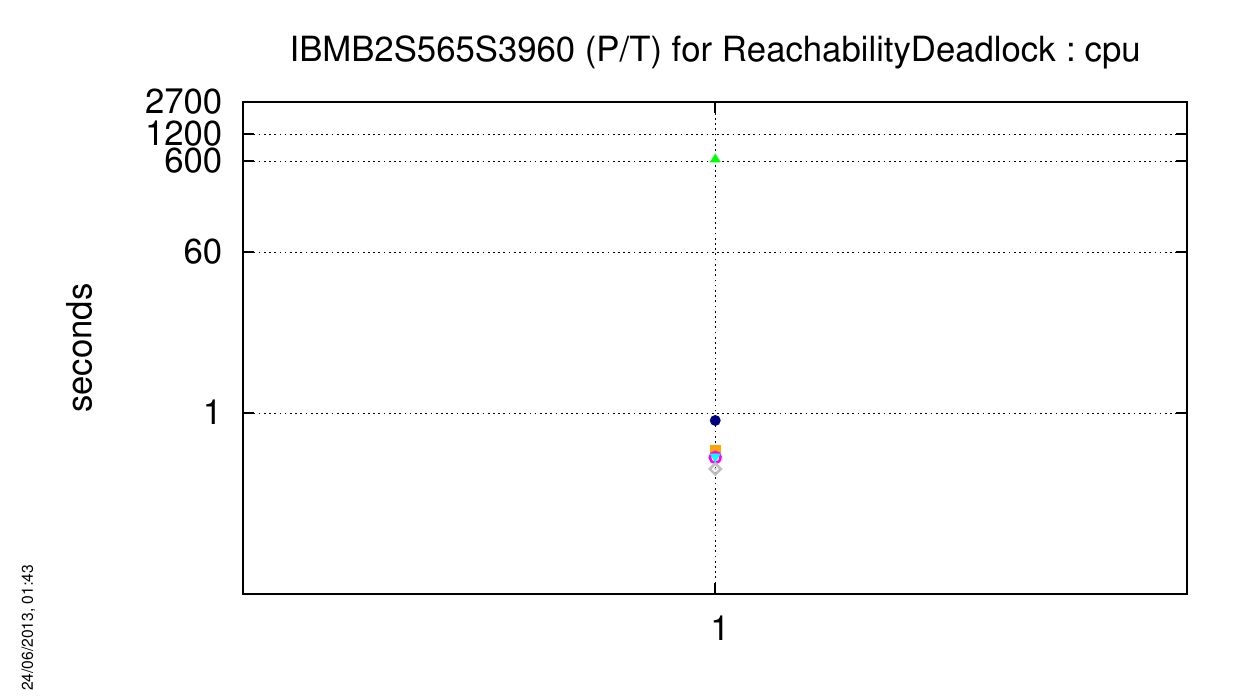}

   \includegraphics[height=1cm]{figures/tools-legend.pdf}
\end{center}

\subsubsection{\acs{QuasiCertifProtocol-COL}}
No instance of this model could be computed for the \textbf{ReachabilityDeadlock} examination.

\subsubsection{\acs{QuasiCertifProtocol-PT}}
The charts below respectively show how tools compete with this ``Suprise'' model (memory and CPU).

\index{Performances!ReachabilityDeadlock!QuasiCertifProtocol (P/T)}
\begin{center}
   \includegraphics[width=7.2cm]{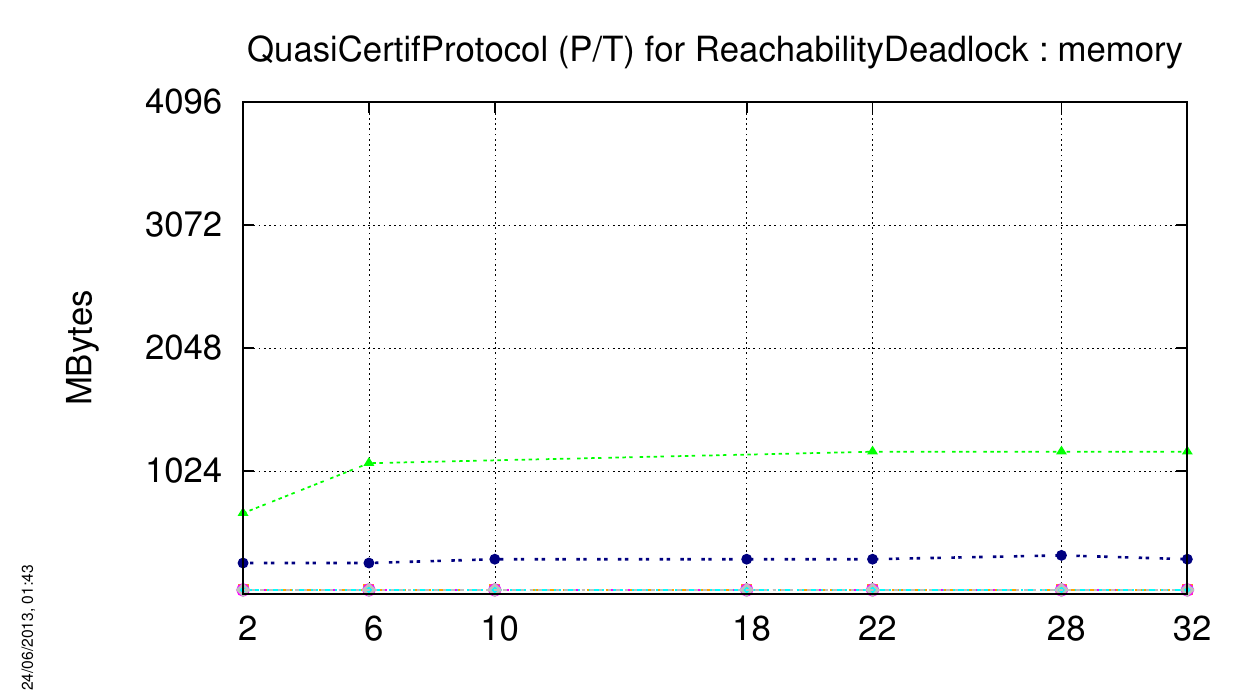}
   \includegraphics[width=7.2cm]{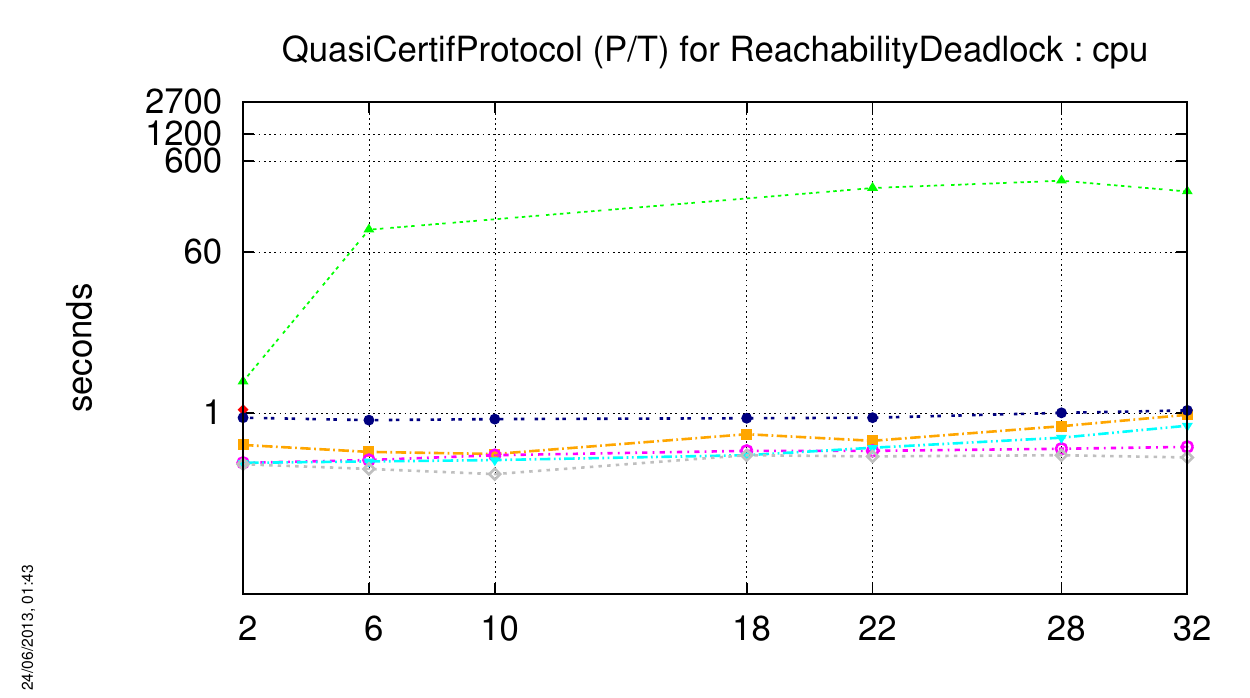}

   \includegraphics[height=1cm]{figures/tools-legend.pdf}
\end{center}

\subsubsection{\acs{Vasy2003-PT}}
The charts below respectively show how tools compete with this ``Suprise'' model (memory and CPU).

\index{Performances!ReachabilityDeadlock!Vasy2003 (P/T)}
\begin{center}
   \includegraphics[width=7.2cm]{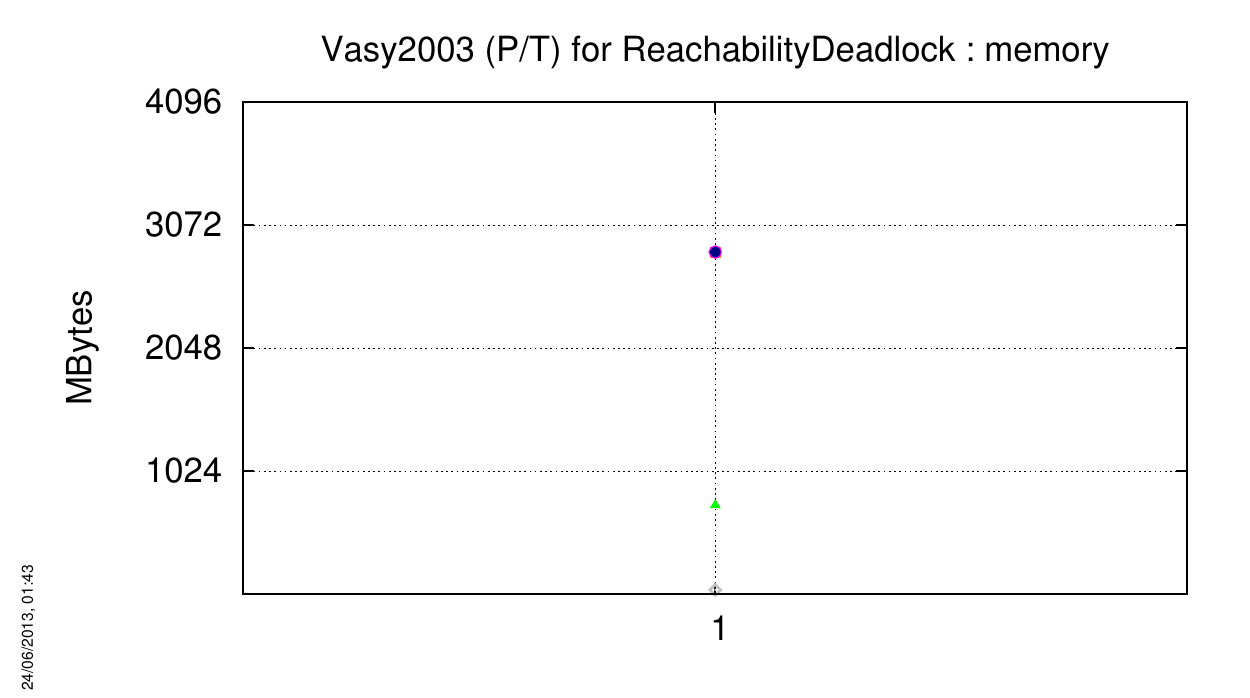}
   \includegraphics[width=7.2cm]{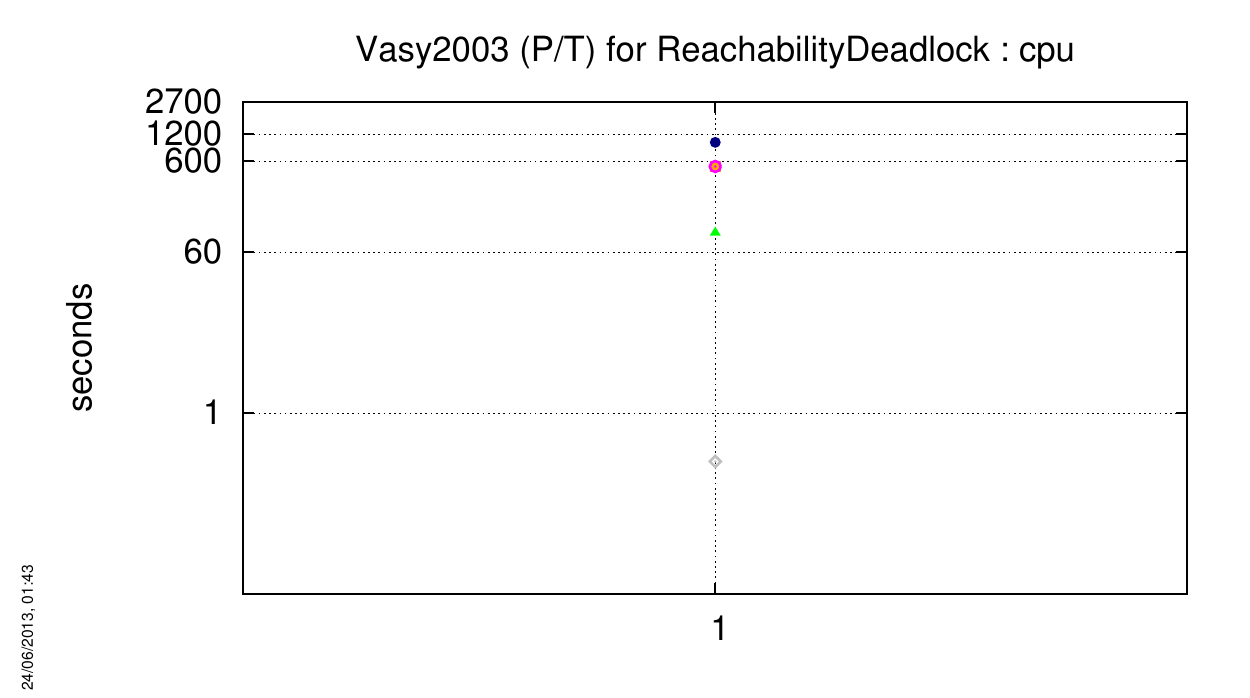}

   \includegraphics[height=1cm]{figures/tools-legend.pdf}
\end{center}

\subsection{Outputs for the ReachabilityDeadlock Examination}
\index{Outputs!ReachabilityDeadlock}

Please find enclosed the brute results for this examination (``Known'' and ``Surprise'' models).
We display only the score of tools that provide a results for at least one instance of one model.
The legend for the values is provided below:
\begin{itemize}
   \item\textbf{nc}: the tool does not compete this examination for this model/instance,
   \item\textbf{cc}: the tool cannot compute this examination for this model/instance,
   \item\textbf{to}: the tool cannot compute this examination for this model/instance within the maximum allowed time,
   \item\textbf{mp}: the tool encountered a memory problem (stack overflow or memory full),
   \item\textbf{nf}: there is no formula available for this type of examination (typically, this concerns P/T nets where
       comparing marking cardinality has no signification when there is no equivalent colored net).
\end{itemize}

\textbf{Note on the display of results for formulas:} each formula is considered as a flag (F if false, T if true, - or ?
when the value cannot be determined). These values are concatenated in the order they appear (we assume it is the order of formulas as they were provided).

\subsubsection{``Known'' Models}

Results are summarized in the table below.
{\footnotesize
\begin{longtable}{|cccccccccc|}
   \hline
      \multicolumn{10}{|c|}{\bf \acs{CSRepetitions-COL}} \\
      \textbf{Instances}\index{Outputs!ReachabilityDeadlock!CSRepetitions (colored)} &
      \acs{cunf} &
      \acs{greatspn} &
      \acs{its-tools} &
      \acs{lola} &
      \acs{lola-optimistic} &
      \acs{lola-optimistic-incomplete} &
      \acs{lola-pessimistic} &
      \acs{marcie} &
      \acs{sara} \\
      02 &
      nc &
      nc &
      \textcolor{red}{\tt \scriptsize F} &
      nc &
      nc &
      nc &
      nc &
      nc &
      nc \\
      03 &
      nc &
      nc &
      \textcolor{red}{\tt \scriptsize F} &
      nc &
      nc &
      nc &
      nc &
      nc &
      nc \\
      04 &
      nc &
      nc &
      \textcolor{red}{\tt \scriptsize mp} &
      nc &
      nc &
      nc &
      nc &
      nc &
      nc \\
      05 &
      nc &
      nc &
      \textcolor{red}{\tt \scriptsize mp} &
      nc &
      nc &
      nc &
      nc &
      nc &
      nc \\
      07 &
      nc &
      nc &
      \textcolor{red}{\tt \scriptsize mp} &
      nc &
      nc &
      nc &
      nc &
      nc &
      nc \\
      10 &
      nc &
      nc &
      \textcolor{red}{\tt \scriptsize mp} &
      nc &
      nc &
      nc &
      nc &
      nc &
      nc \\
\hline
\end{longtable}
}

\subsubsection{``Surprise'' Models}

Results are summarized in the table below.
{\footnotesize
\begin{longtable}{|cccccccc|}
   \hline
      \multicolumn{8}{|c|}{\bf \acs{HouseConstruction-PT}} \\
      instances\index{Outputs!ReachabilityDeadlock!HouseConstruction (P/T)} &
      \acs{cunf} &
      \acs{lola} &
      \acs{lola-optimistic} &
      \acs{lola-optimistic-incomplete} &
      \acs{lola-pessimistic} &
      \acs{marcie} &
      \acs{sara} \\
      002 &
      nc &
      \textcolor{red}{\tt \scriptsize FF} &
      \textcolor{red}{\tt \scriptsize FF} &
      \textcolor{red}{\tt \scriptsize FF} &
      \textcolor{red}{\tt \scriptsize FF} &
      \textcolor{red}{\tt \scriptsize FT} &
      \textcolor{red}{\tt \scriptsize F-} \\
      005 &
      nc &
      \textcolor{red}{\tt \scriptsize FF} &
      \textcolor{red}{\tt \scriptsize FF} &
      \textcolor{red}{\tt \scriptsize FF} &
      \textcolor{red}{\tt \scriptsize FF} &
      \textcolor{red}{\tt \scriptsize FT} &
      \textcolor{red}{\tt \scriptsize F-} \\
      010 &
      nc &
      \textcolor{red}{\tt \scriptsize FF} &
      \textcolor{red}{\tt \scriptsize FF} &
      \textcolor{red}{\tt \scriptsize FF} &
      \textcolor{red}{\tt \scriptsize FF} &
      \textcolor{red}{\tt \scriptsize FT} &
      \textcolor{red}{\tt \scriptsize F-} \\
      020 &
      nc &
      \textcolor{red}{\tt \scriptsize FF} &
      \textcolor{red}{\tt \scriptsize FF} &
      \textcolor{red}{\tt \scriptsize FF} &
      \textcolor{red}{\tt \scriptsize FF} &
      \textcolor{red}{\tt \scriptsize FT} &
      \textcolor{red}{\tt \scriptsize F-} \\
      050 &
      nc &
      \textcolor{red}{\tt \scriptsize FF} &
      \textcolor{red}{\tt \scriptsize FF} &
      \textcolor{red}{\tt \scriptsize FF} &
      \textcolor{red}{\tt \scriptsize FF} &
      \textcolor{red}{\tt \scriptsize cc} &
      \textcolor{red}{\tt \scriptsize F-} \\
\hline
\end{longtable}
}

\subsection{Score for the ReachabilityDeadlock Examination}
\index{Scores!ReachabilityDeadlock}

Please find enclosed the scores for this examination (``Known'' and ``Surprise'' models).
We display only the score of tools that provide a results for at least one instance of one model.
The total is first listed in the table below followed by a detail, for each proposed model.
Meaning of the line labels are:
\begin{itemize}
\item\textbf{1st instance}: the tool gets a bonus for having processed the first instance of this model (+1 point),
\item\textbf{instances}: the tool gets 1 point per instances treated 
(for that, we assume that at least one formula has been successfully computed),
\item\textbf{max reached}: the tool could process all the instances for the model (+2 points),
\item\textbf{best}: the tool is among the ones that processed a maximum of instances within the time and memory confinement (+2 points).
\end{itemize}

\subsubsection{``Known'' Models}

Scores are summarized in the table below.
For each model in this category, a tool may collect up to 411 points.

{\footnotesize
\begin{longtable}{|cccccccccc|}
   \hline
      \multicolumn{10}{|c|}{\bf Total score of the tools for this examination} \\
        &
      \acs{cunf} &
      \acs{greatspn} &
      \acs{its-tools} &
      \acs{lola} &
      \acs{lola-optimistic} &
      \acs{lola-optimistic-incomplete} &
      \acs{lola-pessimistic} &
      \acs{marcie} &
      \acs{sara} \\
      Total Score &
      {\bf 98} &
      {\bf 48} &
      {\bf 52} &
      {\bf 216} &
      {\bf 216} &
      {\bf 202} &
      {\bf 185} &
      {\bf 116} &
      {\bf 251} \\
   \hline
      \multicolumn{10}{|c|}{\bf CSRepetitions (Colored)} \\
      \ \index{Scores!ReachabilityDeadlock!CSRepetitions (colored)} &
      \acs{cunf} &
      \acs{greatspn} &
      \acs{its-tools} &
      \acs{lola} &
      \acs{lola-optimistic} &
      \acs{lola-optimistic-incomplete} &
      \acs{lola-pessimistic} &
      \acs{marcie} &
      \acs{sara} \\
      1st instance &
      \textcolor{red}{0} &
      \textcolor{red}{0} &
      \textcolor{red}{1} &
      \textcolor{red}{0} &
      \textcolor{red}{0} &
      \textcolor{red}{0} &
      \textcolor{red}{0} &
      \textcolor{red}{0} &
      \textcolor{red}{0} \\
      instances &
      \textcolor{red}{0} &
      \textcolor{red}{0} &
      \textcolor{red}{2} &
      \textcolor{red}{0} &
      \textcolor{red}{0} &
      \textcolor{red}{0} &
      \textcolor{red}{0} &
      \textcolor{red}{0} &
      \textcolor{red}{0} \\
      max reached &
      \textcolor{red}{0} &
      \textcolor{red}{0} &
      \textcolor{red}{0} &
      \textcolor{red}{0} &
      \textcolor{red}{0} &
      \textcolor{red}{0} &
      \textcolor{red}{0} &
      \textcolor{red}{0} &
      \textcolor{red}{0} \\
\hline
\end{longtable}
}

\subsubsection{``Surprise'' Models}

Scores are summarized in the table below.
For each model in this category, a tool may collect up to 49 points.

{\footnotesize
\begin{longtable}{|cccccccc|}
   \hline
      \multicolumn{8}{|c|}{\bf Total score of the tools for this examination} \\
        &
      \acs{cunf} &
      \acs{lola} &
      \acs{lola-optimistic} &
      \acs{lola-optimistic-incomplete} &
      \acs{lola-pessimistic} &
      \acs{marcie} &
      \acs{sara} \\
      Total Score &
      {\bf 0} &
      {\bf 37} &
      {\bf 37} &
      {\bf 37} &
      {\bf 31} &
      {\bf 20} &
      {\bf 37} \\
   \hline
      \multicolumn{8}{|c|}{\bf HouseConstruction (P/T)} \\
      \ \index{Scores!ReachabilityDeadlock!HouseConstruction (P/T)} &
      \acs{cunf} &
      \acs{lola} &
      \acs{lola-optimistic} &
      \acs{lola-optimistic-incomplete} &
      \acs{lola-pessimistic} &
      \acs{marcie} &
      \acs{sara} \\
      1st instance &
      \textcolor{red}{0} &
      \textcolor{red}{1} &
      \textcolor{red}{1} &
      \textcolor{red}{1} &
      \textcolor{red}{1} &
      \textcolor{red}{1} &
      \textcolor{red}{1} \\
      instances &
      \textcolor{red}{0} &
      \textcolor{red}{8} &
      \textcolor{red}{8} &
      \textcolor{red}{8} &
      \textcolor{red}{8} &
      \textcolor{red}{4} &
      \textcolor{red}{8} \\
      max reached &
      \textcolor{red}{0} &
      \textcolor{red}{2} &
      \textcolor{red}{2} &
      \textcolor{red}{2} &
      \textcolor{red}{2} &
      \textcolor{red}{0} &
      \textcolor{red}{2} \\
\hline
\end{longtable}
}

\subsection{Trophies for this Examination}
\index{Trophies!ReachabilityDeadlock}

Trophies are divided in three categories: ``Known'' models,
``Surprise'' models, and the global trophies (formula is then
$score_{global} = score_{known} + 2 \times score_{surprise}$).

\subsubsection{For ``Known'' Models} \ \\

\begin{tabular}{c|c|c}
      1 & 2 & 2 \\
   \includegraphics[width=2cm]{figures/gold.jpg} &
   \includegraphics[width=2cm]{figures/silver.jpg} &
   \includegraphics[width=2cm]{figures/silver.jpg} \\
   \acs{sara} &
   \acs{lola} &
   \acs{lola-optimistic} \\
   251 points &
   216 points &
   216 points \\
\end{tabular}

\subsubsection{For ``Surprise'' Models}\  \\

\begin{tabular}{c|c|c|c}
      1 & 1 & 1 & 1 \\
   \includegraphics[width=2cm]{figures/gold.jpg} &
   \includegraphics[width=2cm]{figures/gold.jpg} &
   \includegraphics[width=2cm]{figures/gold.jpg} &
   \includegraphics[width=2cm]{figures/gold.jpg} \\
   \acs{sara} &
   \acs{lola} &
   \acs{lola-optimistic} &
   \acs{lola-optimistic-incomplete} \\
   37 points &
   37 points &
   37 points &
   37 points \\
\end{tabular}

\subsubsection{Global} \ \\

\begin{tabular}{c|c|c}
      1 & 1 & 1 \\
   \includegraphics[width=2cm]{figures/gold.jpg} &
   \includegraphics[width=2cm]{figures/gold.jpg} &
   \includegraphics[width=2cm]{figures/gold.jpg} \\
   \acs{sara} &
   \acs{lola} &
   \acs{lola-optimistic} \\
   325 points &
   290 points &
   290 points \\
\end{tabular}

\newpage

\section{The ReachabilityFireability Examination}
\label{sec:exam:ReachabilityFireability}
\index{Results!ReachabilityFireability}

This examination deals with reachability properties dealing with transition fireability only.
We first show a summary on the handling of models by the participating tools.
Then, we present the computed outputs and the associated scores for this
examination prior to a summary of relevant executions.

\subsection{Handling of Models by Tools}
\index{Performances!ReachabilityFireability}

\subsubsection{\acs{CSRepetitions-COL}}
The charts below respectively show how tools compete with this ``Known'' model (memory and CPU).

\index{Performances!ReachabilityFireability!CSRepetitions (Colored)}
\begin{center}
   \includegraphics[width=7.2cm]{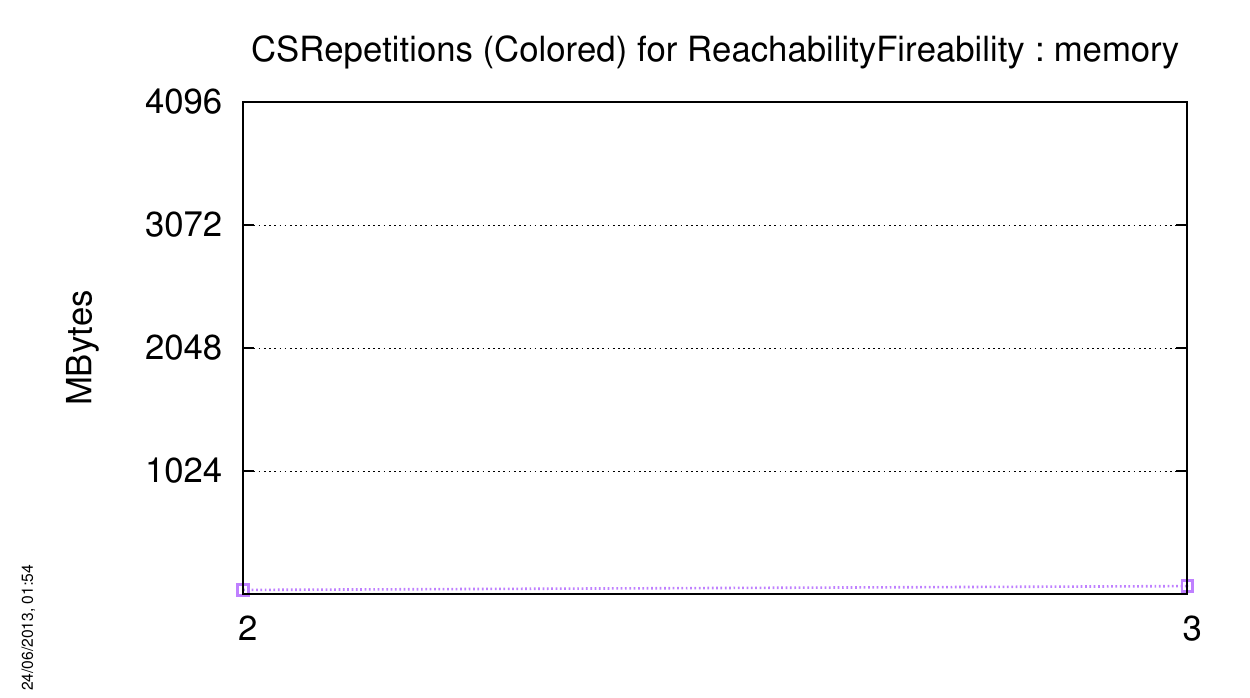}
   \includegraphics[width=7.2cm]{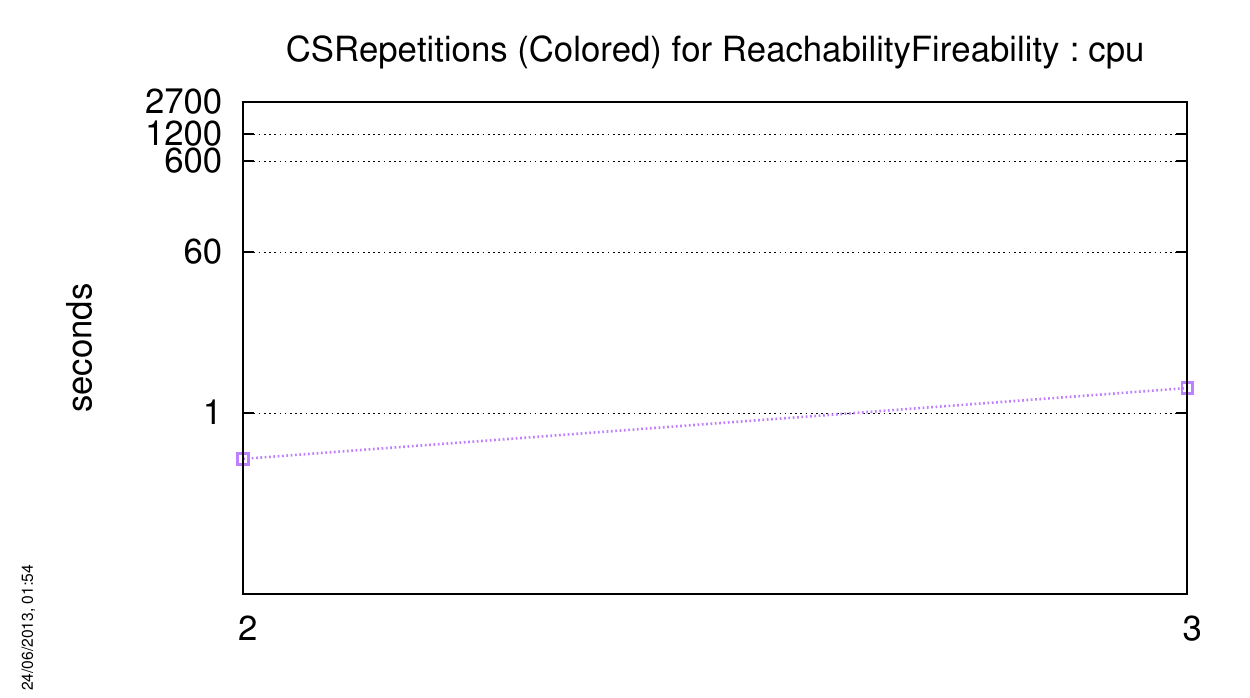}

   \includegraphics[height=1cm]{figures/tools-legend.pdf}
\end{center}

\subsubsection{\acs{CSRepetitions-PT}}
The charts below respectively show how tools compete with this ``Known'' model (memory and CPU).

\index{Performances!ReachabilityFireability!CSRepetitions (P/T)}
\begin{center}
   \includegraphics[width=7.2cm]{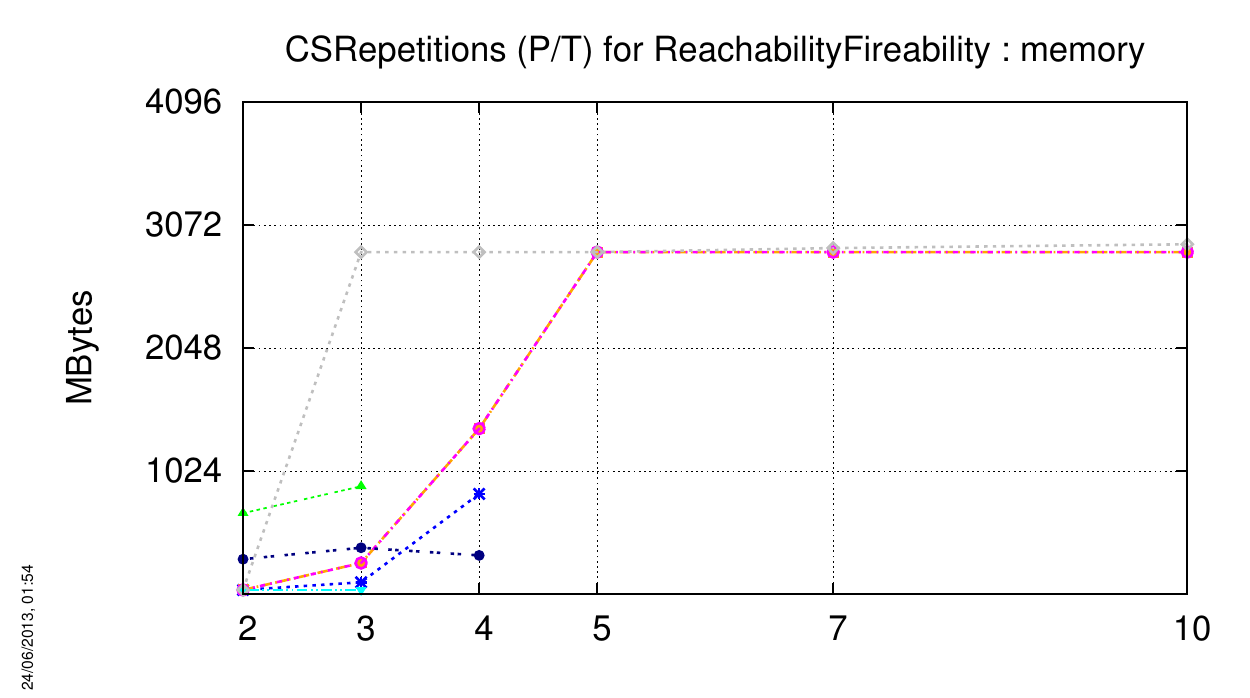}
   \includegraphics[width=7.2cm]{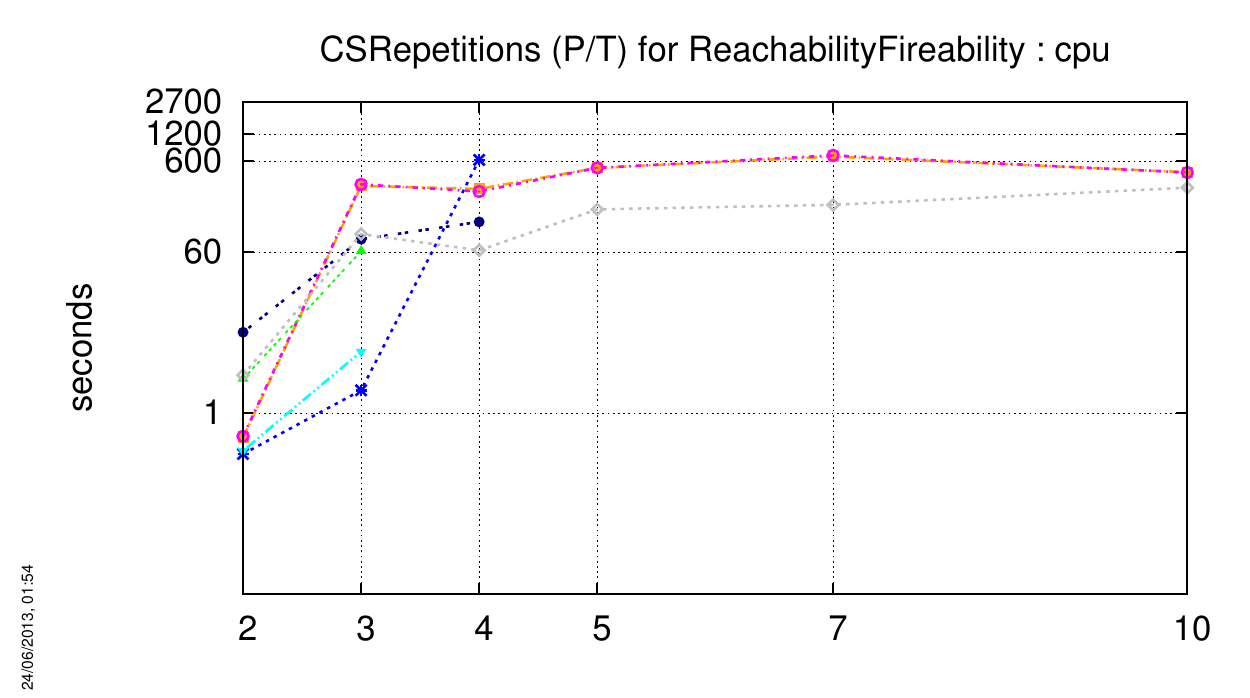}

   \includegraphics[height=1cm]{figures/tools-legend.pdf}
\end{center}

\subsubsection{\acs{Dekker-PT}}
The charts below respectively show how tools compete with this ``Known'' model (memory and CPU).

\index{Performances!ReachabilityFireability!Dekker (P/T)}
\begin{center}
   \includegraphics[width=7.2cm]{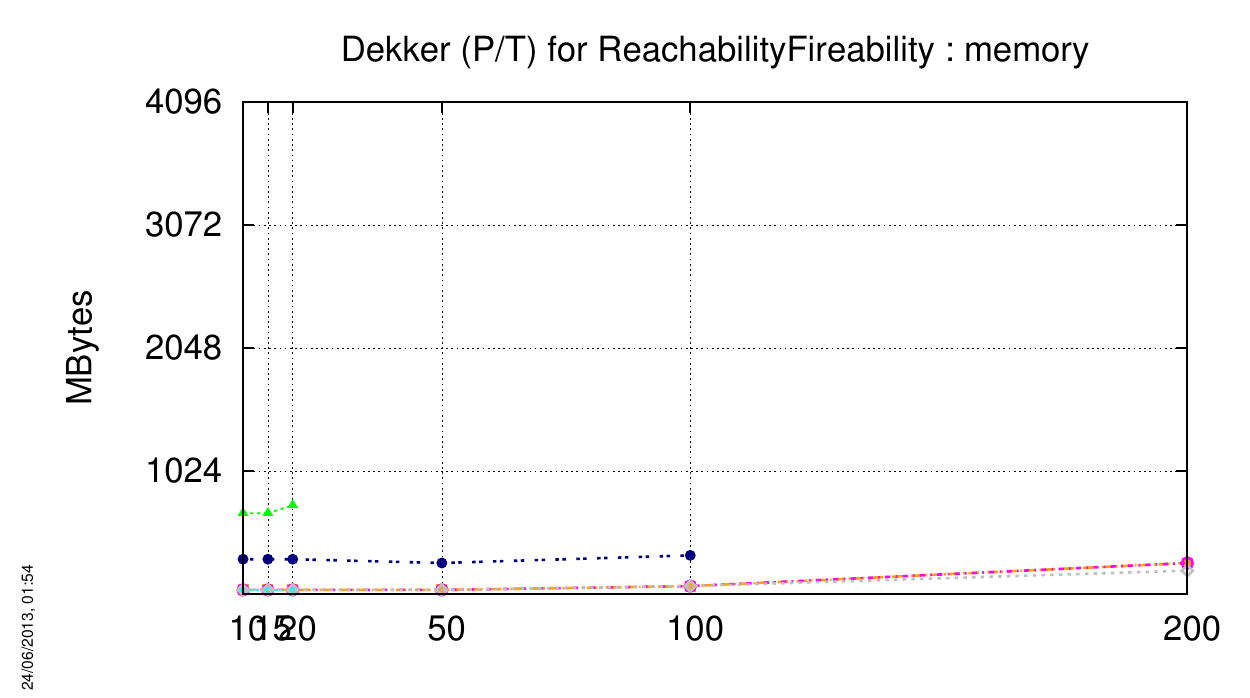}
   \includegraphics[width=7.2cm]{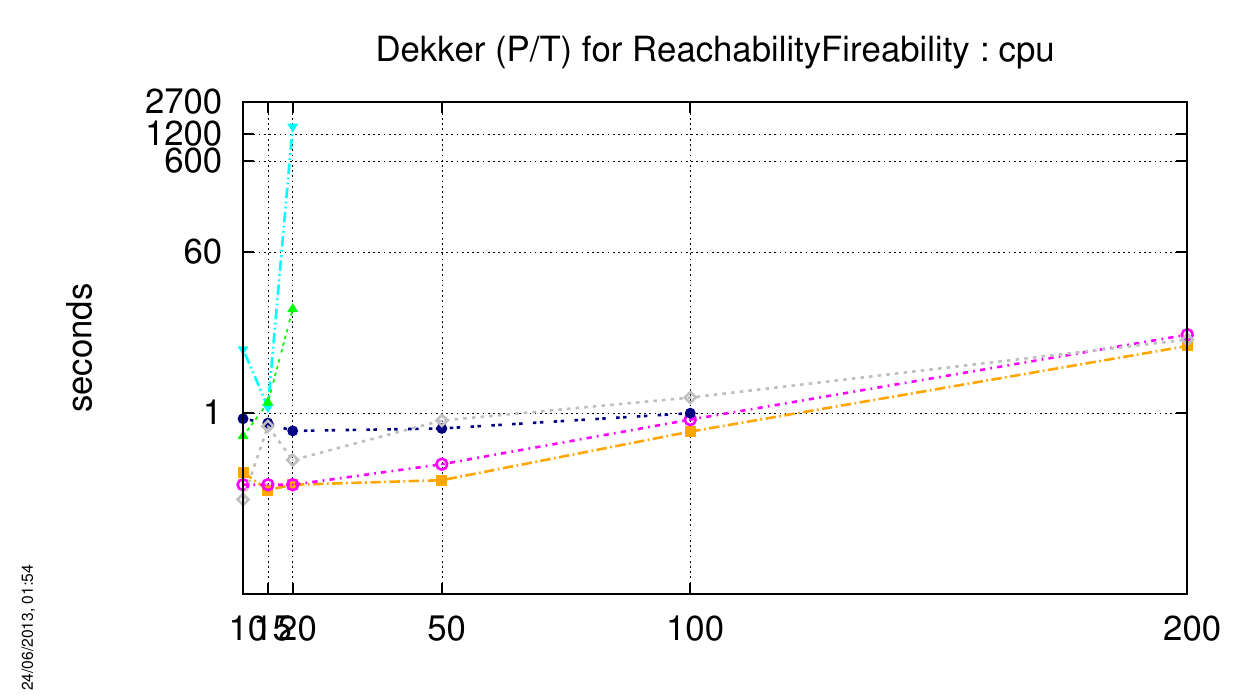}

   \includegraphics[height=1cm]{figures/tools-legend.pdf}
\end{center}

\subsubsection{\acs{DotAndBoxes-COL}}
The charts below respectively show how tools compete with this ``Known'' model (memory and CPU).

\index{Performances!ReachabilityFireability!DotAndBoxes (Colored)}
\begin{center}
   \includegraphics[width=7.2cm]{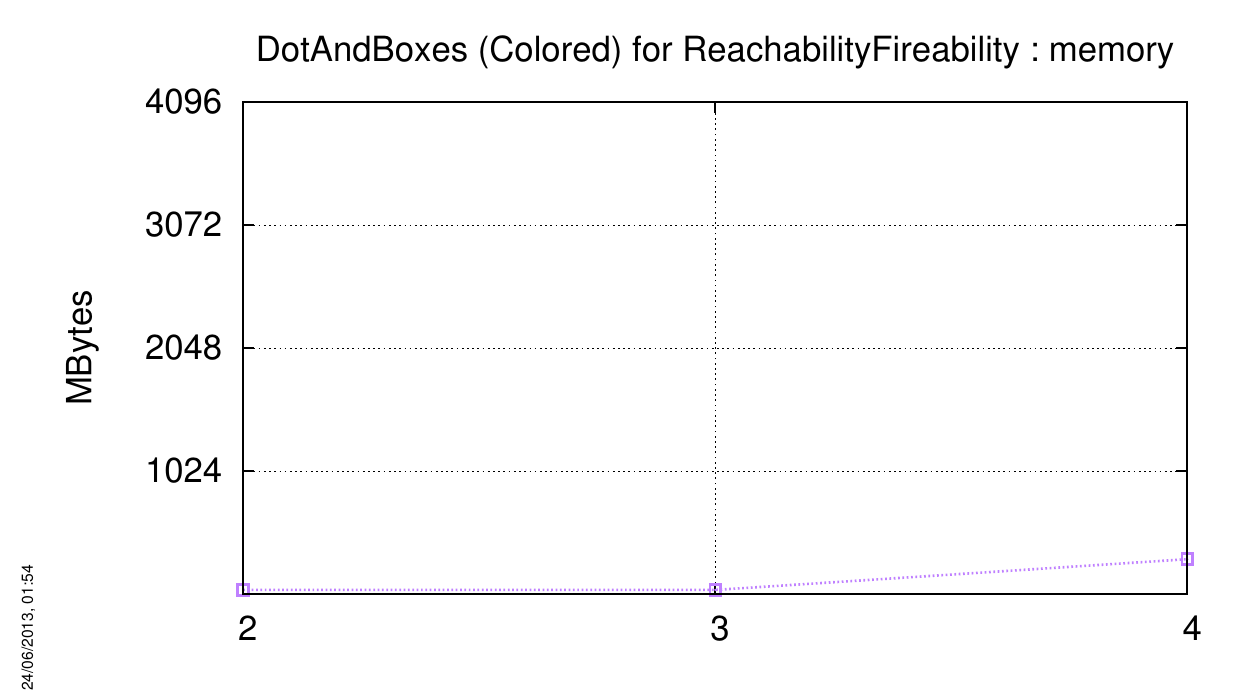}
   \includegraphics[width=7.2cm]{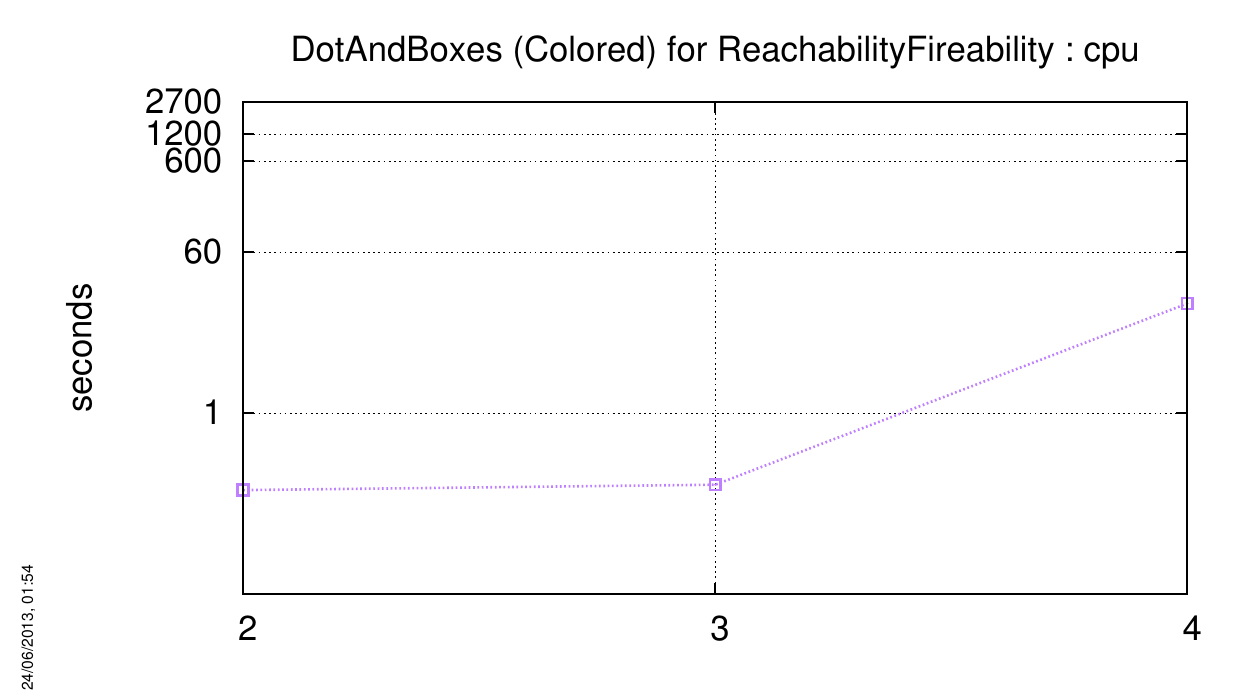}

   \includegraphics[height=1cm]{figures/tools-legend.pdf}
\end{center}

\subsubsection{\acs{DrinkVendingMachine-COL}}
The charts below respectively show how tools compete with this ``Known'' model (memory and CPU).

\index{Performances!ReachabilityFireability!DrinkVendingMachine (Colored)}
\begin{center}
   \includegraphics[width=7.2cm]{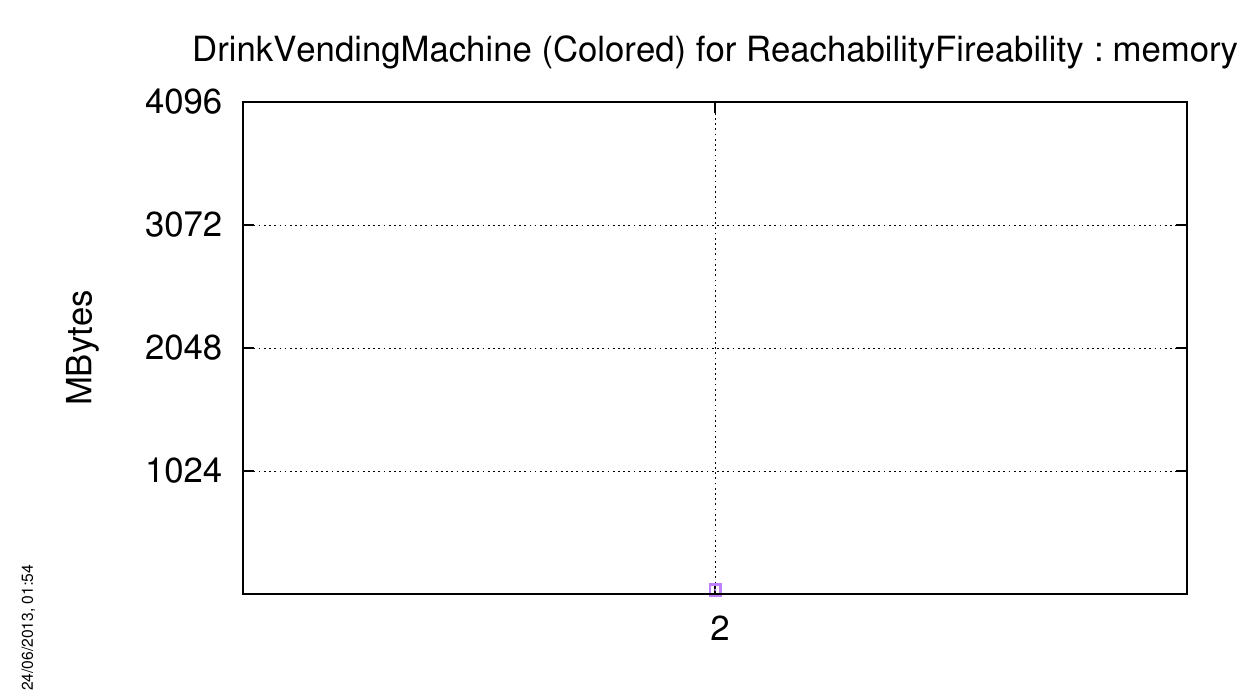}
   \includegraphics[width=7.2cm]{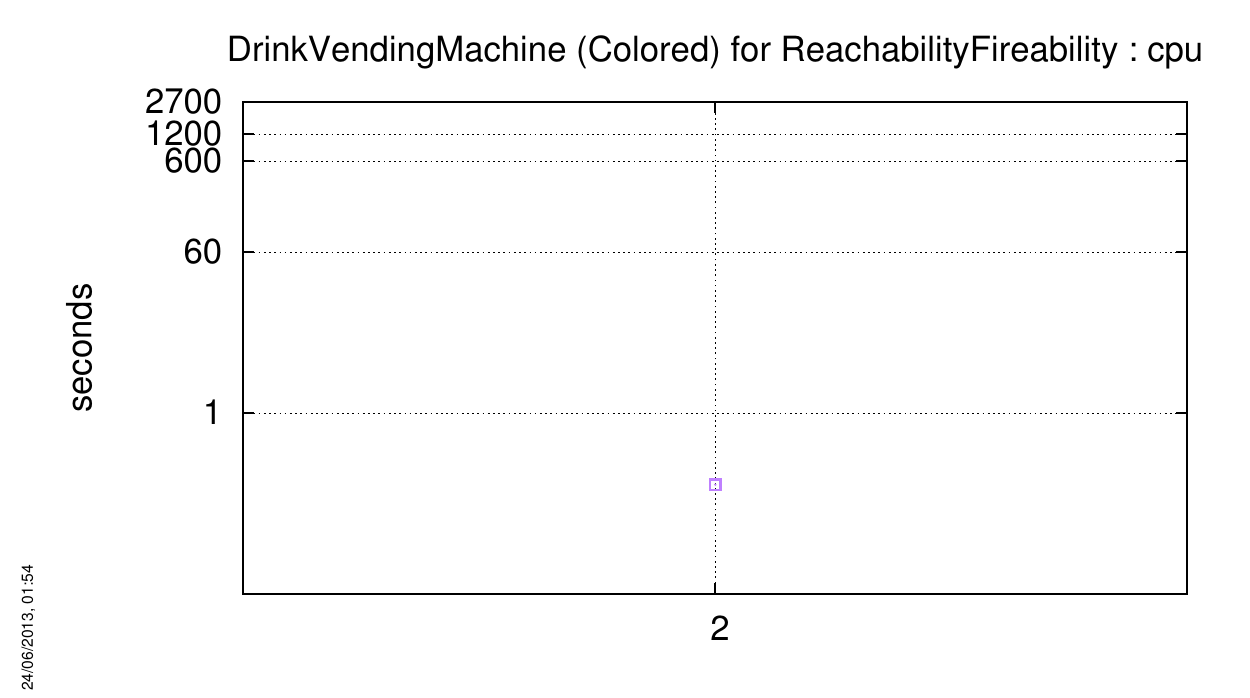}

   \includegraphics[height=1cm]{figures/tools-legend.pdf}
\end{center}

\subsubsection{\acs{DrinkVendingMachine-PT}}
The charts below respectively show how tools compete with this ``Known'' model (memory and CPU).

\index{Performances!ReachabilityFireability!DrinkVendingMachine (P/T)}
\begin{center}
   \includegraphics[width=7.2cm]{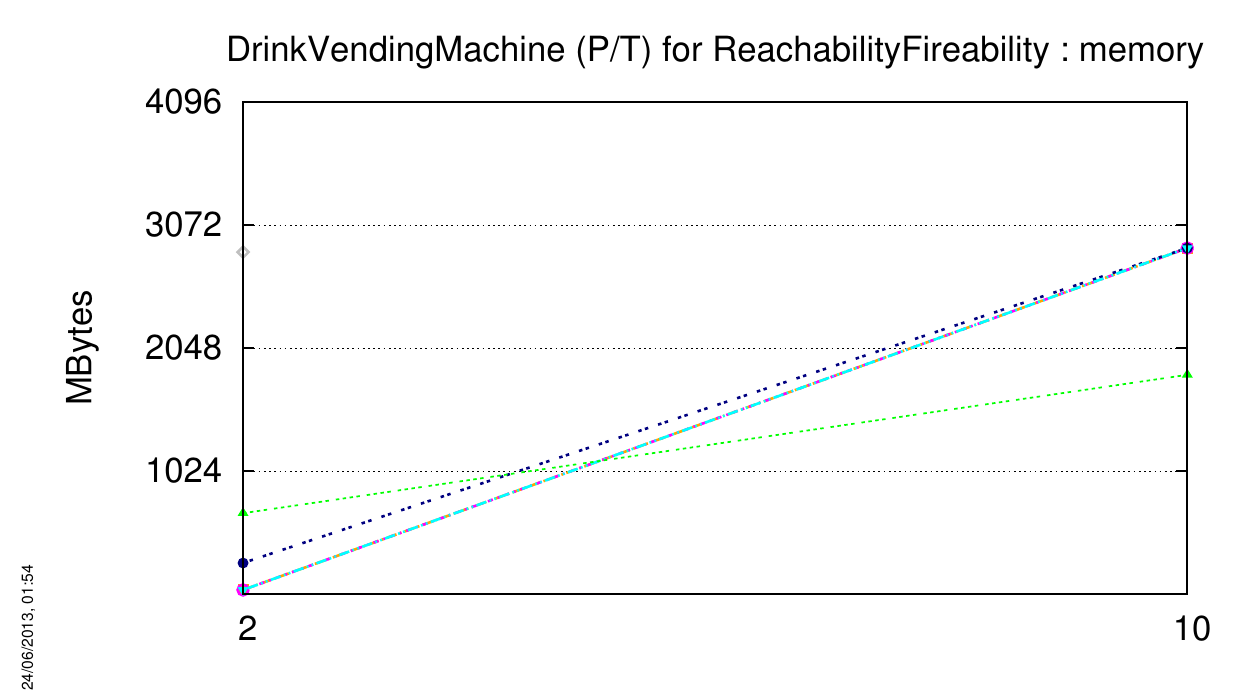}
   \includegraphics[width=7.2cm]{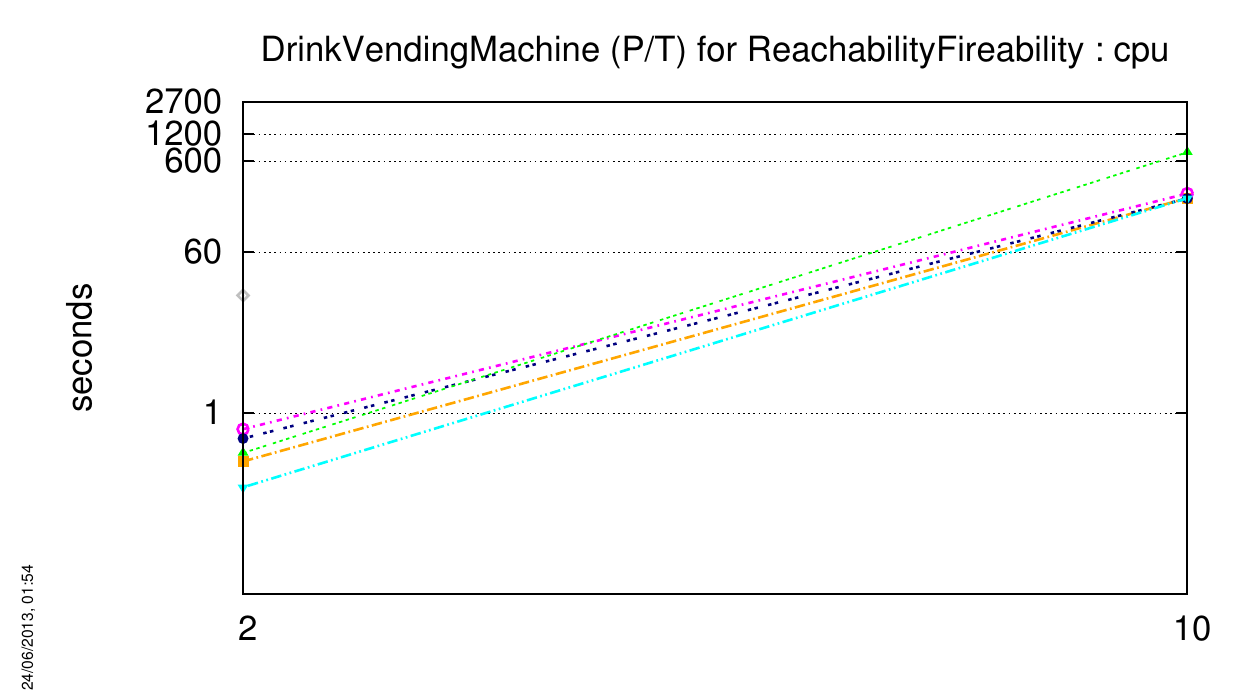}

   \includegraphics[height=1cm]{figures/tools-legend.pdf}
\end{center}

\subsubsection{\acs{Echo-PT}}
The charts below respectively show how tools compete with this ``Known'' model (memory and CPU).

\index{Performances!ReachabilityFireability!Echo (P/T)}
\begin{center}
   \includegraphics[width=7.2cm]{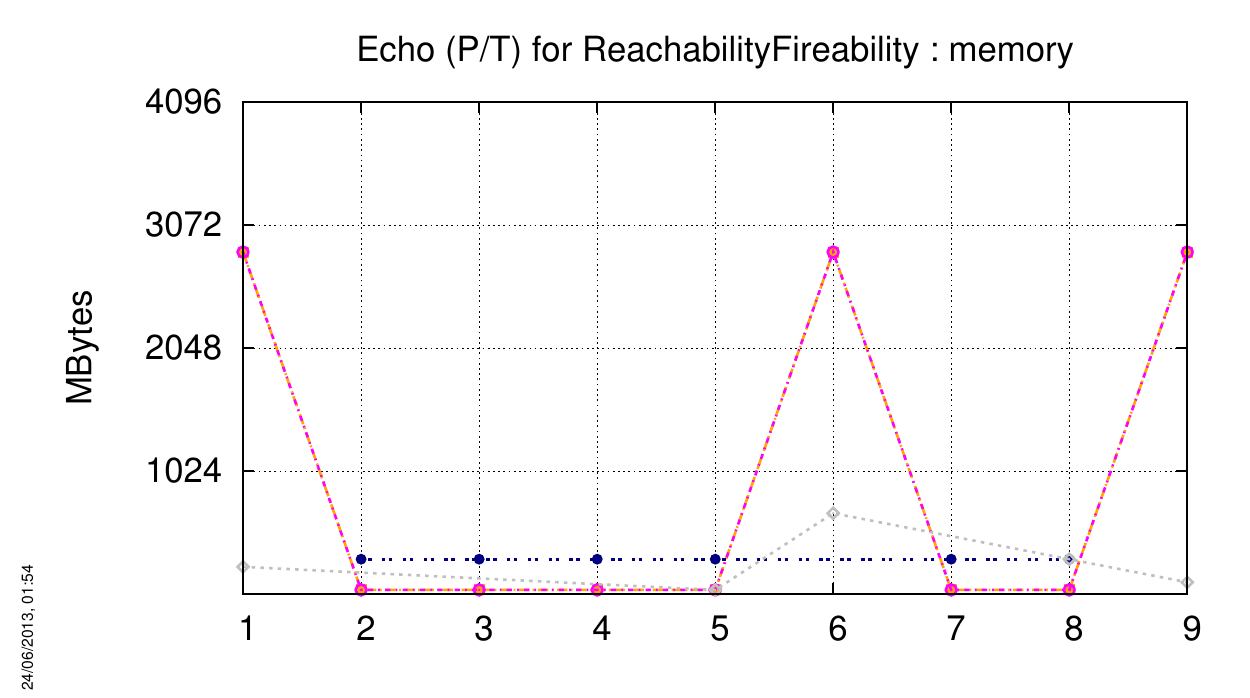}
   \includegraphics[width=7.2cm]{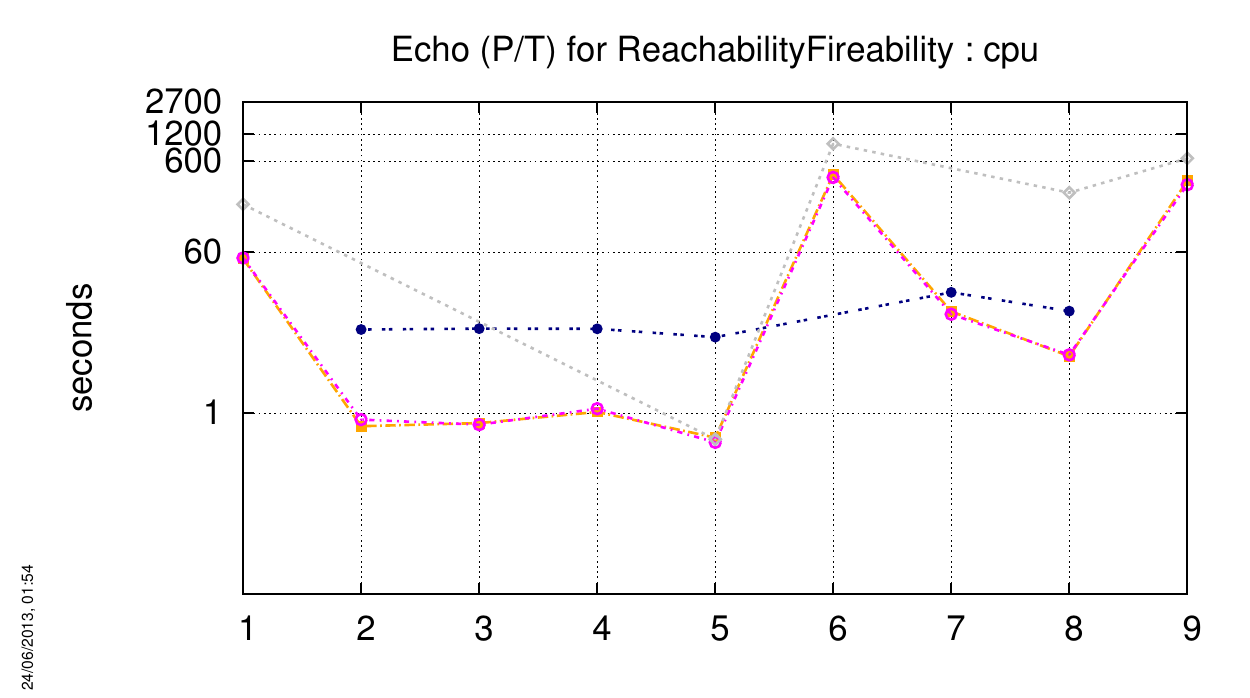}

   \includegraphics[height=1cm]{figures/tools-legend.pdf}
\end{center}

\subsubsection{\acs{Eratosthenes-PT}}
The charts below respectively show how tools compete with this ``Known'' model (memory and CPU).

\index{Performances!ReachabilityFireability!Eratosthenes (P/T)}
\begin{center}
   \includegraphics[width=7.2cm]{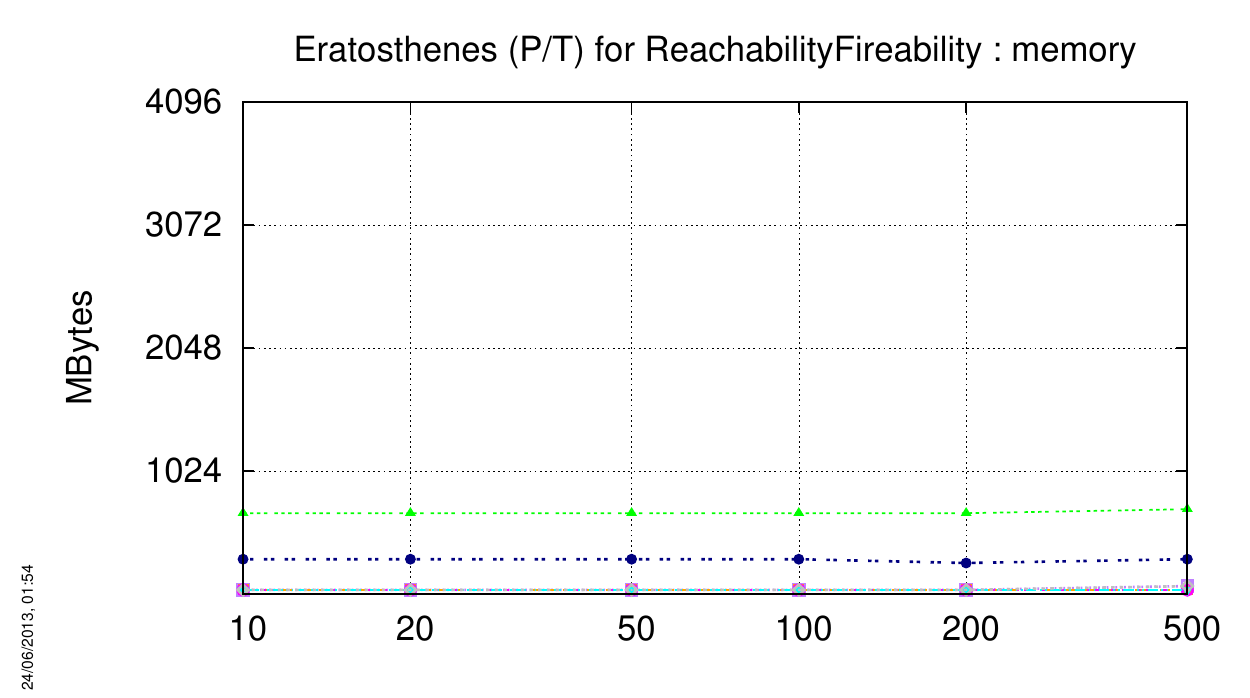}
   \includegraphics[width=7.2cm]{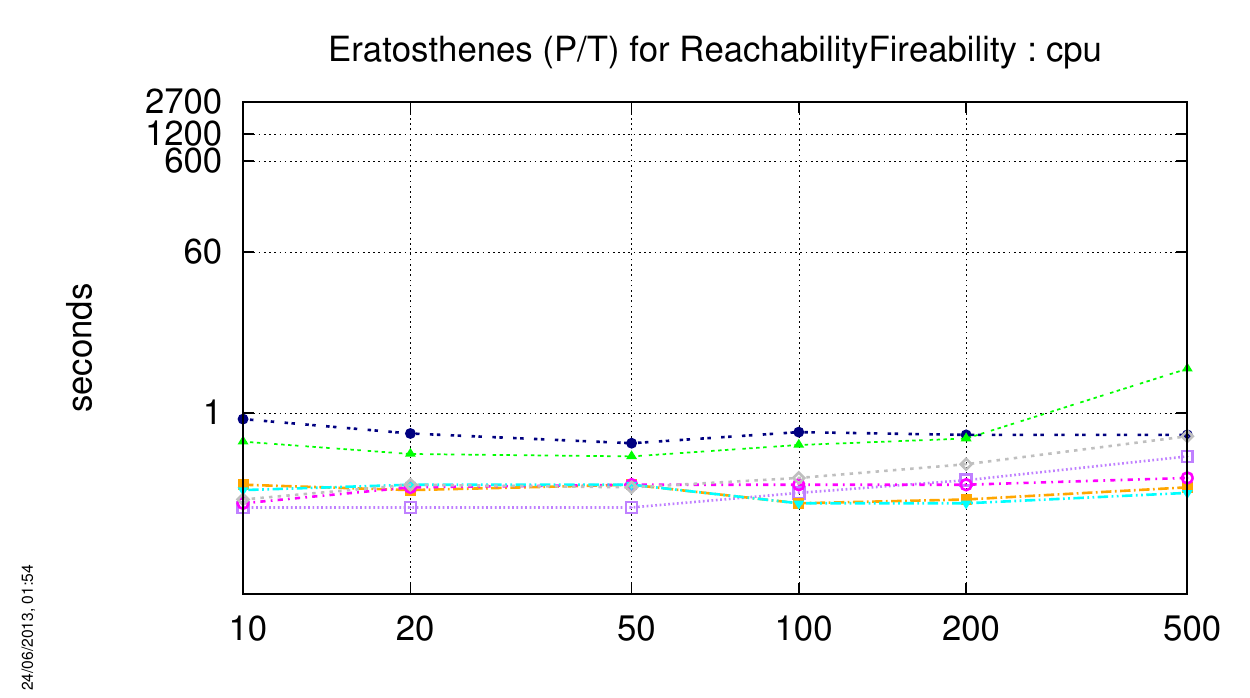}

   \includegraphics[height=1cm]{figures/tools-legend.pdf}
\end{center}

\subsubsection{\acs{FMS-PT}}
The charts below respectively show how tools compete with this ``Known'' model (memory and CPU).

\index{Performances!ReachabilityFireability!FMS (P/T)}
\begin{center}
   \includegraphics[width=7.2cm]{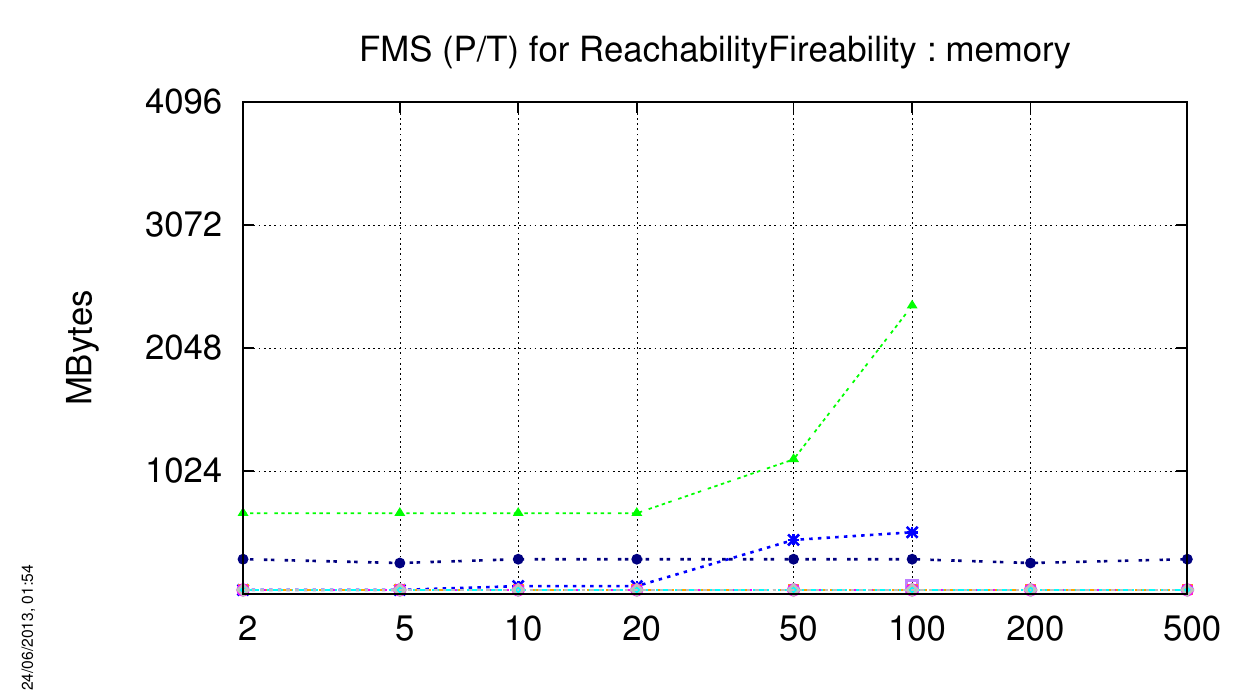}
   \includegraphics[width=7.2cm]{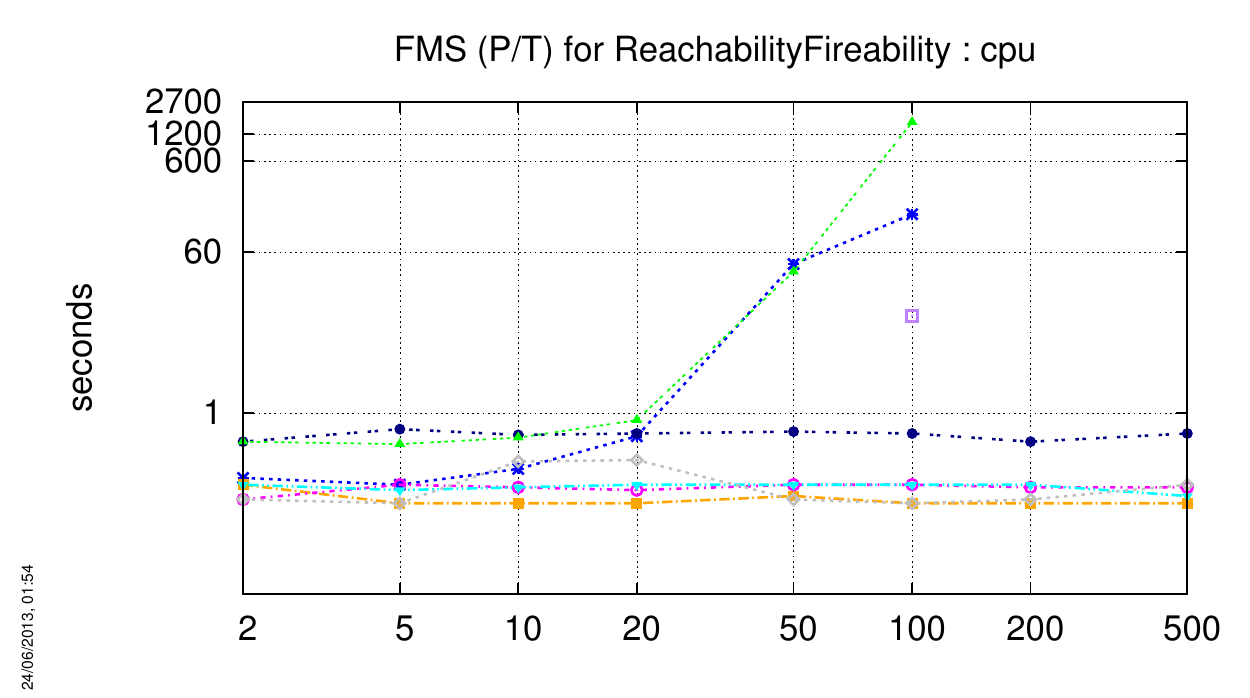}

   \includegraphics[height=1cm]{figures/tools-legend.pdf}
\end{center}

\subsubsection{\acs{GlobalRessAlloc-COL}}
The charts below respectively show how tools compete with this ``Known'' model (memory and CPU).

\index{Performances!ReachabilityFireability!GlobalRessAlloc (Colored)}
\begin{center}
   \includegraphics[width=7.2cm]{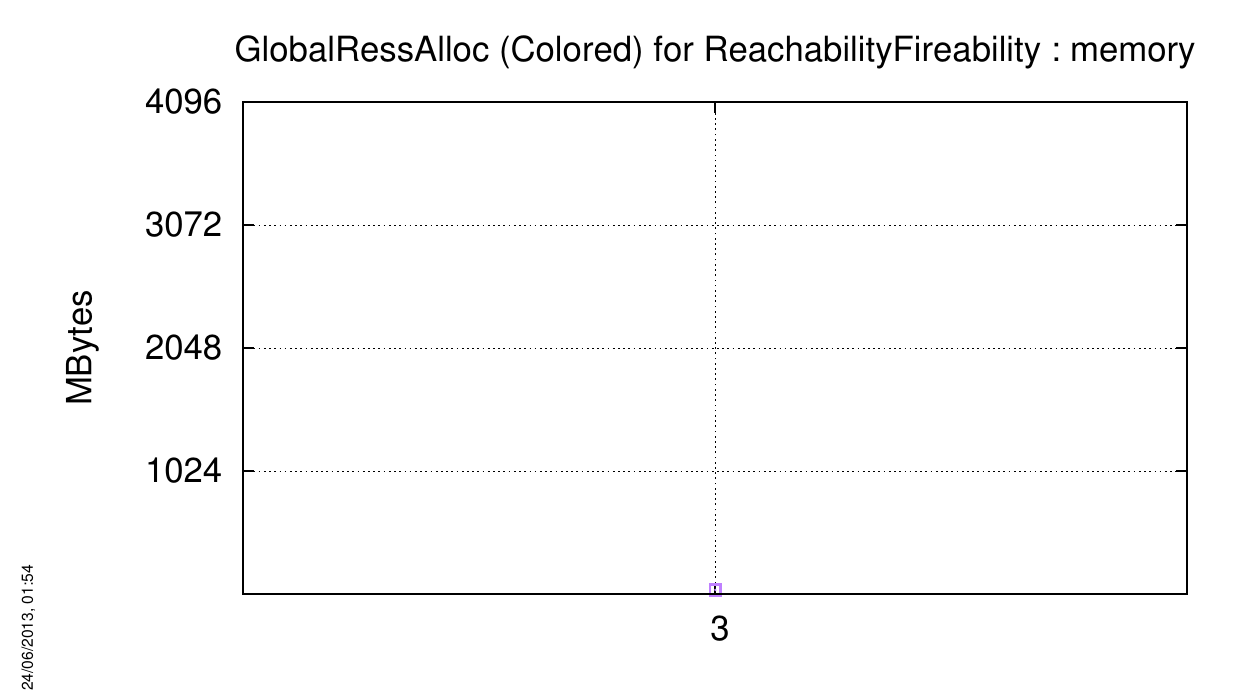}
   \includegraphics[width=7.2cm]{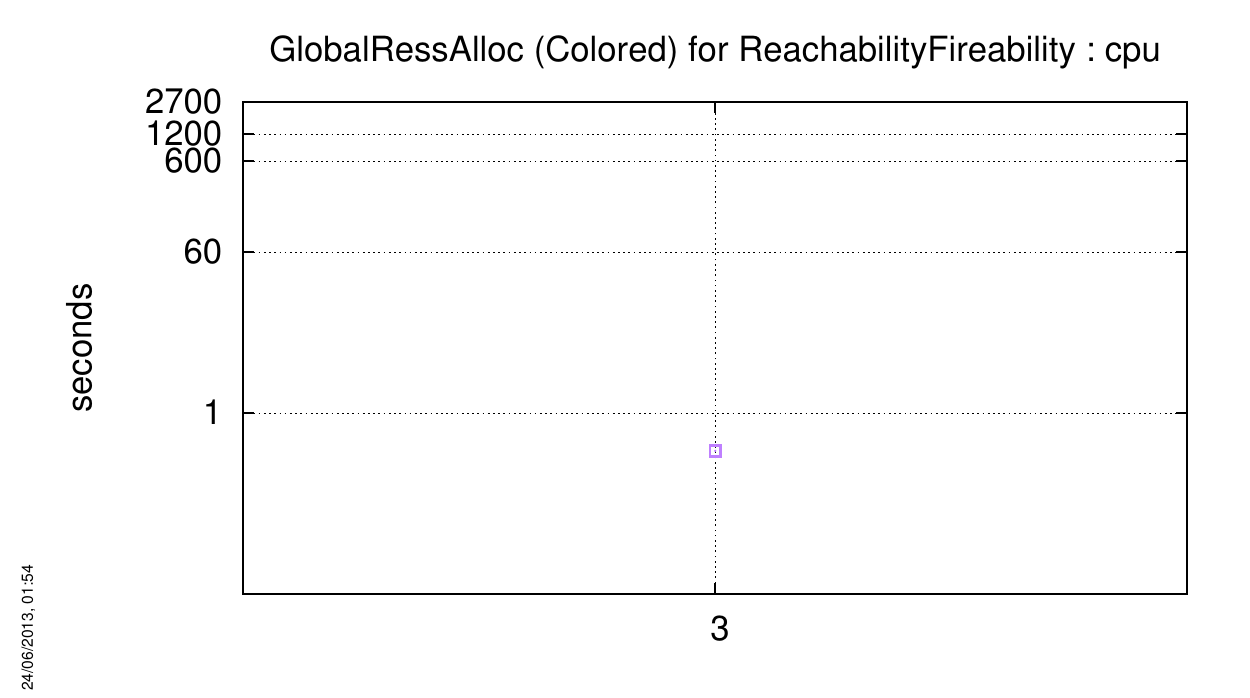}

   \includegraphics[height=1cm]{figures/tools-legend.pdf}
\end{center}

\subsubsection{\acs{GlobalRessAlloc-PT}}
The charts below respectively show how tools compete with this ``Known'' model (memory and CPU).

\index{Performances!ReachabilityFireability!GlobalRessAlloc (P/T)}
\begin{center}
   \includegraphics[width=7.2cm]{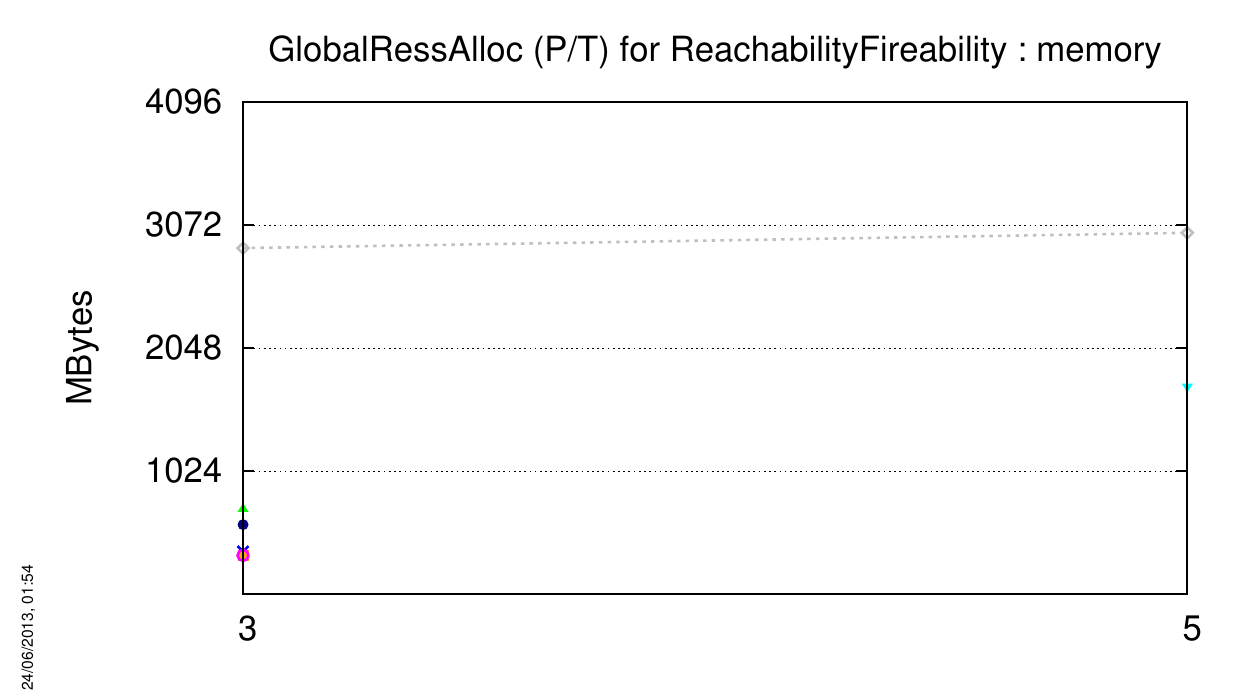}
   \includegraphics[width=7.2cm]{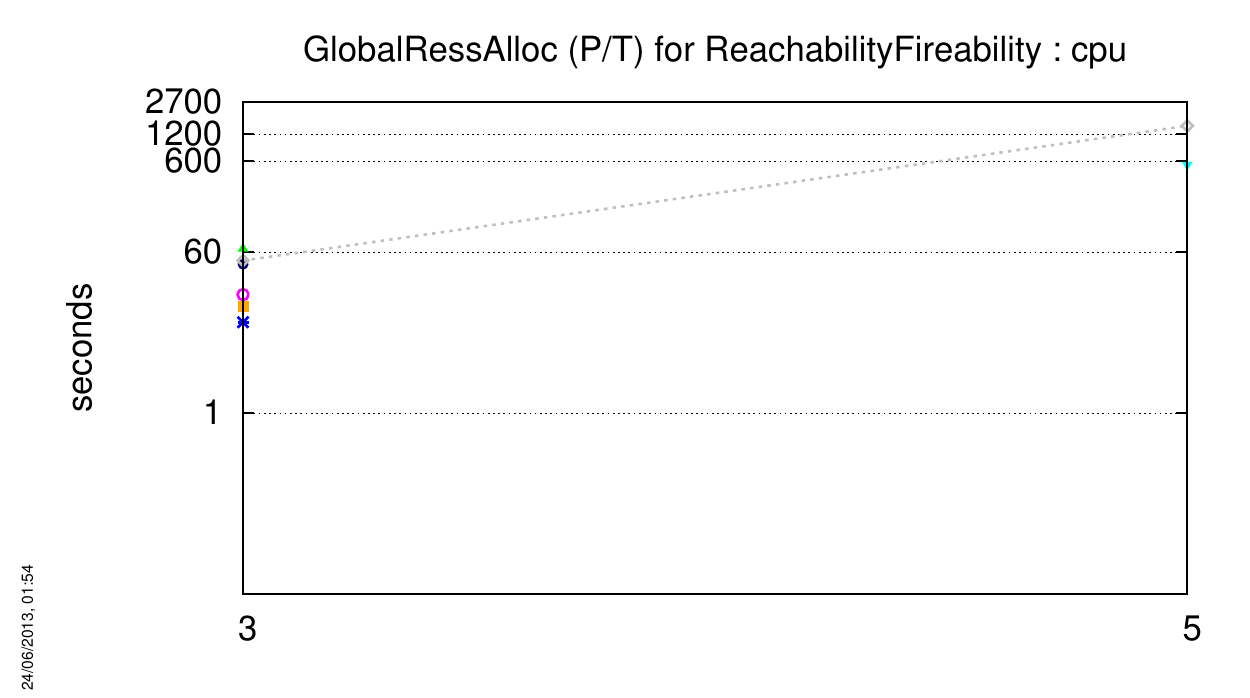}

   \includegraphics[height=1cm]{figures/tools-legend.pdf}
\end{center}

\subsubsection{\acs{Kanban-PT}}
The charts below respectively show how tools compete with this ``Known'' model (memory and CPU).

\index{Performances!ReachabilityFireability!Kanban (P/T)}
\begin{center}
   \includegraphics[width=7.2cm]{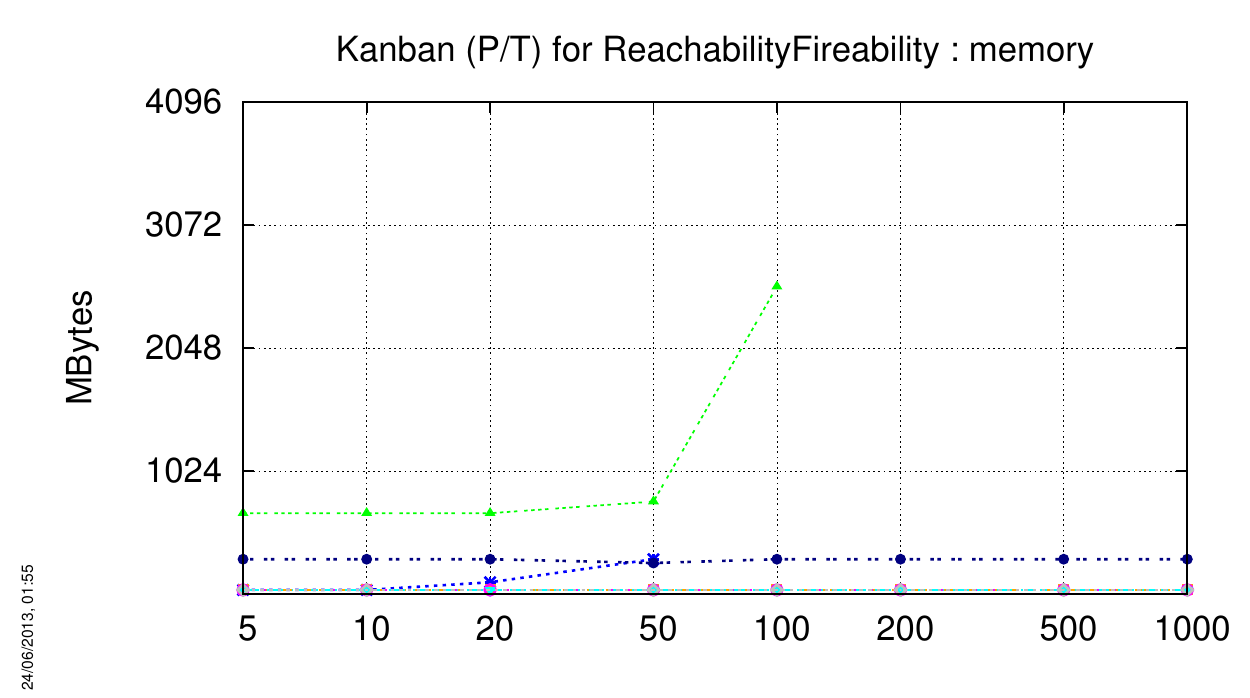}
   \includegraphics[width=7.2cm]{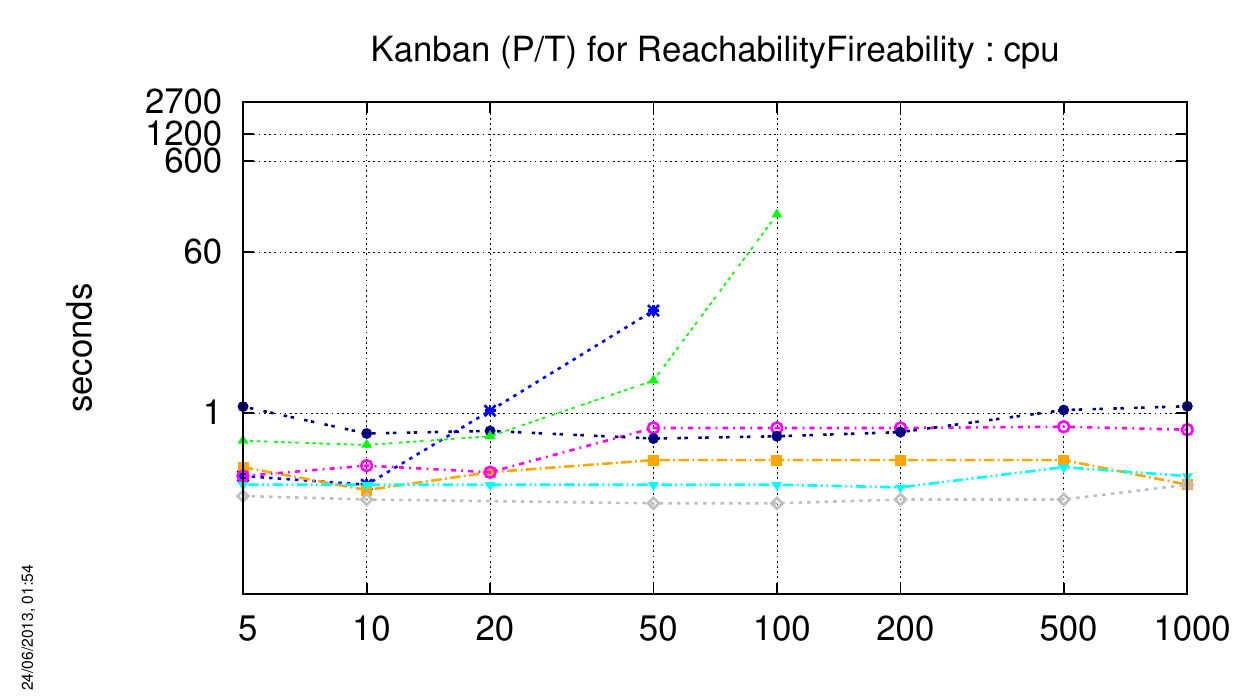}

   \includegraphics[height=1cm]{figures/tools-legend.pdf}
\end{center}

\subsubsection{\acs{LamportFastMutEx-COL}}
The charts below respectively show how tools compete with this ``Known'' model (memory and CPU).

\index{Performances!ReachabilityFireability!LamportFastMutEx (Colored)}
\begin{center}
   \includegraphics[width=7.2cm]{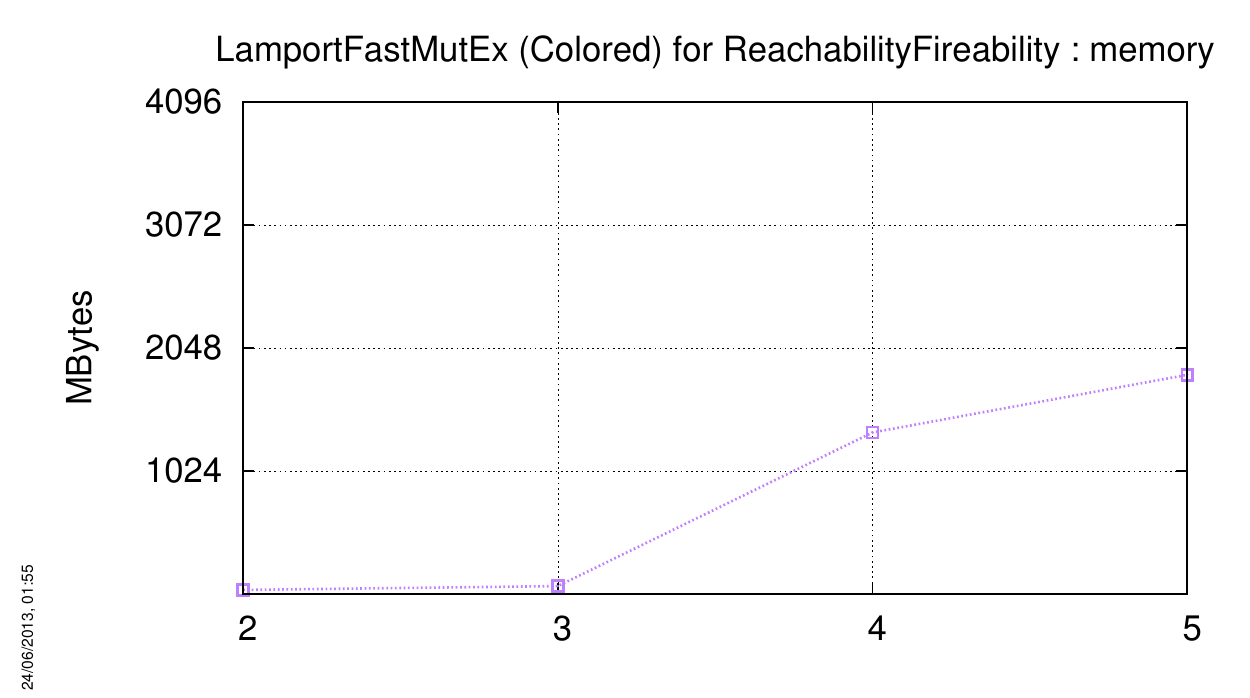}
   \includegraphics[width=7.2cm]{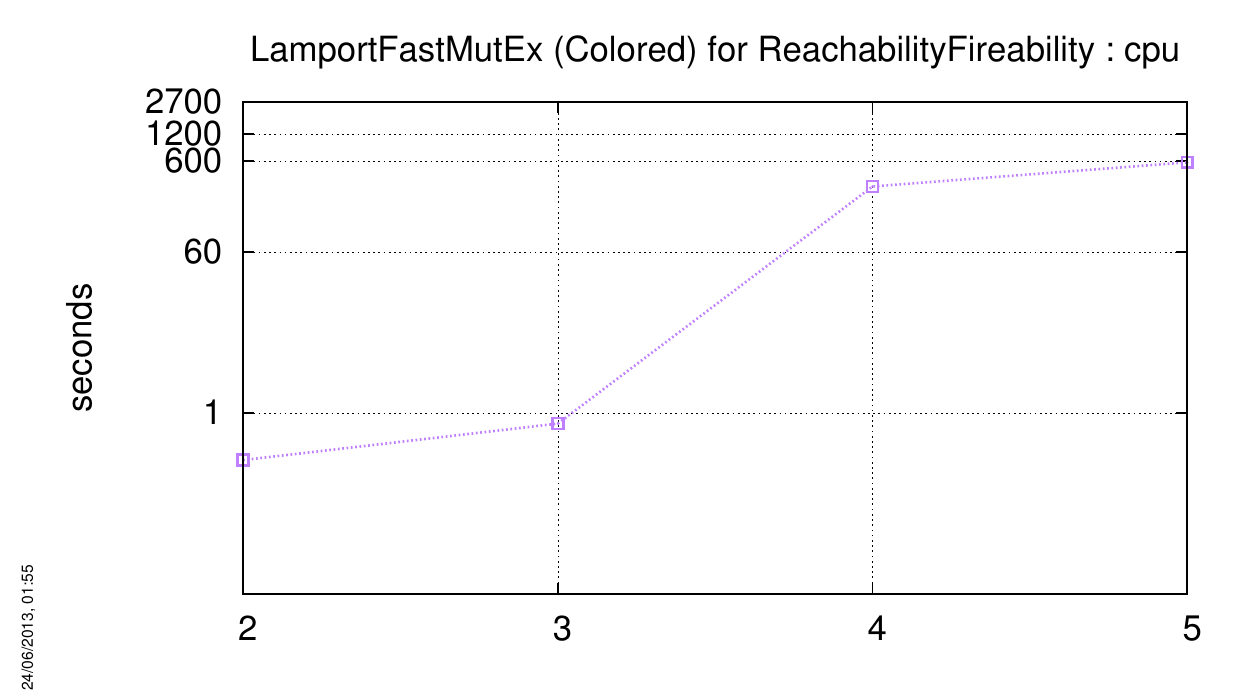}

   \includegraphics[height=1cm]{figures/tools-legend.pdf}
\end{center}

\subsubsection{\acs{LamportFastMutEx-PT}}
The charts below respectively show how tools compete with this ``Known'' model (memory and CPU).

\index{Performances!ReachabilityFireability!LamportFastMutEx (P/T)}
\begin{center}
   \includegraphics[width=7.2cm]{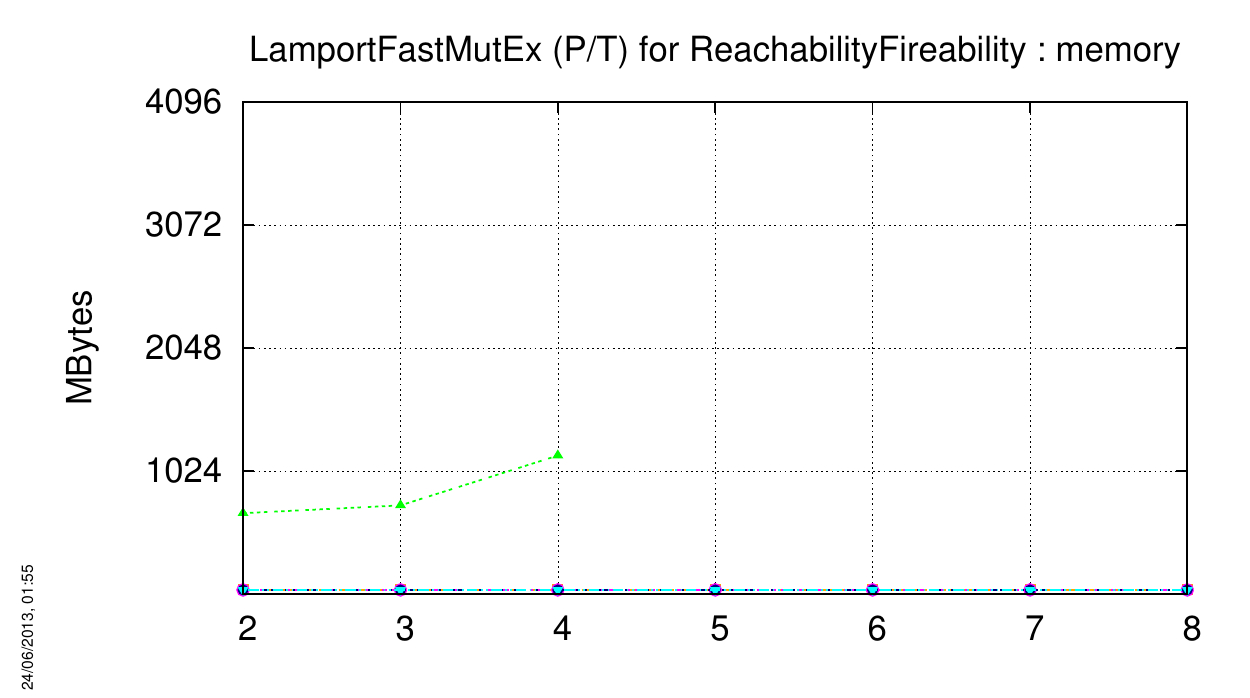}
   \includegraphics[width=7.2cm]{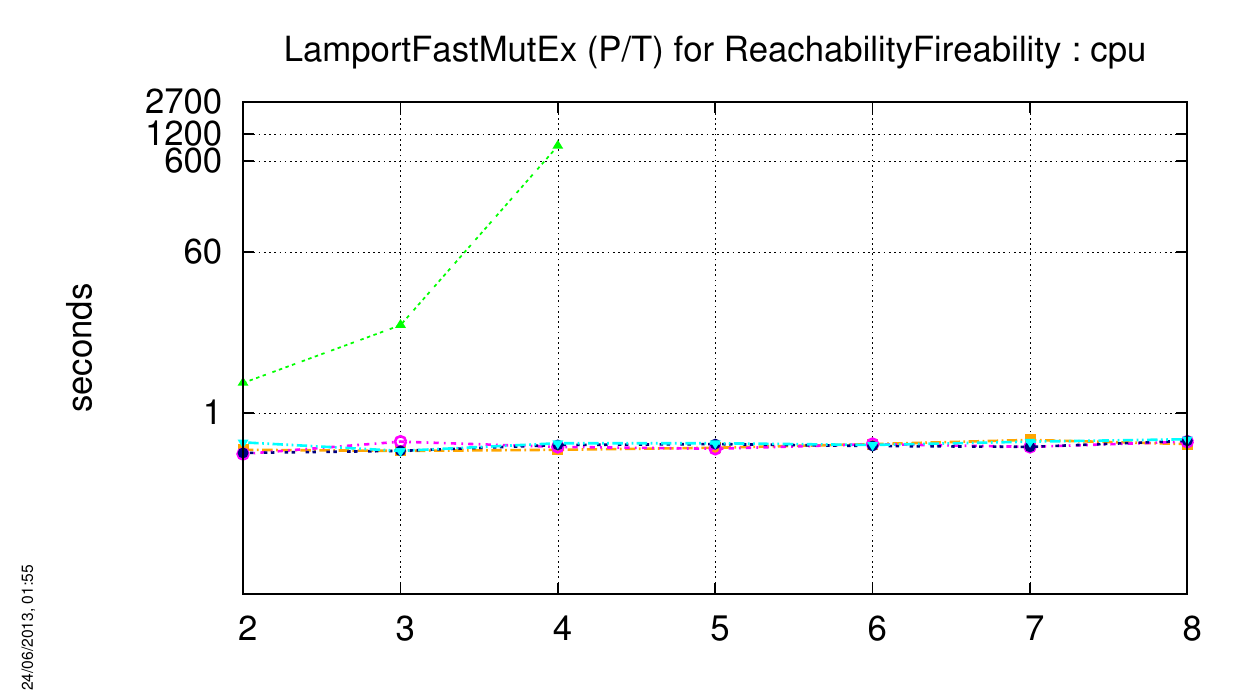}

   \includegraphics[height=1cm]{figures/tools-legend.pdf}
\end{center}

\subsubsection{\acs{MAPK-PT}}
The charts below respectively show how tools compete with this ``Known'' model (memory and CPU).

\index{Performances!ReachabilityFireability!MAPK (P/T)}
\begin{center}
   \includegraphics[width=7.2cm]{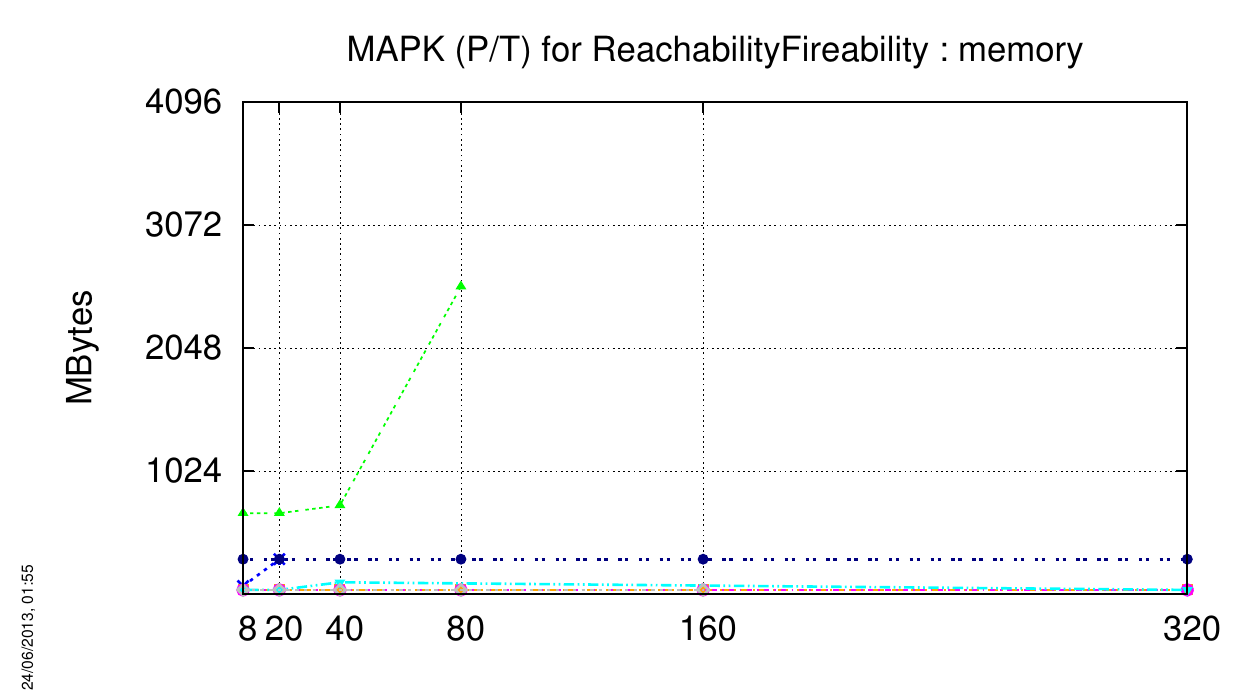}
   \includegraphics[width=7.2cm]{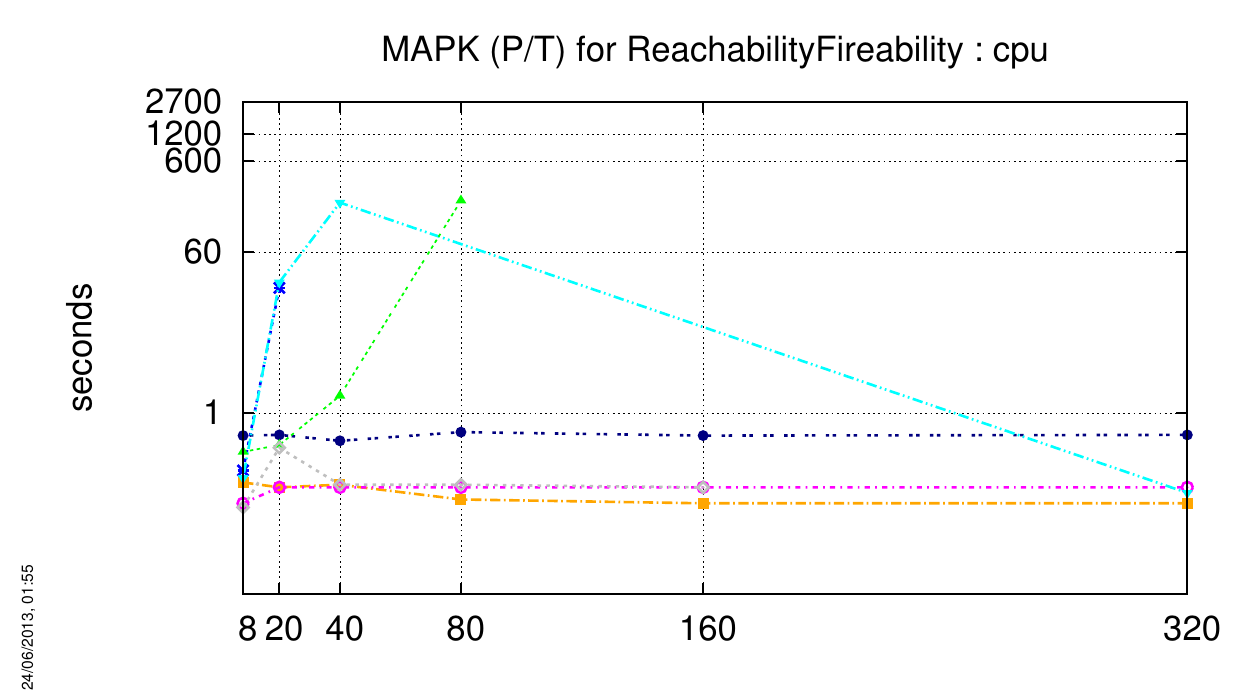}

   \includegraphics[height=1cm]{figures/tools-legend.pdf}
\end{center}

\subsubsection{\acs{NeoElection-COL}}
The charts below respectively show how tools compete with this ``Known'' model (memory and CPU).

\index{Performances!ReachabilityFireability!NeoElection (Colored)}
\begin{center}
   \includegraphics[width=7.2cm]{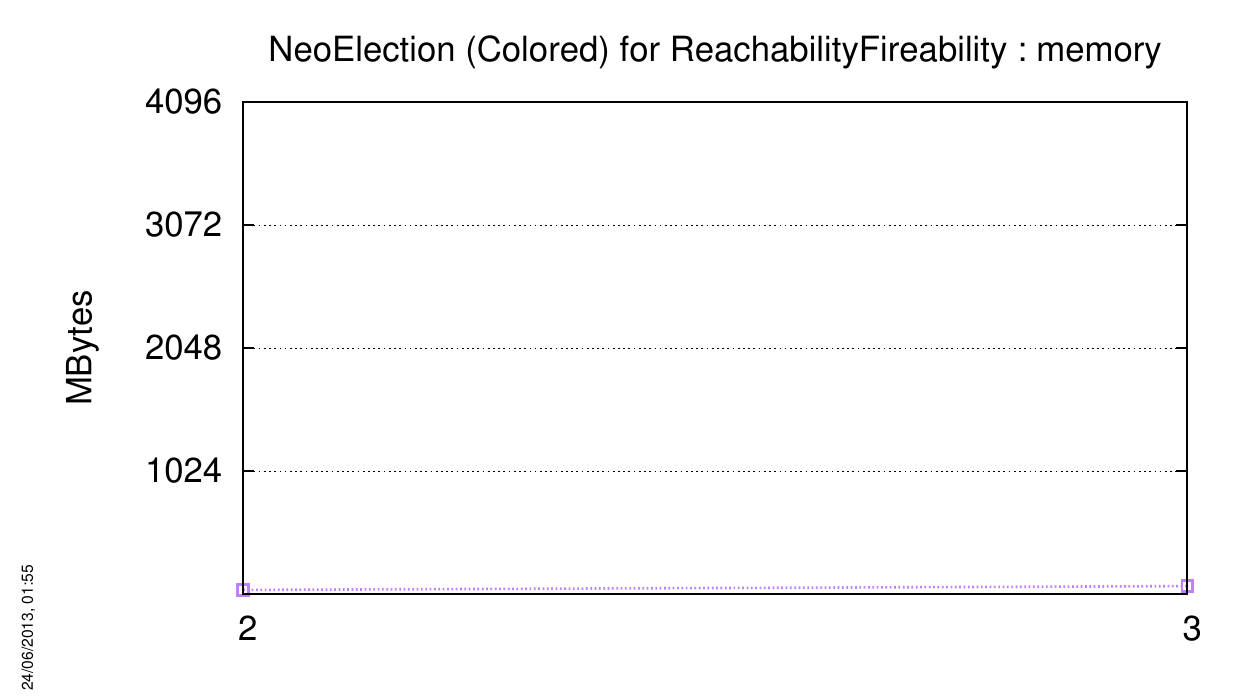}
   \includegraphics[width=7.2cm]{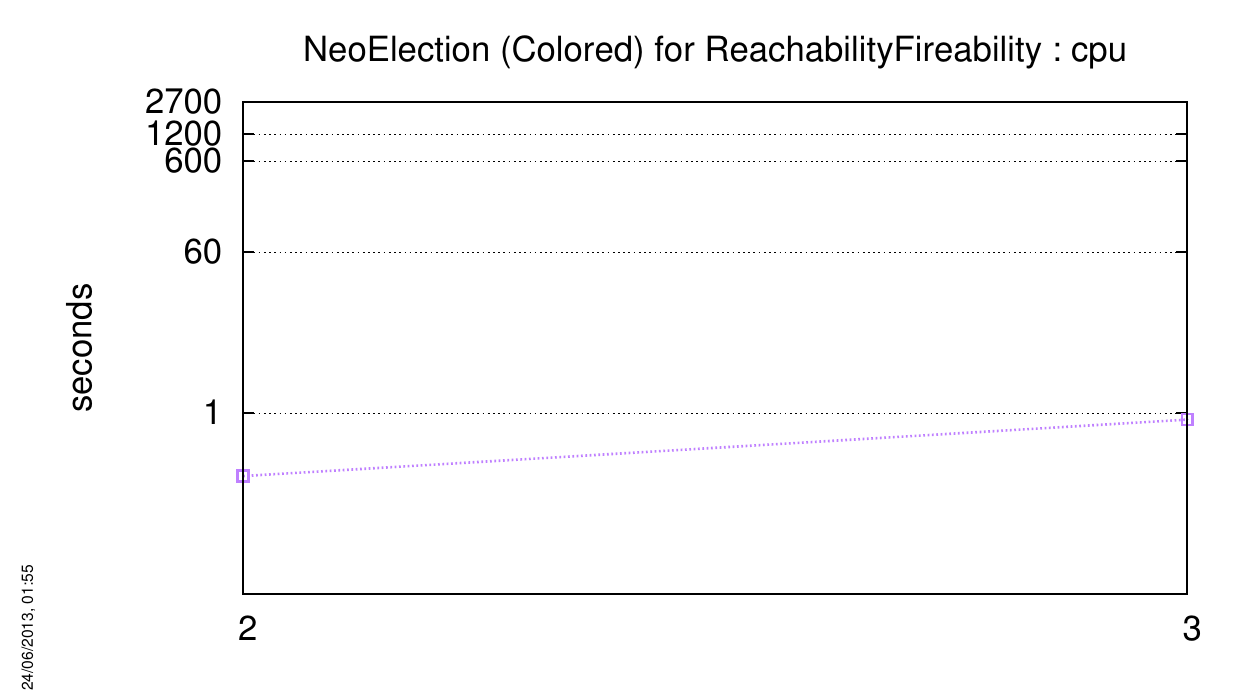}

   \includegraphics[height=1cm]{figures/tools-legend.pdf}
\end{center}

\subsubsection{\acs{NeoElection-PT}}
The charts below respectively show how tools compete with this ``Known'' model (memory and CPU).

\index{Performances!ReachabilityFireability!NeoElection (P/T)}
\begin{center}
   \includegraphics[width=7.2cm]{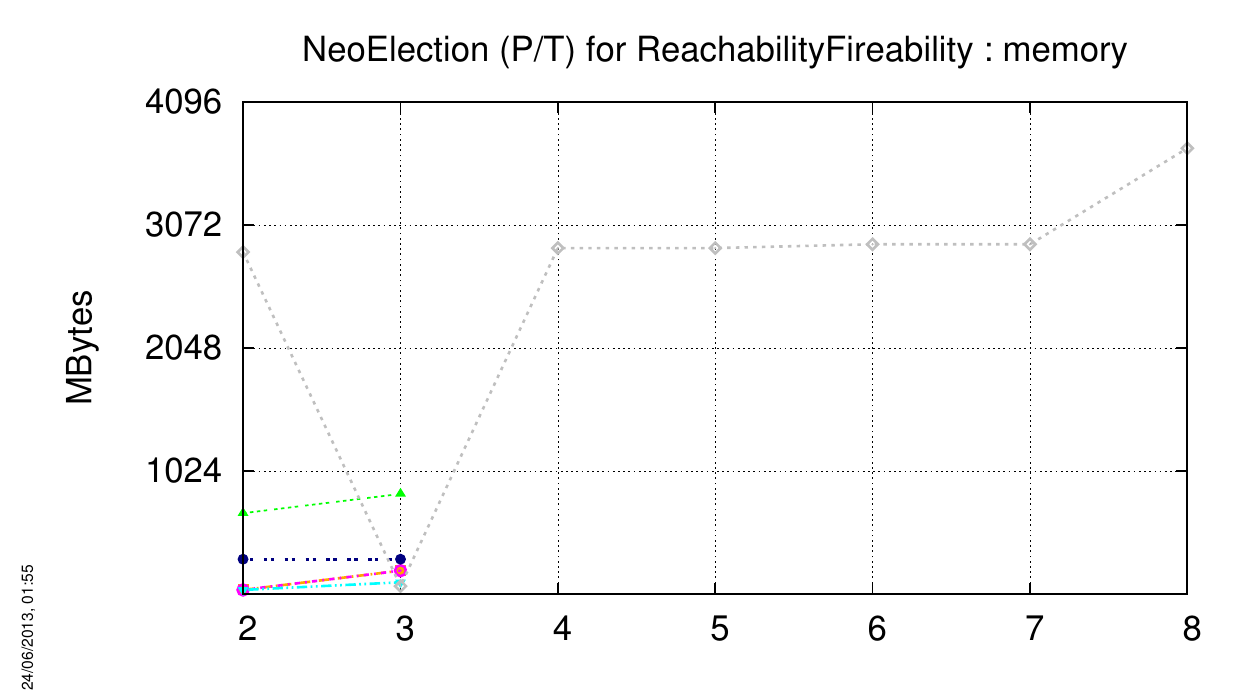}
   \includegraphics[width=7.2cm]{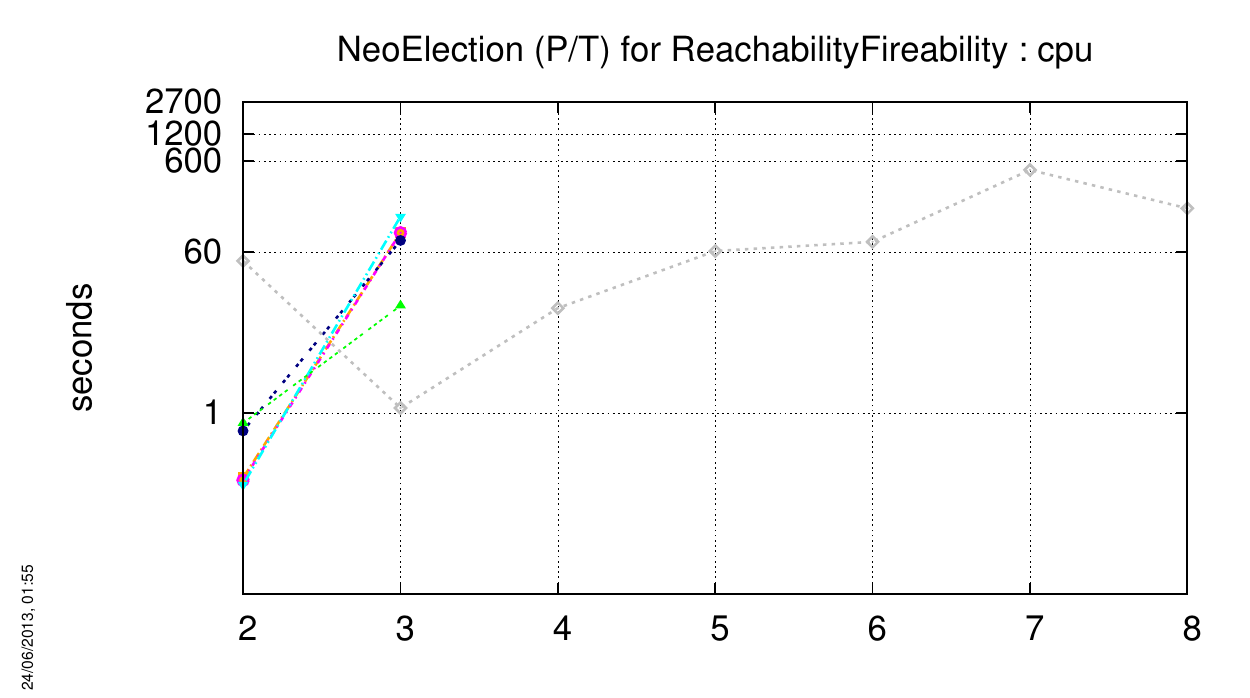}

   \includegraphics[height=1cm]{figures/tools-legend.pdf}
\end{center}

\subsubsection{\acs{PermAdmissibility-COL}}
The charts below respectively show how tools compete with this ``Known'' model (memory and CPU).

\index{Performances!ReachabilityFireability!PermAdmissibility (Colored)}
\begin{center}
   \includegraphics[width=7.2cm]{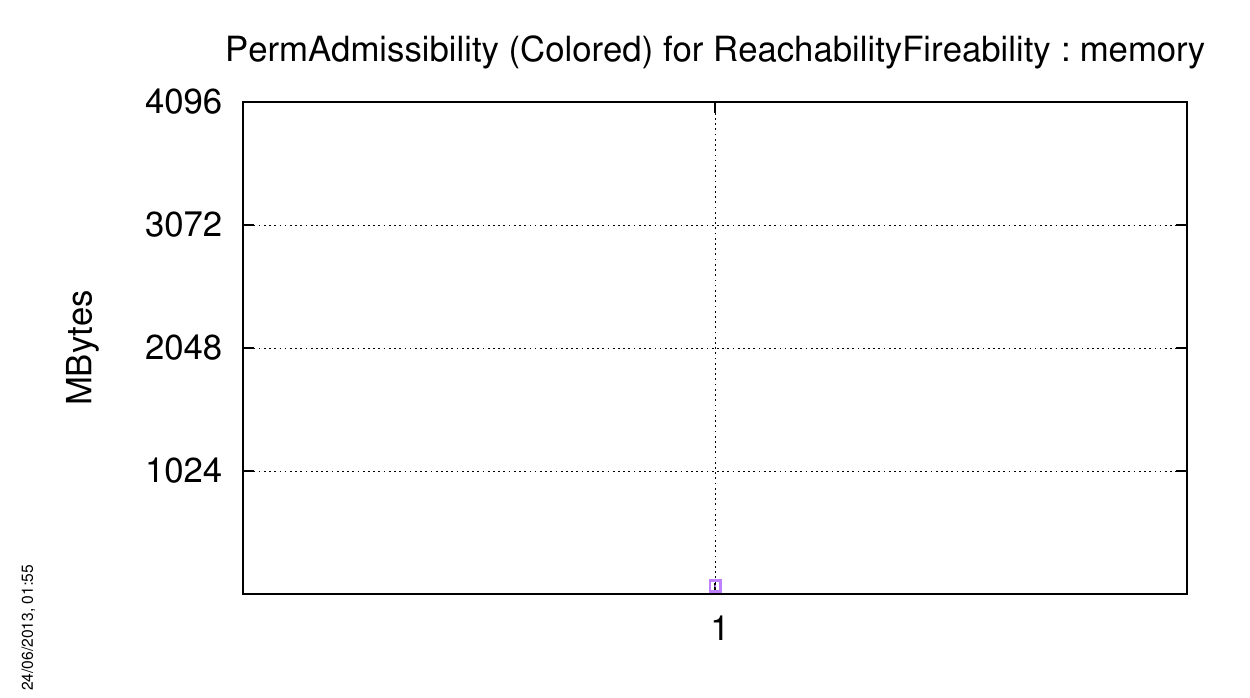}
   \includegraphics[width=7.2cm]{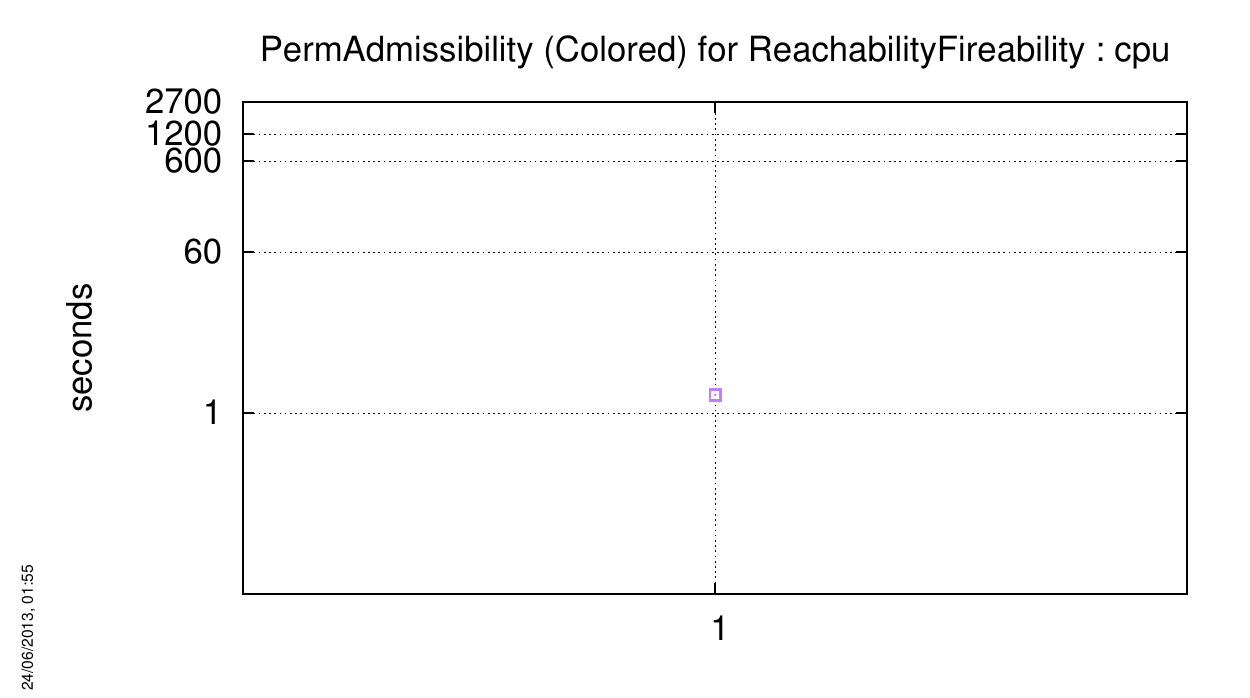}

   \includegraphics[height=1cm]{figures/tools-legend.pdf}
\end{center}

\subsubsection{\acs{PermAdmissibility-PT}}
The charts below respectively show how tools compete with this ``Known'' model (memory and CPU).

\index{Performances!ReachabilityFireability!PermAdmissibility (P/T)}
\begin{center}
   \includegraphics[width=7.2cm]{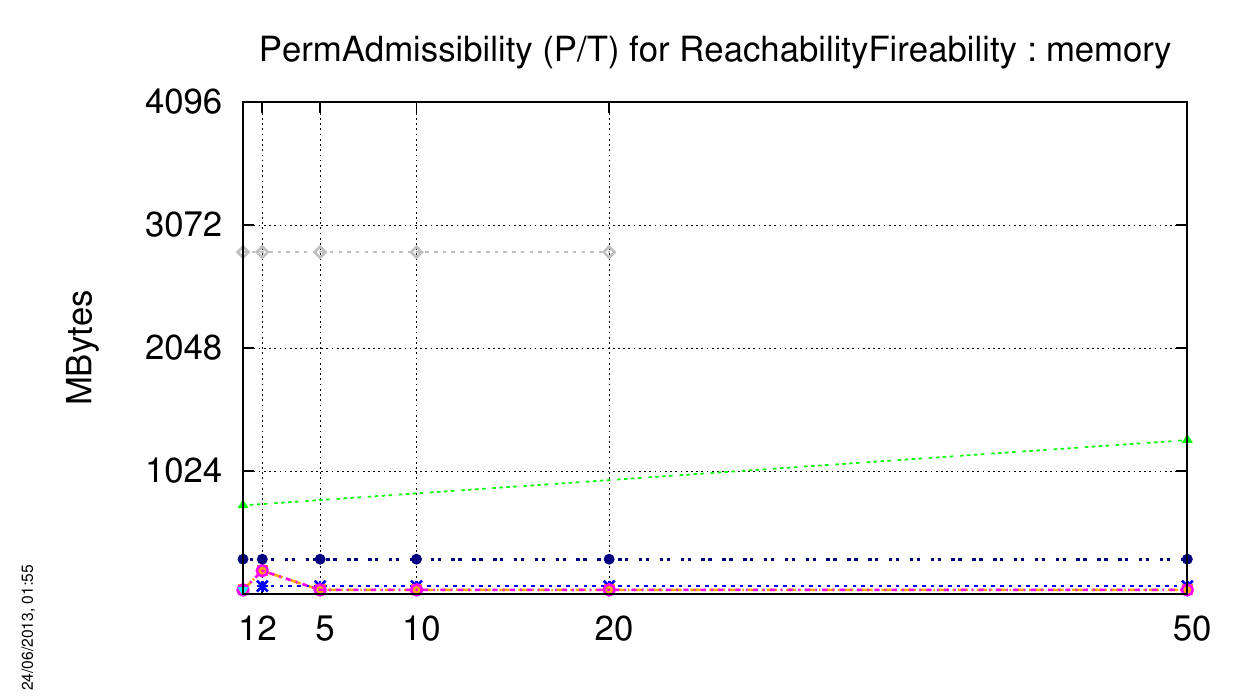}
   \includegraphics[width=7.2cm]{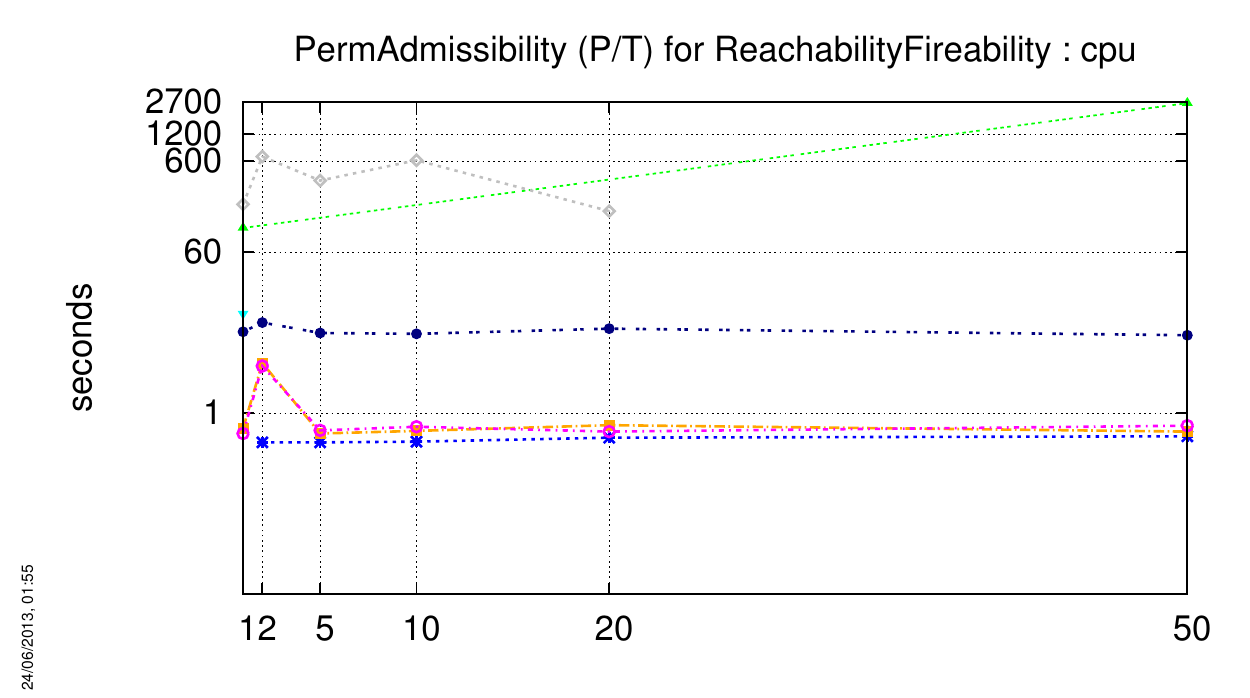}

   \includegraphics[height=1cm]{figures/tools-legend.pdf}
\end{center}

\subsubsection{\acs{Peterson-COL}}
The charts below respectively show how tools compete with this ``Known'' model (memory and CPU).

\index{Performances!ReachabilityFireability!Peterson (Colored)}
\begin{center}
   \includegraphics[width=7.2cm]{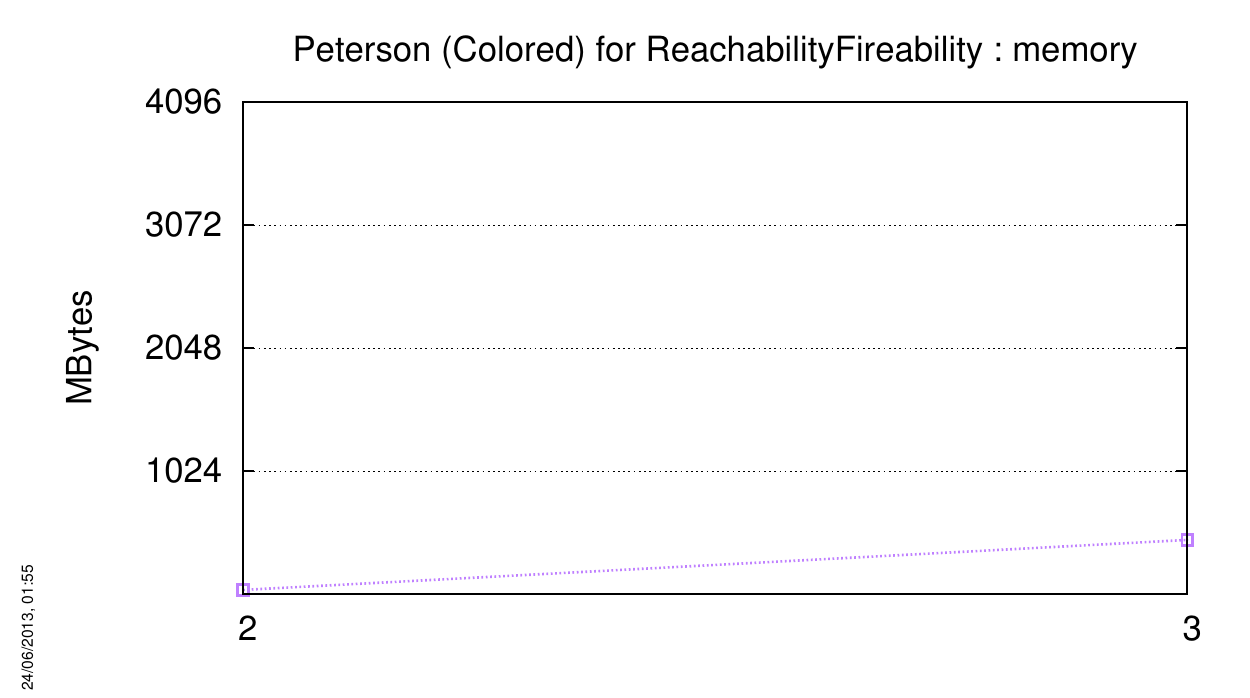}
   \includegraphics[width=7.2cm]{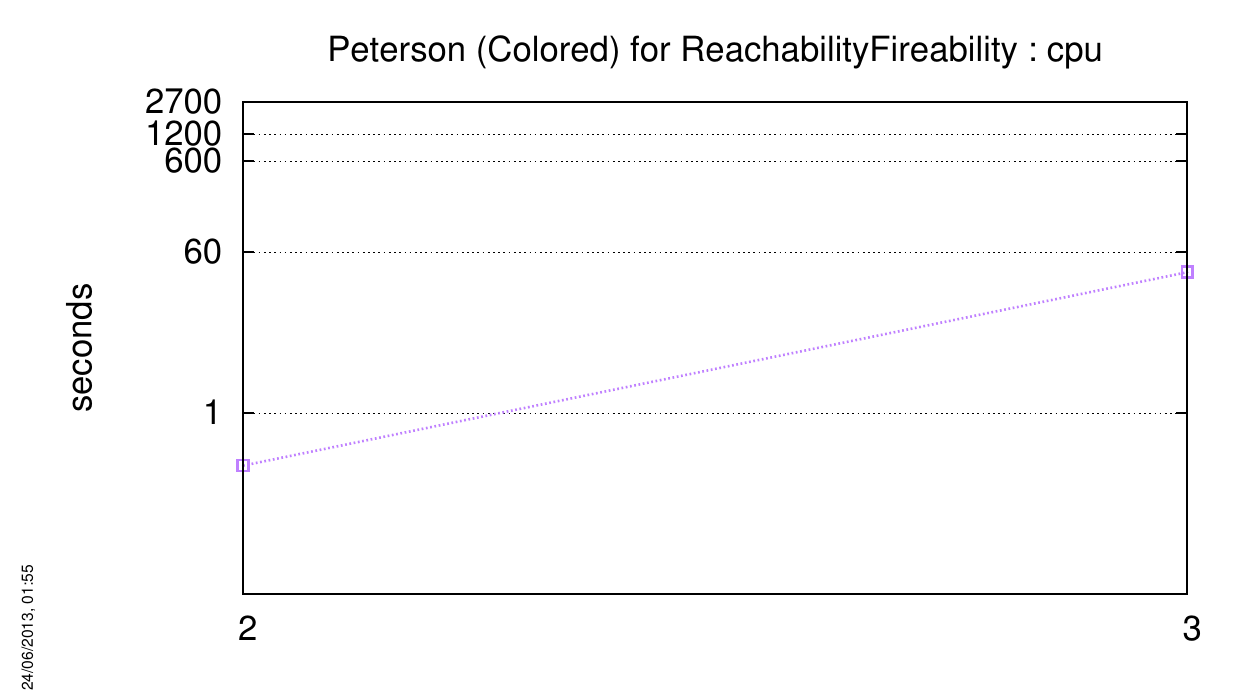}

   \includegraphics[height=1cm]{figures/tools-legend.pdf}
\end{center}

\subsubsection{\acs{Peterson-PT}}
The charts below respectively show how tools compete with this ``Known'' model (memory and CPU).

\index{Performances!ReachabilityFireability!Peterson (P/T)}
\begin{center}
   \includegraphics[width=7.2cm]{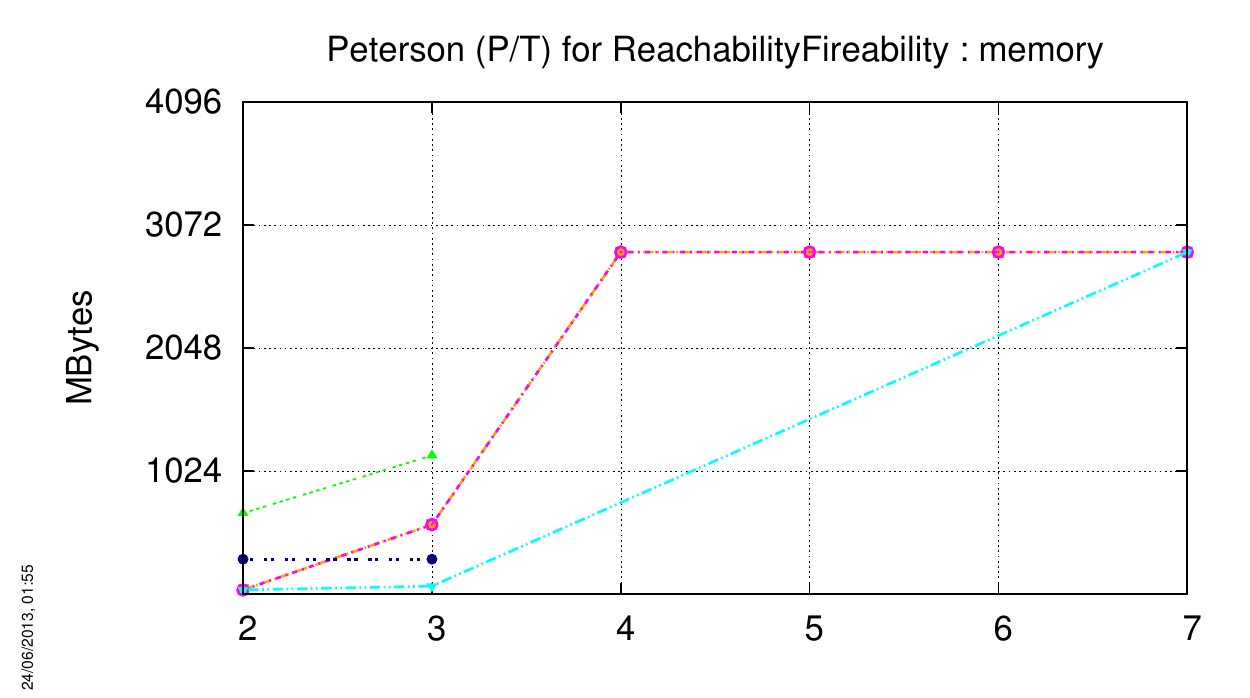}
   \includegraphics[width=7.2cm]{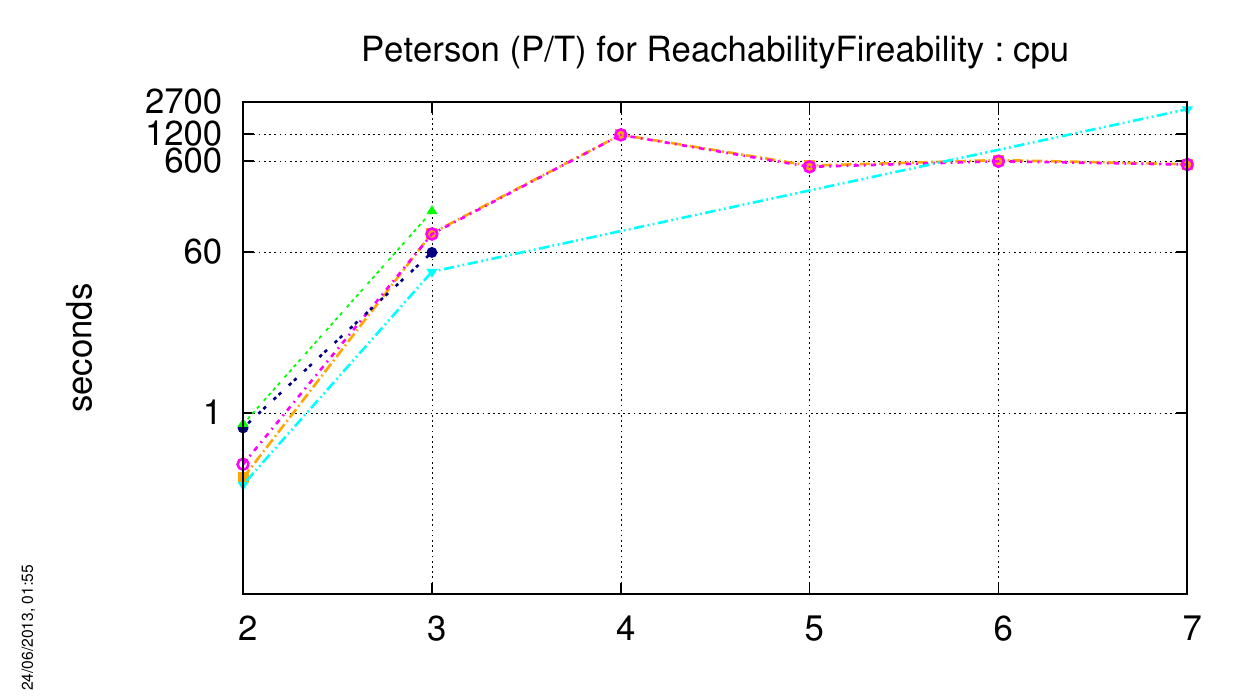}

   \includegraphics[height=1cm]{figures/tools-legend.pdf}
\end{center}

\subsubsection{\acs{Philosophers-COL}}
No instance of this model could be computed for the \textbf{ReachabilityFireability} examination.

\subsubsection{\acs{Philosophers-PT}}
The charts below respectively show how tools compete with this ``Known'' model (memory and CPU).

\index{Performances!ReachabilityFireability!Philosophers (P/T)}
\begin{center}
   \includegraphics[width=7.2cm]{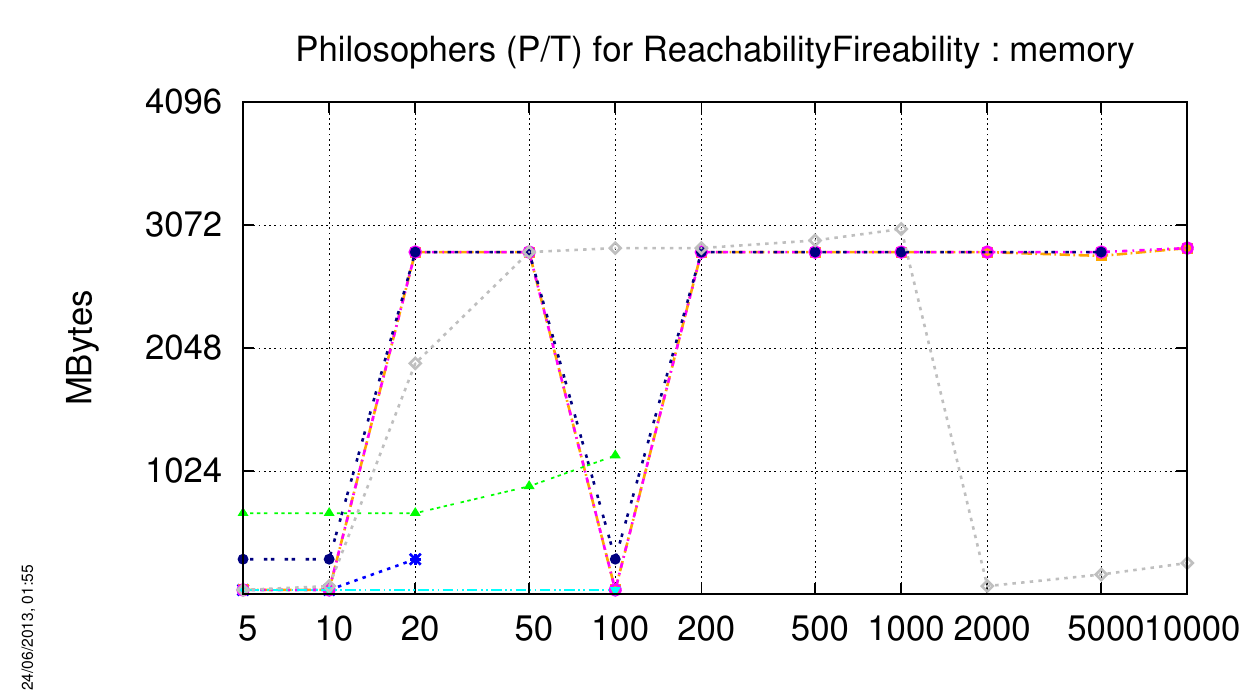}
   \includegraphics[width=7.2cm]{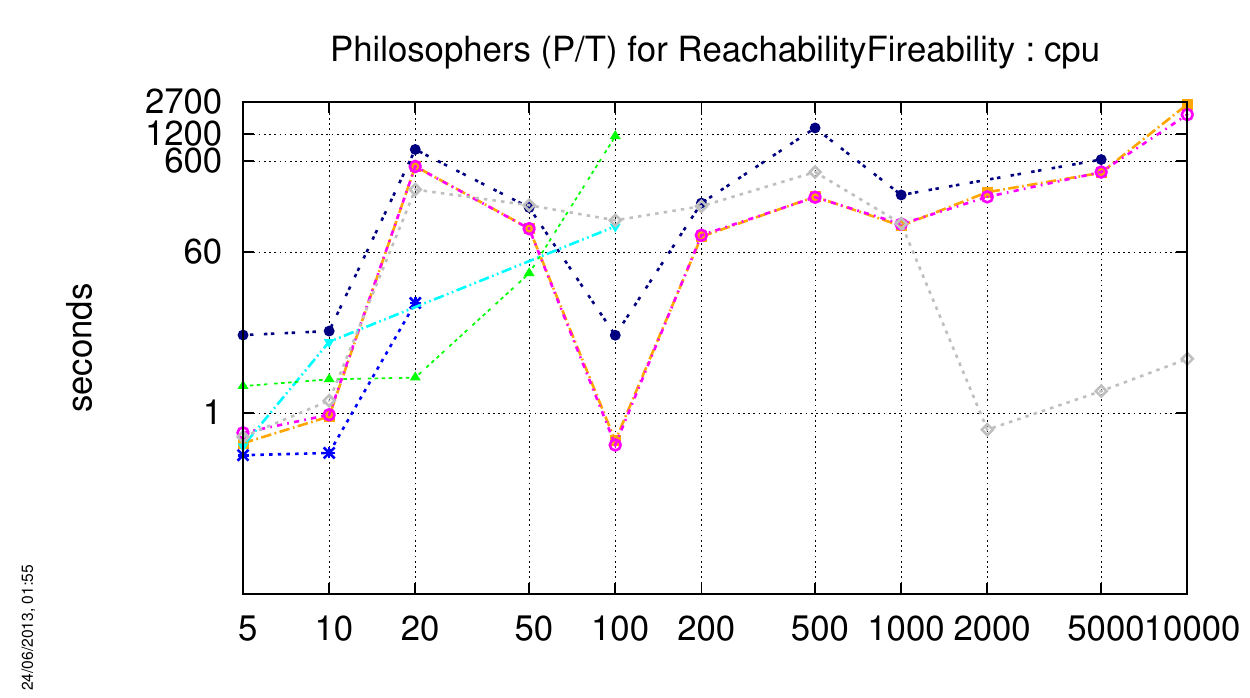}

   \includegraphics[height=1cm]{figures/tools-legend.pdf}
\end{center}

\subsubsection{\acs{PhilosophersDyn-COL}}
No instance of this model could be computed for the \textbf{ReachabilityFireability} examination.

\subsubsection{\acs{PhilosophersDyn-PT}}
The charts below respectively show how tools compete with this ``Known'' model (memory and CPU).

\index{Performances!ReachabilityFireability!PhilosophersDyn (P/T)}
\begin{center}
   \includegraphics[width=7.2cm]{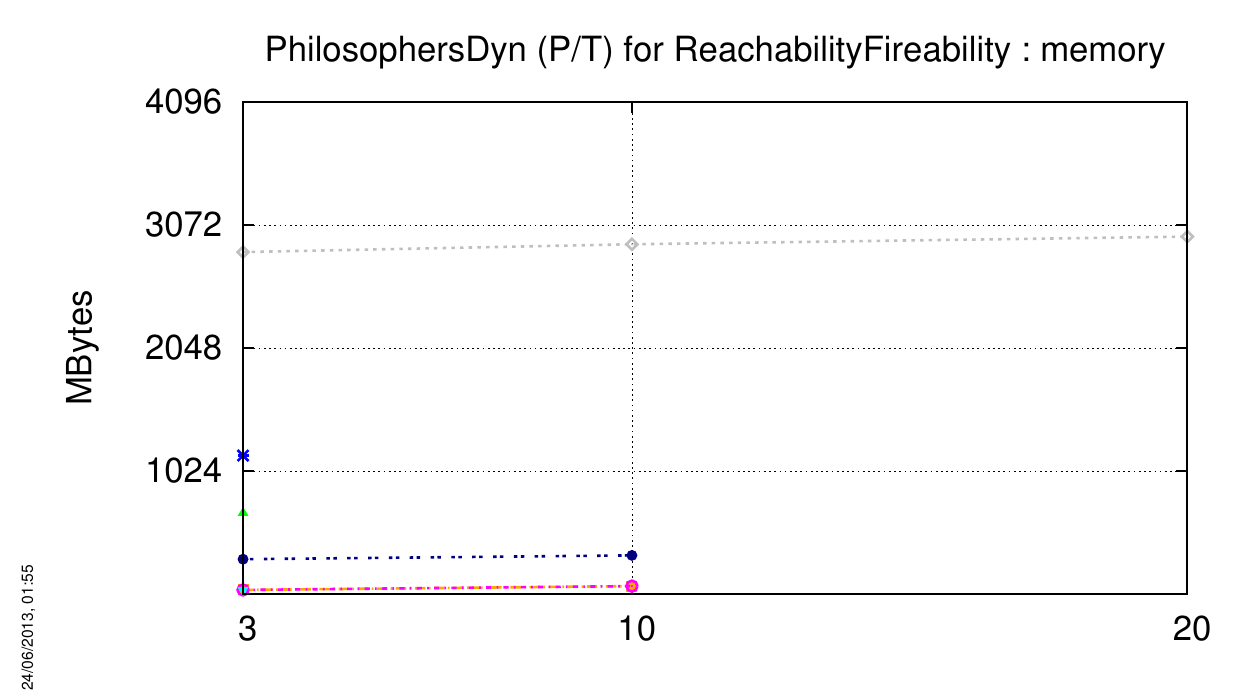}
   \includegraphics[width=7.2cm]{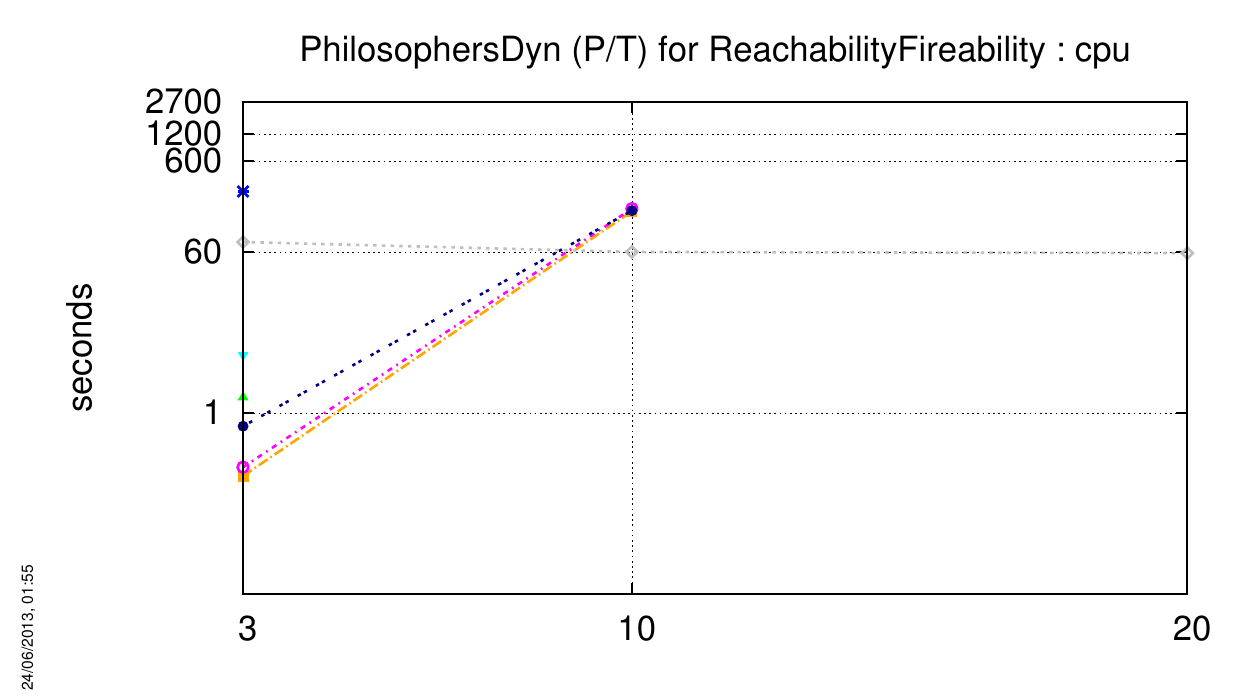}

   \includegraphics[height=1cm]{figures/tools-legend.pdf}
\end{center}

\subsubsection{\acs{Planning-PT}}
No instance of this model could be computed for the \textbf{ReachabilityFireability} examination.

\subsubsection{\acs{Railroad-PT}}
The charts below respectively show how tools compete with this ``Known'' model (memory and CPU).

\index{Performances!ReachabilityFireability!Railroad (P/T)}
\begin{center}
   \includegraphics[width=7.2cm]{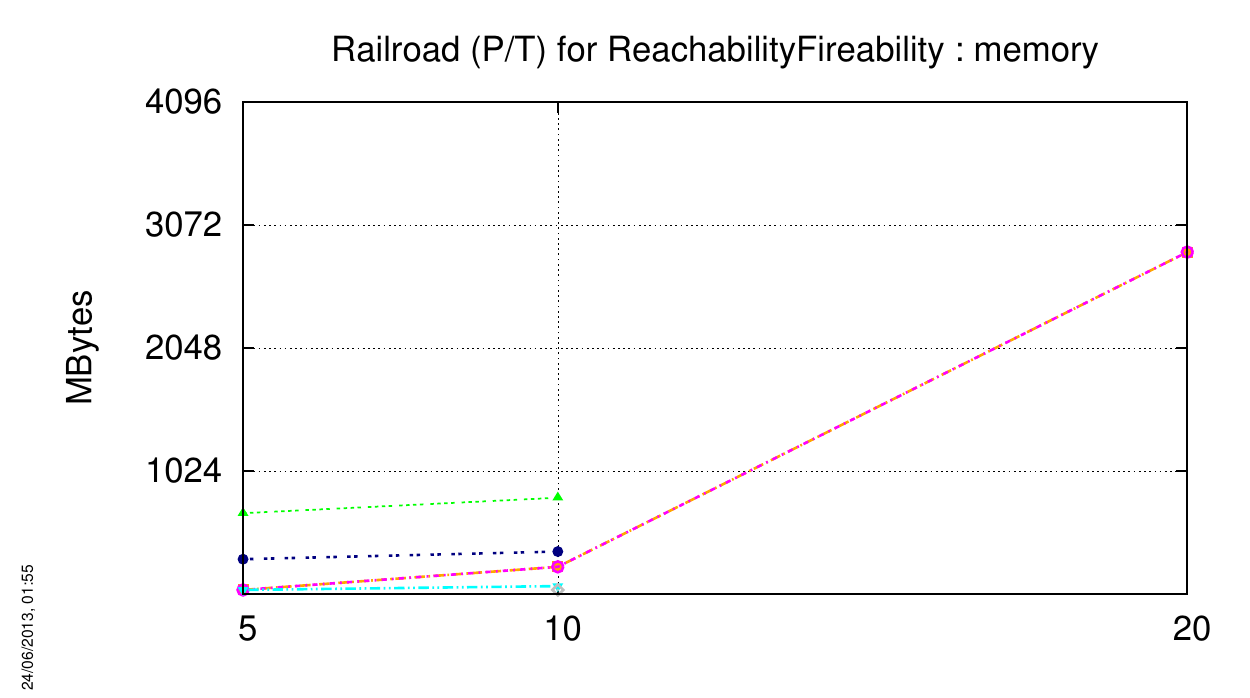}
   \includegraphics[width=7.2cm]{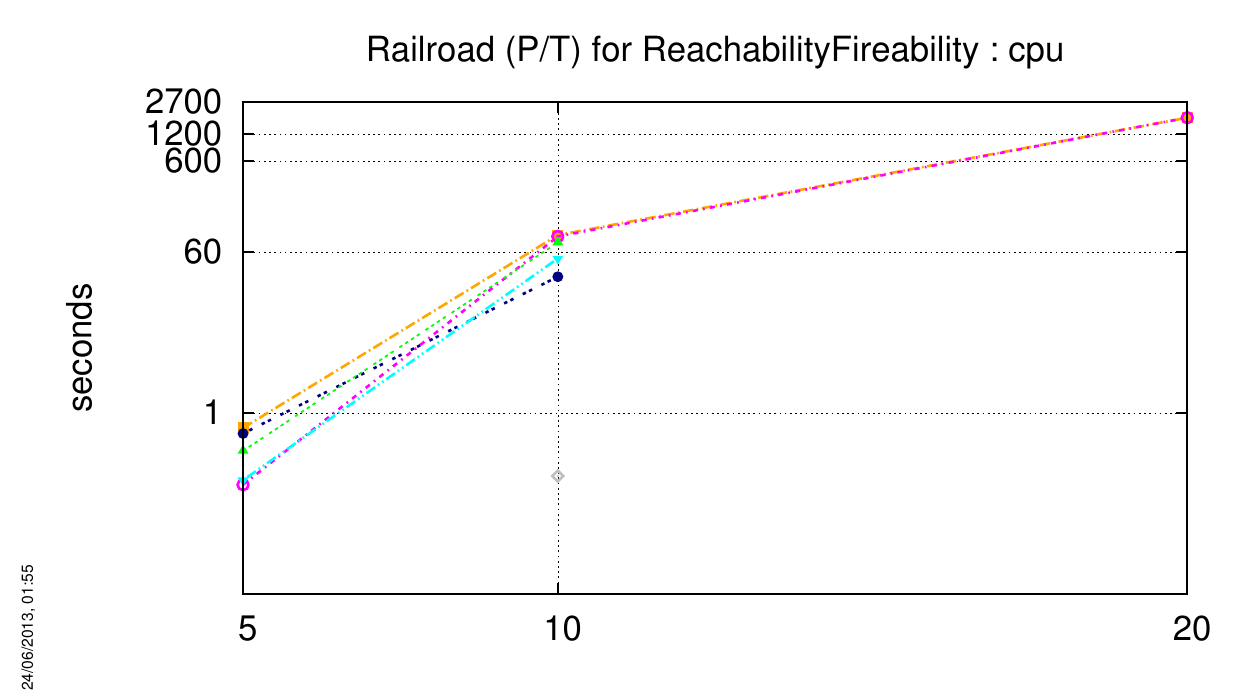}

   \includegraphics[height=1cm]{figures/tools-legend.pdf}
\end{center}

\subsubsection{\acs{RessAllocation-PT}}
The charts below respectively show how tools compete with this ``Known'' model (memory and CPU).

\index{Performances!ReachabilityFireability!RessAllocation (P/T)}
\begin{center}
   \includegraphics[width=7.2cm]{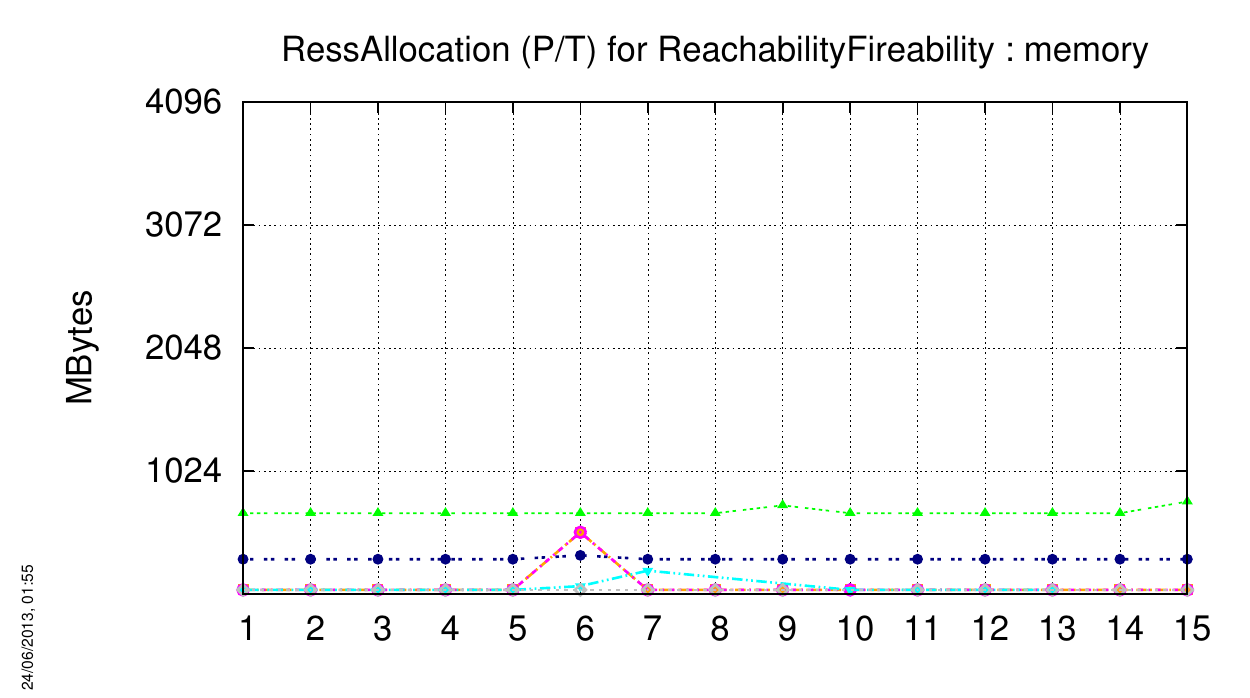}
   \includegraphics[width=7.2cm]{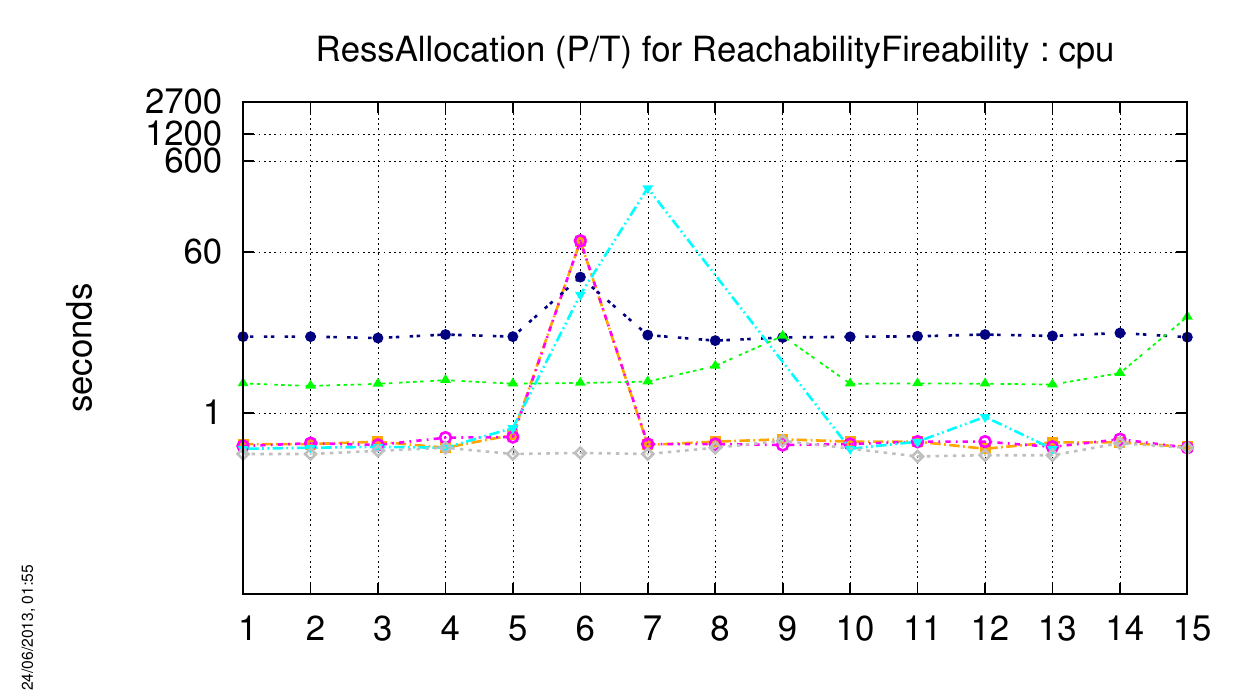}

   \includegraphics[height=1cm]{figures/tools-legend.pdf}
\end{center}

\subsubsection{\acs{Ring-PT}}
The charts below respectively show how tools compete with this ``Known'' model (memory and CPU).

\index{Performances!ReachabilityFireability!Ring (P/T)}
\begin{center}
   \includegraphics[width=7.2cm]{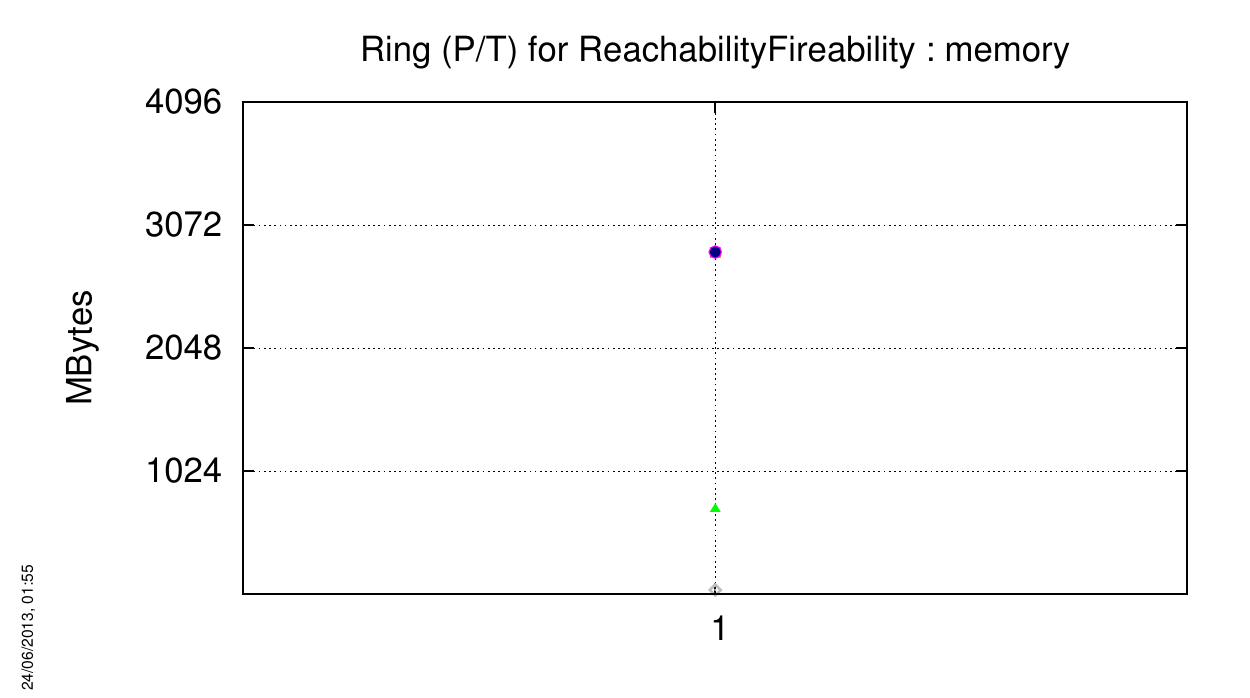}
   \includegraphics[width=7.2cm]{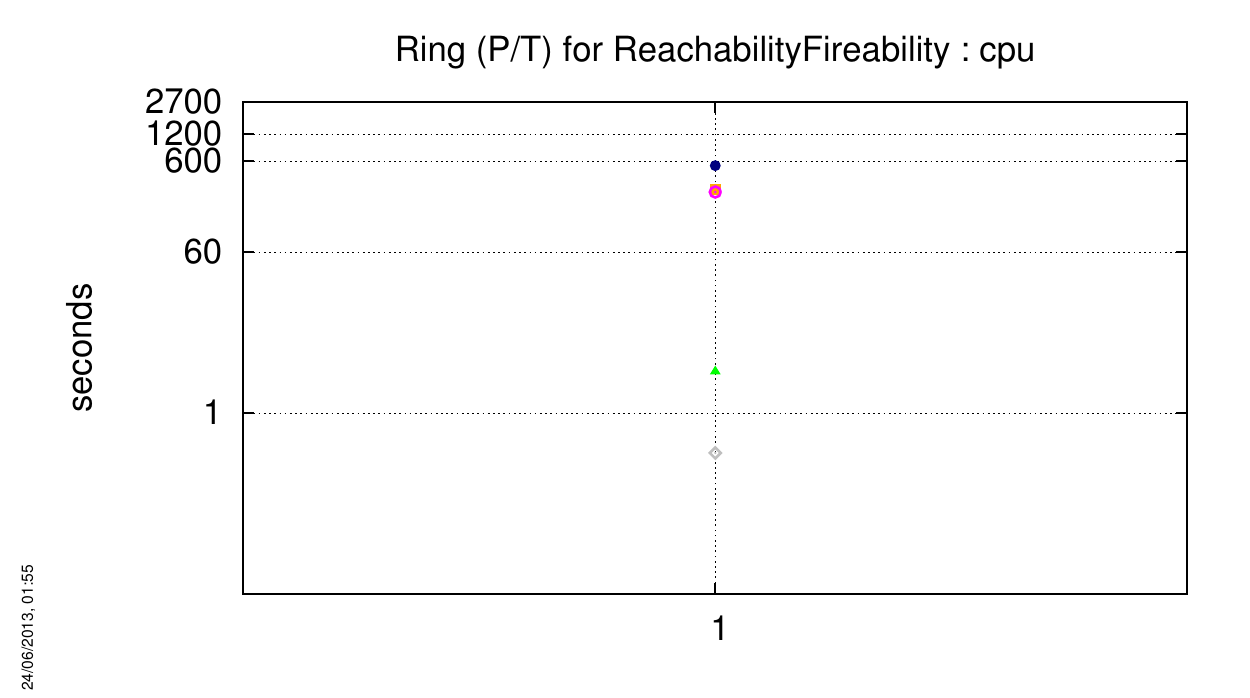}

   \includegraphics[height=1cm]{figures/tools-legend.pdf}
\end{center}

\subsubsection{\acs{RwMutex-PT}}
The charts below respectively show how tools compete with this ``Known'' model (memory and CPU).

\index{Performances!ReachabilityFireability!RwMutex (P/T)}
\begin{center}
   \includegraphics[width=7.2cm]{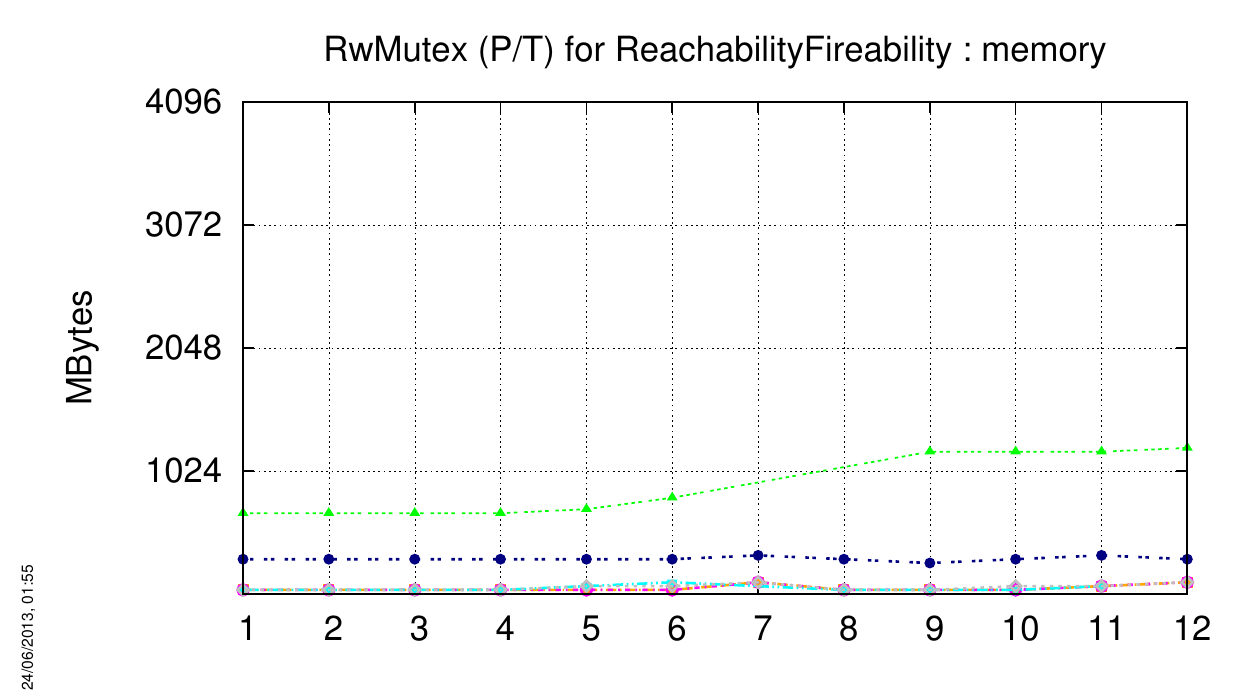}
   \includegraphics[width=7.2cm]{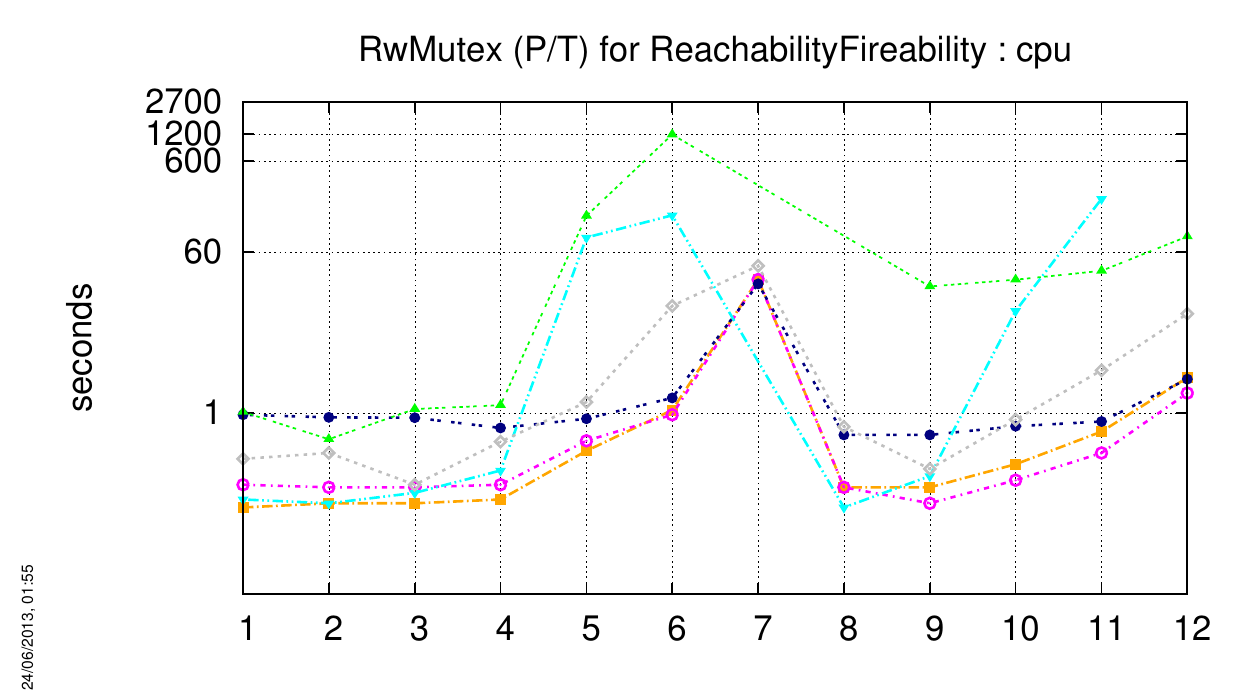}

   \includegraphics[height=1cm]{figures/tools-legend.pdf}
\end{center}

\subsubsection{\acs{SharedMemory-COL}}
No instance of this model could be computed for the \textbf{ReachabilityFireability} examination.

\subsubsection{\acs{SharedMemory-PT}}
The charts below respectively show how tools compete with this ``Known'' model (memory and CPU).

\index{Performances!ReachabilityFireability!SharedMemory (P/T)}
\begin{center}
   \includegraphics[width=7.2cm]{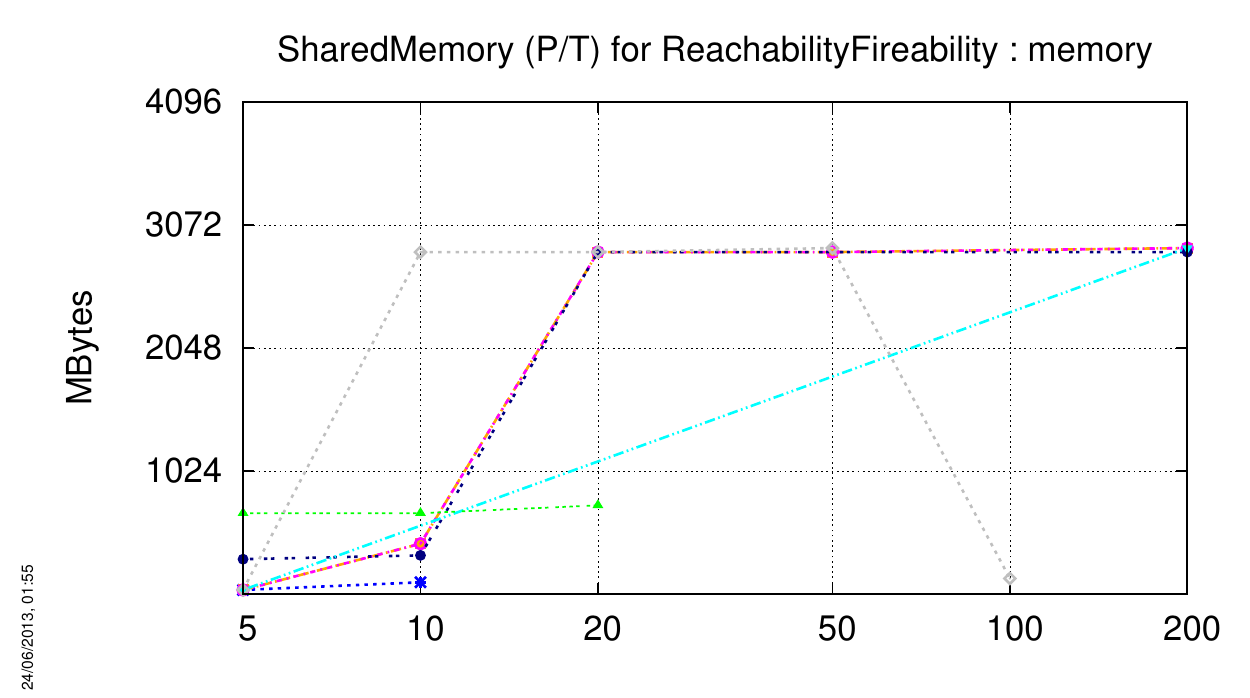}
   \includegraphics[width=7.2cm]{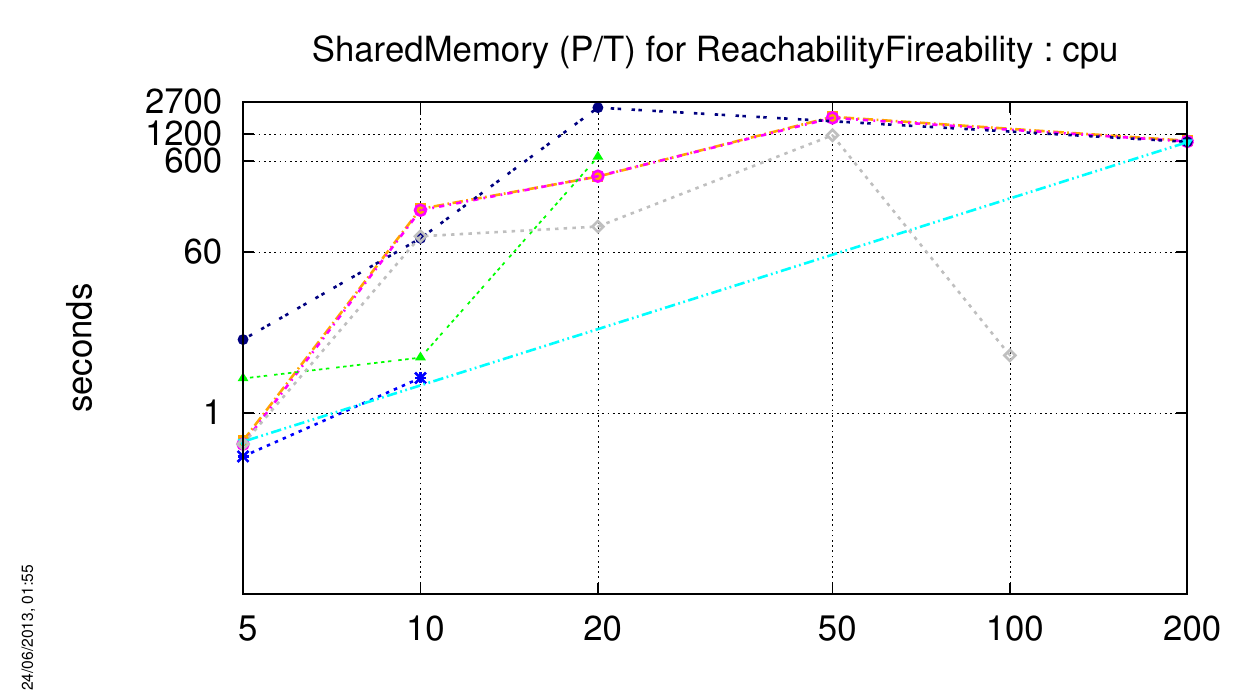}

   \includegraphics[height=1cm]{figures/tools-legend.pdf}
\end{center}

\subsubsection{\acs{SimpleLoadBal-COL}}
The charts below respectively show how tools compete with this ``Known'' model (memory and CPU).

\index{Performances!ReachabilityFireability!SimpleLoadBal (Colored)}
\begin{center}
   \includegraphics[width=7.2cm]{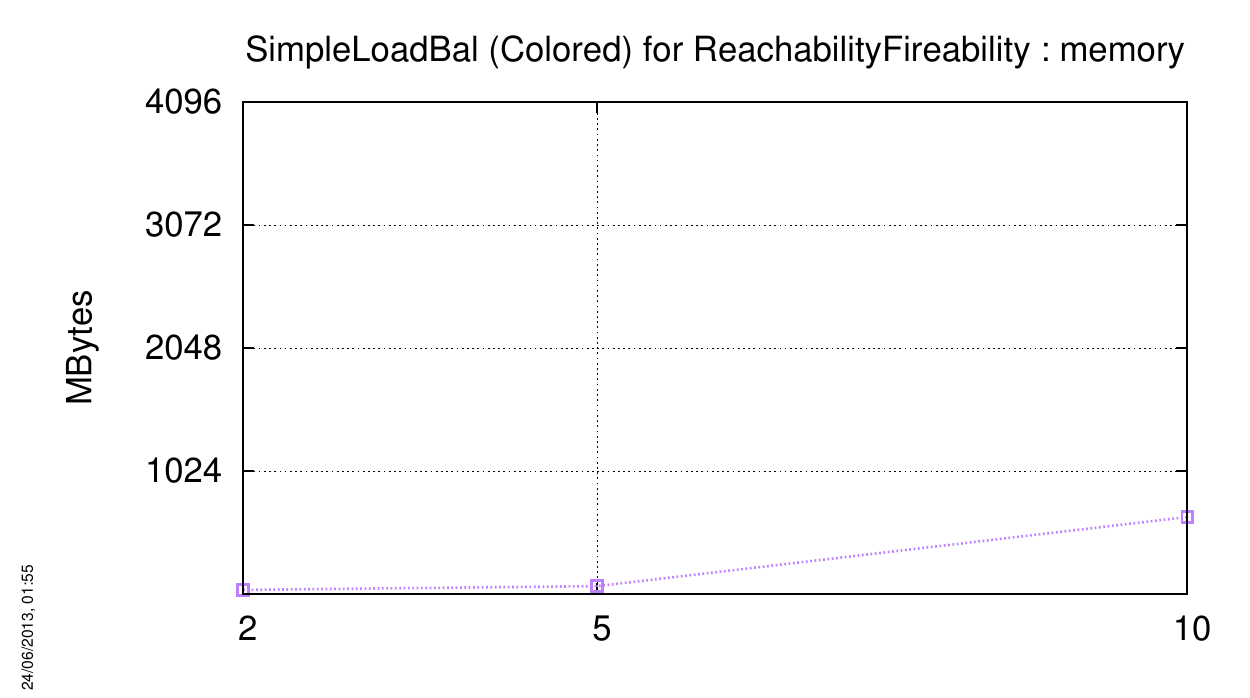}
   \includegraphics[width=7.2cm]{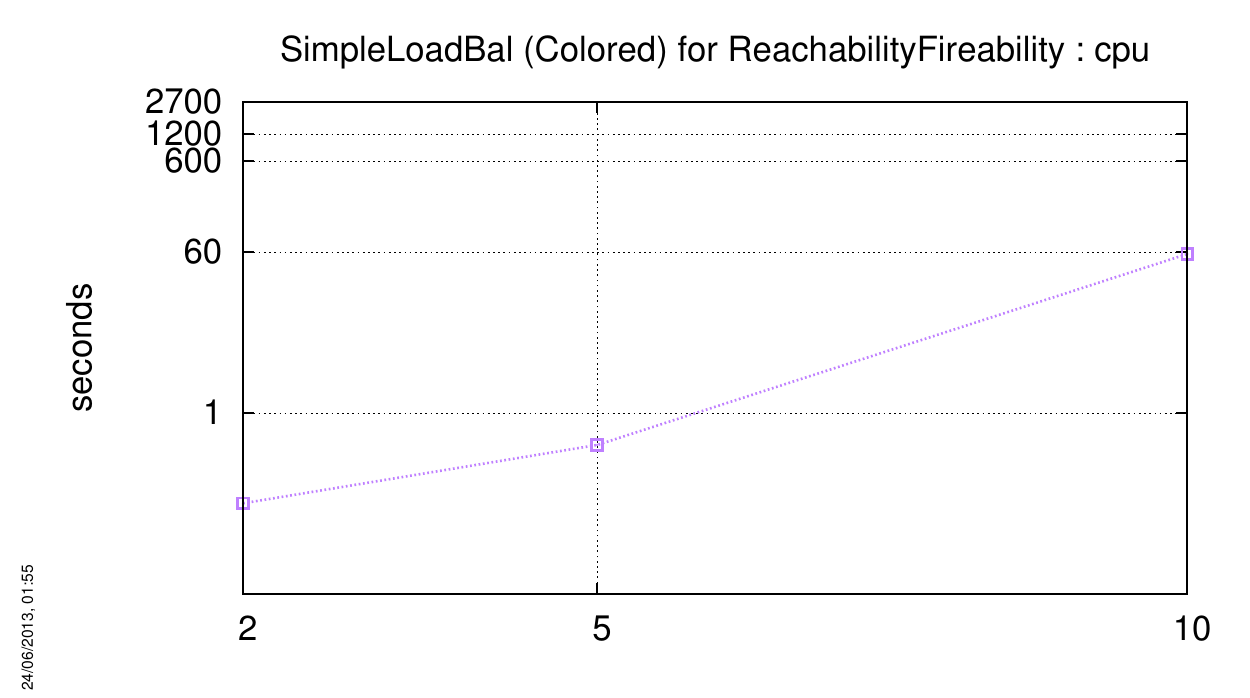}

   \includegraphics[height=1cm]{figures/tools-legend.pdf}
\end{center}

\subsubsection{\acs{SimpleLoadBal-PT}}
The charts below respectively show how tools compete with this ``Known'' model (memory and CPU).

\index{Performances!ReachabilityFireability!SimpleLoadBal (P/T)}
\begin{center}
   \includegraphics[width=7.2cm]{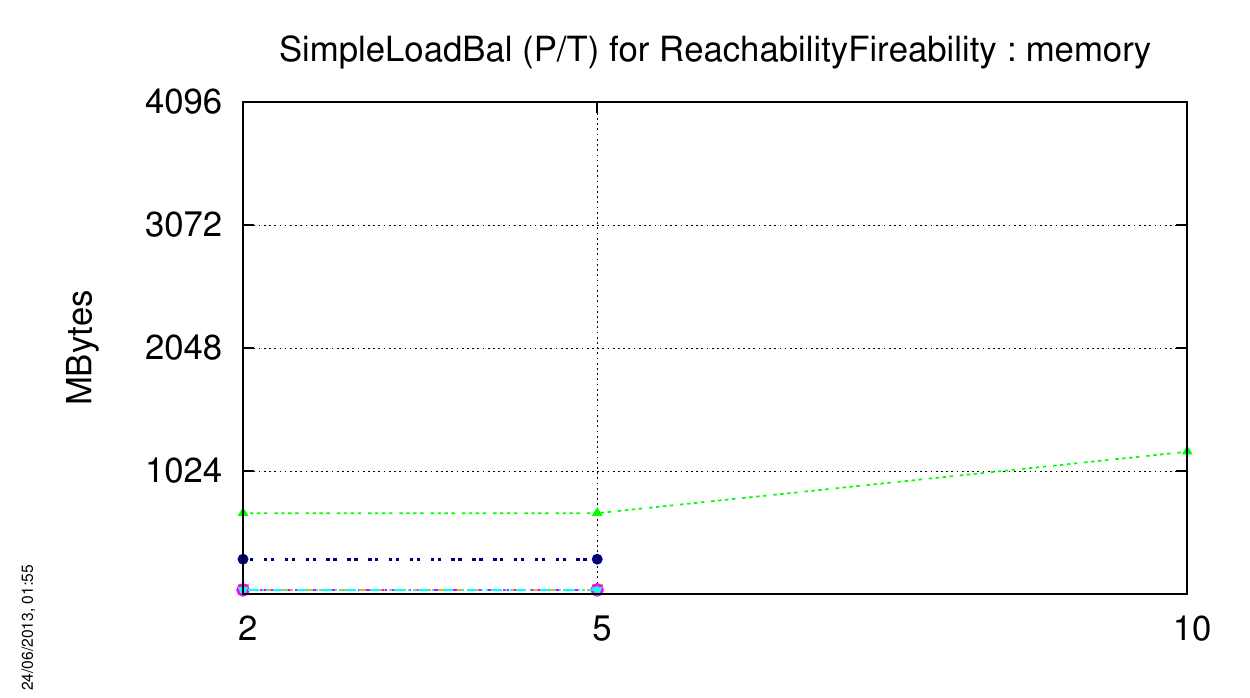}
   \includegraphics[width=7.2cm]{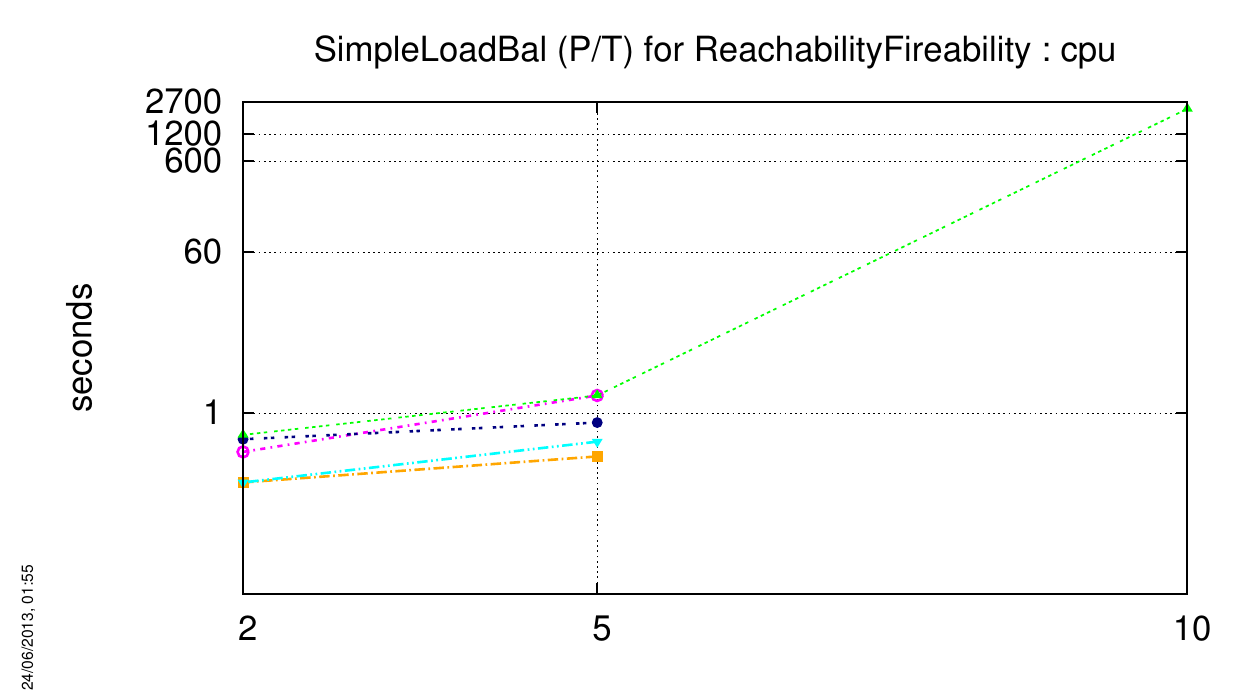}

   \includegraphics[height=1cm]{figures/tools-legend.pdf}
\end{center}

\subsubsection{\acs{TokenRing-COL}}
No instance of this model could be computed for the \textbf{ReachabilityFireability} examination.

\subsubsection{\acs{TokenRing-PT}}
The charts below respectively show how tools compete with this ``Known'' model (memory and CPU).

\index{Performances!ReachabilityFireability!TokenRing (P/T)}
\begin{center}
   \includegraphics[width=7.2cm]{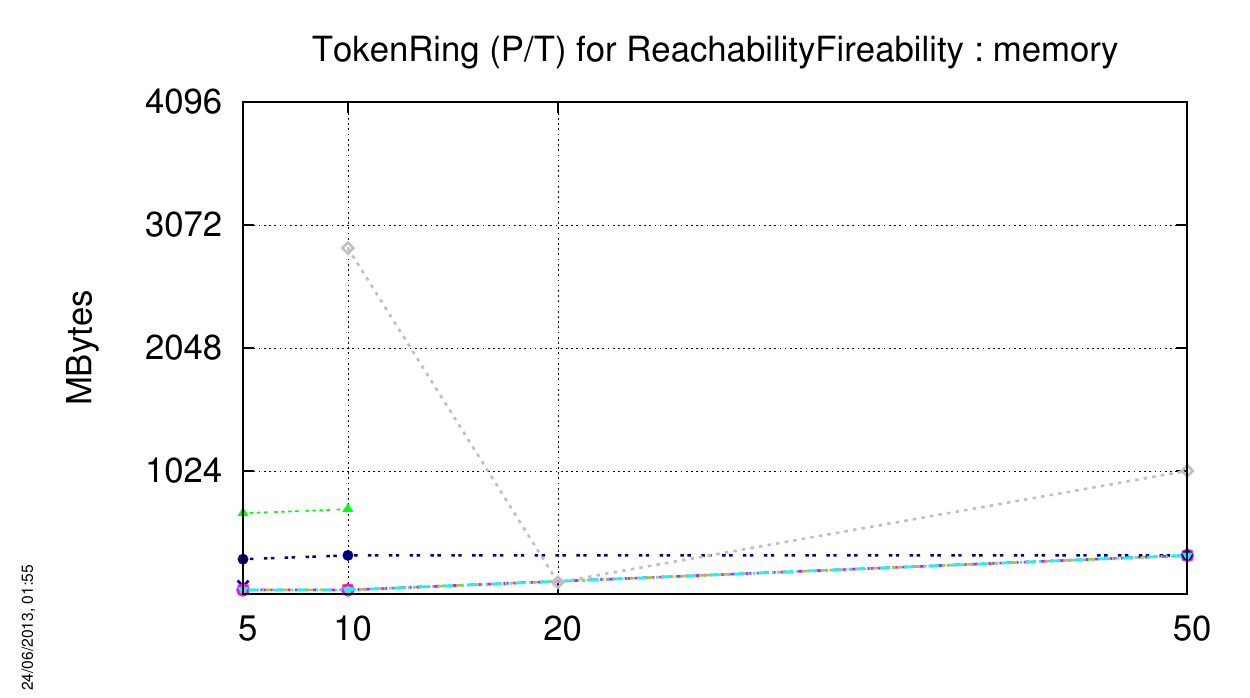}
   \includegraphics[width=7.2cm]{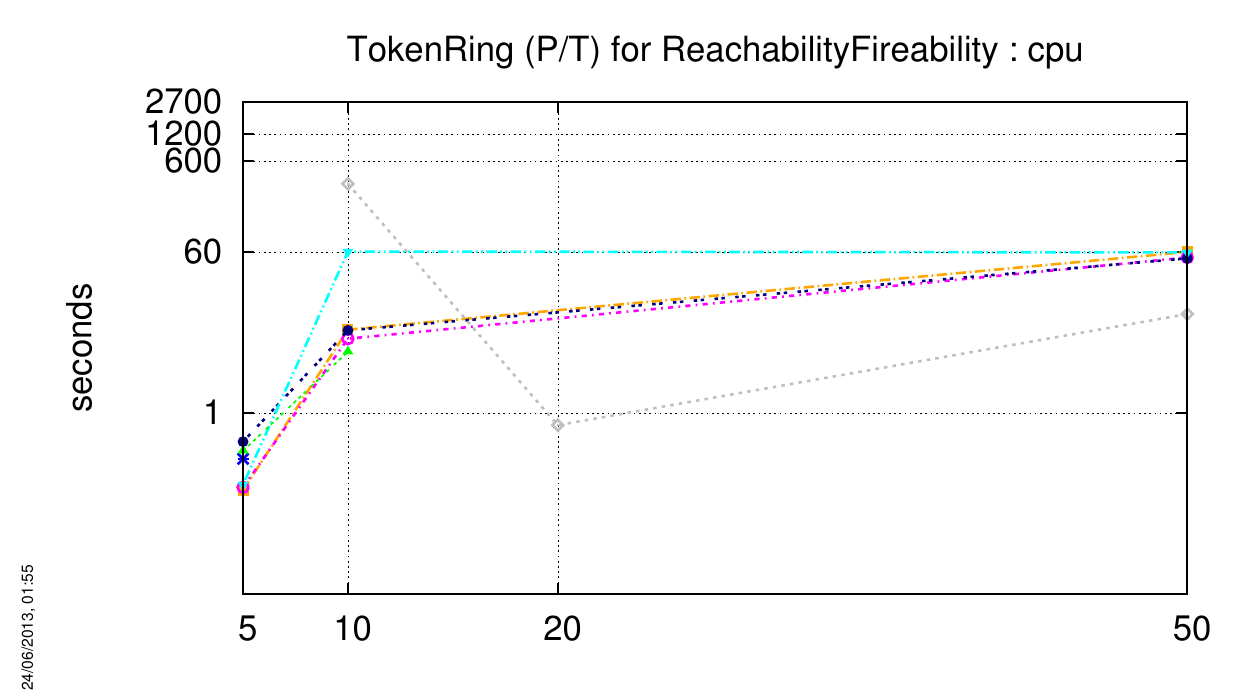}

   \includegraphics[height=1cm]{figures/tools-legend.pdf}
\end{center}

\subsubsection{\acs{HouseConstruction-PT}}
The charts below respectively show how tools compete with this ``Suprise'' model (memory and CPU).

\index{Performances!ReachabilityFireability!HouseConstruction (P/T)}
\begin{center}
   \includegraphics[width=7.2cm]{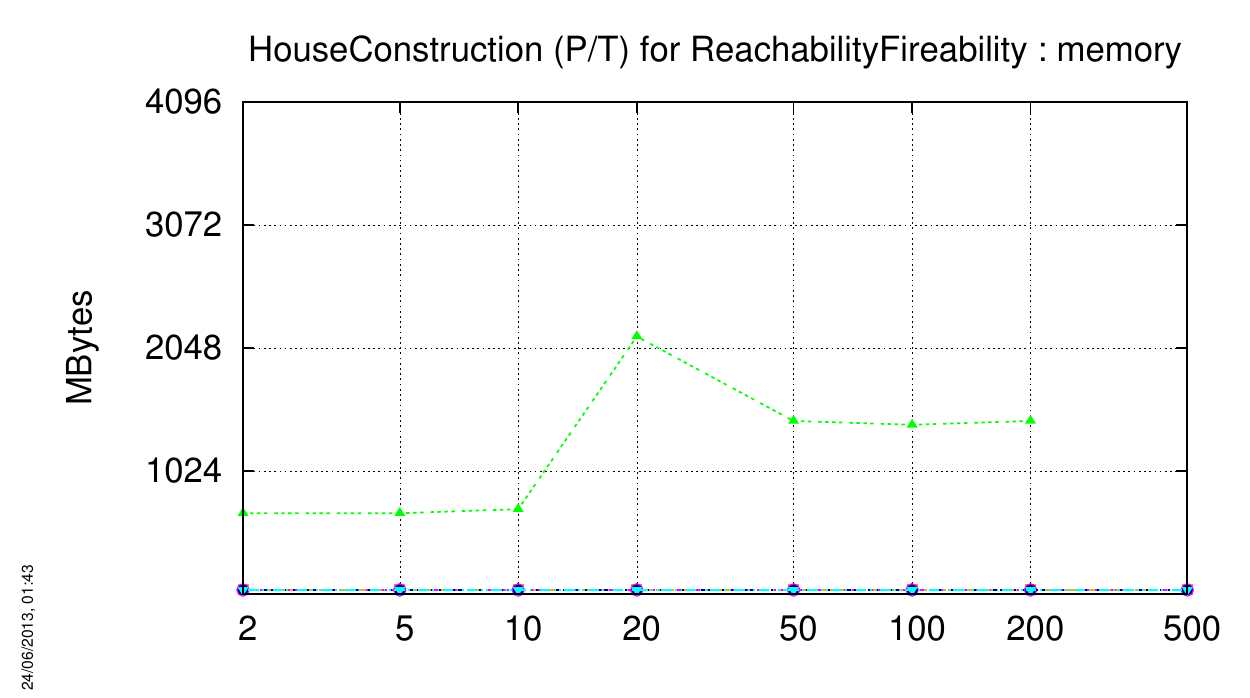}
   \includegraphics[width=7.2cm]{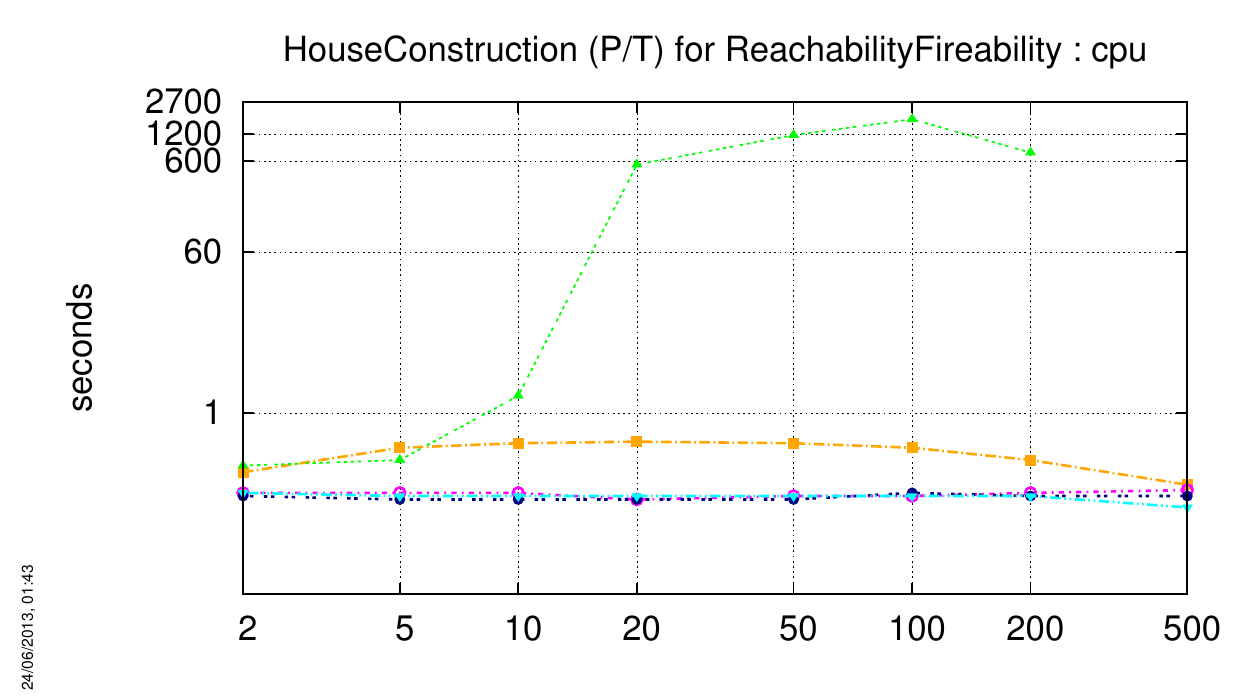}

   \includegraphics[height=1cm]{figures/tools-legend.pdf}
\end{center}

\subsubsection{\acs{IBMB2S565S3960-PT}}
The charts below respectively show how tools compete with this ``Suprise'' model (memory and CPU).

\index{Performances!ReachabilityFireability!IBMB2S565S3960 (P/T)}
\begin{center}
   \includegraphics[width=7.2cm]{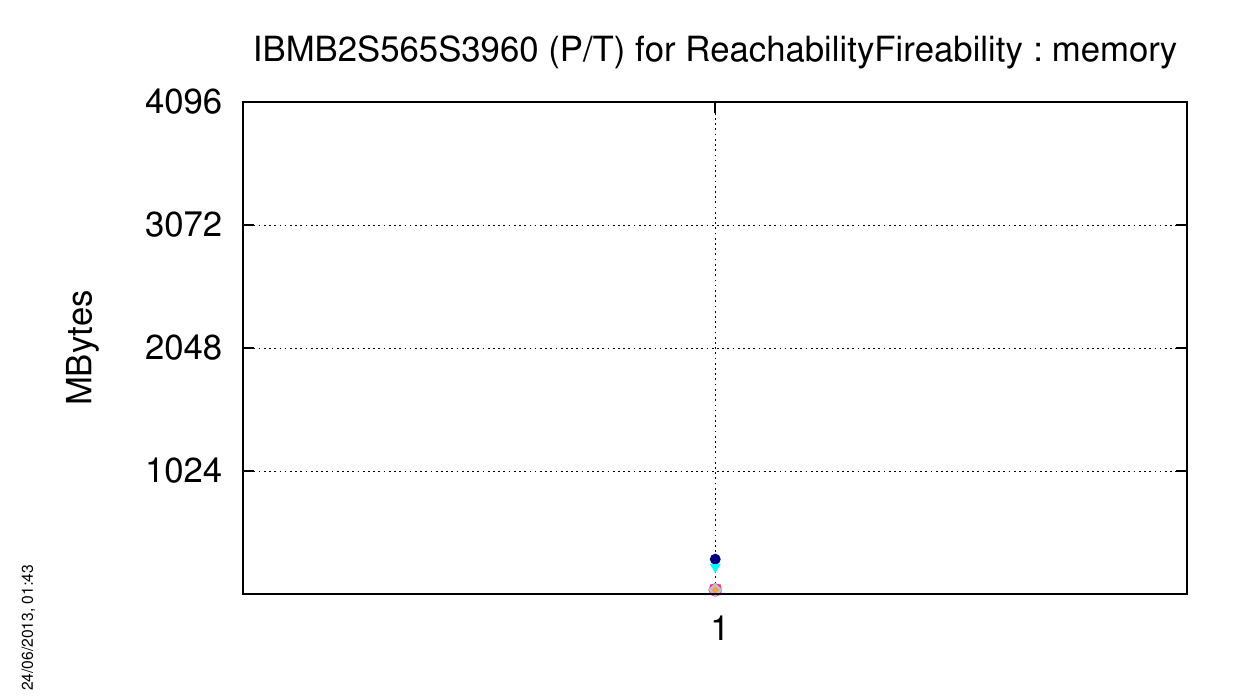}
   \includegraphics[width=7.2cm]{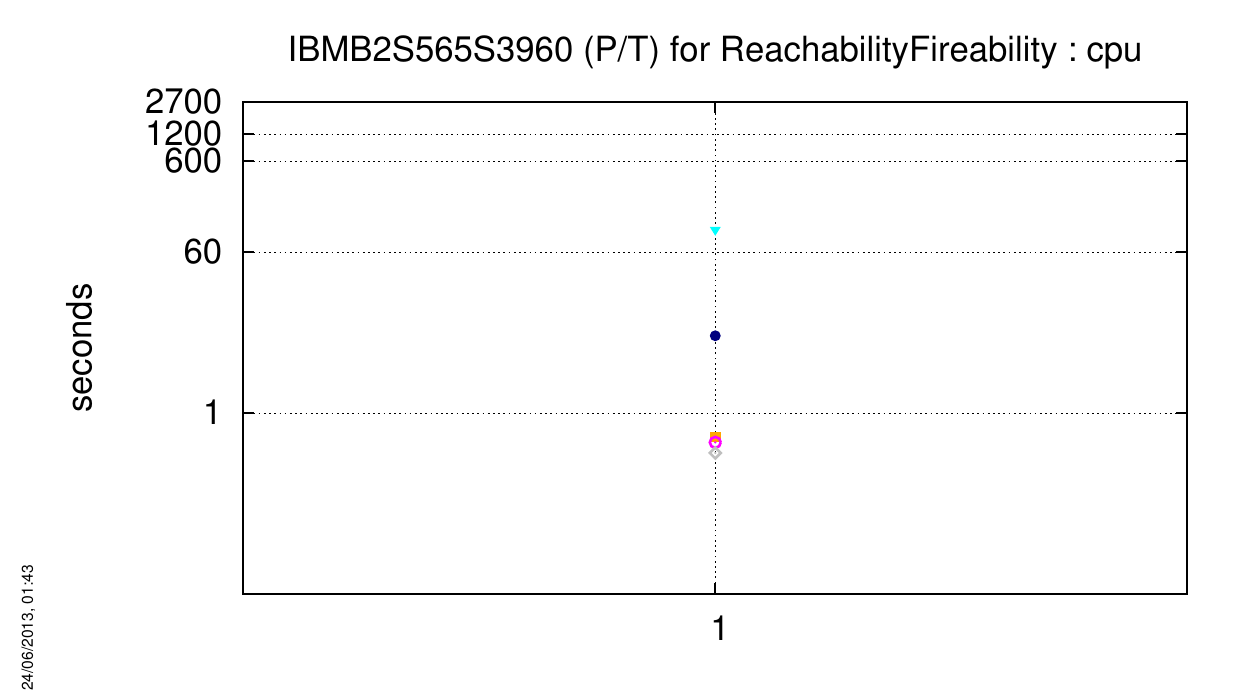}

   \includegraphics[height=1cm]{figures/tools-legend.pdf}
\end{center}

\subsubsection{\acs{QuasiCertifProtocol-COL}}
No instance of this model could be computed for the \textbf{ReachabilityFireability} examination.

\subsubsection{\acs{QuasiCertifProtocol-PT}}
The charts below respectively show how tools compete with this ``Suprise'' model (memory and CPU).

\index{Performances!ReachabilityFireability!QuasiCertifProtocol (P/T)}
\begin{center}
   \includegraphics[width=7.2cm]{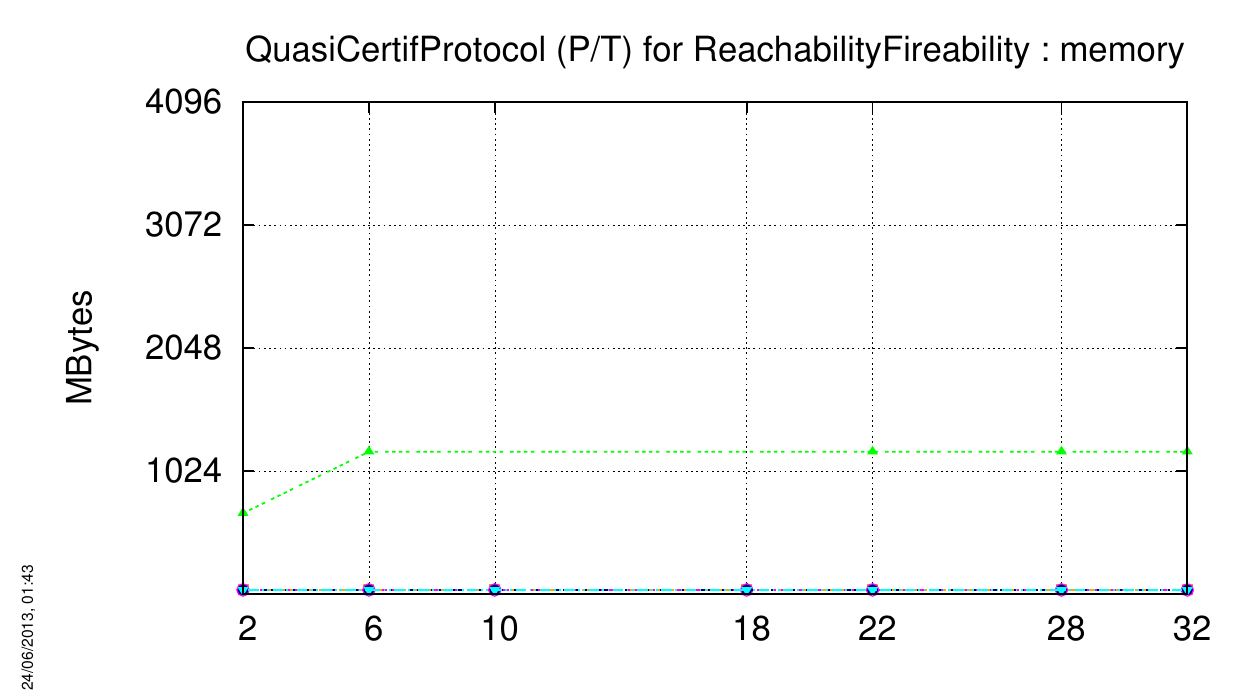}
   \includegraphics[width=7.2cm]{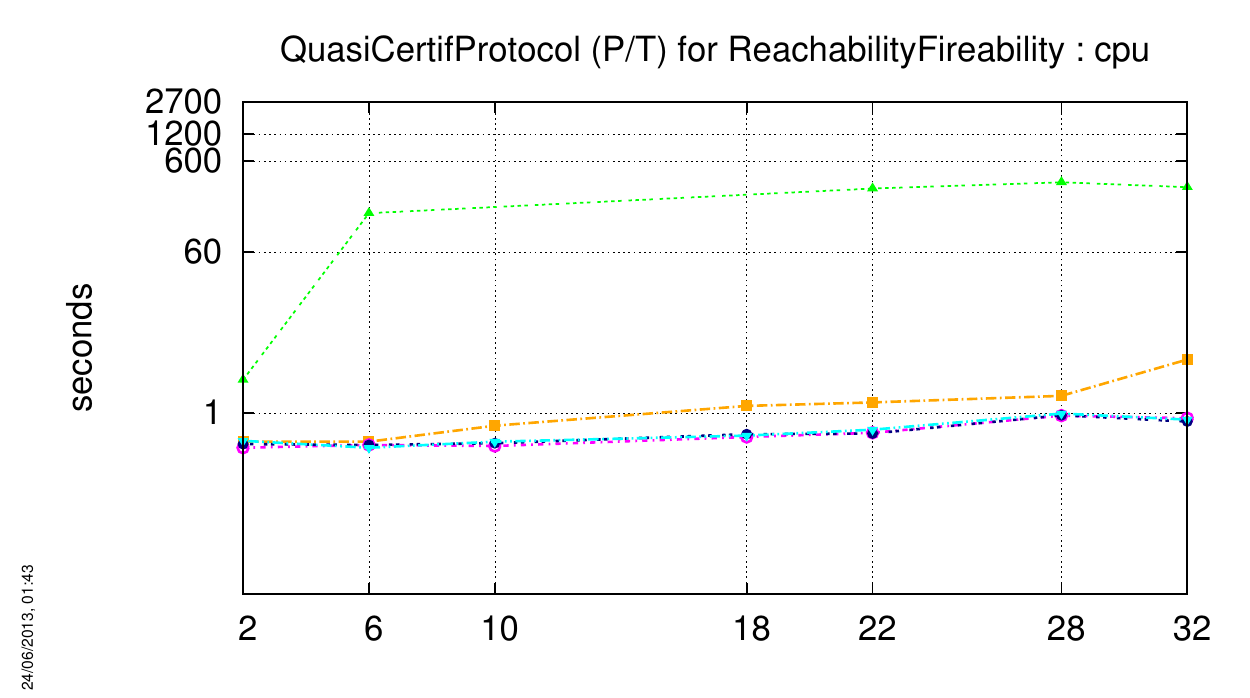}

   \includegraphics[height=1cm]{figures/tools-legend.pdf}
\end{center}

\subsubsection{\acs{Vasy2003-PT}}
The charts below respectively show how tools compete with this ``Suprise'' model (memory and CPU).

\index{Performances!ReachabilityFireability!Vasy2003 (P/T)}
\begin{center}
   \includegraphics[width=7.2cm]{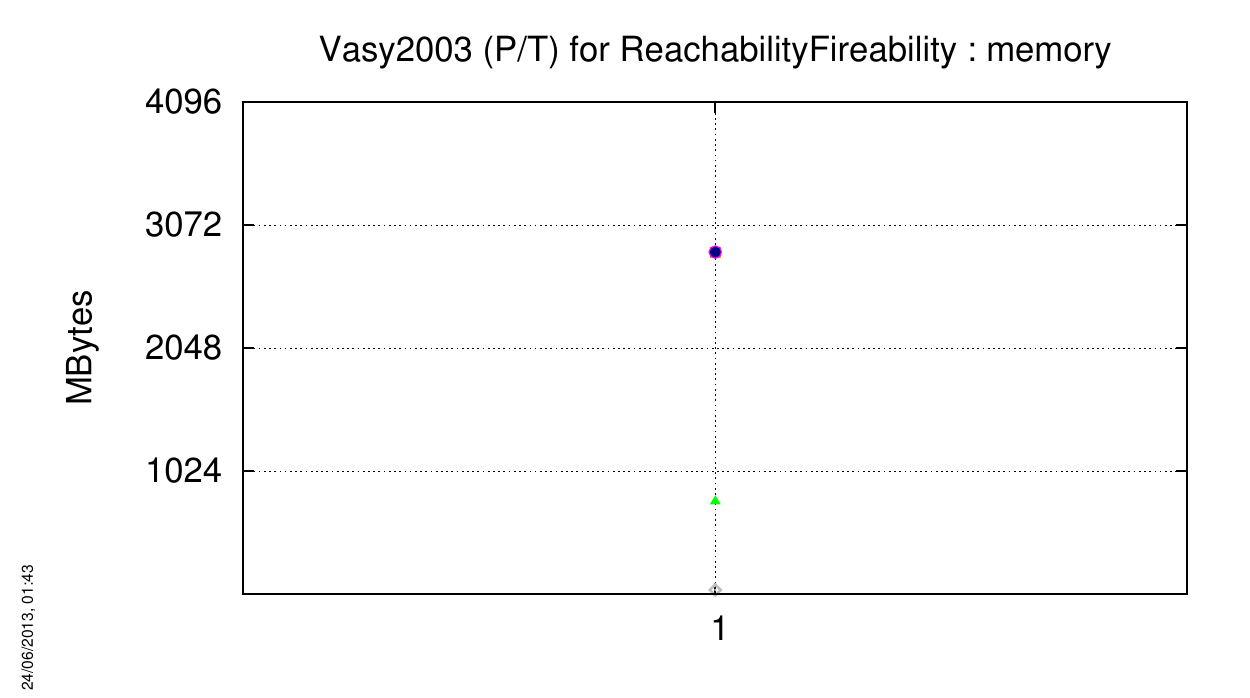}
   \includegraphics[width=7.2cm]{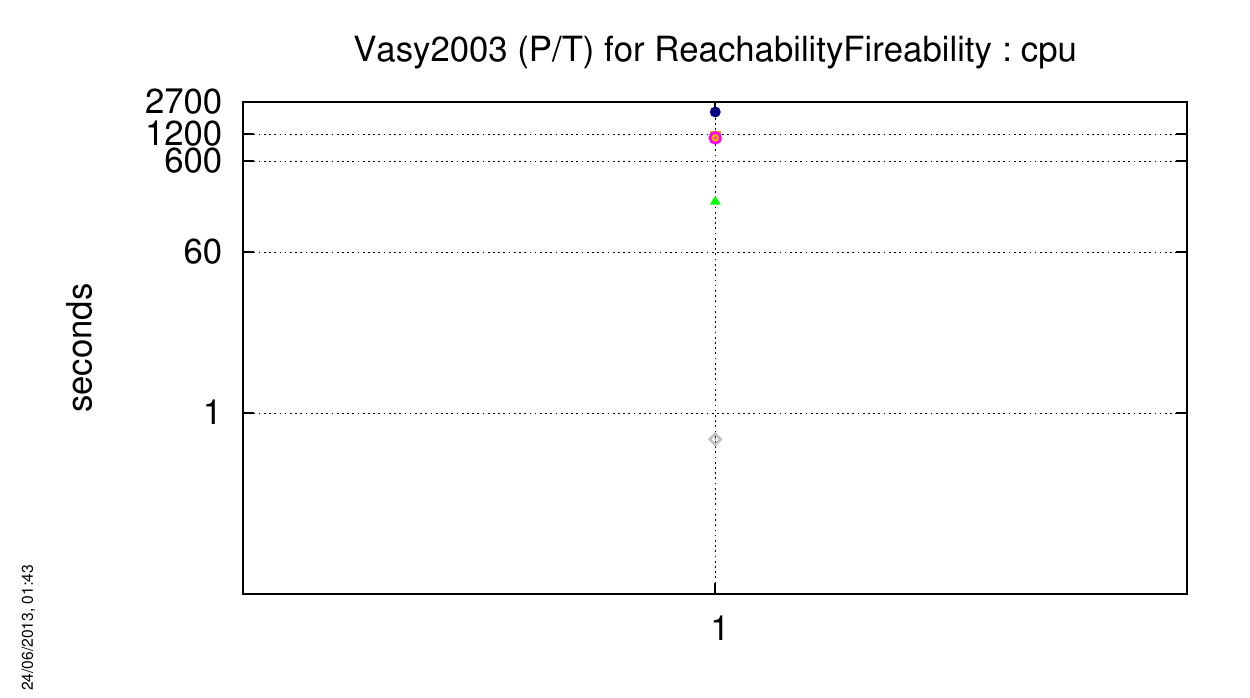}

   \includegraphics[height=1cm]{figures/tools-legend.pdf}
\end{center}

\subsection{Outputs for the ReachabilityFireability Examination}
\index{Outputs!ReachabilityFireability}

Please find enclosed the brute results for this examination (``Known'' and ``Surprise'' models).
We display only the score of tools that provide a results for at least one instance of one model.
The legend for the values is provided below:
\begin{itemize}
   \item\textbf{nc}: the tool does not compete this examination for this model/instance,
   \item\textbf{cc}: the tool cannot compute this examination for this model/instance,
   \item\textbf{to}: the tool cannot compute this examination for this model/instance within the maximum allowed time,
   \item\textbf{mp}: the tool encountered a memory problem (stack overflow or memory full),
   \item\textbf{nf}: there is no formula available for this type of examination (typically, this concerns P/T nets where
       comparing marking cardinality has no signification when there is no equivalent colored net).
\end{itemize}

\textbf{Note on the display of results for formulas:} each formula is considered as a flag (F if false, T if true, - or ?
when the value cannot be determined). These values are concatenated in the order they appear (we assume it is the order of formulas as they were provided).

\subsubsection{``Known'' Models}

Results are summarized in the table below.
{\footnotesize
\begin{longtable}{|ccccccccc|}
   \hline
      \multicolumn{9}{|c|}{\bf \acs{CSRepetitions-COL}} \\
      \textbf{Instances}\index{Outputs!ReachabilityFireability!CSRepetitions (colored)} &
      \acs{greatspn} &
      \acs{its-tools} &
      \acs{lola} &
      \acs{lola-optimistic} &
      \acs{lola-optimistic-incomplete} &
      \acs{lola-pessimistic} &
      \acs{marcie} &
      \acs{sara} \\
      02 &
      nc &
      \textcolor{red}{\tt \scriptsize FFFFFFFF} &
      nc &
      nc &
      nc &
      nc &
      nc &
      nc \\
      03 &
      nc &
      \textcolor{red}{\tt \scriptsize FFFFFFFF} &
      nc &
      nc &
      nc &
      nc &
      nc &
      nc \\
      04 &
      nc &
      \textcolor{red}{\tt \scriptsize mp} &
      nc &
      nc &
      nc &
      nc &
      nc &
      nc \\
      05 &
      nc &
      \textcolor{red}{\tt \scriptsize mp} &
      nc &
      nc &
      nc &
      nc &
      nc &
      nc \\
      07 &
      nc &
      \textcolor{red}{\tt \scriptsize mp} &
      nc &
      nc &
      nc &
      nc &
      nc &
      nc \\
      10 &
      nc &
      \textcolor{red}{\tt \scriptsize mp} &
      nc &
      nc &
      nc &
      nc &
      nc &
      nc \\
\hline
\end{longtable}
}

\subsubsection{``Surprise'' Models}

Results are summarized in the table below.
{\footnotesize
\begin{longtable}{|ccccccc|}
   \hline
      \multicolumn{7}{|c|}{\bf \acs{HouseConstruction-PT}} \\
      instances\index{Outputs!ReachabilityFireability!HouseConstruction (P/T)} &
      \acs{lola} &
      \acs{lola-optimistic} &
      \acs{lola-optimistic-incomplete} &
      \acs{lola-pessimistic} &
      \acs{marcie} &
      \acs{sara} \\
      002 &
      \textcolor{red}{\tt \scriptsize cc} &
      \textcolor{red}{\tt \scriptsize cc} &
      \textcolor{red}{\tt \scriptsize cc} &
      \textcolor{red}{\tt \scriptsize cc} &
      \textcolor{red}{\tt \scriptsize FFFFFFFF} &
      \textcolor{red}{\tt \scriptsize cc} \\
      005 &
      \textcolor{red}{\tt \scriptsize cc} &
      \textcolor{red}{\tt \scriptsize cc} &
      \textcolor{red}{\tt \scriptsize cc} &
      \textcolor{red}{\tt \scriptsize cc} &
      \textcolor{red}{\tt \scriptsize FFFFFFFF} &
      \textcolor{red}{\tt \scriptsize cc} \\
      010 &
      \textcolor{red}{\tt \scriptsize cc} &
      \textcolor{red}{\tt \scriptsize cc} &
      \textcolor{red}{\tt \scriptsize cc} &
      \textcolor{red}{\tt \scriptsize cc} &
      \textcolor{red}{\tt \scriptsize FFFFFFFF} &
      \textcolor{red}{\tt \scriptsize cc} \\
      020 &
      \textcolor{red}{\tt \scriptsize cc} &
      \textcolor{red}{\tt \scriptsize cc} &
      \textcolor{red}{\tt \scriptsize cc} &
      \textcolor{red}{\tt \scriptsize cc} &
      \textcolor{red}{\tt \scriptsize FFFFFFFF} &
      \textcolor{red}{\tt \scriptsize cc} \\
      050 &
      \textcolor{red}{\tt \scriptsize cc} &
      \textcolor{red}{\tt \scriptsize cc} &
      \textcolor{red}{\tt \scriptsize cc} &
      \textcolor{red}{\tt \scriptsize cc} &
      \textcolor{red}{\tt \scriptsize cc} &
      \textcolor{red}{\tt \scriptsize cc} \\
\hline
\end{longtable}
}

\subsection{Score for the ReachabilityFireability Examination}
\index{Scores!ReachabilityFireability}

Please find enclosed the scores for this examination (``Known'' and ``Surprise'' models).
We display only the score of tools that provide a results for at least one instance of one model.
The total is first listed in the table below followed by a detail, for each proposed model.
Meaning of the line labels are:
\begin{itemize}
\item\textbf{1st instance}: the tool gets a bonus for having processed the first instance of this model (+1 point),
\item\textbf{instances}: the tool gets 1 point per instances treated 
(for that, we assume that at least one formula has been successfully computed),
\item\textbf{max reached}: the tool could process all the instances for the model (+2 points),
\item\textbf{best}: the tool is among the ones that processed a maximum of instances within the time and memory confinement (+2 points).
\end{itemize}

\subsubsection{``Known'' Models}

Scores are summarized in the table below.
For each model in this category, a tool may collect up to 411 points.

{\footnotesize
\begin{longtable}{|ccccccccc|}
   \hline
      \multicolumn{9}{|c|}{\bf Total score of the tools for this examination} \\
        &
      \acs{greatspn} &
      \acs{its-tools} &
      \acs{lola} &
      \acs{lola-optimistic} &
      \acs{lola-optimistic-incomplete} &
      \acs{lola-pessimistic} &
      \acs{marcie} &
      \acs{sara} \\
      Total Score &
      {\bf 44} &
      {\bf 59} &
      {\bf 200} &
      {\bf 200} &
      {\bf 161} &
      {\bf 106} &
      {\bf 120} &
      {\bf 151} \\
   \hline
      \multicolumn{9}{|c|}{\bf CSRepetitions (Colored)} \\
      \ \index{Scores!ReachabilityFireability!CSRepetitions (colored)} &
      \acs{greatspn} &
      \acs{its-tools} &
      \acs{lola} &
      \acs{lola-optimistic} &
      \acs{lola-optimistic-incomplete} &
      \acs{lola-pessimistic} &
      \acs{marcie} &
      \acs{sara} \\
      1st instance &
      \textcolor{red}{0} &
      \textcolor{red}{1} &
      \textcolor{red}{0} &
      \textcolor{red}{0} &
      \textcolor{red}{0} &
      \textcolor{red}{0} &
      \textcolor{red}{0} &
      \textcolor{red}{0} \\
      instances &
      \textcolor{red}{0} &
      \textcolor{red}{2} &
      \textcolor{red}{0} &
      \textcolor{red}{0} &
      \textcolor{red}{0} &
      \textcolor{red}{0} &
      \textcolor{red}{0} &
      \textcolor{red}{0} \\
      max reached &
      \textcolor{red}{0} &
      \textcolor{red}{0} &
      \textcolor{red}{0} &
      \textcolor{red}{0} &
      \textcolor{red}{0} &
      \textcolor{red}{0} &
      \textcolor{red}{0} &
      \textcolor{red}{0} \\
\hline
\end{longtable}
}

\subsubsection{``Surprise'' Models}

Scores are summarized in the table below.
For each model in this category, a tool may collect up to 49 points.

{\footnotesize
\begin{longtable}{|ccccccc|}
   \hline
      \multicolumn{7}{|c|}{\bf Total score of the tools for this examination} \\
        &
      \acs{lola} &
      \acs{lola-optimistic} &
      \acs{lola-optimistic-incomplete} &
      \acs{lola-pessimistic} &
      \acs{marcie} &
      \acs{sara} \\
      Total Score &
      {\bf 12} &
      {\bf 12} &
      {\bf 12} &
      {\bf 6} &
      {\bf 18} &
      {\bf 6} \\
   \hline
      \multicolumn{7}{|c|}{\bf HouseConstruction (P/T)} \\
      \ \index{Scores!ReachabilityFireability!HouseConstruction (P/T)} &
      \acs{lola} &
      \acs{lola-optimistic} &
      \acs{lola-optimistic-incomplete} &
      \acs{lola-pessimistic} &
      \acs{marcie} &
      \acs{sara} \\
      1st instance &
      \textcolor{red}{0} &
      \textcolor{red}{0} &
      \textcolor{red}{0} &
      \textcolor{red}{0} &
      \textcolor{red}{1} &
      \textcolor{red}{0} \\
      instances &
      \textcolor{red}{0} &
      \textcolor{red}{0} &
      \textcolor{red}{0} &
      \textcolor{red}{0} &
      \textcolor{red}{4} &
      \textcolor{red}{0} \\
      max reached &
      \textcolor{red}{0} &
      \textcolor{red}{0} &
      \textcolor{red}{0} &
      \textcolor{red}{0} &
      \textcolor{red}{0} &
      \textcolor{red}{0} \\
\hline
\end{longtable}
}

\subsection{Trophies for this Examination}
\index{Trophies!ReachabilityFireability}

Trophies are divided in three categories: ``Known'' models,
``Surprise'' models, and the global trophies (formula is then
$score_{global} = score_{known} + 2 \times score_{surprise}$).

\subsubsection{For ``Known'' Models} \ \\

\begin{tabular}{c|c|c}
      1 & 1 & 3 \\
   \includegraphics[width=2cm]{figures/gold.jpg} &
   \includegraphics[width=2cm]{figures/gold.jpg} &
   \includegraphics[width=2cm]{figures/bronse.jpg} \\
   \acs{lola} &
   \acs{lola-optimistic} &
   \acs{lola-optimistic-incomplete} \\
   200 points &
   200 points &
   161 points \\
\end{tabular}

\subsubsection{For ``Surprise'' Models}\  \\

\begin{tabular}{c|c|c|c|c}
      1 & 2 & 2 & 2 & 2 \\
   \includegraphics[width=2cm]{figures/gold.jpg} &
   \includegraphics[width=2cm]{figures/silver.jpg} &
   \includegraphics[width=2cm]{figures/silver.jpg} &
   \includegraphics[width=2cm]{figures/silver.jpg} &
   \includegraphics[width=2cm]{figures/silver.jpg} \\
   \acs{marcie} &
   \acs{its-tools} &
   \acs{lola} &
   \acs{lola-optimistic} &
   \acs{lola-optimistic-incomplete} \\
   18 points &
   12 points &
   12 points &
   12 points &
   12 points \\
\end{tabular}

\subsubsection{Global} \ \\

\begin{tabular}{c|c|c}
      1 & 1 & 3 \\
   \includegraphics[width=2cm]{figures/gold.jpg} &
   \includegraphics[width=2cm]{figures/gold.jpg} &
   \includegraphics[width=2cm]{figures/bronse.jpg} \\
   \acs{lola} &
   \acs{lola-optimistic} &
   \acs{lola-optimistic-incomplete} \\
   224 points &
   224 points &
   185 points \\
\end{tabular}

\newpage

\section{The ReachabilityMarkingComparison Examination}
\label{sec:exam:ReachabilityMarkingComparison}
\index{Results!ReachabilityMarkingComparison}

This examination deals with reachability properties dealing with marking comparison only.
We first show a summary on the handling of models by the participating tools.
Then, we present the computed outputs and the associated scores for this
examination prior to a summary of relevant executions.

\subsection{Handling of Models by Tools}
\index{Performances!ReachabilityMarkingComparison}

\subsubsection{\acs{CSRepetitions-COL}}
No instance of this model could be computed for the \textbf{ReachabilityMarkingComparison} examination.

\subsubsection{\acs{CSRepetitions-PT}}
The charts below respectively show how tools compete with this ``Known'' model (memory and CPU).

\index{Performances!ReachabilityMarkingComparison!CSRepetitions (P/T)}
\begin{center}
   \includegraphics[width=7.2cm]{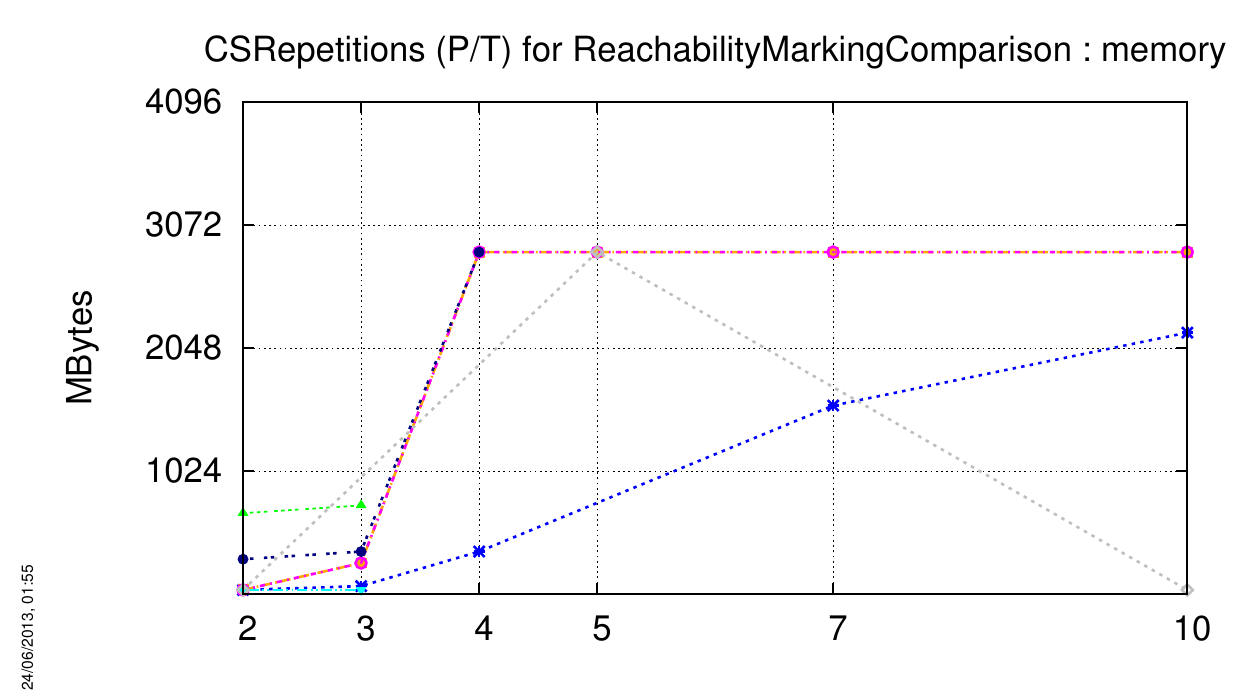}
   \includegraphics[width=7.2cm]{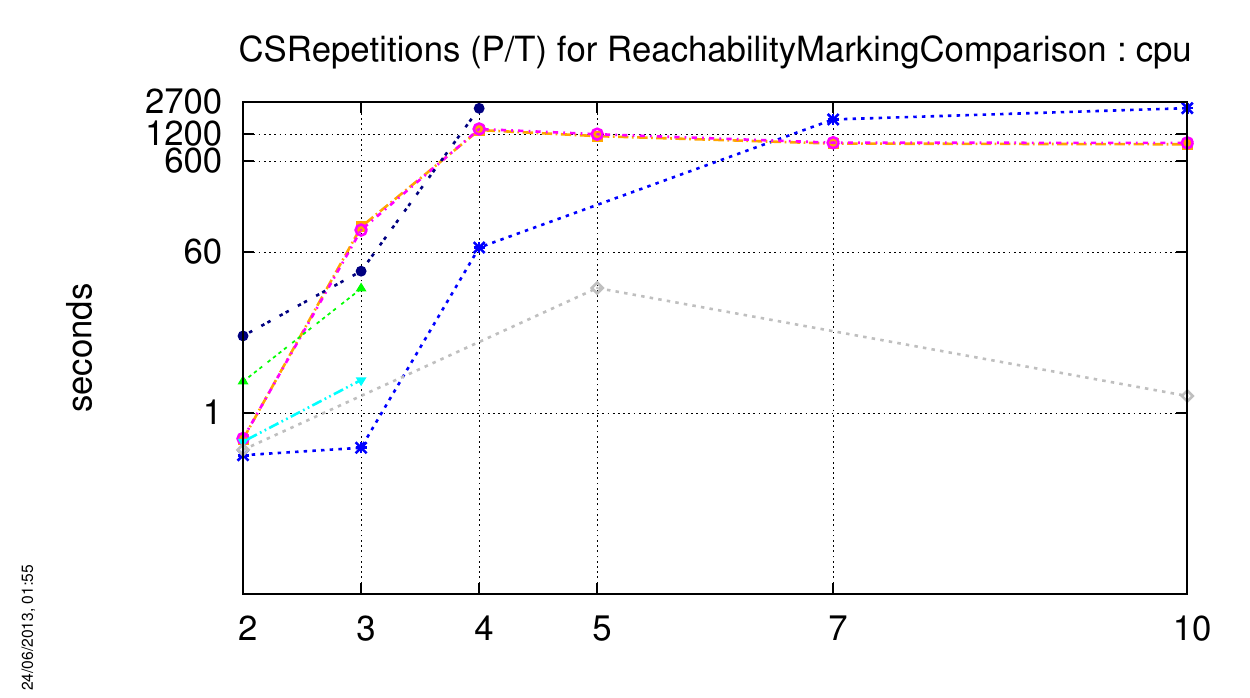}

   \includegraphics[height=1cm]{figures/tools-legend.pdf}
\end{center}

\subsubsection{\acs{Dekker-PT}}
No instance of this model could be computed for the \textbf{ReachabilityMarkingComparison} examination.

\subsubsection{\acs{DotAndBoxes-COL}}
No instance of this model could be computed for the \textbf{ReachabilityMarkingComparison} examination.

\subsubsection{\acs{DrinkVendingMachine-COL}}
No instance of this model could be computed for the \textbf{ReachabilityMarkingComparison} examination.

\subsubsection{\acs{DrinkVendingMachine-PT}}
The charts below respectively show how tools compete with this ``Known'' model (memory and CPU).

\index{Performances!ReachabilityMarkingComparison!DrinkVendingMachine (P/T)}
\begin{center}
   \includegraphics[width=7.2cm]{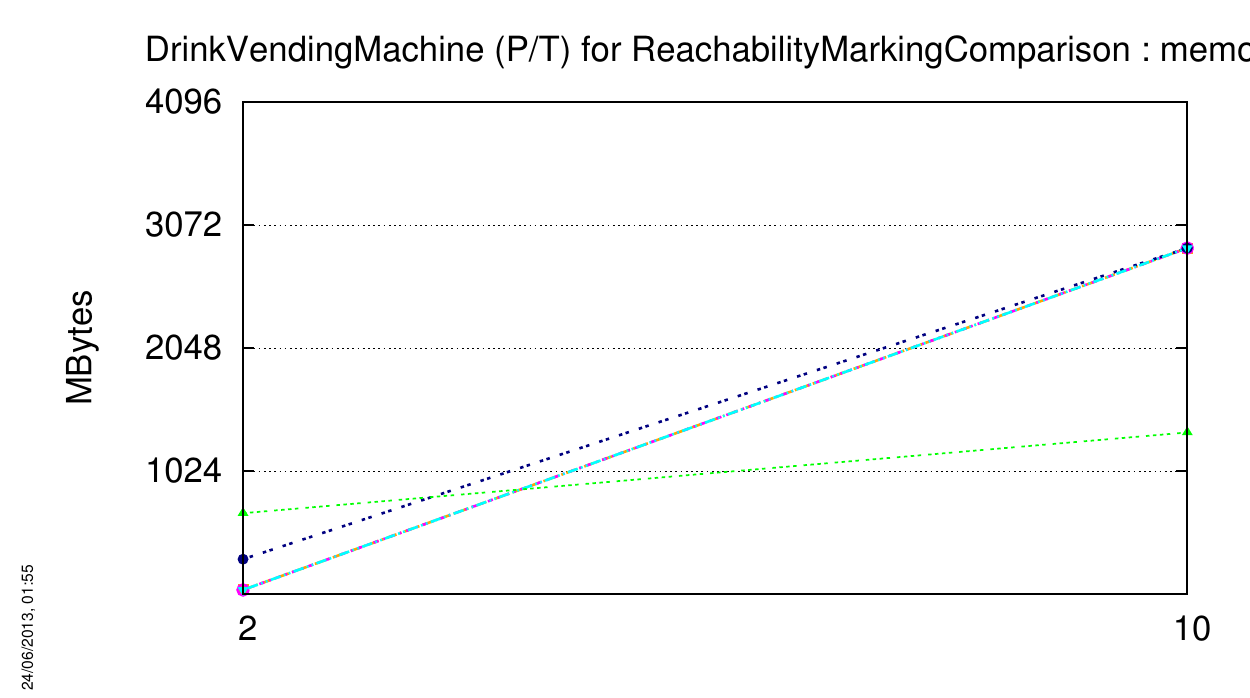}
   \includegraphics[width=7.2cm]{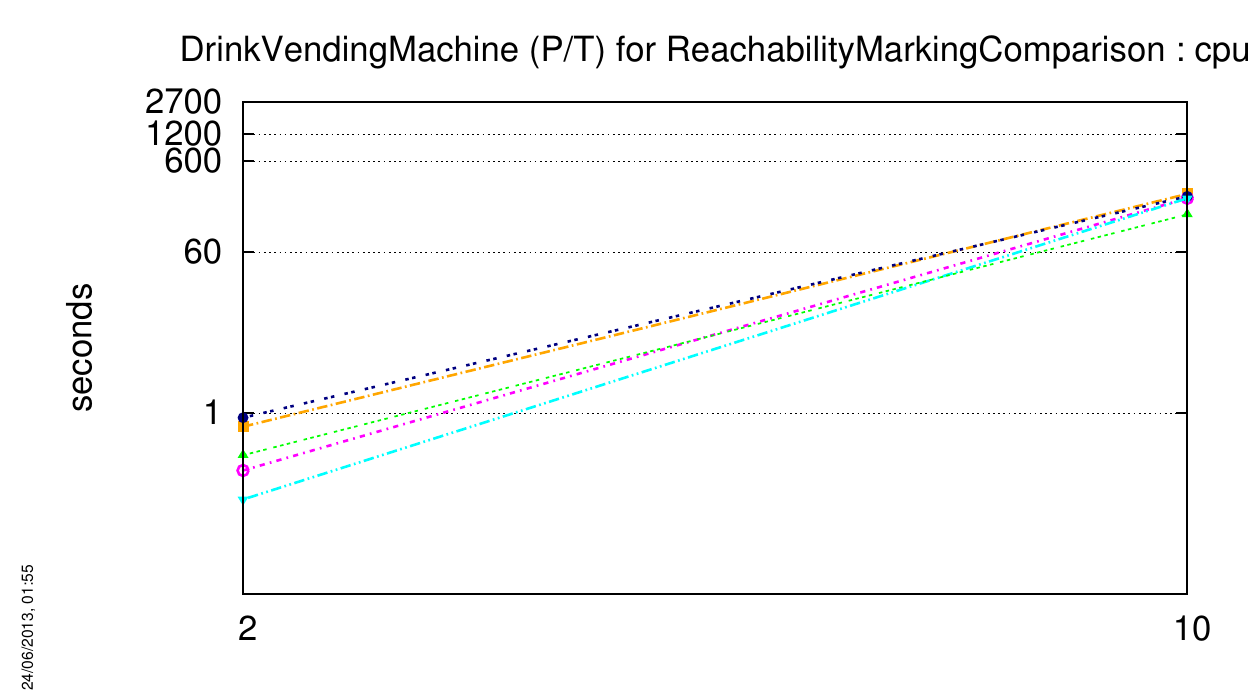}

   \includegraphics[height=1cm]{figures/tools-legend.pdf}
\end{center}

\subsubsection{\acs{Echo-PT}}
No instance of this model could be computed for the \textbf{ReachabilityMarkingComparison} examination.

\subsubsection{\acs{Eratosthenes-PT}}
No instance of this model could be computed for the \textbf{ReachabilityMarkingComparison} examination.

\subsubsection{\acs{FMS-PT}}
No instance of this model could be computed for the \textbf{ReachabilityMarkingComparison} examination.

\subsubsection{\acs{GlobalRessAlloc-COL}}
No instance of this model could be computed for the \textbf{ReachabilityMarkingComparison} examination.

\subsubsection{\acs{GlobalRessAlloc-PT}}
The charts below respectively show how tools compete with this ``Known'' model (memory and CPU).

\index{Performances!ReachabilityMarkingComparison!GlobalRessAlloc (P/T)}
\begin{center}
   \includegraphics[width=7.2cm]{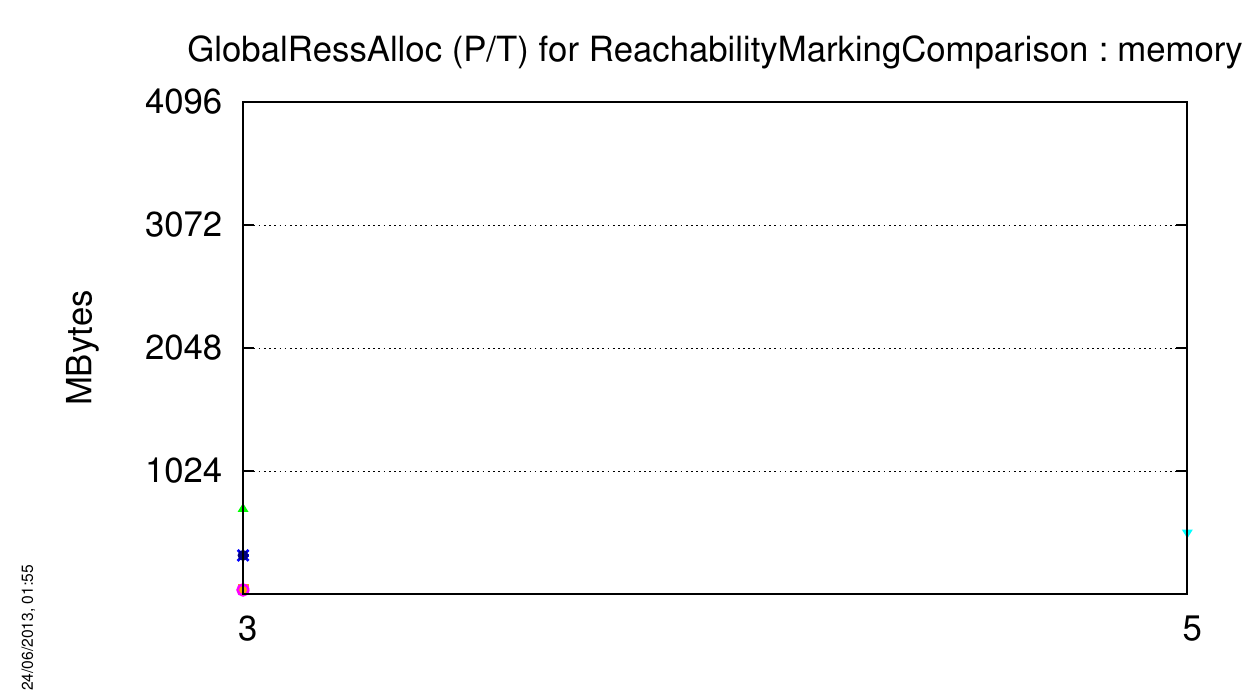}
   \includegraphics[width=7.2cm]{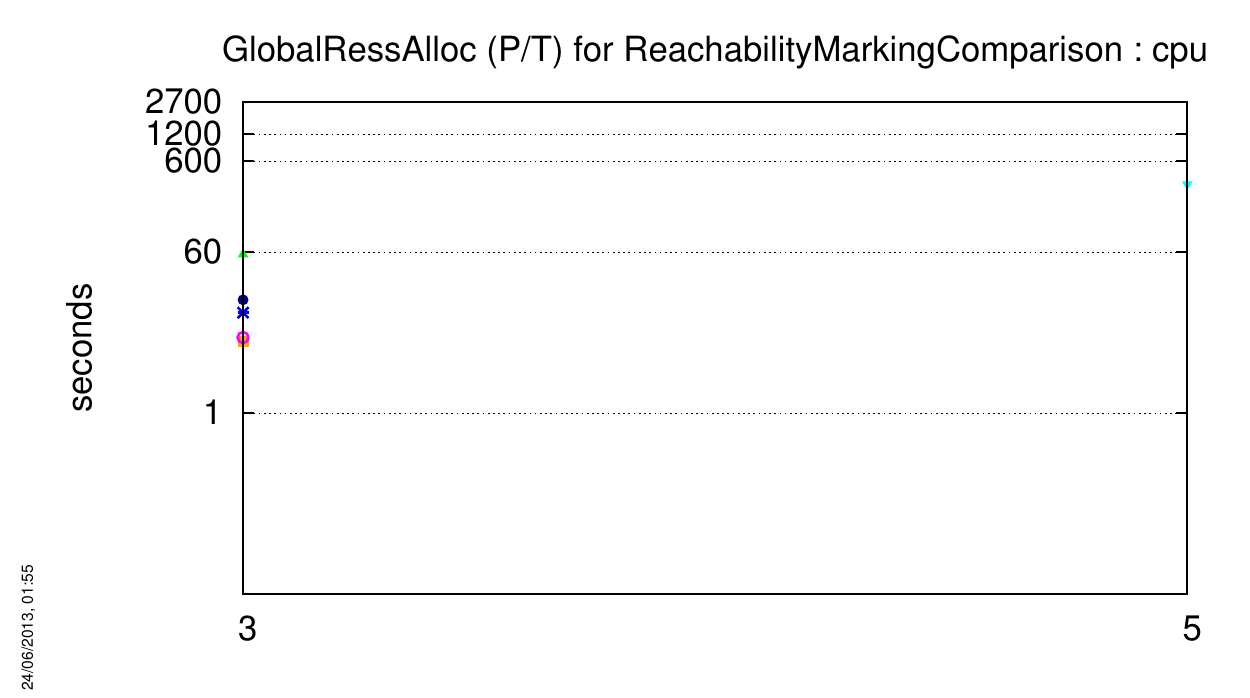}

   \includegraphics[height=1cm]{figures/tools-legend.pdf}
\end{center}

\subsubsection{\acs{Kanban-PT}}
No instance of this model could be computed for the \textbf{ReachabilityMarkingComparison} examination.

\subsubsection{\acs{LamportFastMutEx-COL}}
No instance of this model could be computed for the \textbf{ReachabilityMarkingComparison} examination.

\subsubsection{\acs{LamportFastMutEx-PT}}
The charts below respectively show how tools compete with this ``Known'' model (memory and CPU).

\index{Performances!ReachabilityMarkingComparison!LamportFastMutEx (P/T)}
\begin{center}
   \includegraphics[width=7.2cm]{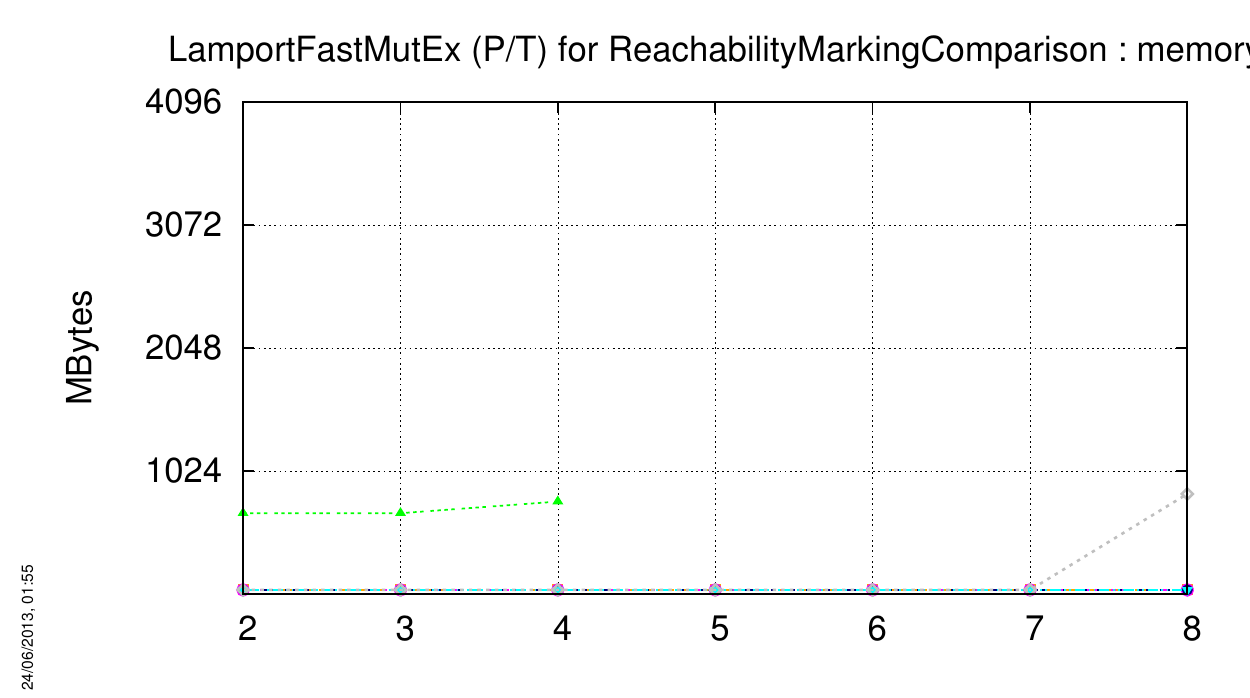}
   \includegraphics[width=7.2cm]{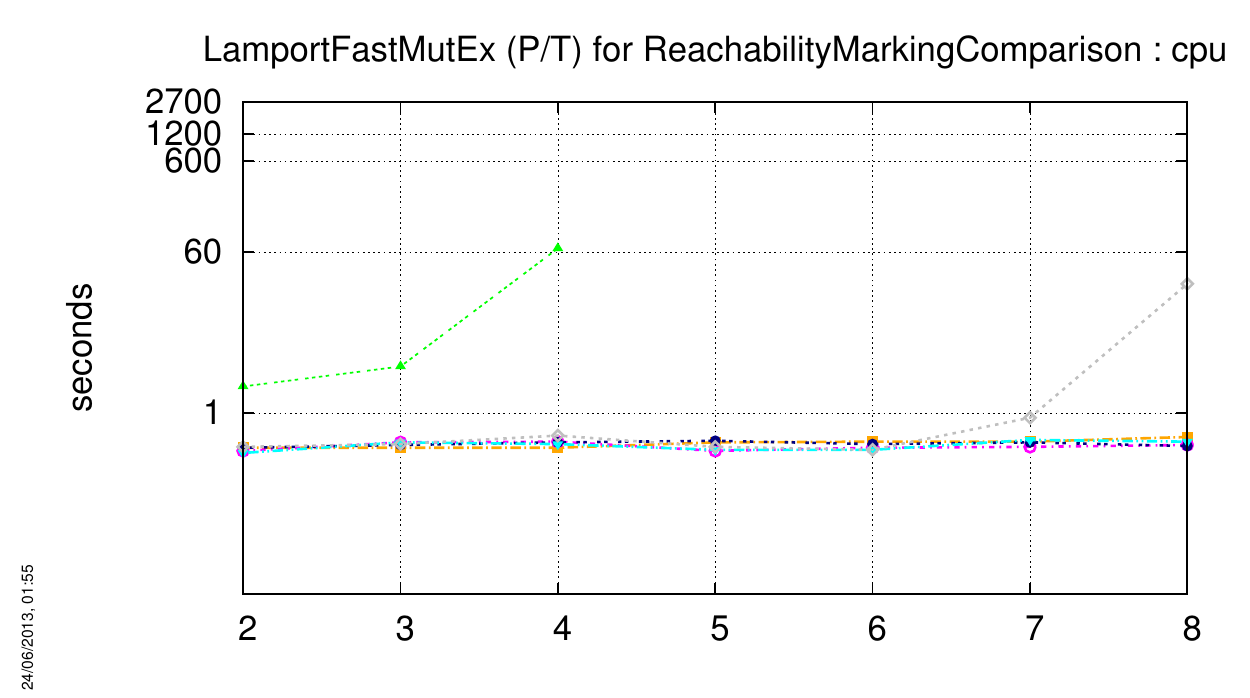}

   \includegraphics[height=1cm]{figures/tools-legend.pdf}
\end{center}

\subsubsection{\acs{MAPK-PT}}
No instance of this model could be computed for the \textbf{ReachabilityMarkingComparison} examination.

\subsubsection{\acs{NeoElection-COL}}
No instance of this model could be computed for the \textbf{ReachabilityMarkingComparison} examination.

\subsubsection{\acs{NeoElection-PT}}
The charts below respectively show how tools compete with this ``Known'' model (memory and CPU).

\index{Performances!ReachabilityMarkingComparison!NeoElection (P/T)}
\begin{center}
   \includegraphics[width=7.2cm]{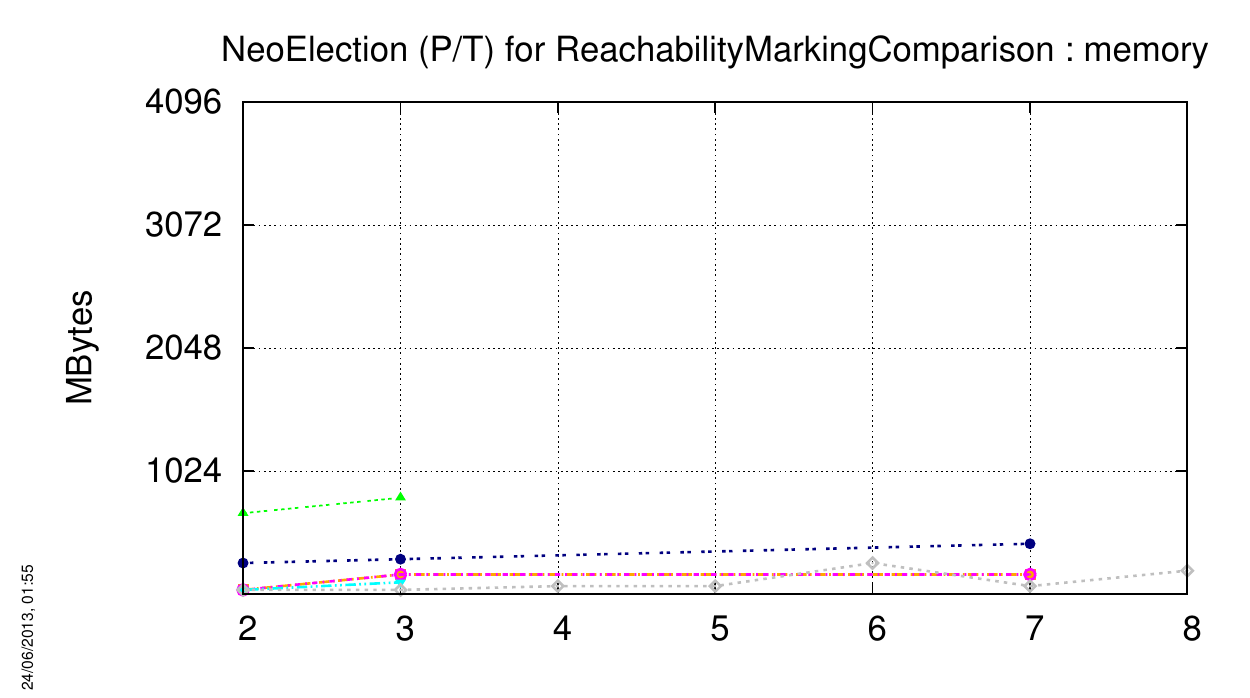}
   \includegraphics[width=7.2cm]{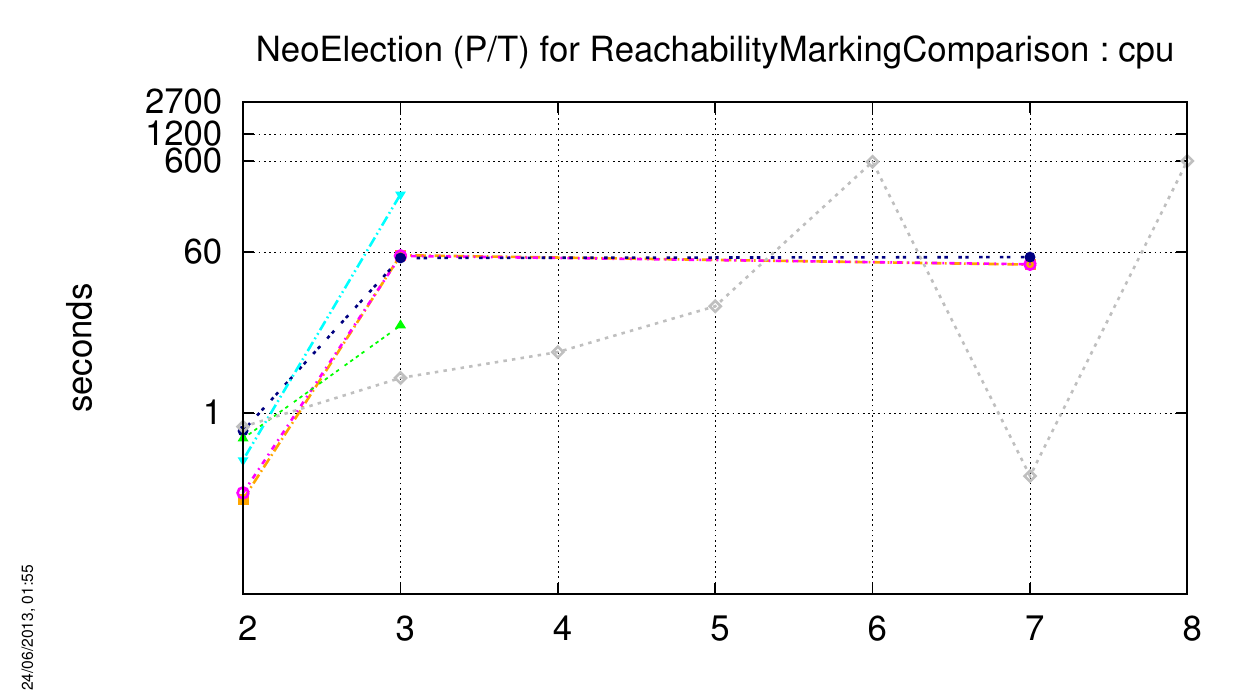}

   \includegraphics[height=1cm]{figures/tools-legend.pdf}
\end{center}

\subsubsection{\acs{PermAdmissibility-COL}}
No instance of this model could be computed for the \textbf{ReachabilityMarkingComparison} examination.

\subsubsection{\acs{PermAdmissibility-PT}}
The charts below respectively show how tools compete with this ``Known'' model (memory and CPU).

\index{Performances!ReachabilityMarkingComparison!PermAdmissibility (P/T)}
\begin{center}
   \includegraphics[width=7.2cm]{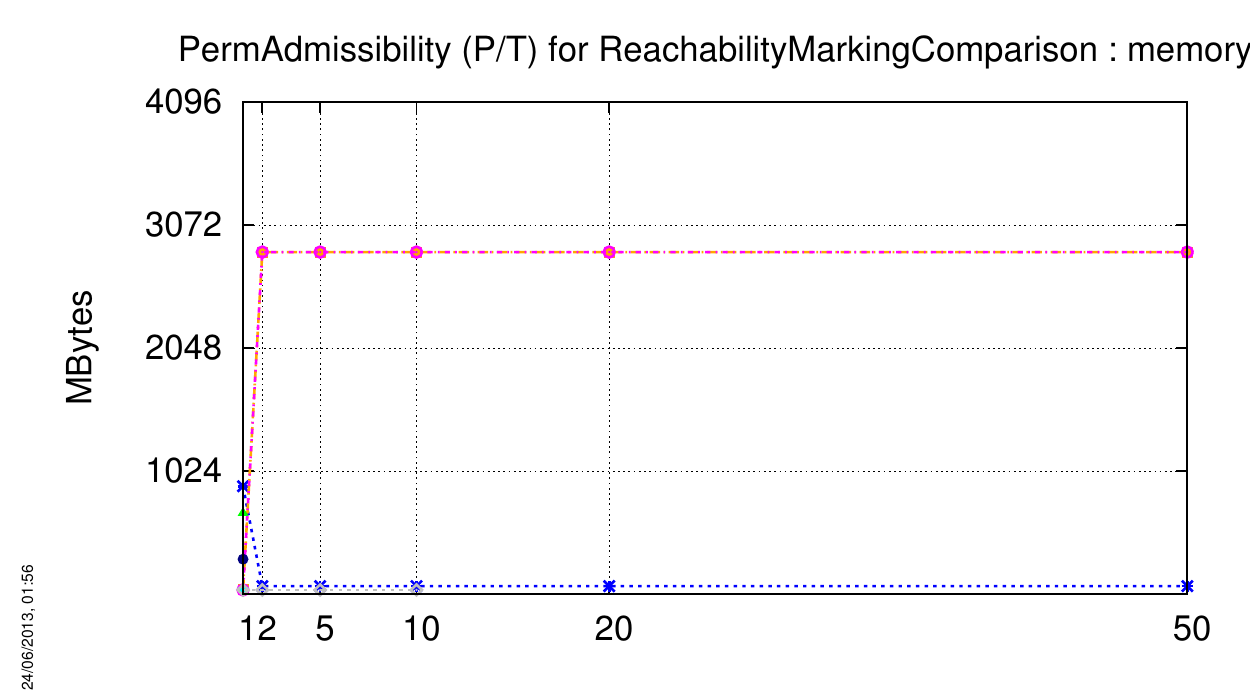}
   \includegraphics[width=7.2cm]{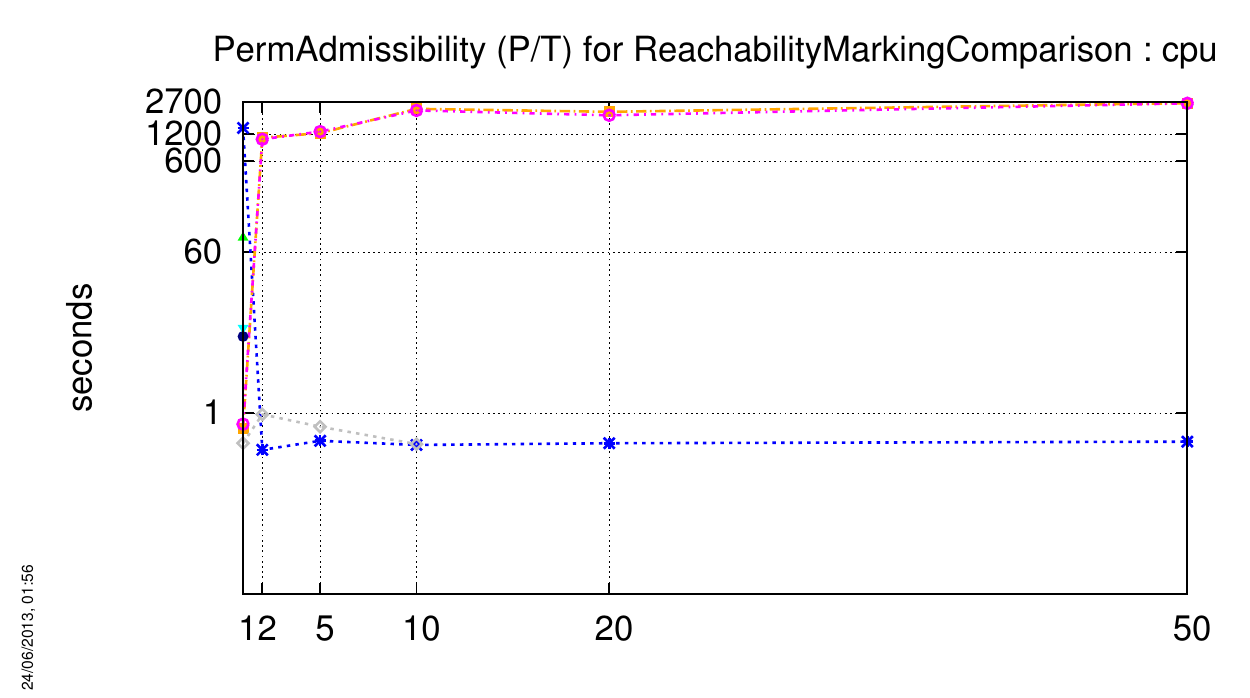}

   \includegraphics[height=1cm]{figures/tools-legend.pdf}
\end{center}

\subsubsection{\acs{Peterson-COL}}
No instance of this model could be computed for the \textbf{ReachabilityMarkingComparison} examination.

\subsubsection{\acs{Peterson-PT}}
The charts below respectively show how tools compete with this ``Known'' model (memory and CPU).

\index{Performances!ReachabilityMarkingComparison!Peterson (P/T)}
\begin{center}
   \includegraphics[width=7.2cm]{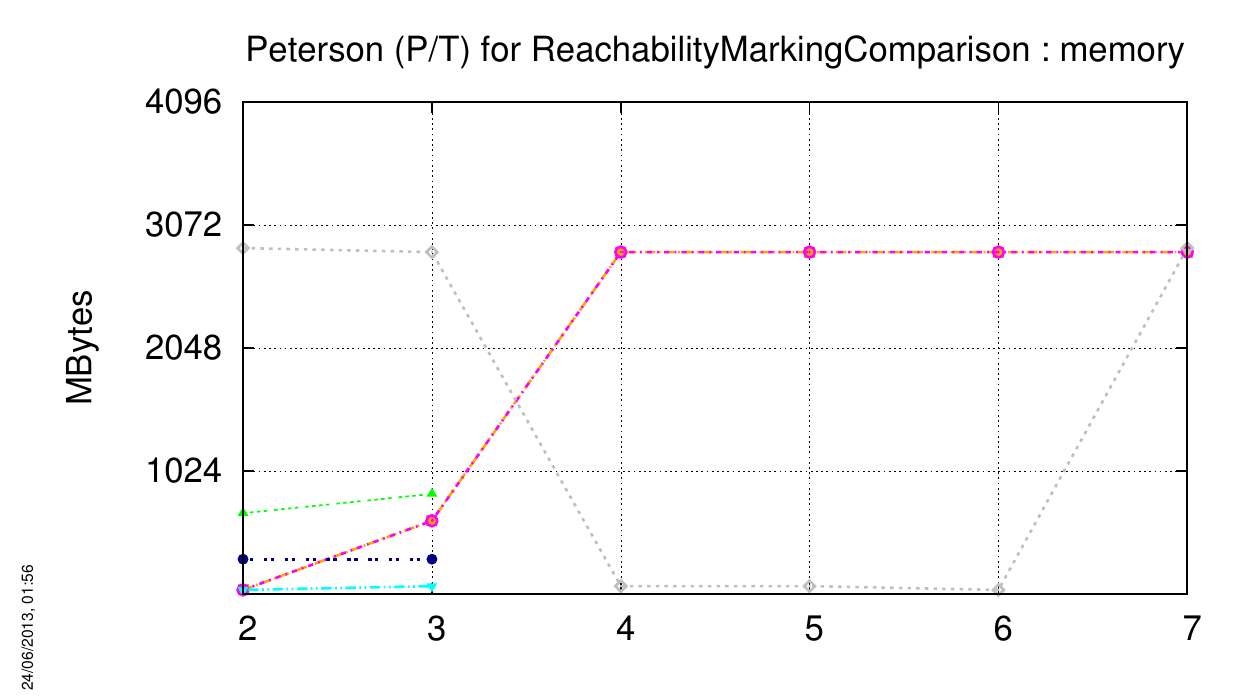}
   \includegraphics[width=7.2cm]{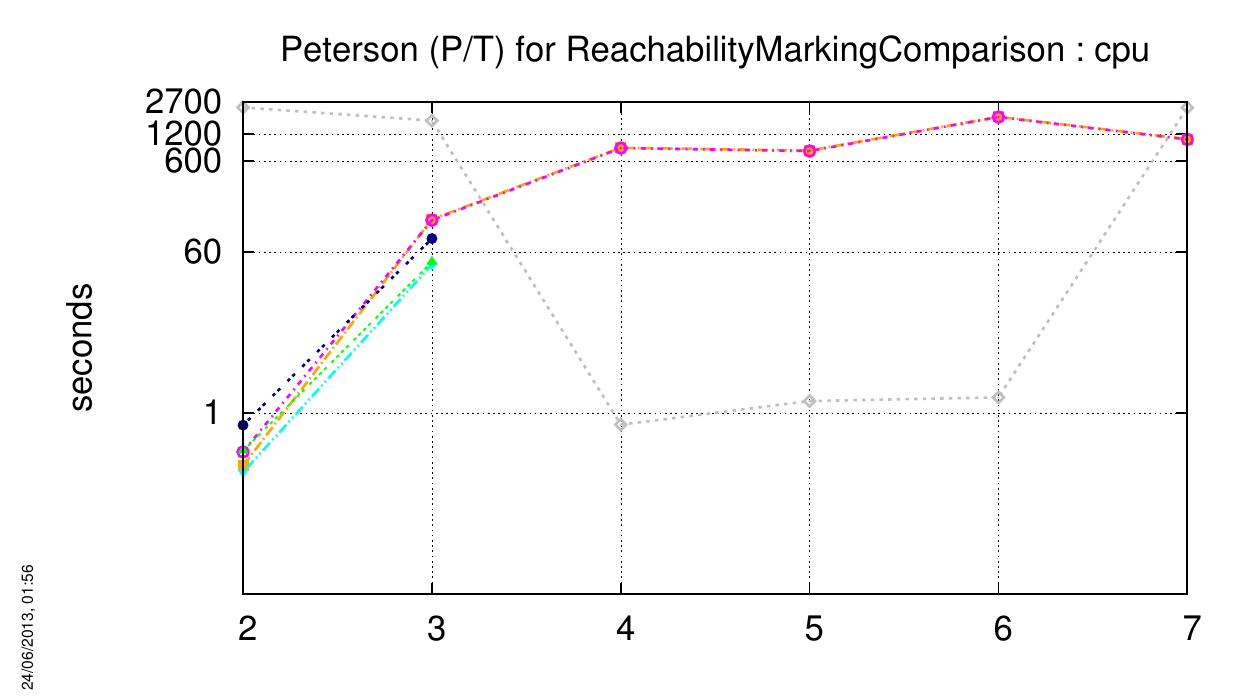}

   \includegraphics[height=1cm]{figures/tools-legend.pdf}
\end{center}

\subsubsection{\acs{Philosophers-COL}}
No instance of this model could be computed for the \textbf{ReachabilityMarkingComparison} examination.

\subsubsection{\acs{Philosophers-PT}}
The charts below respectively show how tools compete with this ``Known'' model (memory and CPU).

\index{Performances!ReachabilityMarkingComparison!Philosophers (P/T)}
\begin{center}
   \includegraphics[width=7.2cm]{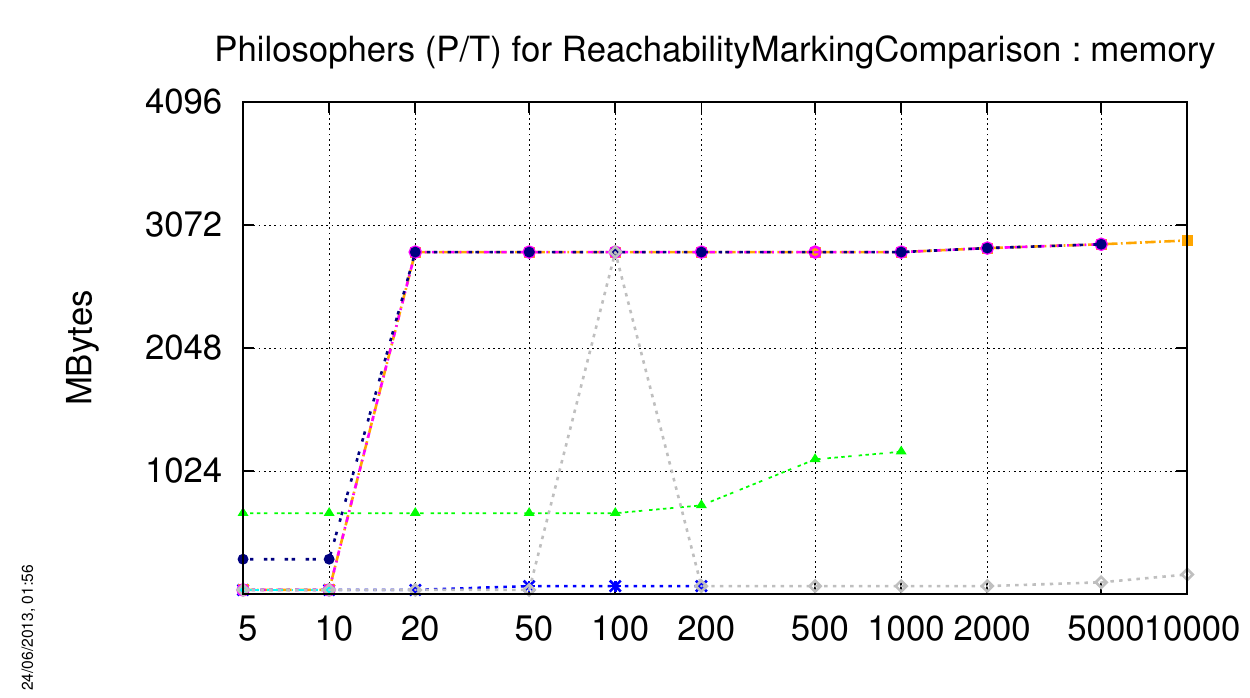}
   \includegraphics[width=7.2cm]{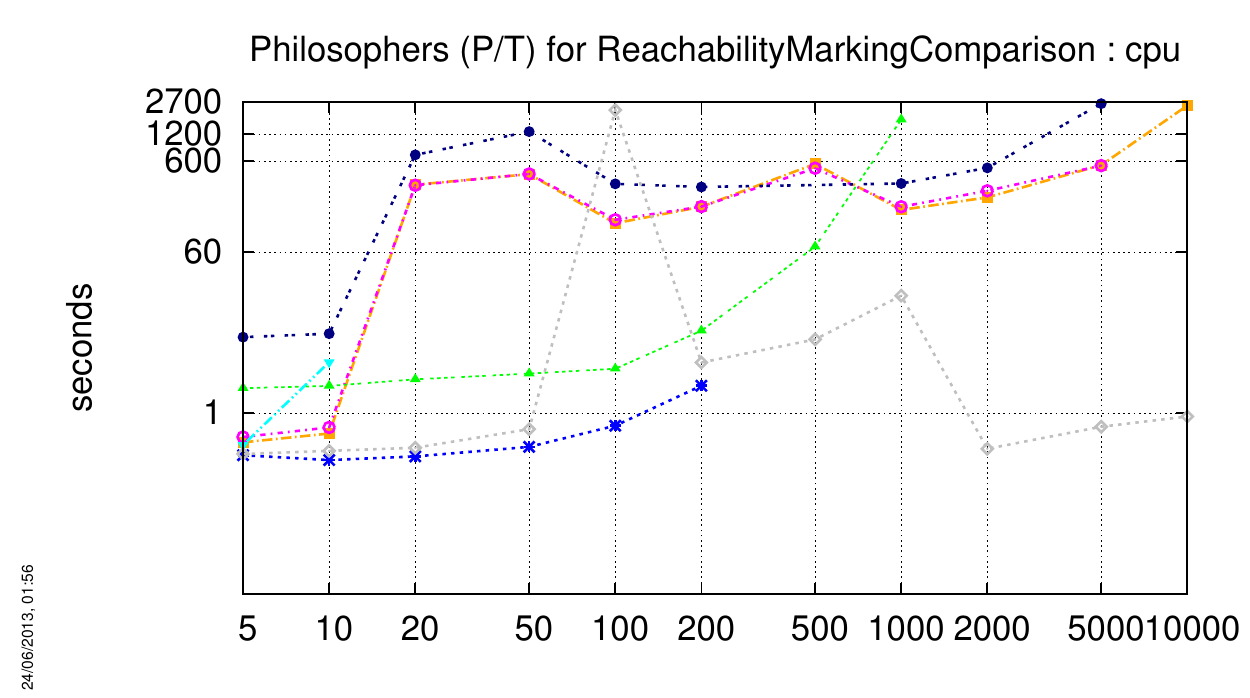}

   \includegraphics[height=1cm]{figures/tools-legend.pdf}
\end{center}

\subsubsection{\acs{PhilosophersDyn-COL}}
No instance of this model could be computed for the \textbf{ReachabilityMarkingComparison} examination.

\subsubsection{\acs{PhilosophersDyn-PT}}
The charts below respectively show how tools compete with this ``Known'' model (memory and CPU).

\index{Performances!ReachabilityMarkingComparison!PhilosophersDyn (P/T)}
\begin{center}
   \includegraphics[width=7.2cm]{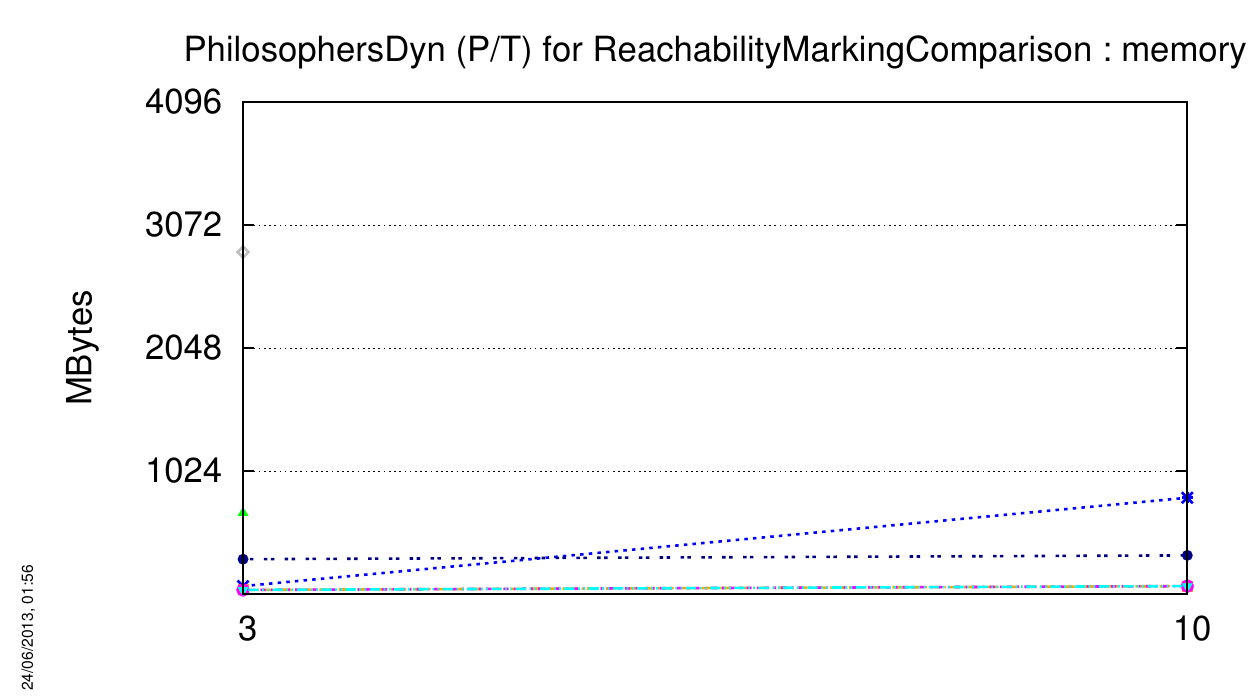}
   \includegraphics[width=7.2cm]{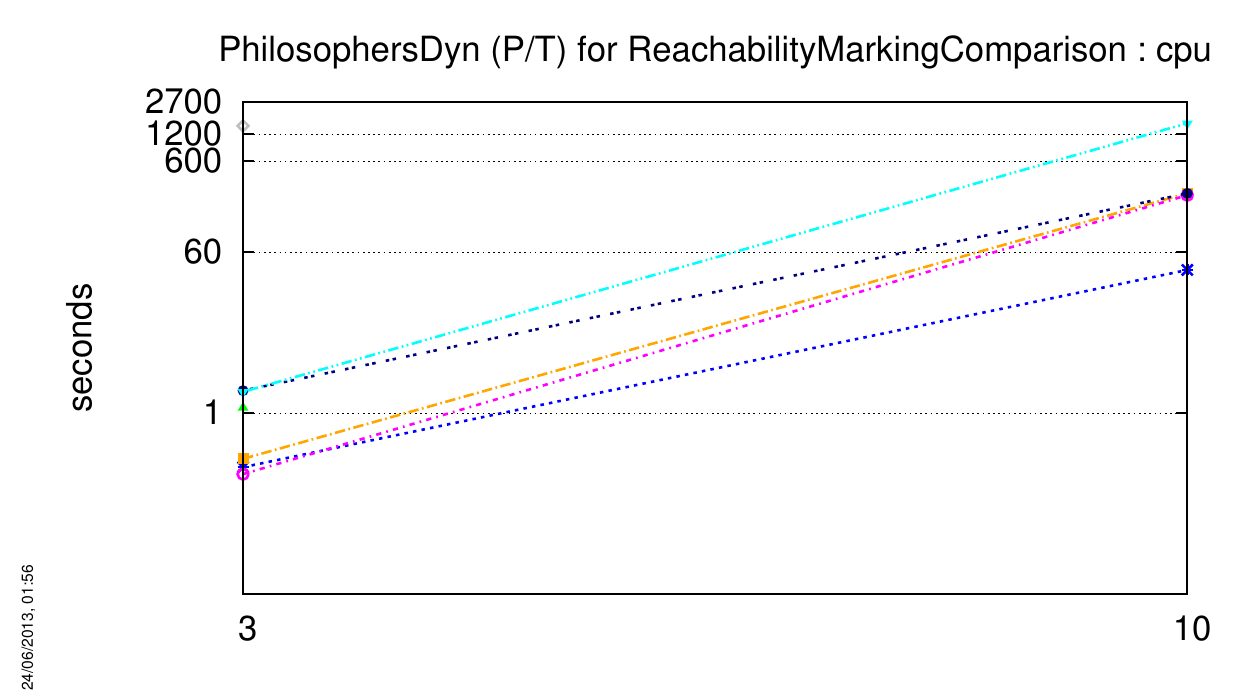}

   \includegraphics[height=1cm]{figures/tools-legend.pdf}
\end{center}

\subsubsection{\acs{Planning-PT}}
No instance of this model could be computed for the \textbf{ReachabilityMarkingComparison} examination.

\subsubsection{\acs{Railroad-PT}}
No instance of this model could be computed for the \textbf{ReachabilityMarkingComparison} examination.

\subsubsection{\acs{RessAllocation-PT}}
No instance of this model could be computed for the \textbf{ReachabilityMarkingComparison} examination.

\subsubsection{\acs{Ring-PT}}
No instance of this model could be computed for the \textbf{ReachabilityMarkingComparison} examination.

\subsubsection{\acs{RwMutex-PT}}
No instance of this model could be computed for the \textbf{ReachabilityMarkingComparison} examination.

\subsubsection{\acs{SharedMemory-COL}}
No instance of this model could be computed for the \textbf{ReachabilityMarkingComparison} examination.

\subsubsection{\acs{SharedMemory-PT}}
The charts below respectively show how tools compete with this ``Known'' model (memory and CPU).

\index{Performances!ReachabilityMarkingComparison!SharedMemory (P/T)}
\begin{center}
   \includegraphics[width=7.2cm]{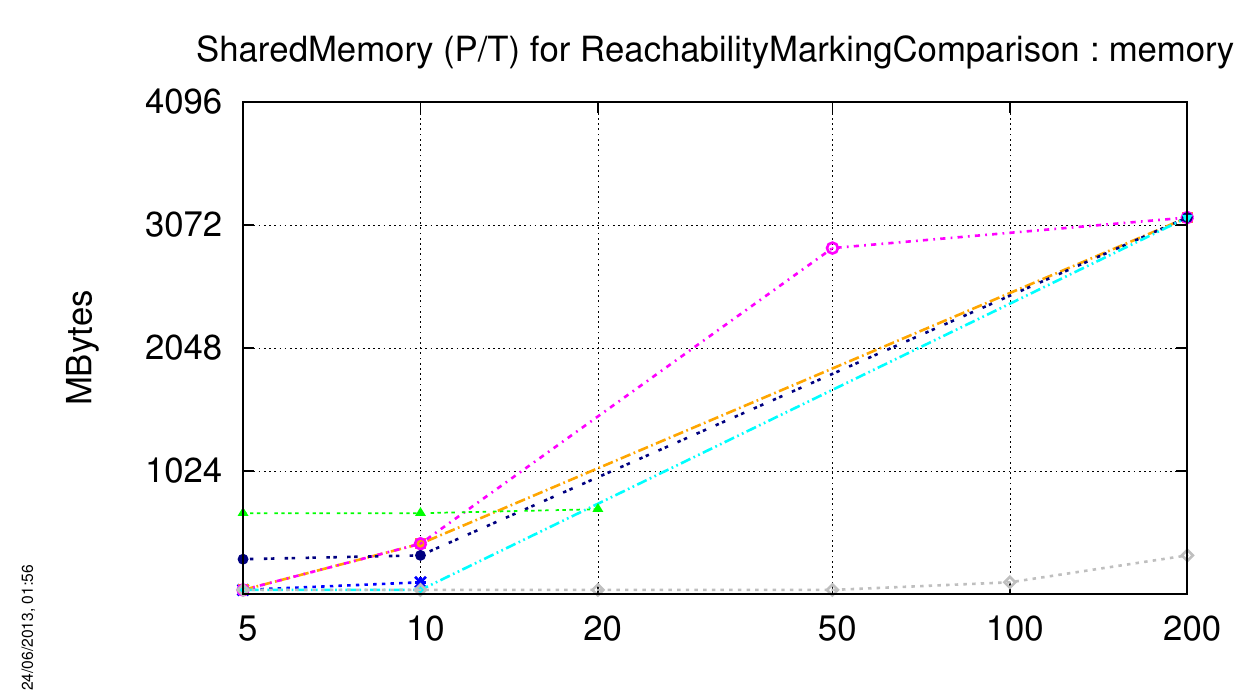}
   \includegraphics[width=7.2cm]{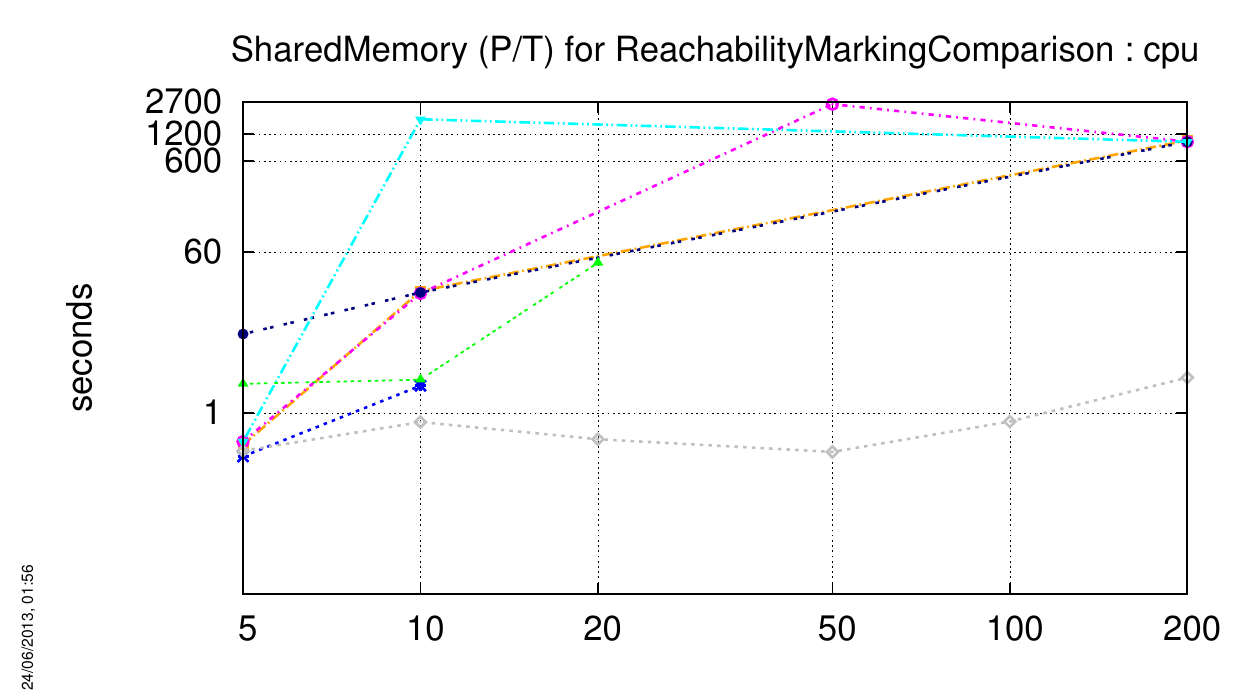}

   \includegraphics[height=1cm]{figures/tools-legend.pdf}
\end{center}

\subsubsection{\acs{SimpleLoadBal-COL}}
No instance of this model could be computed for the \textbf{ReachabilityMarkingComparison} examination.

\subsubsection{\acs{SimpleLoadBal-PT}}
The charts below respectively show how tools compete with this ``Known'' model (memory and CPU).

\index{Performances!ReachabilityMarkingComparison!SimpleLoadBal (P/T)}
\begin{center}
   \includegraphics[width=7.2cm]{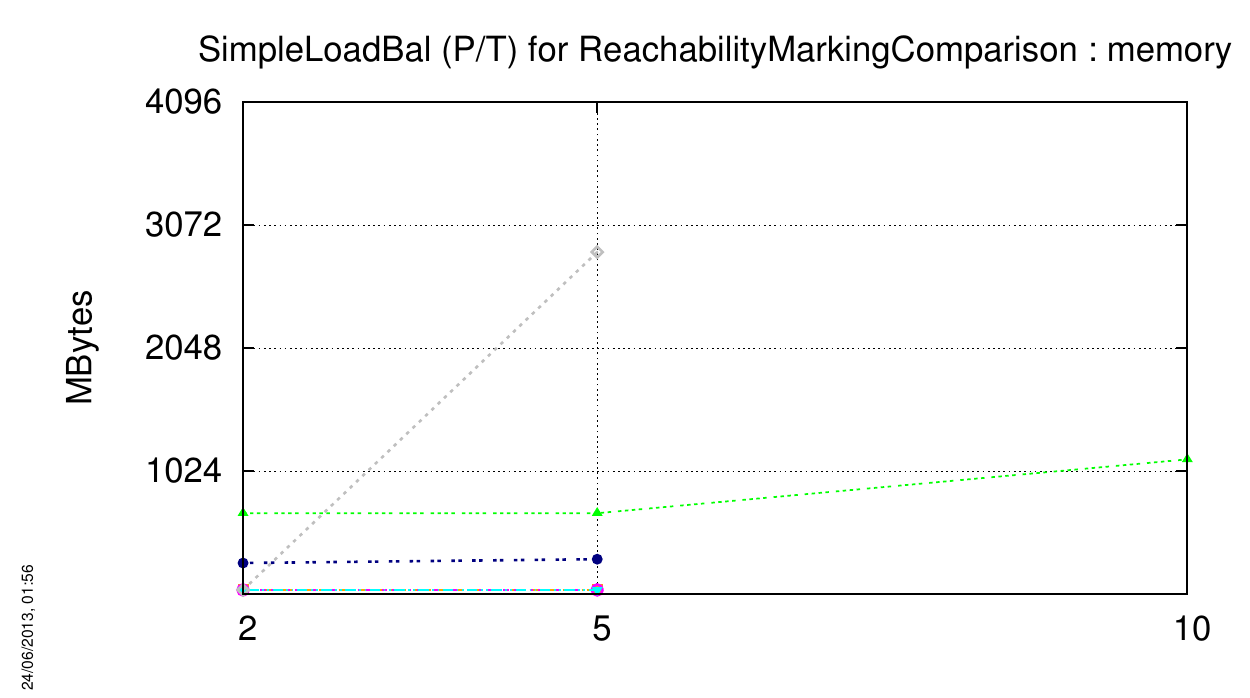}
   \includegraphics[width=7.2cm]{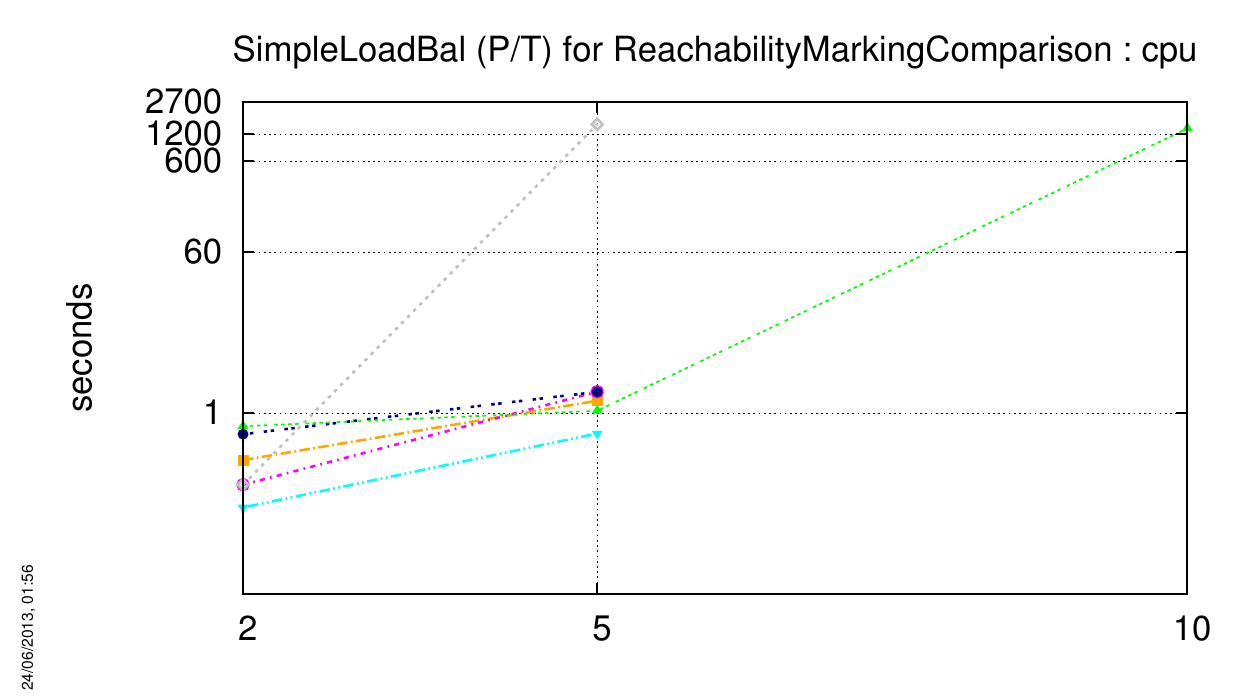}

   \includegraphics[height=1cm]{figures/tools-legend.pdf}
\end{center}

\subsubsection{\acs{TokenRing-COL}}
No instance of this model could be computed for the \textbf{ReachabilityMarkingComparison} examination.

\subsubsection{\acs{TokenRing-PT}}
The charts below respectively show how tools compete with this ``Known'' model (memory and CPU).

\index{Performances!ReachabilityMarkingComparison!TokenRing (P/T)}
\begin{center}
   \includegraphics[width=7.2cm]{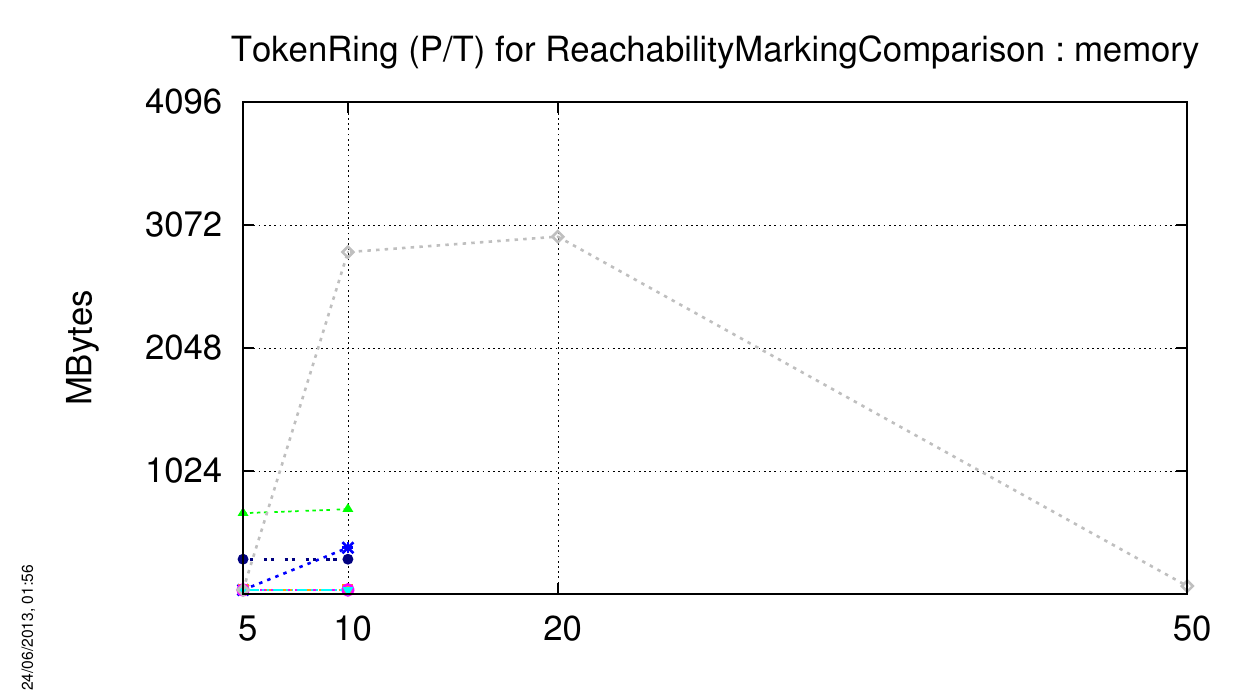}
   \includegraphics[width=7.2cm]{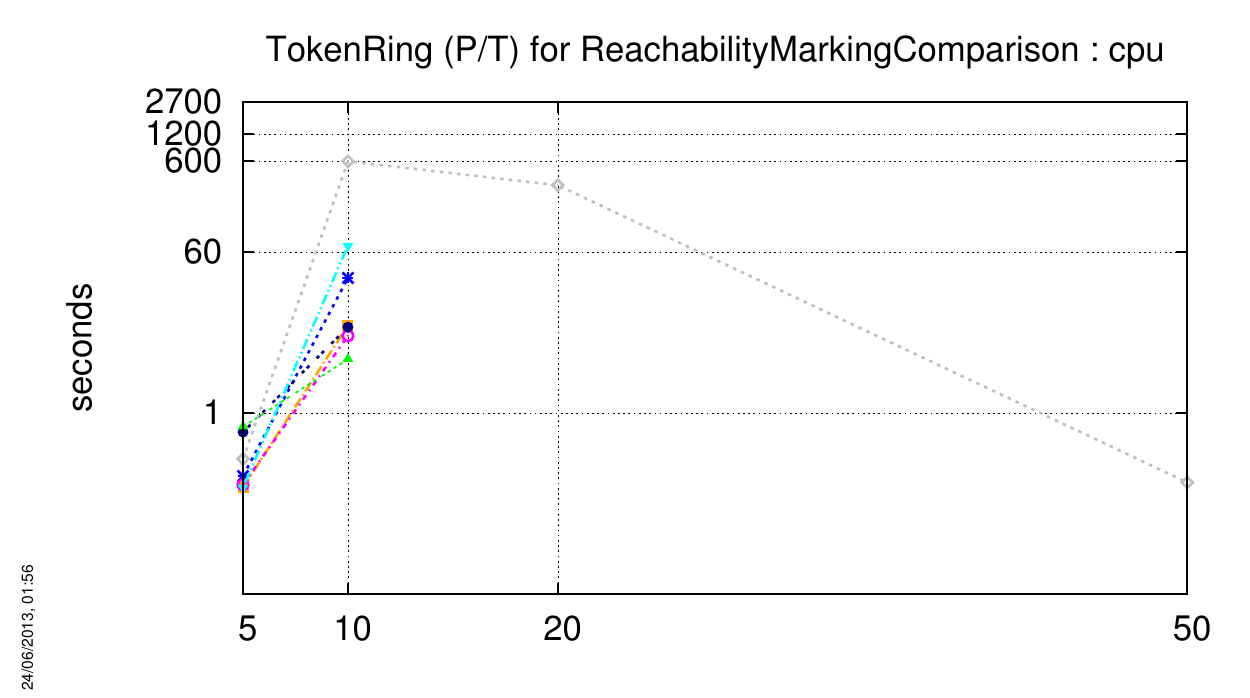}

   \includegraphics[height=1cm]{figures/tools-legend.pdf}
\end{center}

\subsubsection{\acs{HouseConstruction-PT}}
No instance of this model could be computed for the \textbf{ReachabilityMarkingComparison} examination.

\subsubsection{\acs{IBMB2S565S3960-PT}}
No instance of this model could be computed for the \textbf{ReachabilityMarkingComparison} examination.

\subsubsection{\acs{QuasiCertifProtocol-COL}}
No instance of this model could be computed for the \textbf{ReachabilityMarkingComparison} examination.

\subsubsection{\acs{QuasiCertifProtocol-PT}}
The charts below respectively show how tools compete with this ``Suprise'' model (memory and CPU).

\index{Performances!ReachabilityMarkingComparison!QuasiCertifProtocol (P/T)}
\begin{center}
   \includegraphics[width=7.2cm]{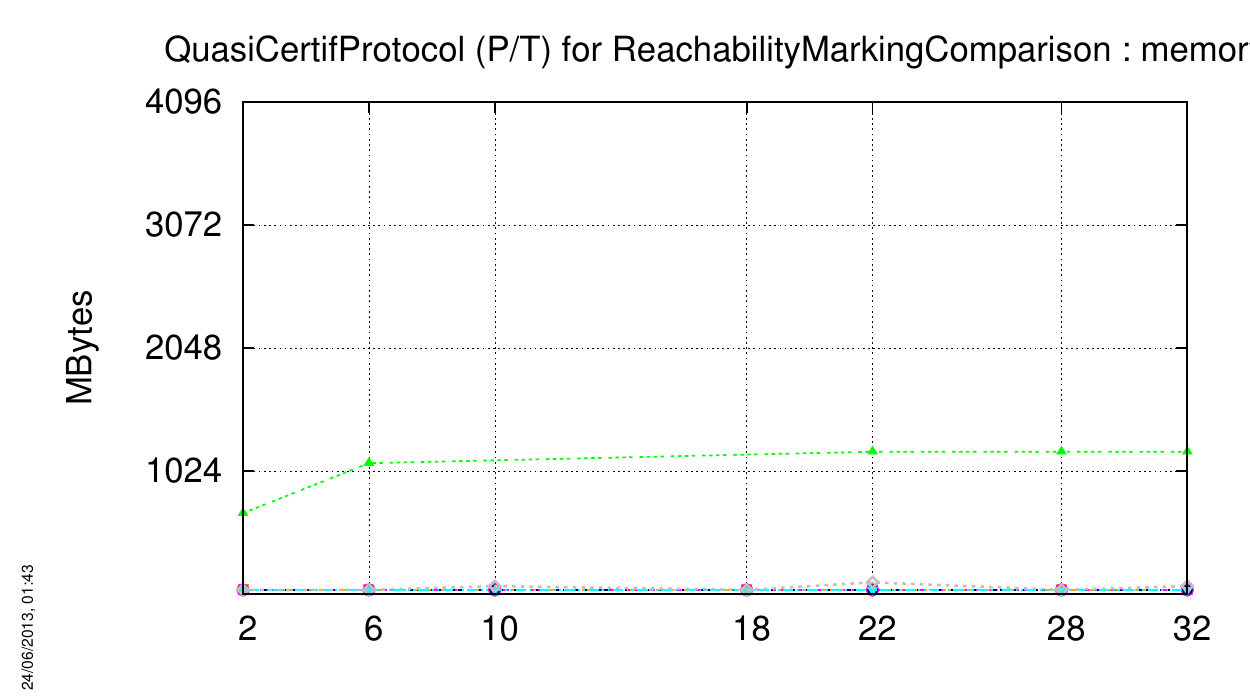}
   \includegraphics[width=7.2cm]{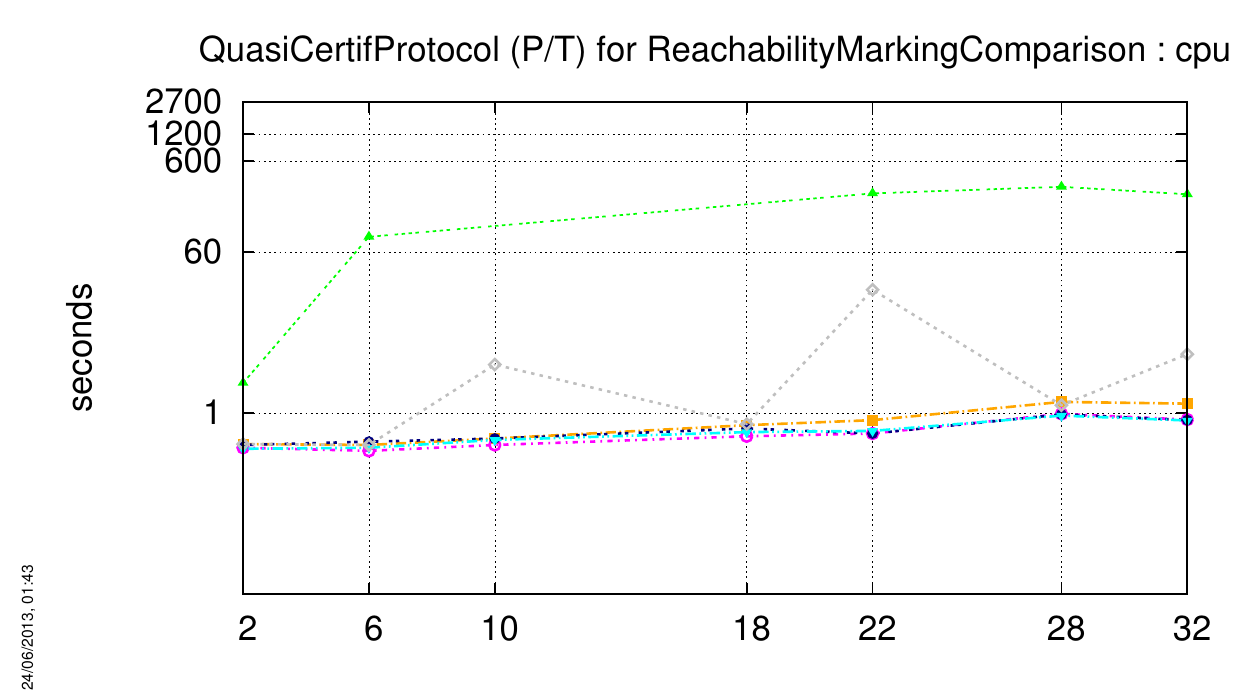}

   \includegraphics[height=1cm]{figures/tools-legend.pdf}
\end{center}

\subsubsection{\acs{Vasy2003-PT}}
No instance of this model could be computed for the \textbf{ReachabilityMarkingComparison} examination.

\subsection{Outputs for the ReachabilityMarkingComparison Examination}
\index{Outputs!ReachabilityMarkingComparison}

Please find enclosed the brute results for this examination (``Known'' and ``Surprise'' models).
We display only the score of tools that provide a results for at least one instance of one model.
The legend for the values is provided below:
\begin{itemize}
   \item\textbf{nc}: the tool does not compete this examination for this model/instance,
   \item\textbf{cc}: the tool cannot compute this examination for this model/instance,
   \item\textbf{to}: the tool cannot compute this examination for this model/instance within the maximum allowed time,
   \item\textbf{mp}: the tool encountered a memory problem (stack overflow or memory full),
   \item\textbf{nf}: there is no formula available for this type of examination (typically, this concerns P/T nets where
       comparing marking cardinality has no signification when there is no equivalent colored net).
\end{itemize}

\textbf{Note on the display of results for formulas:} each formula is considered as a flag (F if false, T if true, - or ?
when the value cannot be determined). These values are concatenated in the order they appear (we assume it is the order of formulas as they were provided).

\subsubsection{``Known'' Models}

Results are summarized in the table below.
{\footnotesize
\begin{longtable}{|cccccccc|}
   \hline
      \multicolumn{8}{|c|}{\bf \acs{CSRepetitions-COL}} \\
      \textbf{Instances}\index{Outputs!ReachabilityMarkingComparison!CSRepetitions (colored)} &
      \acs{greatspn} &
      \acs{lola} &
      \acs{lola-optimistic} &
      \acs{lola-optimistic-incomplete} &
      \acs{lola-pessimistic} &
      \acs{marcie} &
      \acs{sara} \\
      02 &
      nc &
      nc &
      nc &
      nc &
      nc &
      nc &
      nc \\
      03 &
      nc &
      nc &
      nc &
      nc &
      nc &
      nc &
      nc \\
      04 &
      nc &
      nc &
      nc &
      nc &
      nc &
      nc &
      nc \\
      05 &
      nc &
      nc &
      nc &
      nc &
      nc &
      nc &
      nc \\
      07 &
      nc &
      nc &
      nc &
      nc &
      nc &
      nc &
      nc \\
      10 &
      nc &
      nc &
      nc &
      nc &
      nc &
      nc &
      nc \\
\hline
\end{longtable}
}

\subsubsection{``Surprise'' Models}

Results are summarized in the table below.
{\footnotesize
\begin{longtable}{|ccccccc|}
   \hline
      \multicolumn{7}{|c|}{\bf \acs{HouseConstruction-PT}} \\
      instances\index{Outputs!ReachabilityMarkingComparison!HouseConstruction (P/T)} &
      \acs{lola} &
      \acs{lola-optimistic} &
      \acs{lola-optimistic-incomplete} &
      \acs{lola-pessimistic} &
      \acs{marcie} &
      \acs{sara} \\
      002 &
      nf &
      nf &
      nf &
      nf &
      nf &
      nf \\
      005 &
      nf &
      nf &
      nf &
      nf &
      nf &
      nf \\
      010 &
      nf &
      nf &
      nf &
      nf &
      nf &
      nf \\
      020 &
      nf &
      nf &
      nf &
      nf &
      nf &
      nf \\
      050 &
      nf &
      nf &
      nf &
      nf &
      nf &
      nf \\
      100 &
      nf &
      nf &
      nf &
      nf &
      nf &
      nf \\
\hline
\end{longtable}
}

\subsection{Score for the ReachabilityMarkingComparison Examination}
\index{Scores!ReachabilityMarkingComparison}

Please find enclosed the scores for this examination (``Known'' and ``Surprise'' models).
We display only the score of tools that provide a results for at least one instance of one model.
The total is first listed in the table below followed by a detail, for each proposed model.
Meaning of the line labels are:
\begin{itemize}
\item\textbf{1st instance}: the tool gets a bonus for having processed the first instance of this model (+1 point),
\item\textbf{instances}: the tool gets 1 point per instances treated 
(for that, we assume that at least one formula has been successfully computed),
\item\textbf{max reached}: the tool could process all the instances for the model (+2 points),
\item\textbf{best}: the tool is among the ones that processed a maximum of instances within the time and memory confinement (+2 points).
\end{itemize}

\subsubsection{``Known'' Models}

Scores are summarized in the table below.
For each model in this category, a tool may collect up to 411 points.

{\footnotesize
\begin{longtable}{|cccccccc|}
   \hline
      \multicolumn{8}{|c|}{\bf Total score of the tools for this examination} \\
        &
      \acs{greatspn} &
      \acs{lola} &
      \acs{lola-optimistic} &
      \acs{lola-optimistic-incomplete} &
      \acs{lola-pessimistic} &
      \acs{marcie} &
      \acs{sara} \\
      Total Score &
      {\bf 5} &
      {\bf 71} &
      {\bf 69} &
      {\bf 41} &
      {\bf 34} &
      {\bf 48} &
      {\bf 71} \\
   \hline
      \multicolumn{8}{|c|}{\bf CSRepetitions (Colored)} \\
      \ \index{Scores!ReachabilityMarkingComparison!CSRepetitions (colored)} &
      \acs{greatspn} &
      \acs{lola} &
      \acs{lola-optimistic} &
      \acs{lola-optimistic-incomplete} &
      \acs{lola-pessimistic} &
      \acs{marcie} &
      \acs{sara} \\
      1st instance &
      \textcolor{red}{0} &
      \textcolor{red}{0} &
      \textcolor{red}{0} &
      \textcolor{red}{0} &
      \textcolor{red}{0} &
      \textcolor{red}{0} &
      \textcolor{red}{0} \\
      instances &
      \textcolor{red}{0} &
      \textcolor{red}{0} &
      \textcolor{red}{0} &
      \textcolor{red}{0} &
      \textcolor{red}{0} &
      \textcolor{red}{0} &
      \textcolor{red}{0} \\
      max reached &
      \textcolor{red}{0} &
      \textcolor{red}{0} &
      \textcolor{red}{0} &
      \textcolor{red}{0} &
      \textcolor{red}{0} &
      \textcolor{red}{0} &
      \textcolor{red}{0} \\
\hline
\end{longtable}
}

\subsubsection{``Surprise'' Models}

Scores are summarized in the table below.
For each model in this category, a tool may collect up to 49 points.

{\footnotesize
\begin{longtable}{|ccccccc|}
   \hline
      \multicolumn{7}{|c|}{\bf Total score of the tools for this examination} \\
        &
      \acs{lola} &
      \acs{lola-optimistic} &
      \acs{lola-optimistic-incomplete} &
      \acs{lola-pessimistic} &
      \acs{marcie} &
      \acs{sara} \\
      Total Score &
      {\bf 0} &
      {\bf 0} &
      {\bf 0} &
      {\bf 0} &
      {\bf 3} &
      {\bf 12} \\
   \hline
      \multicolumn{7}{|c|}{\bf HouseConstruction (P/T)} \\
      \ \index{Scores!ReachabilityMarkingComparison!HouseConstruction (P/T)} &
      \acs{lola} &
      \acs{lola-optimistic} &
      \acs{lola-optimistic-incomplete} &
      \acs{lola-pessimistic} &
      \acs{marcie} &
      \acs{sara} \\
      1st instance &
      \textcolor{red}{0} &
      \textcolor{red}{0} &
      \textcolor{red}{0} &
      \textcolor{red}{0} &
      \textcolor{red}{0} &
      \textcolor{red}{0} \\
      instances &
      \textcolor{red}{0} &
      \textcolor{red}{0} &
      \textcolor{red}{0} &
      \textcolor{red}{0} &
      \textcolor{red}{0} &
      \textcolor{red}{0} \\
      max reached &
      \textcolor{red}{0} &
      \textcolor{red}{0} &
      \textcolor{red}{0} &
      \textcolor{red}{0} &
      \textcolor{red}{0} &
      \textcolor{red}{0} \\
\hline
\end{longtable}
}

\subsection{Trophies for this Examination}
\index{Trophies!ReachabilityMarkingComparison}

Trophies are divided in three categories: ``Known'' models,
``Surprise'' models, and the global trophies (formula is then
$score_{global} = score_{known} + 2 \times score_{surprise}$).

\subsubsection{For ``Known'' Models} \ \\

\begin{tabular}{c|c|c}
      1 & 1 & 3 \\
   \includegraphics[width=2cm]{figures/gold.jpg} &
   \includegraphics[width=2cm]{figures/gold.jpg} &
   \includegraphics[width=2cm]{figures/bronse.jpg} \\
   \acs{sara} &
   \acs{lola} &
   \acs{lola-optimistic} \\
   71 points &
   71 points &
   69 points \\
\end{tabular}

\subsubsection{For ``Surprise'' Models}\  \\

\begin{tabular}{c|c}
      1 & 2 \\
   \includegraphics[width=2cm]{figures/gold.jpg} &
   \includegraphics[width=2cm]{figures/silver.jpg} \\
   \acs{sara} &
   \acs{marcie} \\
   12 points &
   3 points \\
\end{tabular}

\subsubsection{Global} \ \\

\begin{tabular}{c|c|c}
      1 & 2 & 3 \\
   \includegraphics[width=2cm]{figures/gold.jpg} &
   \includegraphics[width=2cm]{figures/silver.jpg} &
   \includegraphics[width=2cm]{figures/bronse.jpg} \\
   \acs{sara} &
   \acs{lola} &
   \acs{lola-optimistic} \\
   95 points &
   71 points &
   69 points \\
\end{tabular}

\newpage

\section{The ReachabilityPlaceComparison Examination}
\label{sec:exam:ReachabilityPlaceComparison}
\index{Results!ReachabilityPlaceComparison}

This examination deals with reachability properties dealing with the comparison of places marking only.
We first show a summary on the handling of models by the participating tools.
Then, we present the computed outputs and the associated scores for this
examination prior to a summary of relevant executions.

\subsection{Handling of Models by Tools}
\index{Performances!ReachabilityPlaceComparison}

\subsubsection{\acs{CSRepetitions-COL}}
The charts below respectively show how tools compete with this ``Known'' model (memory and CPU).

\index{Performances!ReachabilityPlaceComparison!CSRepetitions (Colored)}
\begin{center}
   \includegraphics[width=7.2cm]{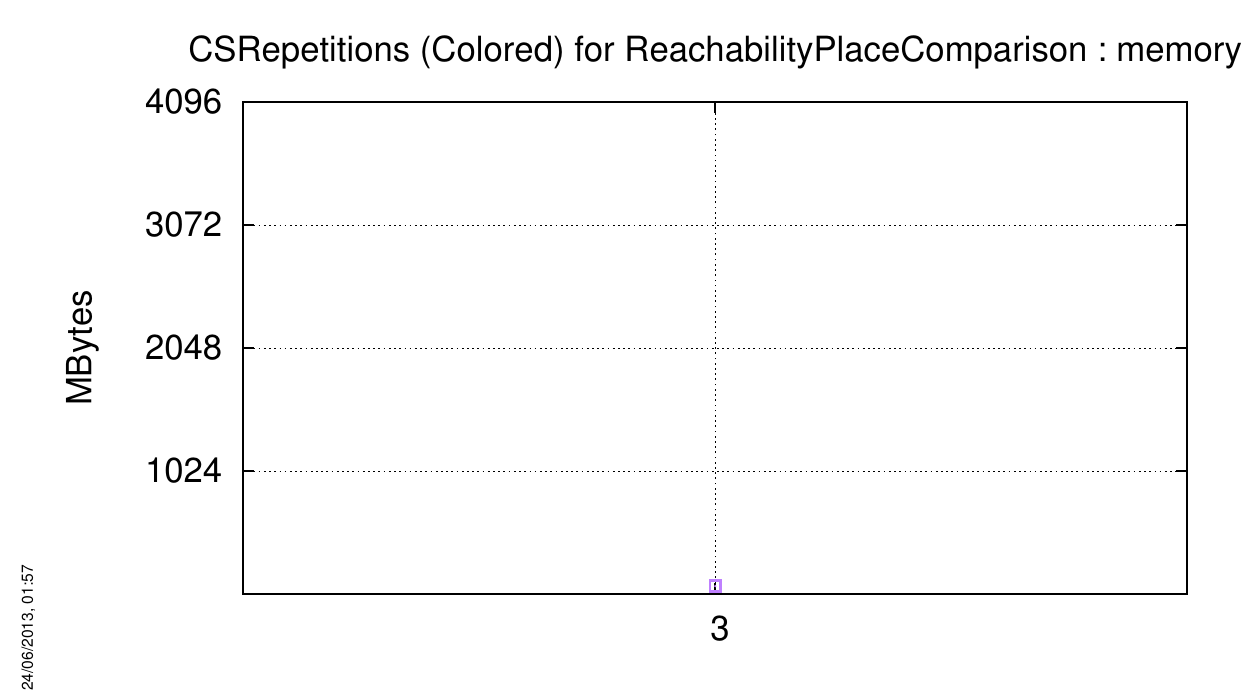}
   \includegraphics[width=7.2cm]{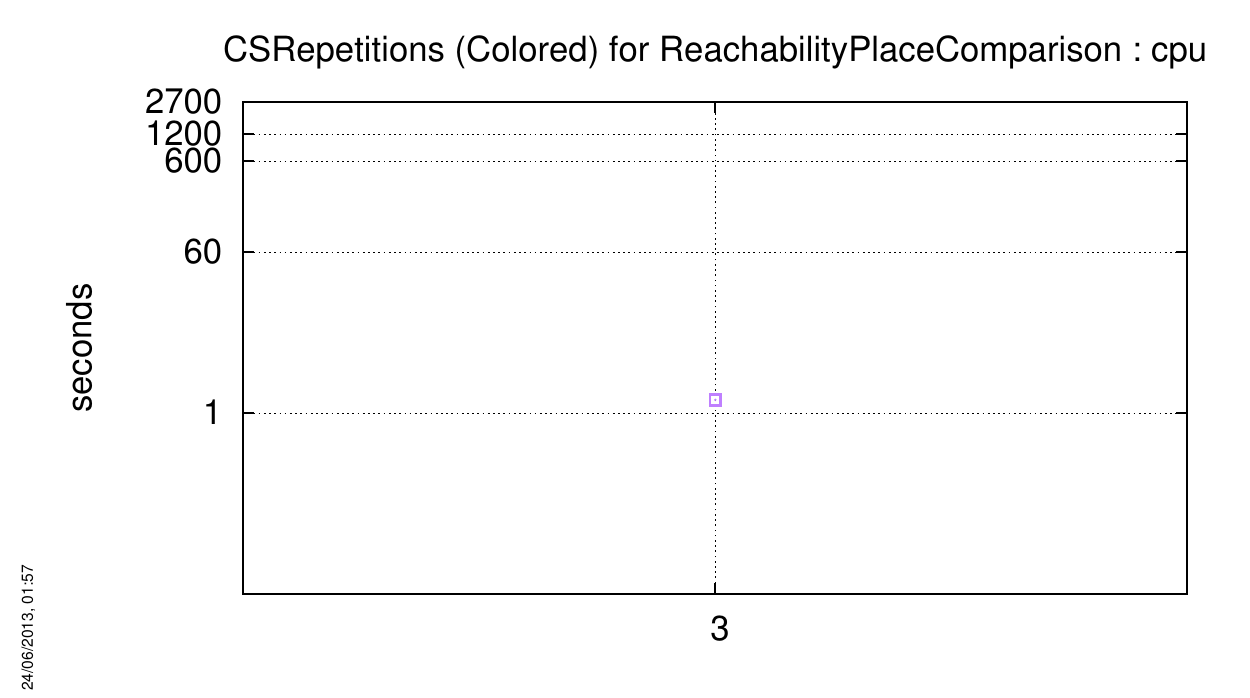}

   \includegraphics[height=1cm]{figures/tools-legend.pdf}
\end{center}

\subsubsection{\acs{CSRepetitions-PT}}
The charts below respectively show how tools compete with this ``Known'' model (memory and CPU).

\index{Performances!ReachabilityPlaceComparison!CSRepetitions (P/T)}
\begin{center}
   \includegraphics[width=7.2cm]{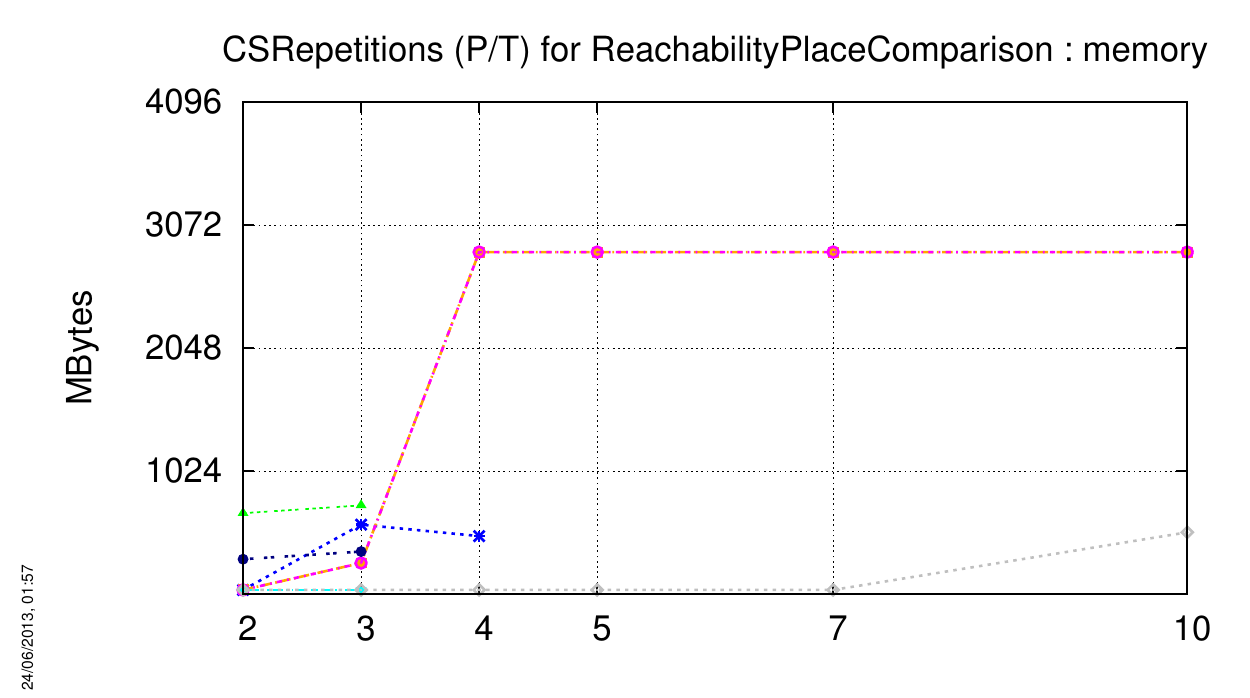}
   \includegraphics[width=7.2cm]{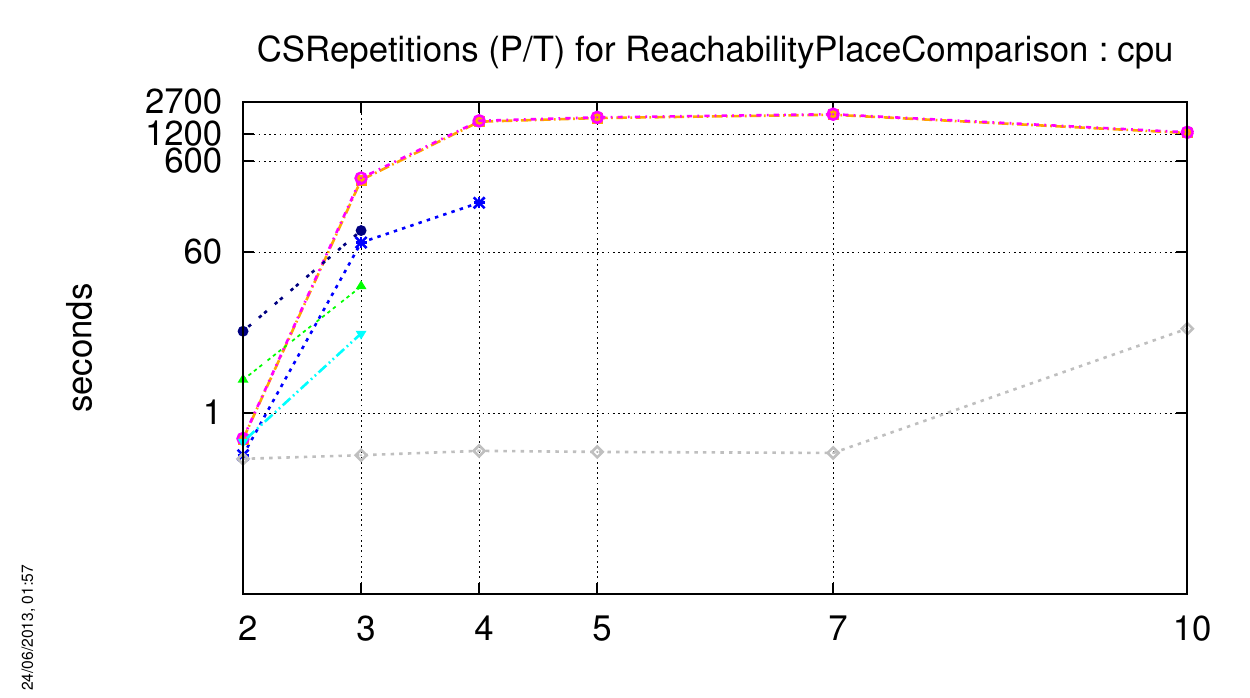}

   \includegraphics[height=1cm]{figures/tools-legend.pdf}
\end{center}

\subsubsection{\acs{Dekker-PT}}
The charts below respectively show how tools compete with this ``Known'' model (memory and CPU).

\index{Performances!ReachabilityPlaceComparison!Dekker (P/T)}
\begin{center}
   \includegraphics[width=7.2cm]{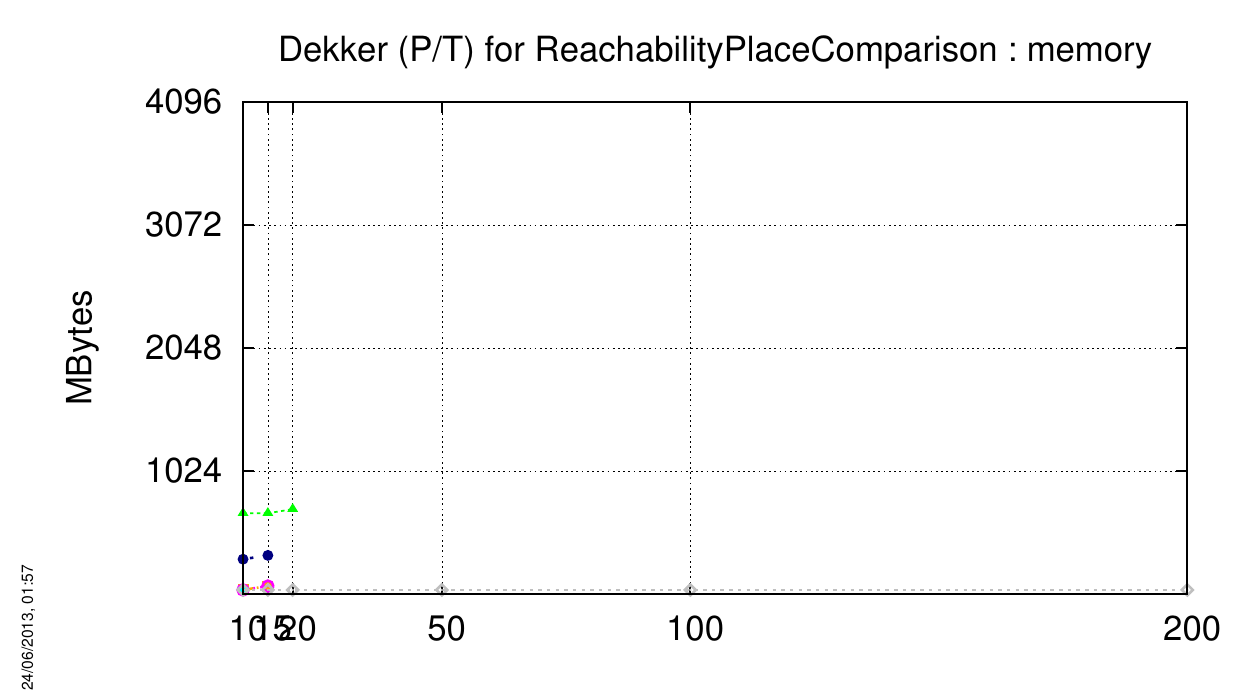}
   \includegraphics[width=7.2cm]{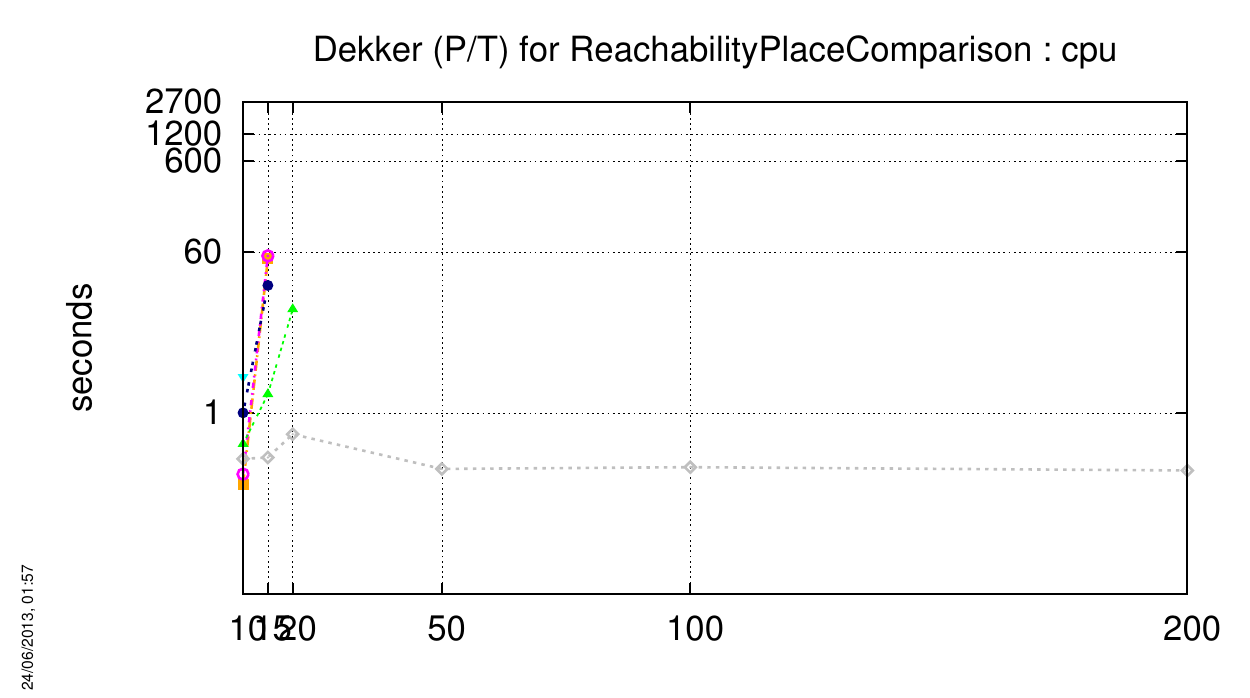}

   \includegraphics[height=1cm]{figures/tools-legend.pdf}
\end{center}

\subsubsection{\acs{DotAndBoxes-COL}}
The charts below respectively show how tools compete with this ``Known'' model (memory and CPU).

\index{Performances!ReachabilityPlaceComparison!DotAndBoxes (Colored)}
\begin{center}
   \includegraphics[width=7.2cm]{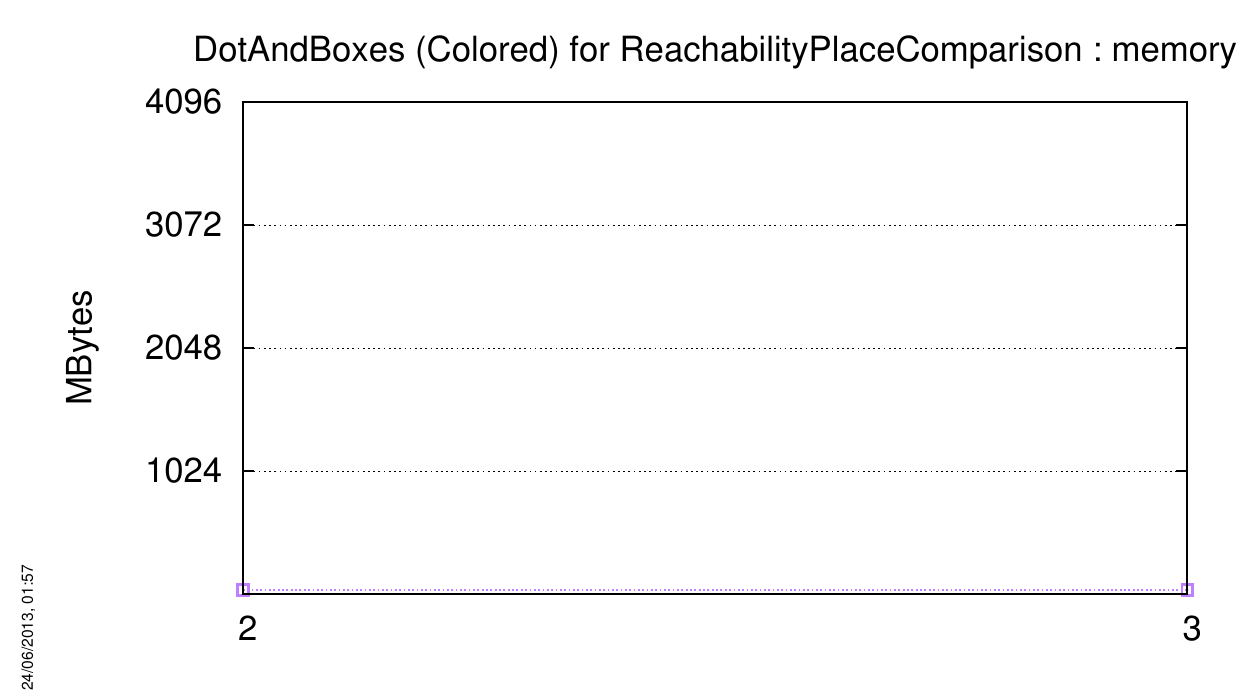}
   \includegraphics[width=7.2cm]{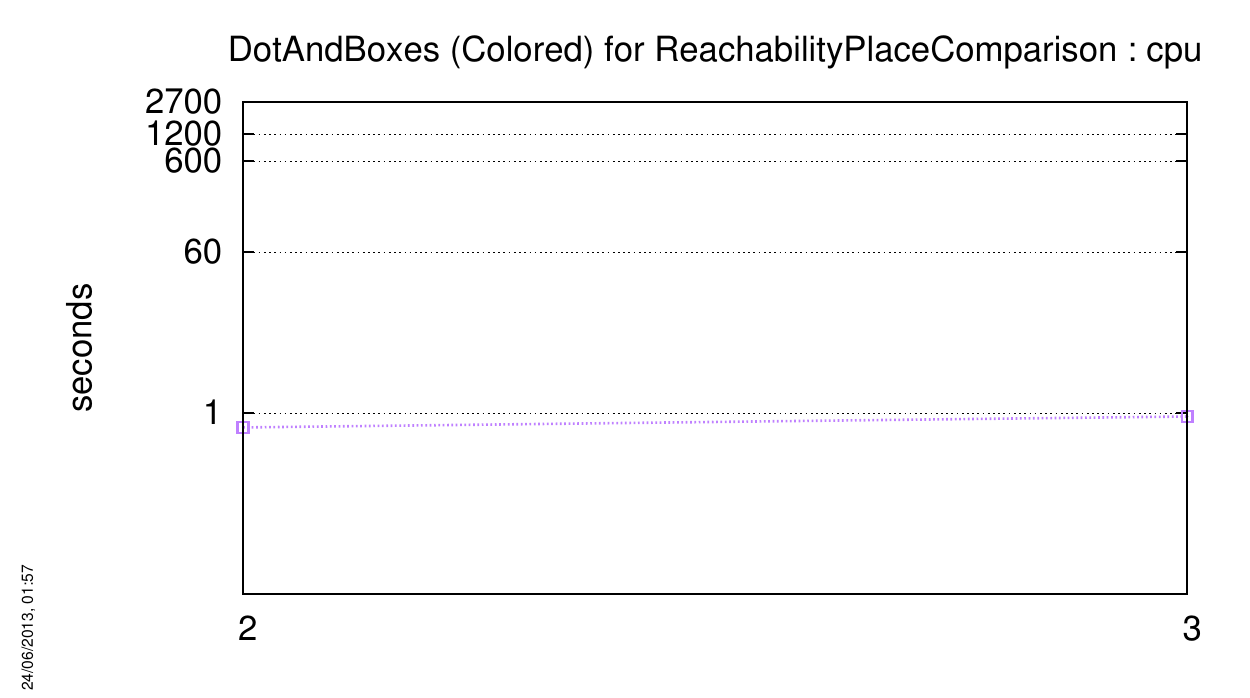}

   \includegraphics[height=1cm]{figures/tools-legend.pdf}
\end{center}

\subsubsection{\acs{DrinkVendingMachine-COL}}
The charts below respectively show how tools compete with this ``Known'' model (memory and CPU).

\index{Performances!ReachabilityPlaceComparison!DrinkVendingMachine (Colored)}
\begin{center}
   \includegraphics[width=7.2cm]{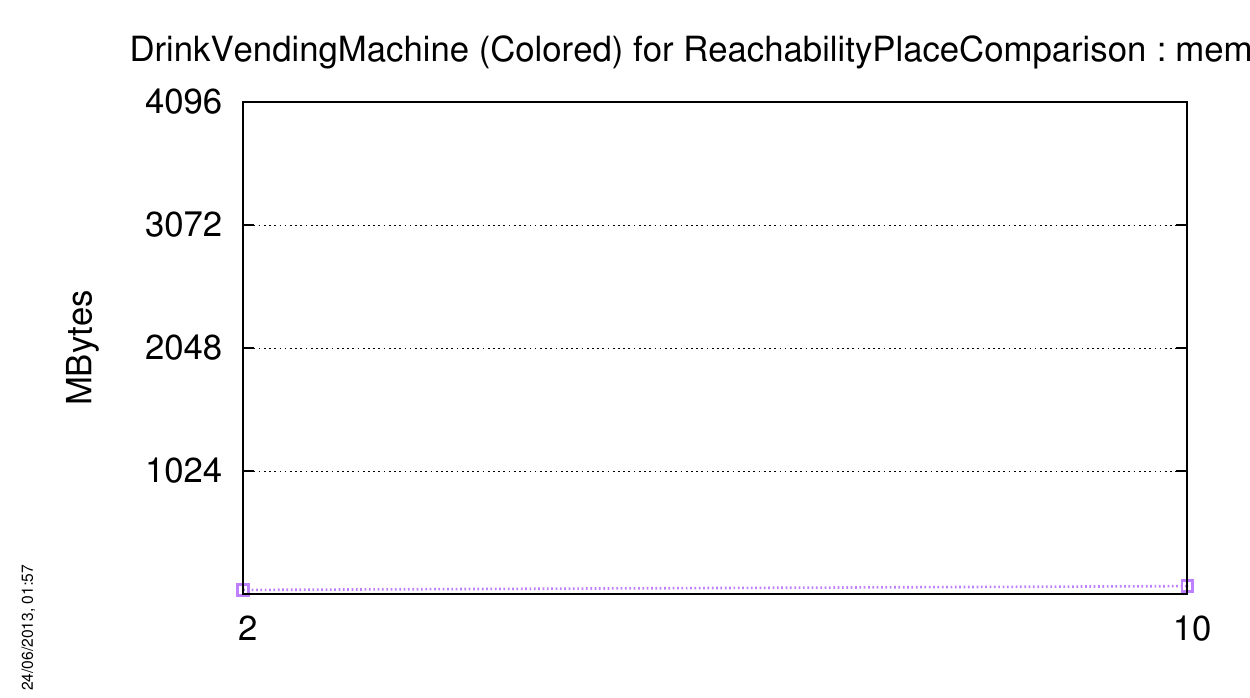}
   \includegraphics[width=7.2cm]{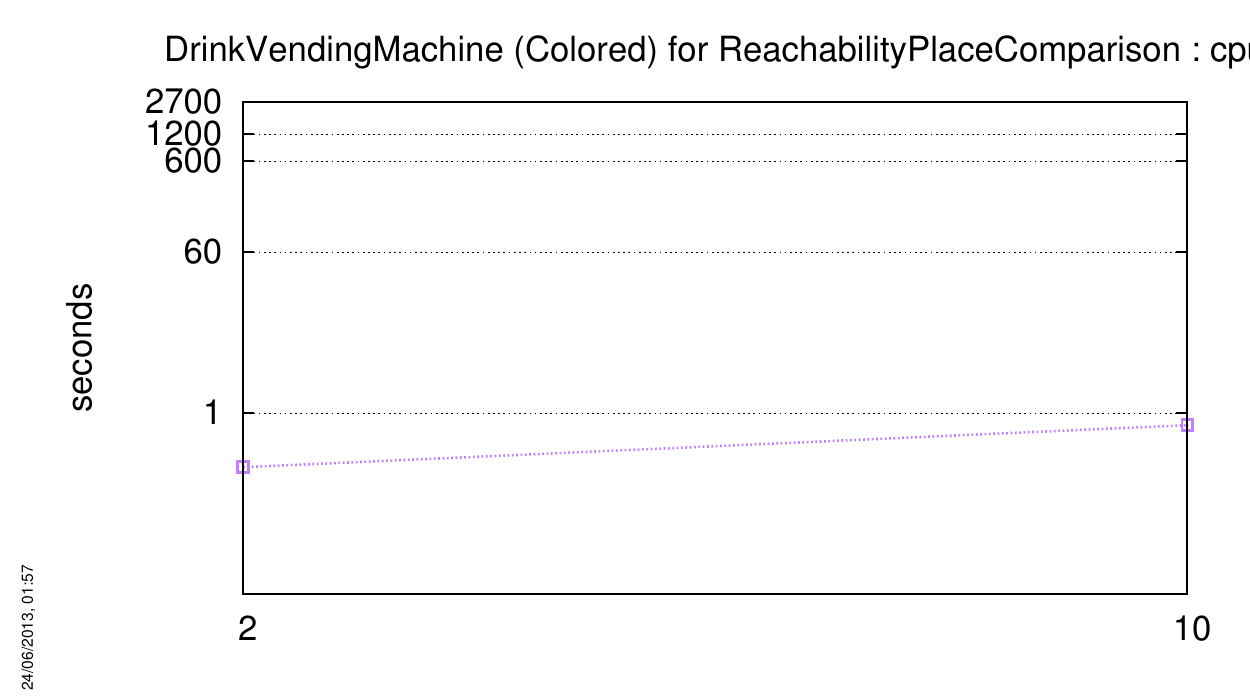}

   \includegraphics[height=1cm]{figures/tools-legend.pdf}
\end{center}

\subsubsection{\acs{DrinkVendingMachine-PT}}
The charts below respectively show how tools compete with this ``Known'' model (memory and CPU).

\index{Performances!ReachabilityPlaceComparison!DrinkVendingMachine (P/T)}
\begin{center}
   \includegraphics[width=7.2cm]{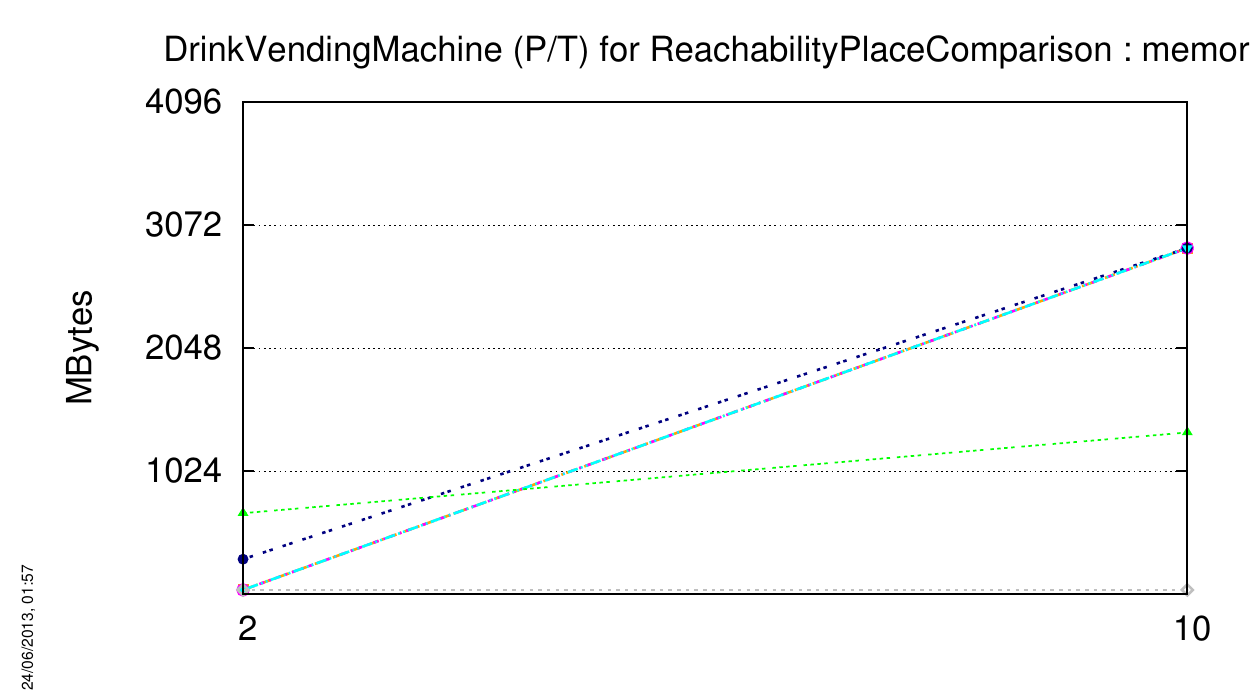}
   \includegraphics[width=7.2cm]{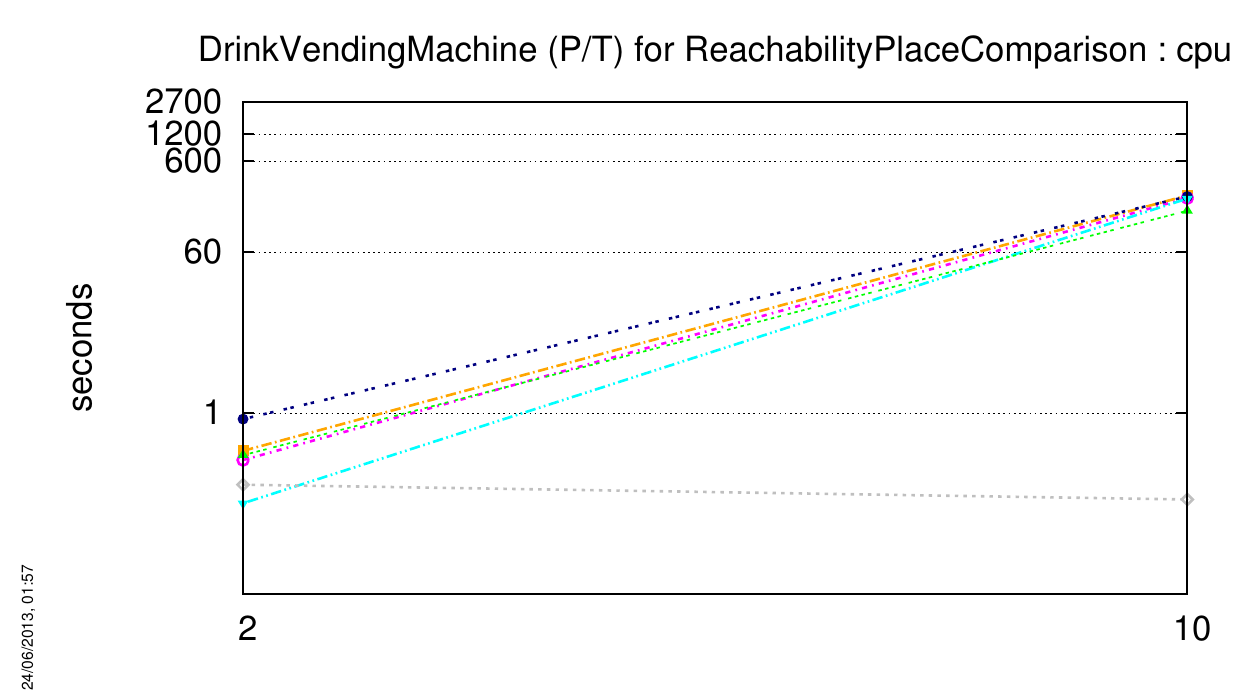}

   \includegraphics[height=1cm]{figures/tools-legend.pdf}
\end{center}

\subsubsection{\acs{Echo-PT}}
The charts below respectively show how tools compete with this ``Known'' model (memory and CPU).

\index{Performances!ReachabilityPlaceComparison!Echo (P/T)}
\begin{center}
   \includegraphics[width=7.2cm]{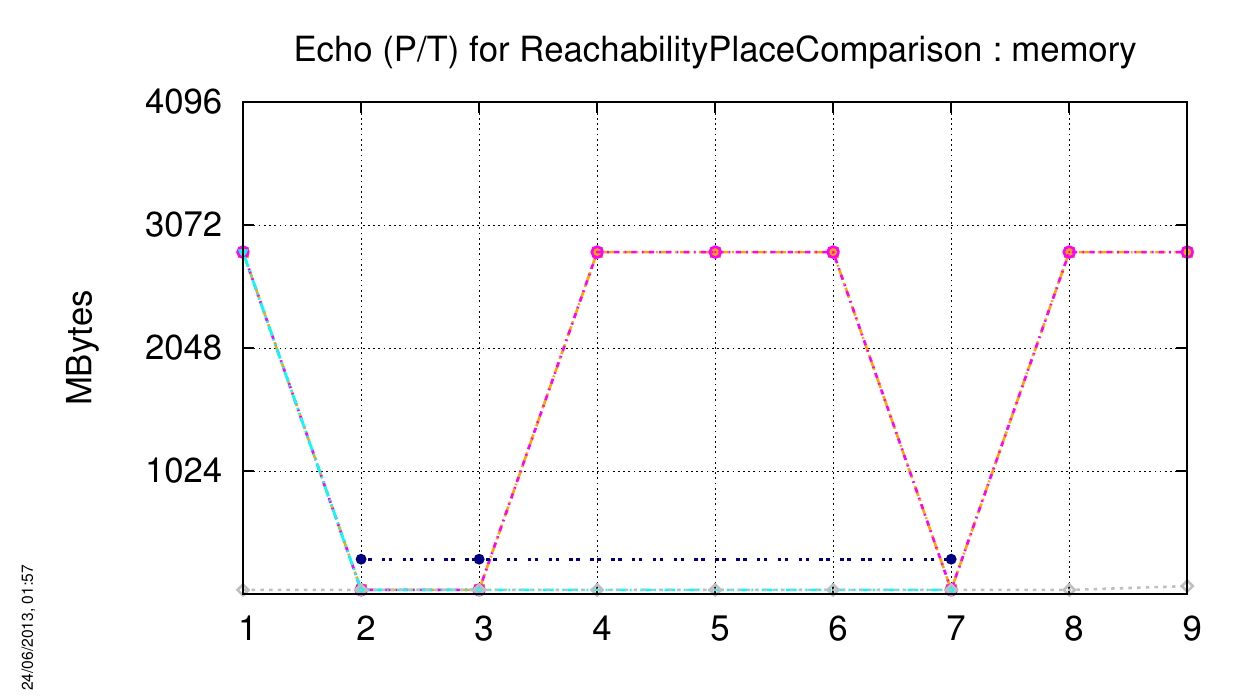}
   \includegraphics[width=7.2cm]{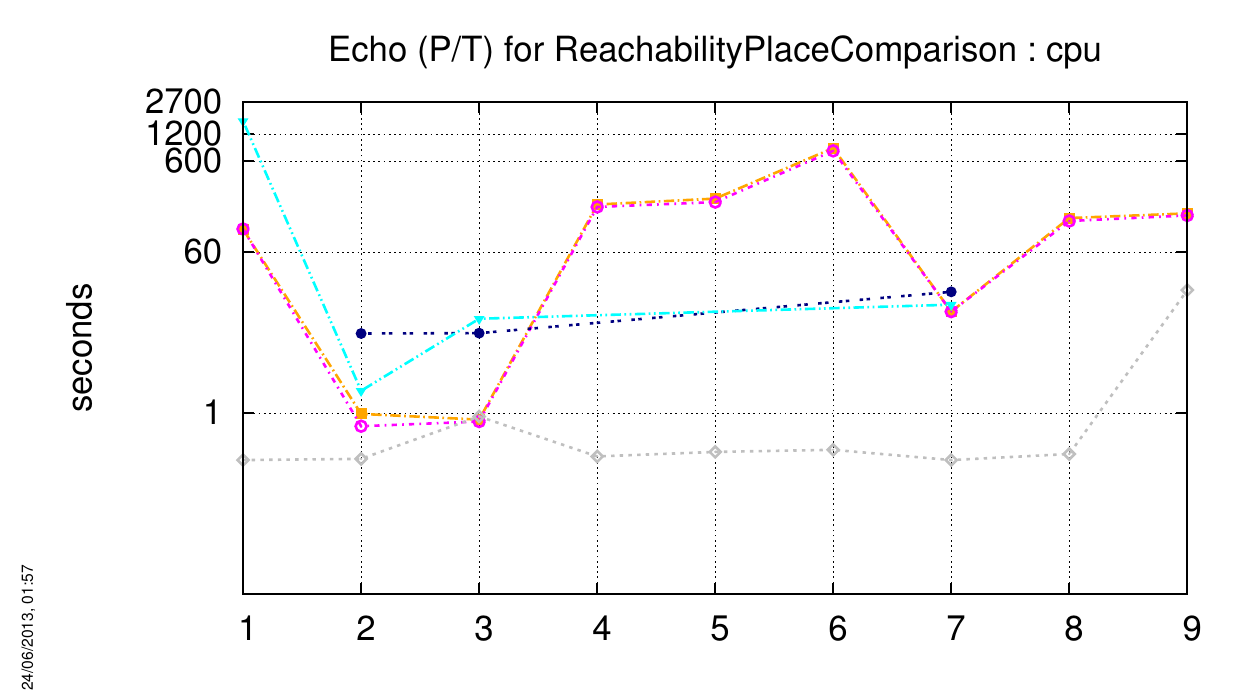}

   \includegraphics[height=1cm]{figures/tools-legend.pdf}
\end{center}

\subsubsection{\acs{Eratosthenes-PT}}
The charts below respectively show how tools compete with this ``Known'' model (memory and CPU).

\index{Performances!ReachabilityPlaceComparison!Eratosthenes (P/T)}
\begin{center}
   \includegraphics[width=7.2cm]{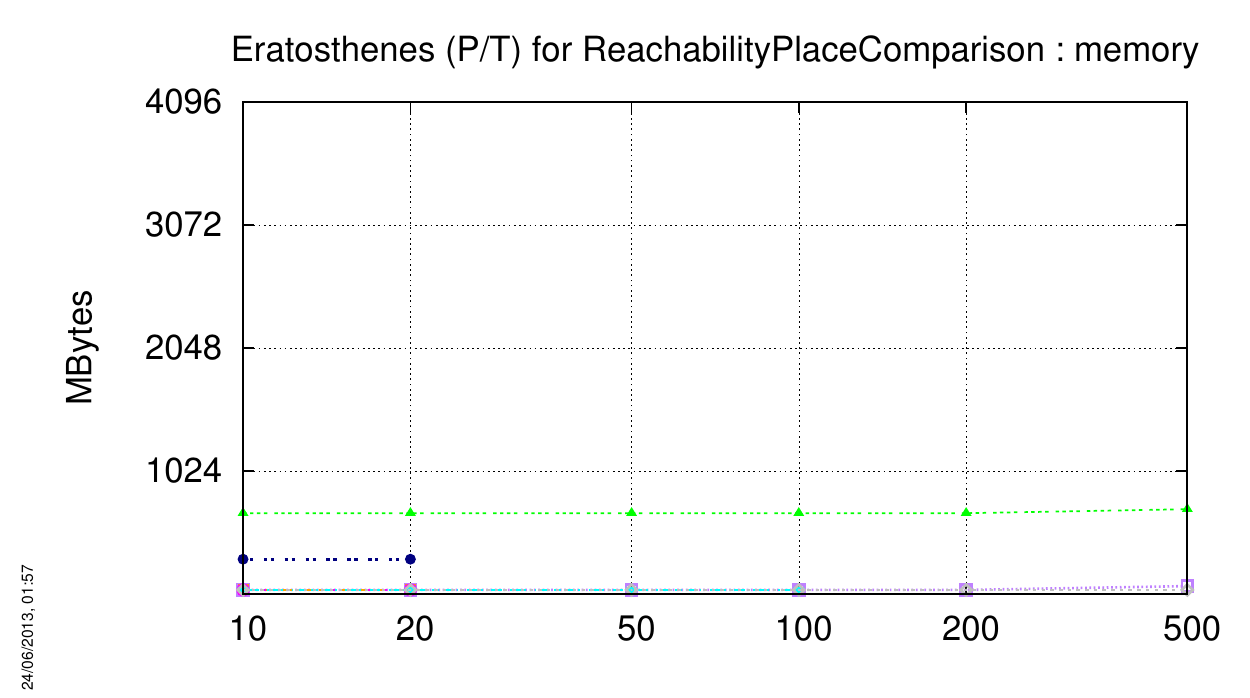}
   \includegraphics[width=7.2cm]{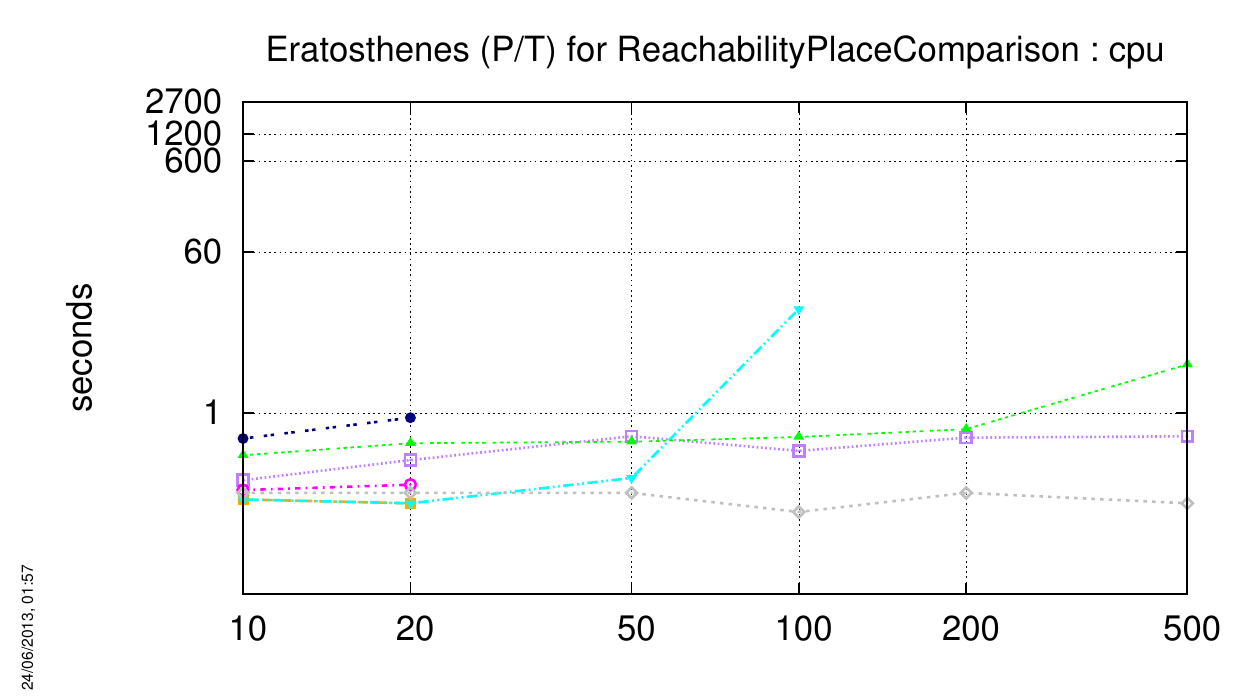}

   \includegraphics[height=1cm]{figures/tools-legend.pdf}
\end{center}

\subsubsection{\acs{FMS-PT}}
The charts below respectively show how tools compete with this ``Known'' model (memory and CPU).

\index{Performances!ReachabilityPlaceComparison!FMS (P/T)}
\begin{center}
   \includegraphics[width=7.2cm]{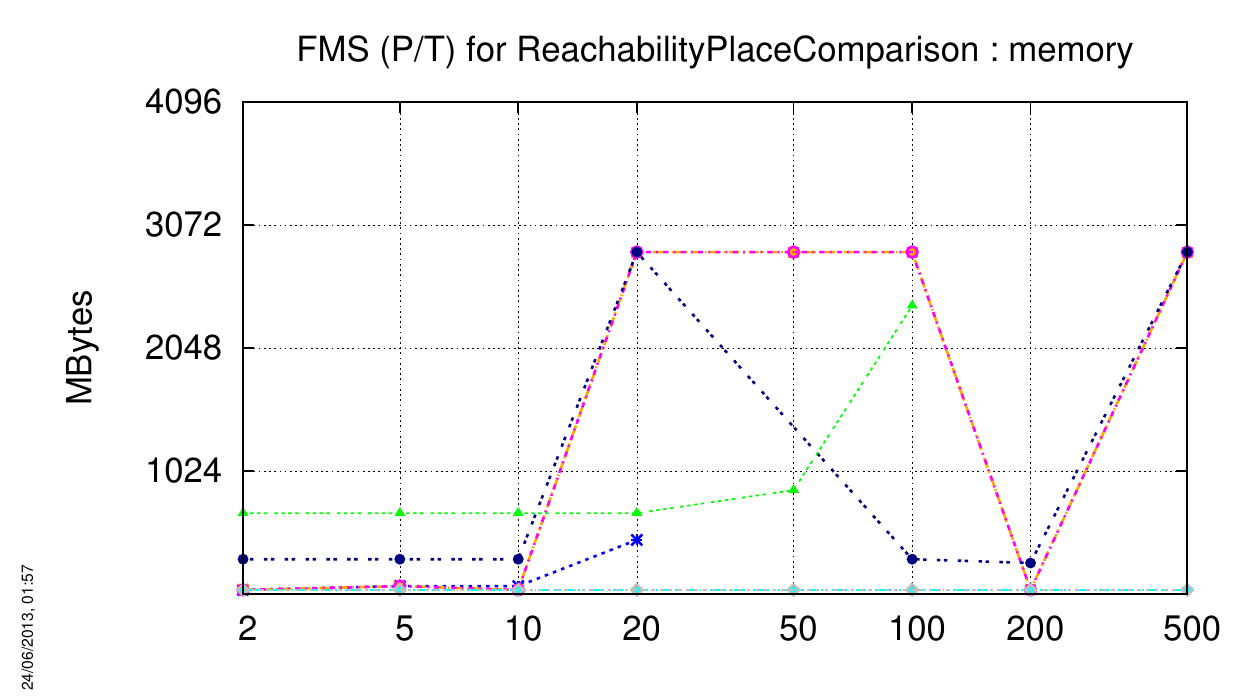}
   \includegraphics[width=7.2cm]{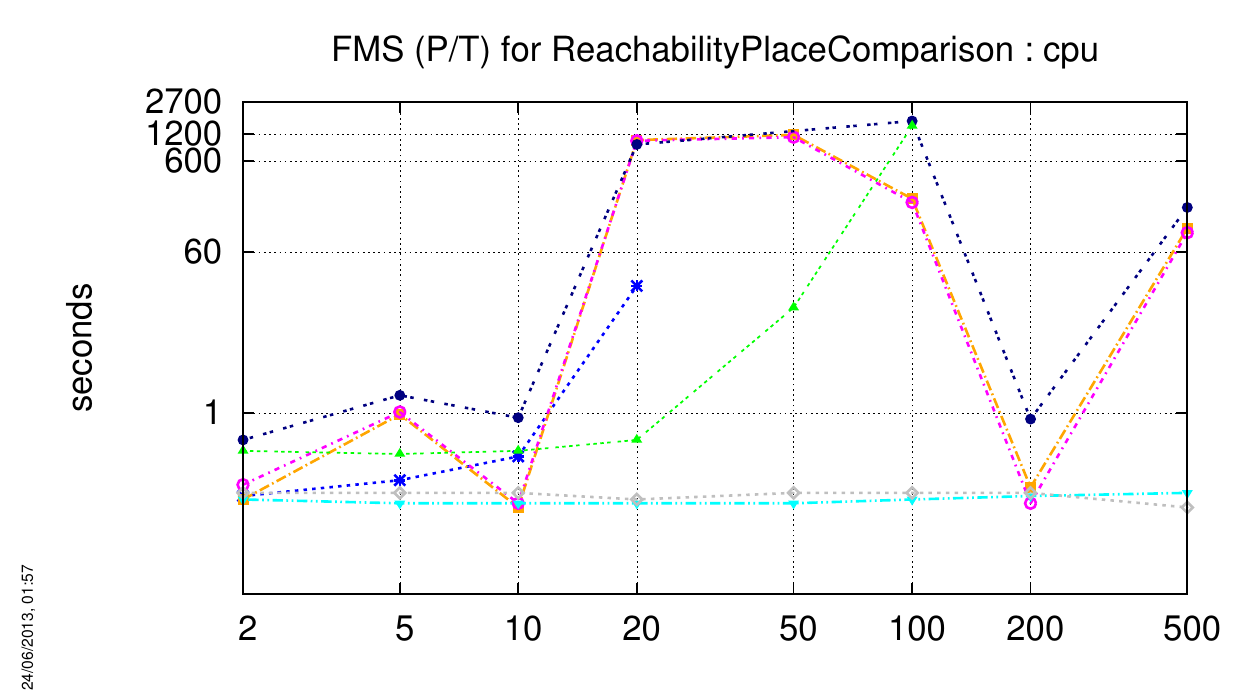}

   \includegraphics[height=1cm]{figures/tools-legend.pdf}
\end{center}

\subsubsection{\acs{GlobalRessAlloc-COL}}
The charts below respectively show how tools compete with this ``Known'' model (memory and CPU).

\index{Performances!ReachabilityPlaceComparison!GlobalRessAlloc (Colored)}
\begin{center}
   \includegraphics[width=7.2cm]{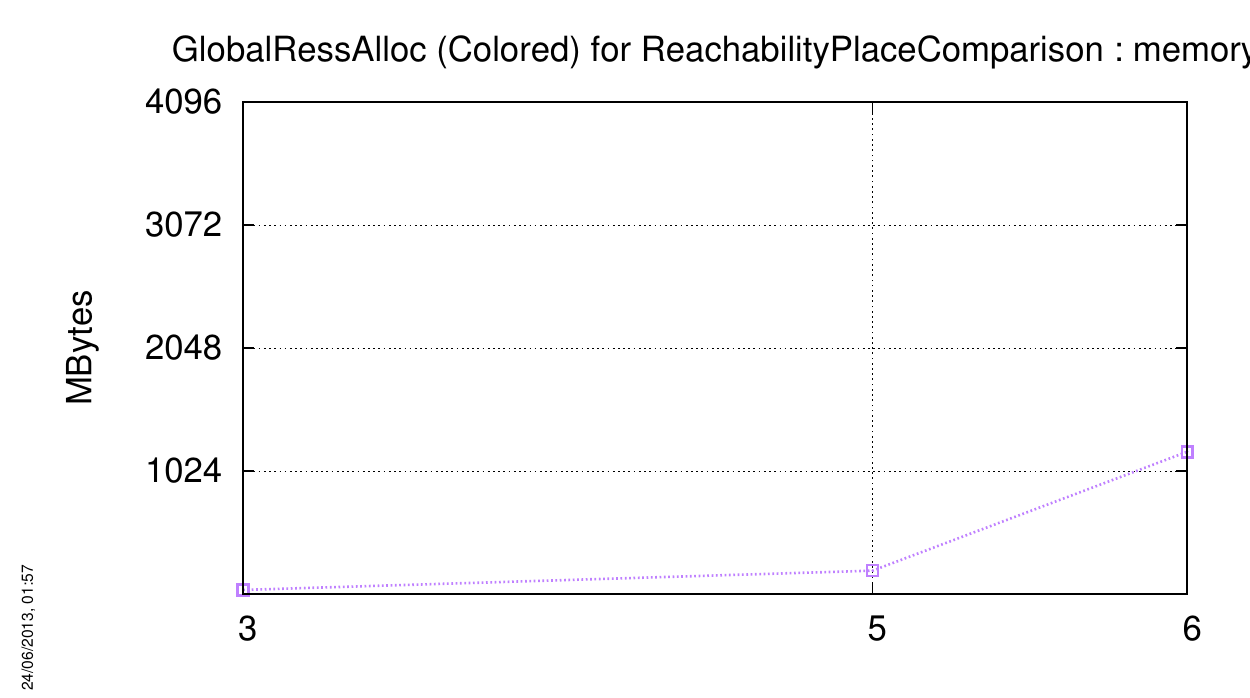}
   \includegraphics[width=7.2cm]{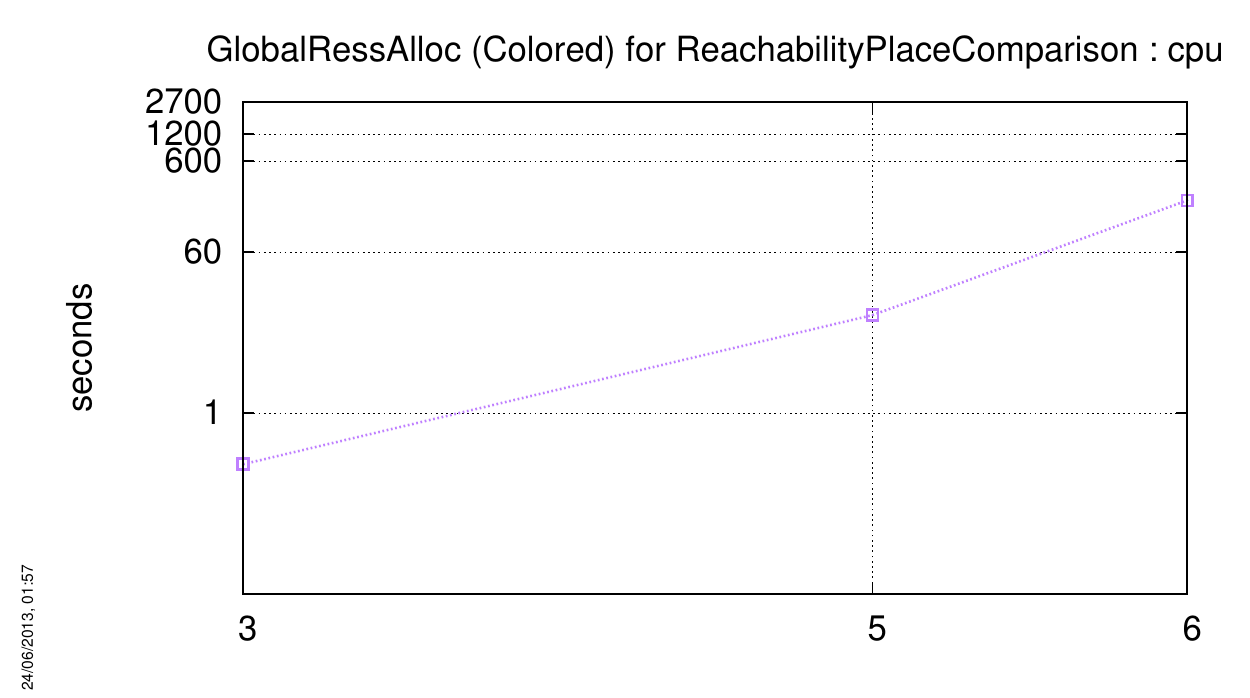}

   \includegraphics[height=1cm]{figures/tools-legend.pdf}
\end{center}

\subsubsection{\acs{GlobalRessAlloc-PT}}
The charts below respectively show how tools compete with this ``Known'' model (memory and CPU).

\index{Performances!ReachabilityPlaceComparison!GlobalRessAlloc (P/T)}
\begin{center}
   \includegraphics[width=7.2cm]{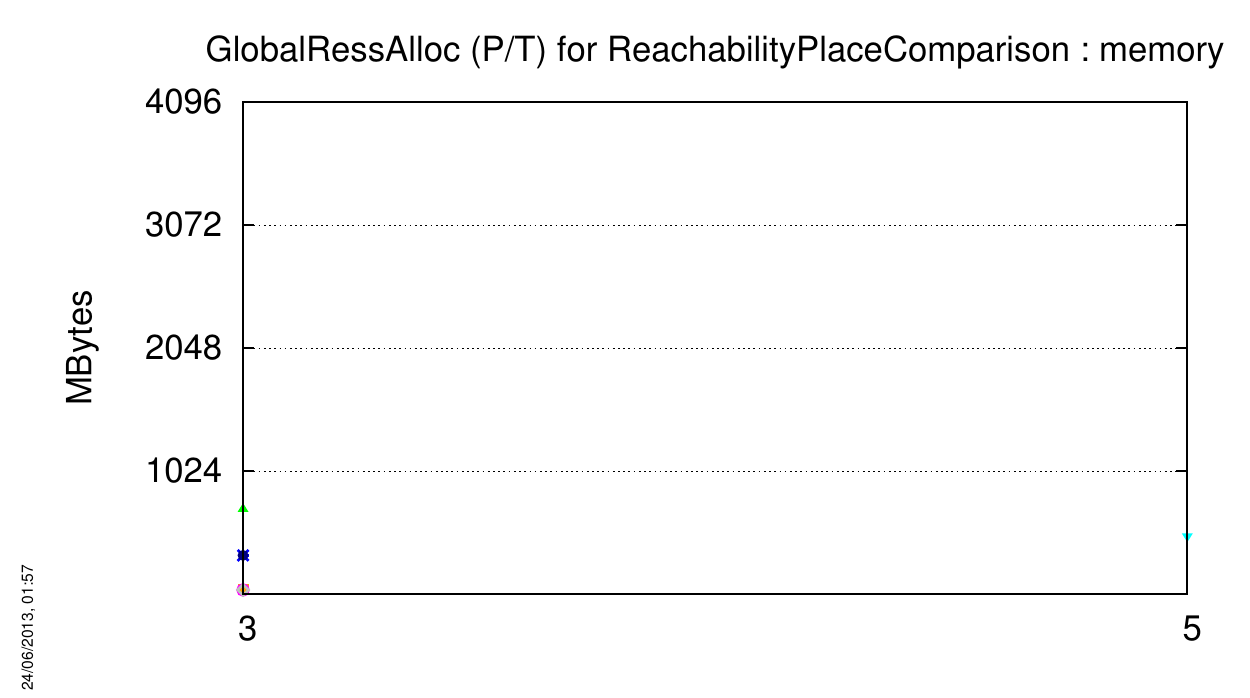}
   \includegraphics[width=7.2cm]{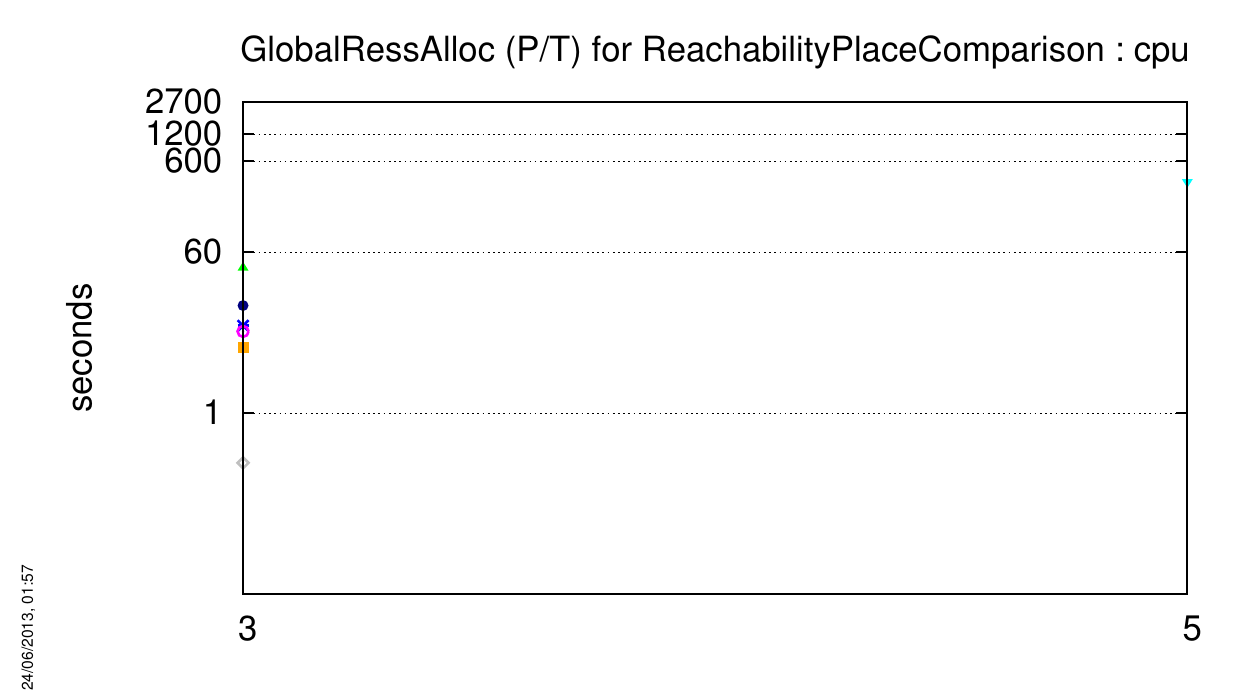}

   \includegraphics[height=1cm]{figures/tools-legend.pdf}
\end{center}

\subsubsection{\acs{Kanban-PT}}
The charts below respectively show how tools compete with this ``Known'' model (memory and CPU).

\index{Performances!ReachabilityPlaceComparison!Kanban (P/T)}
\begin{center}
   \includegraphics[width=7.2cm]{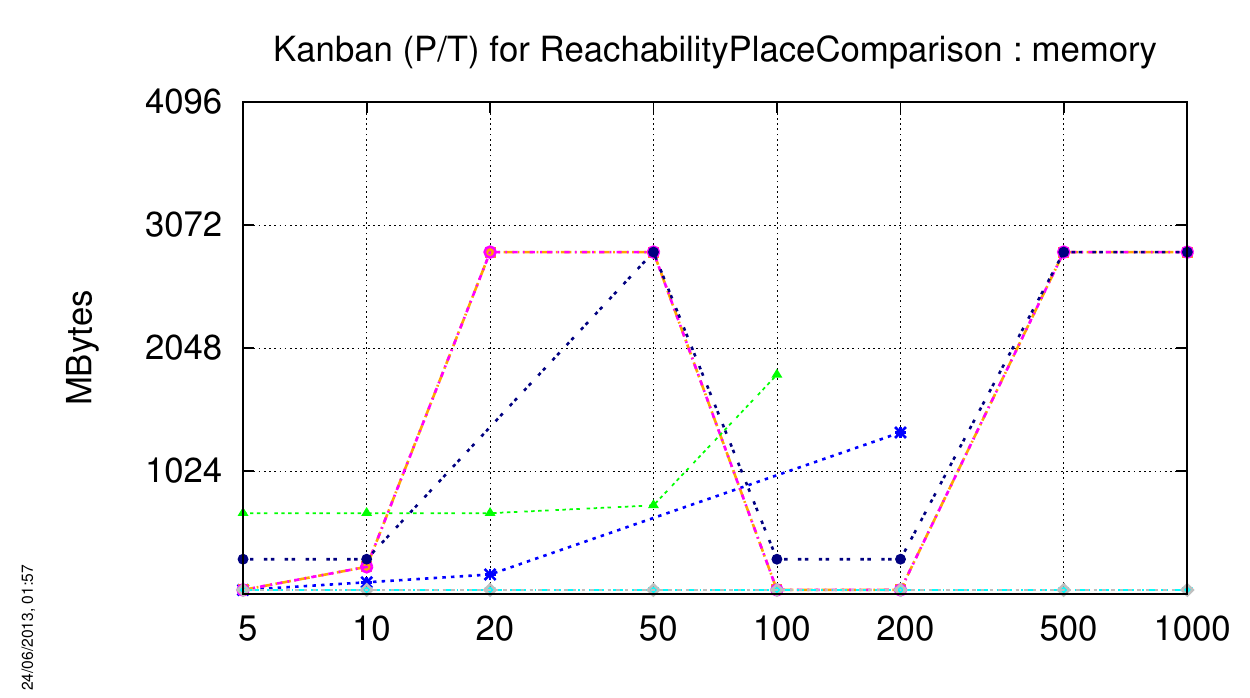}
   \includegraphics[width=7.2cm]{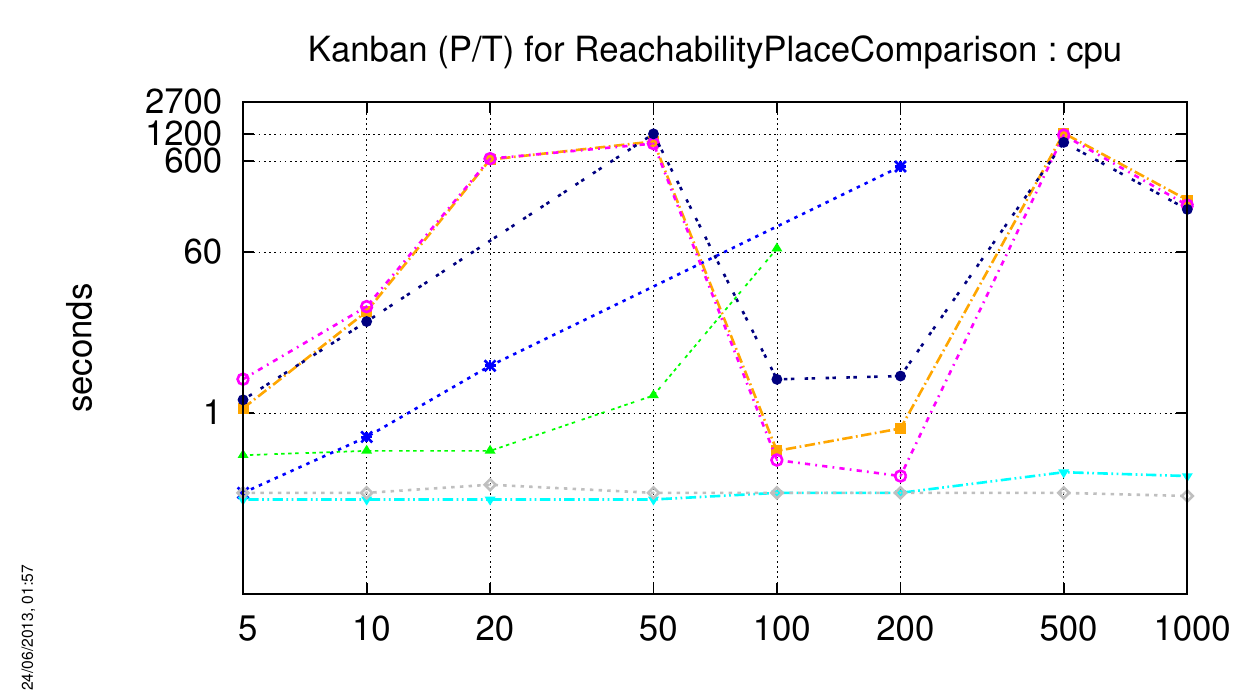}

   \includegraphics[height=1cm]{figures/tools-legend.pdf}
\end{center}

\subsubsection{\acs{LamportFastMutEx-COL}}
The charts below respectively show how tools compete with this ``Known'' model (memory and CPU).

\index{Performances!ReachabilityPlaceComparison!LamportFastMutEx (Colored)}
\begin{center}
   \includegraphics[width=7.2cm]{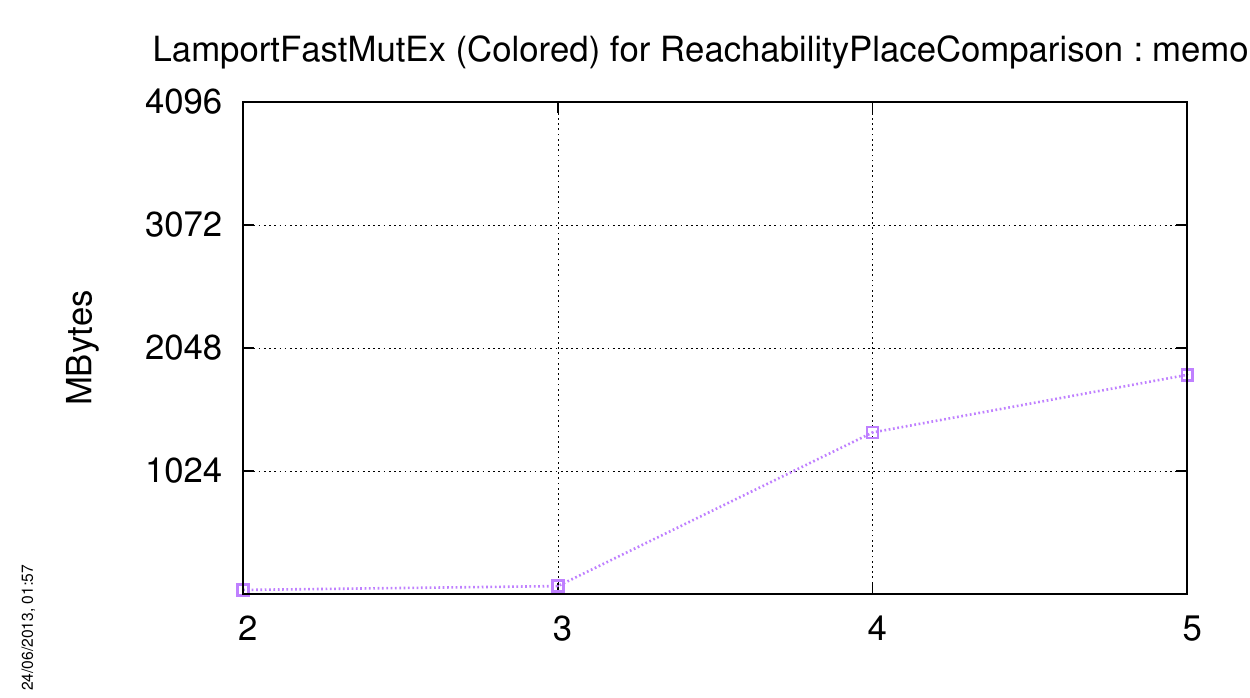}
   \includegraphics[width=7.2cm]{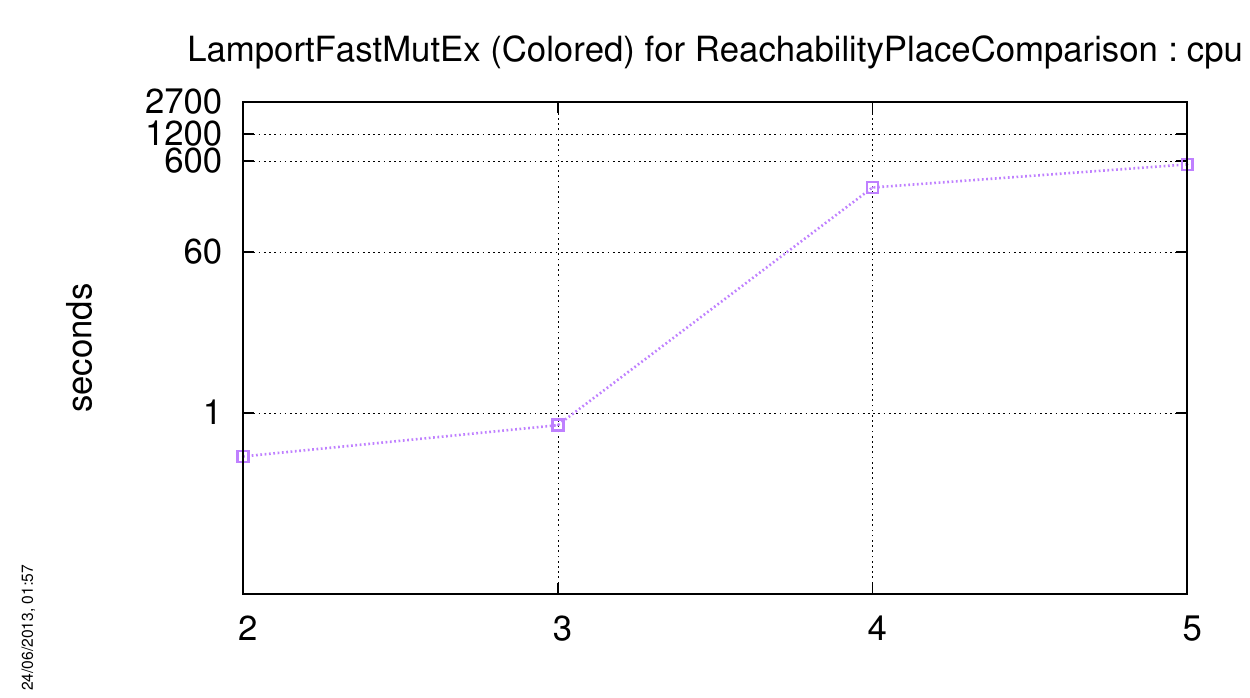}

   \includegraphics[height=1cm]{figures/tools-legend.pdf}
\end{center}

\subsubsection{\acs{LamportFastMutEx-PT}}
The charts below respectively show how tools compete with this ``Known'' model (memory and CPU).

\index{Performances!ReachabilityPlaceComparison!LamportFastMutEx (P/T)}
\begin{center}
   \includegraphics[width=7.2cm]{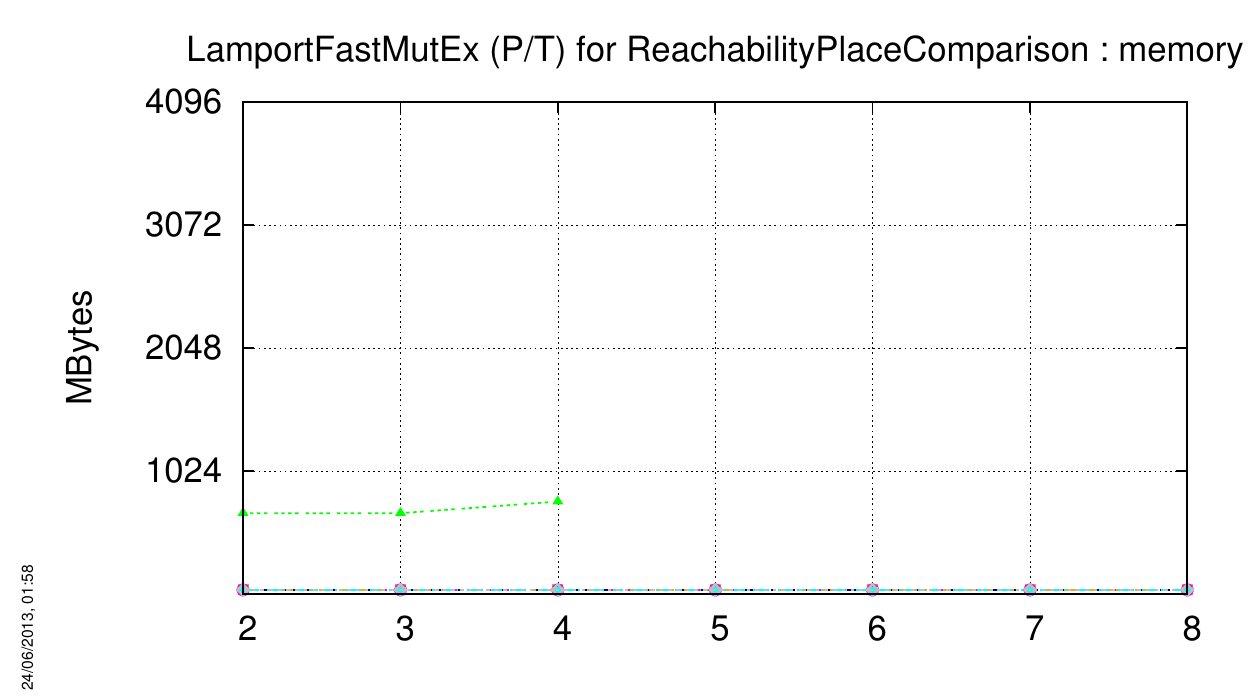}
   \includegraphics[width=7.2cm]{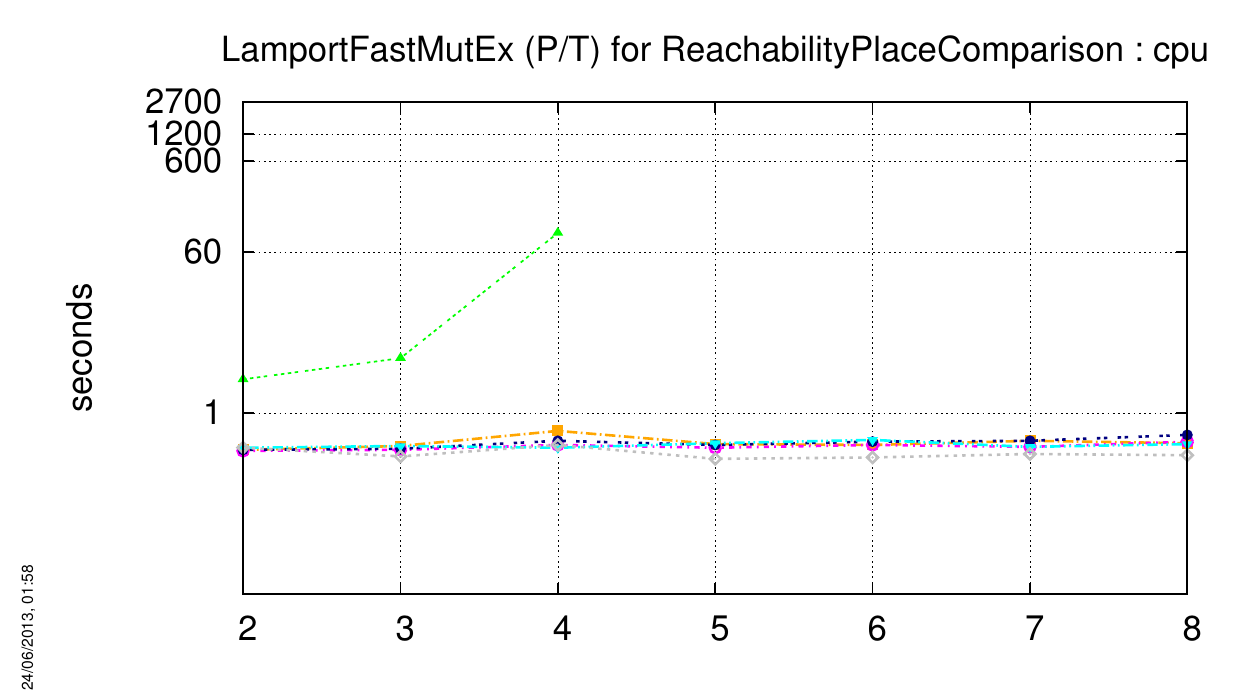}

   \includegraphics[height=1cm]{figures/tools-legend.pdf}
\end{center}

\subsubsection{\acs{MAPK-PT}}
The charts below respectively show how tools compete with this ``Known'' model (memory and CPU).

\index{Performances!ReachabilityPlaceComparison!MAPK (P/T)}
\begin{center}
   \includegraphics[width=7.2cm]{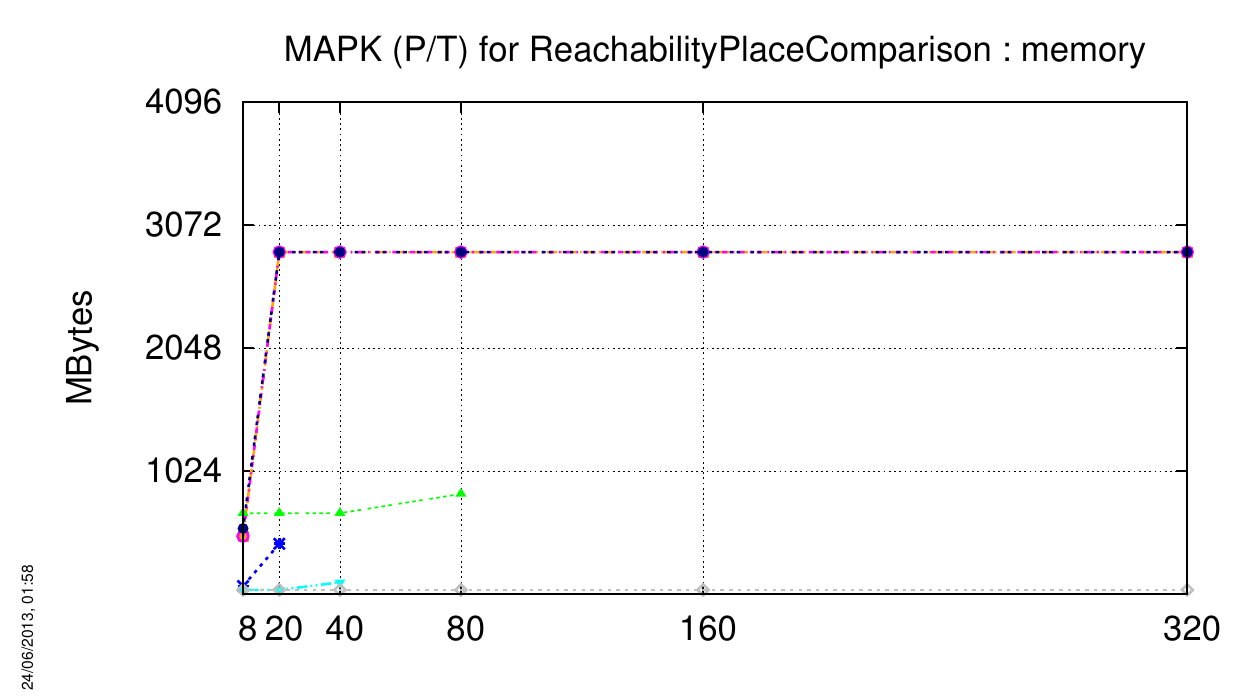}
   \includegraphics[width=7.2cm]{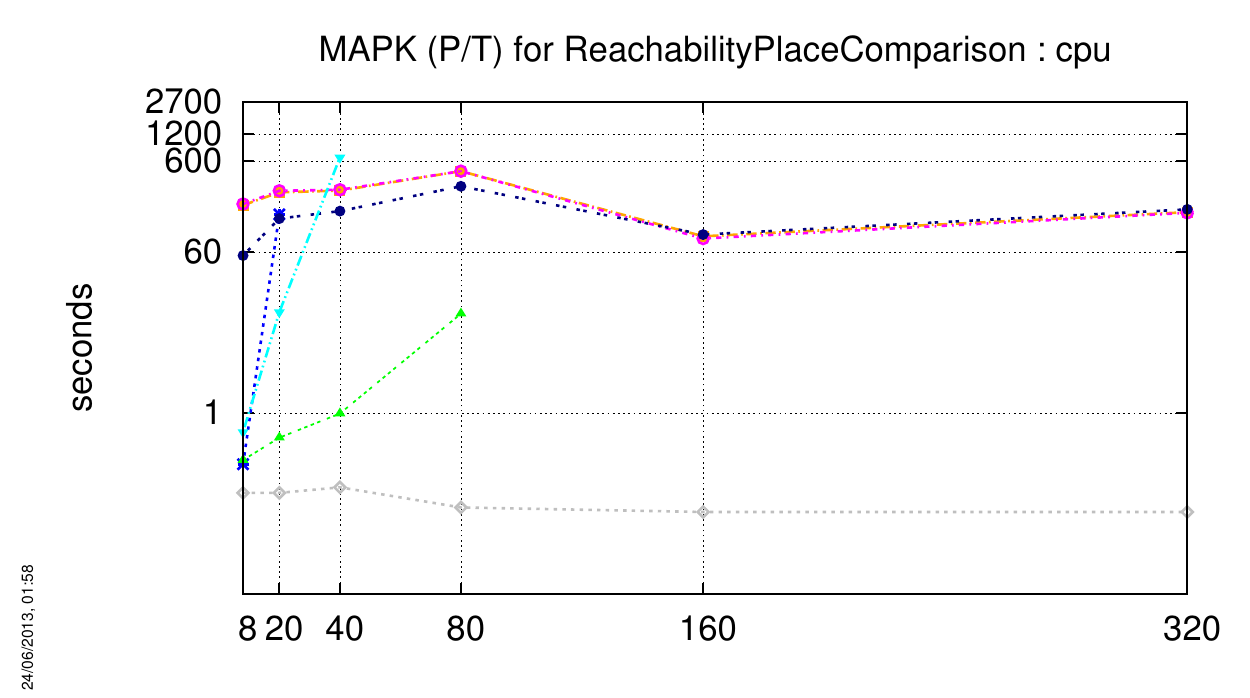}

   \includegraphics[height=1cm]{figures/tools-legend.pdf}
\end{center}

\subsubsection{\acs{NeoElection-COL}}
The charts below respectively show how tools compete with this ``Known'' model (memory and CPU).

\index{Performances!ReachabilityPlaceComparison!NeoElection (Colored)}
\begin{center}
   \includegraphics[width=7.2cm]{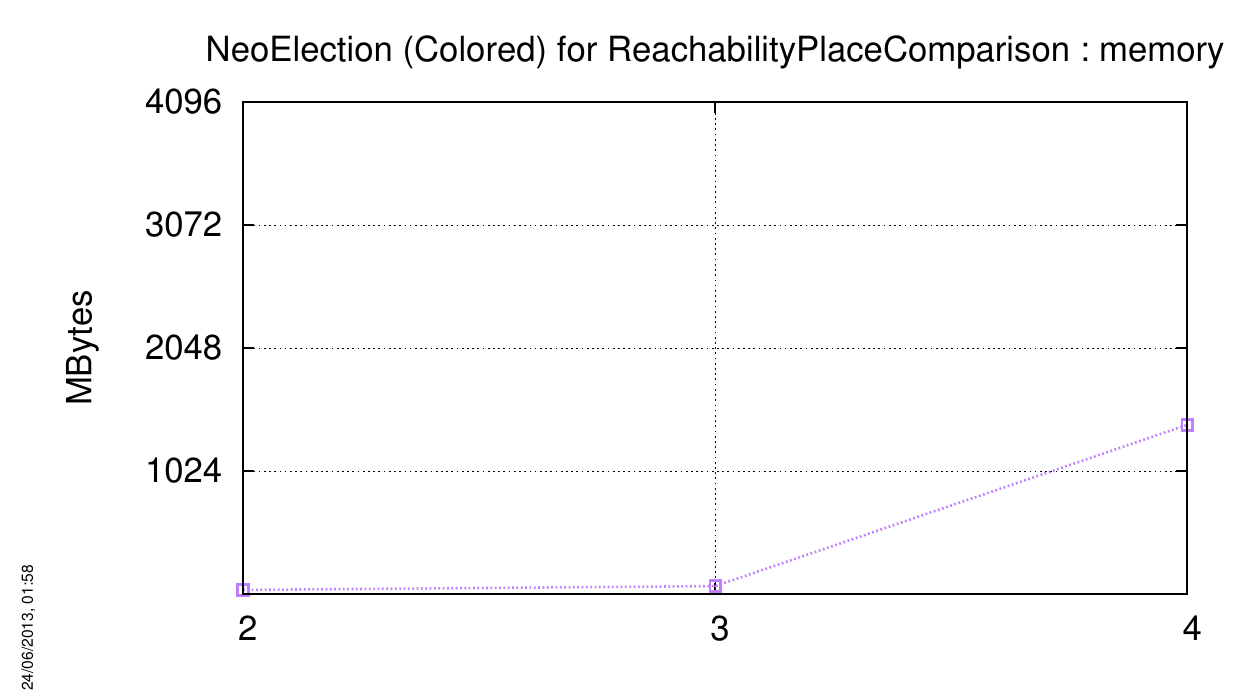}
   \includegraphics[width=7.2cm]{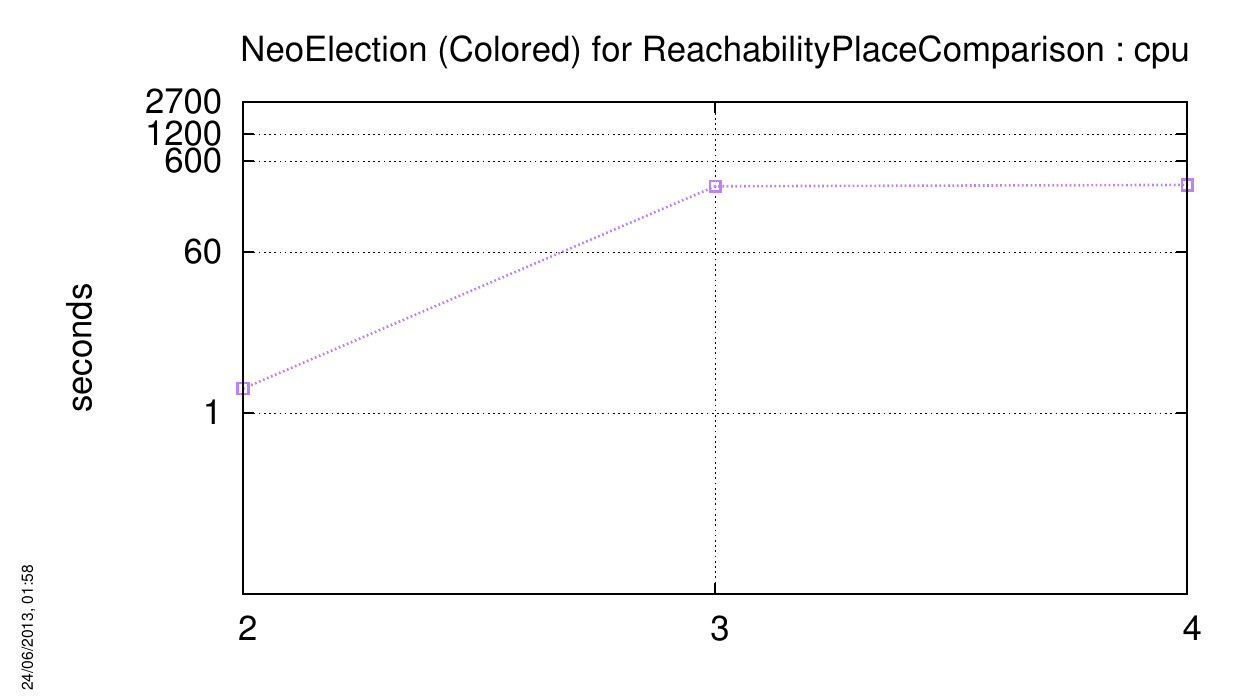}

   \includegraphics[height=1cm]{figures/tools-legend.pdf}
\end{center}

\subsubsection{\acs{NeoElection-PT}}
The charts below respectively show how tools compete with this ``Known'' model (memory and CPU).

\index{Performances!ReachabilityPlaceComparison!NeoElection (P/T)}
\begin{center}
   \includegraphics[width=7.2cm]{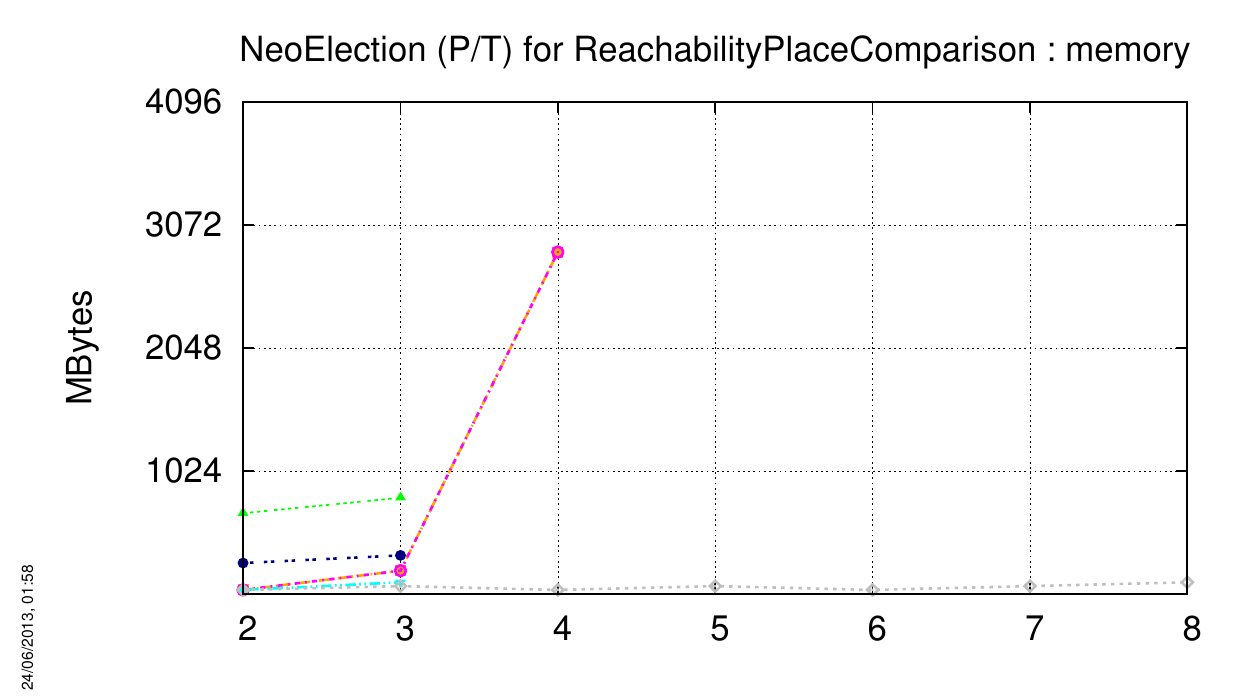}
   \includegraphics[width=7.2cm]{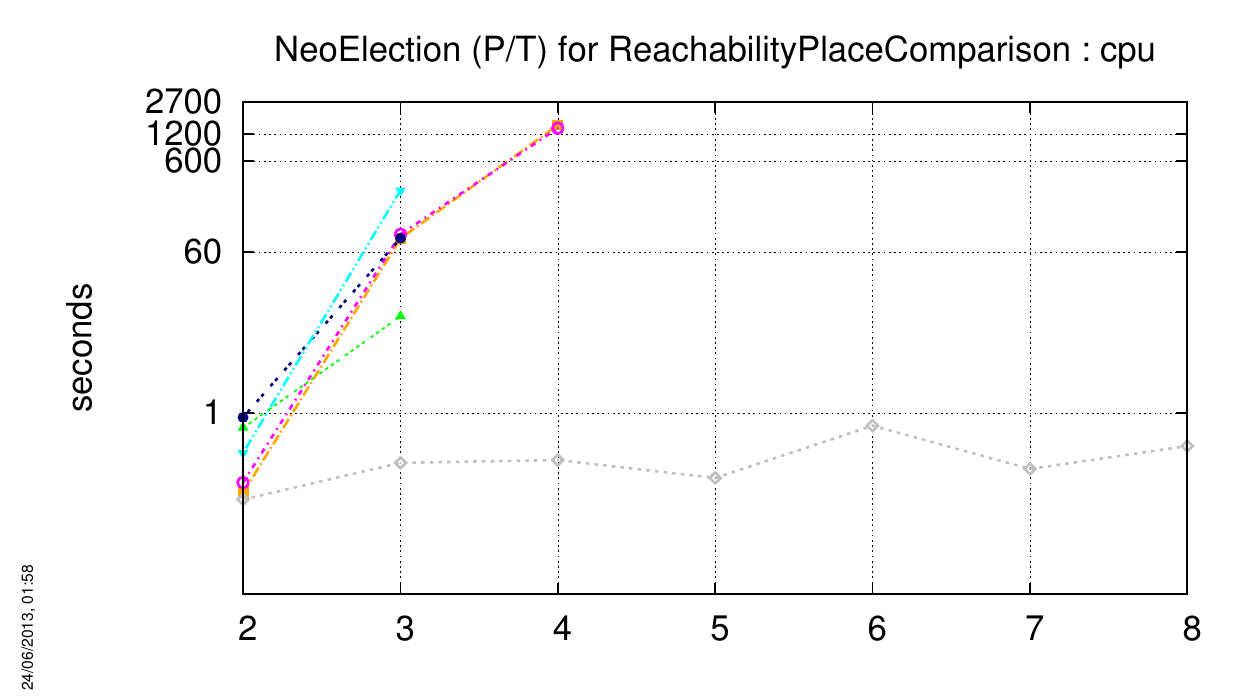}

   \includegraphics[height=1cm]{figures/tools-legend.pdf}
\end{center}

\subsubsection{\acs{PermAdmissibility-COL}}
The charts below respectively show how tools compete with this ``Known'' model (memory and CPU).

\index{Performances!ReachabilityPlaceComparison!PermAdmissibility (Colored)}
\begin{center}
   \includegraphics[width=7.2cm]{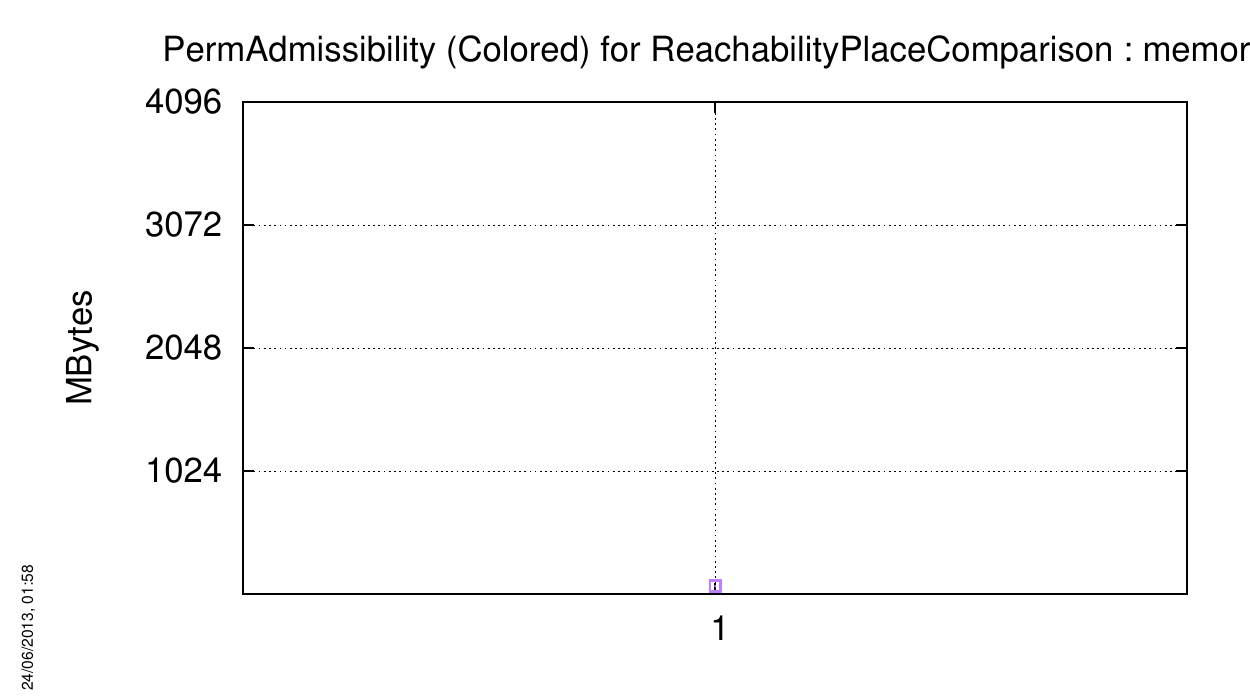}
   \includegraphics[width=7.2cm]{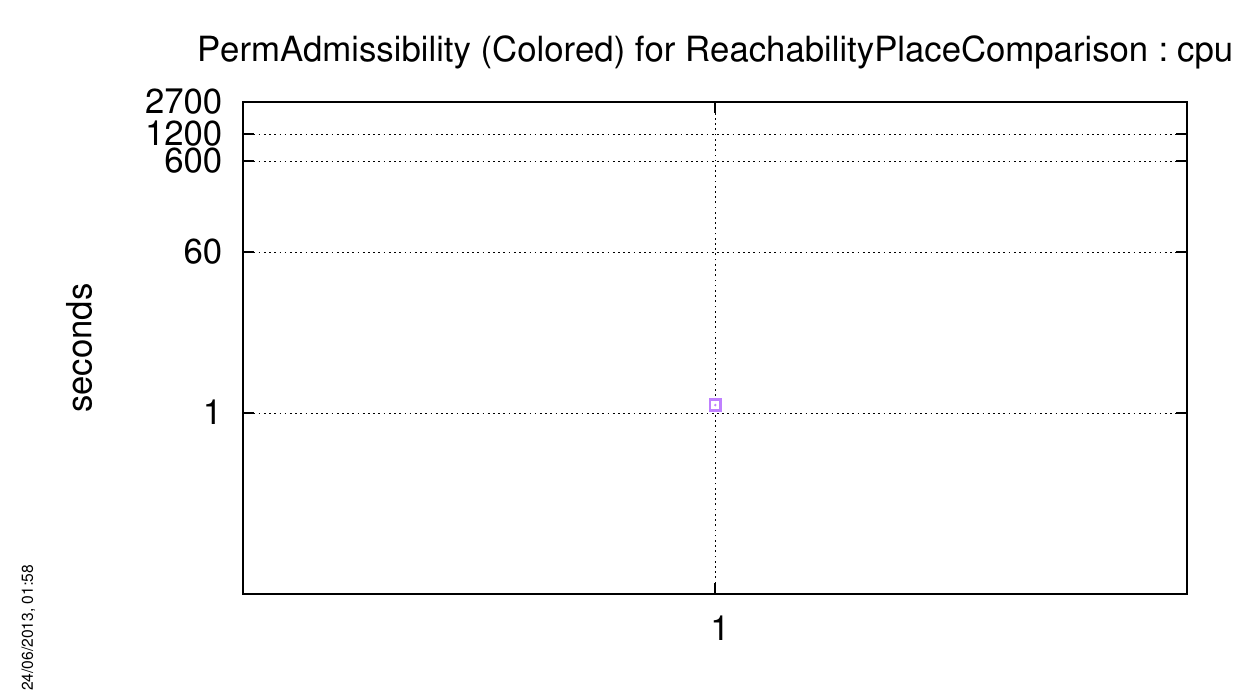}

   \includegraphics[height=1cm]{figures/tools-legend.pdf}
\end{center}

\subsubsection{\acs{PermAdmissibility-PT}}
The charts below respectively show how tools compete with this ``Known'' model (memory and CPU).

\index{Performances!ReachabilityPlaceComparison!PermAdmissibility (P/T)}
\begin{center}
   \includegraphics[width=7.2cm]{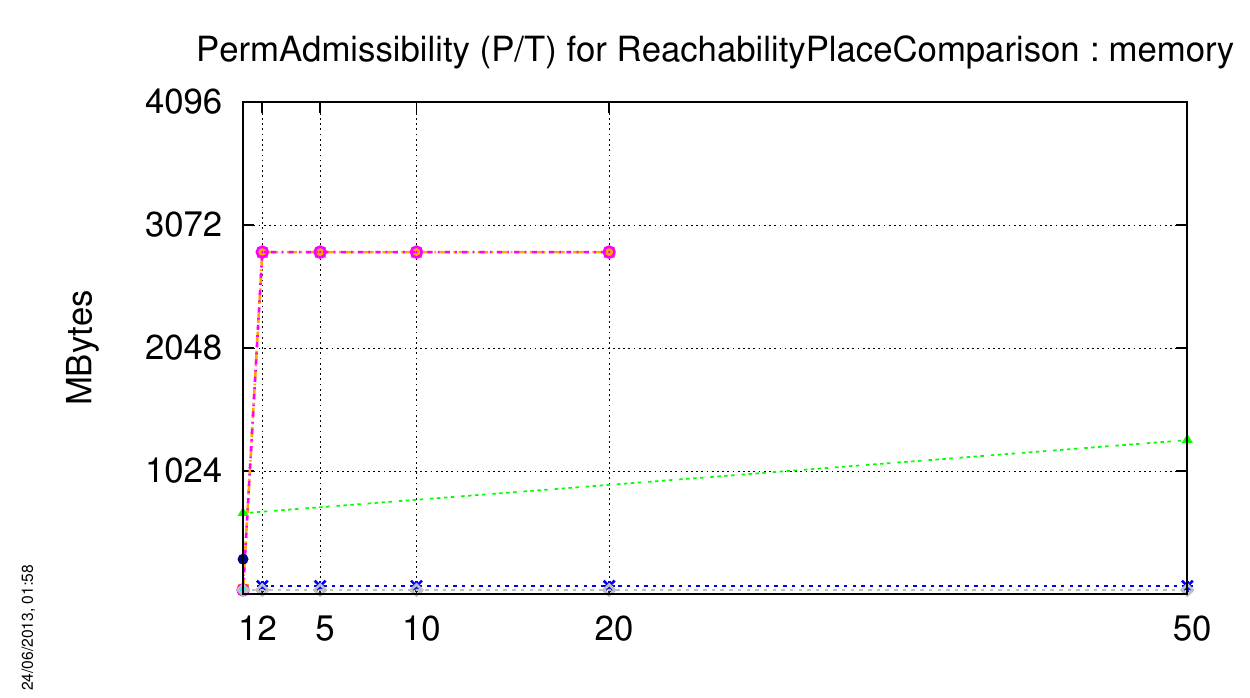}
   \includegraphics[width=7.2cm]{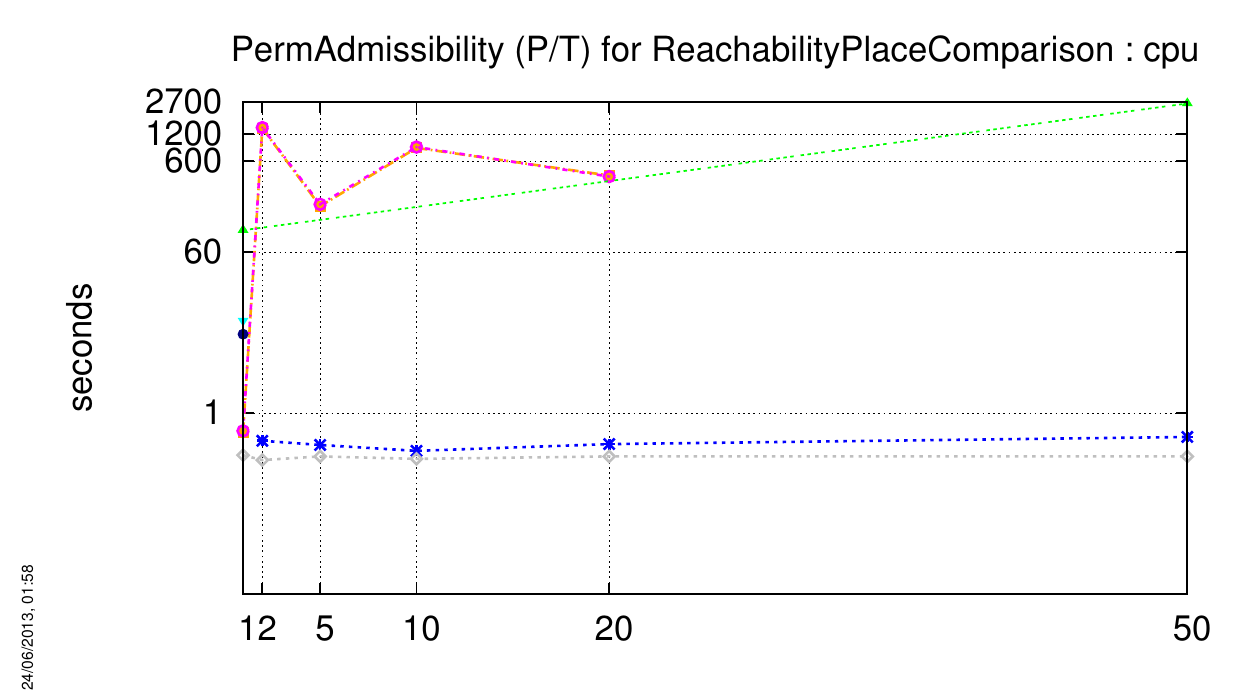}

   \includegraphics[height=1cm]{figures/tools-legend.pdf}
\end{center}

\subsubsection{\acs{Peterson-COL}}
The charts below respectively show how tools compete with this ``Known'' model (memory and CPU).

\index{Performances!ReachabilityPlaceComparison!Peterson (Colored)}
\begin{center}
   \includegraphics[width=7.2cm]{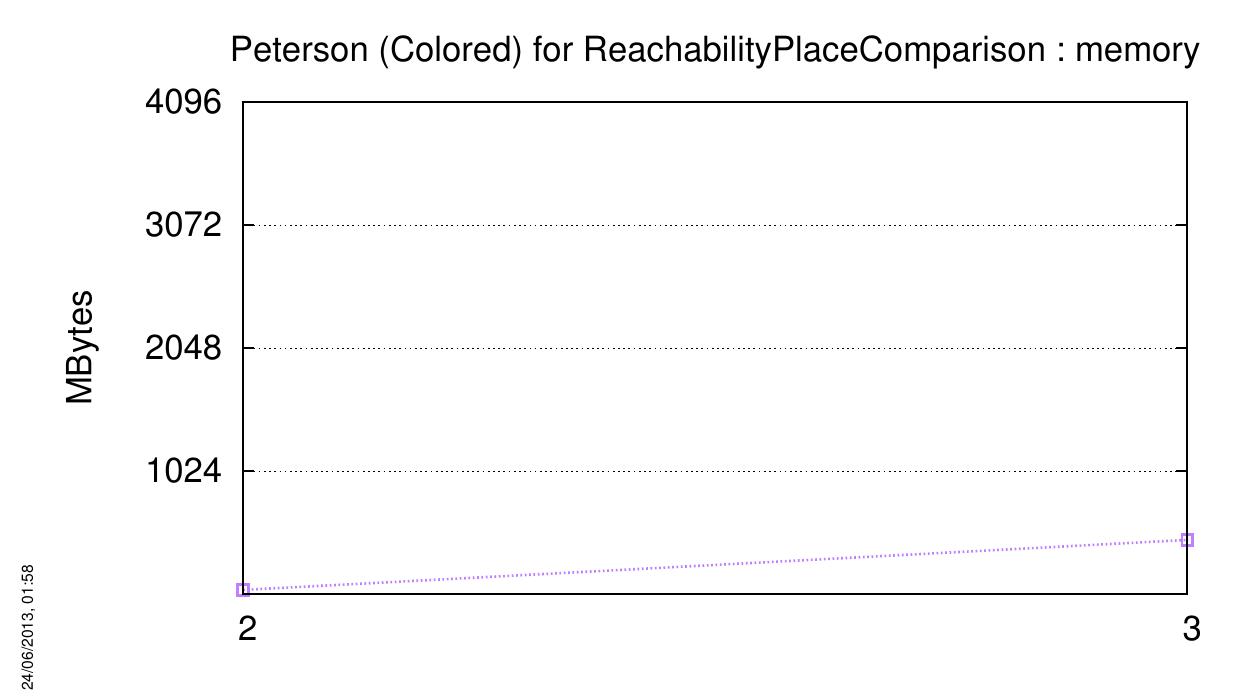}
   \includegraphics[width=7.2cm]{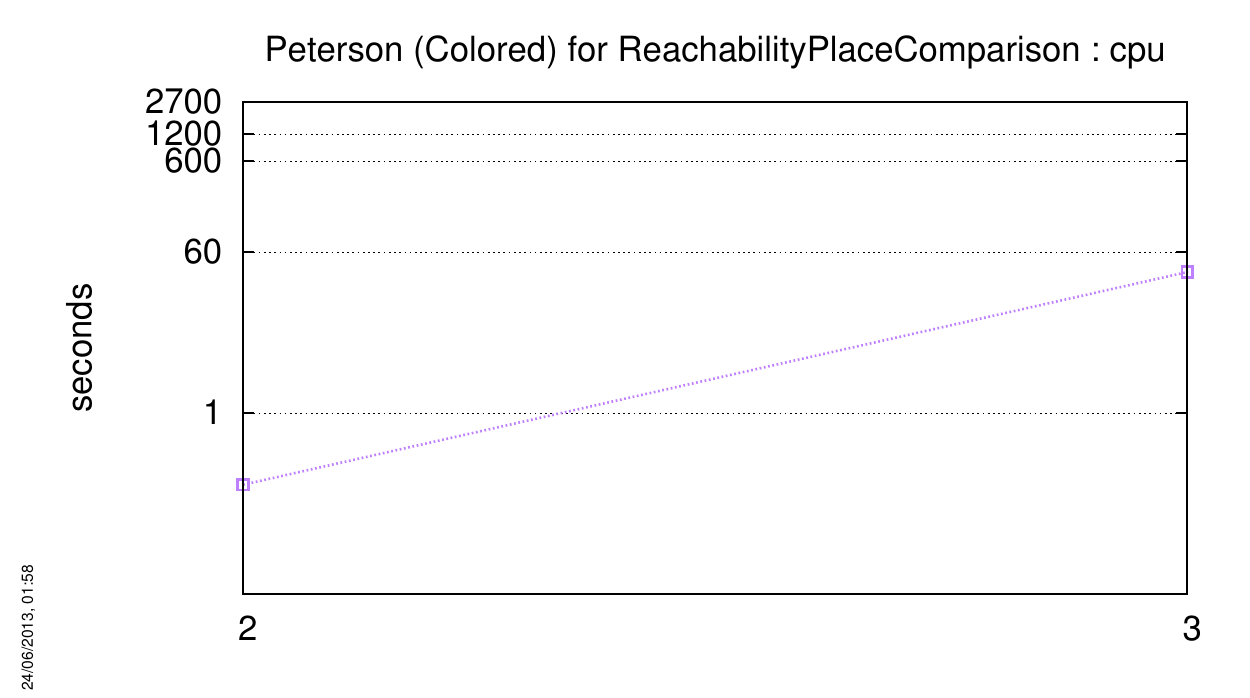}

   \includegraphics[height=1cm]{figures/tools-legend.pdf}
\end{center}

\subsubsection{\acs{Peterson-PT}}
The charts below respectively show how tools compete with this ``Known'' model (memory and CPU).

\index{Performances!ReachabilityPlaceComparison!Peterson (P/T)}
\begin{center}
   \includegraphics[width=7.2cm]{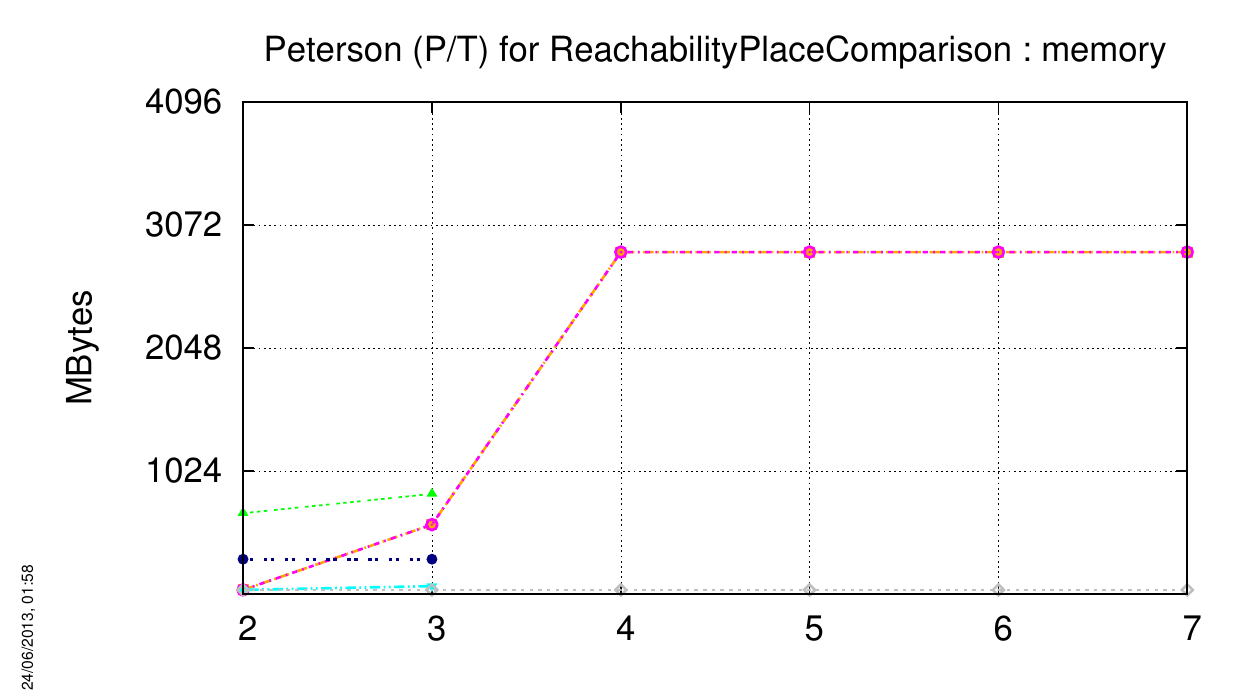}
   \includegraphics[width=7.2cm]{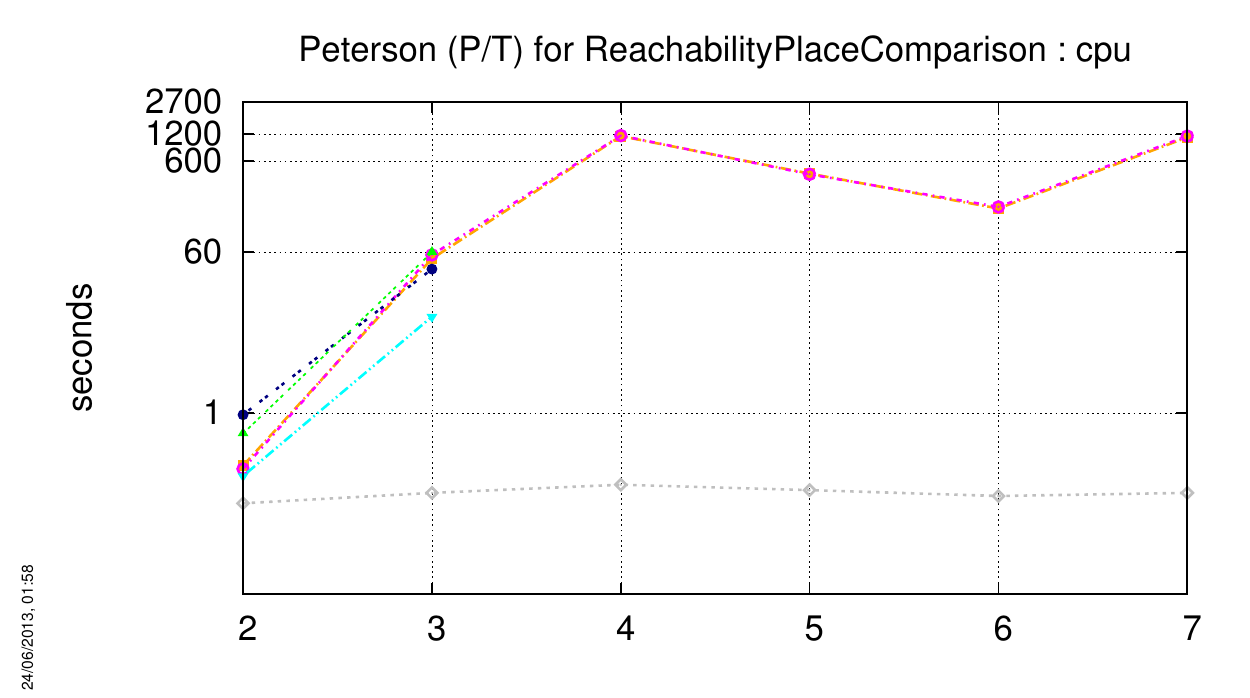}

   \includegraphics[height=1cm]{figures/tools-legend.pdf}
\end{center}

\subsubsection{\acs{Philosophers-COL}}
No instance of this model could be computed for the \textbf{ReachabilityPlaceComparison} examination.

\subsubsection{\acs{Philosophers-PT}}
The charts below respectively show how tools compete with this ``Known'' model (memory and CPU).

\index{Performances!ReachabilityPlaceComparison!Philosophers (P/T)}
\begin{center}
   \includegraphics[width=7.2cm]{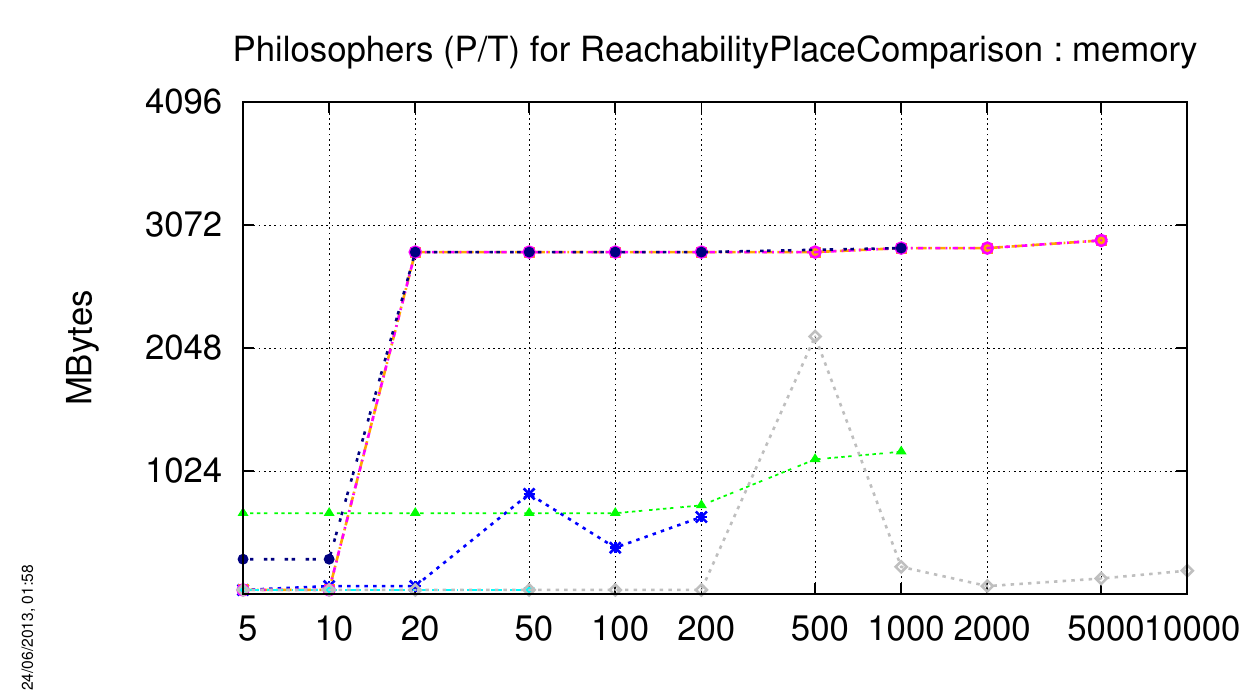}
   \includegraphics[width=7.2cm]{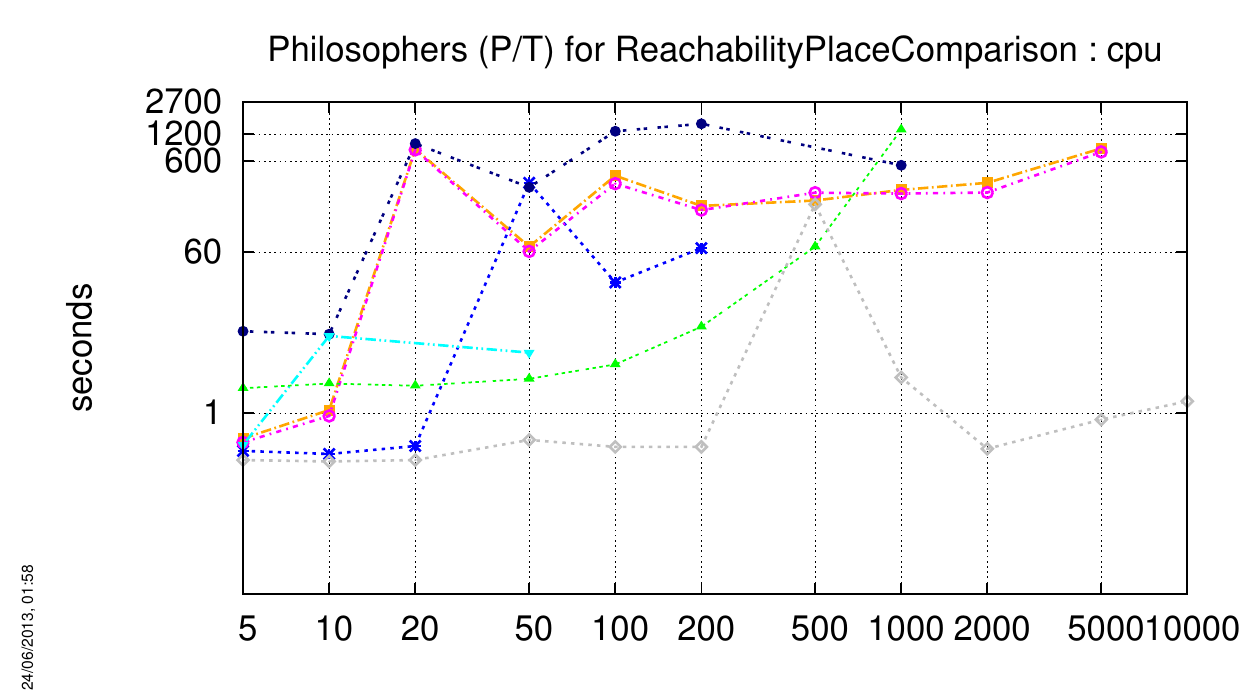}

   \includegraphics[height=1cm]{figures/tools-legend.pdf}
\end{center}

\subsubsection{\acs{PhilosophersDyn-COL}}
No instance of this model could be computed for the \textbf{ReachabilityPlaceComparison} examination.

\subsubsection{\acs{PhilosophersDyn-PT}}
The charts below respectively show how tools compete with this ``Known'' model (memory and CPU).

\index{Performances!ReachabilityPlaceComparison!PhilosophersDyn (P/T)}
\begin{center}
   \includegraphics[width=7.2cm]{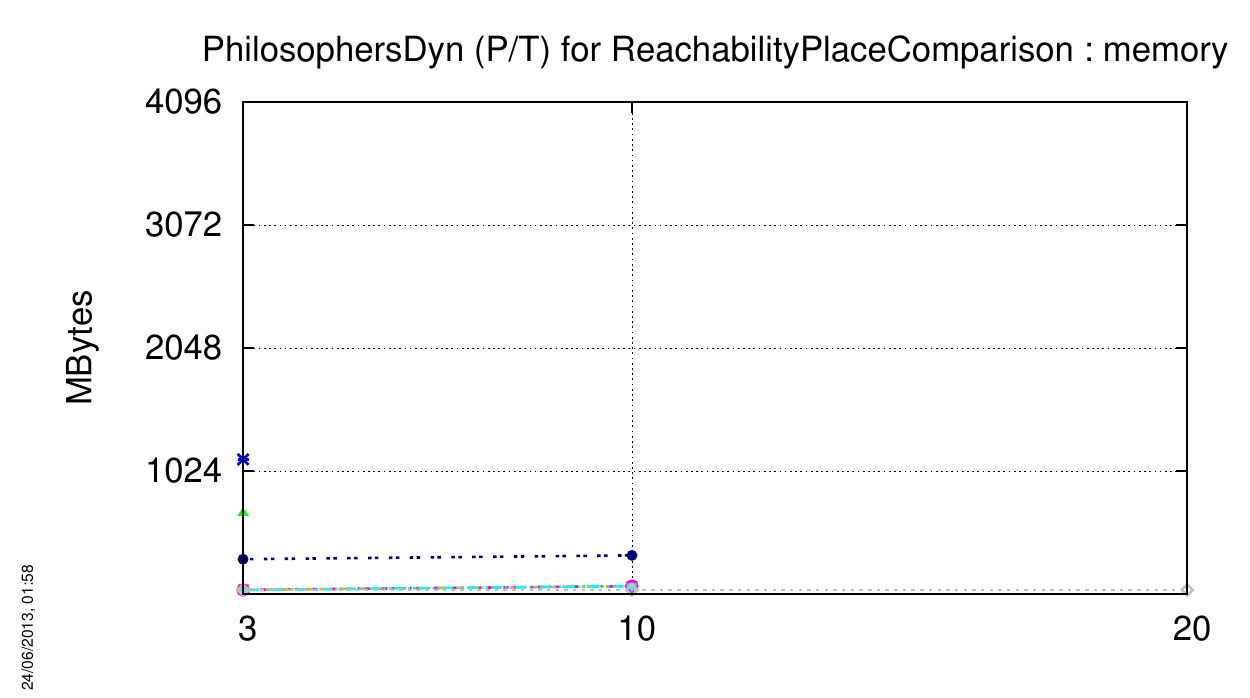}
   \includegraphics[width=7.2cm]{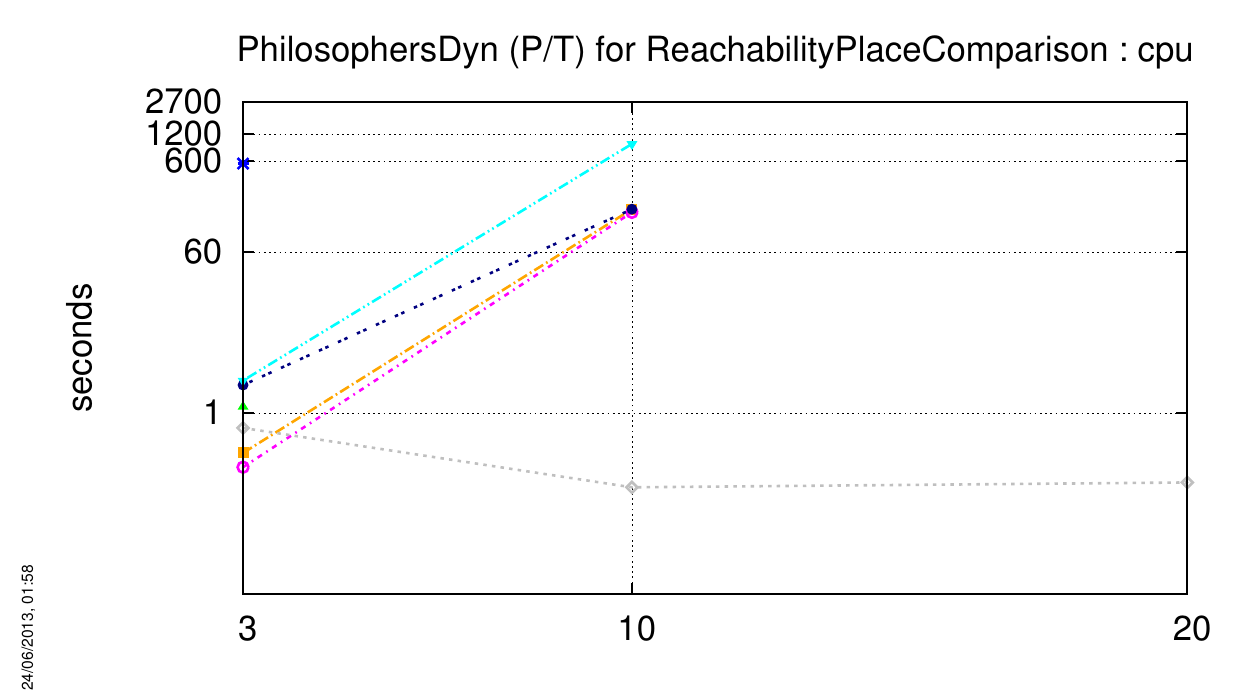}

   \includegraphics[height=1cm]{figures/tools-legend.pdf}
\end{center}

\subsubsection{\acs{Planning-PT}}
No instance of this model could be computed for the \textbf{ReachabilityPlaceComparison} examination.

\subsubsection{\acs{Railroad-PT}}
The charts below respectively show how tools compete with this ``Known'' model (memory and CPU).

\index{Performances!ReachabilityPlaceComparison!Railroad (P/T)}
\begin{center}
   \includegraphics[width=7.2cm]{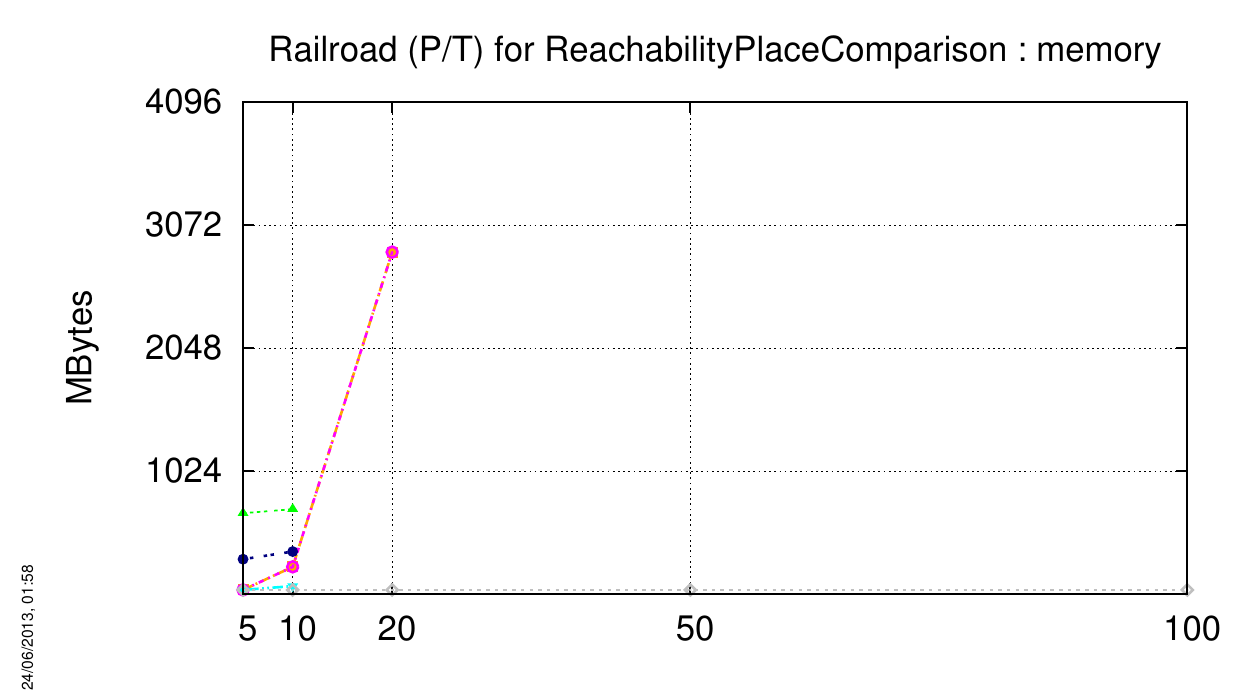}
   \includegraphics[width=7.2cm]{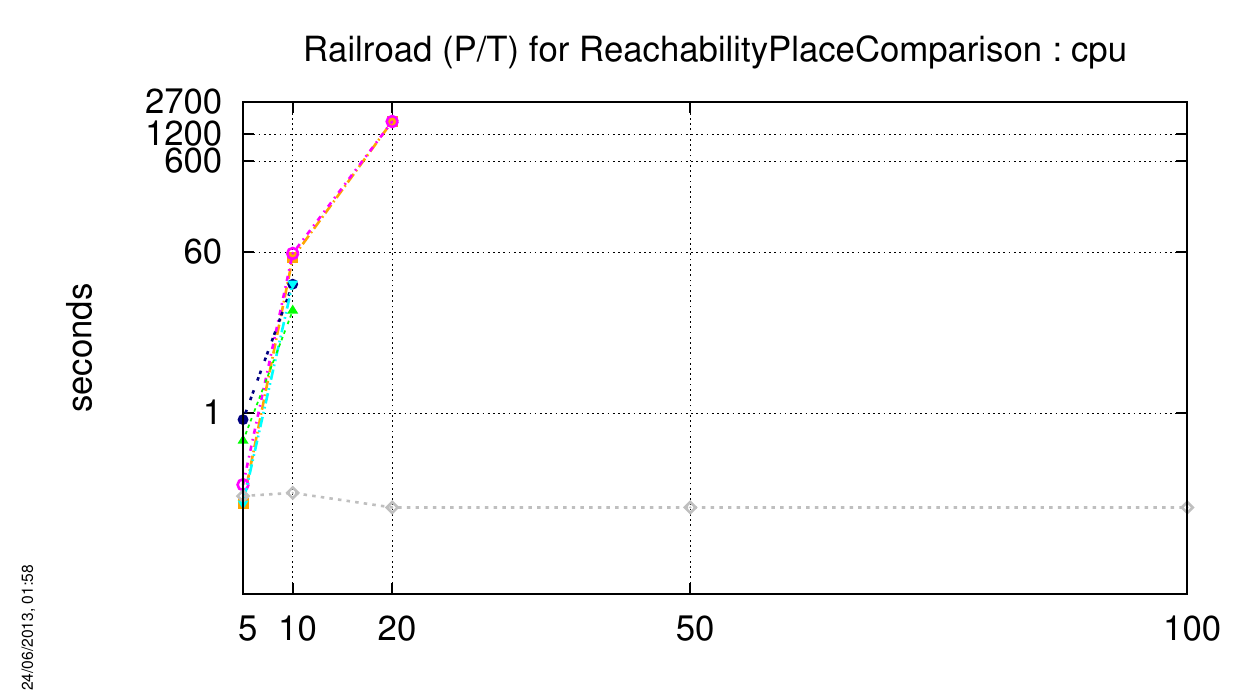}

   \includegraphics[height=1cm]{figures/tools-legend.pdf}
\end{center}

\subsubsection{\acs{RessAllocation-PT}}
The charts below respectively show how tools compete with this ``Known'' model (memory and CPU).

\index{Performances!ReachabilityPlaceComparison!RessAllocation (P/T)}
\begin{center}
   \includegraphics[width=7.2cm]{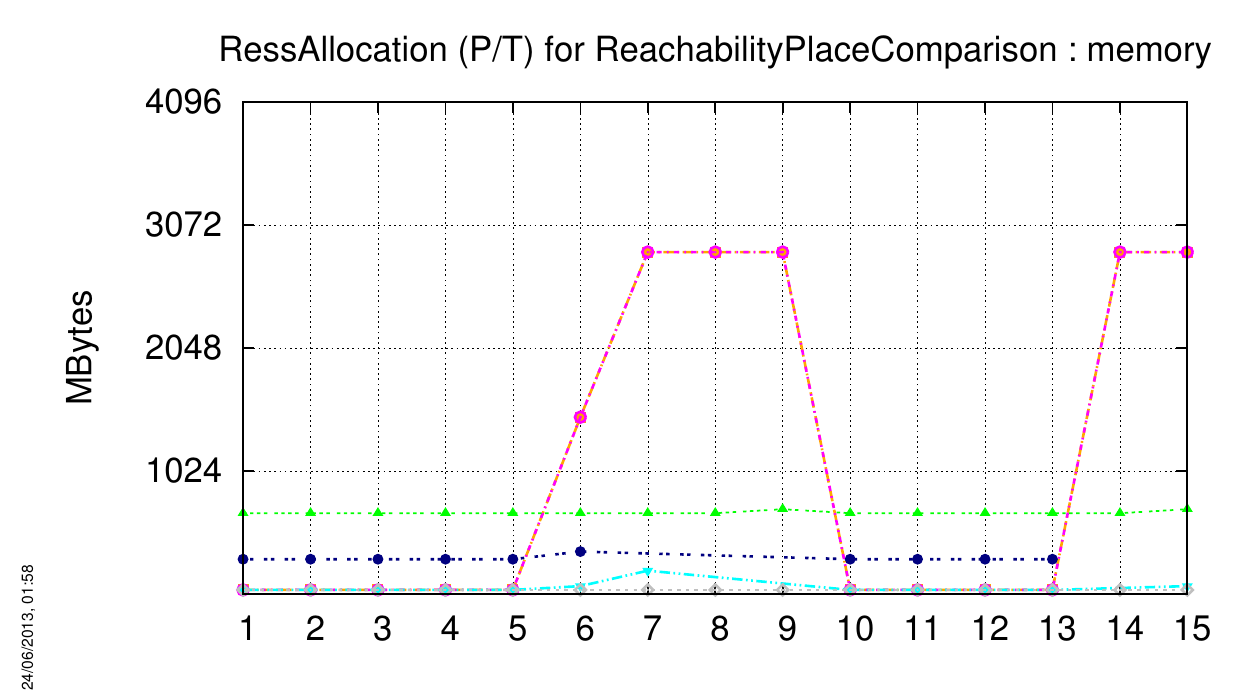}
   \includegraphics[width=7.2cm]{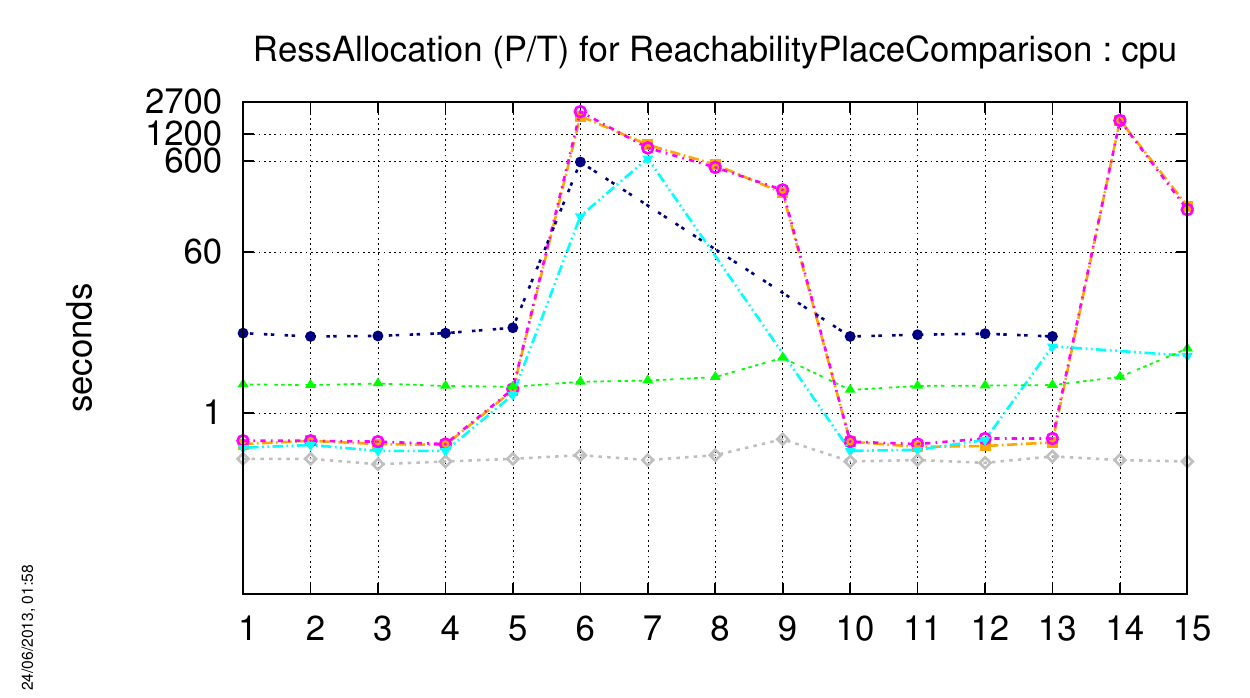}

   \includegraphics[height=1cm]{figures/tools-legend.pdf}
\end{center}

\subsubsection{\acs{Ring-PT}}
The charts below respectively show how tools compete with this ``Known'' model (memory and CPU).

\index{Performances!ReachabilityPlaceComparison!Ring (P/T)}
\begin{center}
   \includegraphics[width=7.2cm]{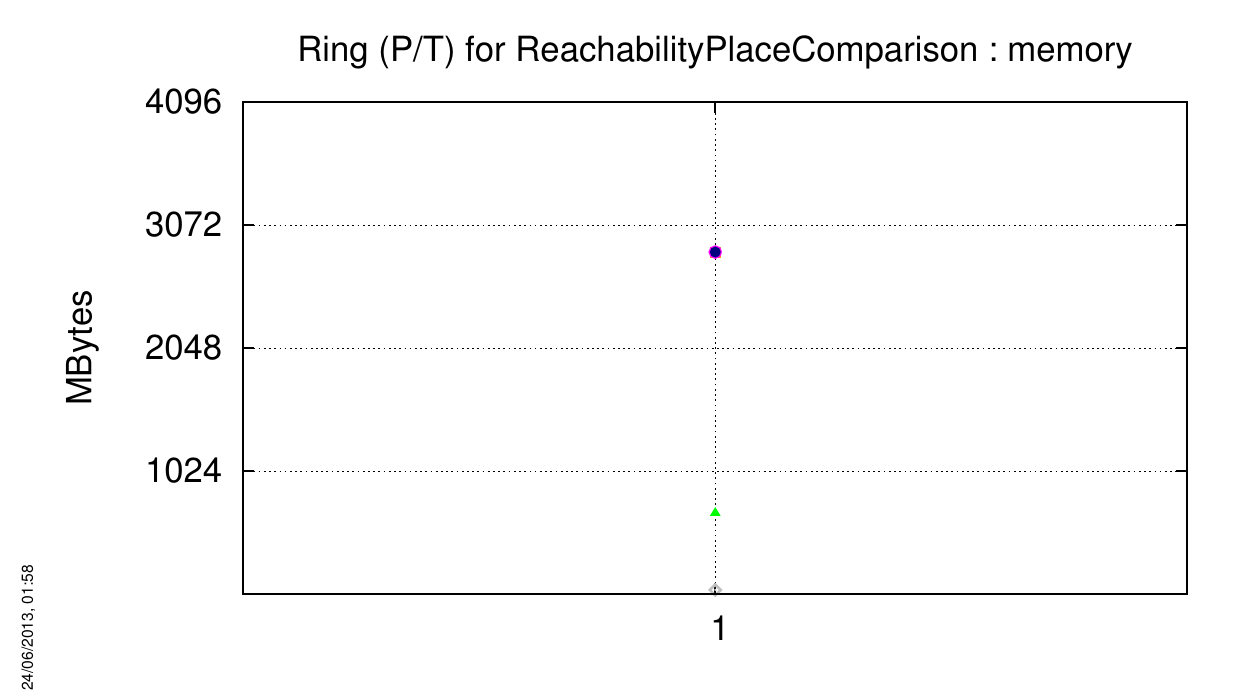}
   \includegraphics[width=7.2cm]{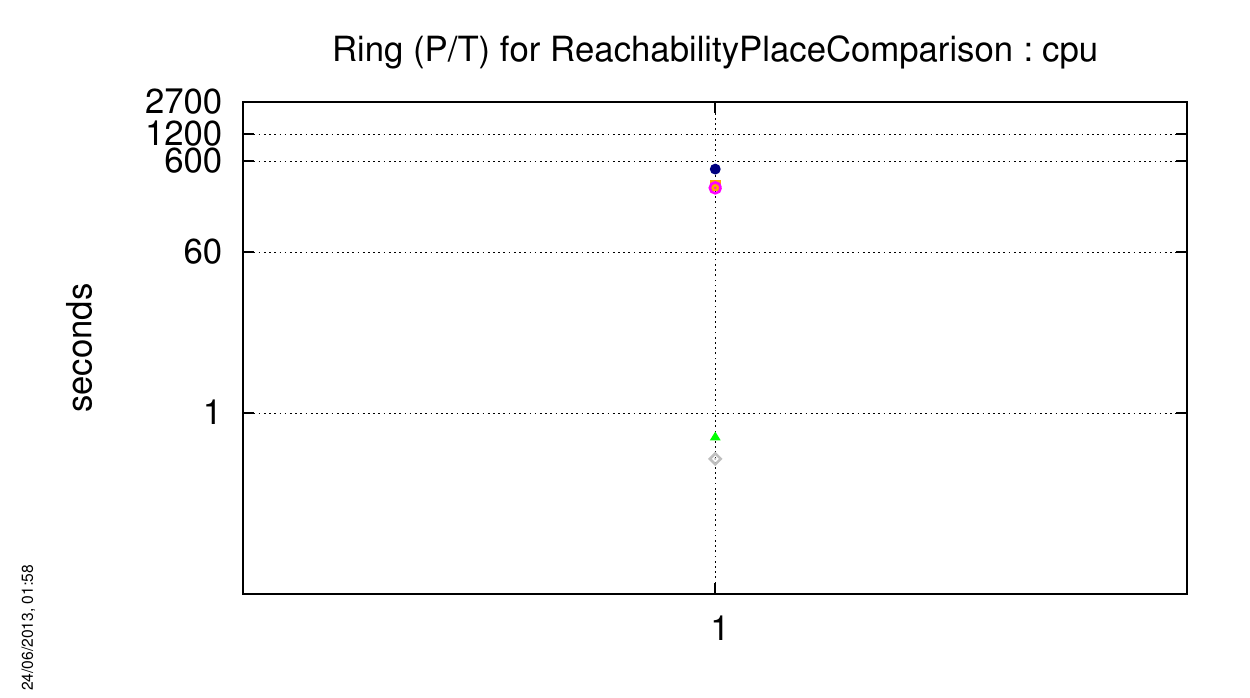}

   \includegraphics[height=1cm]{figures/tools-legend.pdf}
\end{center}

\subsubsection{\acs{RwMutex-PT}}
The charts below respectively show how tools compete with this ``Known'' model (memory and CPU).

\index{Performances!ReachabilityPlaceComparison!RwMutex (P/T)}
\begin{center}
   \includegraphics[width=7.2cm]{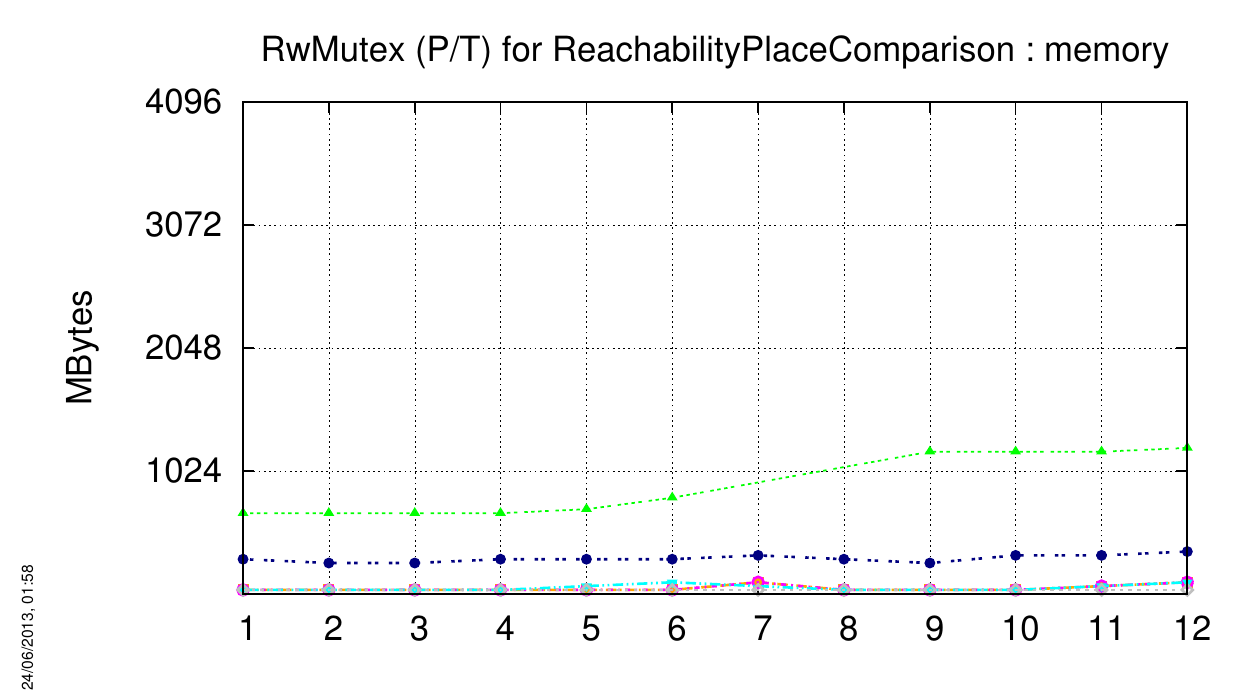}
   \includegraphics[width=7.2cm]{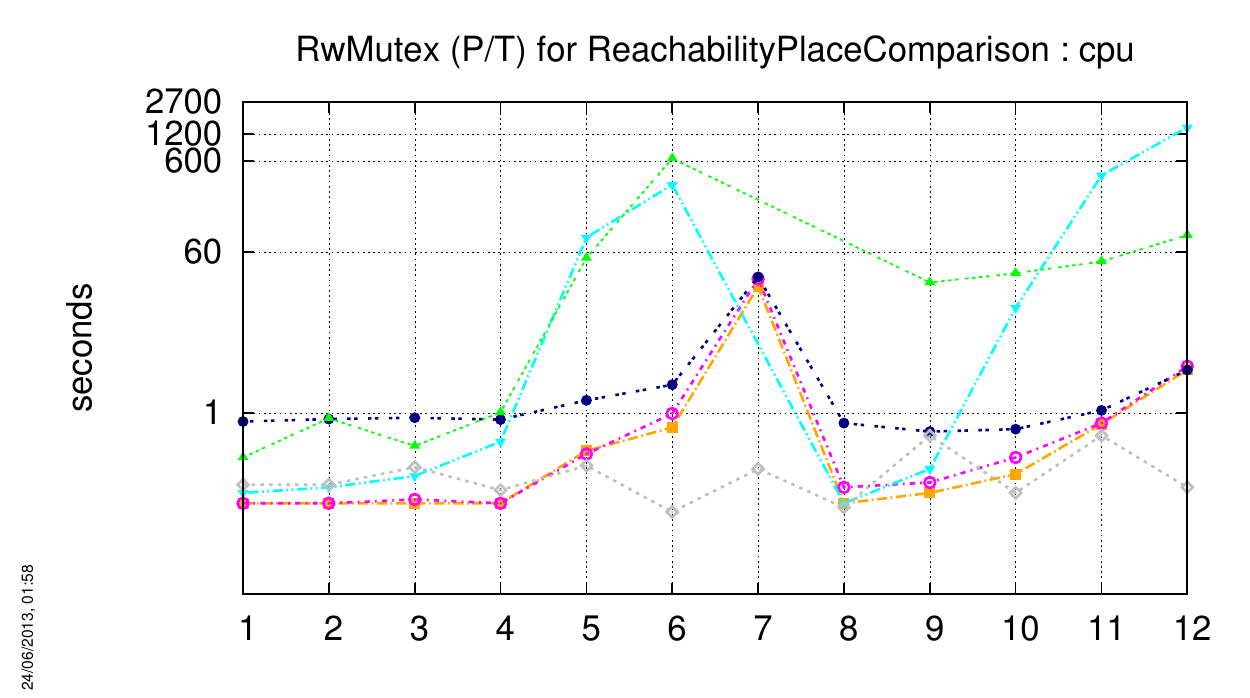}

   \includegraphics[height=1cm]{figures/tools-legend.pdf}
\end{center}

\subsubsection{\acs{SharedMemory-COL}}
No instance of this model could be computed for the \textbf{ReachabilityPlaceComparison} examination.

\subsubsection{\acs{SharedMemory-PT}}
The charts below respectively show how tools compete with this ``Known'' model (memory and CPU).

\index{Performances!ReachabilityPlaceComparison!SharedMemory (P/T)}
\begin{center}
   \includegraphics[width=7.2cm]{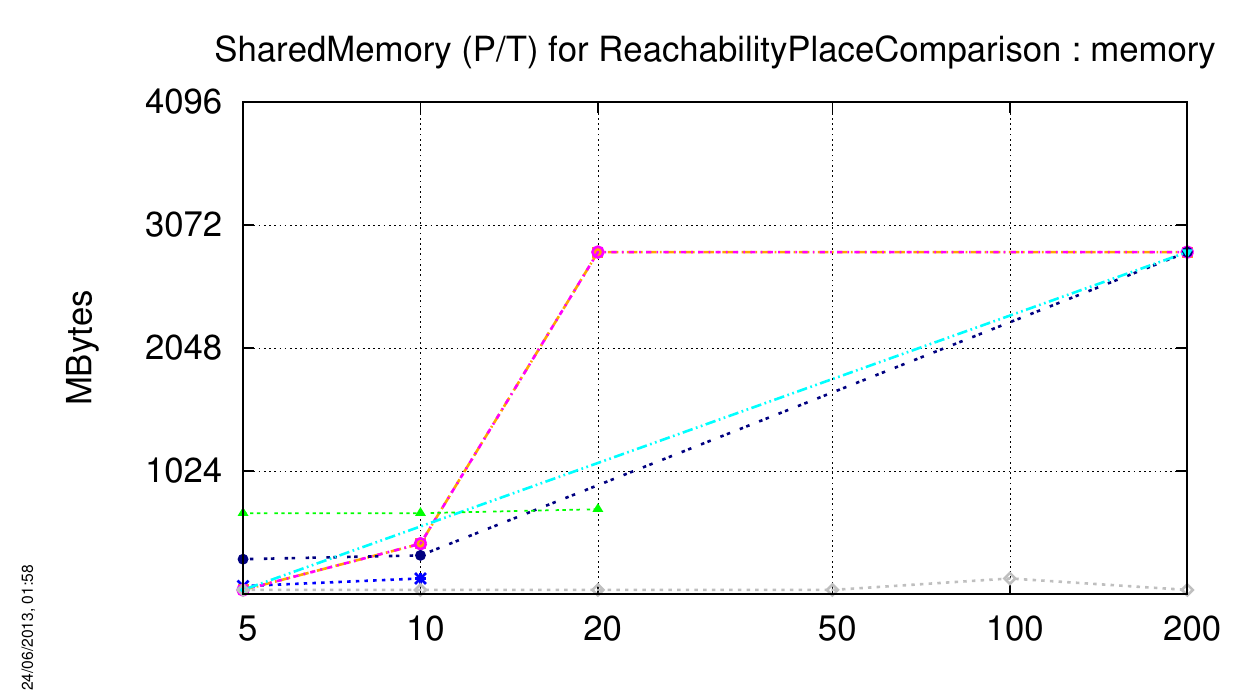}
   \includegraphics[width=7.2cm]{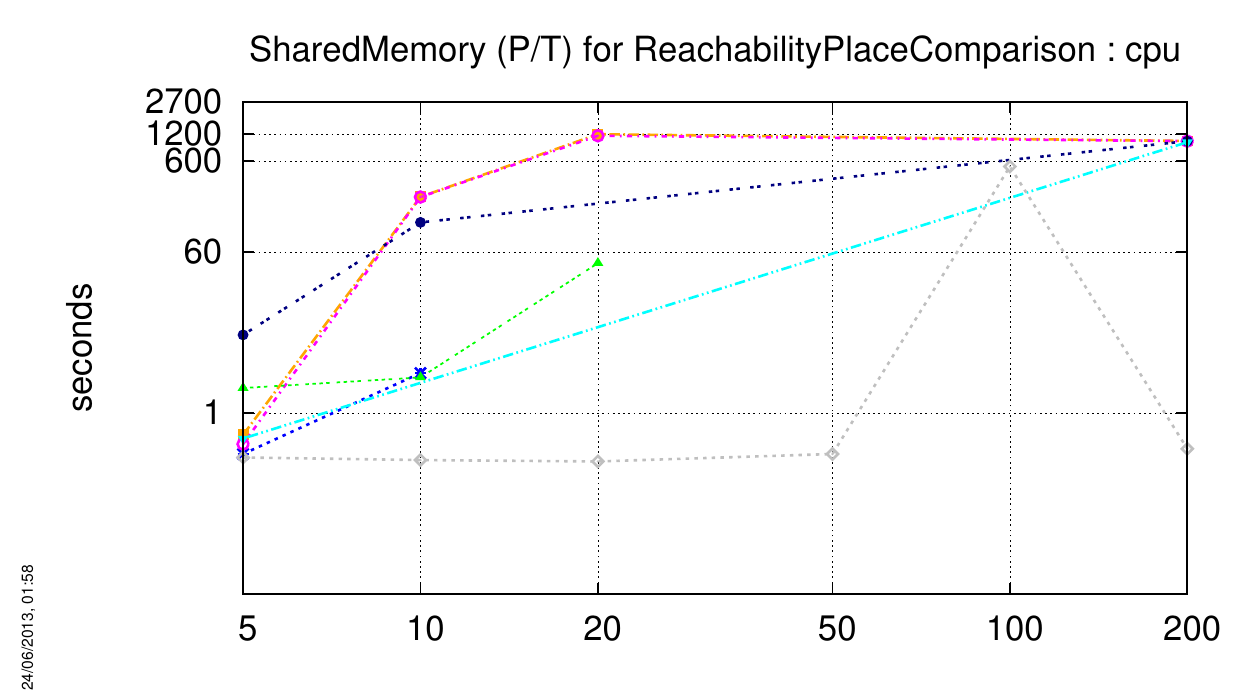}

   \includegraphics[height=1cm]{figures/tools-legend.pdf}
\end{center}

\subsubsection{\acs{SimpleLoadBal-COL}}
The charts below respectively show how tools compete with this ``Known'' model (memory and CPU).

\index{Performances!ReachabilityPlaceComparison!SimpleLoadBal (Colored)}
\begin{center}
   \includegraphics[width=7.2cm]{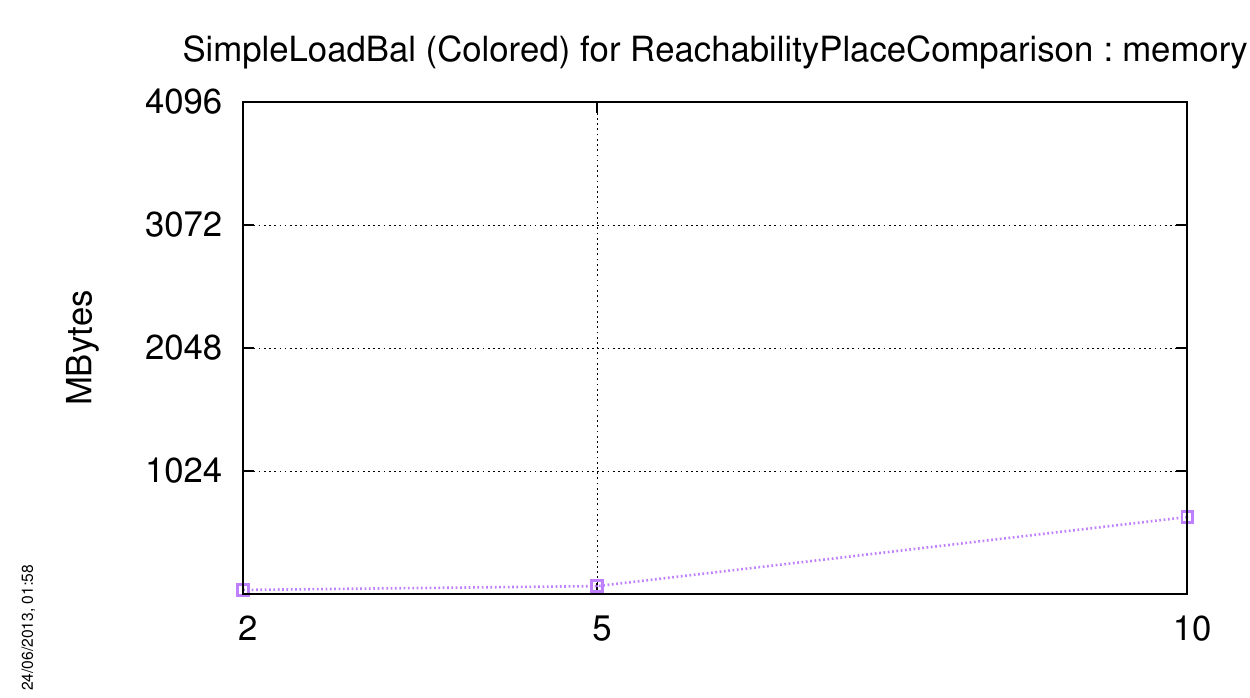}
   \includegraphics[width=7.2cm]{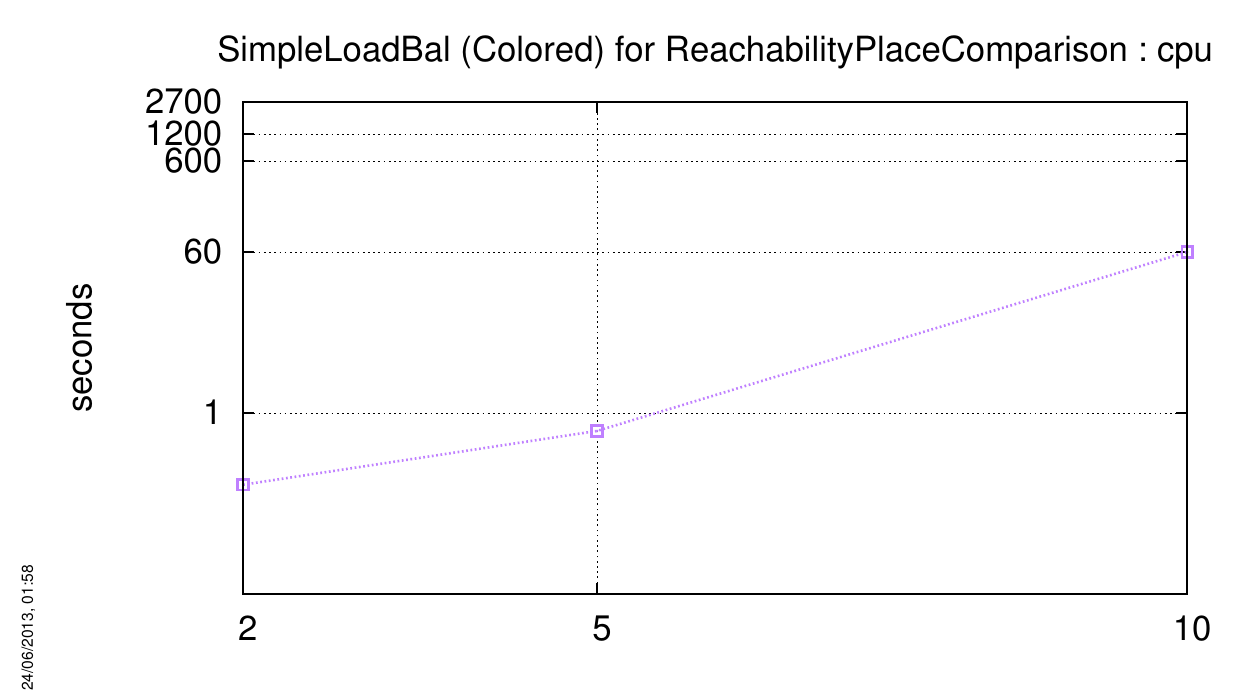}

   \includegraphics[height=1cm]{figures/tools-legend.pdf}
\end{center}

\subsubsection{\acs{SimpleLoadBal-PT}}
The charts below respectively show how tools compete with this ``Known'' model (memory and CPU).

\index{Performances!ReachabilityPlaceComparison!SimpleLoadBal (P/T)}
\begin{center}
   \includegraphics[width=7.2cm]{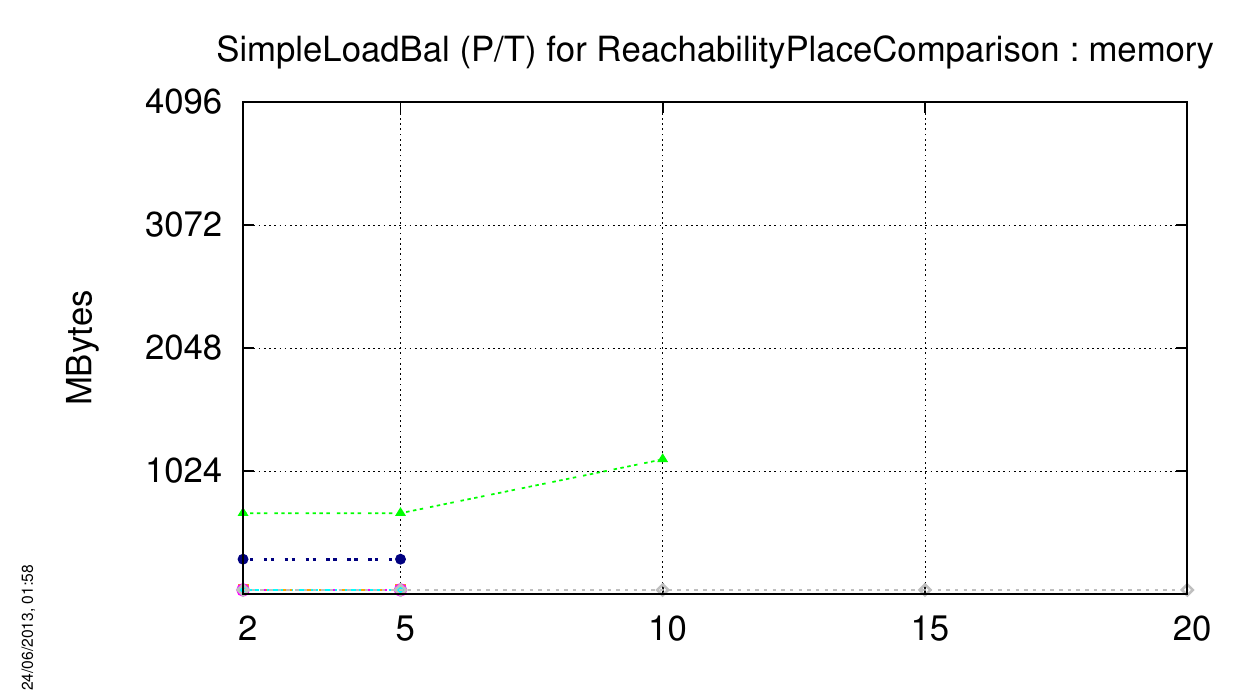}
   \includegraphics[width=7.2cm]{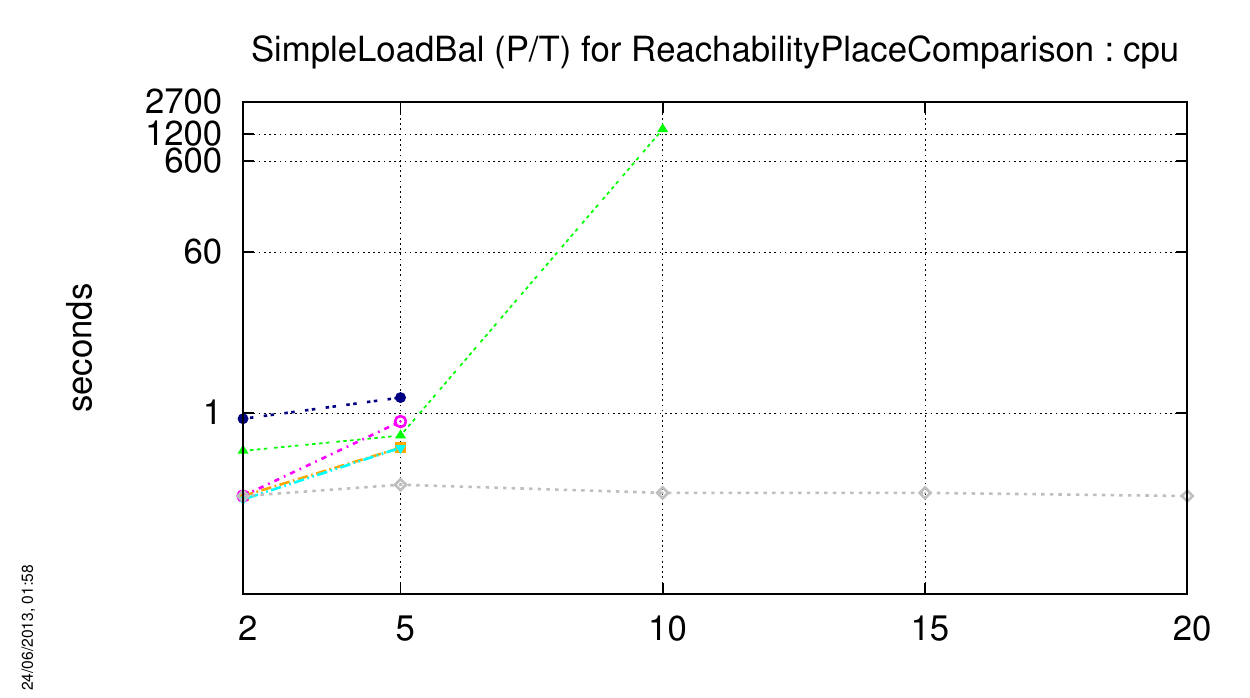}

   \includegraphics[height=1cm]{figures/tools-legend.pdf}
\end{center}

\subsubsection{\acs{TokenRing-COL}}
No instance of this model could be computed for the \textbf{ReachabilityPlaceComparison} examination.

\subsubsection{\acs{TokenRing-PT}}
No instance of this model could be computed for the \textbf{ReachabilityPlaceComparison} examination.

\subsubsection{\acs{HouseConstruction-PT}}
The charts below respectively show how tools compete with this ``Suprise'' model (memory and CPU).

\index{Performances!ReachabilityPlaceComparison!HouseConstruction (P/T)}
\begin{center}
   \includegraphics[width=7.2cm]{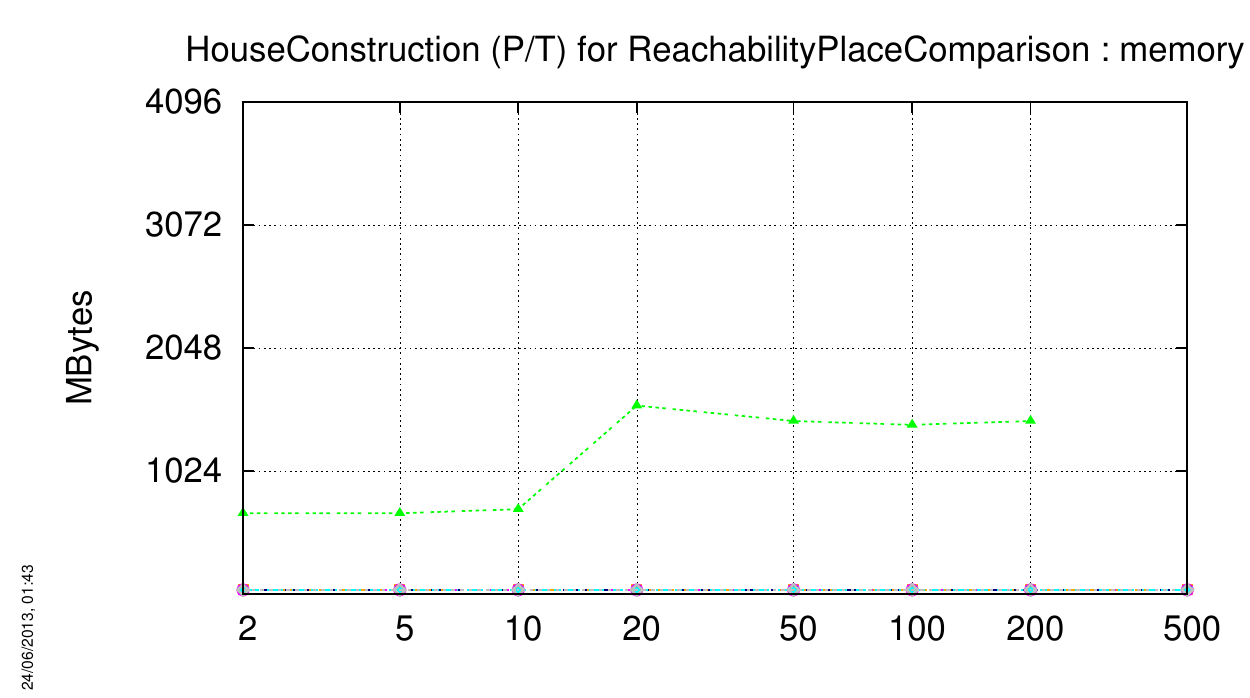}
   \includegraphics[width=7.2cm]{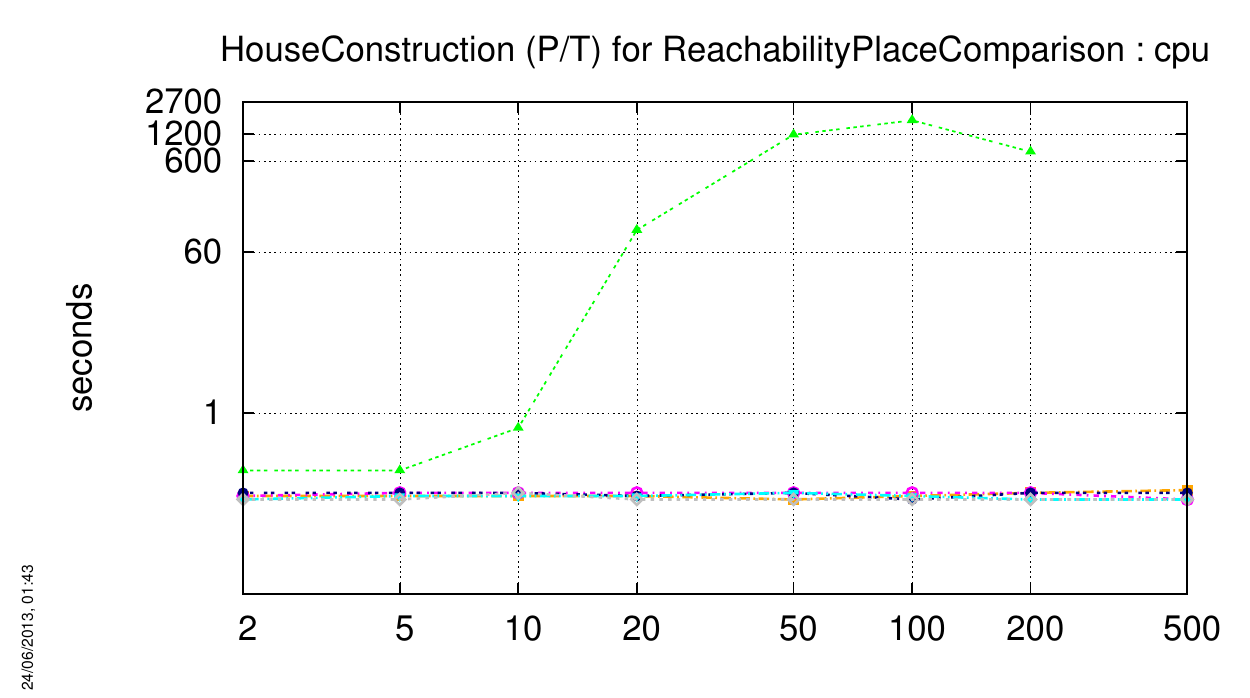}

   \includegraphics[height=1cm]{figures/tools-legend.pdf}
\end{center}

\subsubsection{\acs{IBMB2S565S3960-PT}}
The charts below respectively show how tools compete with this ``Suprise'' model (memory and CPU).

\index{Performances!ReachabilityPlaceComparison!IBMB2S565S3960 (P/T)}
\begin{center}
   \includegraphics[width=7.2cm]{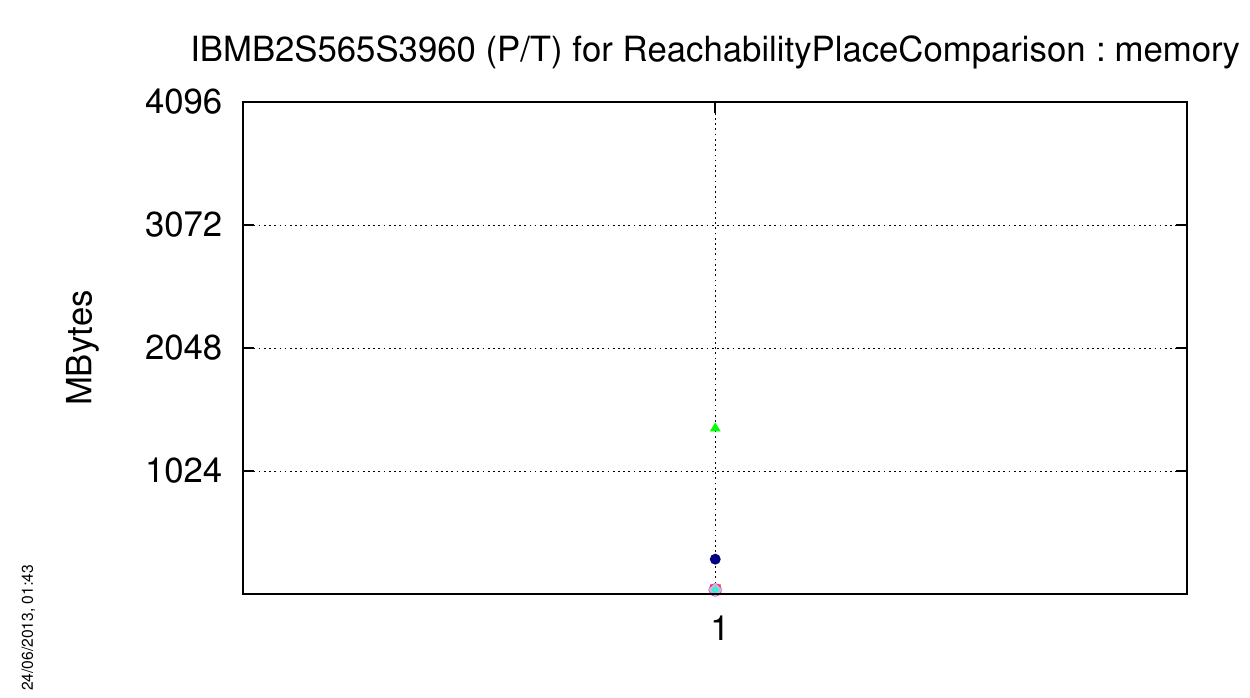}
   \includegraphics[width=7.2cm]{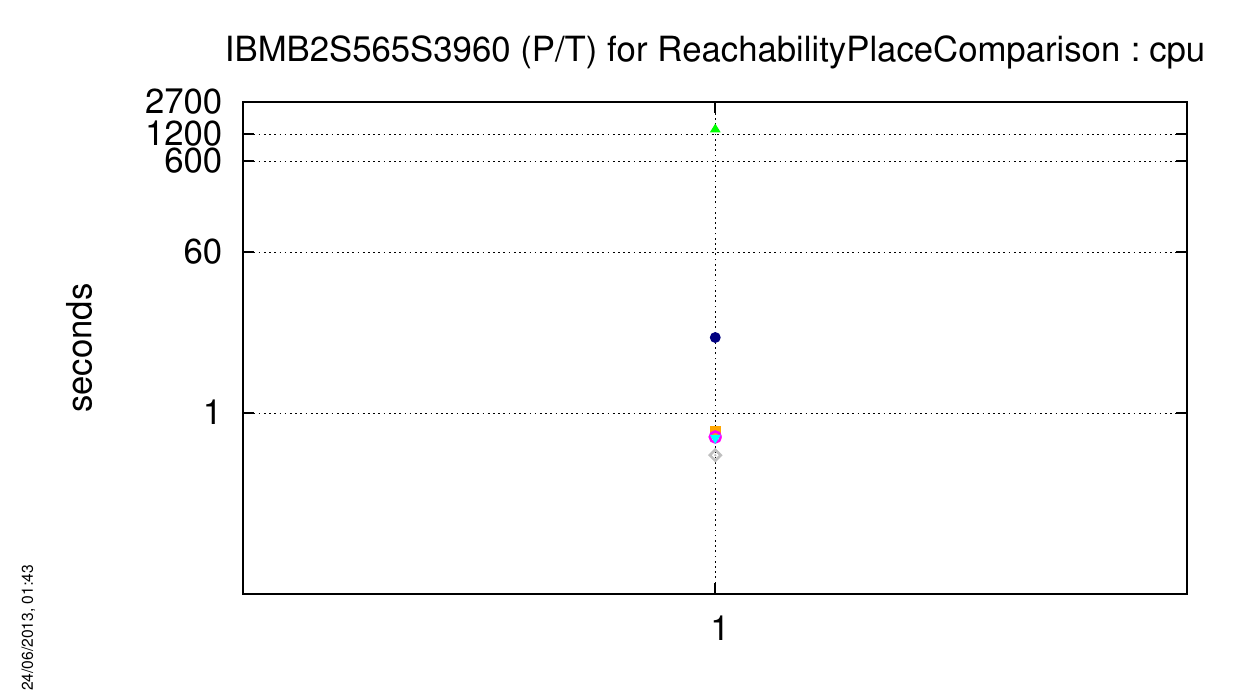}

   \includegraphics[height=1cm]{figures/tools-legend.pdf}
\end{center}

\subsubsection{\acs{QuasiCertifProtocol-COL}}
No instance of this model could be computed for the \textbf{ReachabilityPlaceComparison} examination.

\subsubsection{\acs{QuasiCertifProtocol-PT}}
The charts below respectively show how tools compete with this ``Suprise'' model (memory and CPU).

\index{Performances!ReachabilityPlaceComparison!QuasiCertifProtocol (P/T)}
\begin{center}
   \includegraphics[width=7.2cm]{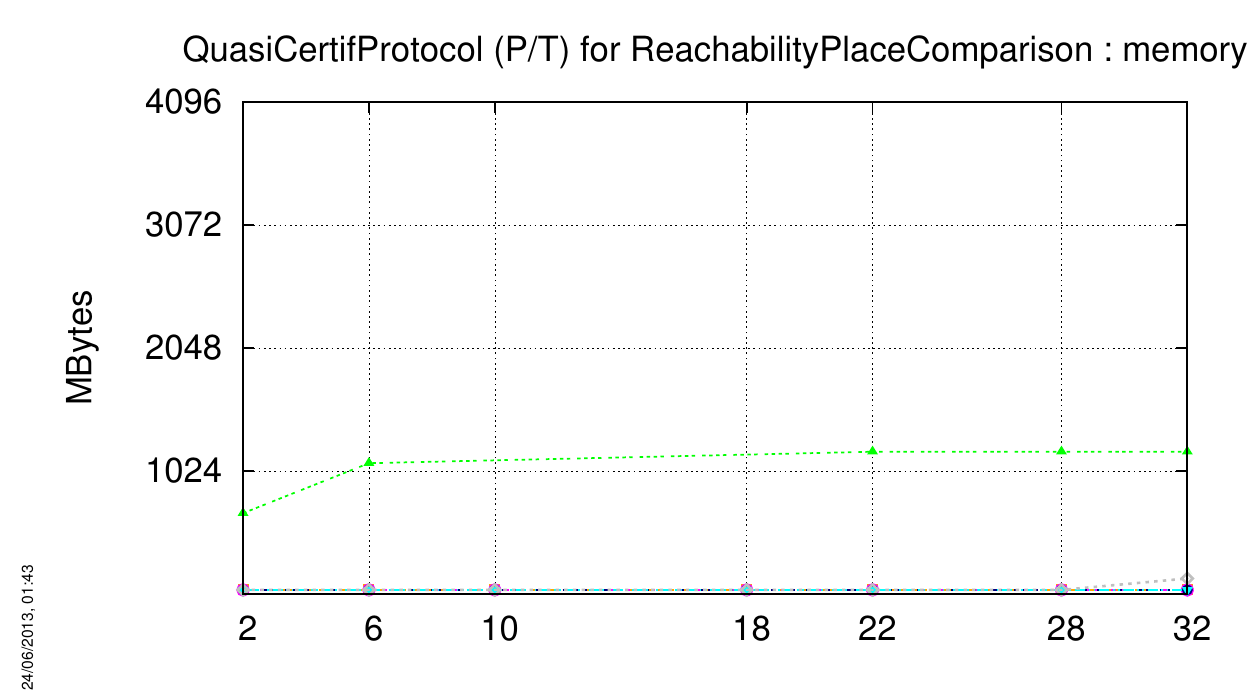}
   \includegraphics[width=7.2cm]{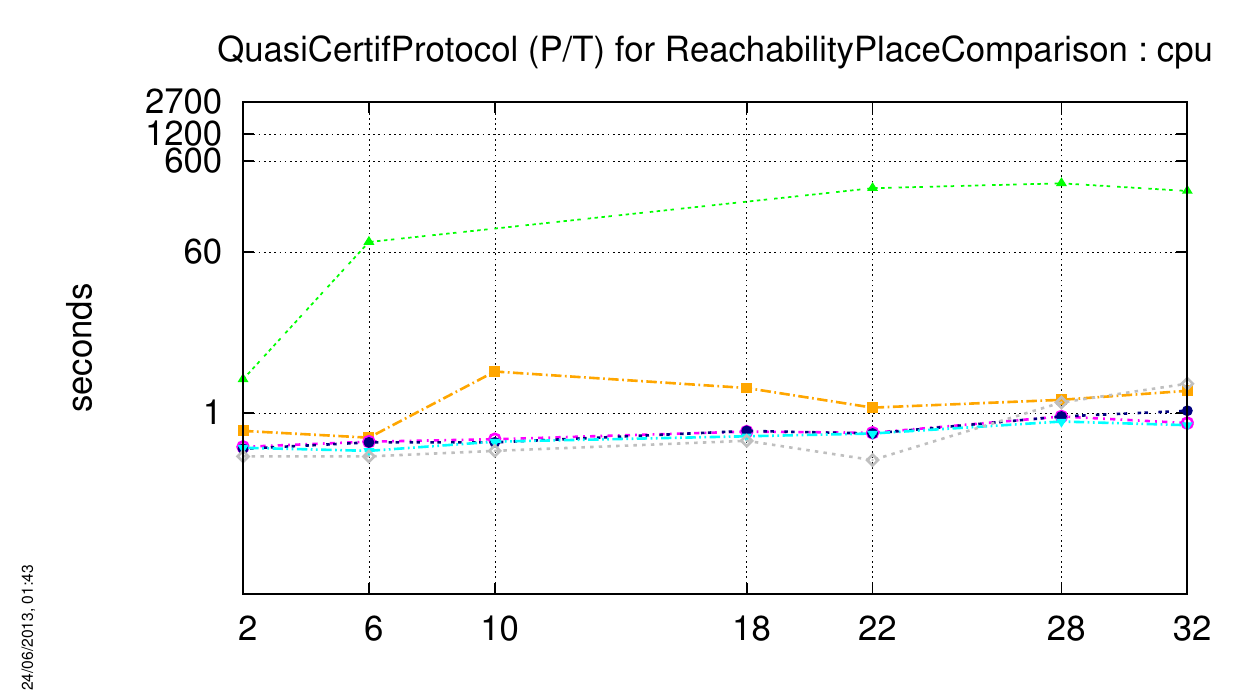}

   \includegraphics[height=1cm]{figures/tools-legend.pdf}
\end{center}

\subsubsection{\acs{Vasy2003-PT}}
The charts below respectively show how tools compete with this ``Suprise'' model (memory and CPU).

\index{Performances!ReachabilityPlaceComparison!Vasy2003 (P/T)}
\begin{center}
   \includegraphics[width=7.2cm]{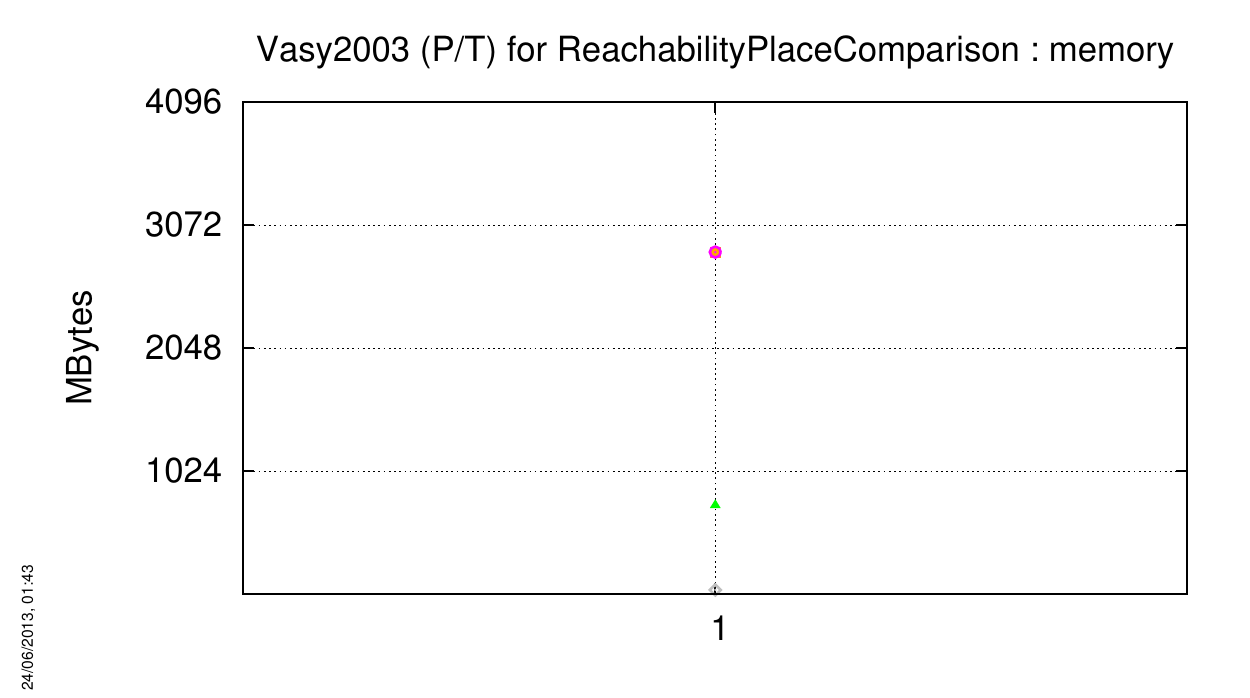}
   \includegraphics[width=7.2cm]{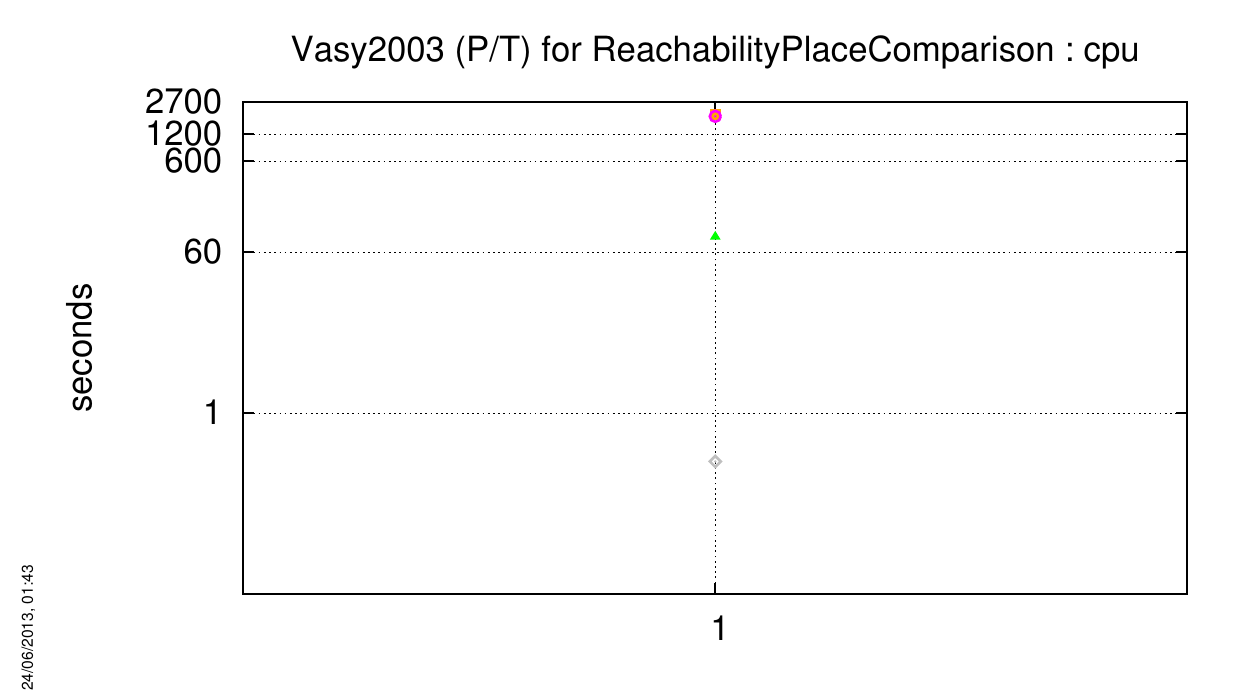}

   \includegraphics[height=1cm]{figures/tools-legend.pdf}
\end{center}

\subsection{Outputs for the ReachabilityPlaceComparison Examination}
\index{Outputs!ReachabilityPlaceComparison}

Please find enclosed the brute results for this examination (``Known'' and ``Surprise'' models).
We display only the score of tools that provide a results for at least one instance of one model.
The legend for the values is provided below:
\begin{itemize}
   \item\textbf{nc}: the tool does not compete this examination for this model/instance,
   \item\textbf{cc}: the tool cannot compute this examination for this model/instance,
   \item\textbf{to}: the tool cannot compute this examination for this model/instance within the maximum allowed time,
   \item\textbf{mp}: the tool encountered a memory problem (stack overflow or memory full),
   \item\textbf{nf}: there is no formula available for this type of examination (typically, this concerns P/T nets where
       comparing marking cardinality has no signification when there is no equivalent colored net).
\end{itemize}

\textbf{Note on the display of results for formulas:} each formula is considered as a flag (F if false, T if true, - or ?
when the value cannot be determined). These values are concatenated in the order they appear (we assume it is the order of formulas as they were provided).

\subsubsection{``Known'' Models}

Results are summarized in the table below.
{\footnotesize
\begin{longtable}{|ccccccccc|}
   \hline
      \multicolumn{9}{|c|}{\bf \acs{CSRepetitions-COL}} \\
      \textbf{Instances}\index{Outputs!ReachabilityPlaceComparison!CSRepetitions (colored)} &
      \acs{greatspn} &
      \acs{its-tools} &
      \acs{lola} &
      \acs{lola-optimistic} &
      \acs{lola-optimistic-incomplete} &
      \acs{lola-pessimistic} &
      \acs{marcie} &
      \acs{sara} \\
      02 &
      nc &
      nc &
      nc &
      nc &
      nc &
      nc &
      nc &
      nc \\
      03 &
      nc &
      \textcolor{red}{\tt \scriptsize FFFFFFFF} &
      nc &
      nc &
      nc &
      nc &
      nc &
      nc \\
      04 &
      nc &
      \textcolor{red}{\tt \scriptsize mp} &
      nc &
      nc &
      nc &
      nc &
      nc &
      nc \\
      05 &
      nc &
      \textcolor{red}{\tt \scriptsize mp} &
      nc &
      nc &
      nc &
      nc &
      nc &
      nc \\
      07 &
      nc &
      \textcolor{red}{\tt \scriptsize mp} &
      nc &
      nc &
      nc &
      nc &
      nc &
      nc \\
      10 &
      nc &
      \textcolor{red}{\tt \scriptsize mp} &
      nc &
      nc &
      nc &
      nc &
      nc &
      nc \\
\hline
\end{longtable}
}

\subsubsection{``Surprise'' Models}

Results are summarized in the table below.
{\footnotesize
\begin{longtable}{|ccccccc|}
   \hline
      \multicolumn{7}{|c|}{\bf \acs{HouseConstruction-PT}} \\
      instances\index{Outputs!ReachabilityPlaceComparison!HouseConstruction (P/T)} &
      \acs{lola} &
      \acs{lola-optimistic} &
      \acs{lola-optimistic-incomplete} &
      \acs{lola-pessimistic} &
      \acs{marcie} &
      \acs{sara} \\
      002 &
      \textcolor{red}{\tt \scriptsize cc} &
      \textcolor{red}{\tt \scriptsize cc} &
      \textcolor{red}{\tt \scriptsize cc} &
      \textcolor{red}{\tt \scriptsize cc} &
      \textcolor{red}{\tt \scriptsize FFFTTFFF} &
      \textcolor{red}{\tt \scriptsize cc} \\
      005 &
      \textcolor{red}{\tt \scriptsize cc} &
      \textcolor{red}{\tt \scriptsize cc} &
      \textcolor{red}{\tt \scriptsize cc} &
      \textcolor{red}{\tt \scriptsize cc} &
      \textcolor{red}{\tt \scriptsize FFFTTFFF} &
      \textcolor{red}{\tt \scriptsize cc} \\
      010 &
      \textcolor{red}{\tt \scriptsize cc} &
      \textcolor{red}{\tt \scriptsize cc} &
      \textcolor{red}{\tt \scriptsize cc} &
      \textcolor{red}{\tt \scriptsize cc} &
      \textcolor{red}{\tt \scriptsize FFFTFFFF} &
      \textcolor{red}{\tt \scriptsize cc} \\
      020 &
      \textcolor{red}{\tt \scriptsize cc} &
      \textcolor{red}{\tt \scriptsize cc} &
      \textcolor{red}{\tt \scriptsize cc} &
      \textcolor{red}{\tt \scriptsize cc} &
      \textcolor{red}{\tt \scriptsize FFFTFFFF} &
      \textcolor{red}{\tt \scriptsize cc} \\
      050 &
      \textcolor{red}{\tt \scriptsize cc} &
      \textcolor{red}{\tt \scriptsize cc} &
      \textcolor{red}{\tt \scriptsize cc} &
      \textcolor{red}{\tt \scriptsize cc} &
      \textcolor{red}{\tt \scriptsize cc} &
      \textcolor{red}{\tt \scriptsize cc} \\
\hline
\end{longtable}
}

\subsection{Score for the ReachabilityPlaceComparison Examination}
\index{Scores!ReachabilityPlaceComparison}

Please find enclosed the scores for this examination (``Known'' and ``Surprise'' models).
We display only the score of tools that provide a results for at least one instance of one model.
The total is first listed in the table below followed by a detail, for each proposed model.
Meaning of the line labels are:
\begin{itemize}
\item\textbf{1st instance}: the tool gets a bonus for having processed the first instance of this model (+1 point),
\item\textbf{instances}: the tool gets 1 point per instances treated 
(for that, we assume that at least one formula has been successfully computed),
\item\textbf{max reached}: the tool could process all the instances for the model (+2 points),
\item\textbf{best}: the tool is among the ones that processed a maximum of instances within the time and memory confinement (+2 points).
\end{itemize}

\subsubsection{``Known'' Models}

Scores are summarized in the table below.
For each model in this category, a tool may collect up to 411 points.

{\footnotesize
\begin{longtable}{|ccccccccc|}
   \hline
      \multicolumn{9}{|c|}{\bf Total score of the tools for this examination} \\
        &
      \acs{greatspn} &
      \acs{its-tools} &
      \acs{lola} &
      \acs{lola-optimistic} &
      \acs{lola-optimistic-incomplete} &
      \acs{lola-pessimistic} &
      \acs{marcie} &
      \acs{sara} \\
      Total Score &
      {\bf 43} &
      {\bf 61} &
      {\bf 170} &
      {\bf 170} &
      {\bf 116} &
      {\bf 98} &
      {\bf 120} &
      {\bf 55} \\
   \hline
      \multicolumn{9}{|c|}{\bf CSRepetitions (Colored)} \\
      \ \index{Scores!ReachabilityPlaceComparison!CSRepetitions (colored)} &
      \acs{greatspn} &
      \acs{its-tools} &
      \acs{lola} &
      \acs{lola-optimistic} &
      \acs{lola-optimistic-incomplete} &
      \acs{lola-pessimistic} &
      \acs{marcie} &
      \acs{sara} \\
      1st instance &
      \textcolor{red}{0} &
      \textcolor{red}{1} &
      \textcolor{red}{0} &
      \textcolor{red}{0} &
      \textcolor{red}{0} &
      \textcolor{red}{0} &
      \textcolor{red}{0} &
      \textcolor{red}{0} \\
      instances &
      \textcolor{red}{0} &
      \textcolor{red}{1} &
      \textcolor{red}{0} &
      \textcolor{red}{0} &
      \textcolor{red}{0} &
      \textcolor{red}{0} &
      \textcolor{red}{0} &
      \textcolor{red}{0} \\
      max reached &
      \textcolor{red}{0} &
      \textcolor{red}{0} &
      \textcolor{red}{0} &
      \textcolor{red}{0} &
      \textcolor{red}{0} &
      \textcolor{red}{0} &
      \textcolor{red}{0} &
      \textcolor{red}{0} \\
\hline
\end{longtable}
}

\subsubsection{``Surprise'' Models}

Scores are summarized in the table below.
For each model in this category, a tool may collect up to 49 points.

{\footnotesize
\begin{longtable}{|ccccccc|}
   \hline
      \multicolumn{7}{|c|}{\bf Total score of the tools for this examination} \\
        &
      \acs{lola} &
      \acs{lola-optimistic} &
      \acs{lola-optimistic-incomplete} &
      \acs{lola-pessimistic} &
      \acs{marcie} &
      \acs{sara} \\
      Total Score &
      {\bf 12} &
      {\bf 12} &
      {\bf 6} &
      {\bf 6} &
      {\bf 22} &
      {\bf 4} \\
   \hline
      \multicolumn{7}{|c|}{\bf HouseConstruction (P/T)} \\
      \ \index{Scores!ReachabilityPlaceComparison!HouseConstruction (P/T)} &
      \acs{lola} &
      \acs{lola-optimistic} &
      \acs{lola-optimistic-incomplete} &
      \acs{lola-pessimistic} &
      \acs{marcie} &
      \acs{sara} \\
      1st instance &
      \textcolor{red}{0} &
      \textcolor{red}{0} &
      \textcolor{red}{0} &
      \textcolor{red}{0} &
      \textcolor{red}{1} &
      \textcolor{red}{0} \\
      instances &
      \textcolor{red}{0} &
      \textcolor{red}{0} &
      \textcolor{red}{0} &
      \textcolor{red}{0} &
      \textcolor{red}{4} &
      \textcolor{red}{0} \\
      max reached &
      \textcolor{red}{0} &
      \textcolor{red}{0} &
      \textcolor{red}{0} &
      \textcolor{red}{0} &
      \textcolor{red}{0} &
      \textcolor{red}{0} \\
\hline
\end{longtable}
}

\subsection{Trophies for this Examination}
\index{Trophies!ReachabilityPlaceComparison}

Trophies are divided in three categories: ``Known'' models,
``Surprise'' models, and the global trophies (formula is then
$score_{global} = score_{known} + 2 \times score_{surprise}$).

\subsubsection{For ``Known'' Models} \ \\

\begin{tabular}{c|c|c|c}
      1 & 1 & 3 & 3 \\
   \includegraphics[width=2cm]{figures/gold.jpg} &
   \includegraphics[width=2cm]{figures/gold.jpg} &
   \includegraphics[width=2cm]{figures/bronse.jpg} &
   \includegraphics[width=2cm]{figures/bronse.jpg} \\
   \acs{lola} &
   \acs{lola-optimistic} &
   \acs{lola-pessimistic} &
   \acs{marcie} \\
   170 points &
   170 points &
   120 points &
   120 points \\
\end{tabular}

\subsubsection{For ``Surprise'' Models}\  \\

\begin{tabular}{c|c|c}
      1 & 2 & 2 \\
   \includegraphics[width=2cm]{figures/gold.jpg} &
   \includegraphics[width=2cm]{figures/silver.jpg} &
   \includegraphics[width=2cm]{figures/silver.jpg} \\
   \acs{marcie} &
   \acs{lola} &
   \acs{lola-optimistic} \\
   22 points &
   12 points &
   12 points \\
\end{tabular}

\subsubsection{Global} \ \\

\begin{tabular}{c|c|c}
      1 & 1 & 3 \\
   \includegraphics[width=2cm]{figures/gold.jpg} &
   \includegraphics[width=2cm]{figures/gold.jpg} &
   \includegraphics[width=2cm]{figures/bronse.jpg} \\
   \acs{lola} &
   \acs{lola-optimistic} &
   \acs{marcie} \\
   194 points &
   194 points &
   164 points \\
\end{tabular}

\newpage

\section{The ReachabilityMix Examination}
\label{sec:exam:ReachabilityMix}
\index{Results!ReachabilityMix}

This examination deals with reachability properties dealing with all the previous type of atomic proposition.
We first show a summary on the handling of models by the participating tools.
Then, we present the computed outputs and the associated scores for this
examination prior to a summary of relevant executions.

\subsection{Handling of Models by Tools}
\index{Performances!ReachabilityMix}

\subsubsection{\acs{CSRepetitions-COL}}
The charts below respectively show how tools compete with this ``Known'' model (memory and CPU).

\index{Performances!ReachabilityMix!CSRepetitions (Colored)}
\begin{center}
   \includegraphics[width=7.2cm]{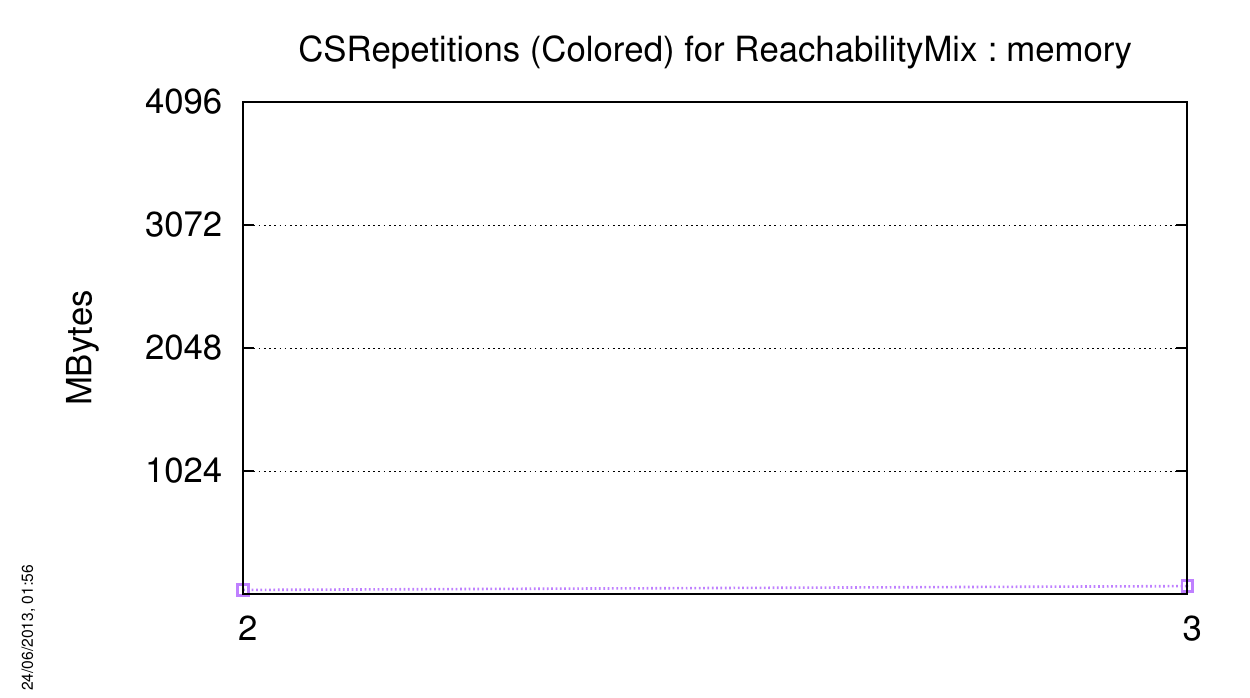}
   \includegraphics[width=7.2cm]{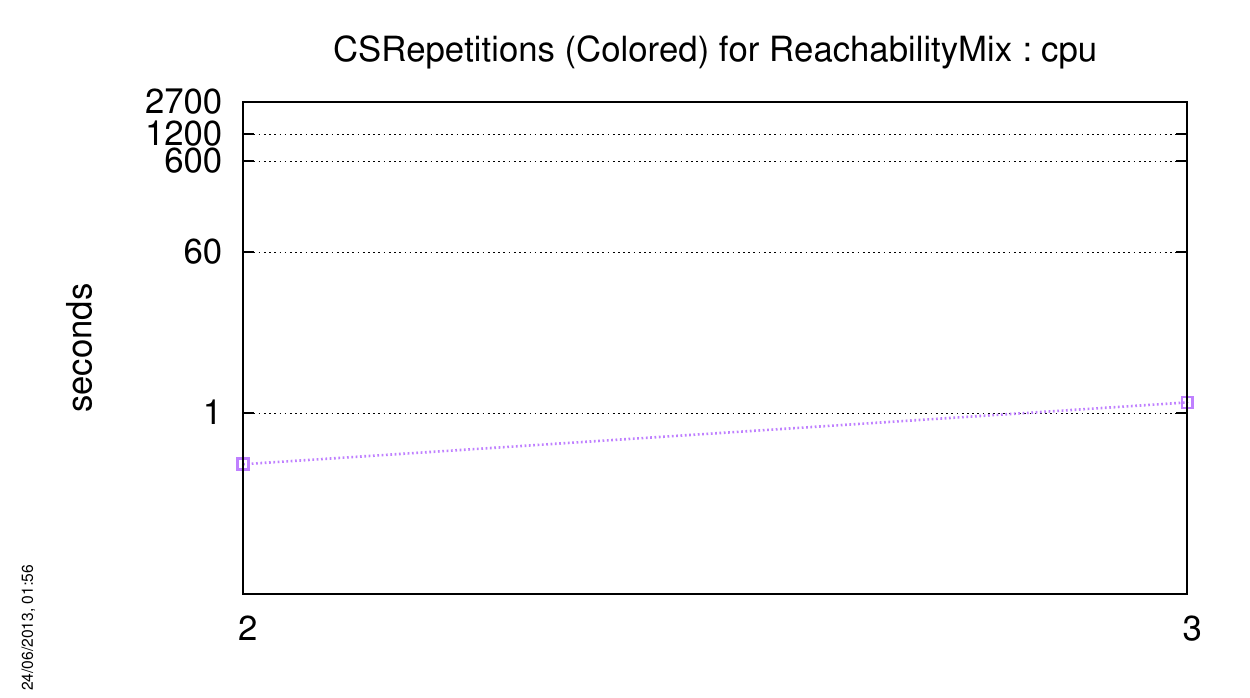}

   \includegraphics[height=1cm]{figures/tools-legend.pdf}
\end{center}

\subsubsection{\acs{CSRepetitions-PT}}
The charts below respectively show how tools compete with this ``Known'' model (memory and CPU).

\index{Performances!ReachabilityMix!CSRepetitions (P/T)}
\begin{center}
   \includegraphics[width=7.2cm]{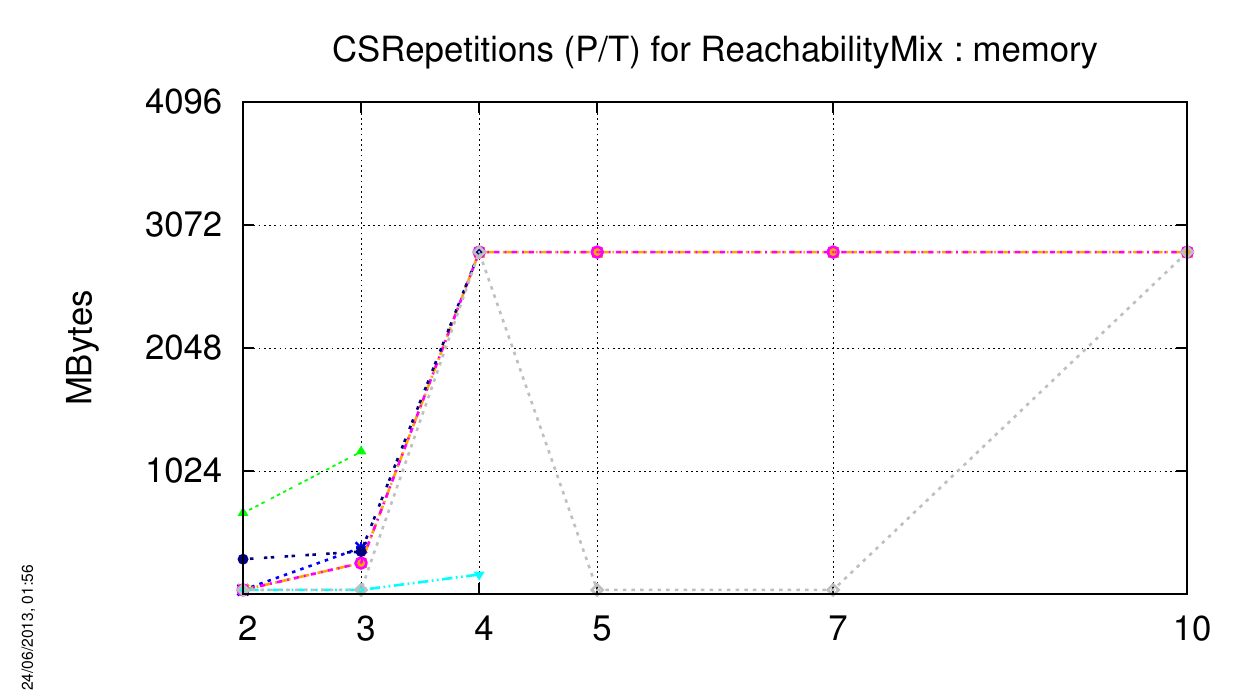}
   \includegraphics[width=7.2cm]{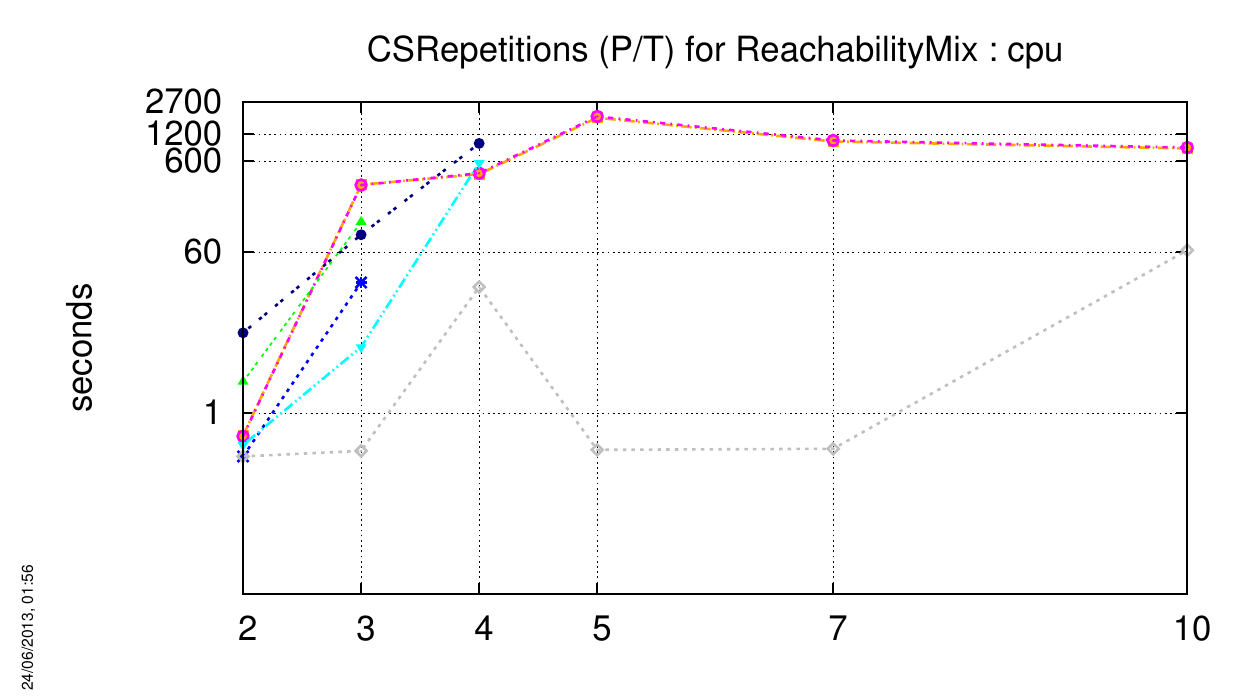}

   \includegraphics[height=1cm]{figures/tools-legend.pdf}
\end{center}

\subsubsection{\acs{Dekker-PT}}
The charts below respectively show how tools compete with this ``Known'' model (memory and CPU).

\index{Performances!ReachabilityMix!Dekker (P/T)}
\begin{center}
   \includegraphics[width=7.2cm]{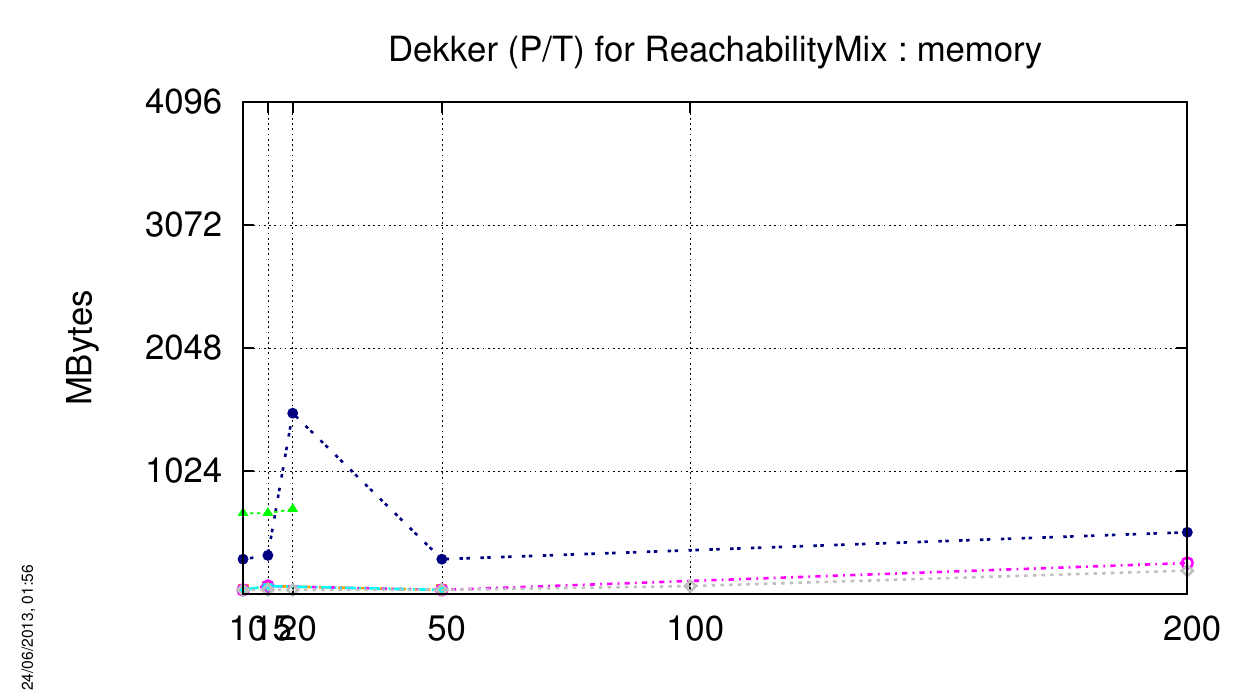}
   \includegraphics[width=7.2cm]{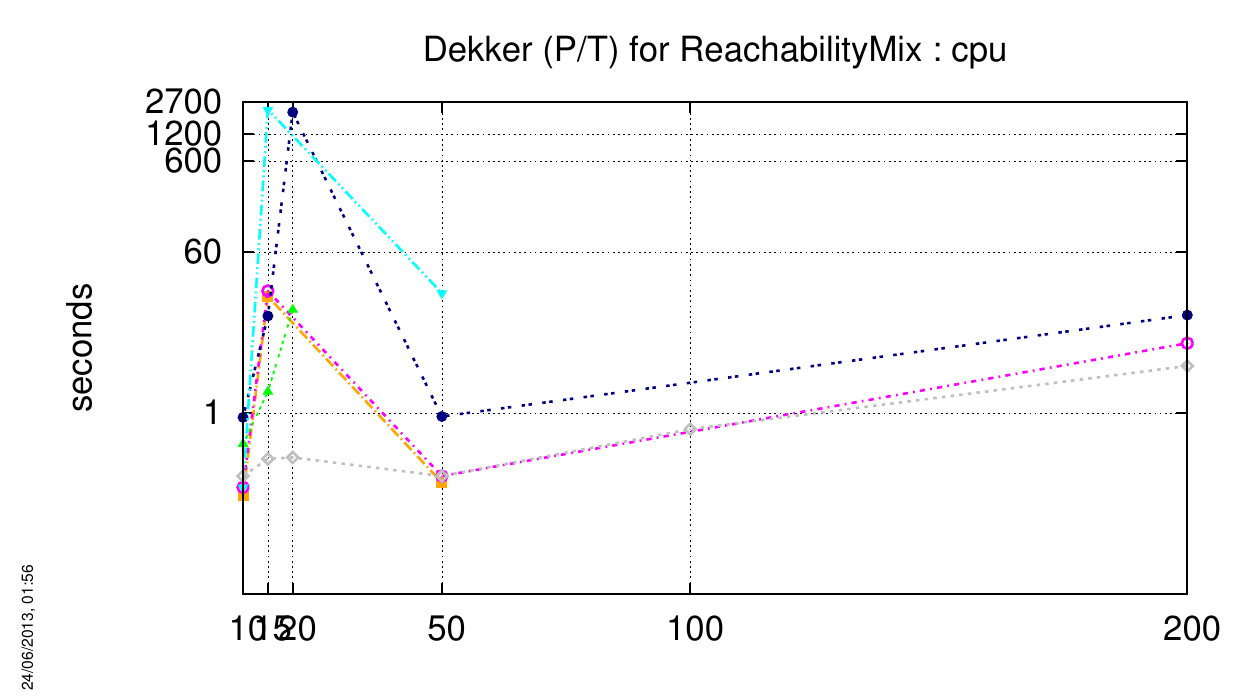}

   \includegraphics[height=1cm]{figures/tools-legend.pdf}
\end{center}

\subsubsection{\acs{DotAndBoxes-COL}}
The charts below respectively show how tools compete with this ``Known'' model (memory and CPU).

\index{Performances!ReachabilityMix!DotAndBoxes (Colored)}
\begin{center}
   \includegraphics[width=7.2cm]{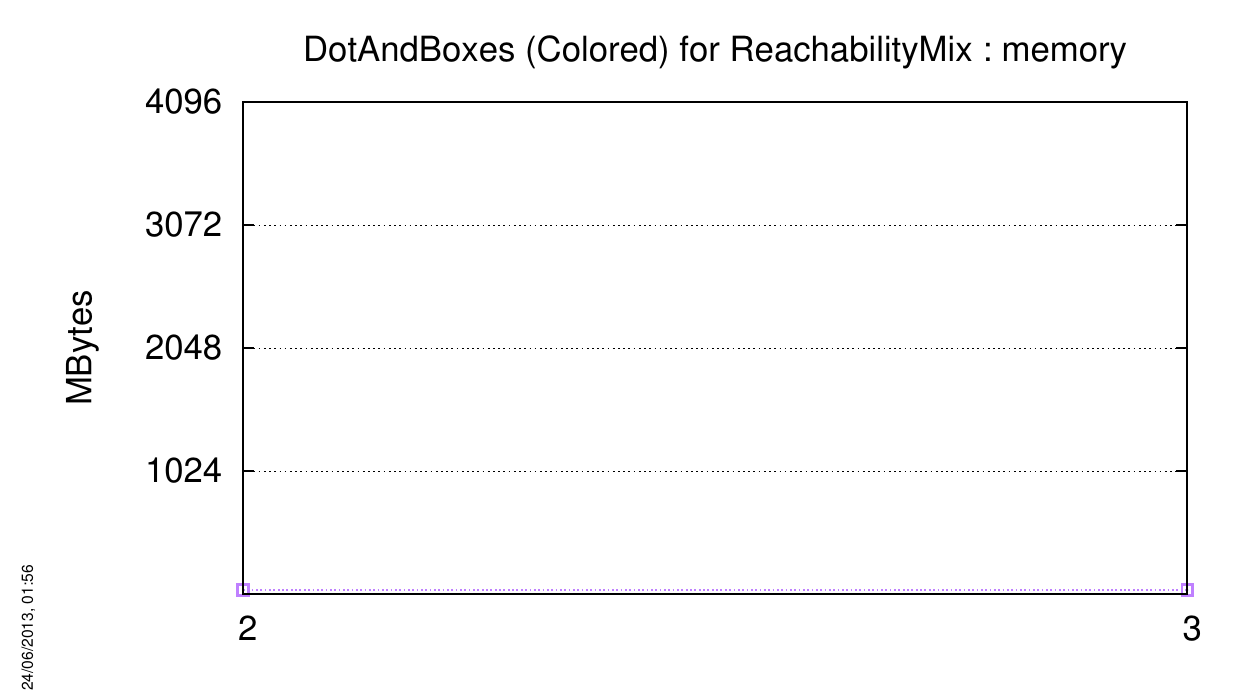}
   \includegraphics[width=7.2cm]{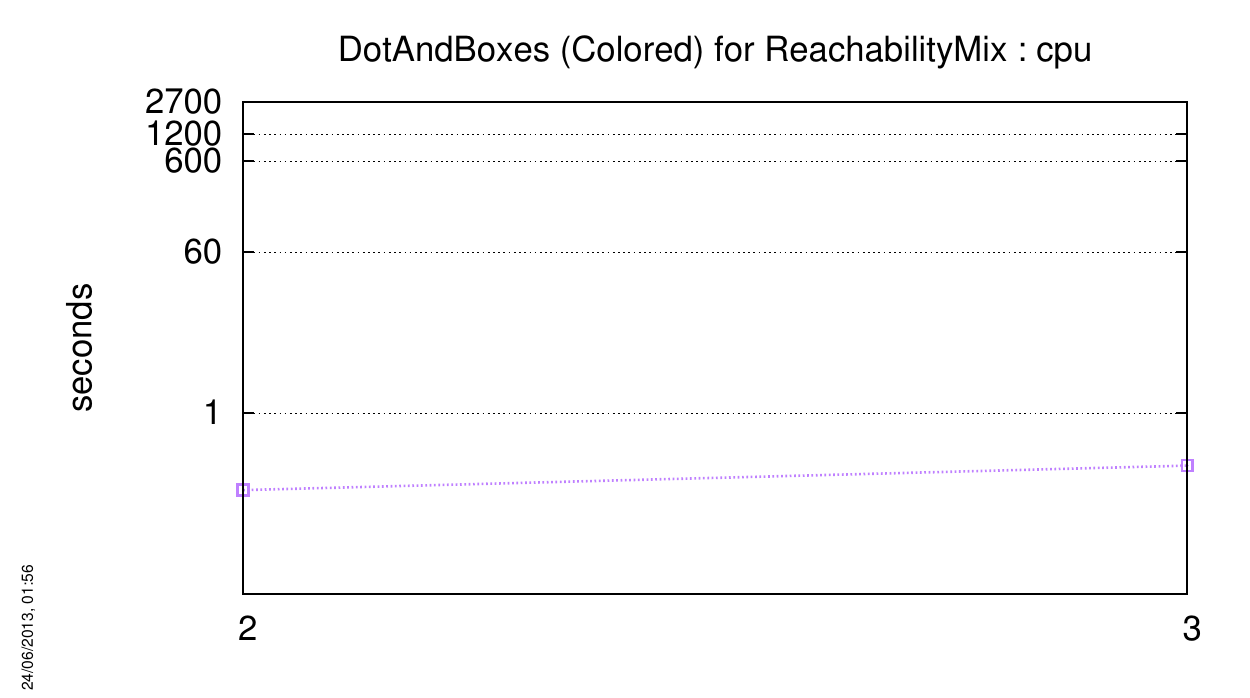}

   \includegraphics[height=1cm]{figures/tools-legend.pdf}
\end{center}

\subsubsection{\acs{DrinkVendingMachine-COL}}
The charts below respectively show how tools compete with this ``Known'' model (memory and CPU).

\index{Performances!ReachabilityMix!DrinkVendingMachine (Colored)}
\begin{center}
   \includegraphics[width=7.2cm]{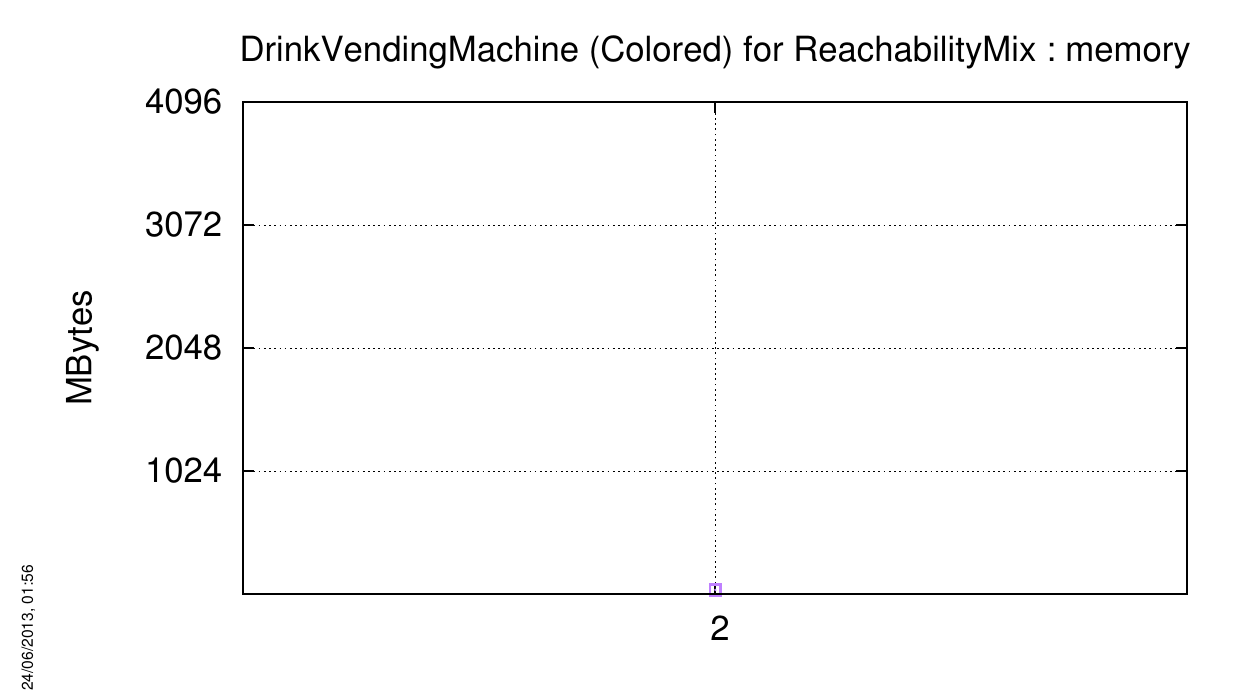}
   \includegraphics[width=7.2cm]{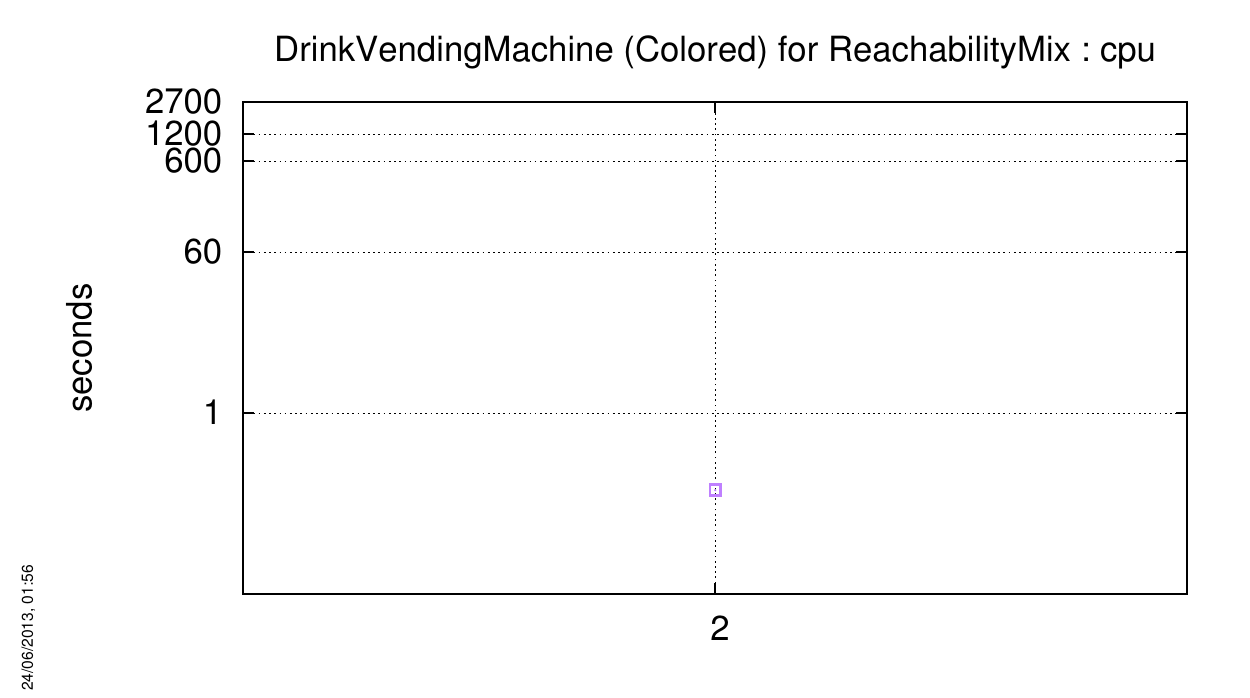}

   \includegraphics[height=1cm]{figures/tools-legend.pdf}
\end{center}

\subsubsection{\acs{DrinkVendingMachine-PT}}
The charts below respectively show how tools compete with this ``Known'' model (memory and CPU).

\index{Performances!ReachabilityMix!DrinkVendingMachine (P/T)}
\begin{center}
   \includegraphics[width=7.2cm]{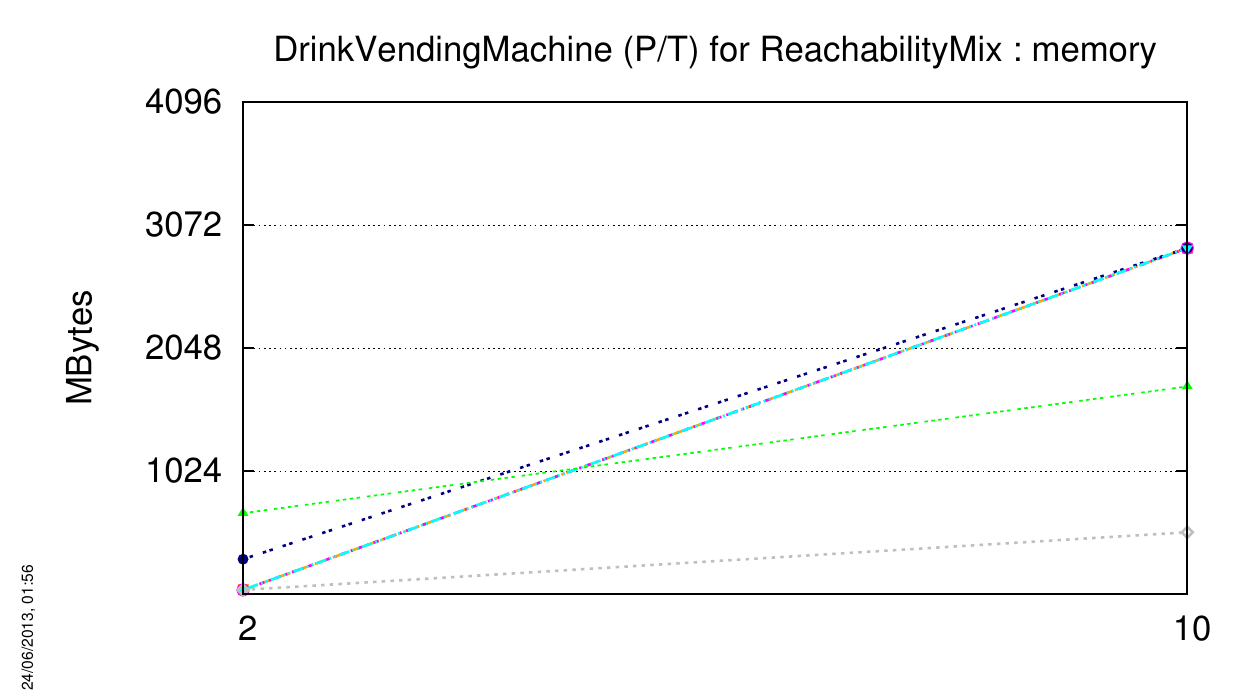}
   \includegraphics[width=7.2cm]{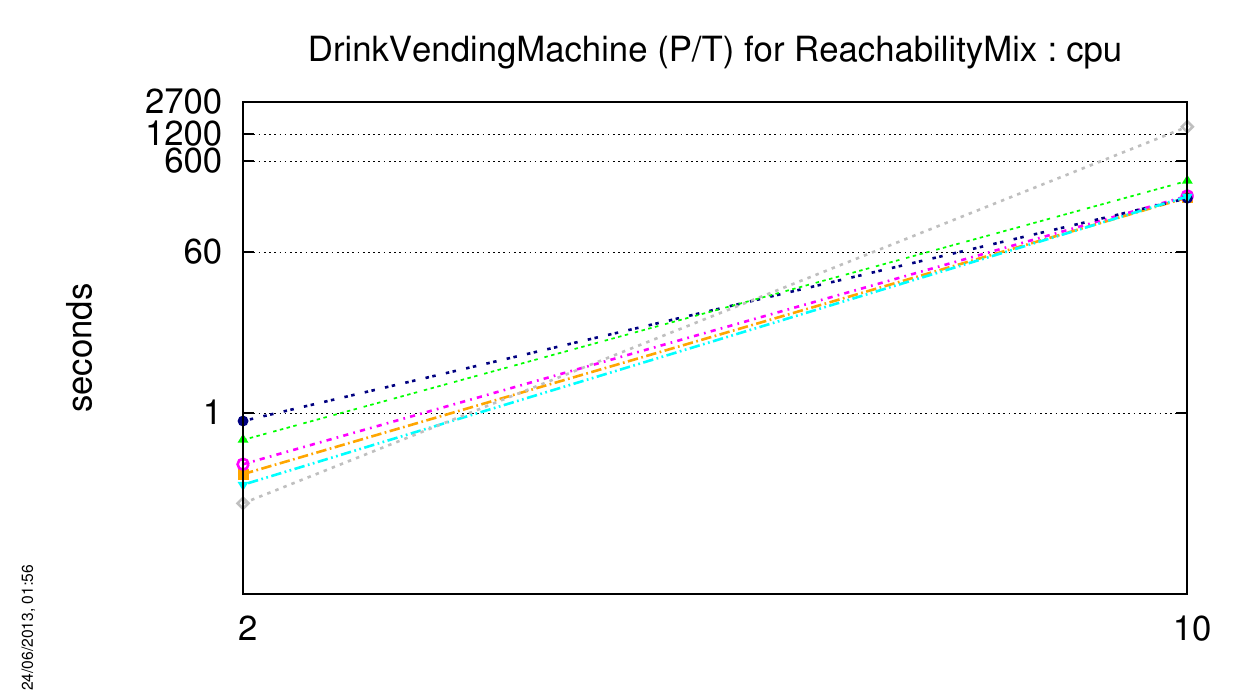}

   \includegraphics[height=1cm]{figures/tools-legend.pdf}
\end{center}

\subsubsection{\acs{Echo-PT}}
The charts below respectively show how tools compete with this ``Known'' model (memory and CPU).

\index{Performances!ReachabilityMix!Echo (P/T)}
\begin{center}
   \includegraphics[width=7.2cm]{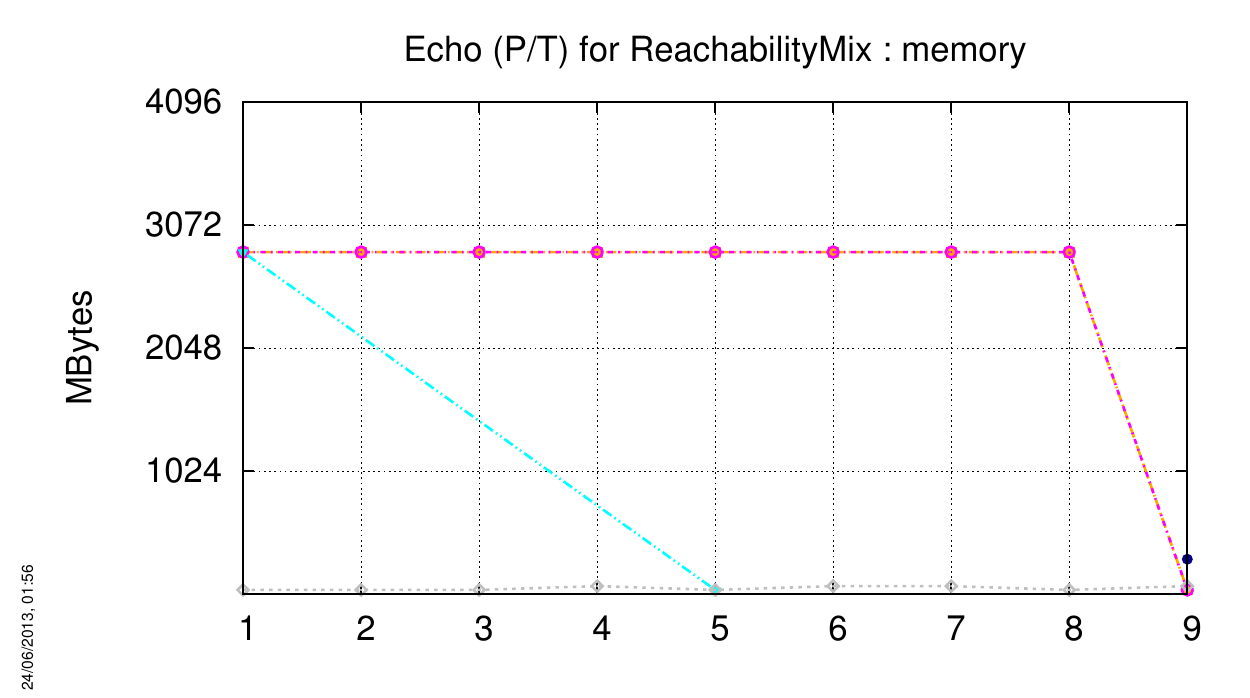}
   \includegraphics[width=7.2cm]{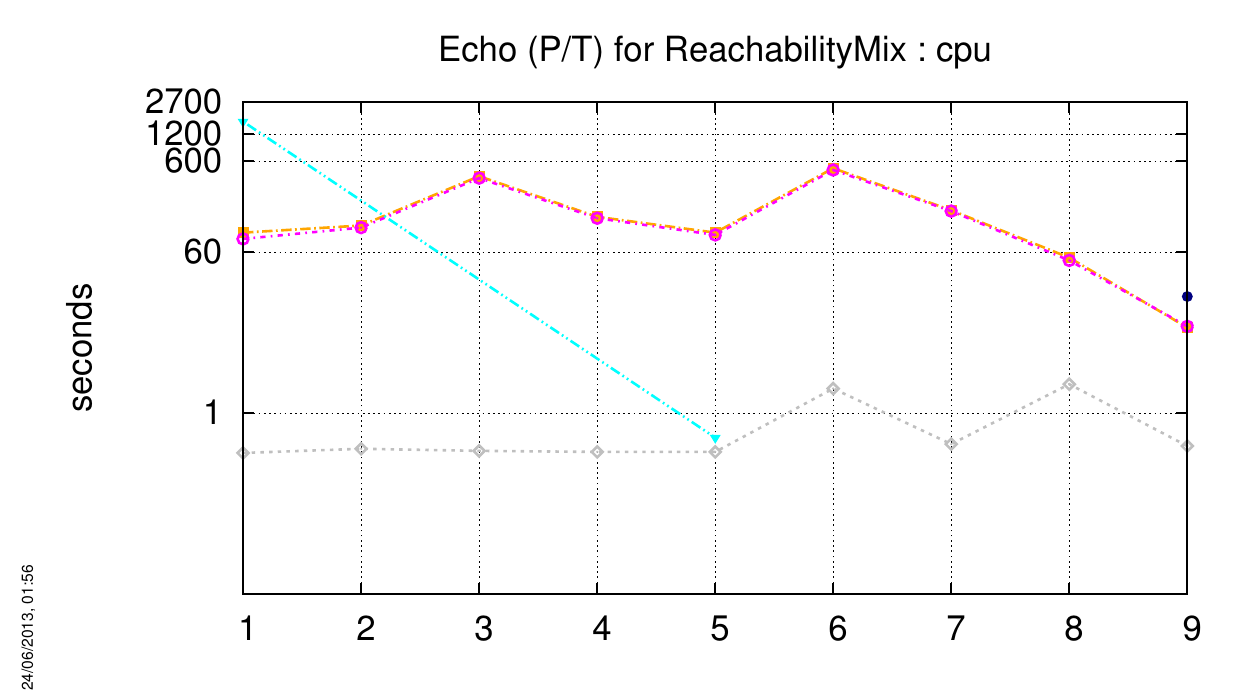}

   \includegraphics[height=1cm]{figures/tools-legend.pdf}
\end{center}

\subsubsection{\acs{Eratosthenes-PT}}
The charts below respectively show how tools compete with this ``Known'' model (memory and CPU).

\index{Performances!ReachabilityMix!Eratosthenes (P/T)}
\begin{center}
   \includegraphics[width=7.2cm]{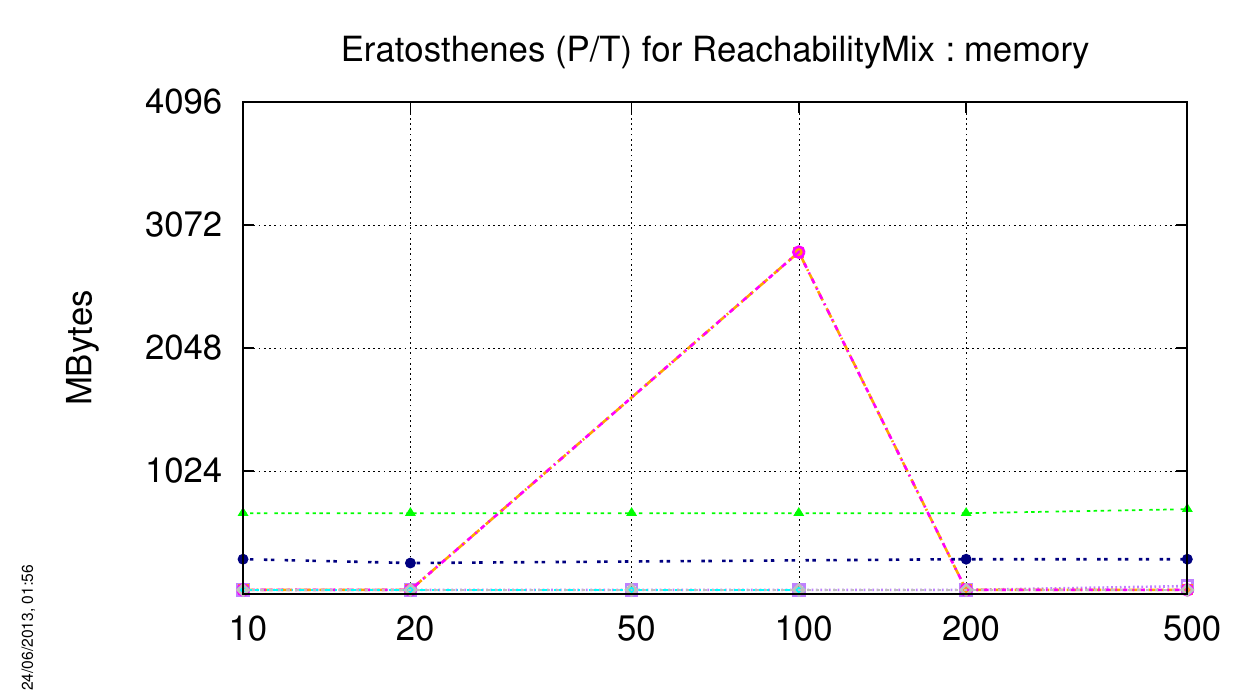}
   \includegraphics[width=7.2cm]{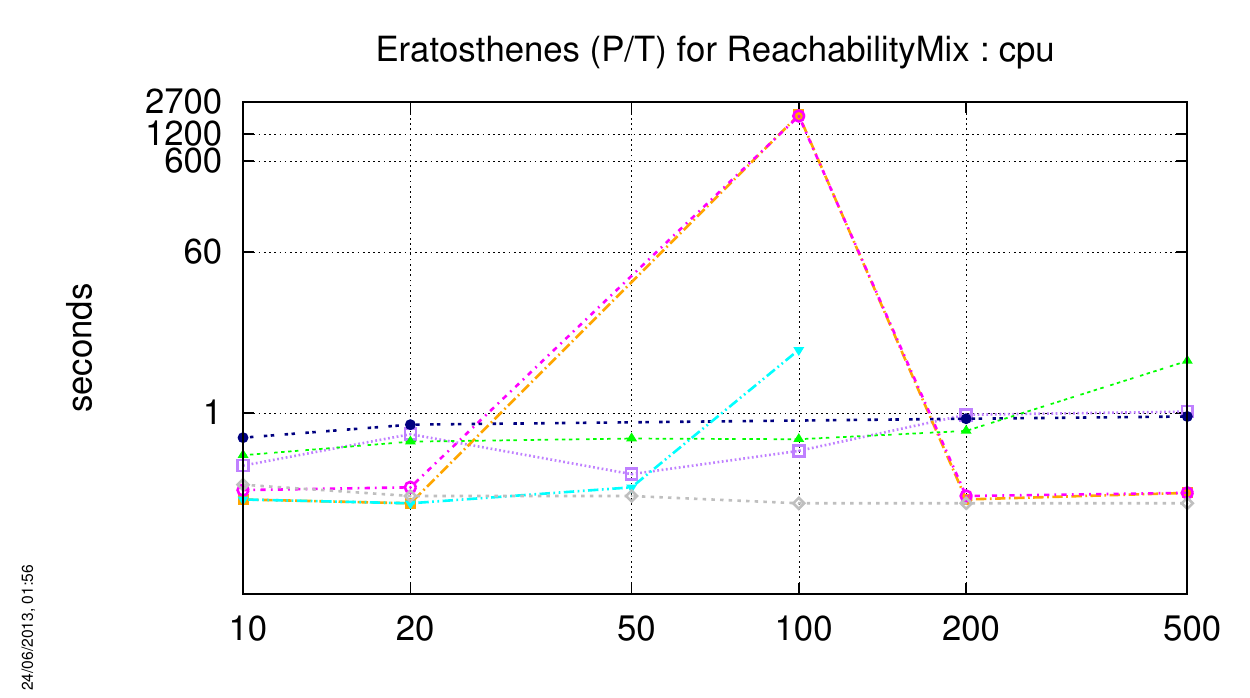}

   \includegraphics[height=1cm]{figures/tools-legend.pdf}
\end{center}

\subsubsection{\acs{FMS-PT}}
The charts below respectively show how tools compete with this ``Known'' model (memory and CPU).

\index{Performances!ReachabilityMix!FMS (P/T)}
\begin{center}
   \includegraphics[width=7.2cm]{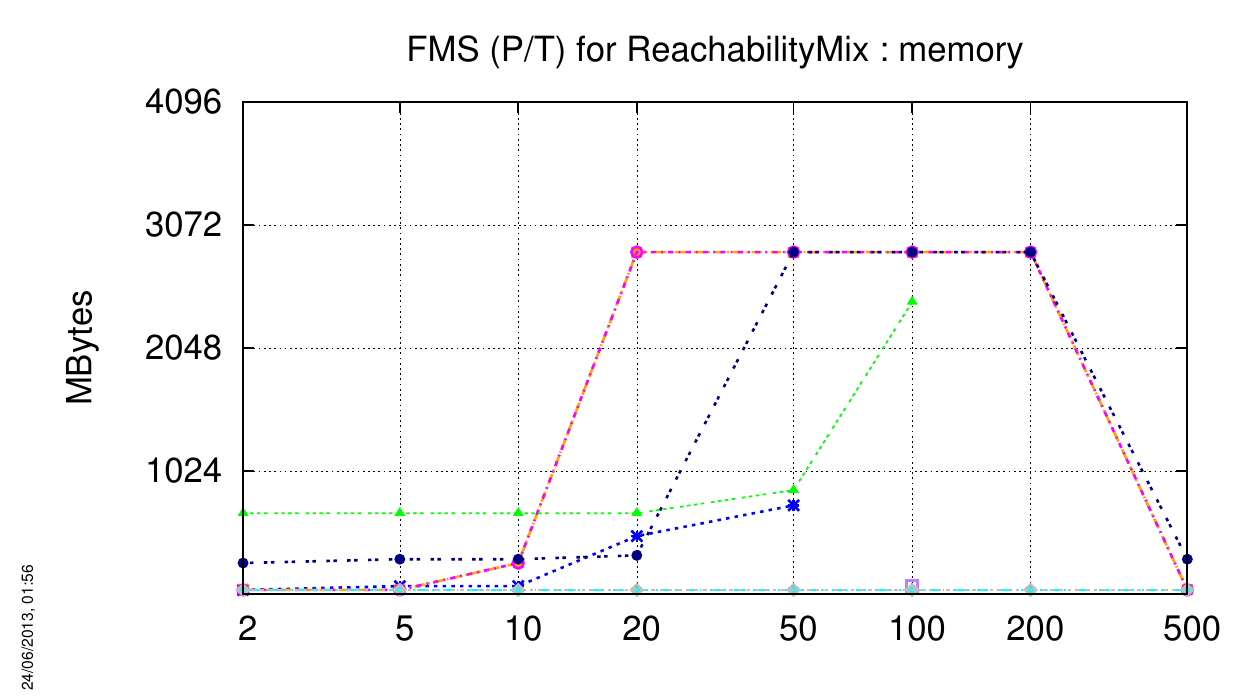}
   \includegraphics[width=7.2cm]{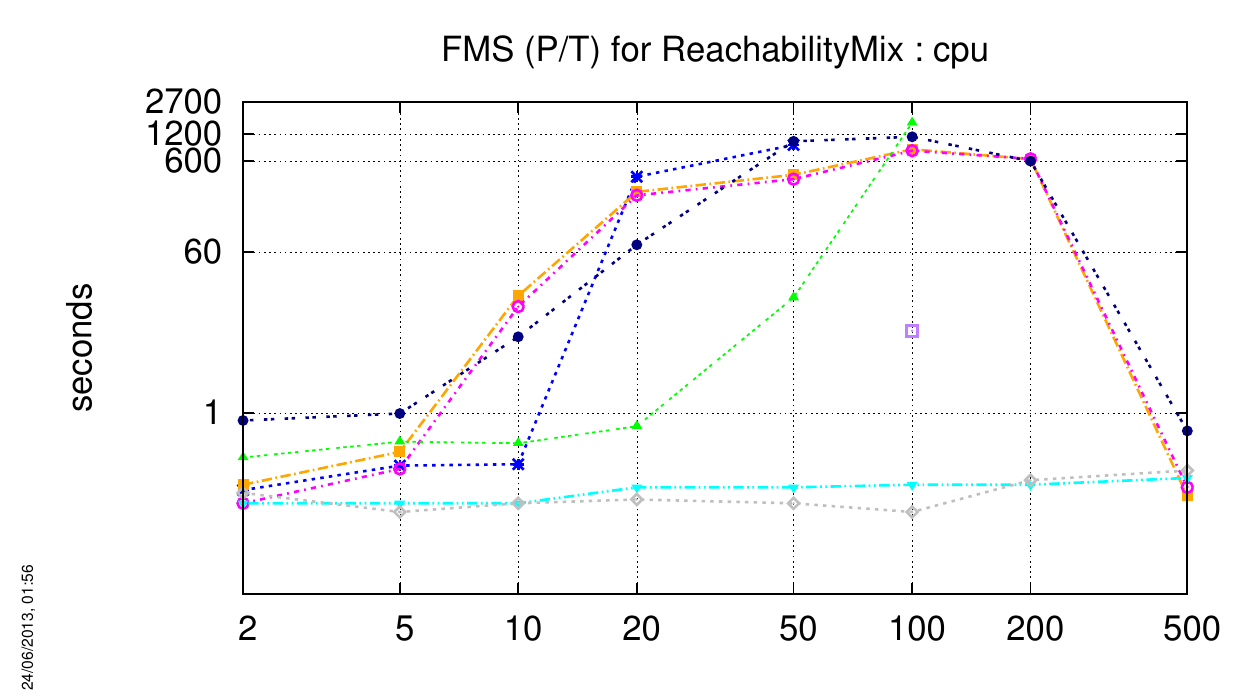}

   \includegraphics[height=1cm]{figures/tools-legend.pdf}
\end{center}

\subsubsection{\acs{GlobalRessAlloc-COL}}
No instance of this model could be computed for the \textbf{ReachabilityMix} examination.

\subsubsection{\acs{GlobalRessAlloc-PT}}
The charts below respectively show how tools compete with this ``Known'' model (memory and CPU).

\index{Performances!ReachabilityMix!GlobalRessAlloc (P/T)}
\begin{center}
   \includegraphics[width=7.2cm]{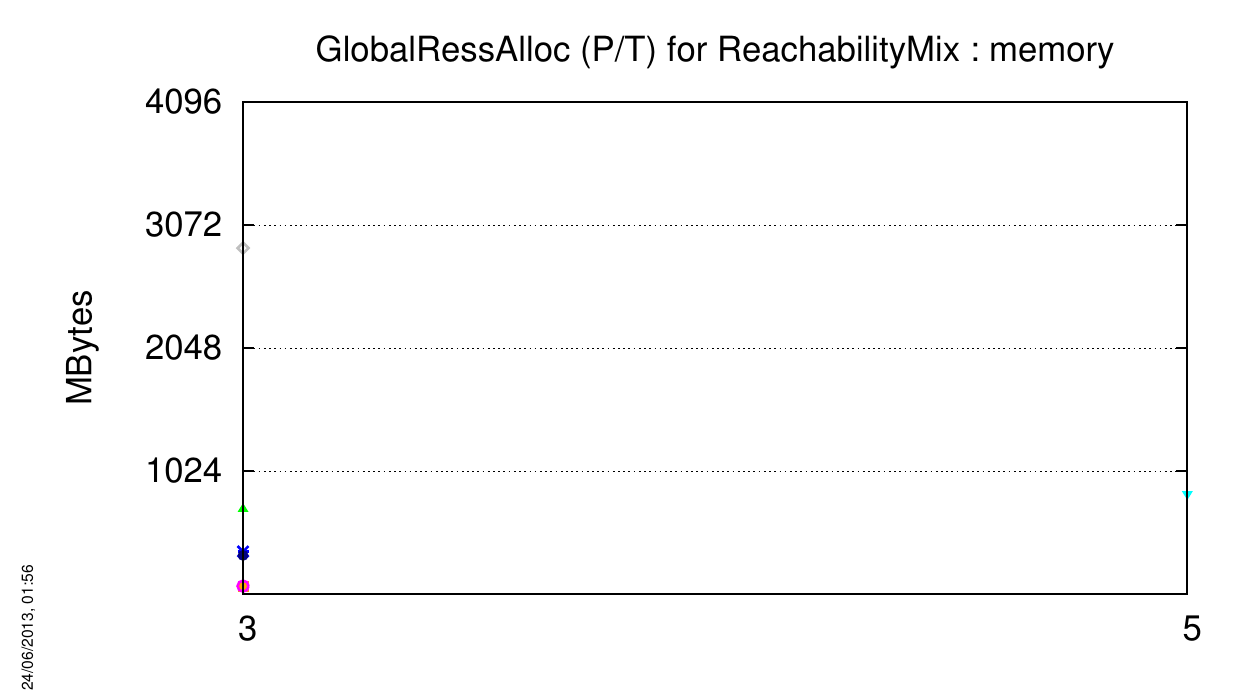}
   \includegraphics[width=7.2cm]{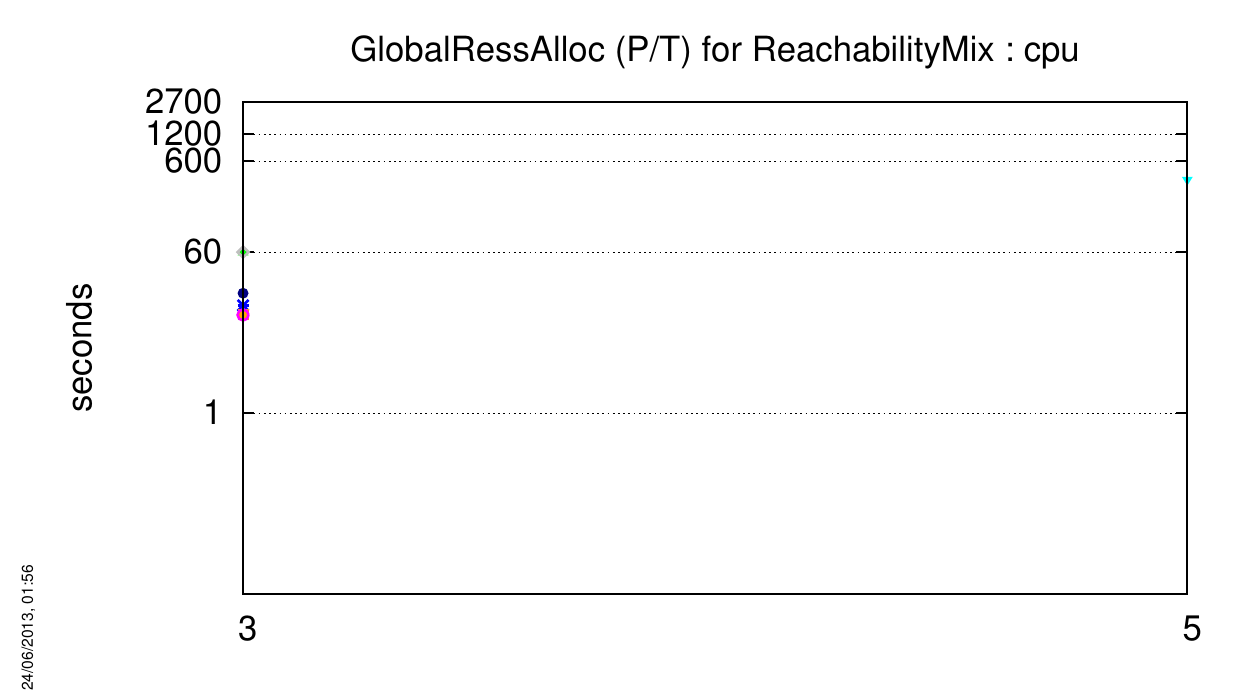}

   \includegraphics[height=1cm]{figures/tools-legend.pdf}
\end{center}

\subsubsection{\acs{Kanban-PT}}
The charts below respectively show how tools compete with this ``Known'' model (memory and CPU).

\index{Performances!ReachabilityMix!Kanban (P/T)}
\begin{center}
   \includegraphics[width=7.2cm]{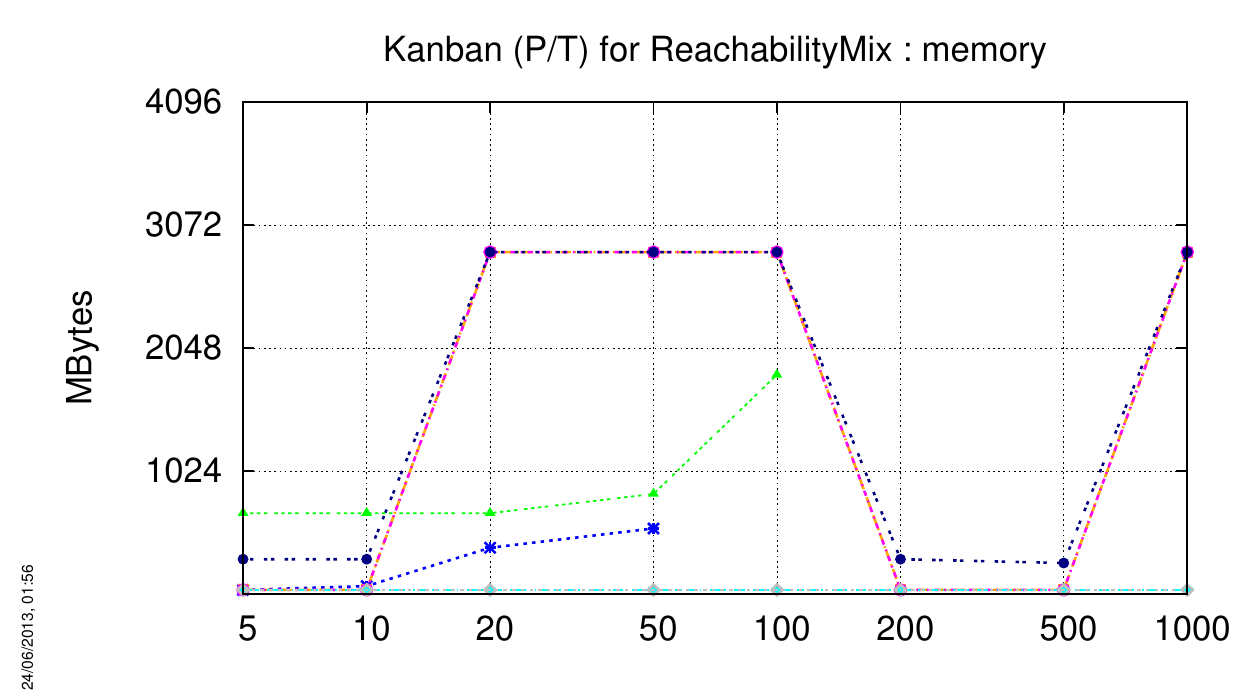}
   \includegraphics[width=7.2cm]{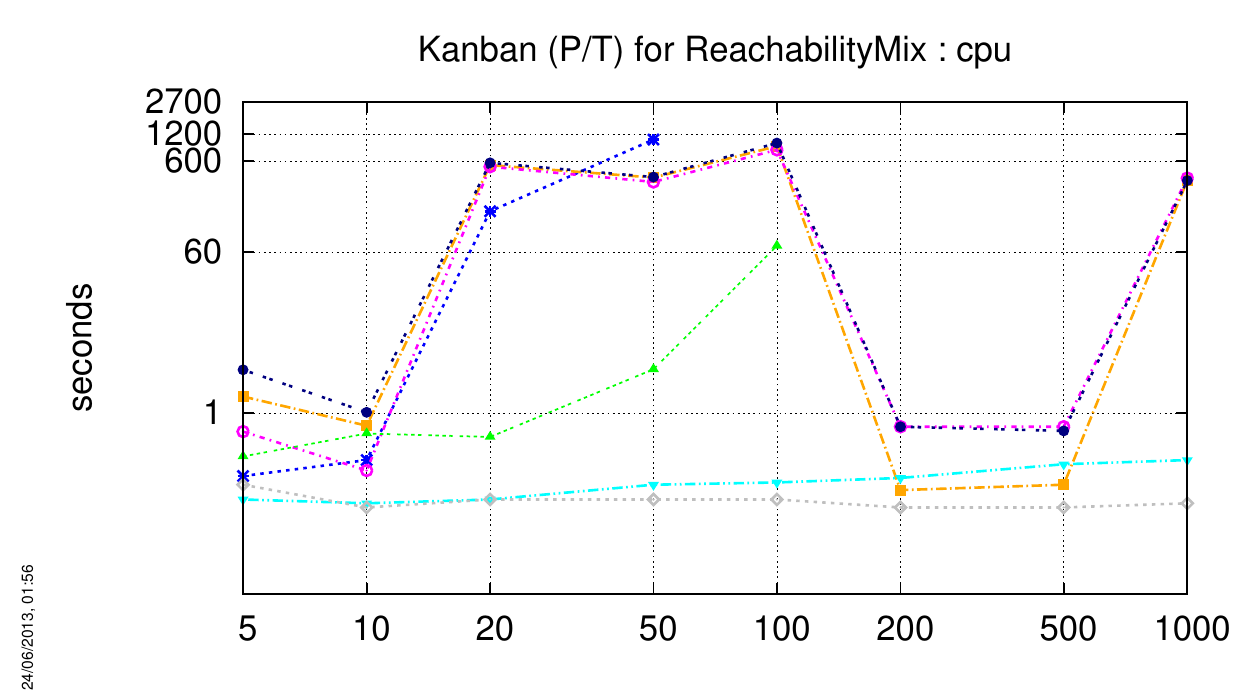}

   \includegraphics[height=1cm]{figures/tools-legend.pdf}
\end{center}

\subsubsection{\acs{LamportFastMutEx-COL}}
The charts below respectively show how tools compete with this ``Known'' model (memory and CPU).

\index{Performances!ReachabilityMix!LamportFastMutEx (Colored)}
\begin{center}
   \includegraphics[width=7.2cm]{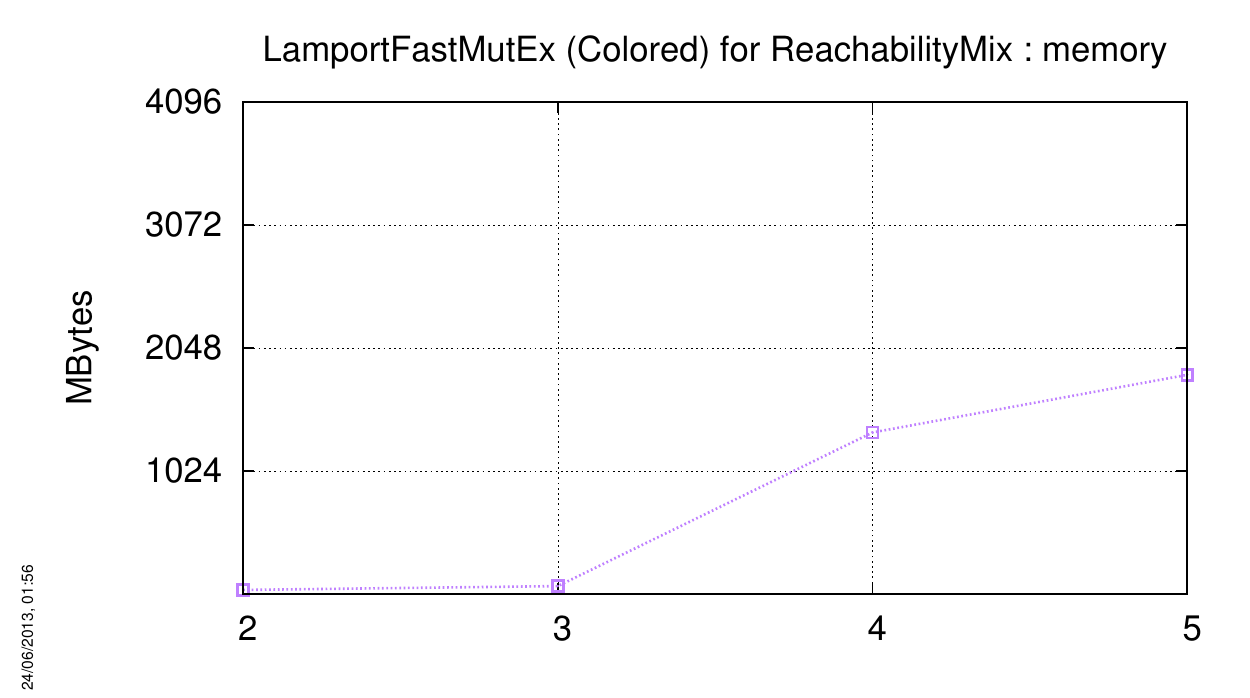}
   \includegraphics[width=7.2cm]{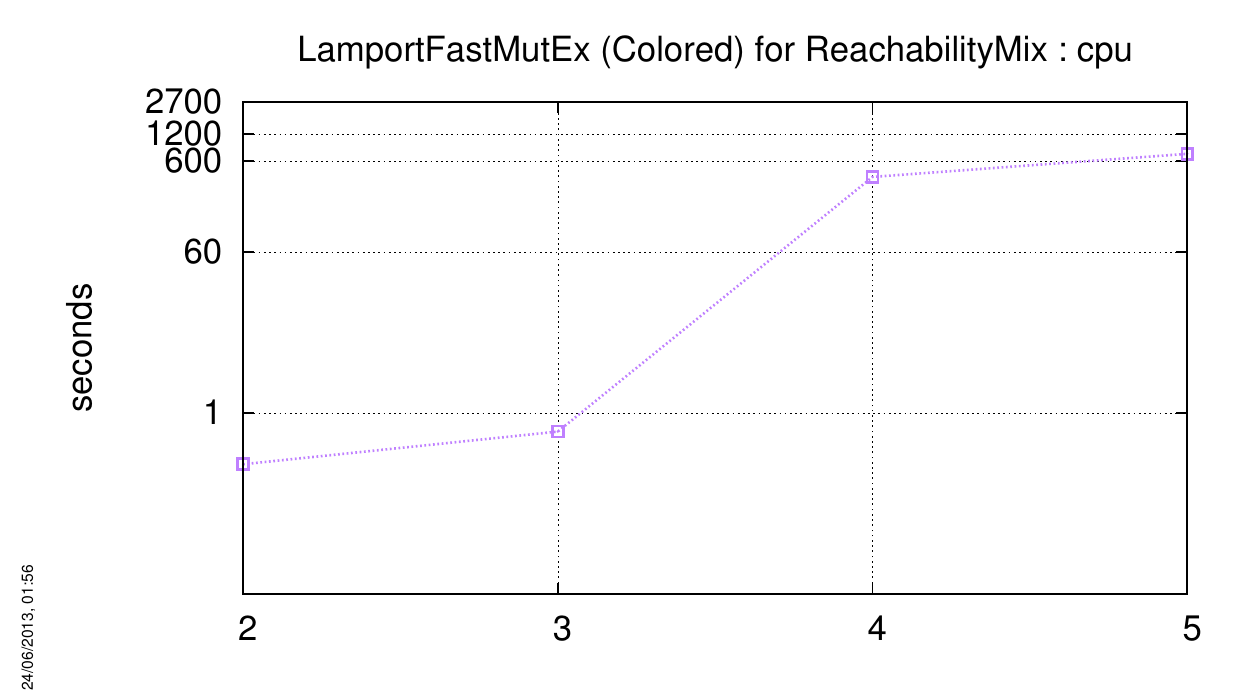}

   \includegraphics[height=1cm]{figures/tools-legend.pdf}
\end{center}

\subsubsection{\acs{LamportFastMutEx-PT}}
The charts below respectively show how tools compete with this ``Known'' model (memory and CPU).

\index{Performances!ReachabilityMix!LamportFastMutEx (P/T)}
\begin{center}
   \includegraphics[width=7.2cm]{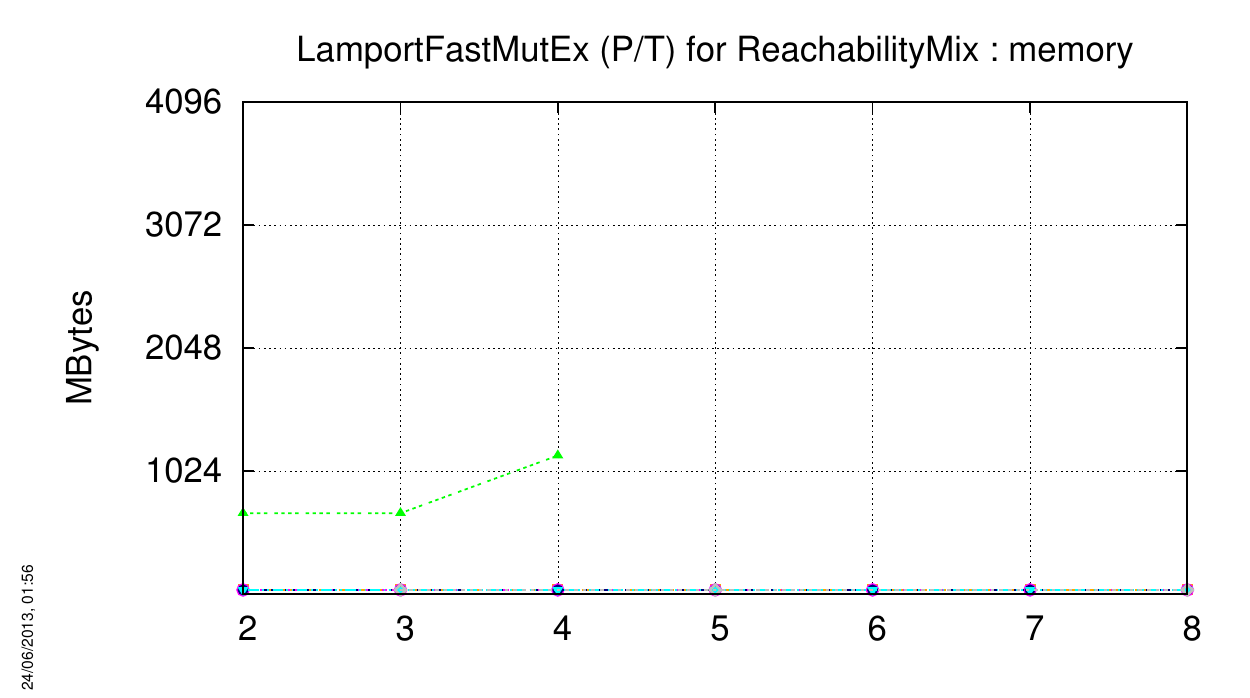}
   \includegraphics[width=7.2cm]{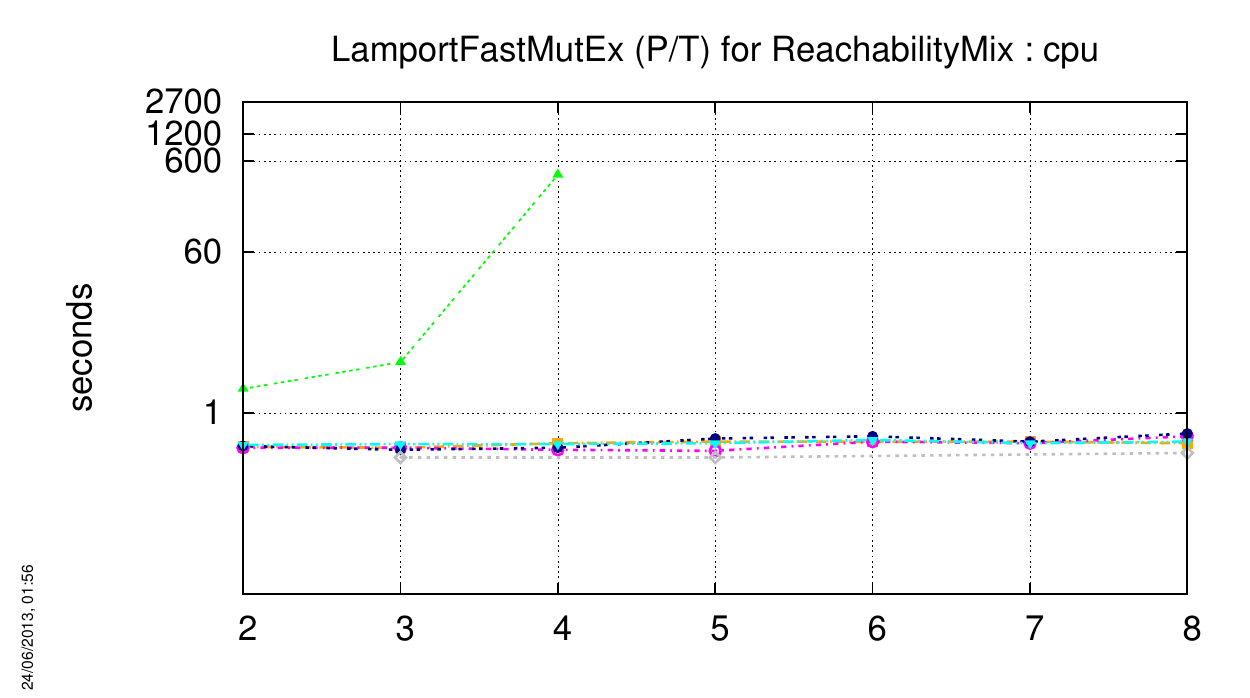}

   \includegraphics[height=1cm]{figures/tools-legend.pdf}
\end{center}

\subsubsection{\acs{MAPK-PT}}
The charts below respectively show how tools compete with this ``Known'' model (memory and CPU).

\index{Performances!ReachabilityMix!MAPK (P/T)}
\begin{center}
   \includegraphics[width=7.2cm]{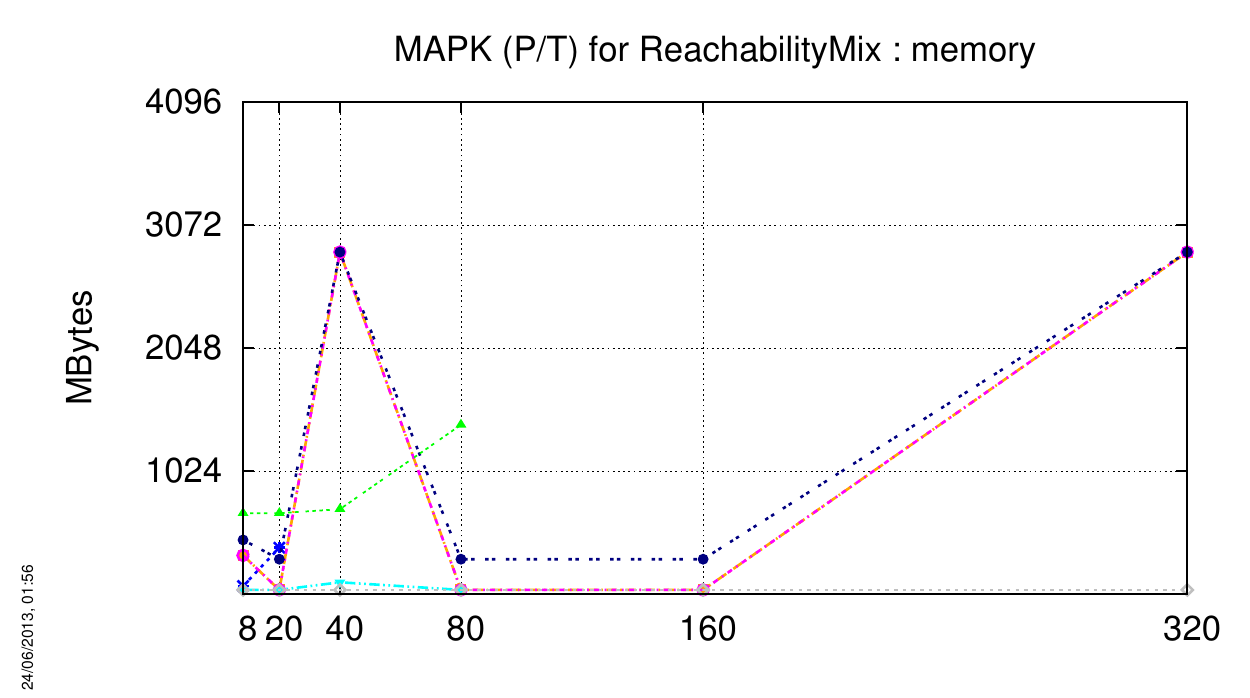}
   \includegraphics[width=7.2cm]{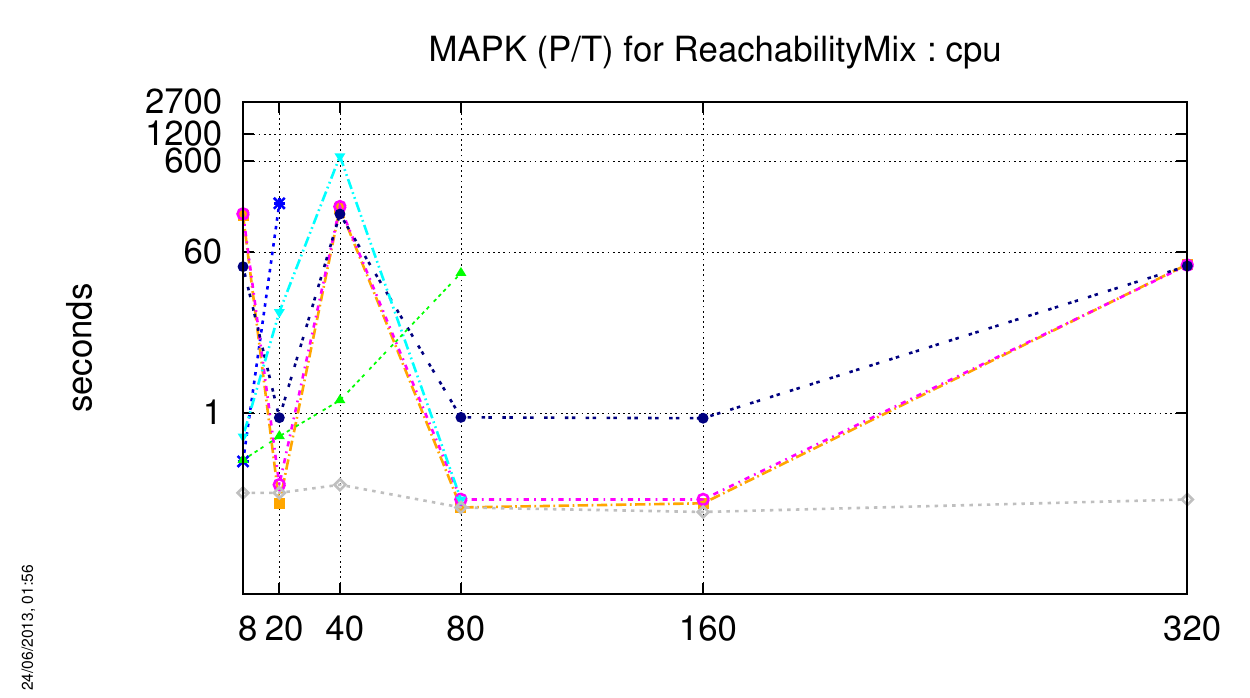}

   \includegraphics[height=1cm]{figures/tools-legend.pdf}
\end{center}

\subsubsection{\acs{NeoElection-COL}}
The charts below respectively show how tools compete with this ``Known'' model (memory and CPU).

\index{Performances!ReachabilityMix!NeoElection (Colored)}
\begin{center}
   \includegraphics[width=7.2cm]{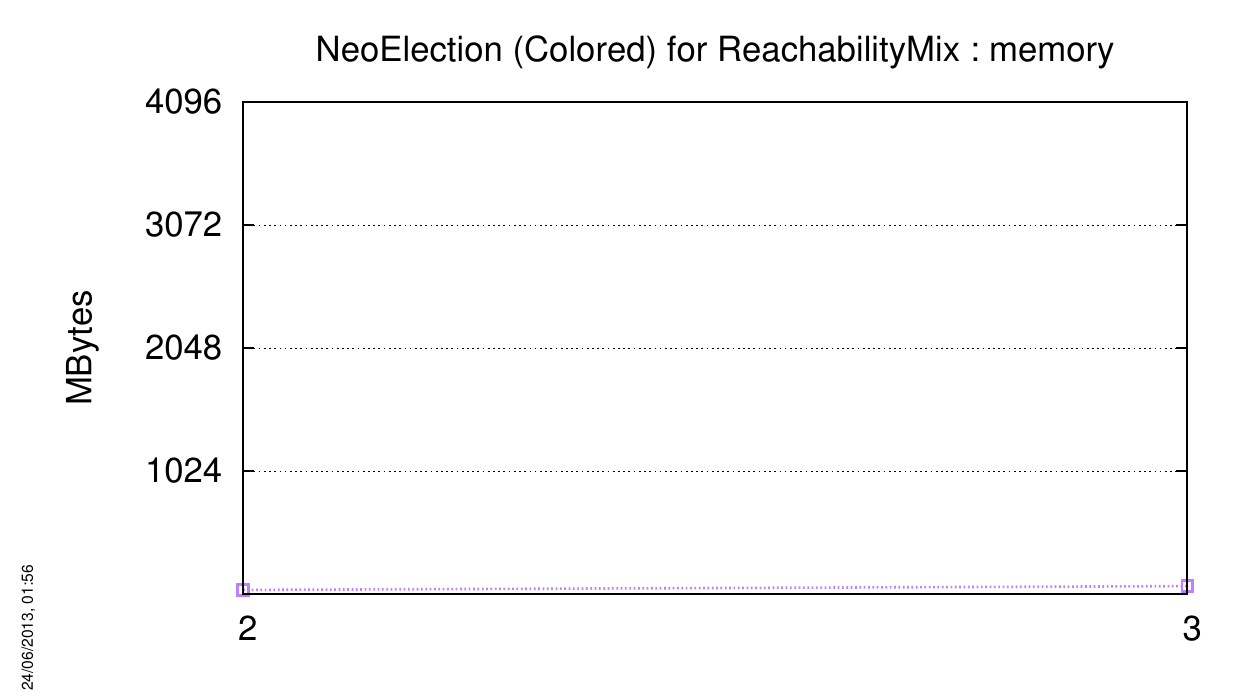}
   \includegraphics[width=7.2cm]{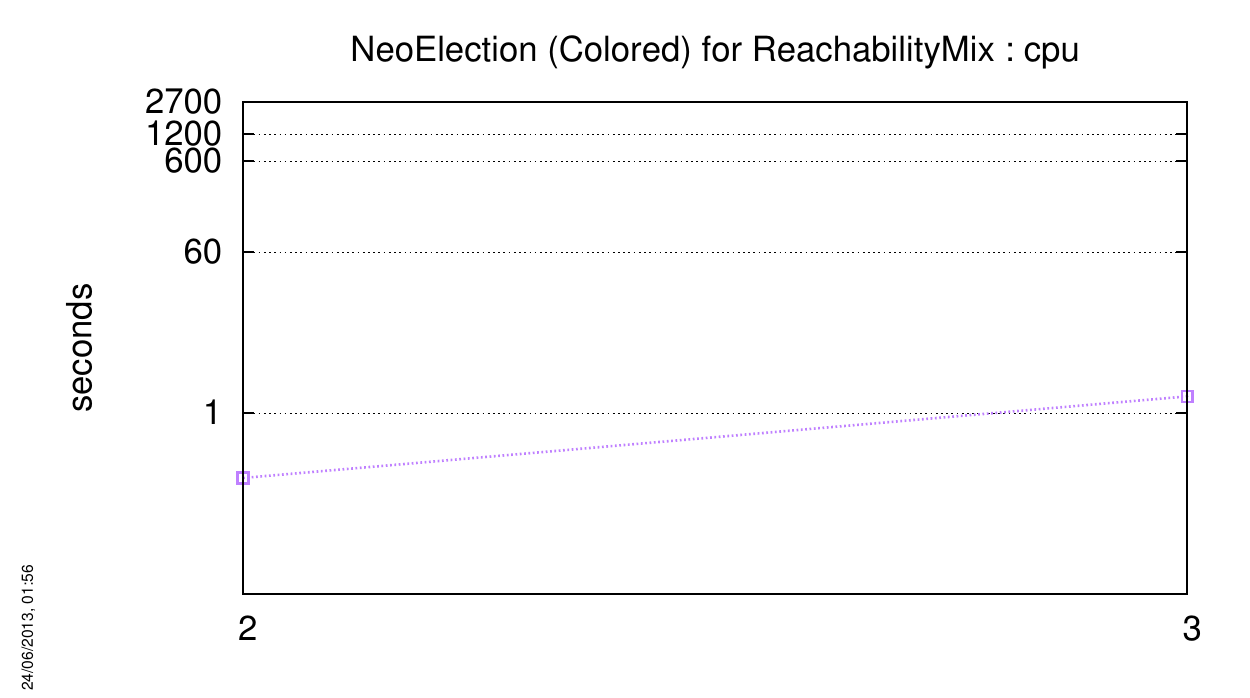}

   \includegraphics[height=1cm]{figures/tools-legend.pdf}
\end{center}

\subsubsection{\acs{NeoElection-PT}}
The charts below respectively show how tools compete with this ``Known'' model (memory and CPU).

\index{Performances!ReachabilityMix!NeoElection (P/T)}
\begin{center}
   \includegraphics[width=7.2cm]{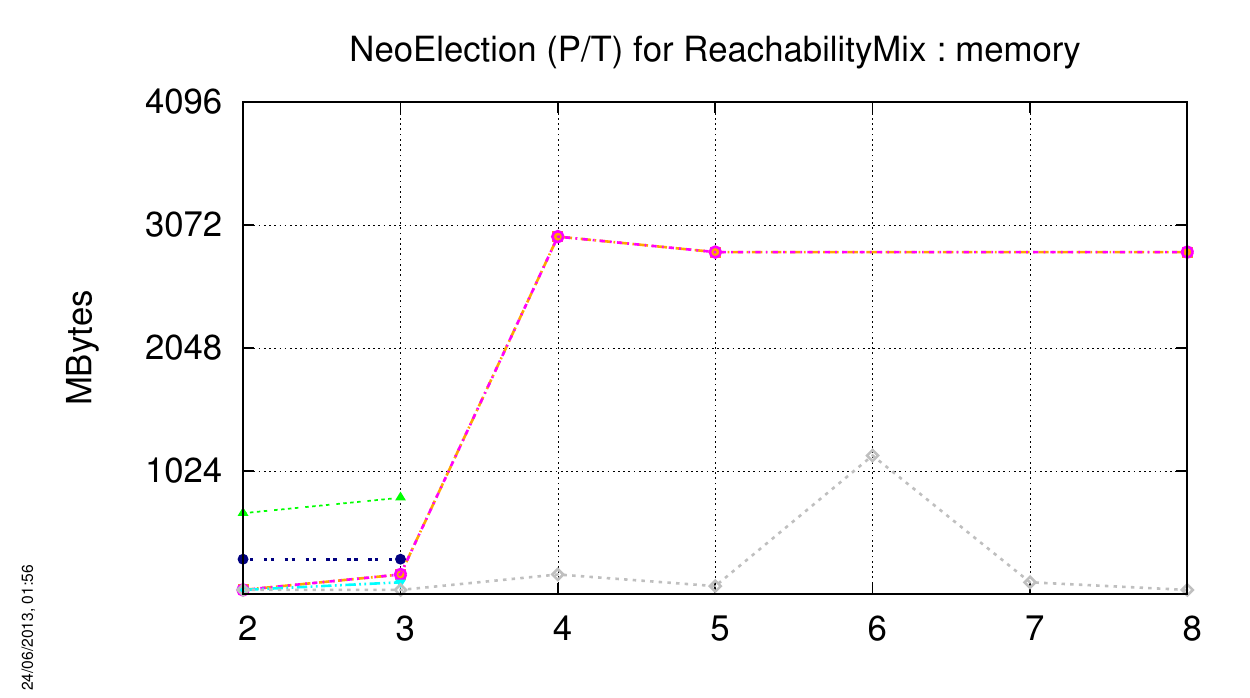}
   \includegraphics[width=7.2cm]{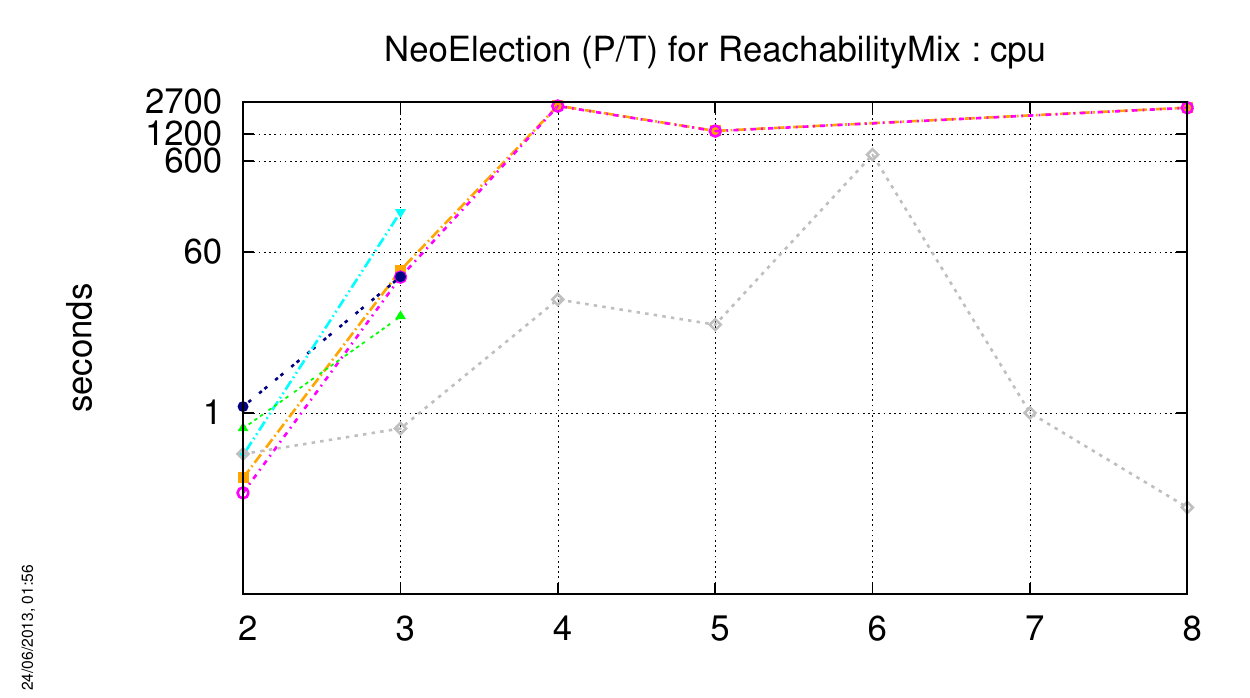}

   \includegraphics[height=1cm]{figures/tools-legend.pdf}
\end{center}

\subsubsection{\acs{PermAdmissibility-COL}}
The charts below respectively show how tools compete with this ``Known'' model (memory and CPU).

\index{Performances!ReachabilityMix!PermAdmissibility (Colored)}
\begin{center}
   \includegraphics[width=7.2cm]{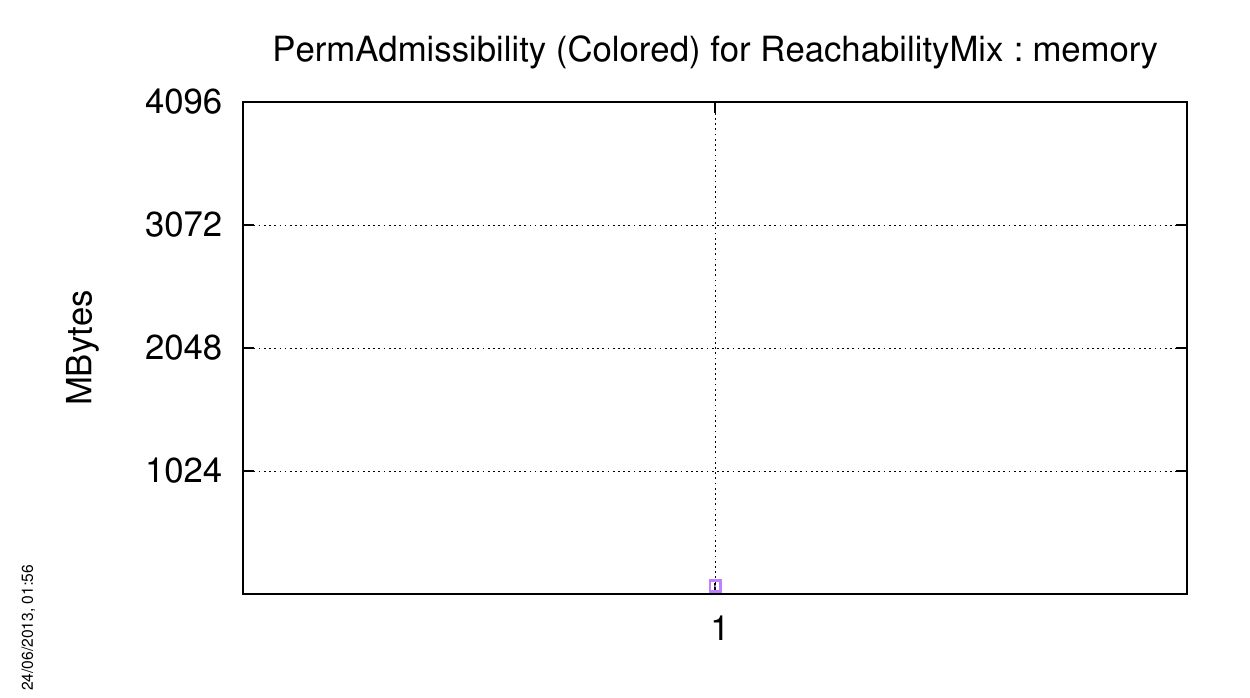}
   \includegraphics[width=7.2cm]{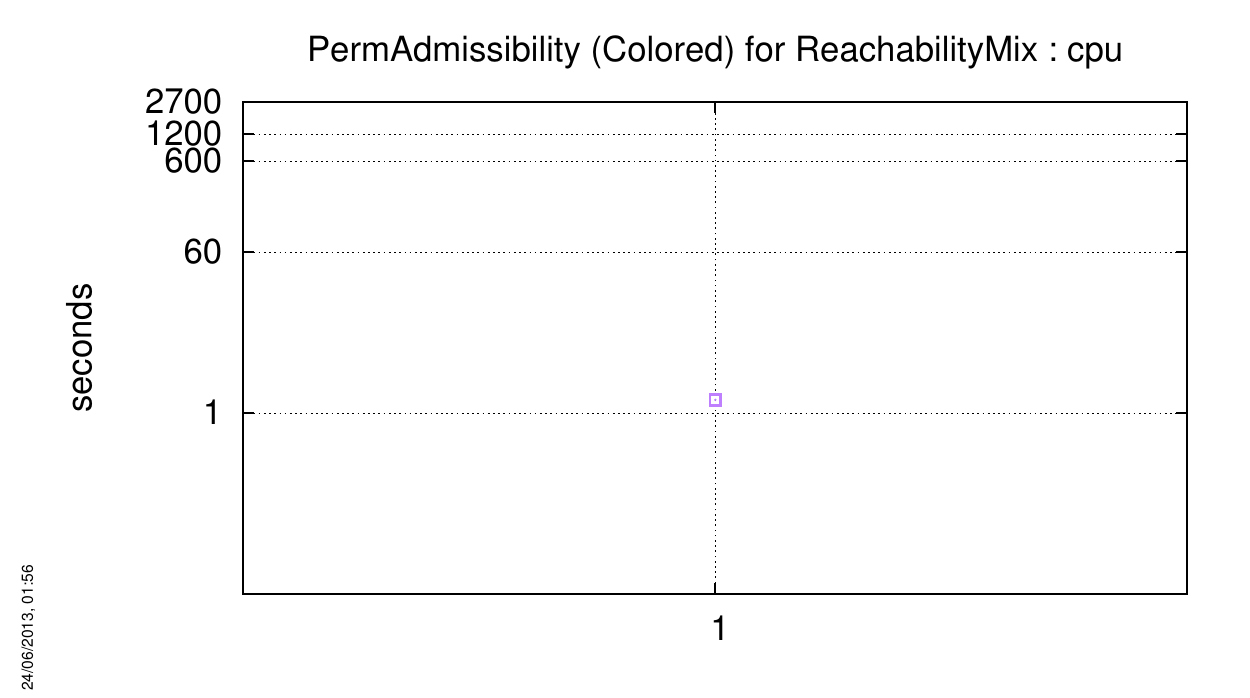}

   \includegraphics[height=1cm]{figures/tools-legend.pdf}
\end{center}

\subsubsection{\acs{PermAdmissibility-PT}}
The charts below respectively show how tools compete with this ``Known'' model (memory and CPU).

\index{Performances!ReachabilityMix!PermAdmissibility (P/T)}
\begin{center}
   \includegraphics[width=7.2cm]{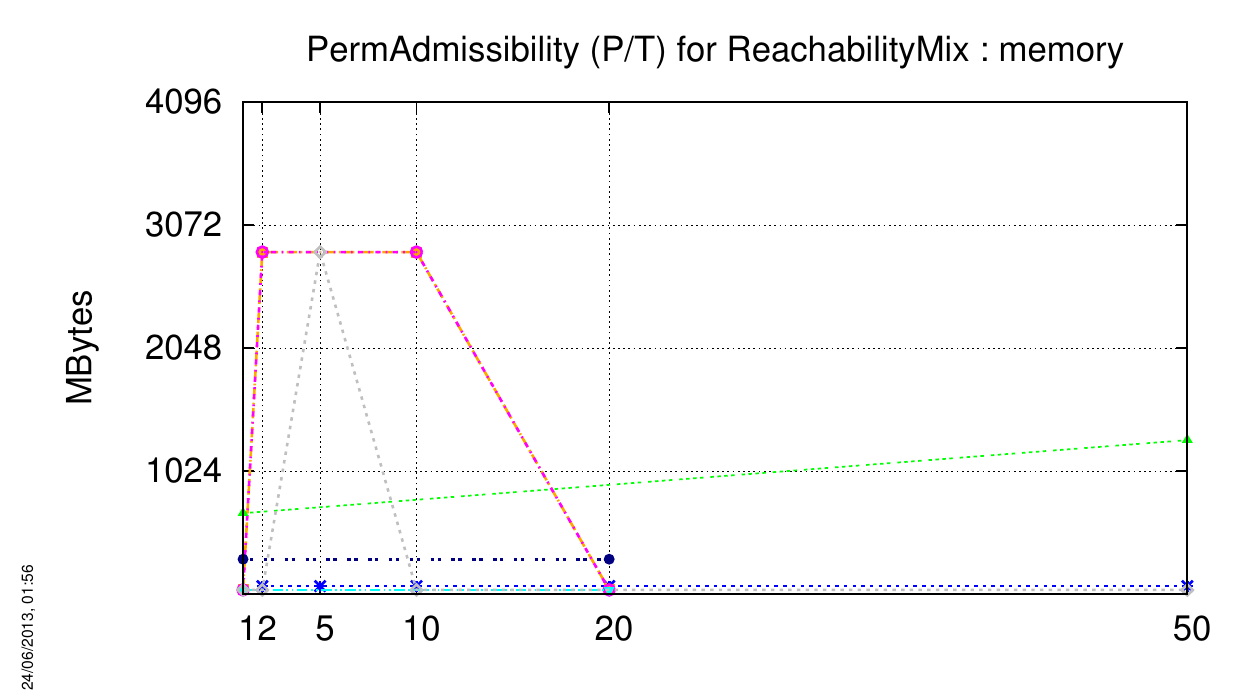}
   \includegraphics[width=7.2cm]{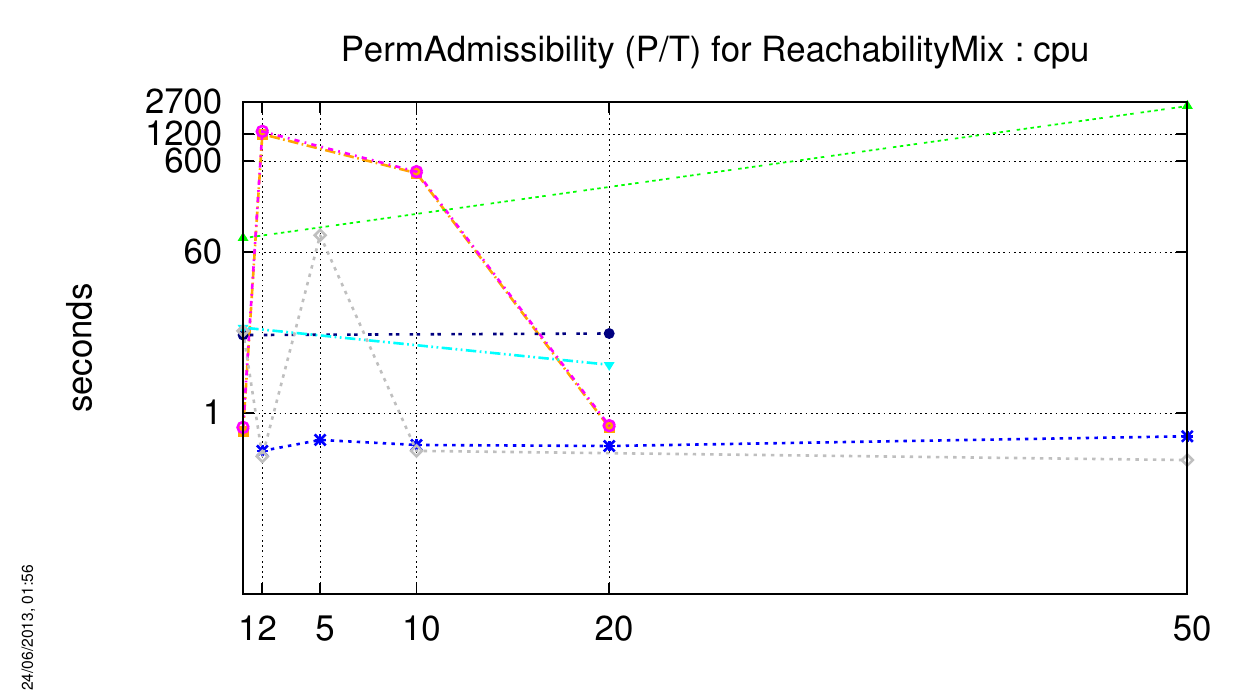}

   \includegraphics[height=1cm]{figures/tools-legend.pdf}
\end{center}

\subsubsection{\acs{Peterson-COL}}
The charts below respectively show how tools compete with this ``Known'' model (memory and CPU).

\index{Performances!ReachabilityMix!Peterson (Colored)}
\begin{center}
   \includegraphics[width=7.2cm]{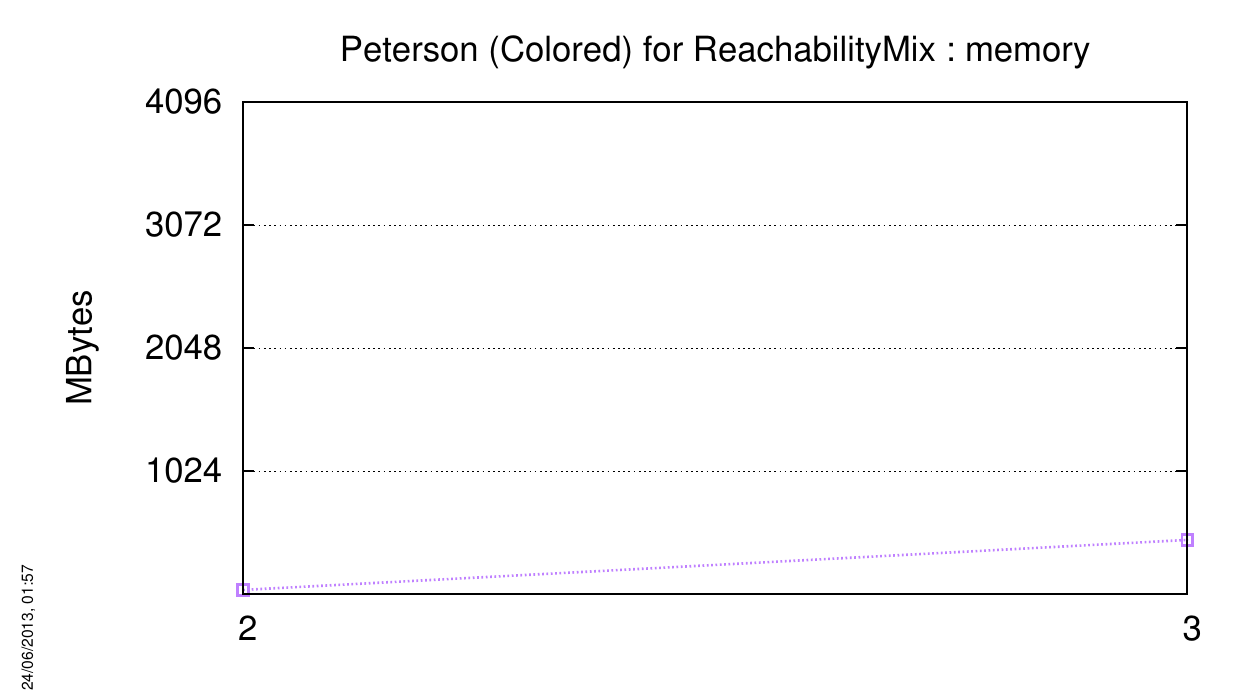}
   \includegraphics[width=7.2cm]{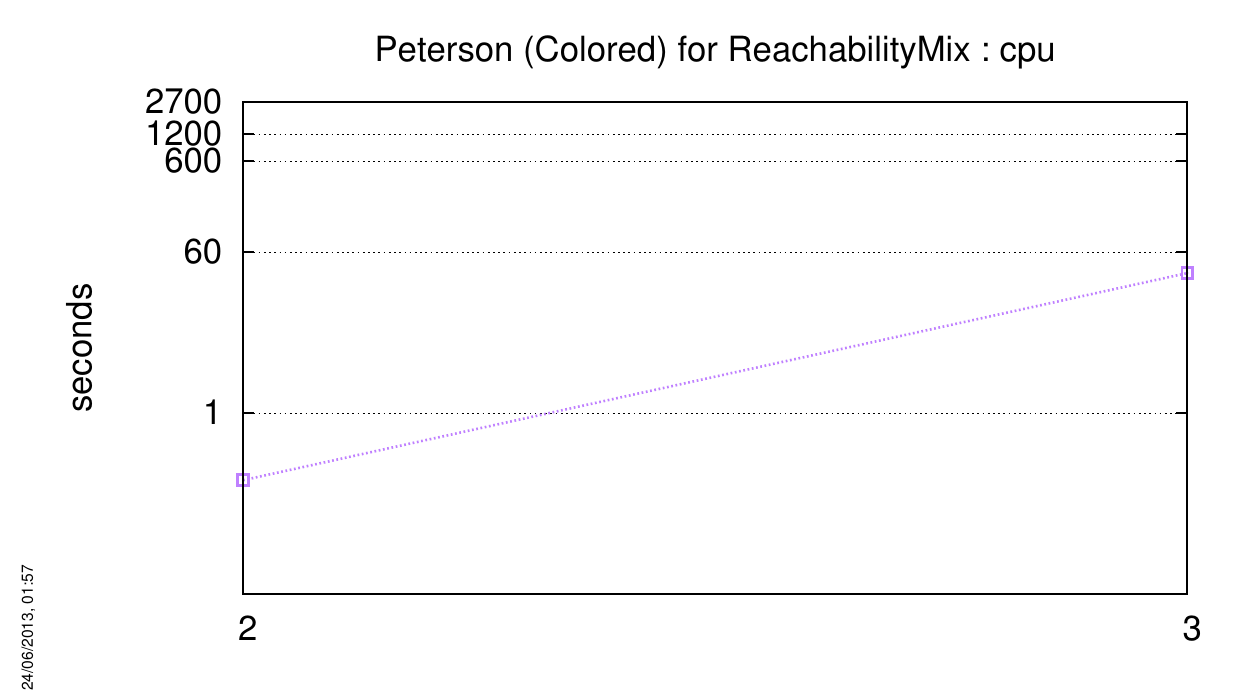}

   \includegraphics[height=1cm]{figures/tools-legend.pdf}
\end{center}

\subsubsection{\acs{Peterson-PT}}
The charts below respectively show how tools compete with this ``Known'' model (memory and CPU).

\index{Performances!ReachabilityMix!Peterson (P/T)}
\begin{center}
   \includegraphics[width=7.2cm]{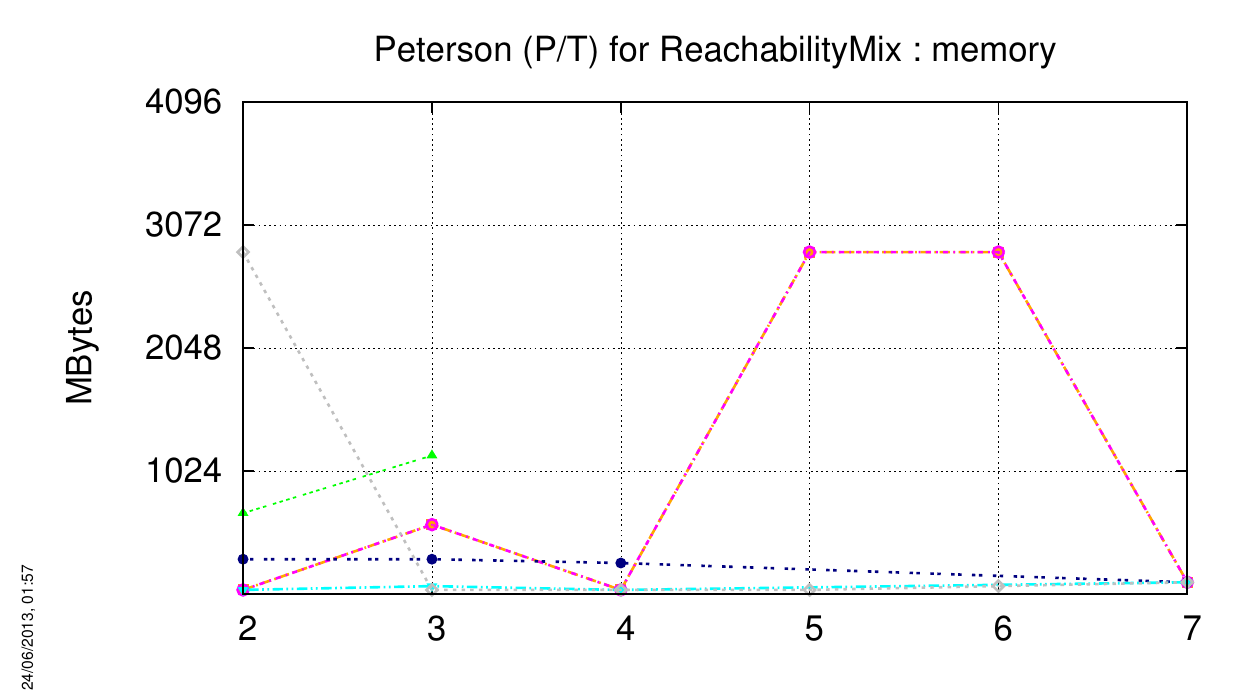}
   \includegraphics[width=7.2cm]{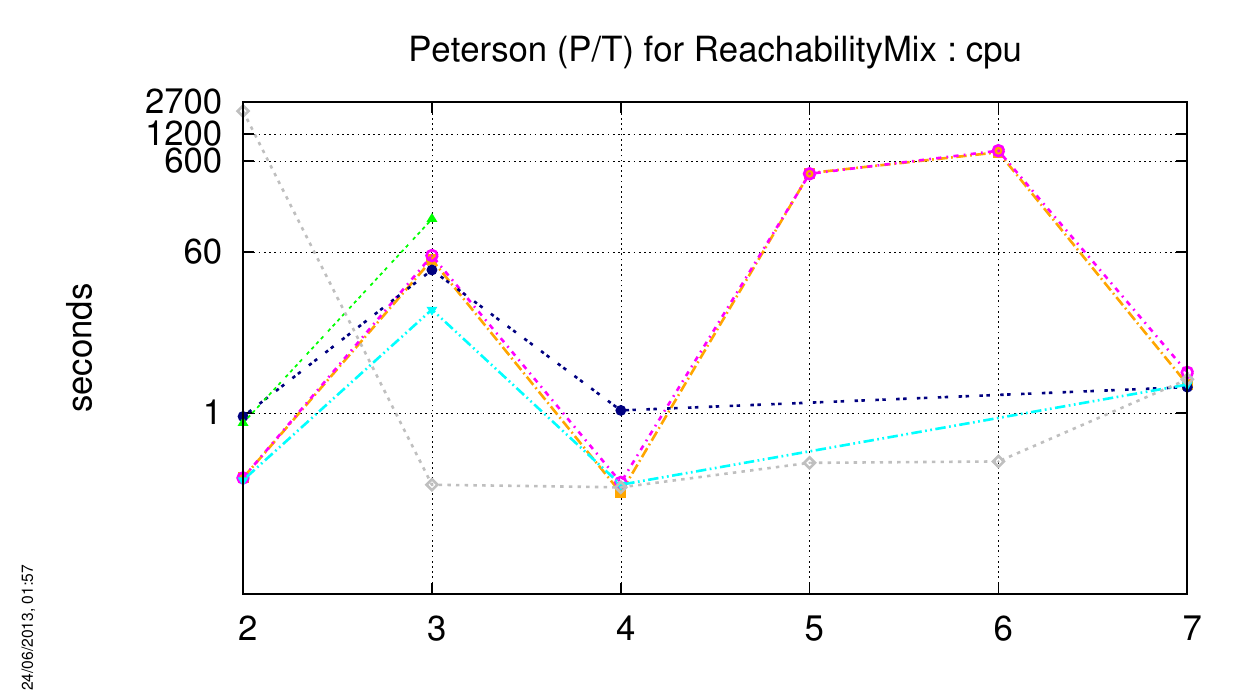}

   \includegraphics[height=1cm]{figures/tools-legend.pdf}
\end{center}

\subsubsection{\acs{Philosophers-COL}}
No instance of this model could be computed for the \textbf{ReachabilityMix} examination.

\subsubsection{\acs{Philosophers-PT}}
The charts below respectively show how tools compete with this ``Known'' model (memory and CPU).

\index{Performances!ReachabilityMix!Philosophers (P/T)}
\begin{center}
   \includegraphics[width=7.2cm]{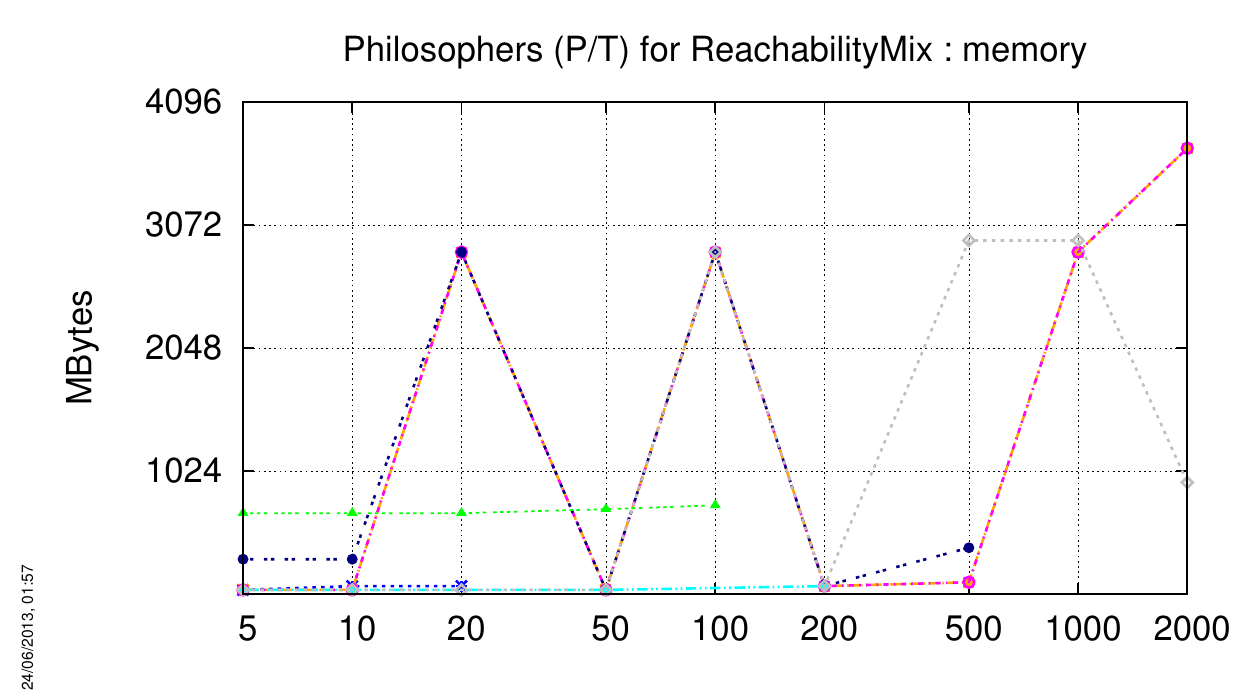}
   \includegraphics[width=7.2cm]{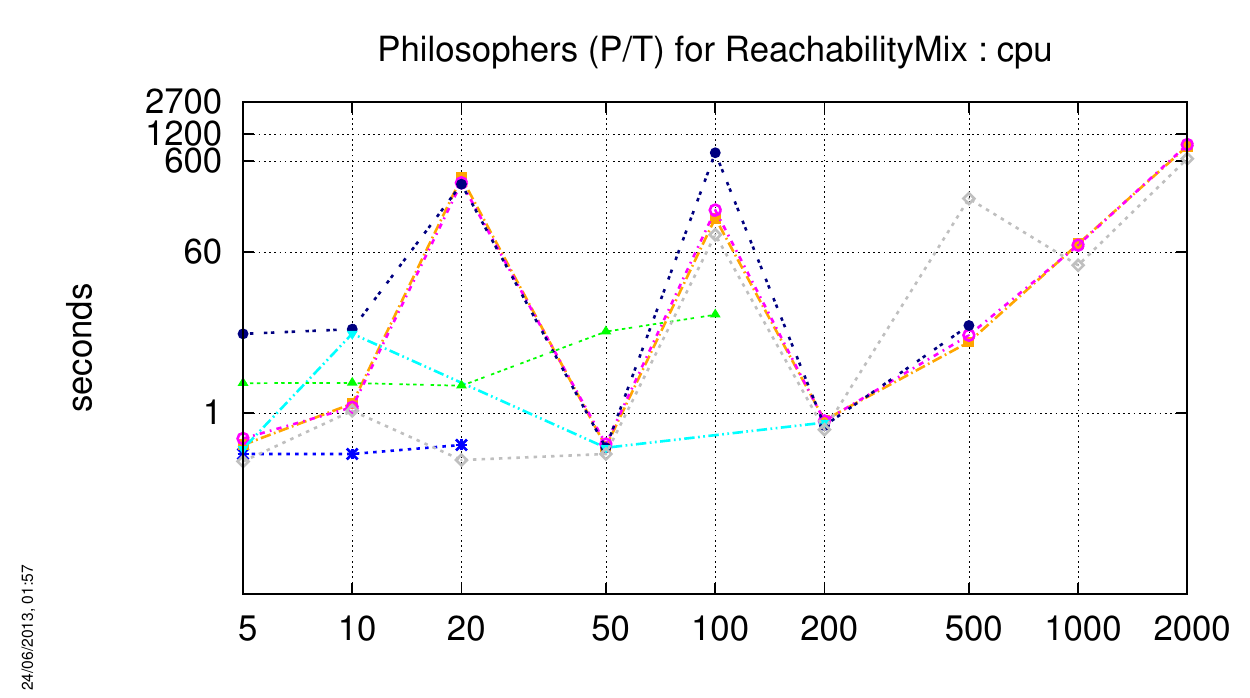}

   \includegraphics[height=1cm]{figures/tools-legend.pdf}
\end{center}

\subsubsection{\acs{PhilosophersDyn-COL}}
No instance of this model could be computed for the \textbf{ReachabilityMix} examination.

\subsubsection{\acs{PhilosophersDyn-PT}}
The charts below respectively show how tools compete with this ``Known'' model (memory and CPU).

\index{Performances!ReachabilityMix!PhilosophersDyn (P/T)}
\begin{center}
   \includegraphics[width=7.2cm]{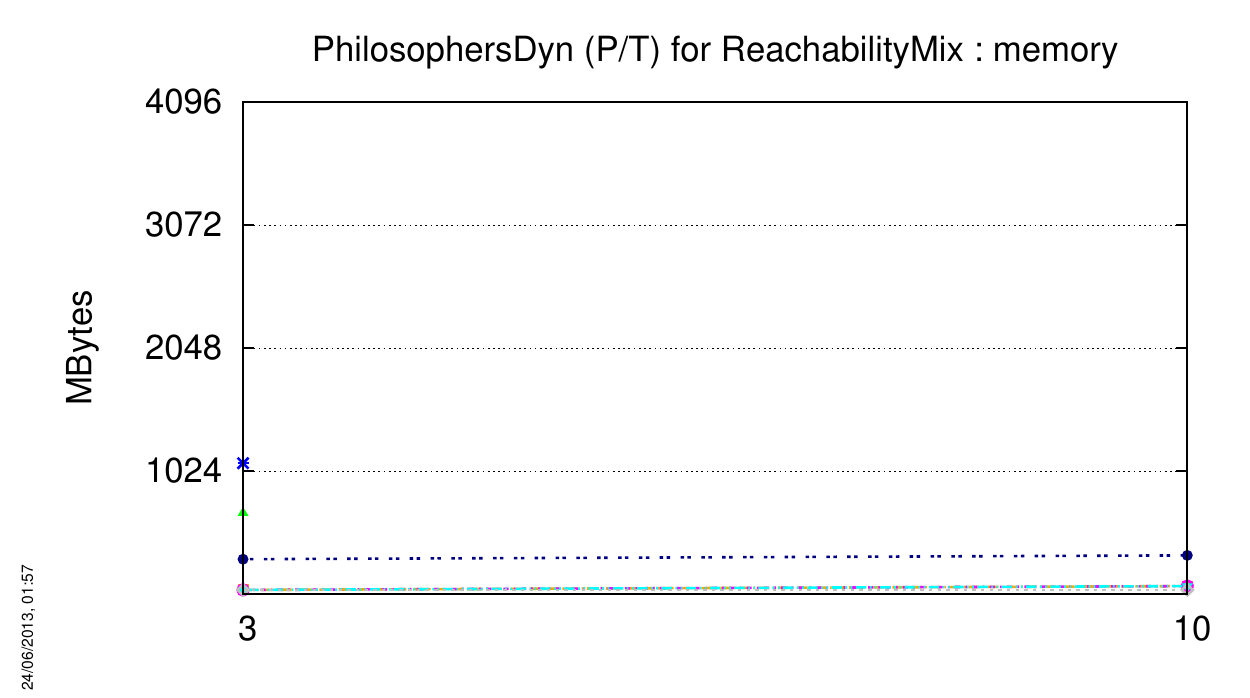}
   \includegraphics[width=7.2cm]{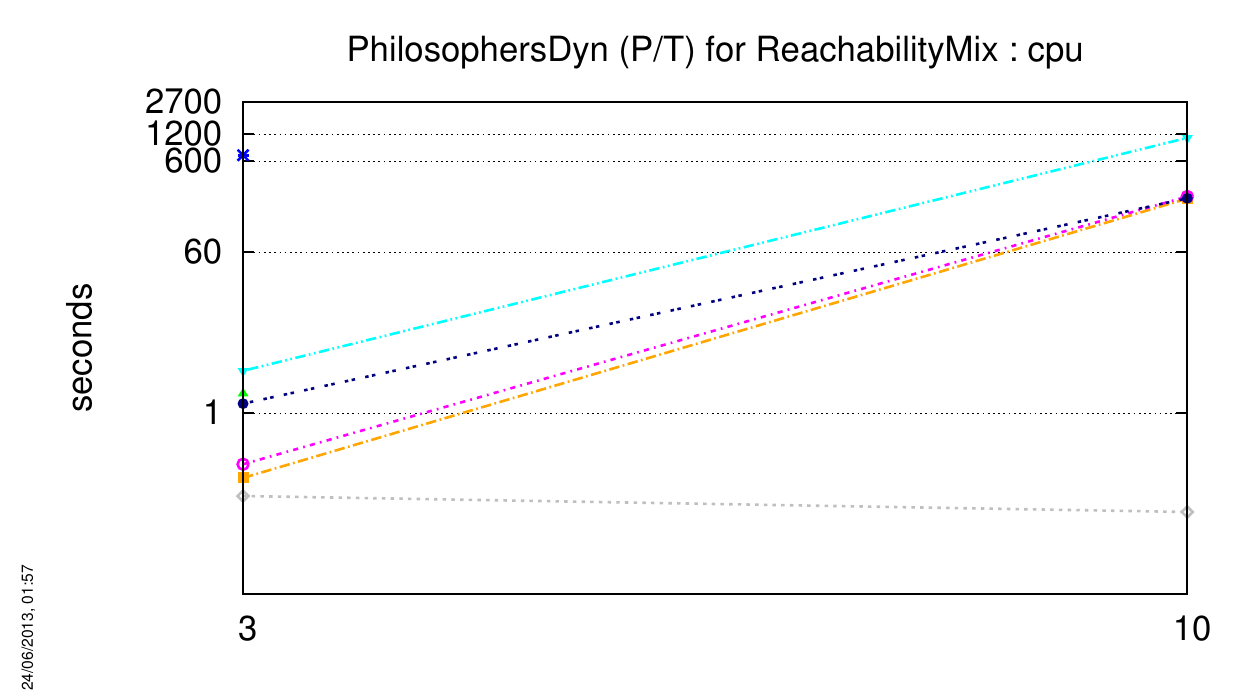}

   \includegraphics[height=1cm]{figures/tools-legend.pdf}
\end{center}

\subsubsection{\acs{Planning-PT}}
No instance of this model could be computed for the \textbf{ReachabilityMix} examination.

\subsubsection{\acs{Railroad-PT}}
The charts below respectively show how tools compete with this ``Known'' model (memory and CPU).

\index{Performances!ReachabilityMix!Railroad (P/T)}
\begin{center}
   \includegraphics[width=7.2cm]{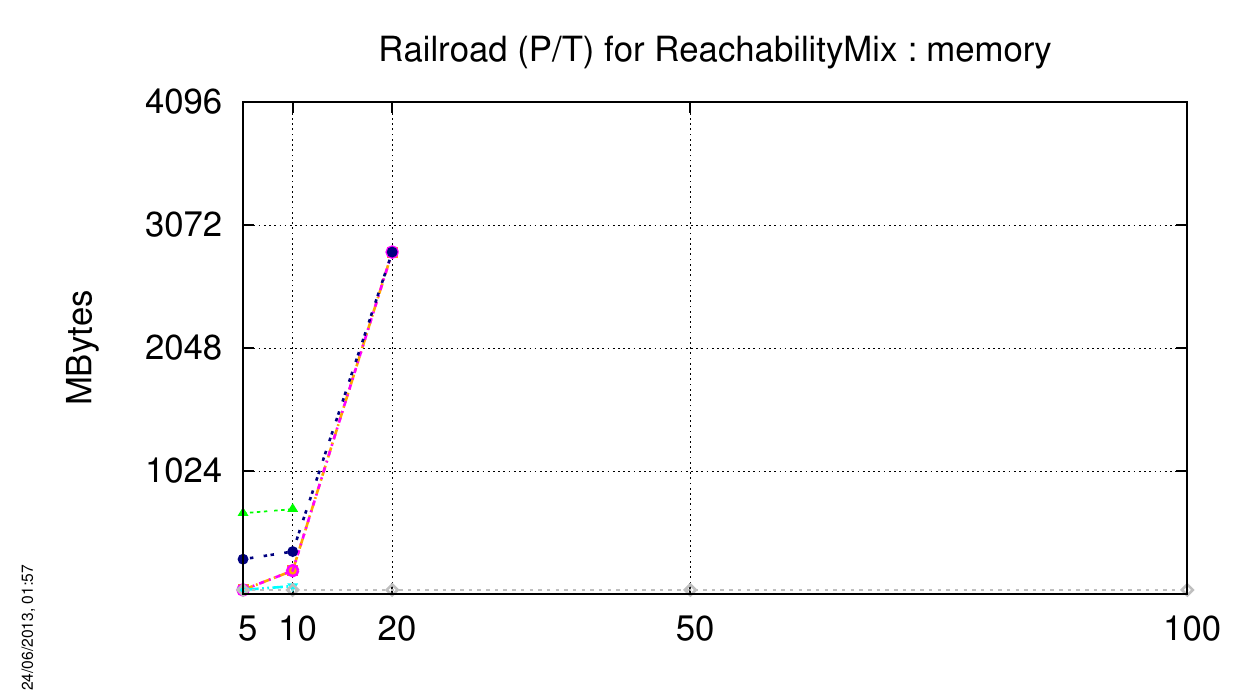}
   \includegraphics[width=7.2cm]{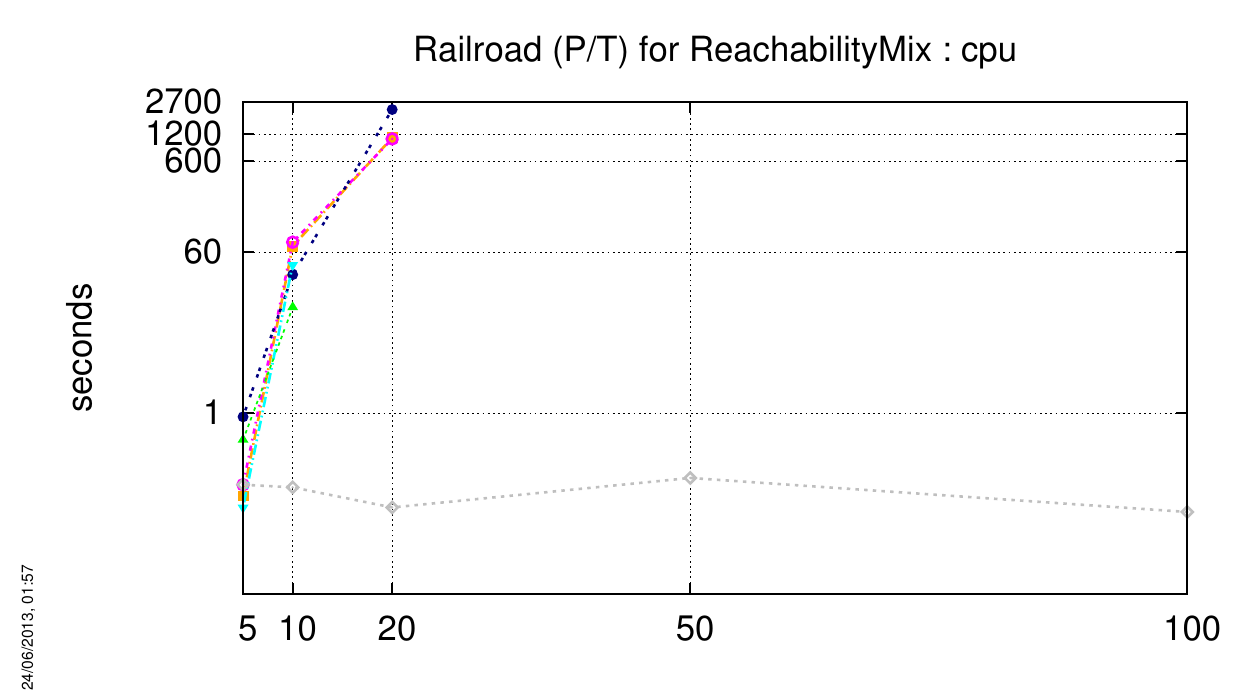}

   \includegraphics[height=1cm]{figures/tools-legend.pdf}
\end{center}

\subsubsection{\acs{RessAllocation-PT}}
The charts below respectively show how tools compete with this ``Known'' model (memory and CPU).

\index{Performances!ReachabilityMix!RessAllocation (P/T)}
\begin{center}
   \includegraphics[width=7.2cm]{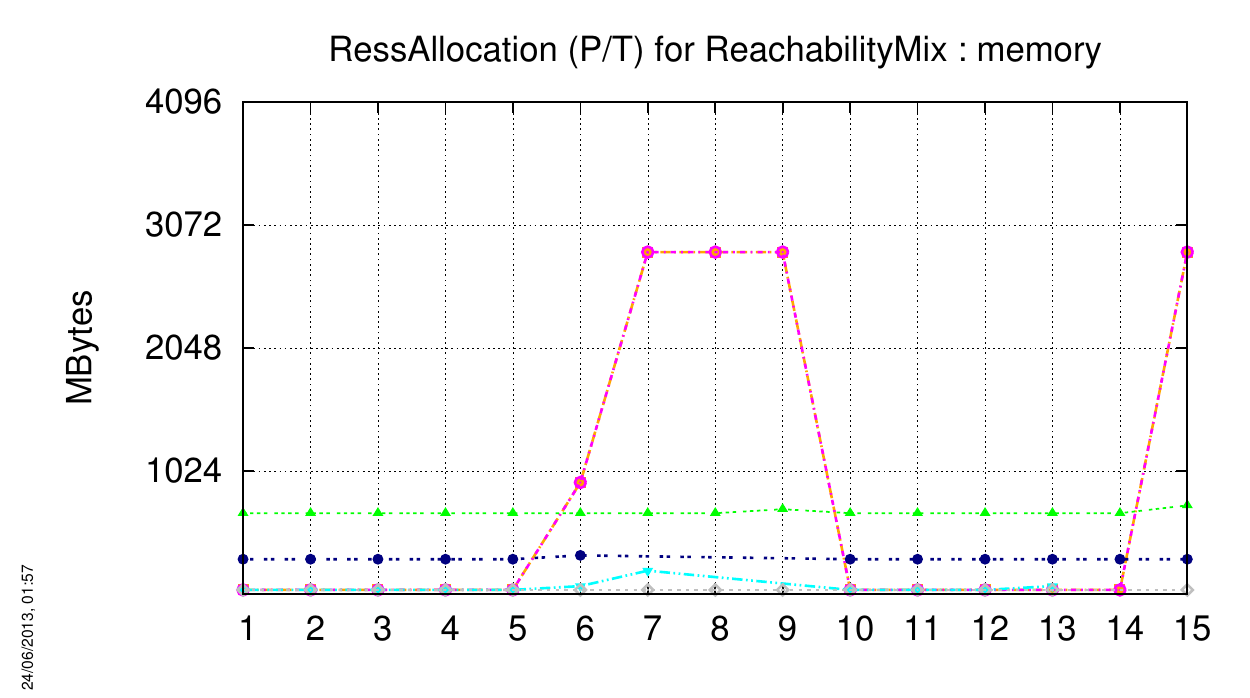}
   \includegraphics[width=7.2cm]{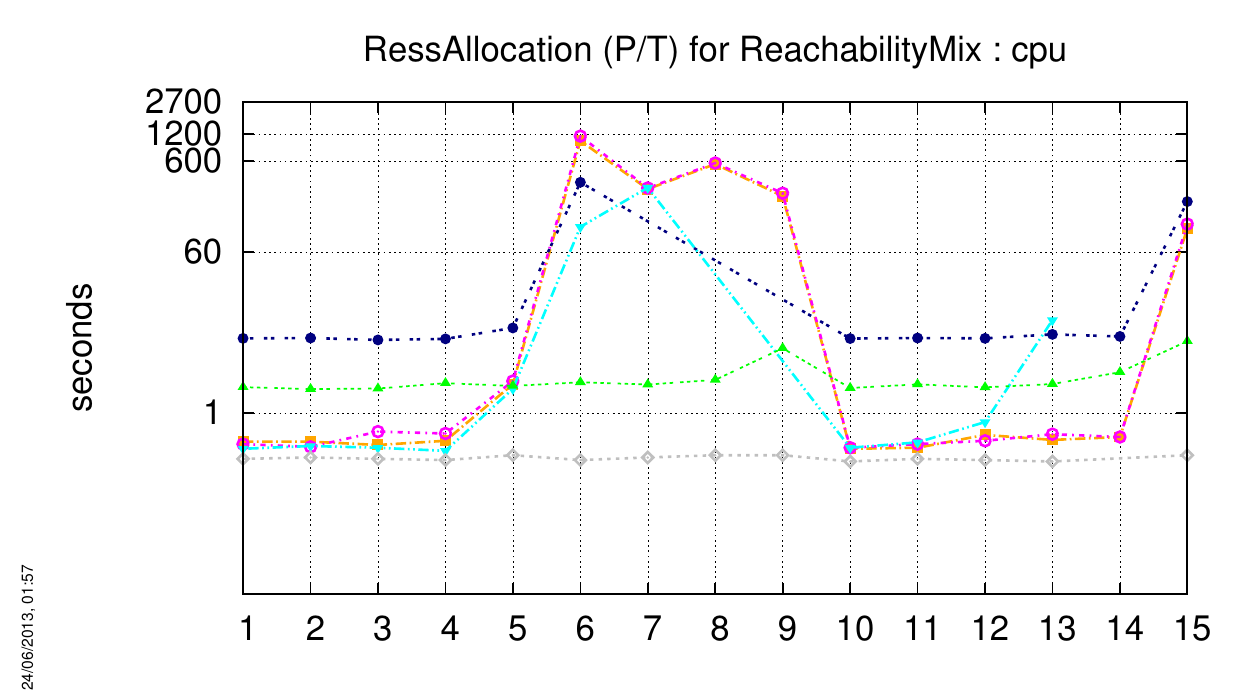}

   \includegraphics[height=1cm]{figures/tools-legend.pdf}
\end{center}

\subsubsection{\acs{Ring-PT}}
The charts below respectively show how tools compete with this ``Known'' model (memory and CPU).

\index{Performances!ReachabilityMix!Ring (P/T)}
\begin{center}
   \includegraphics[width=7.2cm]{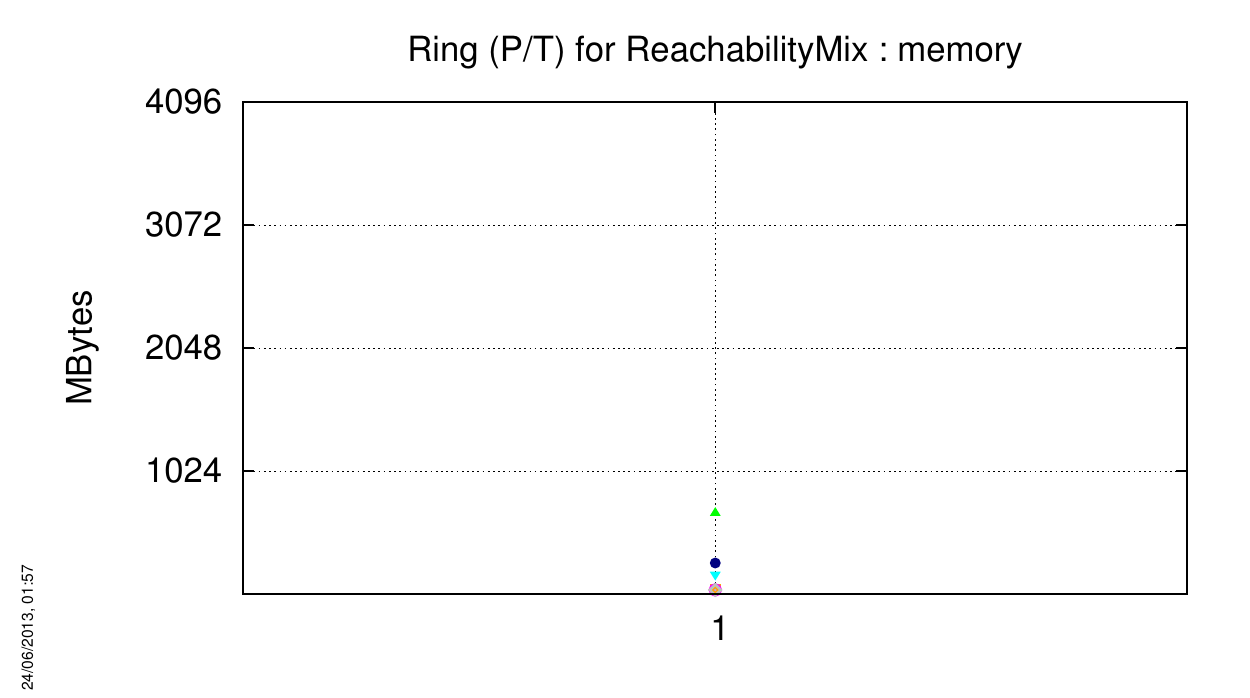}
   \includegraphics[width=7.2cm]{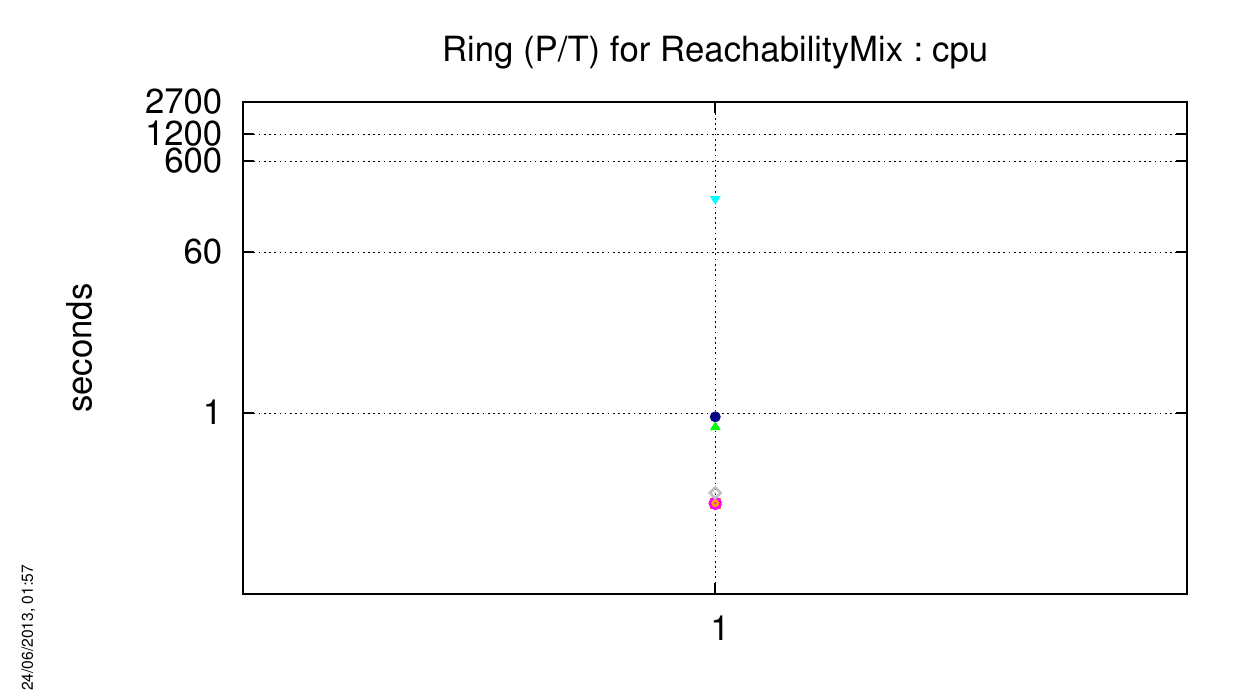}

   \includegraphics[height=1cm]{figures/tools-legend.pdf}
\end{center}

\subsubsection{\acs{RwMutex-PT}}
The charts below respectively show how tools compete with this ``Known'' model (memory and CPU).

\index{Performances!ReachabilityMix!RwMutex (P/T)}
\begin{center}
   \includegraphics[width=7.2cm]{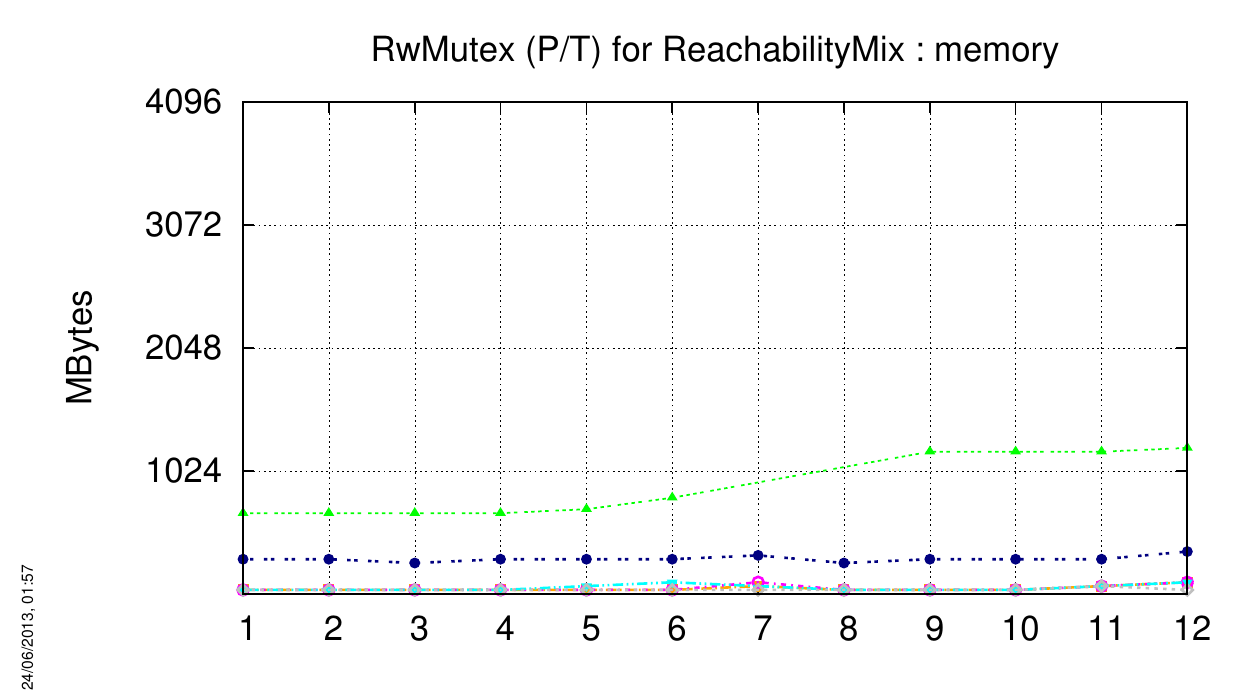}
   \includegraphics[width=7.2cm]{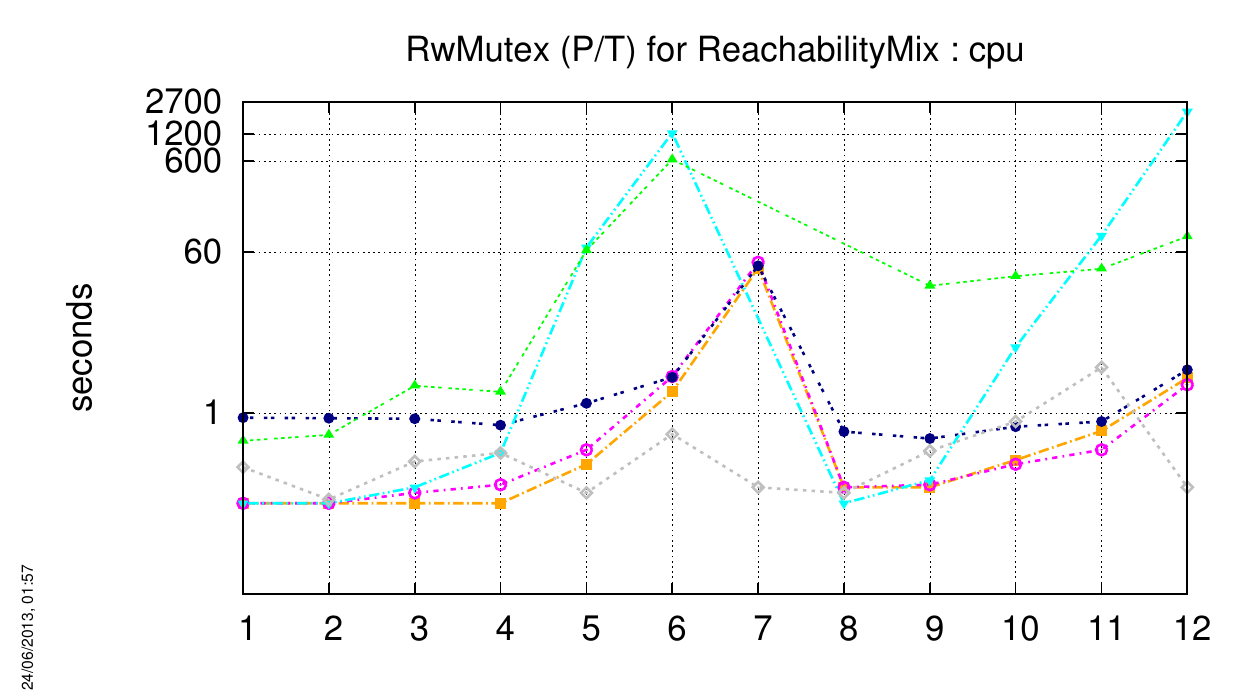}

   \includegraphics[height=1cm]{figures/tools-legend.pdf}
\end{center}

\subsubsection{\acs{SharedMemory-COL}}
No instance of this model could be computed for the \textbf{ReachabilityMix} examination.

\subsubsection{\acs{SharedMemory-PT}}
The charts below respectively show how tools compete with this ``Known'' model (memory and CPU).

\index{Performances!ReachabilityMix!SharedMemory (P/T)}
\begin{center}
   \includegraphics[width=7.2cm]{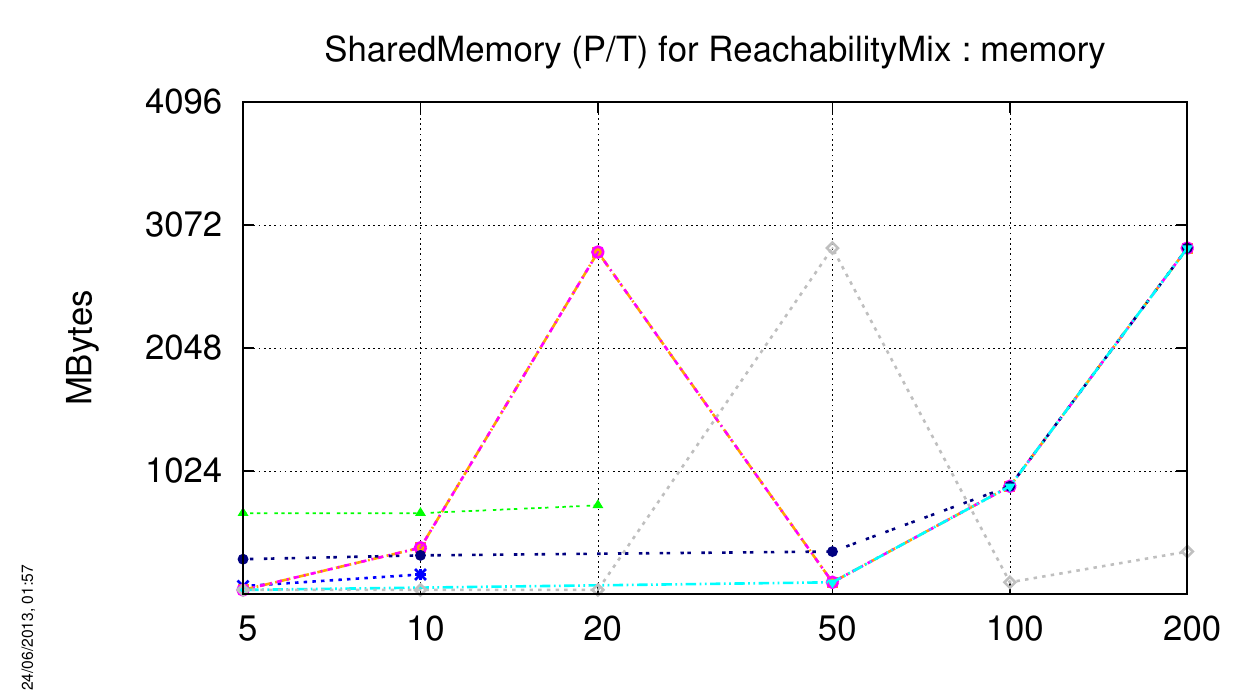}
   \includegraphics[width=7.2cm]{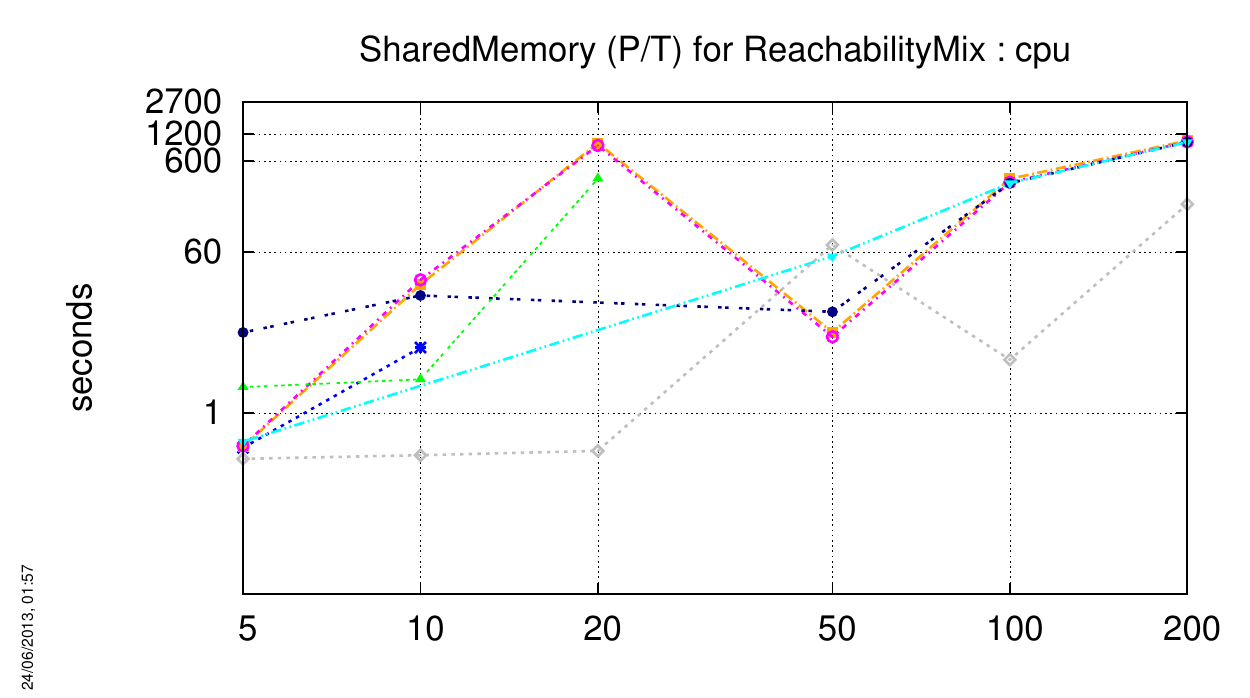}

   \includegraphics[height=1cm]{figures/tools-legend.pdf}
\end{center}

\subsubsection{\acs{SimpleLoadBal-COL}}
The charts below respectively show how tools compete with this ``Known'' model (memory and CPU).

\index{Performances!ReachabilityMix!SimpleLoadBal (Colored)}
\begin{center}
   \includegraphics[width=7.2cm]{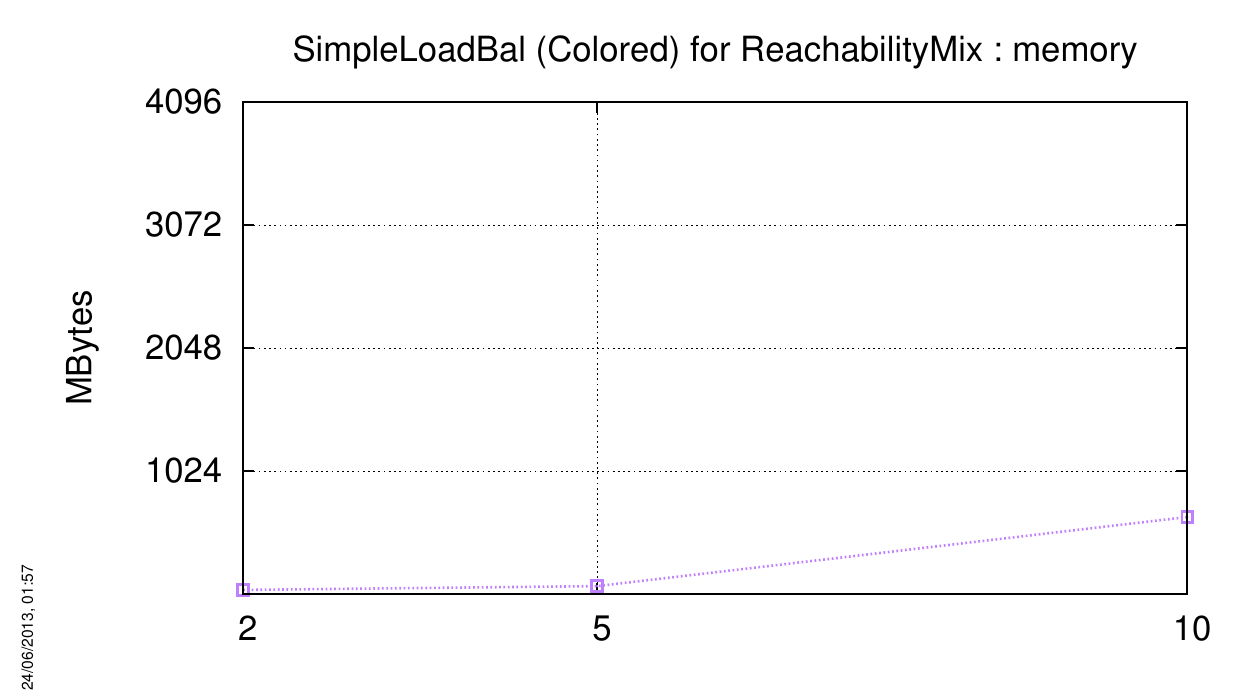}
   \includegraphics[width=7.2cm]{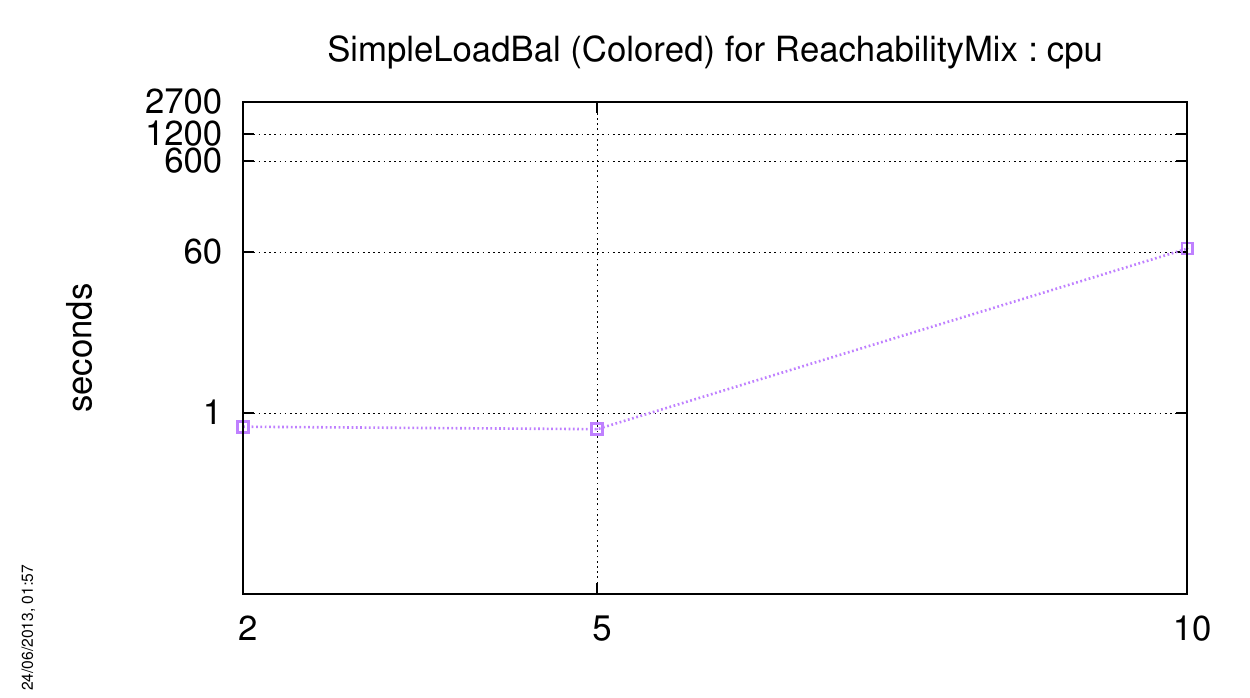}

   \includegraphics[height=1cm]{figures/tools-legend.pdf}
\end{center}

\subsubsection{\acs{SimpleLoadBal-PT}}
The charts below respectively show how tools compete with this ``Known'' model (memory and CPU).

\index{Performances!ReachabilityMix!SimpleLoadBal (P/T)}
\begin{center}
   \includegraphics[width=7.2cm]{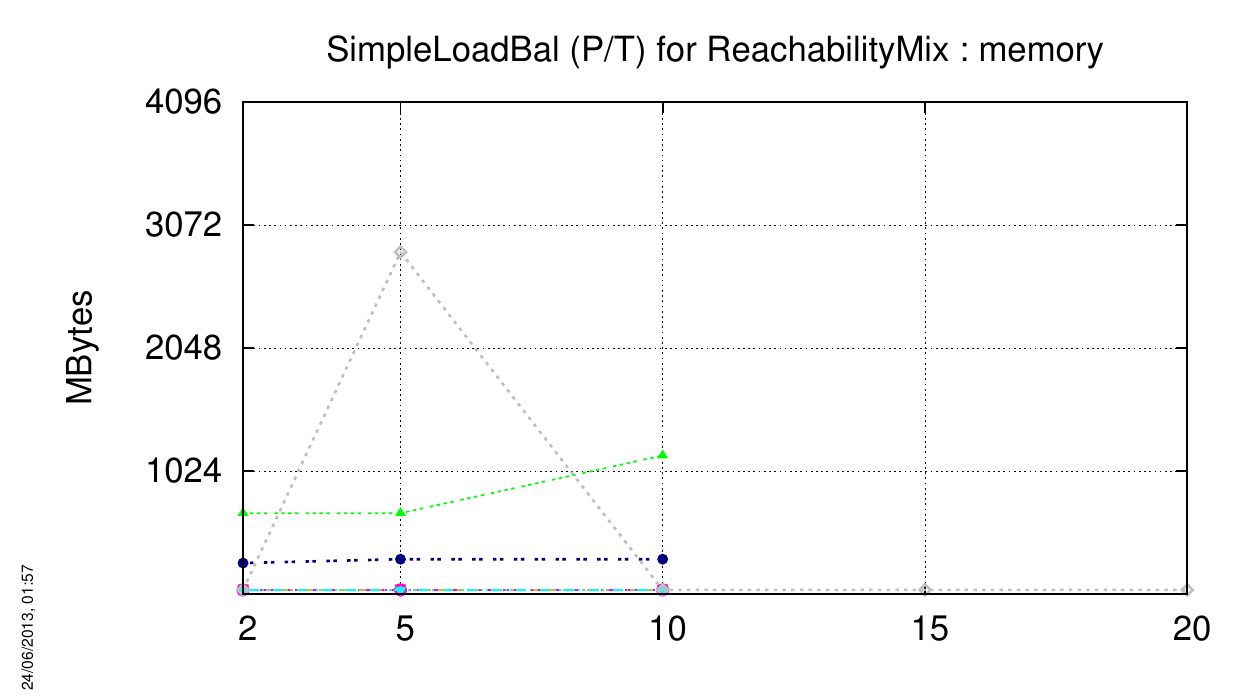}
   \includegraphics[width=7.2cm]{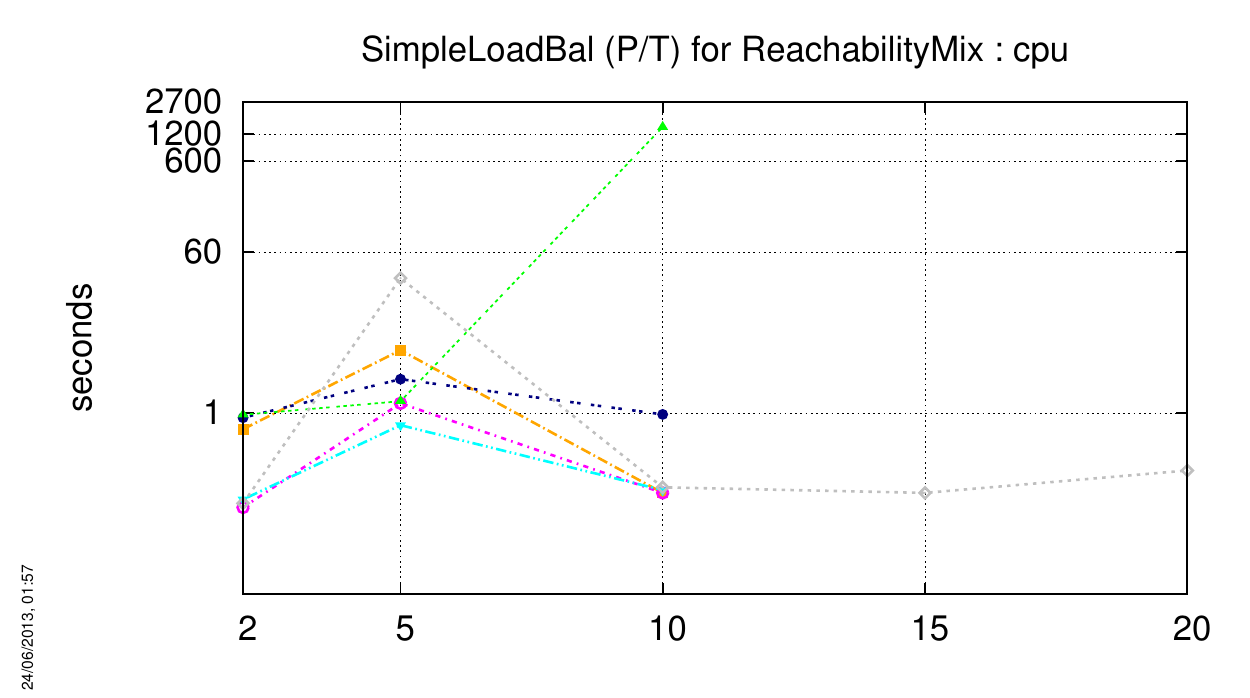}

   \includegraphics[height=1cm]{figures/tools-legend.pdf}
\end{center}

\subsubsection{\acs{TokenRing-COL}}
No instance of this model could be computed for the \textbf{ReachabilityMix} examination.

\subsubsection{\acs{TokenRing-PT}}
The charts below respectively show how tools compete with this ``Known'' model (memory and CPU).

\index{Performances!ReachabilityMix!TokenRing (P/T)}
\begin{center}
   \includegraphics[width=7.2cm]{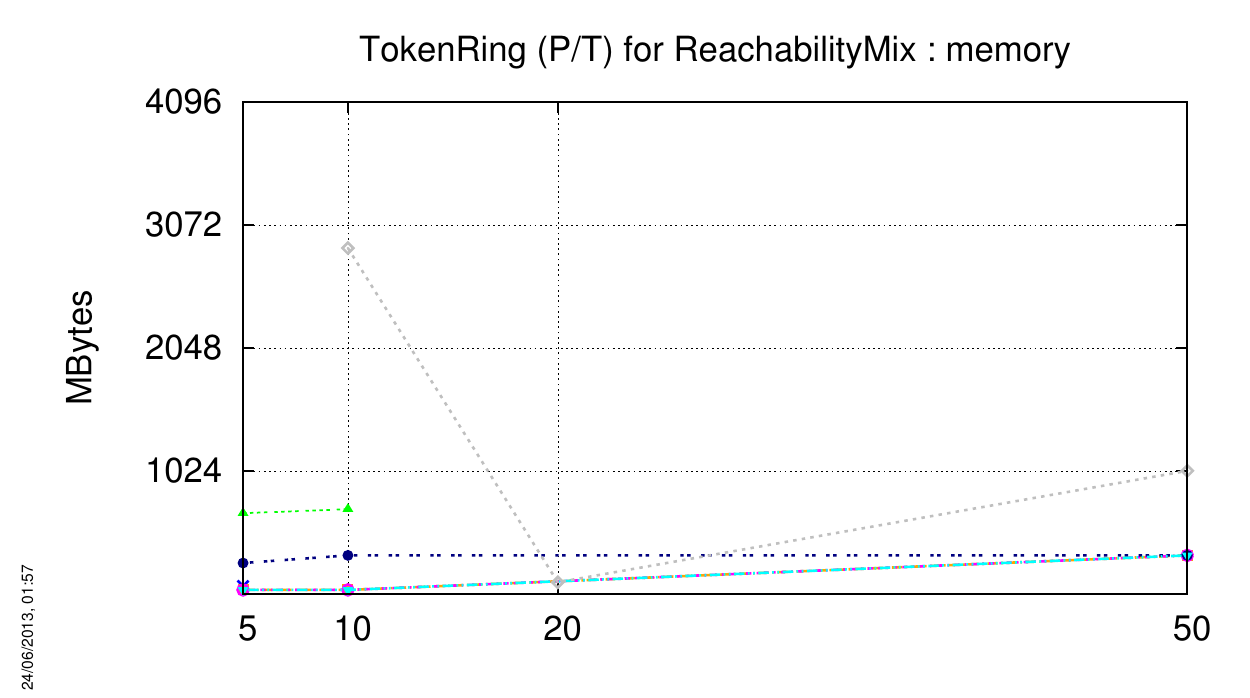}
   \includegraphics[width=7.2cm]{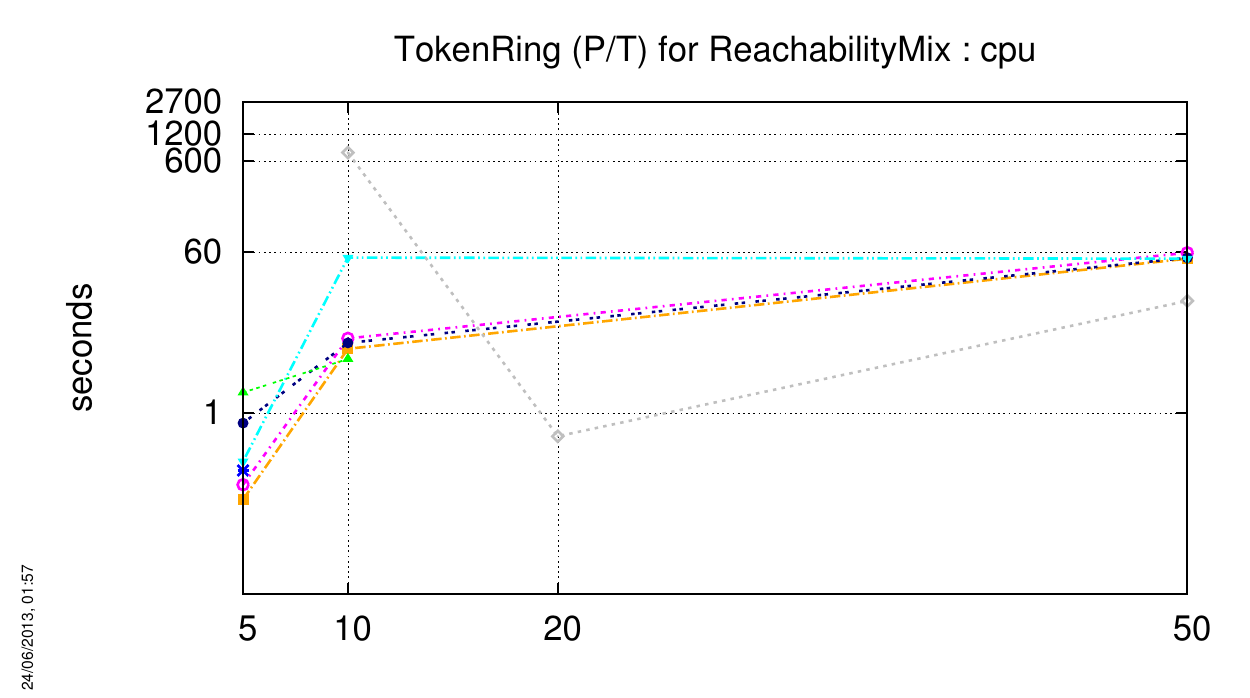}

   \includegraphics[height=1cm]{figures/tools-legend.pdf}
\end{center}

\subsubsection{\acs{HouseConstruction-PT}}
The charts below respectively show how tools compete with this ``Suprise'' model (memory and CPU).

\index{Performances!ReachabilityMix!HouseConstruction (P/T)}
\begin{center}
   \includegraphics[width=7.2cm]{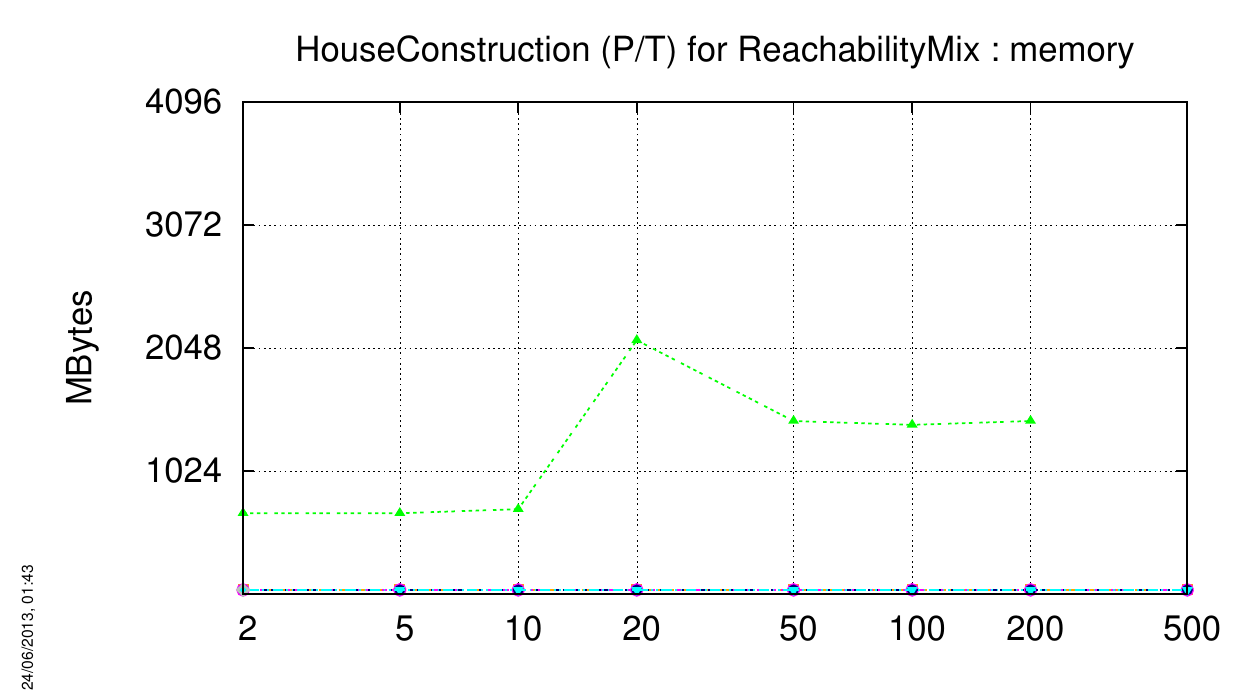}
   \includegraphics[width=7.2cm]{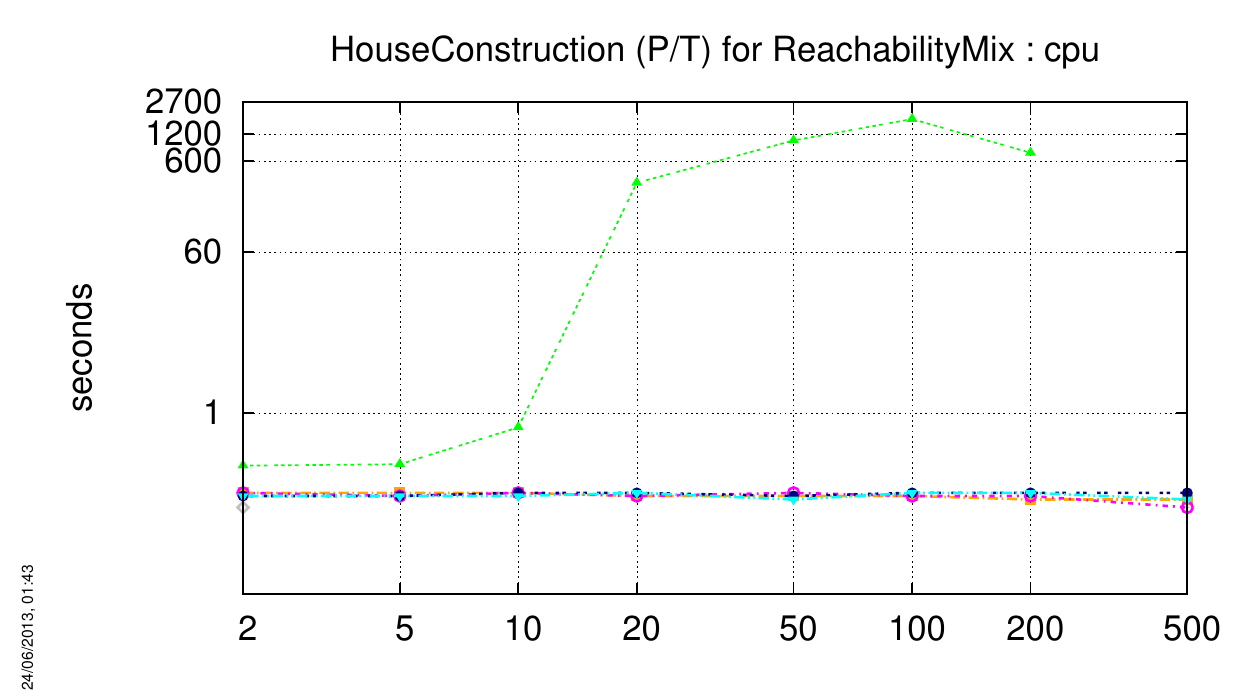}

   \includegraphics[height=1cm]{figures/tools-legend.pdf}
\end{center}

\subsubsection{\acs{IBMB2S565S3960-PT}}
The charts below respectively show how tools compete with this ``Suprise'' model (memory and CPU).

\index{Performances!ReachabilityMix!IBMB2S565S3960 (P/T)}
\begin{center}
   \includegraphics[width=7.2cm]{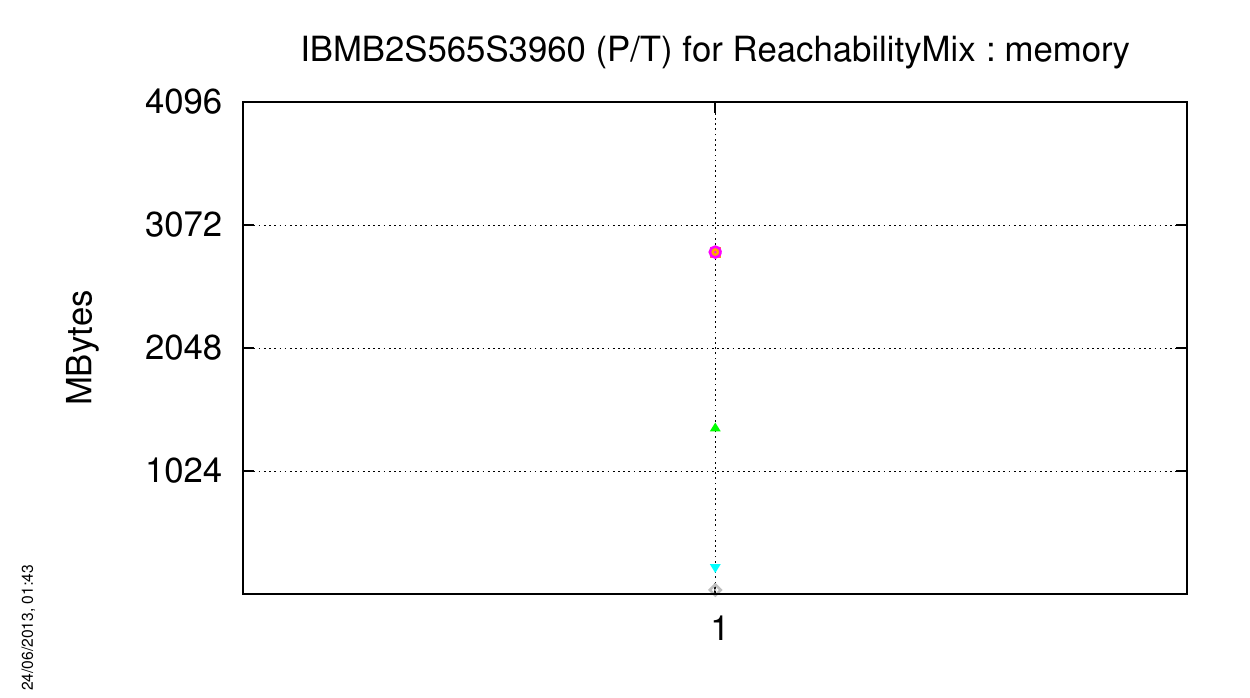}
   \includegraphics[width=7.2cm]{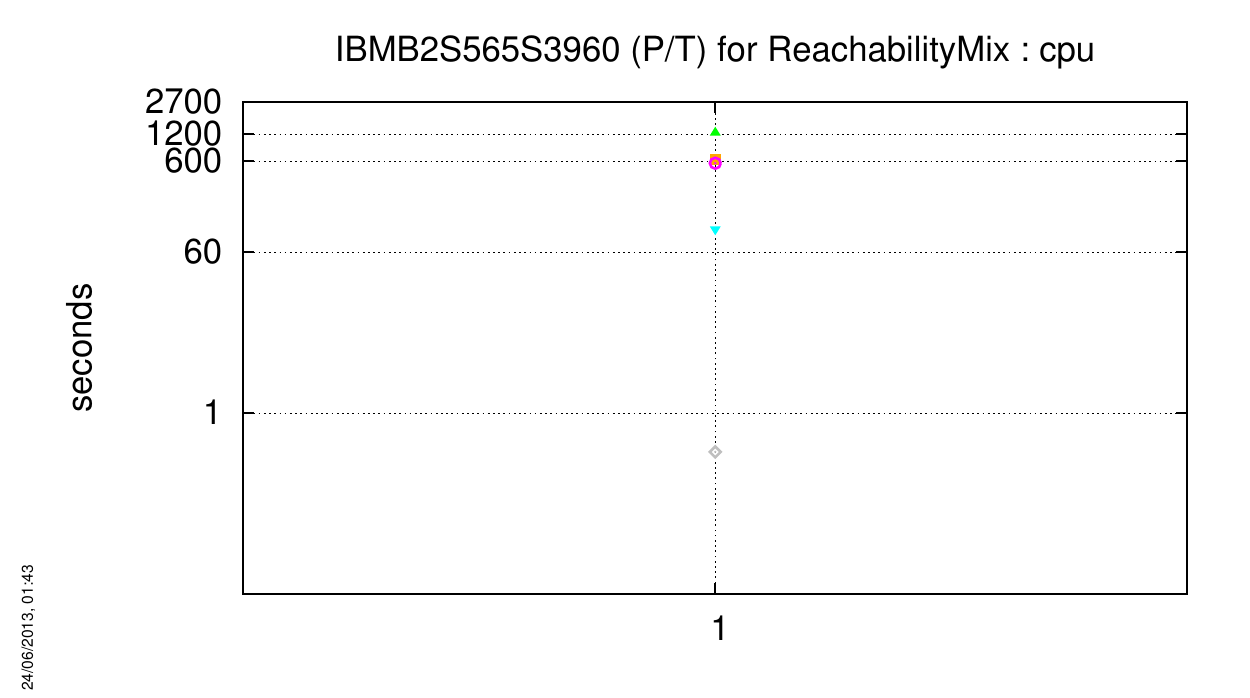}

   \includegraphics[height=1cm]{figures/tools-legend.pdf}
\end{center}

\subsubsection{\acs{QuasiCertifProtocol-COL}}
No instance of this model could be computed for the \textbf{ReachabilityMix} examination.

\subsubsection{\acs{QuasiCertifProtocol-PT}}
The charts below respectively show how tools compete with this ``Suprise'' model (memory and CPU).

\index{Performances!ReachabilityMix!QuasiCertifProtocol (P/T)}
\begin{center}
   \includegraphics[width=7.2cm]{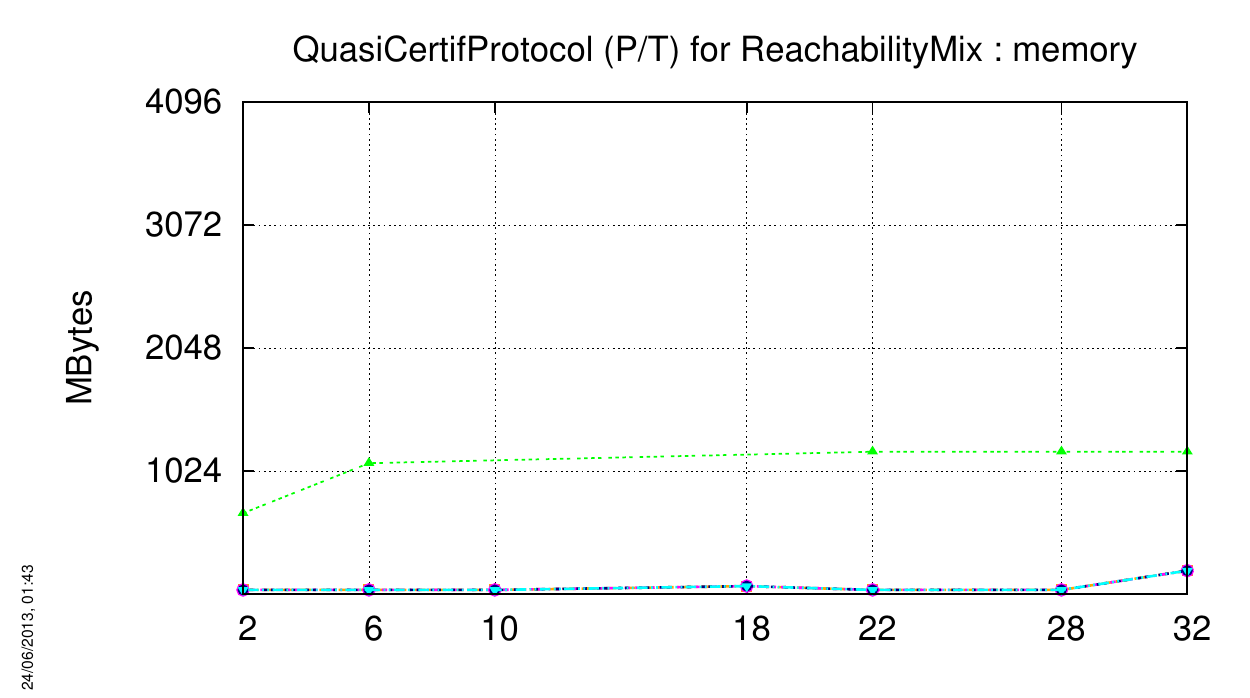}
   \includegraphics[width=7.2cm]{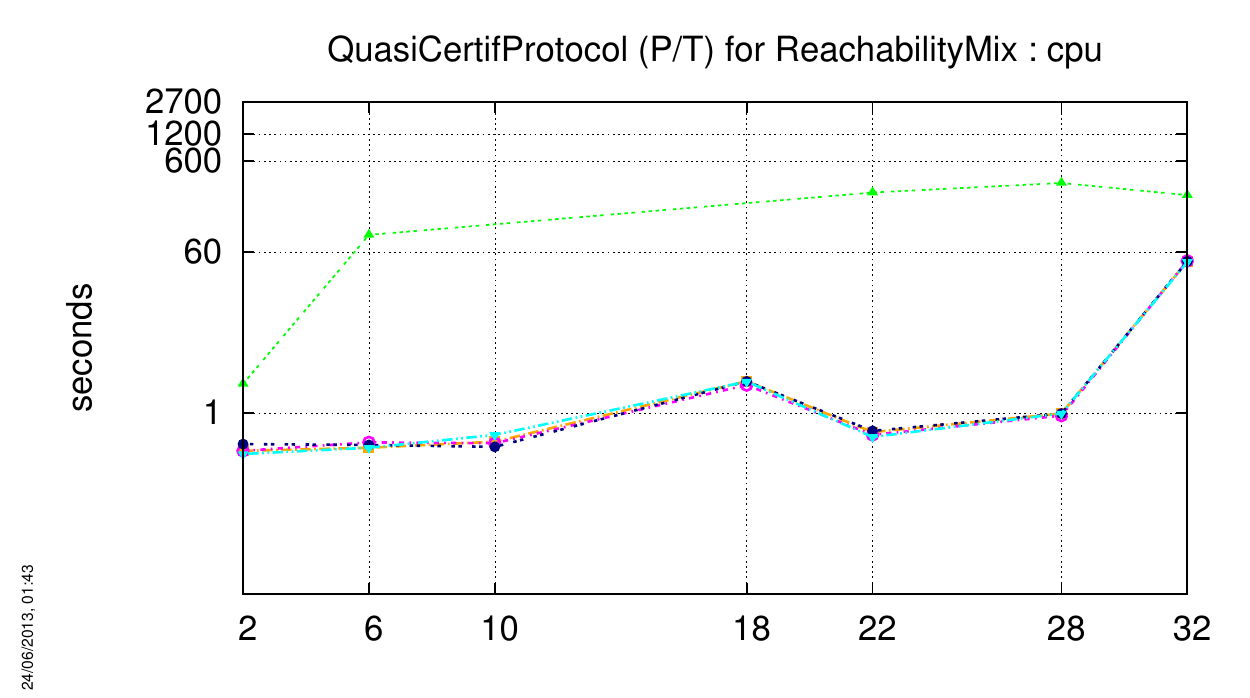}

   \includegraphics[height=1cm]{figures/tools-legend.pdf}
\end{center}

\subsubsection{\acs{Vasy2003-PT}}
The charts below respectively show how tools compete with this ``Suprise'' model (memory and CPU).

\index{Performances!ReachabilityMix!Vasy2003 (P/T)}
\begin{center}
   \includegraphics[width=7.2cm]{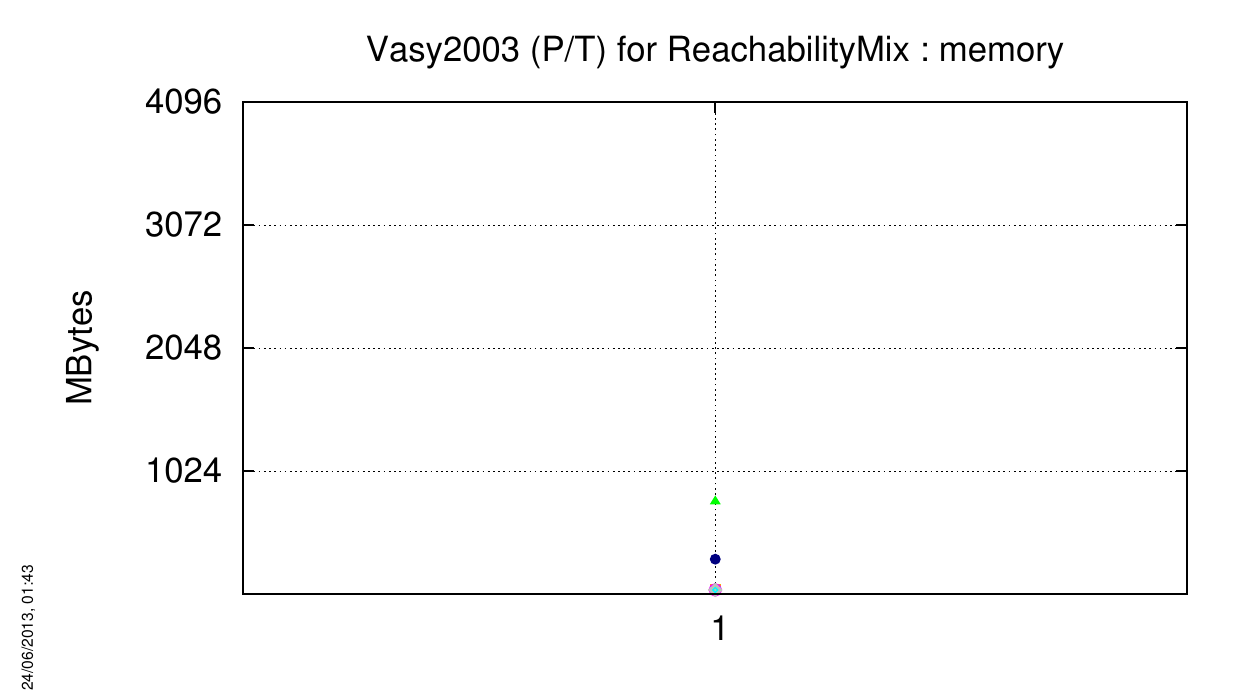}
   \includegraphics[width=7.2cm]{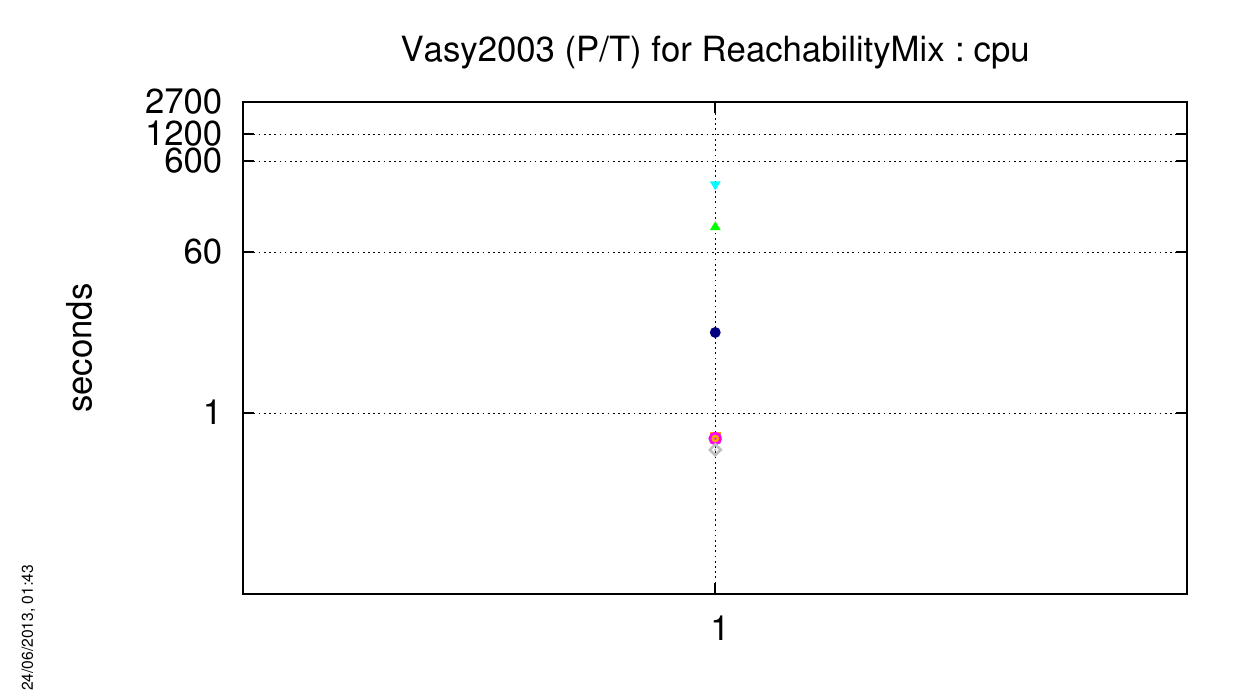}

   \includegraphics[height=1cm]{figures/tools-legend.pdf}
\end{center}

\subsection{Outputs for the ReachabilityMix Examination}
\index{Outputs!ReachabilityMix}

Please find enclosed the brute results for this examination (``Known'' and ``Surprise'' models).
We display only the score of tools that provide a results for at least one instance of one model.
The legend for the values is provided below:
\begin{itemize}
   \item\textbf{nc}: the tool does not compete this examination for this model/instance,
   \item\textbf{cc}: the tool cannot compute this examination for this model/instance,
   \item\textbf{to}: the tool cannot compute this examination for this model/instance within the maximum allowed time,
   \item\textbf{mp}: the tool encountered a memory problem (stack overflow or memory full),
   \item\textbf{nf}: there is no formula available for this type of examination (typically, this concerns P/T nets where
       comparing marking cardinality has no signification when there is no equivalent colored net).
\end{itemize}

\textbf{Note on the display of results for formulas:} each formula is considered as a flag (F if false, T if true, - or ?
when the value cannot be determined). These values are concatenated in the order they appear (we assume it is the order of formulas as they were provided).

\subsubsection{``Known'' Models}

Results are summarized in the table below.
{\footnotesize
\begin{longtable}{|ccccccccc|}
   \hline
      \multicolumn{9}{|c|}{\bf \acs{CSRepetitions-COL}} \\
      \textbf{Instances}\index{Outputs!ReachabilityMix!CSRepetitions (colored)} &
      \acs{greatspn} &
      \acs{its-tools} &
      \acs{lola} &
      \acs{lola-optimistic} &
      \acs{lola-optimistic-incomplete} &
      \acs{lola-pessimistic} &
      \acs{marcie} &
      \acs{sara} \\
      02 &
      nc &
      \textcolor{red}{\tt \scriptsize FFFFFFFF} &
      nc &
      nc &
      nc &
      nc &
      nc &
      nc \\
      03 &
      nc &
      \textcolor{red}{\tt \scriptsize FTFTFFFF} &
      nc &
      nc &
      nc &
      nc &
      nc &
      nc \\
      04 &
      nc &
      \textcolor{red}{\tt \scriptsize mp} &
      nc &
      nc &
      nc &
      nc &
      nc &
      nc \\
      05 &
      nc &
      \textcolor{red}{\tt \scriptsize mp} &
      nc &
      nc &
      nc &
      nc &
      nc &
      nc \\
      07 &
      nc &
      \textcolor{red}{\tt \scriptsize mp} &
      nc &
      nc &
      nc &
      nc &
      nc &
      nc \\
      10 &
      nc &
      \textcolor{red}{\tt \scriptsize mp} &
      nc &
      nc &
      nc &
      nc &
      nc &
      nc \\
\hline
\end{longtable}
}

\subsubsection{``Surprise'' Models}

Results are summarized in the table below.
{\footnotesize
\begin{longtable}{|ccccccc|}
   \hline
      \multicolumn{7}{|c|}{\bf \acs{HouseConstruction-PT}} \\
      instances\index{Outputs!ReachabilityMix!HouseConstruction (P/T)} &
      \acs{lola} &
      \acs{lola-optimistic} &
      \acs{lola-optimistic-incomplete} &
      \acs{lola-pessimistic} &
      \acs{marcie} &
      \acs{sara} \\
      002 &
      \textcolor{red}{\tt \scriptsize cc} &
      \textcolor{red}{\tt \scriptsize cc} &
      \textcolor{red}{\tt \scriptsize cc} &
      \textcolor{red}{\tt \scriptsize cc} &
      \textcolor{red}{\tt \scriptsize FFFFTFFT} &
      \textcolor{red}{\tt \scriptsize cc} \\
      005 &
      \textcolor{red}{\tt \scriptsize cc} &
      \textcolor{red}{\tt \scriptsize cc} &
      \textcolor{red}{\tt \scriptsize cc} &
      \textcolor{red}{\tt \scriptsize cc} &
      \textcolor{red}{\tt \scriptsize FFFFFFFF} &
      \textcolor{red}{\tt \scriptsize cc} \\
      010 &
      \textcolor{red}{\tt \scriptsize cc} &
      \textcolor{red}{\tt \scriptsize cc} &
      \textcolor{red}{\tt \scriptsize cc} &
      \textcolor{red}{\tt \scriptsize cc} &
      \textcolor{red}{\tt \scriptsize FFFFTFFF} &
      \textcolor{red}{\tt \scriptsize cc} \\
      020 &
      \textcolor{red}{\tt \scriptsize cc} &
      \textcolor{red}{\tt \scriptsize cc} &
      \textcolor{red}{\tt \scriptsize cc} &
      \textcolor{red}{\tt \scriptsize cc} &
      \textcolor{red}{\tt \scriptsize FFFTFFFF} &
      \textcolor{red}{\tt \scriptsize cc} \\
      050 &
      \textcolor{red}{\tt \scriptsize cc} &
      \textcolor{red}{\tt \scriptsize cc} &
      \textcolor{red}{\tt \scriptsize cc} &
      \textcolor{red}{\tt \scriptsize cc} &
      \textcolor{red}{\tt \scriptsize cc} &
      \textcolor{red}{\tt \scriptsize cc} \\
\hline
\end{longtable}
}

\subsection{Score for the ReachabilityMix Examination}
\index{Scores!ReachabilityMix}

Please find enclosed the scores for this examination (``Known'' and ``Surprise'' models).
We display only the score of tools that provide a results for at least one instance of one model.
The total is first listed in the table below followed by a detail, for each proposed model.
Meaning of the line labels are:
\begin{itemize}
\item\textbf{1st instance}: the tool gets a bonus for having processed the first instance of this model (+1 point),
\item\textbf{instances}: the tool gets 1 point per instances treated 
(for that, we assume that at least one formula has been successfully computed),
\item\textbf{max reached}: the tool could process all the instances for the model (+2 points),
\item\textbf{best}: the tool is among the ones that processed a maximum of instances within the time and memory confinement (+2 points).
\end{itemize}

\subsubsection{``Known'' Models}

Scores are summarized in the table below.
For each model in this category, a tool may collect up to 411 points.

{\footnotesize
\begin{longtable}{|ccccccccc|}
   \hline
      \multicolumn{9}{|c|}{\bf Total score of the tools for this examination} \\
        &
      \acs{greatspn} &
      \acs{its-tools} &
      \acs{lola} &
      \acs{lola-optimistic} &
      \acs{lola-optimistic-incomplete} &
      \acs{lola-pessimistic} &
      \acs{marcie} &
      \acs{sara} \\
      Total Score &
      {\bf 40} &
      {\bf 54} &
      {\bf 184} &
      {\bf 189} &
      {\bf 154} &
      {\bf 111} &
      {\bf 118} &
      {\bf 80} \\
   \hline
      \multicolumn{9}{|c|}{\bf CSRepetitions (Colored)} \\
      \ \index{Scores!ReachabilityMix!CSRepetitions (colored)} &
      \acs{greatspn} &
      \acs{its-tools} &
      \acs{lola} &
      \acs{lola-optimistic} &
      \acs{lola-optimistic-incomplete} &
      \acs{lola-pessimistic} &
      \acs{marcie} &
      \acs{sara} \\
      1st instance &
      \textcolor{red}{0} &
      \textcolor{red}{1} &
      \textcolor{red}{0} &
      \textcolor{red}{0} &
      \textcolor{red}{0} &
      \textcolor{red}{0} &
      \textcolor{red}{0} &
      \textcolor{red}{0} \\
      instances &
      \textcolor{red}{0} &
      \textcolor{red}{2} &
      \textcolor{red}{0} &
      \textcolor{red}{0} &
      \textcolor{red}{0} &
      \textcolor{red}{0} &
      \textcolor{red}{0} &
      \textcolor{red}{0} \\
      max reached &
      \textcolor{red}{0} &
      \textcolor{red}{0} &
      \textcolor{red}{0} &
      \textcolor{red}{0} &
      \textcolor{red}{0} &
      \textcolor{red}{0} &
      \textcolor{red}{0} &
      \textcolor{red}{0} \\
\hline
\end{longtable}
}

\subsubsection{``Surprise'' Models}

Scores are summarized in the table below.
For each model in this category, a tool may collect up to 49 points.

{\footnotesize
\begin{longtable}{|ccccccc|}
   \hline
      \multicolumn{7}{|c|}{\bf Total score of the tools for this examination} \\
        &
      \acs{lola} &
      \acs{lola-optimistic} &
      \acs{lola-optimistic-incomplete} &
      \acs{lola-pessimistic} &
      \acs{marcie} &
      \acs{sara} \\
      Total Score &
      {\bf 12} &
      {\bf 12} &
      {\bf 6} &
      {\bf 12} &
      {\bf 24} &
      {\bf 0} \\
   \hline
      \multicolumn{7}{|c|}{\bf HouseConstruction (P/T)} \\
      \ \index{Scores!ReachabilityMix!HouseConstruction (P/T)} &
      \acs{lola} &
      \acs{lola-optimistic} &
      \acs{lola-optimistic-incomplete} &
      \acs{lola-pessimistic} &
      \acs{marcie} &
      \acs{sara} \\
      1st instance &
      \textcolor{red}{0} &
      \textcolor{red}{0} &
      \textcolor{red}{0} &
      \textcolor{red}{0} &
      \textcolor{red}{1} &
      \textcolor{red}{0} \\
      instances &
      \textcolor{red}{0} &
      \textcolor{red}{0} &
      \textcolor{red}{0} &
      \textcolor{red}{0} &
      \textcolor{red}{4} &
      \textcolor{red}{0} \\
      max reached &
      \textcolor{red}{0} &
      \textcolor{red}{0} &
      \textcolor{red}{0} &
      \textcolor{red}{0} &
      \textcolor{red}{0} &
      \textcolor{red}{0} \\
\hline
\end{longtable}
}

\subsection{Trophies for this Examination}
\index{Trophies!ReachabilityMix}

Trophies are divided in three categories: ``Known'' models,
``Surprise'' models, and the global trophies (formula is then
$score_{global} = score_{known} + 2 \times score_{surprise}$).

\subsubsection{For ``Known'' Models} \ \\

\begin{tabular}{c|c|c}
      1 & 2 & 3 \\
   \includegraphics[width=2cm]{figures/gold.jpg} &
   \includegraphics[width=2cm]{figures/silver.jpg} &
   \includegraphics[width=2cm]{figures/bronse.jpg} \\
   \acs{lola-optimistic} &
   \acs{lola} &
   \acs{lola-optimistic-incomplete} \\
   189 points &
   184 points &
   154 points \\
\end{tabular}

\subsubsection{For ``Surprise'' Models}\  \\

\begin{tabular}{c|c|c|c}
      1 & 2 & 2 & 2 \\
   \includegraphics[width=2cm]{figures/gold.jpg} &
   \includegraphics[width=2cm]{figures/silver.jpg} &
   \includegraphics[width=2cm]{figures/silver.jpg} &
   \includegraphics[width=2cm]{figures/silver.jpg} \\
   \acs{marcie} &
   \acs{lola-optimistic} &
   \acs{lola} &
   \acs{lola-optimistic-incomplete} \\
   24 points &
   12 points &
   12 points &
   12 points \\
\end{tabular}

\subsubsection{Global} \ \\

\begin{tabular}{c|c|c|c}
      1 & 2 & 3 & 3 \\
   \includegraphics[width=2cm]{figures/gold.jpg} &
   \includegraphics[width=2cm]{figures/silver.jpg} &
   \includegraphics[width=2cm]{figures/bronse.jpg} &
   \includegraphics[width=2cm]{figures/bronse.jpg} \\
   \acs{lola-optimistic} &
   \acs{lola} &
   \acs{marcie} &
   \acs{lola-optimistic-incomplete} \\
   213 points &
   208 points &
   166 points &
   166 points \\
\end{tabular}

\part{CTL-based Analysis}
\label{part:four}
\newpage

\section{The CTLCardinalityComparison Examination}
\label{sec:exam:CTLCardinalityComparison}
\index{Results!CTLCardinalityComparison}

This examination deals with CTL properties dealing with checking cardinality of marking only.
We first show a summary on the handling of models by the participating tools.
Then, we present the computed outputs and the associated scores for this
examination prior to a summary of relevant executions.

\subsection{Handling of Models by Tools}
\index{Performances!CTLCardinalityComparison}

\subsubsection{\acs{CSRepetitions-COL}}
No instance of this model could be computed for the \textbf{CTLCardinalityComparison} examination.

\subsubsection{\acs{CSRepetitions-PT}}
The charts below respectively show how tools compete with this ``Known'' model (memory and CPU).

\index{Performances!CTLCardinalityComparison!CSRepetitions (P/T)}
\begin{center}
   \includegraphics[width=7.2cm]{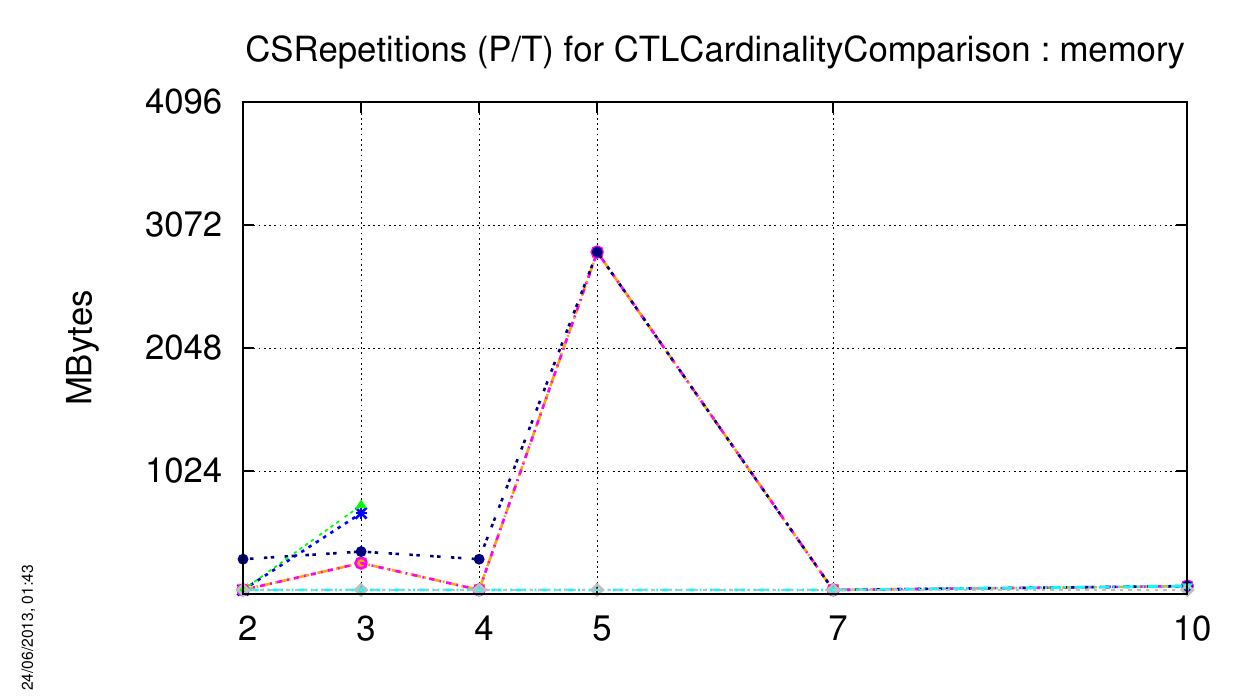}
   \includegraphics[width=7.2cm]{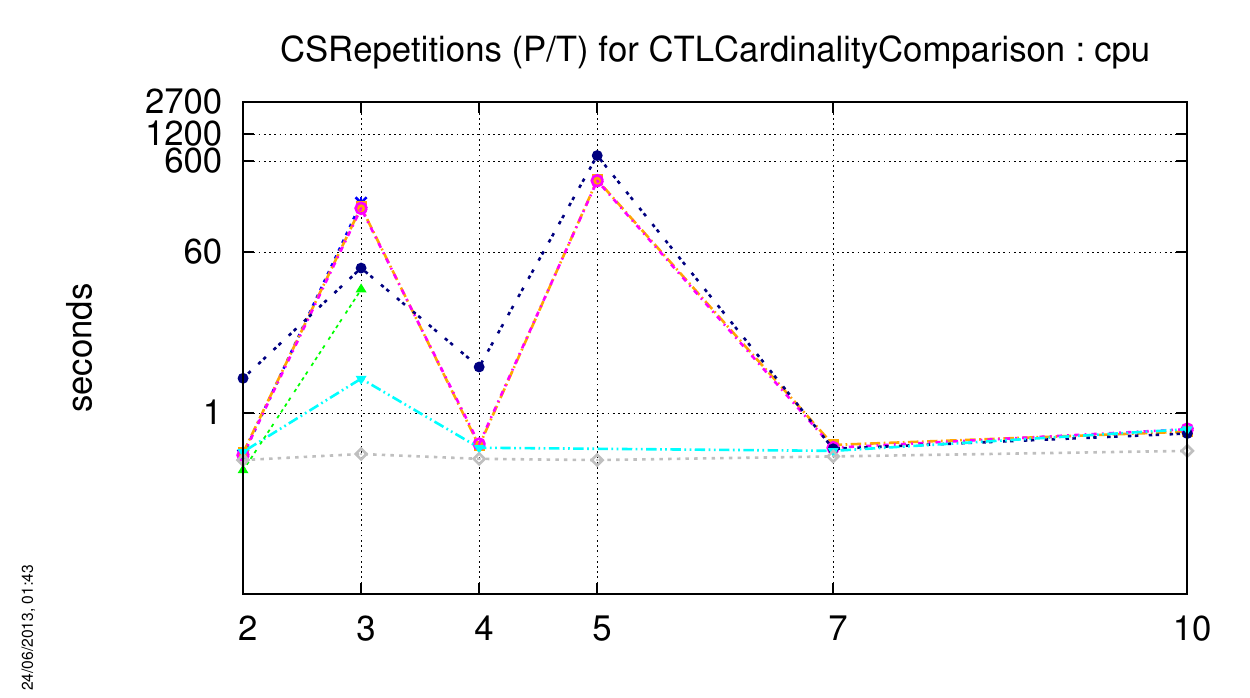}

   \includegraphics[height=1cm]{figures/tools-legend.pdf}
\end{center}

\subsubsection{\acs{Dekker-PT}}
The charts below respectively show how tools compete with this ``Known'' model (memory and CPU).

\index{Performances!CTLCardinalityComparison!Dekker (P/T)}
\begin{center}
   \includegraphics[width=7.2cm]{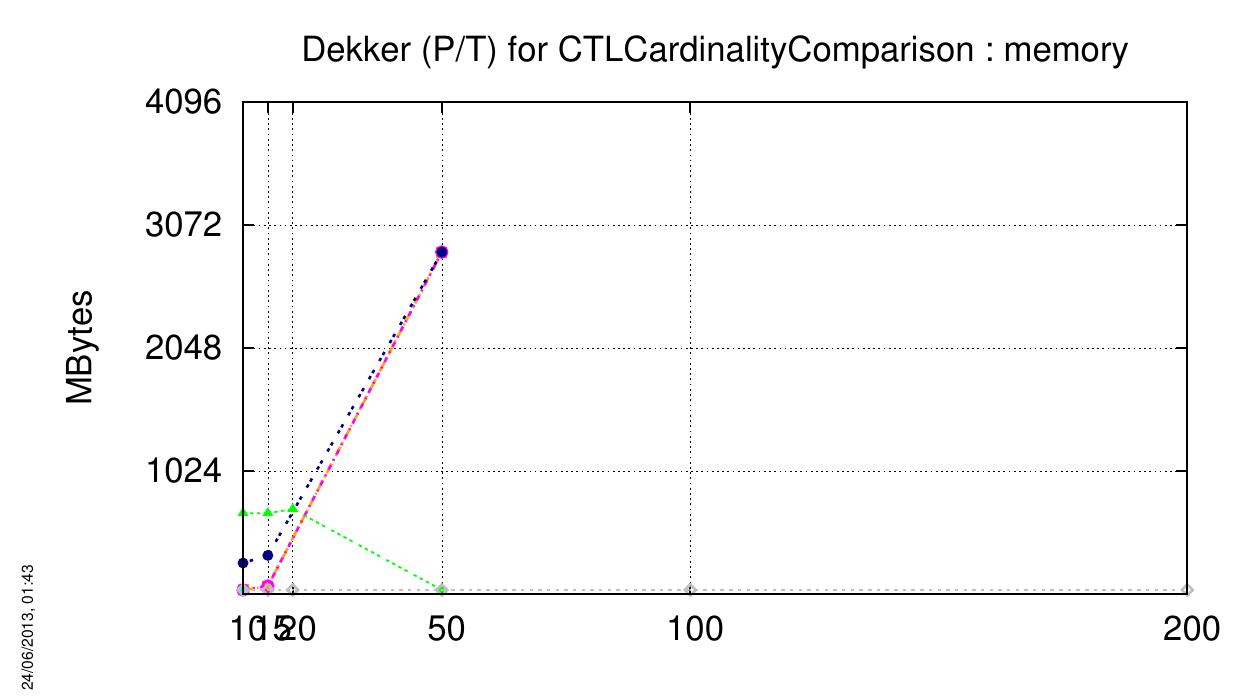}
   \includegraphics[width=7.2cm]{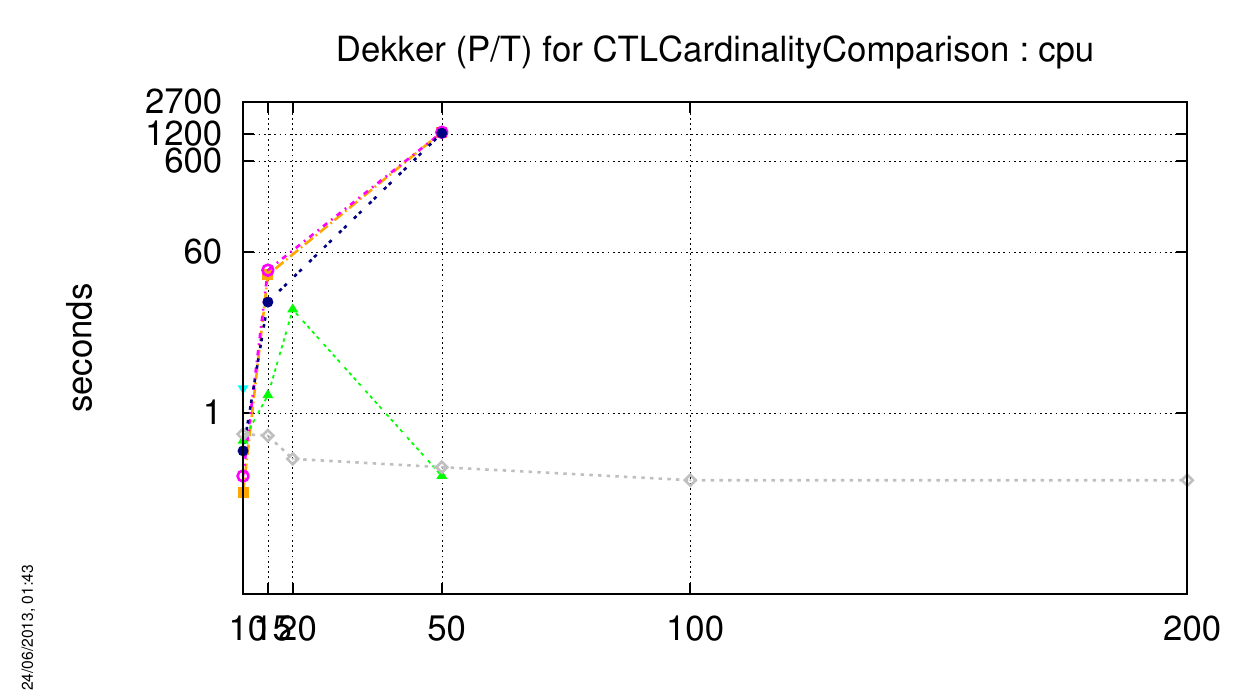}

   \includegraphics[height=1cm]{figures/tools-legend.pdf}
\end{center}

\subsubsection{\acs{DotAndBoxes-COL}}
No instance of this model could be computed for the \textbf{CTLCardinalityComparison} examination.

\subsubsection{\acs{DrinkVendingMachine-COL}}
No instance of this model could be computed for the \textbf{CTLCardinalityComparison} examination.

\subsubsection{\acs{DrinkVendingMachine-PT}}
The charts below respectively show how tools compete with this ``Known'' model (memory and CPU).

\index{Performances!CTLCardinalityComparison!DrinkVendingMachine (P/T)}
\begin{center}
   \includegraphics[width=7.2cm]{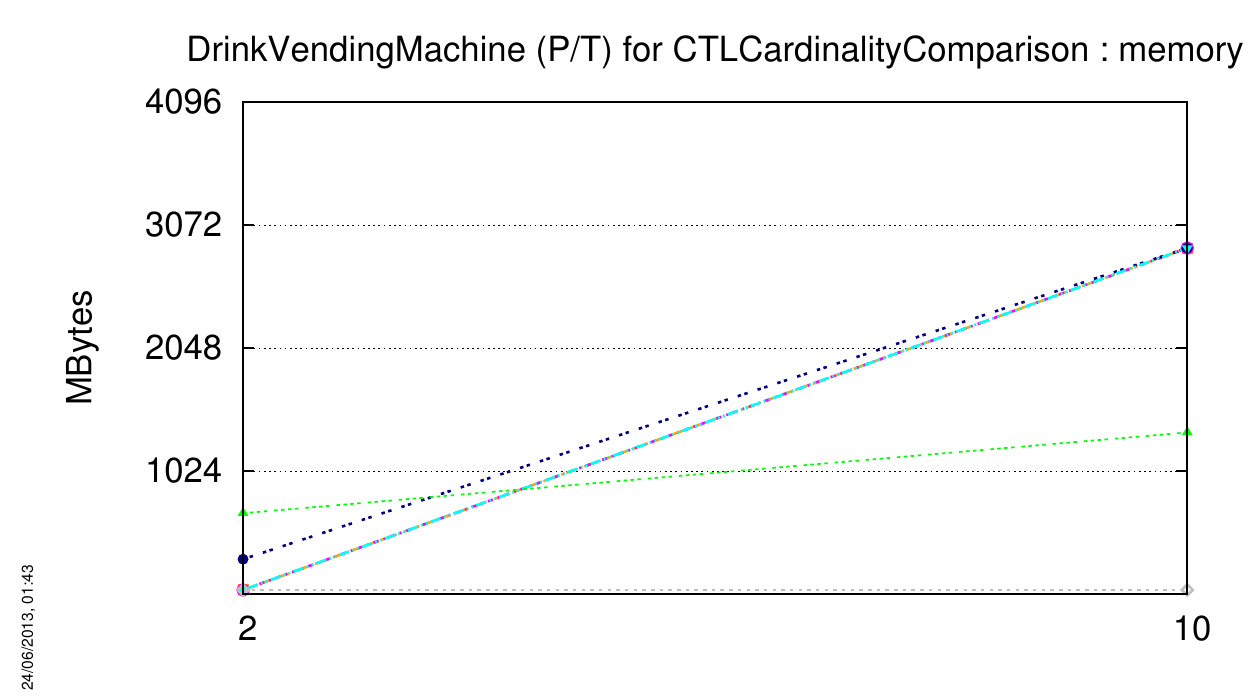}
   \includegraphics[width=7.2cm]{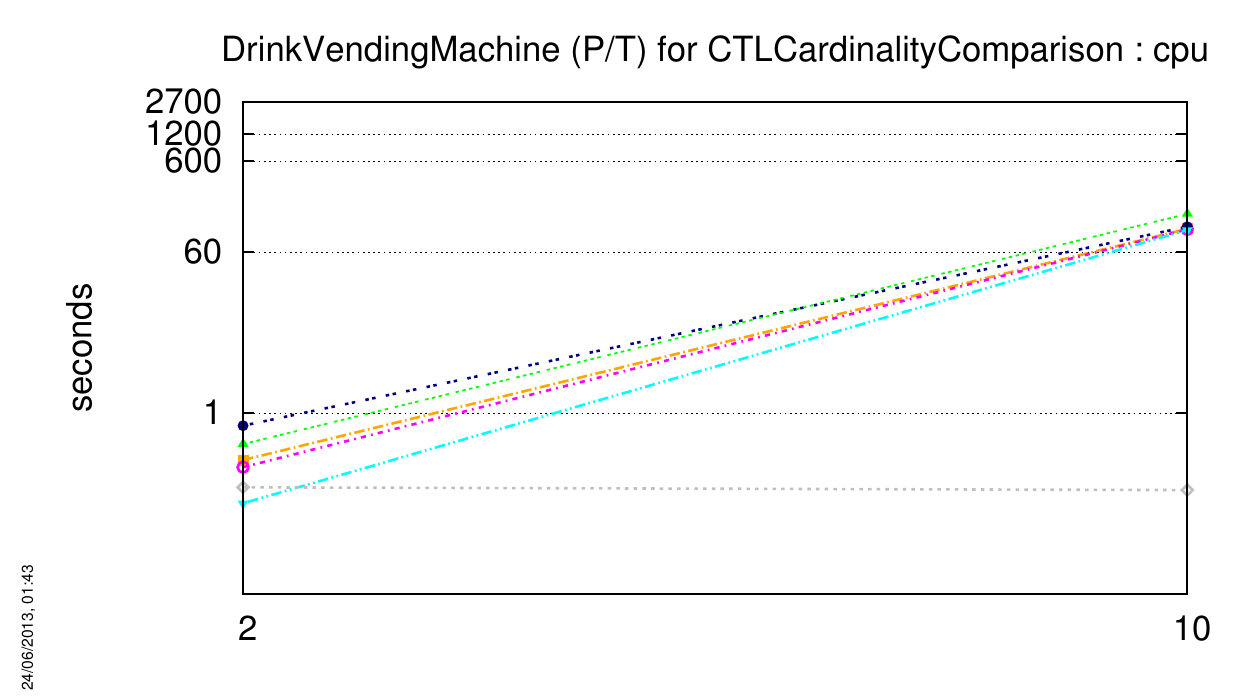}

   \includegraphics[height=1cm]{figures/tools-legend.pdf}
\end{center}

\subsubsection{\acs{Echo-PT}}
The charts below respectively show how tools compete with this ``Known'' model (memory and CPU).

\index{Performances!CTLCardinalityComparison!Echo (P/T)}
\begin{center}
   \includegraphics[width=7.2cm]{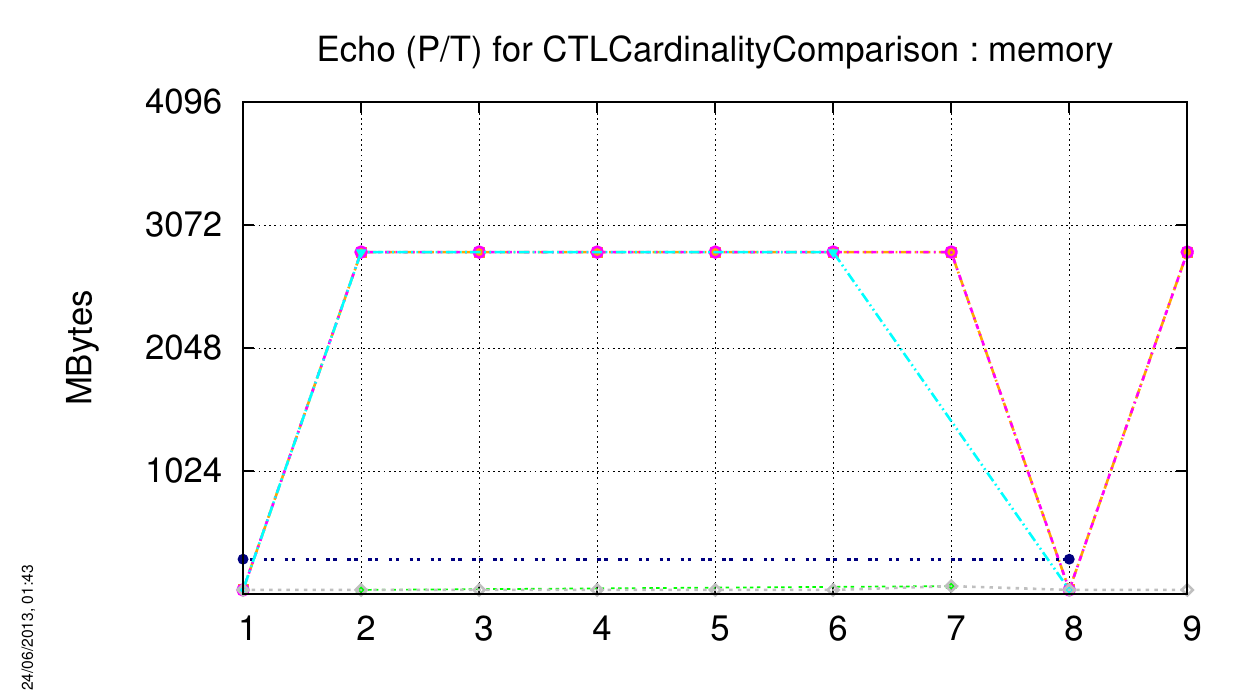}
   \includegraphics[width=7.2cm]{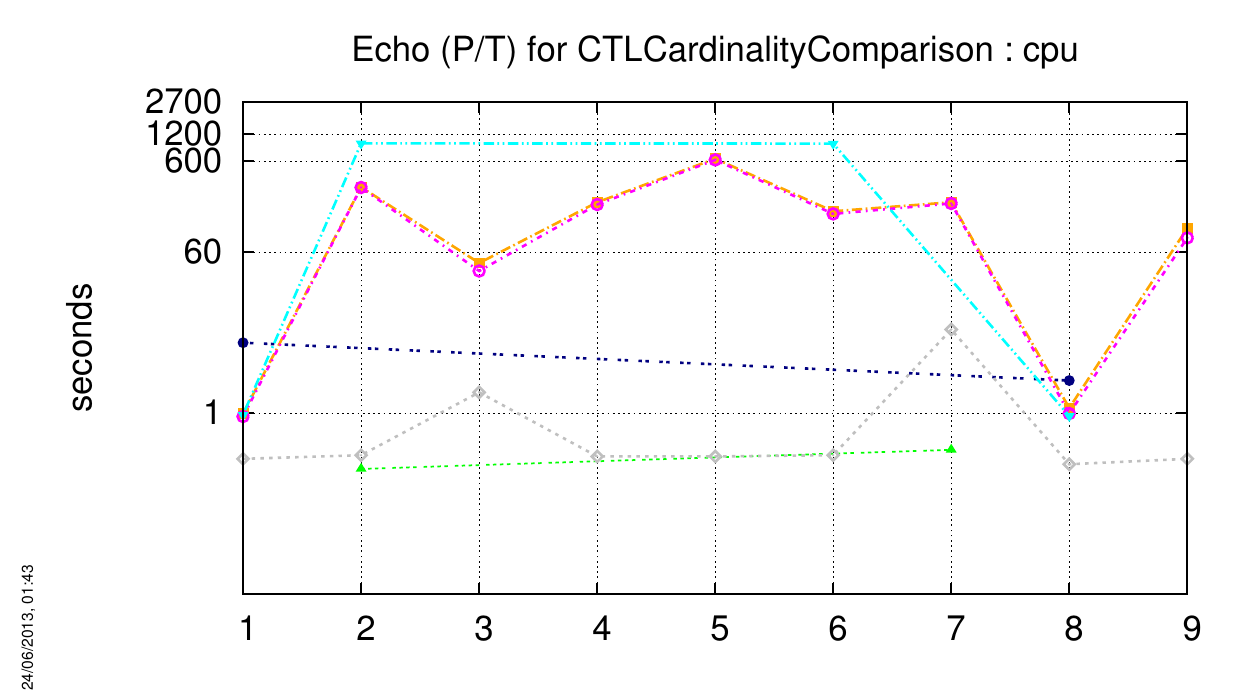}

   \includegraphics[height=1cm]{figures/tools-legend.pdf}
\end{center}

\subsubsection{\acs{Eratosthenes-PT}}
The charts below respectively show how tools compete with this ``Known'' model (memory and CPU).

\index{Performances!CTLCardinalityComparison!Eratosthenes (P/T)}
\begin{center}
   \includegraphics[width=7.2cm]{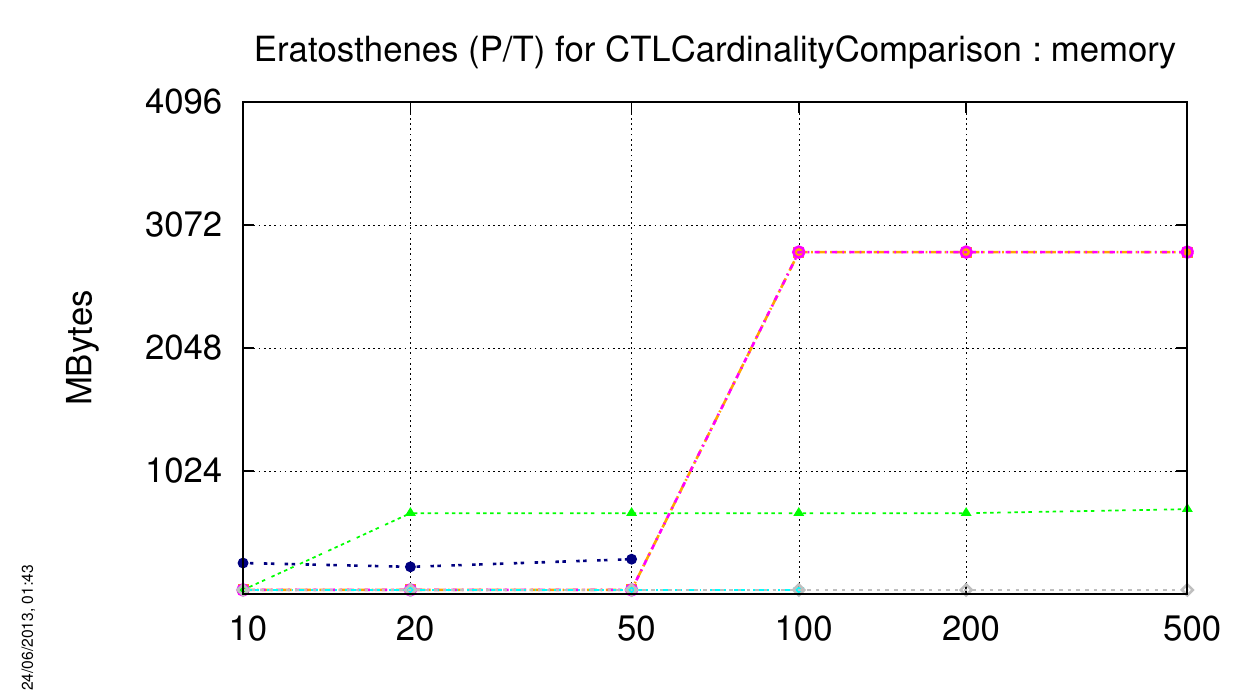}
   \includegraphics[width=7.2cm]{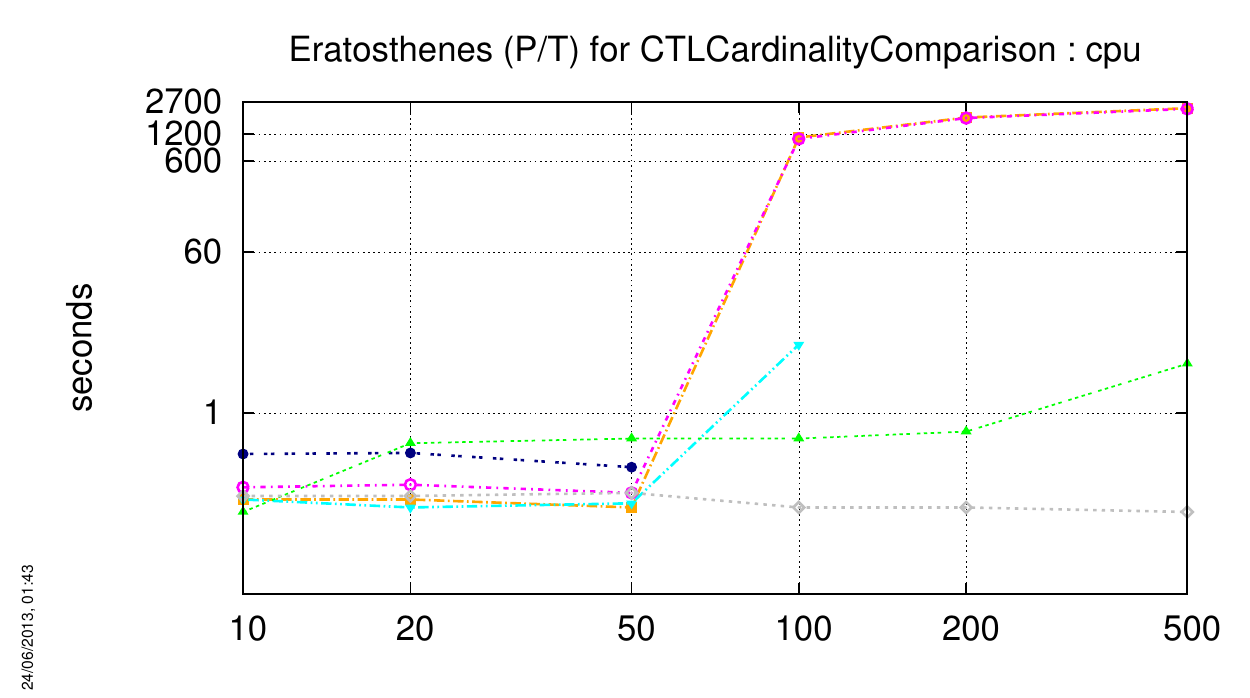}

   \includegraphics[height=1cm]{figures/tools-legend.pdf}
\end{center}

\subsubsection{\acs{FMS-PT}}
The charts below respectively show how tools compete with this ``Known'' model (memory and CPU).

\index{Performances!CTLCardinalityComparison!FMS (P/T)}
\begin{center}
   \includegraphics[width=7.2cm]{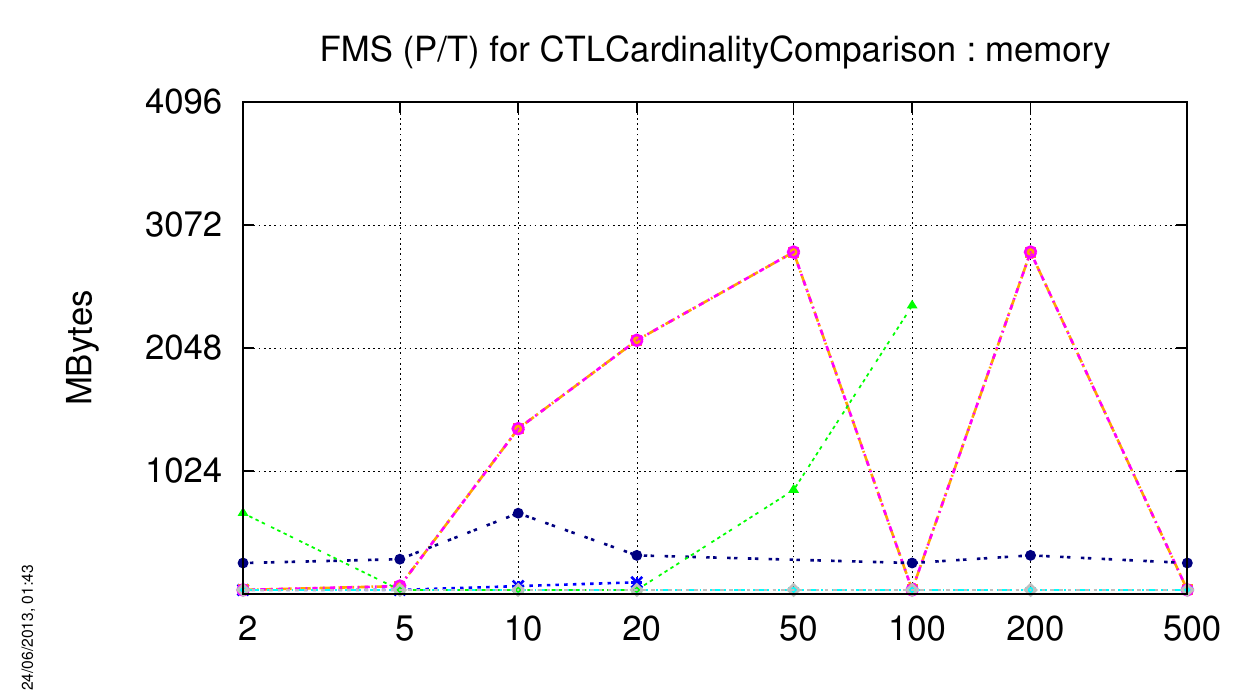}
   \includegraphics[width=7.2cm]{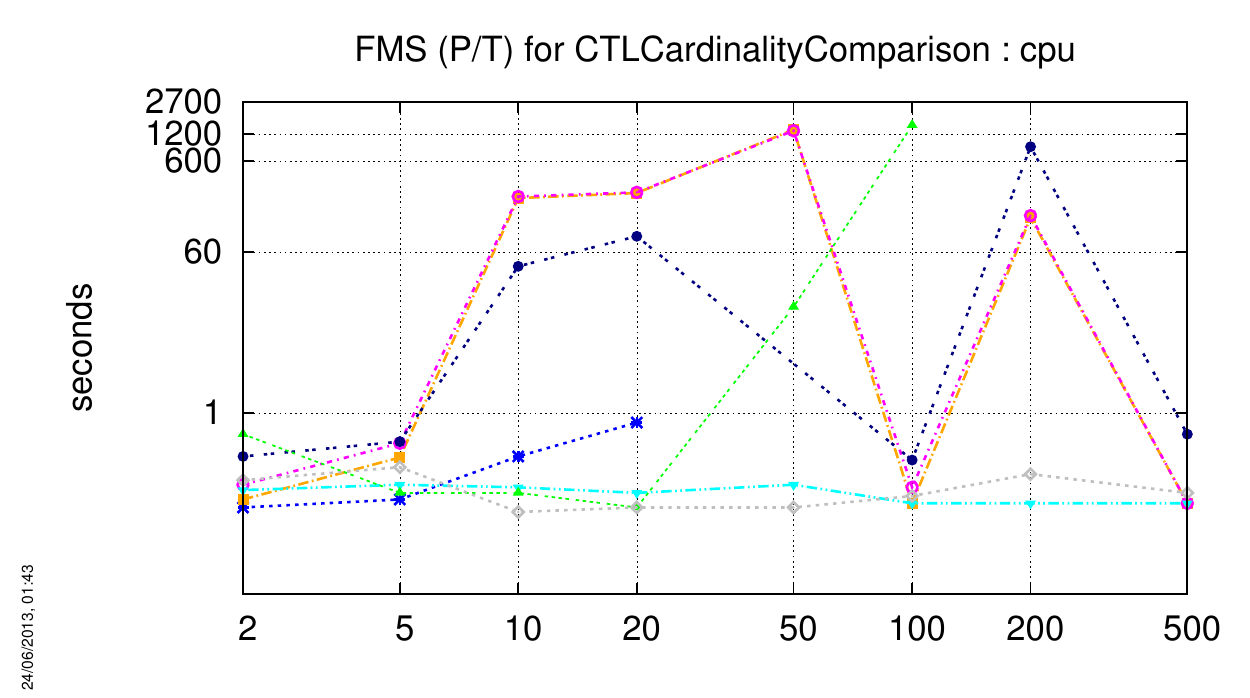}

   \includegraphics[height=1cm]{figures/tools-legend.pdf}
\end{center}

\subsubsection{\acs{GlobalRessAlloc-COL}}
No instance of this model could be computed for the \textbf{CTLCardinalityComparison} examination.

\subsubsection{\acs{GlobalRessAlloc-PT}}
The charts below respectively show how tools compete with this ``Known'' model (memory and CPU).

\index{Performances!CTLCardinalityComparison!GlobalRessAlloc (P/T)}
\begin{center}
   \includegraphics[width=7.2cm]{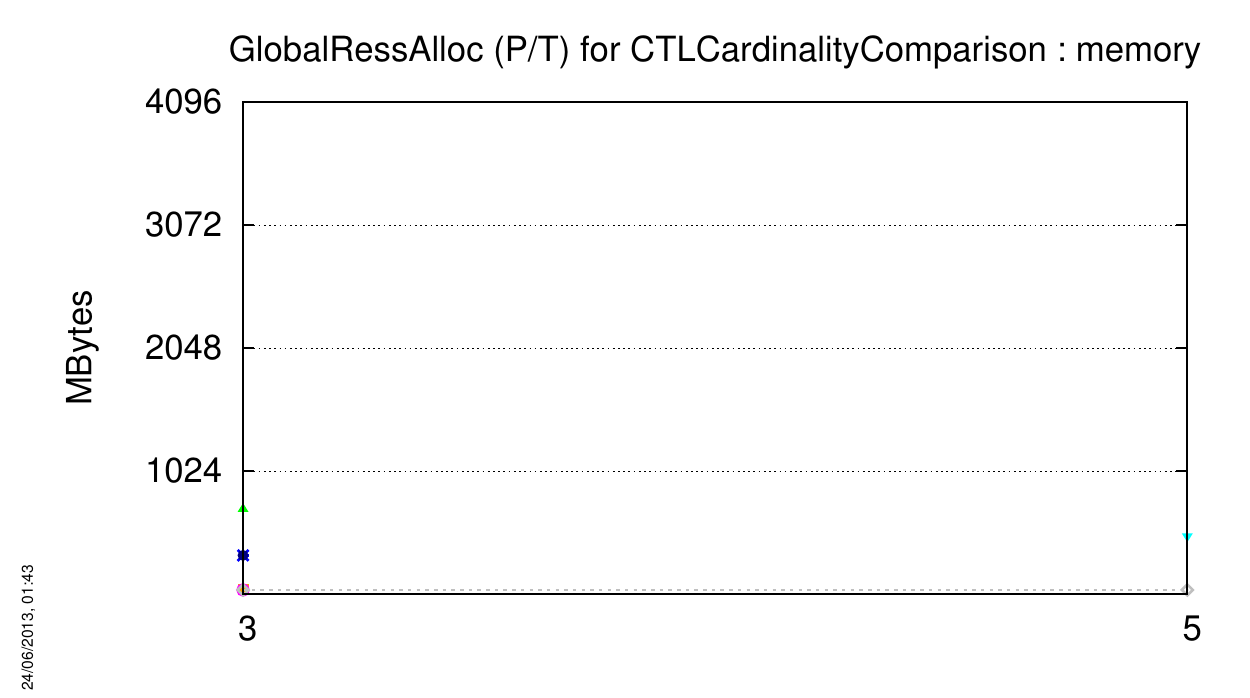}
   \includegraphics[width=7.2cm]{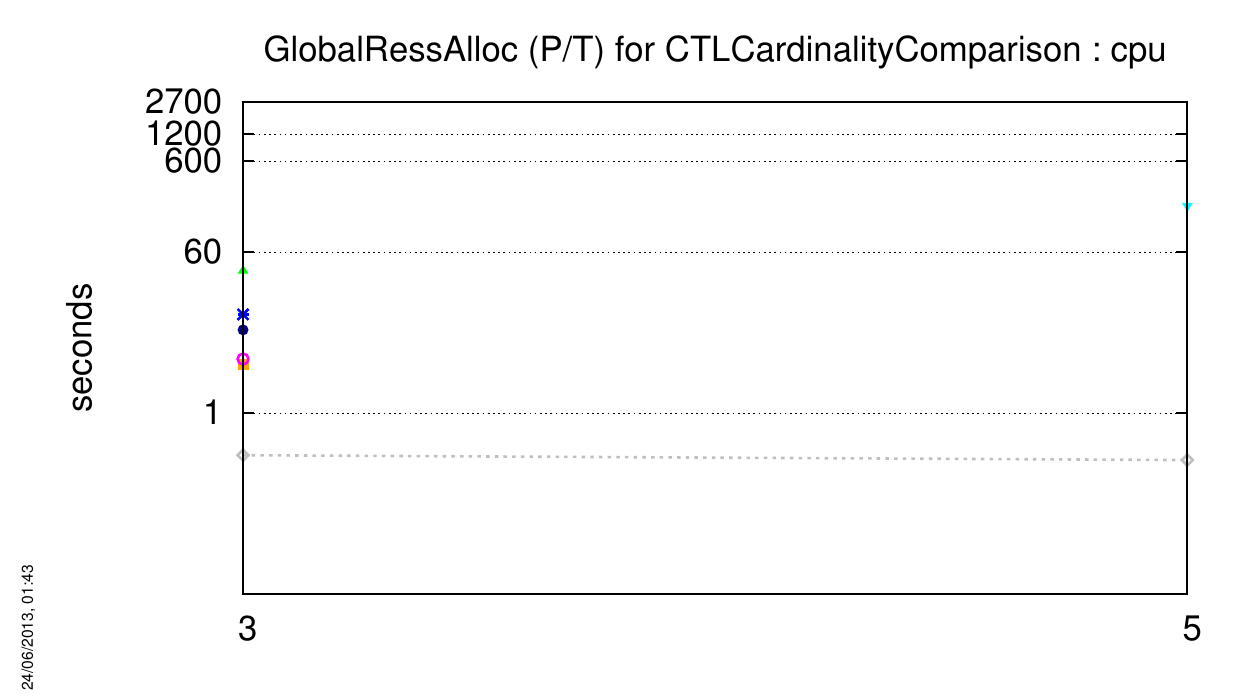}

   \includegraphics[height=1cm]{figures/tools-legend.pdf}
\end{center}

\subsubsection{\acs{Kanban-PT}}
The charts below respectively show how tools compete with this ``Known'' model (memory and CPU).

\index{Performances!CTLCardinalityComparison!Kanban (P/T)}
\begin{center}
   \includegraphics[width=7.2cm]{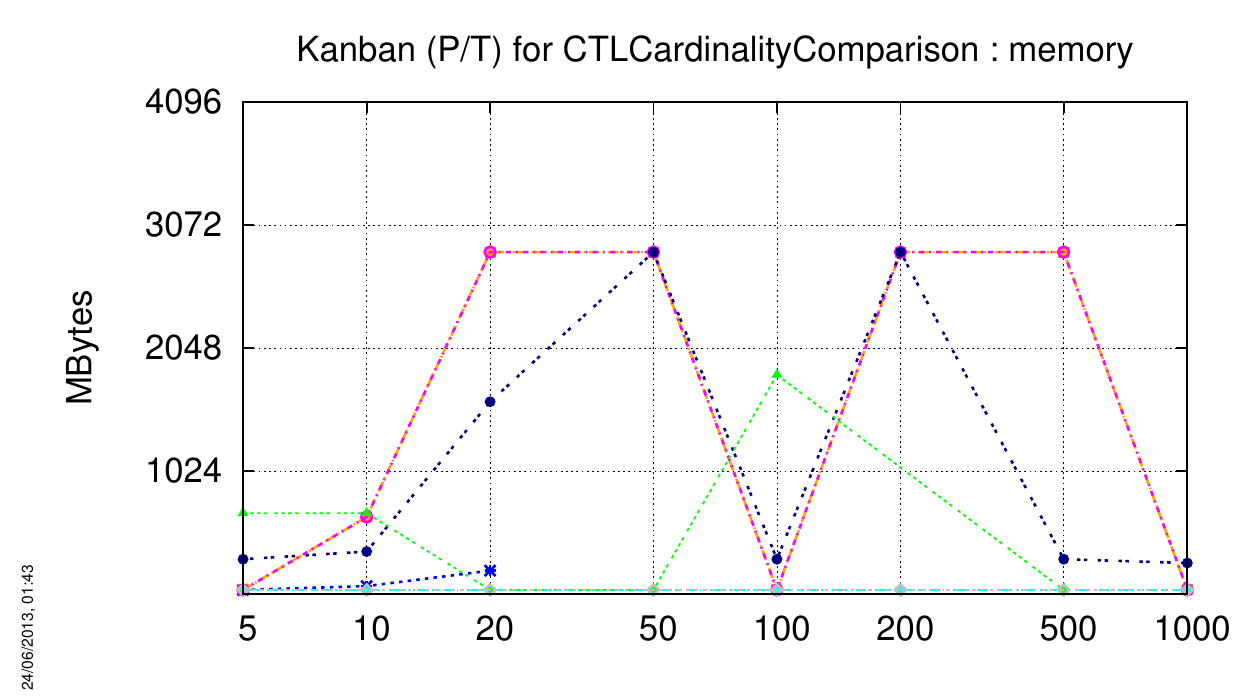}
   \includegraphics[width=7.2cm]{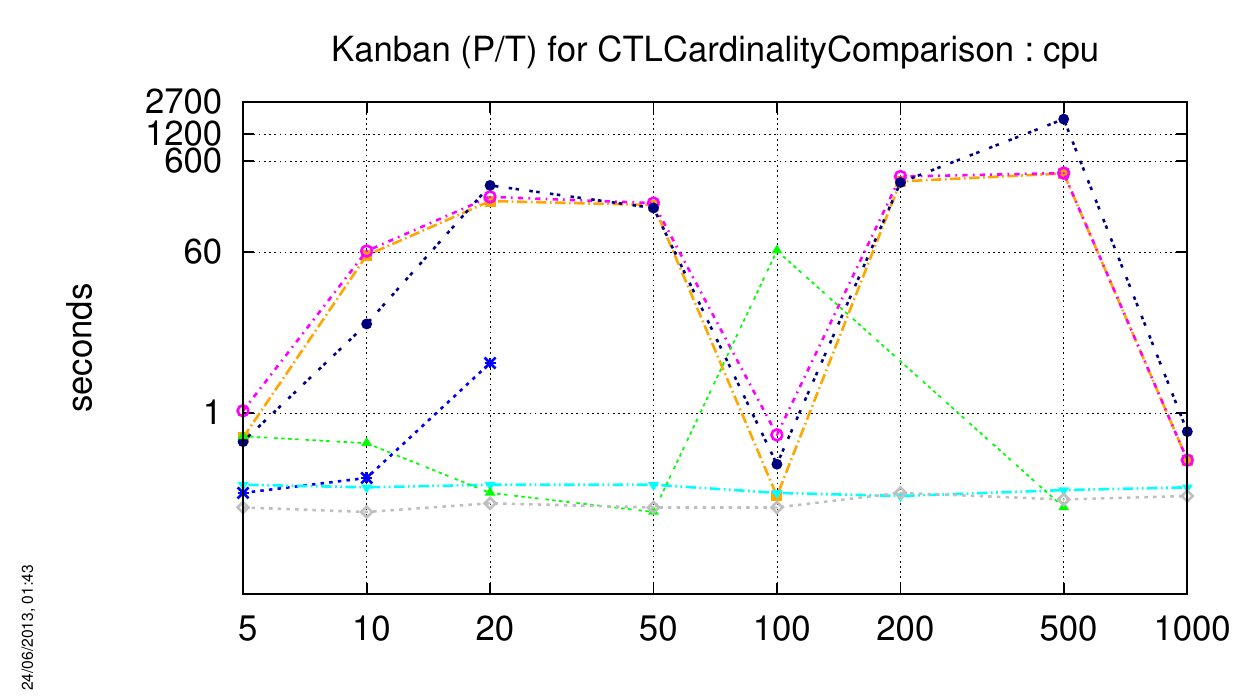}

   \includegraphics[height=1cm]{figures/tools-legend.pdf}
\end{center}

\subsubsection{\acs{LamportFastMutEx-COL}}
No instance of this model could be computed for the \textbf{CTLCardinalityComparison} examination.

\subsubsection{\acs{LamportFastMutEx-PT}}
The charts below respectively show how tools compete with this ``Known'' model (memory and CPU).

\index{Performances!CTLCardinalityComparison!LamportFastMutEx (P/T)}
\begin{center}
   \includegraphics[width=7.2cm]{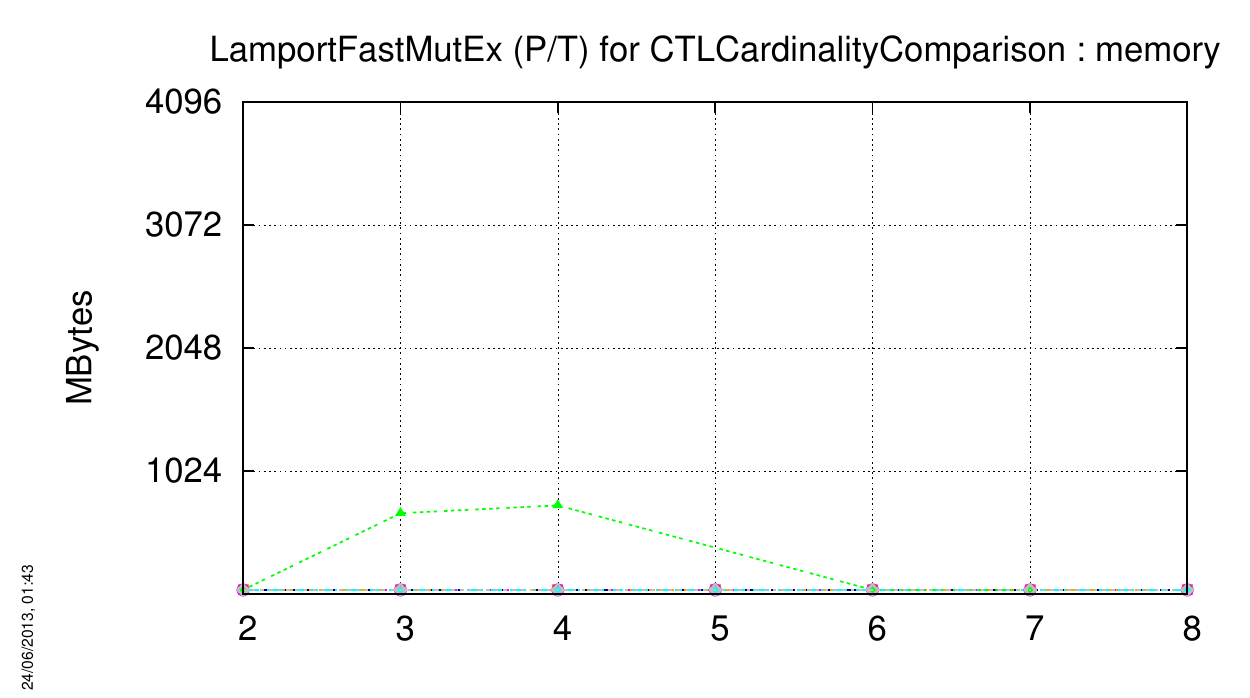}
   \includegraphics[width=7.2cm]{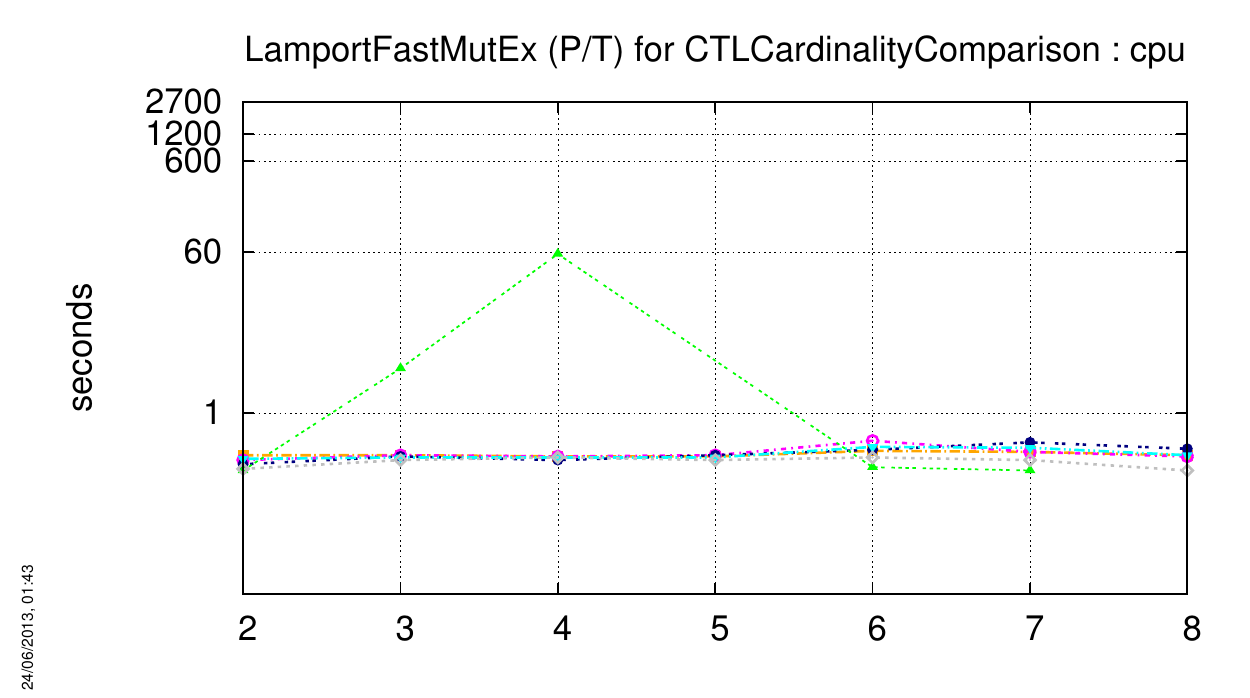}

   \includegraphics[height=1cm]{figures/tools-legend.pdf}
\end{center}

\subsubsection{\acs{MAPK-PT}}
The charts below respectively show how tools compete with this ``Known'' model (memory and CPU).

\index{Performances!CTLCardinalityComparison!MAPK (P/T)}
\begin{center}
   \includegraphics[width=7.2cm]{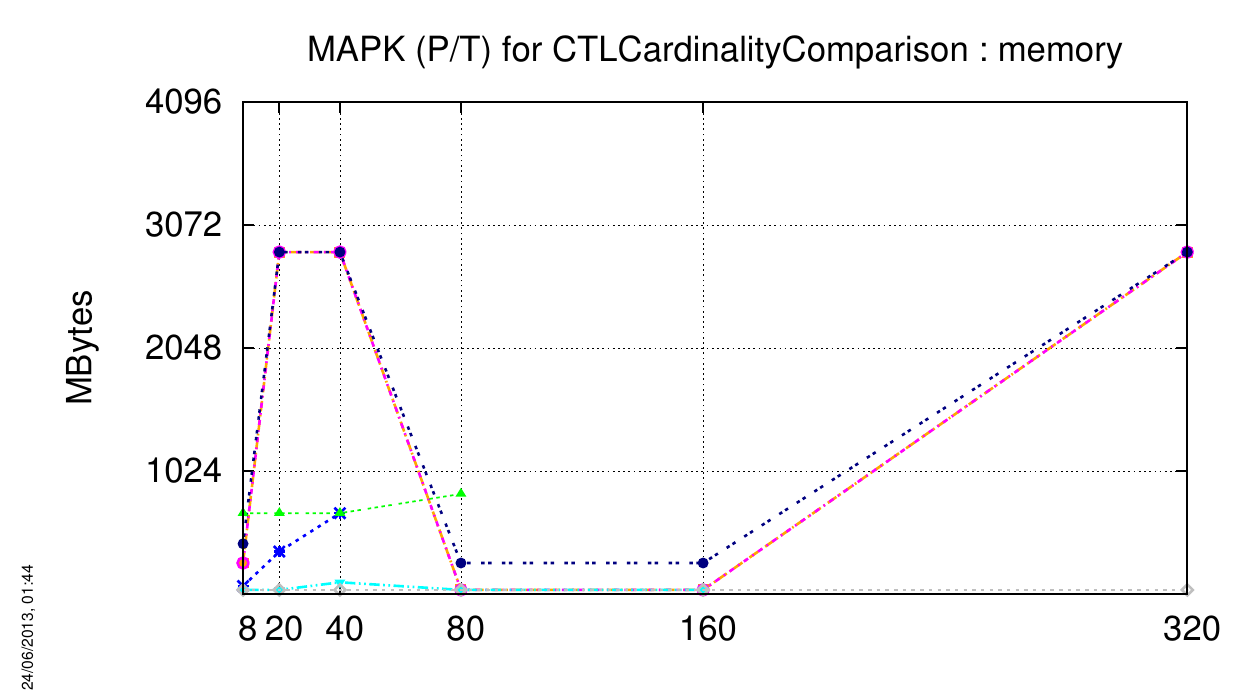}
   \includegraphics[width=7.2cm]{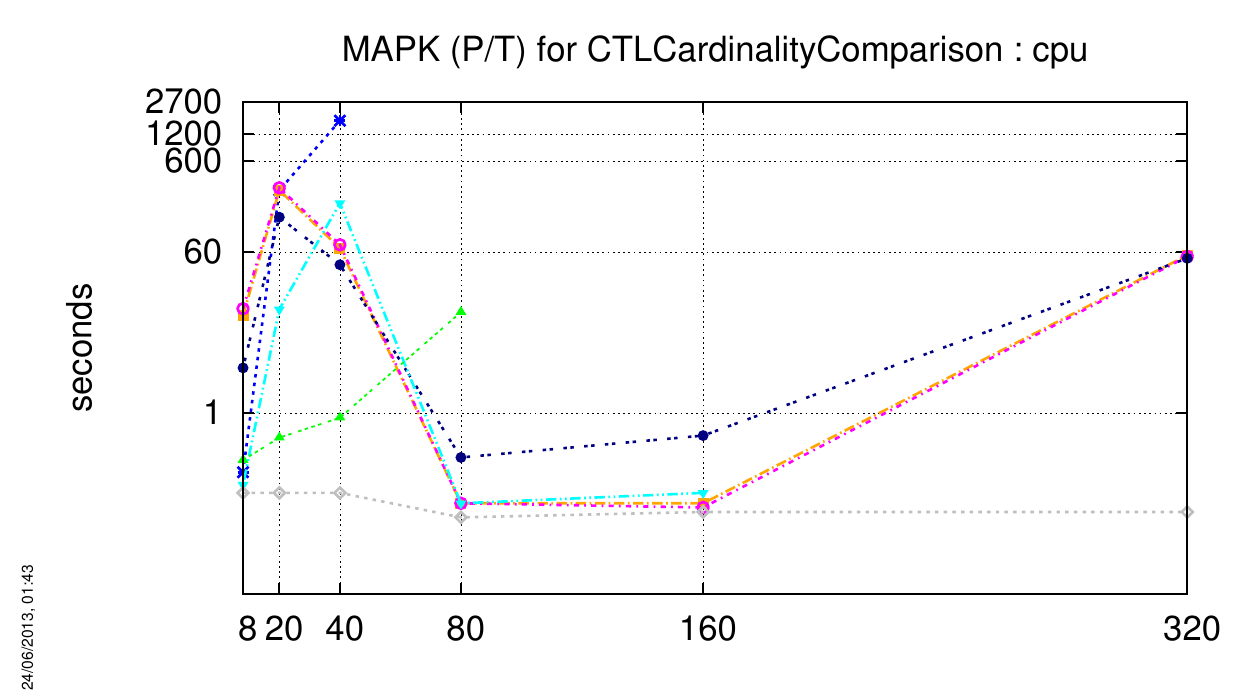}

   \includegraphics[height=1cm]{figures/tools-legend.pdf}
\end{center}

\subsubsection{\acs{NeoElection-COL}}
No instance of this model could be computed for the \textbf{CTLCardinalityComparison} examination.

\subsubsection{\acs{NeoElection-PT}}
The charts below respectively show how tools compete with this ``Known'' model (memory and CPU).

\index{Performances!CTLCardinalityComparison!NeoElection (P/T)}
\begin{center}
   \includegraphics[width=7.2cm]{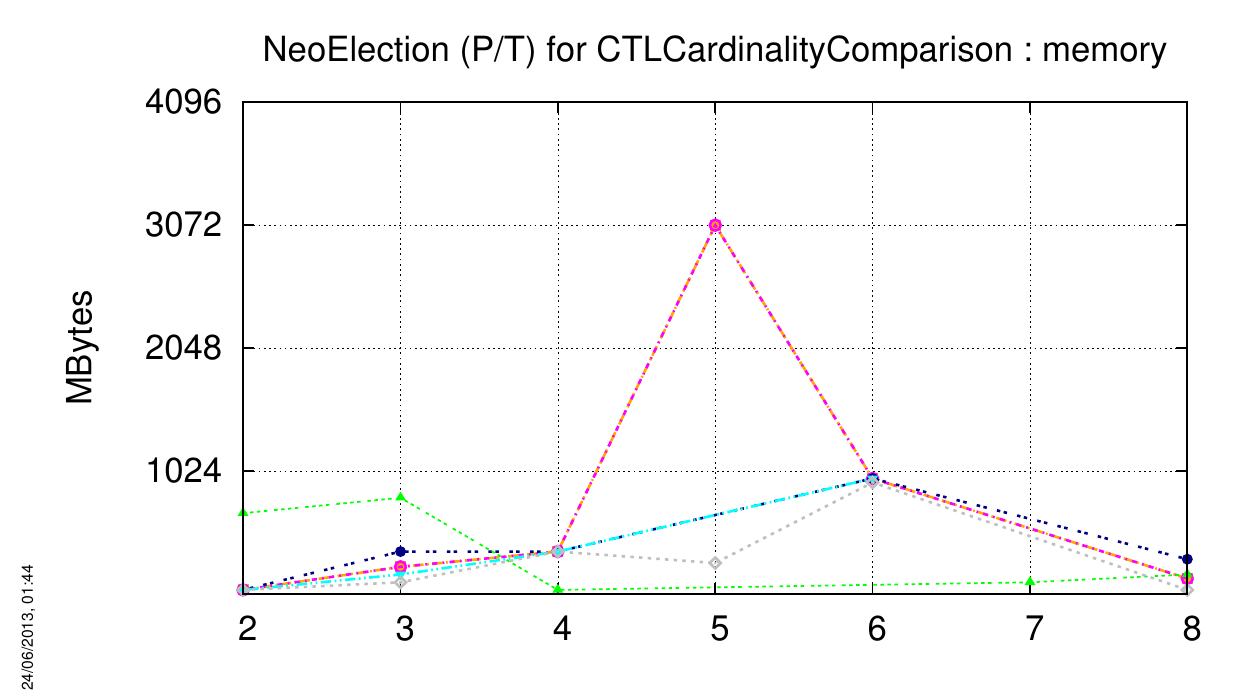}
   \includegraphics[width=7.2cm]{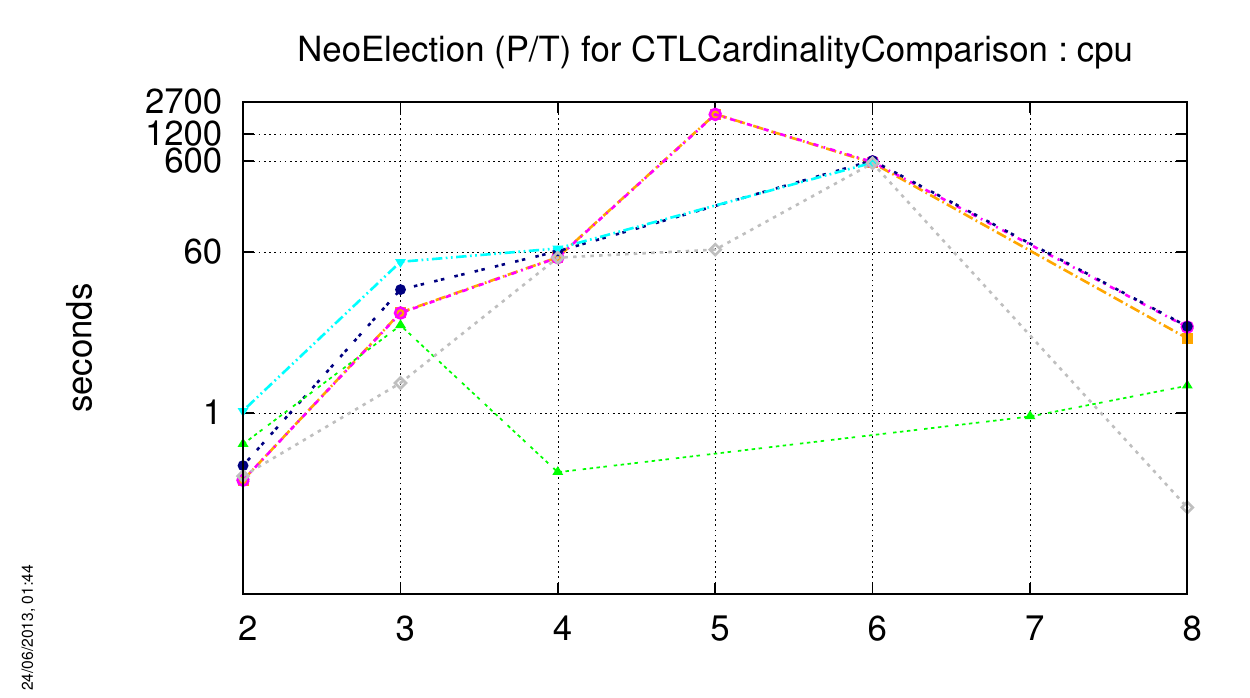}

   \includegraphics[height=1cm]{figures/tools-legend.pdf}
\end{center}

\subsubsection{\acs{PermAdmissibility-COL}}
No instance of this model could be computed for the \textbf{CTLCardinalityComparison} examination.

\subsubsection{\acs{PermAdmissibility-PT}}
The charts below respectively show how tools compete with this ``Known'' model (memory and CPU).

\index{Performances!CTLCardinalityComparison!PermAdmissibility (P/T)}
\begin{center}
   \includegraphics[width=7.2cm]{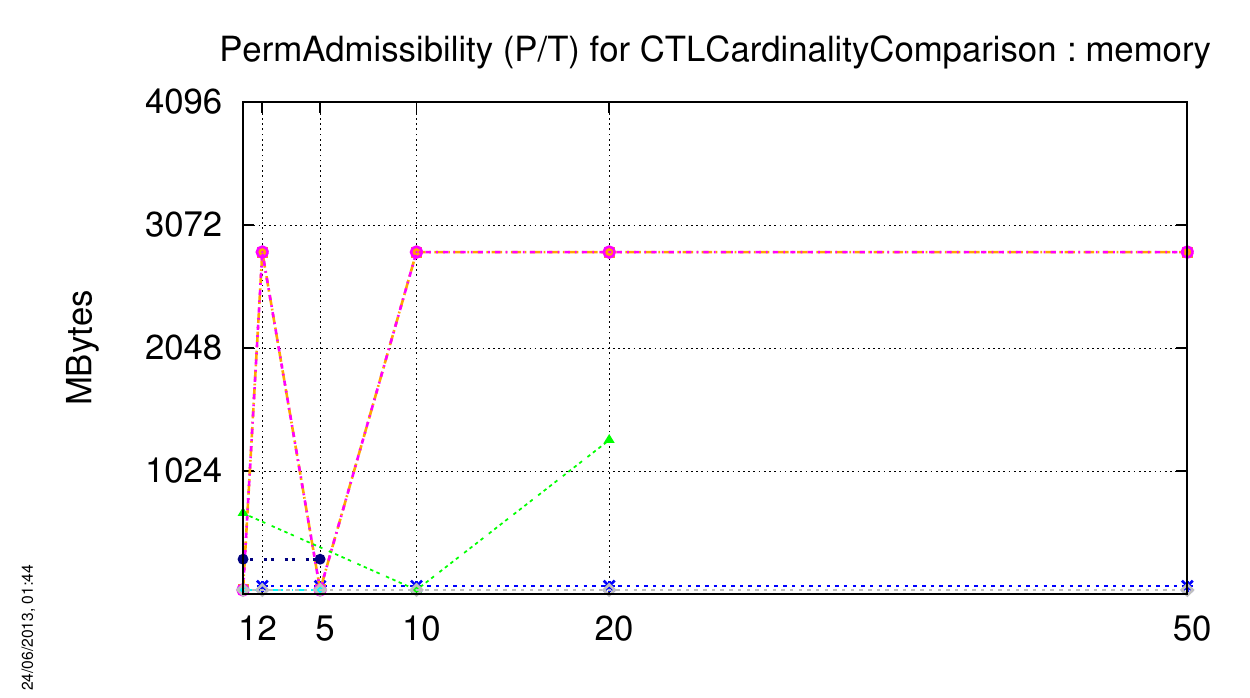}
   \includegraphics[width=7.2cm]{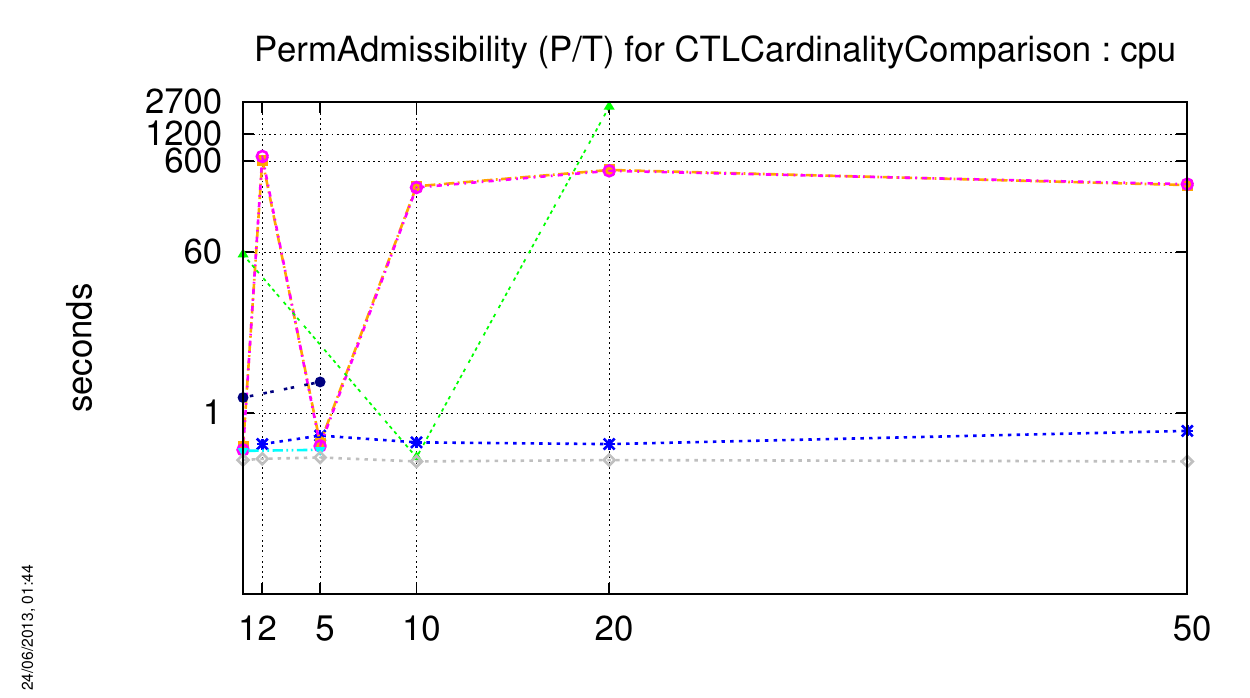}

   \includegraphics[height=1cm]{figures/tools-legend.pdf}
\end{center}

\subsubsection{\acs{Peterson-COL}}
No instance of this model could be computed for the \textbf{CTLCardinalityComparison} examination.

\subsubsection{\acs{Peterson-PT}}
The charts below respectively show how tools compete with this ``Known'' model (memory and CPU).

\index{Performances!CTLCardinalityComparison!Peterson (P/T)}
\begin{center}
   \includegraphics[width=7.2cm]{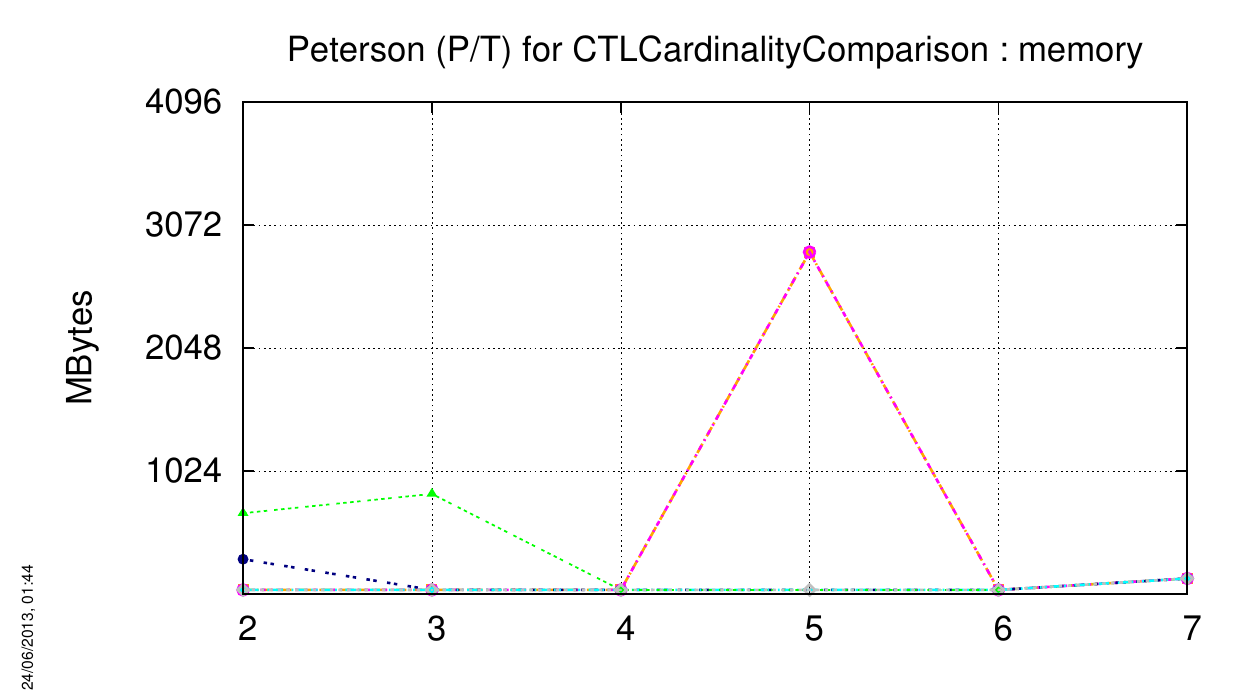}
   \includegraphics[width=7.2cm]{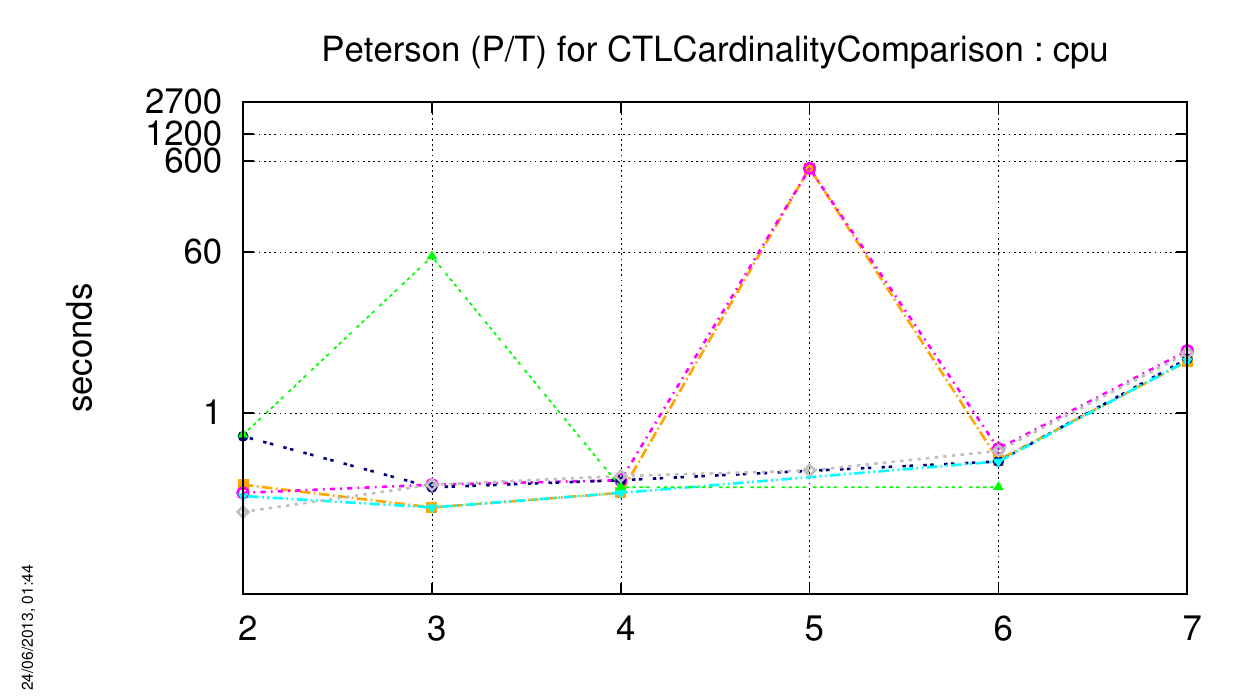}

   \includegraphics[height=1cm]{figures/tools-legend.pdf}
\end{center}

\subsubsection{\acs{Philosophers-COL}}
No instance of this model could be computed for the \textbf{CTLCardinalityComparison} examination.

\subsubsection{\acs{Philosophers-PT}}
The charts below respectively show how tools compete with this ``Known'' model (memory and CPU).

\index{Performances!CTLCardinalityComparison!Philosophers (P/T)}
\begin{center}
   \includegraphics[width=7.2cm]{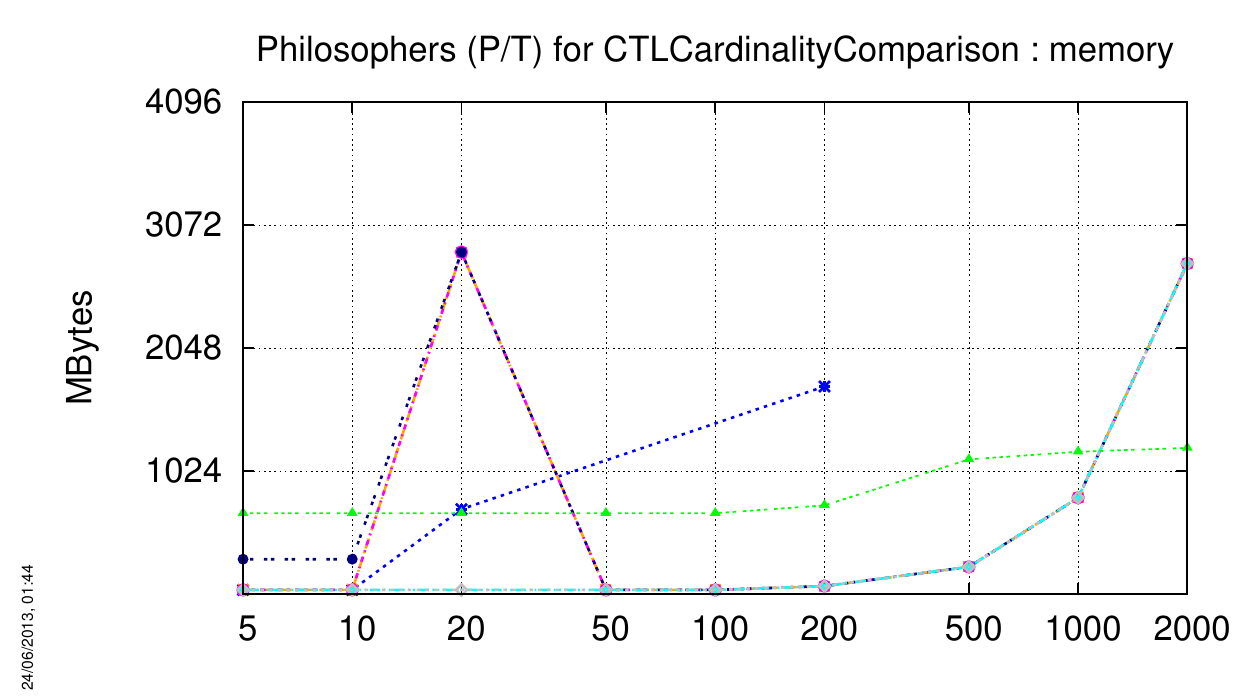}
   \includegraphics[width=7.2cm]{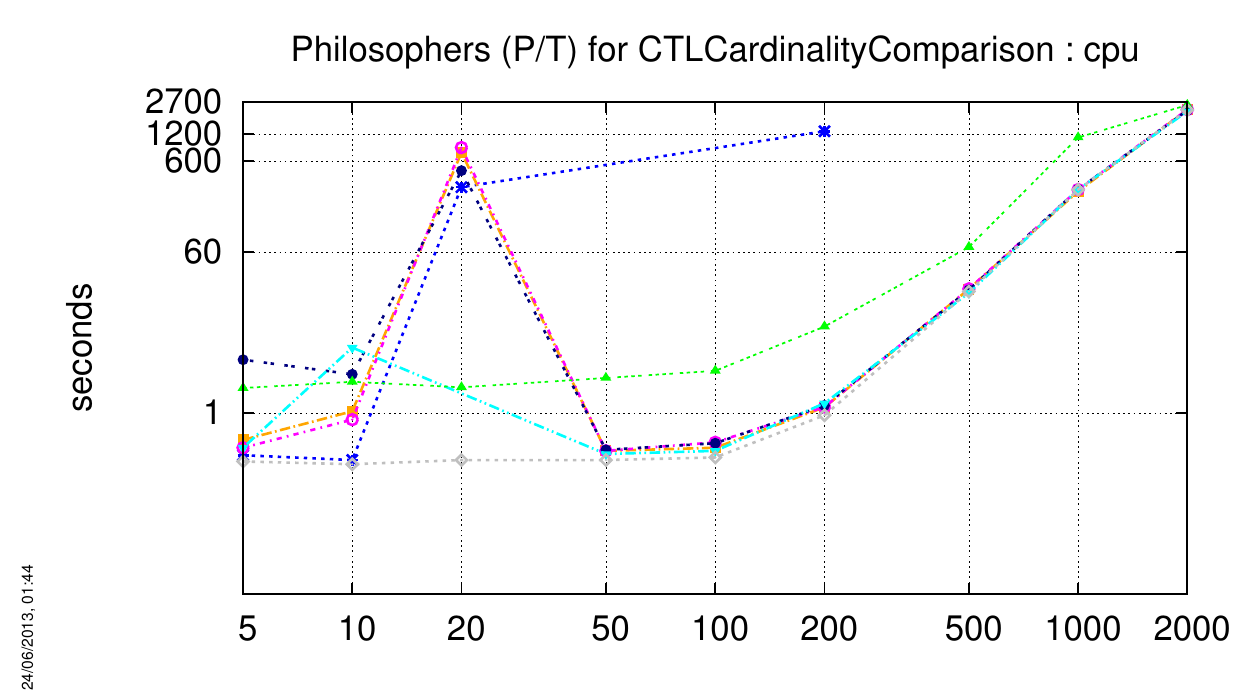}

   \includegraphics[height=1cm]{figures/tools-legend.pdf}
\end{center}

\subsubsection{\acs{PhilosophersDyn-COL}}
No instance of this model could be computed for the \textbf{CTLCardinalityComparison} examination.

\subsubsection{\acs{PhilosophersDyn-PT}}
The charts below respectively show how tools compete with this ``Known'' model (memory and CPU).

\index{Performances!CTLCardinalityComparison!PhilosophersDyn (P/T)}
\begin{center}
   \includegraphics[width=7.2cm]{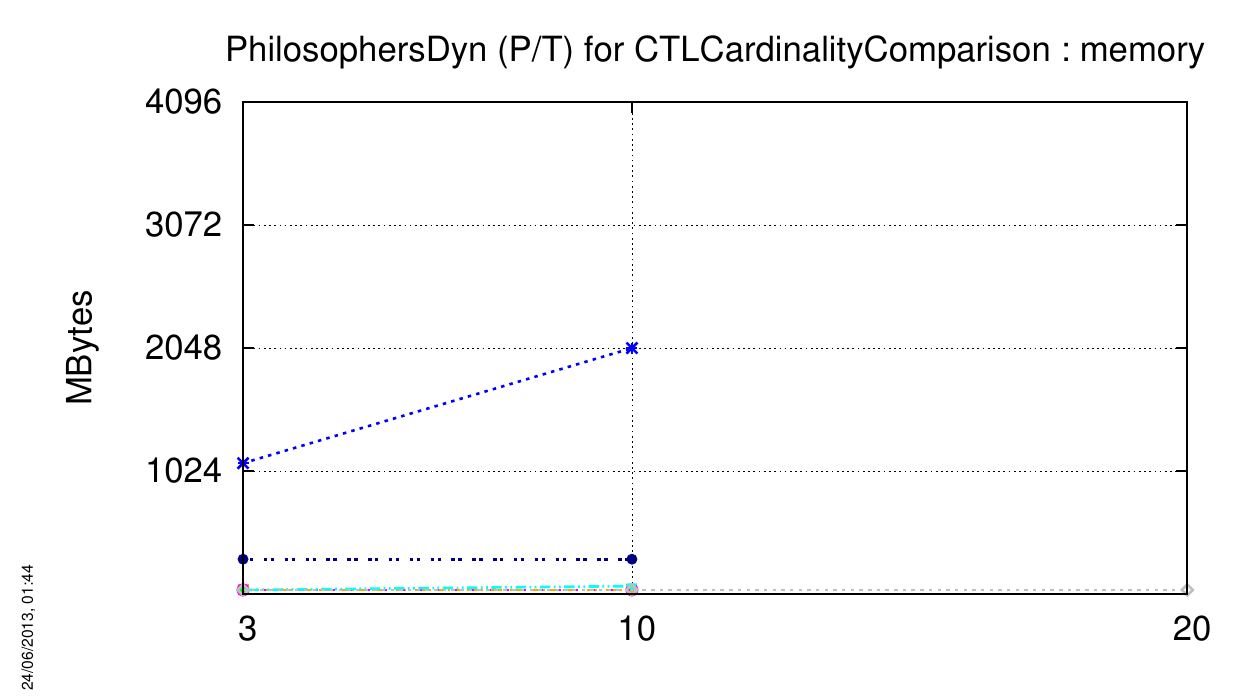}
   \includegraphics[width=7.2cm]{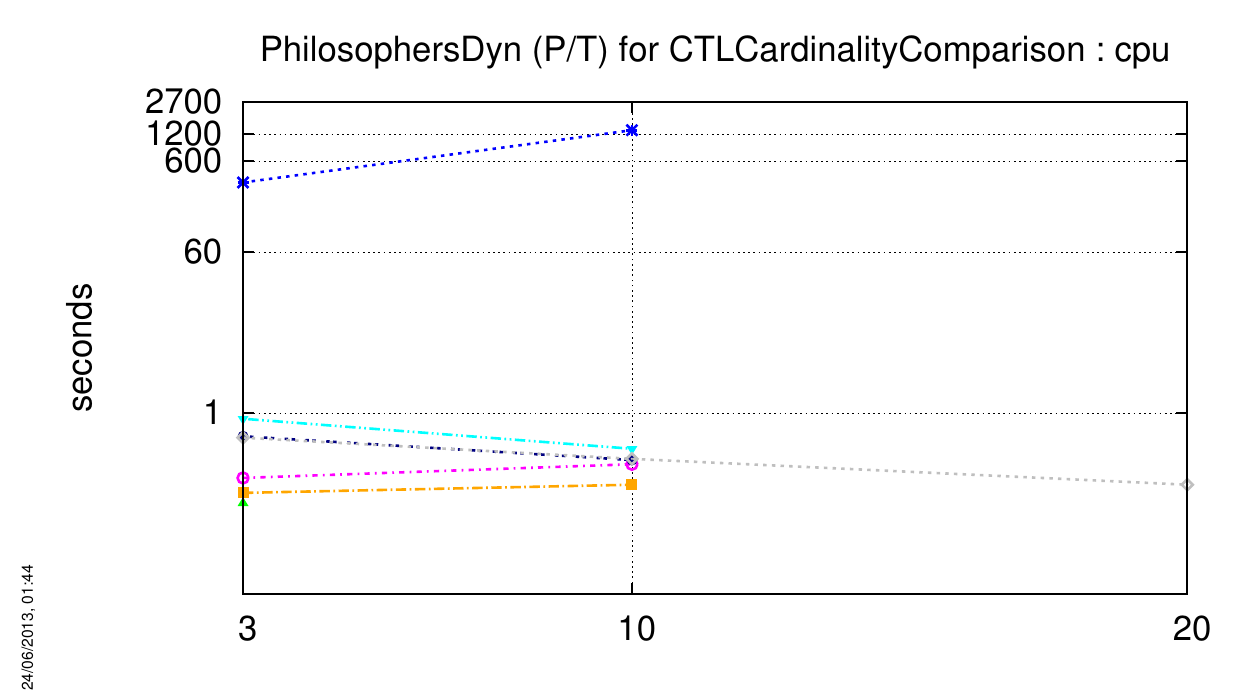}

   \includegraphics[height=1cm]{figures/tools-legend.pdf}
\end{center}

\subsubsection{\acs{Planning-PT}}
No instance of this model could be computed for the \textbf{CTLCardinalityComparison} examination.

\subsubsection{\acs{Railroad-PT}}
The charts below respectively show how tools compete with this ``Known'' model (memory and CPU).

\index{Performances!CTLCardinalityComparison!Railroad (P/T)}
\begin{center}
   \includegraphics[width=7.2cm]{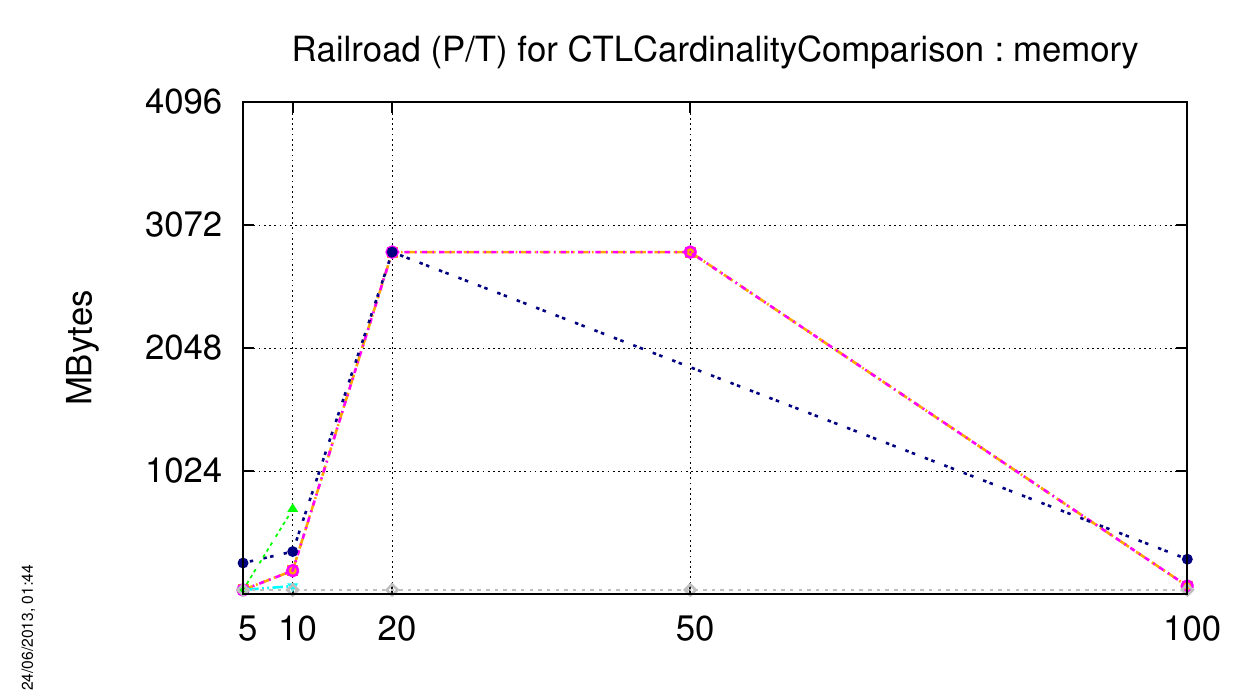}
   \includegraphics[width=7.2cm]{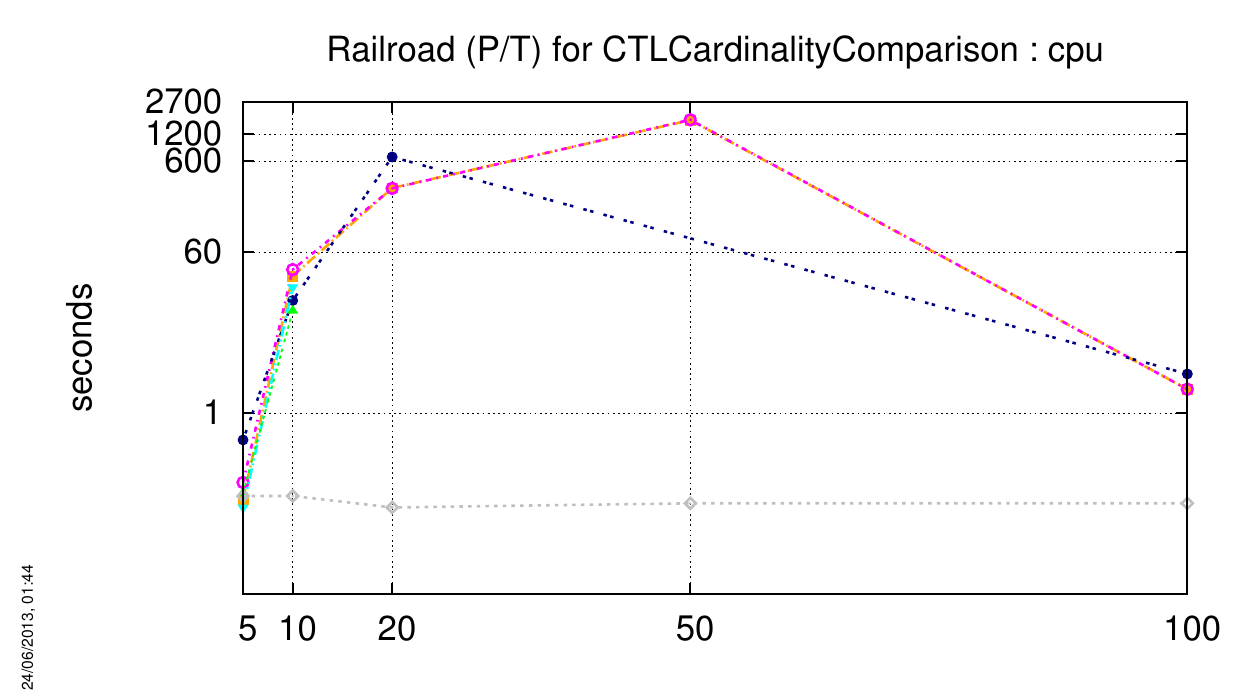}

   \includegraphics[height=1cm]{figures/tools-legend.pdf}
\end{center}

\subsubsection{\acs{RessAllocation-PT}}
The charts below respectively show how tools compete with this ``Known'' model (memory and CPU).

\index{Performances!CTLCardinalityComparison!RessAllocation (P/T)}
\begin{center}
   \includegraphics[width=7.2cm]{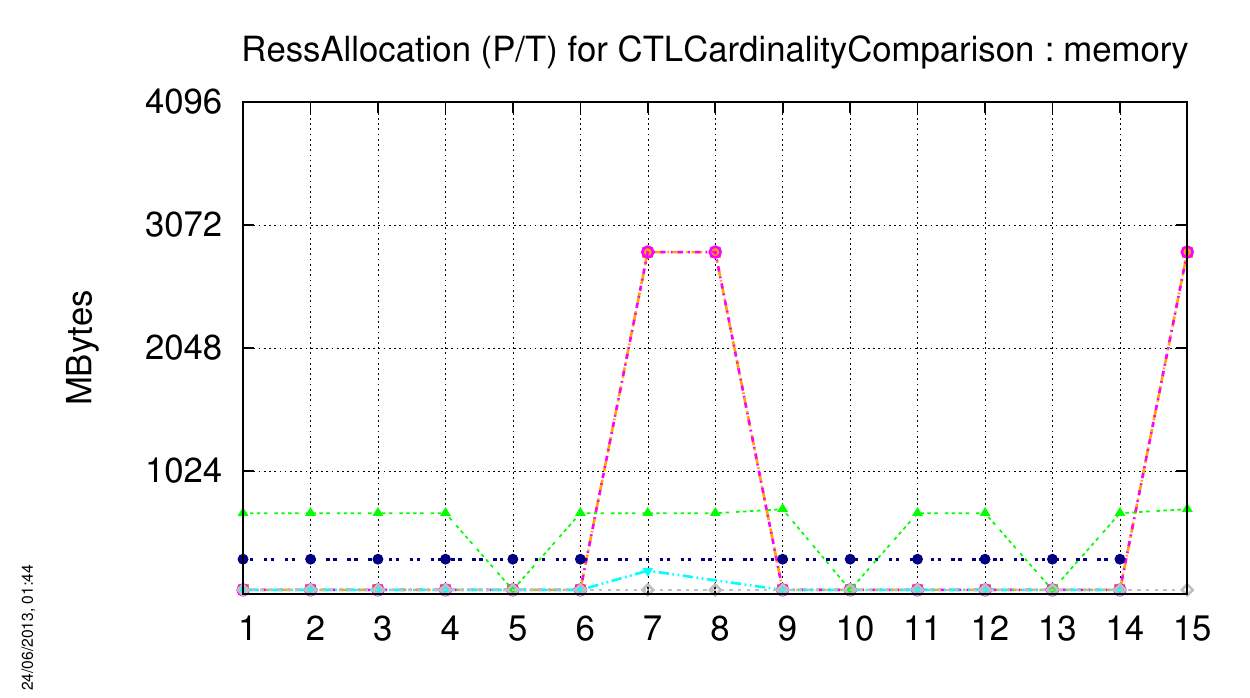}
   \includegraphics[width=7.2cm]{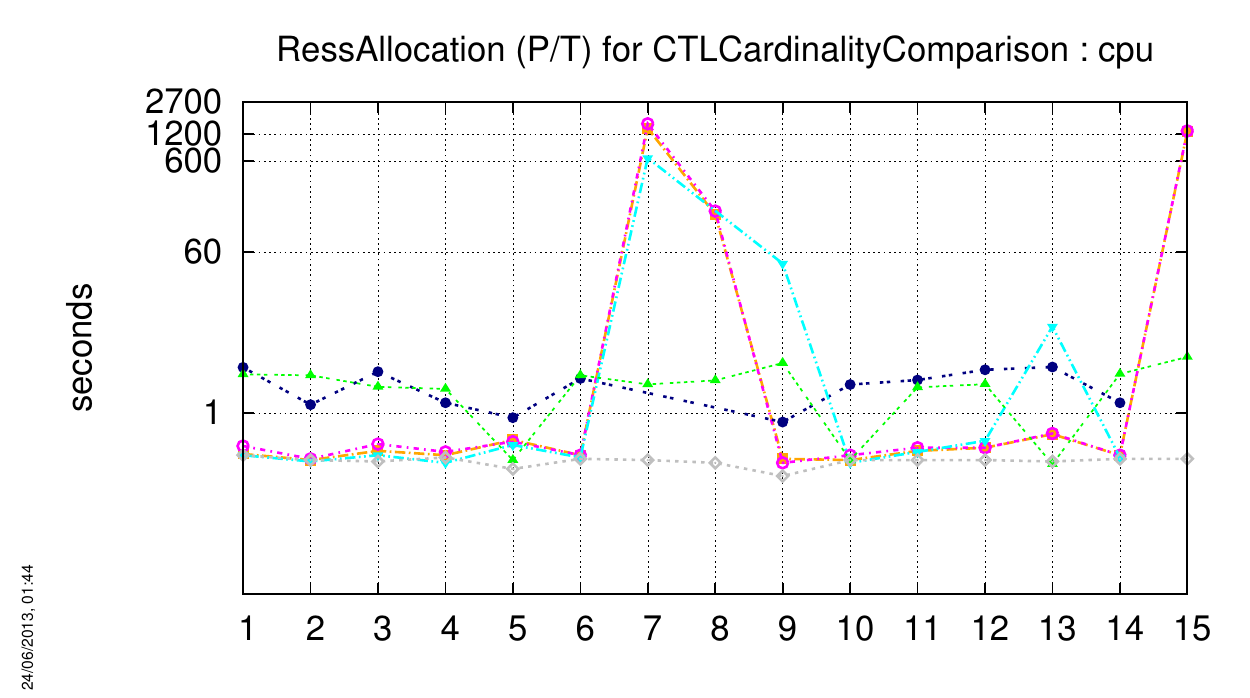}

   \includegraphics[height=1cm]{figures/tools-legend.pdf}
\end{center}

\subsubsection{\acs{Ring-PT}}
The charts below respectively show how tools compete with this ``Known'' model (memory and CPU).

\index{Performances!CTLCardinalityComparison!Ring (P/T)}
\begin{center}
   \includegraphics[width=7.2cm]{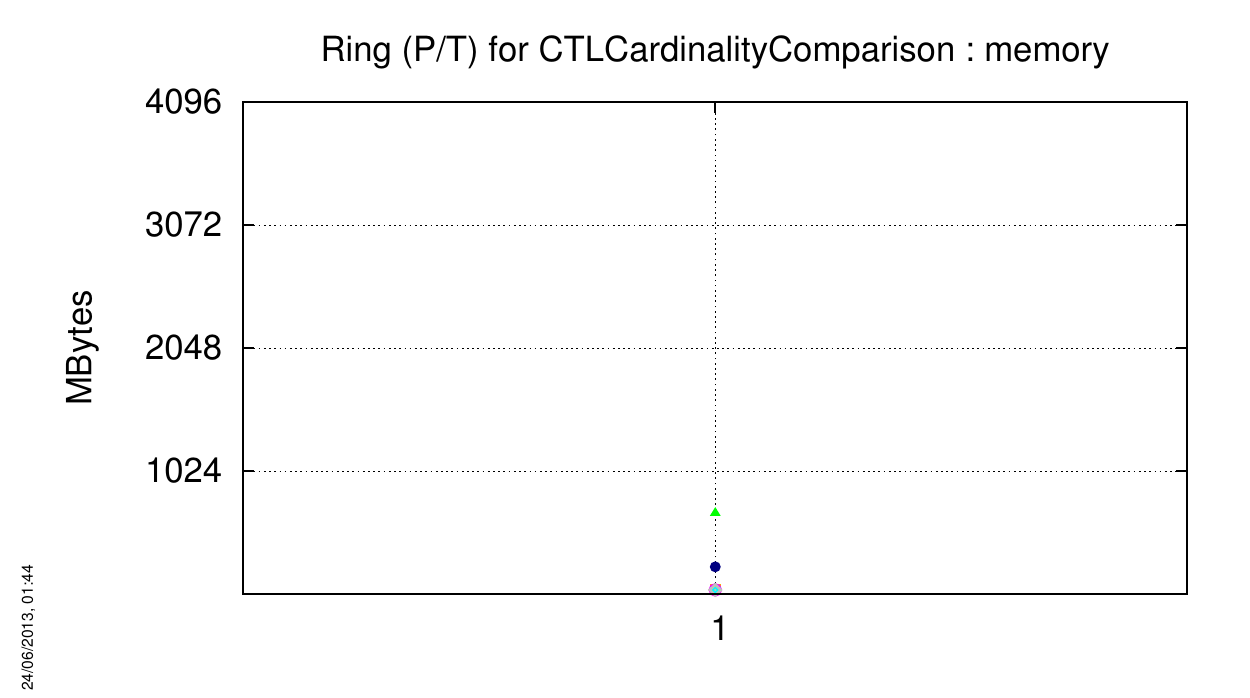}
   \includegraphics[width=7.2cm]{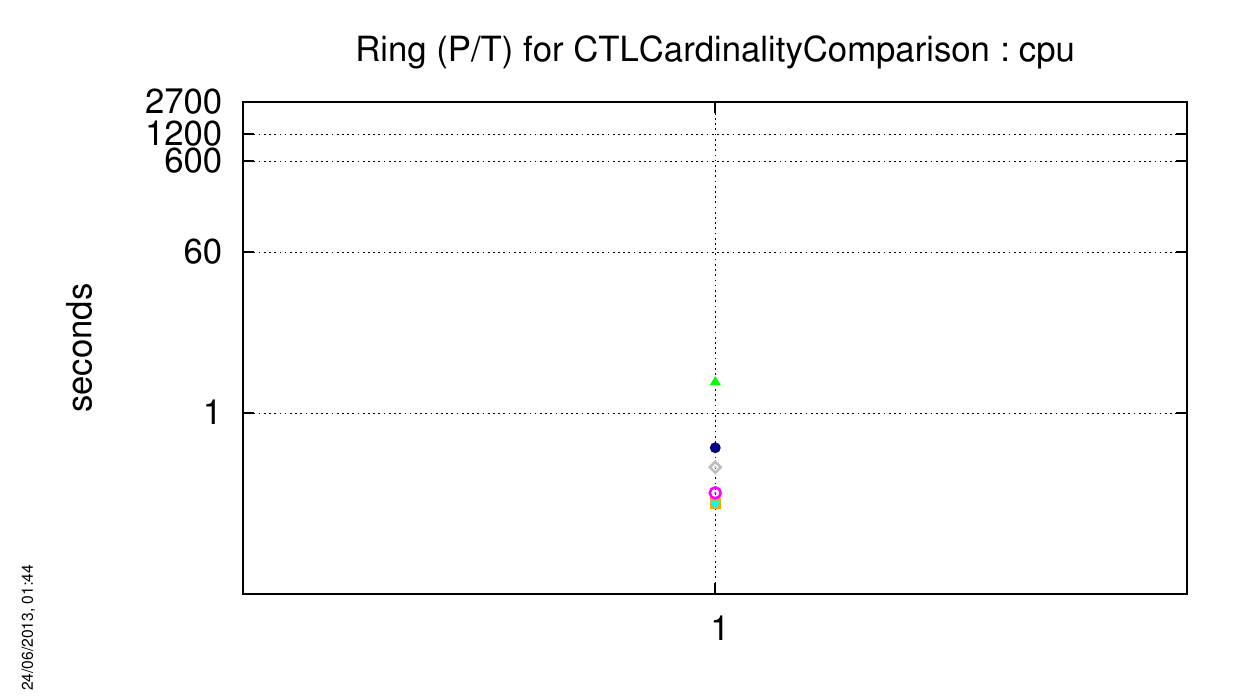}

   \includegraphics[height=1cm]{figures/tools-legend.pdf}
\end{center}

\subsubsection{\acs{RwMutex-PT}}
The charts below respectively show how tools compete with this ``Known'' model (memory and CPU).

\index{Performances!CTLCardinalityComparison!RwMutex (P/T)}
\begin{center}
   \includegraphics[width=7.2cm]{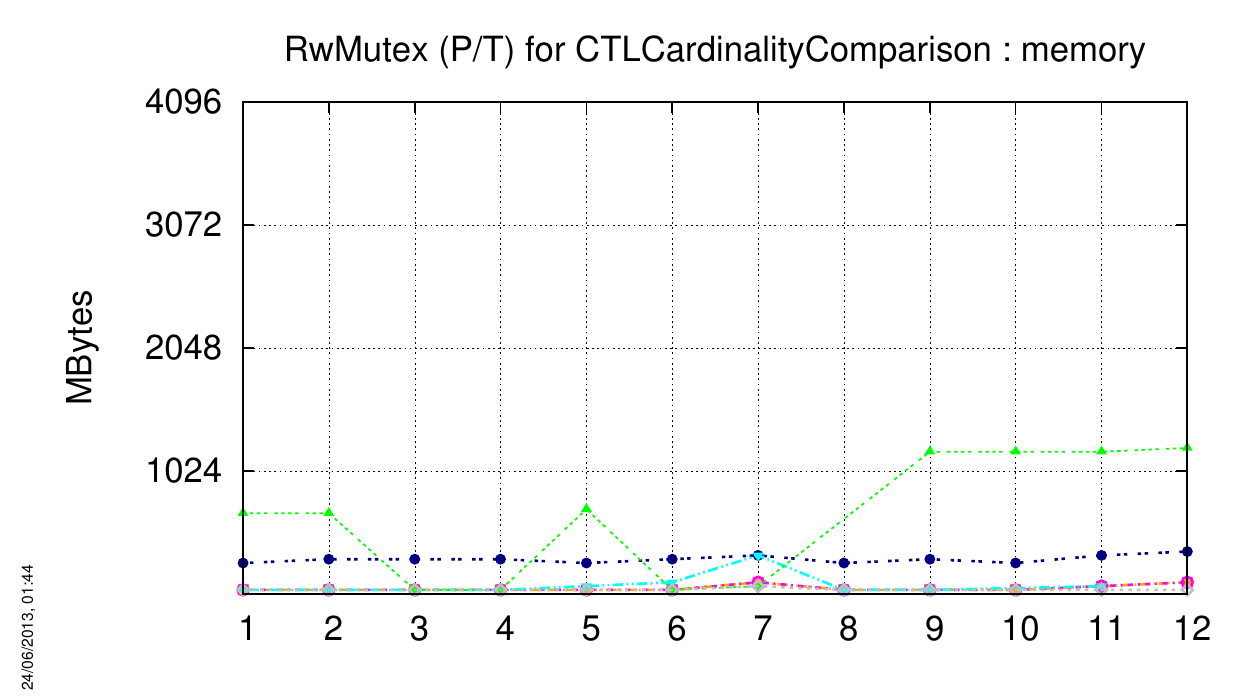}
   \includegraphics[width=7.2cm]{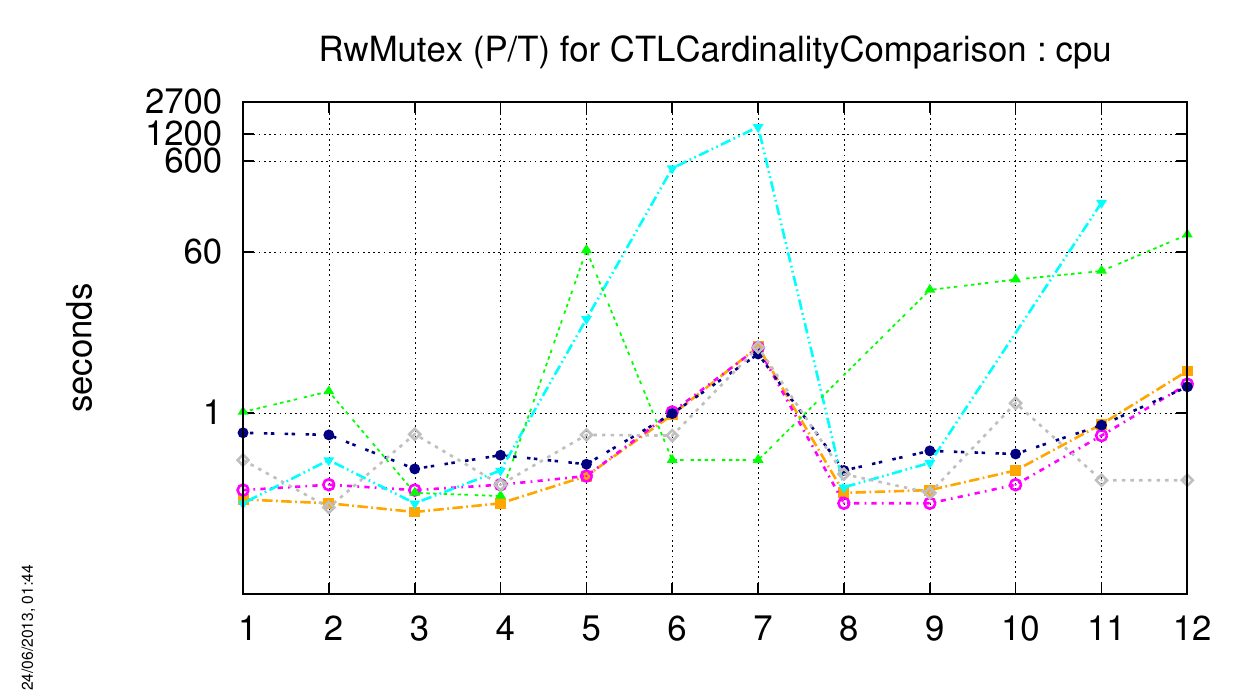}

   \includegraphics[height=1cm]{figures/tools-legend.pdf}
\end{center}

\subsubsection{\acs{SharedMemory-COL}}
No instance of this model could be computed for the \textbf{CTLCardinalityComparison} examination.

\subsubsection{\acs{SharedMemory-PT}}
The charts below respectively show how tools compete with this ``Known'' model (memory and CPU).

\index{Performances!CTLCardinalityComparison!SharedMemory (P/T)}
\begin{center}
   \includegraphics[width=7.2cm]{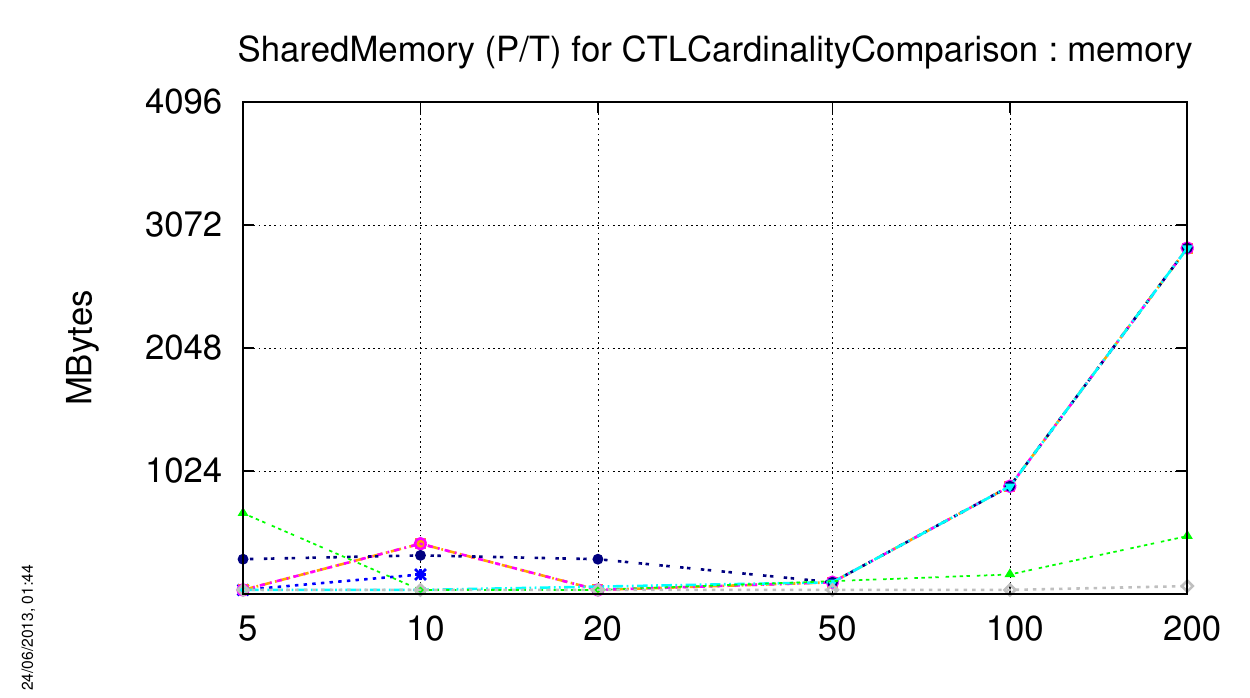}
   \includegraphics[width=7.2cm]{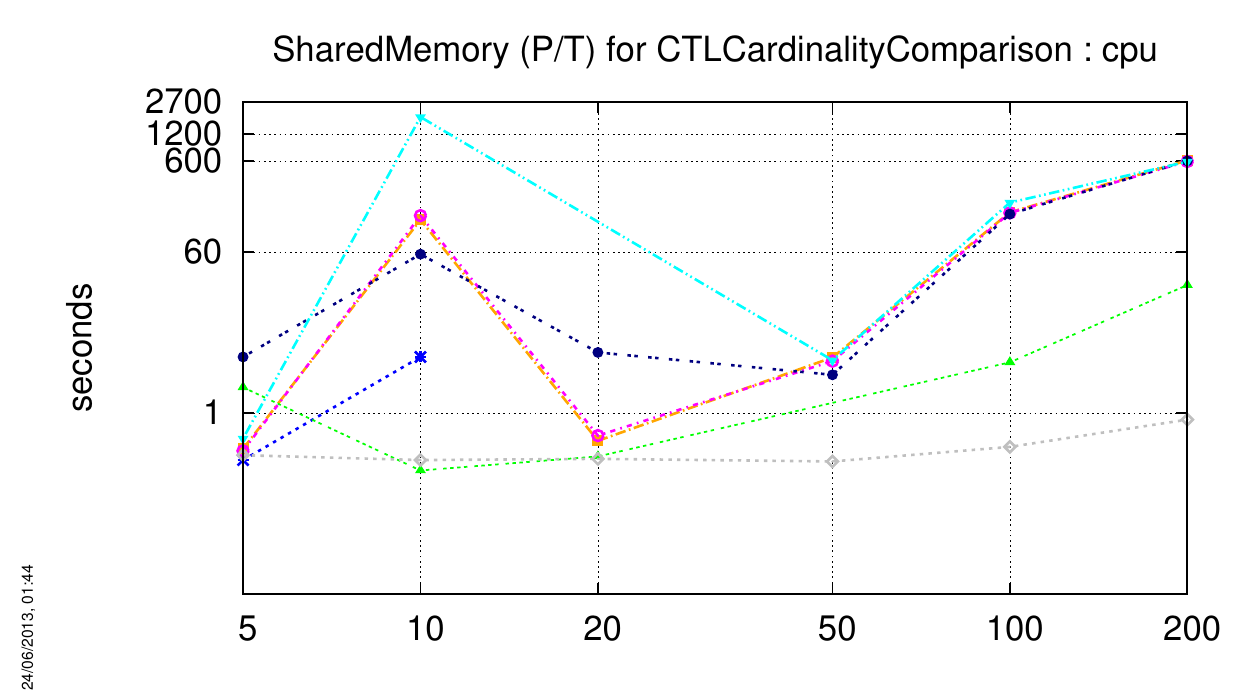}

   \includegraphics[height=1cm]{figures/tools-legend.pdf}
\end{center}

\subsubsection{\acs{SimpleLoadBal-COL}}
No instance of this model could be computed for the \textbf{CTLCardinalityComparison} examination.

\subsubsection{\acs{SimpleLoadBal-PT}}
The charts below respectively show how tools compete with this ``Known'' model (memory and CPU).

\index{Performances!CTLCardinalityComparison!SimpleLoadBal (P/T)}
\begin{center}
   \includegraphics[width=7.2cm]{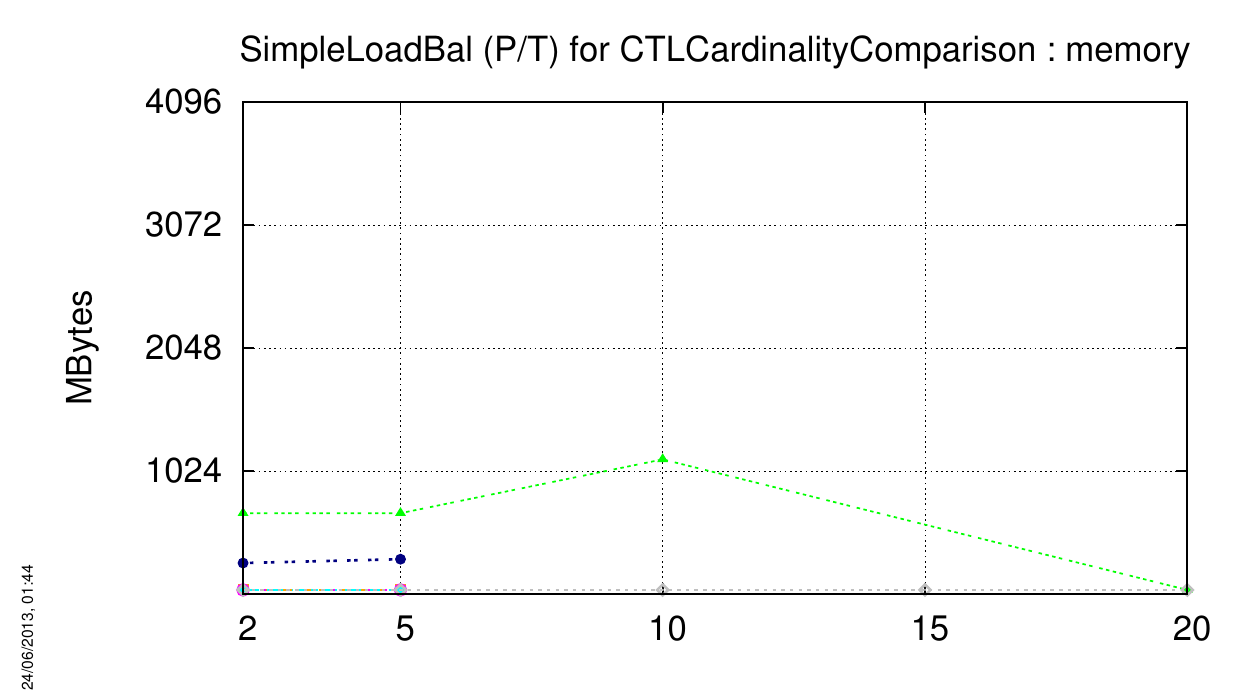}
   \includegraphics[width=7.2cm]{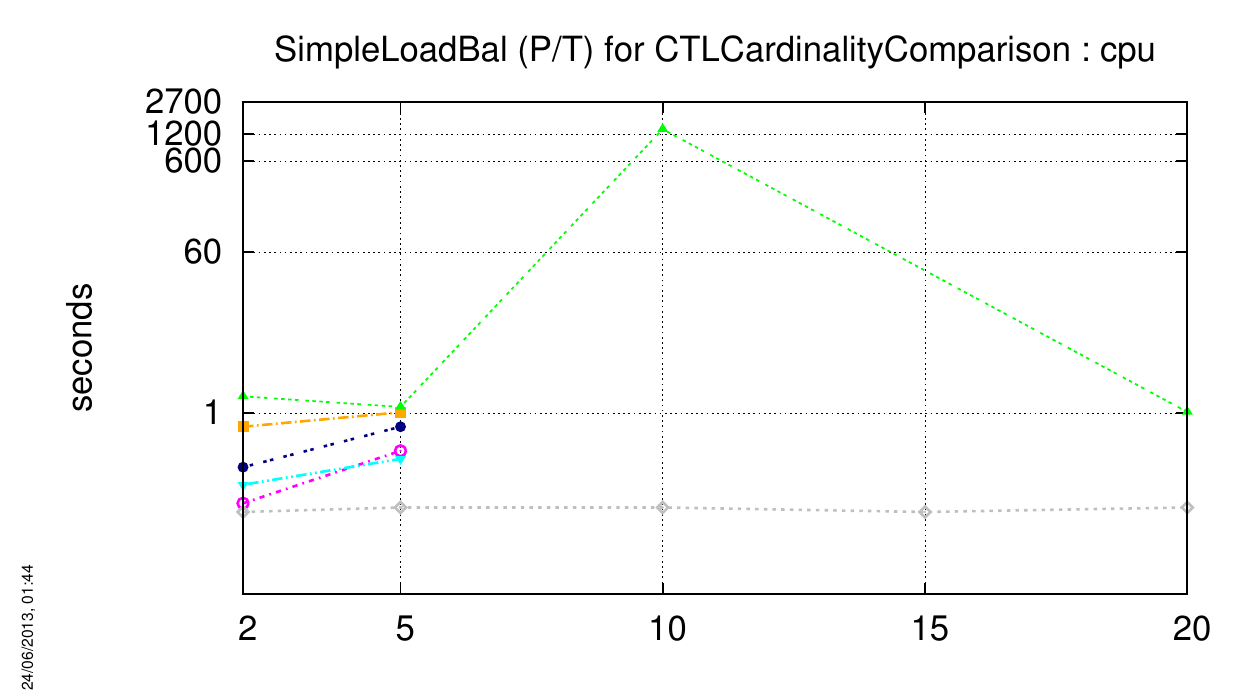}

   \includegraphics[height=1cm]{figures/tools-legend.pdf}
\end{center}

\subsubsection{\acs{TokenRing-COL}}
No instance of this model could be computed for the \textbf{CTLCardinalityComparison} examination.

\subsubsection{\acs{TokenRing-PT}}
No instance of this model could be computed for the \textbf{CTLCardinalityComparison} examination.

\subsubsection{\acs{HouseConstruction-PT}}
The charts below respectively show how tools compete with this ``Suprise'' model (memory and CPU).

\index{Performances!CTLCardinalityComparison!HouseConstruction (P/T)}
\begin{center}
   \includegraphics[width=7.2cm]{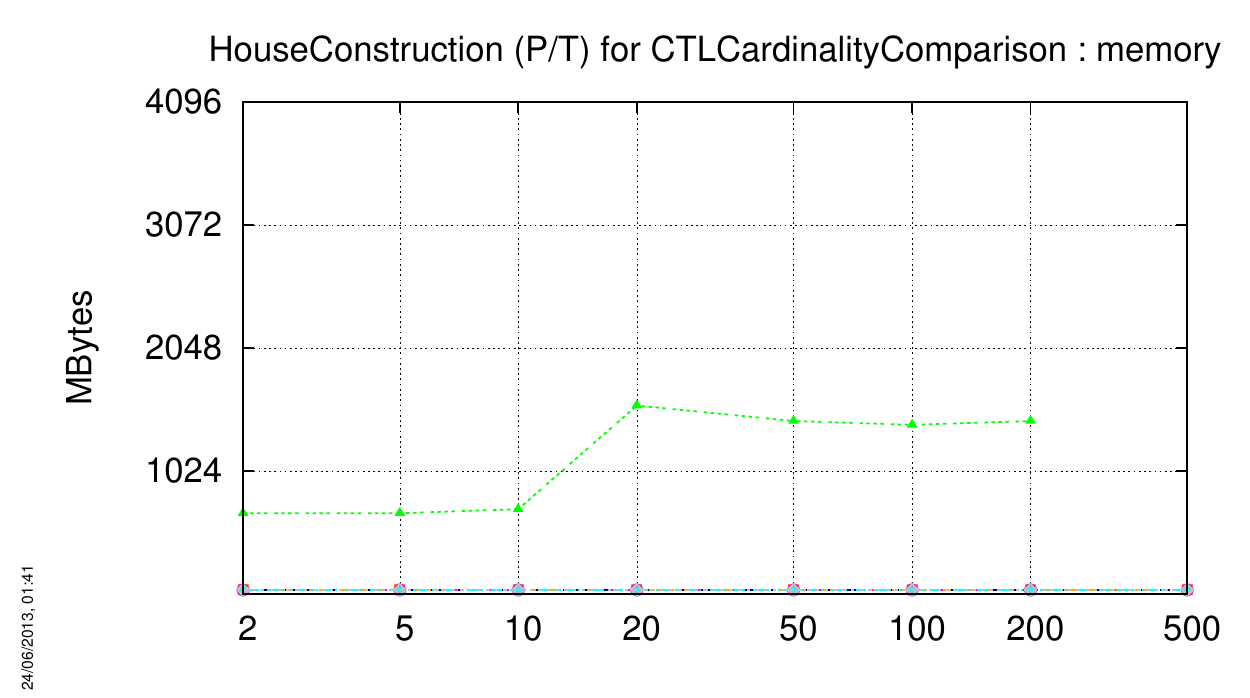}
   \includegraphics[width=7.2cm]{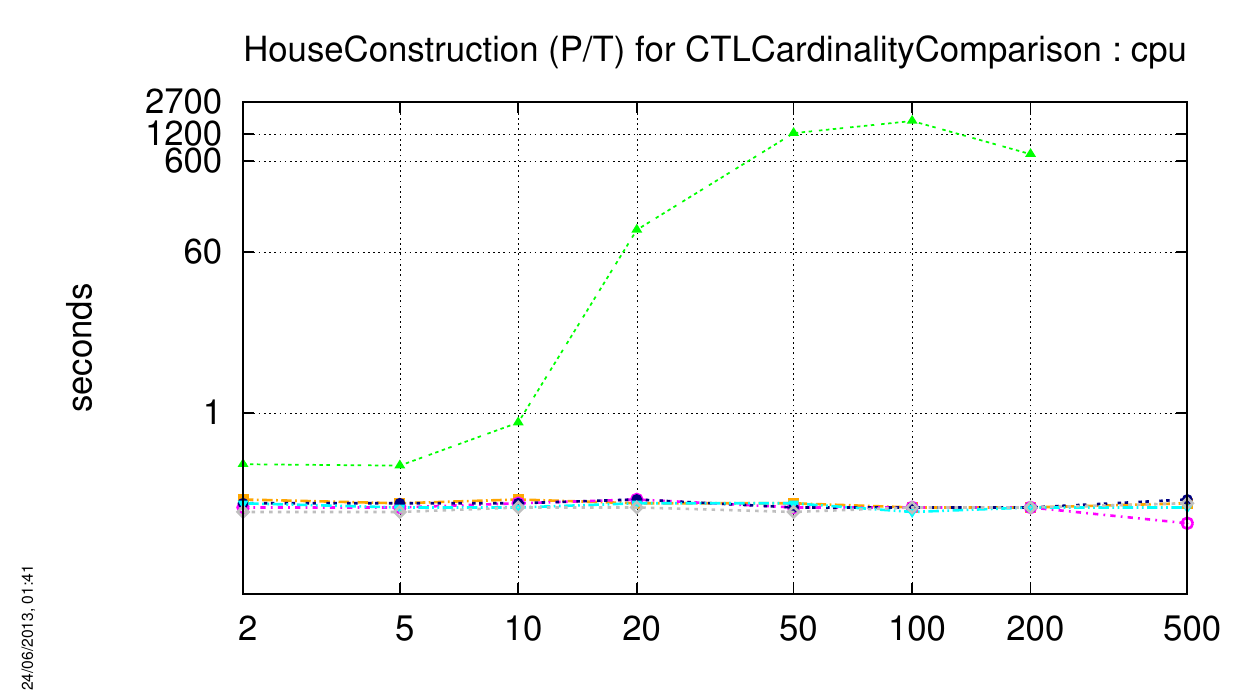}

   \includegraphics[height=1cm]{figures/tools-legend.pdf}
\end{center}

\subsubsection{\acs{IBMB2S565S3960-PT}}
The charts below respectively show how tools compete with this ``Suprise'' model (memory and CPU).

\index{Performances!CTLCardinalityComparison!IBMB2S565S3960 (P/T)}
\begin{center}
   \includegraphics[width=7.2cm]{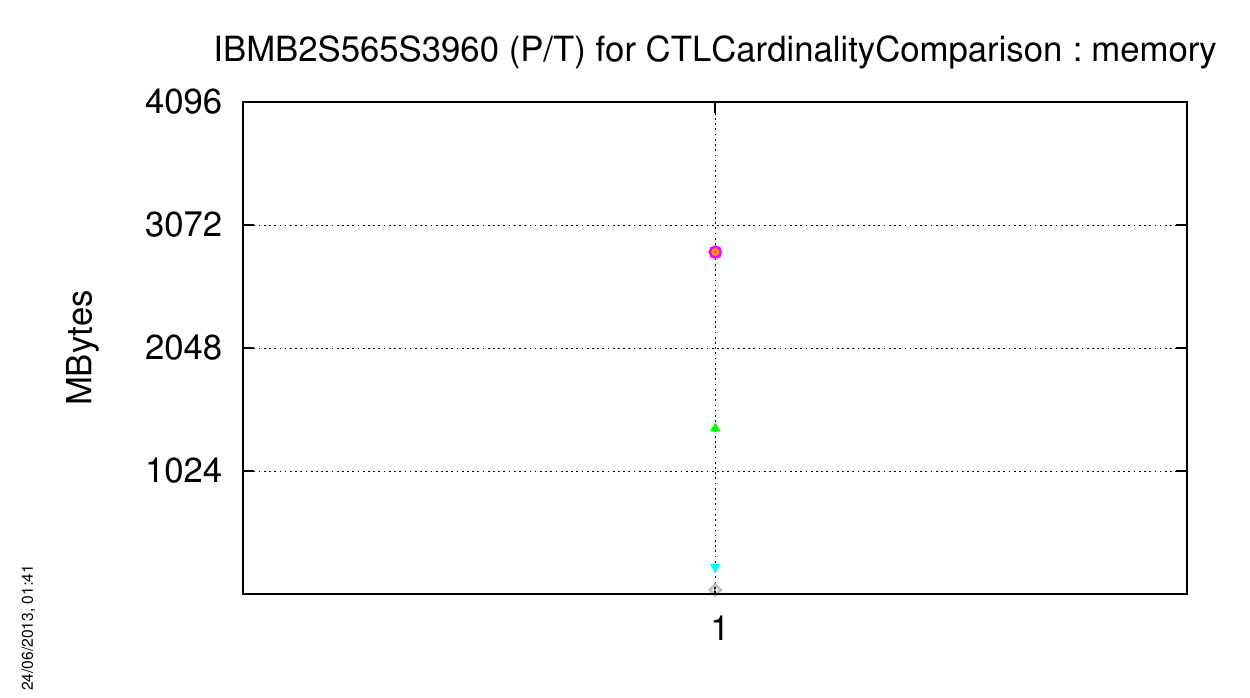}
   \includegraphics[width=7.2cm]{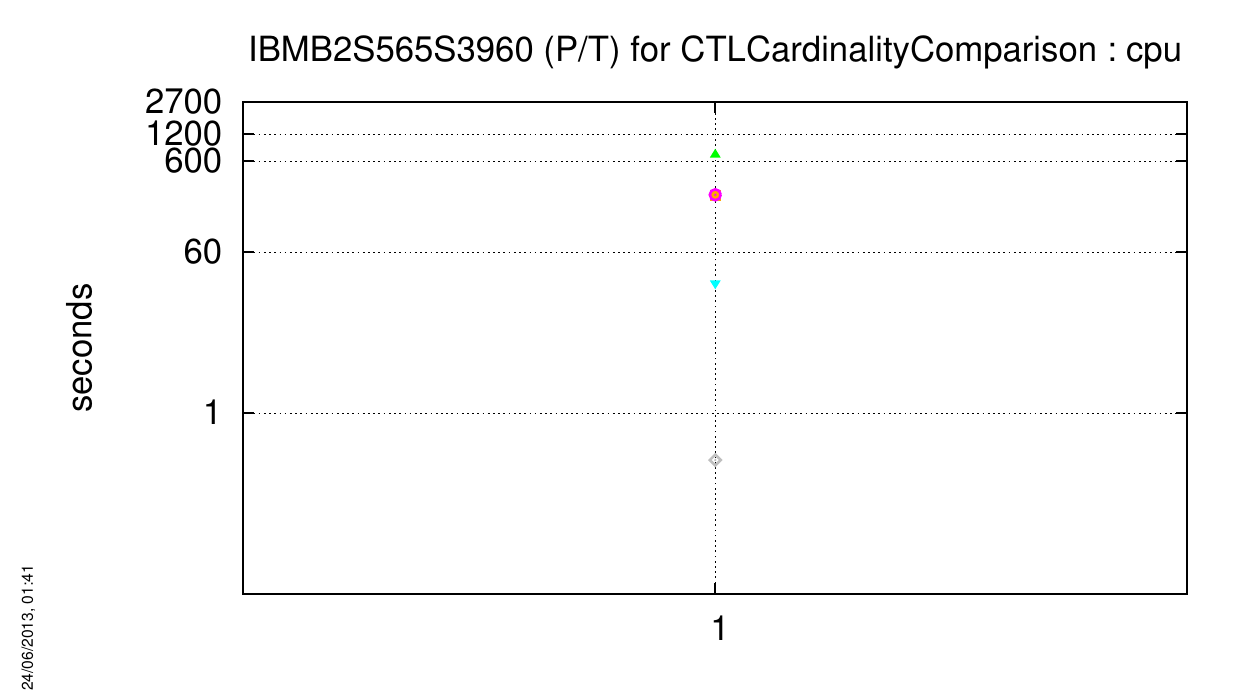}

   \includegraphics[height=1cm]{figures/tools-legend.pdf}
\end{center}

\subsubsection{\acs{QuasiCertifProtocol-COL}}
No instance of this model could be computed for the \textbf{CTLCardinalityComparison} examination.

\subsubsection{\acs{QuasiCertifProtocol-PT}}
The charts below respectively show how tools compete with this ``Suprise'' model (memory and CPU).

\index{Performances!CTLCardinalityComparison!QuasiCertifProtocol (P/T)}
\begin{center}
   \includegraphics[width=7.2cm]{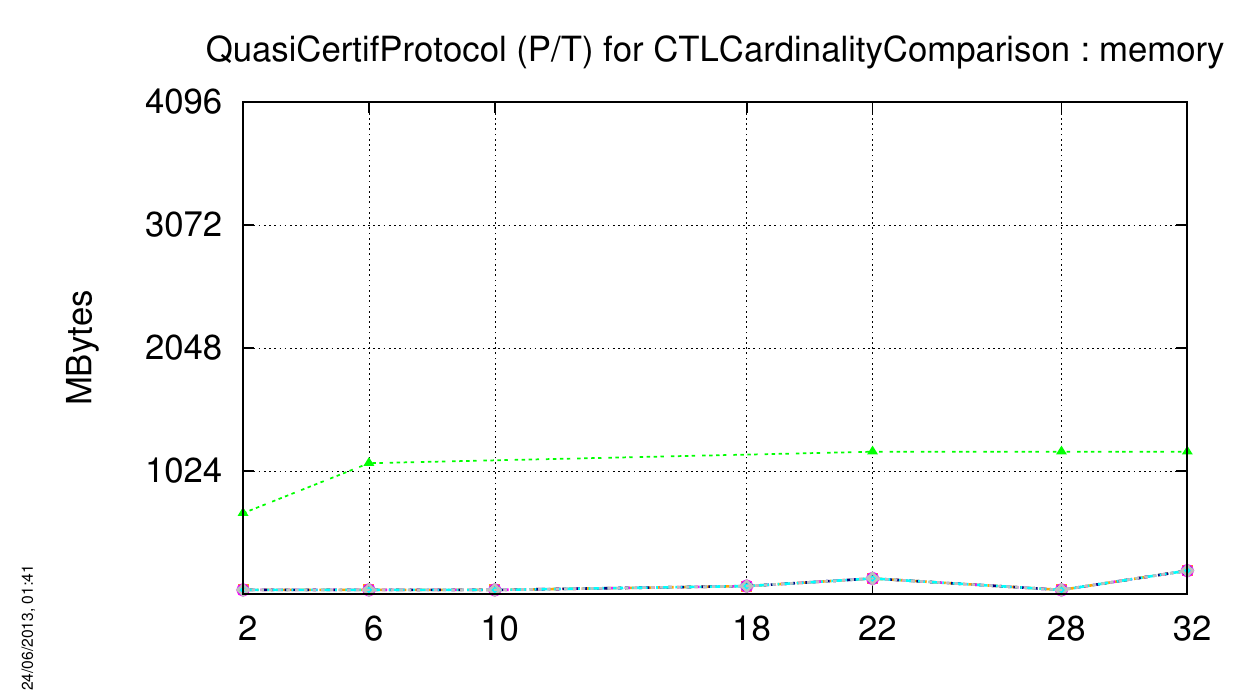}
   \includegraphics[width=7.2cm]{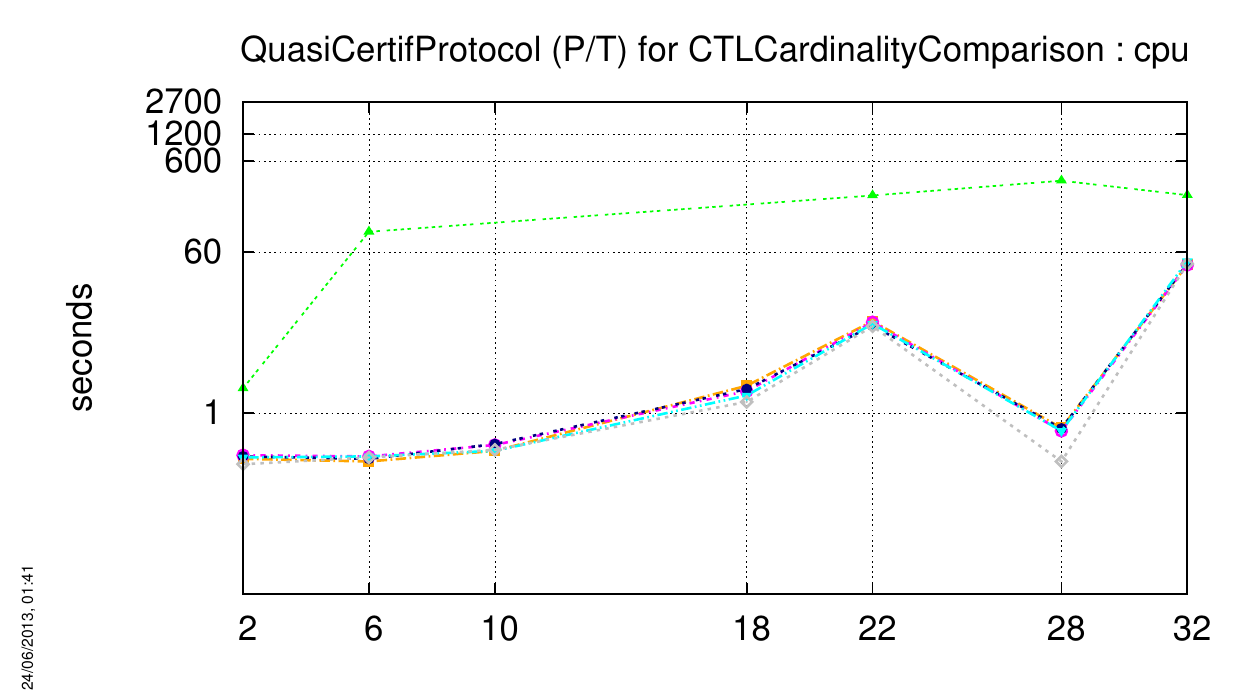}

   \includegraphics[height=1cm]{figures/tools-legend.pdf}
\end{center}

\subsubsection{\acs{Vasy2003-PT}}
The charts below respectively show how tools compete with this ``Suprise'' model (memory and CPU).

\index{Performances!CTLCardinalityComparison!Vasy2003 (P/T)}
\begin{center}
   \includegraphics[width=7.2cm]{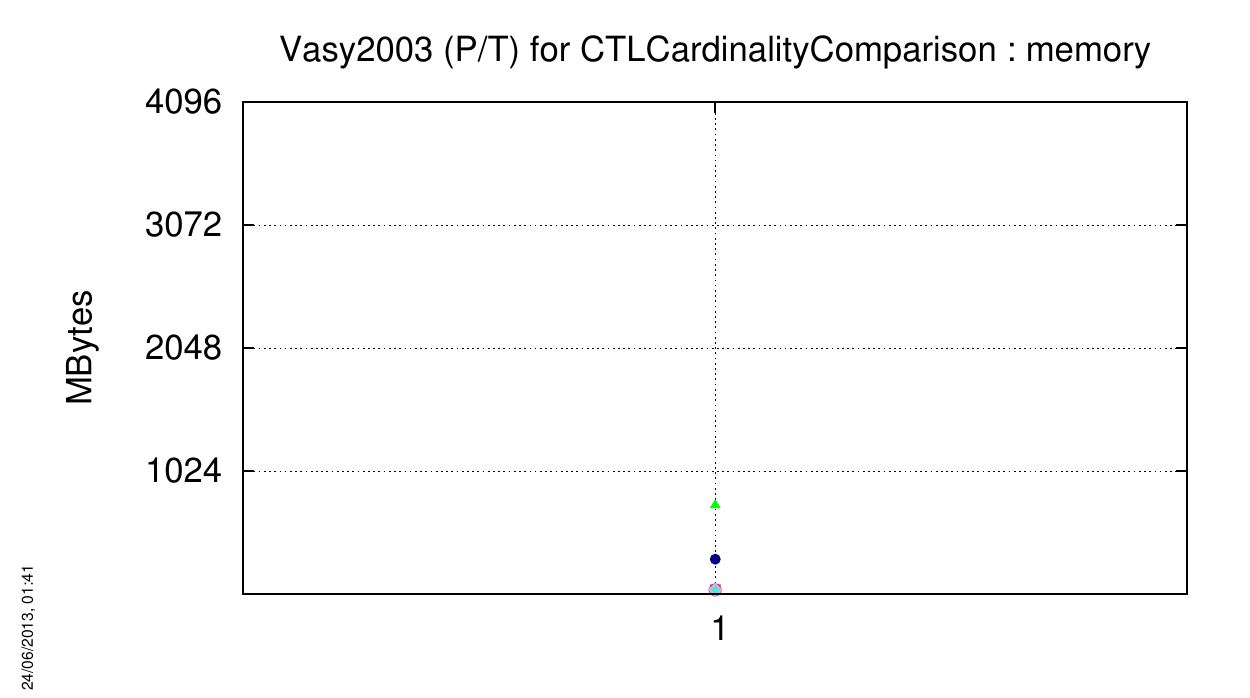}
   \includegraphics[width=7.2cm]{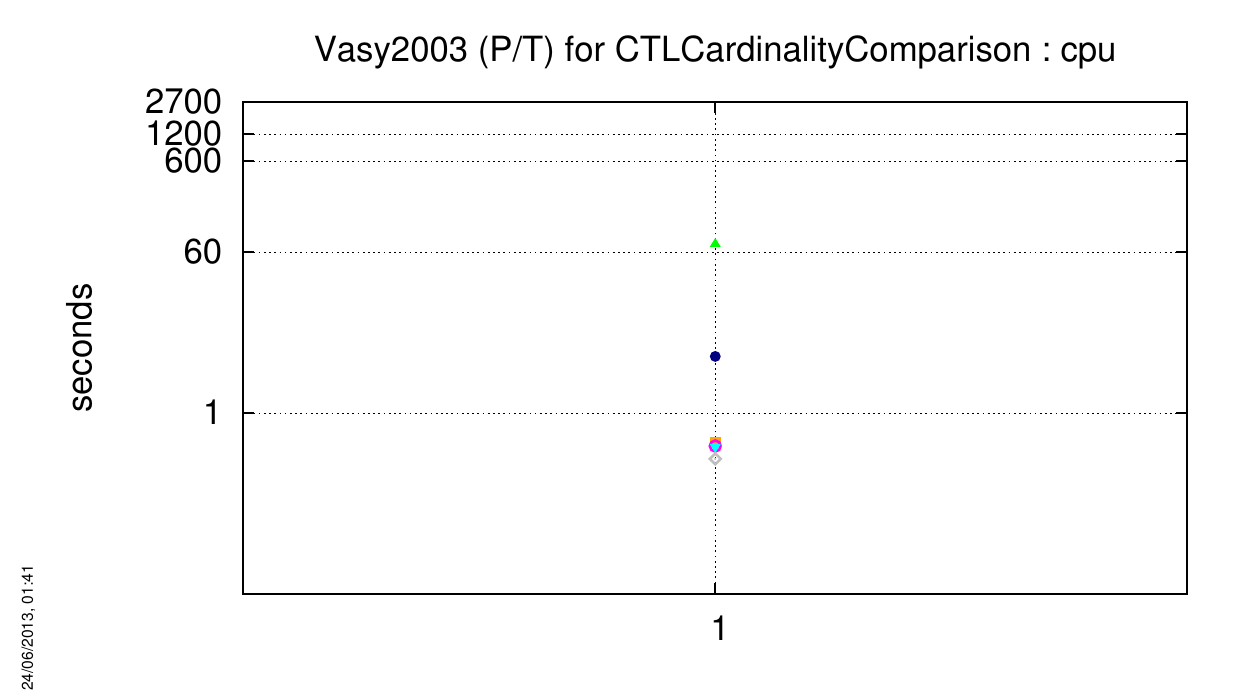}

   \includegraphics[height=1cm]{figures/tools-legend.pdf}
\end{center}

\subsection{Outputs for the CTLCardinalityComparison Examination}
\index{Outputs!CTLCardinalityComparison}

Please find enclosed the brute results for this examination (``Known'' and ``Surprise'' models).
We display only the score of tools that provide a results for at least one instance of one model.
The legend for the values is provided below:
\begin{itemize}
   \item\textbf{nc}: the tool does not compete this examination for this model/instance,
   \item\textbf{cc}: the tool cannot compute this examination for this model/instance,
   \item\textbf{to}: the tool cannot compute this examination for this model/instance within the maximum allowed time,
   \item\textbf{mp}: the tool encountered a memory problem (stack overflow or memory full),
   \item\textbf{nf}: there is no formula available for this type of examination (typically, this concerns P/T nets where
       comparing marking cardinality has no signification when there is no equivalent colored net).
\end{itemize}

\textbf{Note on the display of results for formulas:} each formula is considered as a flag (F if false, T if true, - or ?
when the value cannot be determined). These values are concatenated in the order they appear (we assume it is the order of formulas as they were provided).

\subsubsection{``Known'' Models}

Results are summarized in the table below.
{\footnotesize
\begin{longtable}{|cccccccc|}
   \hline
      \multicolumn{8}{|c|}{\bf \acs{CSRepetitions-COL}} \\
      \textbf{Instances}\index{Outputs!CTLCardinalityComparison!CSRepetitions (colored)} &
      \acs{greatspn} &
      \acs{lola} &
      \acs{lola-optimistic} &
      \acs{lola-optimistic-incomplete} &
      \acs{lola-pessimistic} &
      \acs{marcie} &
      \acs{sara} \\
      02 &
      nc &
      nc &
      nc &
      nc &
      nc &
      nc &
      nc \\
      03 &
      nc &
      nc &
      nc &
      nc &
      nc &
      nc &
      nc \\
      04 &
      nc &
      nc &
      nc &
      nc &
      nc &
      nc &
      nc \\
      05 &
      nc &
      nc &
      nc &
      nc &
      nc &
      nc &
      nc \\
      07 &
      nc &
      nc &
      nc &
      nc &
      nc &
      nc &
      nc \\
      10 &
      nc &
      nc &
      nc &
      nc &
      nc &
      nc &
      nc \\
\hline
\end{longtable}
}

\subsubsection{``Surprise'' Models}

Results are summarized in the table below.
{\footnotesize
\begin{longtable}{|ccccccc|}
   \hline
      \multicolumn{7}{|c|}{\bf \acs{HouseConstruction-PT}} \\
      instances\index{Outputs!CTLCardinalityComparison!HouseConstruction (P/T)} &
      \acs{lola} &
      \acs{lola-optimistic} &
      \acs{lola-optimistic-incomplete} &
      \acs{lola-pessimistic} &
      \acs{marcie} &
      \acs{sara} \\
      002 &
      \textcolor{red}{\tt \scriptsize cc} &
      \textcolor{red}{\tt \scriptsize cc} &
      \textcolor{red}{\tt \scriptsize cc} &
      \textcolor{red}{\tt \scriptsize cc} &
      \textcolor{red}{\tt \scriptsize FFFFTFFF} &
      \textcolor{red}{\tt \scriptsize cc} \\
      005 &
      \textcolor{red}{\tt \scriptsize cc} &
      \textcolor{red}{\tt \scriptsize cc} &
      \textcolor{red}{\tt \scriptsize cc} &
      \textcolor{red}{\tt \scriptsize cc} &
      \textcolor{red}{\tt \scriptsize FFFTFFFF} &
      \textcolor{red}{\tt \scriptsize cc} \\
      010 &
      \textcolor{red}{\tt \scriptsize cc} &
      \textcolor{red}{\tt \scriptsize cc} &
      \textcolor{red}{\tt \scriptsize cc} &
      \textcolor{red}{\tt \scriptsize cc} &
      \textcolor{red}{\tt \scriptsize FFFFTFFF} &
      \textcolor{red}{\tt \scriptsize cc} \\
      020 &
      \textcolor{red}{\tt \scriptsize cc} &
      \textcolor{red}{\tt \scriptsize cc} &
      \textcolor{red}{\tt \scriptsize cc} &
      \textcolor{red}{\tt \scriptsize cc} &
      \textcolor{red}{\tt \scriptsize FFFTFFFF} &
      \textcolor{red}{\tt \scriptsize cc} \\
      050 &
      \textcolor{red}{\tt \scriptsize cc} &
      \textcolor{red}{\tt \scriptsize cc} &
      \textcolor{red}{\tt \scriptsize cc} &
      \textcolor{red}{\tt \scriptsize cc} &
      \textcolor{red}{\tt \scriptsize cc} &
      \textcolor{red}{\tt \scriptsize cc} \\
\hline
\end{longtable}
}

\subsection{Score for the CTLCardinalityComparison Examination}
\index{Scores!CTLCardinalityComparison}

Please find enclosed the scores for this examination (``Known'' and ``Surprise'' models).
We display only the score of tools that provide a results for at least one instance of one model.
The total is first listed in the table below followed by a detail, for each proposed model.
Meaning of the line labels are:
\begin{itemize}
\item\textbf{1st instance}: the tool gets a bonus for having processed the first instance of this model (+1 point),
\item\textbf{instances}: the tool gets 1 point per instances treated 
(for that, we assume that at least one formula has been successfully computed),
\item\textbf{max reached}: the tool could process all the instances for the model (+2 points),
\item\textbf{best}: the tool is among the ones that processed a maximum of instances within the time and memory confinement (+2 points).
\end{itemize}

\subsubsection{``Known'' Models}

Scores are summarized in the table below.
For each model in this category, a tool may collect up to 411 points.

{\footnotesize
\begin{longtable}{|cccccccc|}
   \hline
      \multicolumn{8}{|c|}{\bf Total score of the tools for this examination} \\
        &
      \acs{greatspn} &
      \acs{lola} &
      \acs{lola-optimistic} &
      \acs{lola-optimistic-incomplete} &
      \acs{lola-pessimistic} &
      \acs{marcie} &
      \acs{sara} \\
      Total Score &
      {\bf 43} &
      {\bf 172} &
      {\bf 172} &
      {\bf 123} &
      {\bf 96} &
      {\bf 102} &
      {\bf 17} \\
   \hline
      \multicolumn{8}{|c|}{\bf CSRepetitions (Colored)} \\
      \ \index{Scores!CTLCardinalityComparison!CSRepetitions (colored)} &
      \acs{greatspn} &
      \acs{lola} &
      \acs{lola-optimistic} &
      \acs{lola-optimistic-incomplete} &
      \acs{lola-pessimistic} &
      \acs{marcie} &
      \acs{sara} \\
      1st instance &
      \textcolor{red}{0} &
      \textcolor{red}{0} &
      \textcolor{red}{0} &
      \textcolor{red}{0} &
      \textcolor{red}{0} &
      \textcolor{red}{0} &
      \textcolor{red}{0} \\
      instances &
      \textcolor{red}{0} &
      \textcolor{red}{0} &
      \textcolor{red}{0} &
      \textcolor{red}{0} &
      \textcolor{red}{0} &
      \textcolor{red}{0} &
      \textcolor{red}{0} \\
      max reached &
      \textcolor{red}{0} &
      \textcolor{red}{0} &
      \textcolor{red}{0} &
      \textcolor{red}{0} &
      \textcolor{red}{0} &
      \textcolor{red}{0} &
      \textcolor{red}{0} \\
\hline
\end{longtable}
}

\subsubsection{``Surprise'' Models}

Scores are summarized in the table below.
For each model in this category, a tool may collect up to 49 points.

{\footnotesize
\begin{longtable}{|ccccccc|}
   \hline
      \multicolumn{7}{|c|}{\bf Total score of the tools for this examination} \\
        &
      \acs{lola} &
      \acs{lola-optimistic} &
      \acs{lola-optimistic-incomplete} &
      \acs{lola-pessimistic} &
      \acs{marcie} &
      \acs{sara} \\
      Total Score &
      {\bf 12} &
      {\bf 12} &
      {\bf 6} &
      {\bf 12} &
      {\bf 22} &
      {\bf 6} \\
   \hline
      \multicolumn{7}{|c|}{\bf HouseConstruction (P/T)} \\
      \ \index{Scores!CTLCardinalityComparison!HouseConstruction (P/T)} &
      \acs{lola} &
      \acs{lola-optimistic} &
      \acs{lola-optimistic-incomplete} &
      \acs{lola-pessimistic} &
      \acs{marcie} &
      \acs{sara} \\
      1st instance &
      \textcolor{red}{0} &
      \textcolor{red}{0} &
      \textcolor{red}{0} &
      \textcolor{red}{0} &
      \textcolor{red}{1} &
      \textcolor{red}{1} \\
      instances &
      \textcolor{red}{0} &
      \textcolor{red}{0} &
      \textcolor{red}{0} &
      \textcolor{red}{0} &
      \textcolor{red}{4} &
      \textcolor{red}{1} \\
      max reached &
      \textcolor{red}{0} &
      \textcolor{red}{0} &
      \textcolor{red}{0} &
      \textcolor{red}{0} &
      \textcolor{red}{0} &
      \textcolor{red}{2} \\
\hline
\end{longtable}
}

\subsection{Trophies for this Examination}
\index{Trophies!CTLCardinalityComparison}

Trophies are divided in three categories: ``Known'' models,
``Surprise'' models, and the global trophies (formula is then
$score_{global} = score_{known} + 2 \times score_{surprise}$).

\subsubsection{For ``Known'' Models} \ \\

\begin{tabular}{c|c|c}
      1 & 1 & 3 \\
   \includegraphics[width=2cm]{figures/gold.jpg} &
   \includegraphics[width=2cm]{figures/gold.jpg} &
   \includegraphics[width=2cm]{figures/bronse.jpg} \\
   \acs{lola} &
   \acs{lola-optimistic} &
   \acs{marcie} \\
   172 points &
   172 points &
   102 points \\
\end{tabular}

\subsubsection{For ``Surprise'' Models}\  \\

\begin{tabular}{c|c|c}
      1 & 2 & 2 \\
   \includegraphics[width=2cm]{figures/gold.jpg} &
   \includegraphics[width=2cm]{figures/silver.jpg} &
   \includegraphics[width=2cm]{figures/silver.jpg} \\
   \acs{marcie} &
   \acs{lola} &
   \acs{lola-optimistic} \\
   22 points &
   12 points &
   12 points \\
\end{tabular}

\subsubsection{Global} \ \\

\begin{tabular}{c|c|c}
      1 & 1 & 3 \\
   \includegraphics[width=2cm]{figures/gold.jpg} &
   \includegraphics[width=2cm]{figures/gold.jpg} &
   \includegraphics[width=2cm]{figures/bronse.jpg} \\
   \acs{lola} &
   \acs{lola-optimistic} &
   \acs{marcie} \\
   196 points &
   196 points &
   146 points \\
\end{tabular}

\newpage

\section{The CTLFireability Examination}
\label{sec:exam:CTLFireability}
\index{Results!CTLFireability}

This examination deals with CTL properties dealing with transition fireability only.
We first show a summary on the handling of models by the participating tools.
Then, we present the computed outputs and the associated scores for this
examination prior to a summary of relevant executions.

\subsection{Handling of Models by Tools}
\index{Performances!CTLFireability}

\subsubsection{\acs{CSRepetitions-COL}}
No instance of this model could be computed for the \textbf{CTLFireability} examination.

\subsubsection{\acs{CSRepetitions-PT}}
The charts below respectively show how tools compete with this ``Known'' model (memory and CPU).

\index{Performances!CTLFireability!CSRepetitions (P/T)}
\begin{center}
   \includegraphics[width=7.2cm]{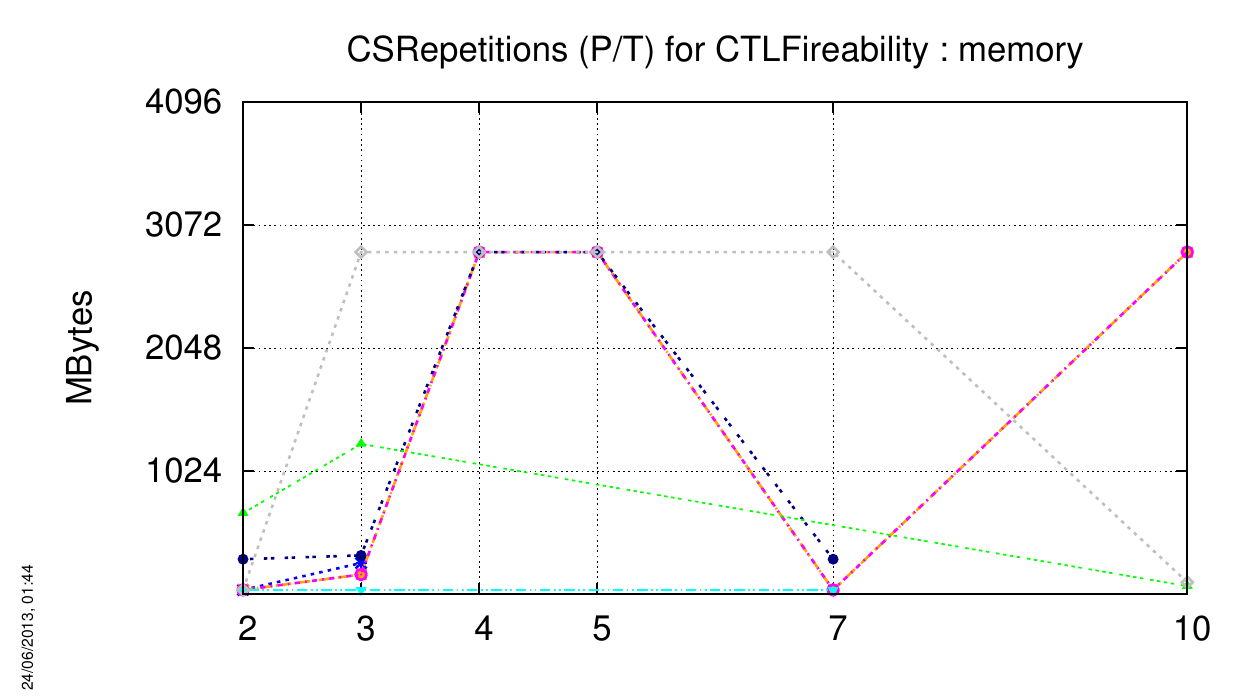}
   \includegraphics[width=7.2cm]{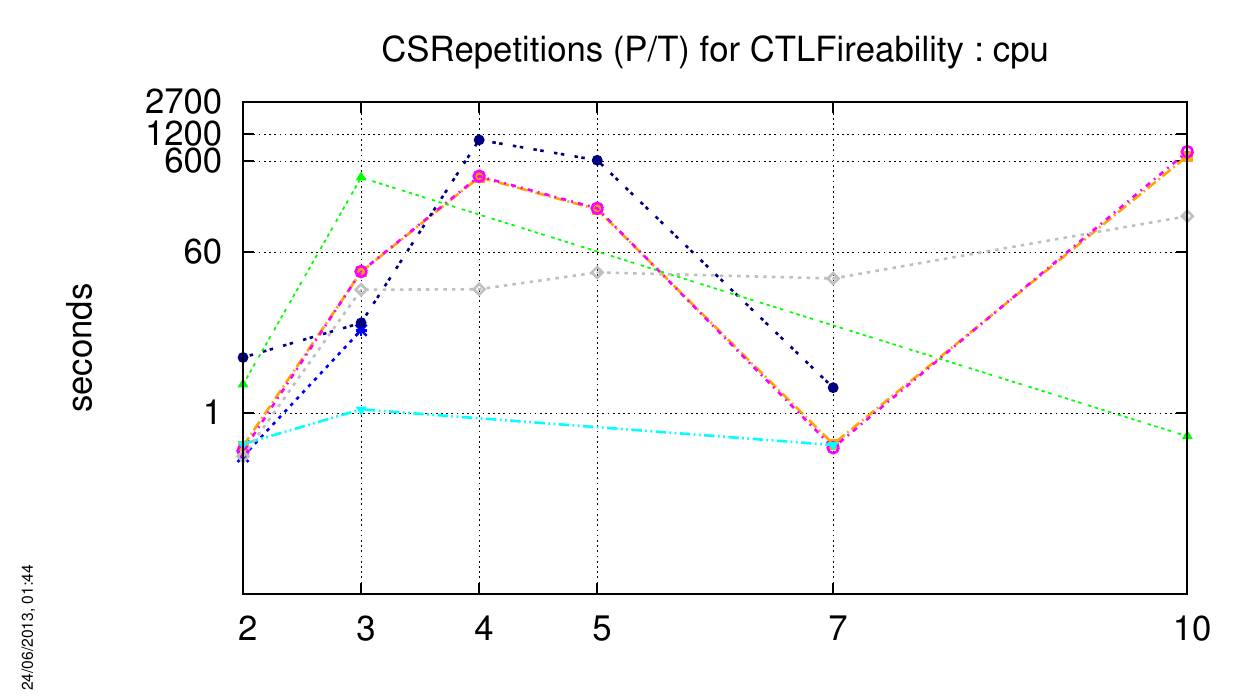}

   \includegraphics[height=1cm]{figures/tools-legend.pdf}
\end{center}

\subsubsection{\acs{Dekker-PT}}
The charts below respectively show how tools compete with this ``Known'' model (memory and CPU).

\index{Performances!CTLFireability!Dekker (P/T)}
\begin{center}
   \includegraphics[width=7.2cm]{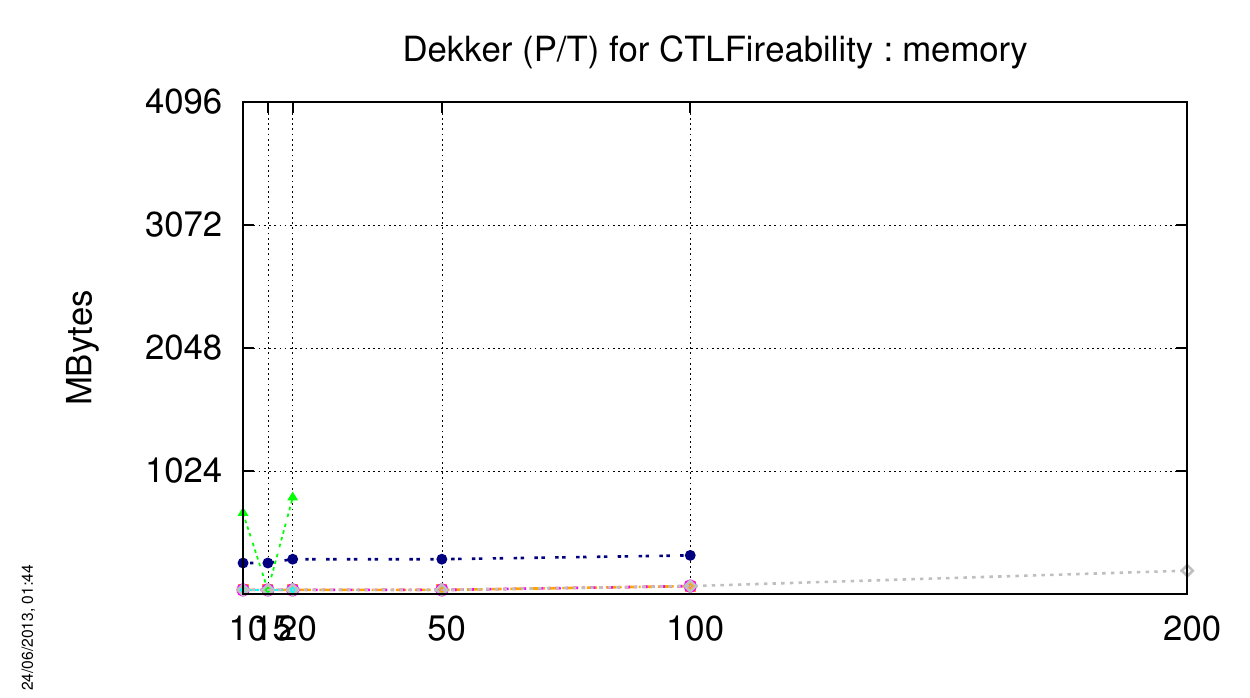}
   \includegraphics[width=7.2cm]{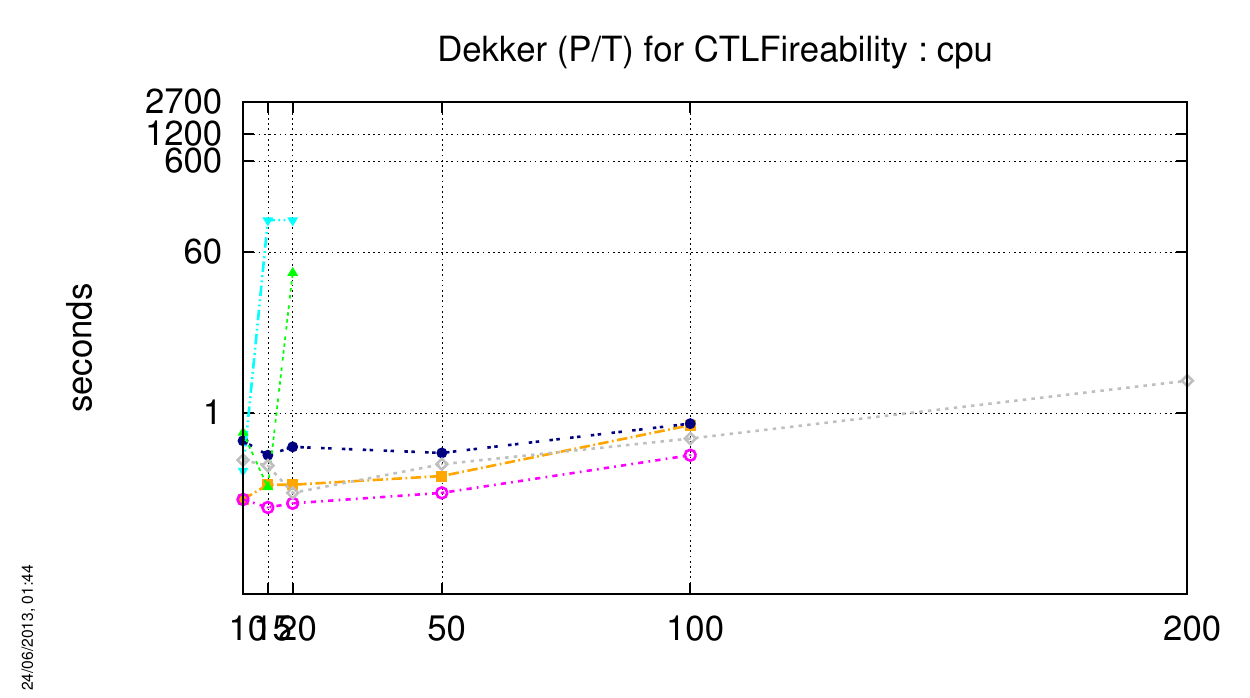}

   \includegraphics[height=1cm]{figures/tools-legend.pdf}
\end{center}

\subsubsection{\acs{DotAndBoxes-COL}}
No instance of this model could be computed for the \textbf{CTLFireability} examination.

\subsubsection{\acs{DrinkVendingMachine-COL}}
No instance of this model could be computed for the \textbf{CTLFireability} examination.

\subsubsection{\acs{DrinkVendingMachine-PT}}
The charts below respectively show how tools compete with this ``Known'' model (memory and CPU).

\index{Performances!CTLFireability!DrinkVendingMachine (P/T)}
\begin{center}
   \includegraphics[width=7.2cm]{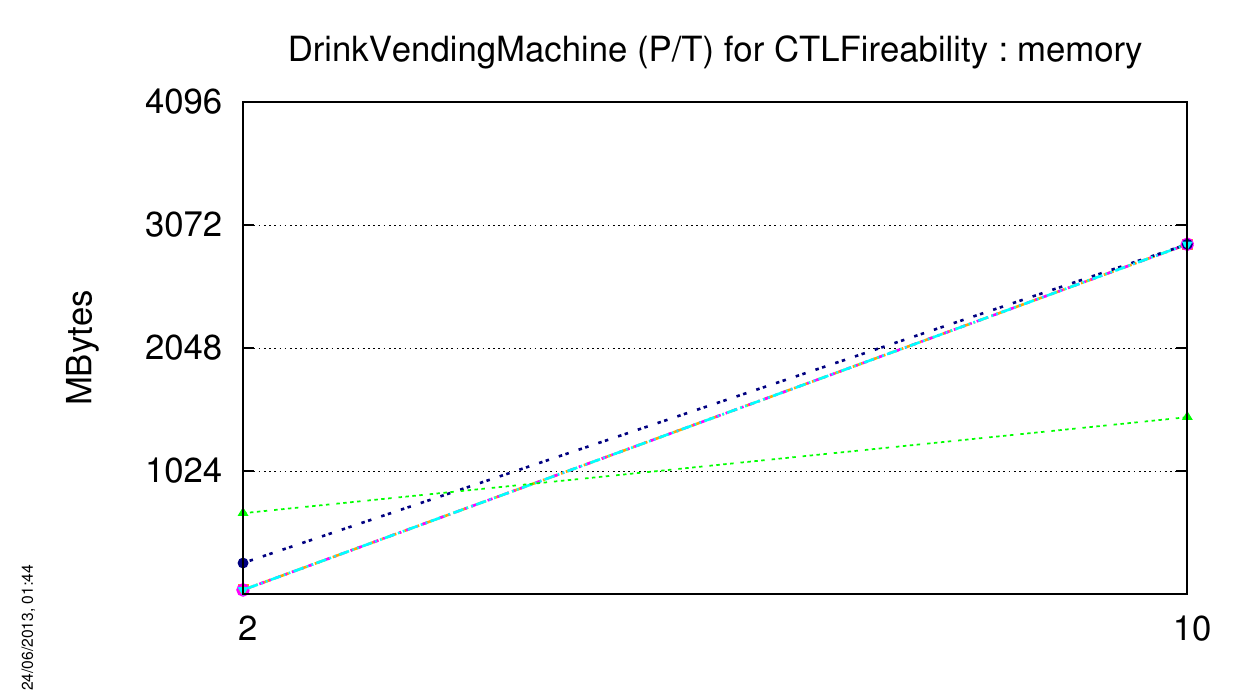}
   \includegraphics[width=7.2cm]{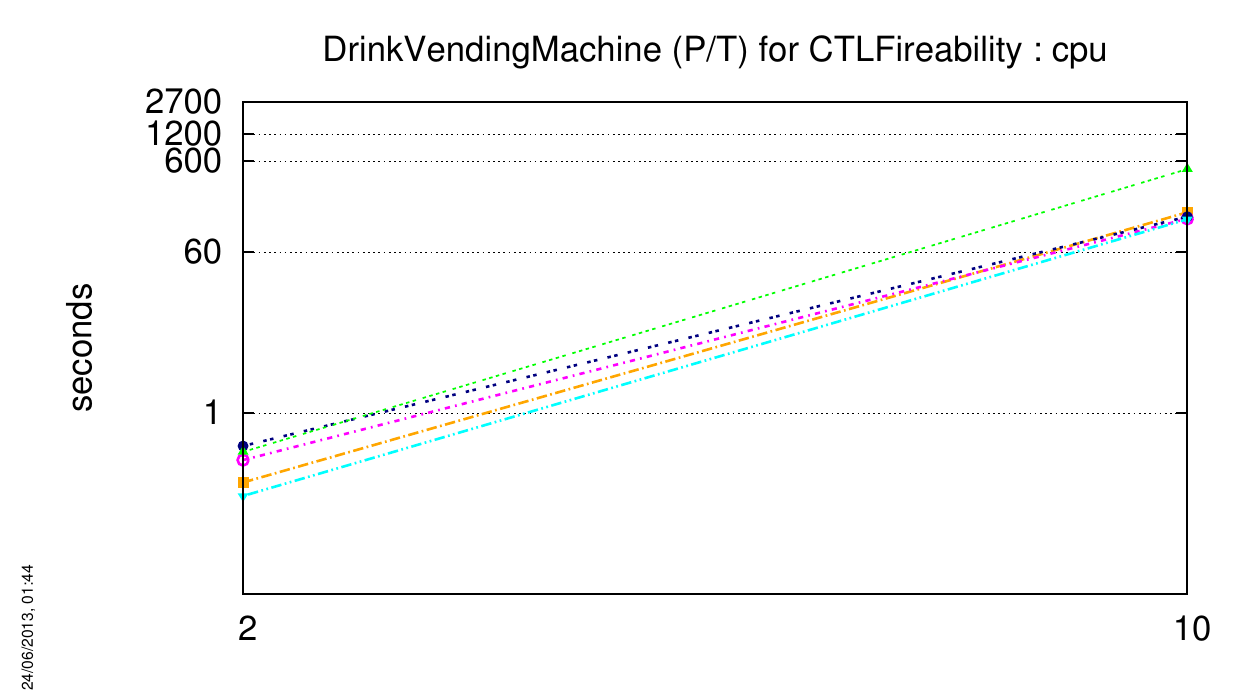}

   \includegraphics[height=1cm]{figures/tools-legend.pdf}
\end{center}

\subsubsection{\acs{Echo-PT}}
The charts below respectively show how tools compete with this ``Known'' model (memory and CPU).

\index{Performances!CTLFireability!Echo (P/T)}
\begin{center}
   \includegraphics[width=7.2cm]{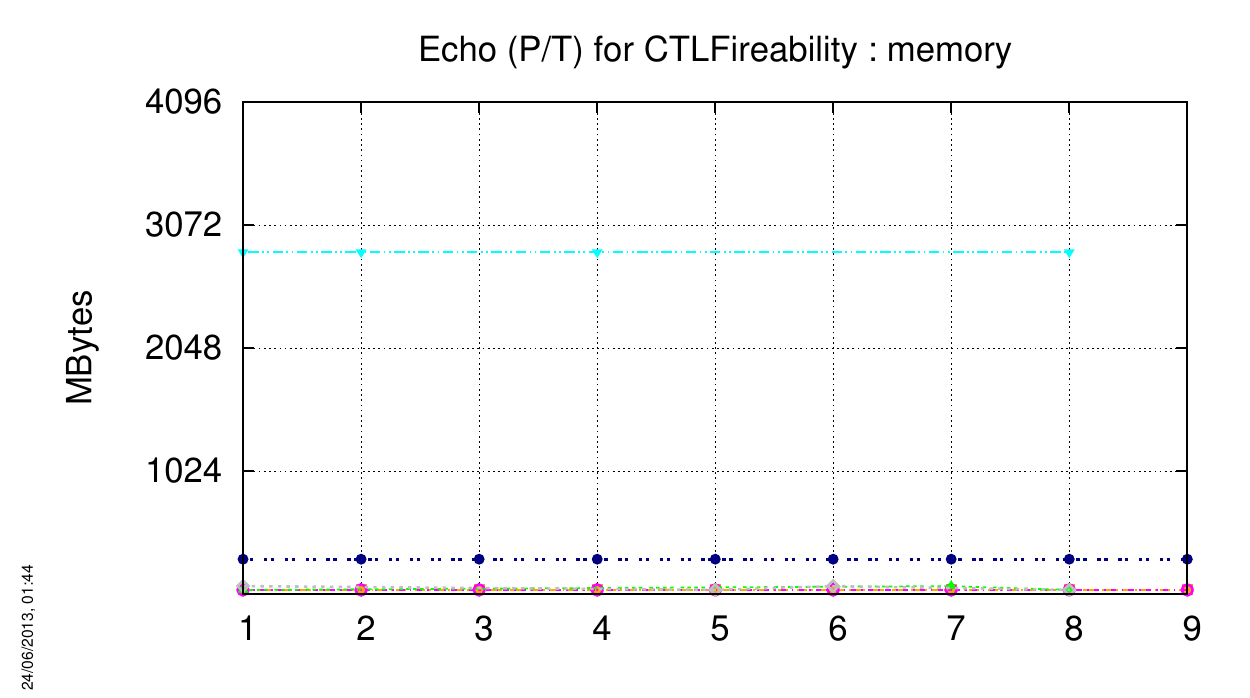}
   \includegraphics[width=7.2cm]{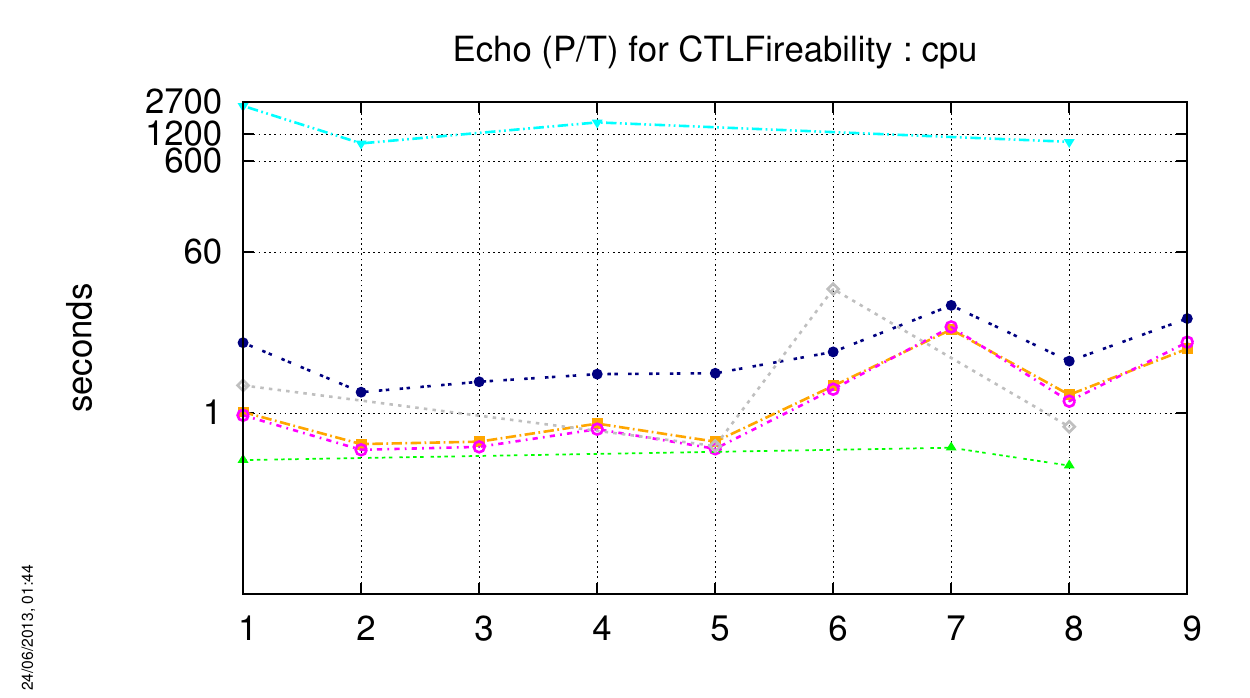}

   \includegraphics[height=1cm]{figures/tools-legend.pdf}
\end{center}

\subsubsection{\acs{Eratosthenes-PT}}
The charts below respectively show how tools compete with this ``Known'' model (memory and CPU).

\index{Performances!CTLFireability!Eratosthenes (P/T)}
\begin{center}
   \includegraphics[width=7.2cm]{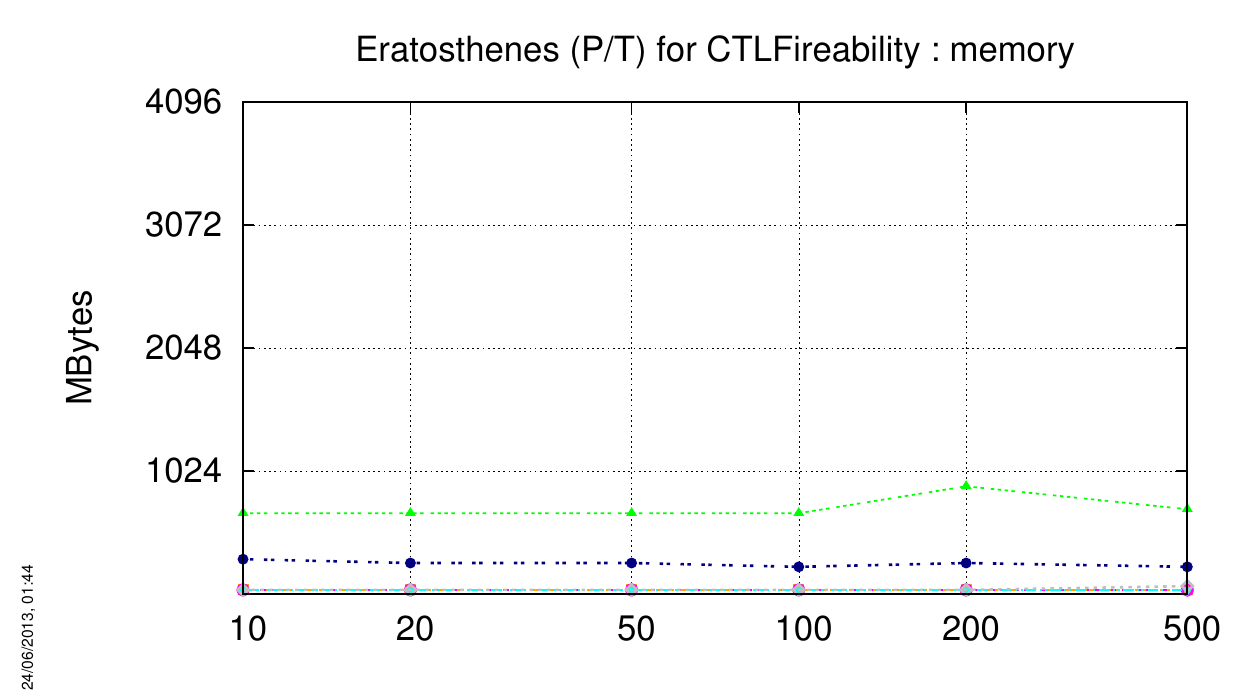}
   \includegraphics[width=7.2cm]{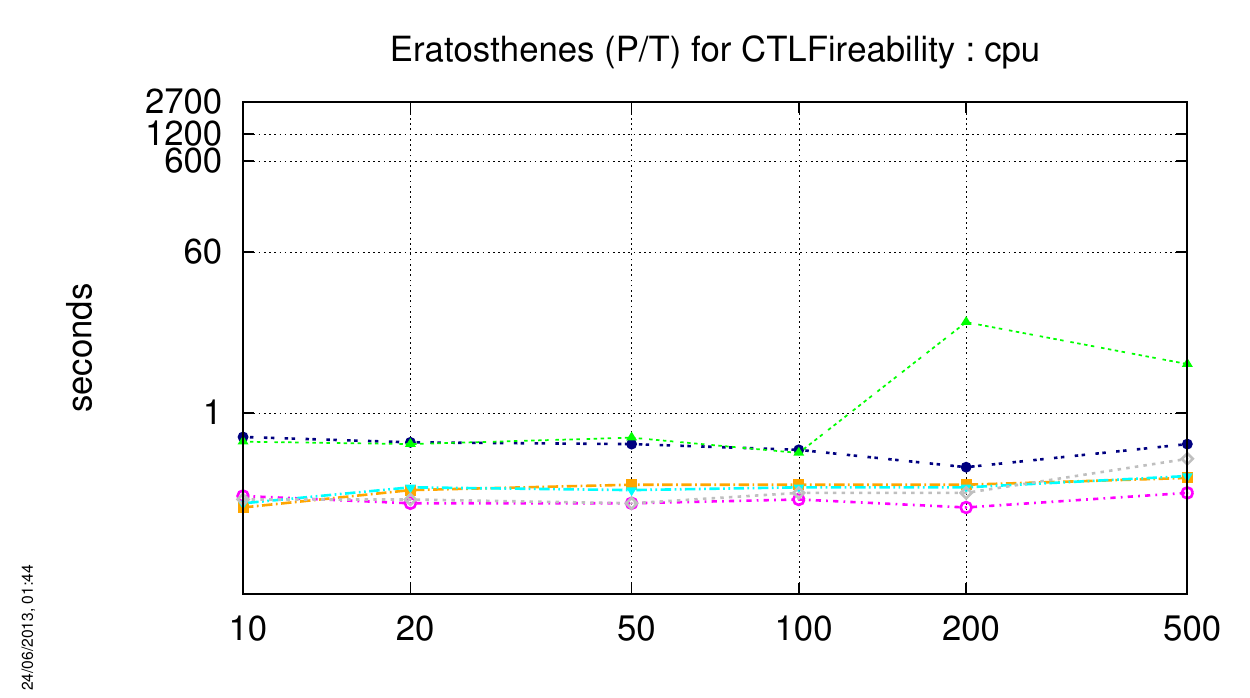}

   \includegraphics[height=1cm]{figures/tools-legend.pdf}
\end{center}

\subsubsection{\acs{FMS-PT}}
The charts below respectively show how tools compete with this ``Known'' model (memory and CPU).

\index{Performances!CTLFireability!FMS (P/T)}
\begin{center}
   \includegraphics[width=7.2cm]{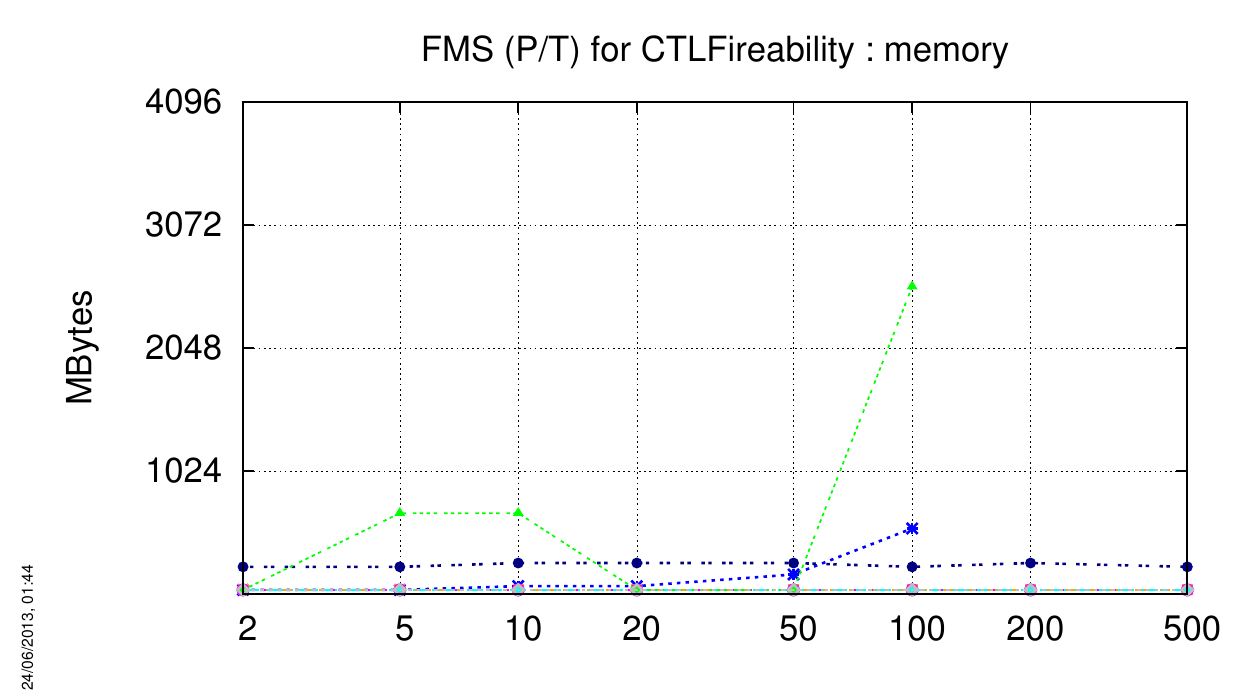}
   \includegraphics[width=7.2cm]{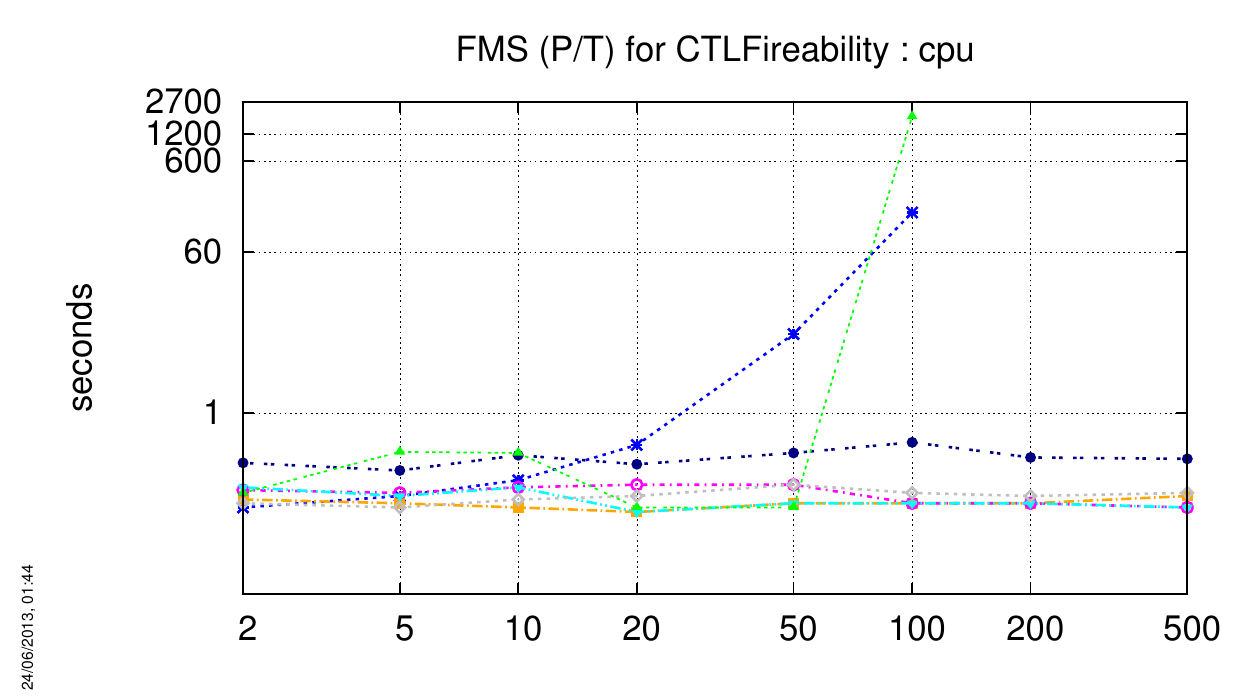}

   \includegraphics[height=1cm]{figures/tools-legend.pdf}
\end{center}

\subsubsection{\acs{GlobalRessAlloc-COL}}
No instance of this model could be computed for the \textbf{CTLFireability} examination.

\subsubsection{\acs{GlobalRessAlloc-PT}}
The charts below respectively show how tools compete with this ``Known'' model (memory and CPU).

\index{Performances!CTLFireability!GlobalRessAlloc (P/T)}
\begin{center}
   \includegraphics[width=7.2cm]{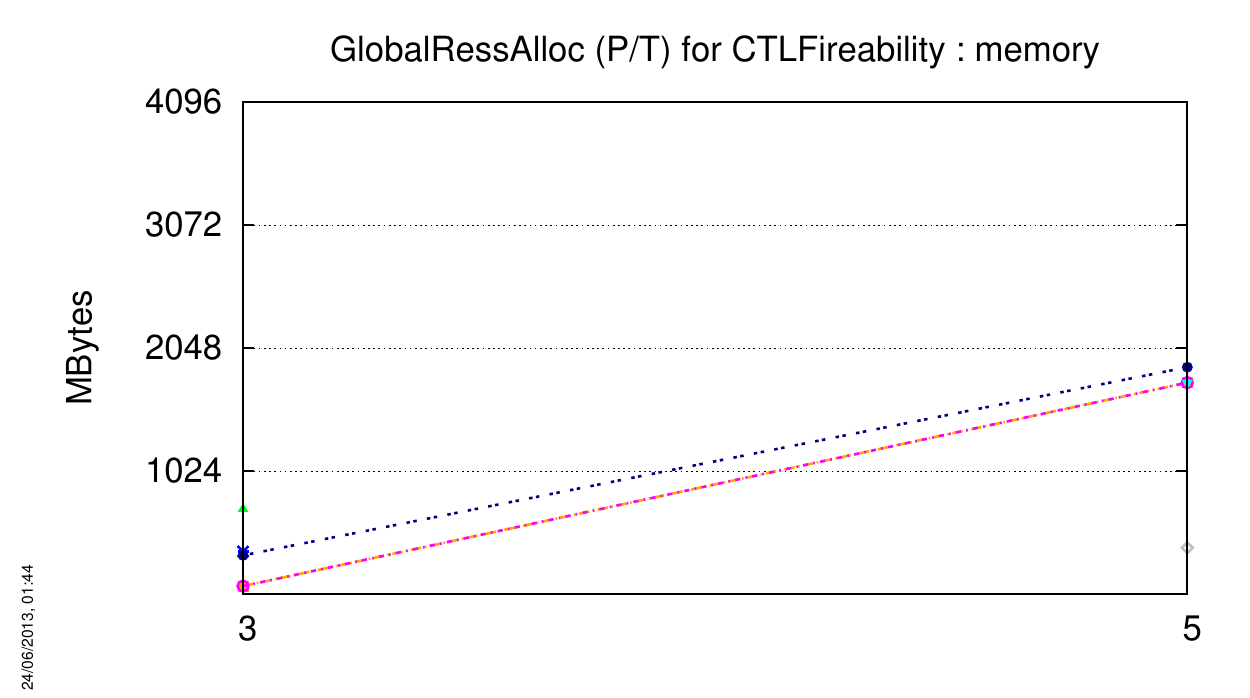}
   \includegraphics[width=7.2cm]{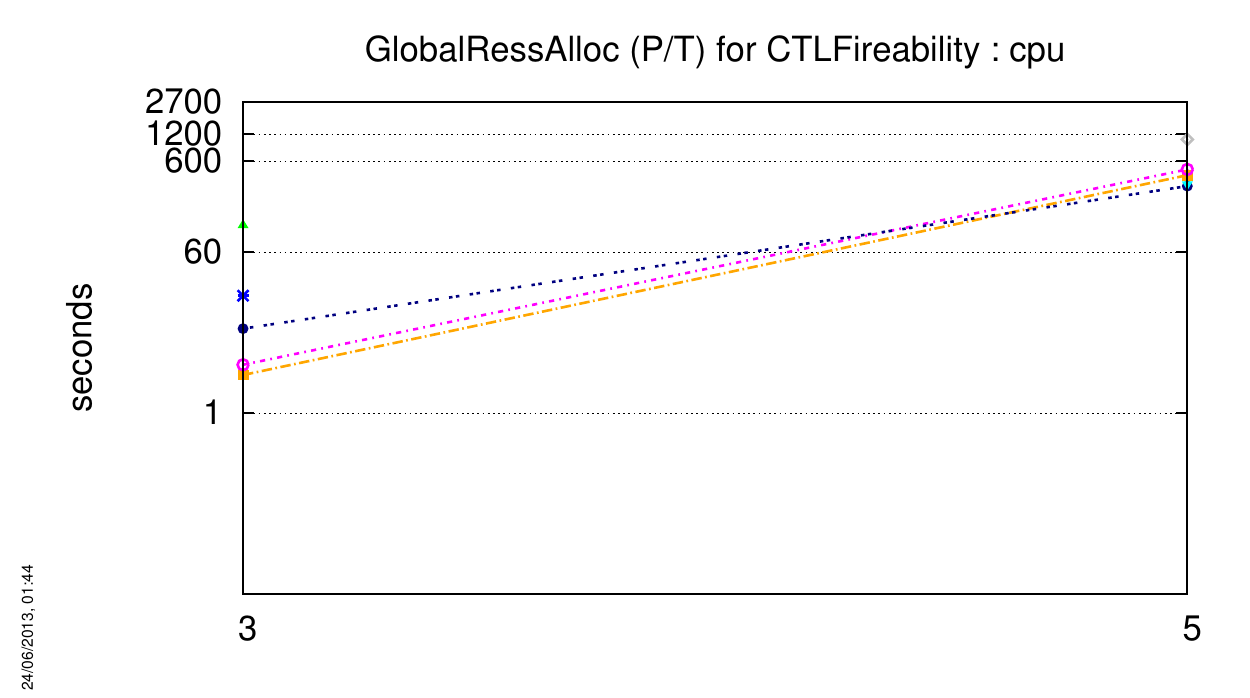}

   \includegraphics[height=1cm]{figures/tools-legend.pdf}
\end{center}

\subsubsection{\acs{Kanban-PT}}
The charts below respectively show how tools compete with this ``Known'' model (memory and CPU).

\index{Performances!CTLFireability!Kanban (P/T)}
\begin{center}
   \includegraphics[width=7.2cm]{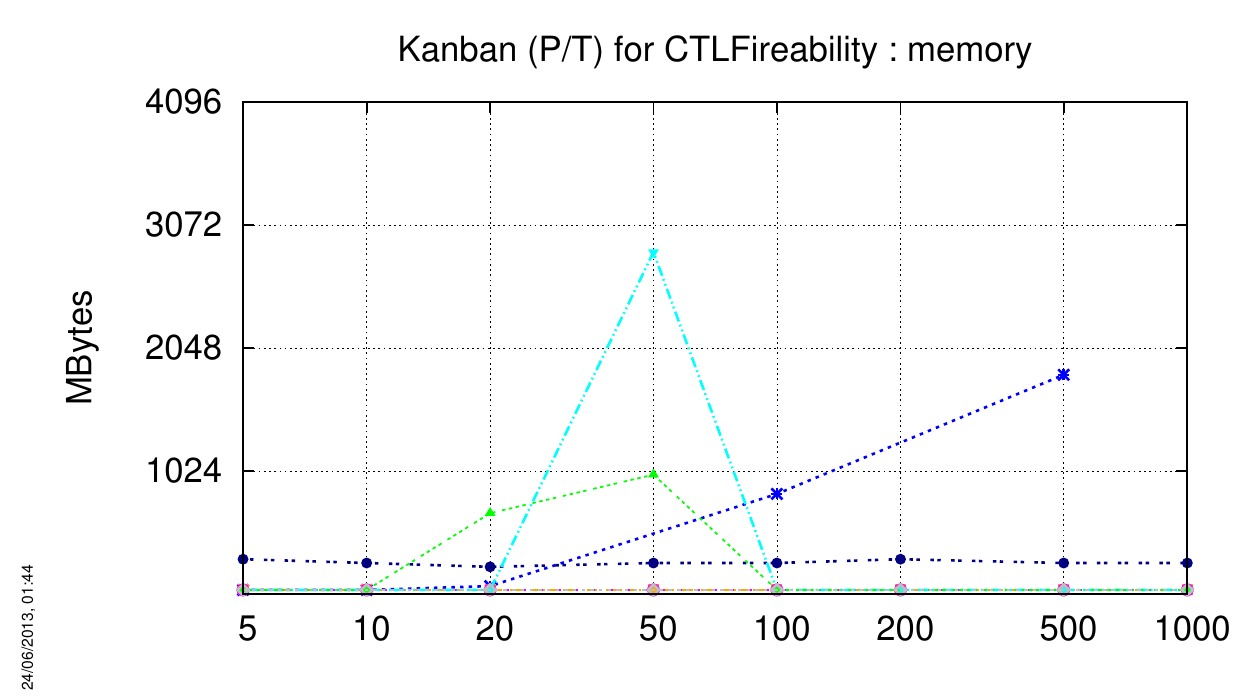}
   \includegraphics[width=7.2cm]{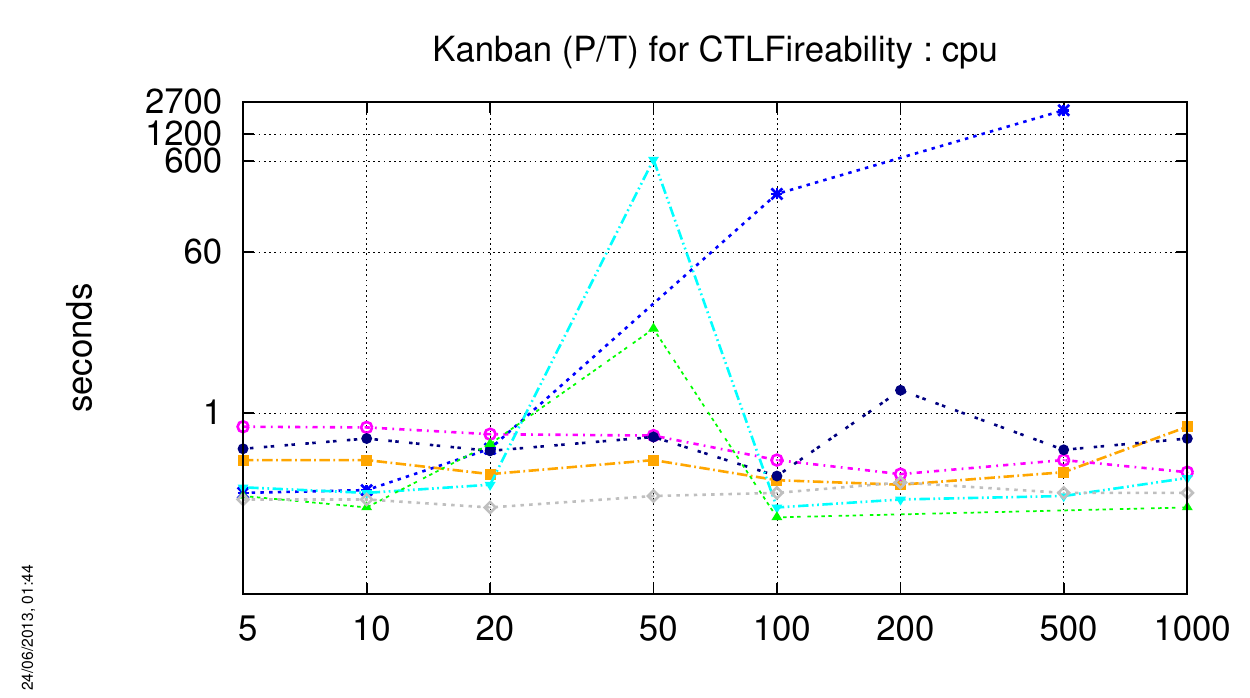}

   \includegraphics[height=1cm]{figures/tools-legend.pdf}
\end{center}

\subsubsection{\acs{LamportFastMutEx-COL}}
No instance of this model could be computed for the \textbf{CTLFireability} examination.

\subsubsection{\acs{LamportFastMutEx-PT}}
The charts below respectively show how tools compete with this ``Known'' model (memory and CPU).

\index{Performances!CTLFireability!LamportFastMutEx (P/T)}
\begin{center}
   \includegraphics[width=7.2cm]{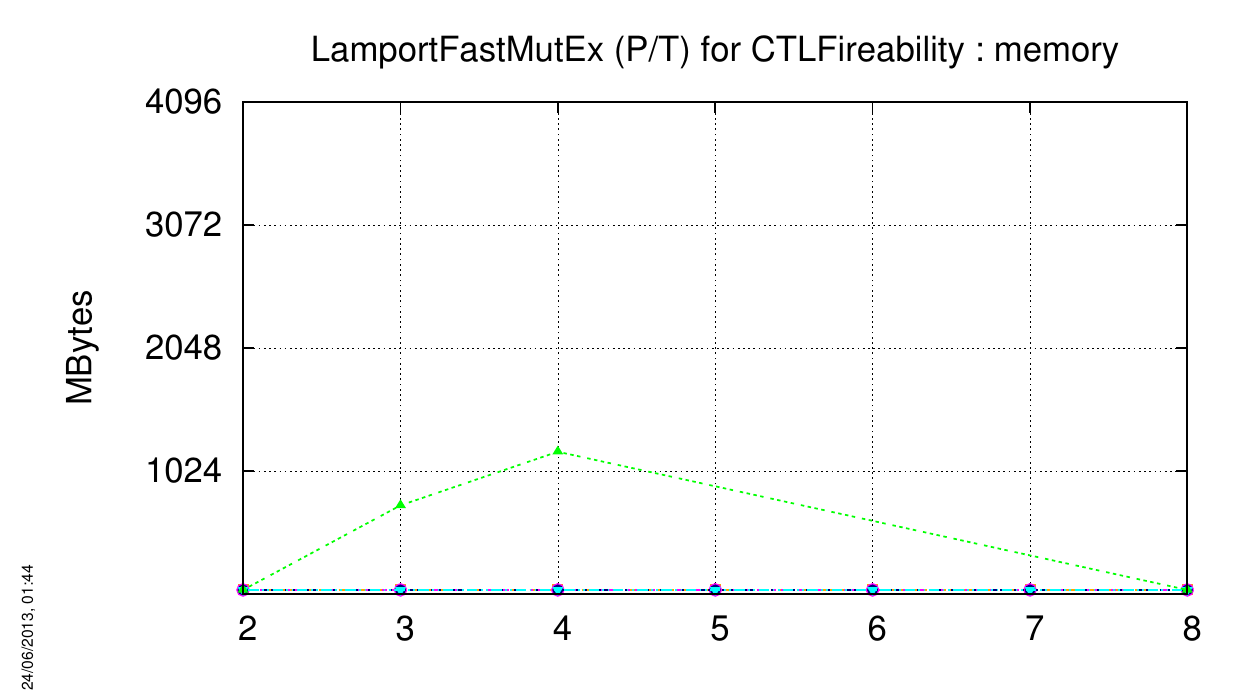}
   \includegraphics[width=7.2cm]{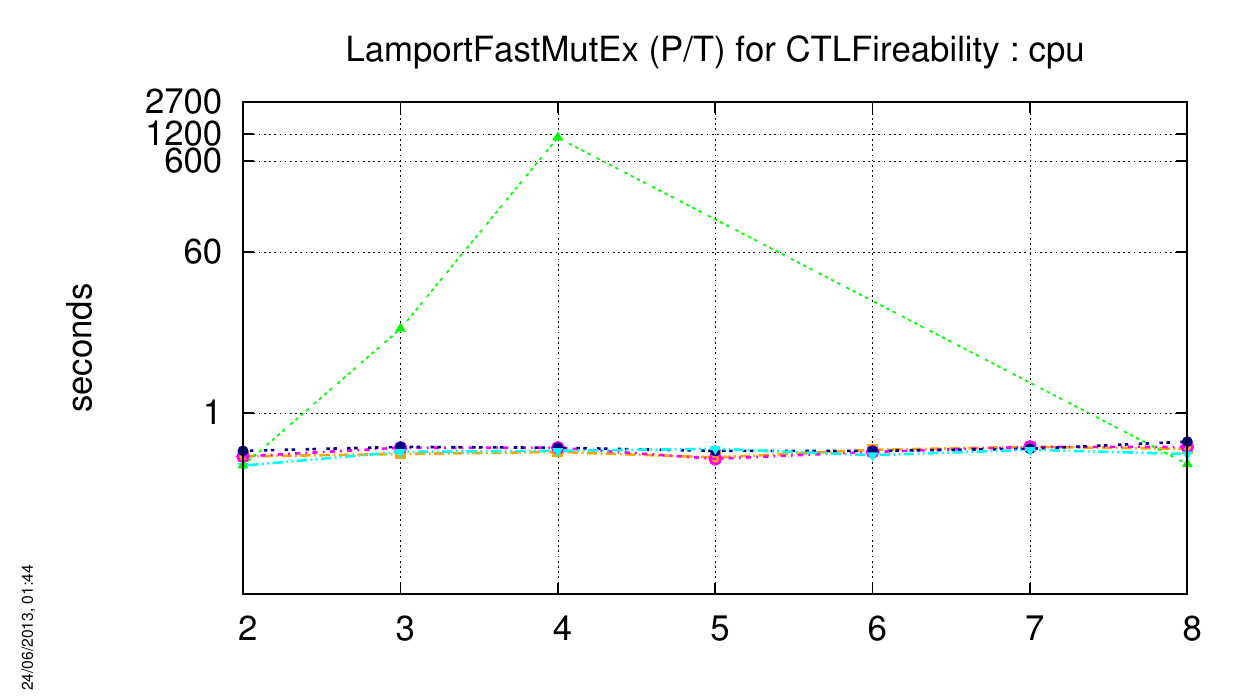}

   \includegraphics[height=1cm]{figures/tools-legend.pdf}
\end{center}

\subsubsection{\acs{MAPK-PT}}
The charts below respectively show how tools compete with this ``Known'' model (memory and CPU).

\index{Performances!CTLFireability!MAPK (P/T)}
\begin{center}
   \includegraphics[width=7.2cm]{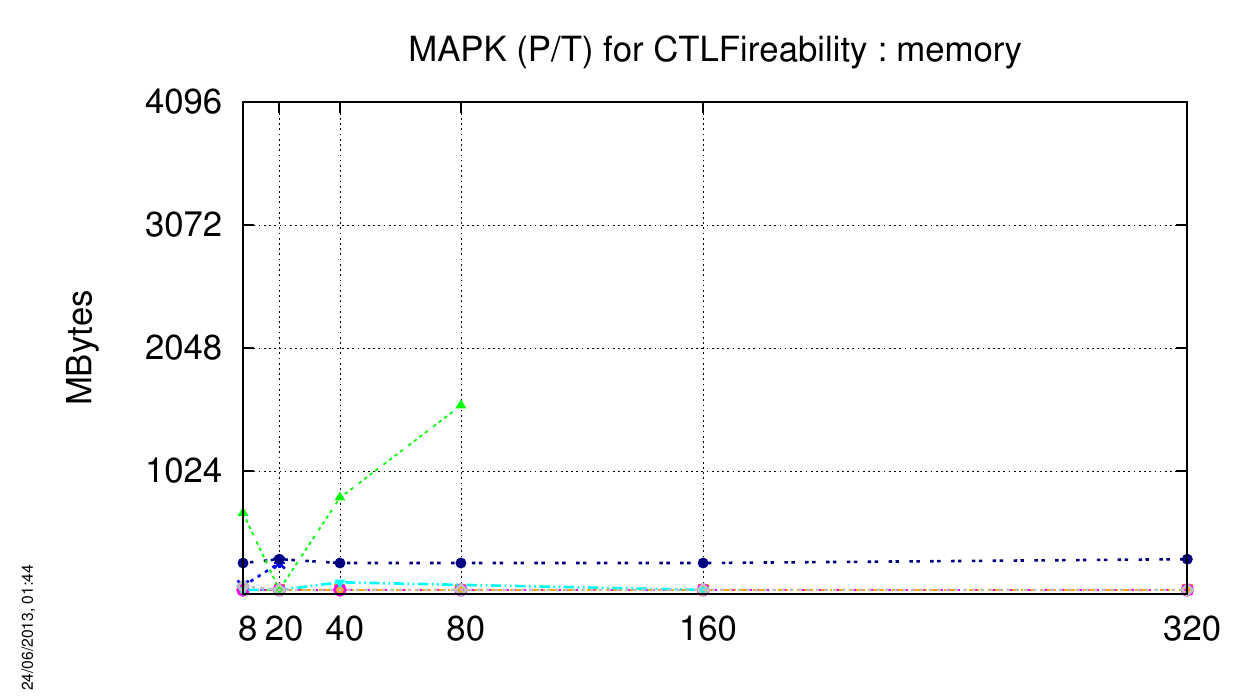}
   \includegraphics[width=7.2cm]{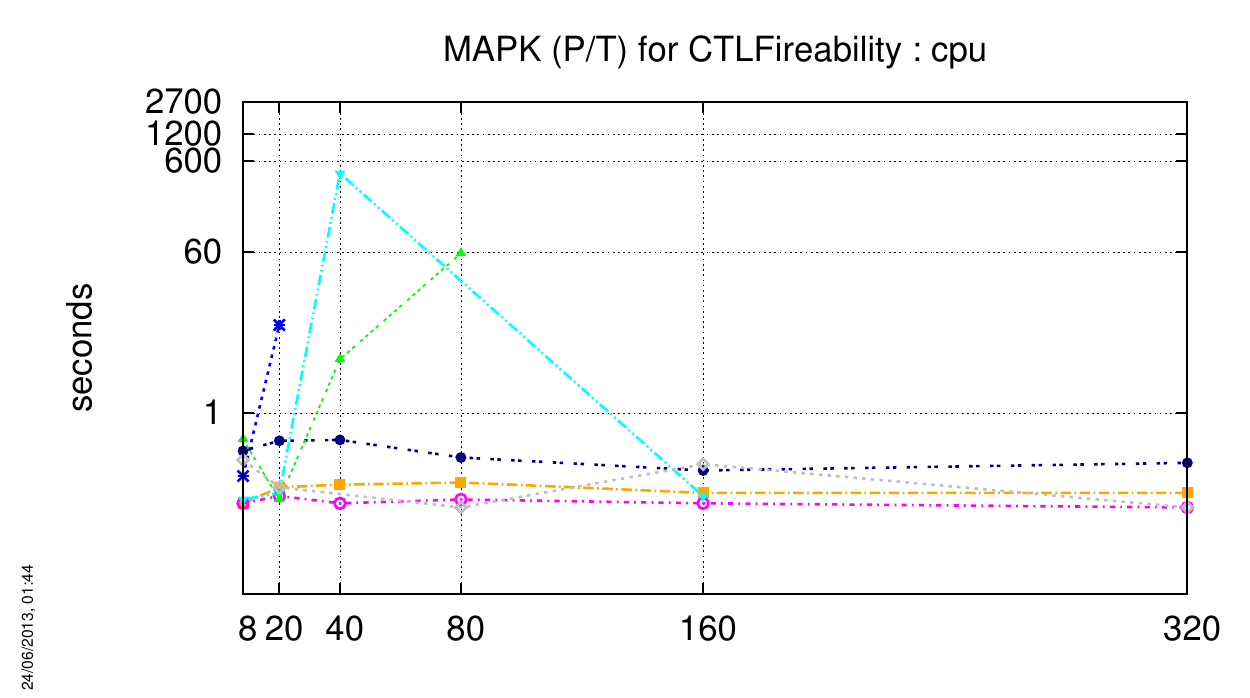}

   \includegraphics[height=1cm]{figures/tools-legend.pdf}
\end{center}

\subsubsection{\acs{NeoElection-COL}}
No instance of this model could be computed for the \textbf{CTLFireability} examination.

\subsubsection{\acs{NeoElection-PT}}
The charts below respectively show how tools compete with this ``Known'' model (memory and CPU).

\index{Performances!CTLFireability!NeoElection (P/T)}
\begin{center}
   \includegraphics[width=7.2cm]{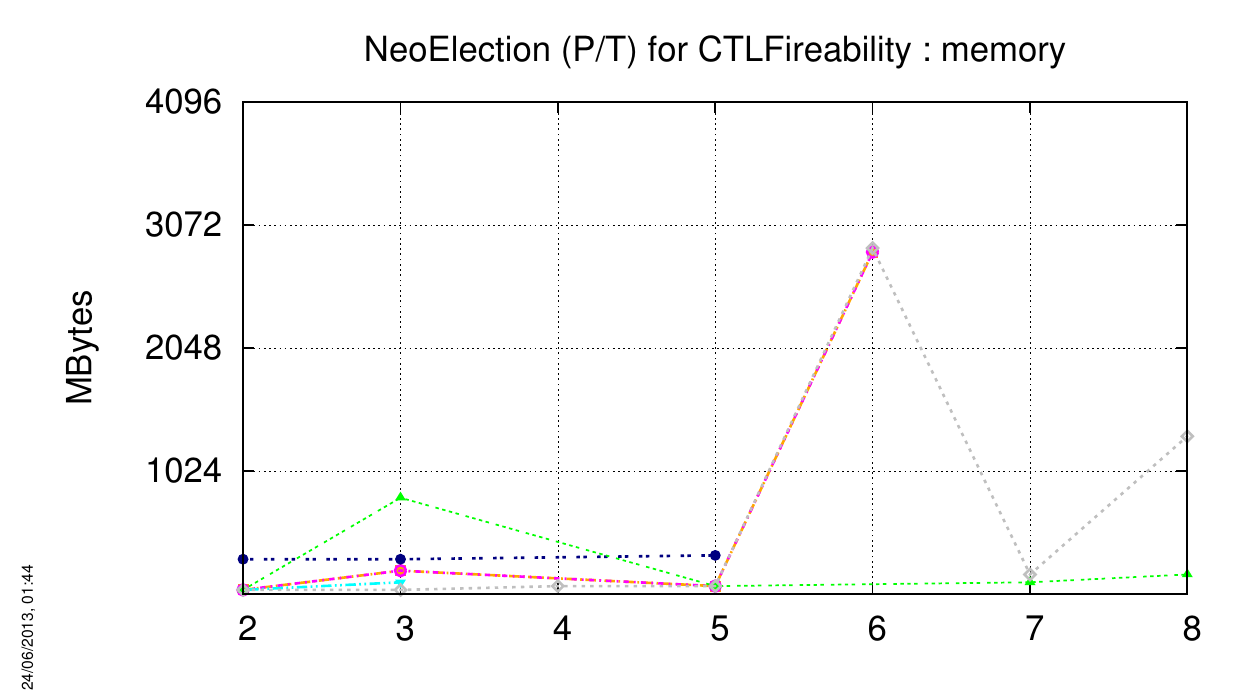}
   \includegraphics[width=7.2cm]{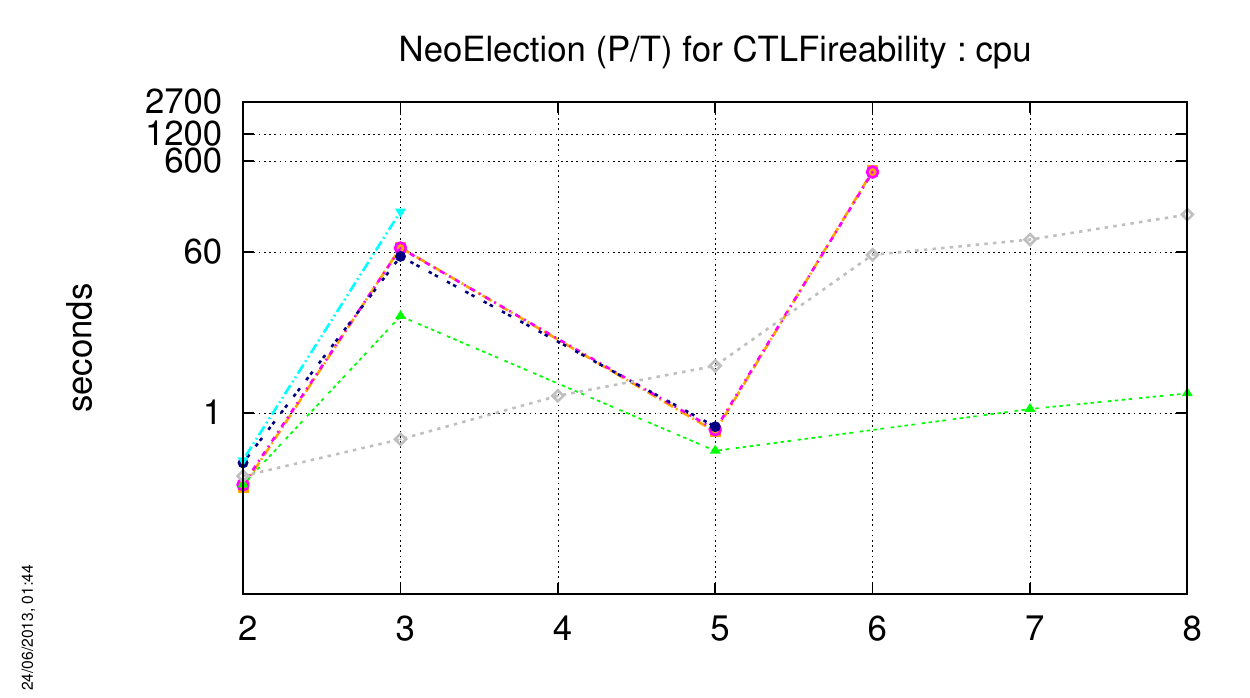}

   \includegraphics[height=1cm]{figures/tools-legend.pdf}
\end{center}

\subsubsection{\acs{PermAdmissibility-COL}}
No instance of this model could be computed for the \textbf{CTLFireability} examination.

\subsubsection{\acs{PermAdmissibility-PT}}
The charts below respectively show how tools compete with this ``Known'' model (memory and CPU).

\index{Performances!CTLFireability!PermAdmissibility (P/T)}
\begin{center}
   \includegraphics[width=7.2cm]{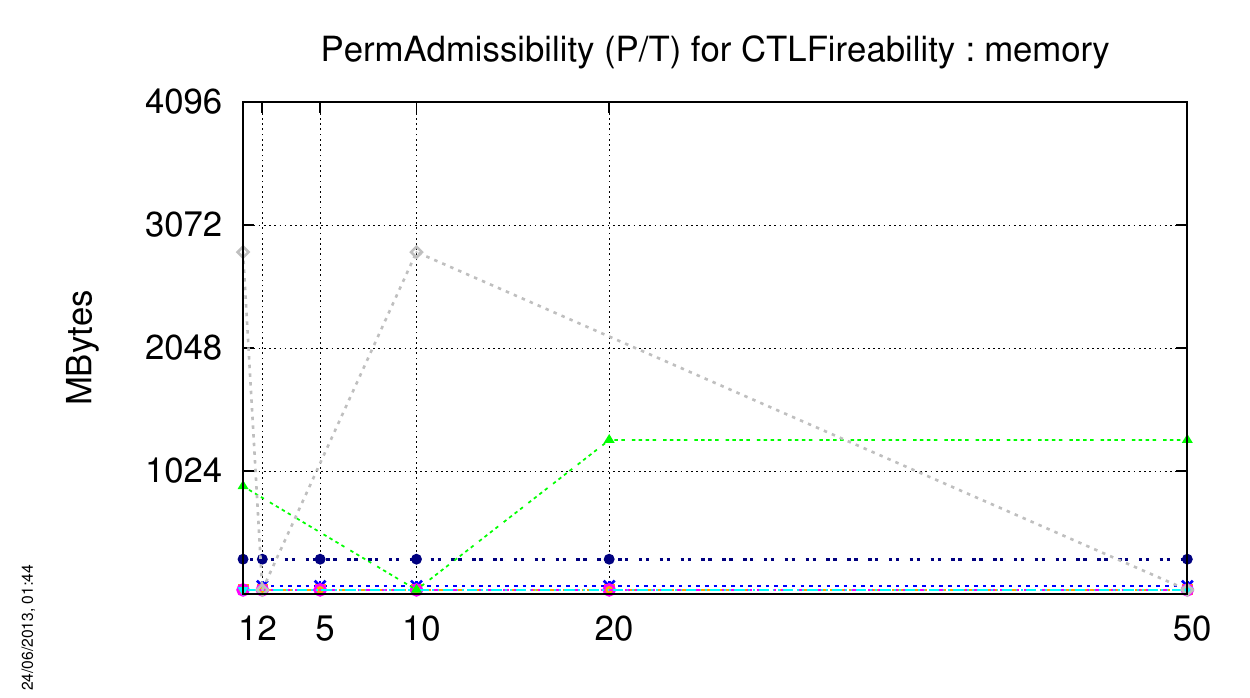}
   \includegraphics[width=7.2cm]{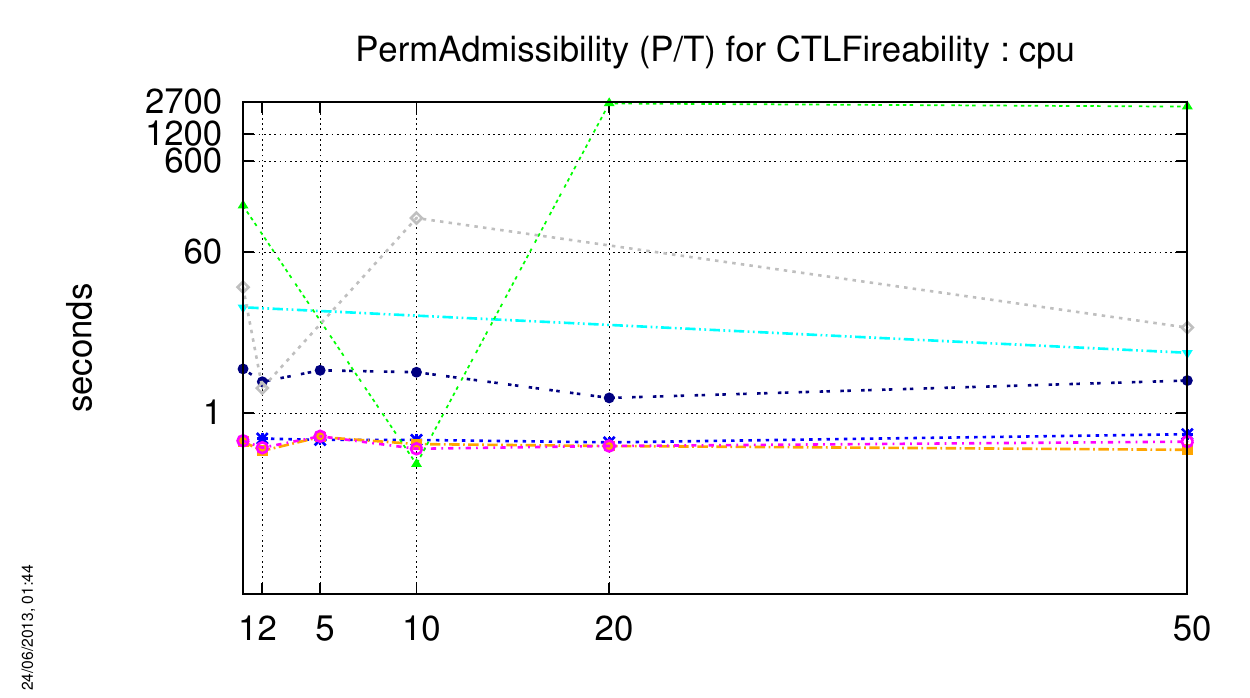}

   \includegraphics[height=1cm]{figures/tools-legend.pdf}
\end{center}

\subsubsection{\acs{Peterson-COL}}
No instance of this model could be computed for the \textbf{CTLFireability} examination.

\subsubsection{\acs{Peterson-PT}}
The charts below respectively show how tools compete with this ``Known'' model (memory and CPU).

\index{Performances!CTLFireability!Peterson (P/T)}
\begin{center}
   \includegraphics[width=7.2cm]{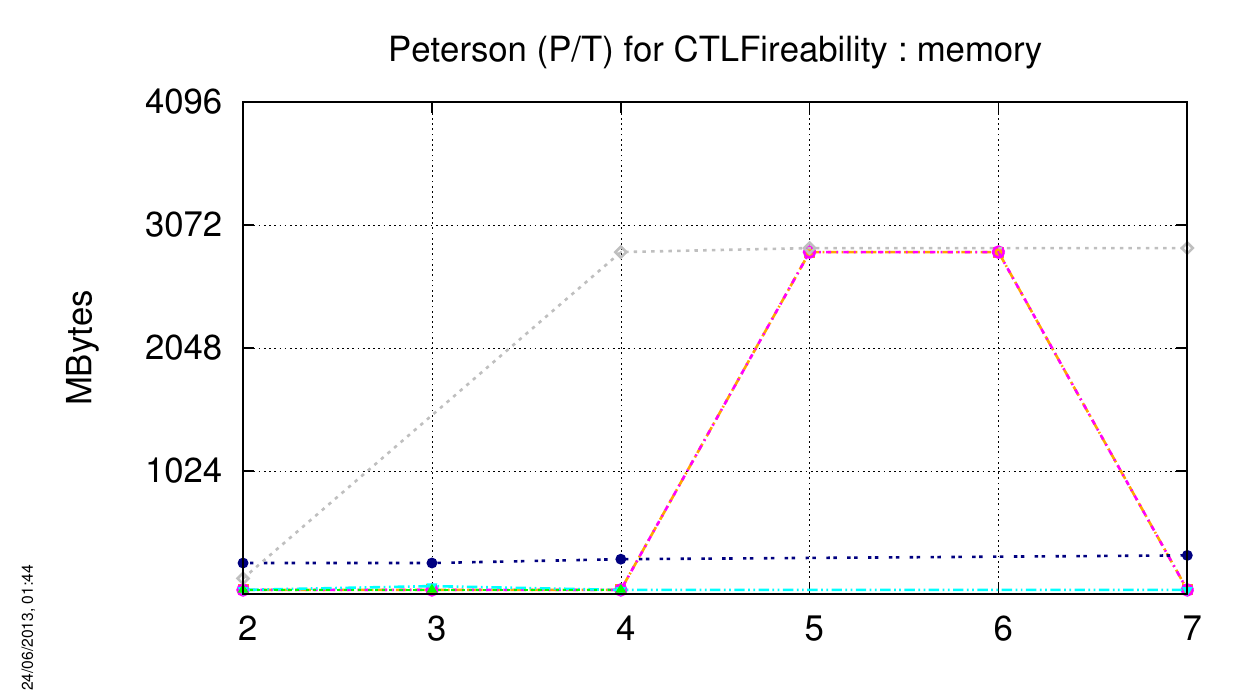}
   \includegraphics[width=7.2cm]{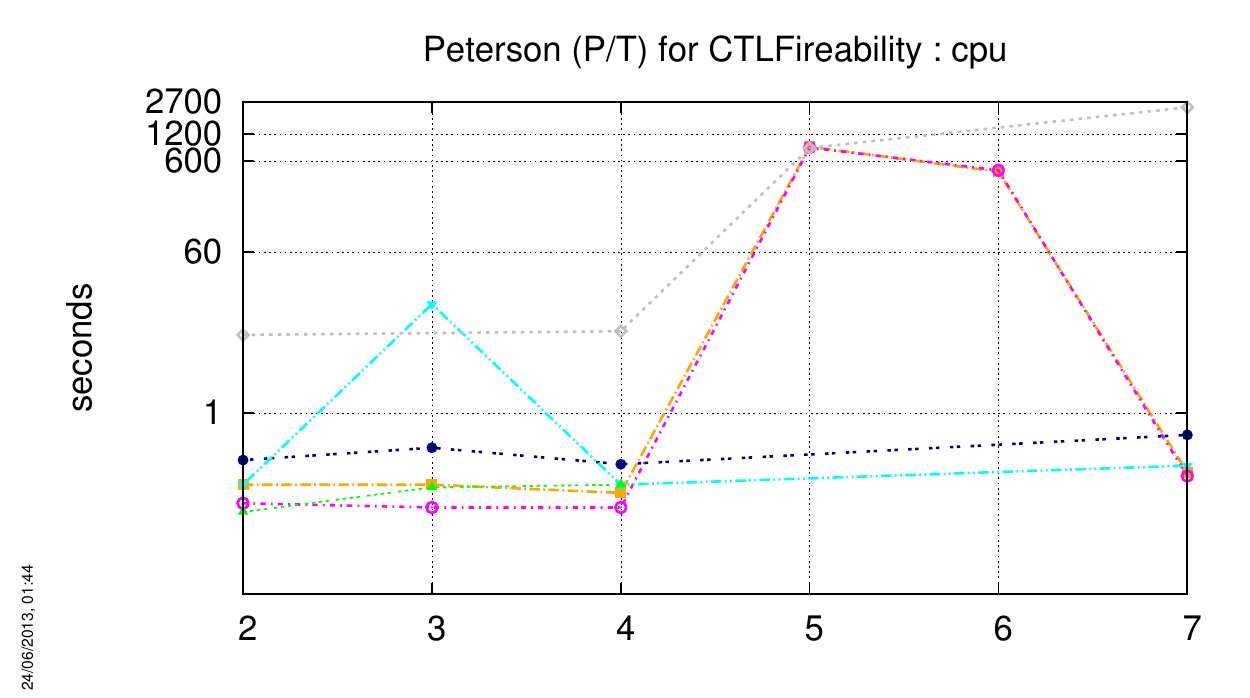}

   \includegraphics[height=1cm]{figures/tools-legend.pdf}
\end{center}

\subsubsection{\acs{Philosophers-COL}}
No instance of this model could be computed for the \textbf{CTLFireability} examination.

\subsubsection{\acs{Philosophers-PT}}
The charts below respectively show how tools compete with this ``Known'' model (memory and CPU).

\index{Performances!CTLFireability!Philosophers (P/T)}
\begin{center}
   \includegraphics[width=7.2cm]{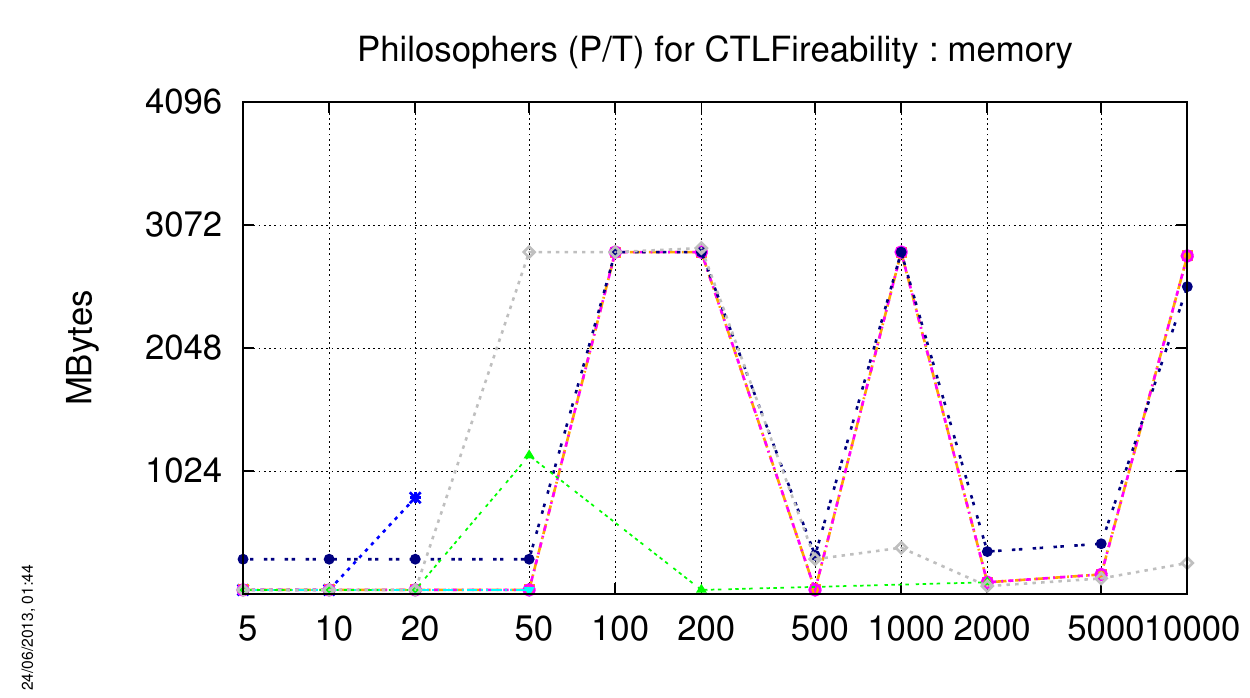}
   \includegraphics[width=7.2cm]{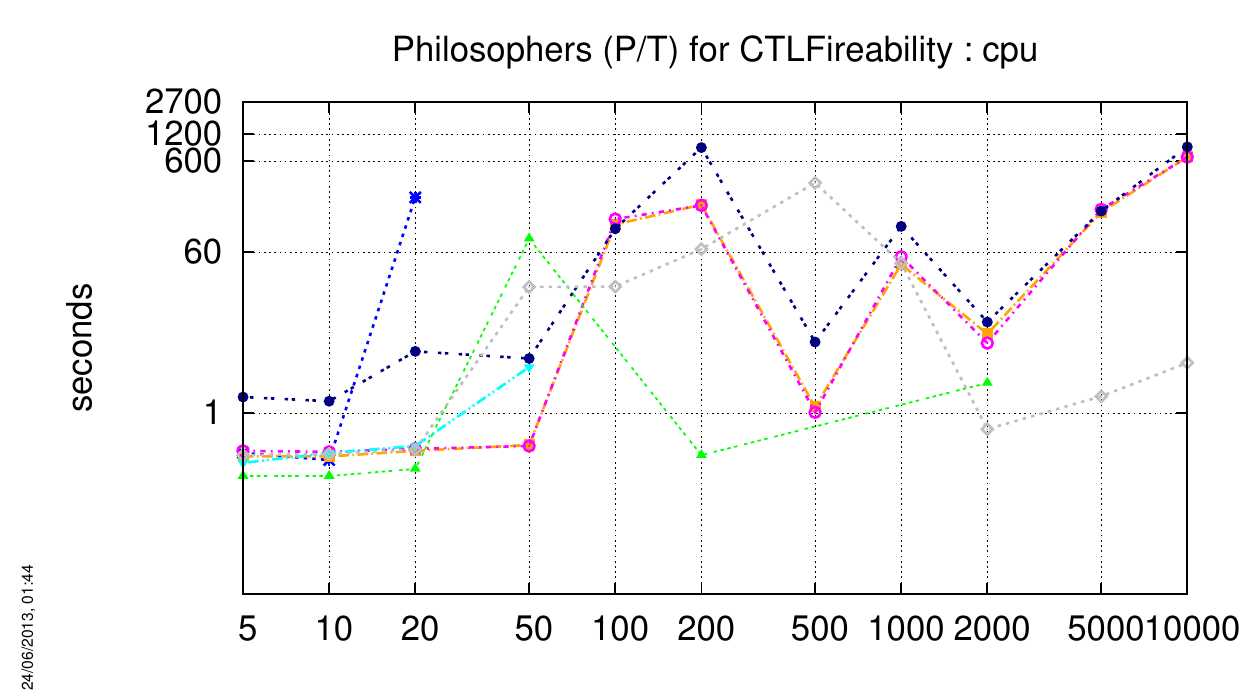}

   \includegraphics[height=1cm]{figures/tools-legend.pdf}
\end{center}

\subsubsection{\acs{PhilosophersDyn-COL}}
No instance of this model could be computed for the \textbf{CTLFireability} examination.

\subsubsection{\acs{PhilosophersDyn-PT}}
The charts below respectively show how tools compete with this ``Known'' model (memory and CPU).

\index{Performances!CTLFireability!PhilosophersDyn (P/T)}
\begin{center}
   \includegraphics[width=7.2cm]{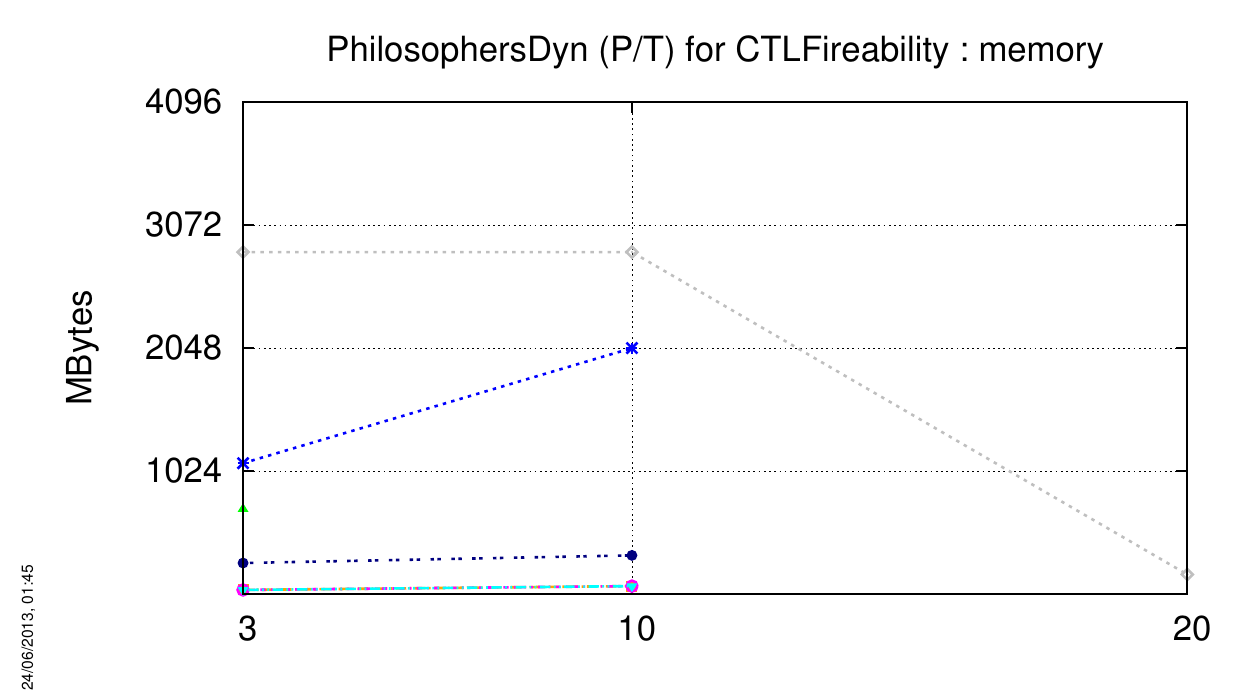}
   \includegraphics[width=7.2cm]{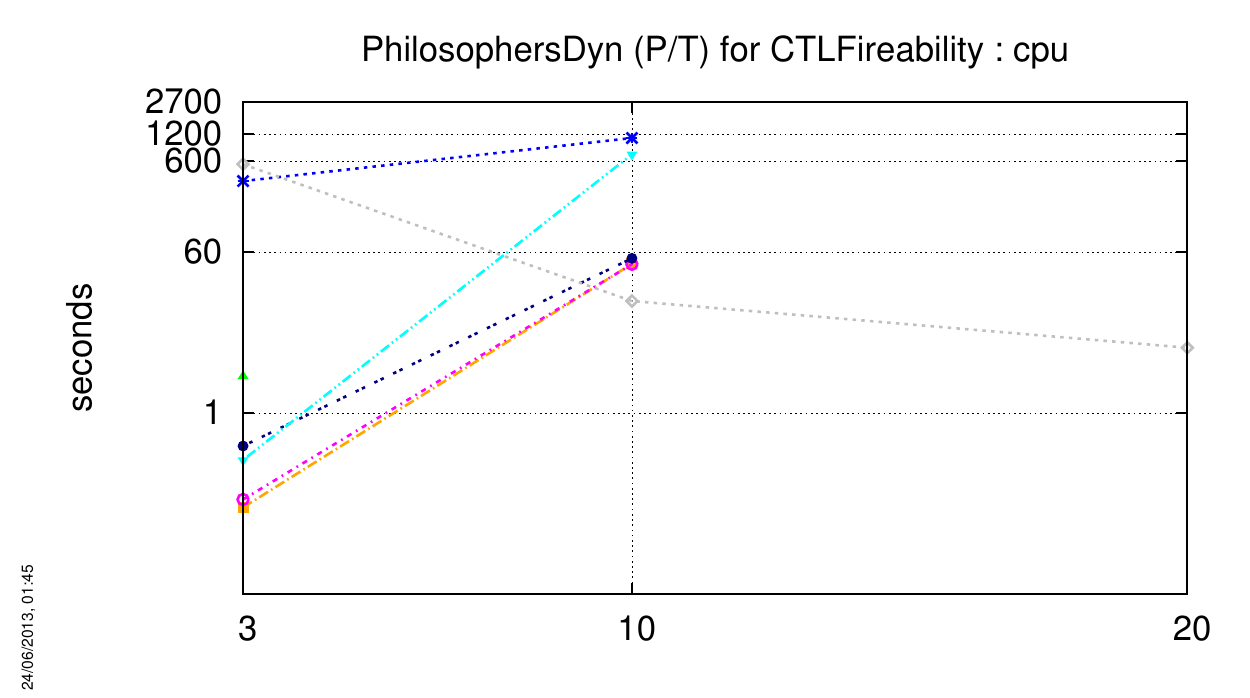}

   \includegraphics[height=1cm]{figures/tools-legend.pdf}
\end{center}

\subsubsection{\acs{Planning-PT}}
No instance of this model could be computed for the \textbf{CTLFireability} examination.

\subsubsection{\acs{Railroad-PT}}
The charts below respectively show how tools compete with this ``Known'' model (memory and CPU).

\index{Performances!CTLFireability!Railroad (P/T)}
\begin{center}
   \includegraphics[width=7.2cm]{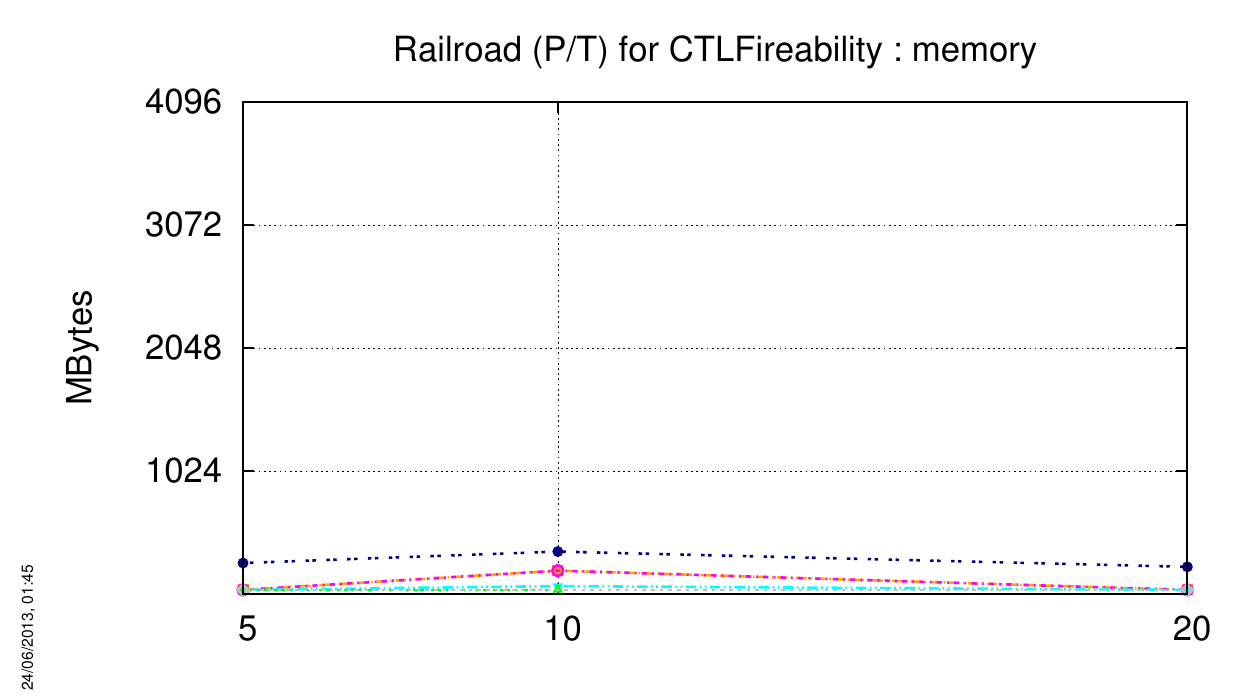}
   \includegraphics[width=7.2cm]{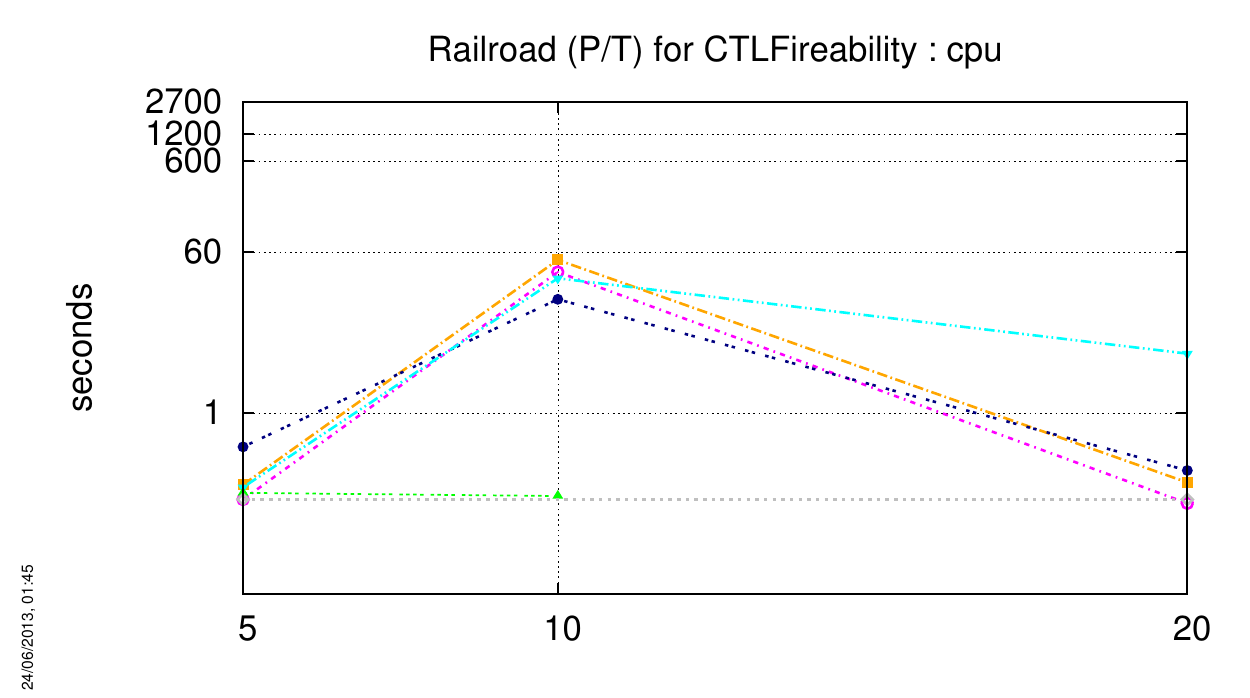}

   \includegraphics[height=1cm]{figures/tools-legend.pdf}
\end{center}

\subsubsection{\acs{RessAllocation-PT}}
The charts below respectively show how tools compete with this ``Known'' model (memory and CPU).

\index{Performances!CTLFireability!RessAllocation (P/T)}
\begin{center}
   \includegraphics[width=7.2cm]{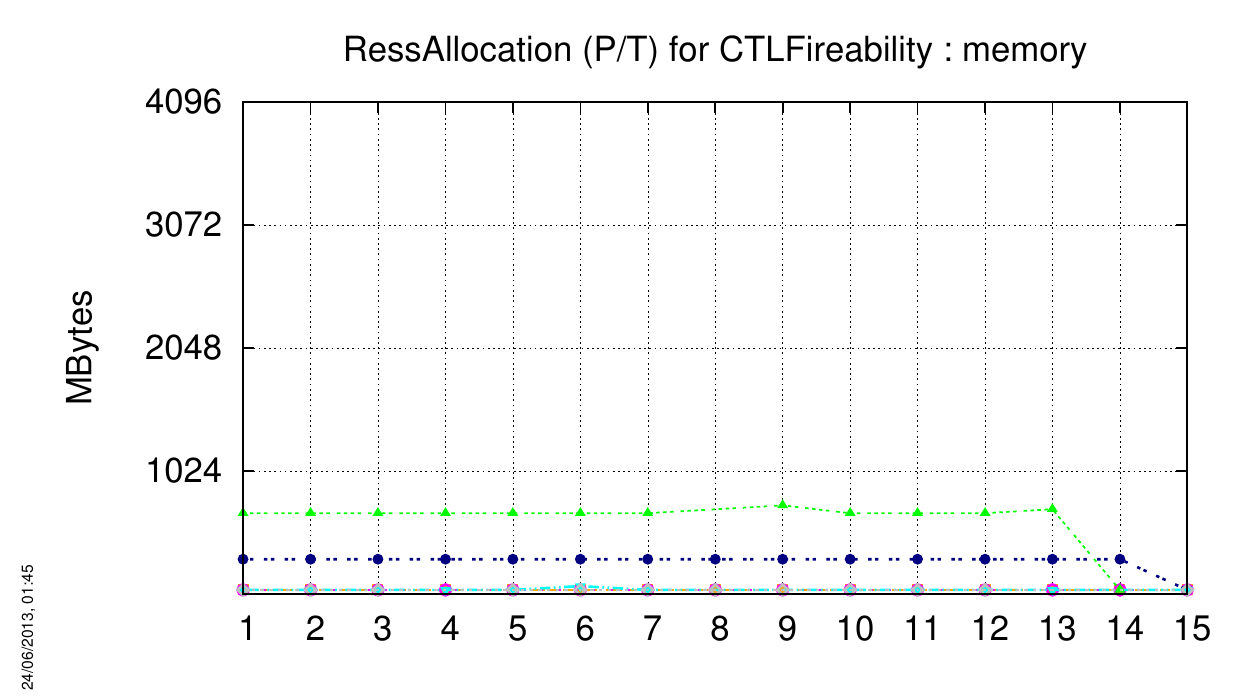}
   \includegraphics[width=7.2cm]{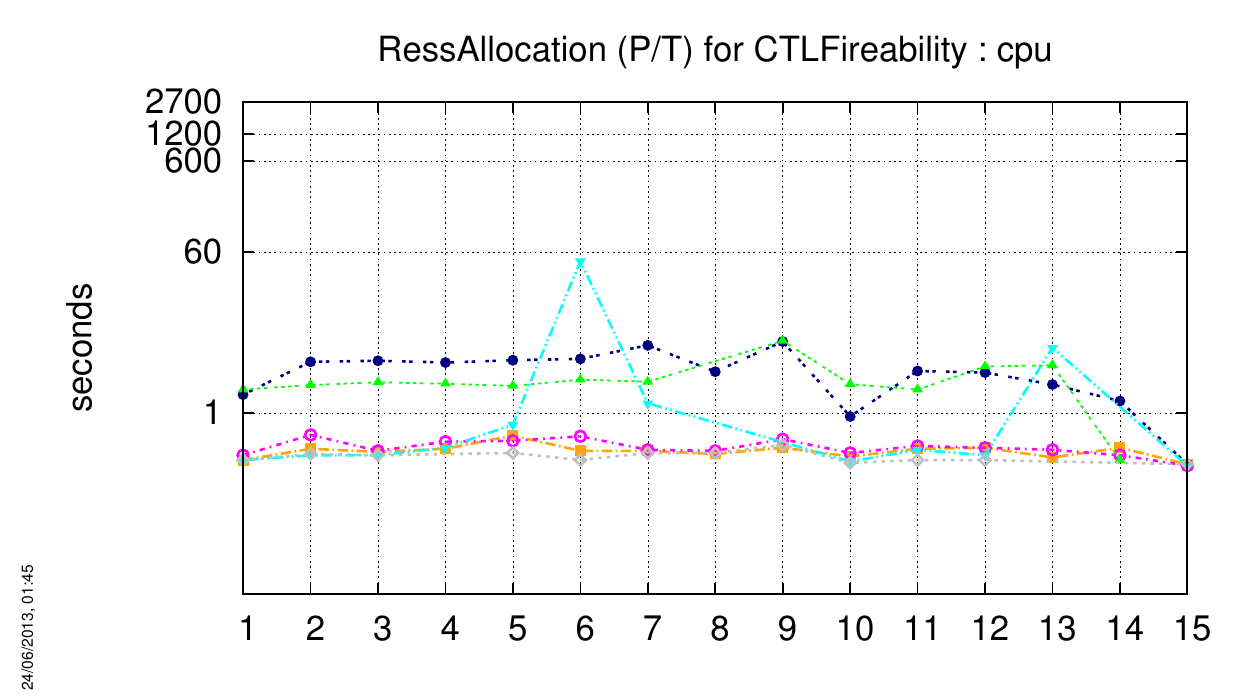}

   \includegraphics[height=1cm]{figures/tools-legend.pdf}
\end{center}

\subsubsection{\acs{Ring-PT}}
The charts below respectively show how tools compete with this ``Known'' model (memory and CPU).

\index{Performances!CTLFireability!Ring (P/T)}
\begin{center}
   \includegraphics[width=7.2cm]{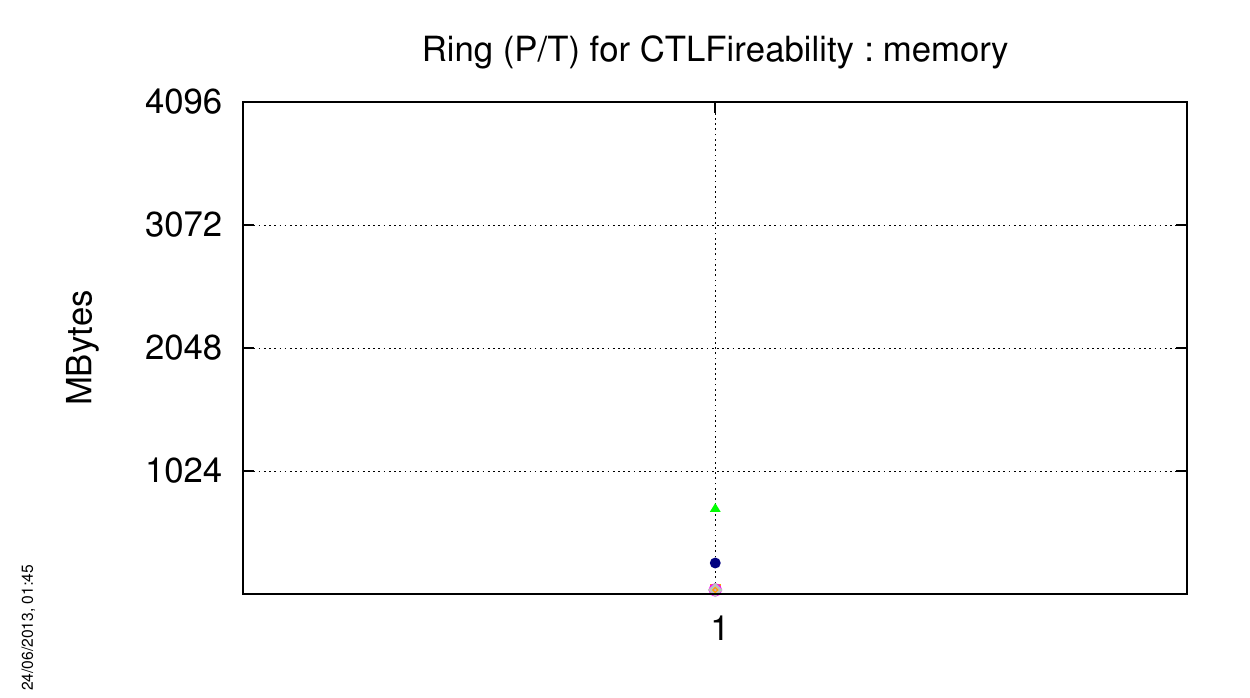}
   \includegraphics[width=7.2cm]{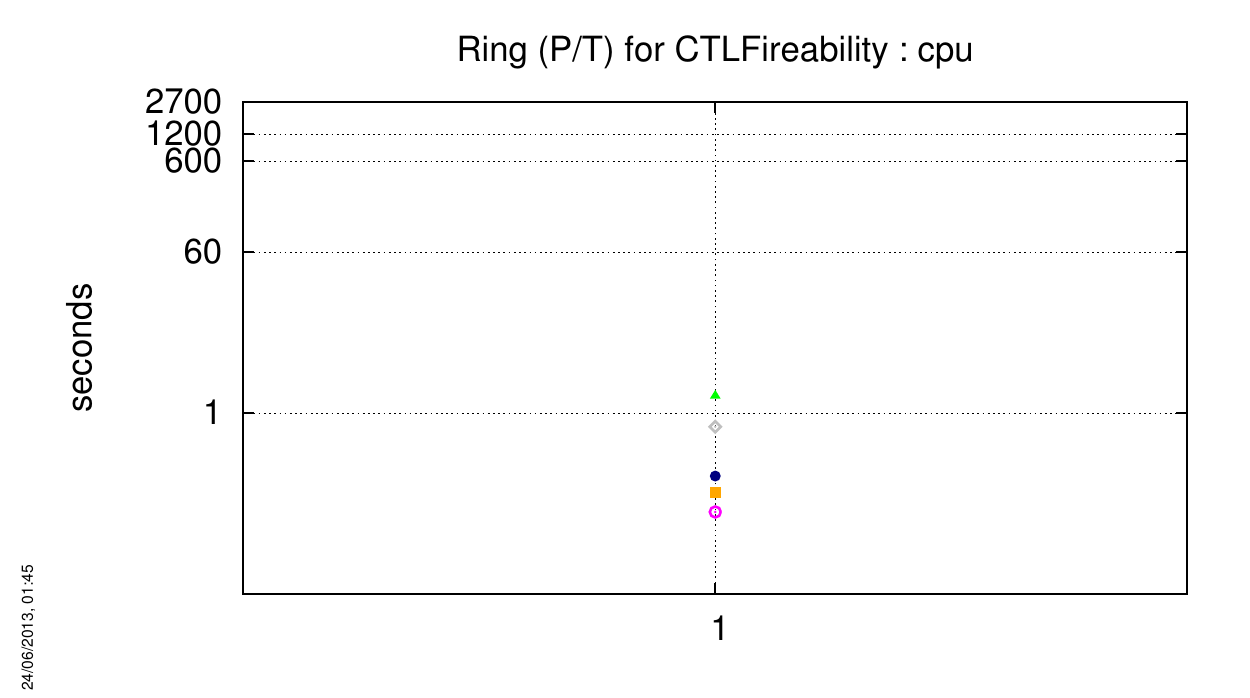}

   \includegraphics[height=1cm]{figures/tools-legend.pdf}
\end{center}

\subsubsection{\acs{RwMutex-PT}}
The charts below respectively show how tools compete with this ``Known'' model (memory and CPU).

\index{Performances!CTLFireability!RwMutex (P/T)}
\begin{center}
   \includegraphics[width=7.2cm]{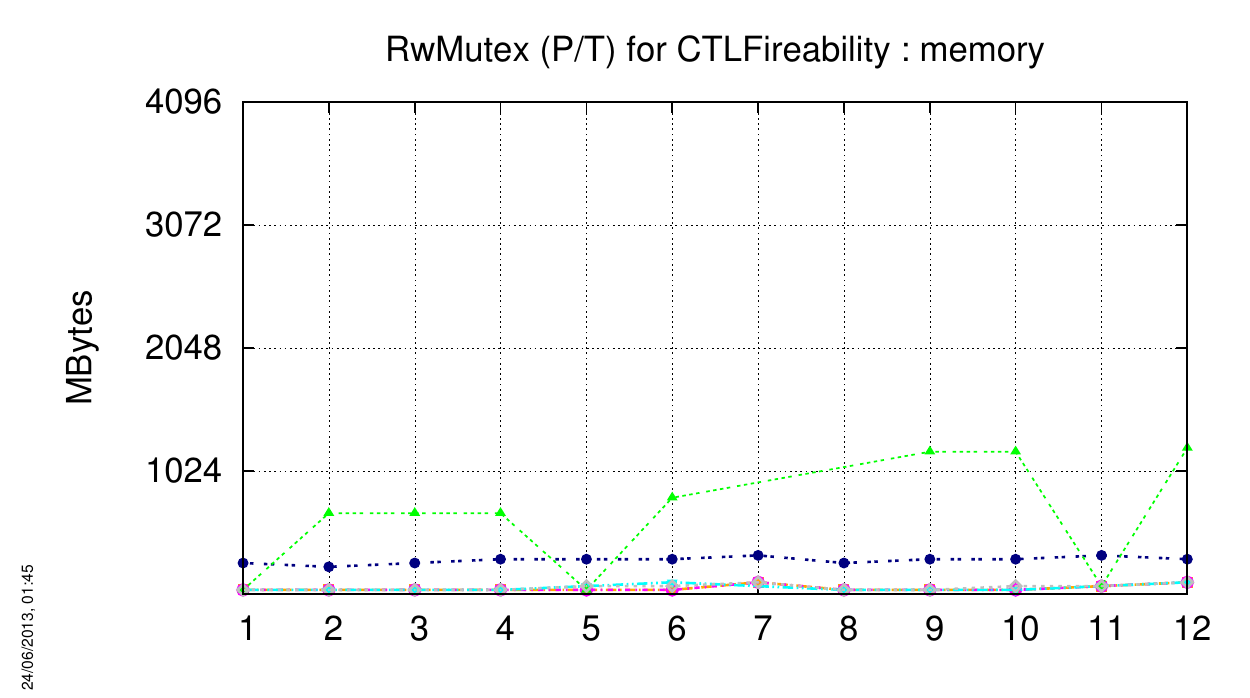}
   \includegraphics[width=7.2cm]{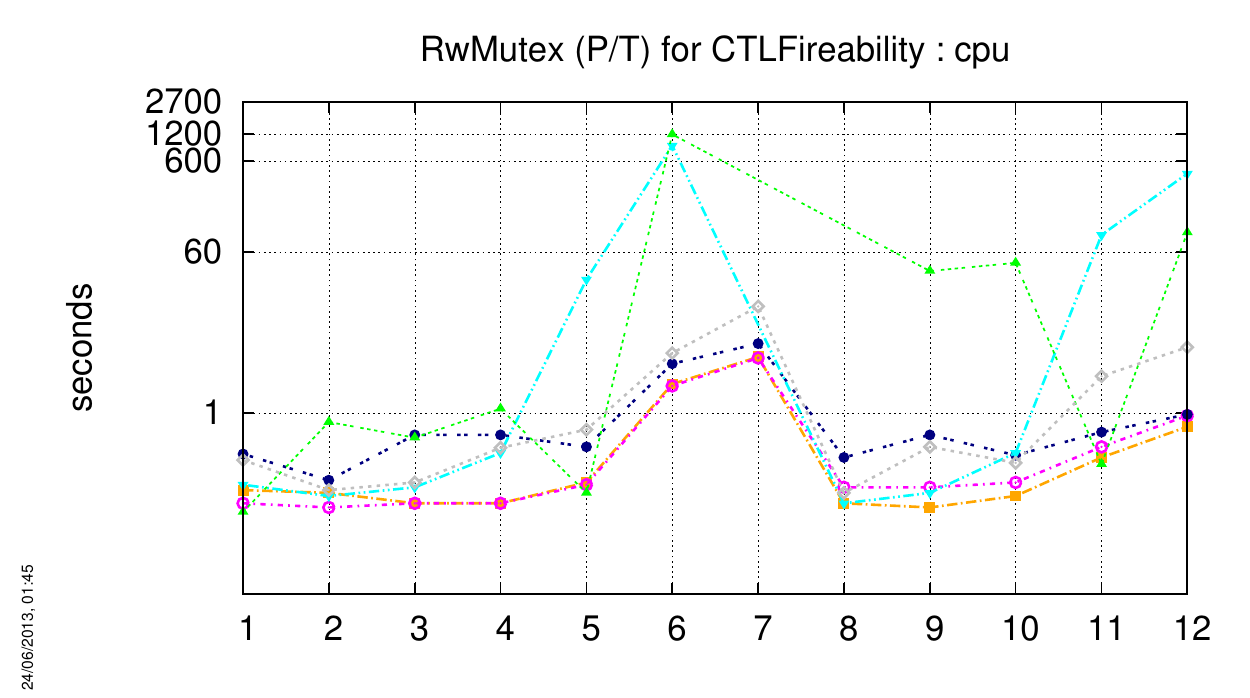}

   \includegraphics[height=1cm]{figures/tools-legend.pdf}
\end{center}

\subsubsection{\acs{SharedMemory-COL}}
No instance of this model could be computed for the \textbf{CTLFireability} examination.

\subsubsection{\acs{SharedMemory-PT}}
The charts below respectively show how tools compete with this ``Known'' model (memory and CPU).

\index{Performances!CTLFireability!SharedMemory (P/T)}
\begin{center}
   \includegraphics[width=7.2cm]{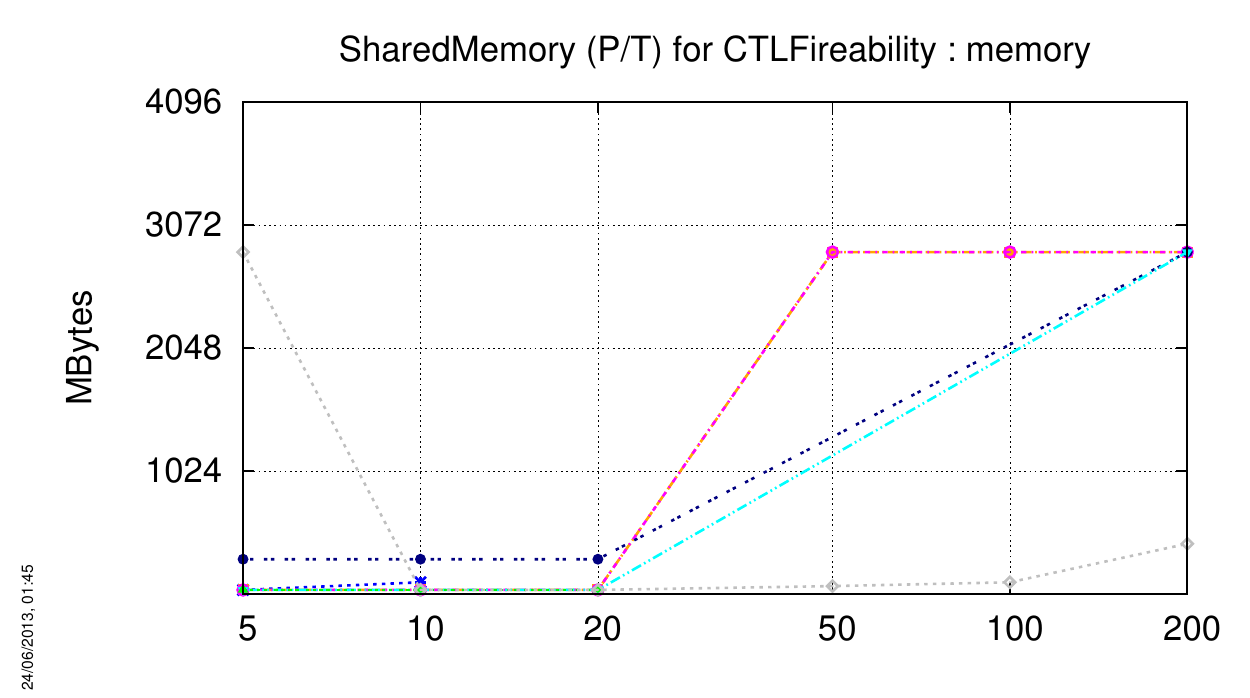}
   \includegraphics[width=7.2cm]{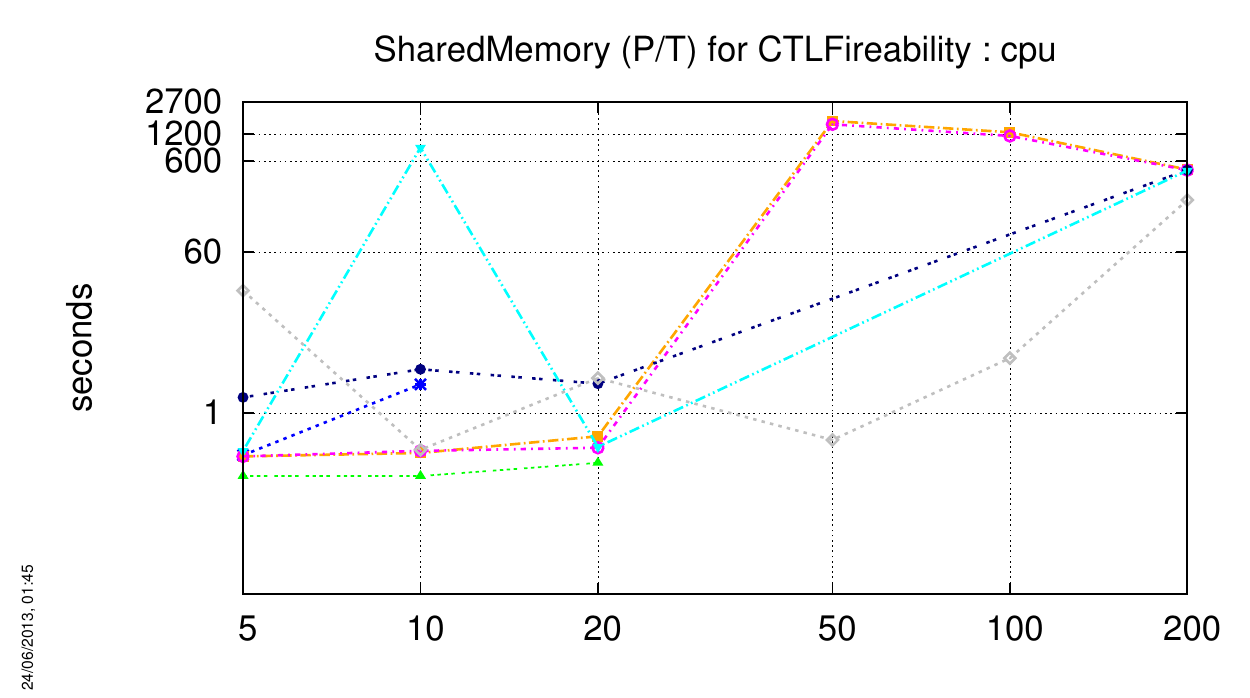}

   \includegraphics[height=1cm]{figures/tools-legend.pdf}
\end{center}

\subsubsection{\acs{SimpleLoadBal-COL}}
No instance of this model could be computed for the \textbf{CTLFireability} examination.

\subsubsection{\acs{SimpleLoadBal-PT}}
The charts below respectively show how tools compete with this ``Known'' model (memory and CPU).

\index{Performances!CTLFireability!SimpleLoadBal (P/T)}
\begin{center}
   \includegraphics[width=7.2cm]{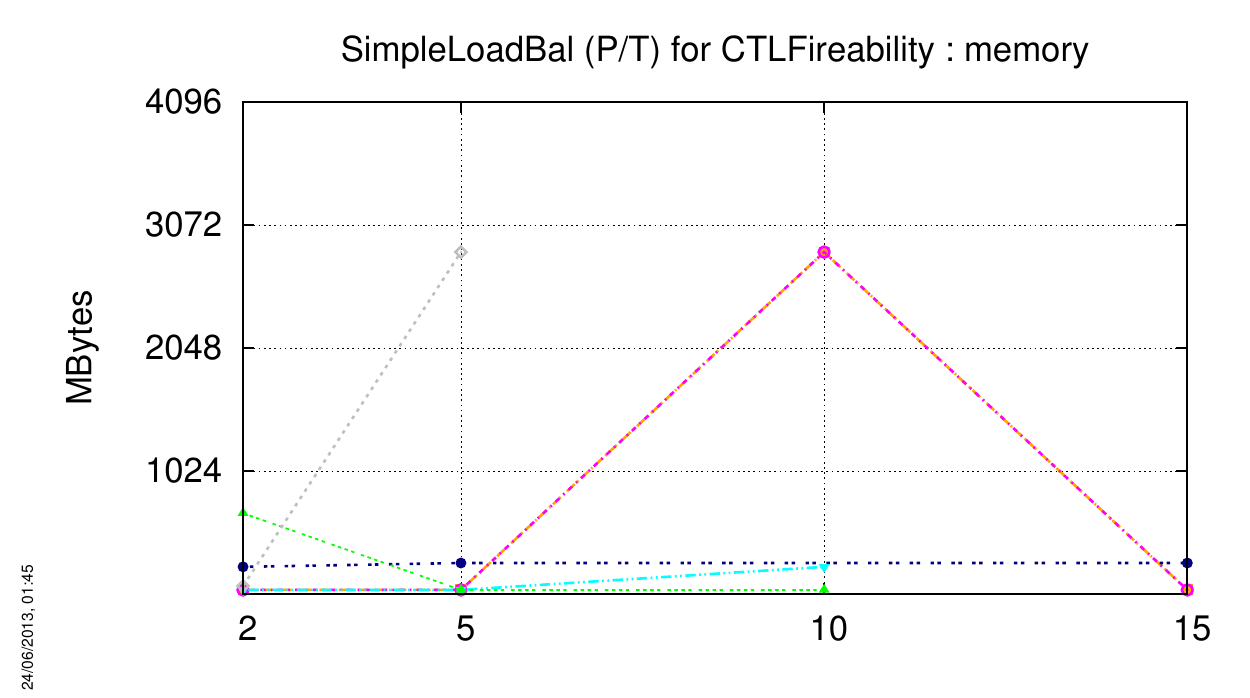}
   \includegraphics[width=7.2cm]{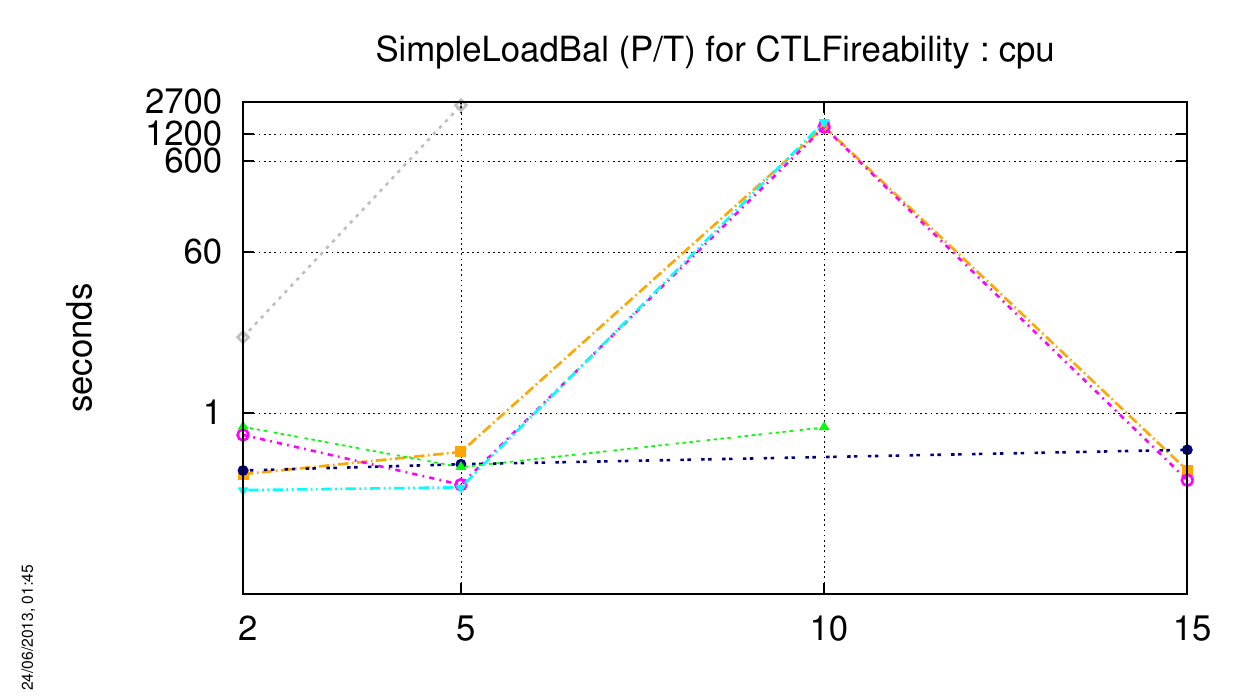}

   \includegraphics[height=1cm]{figures/tools-legend.pdf}
\end{center}

\subsubsection{\acs{TokenRing-COL}}
No instance of this model could be computed for the \textbf{CTLFireability} examination.

\subsubsection{\acs{TokenRing-PT}}
The charts below respectively show how tools compete with this ``Known'' model (memory and CPU).

\index{Performances!CTLFireability!TokenRing (P/T)}
\begin{center}
   \includegraphics[width=7.2cm]{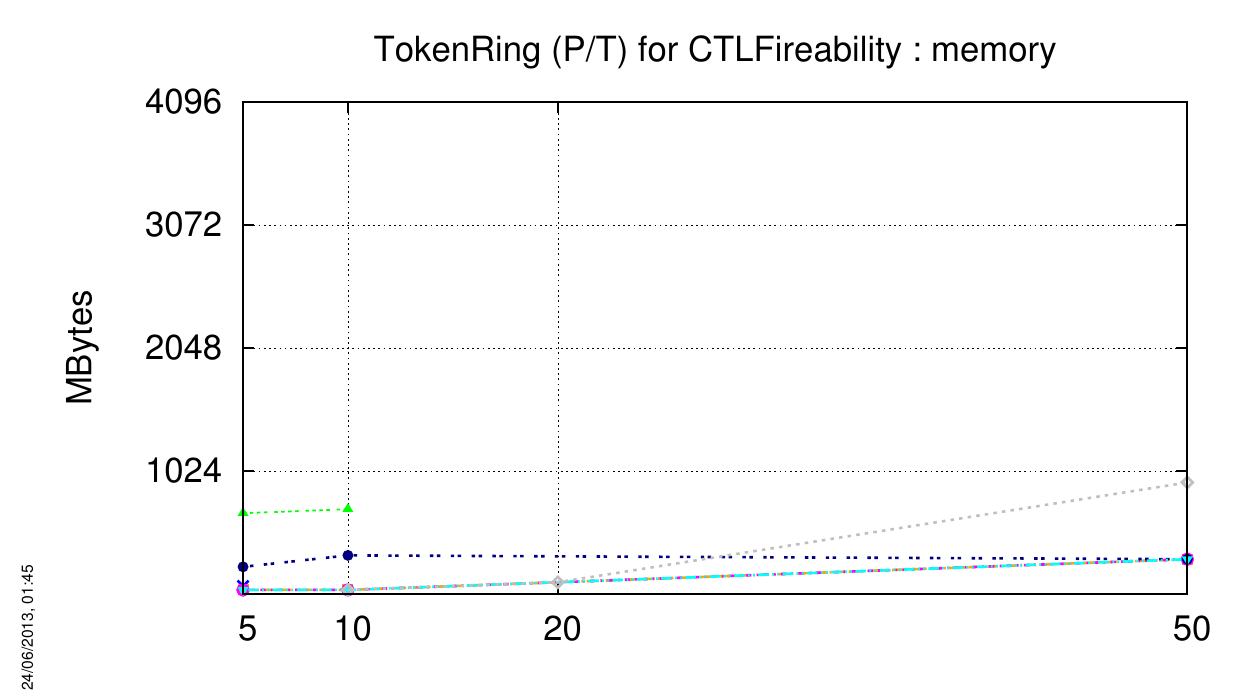}
   \includegraphics[width=7.2cm]{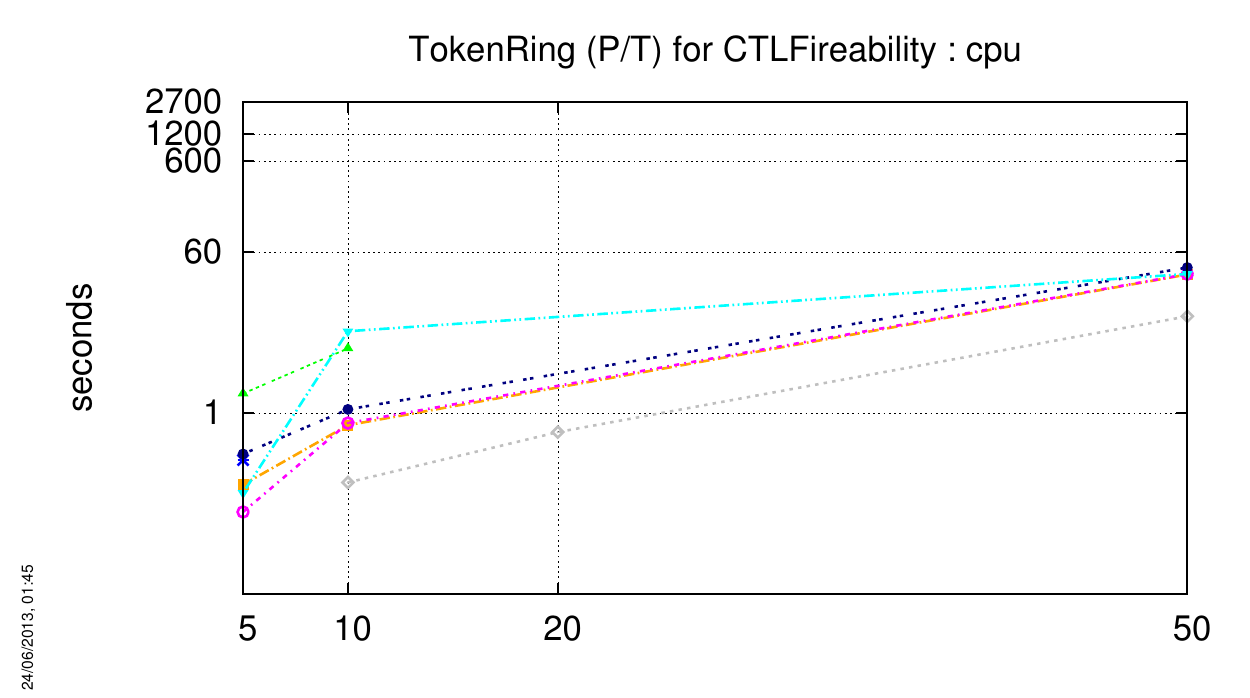}

   \includegraphics[height=1cm]{figures/tools-legend.pdf}
\end{center}

\subsubsection{\acs{HouseConstruction-PT}}
The charts below respectively show how tools compete with this ``Suprise'' model (memory and CPU).

\index{Performances!CTLFireability!HouseConstruction (P/T)}
\begin{center}
   \includegraphics[width=7.2cm]{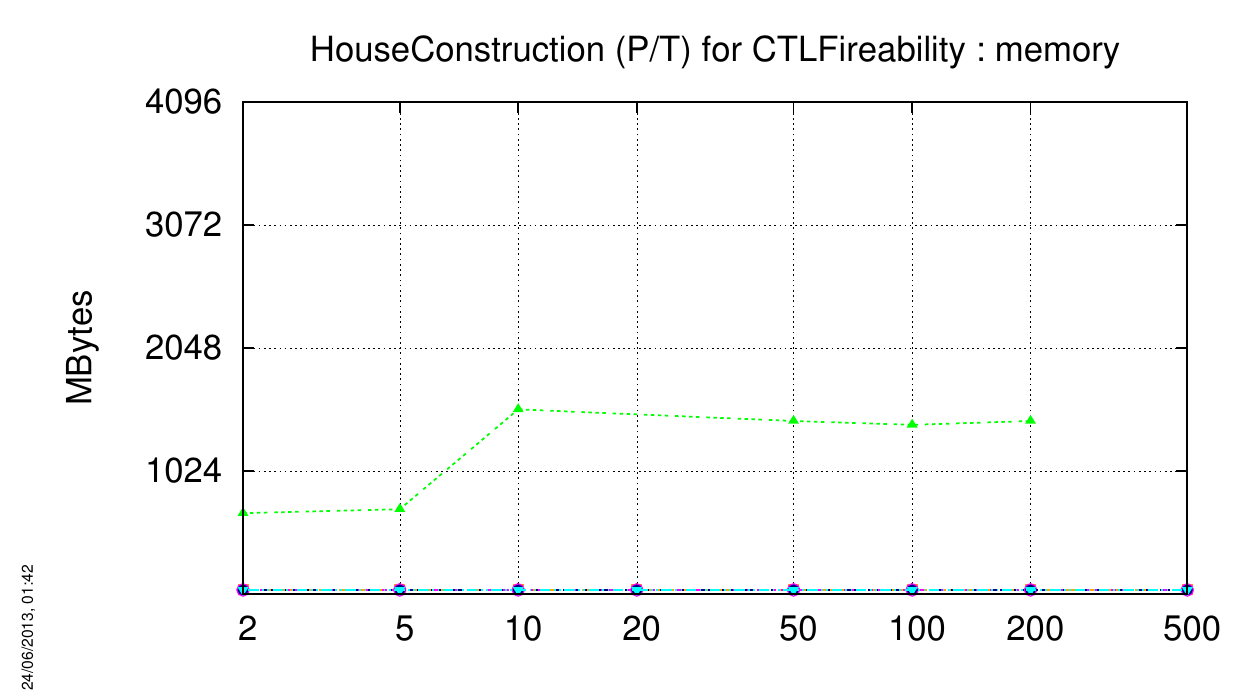}
   \includegraphics[width=7.2cm]{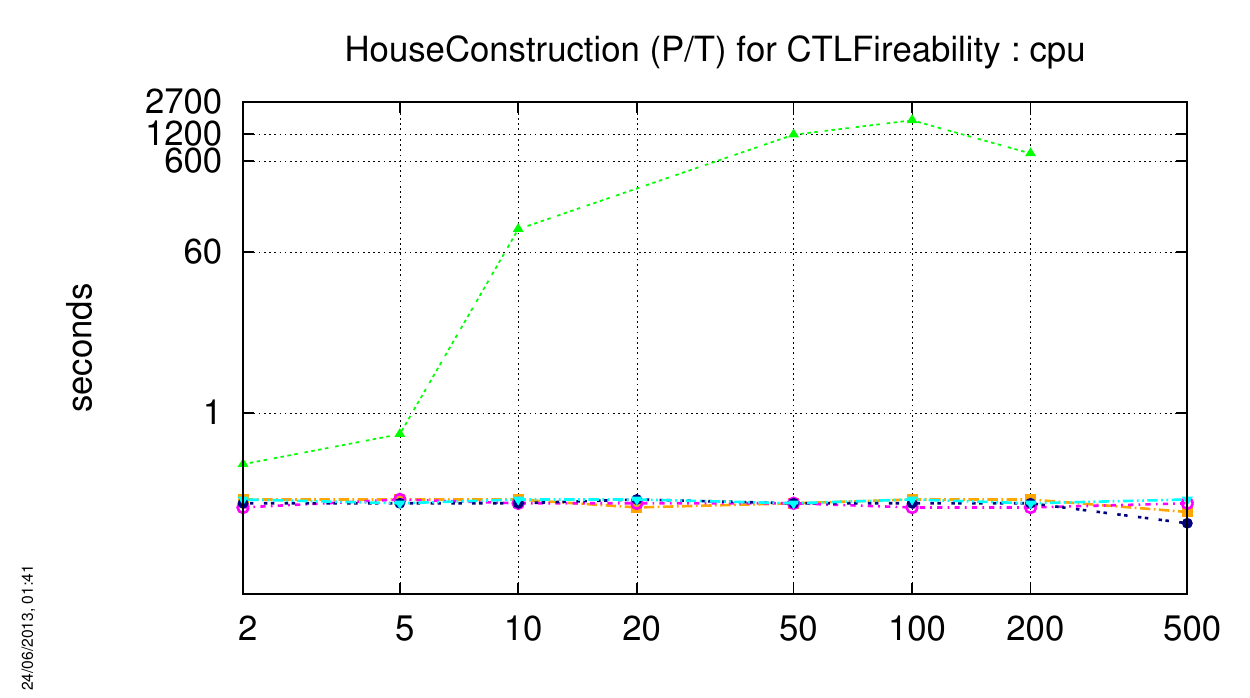}

   \includegraphics[height=1cm]{figures/tools-legend.pdf}
\end{center}

\subsubsection{\acs{IBMB2S565S3960-PT}}
The charts below respectively show how tools compete with this ``Suprise'' model (memory and CPU).

\index{Performances!CTLFireability!IBMB2S565S3960 (P/T)}
\begin{center}
   \includegraphics[width=7.2cm]{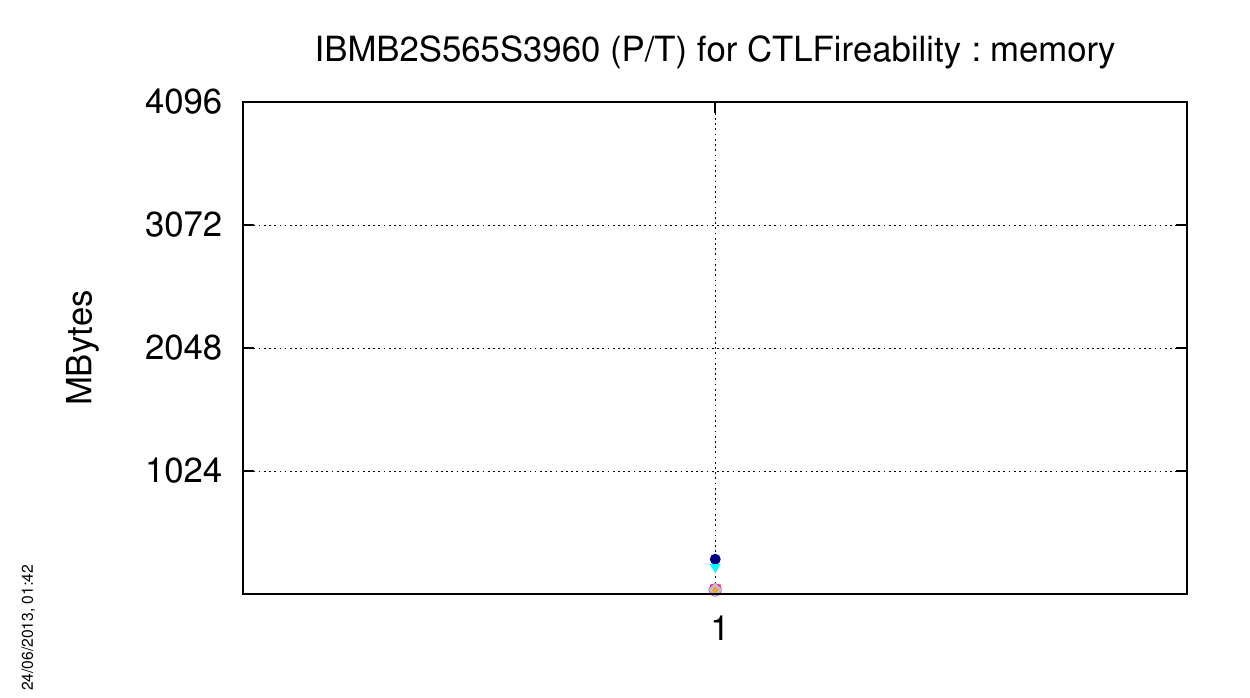}
   \includegraphics[width=7.2cm]{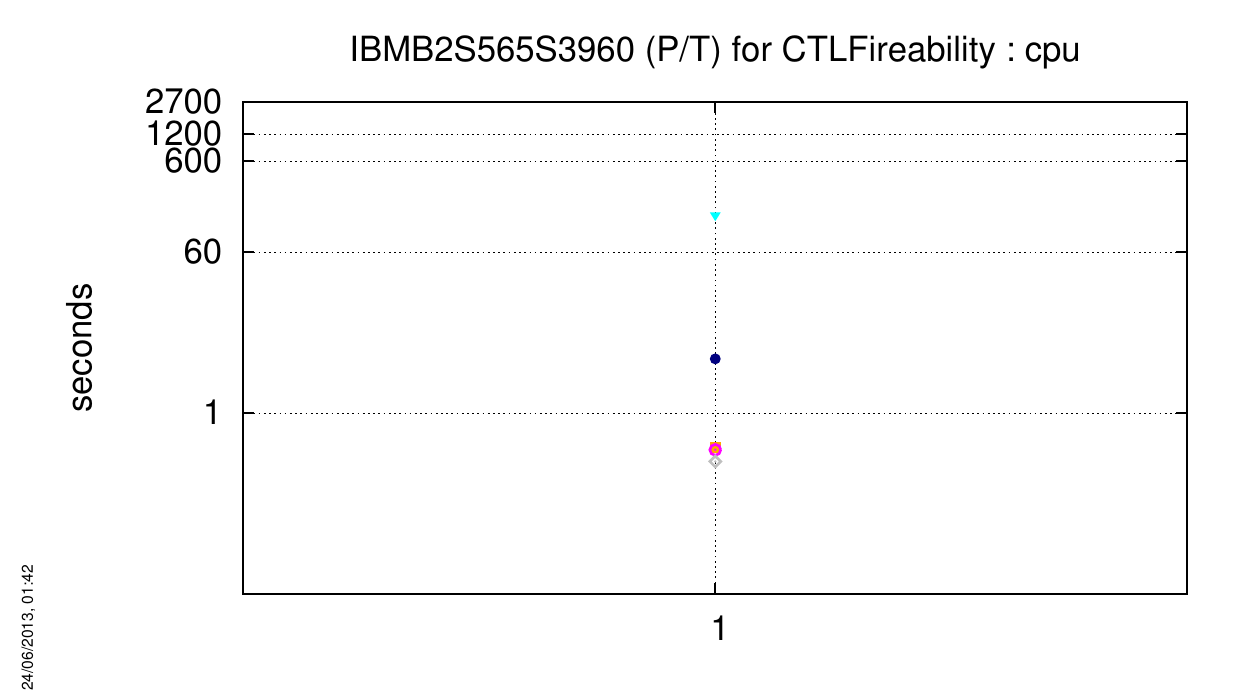}

   \includegraphics[height=1cm]{figures/tools-legend.pdf}
\end{center}

\subsubsection{\acs{QuasiCertifProtocol-COL}}
No instance of this model could be computed for the \textbf{CTLFireability} examination.

\subsubsection{\acs{QuasiCertifProtocol-PT}}
The charts below respectively show how tools compete with this ``Suprise'' model (memory and CPU).

\index{Performances!CTLFireability!QuasiCertifProtocol (P/T)}
\begin{center}
   \includegraphics[width=7.2cm]{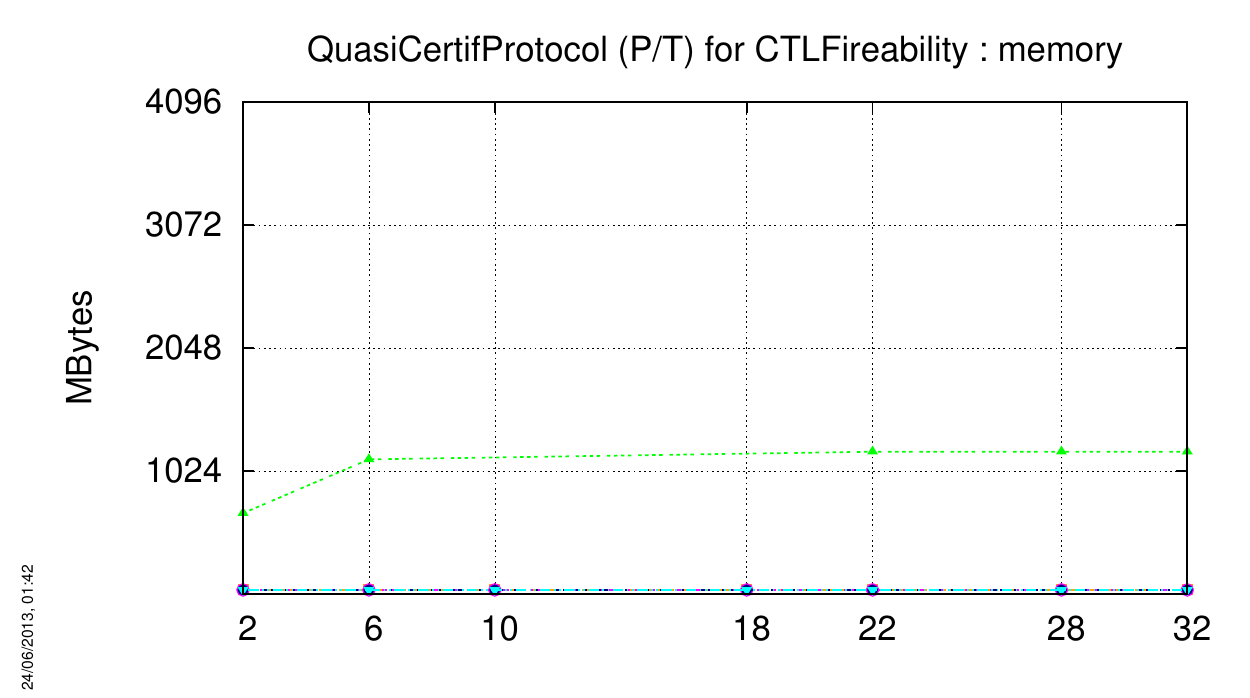}
   \includegraphics[width=7.2cm]{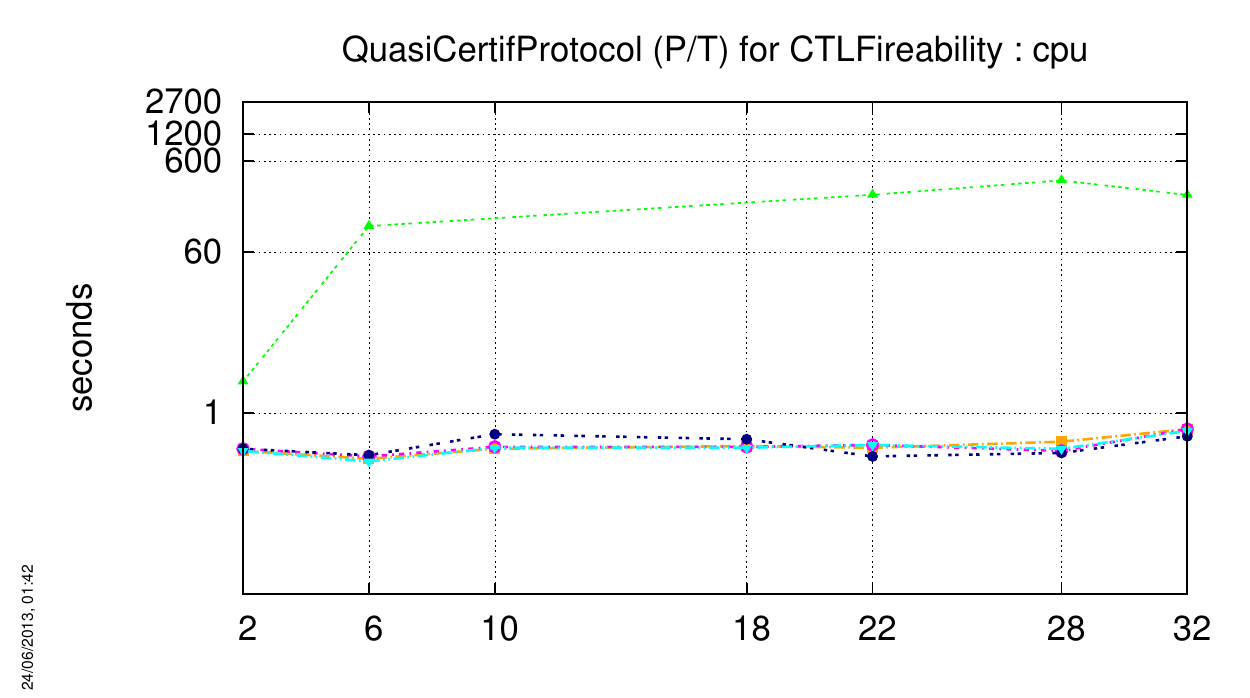}

   \includegraphics[height=1cm]{figures/tools-legend.pdf}
\end{center}

\subsubsection{\acs{Vasy2003-PT}}
The charts below respectively show how tools compete with this ``Suprise'' model (memory and CPU).

\index{Performances!CTLFireability!Vasy2003 (P/T)}
\begin{center}
   \includegraphics[width=7.2cm]{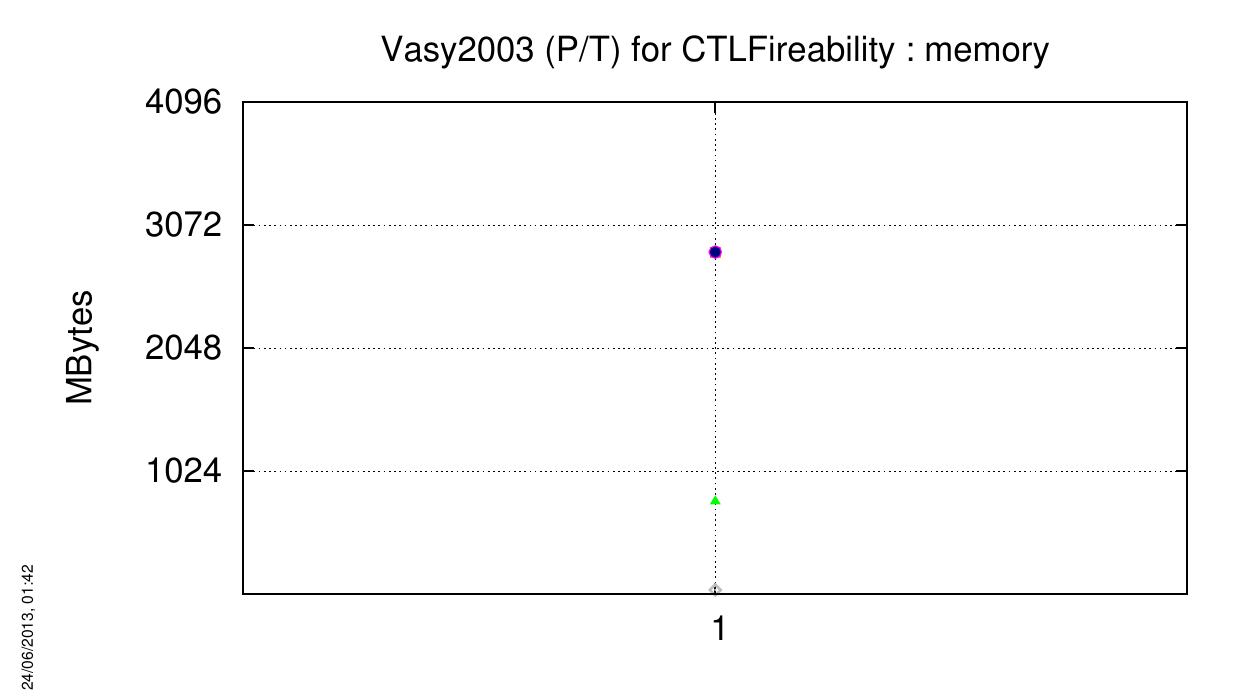}
   \includegraphics[width=7.2cm]{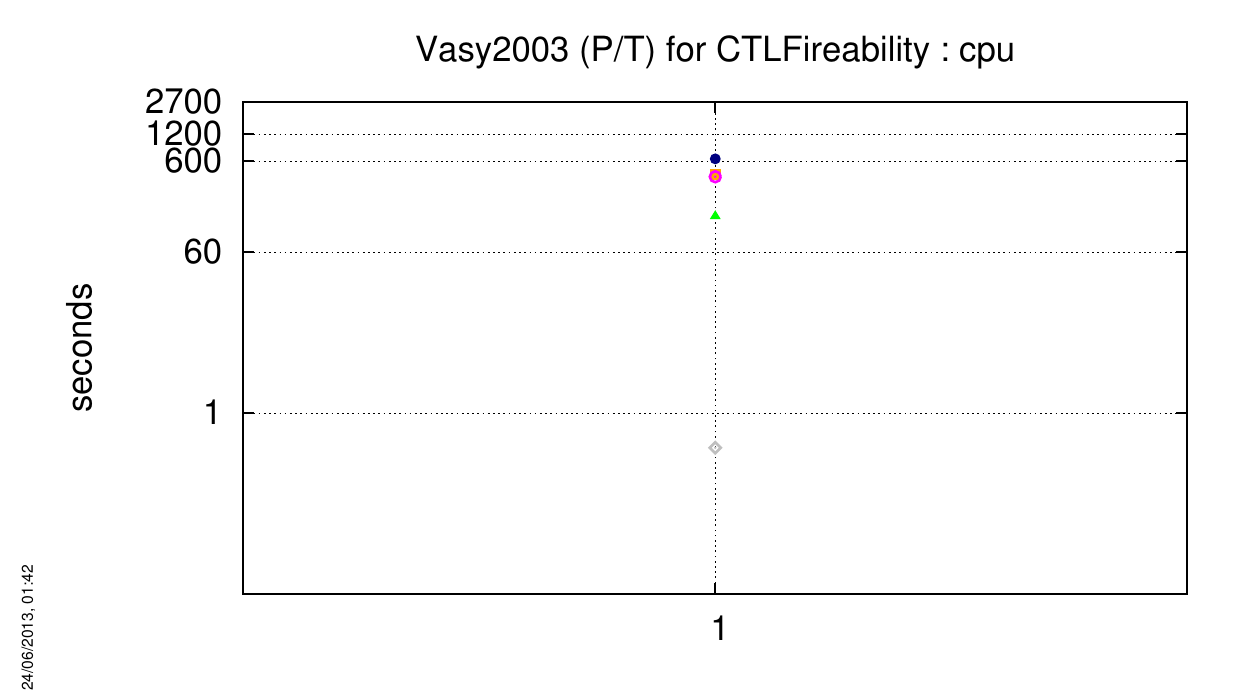}

   \includegraphics[height=1cm]{figures/tools-legend.pdf}
\end{center}

\subsection{Outputs for the CTLFireability Examination}
\index{Outputs!CTLFireability}

Please find enclosed the brute results for this examination (``Known'' and ``Surprise'' models).
We display only the score of tools that provide a results for at least one instance of one model.
The legend for the values is provided below:
\begin{itemize}
   \item\textbf{nc}: the tool does not compete this examination for this model/instance,
   \item\textbf{cc}: the tool cannot compute this examination for this model/instance,
   \item\textbf{to}: the tool cannot compute this examination for this model/instance within the maximum allowed time,
   \item\textbf{mp}: the tool encountered a memory problem (stack overflow or memory full),
   \item\textbf{nf}: there is no formula available for this type of examination (typically, this concerns P/T nets where
       comparing marking cardinality has no signification when there is no equivalent colored net).
\end{itemize}

\textbf{Note on the display of results for formulas:} each formula is considered as a flag (F if false, T if true, - or ?
when the value cannot be determined). These values are concatenated in the order they appear (we assume it is the order of formulas as they were provided).

\subsubsection{``Known'' Models}

Results are summarized in the table below.
{\footnotesize
\begin{longtable}{|cccccccc|}
   \hline
      \multicolumn{8}{|c|}{\bf \acs{CSRepetitions-COL}} \\
      \textbf{Instances}\index{Outputs!CTLFireability!CSRepetitions (colored)} &
      \acs{greatspn} &
      \acs{lola} &
      \acs{lola-optimistic} &
      \acs{lola-optimistic-incomplete} &
      \acs{lola-pessimistic} &
      \acs{marcie} &
      \acs{sara} \\
      02 &
      nc &
      nc &
      nc &
      nc &
      nc &
      nc &
      nc \\
      03 &
      nc &
      nc &
      nc &
      nc &
      nc &
      nc &
      nc \\
      04 &
      nc &
      nc &
      nc &
      nc &
      nc &
      nc &
      nc \\
      05 &
      nc &
      nc &
      nc &
      nc &
      nc &
      nc &
      nc \\
      07 &
      nc &
      nc &
      nc &
      nc &
      nc &
      nc &
      nc \\
      10 &
      nc &
      nc &
      nc &
      nc &
      nc &
      nc &
      nc \\
\hline
\end{longtable}
}

\subsubsection{``Surprise'' Models}

\input{result_surprise_CTLFireability.tex}

\subsection{Score for the CTLFireability Examination}
\index{Scores!CTLFireability}

Please find enclosed the scores for this examination (``Known'' and ``Surprise'' models).
We display only the score of tools that provide a results for at least one instance of one model.
The total is first listed in the table below followed by a detail, for each proposed model.
Meaning of the line labels are:
\begin{itemize}
\item\textbf{1st instance}: the tool gets a bonus for having processed the first instance of this model (+1 point),
\item\textbf{instances}: the tool gets 1 point per instances treated 
(for that, we assume that at least one formula has been successfully computed),
\item\textbf{max reached}: the tool could process all the instances for the model (+2 points),
\item\textbf{best}: the tool is among the ones that processed a maximum of instances within the time and memory confinement (+2 points).
\end{itemize}

\subsubsection{``Known'' Models}

\input{score_known_CTLFireability.tex}

\subsubsection{``Surprise'' Models}

\input{score_surprise_CTLFireability.tex}

\subsection{Trophies for this Examination}
\index{Trophies!CTLFireability}

Trophies are divided in three categories: ``Known'' models,
``Surprise'' models, and the global trophies (formula is then
$score_{global} = score_{known} + 2 \times score_{surprise}$).

\subsubsection{For ``Known'' Models} \ \\

\begin{tabular}{c|c|c}
      1 & 1 & 3 \\
   \includegraphics[width=2cm]{figures/gold.jpg} &
   \includegraphics[width=2cm]{figures/gold.jpg} &
   \includegraphics[width=2cm]{figures/bronse.jpg} \\
   \acs{lola} &
   \acs{lola-optimistic} &
   \acs{lola-optimistic-incomplete} \\
   198 points &
   198 points &
   178 points \\
\end{tabular}

\subsubsection{For ``Surprise'' Models}\  \\

\begin{tabular}{c|c|c}
      1 & 2 & 2 \\
   \includegraphics[width=2cm]{figures/gold.jpg} &
   \includegraphics[width=2cm]{figures/silver.jpg} &
   \includegraphics[width=2cm]{figures/silver.jpg} \\
   \acs{marcie} &
   \acs{lola} &
   \acs{lola-optimistic} \\
   17 points &
   12 points &
   12 points \\
\end{tabular}

\subsubsection{Global} \ \\

\begin{tabular}{c|c|c}
      1 & 1 & 3 \\
   \includegraphics[width=2cm]{figures/gold.jpg} &
   \includegraphics[width=2cm]{figures/gold.jpg} &
   \includegraphics[width=2cm]{figures/bronse.jpg} \\
   \acs{lola} &
   \acs{lola-optimistic} &
   \acs{lola-optimistic-incomplete} \\
   222 points &
   222 points &
   202 points \\
\end{tabular}

\newpage

\section{The CTLMarkingComparison Examination}
\label{sec:exam:CTLMarkingComparison}
\index{Results!CTLMarkingComparison}

This examination deals with CTL properties dealing with marking comparison only.
We first show a summary on the handling of models by the participating tools.
Then, we present the computed outputs and the associated scores for this
examination prior to a summary of relevant executions.

\subsection{Handling of Models by Tools}
\index{Performances!CTLMarkingComparison}

\subsubsection{\acs{CSRepetitions-COL}}
No instance of this model could be computed for the \textbf{CTLMarkingComparison} examination.

\subsubsection{\acs{CSRepetitions-PT}}
The charts below respectively show how tools compete with this ``Known'' model (memory and CPU).

\index{Performances!CTLMarkingComparison!CSRepetitions (P/T)}
\begin{center}
   \includegraphics[width=7.2cm]{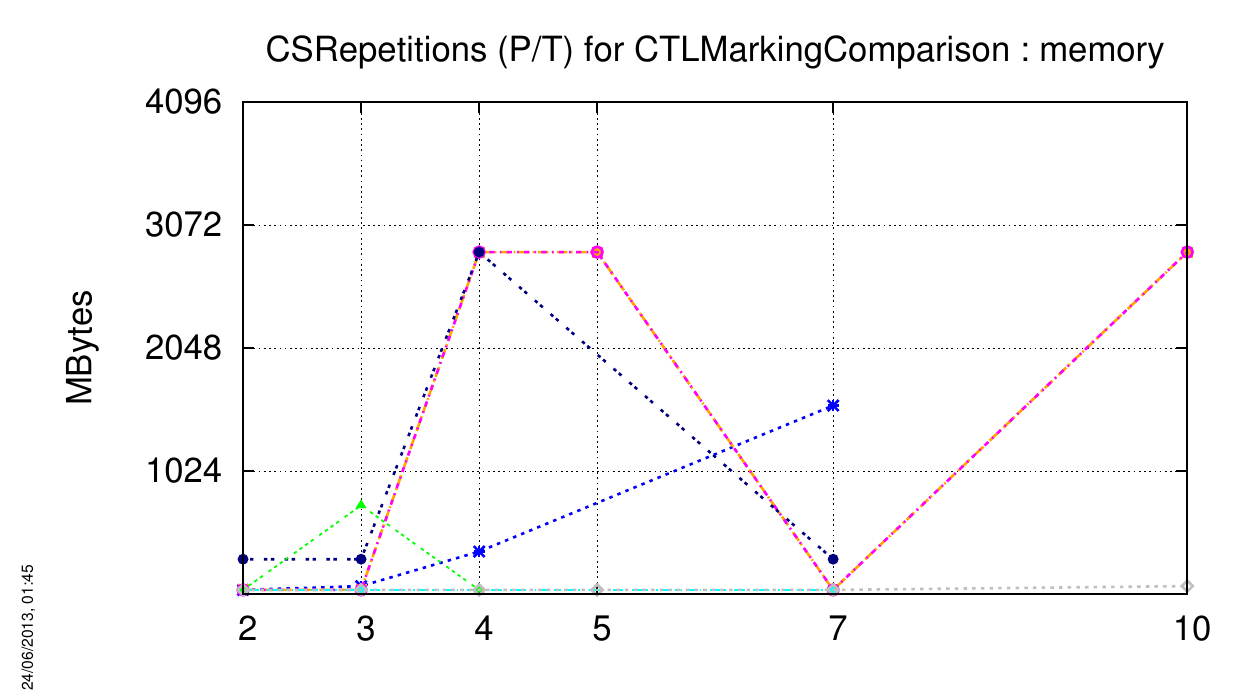}
   \includegraphics[width=7.2cm]{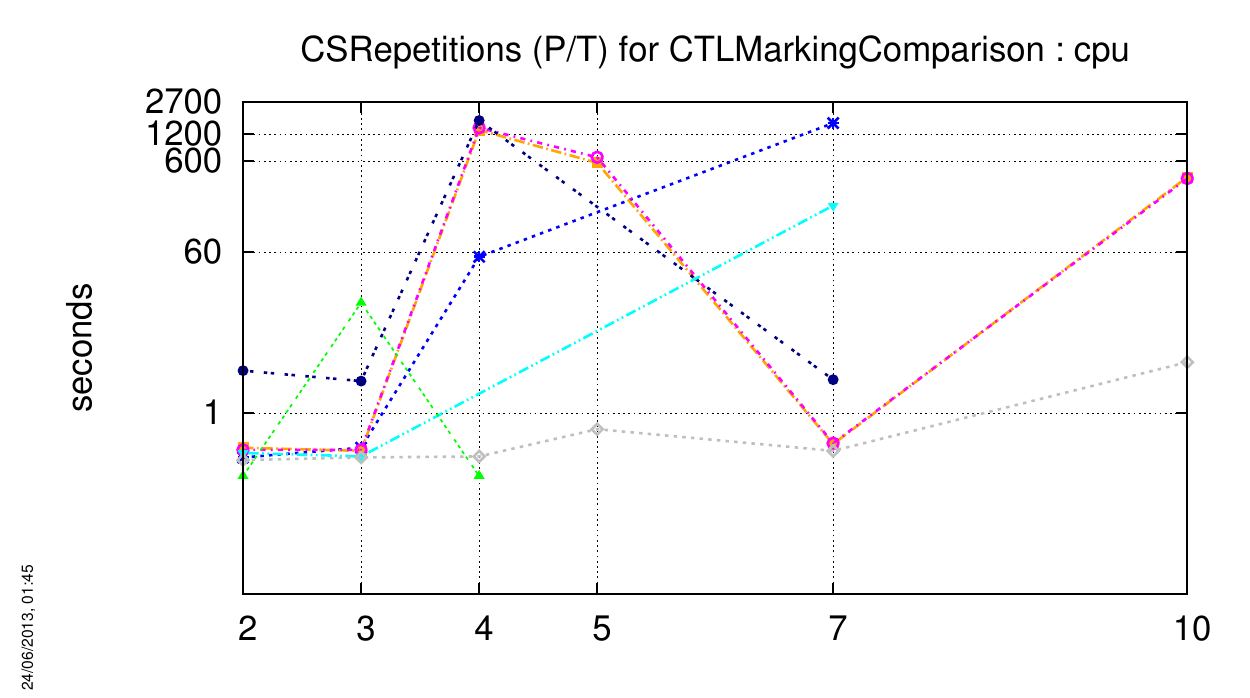}

   \includegraphics[height=1cm]{figures/tools-legend.pdf}
\end{center}

\subsubsection{\acs{Dekker-PT}}
No instance of this model could be computed for the \textbf{CTLMarkingComparison} examination.

\subsubsection{\acs{DotAndBoxes-COL}}
No instance of this model could be computed for the \textbf{CTLMarkingComparison} examination.

\subsubsection{\acs{DrinkVendingMachine-COL}}
No instance of this model could be computed for the \textbf{CTLMarkingComparison} examination.

\subsubsection{\acs{DrinkVendingMachine-PT}}
The charts below respectively show how tools compete with this ``Known'' model (memory and CPU).

\index{Performances!CTLMarkingComparison!DrinkVendingMachine (P/T)}
\begin{center}
   \includegraphics[width=7.2cm]{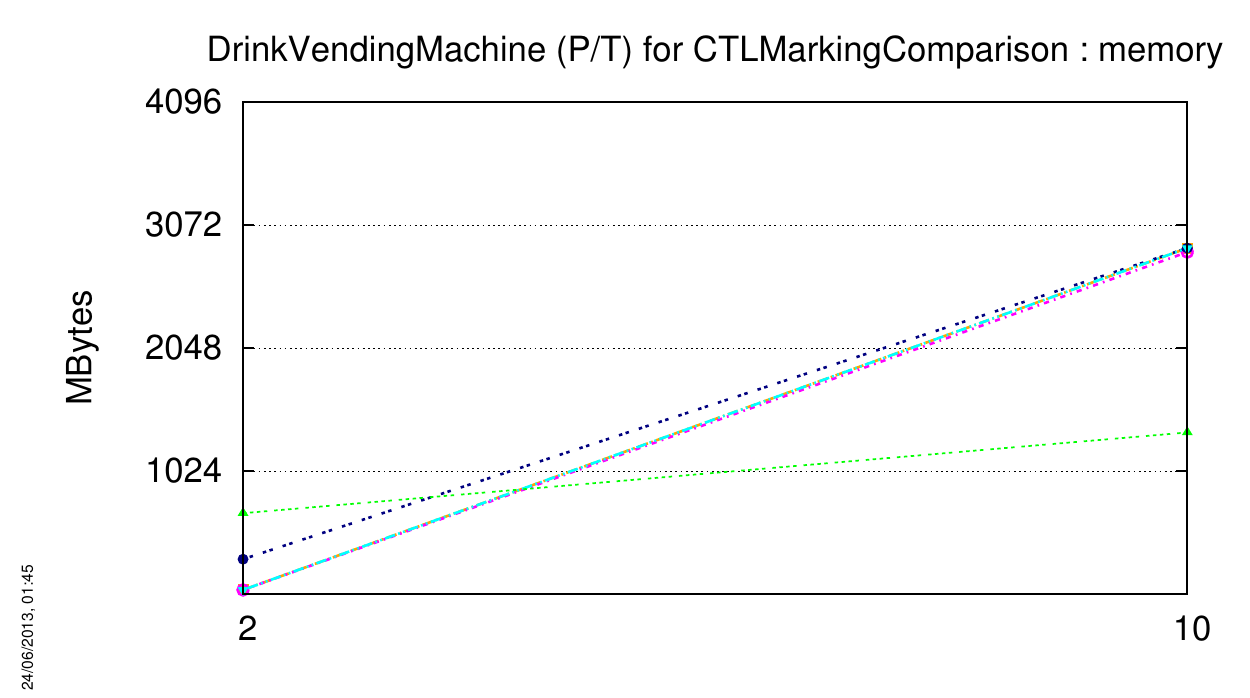}
   \includegraphics[width=7.2cm]{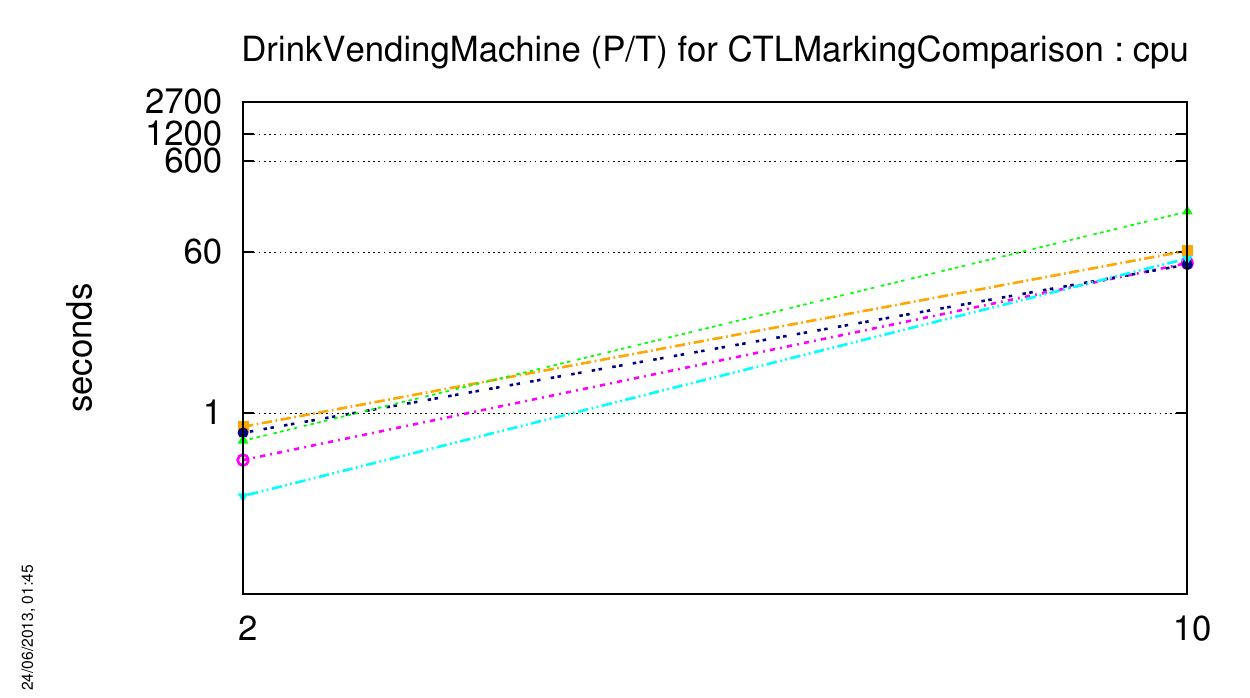}

   \includegraphics[height=1cm]{figures/tools-legend.pdf}
\end{center}

\subsubsection{\acs{Echo-PT}}
No instance of this model could be computed for the \textbf{CTLMarkingComparison} examination.

\subsubsection{\acs{Eratosthenes-PT}}
No instance of this model could be computed for the \textbf{CTLMarkingComparison} examination.

\subsubsection{\acs{FMS-PT}}
No instance of this model could be computed for the \textbf{CTLMarkingComparison} examination.

\subsubsection{\acs{GlobalRessAlloc-COL}}
No instance of this model could be computed for the \textbf{CTLMarkingComparison} examination.

\subsubsection{\acs{GlobalRessAlloc-PT}}
The charts below respectively show how tools compete with this ``Known'' model (memory and CPU).

\index{Performances!CTLMarkingComparison!GlobalRessAlloc (P/T)}
\begin{center}
   \includegraphics[width=7.2cm]{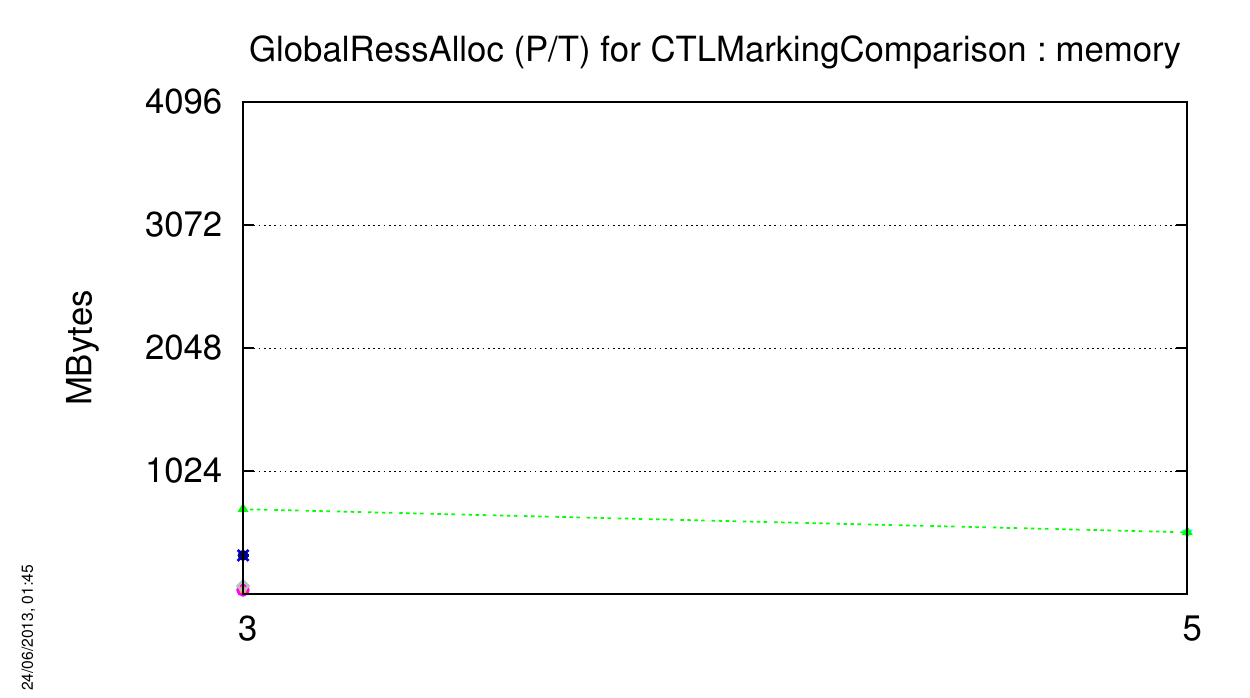}
   \includegraphics[width=7.2cm]{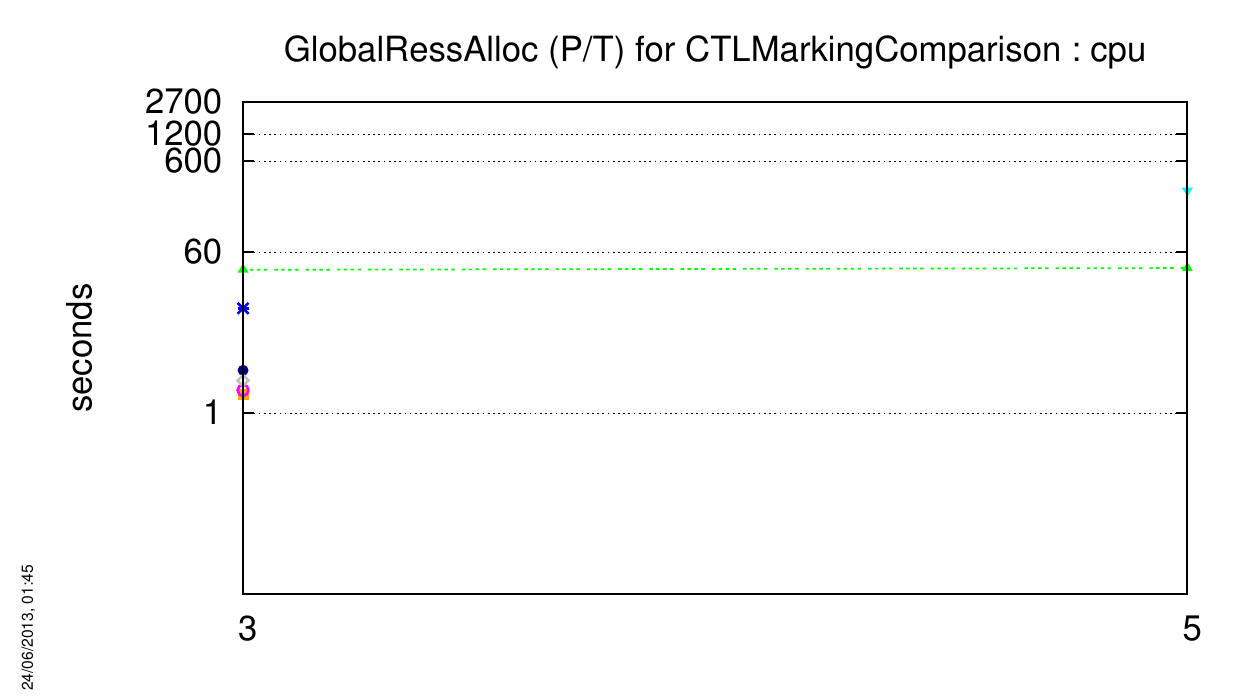}

   \includegraphics[height=1cm]{figures/tools-legend.pdf}
\end{center}

\subsubsection{\acs{Kanban-PT}}
No instance of this model could be computed for the \textbf{CTLMarkingComparison} examination.

\subsubsection{\acs{LamportFastMutEx-COL}}
No instance of this model could be computed for the \textbf{CTLMarkingComparison} examination.

\subsubsection{\acs{LamportFastMutEx-PT}}
The charts below respectively show how tools compete with this ``Known'' model (memory and CPU).

\index{Performances!CTLMarkingComparison!LamportFastMutEx (P/T)}
\begin{center}
   \includegraphics[width=7.2cm]{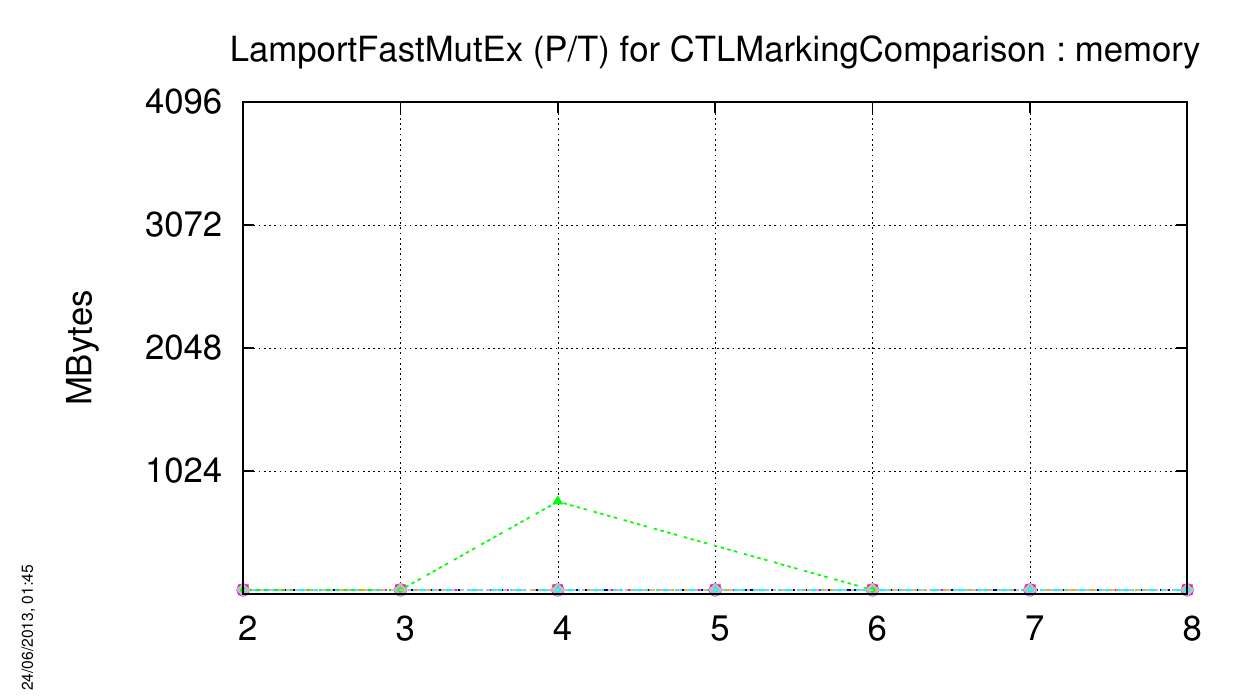}
   \includegraphics[width=7.2cm]{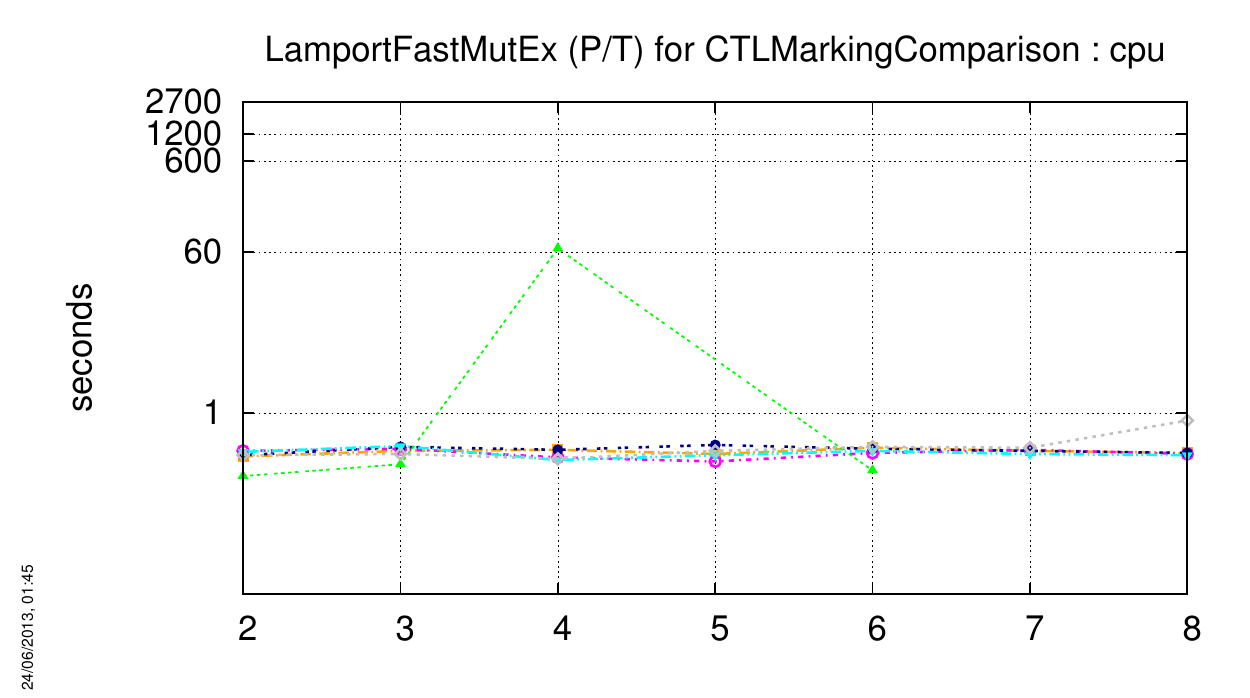}

   \includegraphics[height=1cm]{figures/tools-legend.pdf}
\end{center}

\subsubsection{\acs{MAPK-PT}}
No instance of this model could be computed for the \textbf{CTLMarkingComparison} examination.

\subsubsection{\acs{NeoElection-COL}}
No instance of this model could be computed for the \textbf{CTLMarkingComparison} examination.

\subsubsection{\acs{NeoElection-PT}}
The charts below respectively show how tools compete with this ``Known'' model (memory and CPU).

\index{Performances!CTLMarkingComparison!NeoElection (P/T)}
\begin{center}
   \includegraphics[width=7.2cm]{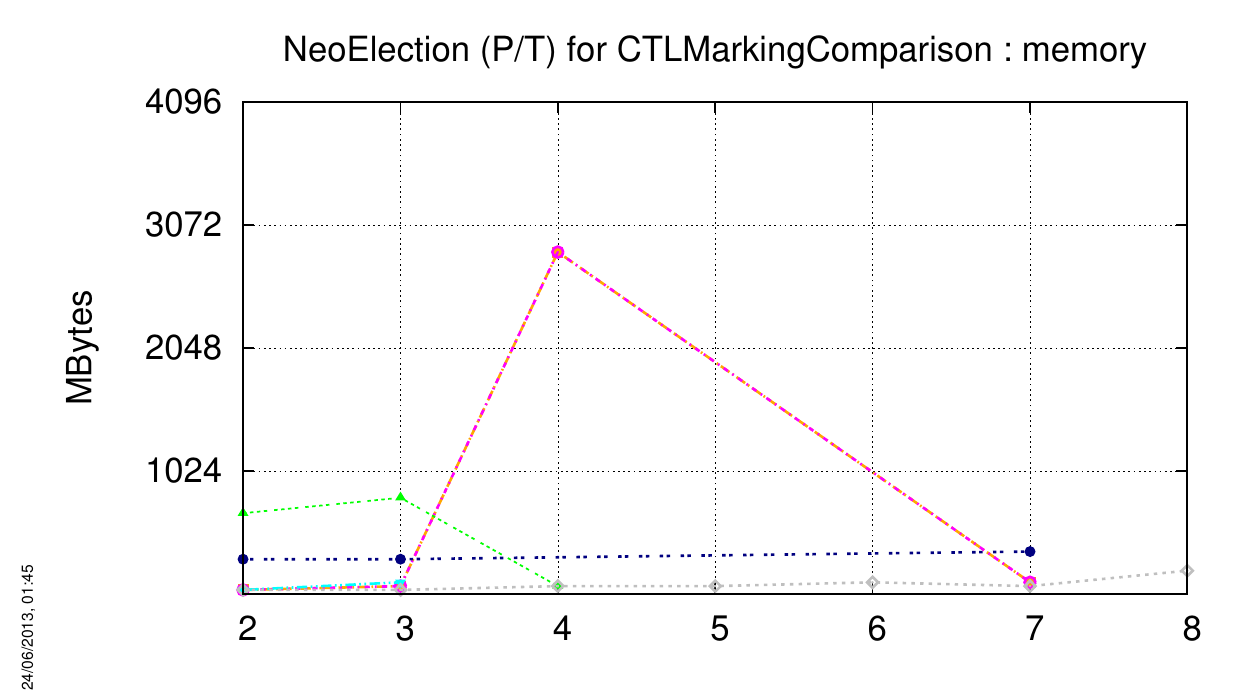}
   \includegraphics[width=7.2cm]{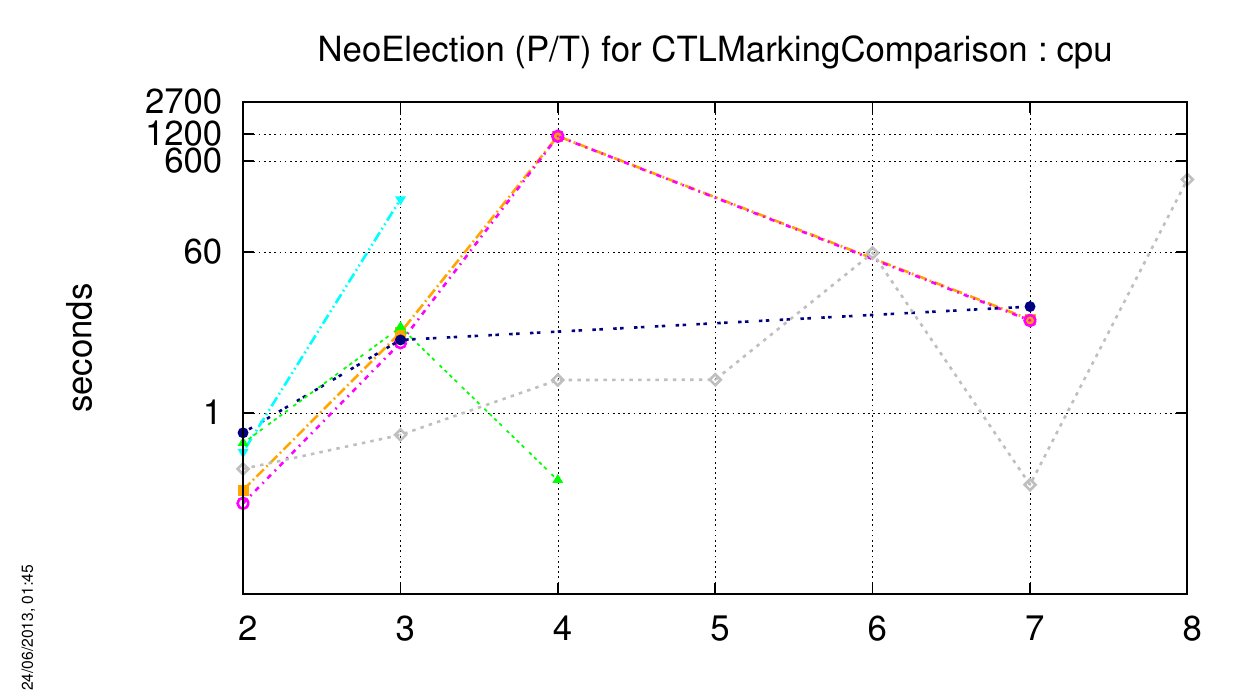}

   \includegraphics[height=1cm]{figures/tools-legend.pdf}
\end{center}

\subsubsection{\acs{PermAdmissibility-COL}}
No instance of this model could be computed for the \textbf{CTLMarkingComparison} examination.

\subsubsection{\acs{PermAdmissibility-PT}}
The charts below respectively show how tools compete with this ``Known'' model (memory and CPU).

\index{Performances!CTLMarkingComparison!PermAdmissibility (P/T)}
\begin{center}
   \includegraphics[width=7.2cm]{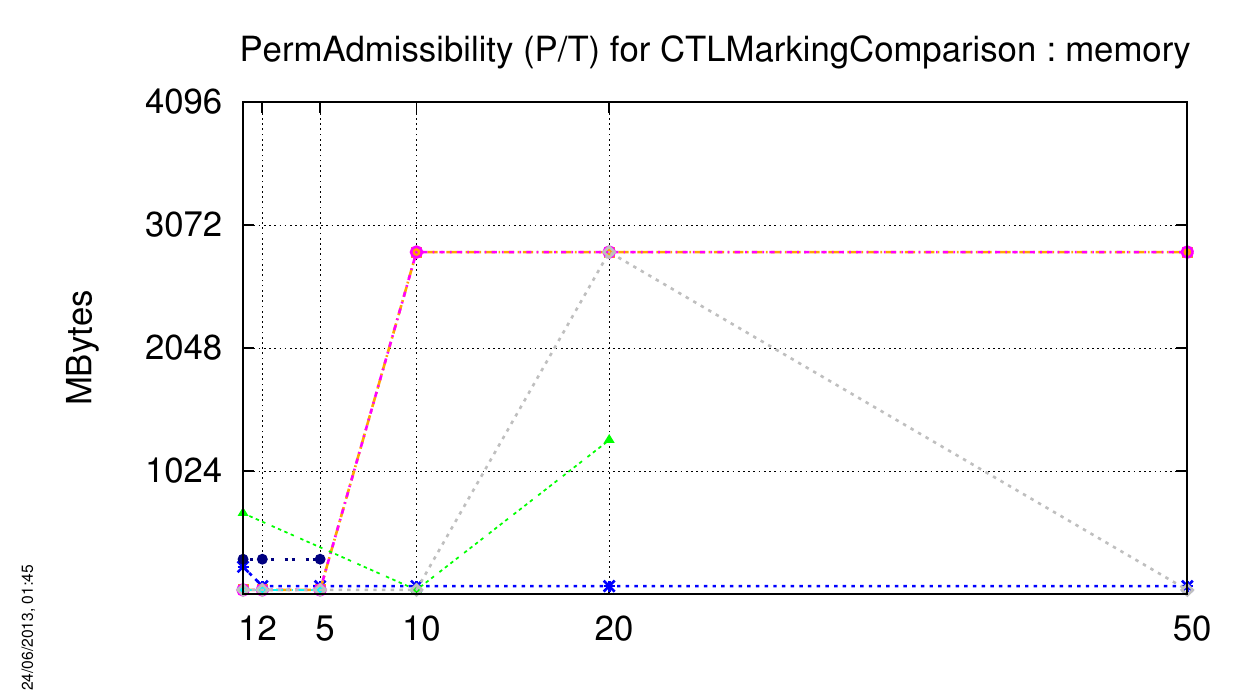}
   \includegraphics[width=7.2cm]{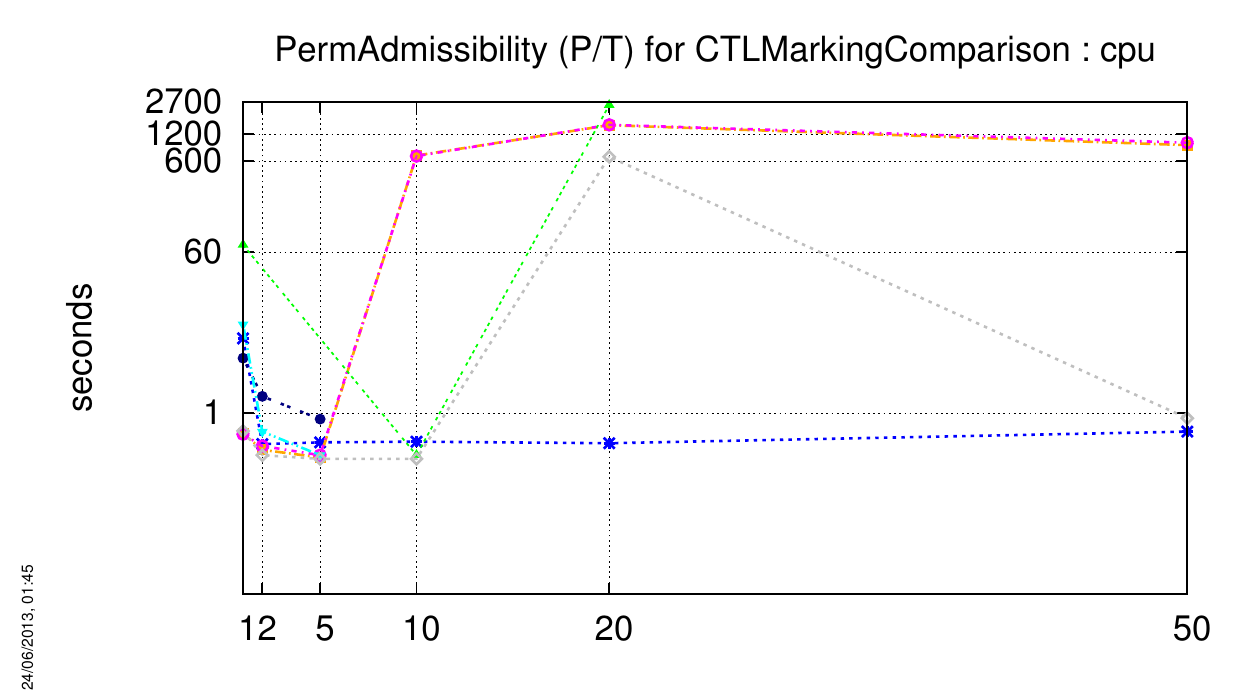}

   \includegraphics[height=1cm]{figures/tools-legend.pdf}
\end{center}

\subsubsection{\acs{Peterson-COL}}
No instance of this model could be computed for the \textbf{CTLMarkingComparison} examination.

\subsubsection{\acs{Peterson-PT}}
The charts below respectively show how tools compete with this ``Known'' model (memory and CPU).

\index{Performances!CTLMarkingComparison!Peterson (P/T)}
\begin{center}
   \includegraphics[width=7.2cm]{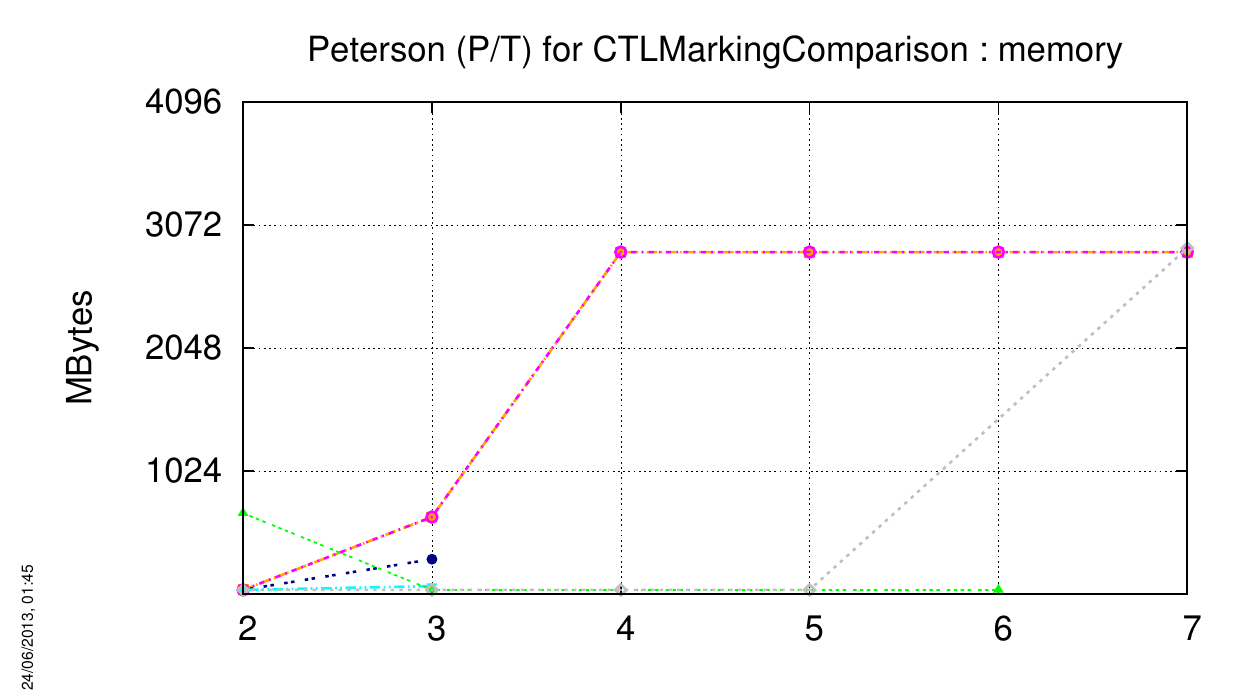}
   \includegraphics[width=7.2cm]{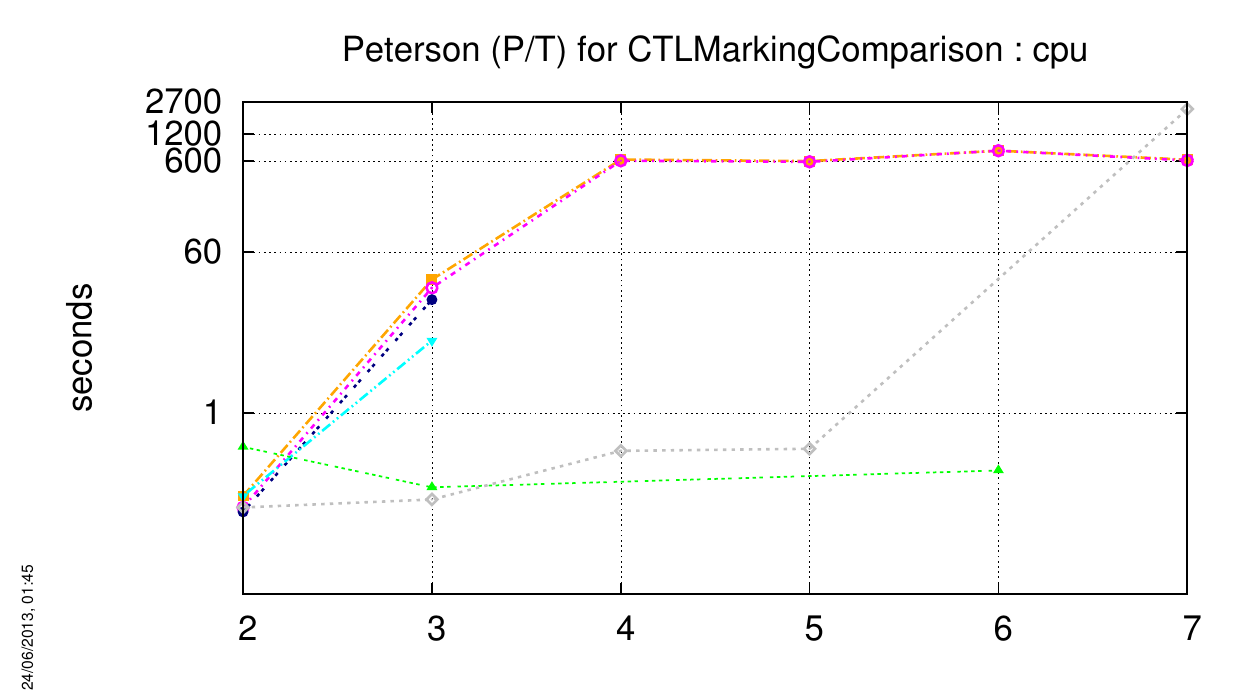}

   \includegraphics[height=1cm]{figures/tools-legend.pdf}
\end{center}

\subsubsection{\acs{Philosophers-COL}}
No instance of this model could be computed for the \textbf{CTLMarkingComparison} examination.

\subsubsection{\acs{Philosophers-PT}}
The charts below respectively show how tools compete with this ``Known'' model (memory and CPU).

\index{Performances!CTLMarkingComparison!Philosophers (P/T)}
\begin{center}
   \includegraphics[width=7.2cm]{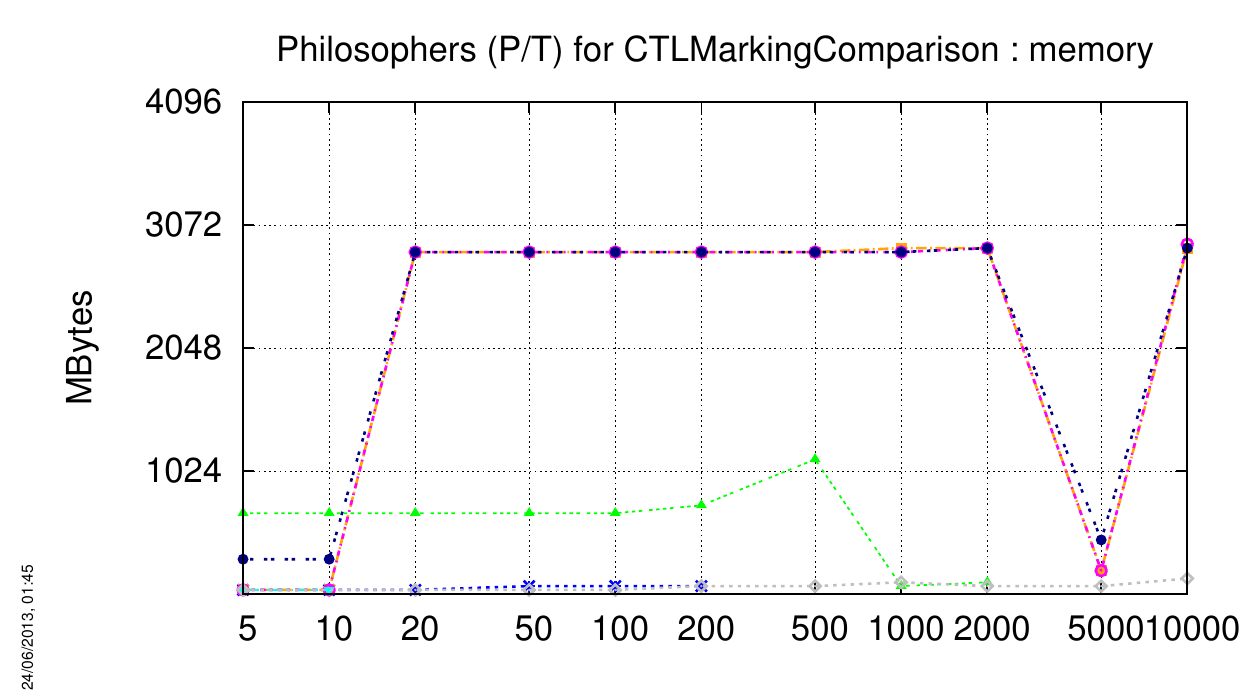}
   \includegraphics[width=7.2cm]{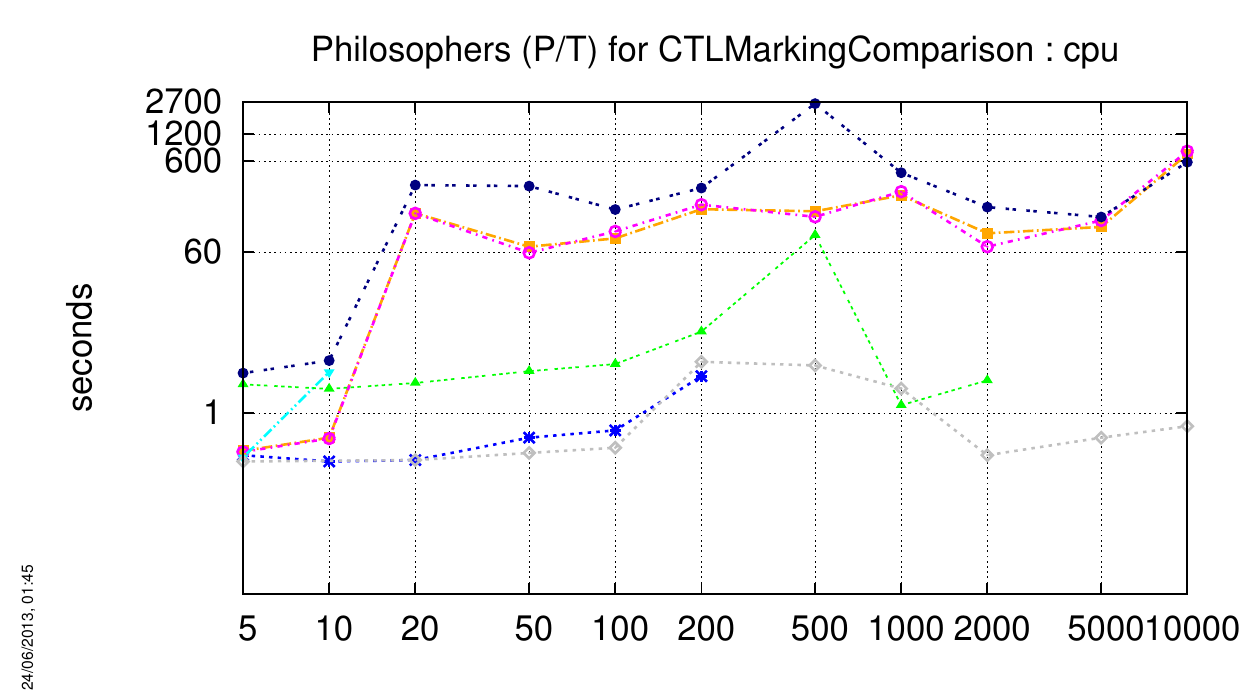}

   \includegraphics[height=1cm]{figures/tools-legend.pdf}
\end{center}

\subsubsection{\acs{PhilosophersDyn-COL}}
No instance of this model could be computed for the \textbf{CTLMarkingComparison} examination.

\subsubsection{\acs{PhilosophersDyn-PT}}
The charts below respectively show how tools compete with this ``Known'' model (memory and CPU).

\index{Performances!CTLMarkingComparison!PhilosophersDyn (P/T)}
\begin{center}
   \includegraphics[width=7.2cm]{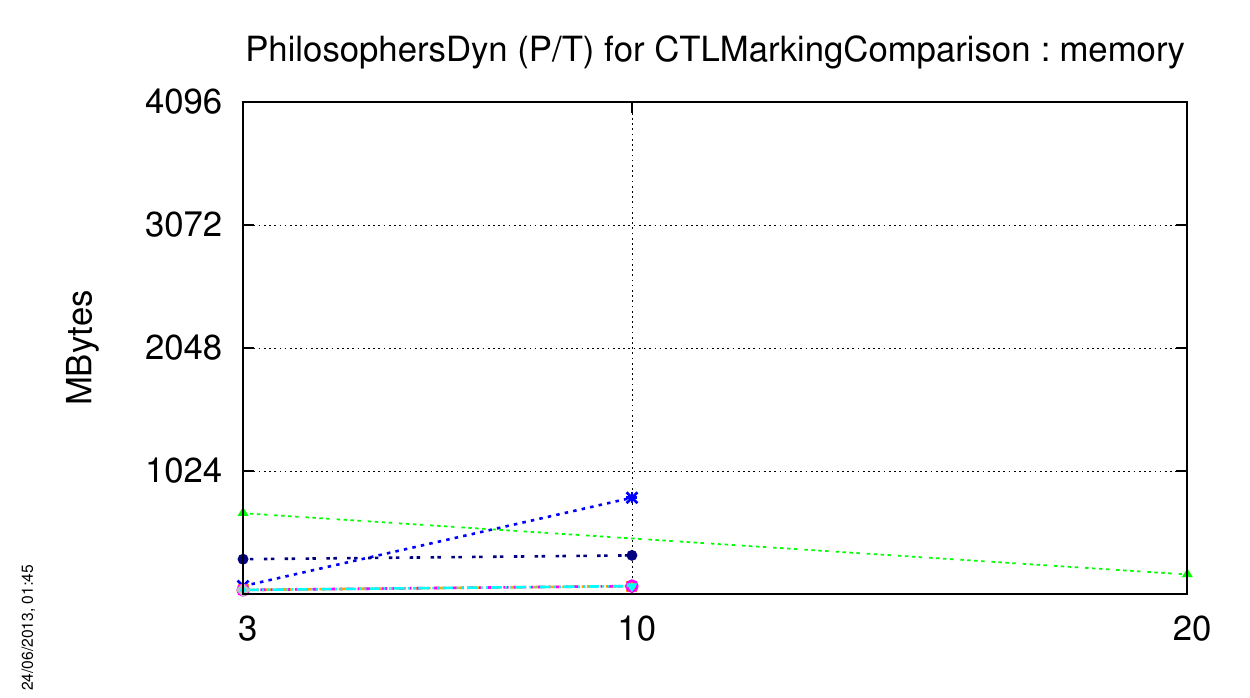}
   \includegraphics[width=7.2cm]{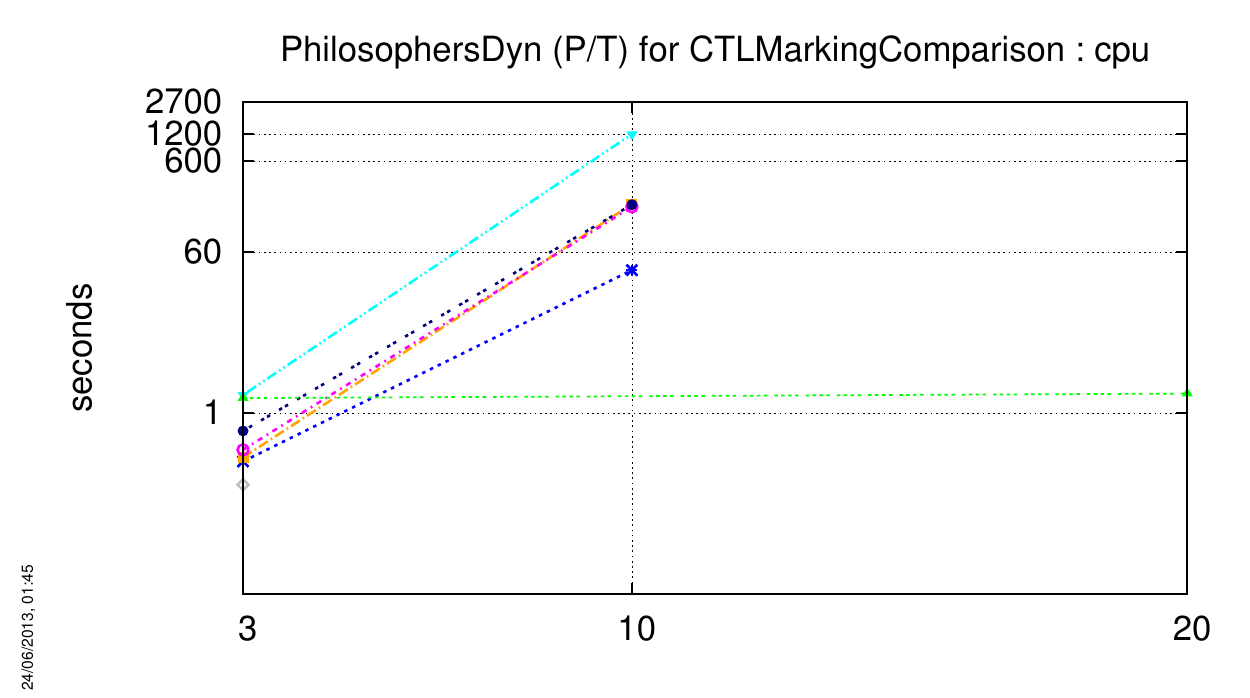}

   \includegraphics[height=1cm]{figures/tools-legend.pdf}
\end{center}

\subsubsection{\acs{Planning-PT}}
No instance of this model could be computed for the \textbf{CTLMarkingComparison} examination.

\subsubsection{\acs{Railroad-PT}}
No instance of this model could be computed for the \textbf{CTLMarkingComparison} examination.

\subsubsection{\acs{RessAllocation-PT}}
No instance of this model could be computed for the \textbf{CTLMarkingComparison} examination.

\subsubsection{\acs{Ring-PT}}
No instance of this model could be computed for the \textbf{CTLMarkingComparison} examination.

\subsubsection{\acs{RwMutex-PT}}
No instance of this model could be computed for the \textbf{CTLMarkingComparison} examination.

\subsubsection{\acs{SharedMemory-COL}}
No instance of this model could be computed for the \textbf{CTLMarkingComparison} examination.

\subsubsection{\acs{SharedMemory-PT}}
The charts below respectively show how tools compete with this ``Known'' model (memory and CPU).

\index{Performances!CTLMarkingComparison!SharedMemory (P/T)}
\begin{center}
   \includegraphics[width=7.2cm]{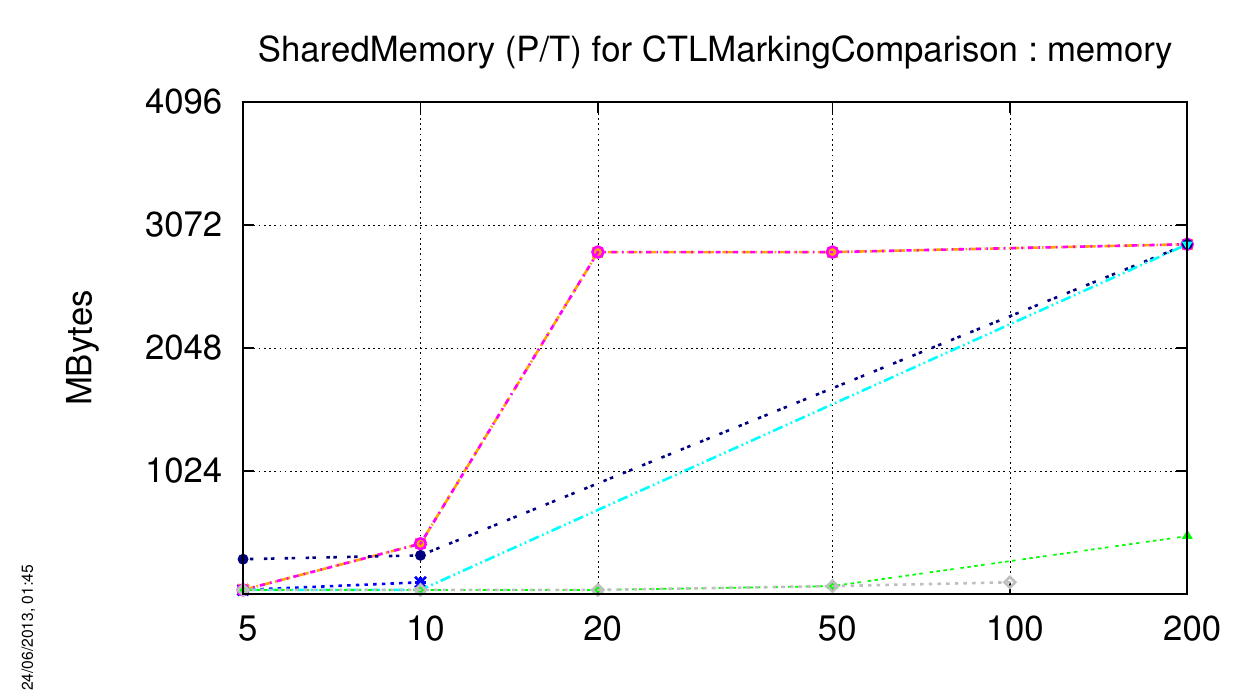}
   \includegraphics[width=7.2cm]{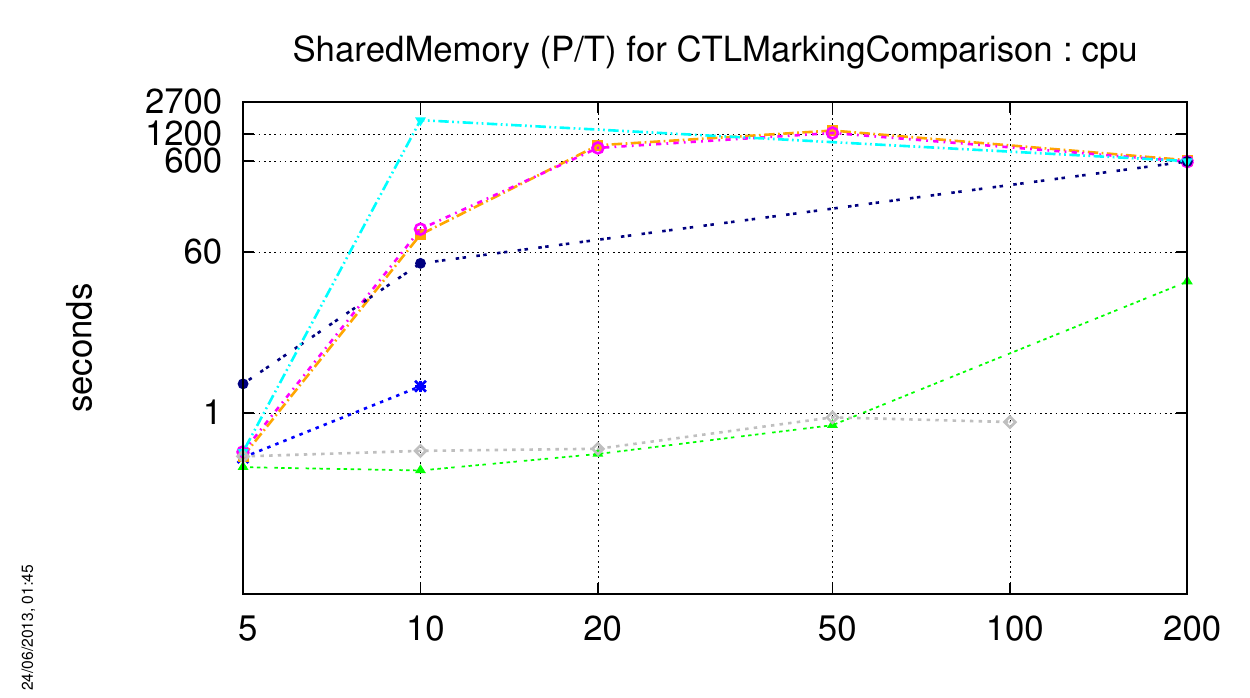}

   \includegraphics[height=1cm]{figures/tools-legend.pdf}
\end{center}

\subsubsection{\acs{SimpleLoadBal-COL}}
No instance of this model could be computed for the \textbf{CTLMarkingComparison} examination.

\subsubsection{\acs{SimpleLoadBal-PT}}
The charts below respectively show how tools compete with this ``Known'' model (memory and CPU).

\index{Performances!CTLMarkingComparison!SimpleLoadBal (P/T)}
\begin{center}
   \includegraphics[width=7.2cm]{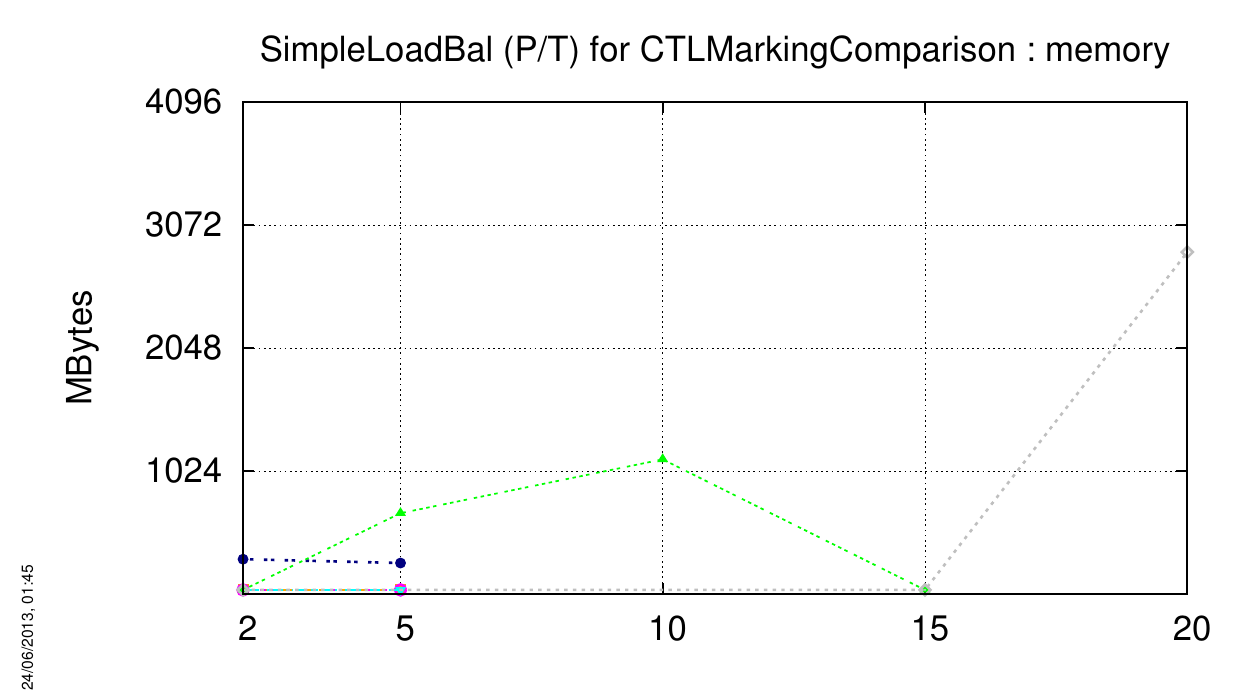}
   \includegraphics[width=7.2cm]{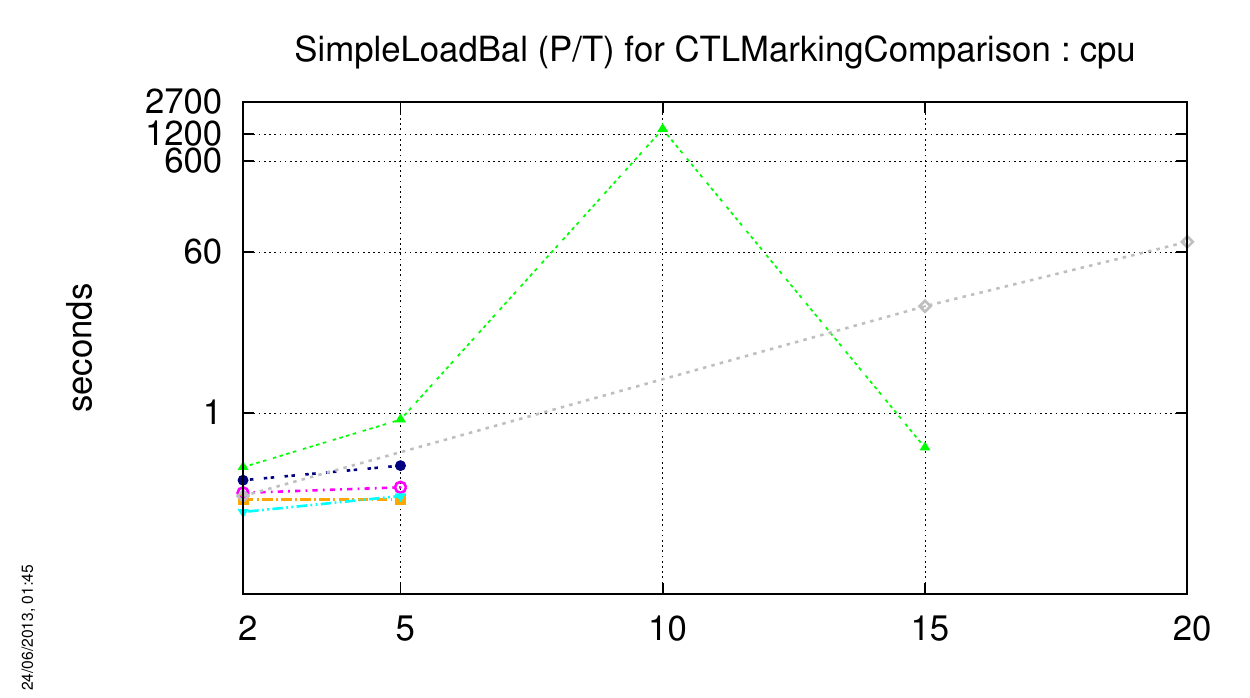}

   \includegraphics[height=1cm]{figures/tools-legend.pdf}
\end{center}

\subsubsection{\acs{TokenRing-COL}}
No instance of this model could be computed for the \textbf{CTLMarkingComparison} examination.

\subsubsection{\acs{TokenRing-PT}}
The charts below respectively show how tools compete with this ``Known'' model (memory and CPU).

\index{Performances!CTLMarkingComparison!TokenRing (P/T)}
\begin{center}
   \includegraphics[width=7.2cm]{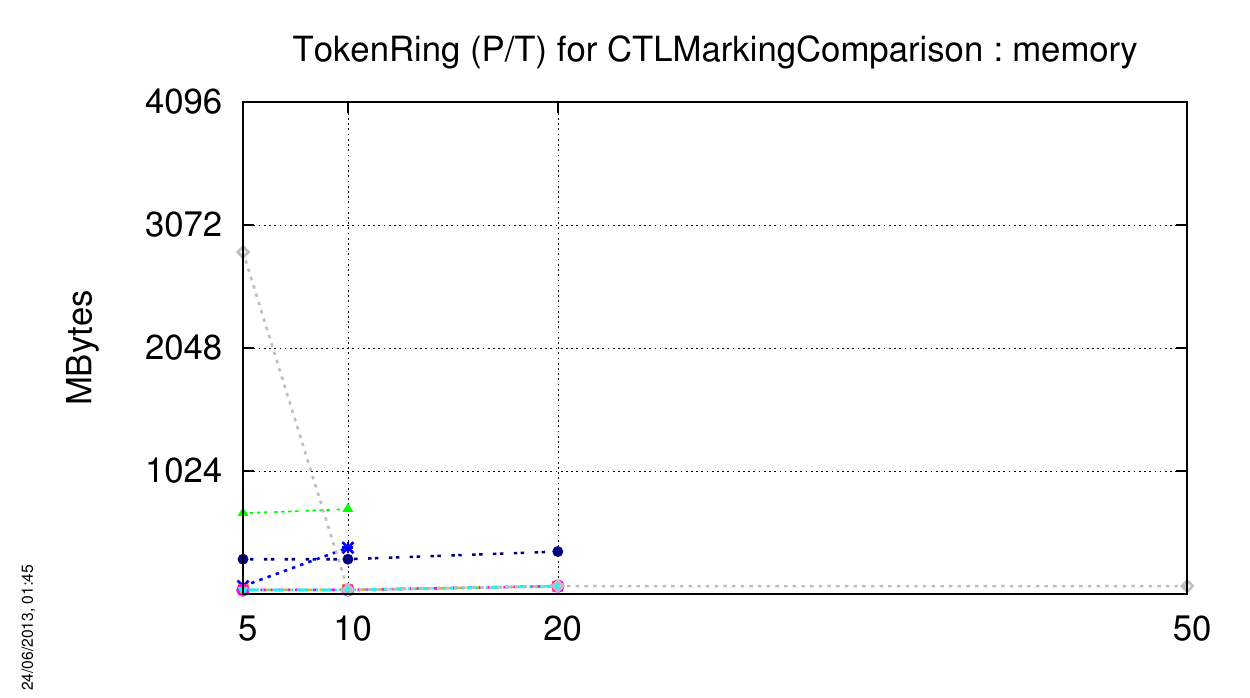}
   \includegraphics[width=7.2cm]{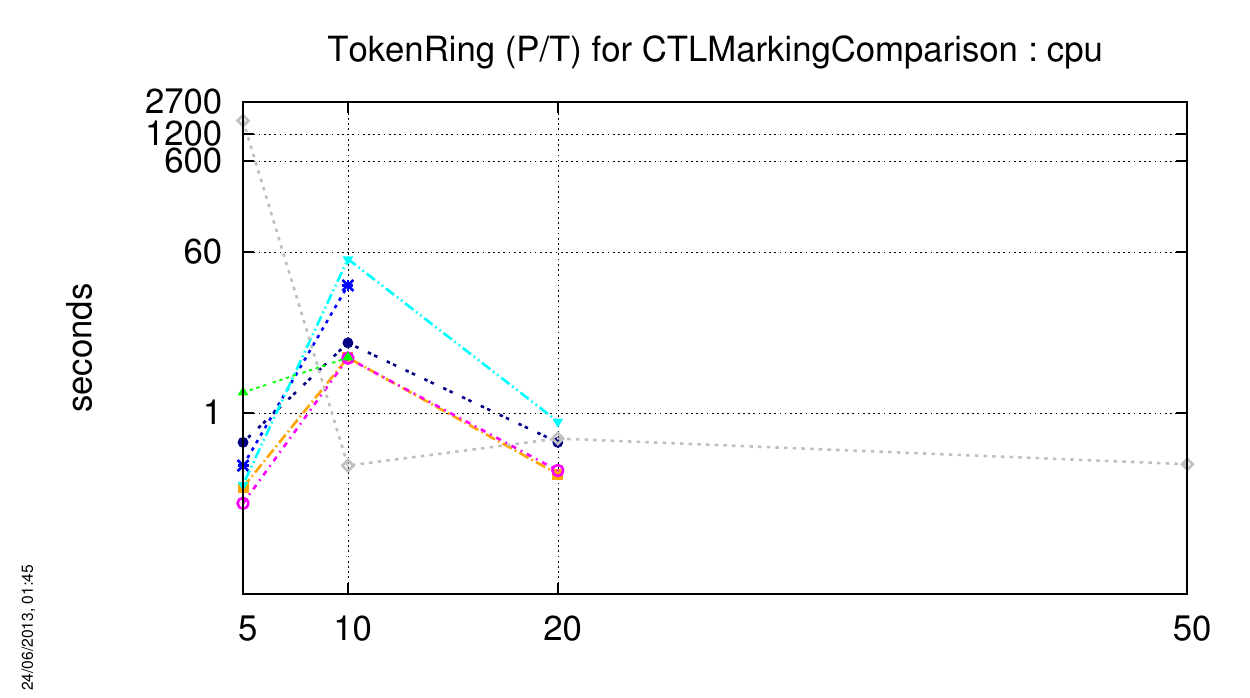}

   \includegraphics[height=1cm]{figures/tools-legend.pdf}
\end{center}

\subsubsection{\acs{HouseConstruction-PT}}
No instance of this model could be computed for the \textbf{CTLMarkingComparison} examination.

\subsubsection{\acs{IBMB2S565S3960-PT}}
No instance of this model could be computed for the \textbf{CTLMarkingComparison} examination.

\subsubsection{\acs{QuasiCertifProtocol-COL}}
No instance of this model could be computed for the \textbf{CTLMarkingComparison} examination.

\subsubsection{\acs{QuasiCertifProtocol-PT}}
The charts below respectively show how tools compete with this ``Suprise'' model (memory and CPU).

\index{Performances!CTLMarkingComparison!QuasiCertifProtocol (P/T)}
\begin{center}
   \includegraphics[width=7.2cm]{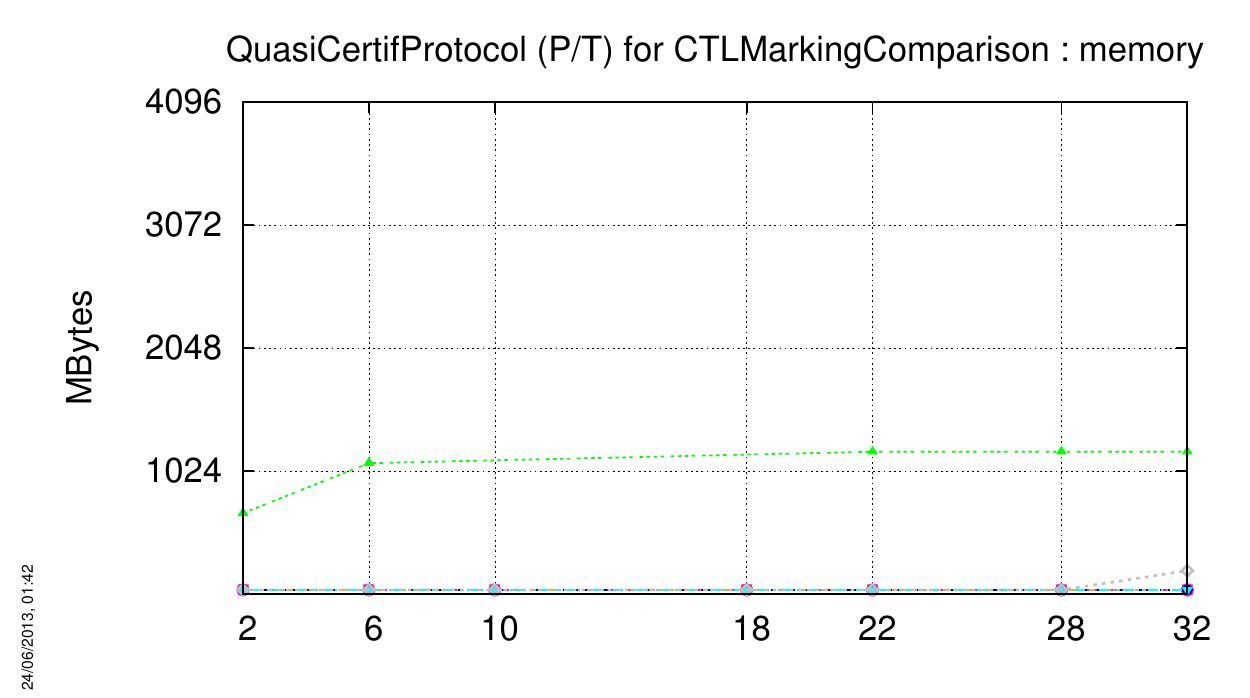}
   \includegraphics[width=7.2cm]{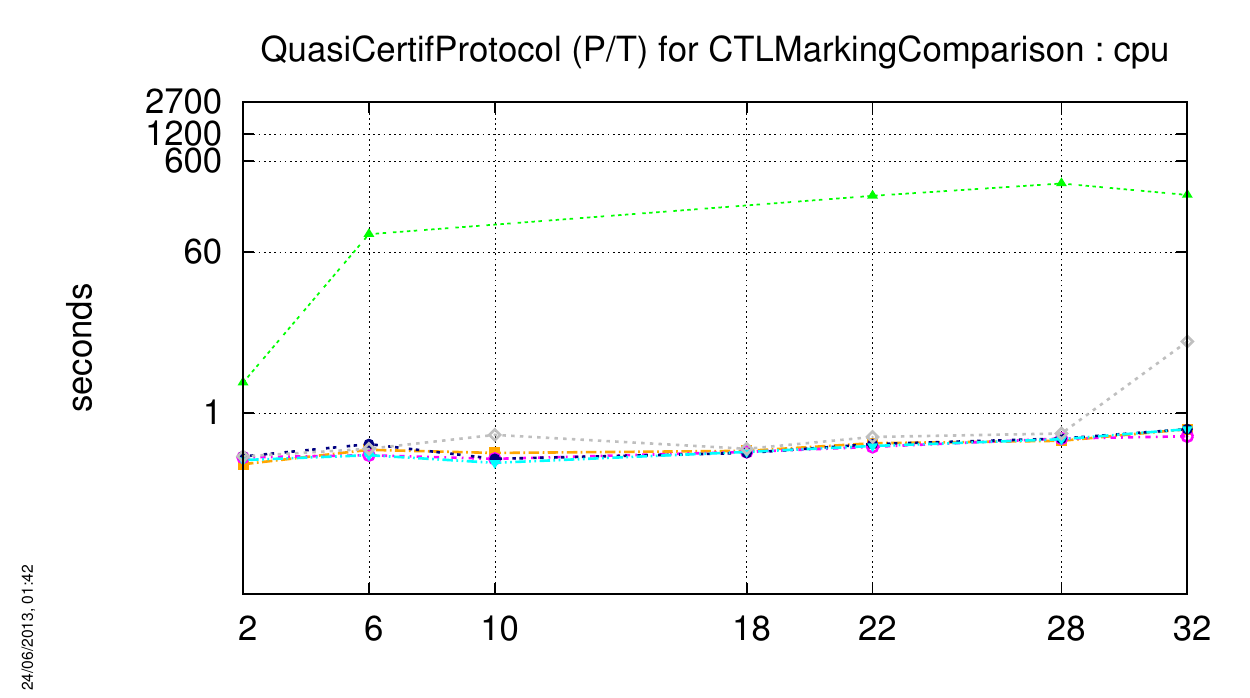}

   \includegraphics[height=1cm]{figures/tools-legend.pdf}
\end{center}

\subsubsection{\acs{Vasy2003-PT}}
No instance of this model could be computed for the \textbf{CTLMarkingComparison} examination.

\subsection{Outputs for the CTLMarkingComparison Examination}
\index{Outputs!CTLMarkingComparison}

Please find enclosed the brute results for this examination (``Known'' and ``Surprise'' models).
We display only the score of tools that provide a results for at least one instance of one model.
The legend for the values is provided below:
\begin{itemize}
   \item\textbf{nc}: the tool does not compete this examination for this model/instance,
   \item\textbf{cc}: the tool cannot compute this examination for this model/instance,
   \item\textbf{to}: the tool cannot compute this examination for this model/instance within the maximum allowed time,
   \item\textbf{mp}: the tool encountered a memory problem (stack overflow or memory full),
   \item\textbf{nf}: there is no formula available for this type of examination (typically, this concerns P/T nets where
       comparing marking cardinality has no signification when there is no equivalent colored net).
\end{itemize}

\textbf{Note on the display of results for formulas:} each formula is considered as a flag (F if false, T if true, - or ?
when the value cannot be determined). These values are concatenated in the order they appear (we assume it is the order of formulas as they were provided).

\subsubsection{``Known'' Models}

\input{result_known_CTLMarkingComparison.tex}

\subsubsection{``Surprise'' Models}

\input{result_surprise_CTLMarkingComparison.tex}

\subsection{Score for the CTLMarkingComparison Examination}
\index{Scores!CTLMarkingComparison}

Please find enclosed the scores for this examination (``Known'' and ``Surprise'' models).
We display only the score of tools that provide a results for at least one instance of one model.
The total is first listed in the table below followed by a detail, for each proposed model.
Meaning of the line labels are:
\begin{itemize}
\item\textbf{1st instance}: the tool gets a bonus for having processed the first instance of this model (+1 point),
\item\textbf{instances}: the tool gets 1 point per instances treated 
(for that, we assume that at least one formula has been successfully computed),
\item\textbf{max reached}: the tool could process all the instances for the model (+2 points),
\item\textbf{best}: the tool is among the ones that processed a maximum of instances within the time and memory confinement (+2 points).
\end{itemize}

\subsubsection{``Known'' Models}

\input{score_known_CTLMarkingComparison.tex}

\subsubsection{``Surprise'' Models}

\input{score_surprise_CTLMarkingComparison.tex}

\subsection{Trophies for this Examination}
\index{Trophies!CTLMarkingComparison}

Trophies are divided in three categories: ``Known'' models,
``Surprise'' models, and the global trophies (formula is then
$score_{global} = score_{known} + 2 \times score_{surprise}$).

\subsubsection{For ``Known'' Models} \ \\

\begin{tabular}{c|c|c}
      1 & 2 & 2 \\
   \includegraphics[width=2cm]{figures/gold.jpg} &
   \includegraphics[width=2cm]{figures/silver.jpg} &
   \includegraphics[width=2cm]{figures/silver.jpg} \\
   \acs{sara} &
   \acs{lola} &
   \acs{lola-optimistic} \\
   87 points &
   71 points &
   71 points \\
\end{tabular}

\subsubsection{For ``Surprise'' Models}\  \\

\begin{tabular}{c|c}
      1 & 2 \\
   \includegraphics[width=2cm]{figures/gold.jpg} &
   \includegraphics[width=2cm]{figures/silver.jpg} \\
   \acs{sara} &
   \acs{marcie} \\
   12 points &
   3 points \\
\end{tabular}

\subsubsection{Global} \ \\

\begin{tabular}{c|c|c}
      1 & 2 & 2 \\
   \includegraphics[width=2cm]{figures/gold.jpg} &
   \includegraphics[width=2cm]{figures/silver.jpg} &
   \includegraphics[width=2cm]{figures/silver.jpg} \\
   \acs{sara} &
   \acs{lola} &
   \acs{lola-optimistic} \\
   111 points &
   71 points &
   71 points \\
\end{tabular}

\newpage

\section{The CTLPlaceComparison Examination}
\label{sec:exam:CTLPlaceComparison}
\index{Results!CTLPlaceComparison}

This examination deals with CTL properties dealing with the comparison of places marking only.
We first show a summary on the handling of models by the participating tools.
Then, we present the computed outputs and the associated scores for this
examination prior to a summary of relevant executions.

\subsection{Handling of Models by Tools}
\index{Performances!CTLPlaceComparison}

\subsubsection{\acs{CSRepetitions-COL}}
No instance of this model could be computed for the \textbf{CTLPlaceComparison} examination.

\subsubsection{\acs{CSRepetitions-PT}}
The charts below respectively show how tools compete with this ``Known'' model (memory and CPU).

\index{Performances!CTLPlaceComparison!CSRepetitions (P/T)}
\begin{center}
   \includegraphics[width=7.2cm]{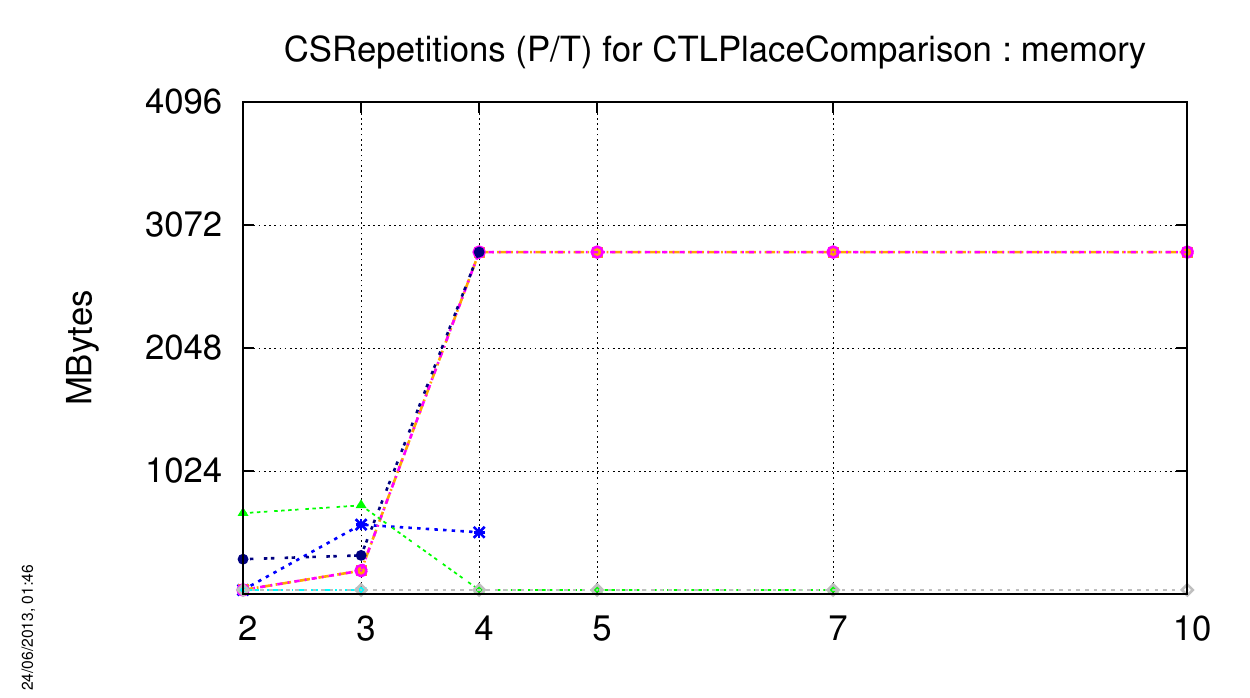}
   \includegraphics[width=7.2cm]{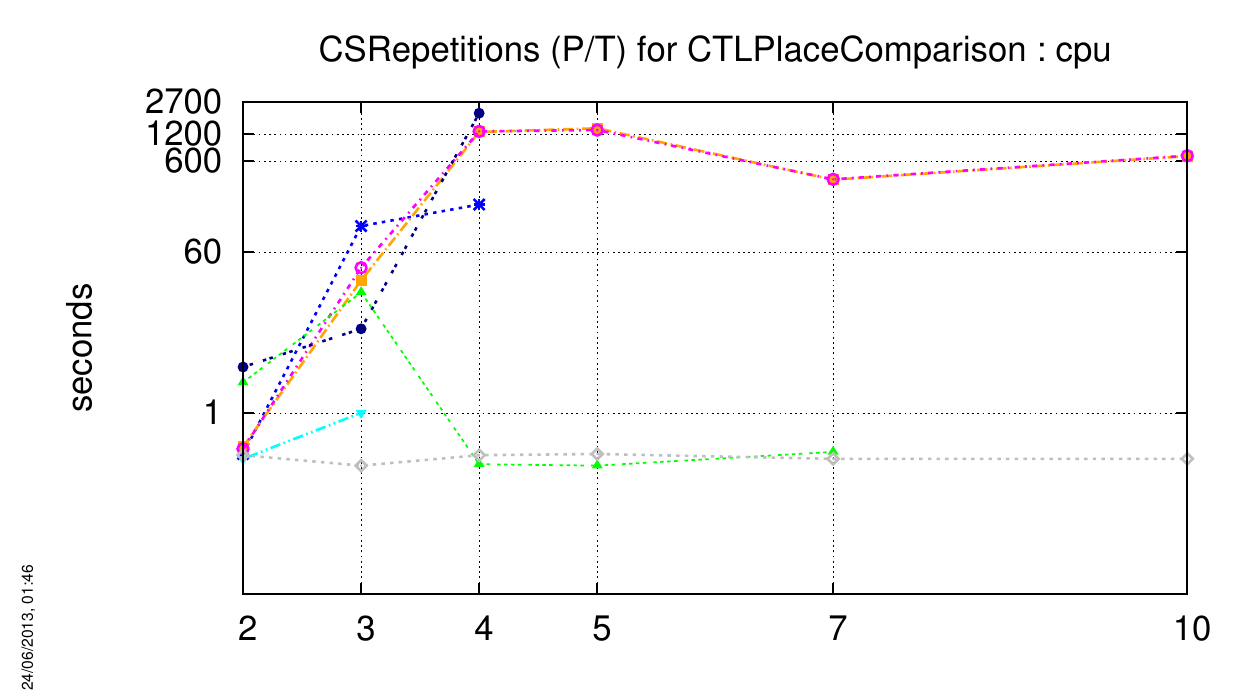}

   \includegraphics[height=1cm]{figures/tools-legend.pdf}
\end{center}

\subsubsection{\acs{Dekker-PT}}
The charts below respectively show how tools compete with this ``Known'' model (memory and CPU).

\index{Performances!CTLPlaceComparison!Dekker (P/T)}
\begin{center}
   \includegraphics[width=7.2cm]{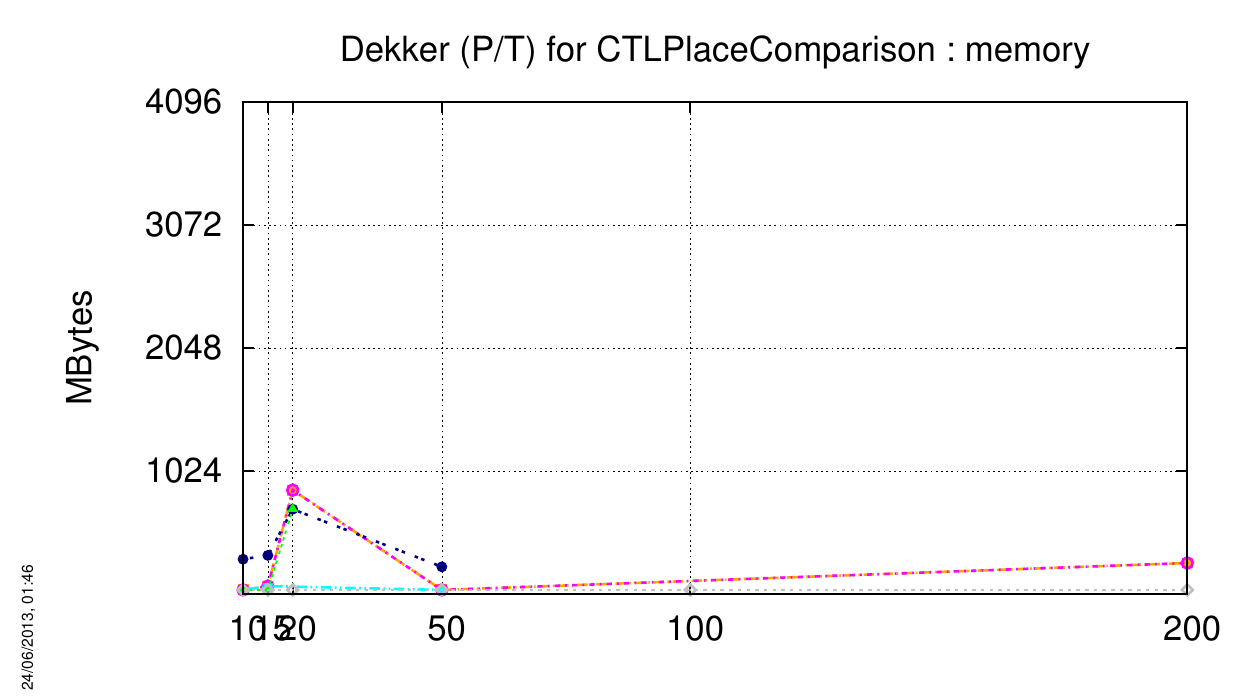}
   \includegraphics[width=7.2cm]{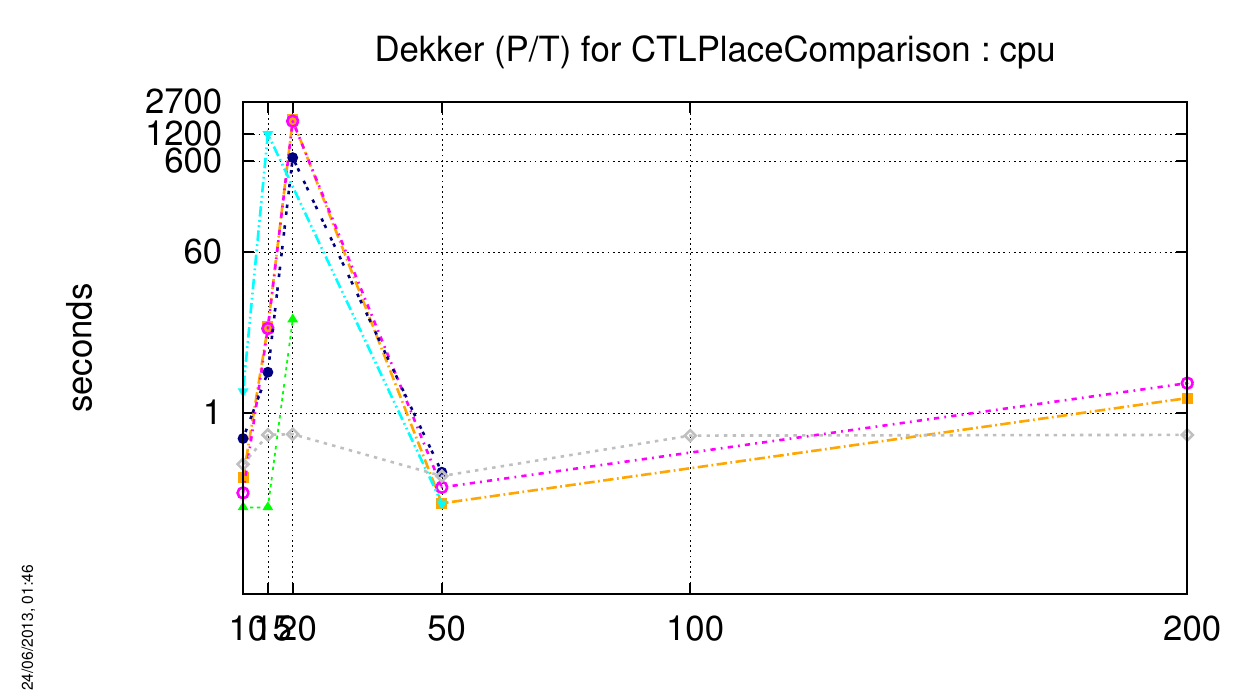}

   \includegraphics[height=1cm]{figures/tools-legend.pdf}
\end{center}

\subsubsection{\acs{DotAndBoxes-COL}}
No instance of this model could be computed for the \textbf{CTLPlaceComparison} examination.

\subsubsection{\acs{DrinkVendingMachine-COL}}
No instance of this model could be computed for the \textbf{CTLPlaceComparison} examination.

\subsubsection{\acs{DrinkVendingMachine-PT}}
The charts below respectively show how tools compete with this ``Known'' model (memory and CPU).

\index{Performances!CTLPlaceComparison!DrinkVendingMachine (P/T)}
\begin{center}
   \includegraphics[width=7.2cm]{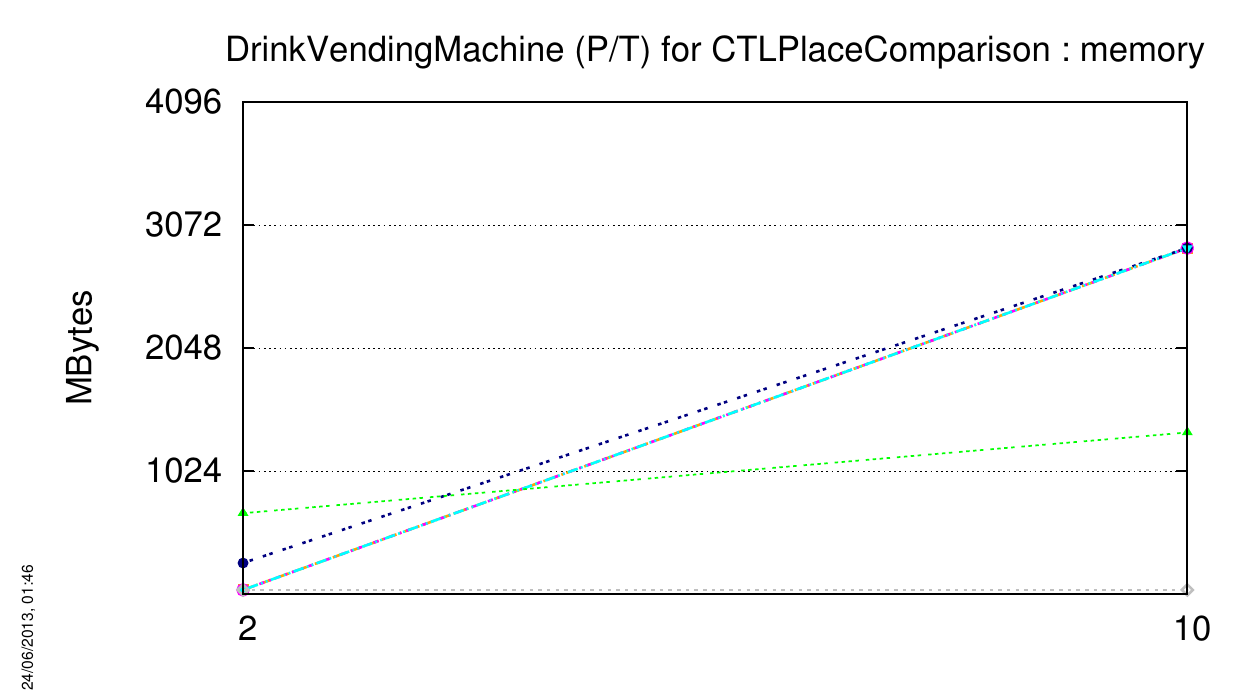}
   \includegraphics[width=7.2cm]{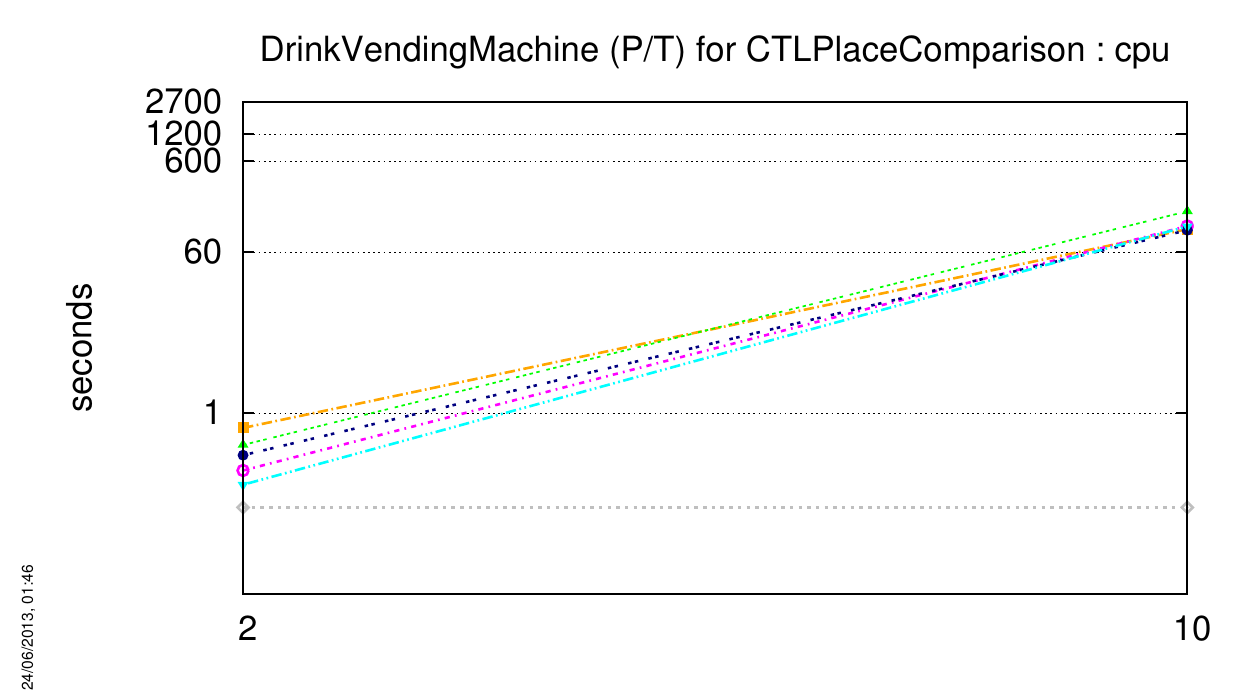}

   \includegraphics[height=1cm]{figures/tools-legend.pdf}
\end{center}

\subsubsection{\acs{Echo-PT}}
The charts below respectively show how tools compete with this ``Known'' model (memory and CPU).

\index{Performances!CTLPlaceComparison!Echo (P/T)}
\begin{center}
   \includegraphics[width=7.2cm]{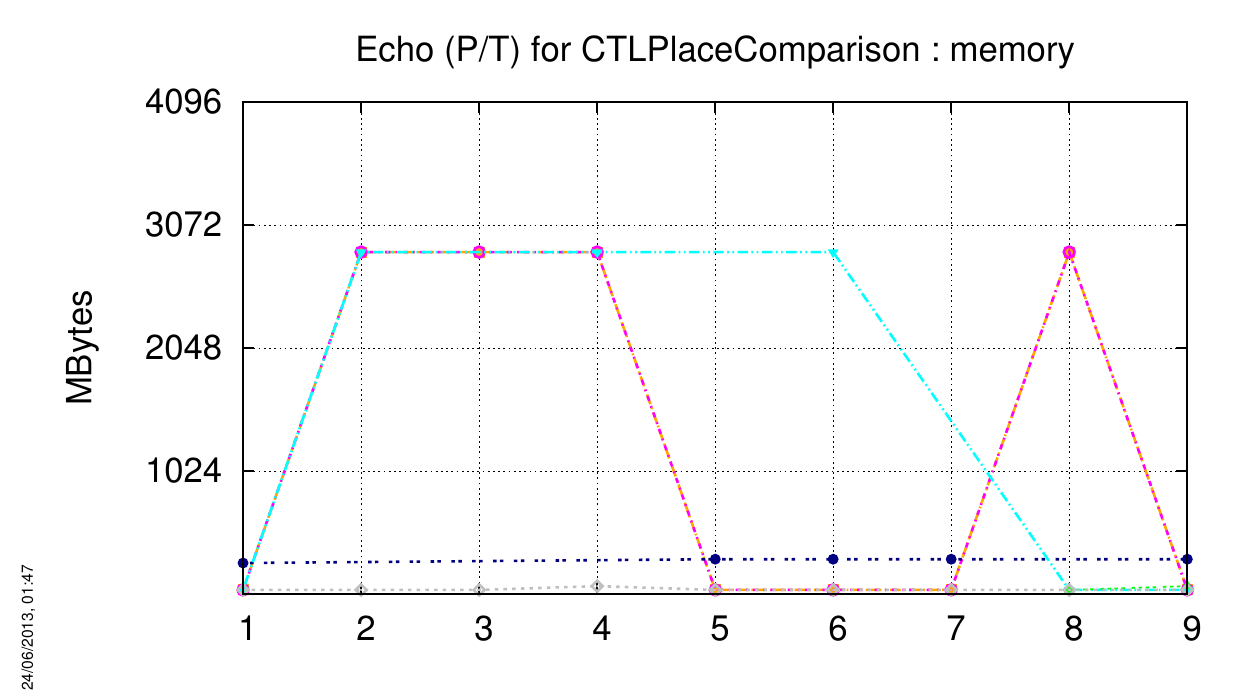}
   \includegraphics[width=7.2cm]{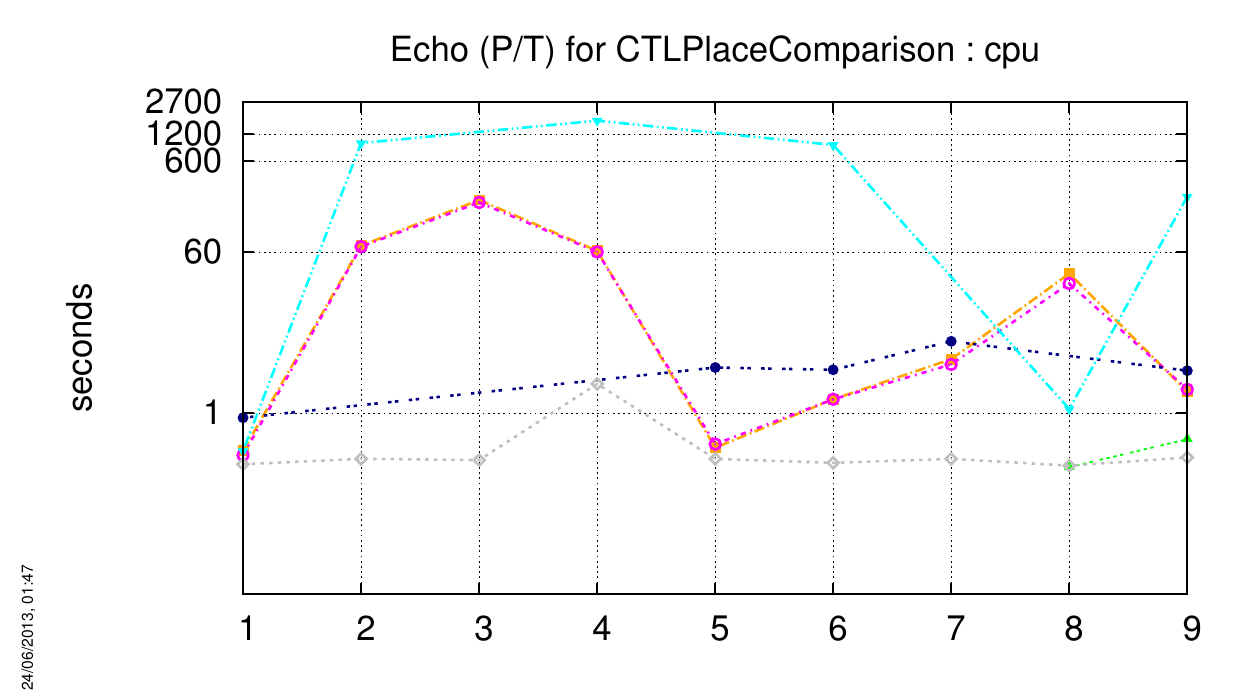}

   \includegraphics[height=1cm]{figures/tools-legend.pdf}
\end{center}

\subsubsection{\acs{Eratosthenes-PT}}
The charts below respectively show how tools compete with this ``Known'' model (memory and CPU).

\index{Performances!CTLPlaceComparison!Eratosthenes (P/T)}
\begin{center}
   \includegraphics[width=7.2cm]{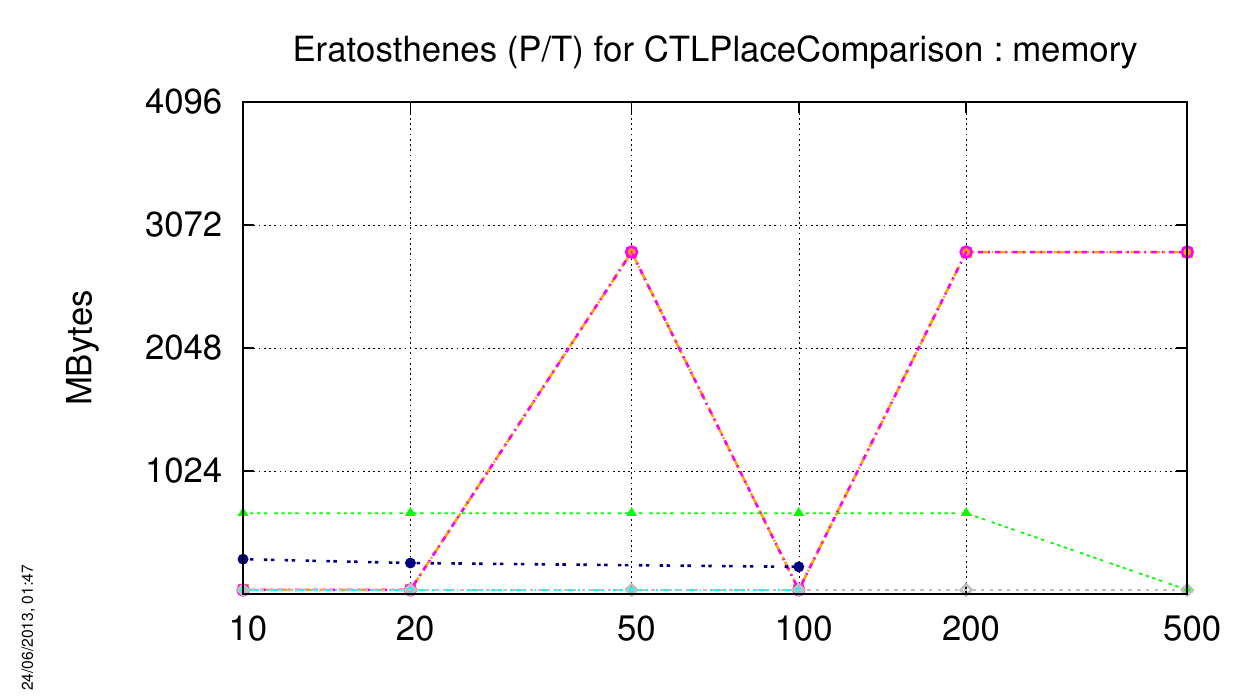}
   \includegraphics[width=7.2cm]{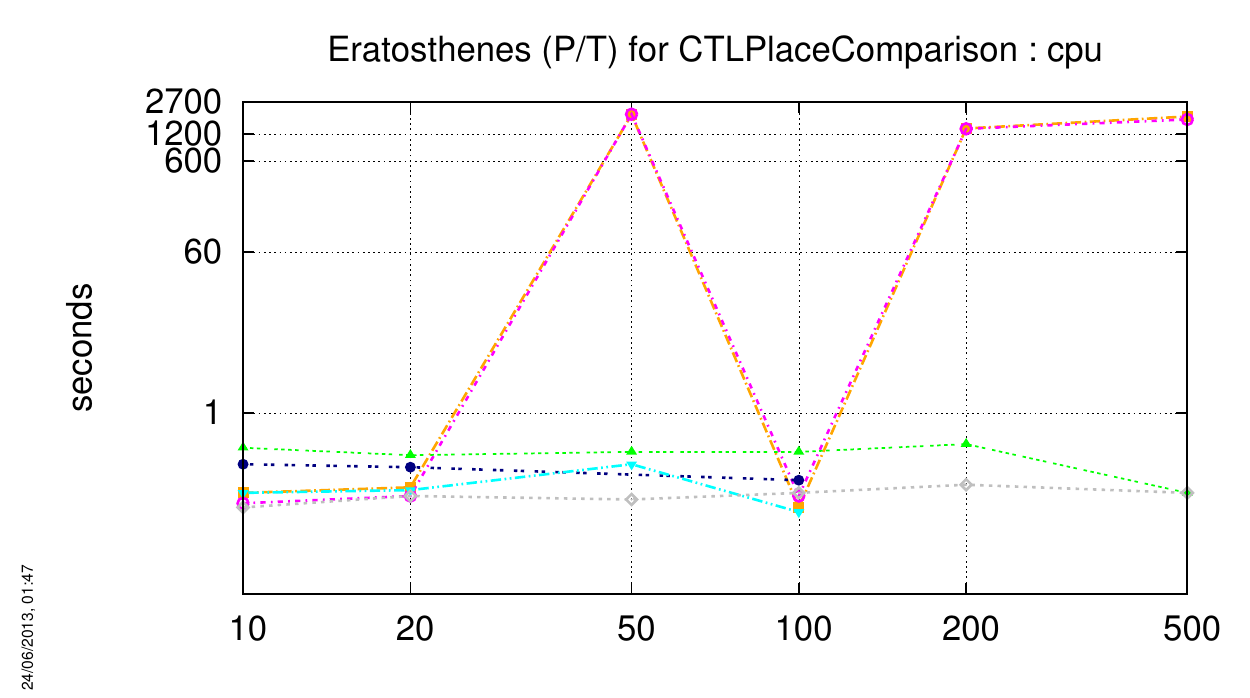}

   \includegraphics[height=1cm]{figures/tools-legend.pdf}
\end{center}

\subsubsection{\acs{FMS-PT}}
The charts below respectively show how tools compete with this ``Known'' model (memory and CPU).

\index{Performances!CTLPlaceComparison!FMS (P/T)}
\begin{center}
   \includegraphics[width=7.2cm]{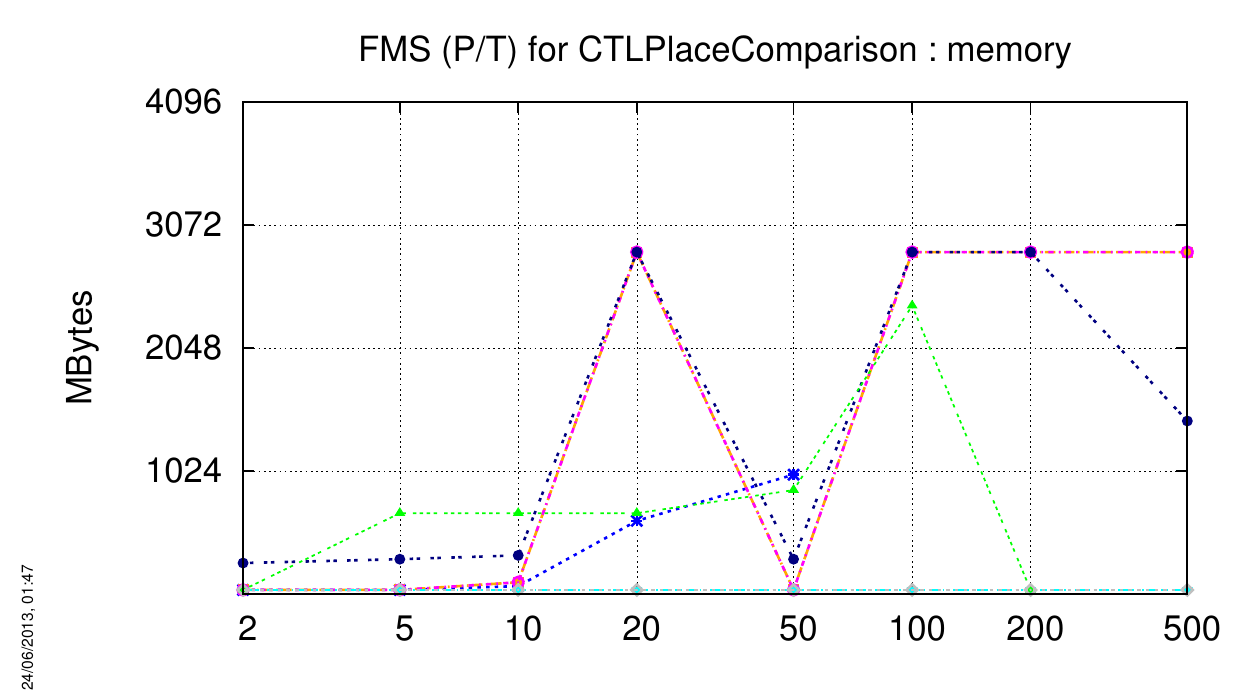}
   \includegraphics[width=7.2cm]{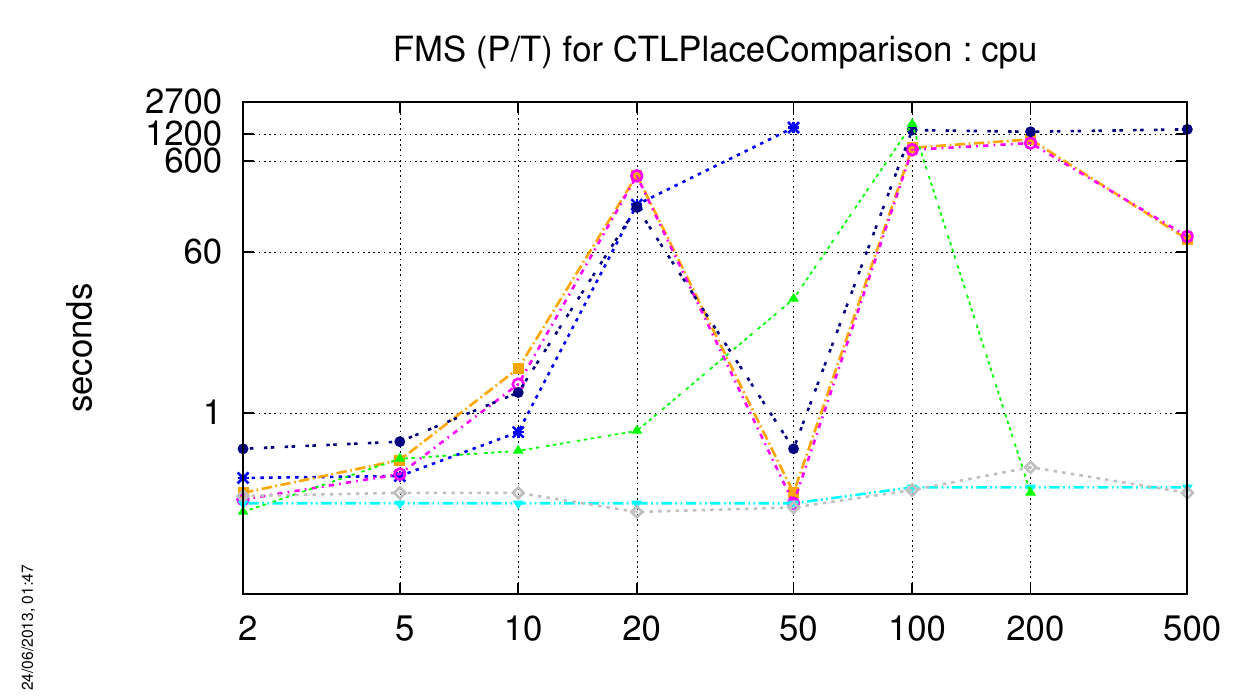}

   \includegraphics[height=1cm]{figures/tools-legend.pdf}
\end{center}

\subsubsection{\acs{GlobalRessAlloc-COL}}
No instance of this model could be computed for the \textbf{CTLPlaceComparison} examination.

\subsubsection{\acs{GlobalRessAlloc-PT}}
The charts below respectively show how tools compete with this ``Known'' model (memory and CPU).

\index{Performances!CTLPlaceComparison!GlobalRessAlloc (P/T)}
\begin{center}
   \includegraphics[width=7.2cm]{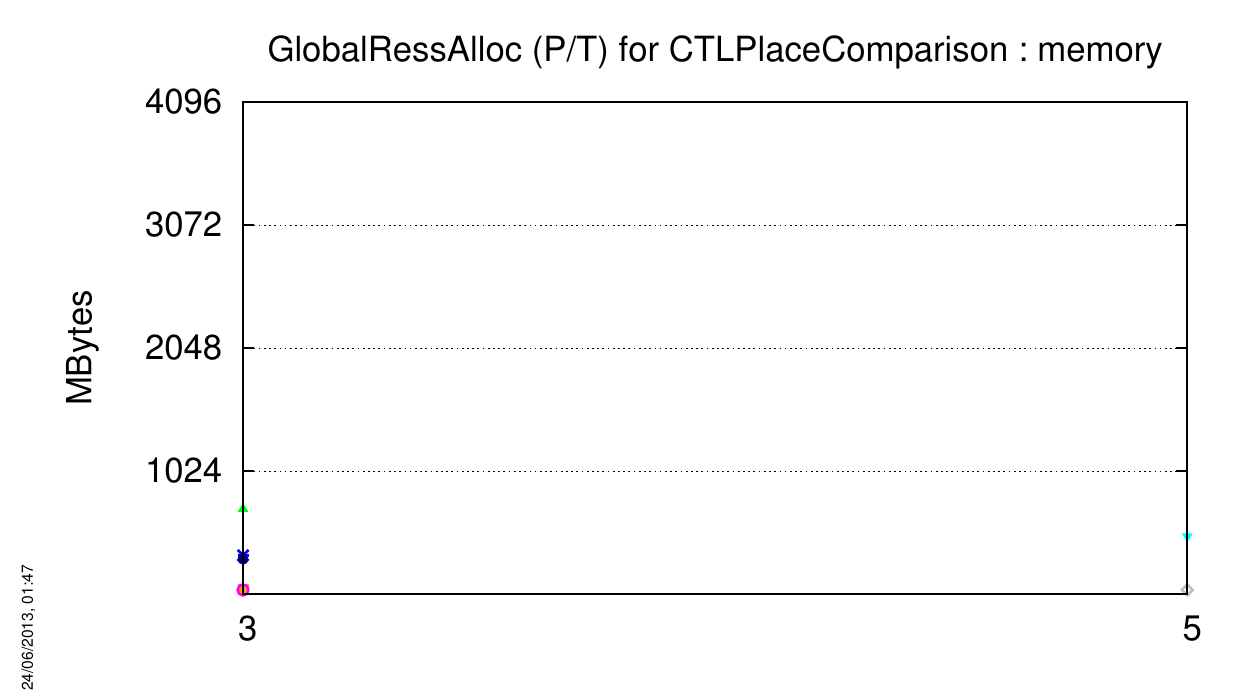}
   \includegraphics[width=7.2cm]{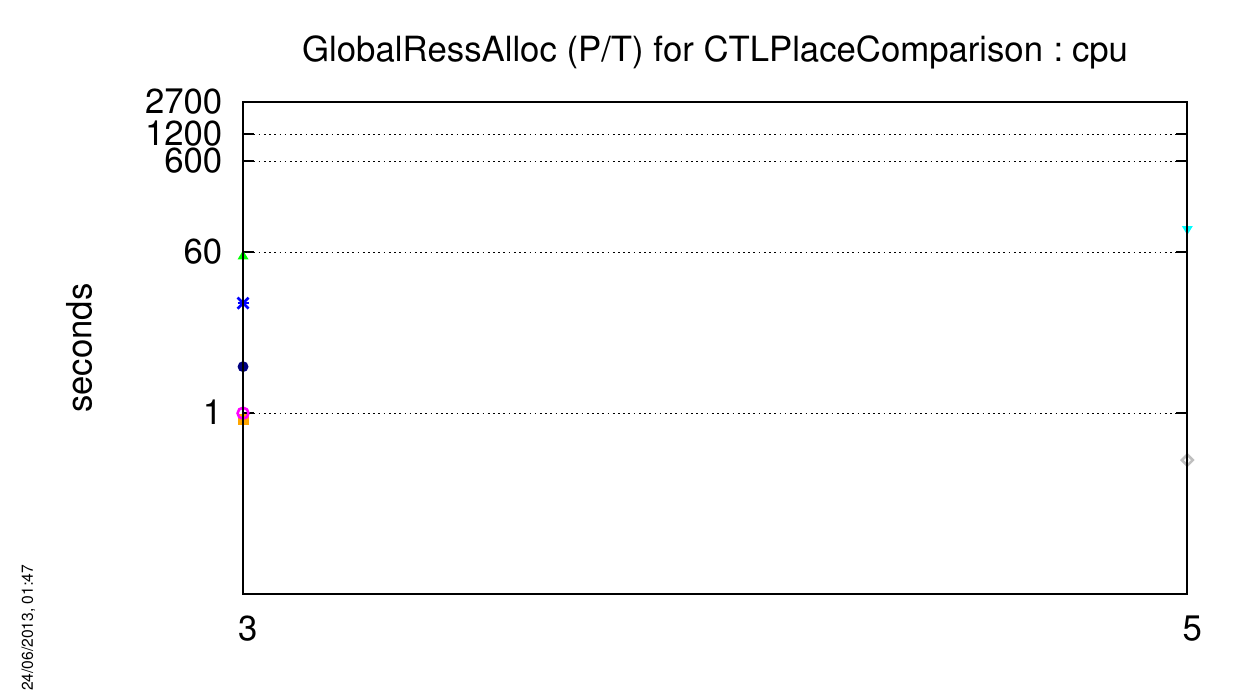}

   \includegraphics[height=1cm]{figures/tools-legend.pdf}
\end{center}

\subsubsection{\acs{Kanban-PT}}
The charts below respectively show how tools compete with this ``Known'' model (memory and CPU).

\index{Performances!CTLPlaceComparison!Kanban (P/T)}
\begin{center}
   \includegraphics[width=7.2cm]{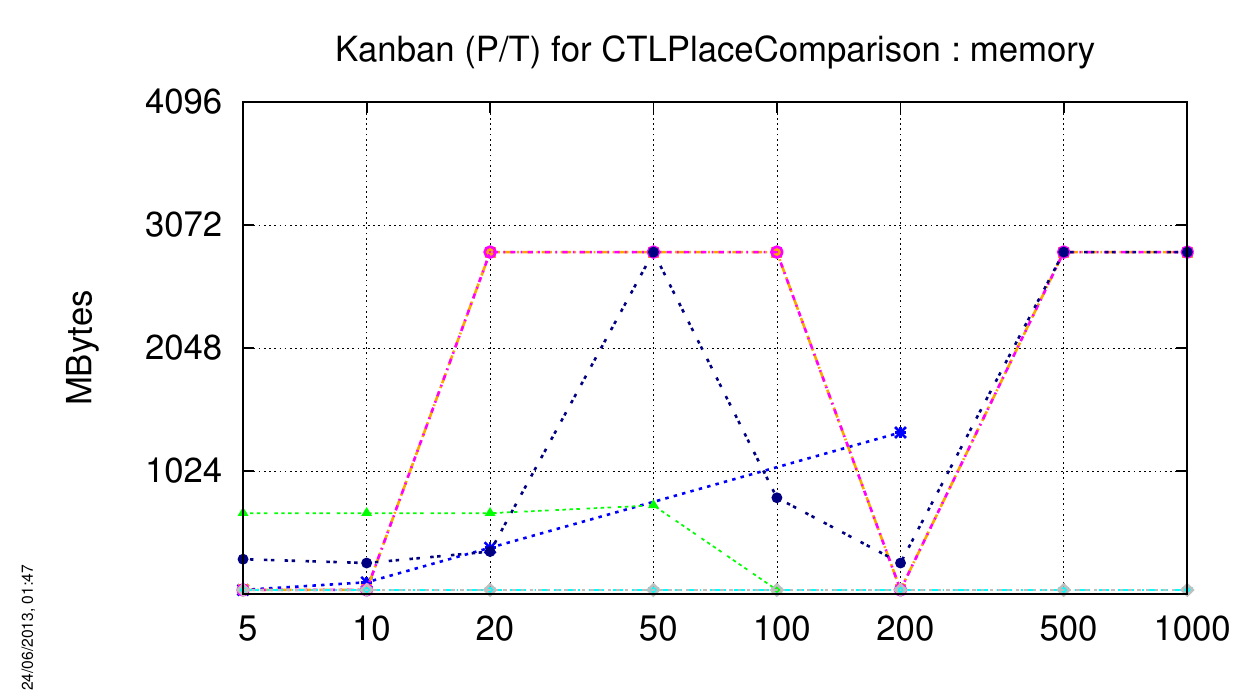}
   \includegraphics[width=7.2cm]{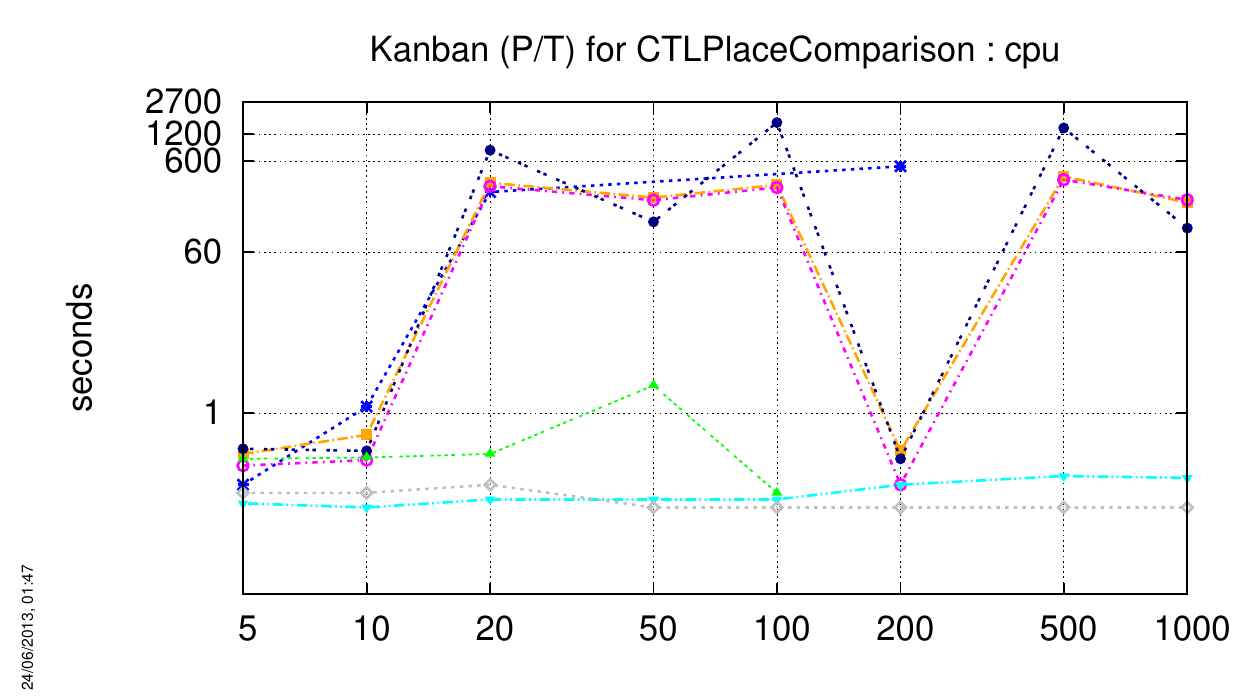}

   \includegraphics[height=1cm]{figures/tools-legend.pdf}
\end{center}

\subsubsection{\acs{LamportFastMutEx-COL}}
No instance of this model could be computed for the \textbf{CTLPlaceComparison} examination.

\subsubsection{\acs{LamportFastMutEx-PT}}
The charts below respectively show how tools compete with this ``Known'' model (memory and CPU).

\index{Performances!CTLPlaceComparison!LamportFastMutEx (P/T)}
\begin{center}
   \includegraphics[width=7.2cm]{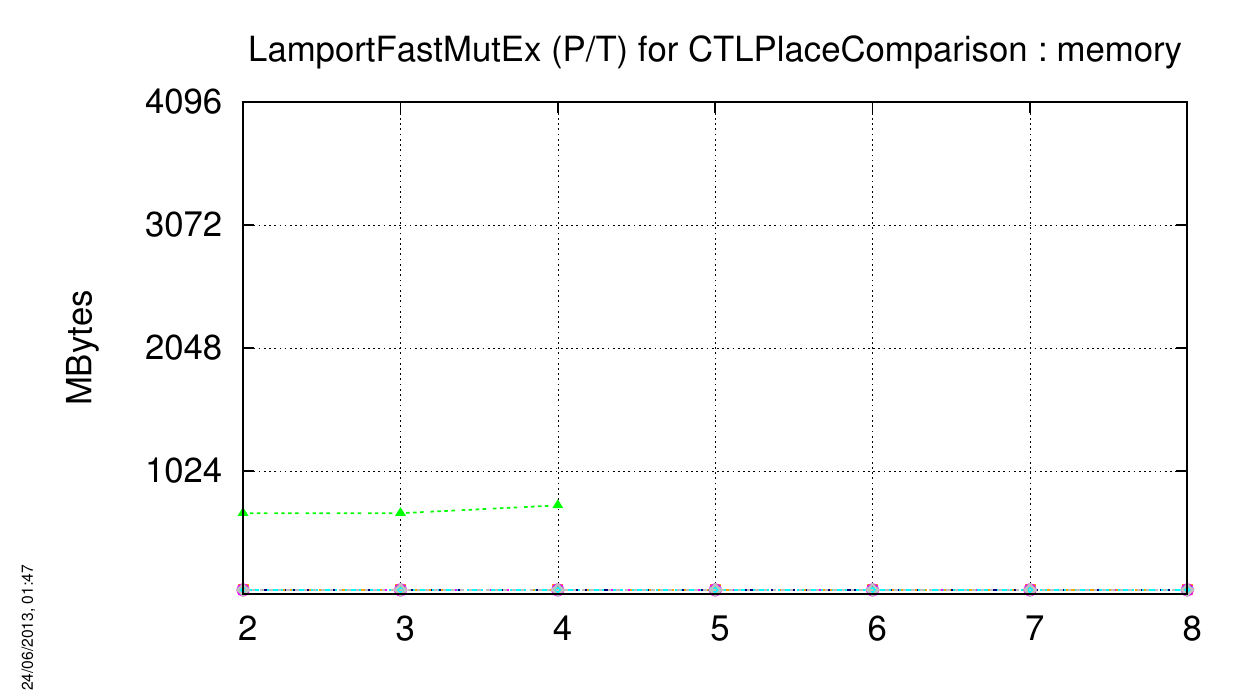}
   \includegraphics[width=7.2cm]{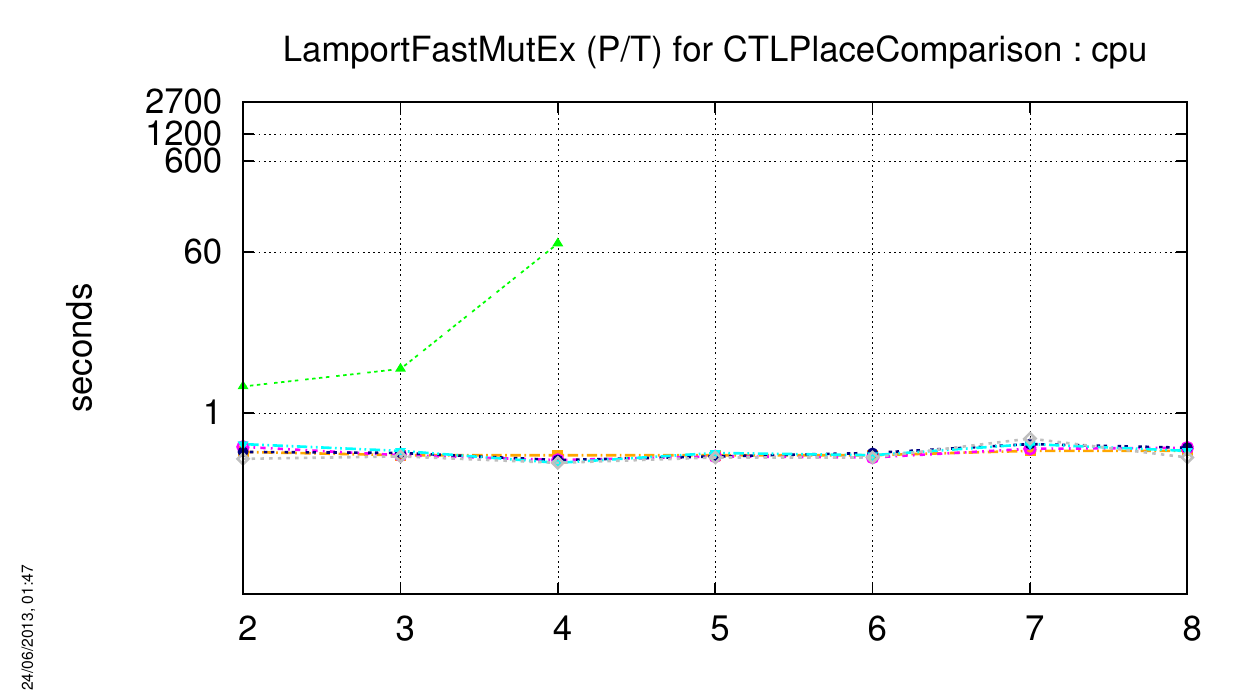}

   \includegraphics[height=1cm]{figures/tools-legend.pdf}
\end{center}

\subsubsection{\acs{MAPK-PT}}
The charts below respectively show how tools compete with this ``Known'' model (memory and CPU).

\index{Performances!CTLPlaceComparison!MAPK (P/T)}
\begin{center}
   \includegraphics[width=7.2cm]{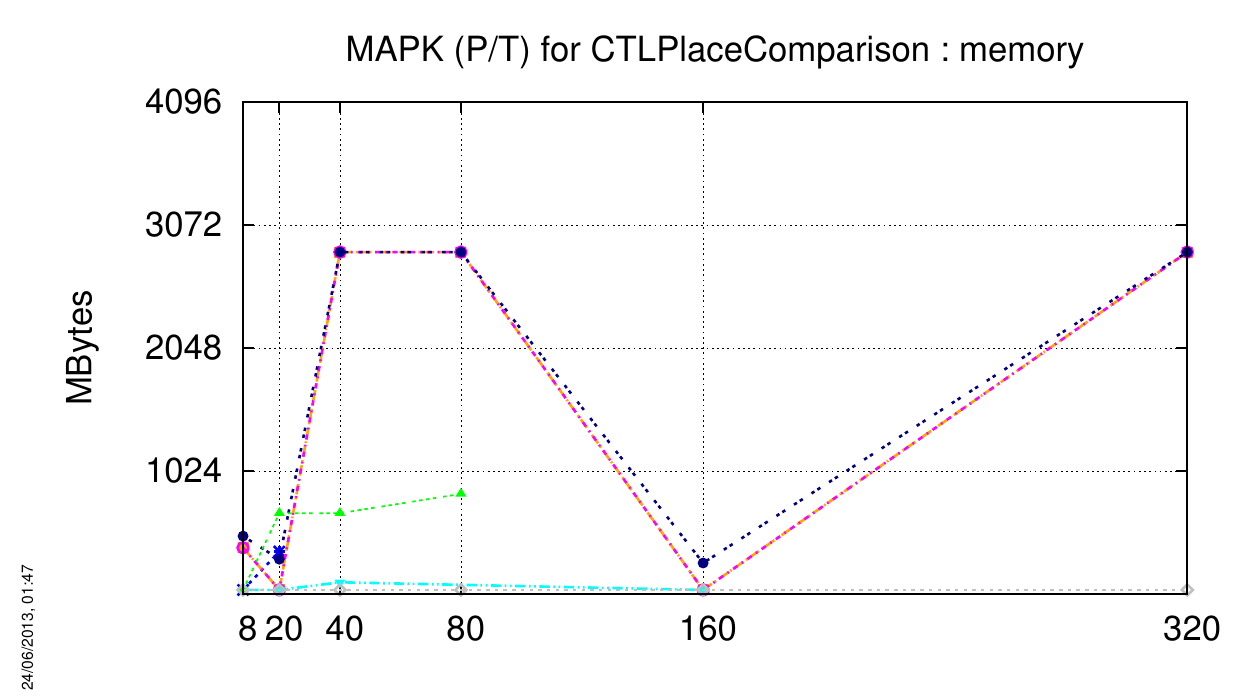}
   \includegraphics[width=7.2cm]{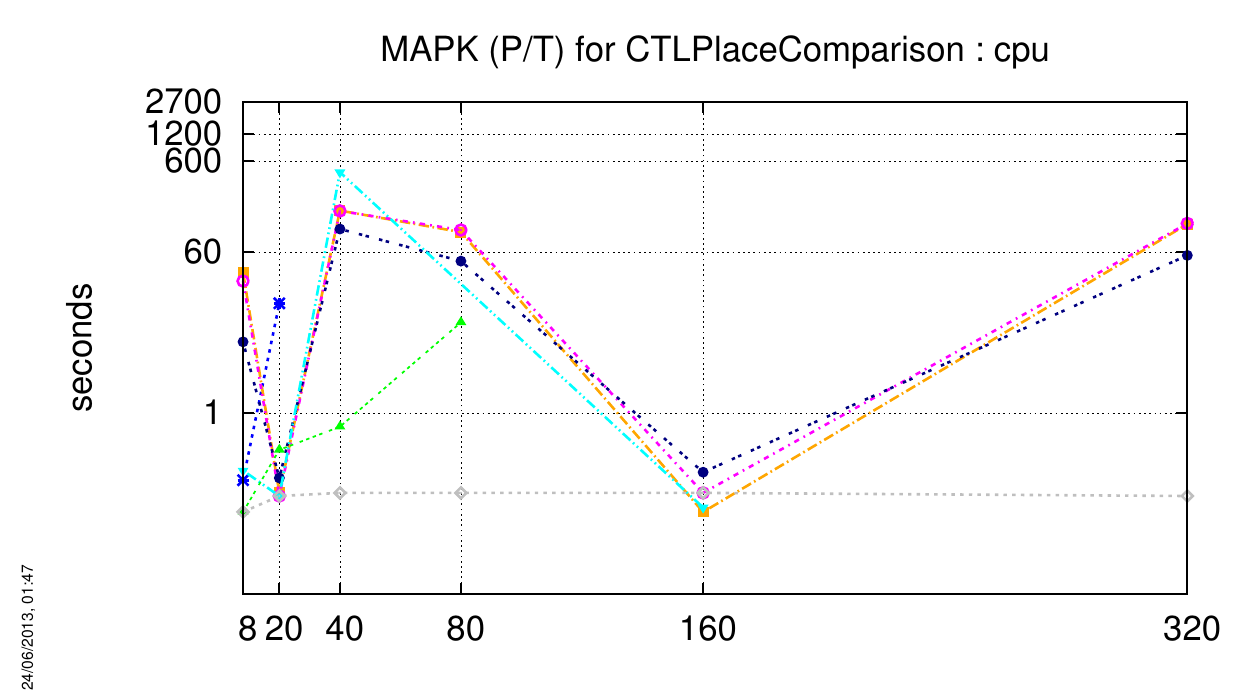}

   \includegraphics[height=1cm]{figures/tools-legend.pdf}
\end{center}

\subsubsection{\acs{NeoElection-COL}}
No instance of this model could be computed for the \textbf{CTLPlaceComparison} examination.

\subsubsection{\acs{NeoElection-PT}}
The charts below respectively show how tools compete with this ``Known'' model (memory and CPU).

\index{Performances!CTLPlaceComparison!NeoElection (P/T)}
\begin{center}
   \includegraphics[width=7.2cm]{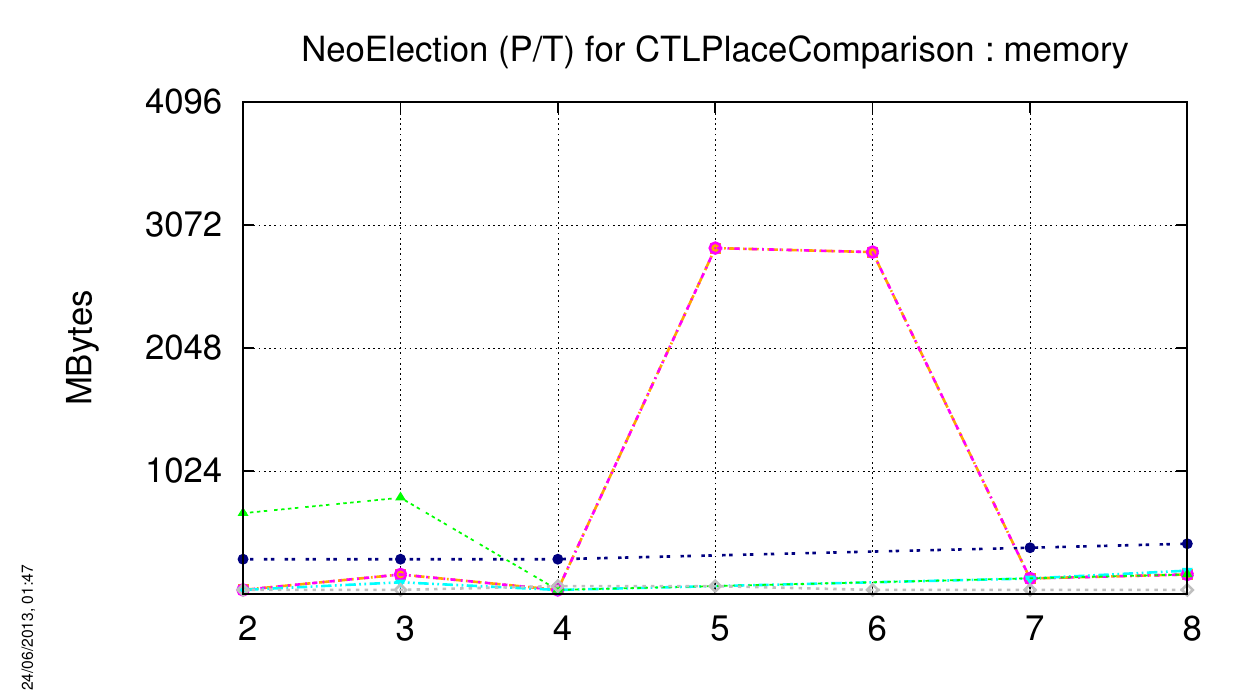}
   \includegraphics[width=7.2cm]{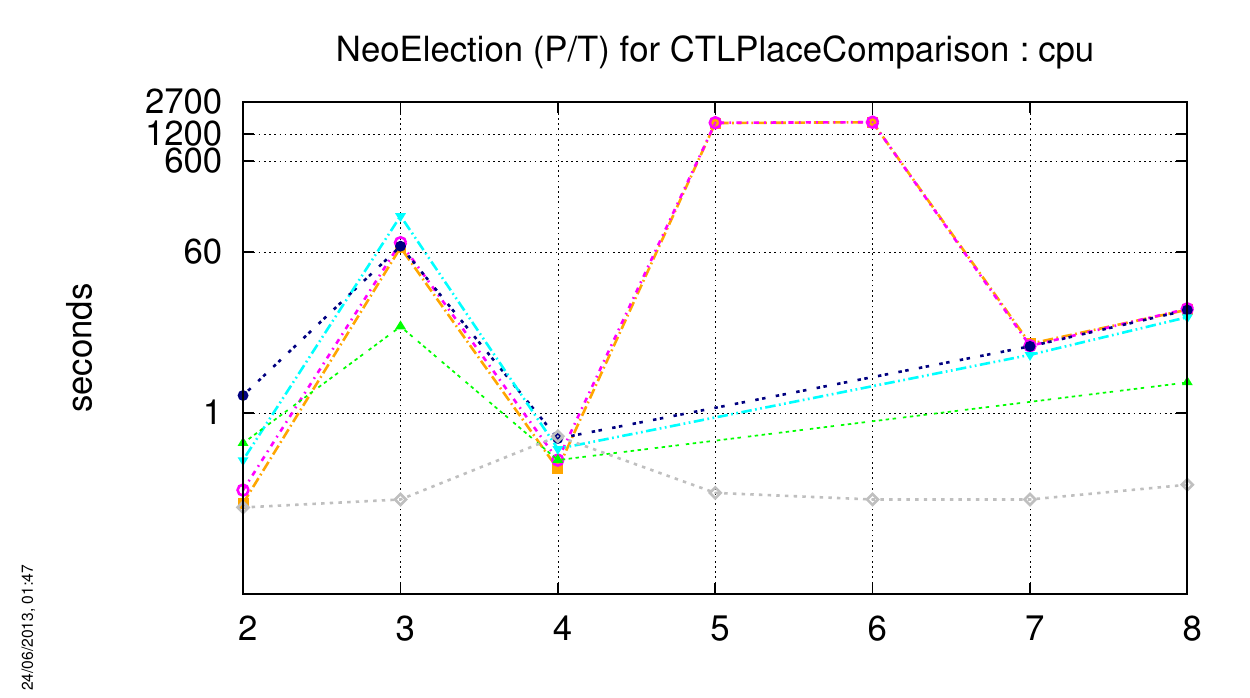}

   \includegraphics[height=1cm]{figures/tools-legend.pdf}
\end{center}

\subsubsection{\acs{PermAdmissibility-COL}}
No instance of this model could be computed for the \textbf{CTLPlaceComparison} examination.

\subsubsection{\acs{PermAdmissibility-PT}}
The charts below respectively show how tools compete with this ``Known'' model (memory and CPU).

\index{Performances!CTLPlaceComparison!PermAdmissibility (P/T)}
\begin{center}
   \includegraphics[width=7.2cm]{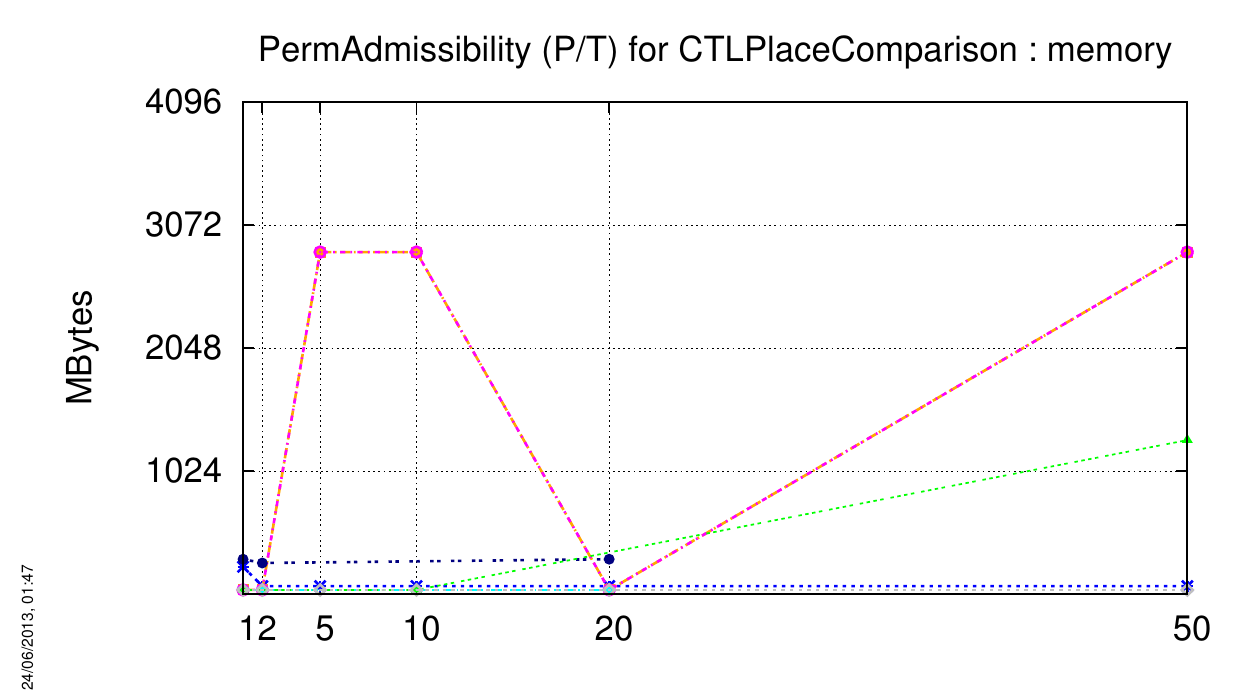}
   \includegraphics[width=7.2cm]{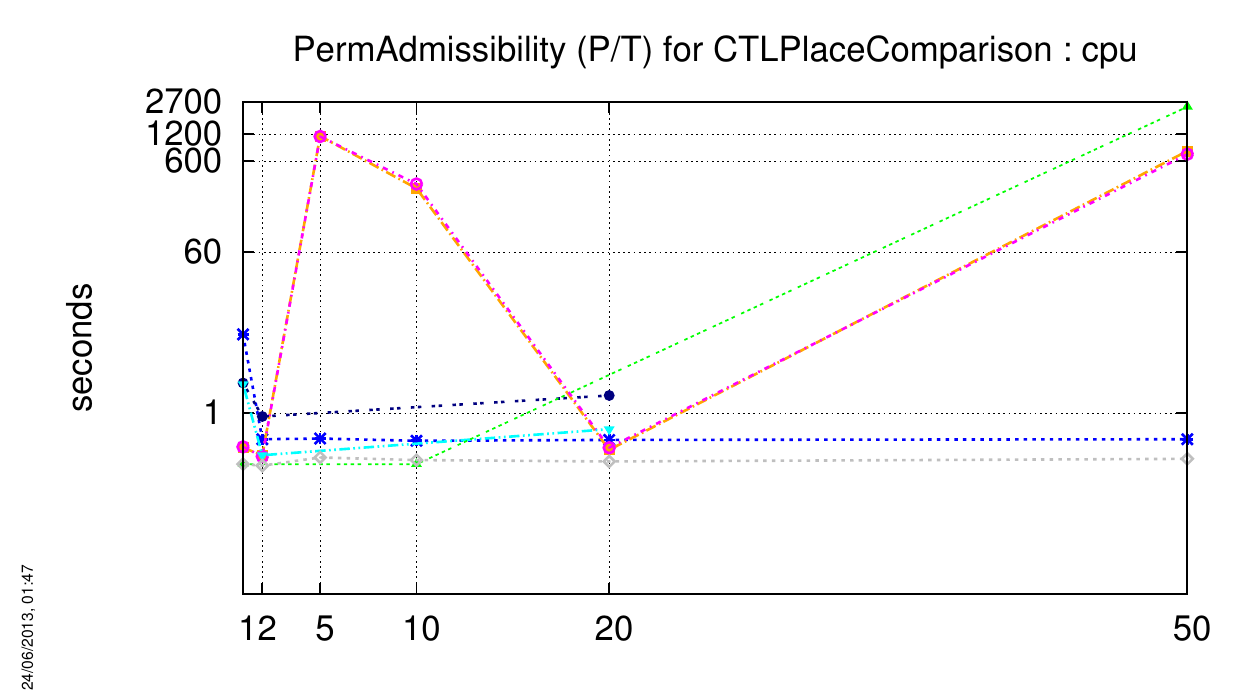}

   \includegraphics[height=1cm]{figures/tools-legend.pdf}
\end{center}

\subsubsection{\acs{Peterson-COL}}
No instance of this model could be computed for the \textbf{CTLPlaceComparison} examination.

\subsubsection{\acs{Peterson-PT}}
The charts below respectively show how tools compete with this ``Known'' model (memory and CPU).

\index{Performances!CTLPlaceComparison!Peterson (P/T)}
\begin{center}
   \includegraphics[width=7.2cm]{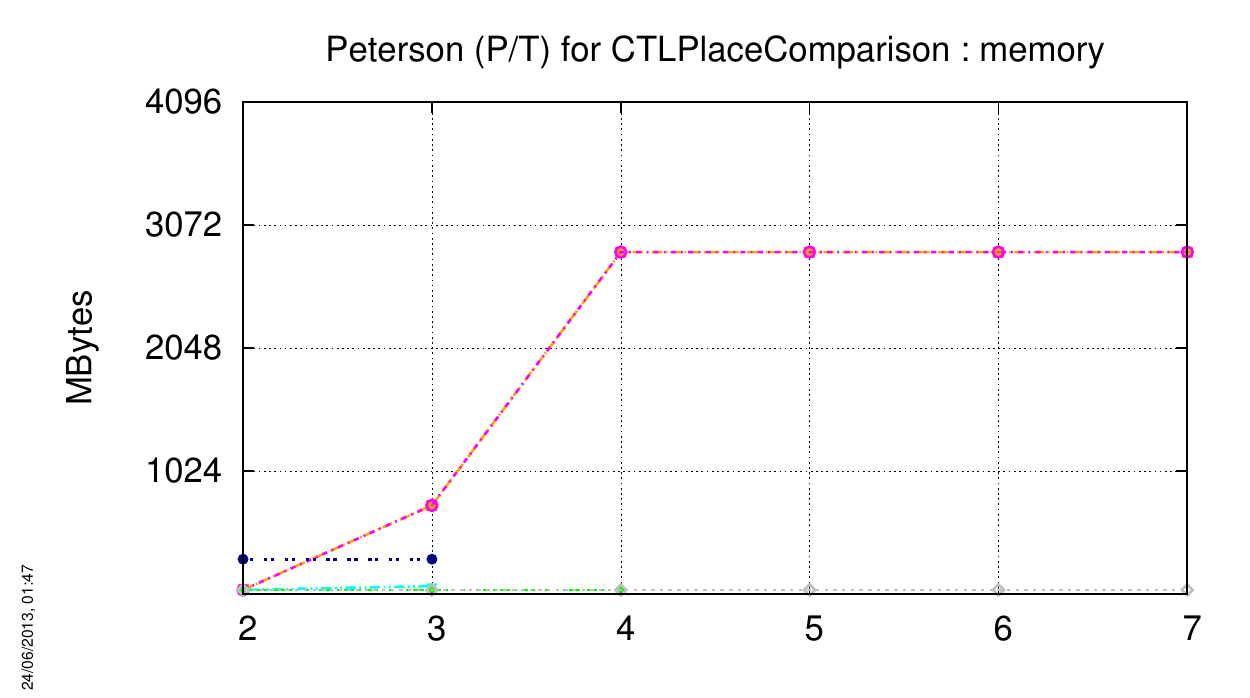}
   \includegraphics[width=7.2cm]{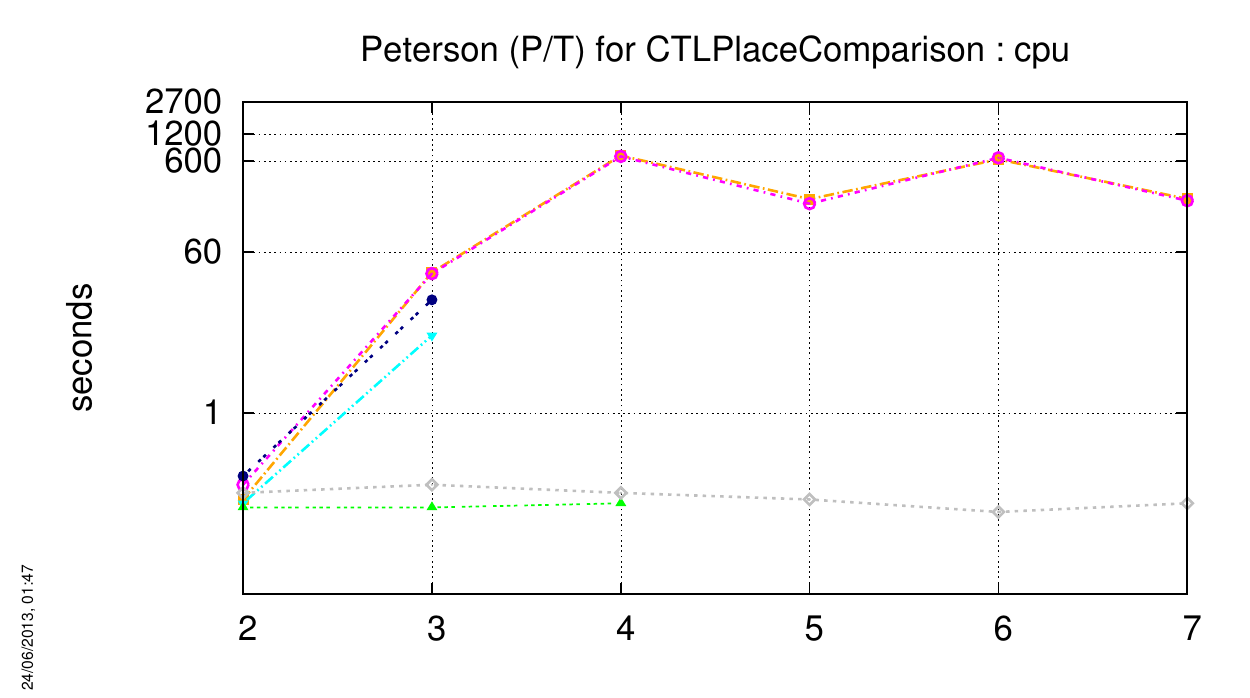}

   \includegraphics[height=1cm]{figures/tools-legend.pdf}
\end{center}

\subsubsection{\acs{Philosophers-COL}}
No instance of this model could be computed for the \textbf{CTLPlaceComparison} examination.

\subsubsection{\acs{Philosophers-PT}}
The charts below respectively show how tools compete with this ``Known'' model (memory and CPU).

\index{Performances!CTLPlaceComparison!Philosophers (P/T)}
\begin{center}
   \includegraphics[width=7.2cm]{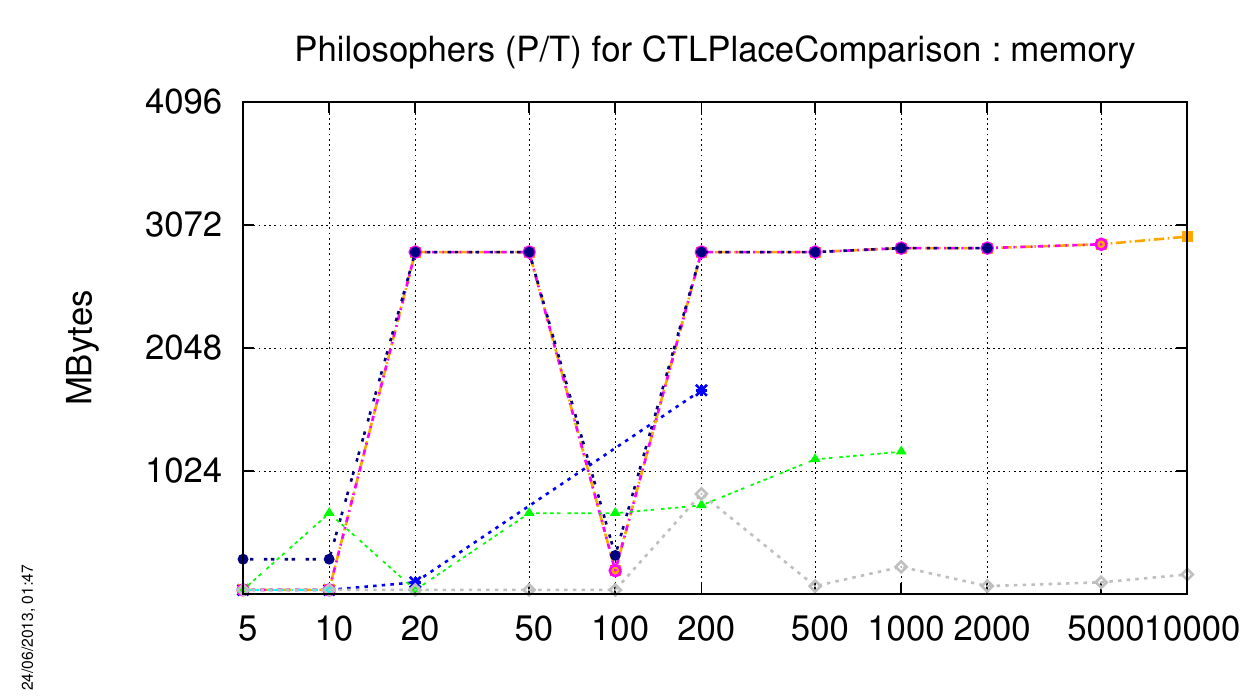}
   \includegraphics[width=7.2cm]{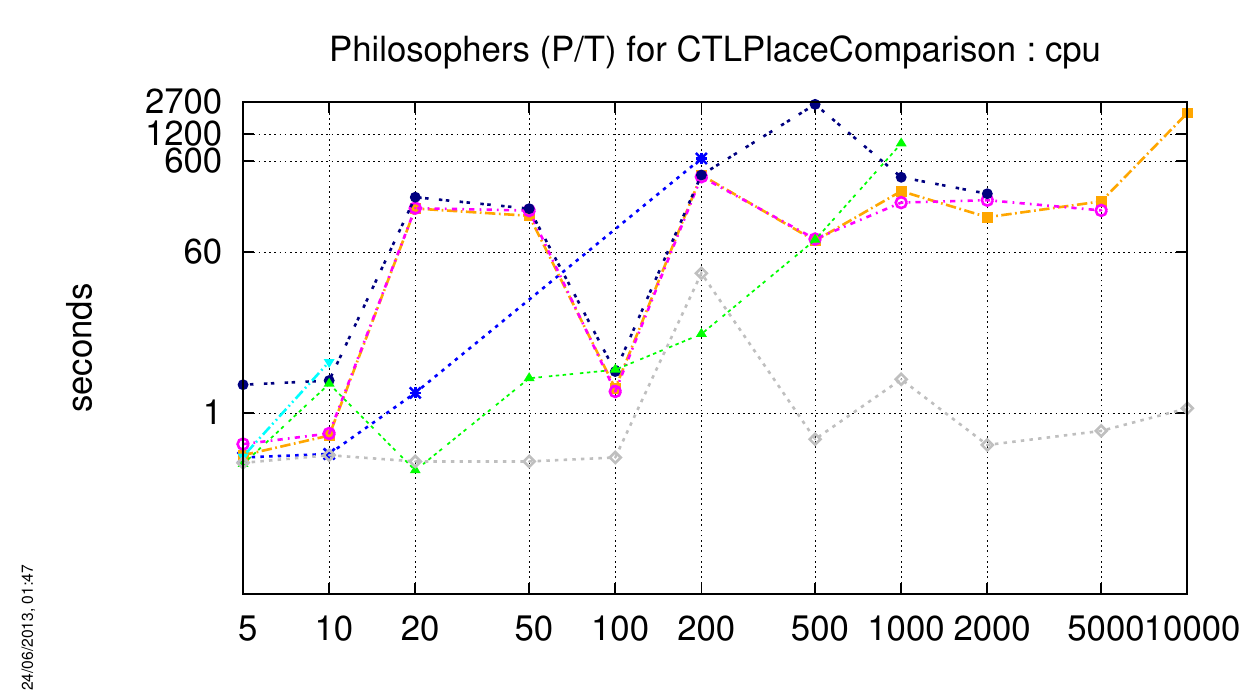}

   \includegraphics[height=1cm]{figures/tools-legend.pdf}
\end{center}

\subsubsection{\acs{PhilosophersDyn-COL}}
No instance of this model could be computed for the \textbf{CTLPlaceComparison} examination.

\subsubsection{\acs{PhilosophersDyn-PT}}
The charts below respectively show how tools compete with this ``Known'' model (memory and CPU).

\index{Performances!CTLPlaceComparison!PhilosophersDyn (P/T)}
\begin{center}
   \includegraphics[width=7.2cm]{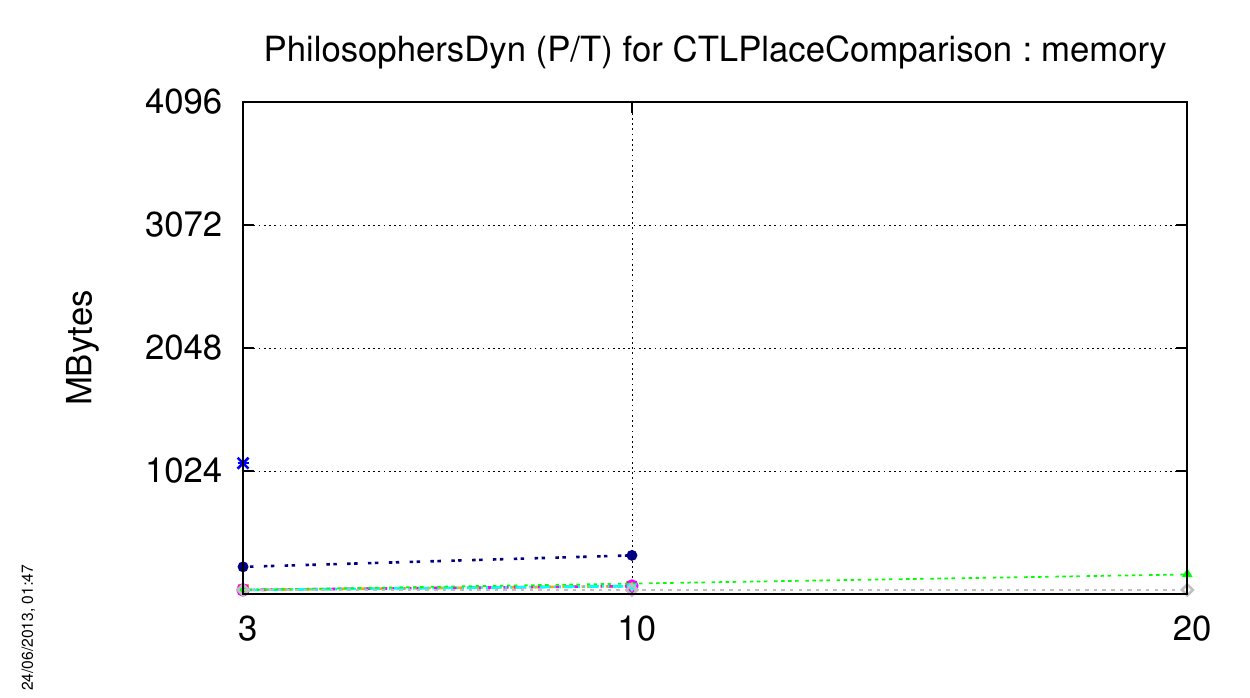}
   \includegraphics[width=7.2cm]{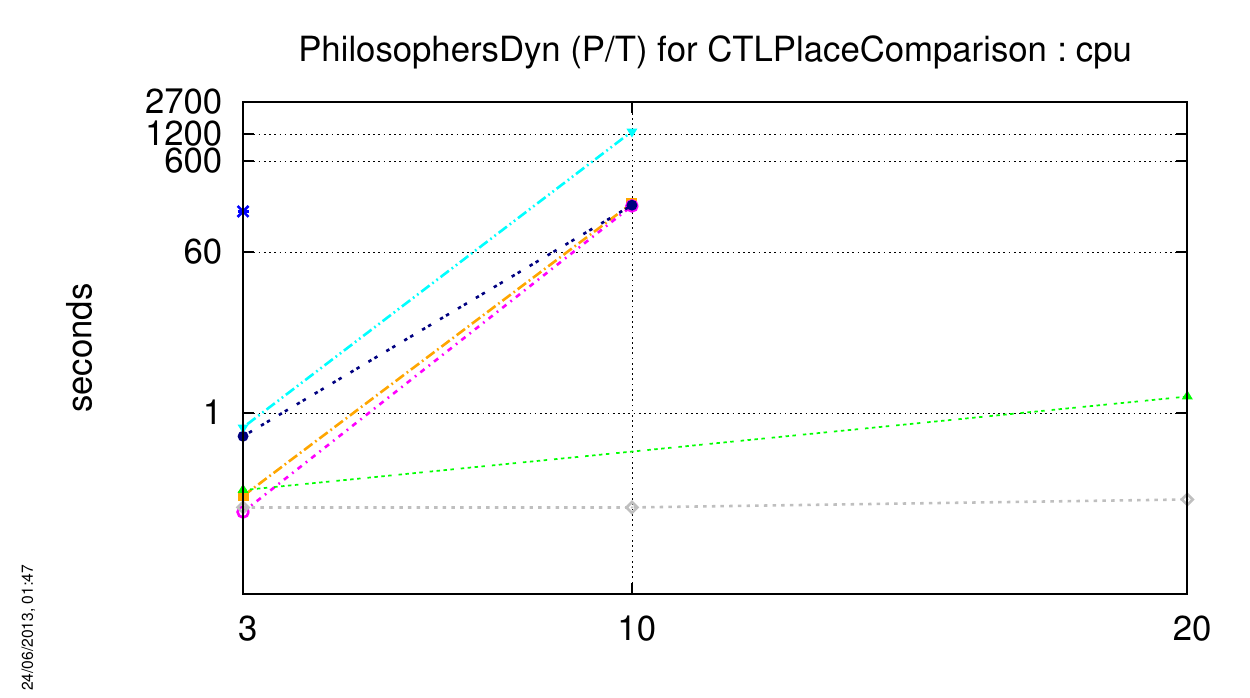}

   \includegraphics[height=1cm]{figures/tools-legend.pdf}
\end{center}

\subsubsection{\acs{Planning-PT}}
No instance of this model could be computed for the \textbf{CTLPlaceComparison} examination.

\subsubsection{\acs{Railroad-PT}}
The charts below respectively show how tools compete with this ``Known'' model (memory and CPU).

\index{Performances!CTLPlaceComparison!Railroad (P/T)}
\begin{center}
   \includegraphics[width=7.2cm]{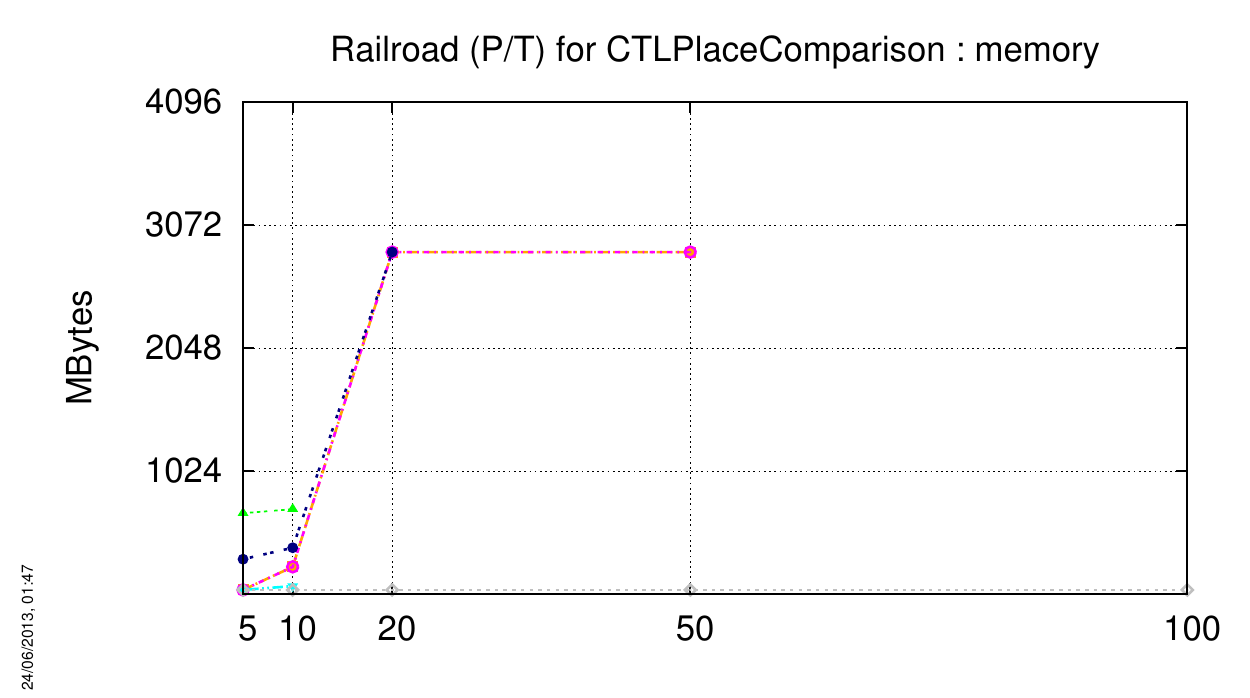}
   \includegraphics[width=7.2cm]{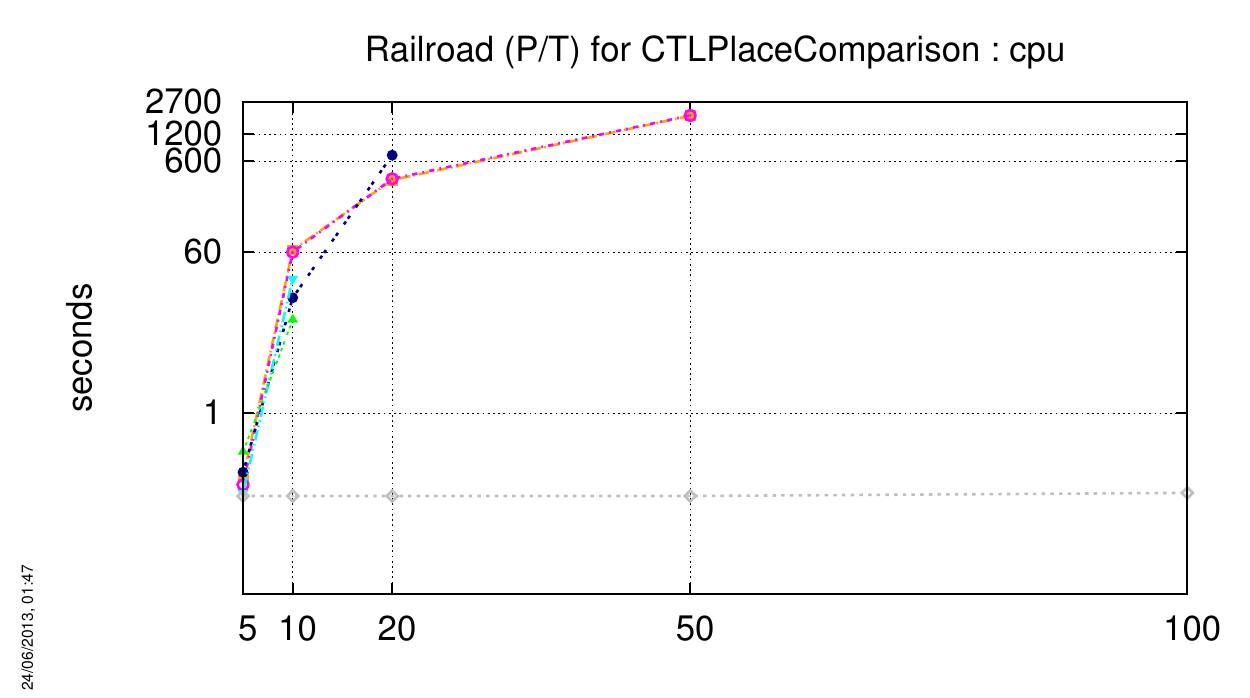}

   \includegraphics[height=1cm]{figures/tools-legend.pdf}
\end{center}

\subsubsection{\acs{RessAllocation-PT}}
The charts below respectively show how tools compete with this ``Known'' model (memory and CPU).

\index{Performances!CTLPlaceComparison!RessAllocation (P/T)}
\begin{center}
   \includegraphics[width=7.2cm]{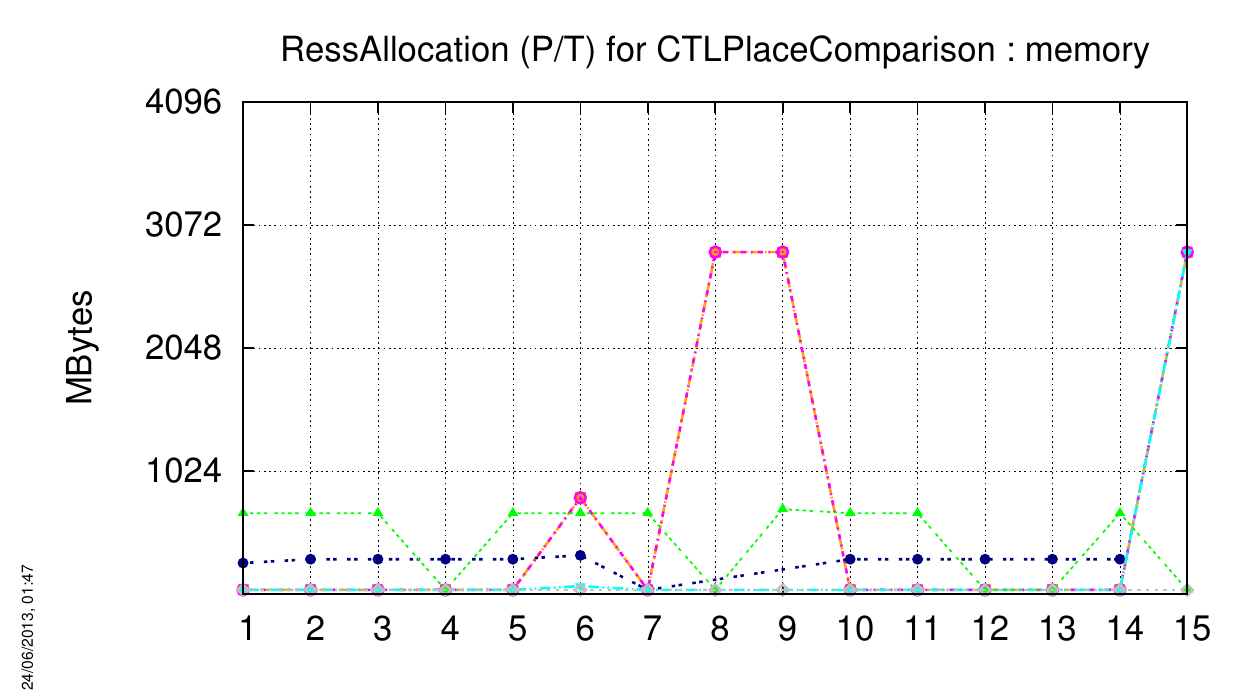}
   \includegraphics[width=7.2cm]{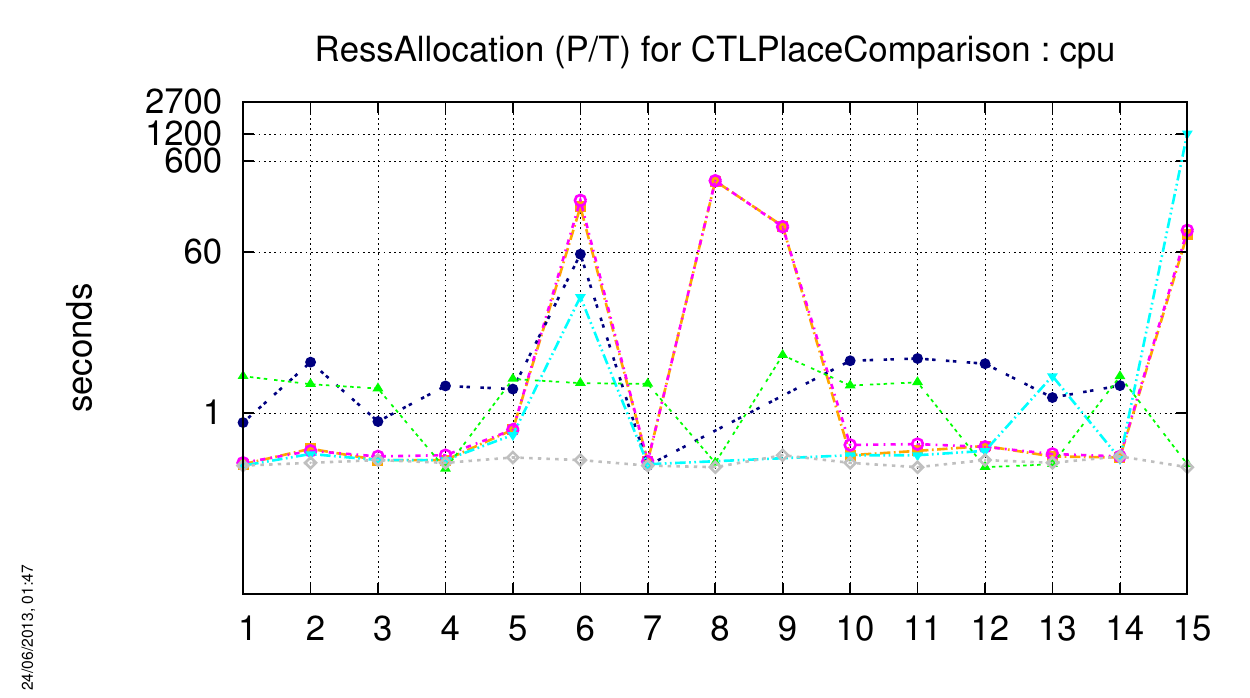}

   \includegraphics[height=1cm]{figures/tools-legend.pdf}
\end{center}

\subsubsection{\acs{Ring-PT}}
The charts below respectively show how tools compete with this ``Known'' model (memory and CPU).

\index{Performances!CTLPlaceComparison!Ring (P/T)}
\begin{center}
   \includegraphics[width=7.2cm]{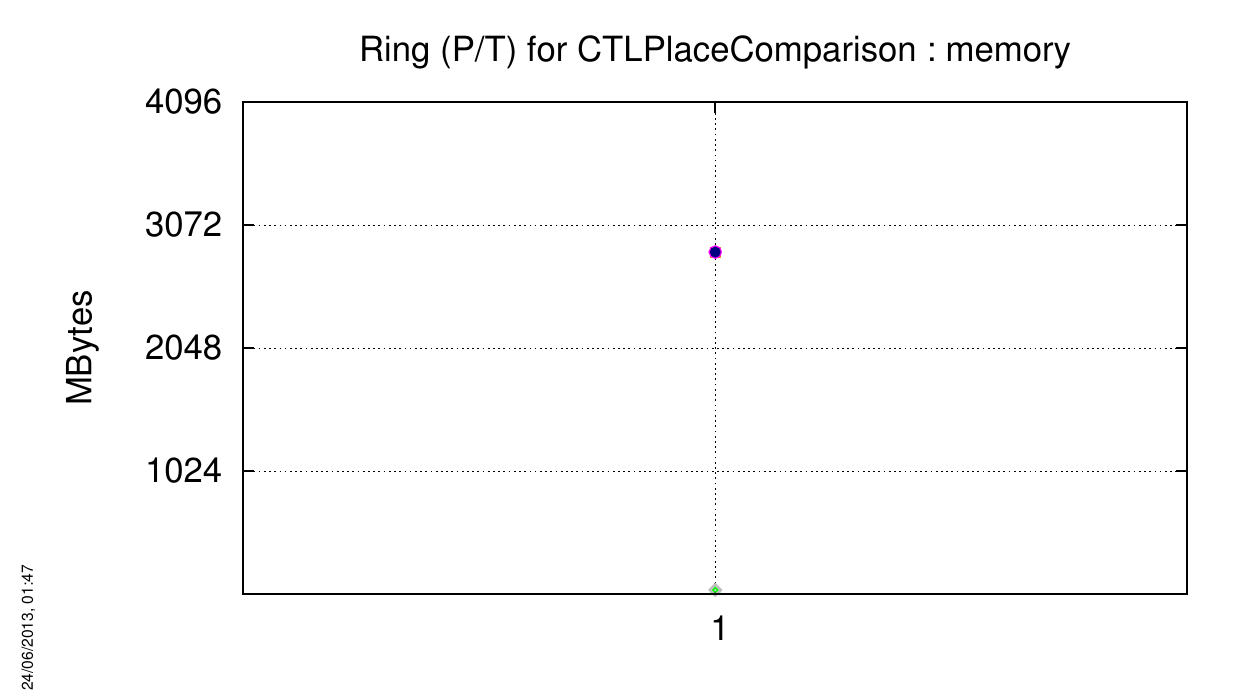}
   \includegraphics[width=7.2cm]{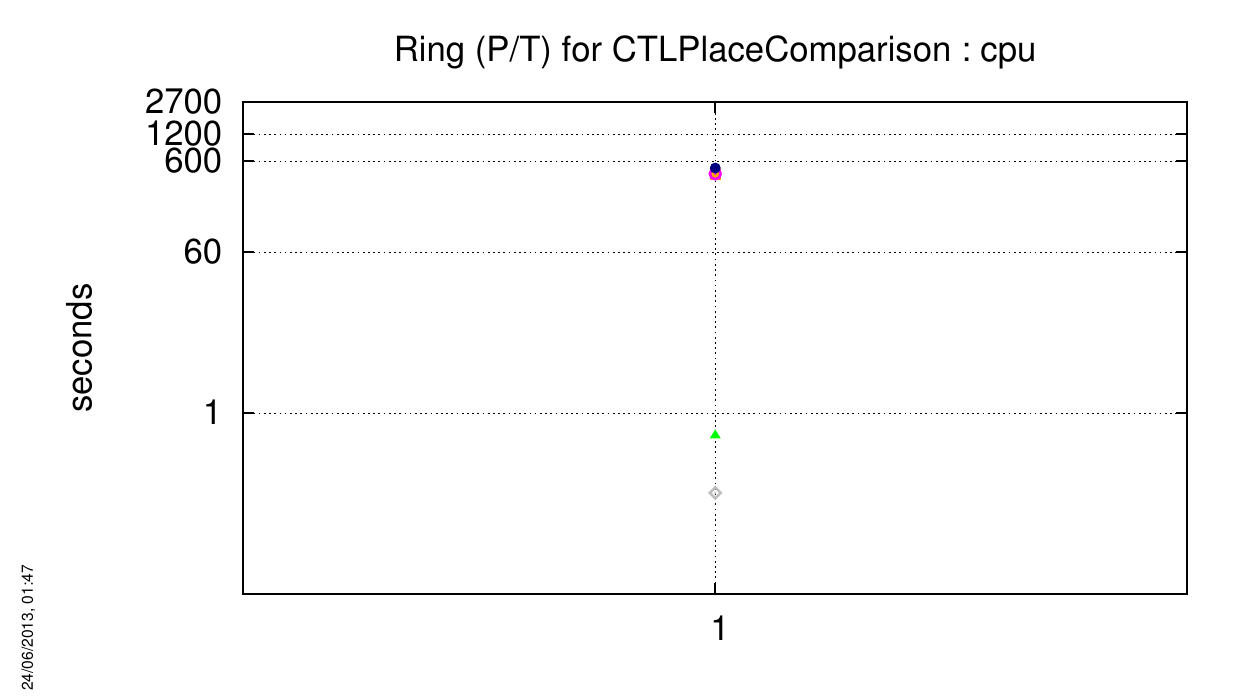}

   \includegraphics[height=1cm]{figures/tools-legend.pdf}
\end{center}

\subsubsection{\acs{RwMutex-PT}}
The charts below respectively show how tools compete with this ``Known'' model (memory and CPU).

\index{Performances!CTLPlaceComparison!RwMutex (P/T)}
\begin{center}
   \includegraphics[width=7.2cm]{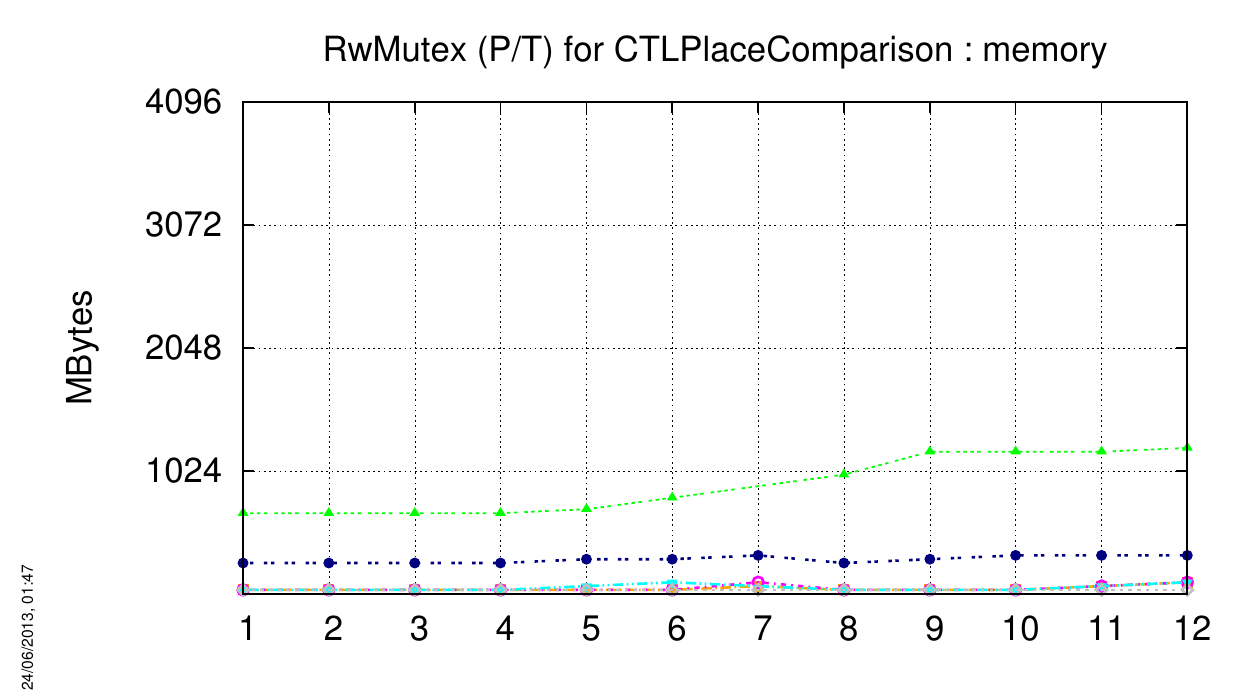}
   \includegraphics[width=7.2cm]{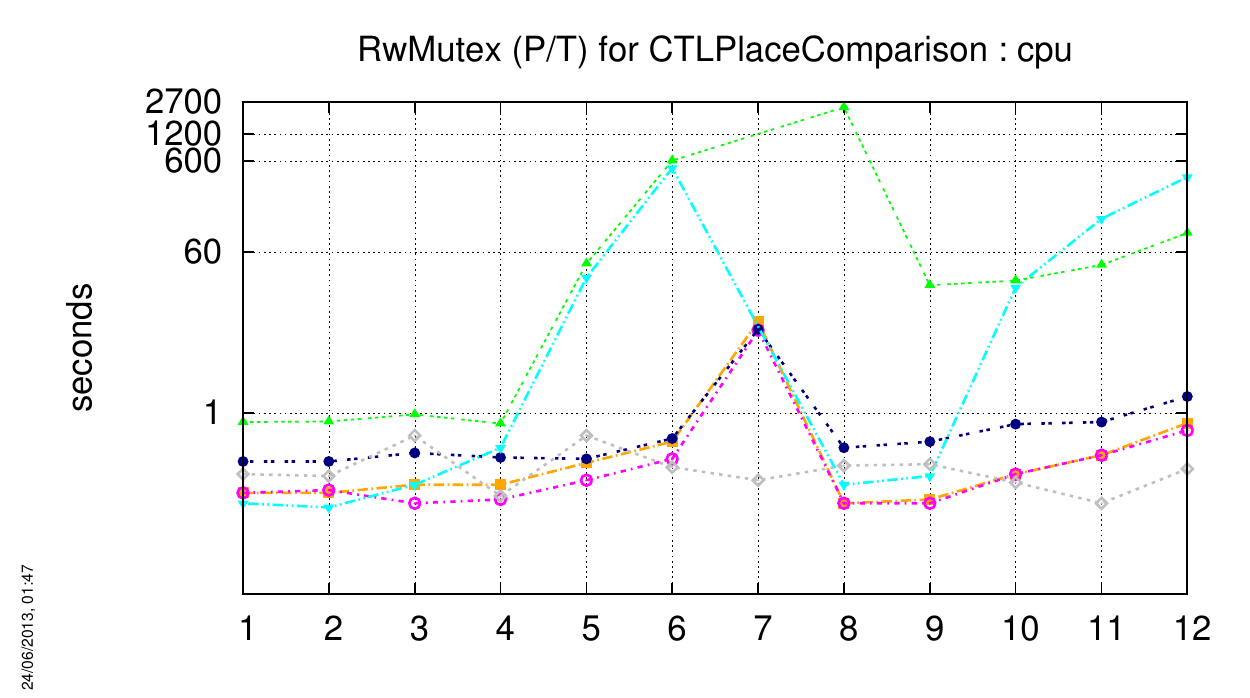}

   \includegraphics[height=1cm]{figures/tools-legend.pdf}
\end{center}

\subsubsection{\acs{SharedMemory-COL}}
No instance of this model could be computed for the \textbf{CTLPlaceComparison} examination.

\subsubsection{\acs{SharedMemory-PT}}
The charts below respectively show how tools compete with this ``Known'' model (memory and CPU).

\index{Performances!CTLPlaceComparison!SharedMemory (P/T)}
\begin{center}
   \includegraphics[width=7.2cm]{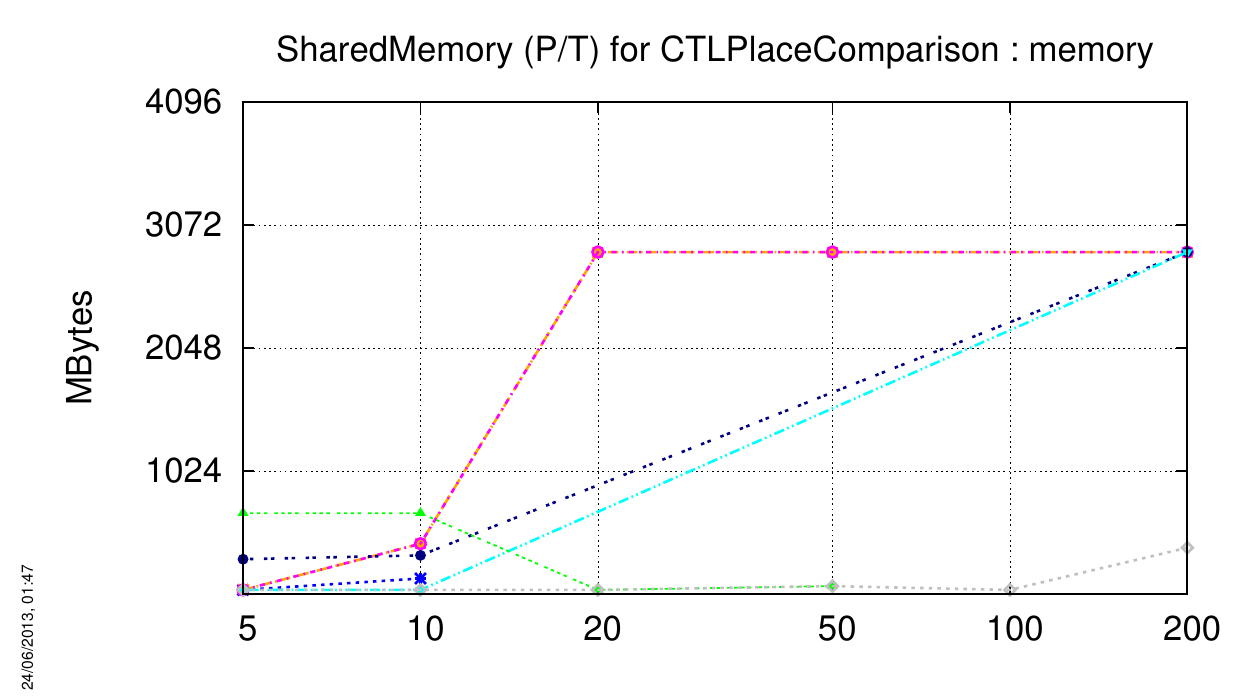}
   \includegraphics[width=7.2cm]{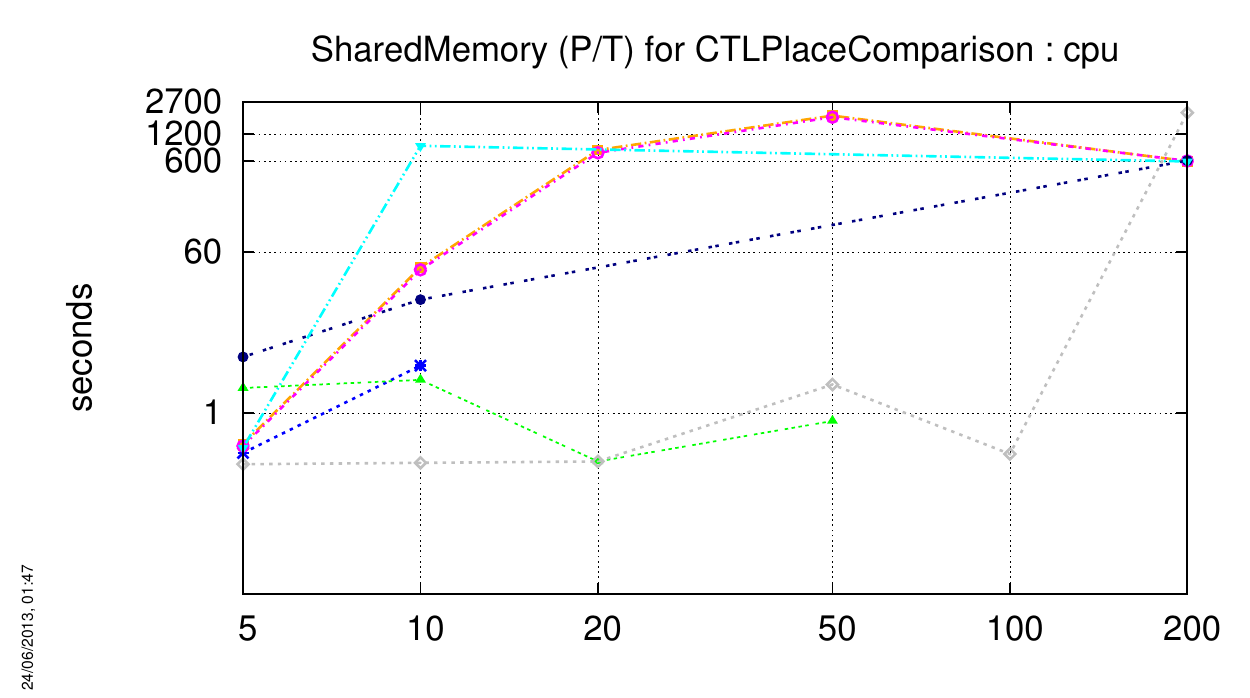}

   \includegraphics[height=1cm]{figures/tools-legend.pdf}
\end{center}

\subsubsection{\acs{SimpleLoadBal-COL}}
No instance of this model could be computed for the \textbf{CTLPlaceComparison} examination.

\subsubsection{\acs{SimpleLoadBal-PT}}
The charts below respectively show how tools compete with this ``Known'' model (memory and CPU).

\index{Performances!CTLPlaceComparison!SimpleLoadBal (P/T)}
\begin{center}
   \includegraphics[width=7.2cm]{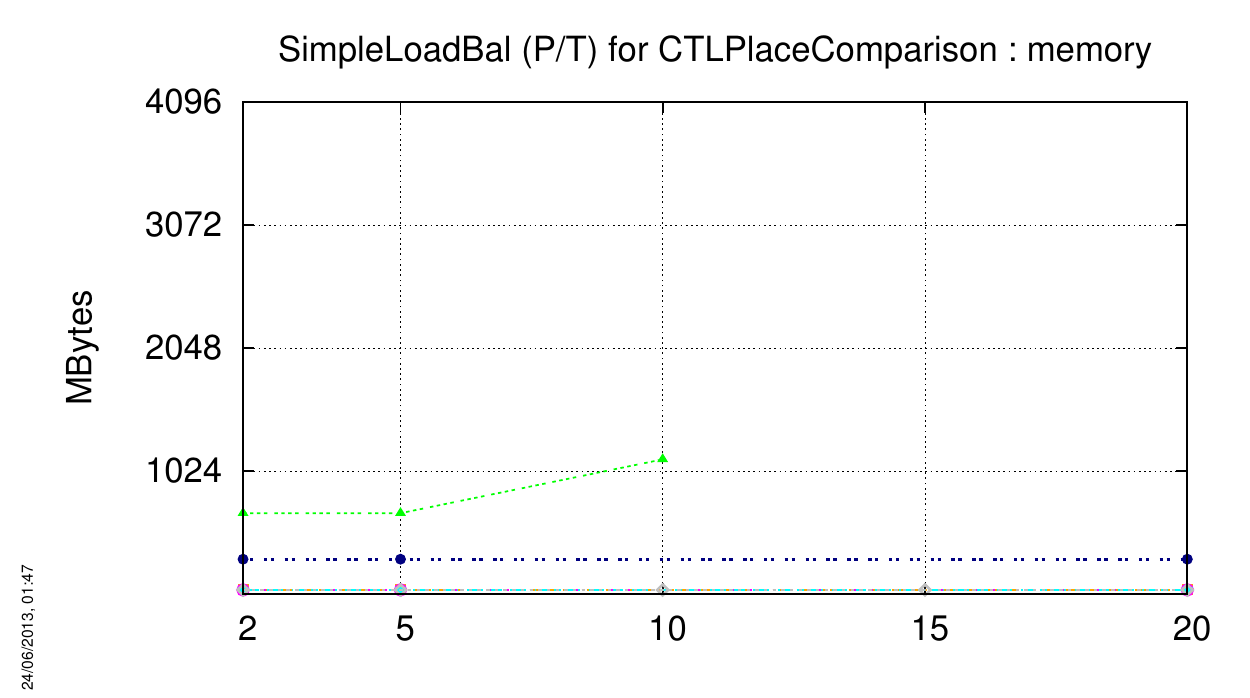}
   \includegraphics[width=7.2cm]{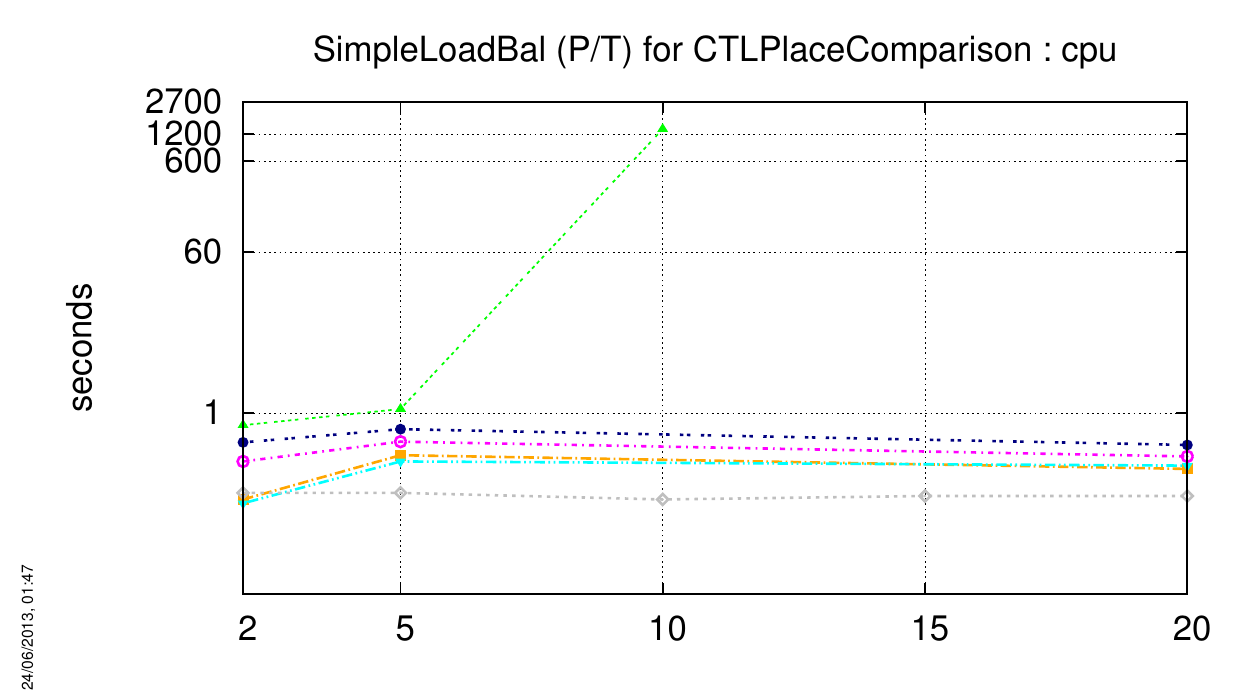}

   \includegraphics[height=1cm]{figures/tools-legend.pdf}
\end{center}

\subsubsection{\acs{TokenRing-COL}}
No instance of this model could be computed for the \textbf{CTLPlaceComparison} examination.

\subsubsection{\acs{TokenRing-PT}}
No instance of this model could be computed for the \textbf{CTLPlaceComparison} examination.

\subsubsection{\acs{HouseConstruction-PT}}
The charts below respectively show how tools compete with this ``Suprise'' model (memory and CPU).

\index{Performances!CTLPlaceComparison!HouseConstruction (P/T)}
\begin{center}
   \includegraphics[width=7.2cm]{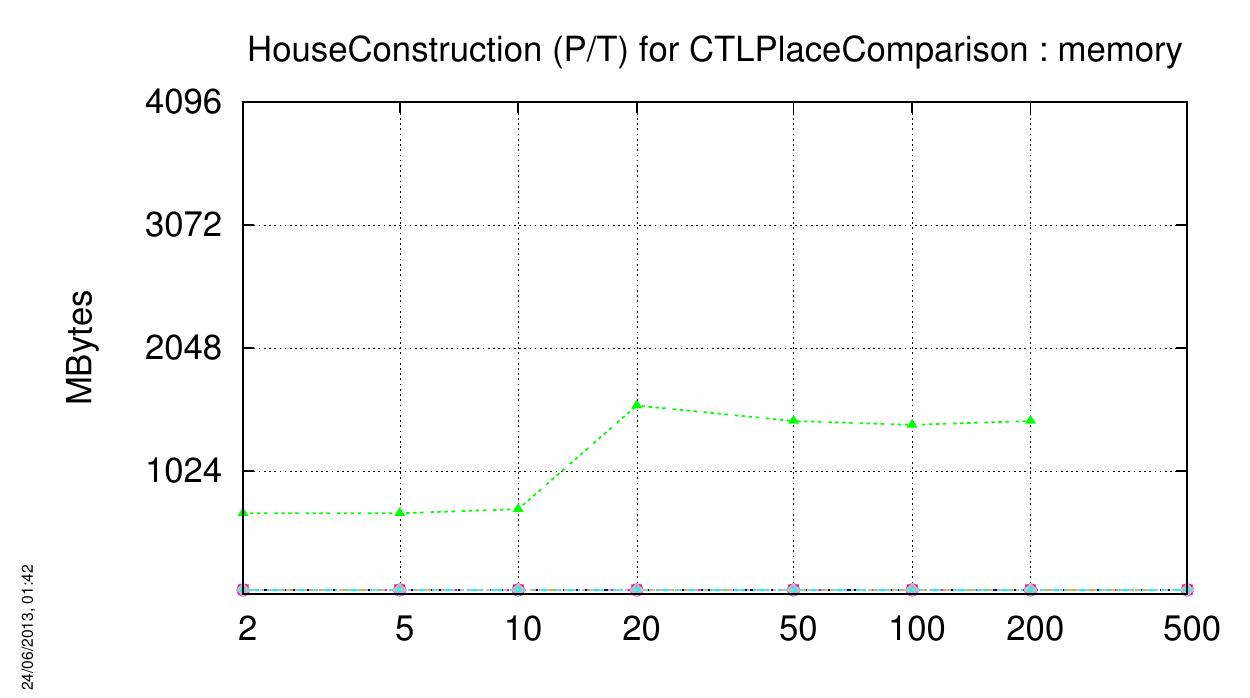}
   \includegraphics[width=7.2cm]{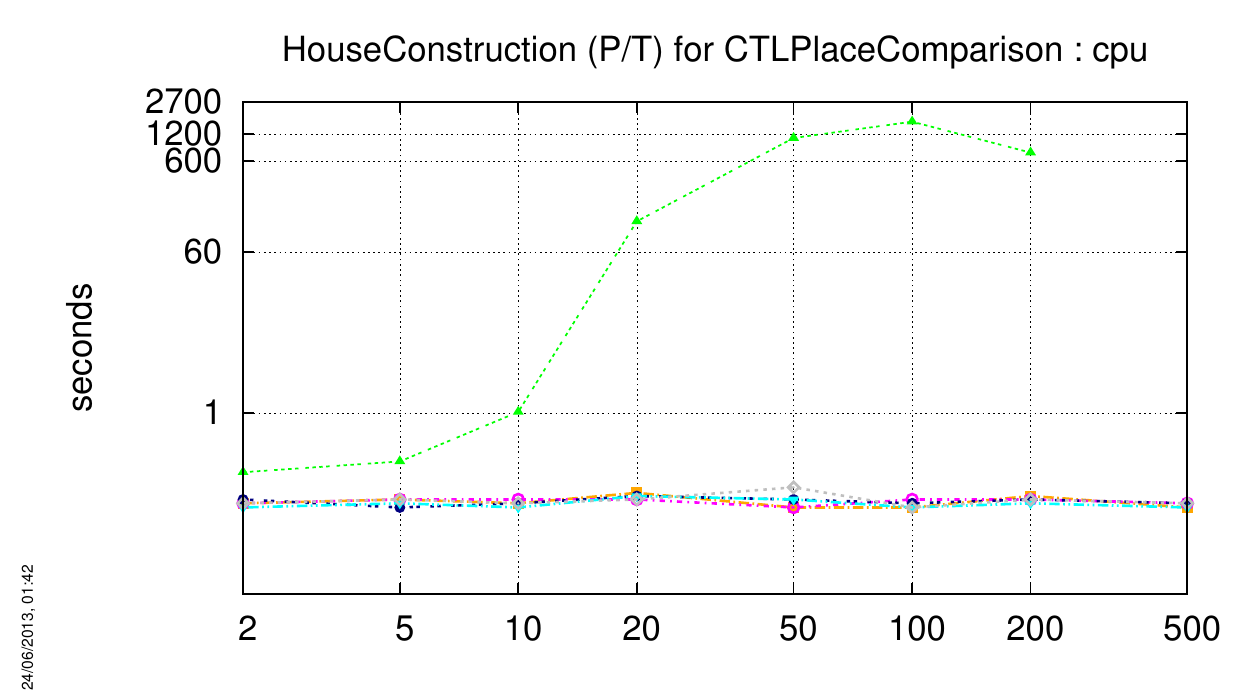}

   \includegraphics[height=1cm]{figures/tools-legend.pdf}
\end{center}

\subsubsection{\acs{IBMB2S565S3960-PT}}
The charts below respectively show how tools compete with this ``Suprise'' model (memory and CPU).

\index{Performances!CTLPlaceComparison!IBMB2S565S3960 (P/T)}
\begin{center}
   \includegraphics[width=7.2cm]{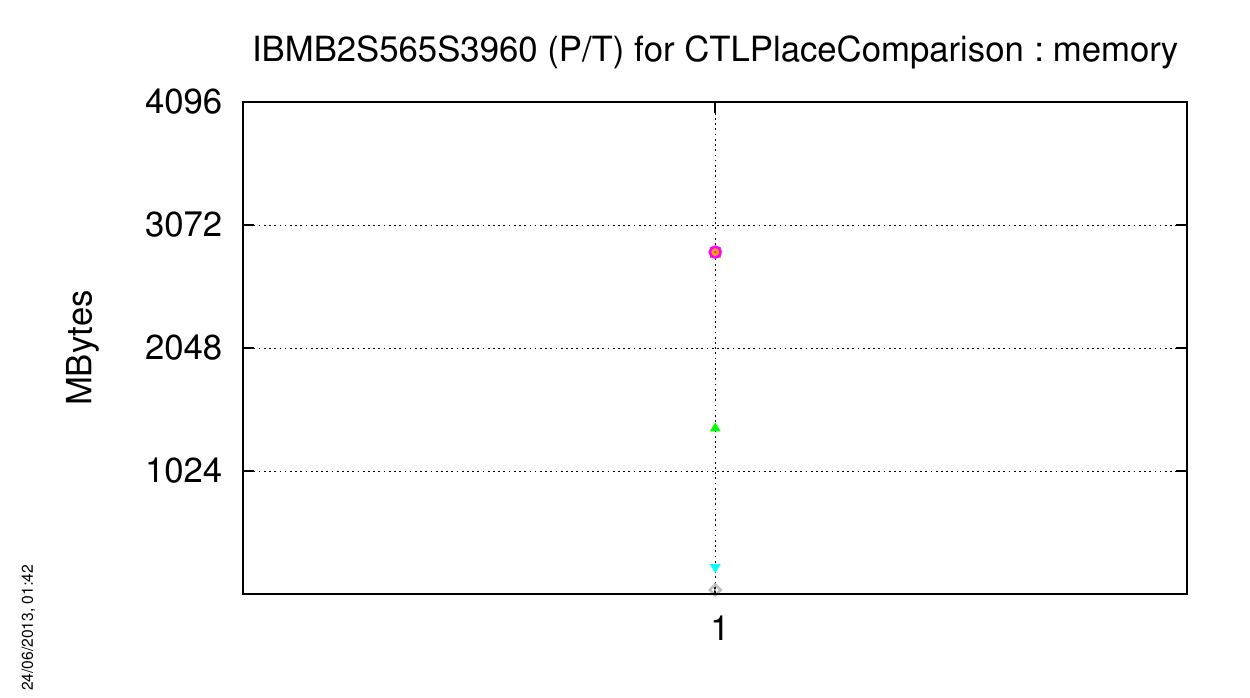}
   \includegraphics[width=7.2cm]{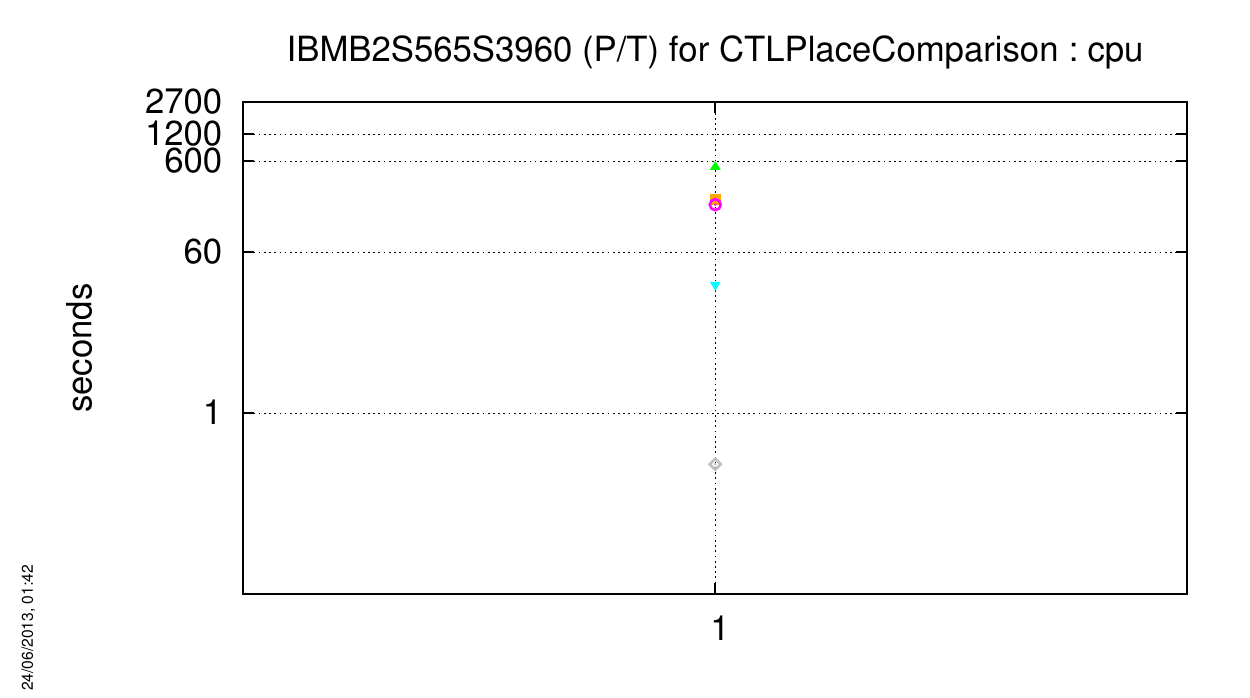}

   \includegraphics[height=1cm]{figures/tools-legend.pdf}
\end{center}

\subsubsection{\acs{QuasiCertifProtocol-COL}}
No instance of this model could be computed for the \textbf{CTLPlaceComparison} examination.

\subsubsection{\acs{QuasiCertifProtocol-PT}}
The charts below respectively show how tools compete with this ``Suprise'' model (memory and CPU).

\index{Performances!CTLPlaceComparison!QuasiCertifProtocol (P/T)}
\begin{center}
   \includegraphics[width=7.2cm]{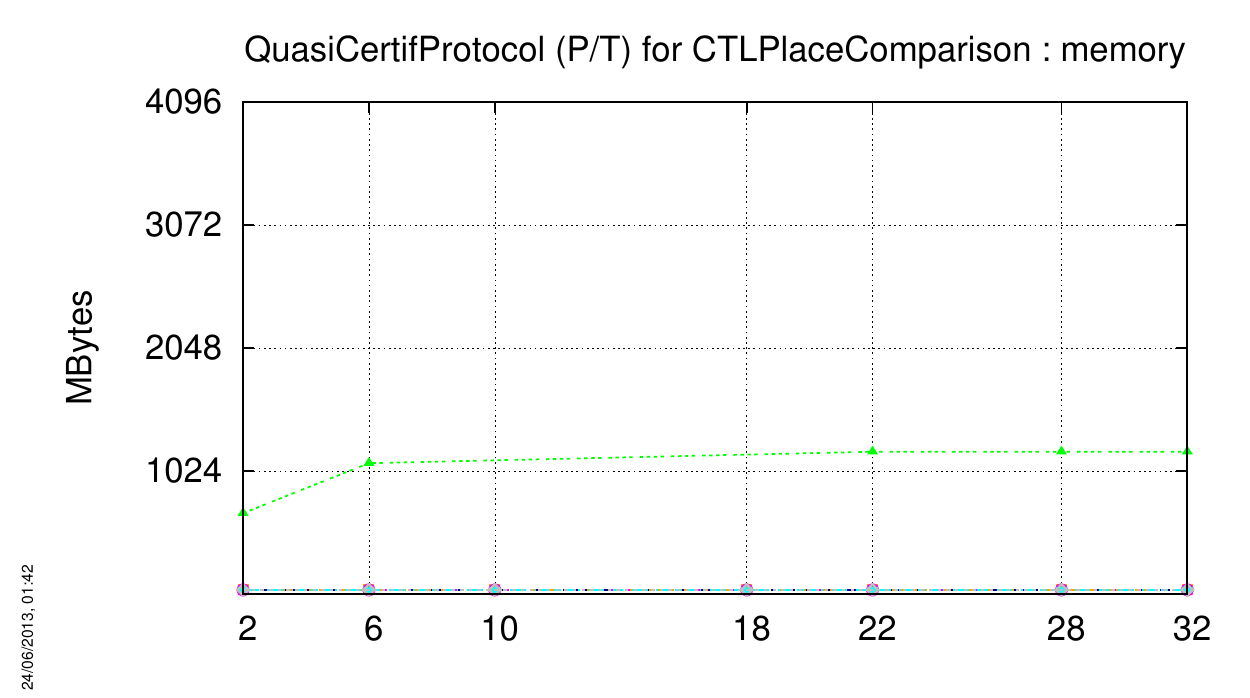}
   \includegraphics[width=7.2cm]{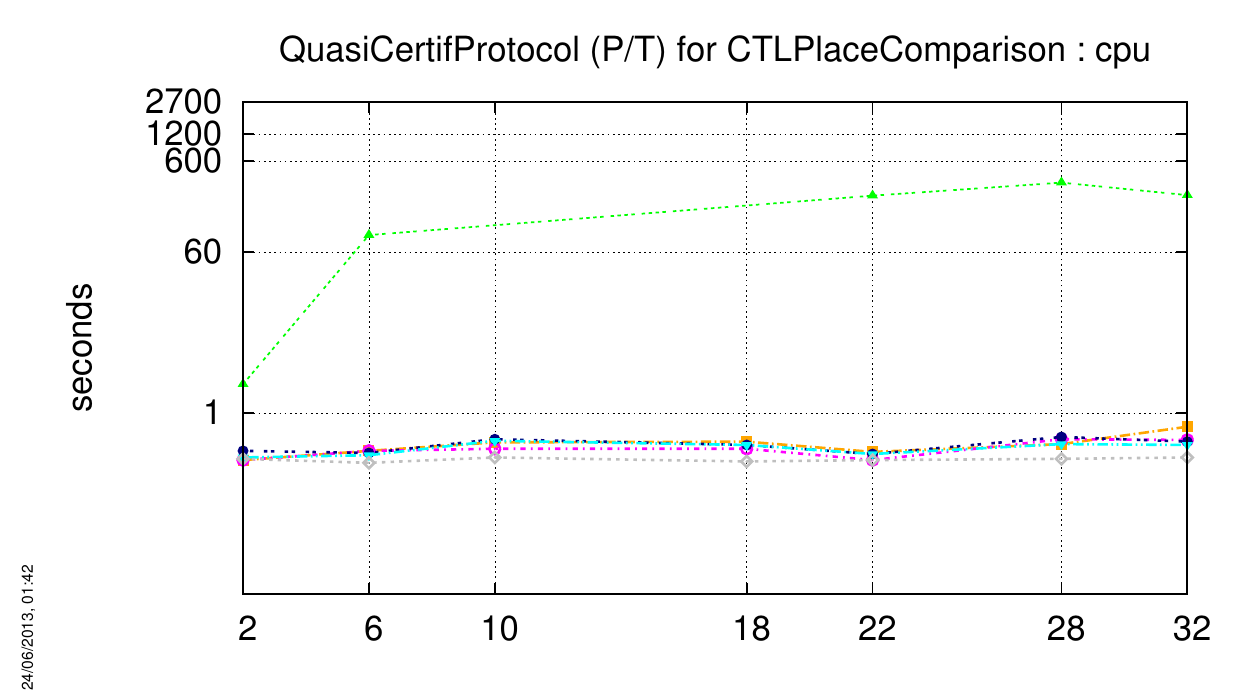}

   \includegraphics[height=1cm]{figures/tools-legend.pdf}
\end{center}

\subsubsection{\acs{Vasy2003-PT}}
The charts below respectively show how tools compete with this ``Suprise'' model (memory and CPU).

\index{Performances!CTLPlaceComparison!Vasy2003 (P/T)}
\begin{center}
   \includegraphics[width=7.2cm]{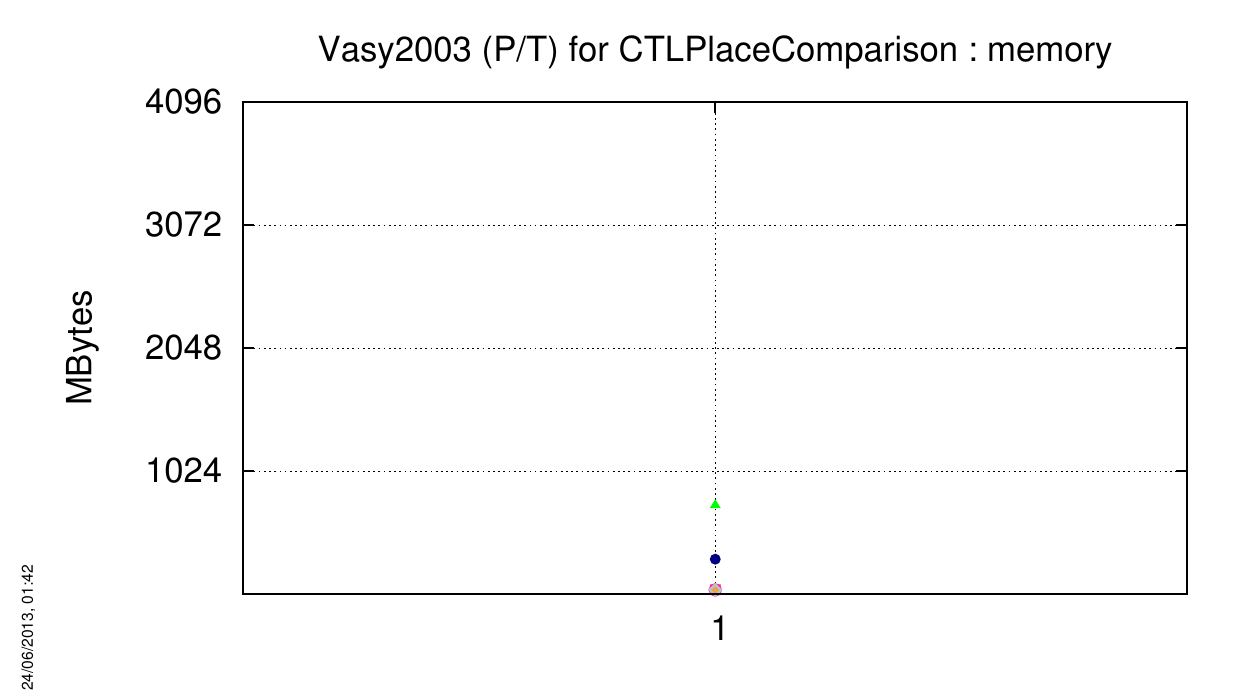}
   \includegraphics[width=7.2cm]{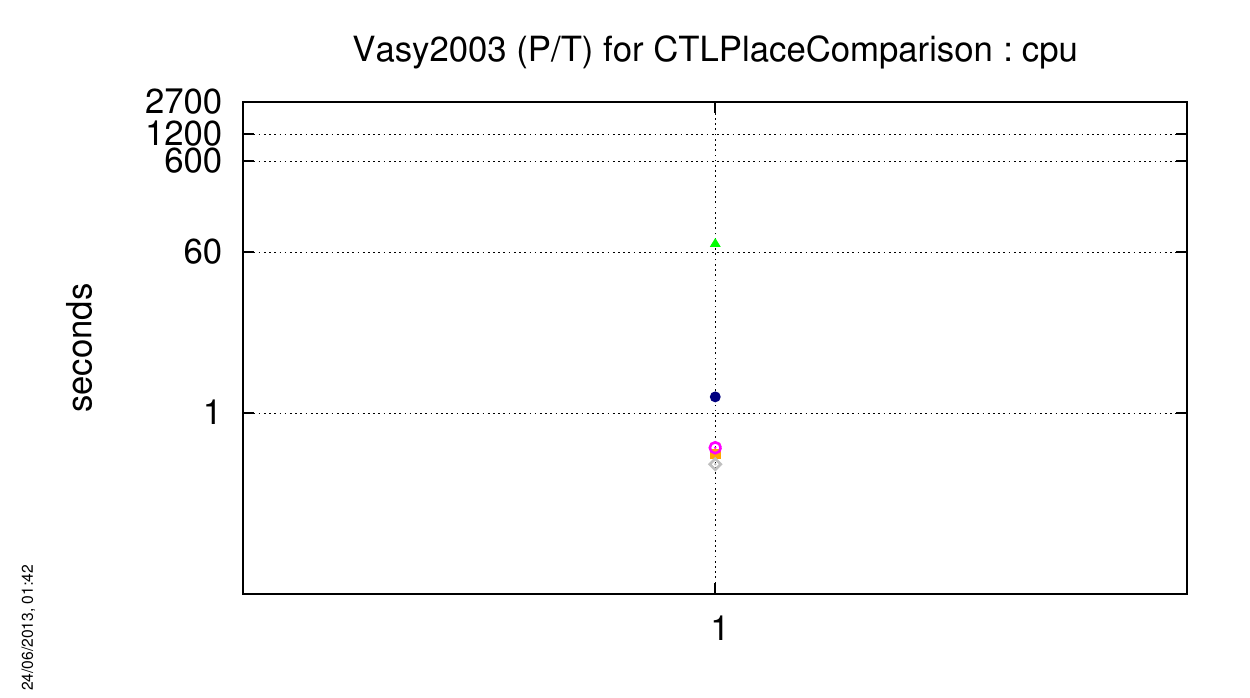}

   \includegraphics[height=1cm]{figures/tools-legend.pdf}
\end{center}

\subsection{Outputs for the CTLPlaceComparison Examination}
\index{Outputs!CTLPlaceComparison}

Please find enclosed the brute results for this examination (``Known'' and ``Surprise'' models).
We display only the score of tools that provide a results for at least one instance of one model.
The legend for the values is provided below:
\begin{itemize}
   \item\textbf{nc}: the tool does not compete this examination for this model/instance,
   \item\textbf{cc}: the tool cannot compute this examination for this model/instance,
   \item\textbf{to}: the tool cannot compute this examination for this model/instance within the maximum allowed time,
   \item\textbf{mp}: the tool encountered a memory problem (stack overflow or memory full),
   \item\textbf{nf}: there is no formula available for this type of examination (typically, this concerns P/T nets where
       comparing marking cardinality has no signification when there is no equivalent colored net).
\end{itemize}

\textbf{Note on the display of results for formulas:} each formula is considered as a flag (F if false, T if true, - or ?
when the value cannot be determined). These values are concatenated in the order they appear (we assume it is the order of formulas as they were provided).

\subsubsection{``Known'' Models}

\input{result_known_CTLPlaceComparison.tex}

\subsubsection{``Surprise'' Models}

\input{result_surprise_CTLPlaceComparison.tex}

\subsection{Score for the CTLPlaceComparison Examination}
\index{Scores!CTLPlaceComparison}

Please find enclosed the scores for this examination (``Known'' and ``Surprise'' models).
We display only the score of tools that provide a results for at least one instance of one model.
The total is first listed in the table below followed by a detail, for each proposed model.
Meaning of the line labels are:
\begin{itemize}
\item\textbf{1st instance}: the tool gets a bonus for having processed the first instance of this model (+1 point),
\item\textbf{instances}: the tool gets 1 point per instances treated 
(for that, we assume that at least one formula has been successfully computed),
\item\textbf{max reached}: the tool could process all the instances for the model (+2 points),
\item\textbf{best}: the tool is among the ones that processed a maximum of instances within the time and memory confinement (+2 points).
\end{itemize}

\subsubsection{``Known'' Models}

\input{score_known_CTLPlaceComparison.tex}

\subsubsection{``Surprise'' Models}

\input{score_surprise_CTLPlaceComparison.tex}

\subsection{Trophies for this Examination}
\index{Trophies!CTLPlaceComparison}

Trophies are divided in three categories: ``Known'' models,
``Surprise'' models, and the global trophies (formula is then
$score_{global} = score_{known} + 2 \times score_{surprise}$).

\subsubsection{For ``Known'' Models} \ \\

\begin{tabular}{c|c|c}
      1 & 2 & 3 \\
   \includegraphics[width=2cm]{figures/gold.jpg} &
   \includegraphics[width=2cm]{figures/silver.jpg} &
   \includegraphics[width=2cm]{figures/bronse.jpg} \\
   \acs{lola} &
   \acs{lola-optimistic} &
   \acs{lola-optimistic-incomplete} \\
   192 points &
   182 points &
   135 points \\
\end{tabular}

\subsubsection{For ``Surprise'' Models}\  \\

\begin{tabular}{c|c|c|c}
      1 & 2 & 2 & 2 \\
   \includegraphics[width=2cm]{figures/gold.jpg} &
   \includegraphics[width=2cm]{figures/silver.jpg} &
   \includegraphics[width=2cm]{figures/silver.jpg} &
   \includegraphics[width=2cm]{figures/silver.jpg} \\
   \acs{marcie} &
   \acs{lola} &
   \acs{lola-optimistic} &
   \acs{lola-optimistic-incomplete} \\
   22 points &
   6 points &
   6 points &
   6 points \\
\end{tabular}

\subsubsection{Global} \ \\

\begin{tabular}{c|c|c}
      1 & 2 & 3 \\
   \includegraphics[width=2cm]{figures/gold.jpg} &
   \includegraphics[width=2cm]{figures/silver.jpg} &
   \includegraphics[width=2cm]{figures/bronse.jpg} \\
   \acs{lola} &
   \acs{lola-optimistic} &
   \acs{lola-optimistic-incomplete} \\
   204 points &
   199 points &
   147 points \\
\end{tabular}

\newpage

\section{The CTLMix Examination}
\label{sec:exam:CTLMix}
\index{Results!CTLMix}

This examination deals with CTL properties dealing with all the previous type of atomic proposition.
We first show a summary on the handling of models by the participating tools.
Then, we present the computed outputs and the associated scores for this
examination prior to a summary of relevant executions.

\subsection{Handling of Models by Tools}
\index{Performances!CTLMix}

\subsubsection{\acs{CSRepetitions-COL}}
No instance of this model could be computed for the \textbf{CTLMix} examination.

\subsubsection{\acs{CSRepetitions-PT}}
The charts below respectively show how tools compete with this ``Known'' model (memory and CPU).

\index{Performances!CTLMix!CSRepetitions (P/T)}
\begin{center}
   \includegraphics[width=7.2cm]{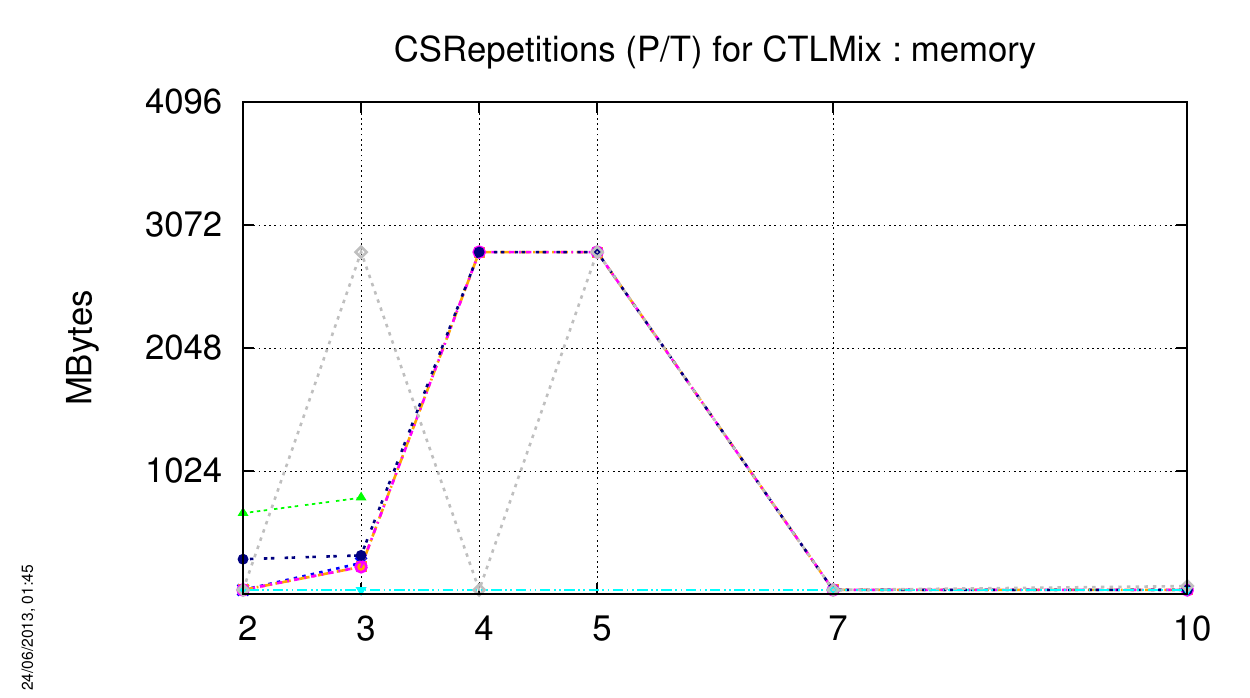}
   \includegraphics[width=7.2cm]{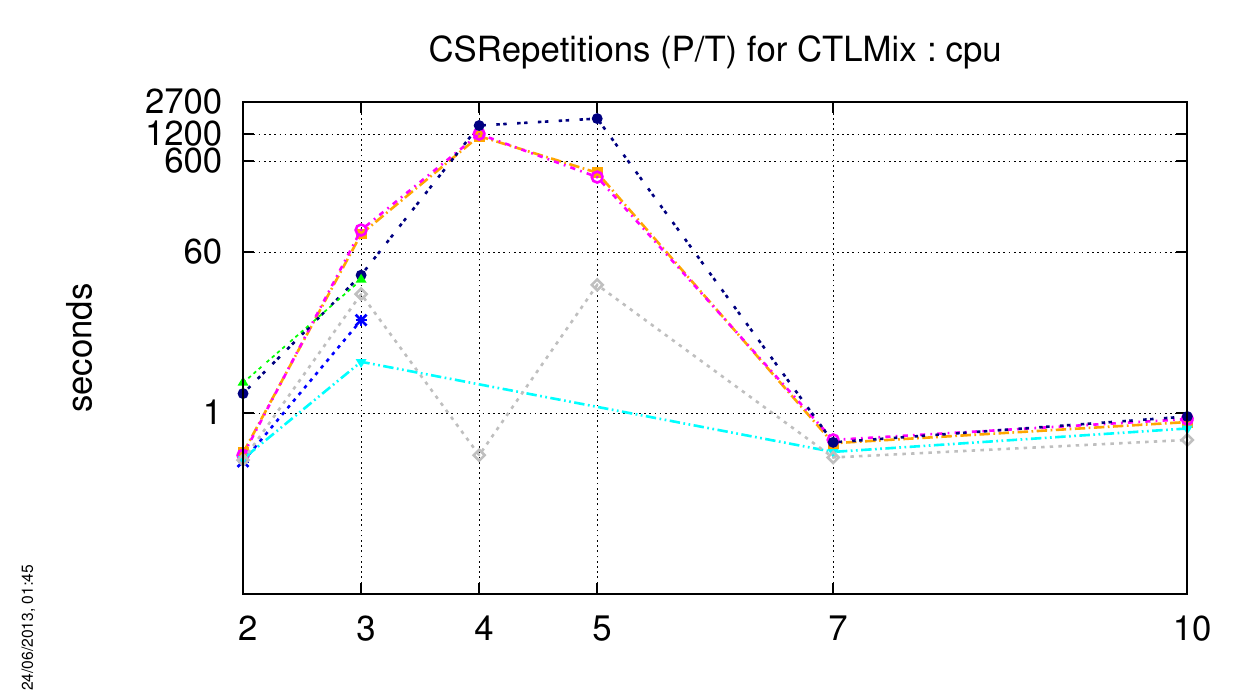}

   \includegraphics[height=1cm]{figures/tools-legend.pdf}
\end{center}

\subsubsection{\acs{Dekker-PT}}
The charts below respectively show how tools compete with this ``Known'' model (memory and CPU).

\index{Performances!CTLMix!Dekker (P/T)}
\begin{center}
   \includegraphics[width=7.2cm]{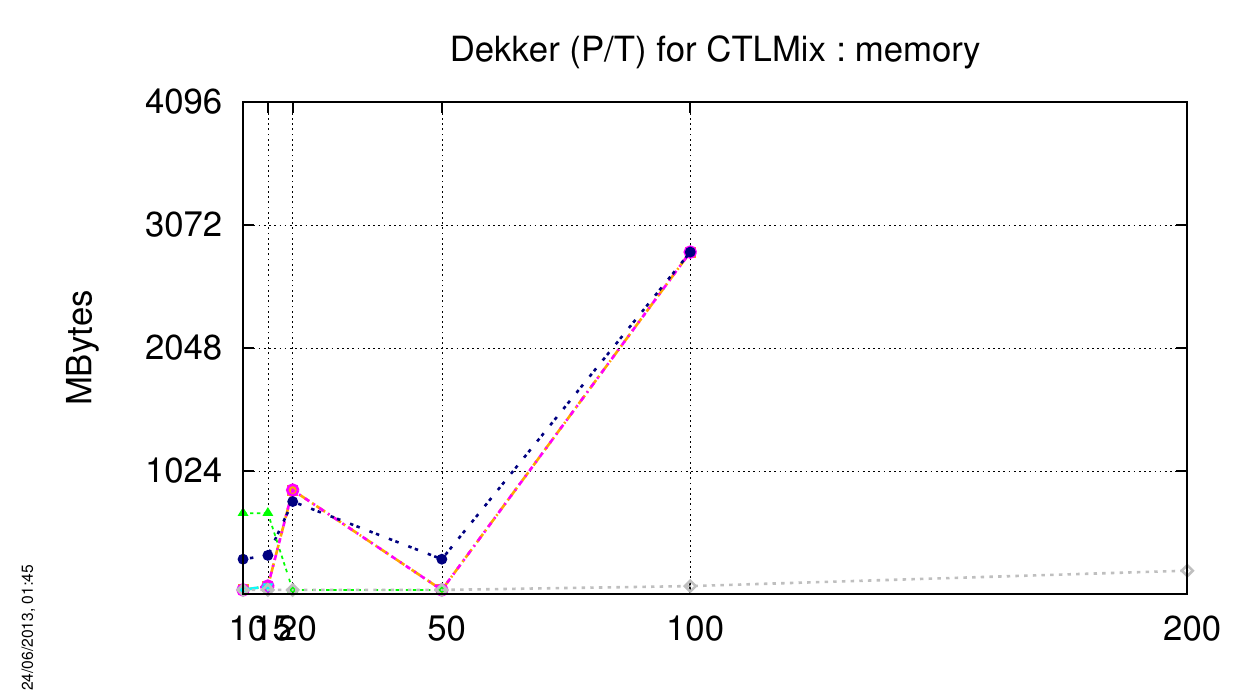}
   \includegraphics[width=7.2cm]{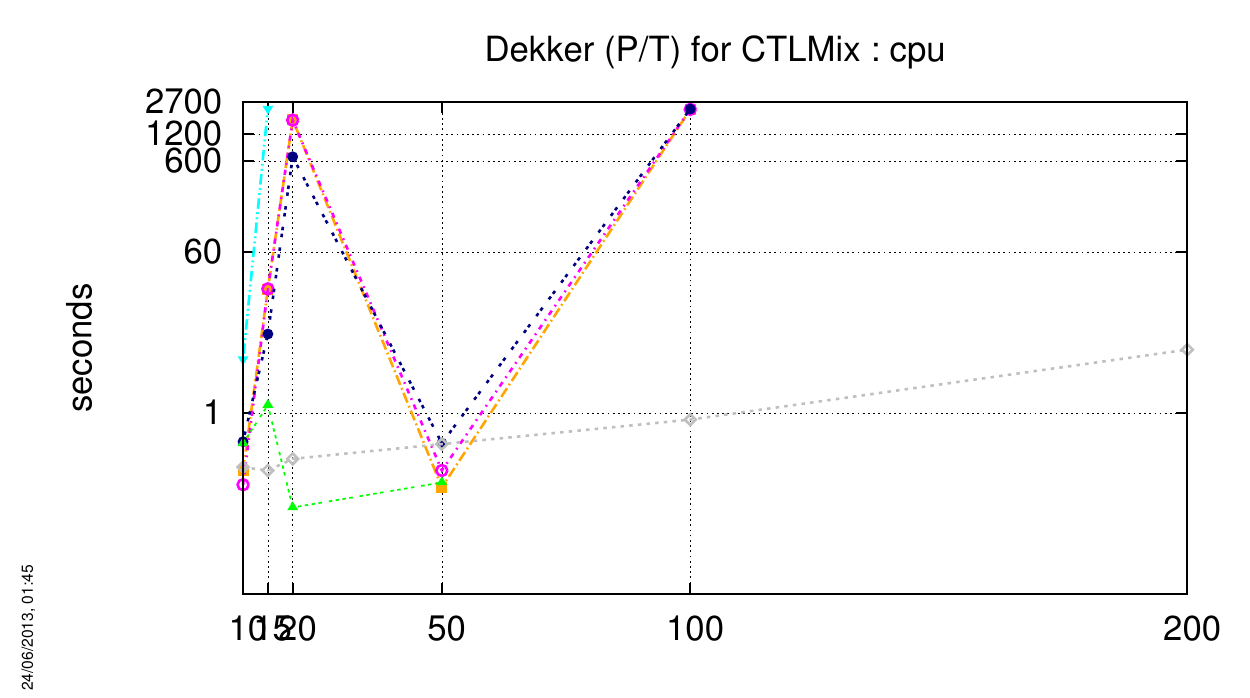}

   \includegraphics[height=1cm]{figures/tools-legend.pdf}
\end{center}

\subsubsection{\acs{DotAndBoxes-COL}}
No instance of this model could be computed for the \textbf{CTLMix} examination.

\subsubsection{\acs{DrinkVendingMachine-COL}}
No instance of this model could be computed for the \textbf{CTLMix} examination.

\subsubsection{\acs{DrinkVendingMachine-PT}}
The charts below respectively show how tools compete with this ``Known'' model (memory and CPU).

\index{Performances!CTLMix!DrinkVendingMachine (P/T)}
\begin{center}
   \includegraphics[width=7.2cm]{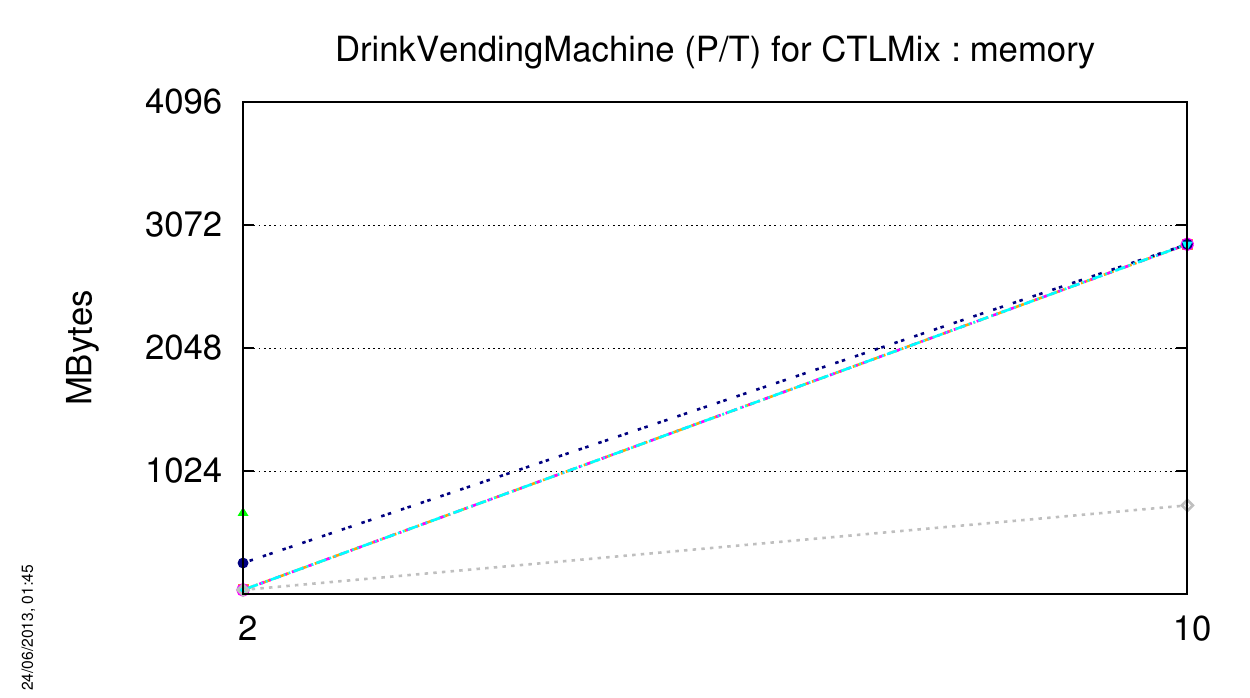}
   \includegraphics[width=7.2cm]{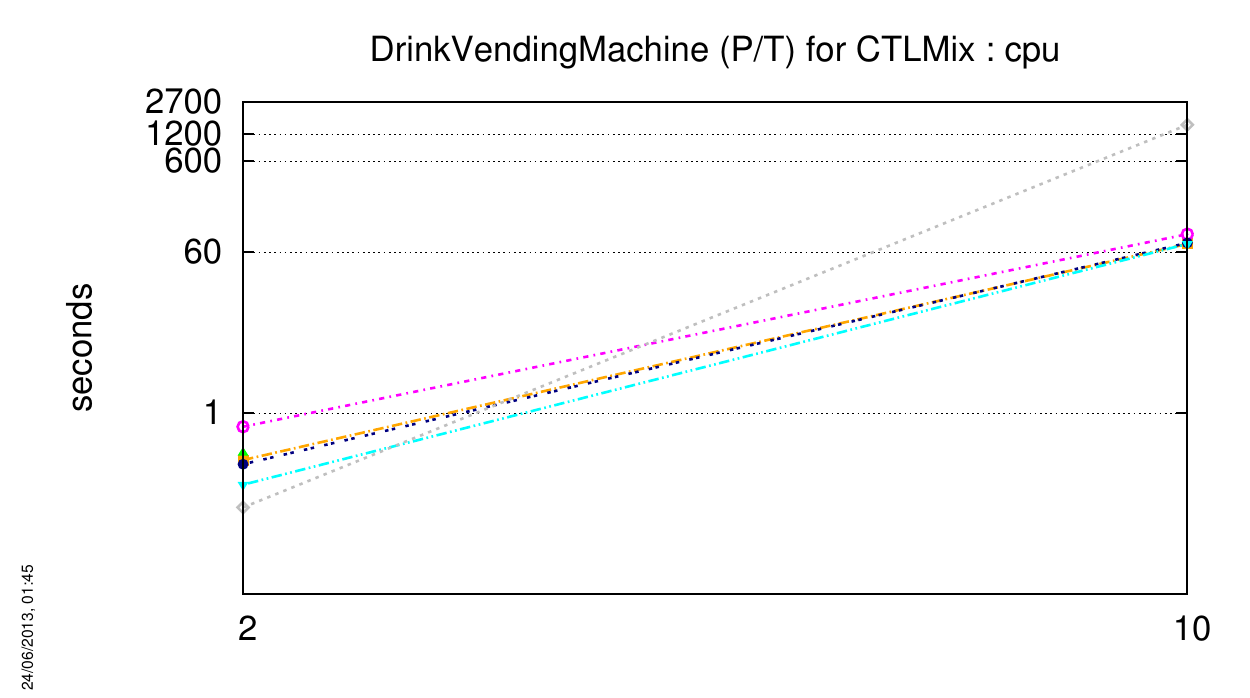}

   \includegraphics[height=1cm]{figures/tools-legend.pdf}
\end{center}

\subsubsection{\acs{Echo-PT}}
The charts below respectively show how tools compete with this ``Known'' model (memory and CPU).

\index{Performances!CTLMix!Echo (P/T)}
\begin{center}
   \includegraphics[width=7.2cm]{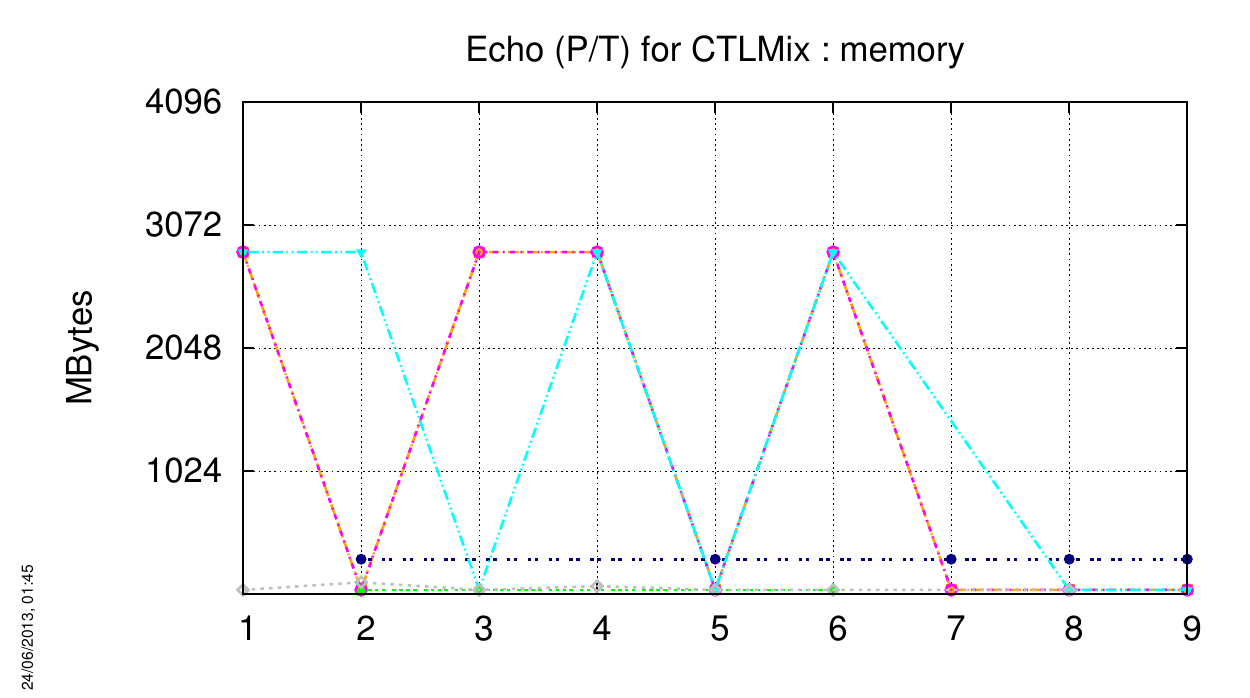}
   \includegraphics[width=7.2cm]{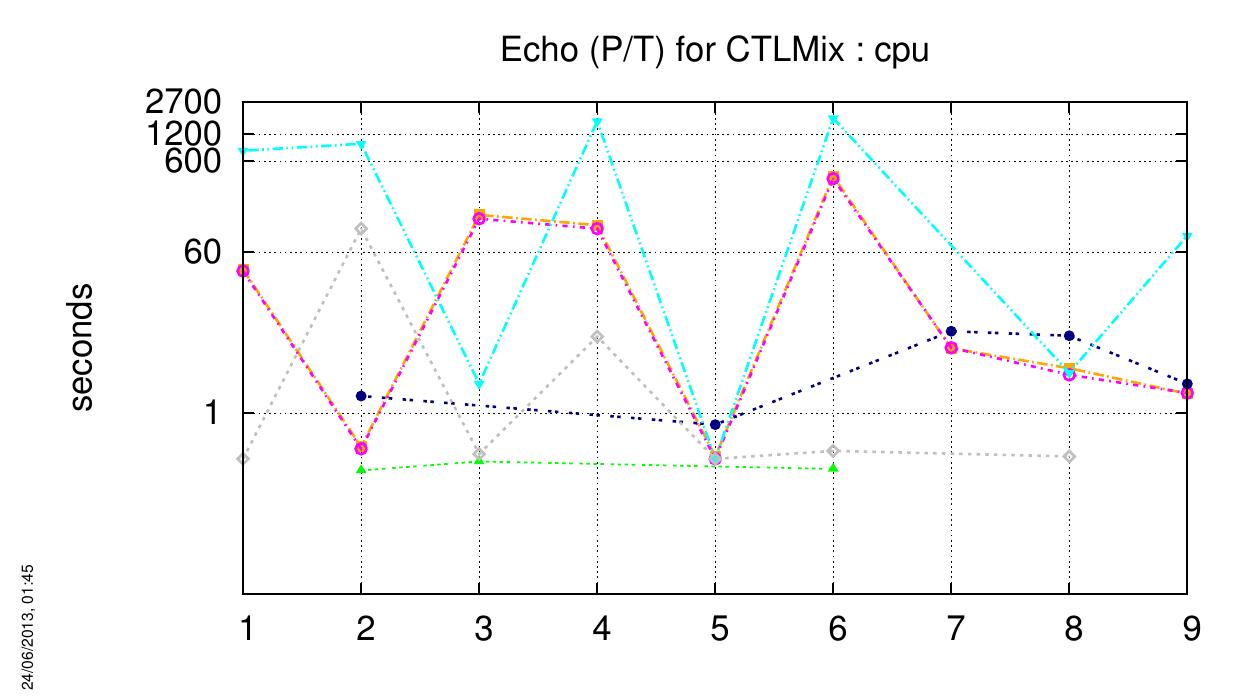}

   \includegraphics[height=1cm]{figures/tools-legend.pdf}
\end{center}

\subsubsection{\acs{Eratosthenes-PT}}
The charts below respectively show how tools compete with this ``Known'' model (memory and CPU).

\index{Performances!CTLMix!Eratosthenes (P/T)}
\begin{center}
   \includegraphics[width=7.2cm]{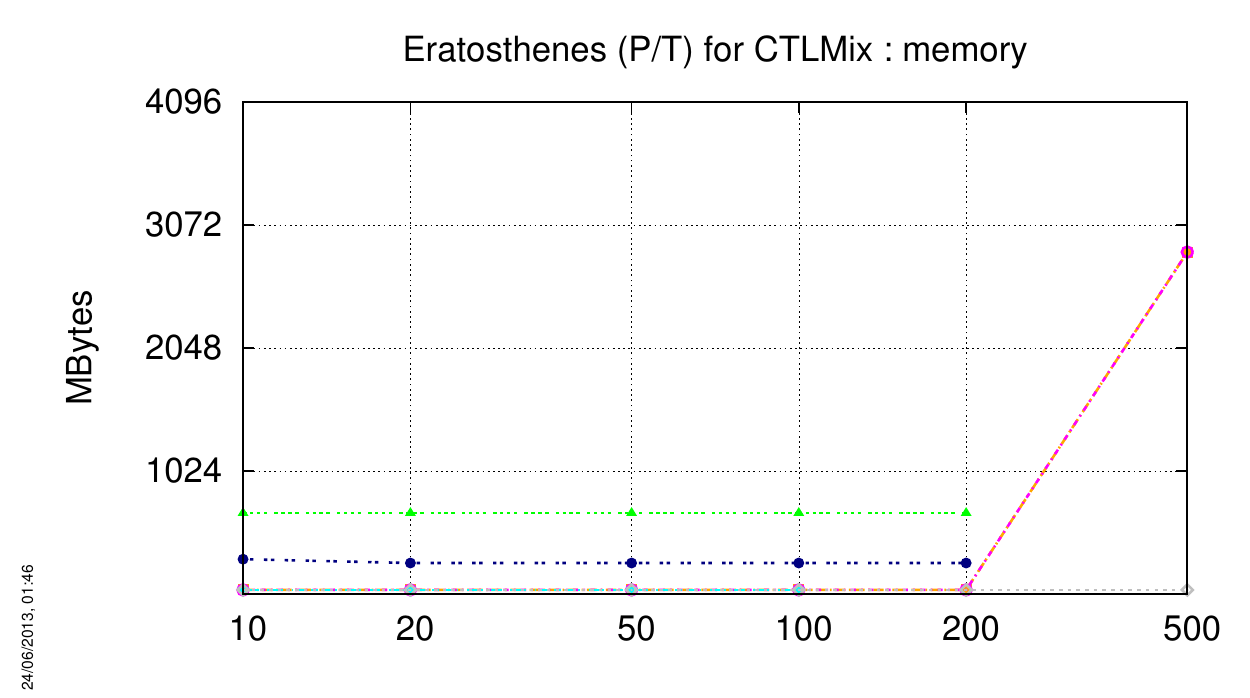}
   \includegraphics[width=7.2cm]{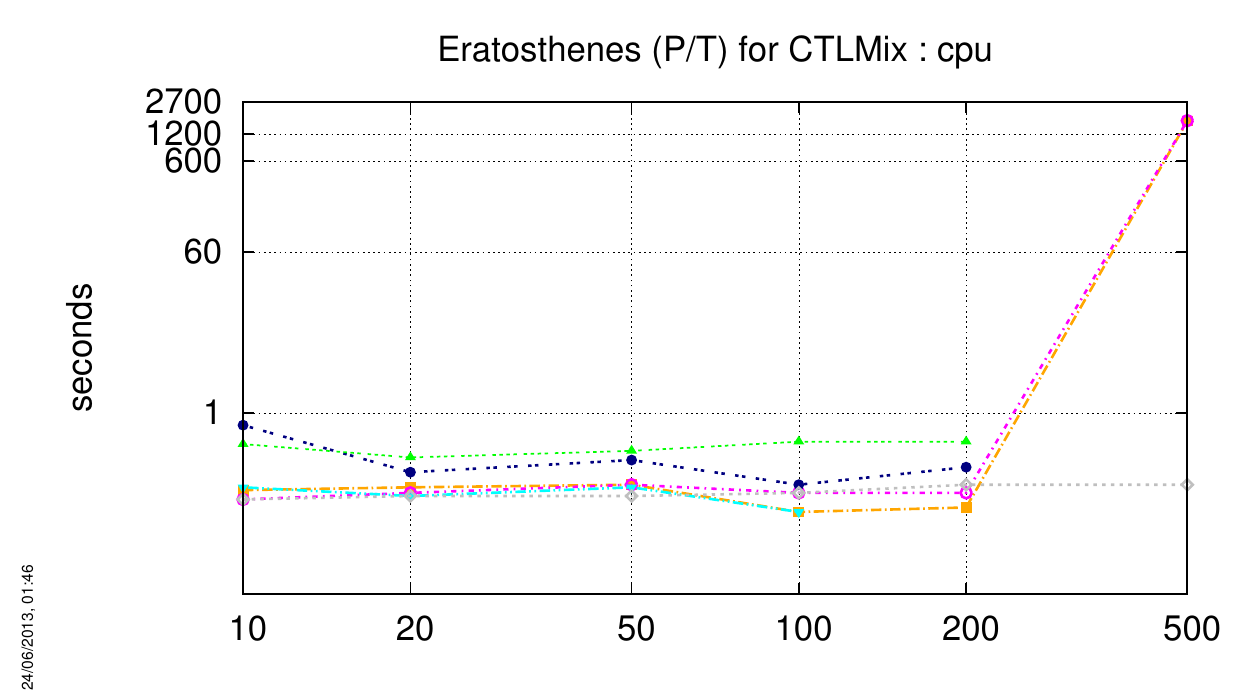}

   \includegraphics[height=1cm]{figures/tools-legend.pdf}
\end{center}

\subsubsection{\acs{FMS-PT}}
The charts below respectively show how tools compete with this ``Known'' model (memory and CPU).

\index{Performances!CTLMix!FMS (P/T)}
\begin{center}
   \includegraphics[width=7.2cm]{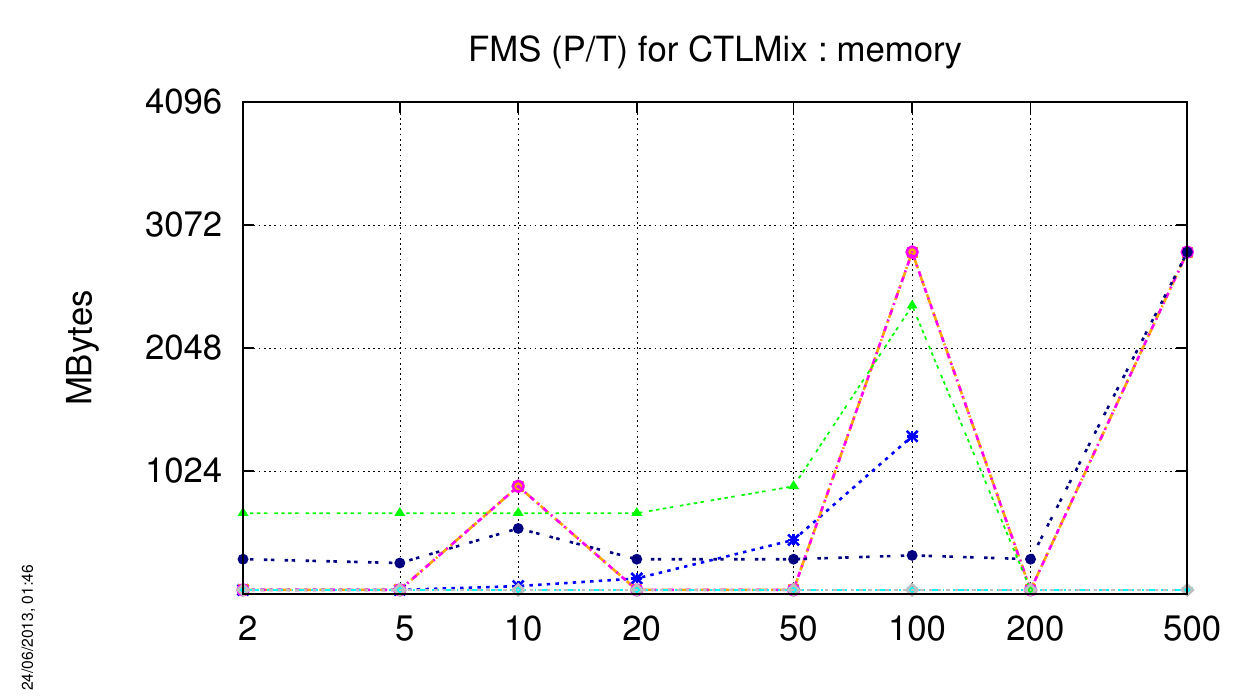}
   \includegraphics[width=7.2cm]{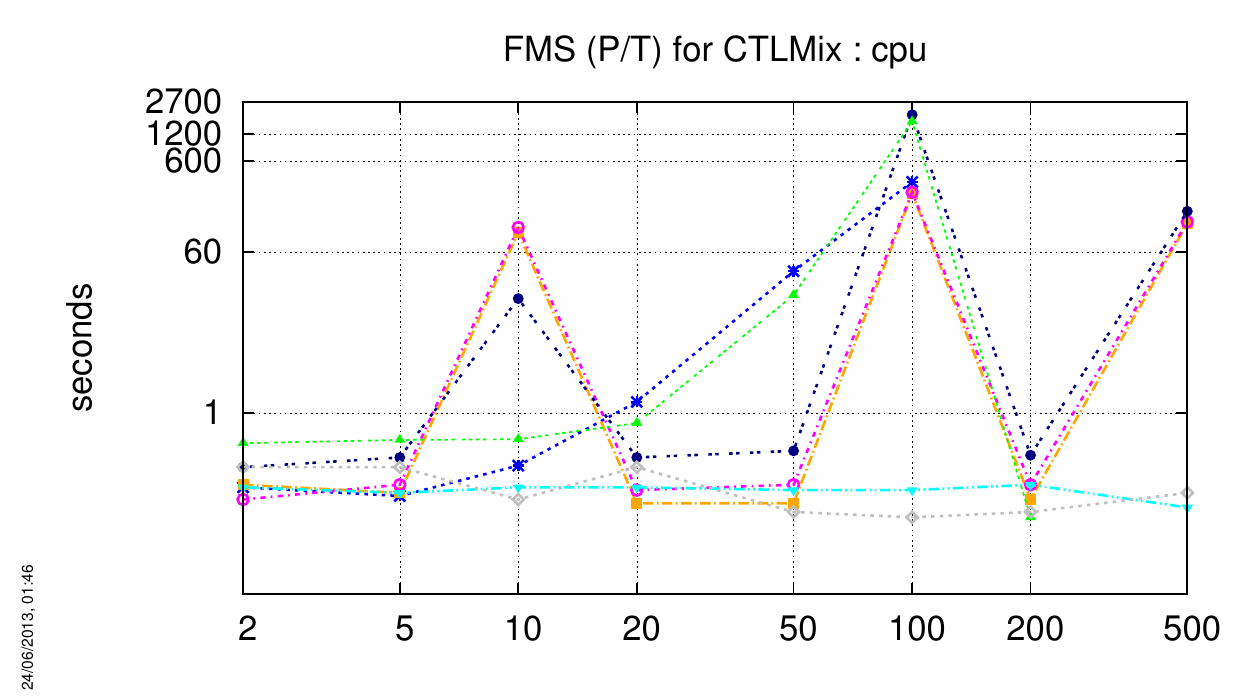}

   \includegraphics[height=1cm]{figures/tools-legend.pdf}
\end{center}

\subsubsection{\acs{GlobalRessAlloc-COL}}
No instance of this model could be computed for the \textbf{CTLMix} examination.

\subsubsection{\acs{GlobalRessAlloc-PT}}
The charts below respectively show how tools compete with this ``Known'' model (memory and CPU).

\index{Performances!CTLMix!GlobalRessAlloc (P/T)}
\begin{center}
   \includegraphics[width=7.2cm]{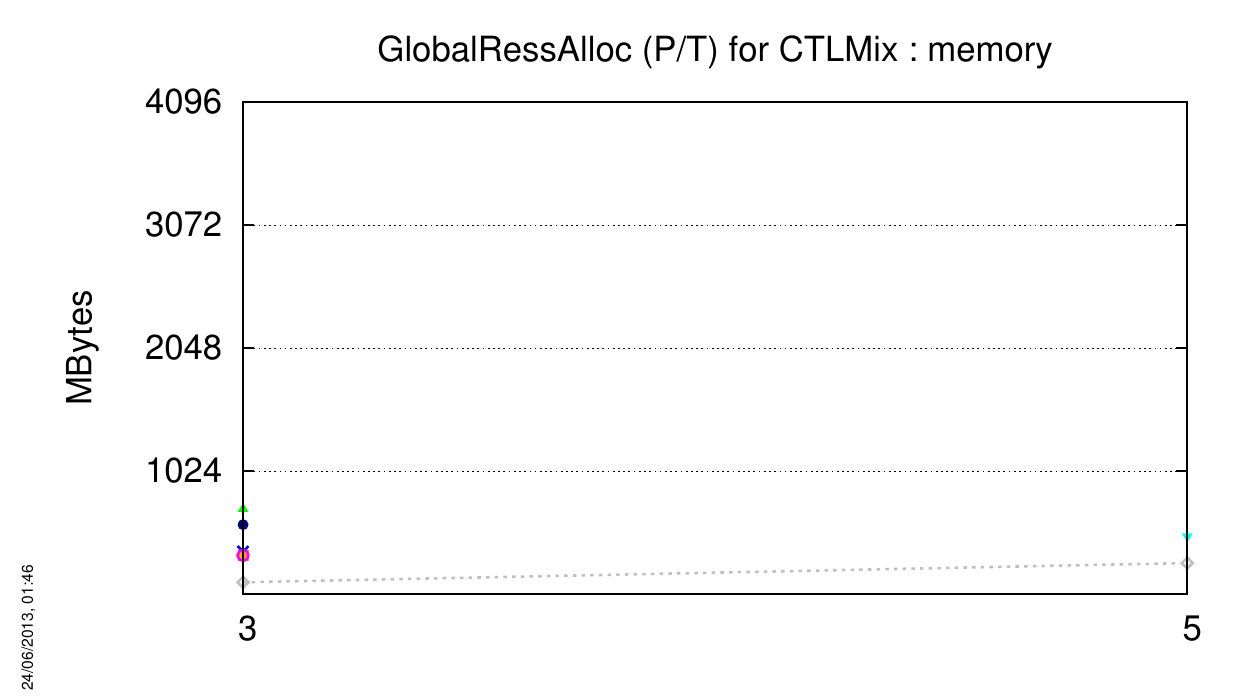}
   \includegraphics[width=7.2cm]{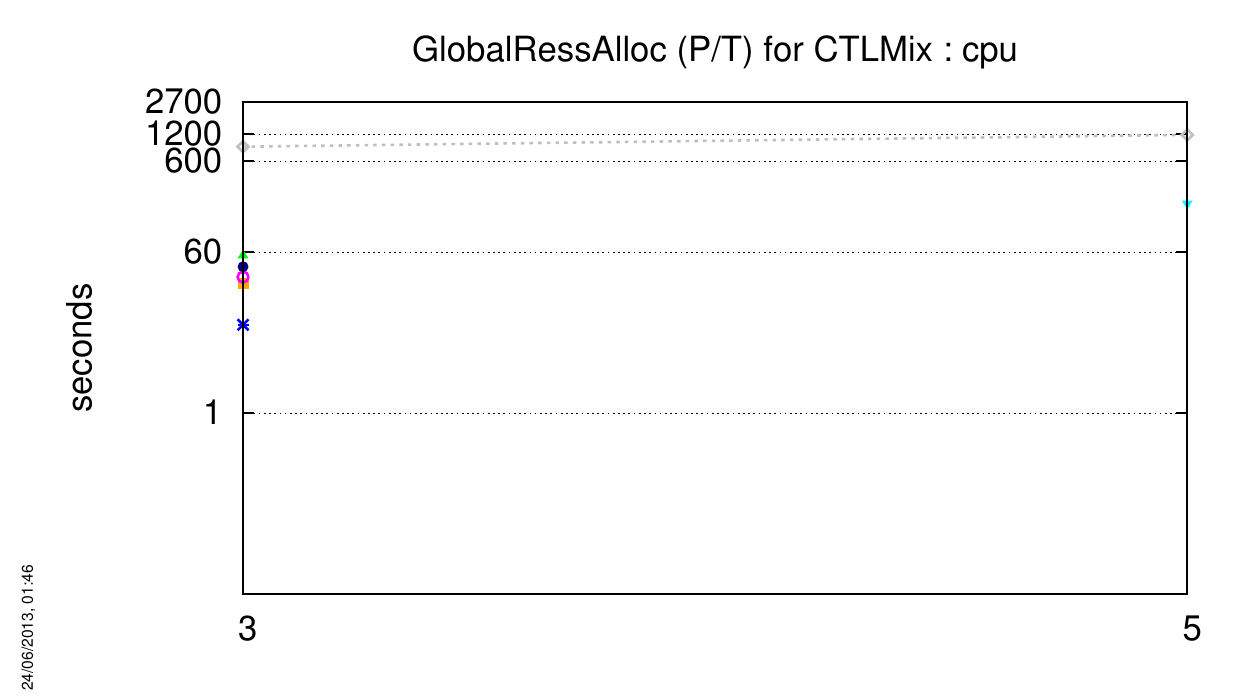}

   \includegraphics[height=1cm]{figures/tools-legend.pdf}
\end{center}

\subsubsection{\acs{Kanban-PT}}
The charts below respectively show how tools compete with this ``Known'' model (memory and CPU).

\index{Performances!CTLMix!Kanban (P/T)}
\begin{center}
   \includegraphics[width=7.2cm]{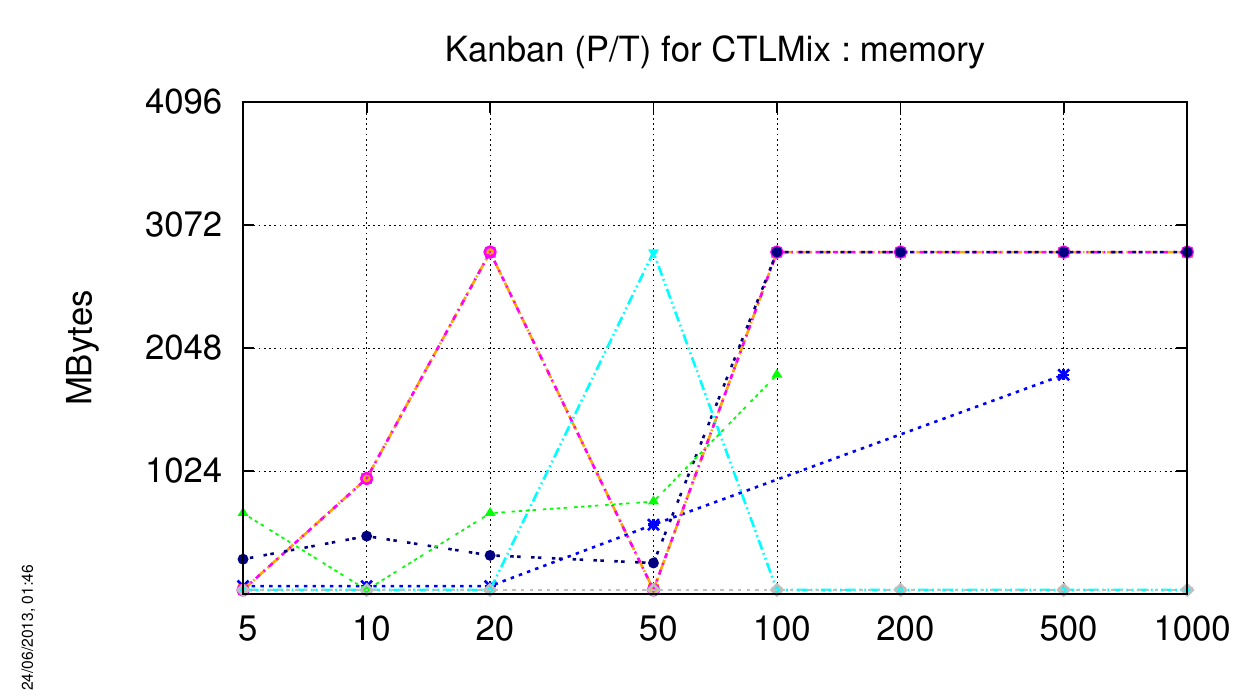}
   \includegraphics[width=7.2cm]{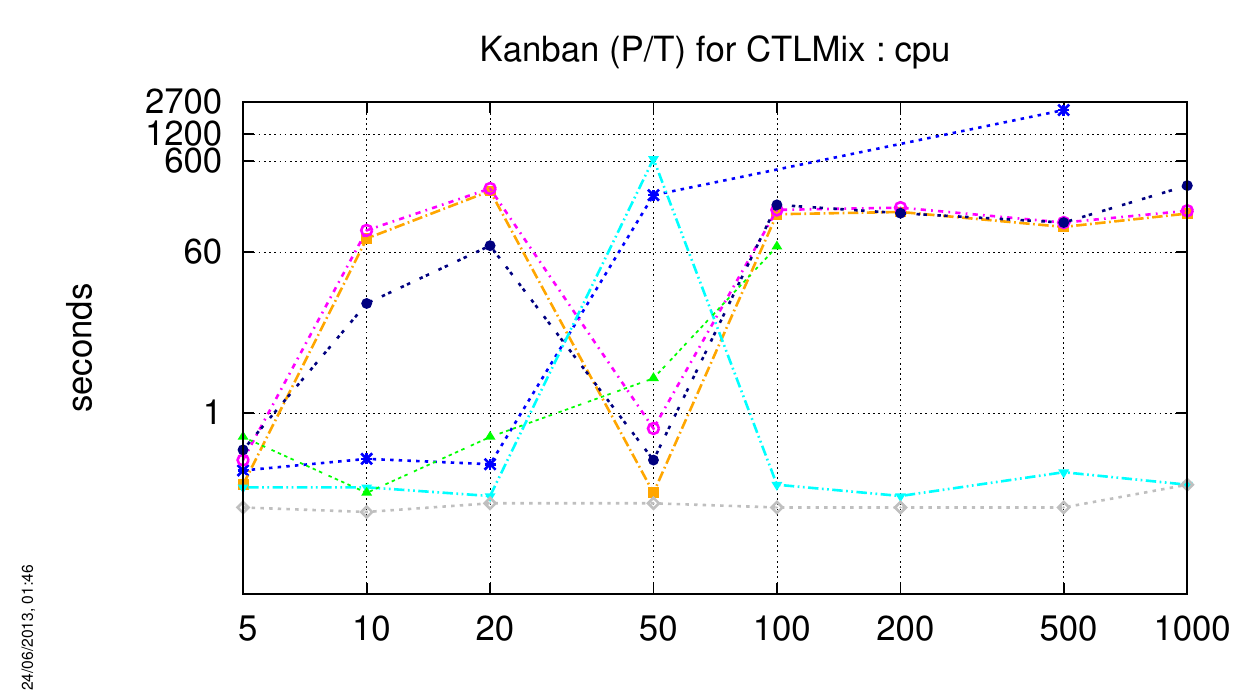}

   \includegraphics[height=1cm]{figures/tools-legend.pdf}
\end{center}

\subsubsection{\acs{LamportFastMutEx-COL}}
No instance of this model could be computed for the \textbf{CTLMix} examination.

\subsubsection{\acs{LamportFastMutEx-PT}}
The charts below respectively show how tools compete with this ``Known'' model (memory and CPU).

\index{Performances!CTLMix!LamportFastMutEx (P/T)}
\begin{center}
   \includegraphics[width=7.2cm]{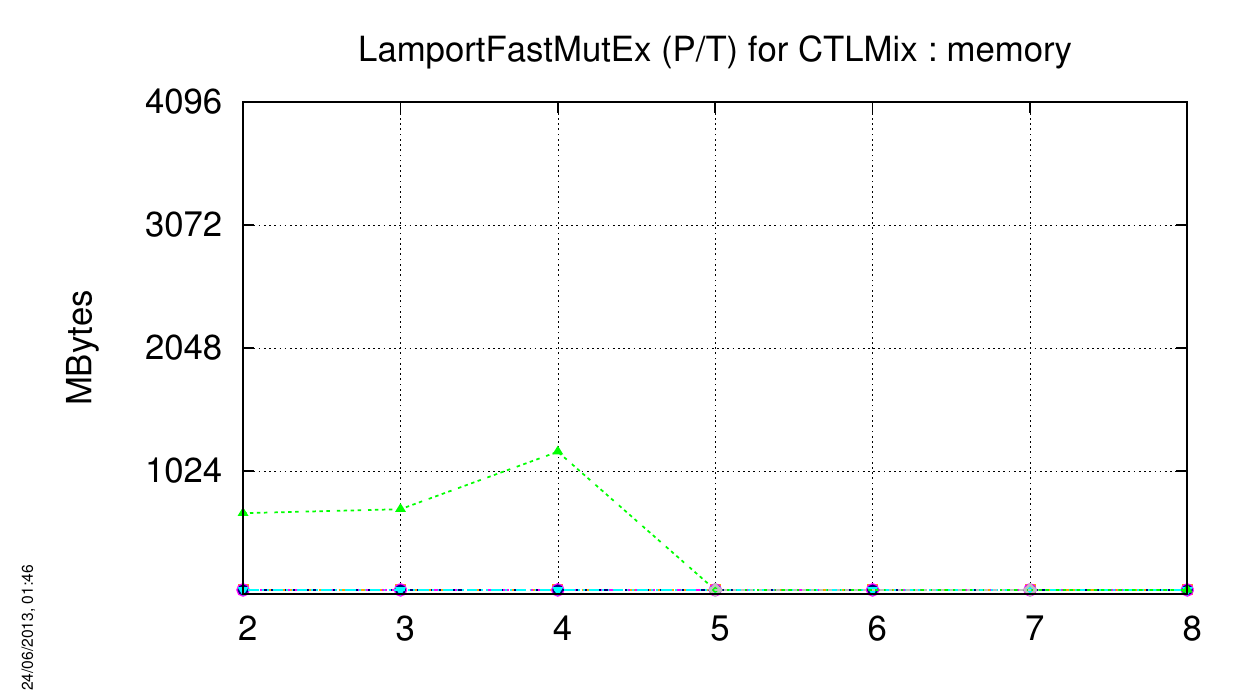}
   \includegraphics[width=7.2cm]{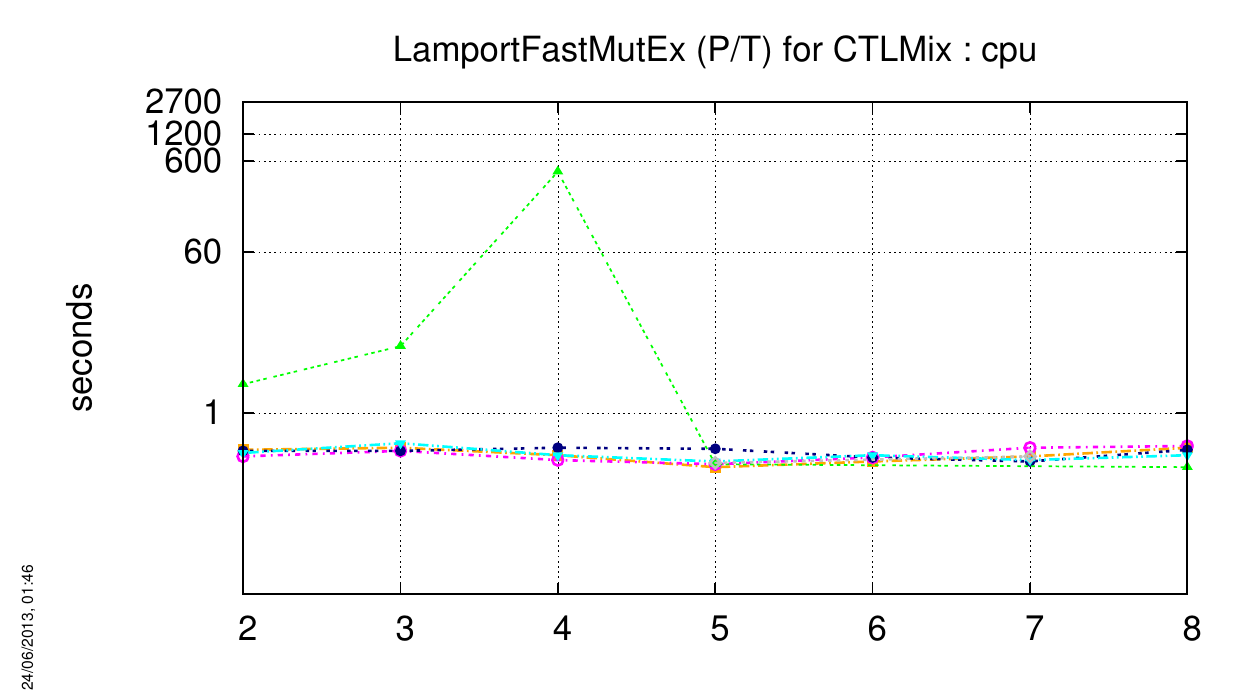}

   \includegraphics[height=1cm]{figures/tools-legend.pdf}
\end{center}

\subsubsection{\acs{MAPK-PT}}
The charts below respectively show how tools compete with this ``Known'' model (memory and CPU).

\index{Performances!CTLMix!MAPK (P/T)}
\begin{center}
   \includegraphics[width=7.2cm]{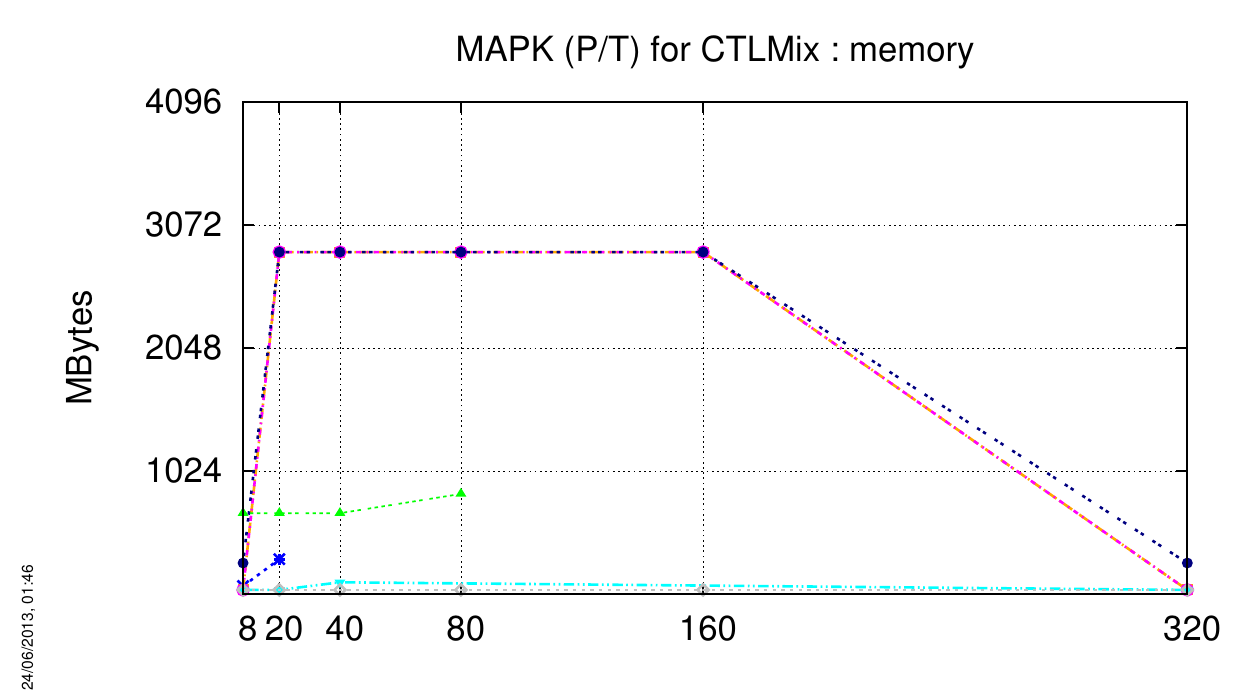}
   \includegraphics[width=7.2cm]{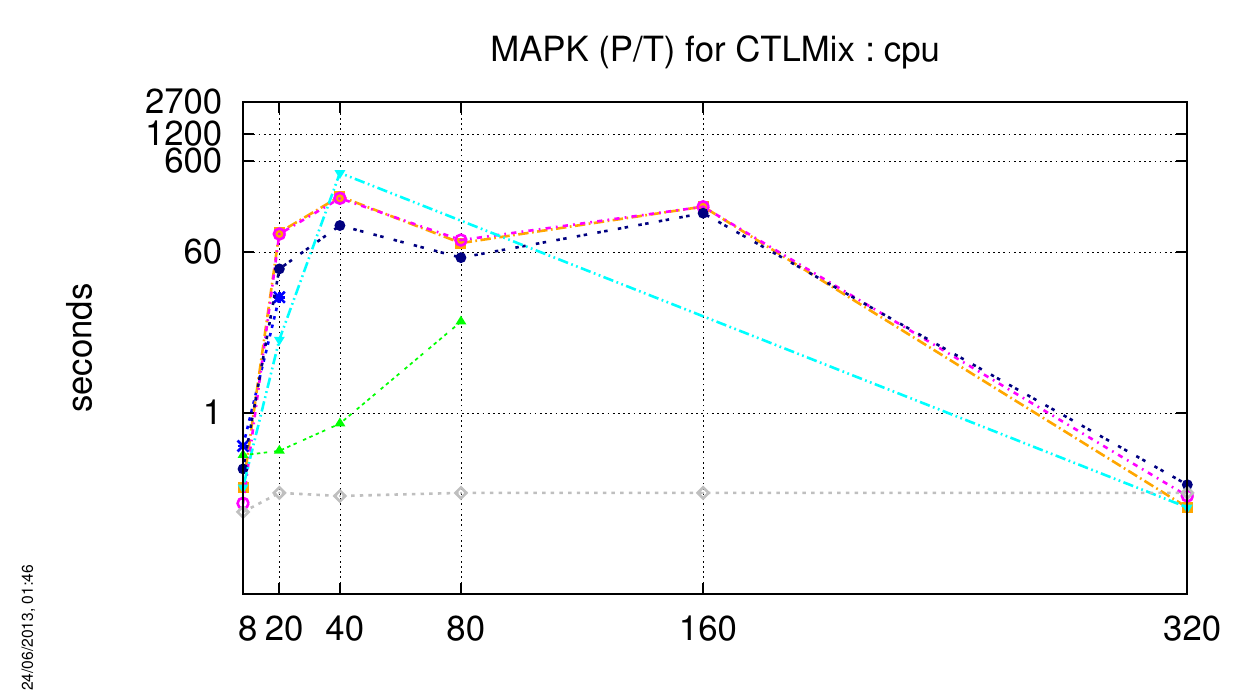}

   \includegraphics[height=1cm]{figures/tools-legend.pdf}
\end{center}

\subsubsection{\acs{NeoElection-COL}}
No instance of this model could be computed for the \textbf{CTLMix} examination.

\subsubsection{\acs{NeoElection-PT}}
The charts below respectively show how tools compete with this ``Known'' model (memory and CPU).

\index{Performances!CTLMix!NeoElection (P/T)}
\begin{center}
   \includegraphics[width=7.2cm]{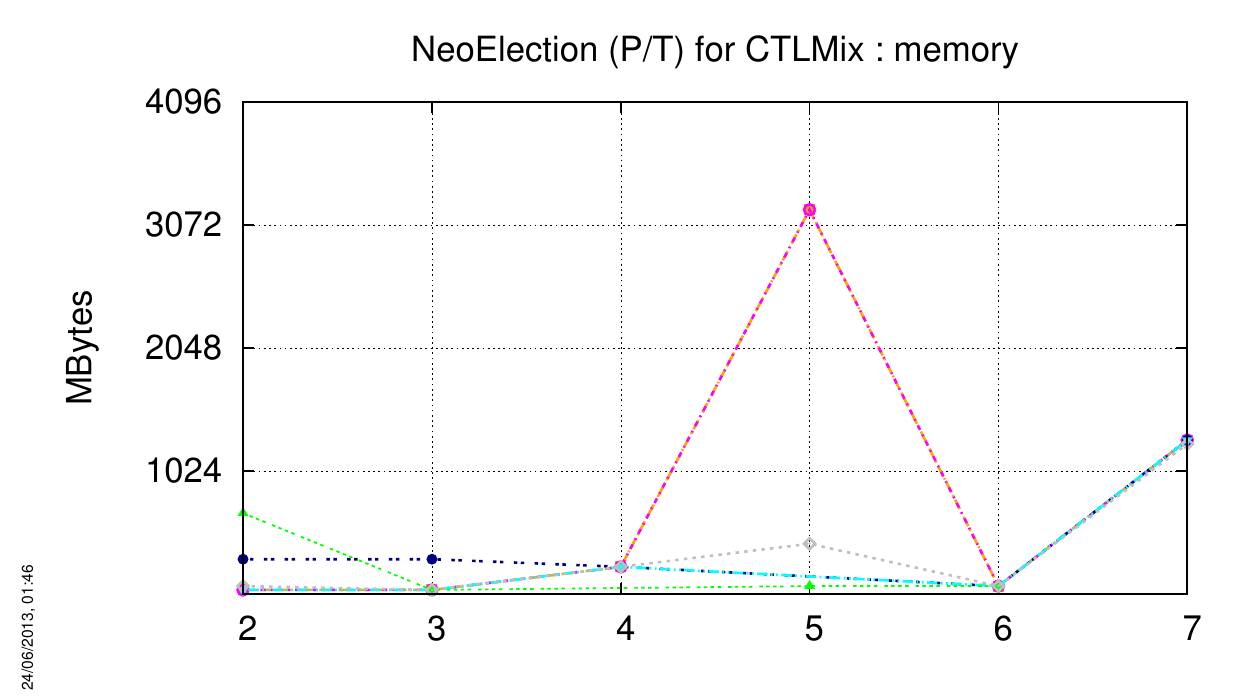}
   \includegraphics[width=7.2cm]{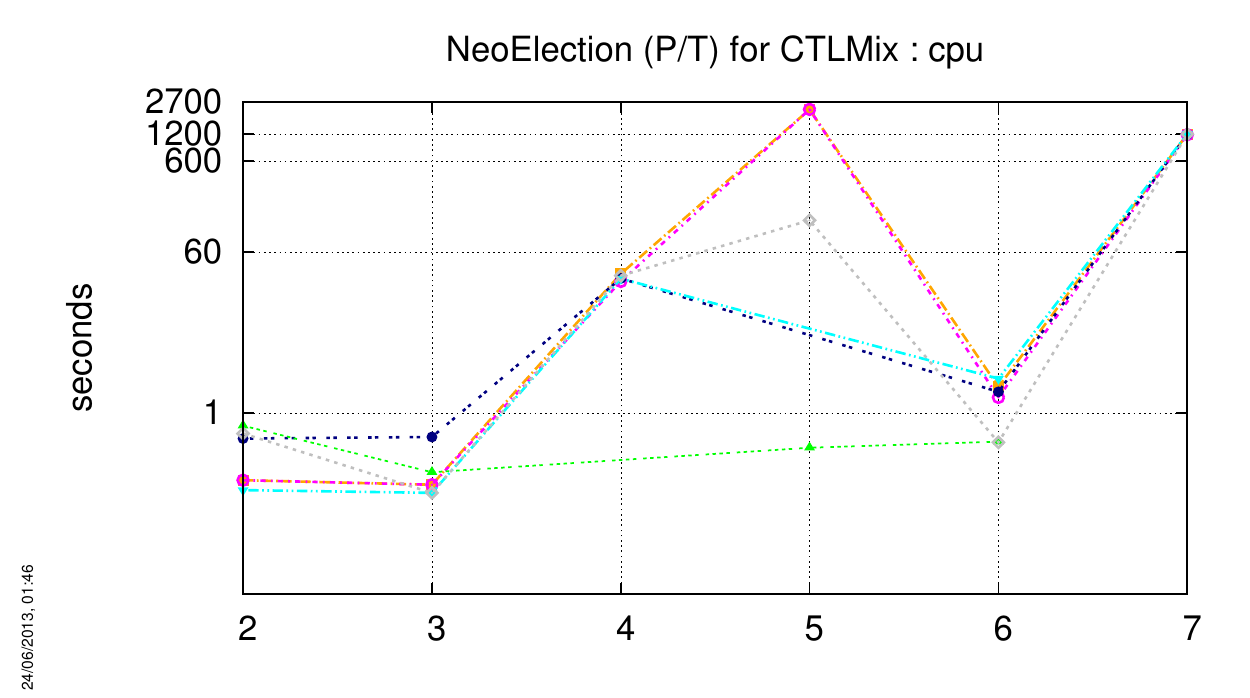}

   \includegraphics[height=1cm]{figures/tools-legend.pdf}
\end{center}

\subsubsection{\acs{PermAdmissibility-COL}}
No instance of this model could be computed for the \textbf{CTLMix} examination.

\subsubsection{\acs{PermAdmissibility-PT}}
The charts below respectively show how tools compete with this ``Known'' model (memory and CPU).

\index{Performances!CTLMix!PermAdmissibility (P/T)}
\begin{center}
   \includegraphics[width=7.2cm]{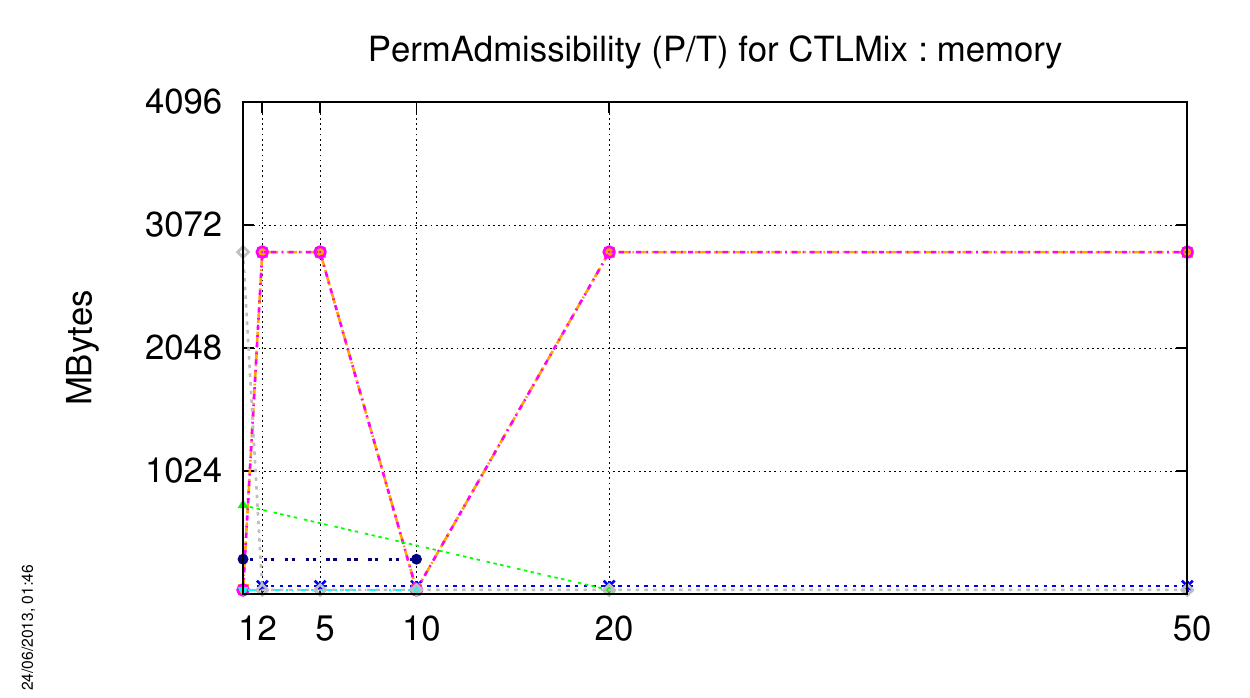}
   \includegraphics[width=7.2cm]{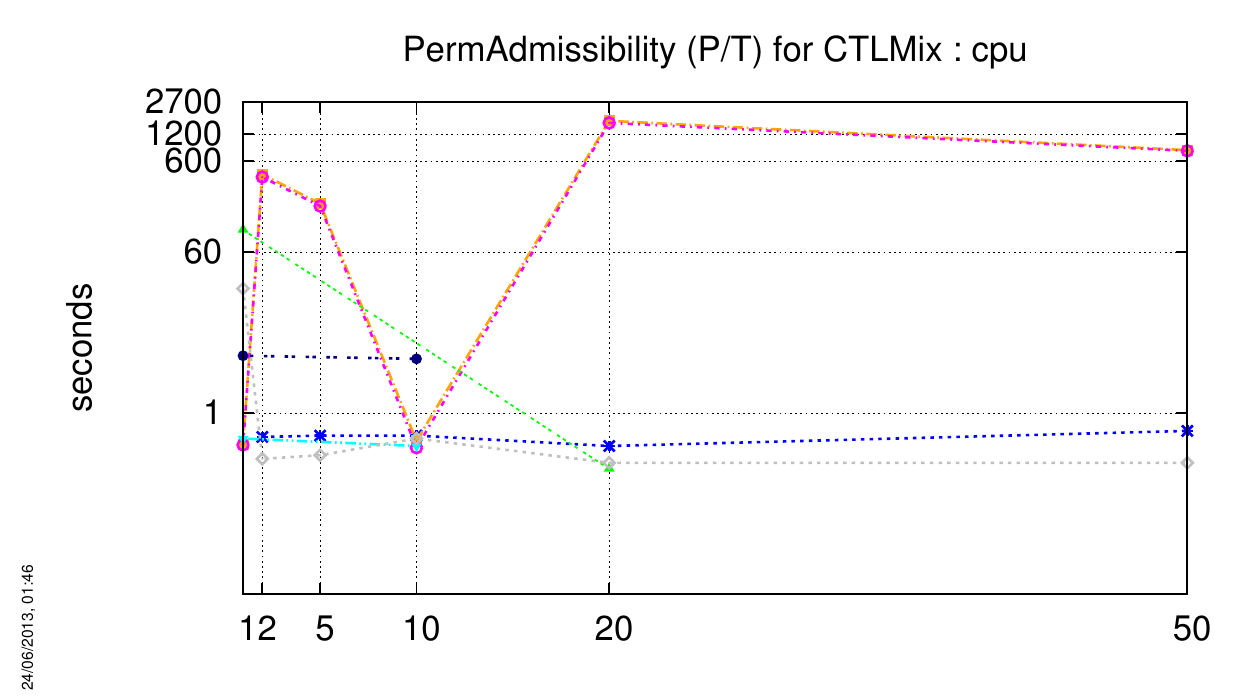}

   \includegraphics[height=1cm]{figures/tools-legend.pdf}
\end{center}

\subsubsection{\acs{Peterson-COL}}
No instance of this model could be computed for the \textbf{CTLMix} examination.

\subsubsection{\acs{Peterson-PT}}
The charts below respectively show how tools compete with this ``Known'' model (memory and CPU).

\index{Performances!CTLMix!Peterson (P/T)}
\begin{center}
   \includegraphics[width=7.2cm]{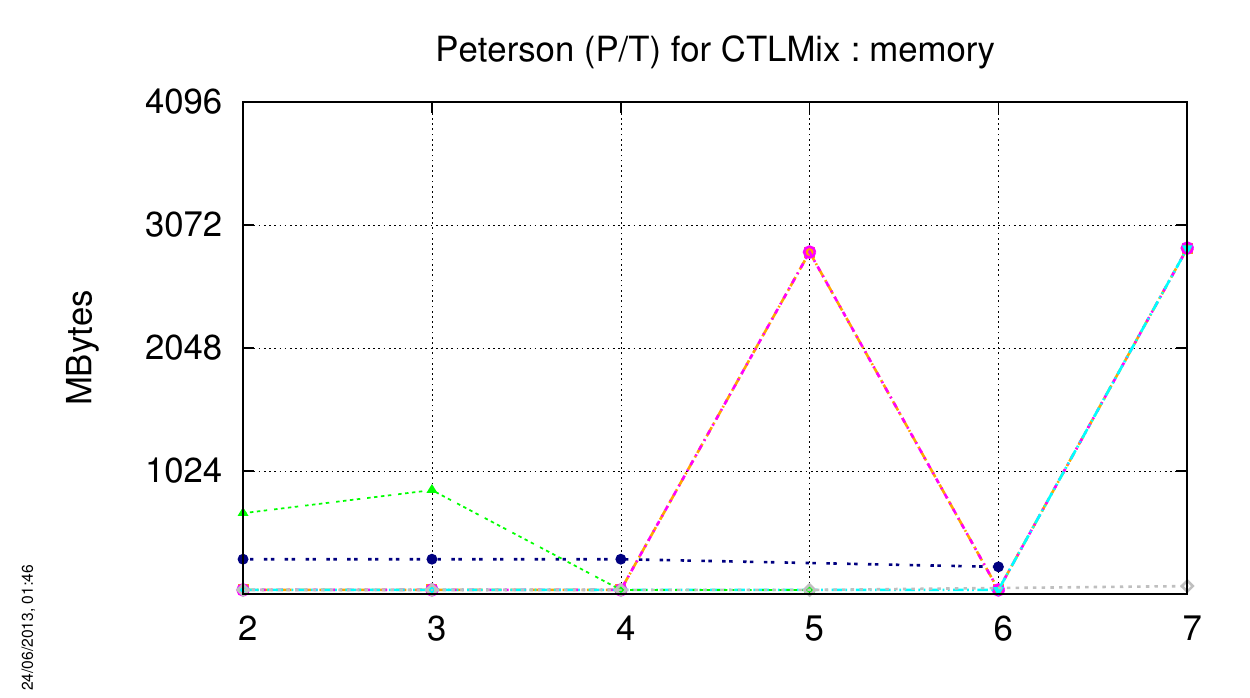}
   \includegraphics[width=7.2cm]{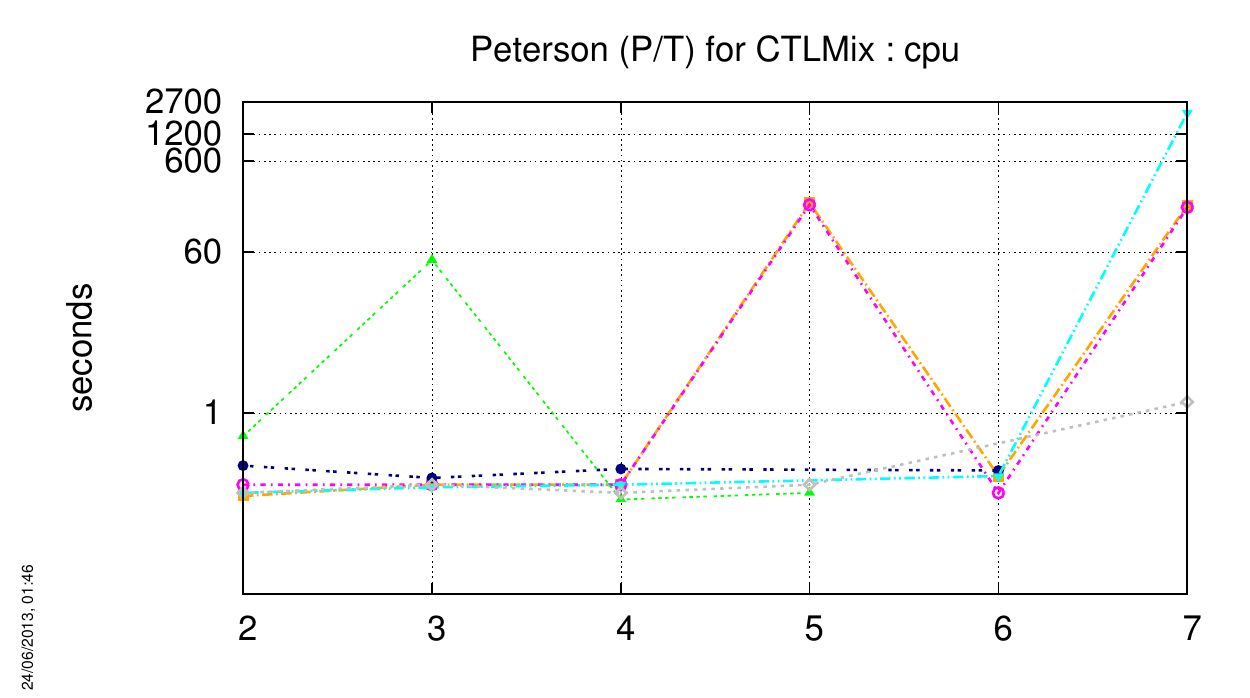}

   \includegraphics[height=1cm]{figures/tools-legend.pdf}
\end{center}

\subsubsection{\acs{Philosophers-COL}}
No instance of this model could be computed for the \textbf{CTLMix} examination.

\subsubsection{\acs{Philosophers-PT}}
The charts below respectively show how tools compete with this ``Known'' model (memory and CPU).

\index{Performances!CTLMix!Philosophers (P/T)}
\begin{center}
   \includegraphics[width=7.2cm]{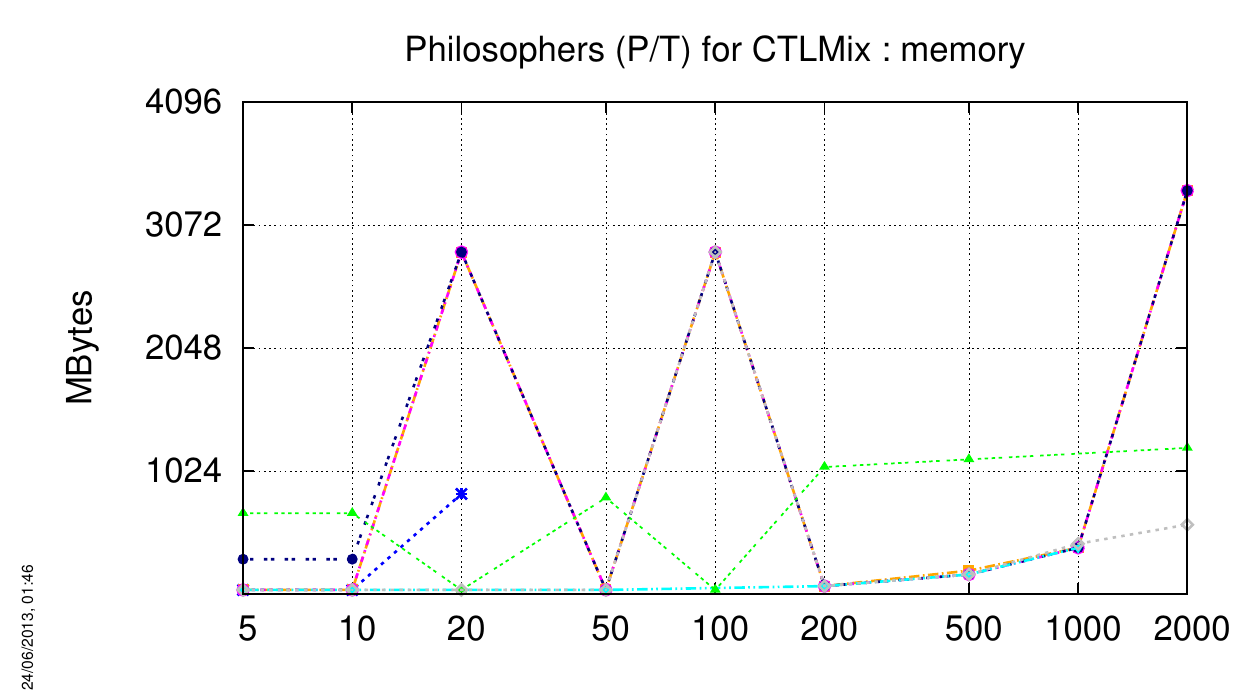}
   \includegraphics[width=7.2cm]{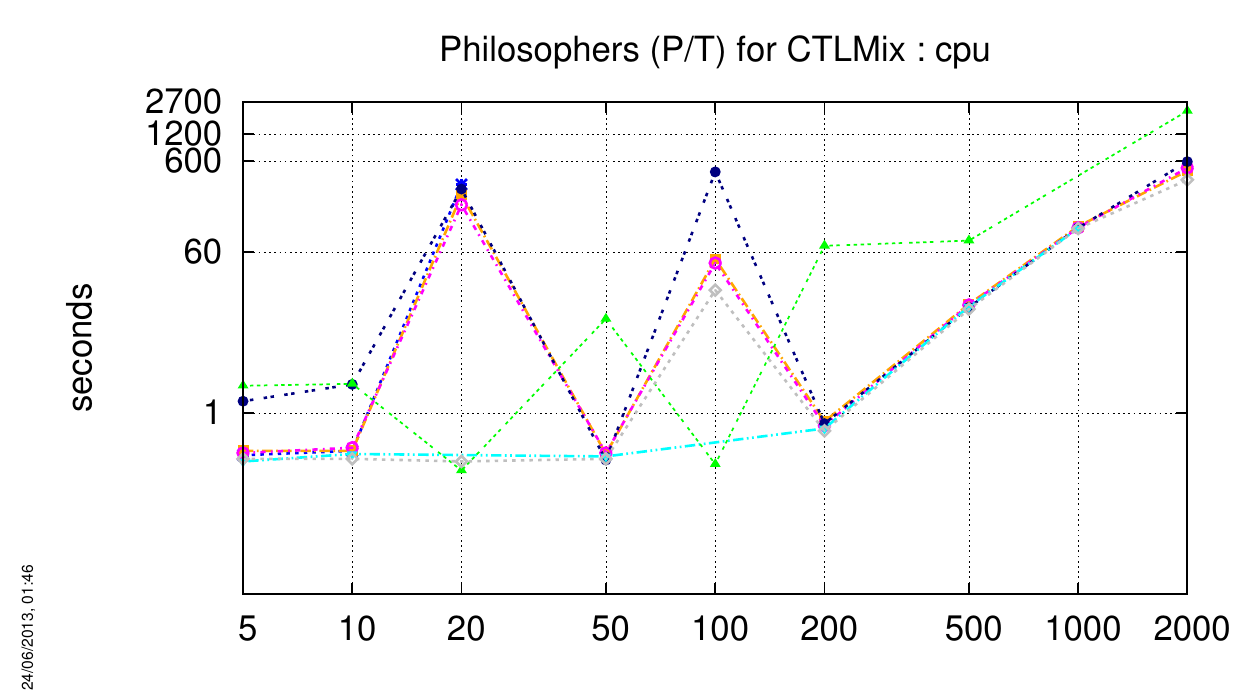}

   \includegraphics[height=1cm]{figures/tools-legend.pdf}
\end{center}

\subsubsection{\acs{PhilosophersDyn-COL}}
No instance of this model could be computed for the \textbf{CTLMix} examination.

\subsubsection{\acs{PhilosophersDyn-PT}}
The charts below respectively show how tools compete with this ``Known'' model (memory and CPU).

\index{Performances!CTLMix!PhilosophersDyn (P/T)}
\begin{center}
   \includegraphics[width=7.2cm]{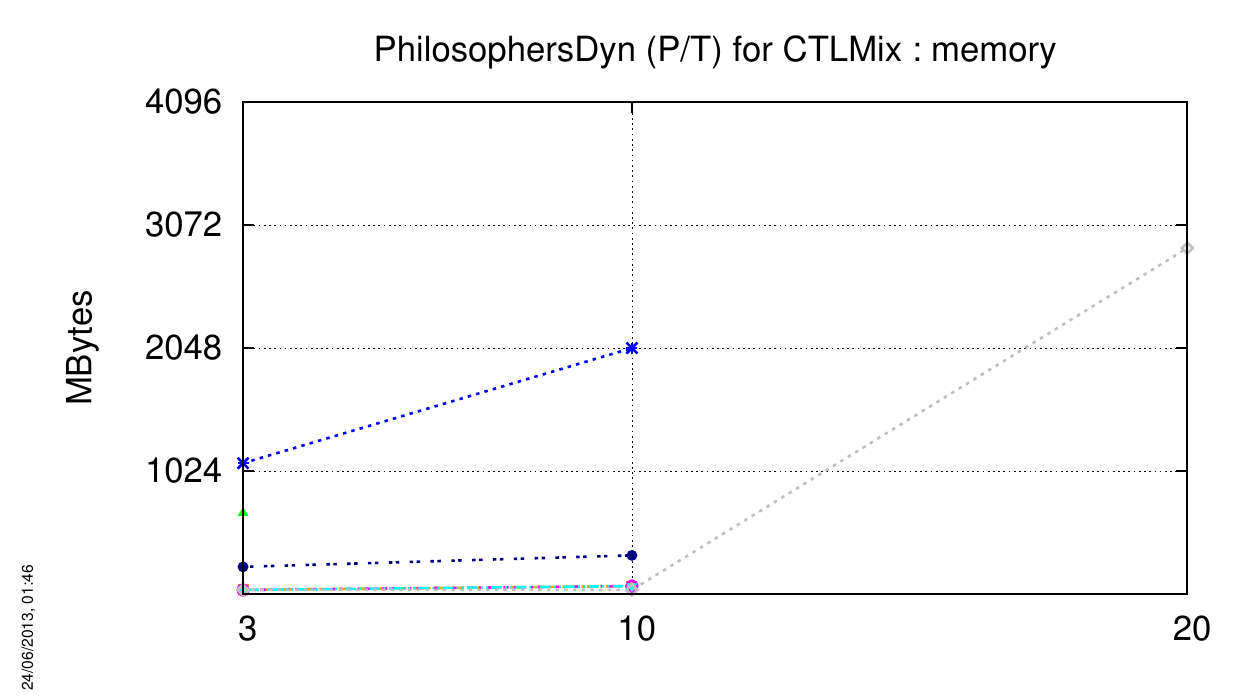}
   \includegraphics[width=7.2cm]{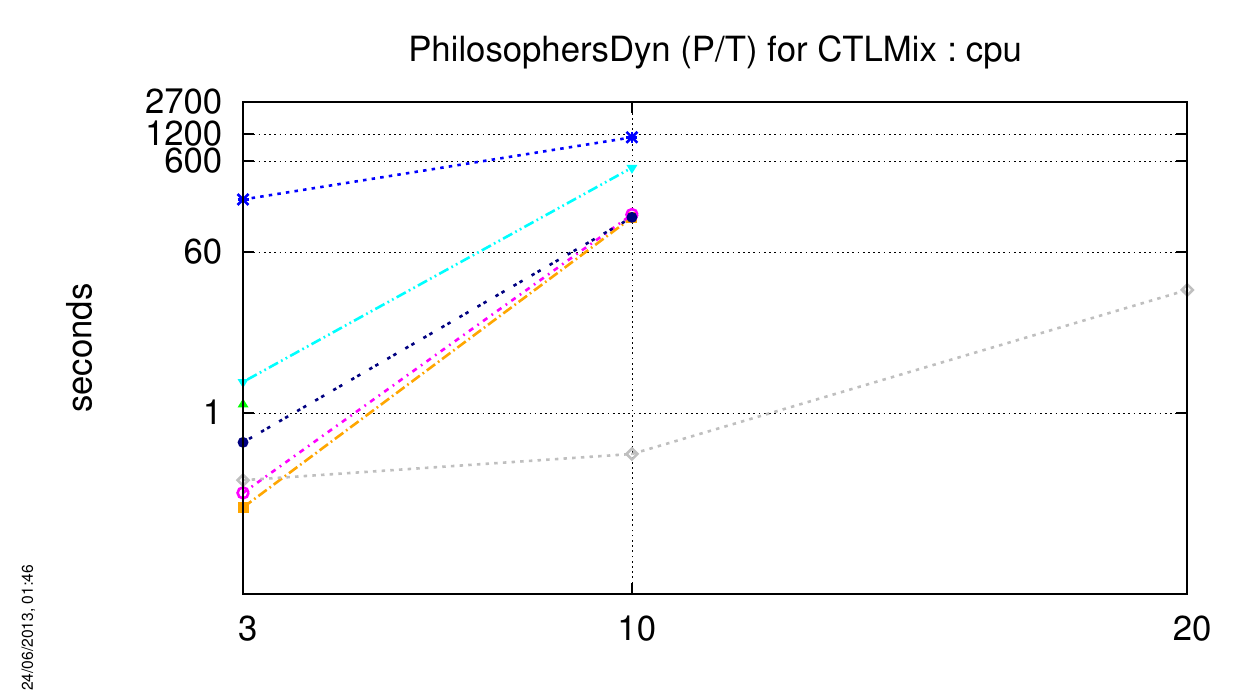}

   \includegraphics[height=1cm]{figures/tools-legend.pdf}
\end{center}

\subsubsection{\acs{Planning-PT}}
No instance of this model could be computed for the \textbf{CTLMix} examination.

\subsubsection{\acs{Railroad-PT}}
The charts below respectively show how tools compete with this ``Known'' model (memory and CPU).

\index{Performances!CTLMix!Railroad (P/T)}
\begin{center}
   \includegraphics[width=7.2cm]{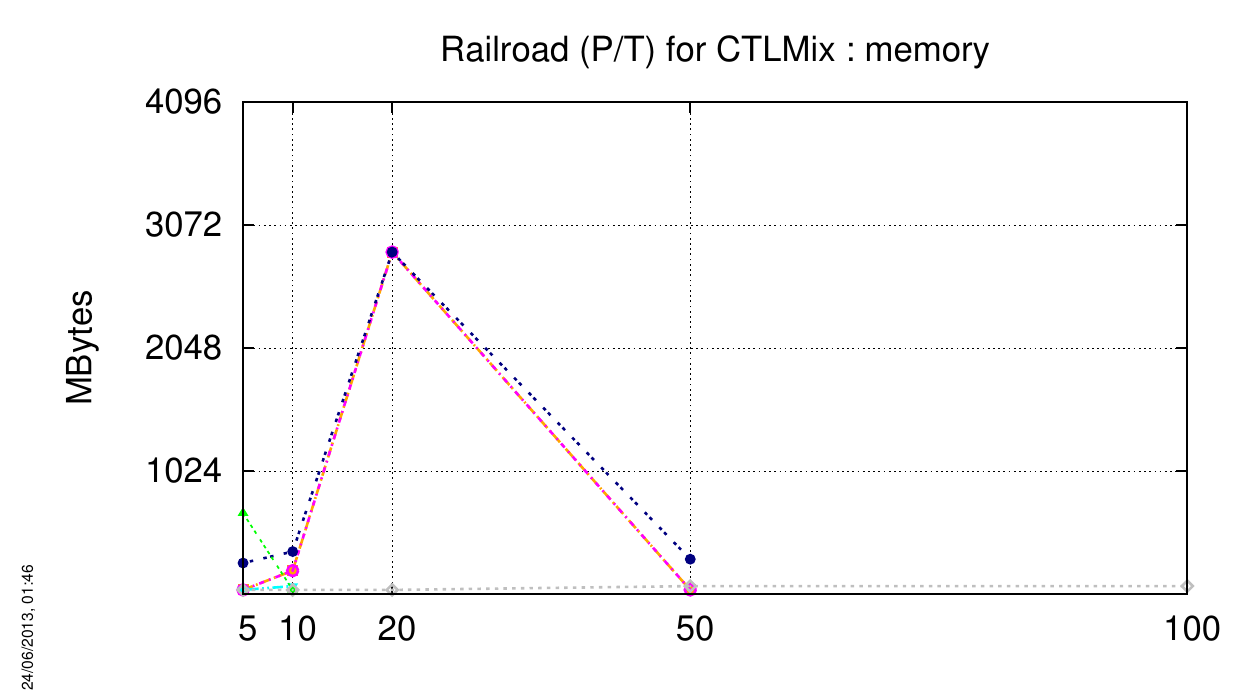}
   \includegraphics[width=7.2cm]{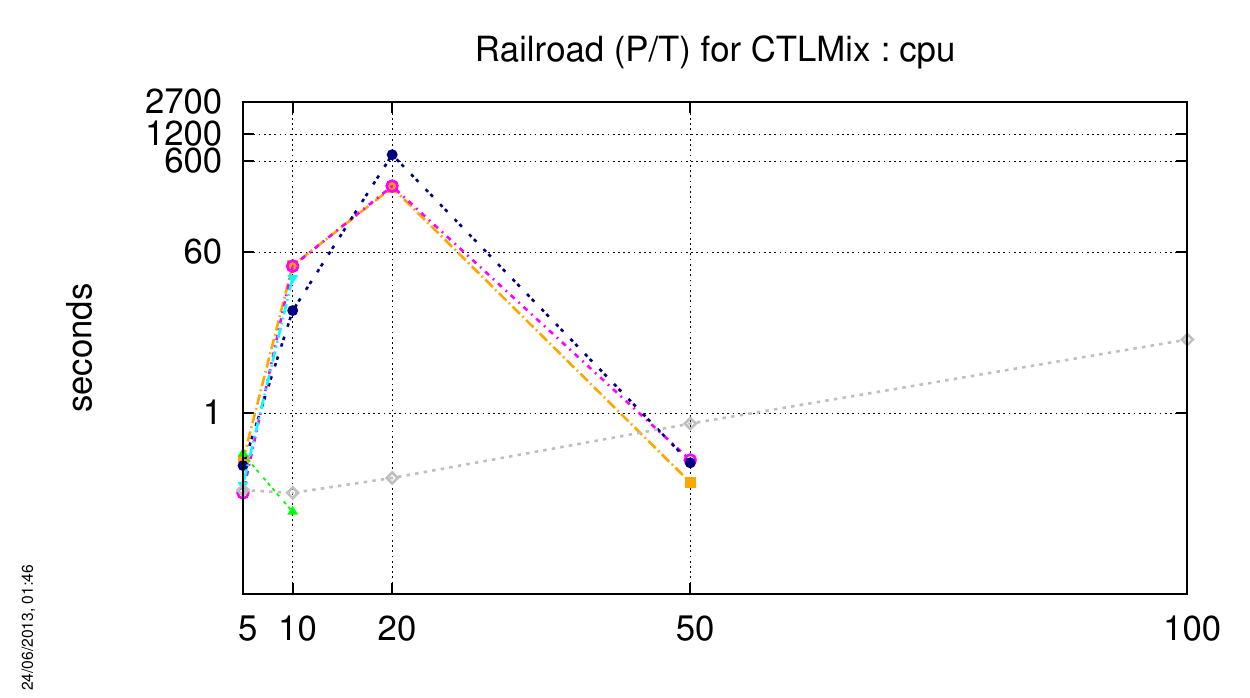}

   \includegraphics[height=1cm]{figures/tools-legend.pdf}
\end{center}

\subsubsection{\acs{RessAllocation-PT}}
The charts below respectively show how tools compete with this ``Known'' model (memory and CPU).

\index{Performances!CTLMix!RessAllocation (P/T)}
\begin{center}
   \includegraphics[width=7.2cm]{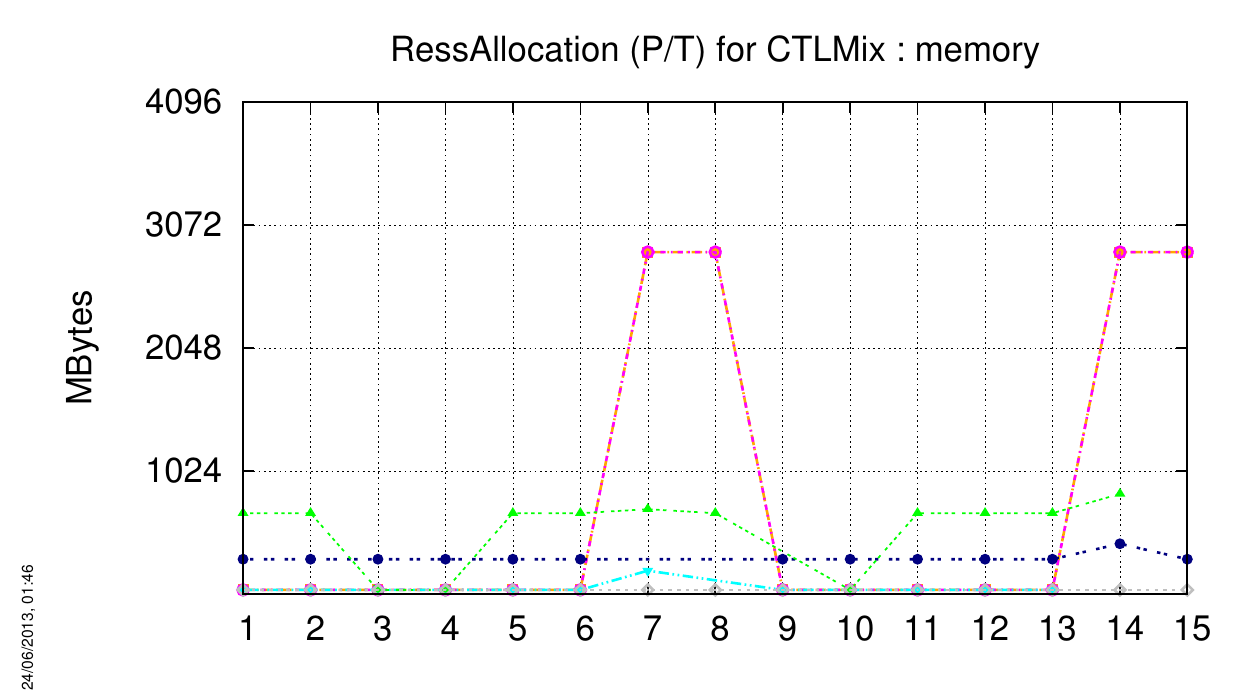}
   \includegraphics[width=7.2cm]{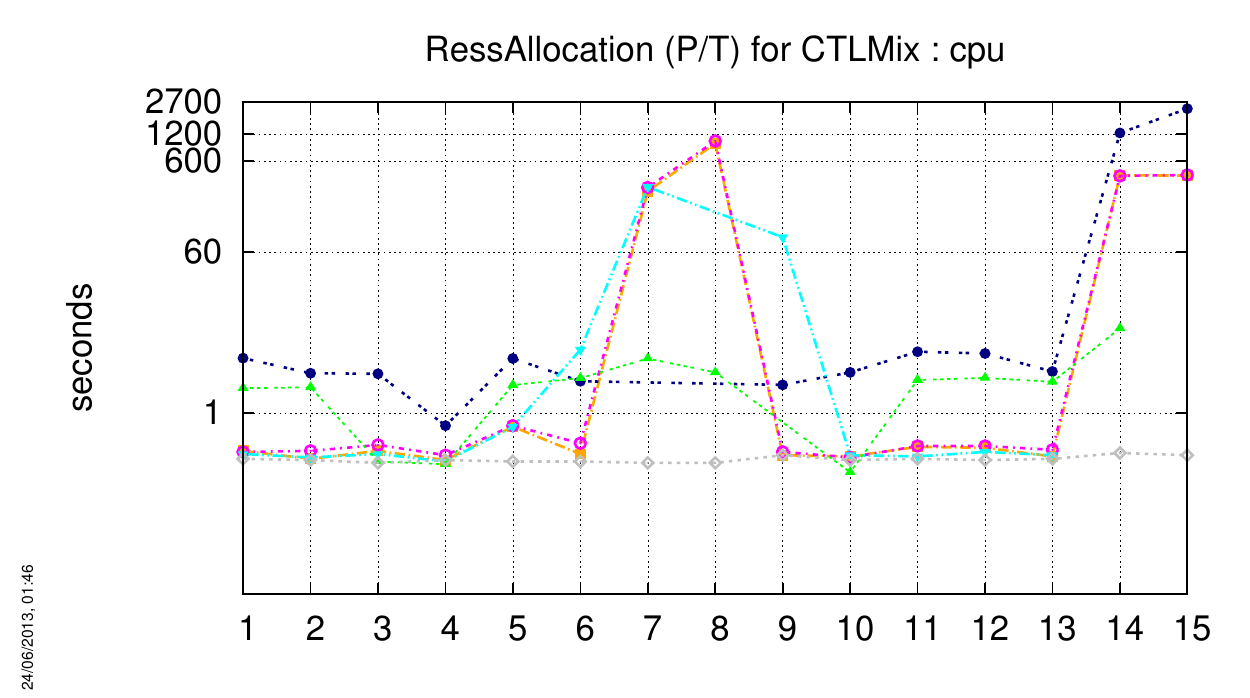}

   \includegraphics[height=1cm]{figures/tools-legend.pdf}
\end{center}

\subsubsection{\acs{Ring-PT}}
The charts below respectively show how tools compete with this ``Known'' model (memory and CPU).

\index{Performances!CTLMix!Ring (P/T)}
\begin{center}
   \includegraphics[width=7.2cm]{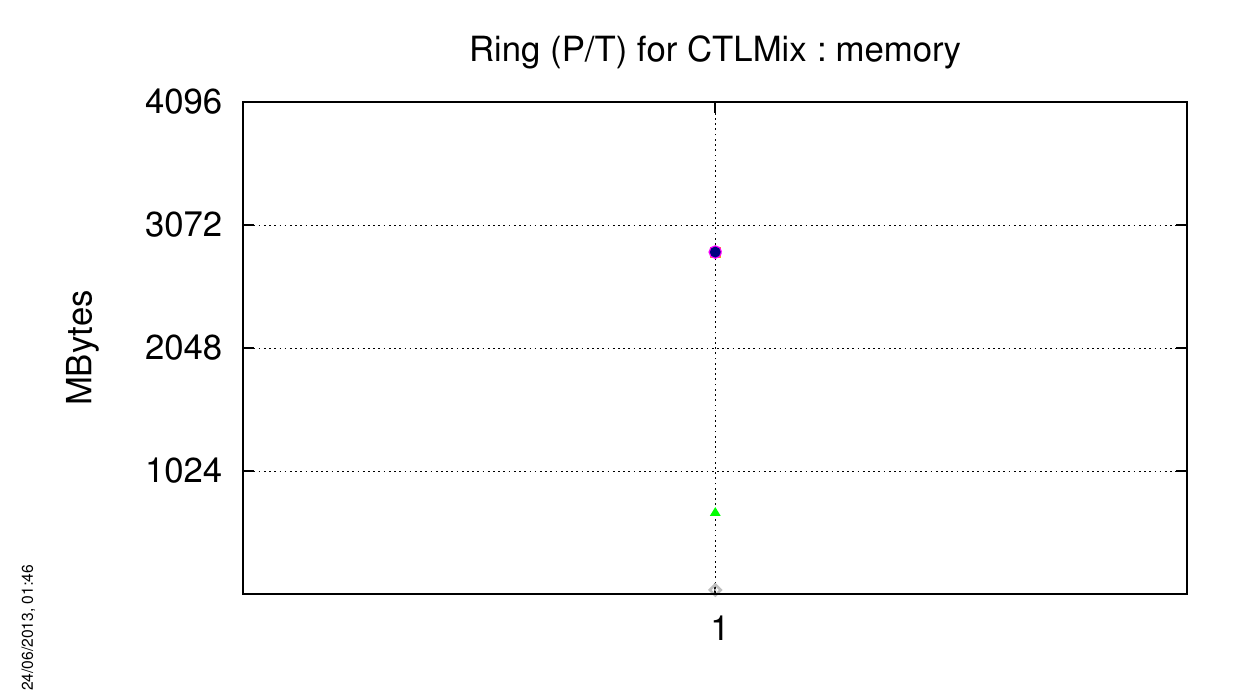}
   \includegraphics[width=7.2cm]{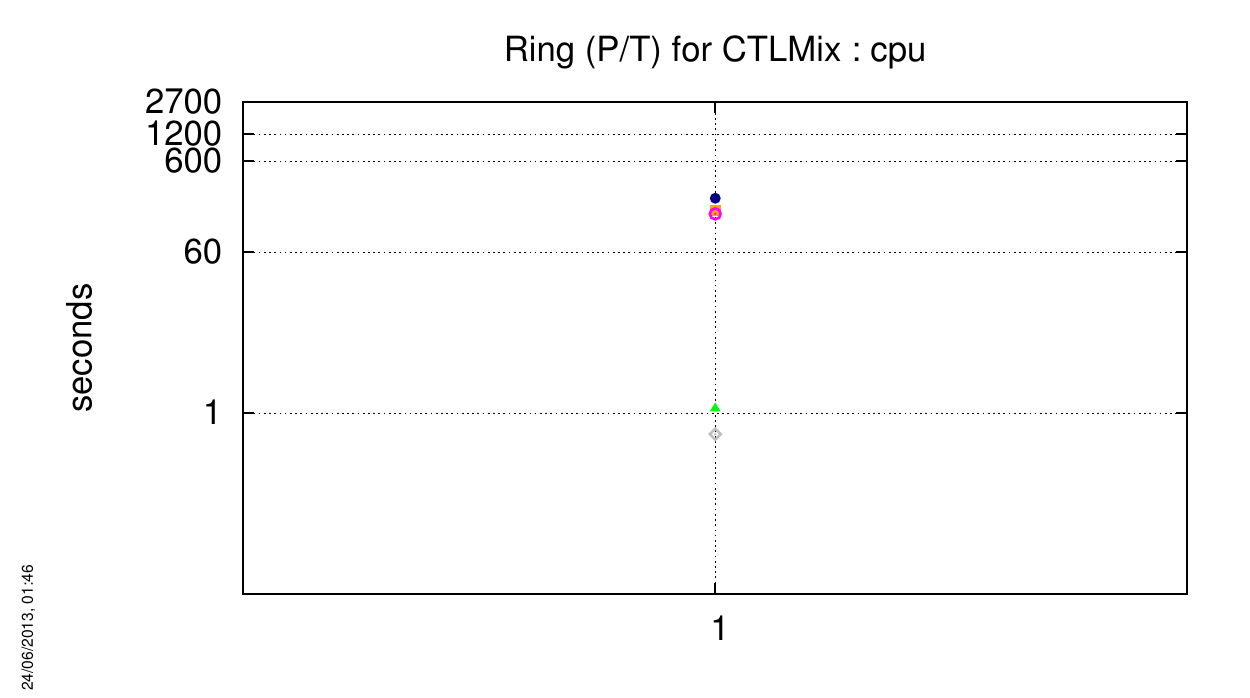}

   \includegraphics[height=1cm]{figures/tools-legend.pdf}
\end{center}

\subsubsection{\acs{RwMutex-PT}}
The charts below respectively show how tools compete with this ``Known'' model (memory and CPU).

\index{Performances!CTLMix!RwMutex (P/T)}
\begin{center}
   \includegraphics[width=7.2cm]{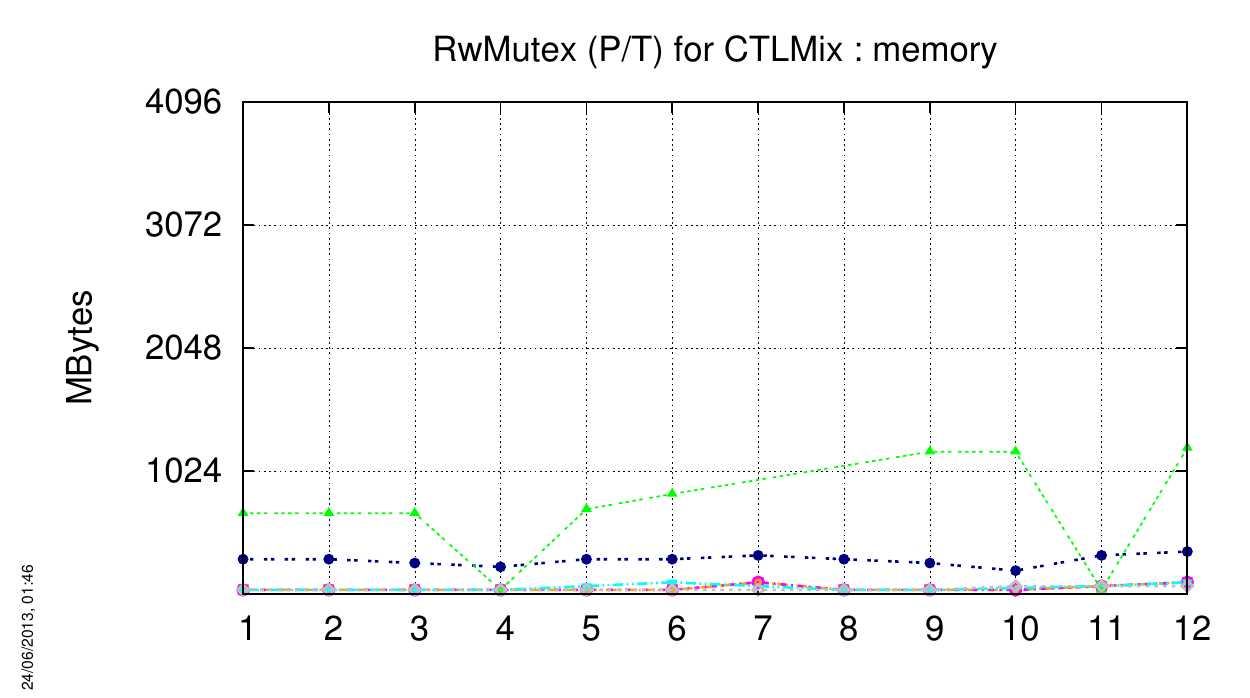}
   \includegraphics[width=7.2cm]{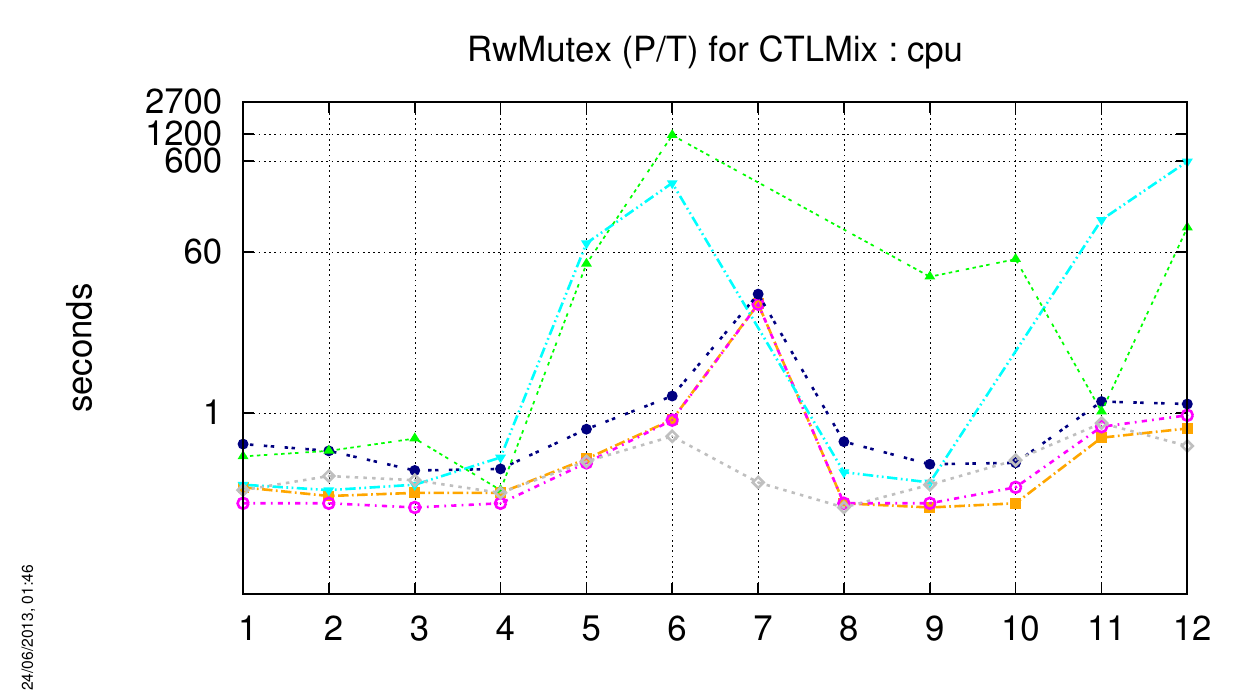}

   \includegraphics[height=1cm]{figures/tools-legend.pdf}
\end{center}

\subsubsection{\acs{SharedMemory-COL}}
No instance of this model could be computed for the \textbf{CTLMix} examination.

\subsubsection{\acs{SharedMemory-PT}}
The charts below respectively show how tools compete with this ``Known'' model (memory and CPU).

\index{Performances!CTLMix!SharedMemory (P/T)}
\begin{center}
   \includegraphics[width=7.2cm]{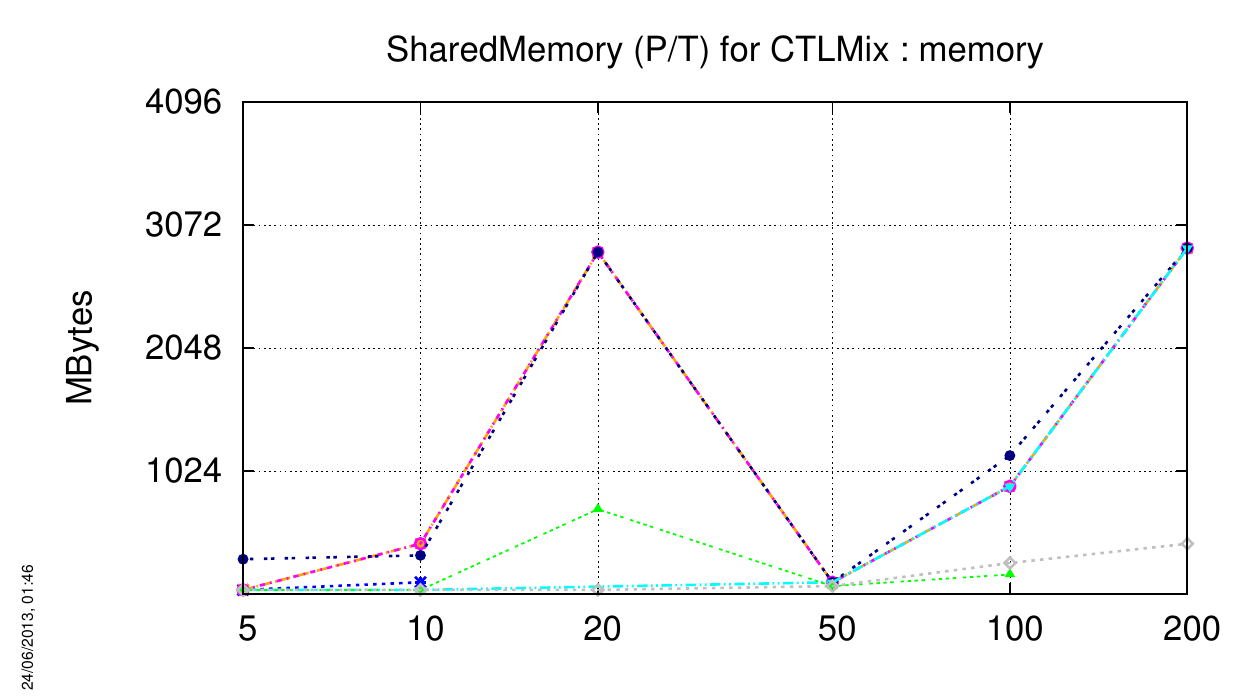}
   \includegraphics[width=7.2cm]{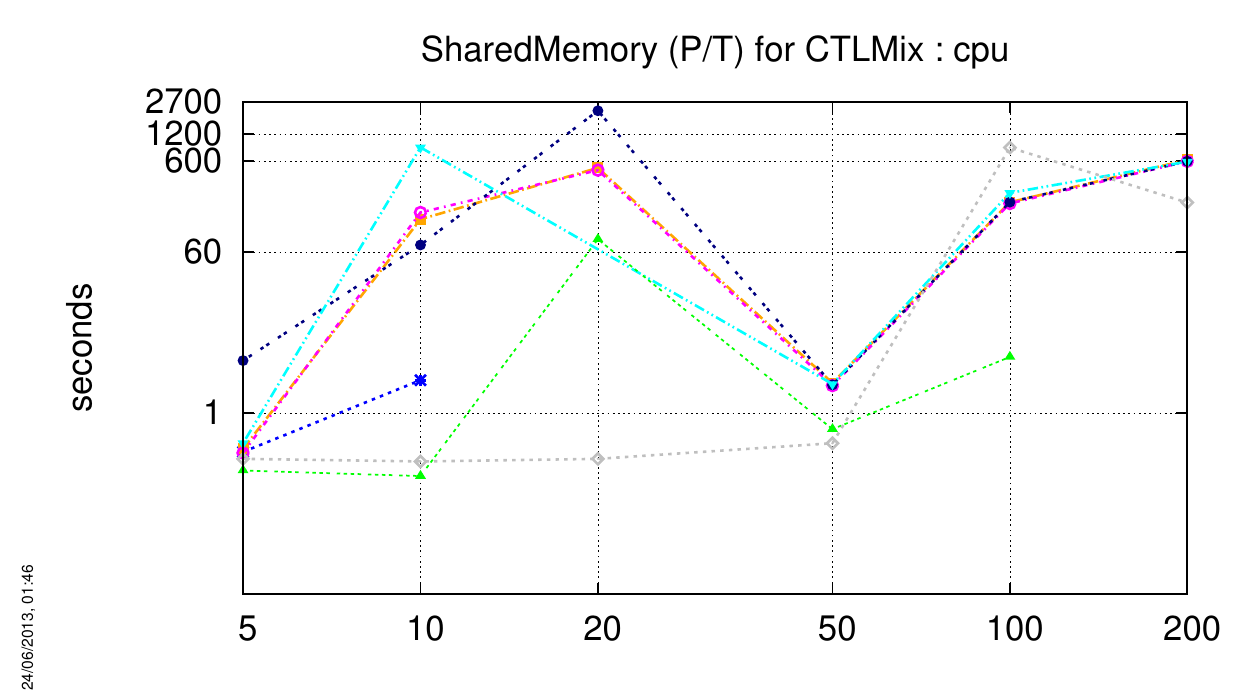}

   \includegraphics[height=1cm]{figures/tools-legend.pdf}
\end{center}

\subsubsection{\acs{SimpleLoadBal-COL}}
No instance of this model could be computed for the \textbf{CTLMix} examination.

\subsubsection{\acs{SimpleLoadBal-PT}}
The charts below respectively show how tools compete with this ``Known'' model (memory and CPU).

\index{Performances!CTLMix!SimpleLoadBal (P/T)}
\begin{center}
   \includegraphics[width=7.2cm]{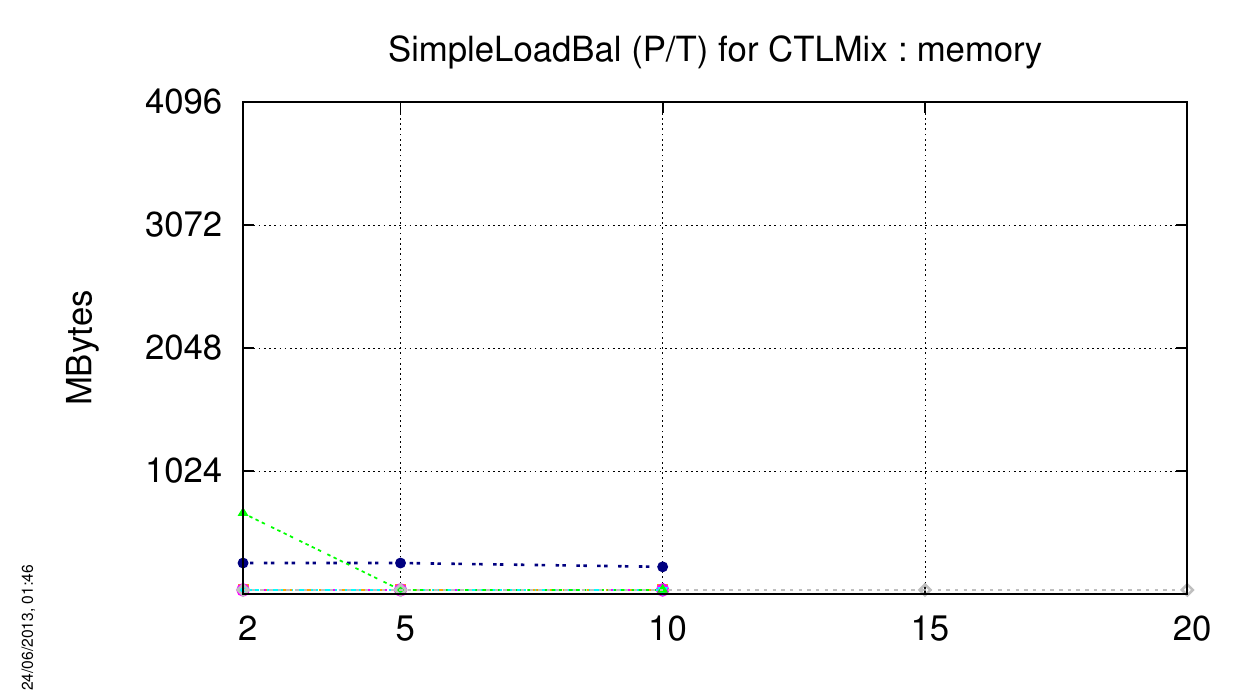}
   \includegraphics[width=7.2cm]{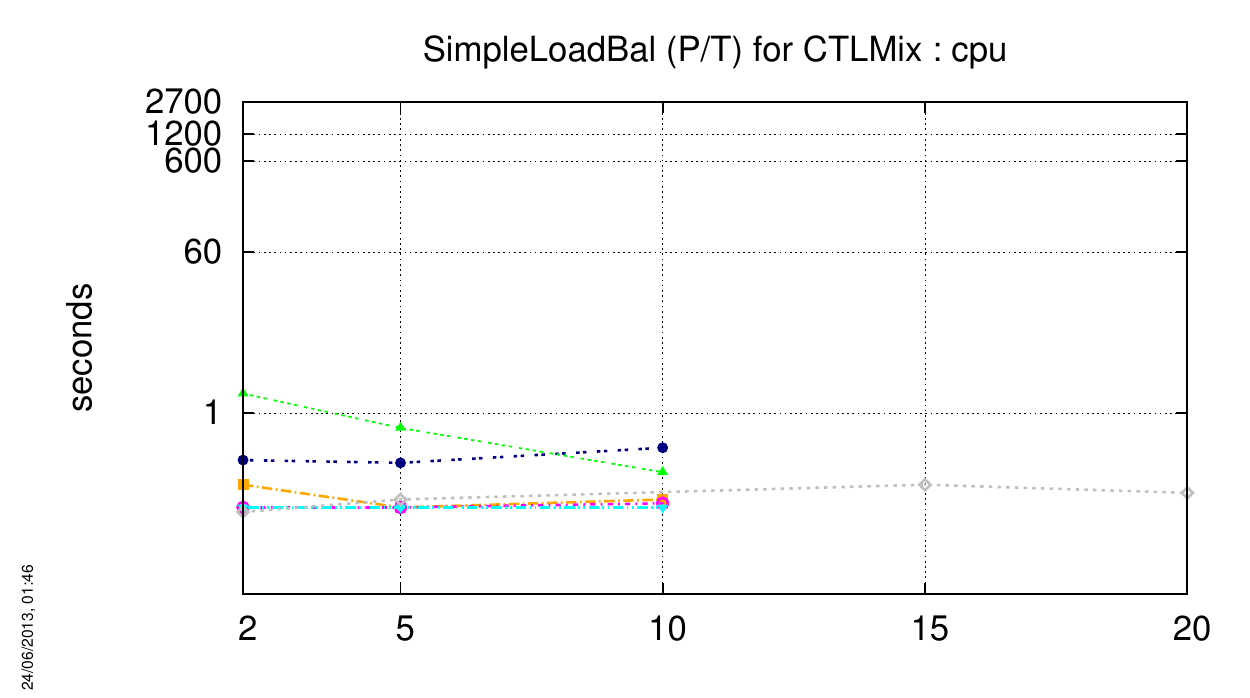}

   \includegraphics[height=1cm]{figures/tools-legend.pdf}
\end{center}

\subsubsection{\acs{TokenRing-COL}}
No instance of this model could be computed for the \textbf{CTLMix} examination.

\subsubsection{\acs{TokenRing-PT}}
The charts below respectively show how tools compete with this ``Known'' model (memory and CPU).

\index{Performances!CTLMix!TokenRing (P/T)}
\begin{center}
   \includegraphics[width=7.2cm]{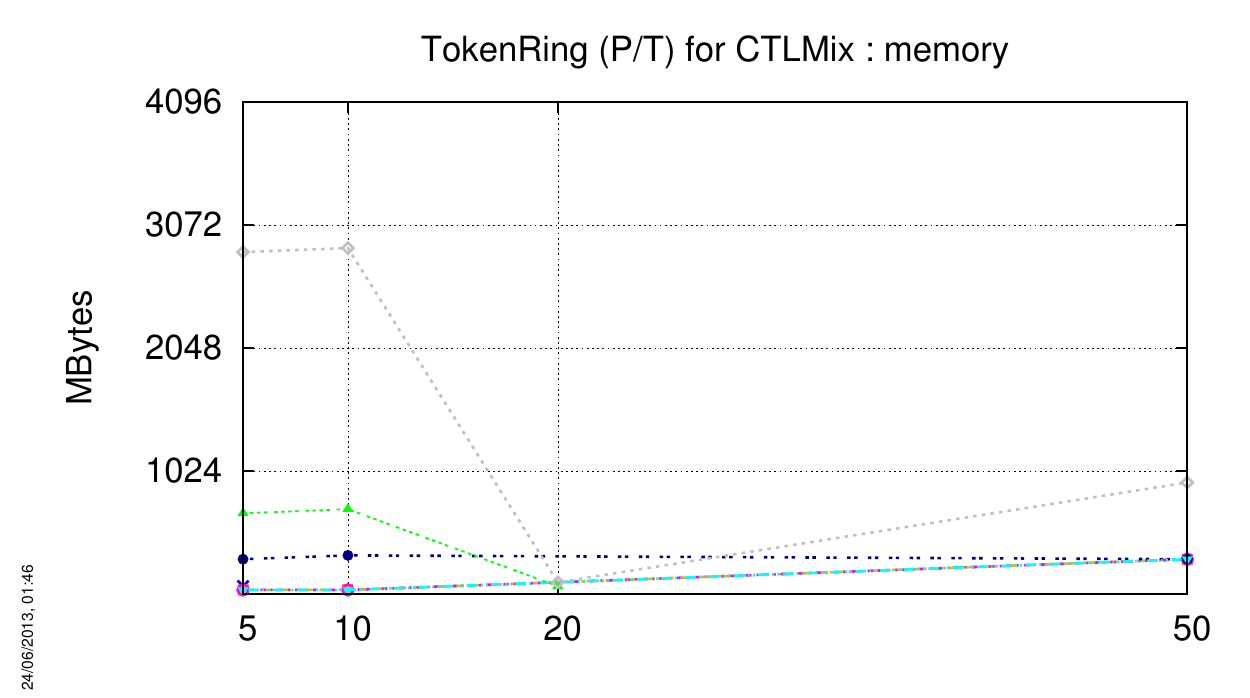}
   \includegraphics[width=7.2cm]{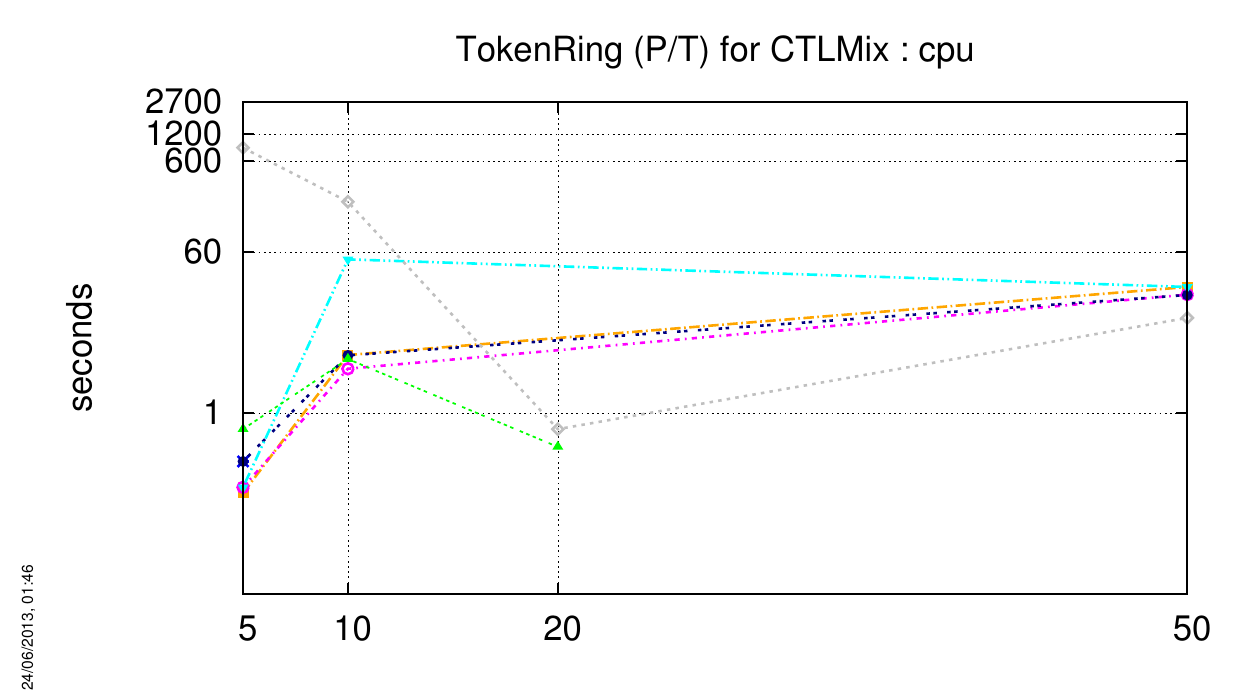}

   \includegraphics[height=1cm]{figures/tools-legend.pdf}
\end{center}

\subsubsection{\acs{HouseConstruction-PT}}
The charts below respectively show how tools compete with this ``Suprise'' model (memory and CPU).

\index{Performances!CTLMix!HouseConstruction (P/T)}
\begin{center}
   \includegraphics[width=7.2cm]{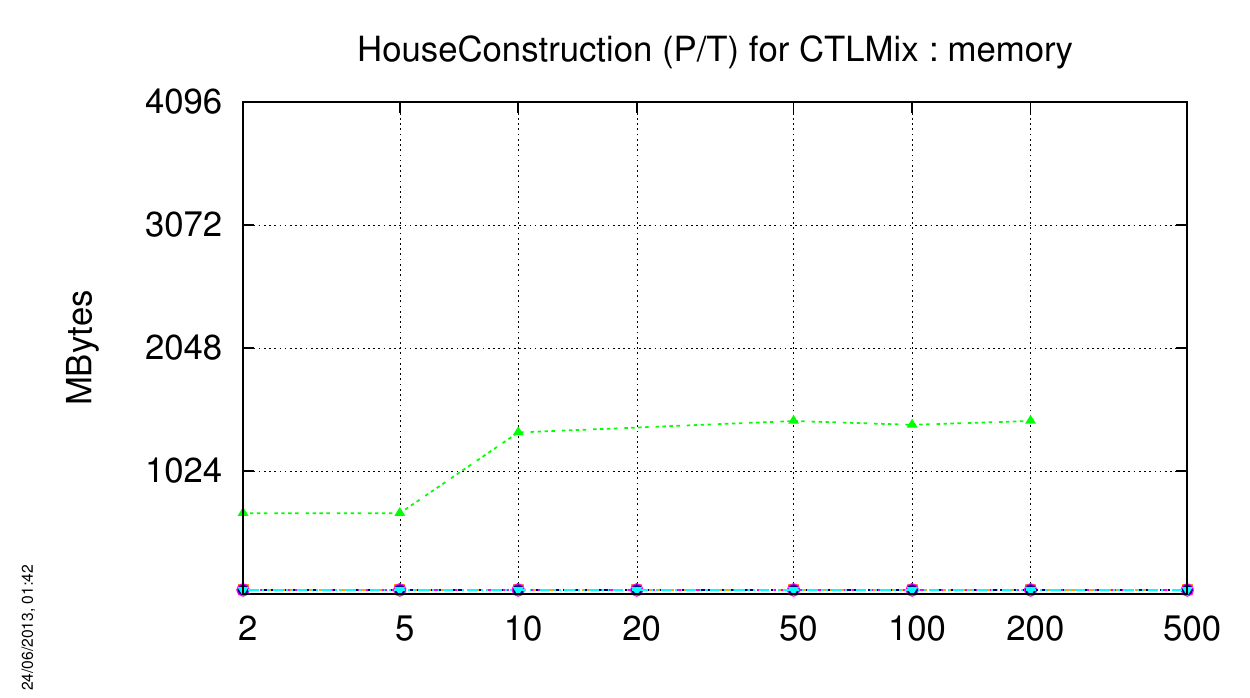}
   \includegraphics[width=7.2cm]{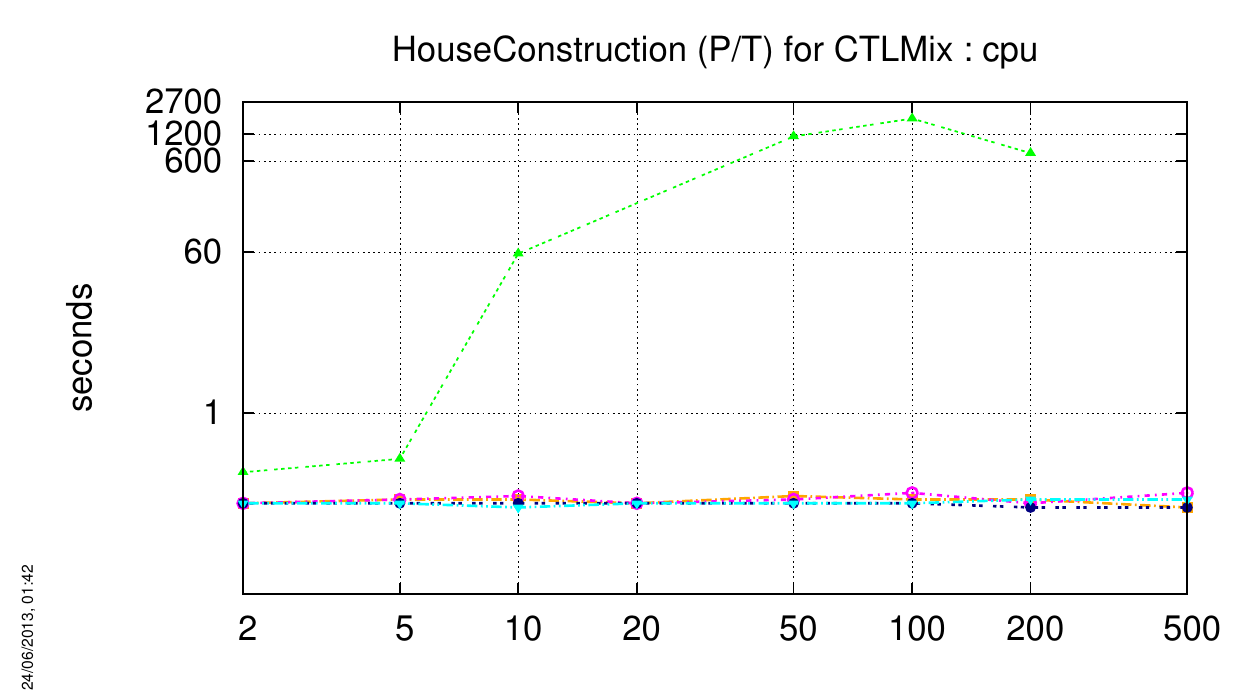}

   \includegraphics[height=1cm]{figures/tools-legend.pdf}
\end{center}

\subsubsection{\acs{IBMB2S565S3960-PT}}
The charts below respectively show how tools compete with this ``Suprise'' model (memory and CPU).

\index{Performances!CTLMix!IBMB2S565S3960 (P/T)}
\begin{center}
   \includegraphics[width=7.2cm]{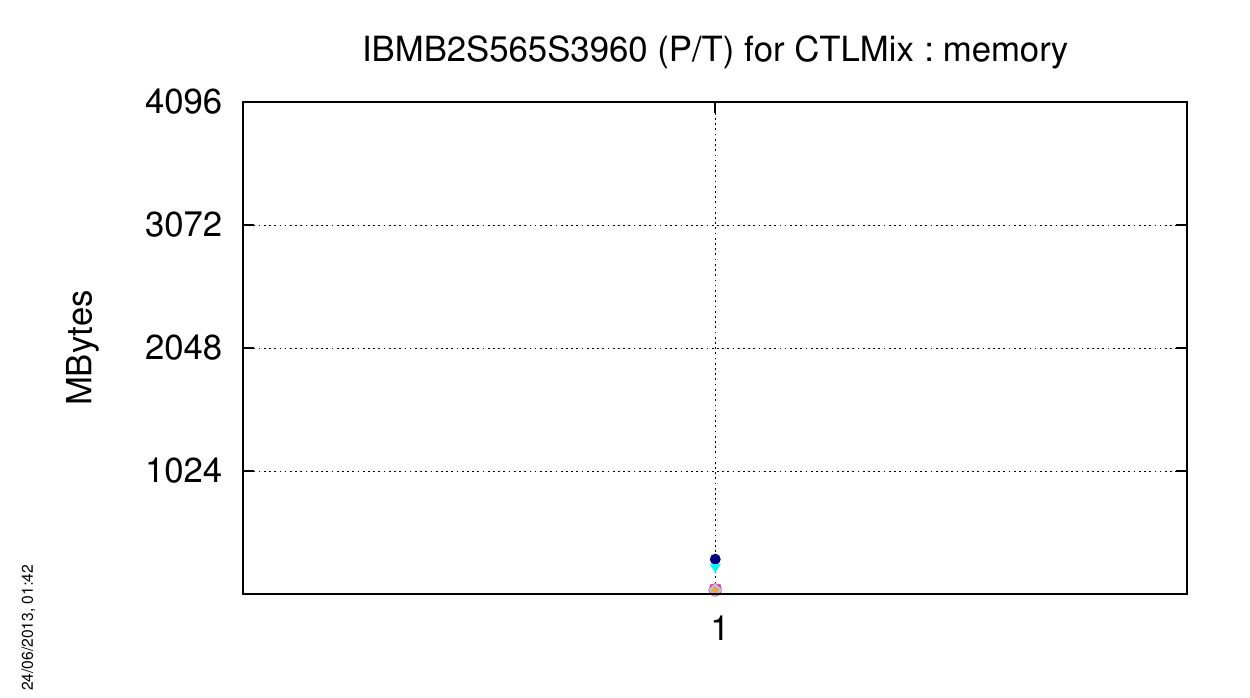}
   \includegraphics[width=7.2cm]{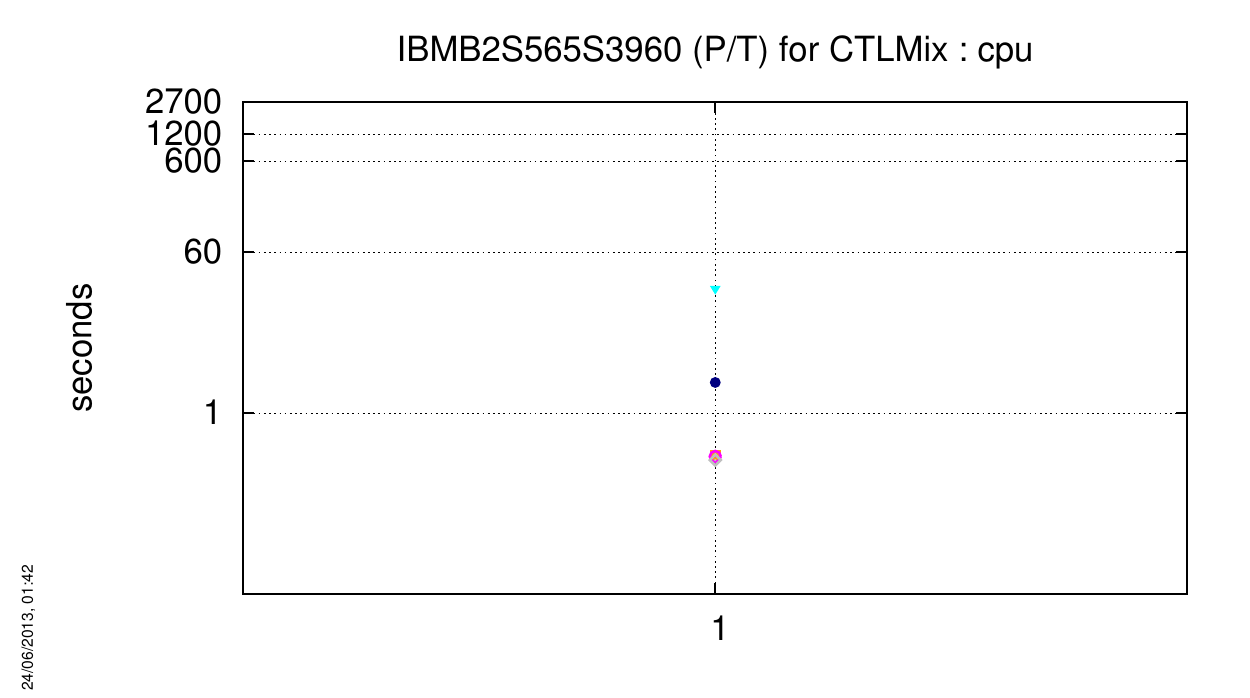}

   \includegraphics[height=1cm]{figures/tools-legend.pdf}
\end{center}

\subsubsection{\acs{QuasiCertifProtocol-COL}}
No instance of this model could be computed for the \textbf{CTLMix} examination.

\subsubsection{\acs{QuasiCertifProtocol-PT}}
The charts below respectively show how tools compete with this ``Suprise'' model (memory and CPU).

\index{Performances!CTLMix!QuasiCertifProtocol (P/T)}
\begin{center}
   \includegraphics[width=7.2cm]{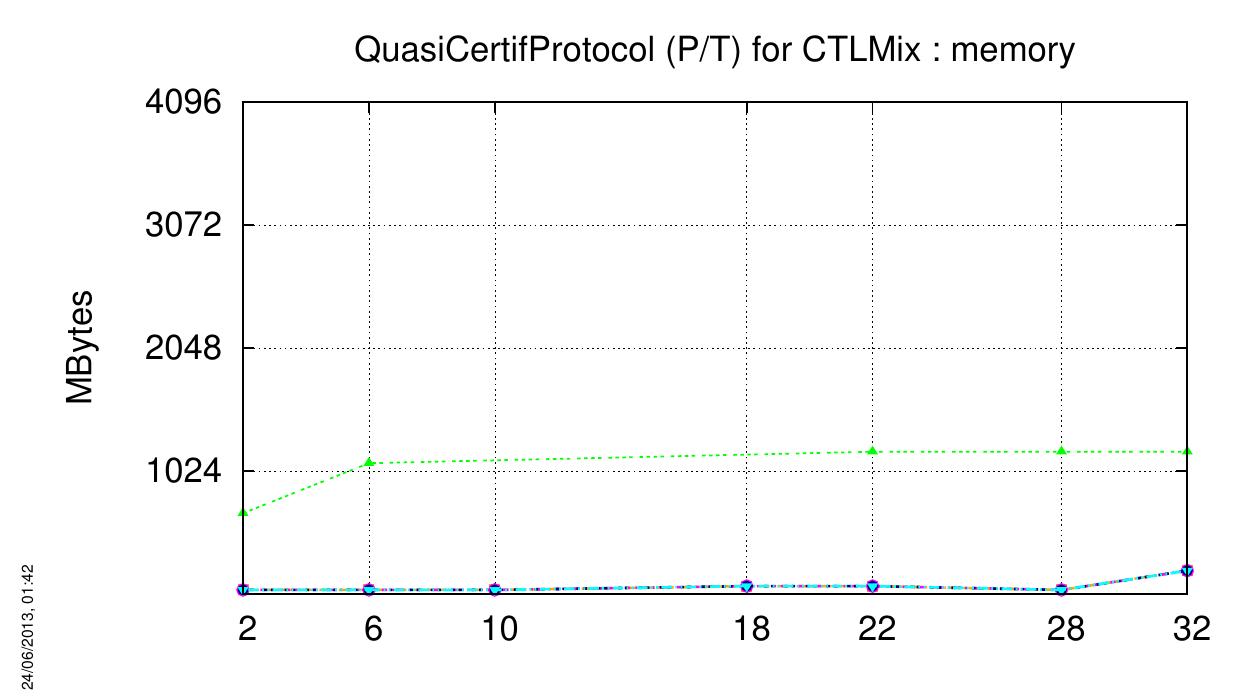}
   \includegraphics[width=7.2cm]{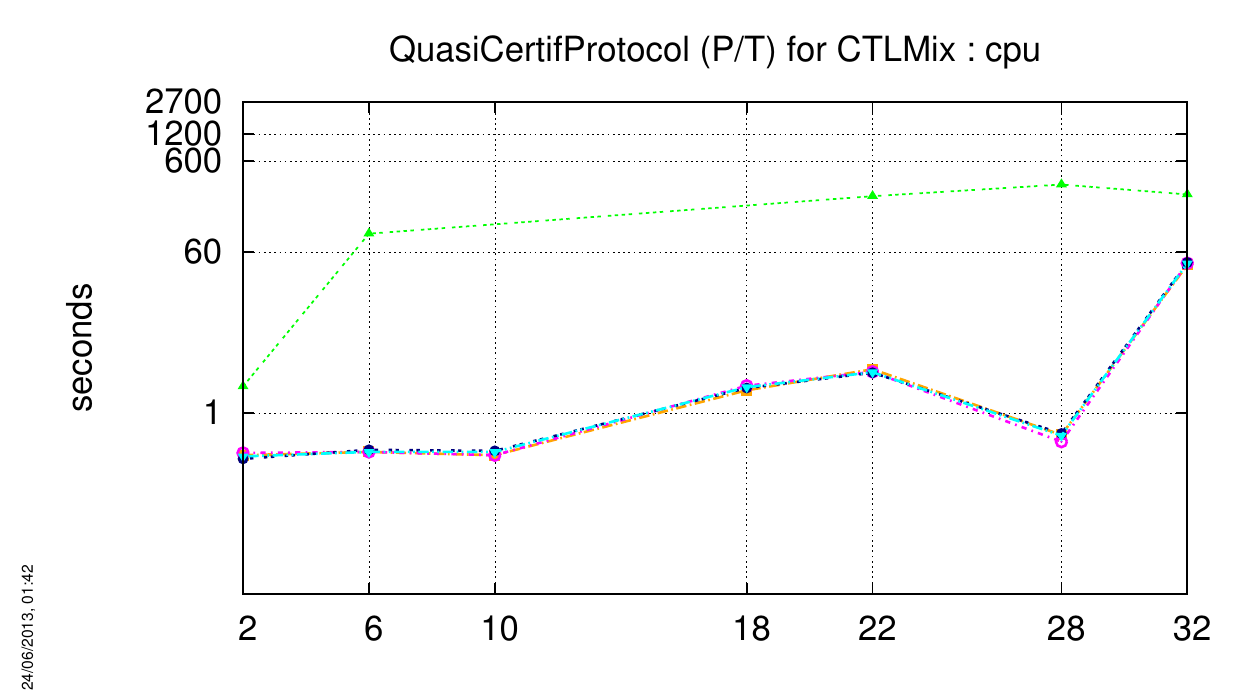}

   \includegraphics[height=1cm]{figures/tools-legend.pdf}
\end{center}

\subsubsection{\acs{Vasy2003-PT}}
The charts below respectively show how tools compete with this ``Suprise'' model (memory and CPU).

\index{Performances!CTLMix!Vasy2003 (P/T)}
\begin{center}
   \includegraphics[width=7.2cm]{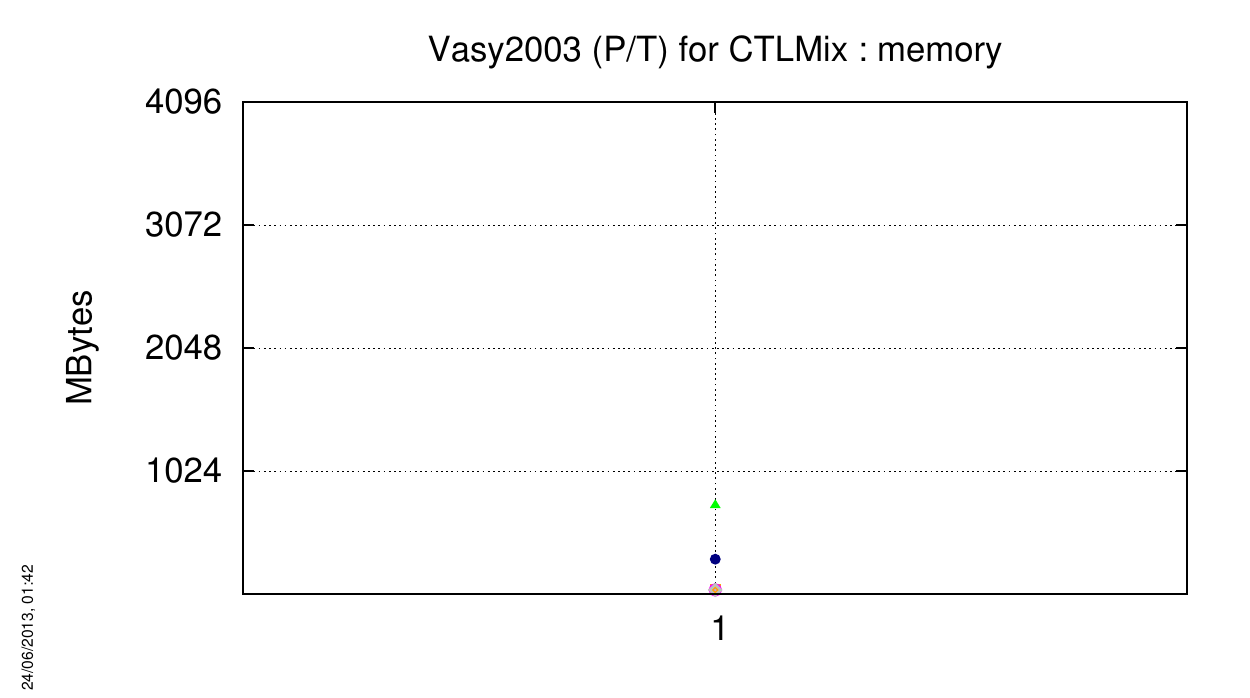}
   \includegraphics[width=7.2cm]{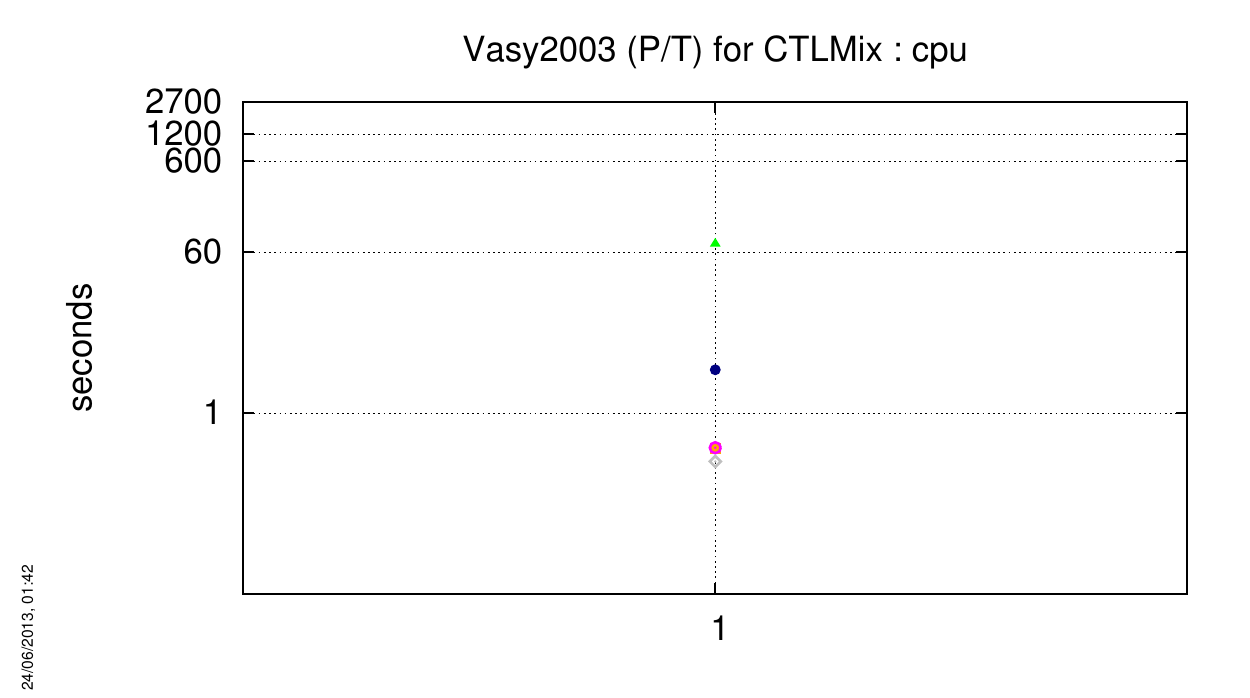}

   \includegraphics[height=1cm]{figures/tools-legend.pdf}
\end{center}

\subsection{Outputs for the CTLMix Examination}
\index{Outputs!CTLMix}

Please find enclosed the brute results for this examination (``Known'' and ``Surprise'' models).
We display only the score of tools that provide a results for at least one instance of one model.
The legend for the values is provided below:
\begin{itemize}
   \item\textbf{nc}: the tool does not compete this examination for this model/instance,
   \item\textbf{cc}: the tool cannot compute this examination for this model/instance,
   \item\textbf{to}: the tool cannot compute this examination for this model/instance within the maximum allowed time,
   \item\textbf{mp}: the tool encountered a memory problem (stack overflow or memory full),
   \item\textbf{nf}: there is no formula available for this type of examination (typically, this concerns P/T nets where
       comparing marking cardinality has no signification when there is no equivalent colored net).
\end{itemize}

\textbf{Note on the display of results for formulas:} each formula is considered as a flag (F if false, T if true, - or ?
when the value cannot be determined). These values are concatenated in the order they appear (we assume it is the order of formulas as they were provided).

\subsubsection{``Known'' Models}

\input{result_known_CTLMix.tex}

\subsubsection{``Surprise'' Models}

\input{result_surprise_CTLMix.tex}

\subsection{Score for the CTLMix Examination}
\index{Scores!CTLMix}

Please find enclosed the scores for this examination (``Known'' and ``Surprise'' models).
We display only the score of tools that provide a results for at least one instance of one model.
The total is first listed in the table below followed by a detail, for each proposed model.
Meaning of the line labels are:
\begin{itemize}
\item\textbf{1st instance}: the tool gets a bonus for having processed the first instance of this model (+1 point),
\item\textbf{instances}: the tool gets 1 point per instances treated 
(for that, we assume that at least one formula has been successfully computed),
\item\textbf{max reached}: the tool could process all the instances for the model (+2 points),
\item\textbf{best}: the tool is among the ones that processed a maximum of instances within the time and memory confinement (+2 points).
\end{itemize}

\subsubsection{``Known'' Models}

\input{score_known_CTLMix.tex}

\subsubsection{``Surprise'' Models}

\input{score_surprise_CTLMix.tex}

\subsection{Trophies for this Examination}
\index{Trophies!CTLMix}

Trophies are divided in three categories: ``Known'' models,
``Surprise'' models, and the global trophies (formula is then
$score_{global} = score_{known} + 2 \times score_{surprise}$).

\subsubsection{For ``Known'' Models} \ \\

\begin{tabular}{c|c|c}
      1 & 1 & 3 \\
   \includegraphics[width=2cm]{figures/gold.jpg} &
   \includegraphics[width=2cm]{figures/gold.jpg} &
   \includegraphics[width=2cm]{figures/bronse.jpg} \\
   \acs{lola} &
   \acs{lola-optimistic} &
   \acs{lola-optimistic-incomplete} \\
   191 points &
   191 points &
   158 points \\
\end{tabular}

\subsubsection{For ``Surprise'' Models}\  \\

\begin{tabular}{c|c|c|c}
      1 & 2 & 2 & 2 \\
   \includegraphics[width=2cm]{figures/gold.jpg} &
   \includegraphics[width=2cm]{figures/silver.jpg} &
   \includegraphics[width=2cm]{figures/silver.jpg} &
   \includegraphics[width=2cm]{figures/silver.jpg} \\
   \acs{marcie} &
   \acs{lola} &
   \acs{lola-optimistic} &
   \acs{lola-optimistic-incomplete} \\
   17 points &
   12 points &
   12 points &
   12 points \\
\end{tabular}

\subsubsection{Global} \ \\

\begin{tabular}{c|c|c}
      1 & 1 & 3 \\
   \includegraphics[width=2cm]{figures/gold.jpg} &
   \includegraphics[width=2cm]{figures/gold.jpg} &
   \includegraphics[width=2cm]{figures/bronse.jpg} \\
   \acs{lola} &
   \acs{lola-optimistic} &
   \acs{lola-optimistic-incomplete} \\
   215 points &
   215 points &
   182 points \\
\end{tabular}

\part{LTL-based Analysis}
\label{part:five}
\newpage

\section{The LTLCardinalityComparison Examination}
\label{sec:exam:LTLCardinalityComparison}
\index{Results!LTLCardinalityComparison}

This examination deals with LTL properties dealing with checking cardinality of marking only.
We first show a summary on the handling of models by the participating tools.
Then, we present the computed outputs and the associated scores for this
examination prior to a summary of relevant executions.

\subsection{Handling of Models by Tools}
\index{Performances!LTLCardinalityComparison}

\subsubsection{\acs{CSRepetitions-COL}}
No instance of this model could be computed for the \textbf{LTLCardinalityComparison} examination.

\subsubsection{\acs{CSRepetitions-PT}}
The charts below respectively show how tools compete with this ``Known'' model (memory and CPU).

\index{Performances!LTLCardinalityComparison!CSRepetitions (P/T)}
\begin{center}
   \includegraphics[width=7.2cm]{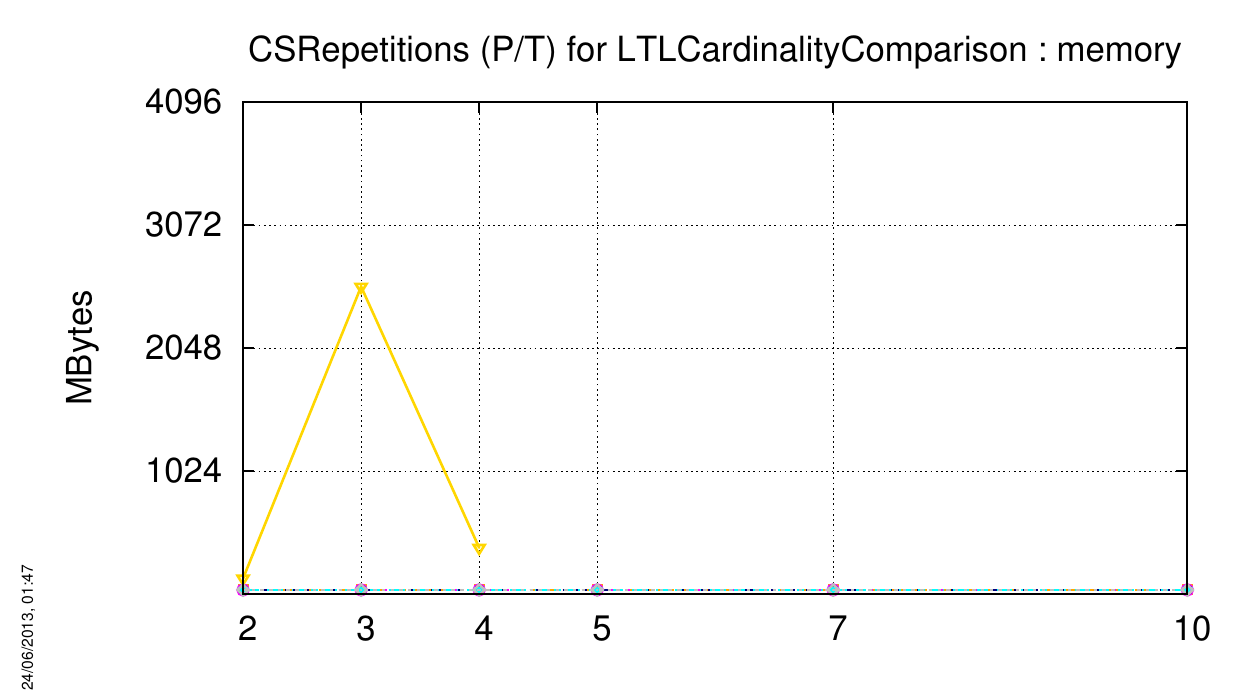}
   \includegraphics[width=7.2cm]{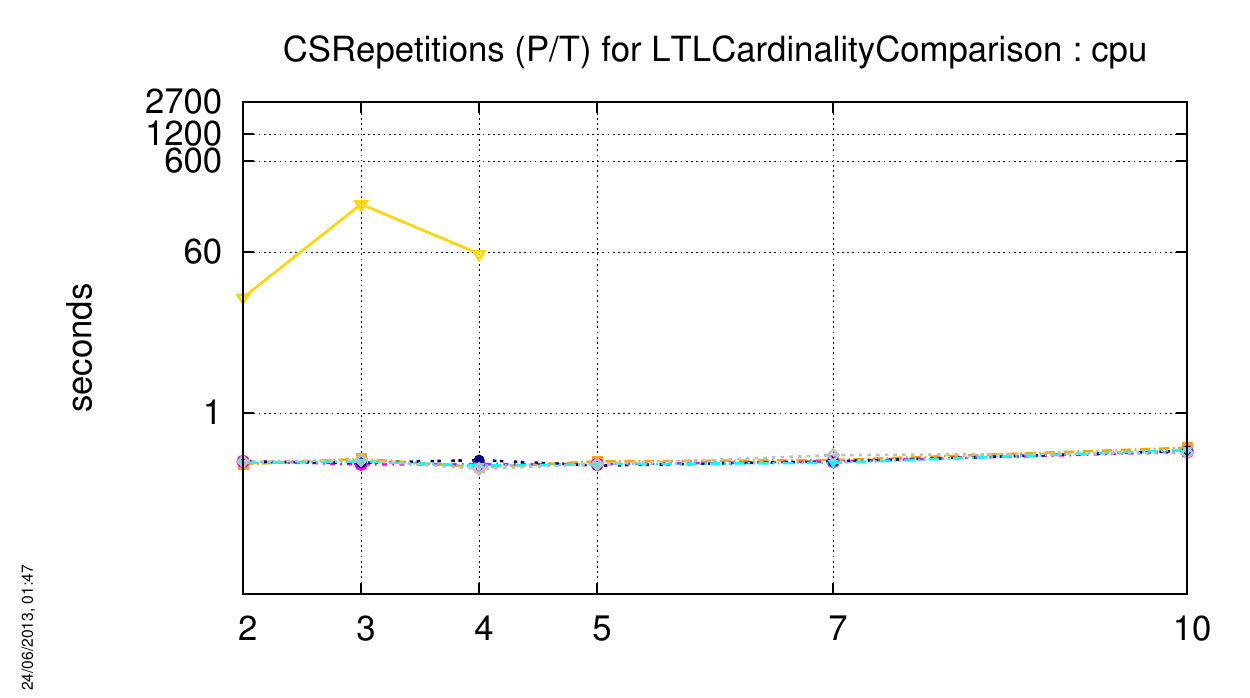}

   \includegraphics[height=1cm]{figures/tools-legend.pdf}
\end{center}

\subsubsection{\acs{Dekker-PT}}
The charts below respectively show how tools compete with this ``Known'' model (memory and CPU).

\index{Performances!LTLCardinalityComparison!Dekker (P/T)}
\begin{center}
   \includegraphics[width=7.2cm]{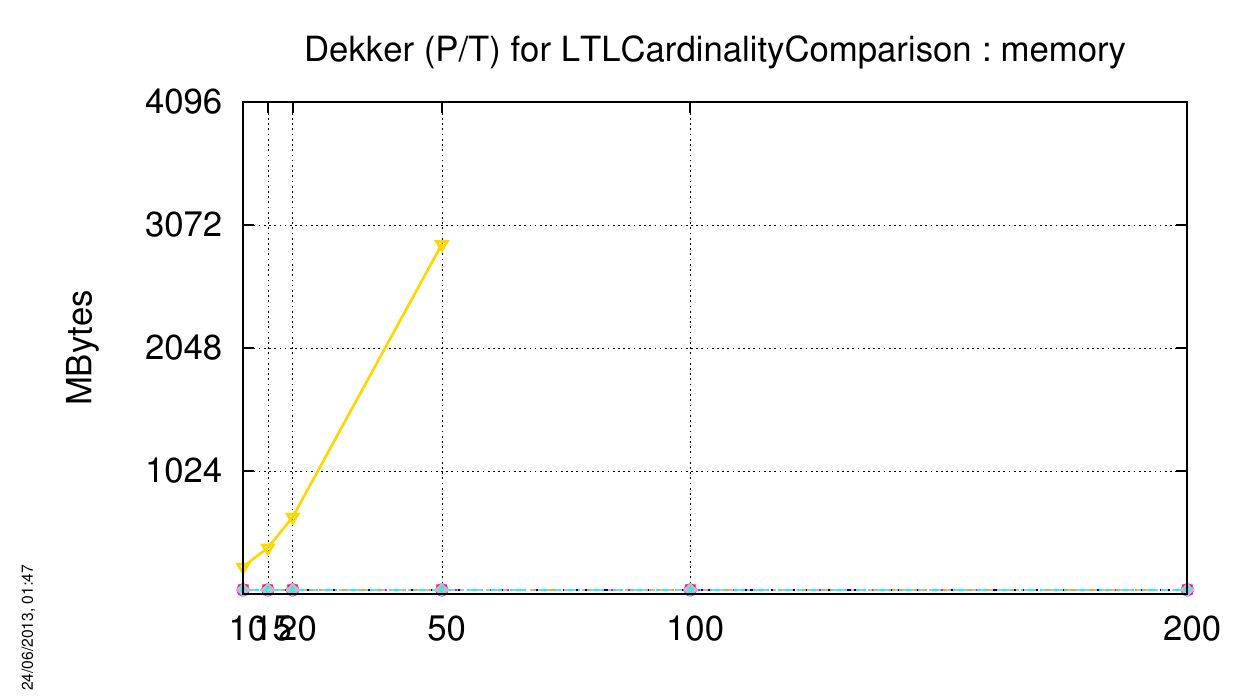}
   \includegraphics[width=7.2cm]{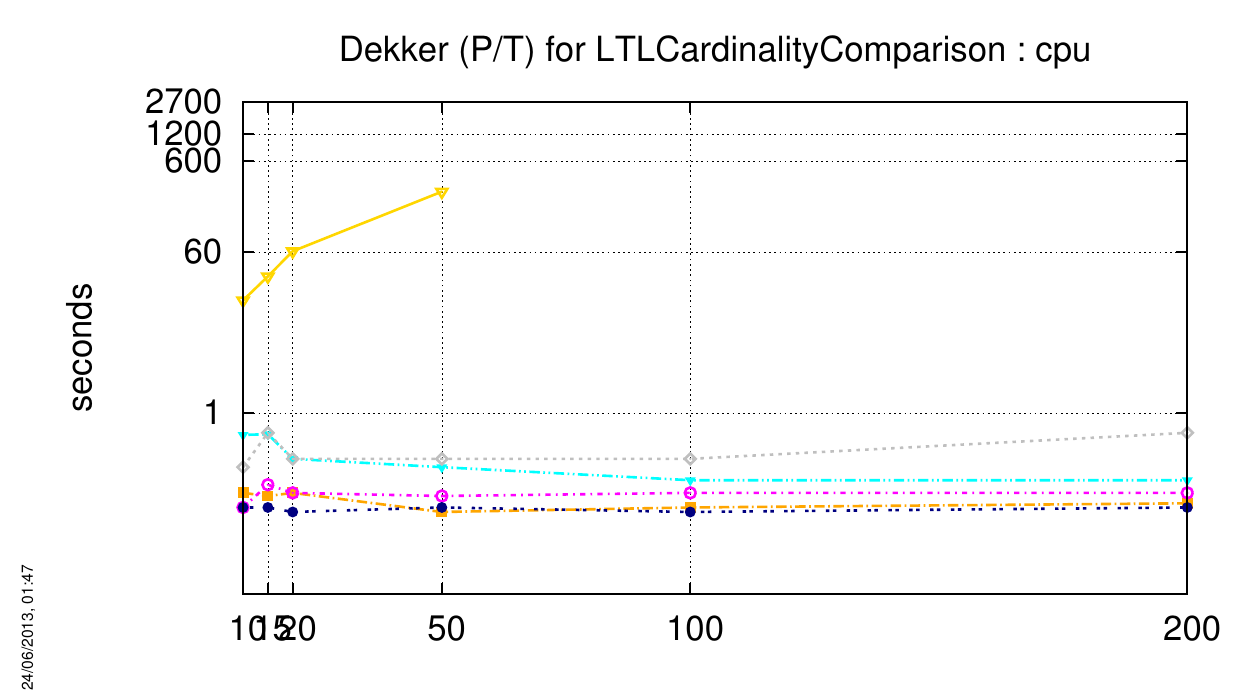}

   \includegraphics[height=1cm]{figures/tools-legend.pdf}
\end{center}

\subsubsection{\acs{DotAndBoxes-COL}}
No instance of this model could be computed for the \textbf{LTLCardinalityComparison} examination.

\subsubsection{\acs{DrinkVendingMachine-COL}}
No instance of this model could be computed for the \textbf{LTLCardinalityComparison} examination.

\subsubsection{\acs{DrinkVendingMachine-PT}}
The charts below respectively show how tools compete with this ``Known'' model (memory and CPU).

\index{Performances!LTLCardinalityComparison!DrinkVendingMachine (P/T)}
\begin{center}
   \includegraphics[width=7.2cm]{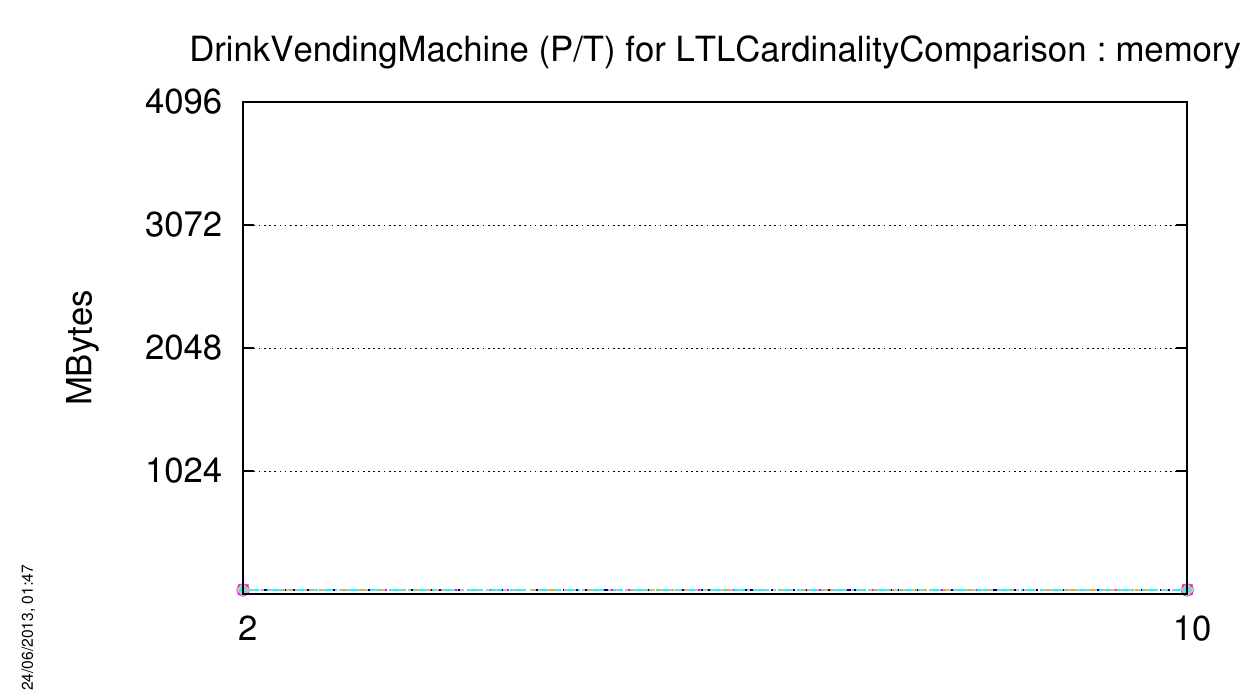}
   \includegraphics[width=7.2cm]{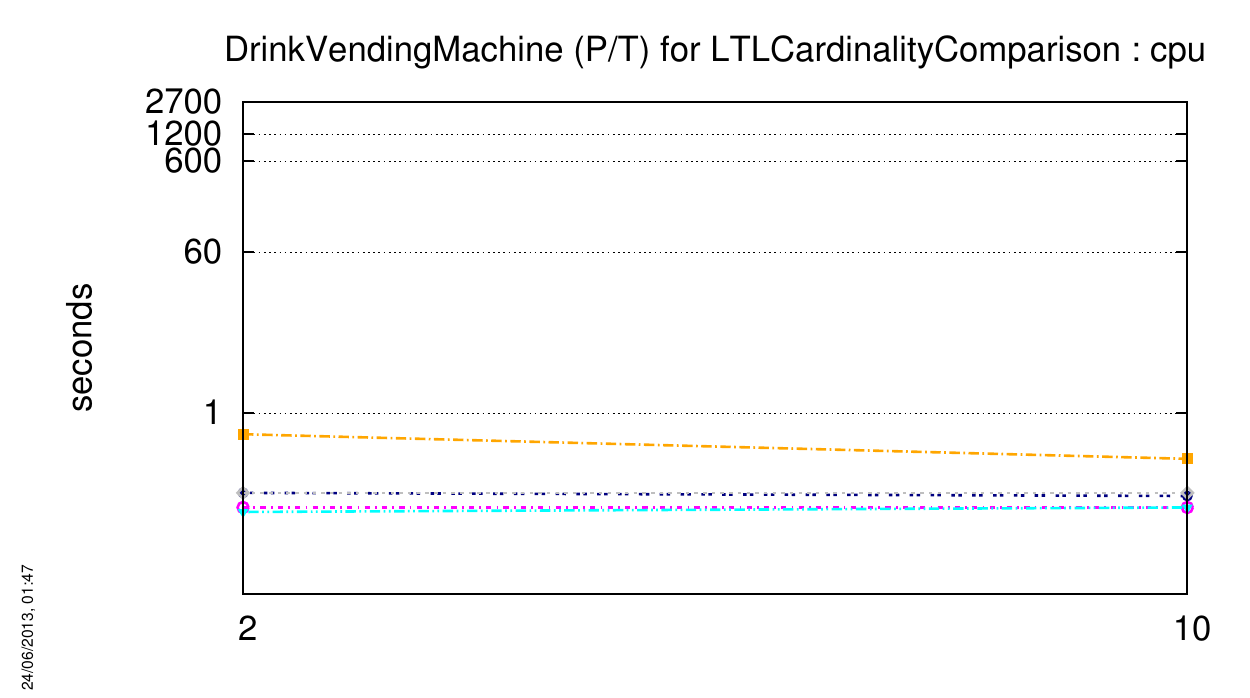}

   \includegraphics[height=1cm]{figures/tools-legend.pdf}
\end{center}

\subsubsection{\acs{Echo-PT}}
The charts below respectively show how tools compete with this ``Known'' model (memory and CPU).

\index{Performances!LTLCardinalityComparison!Echo (P/T)}
\begin{center}
   \includegraphics[width=7.2cm]{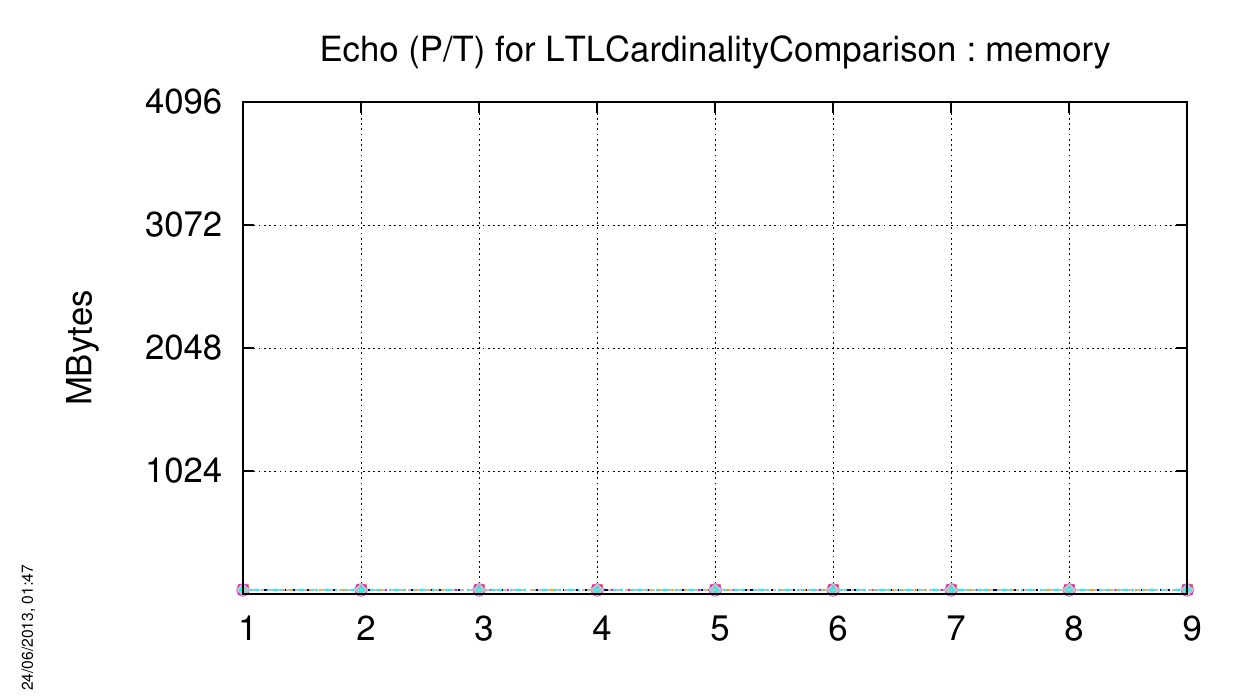}
   \includegraphics[width=7.2cm]{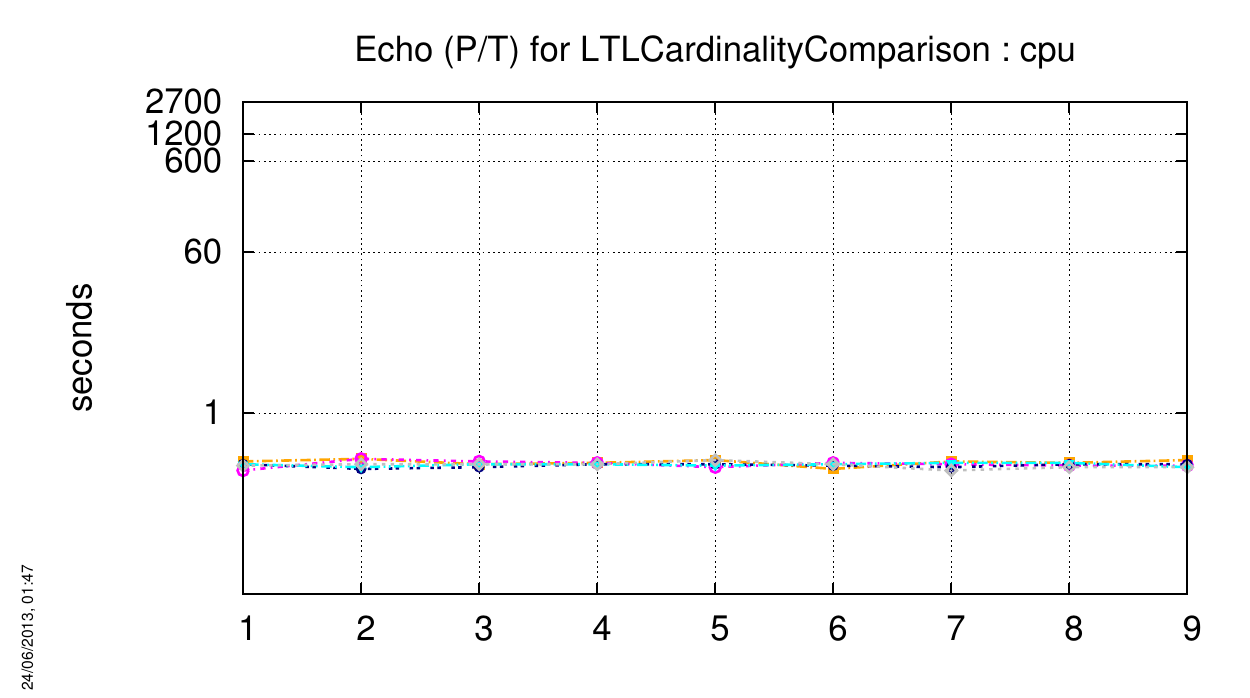}

   \includegraphics[height=1cm]{figures/tools-legend.pdf}
\end{center}

\subsubsection{\acs{Eratosthenes-PT}}
The charts below respectively show how tools compete with this ``Known'' model (memory and CPU).

\index{Performances!LTLCardinalityComparison!Eratosthenes (P/T)}
\begin{center}
   \includegraphics[width=7.2cm]{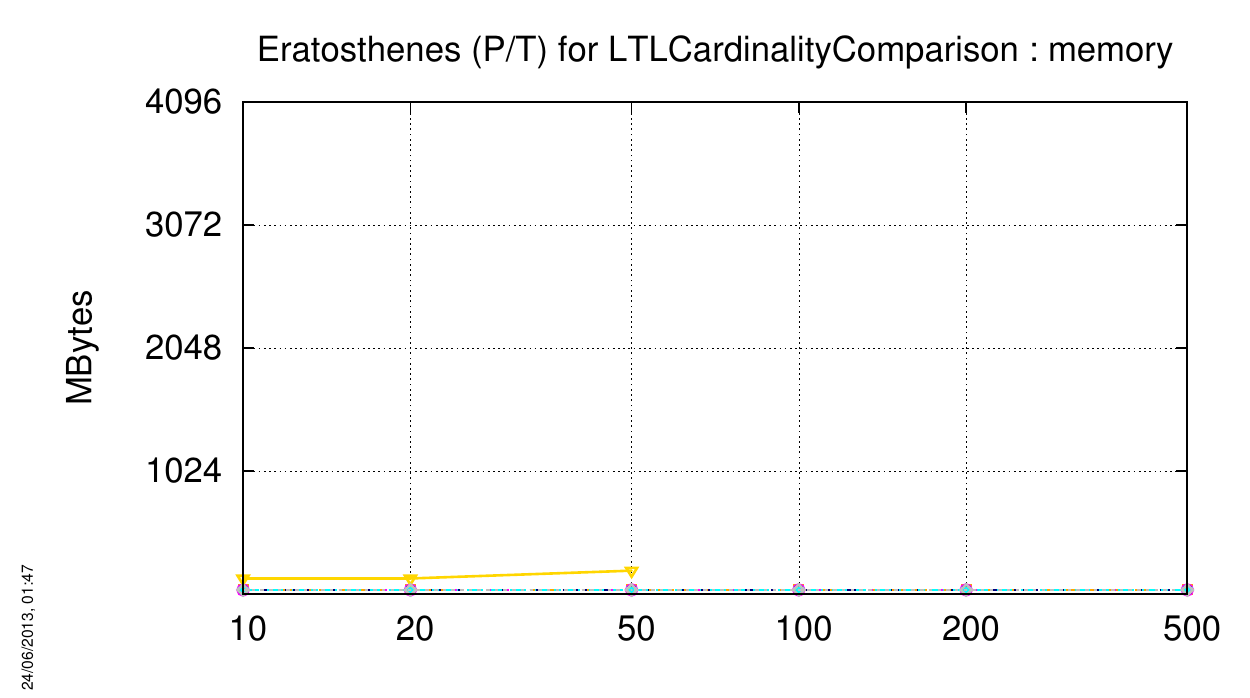}
   \includegraphics[width=7.2cm]{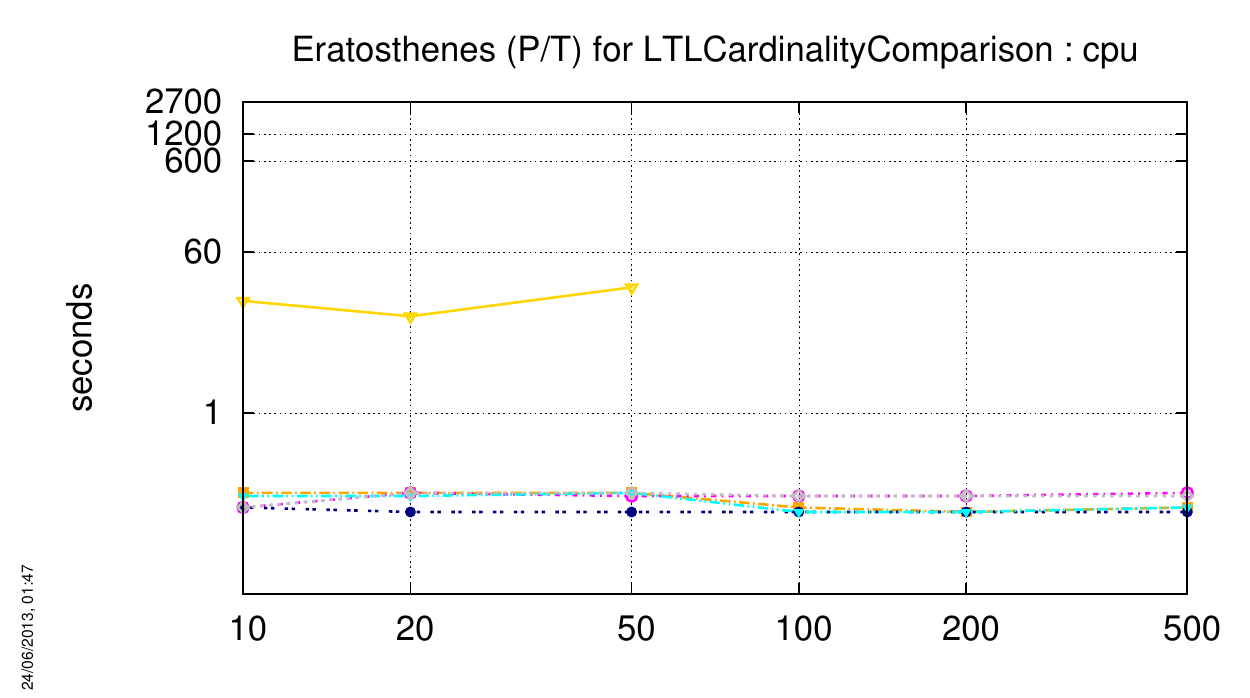}

   \includegraphics[height=1cm]{figures/tools-legend.pdf}
\end{center}

\subsubsection{\acs{FMS-PT}}
The charts below respectively show how tools compete with this ``Known'' model (memory and CPU).

\index{Performances!LTLCardinalityComparison!FMS (P/T)}
\begin{center}
   \includegraphics[width=7.2cm]{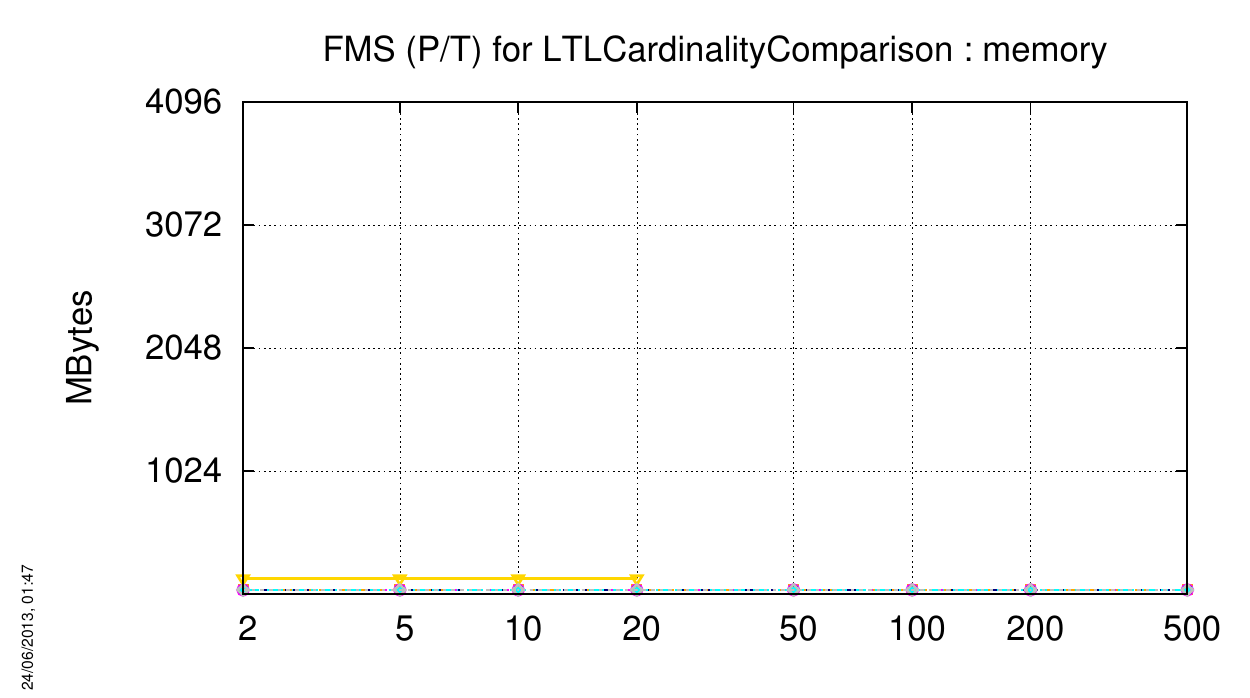}
   \includegraphics[width=7.2cm]{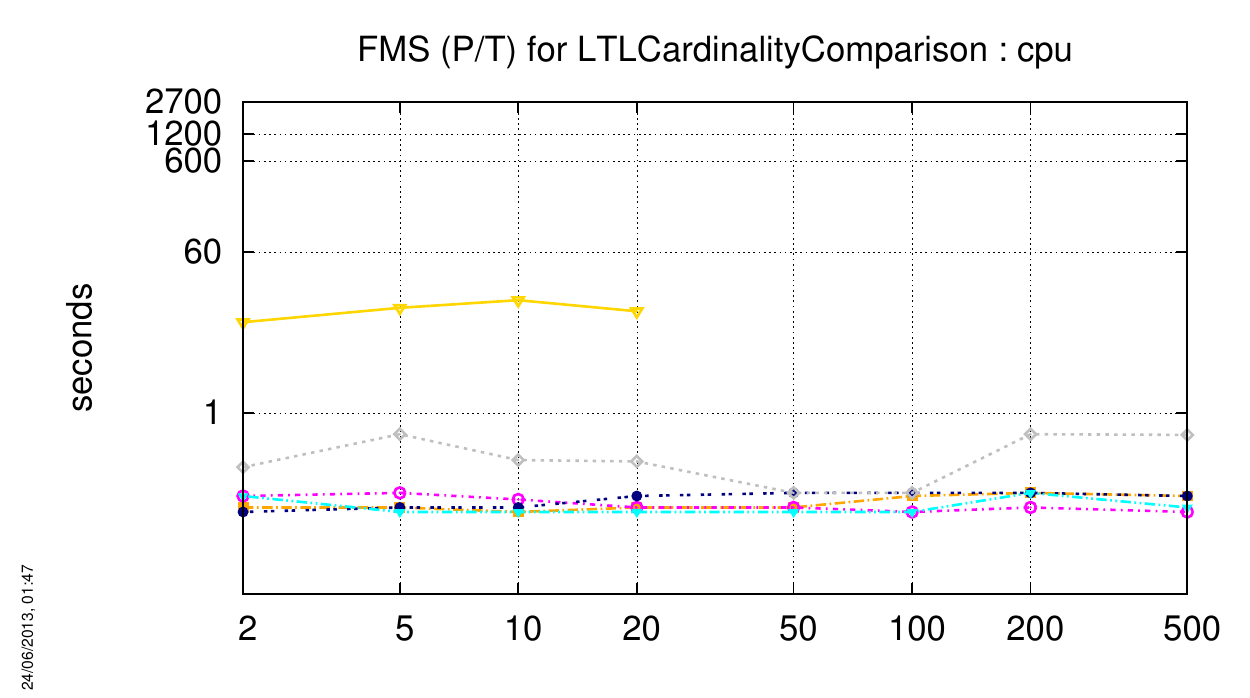}

   \includegraphics[height=1cm]{figures/tools-legend.pdf}
\end{center}

\subsubsection{\acs{GlobalRessAlloc-COL}}
No instance of this model could be computed for the \textbf{LTLCardinalityComparison} examination.

\subsubsection{\acs{GlobalRessAlloc-PT}}
The charts below respectively show how tools compete with this ``Known'' model (memory and CPU).

\index{Performances!LTLCardinalityComparison!GlobalRessAlloc (P/T)}
\begin{center}
   \includegraphics[width=7.2cm]{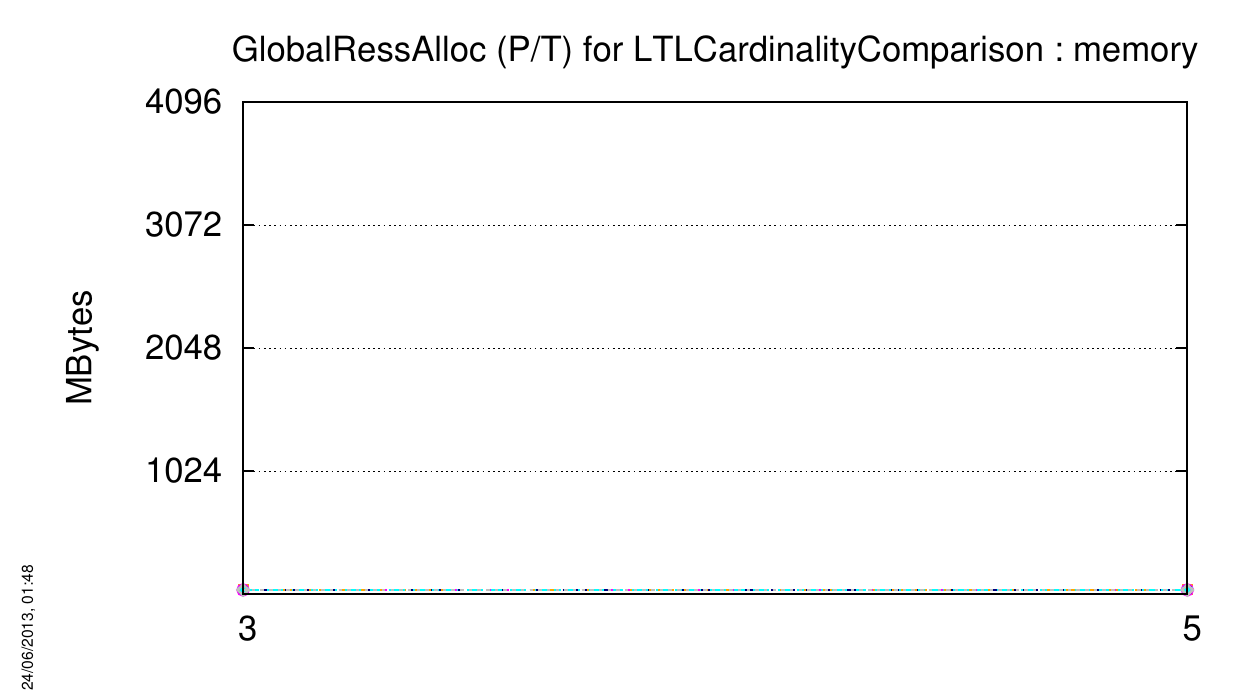}
   \includegraphics[width=7.2cm]{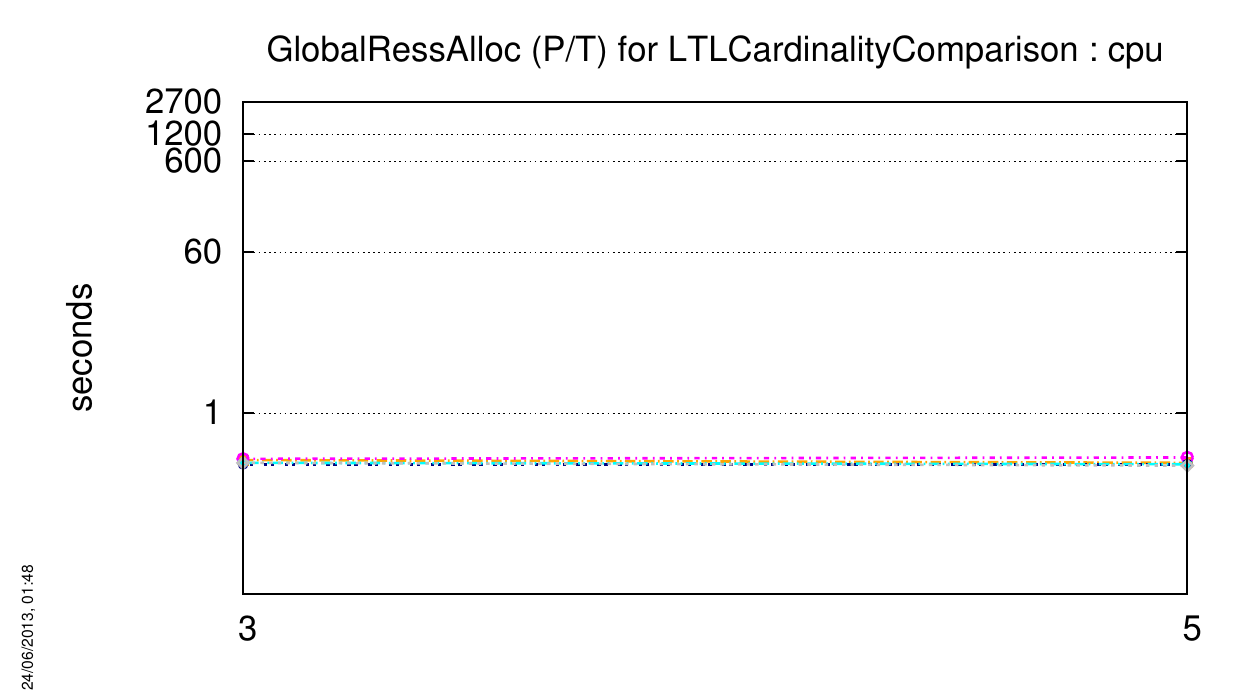}

   \includegraphics[height=1cm]{figures/tools-legend.pdf}
\end{center}

\subsubsection{\acs{Kanban-PT}}
The charts below respectively show how tools compete with this ``Known'' model (memory and CPU).

\index{Performances!LTLCardinalityComparison!Kanban (P/T)}
\begin{center}
   \includegraphics[width=7.2cm]{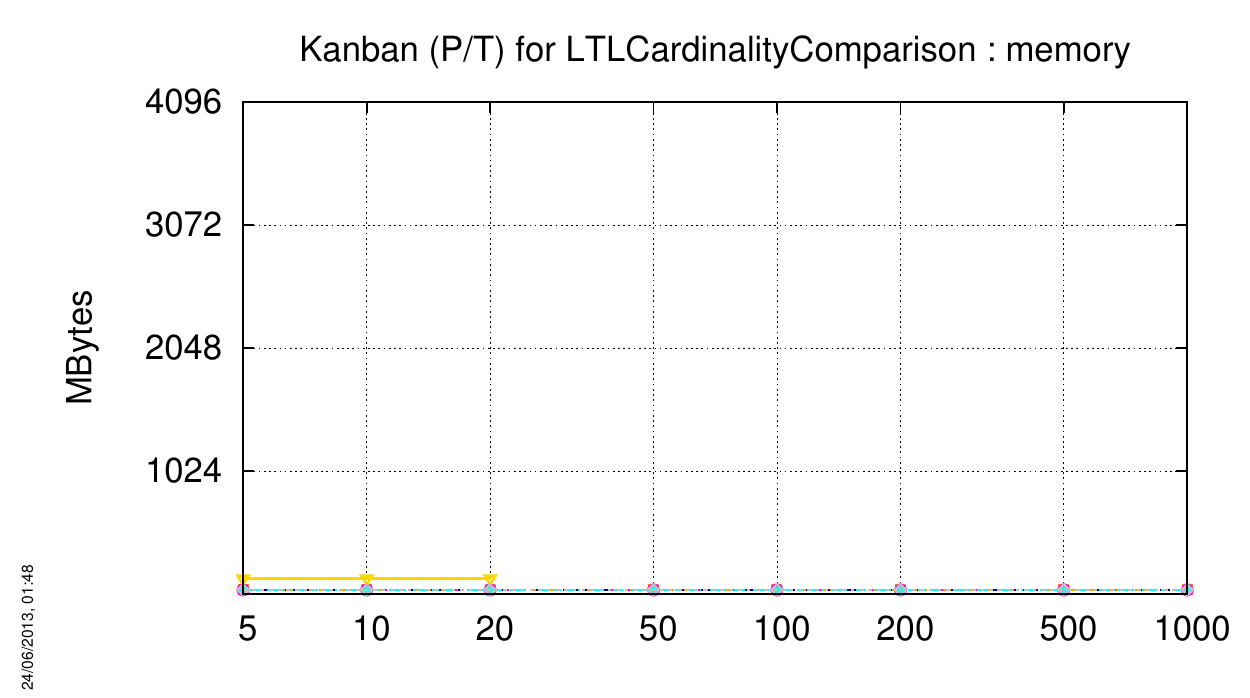}
   \includegraphics[width=7.2cm]{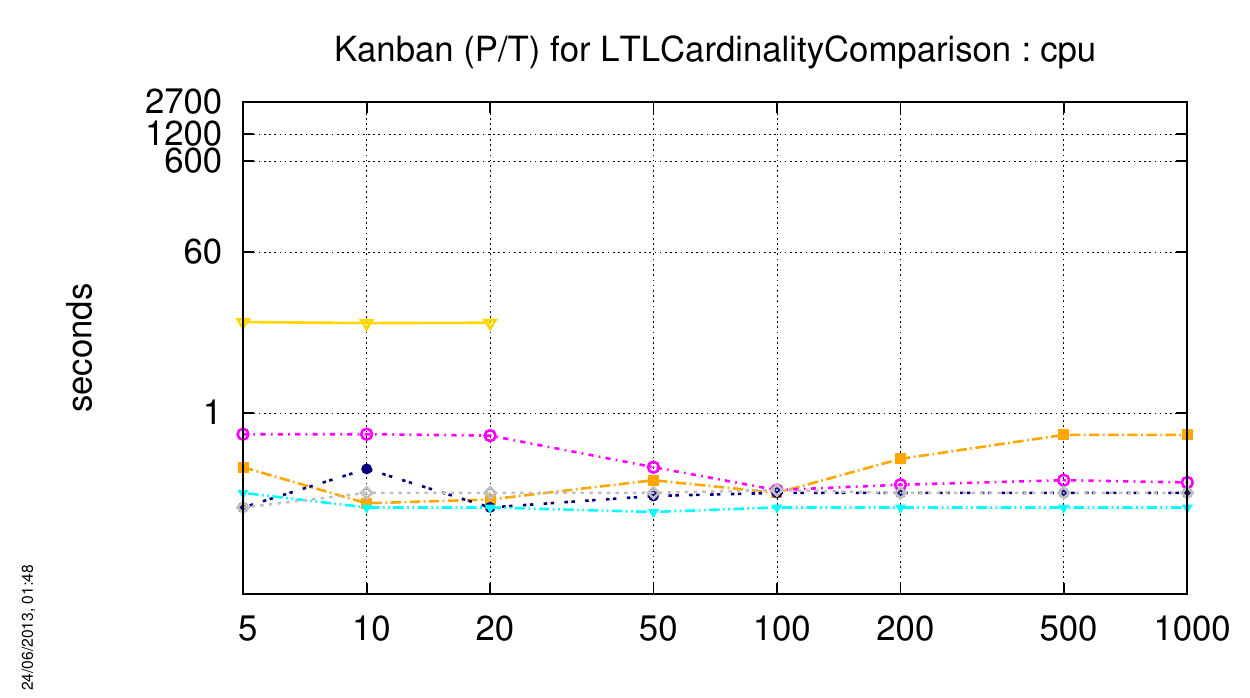}

   \includegraphics[height=1cm]{figures/tools-legend.pdf}
\end{center}

\subsubsection{\acs{LamportFastMutEx-COL}}
No instance of this model could be computed for the \textbf{LTLCardinalityComparison} examination.

\subsubsection{\acs{LamportFastMutEx-PT}}
The charts below respectively show how tools compete with this ``Known'' model (memory and CPU).

\index{Performances!LTLCardinalityComparison!LamportFastMutEx (P/T)}
\begin{center}
   \includegraphics[width=7.2cm]{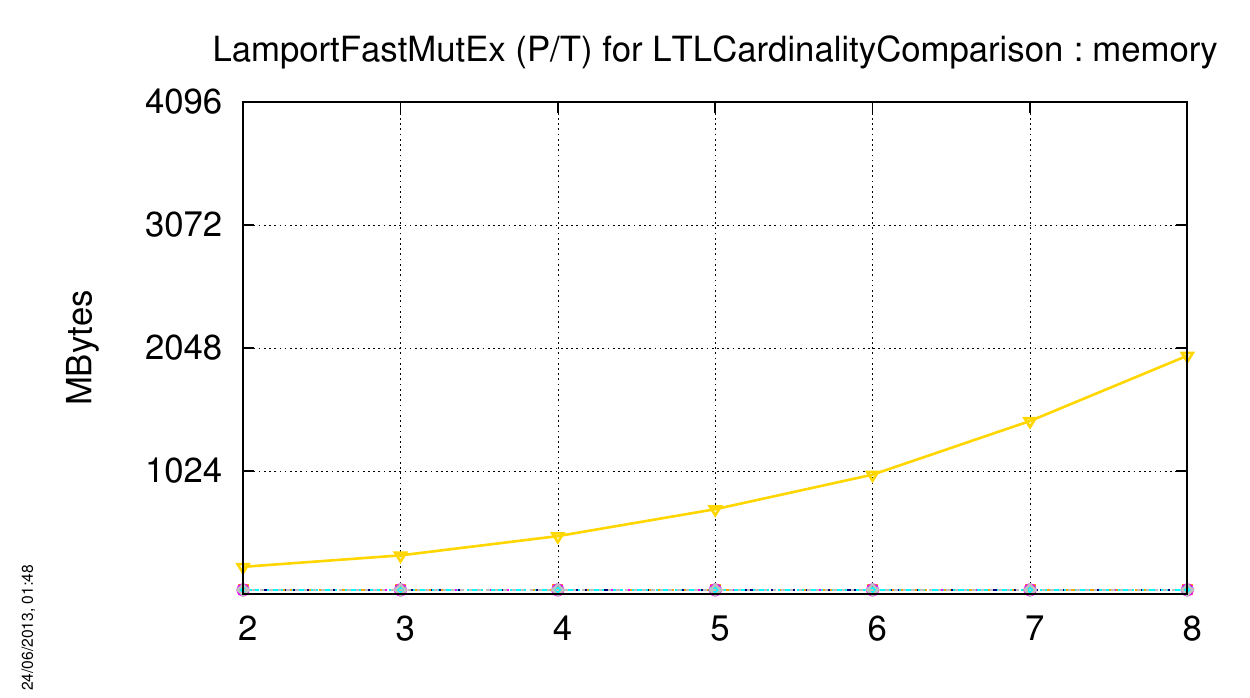}
   \includegraphics[width=7.2cm]{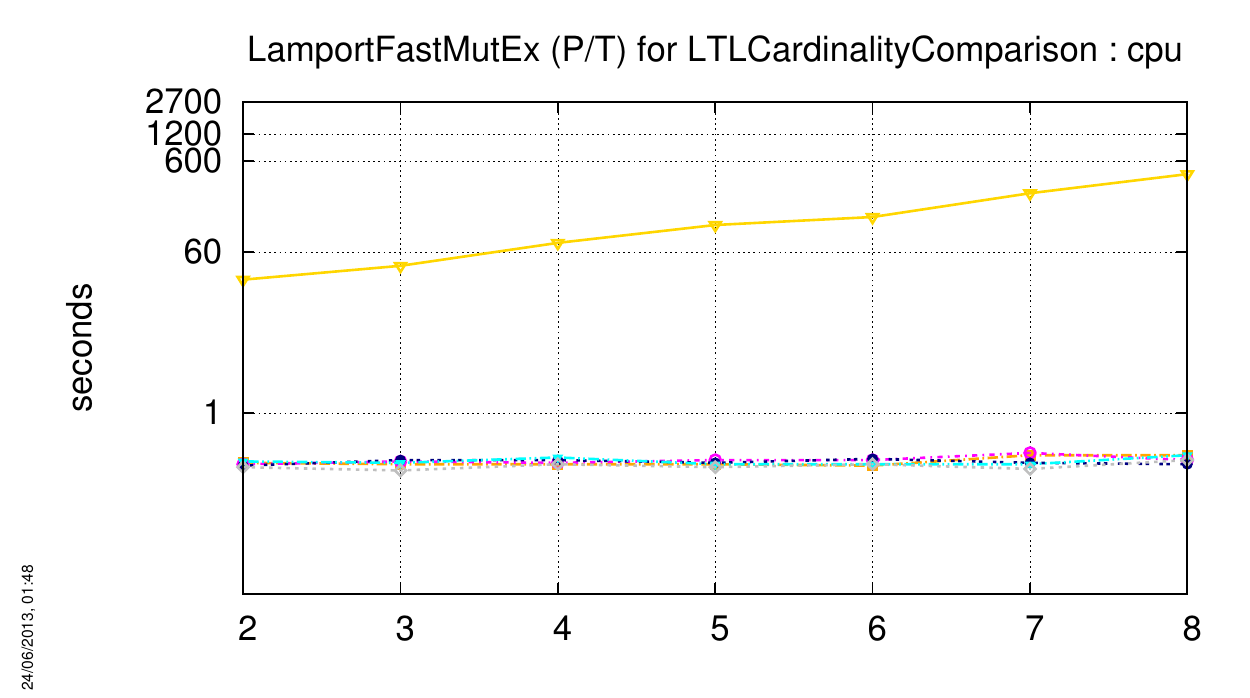}

   \includegraphics[height=1cm]{figures/tools-legend.pdf}
\end{center}

\subsubsection{\acs{MAPK-PT}}
The charts below respectively show how tools compete with this ``Known'' model (memory and CPU).

\index{Performances!LTLCardinalityComparison!MAPK (P/T)}
\begin{center}
   \includegraphics[width=7.2cm]{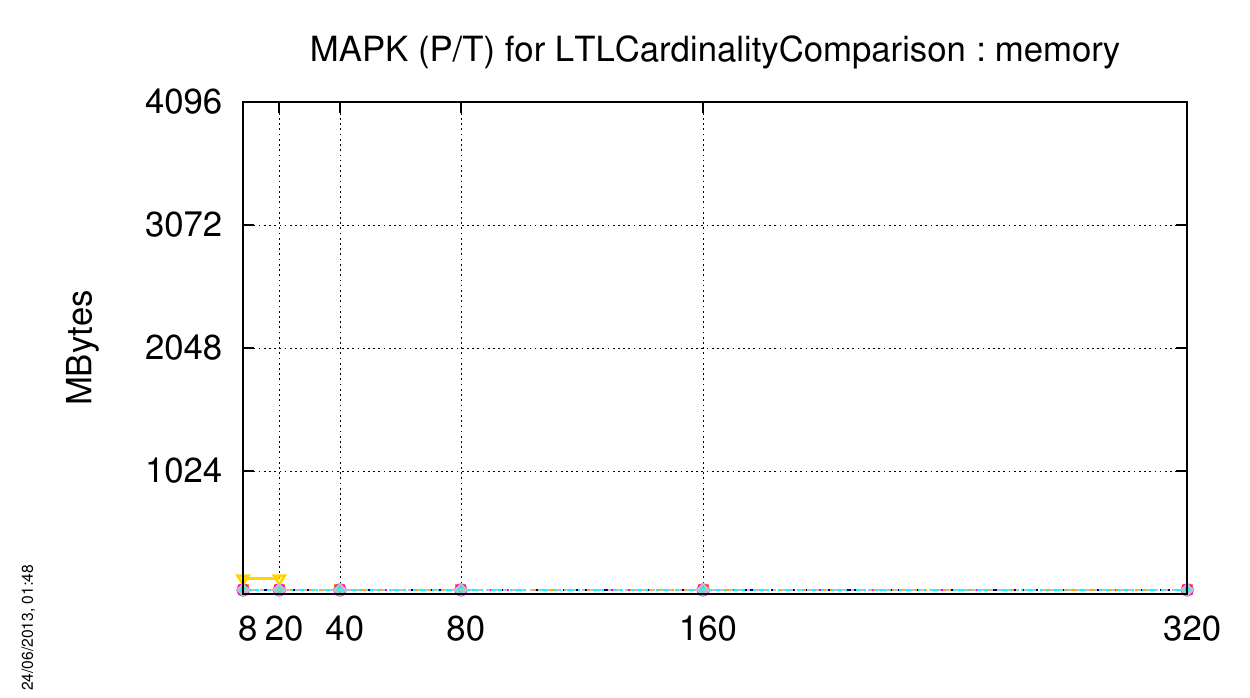}
   \includegraphics[width=7.2cm]{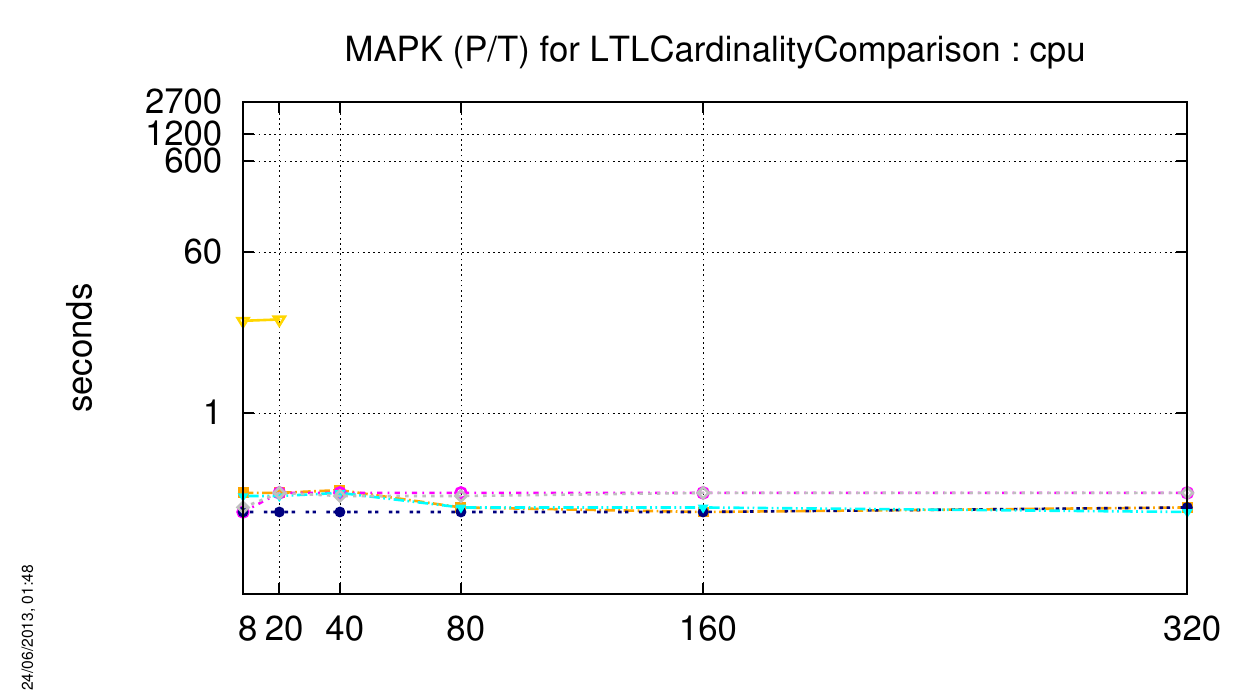}

   \includegraphics[height=1cm]{figures/tools-legend.pdf}
\end{center}

\subsubsection{\acs{NeoElection-COL}}
No instance of this model could be computed for the \textbf{LTLCardinalityComparison} examination.

\subsubsection{\acs{NeoElection-PT}}
The charts below respectively show how tools compete with this ``Known'' model (memory and CPU).

\index{Performances!LTLCardinalityComparison!NeoElection (P/T)}
\begin{center}
   \includegraphics[width=7.2cm]{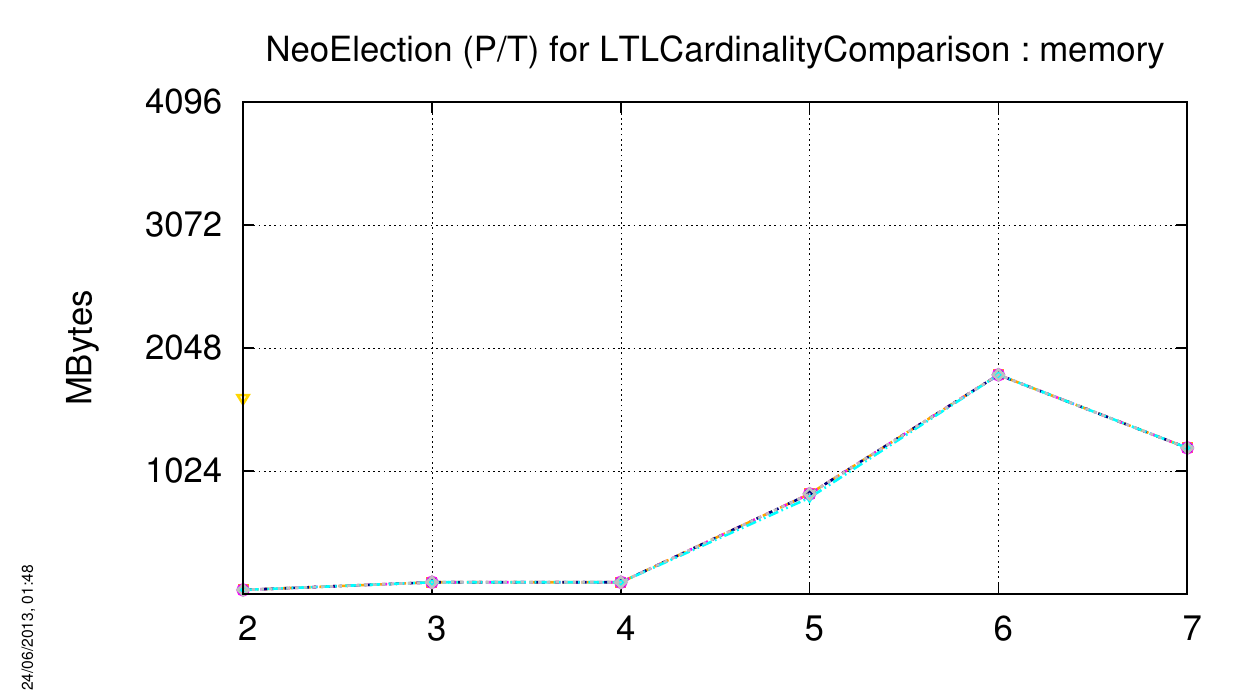}
   \includegraphics[width=7.2cm]{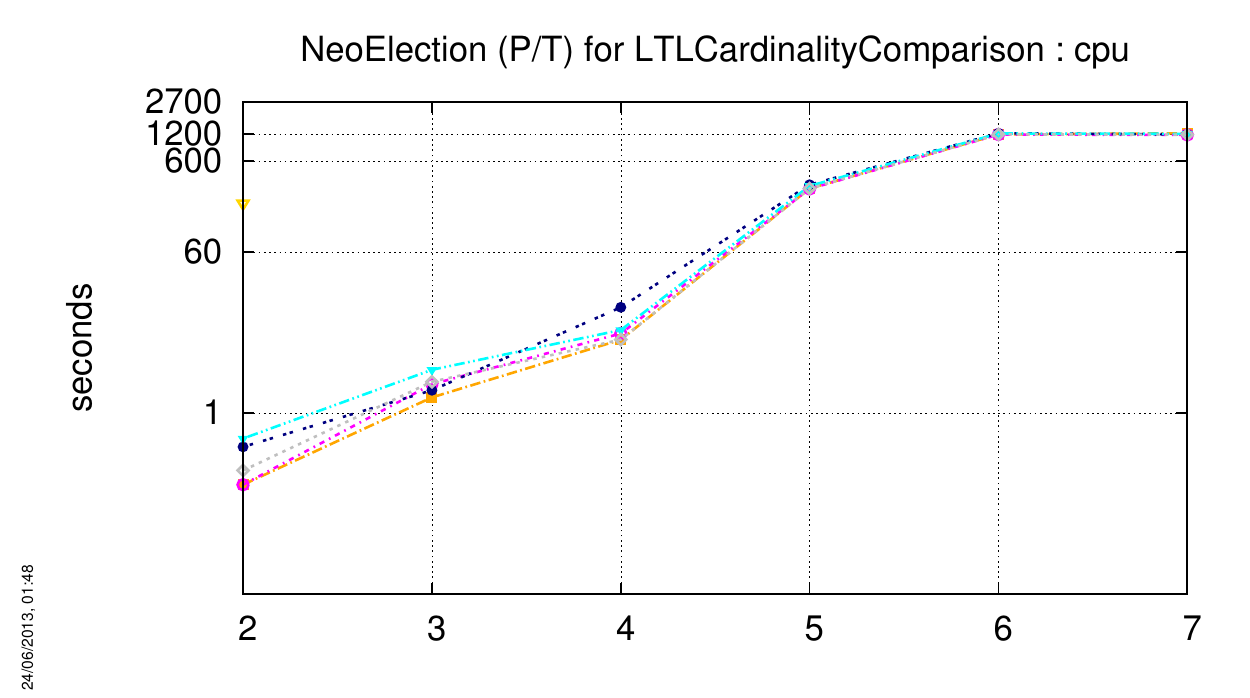}

   \includegraphics[height=1cm]{figures/tools-legend.pdf}
\end{center}

\subsubsection{\acs{PermAdmissibility-COL}}
No instance of this model could be computed for the \textbf{LTLCardinalityComparison} examination.

\subsubsection{\acs{PermAdmissibility-PT}}
The charts below respectively show how tools compete with this ``Known'' model (memory and CPU).

\index{Performances!LTLCardinalityComparison!PermAdmissibility (P/T)}
\begin{center}
   \includegraphics[width=7.2cm]{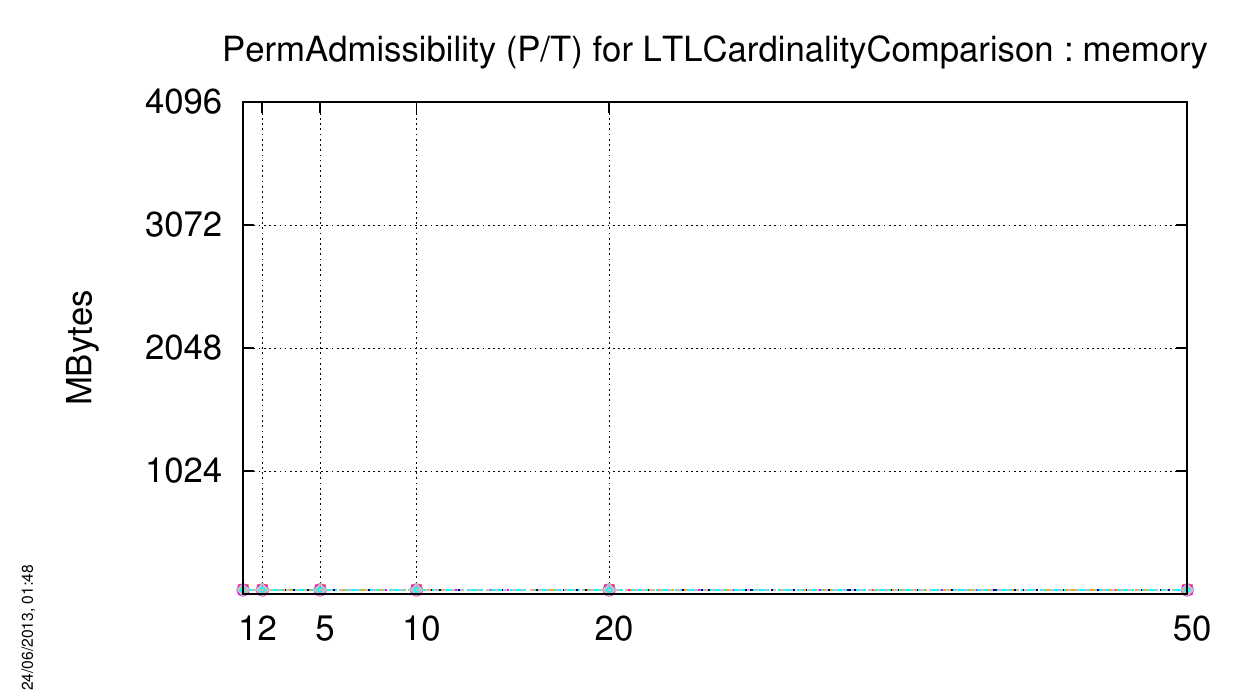}
   \includegraphics[width=7.2cm]{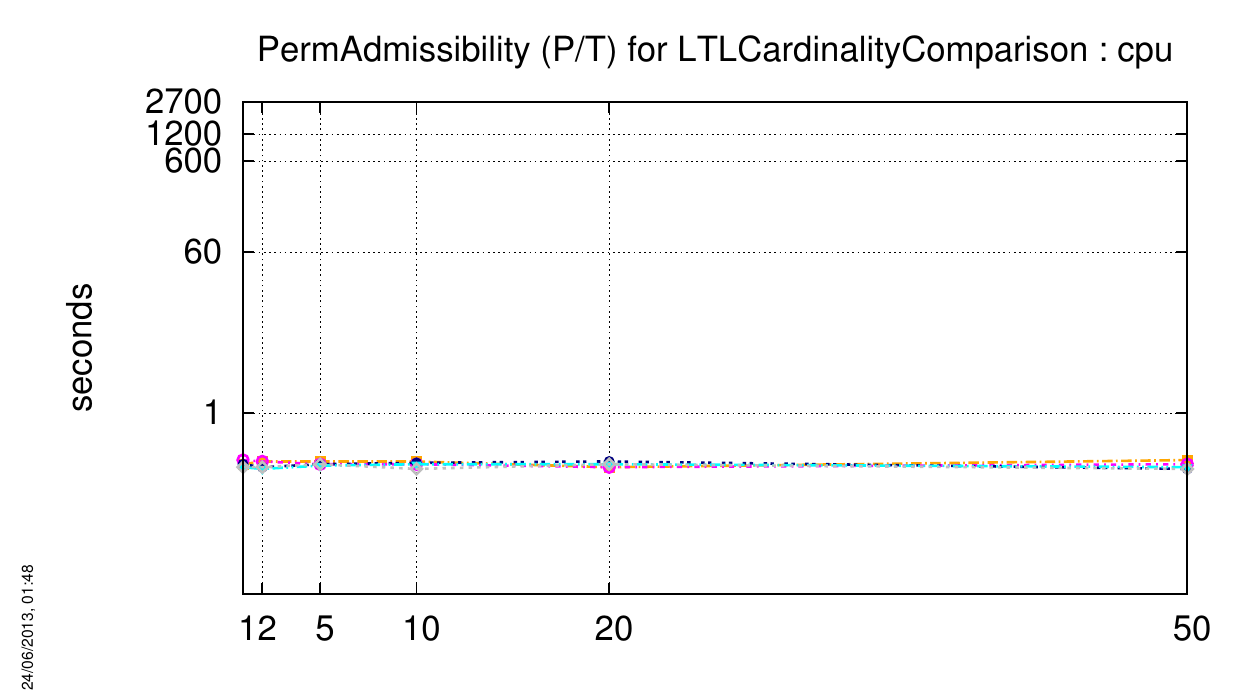}

   \includegraphics[height=1cm]{figures/tools-legend.pdf}
\end{center}

\subsubsection{\acs{Peterson-COL}}
No instance of this model could be computed for the \textbf{LTLCardinalityComparison} examination.

\subsubsection{\acs{Peterson-PT}}
The charts below respectively show how tools compete with this ``Known'' model (memory and CPU).

\index{Performances!LTLCardinalityComparison!Peterson (P/T)}
\begin{center}
   \includegraphics[width=7.2cm]{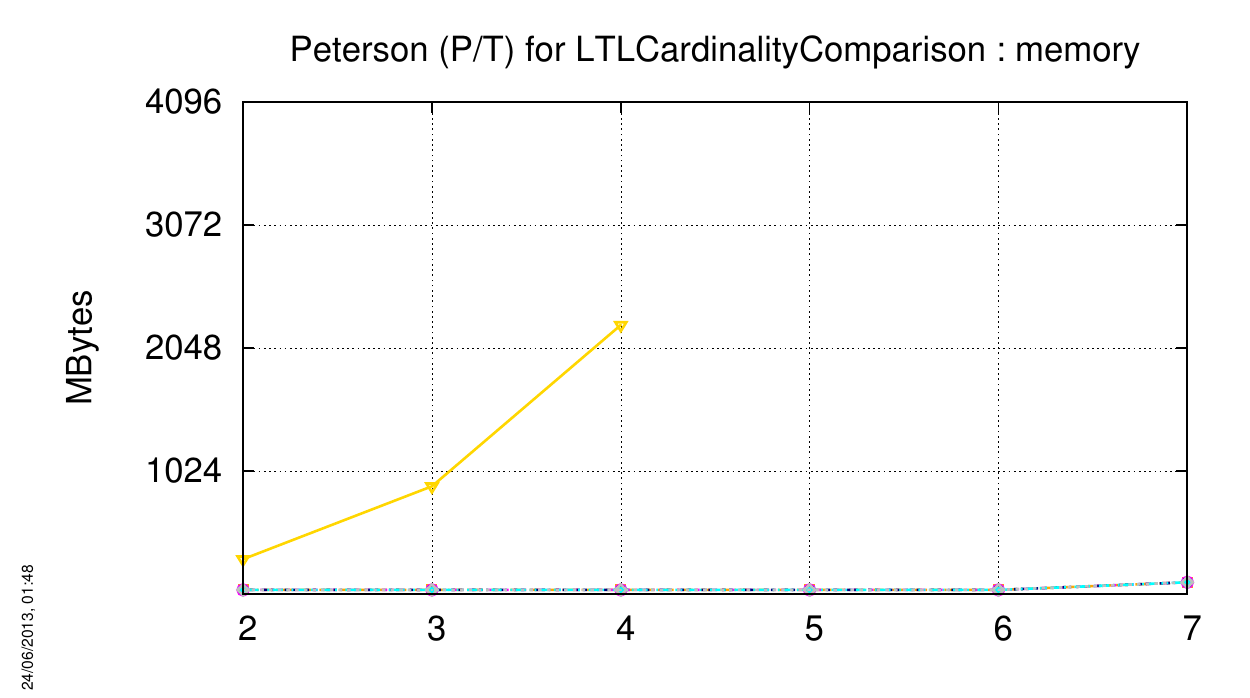}
   \includegraphics[width=7.2cm]{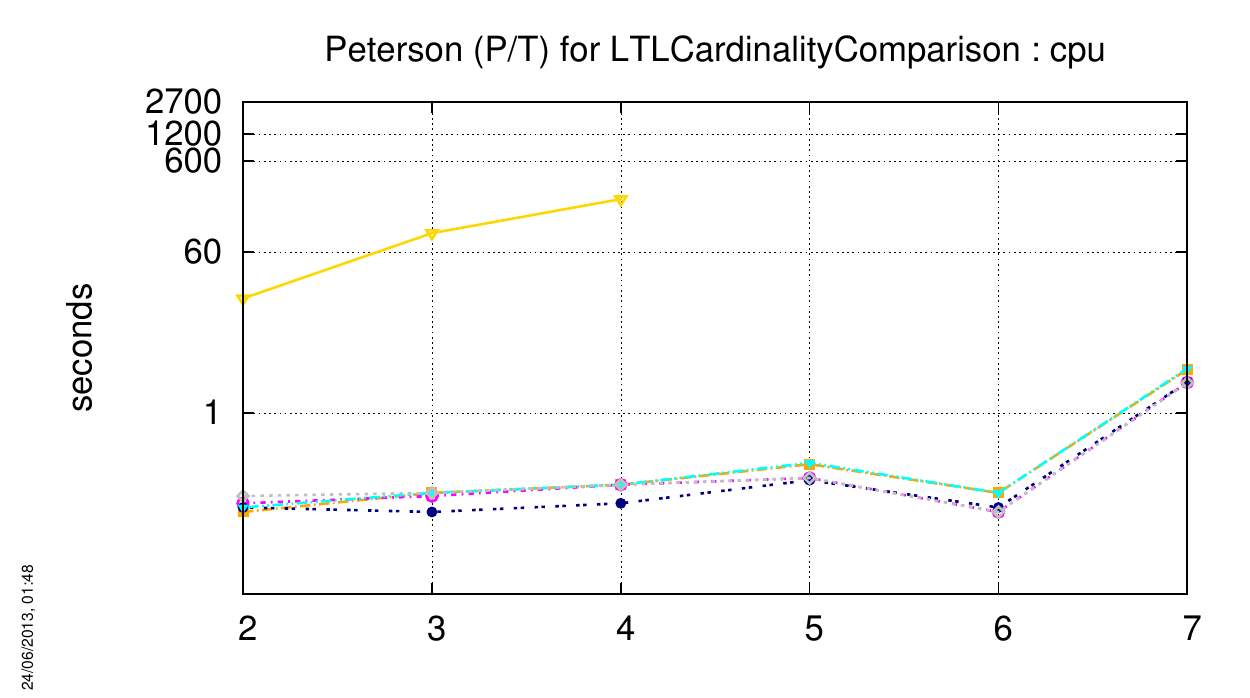}

   \includegraphics[height=1cm]{figures/tools-legend.pdf}
\end{center}

\subsubsection{\acs{Philosophers-COL}}
No instance of this model could be computed for the \textbf{LTLCardinalityComparison} examination.

\subsubsection{\acs{Philosophers-PT}}
The charts below respectively show how tools compete with this ``Known'' model (memory and CPU).

\index{Performances!LTLCardinalityComparison!Philosophers (P/T)}
\begin{center}
   \includegraphics[width=7.2cm]{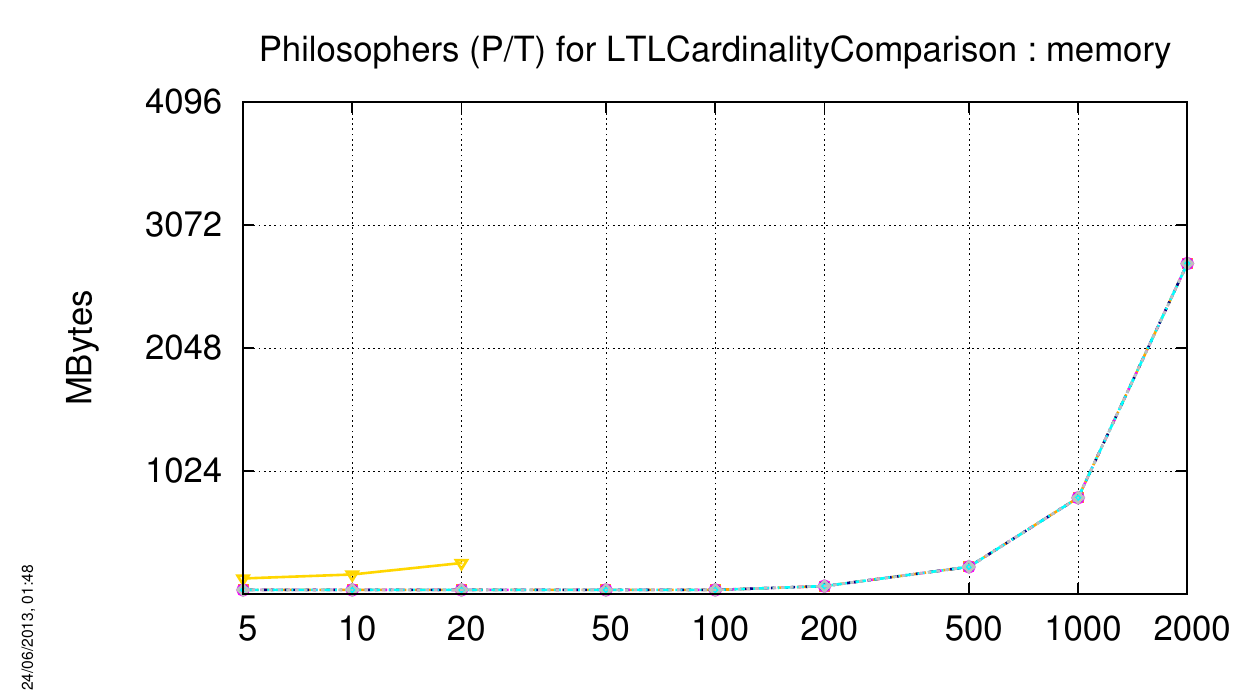}
   \includegraphics[width=7.2cm]{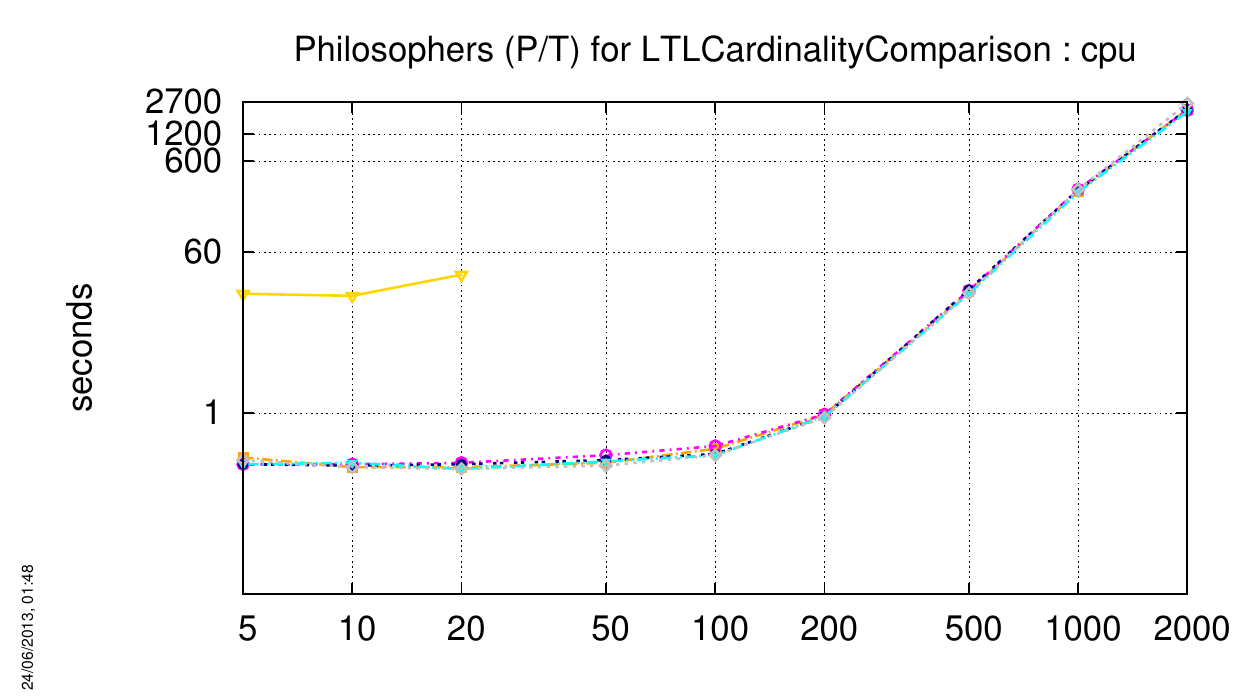}

   \includegraphics[height=1cm]{figures/tools-legend.pdf}
\end{center}

\subsubsection{\acs{PhilosophersDyn-COL}}
No instance of this model could be computed for the \textbf{LTLCardinalityComparison} examination.

\subsubsection{\acs{PhilosophersDyn-PT}}
The charts below respectively show how tools compete with this ``Known'' model (memory and CPU).

\index{Performances!LTLCardinalityComparison!PhilosophersDyn (P/T)}
\begin{center}
   \includegraphics[width=7.2cm]{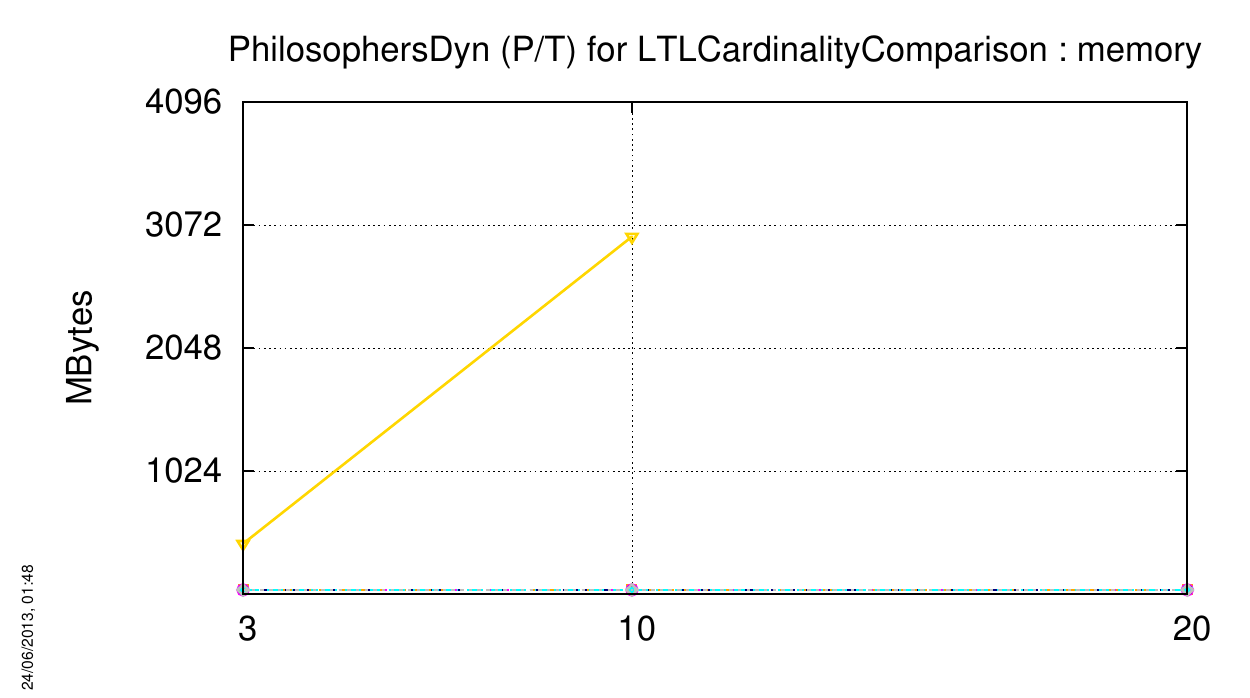}
   \includegraphics[width=7.2cm]{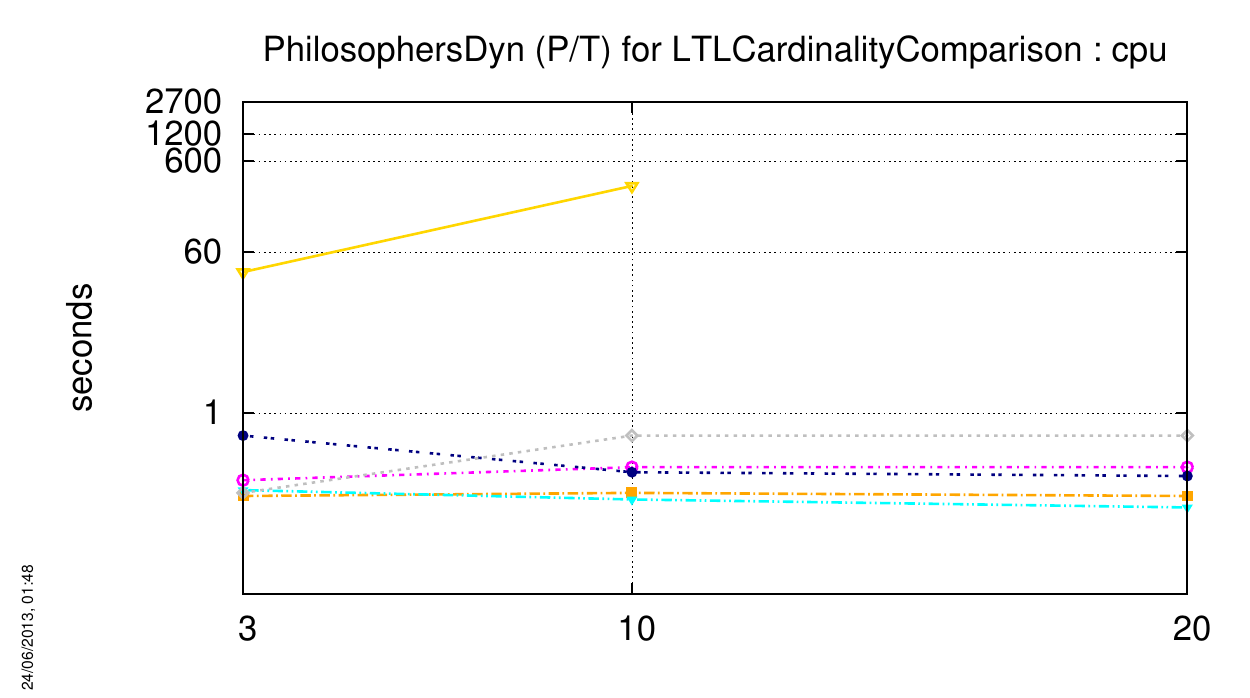}

   \includegraphics[height=1cm]{figures/tools-legend.pdf}
\end{center}

\subsubsection{\acs{Planning-PT}}
No instance of this model could be computed for the \textbf{LTLCardinalityComparison} examination.

\subsubsection{\acs{Railroad-PT}}
The charts below respectively show how tools compete with this ``Known'' model (memory and CPU).

\index{Performances!LTLCardinalityComparison!Railroad (P/T)}
\begin{center}
   \includegraphics[width=7.2cm]{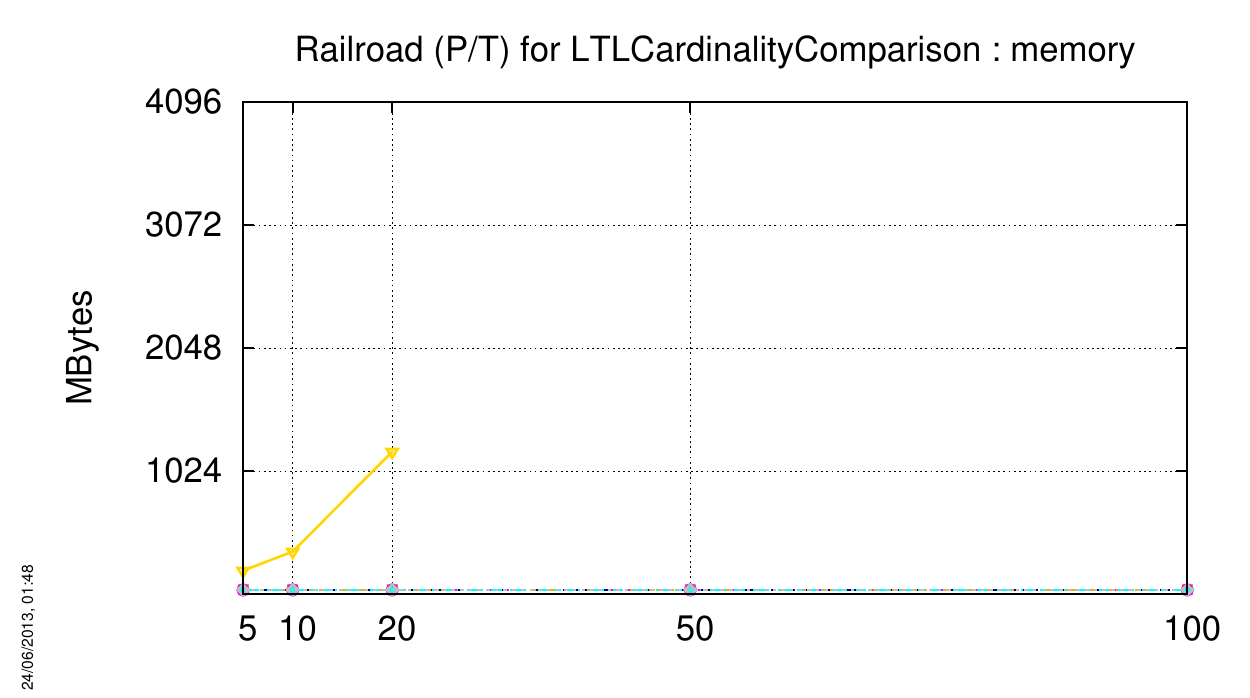}
   \includegraphics[width=7.2cm]{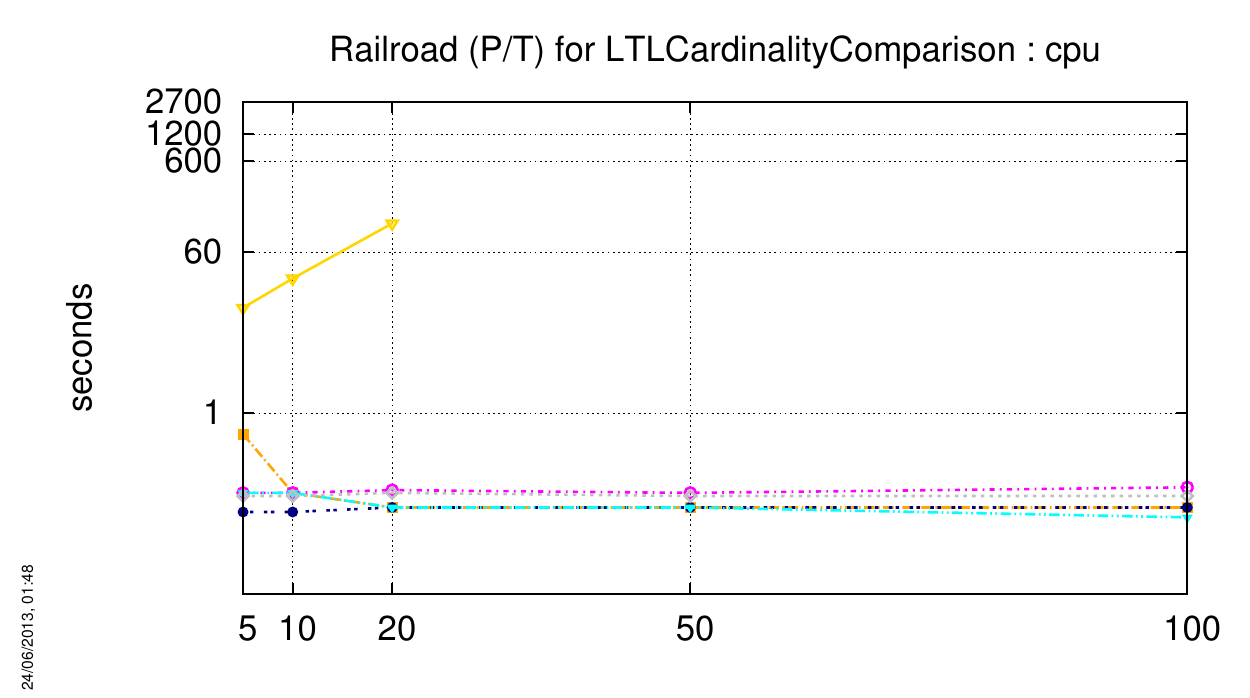}

   \includegraphics[height=1cm]{figures/tools-legend.pdf}
\end{center}

\subsubsection{\acs{RessAllocation-PT}}
The charts below respectively show how tools compete with this ``Known'' model (memory and CPU).

\index{Performances!LTLCardinalityComparison!RessAllocation (P/T)}
\begin{center}
   \includegraphics[width=7.2cm]{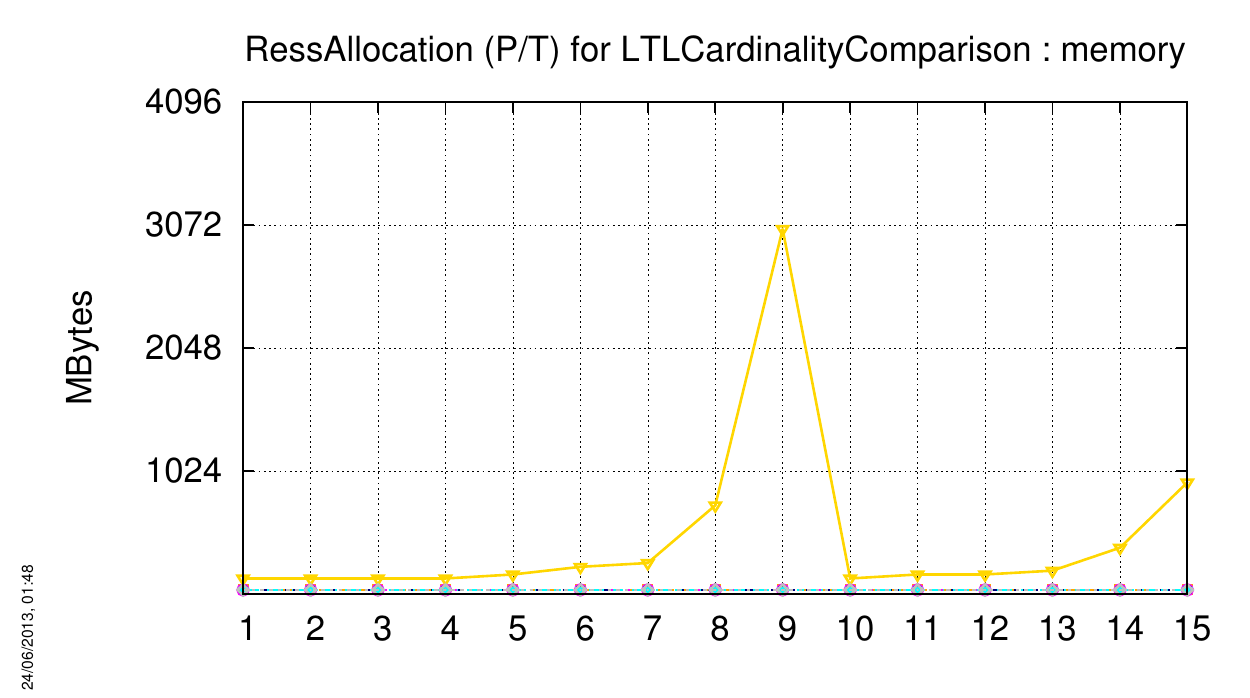}
   \includegraphics[width=7.2cm]{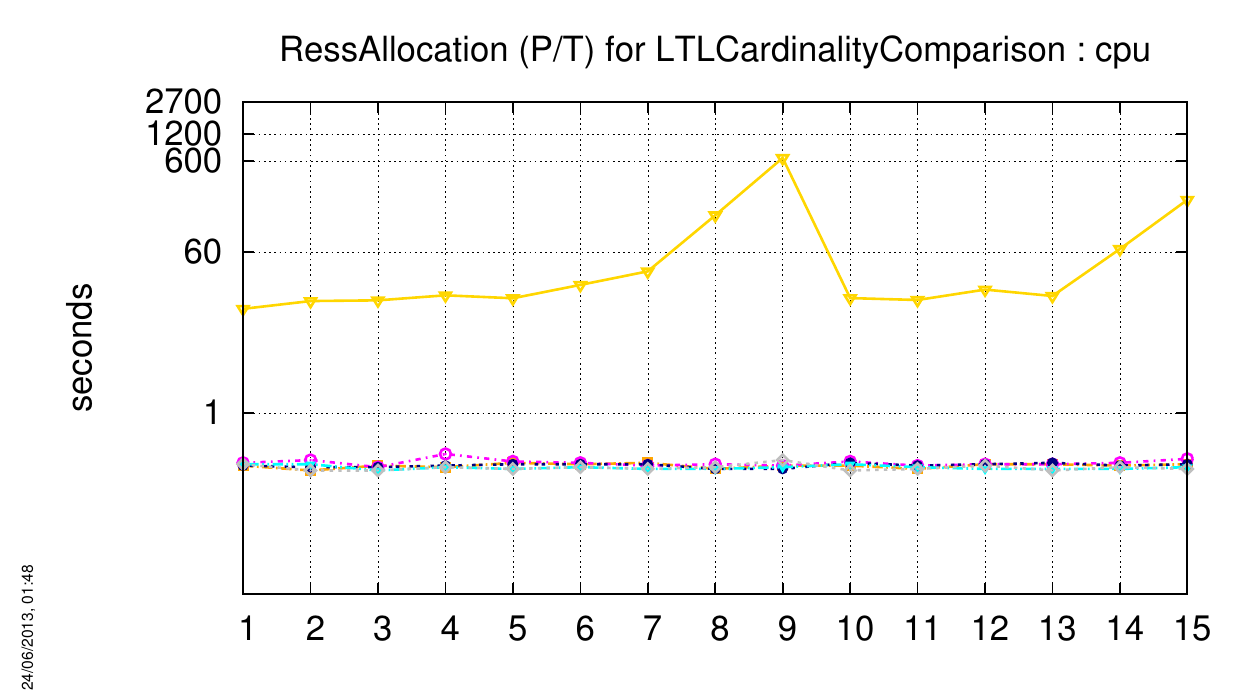}

   \includegraphics[height=1cm]{figures/tools-legend.pdf}
\end{center}

\subsubsection{\acs{Ring-PT}}
The charts below respectively show how tools compete with this ``Known'' model (memory and CPU).

\index{Performances!LTLCardinalityComparison!Ring (P/T)}
\begin{center}
   \includegraphics[width=7.2cm]{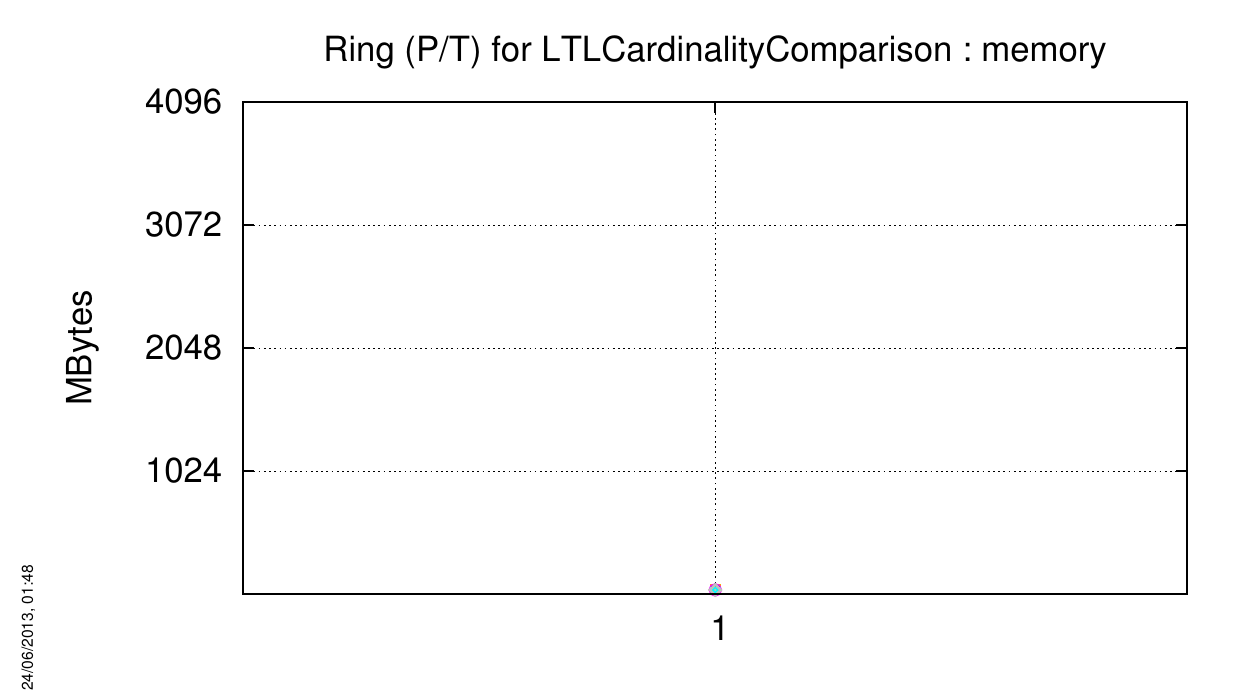}
   \includegraphics[width=7.2cm]{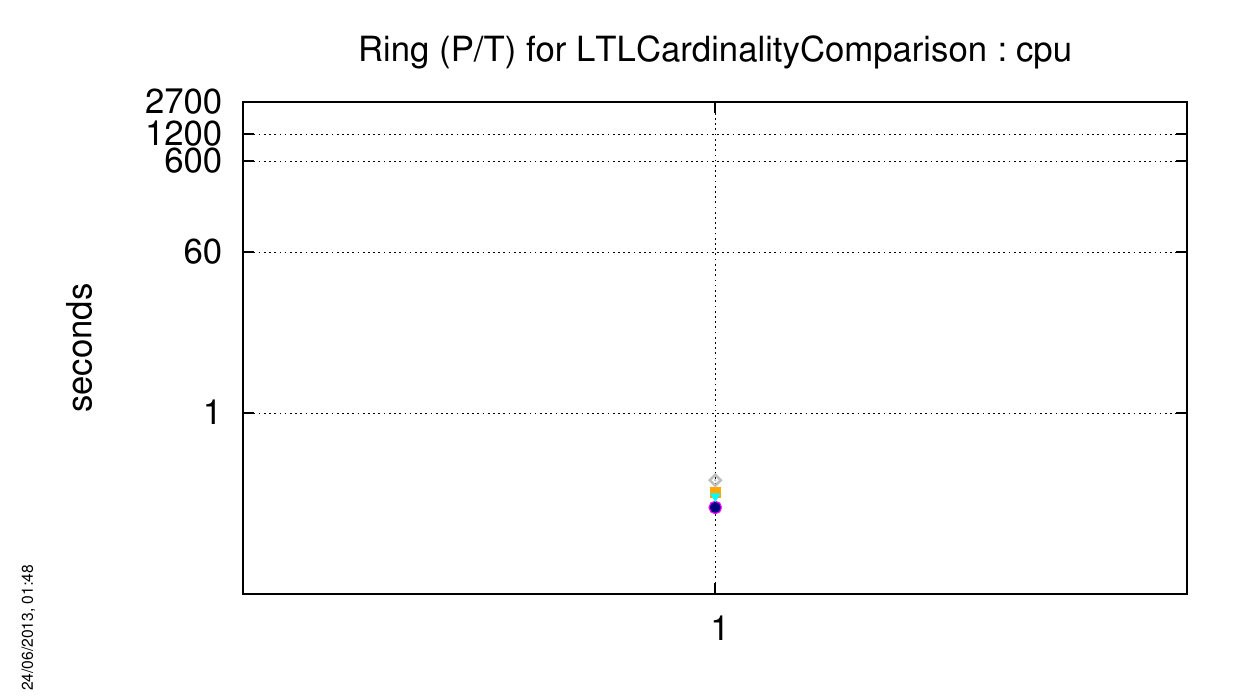}

   \includegraphics[height=1cm]{figures/tools-legend.pdf}
\end{center}

\subsubsection{\acs{RwMutex-PT}}
The charts below respectively show how tools compete with this ``Known'' model (memory and CPU).

\index{Performances!LTLCardinalityComparison!RwMutex (P/T)}
\begin{center}
   \includegraphics[width=7.2cm]{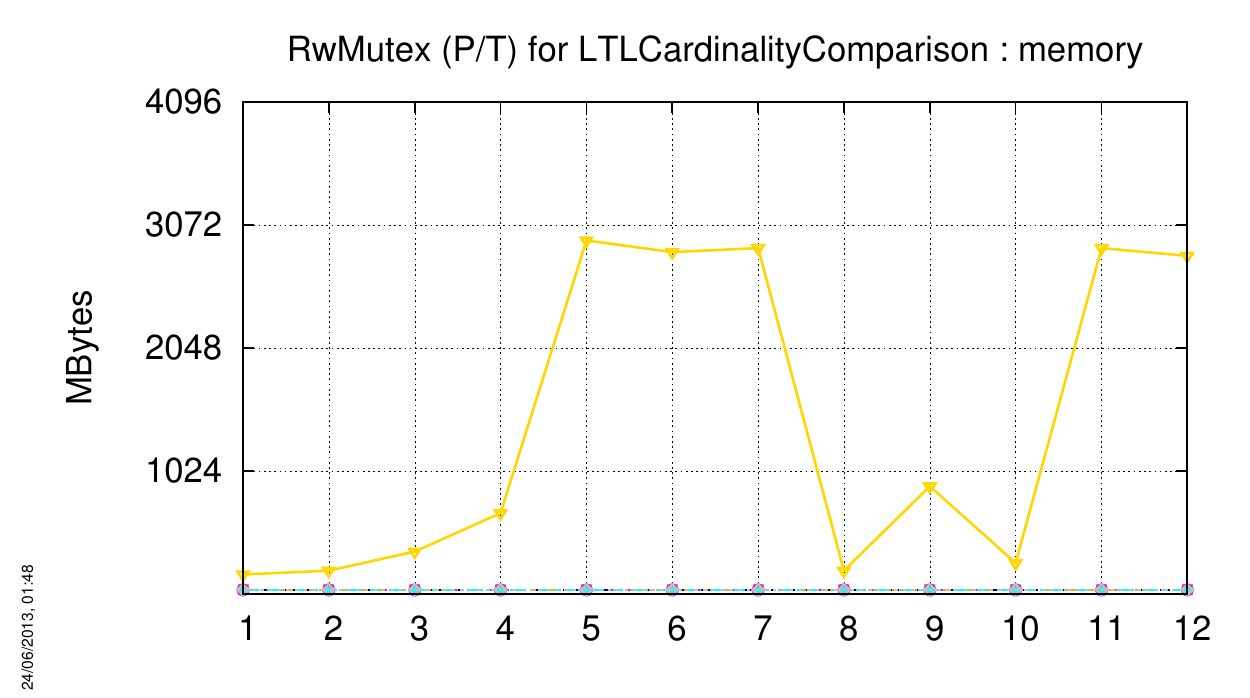}
   \includegraphics[width=7.2cm]{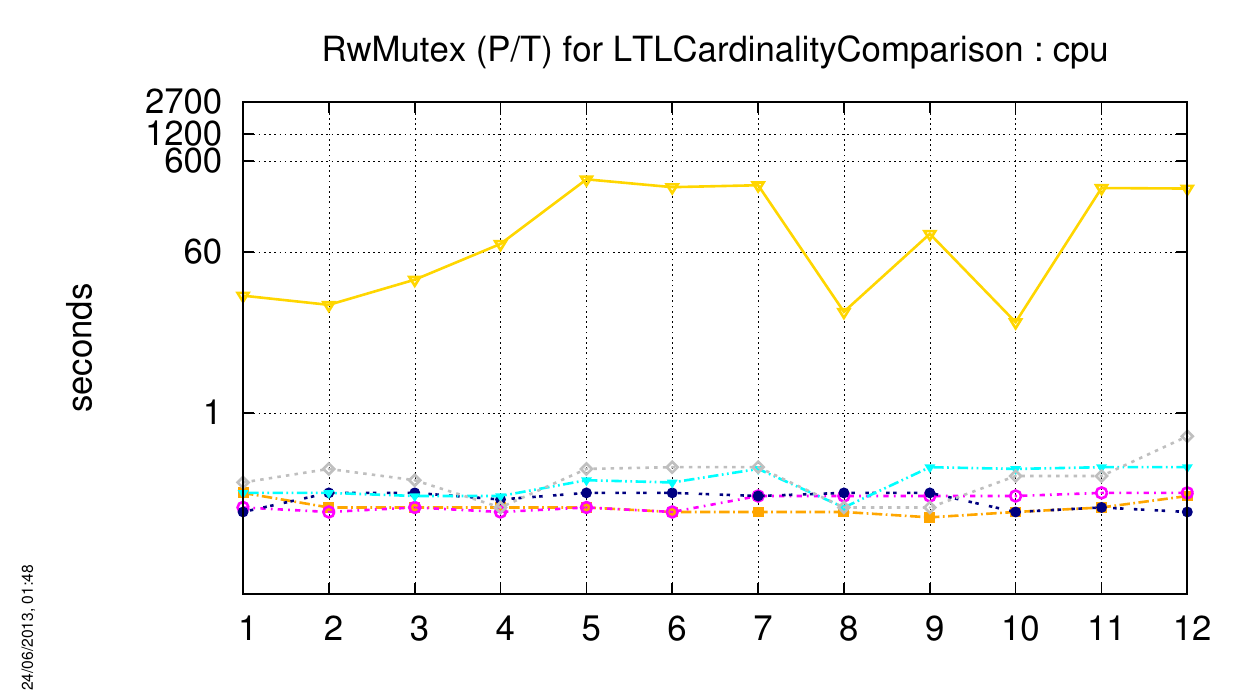}

   \includegraphics[height=1cm]{figures/tools-legend.pdf}
\end{center}

\subsubsection{\acs{SharedMemory-COL}}
No instance of this model could be computed for the \textbf{LTLCardinalityComparison} examination.

\subsubsection{\acs{SharedMemory-PT}}
The charts below respectively show how tools compete with this ``Known'' model (memory and CPU).

\index{Performances!LTLCardinalityComparison!SharedMemory (P/T)}
\begin{center}
   \includegraphics[width=7.2cm]{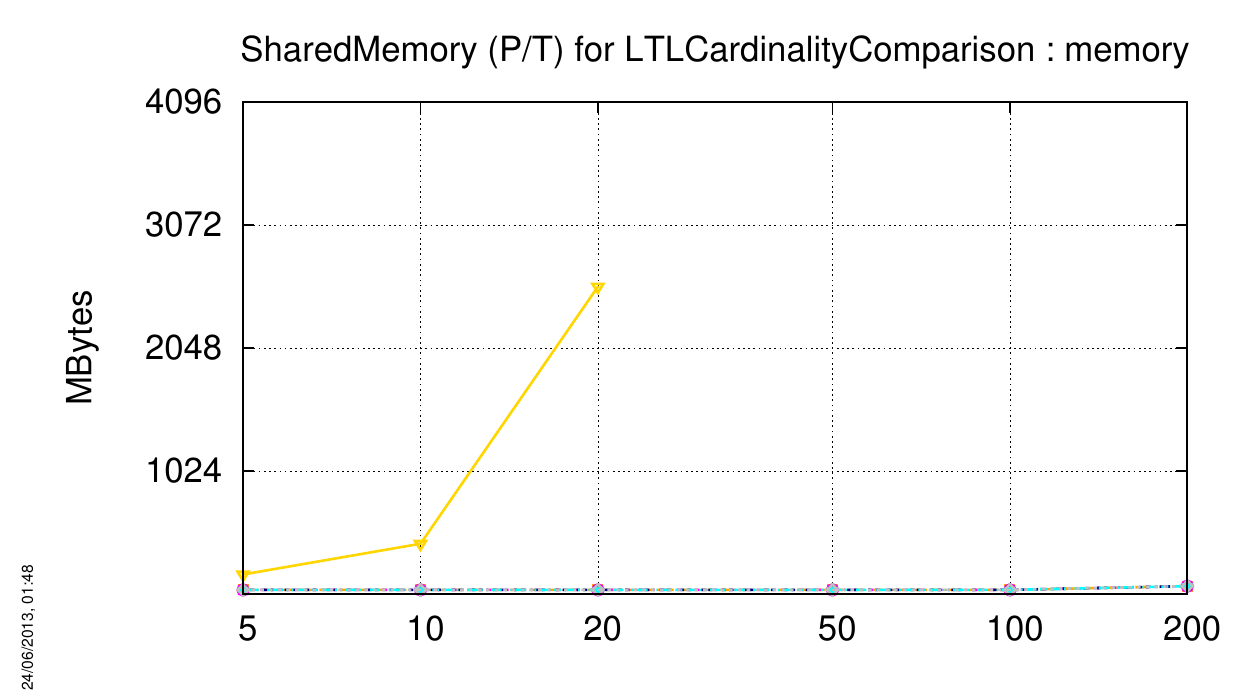}
   \includegraphics[width=7.2cm]{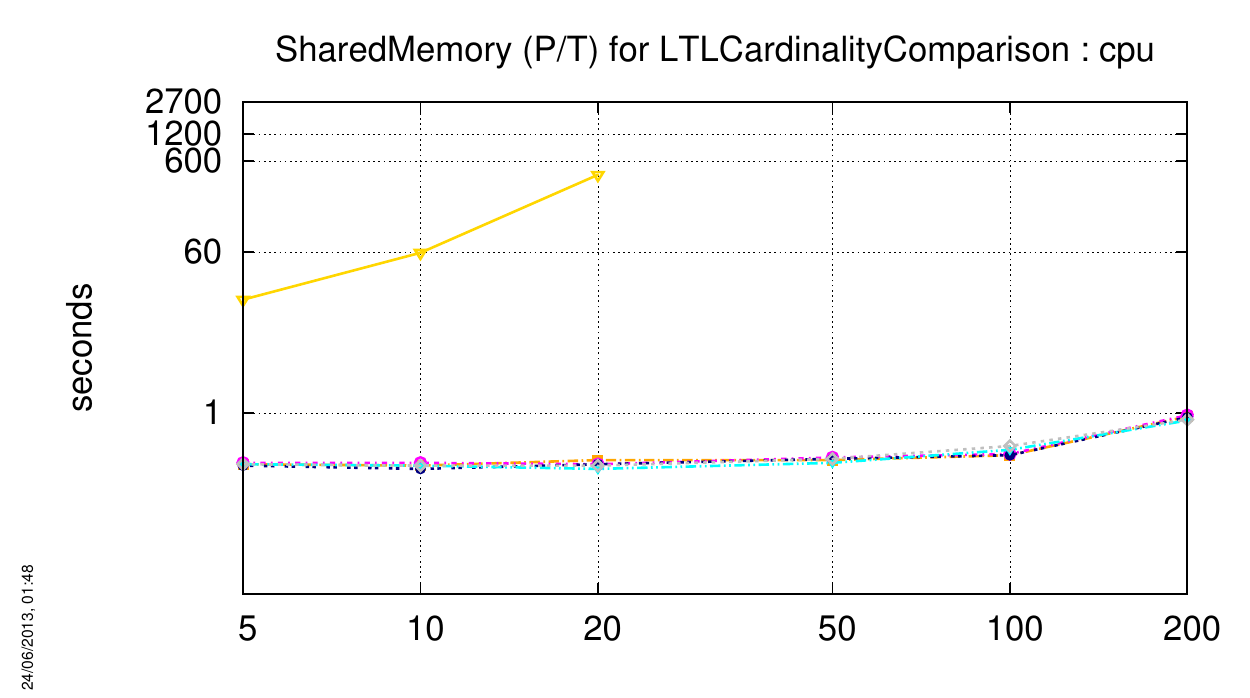}

   \includegraphics[height=1cm]{figures/tools-legend.pdf}
\end{center}

\subsubsection{\acs{SimpleLoadBal-COL}}
No instance of this model could be computed for the \textbf{LTLCardinalityComparison} examination.

\subsubsection{\acs{SimpleLoadBal-PT}}
The charts below respectively show how tools compete with this ``Known'' model (memory and CPU).

\index{Performances!LTLCardinalityComparison!SimpleLoadBal (P/T)}
\begin{center}
   \includegraphics[width=7.2cm]{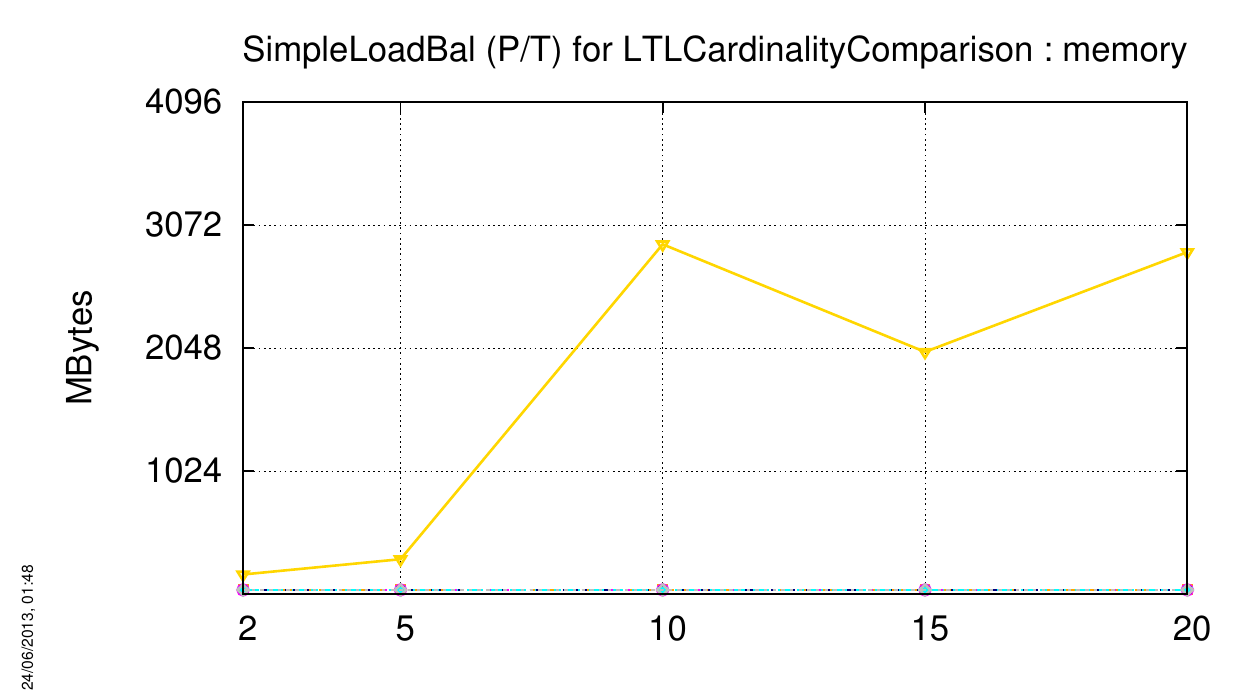}
   \includegraphics[width=7.2cm]{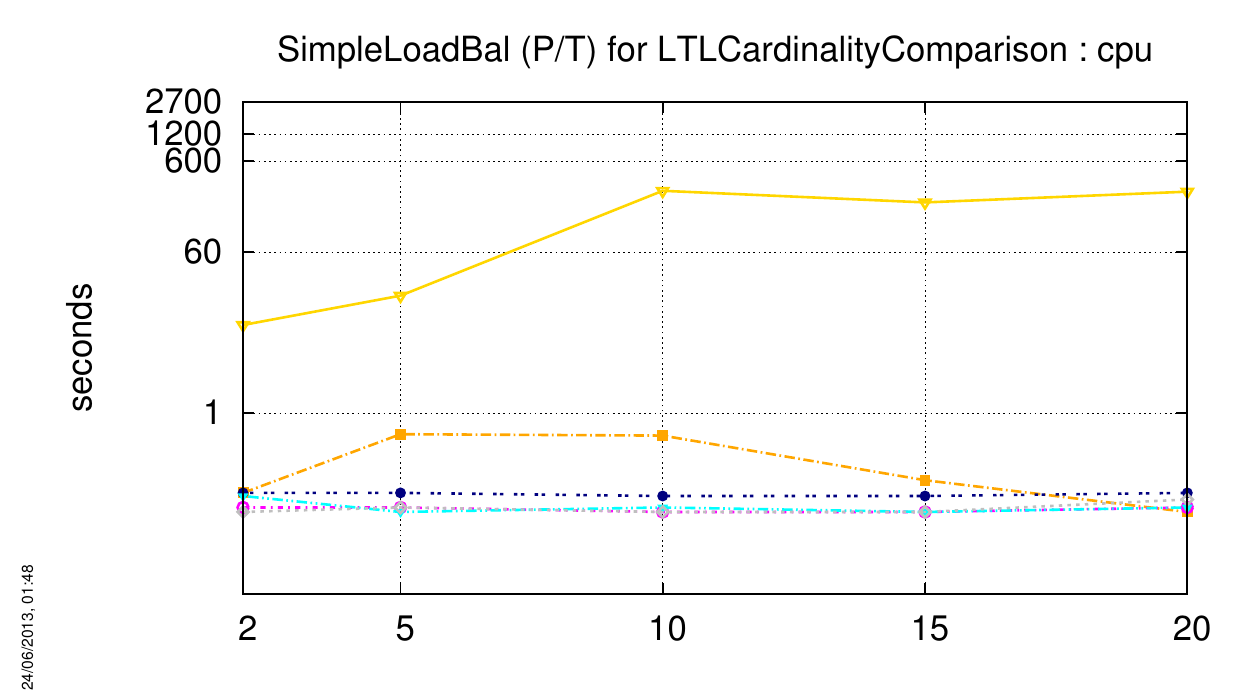}

   \includegraphics[height=1cm]{figures/tools-legend.pdf}
\end{center}

\subsubsection{\acs{TokenRing-COL}}
No instance of this model could be computed for the \textbf{LTLCardinalityComparison} examination.

\subsubsection{\acs{TokenRing-PT}}
No instance of this model could be computed for the \textbf{LTLCardinalityComparison} examination.

\subsubsection{\acs{HouseConstruction-PT}}
The charts below respectively show how tools compete with this ``Suprise'' model (memory and CPU).

\index{Performances!LTLCardinalityComparison!HouseConstruction (P/T)}
\begin{center}
   \includegraphics[width=7.2cm]{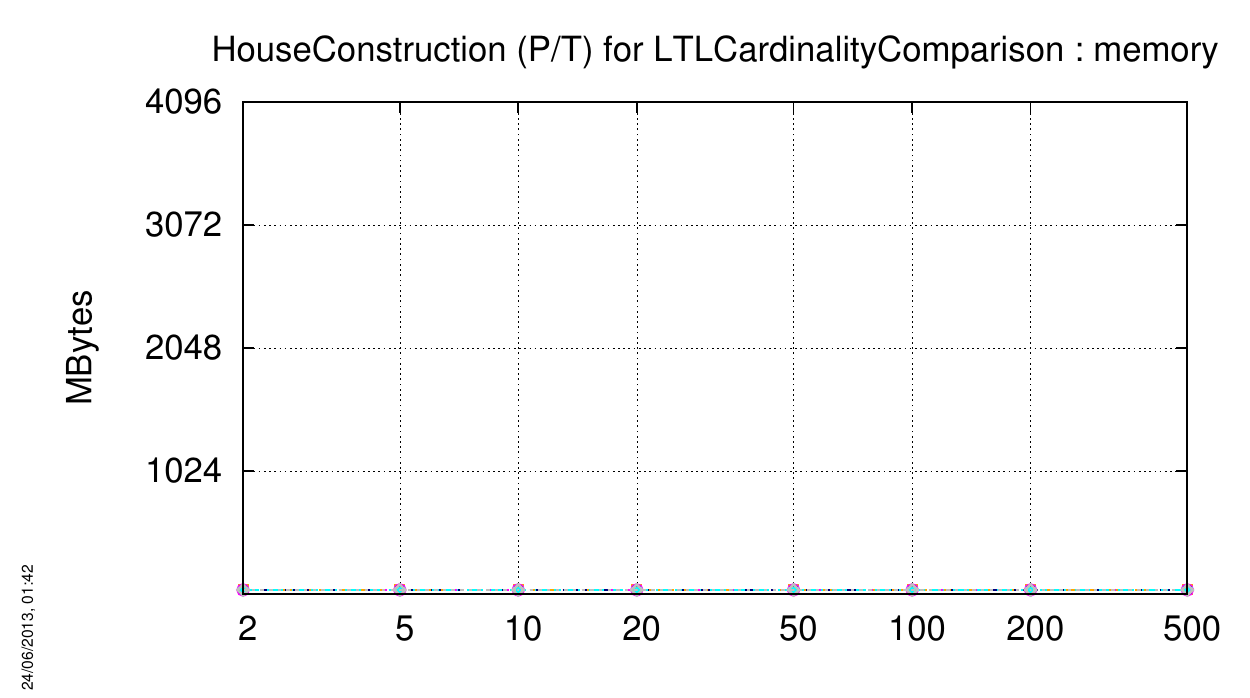}
   \includegraphics[width=7.2cm]{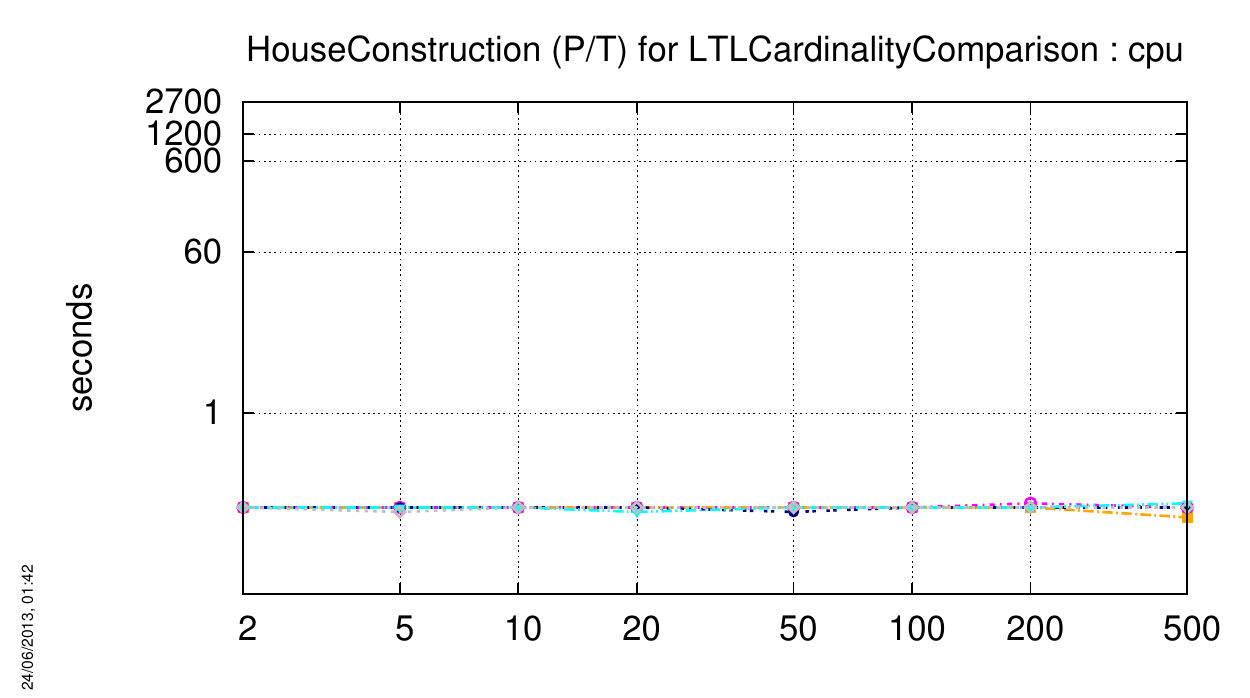}

   \includegraphics[height=1cm]{figures/tools-legend.pdf}
\end{center}

\subsubsection{\acs{IBMB2S565S3960-PT}}
The charts below respectively show how tools compete with this ``Suprise'' model (memory and CPU).

\index{Performances!LTLCardinalityComparison!IBMB2S565S3960 (P/T)}
\begin{center}
   \includegraphics[width=7.2cm]{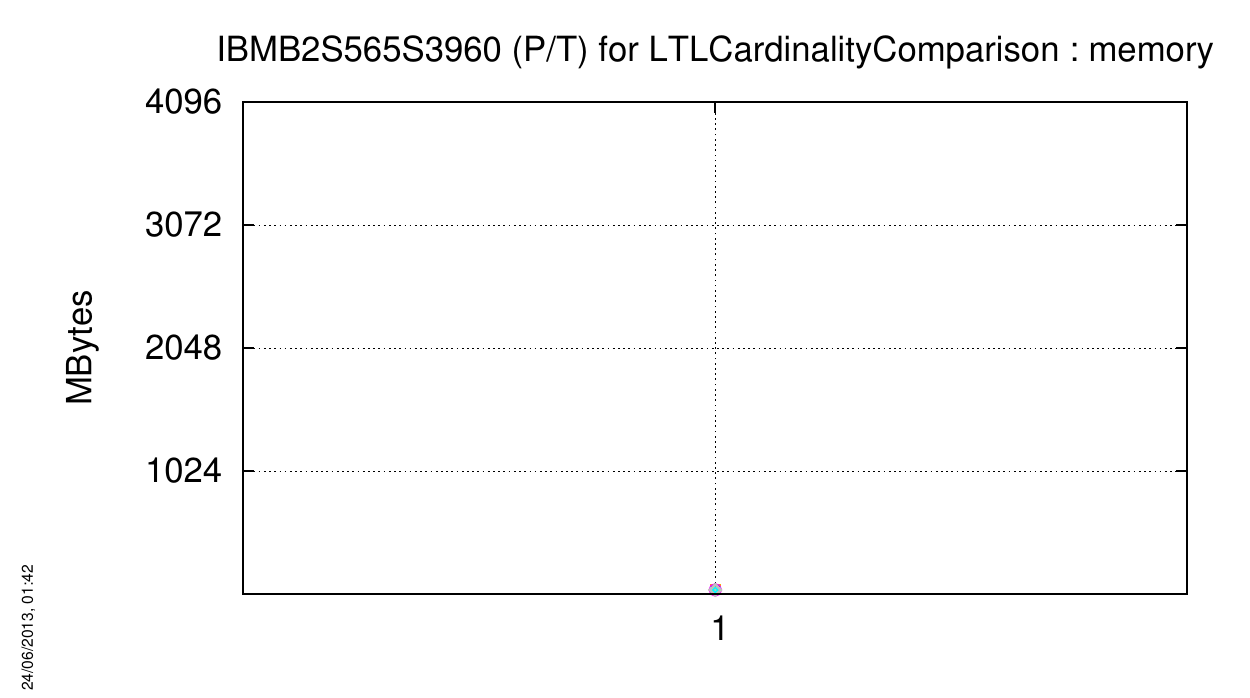}
   \includegraphics[width=7.2cm]{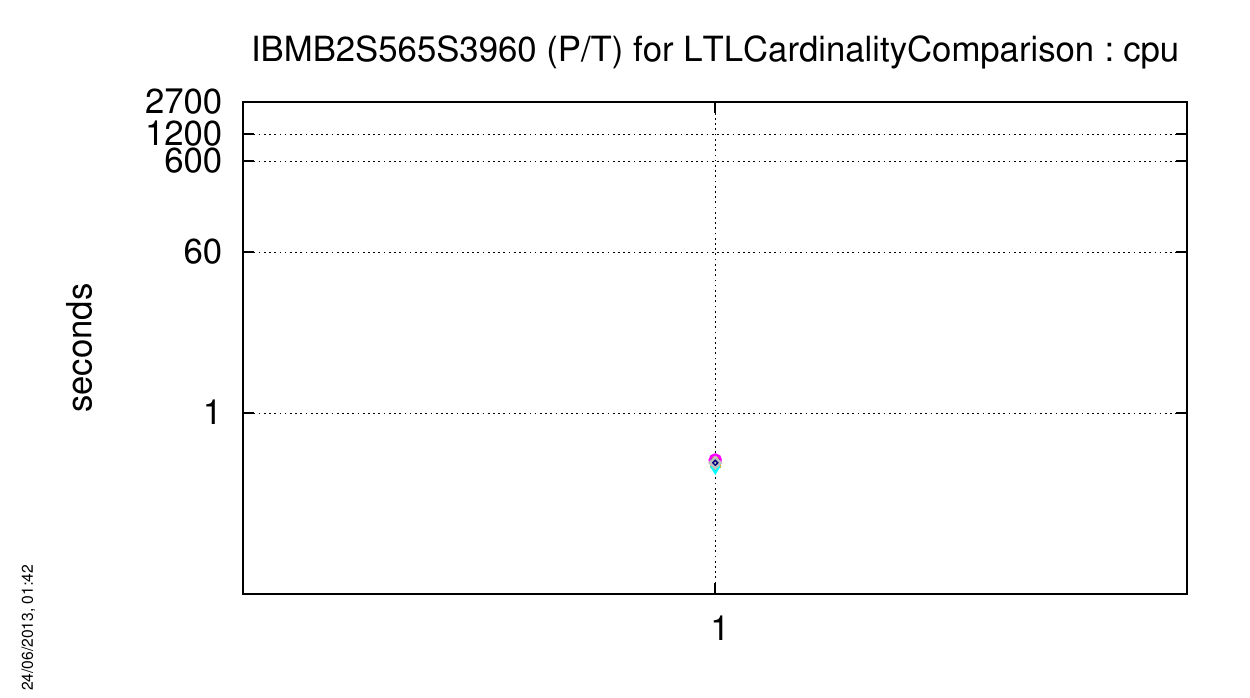}

   \includegraphics[height=1cm]{figures/tools-legend.pdf}
\end{center}

\subsubsection{\acs{QuasiCertifProtocol-COL}}
No instance of this model could be computed for the \textbf{LTLCardinalityComparison} examination.

\subsubsection{\acs{QuasiCertifProtocol-PT}}
The charts below respectively show how tools compete with this ``Suprise'' model (memory and CPU).

\index{Performances!LTLCardinalityComparison!QuasiCertifProtocol (P/T)}
\begin{center}
   \includegraphics[width=7.2cm]{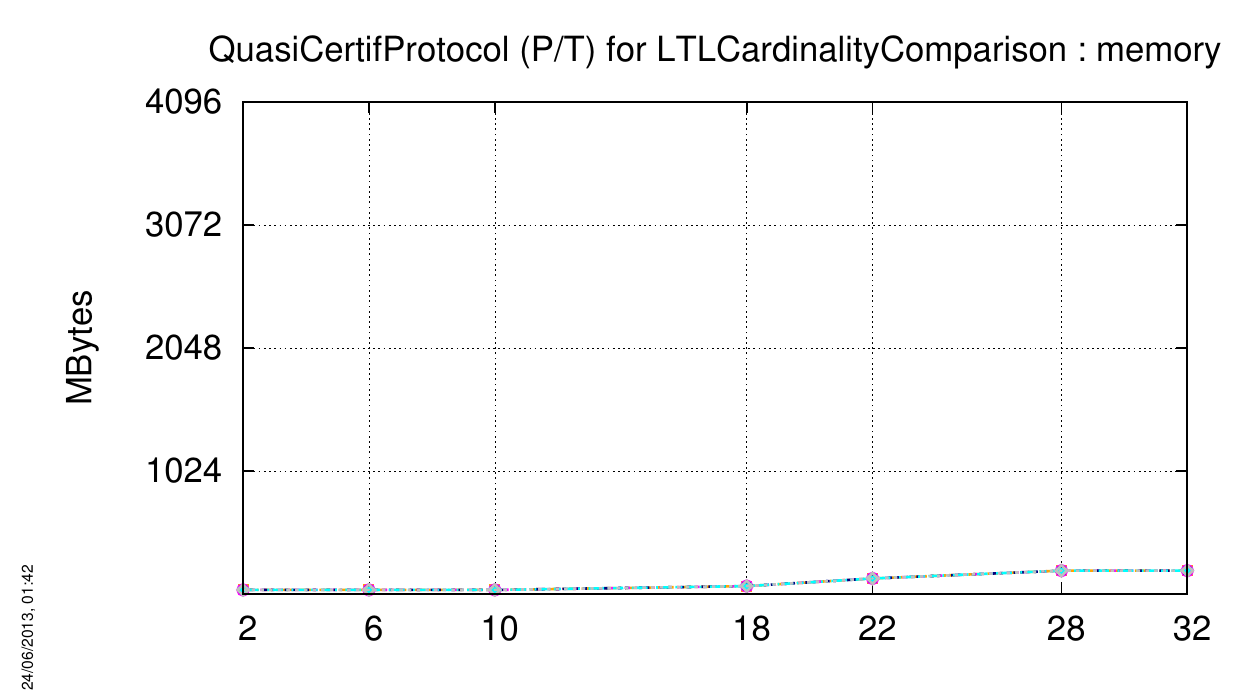}
   \includegraphics[width=7.2cm]{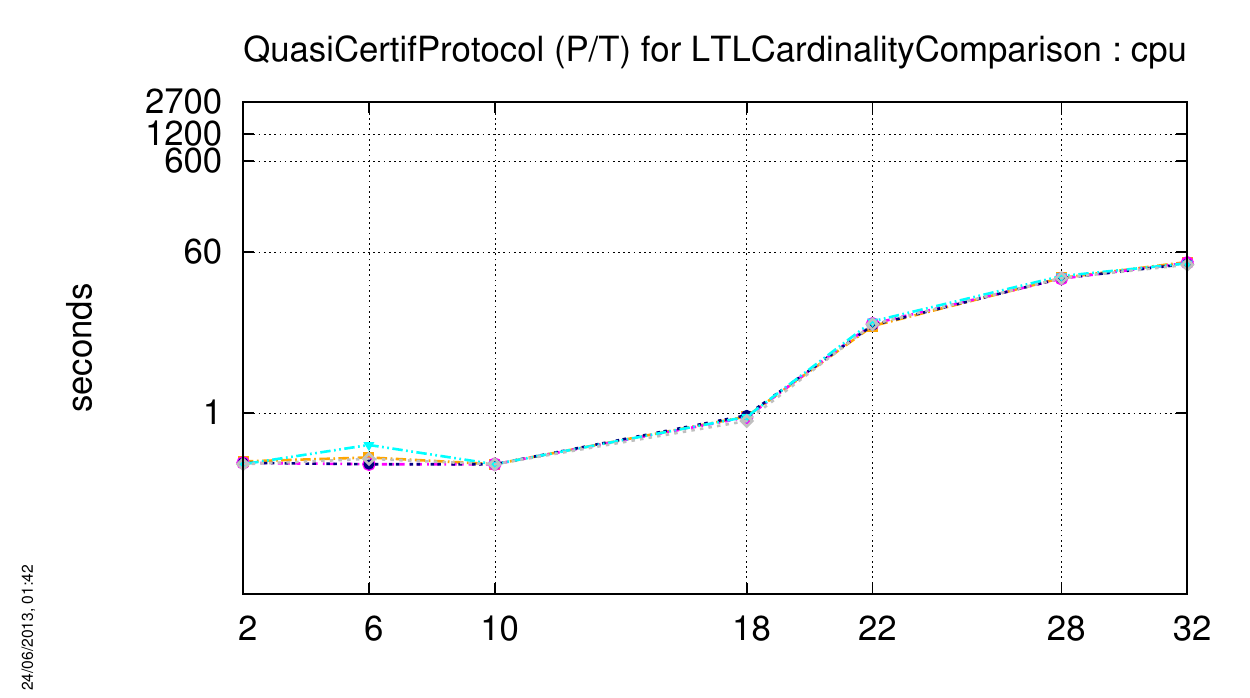}

   \includegraphics[height=1cm]{figures/tools-legend.pdf}
\end{center}

\subsubsection{\acs{Vasy2003-PT}}
The charts below respectively show how tools compete with this ``Suprise'' model (memory and CPU).

\index{Performances!LTLCardinalityComparison!Vasy2003 (P/T)}
\begin{center}
   \includegraphics[width=7.2cm]{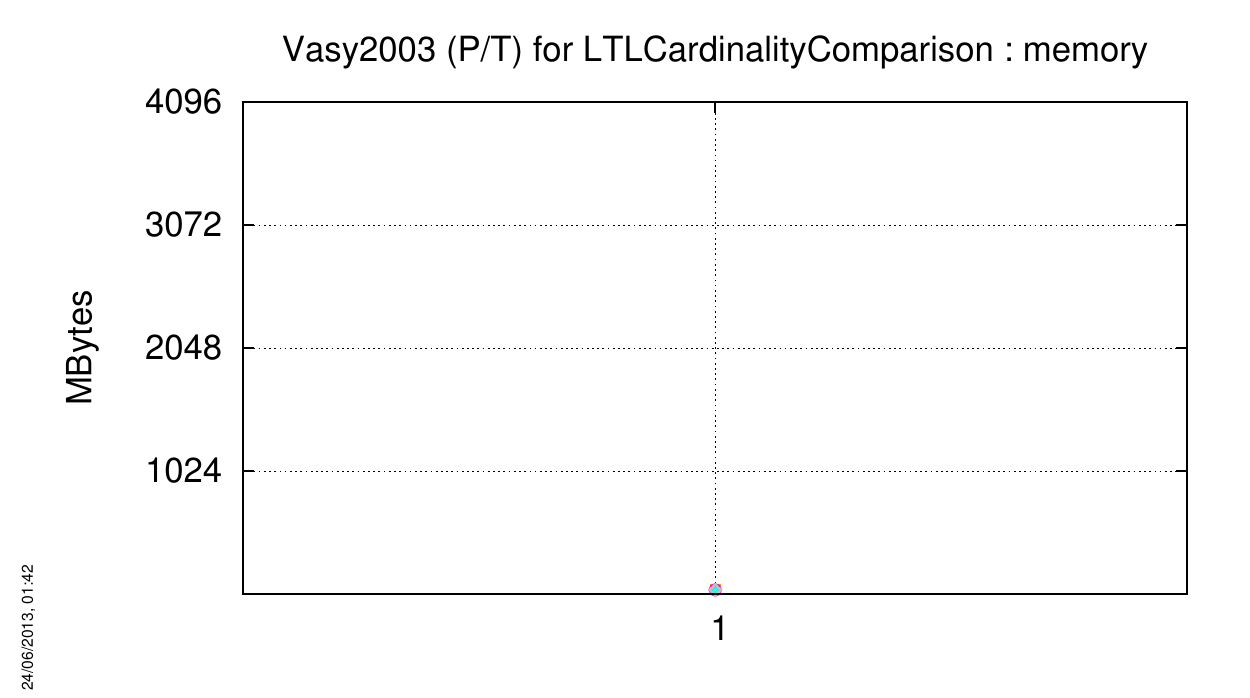}
   \includegraphics[width=7.2cm]{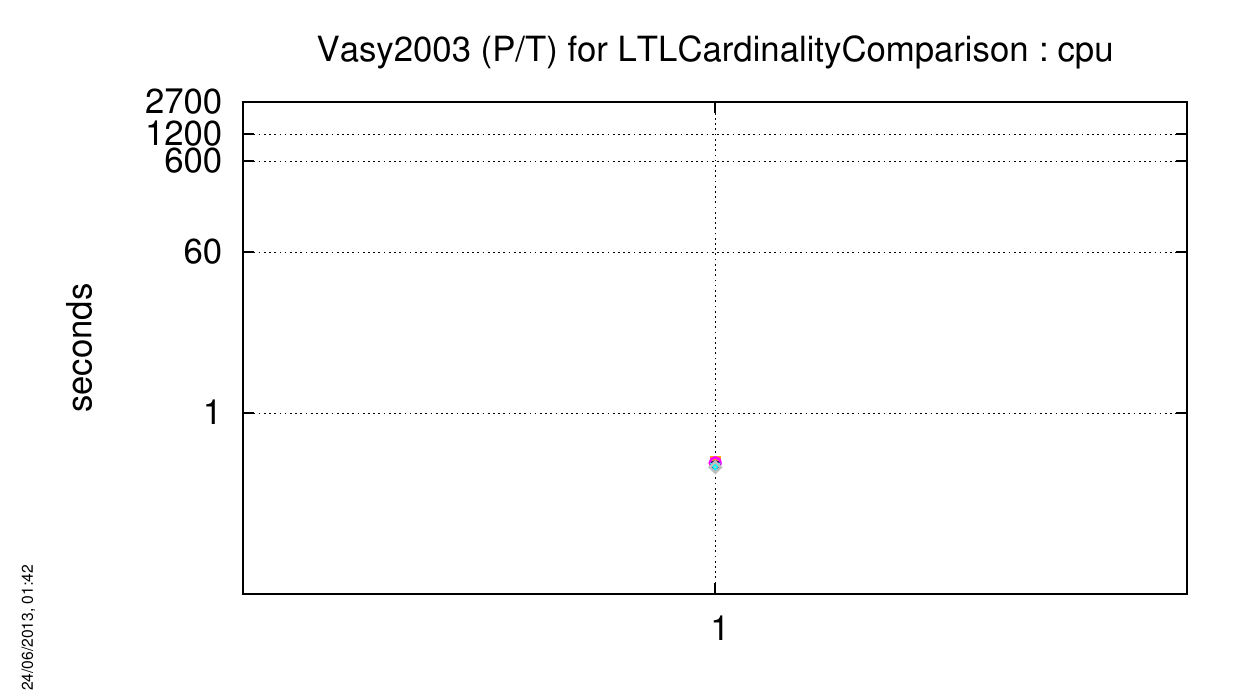}

   \includegraphics[height=1cm]{figures/tools-legend.pdf}
\end{center}

\subsection{Outputs for the LTLCardinalityComparison Examination}
\index{Outputs!LTLCardinalityComparison}

Please find enclosed the brute results for this examination (``Known'' and ``Surprise'' models).
We display only the score of tools that provide a results for at least one instance of one model.
The legend for the values is provided below:
\begin{itemize}
   \item\textbf{nc}: the tool does not compete this examination for this model/instance,
   \item\textbf{cc}: the tool cannot compute this examination for this model/instance,
   \item\textbf{to}: the tool cannot compute this examination for this model/instance within the maximum allowed time,
   \item\textbf{mp}: the tool encountered a memory problem (stack overflow or memory full),
   \item\textbf{nf}: there is no formula available for this type of examination (typically, this concerns P/T nets where
       comparing marking cardinality has no signification when there is no equivalent colored net).
\end{itemize}

\textbf{Note on the display of results for formulas:} each formula is considered as a flag (F if false, T if true, - or ?
when the value cannot be determined). These values are concatenated in the order they appear (we assume it is the order of formulas as they were provided).

\subsubsection{``Known'' Models}

\input{result_known_LTLCardinalityComparison.tex}

\subsubsection{``Surprise'' Models}

\input{result_surprise_LTLCardinalityComparison.tex}

\subsection{Score for the LTLCardinalityComparison Examination}
\index{Scores!LTLCardinalityComparison}

Please find enclosed the scores for this examination (``Known'' and ``Surprise'' models).
We display only the score of tools that provide a results for at least one instance of one model.
The total is first listed in the table below followed by a detail, for each proposed model.
Meaning of the line labels are:
\begin{itemize}
\item\textbf{1st instance}: the tool gets a bonus for having processed the first instance of this model (+1 point),
\item\textbf{instances}: the tool gets 1 point per instances treated 
(for that, we assume that at least one formula has been successfully computed),
\item\textbf{max reached}: the tool could process all the instances for the model (+2 points),
\item\textbf{best}: the tool is among the ones that processed a maximum of instances within the time and memory confinement (+2 points).
\end{itemize}

\subsubsection{``Known'' Models}

\input{score_known_LTLCardinalityComparison.tex}

\subsubsection{``Surprise'' Models}

\input{score_surprise_LTLCardinalityComparison.tex}

\subsection{Trophies for this Examination}
\index{Trophies!LTLCardinalityComparison}

Trophies are divided in three categories: ``Known'' models,
``Surprise'' models, and the global trophies (formula is then
$score_{global} = score_{known} + 2 \times score_{surprise}$).

\subsubsection{For ``Known'' Models} \ \\

\begin{tabular}{c}
      1 \\
   \includegraphics[width=2cm]{figures/gold.jpg} \\
   \acs{neco} \\
   114 points \\
\end{tabular}

\subsubsection{For ``Surprise'' Models}\  \\

No tool could complete this examination.

\subsubsection{Global} \ \\

\begin{tabular}{c}
      1 \\
   \includegraphics[width=2cm]{figures/gold.jpg} \\
   \acs{neco} \\
   114 points \\
\end{tabular}

\newpage

\section{The LTLFireability Examination}
\label{sec:exam:LTLFireability}
\index{Results!LTLFireability}

This examination deals with LTL properties dealing with transition fireability only.
We first show a summary on the handling of models by the participating tools.
Then, we present the computed outputs and the associated scores for this
examination prior to a summary of relevant executions.

\subsection{Handling of Models by Tools}
\index{Performances!LTLFireability}

\subsubsection{\acs{CSRepetitions-COL}}
No instance of this model could be computed for the \textbf{LTLFireability} examination.

\subsubsection{\acs{CSRepetitions-PT}}
The charts below respectively show how tools compete with this ``Known'' model (memory and CPU).

\index{Performances!LTLFireability!CSRepetitions (P/T)}
\begin{center}
   \includegraphics[width=7.2cm]{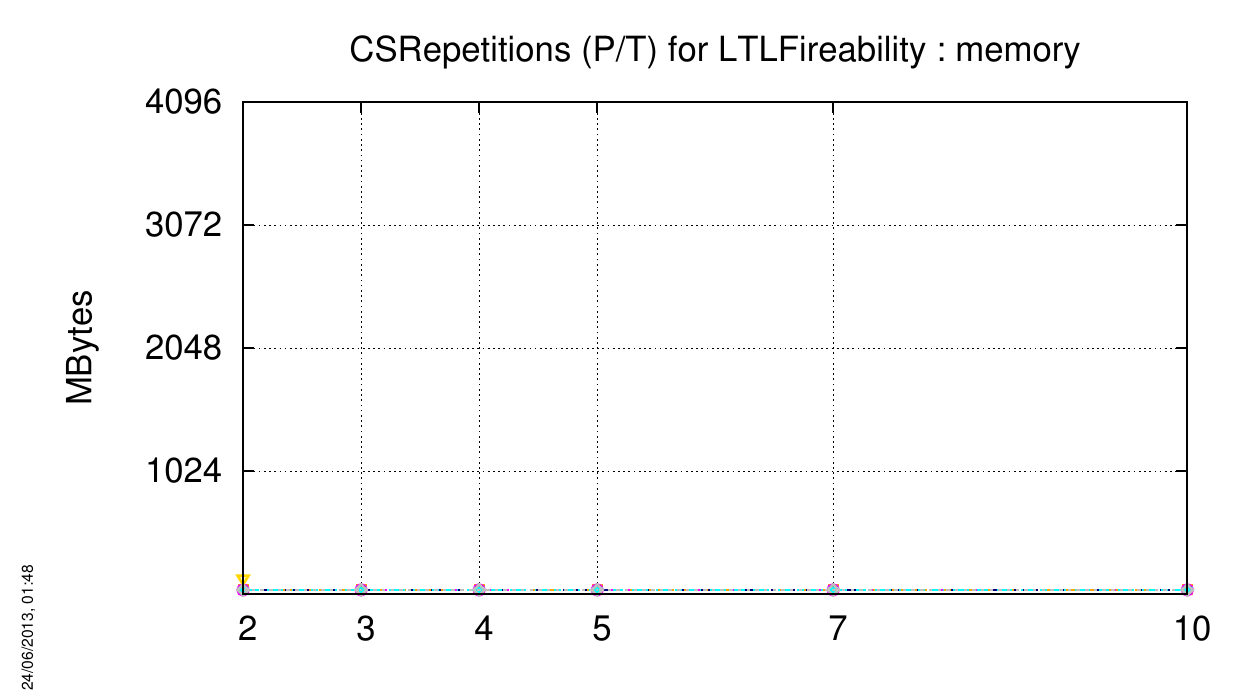}
   \includegraphics[width=7.2cm]{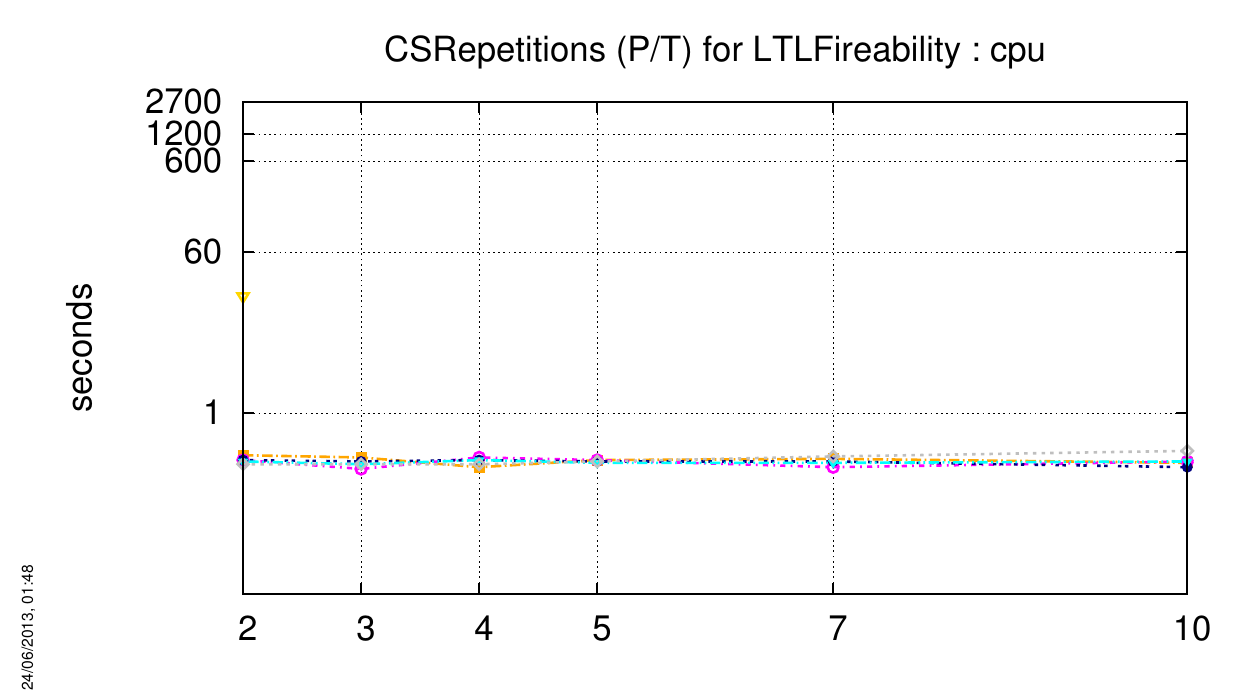}

   \includegraphics[height=1cm]{figures/tools-legend.pdf}
\end{center}

\subsubsection{\acs{Dekker-PT}}
The charts below respectively show how tools compete with this ``Known'' model (memory and CPU).

\index{Performances!LTLFireability!Dekker (P/T)}
\begin{center}
   \includegraphics[width=7.2cm]{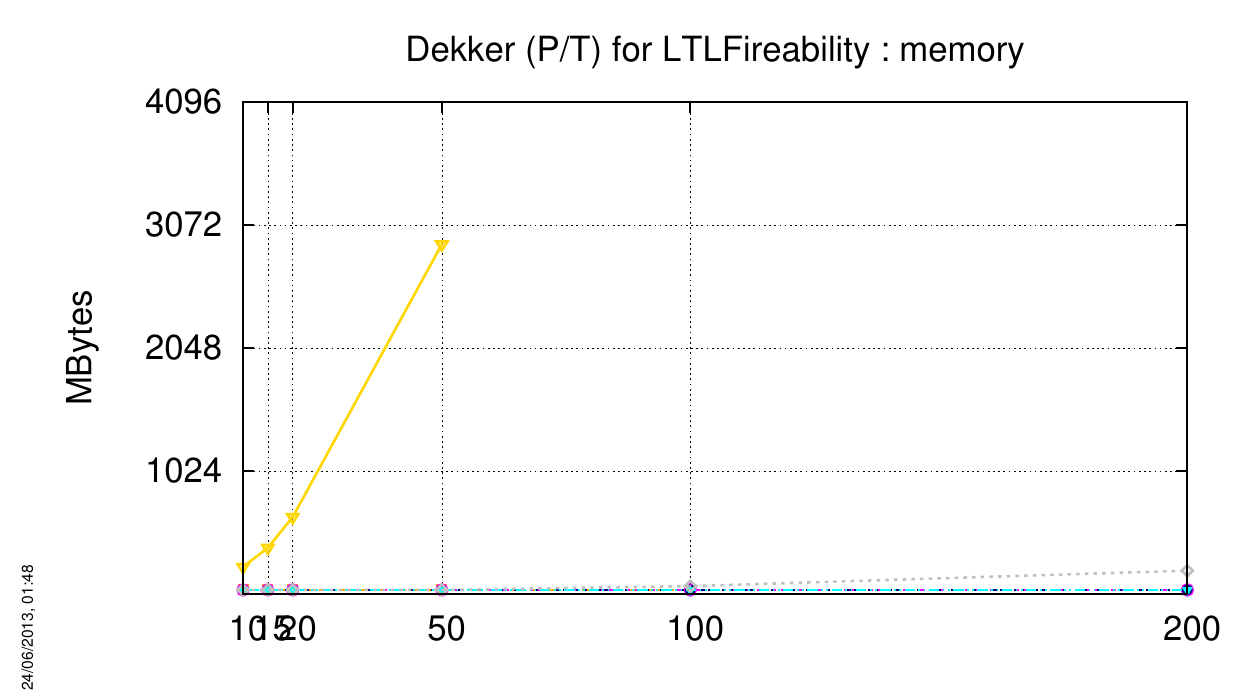}
   \includegraphics[width=7.2cm]{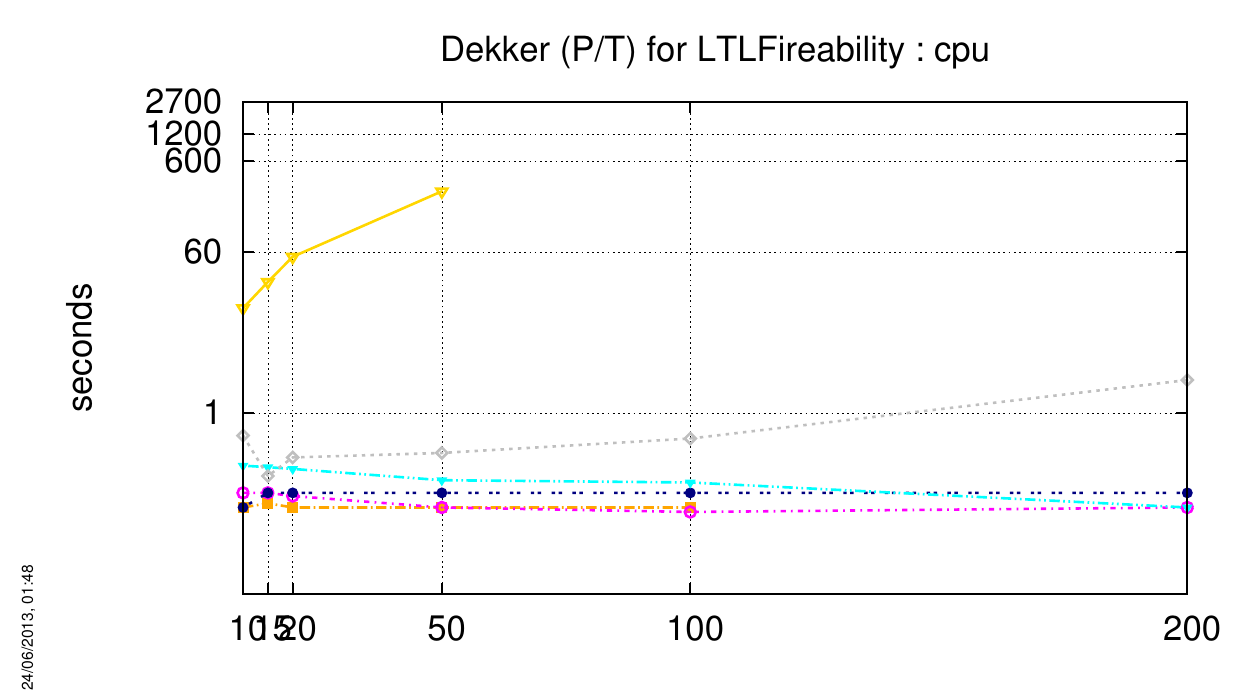}

   \includegraphics[height=1cm]{figures/tools-legend.pdf}
\end{center}

\subsubsection{\acs{DotAndBoxes-COL}}
No instance of this model could be computed for the \textbf{LTLFireability} examination.

\subsubsection{\acs{DrinkVendingMachine-COL}}
No instance of this model could be computed for the \textbf{LTLFireability} examination.

\subsubsection{\acs{DrinkVendingMachine-PT}}
The charts below respectively show how tools compete with this ``Known'' model (memory and CPU).

\index{Performances!LTLFireability!DrinkVendingMachine (P/T)}
\begin{center}
   \includegraphics[width=7.2cm]{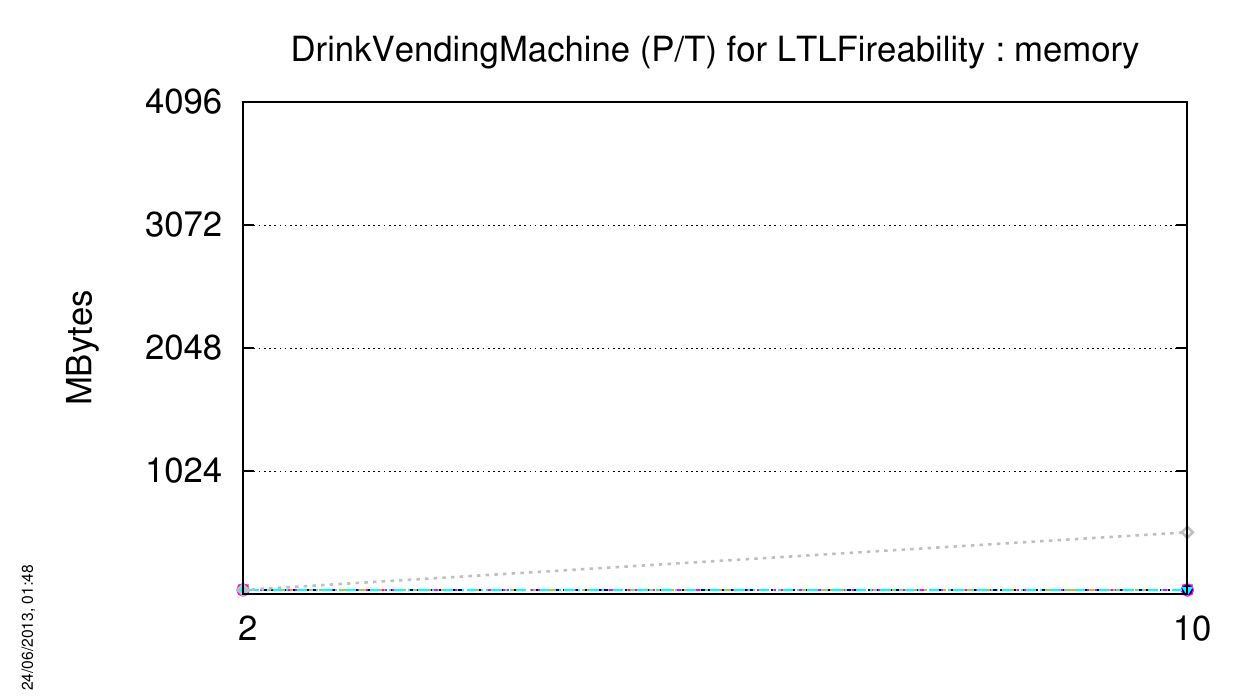}
   \includegraphics[width=7.2cm]{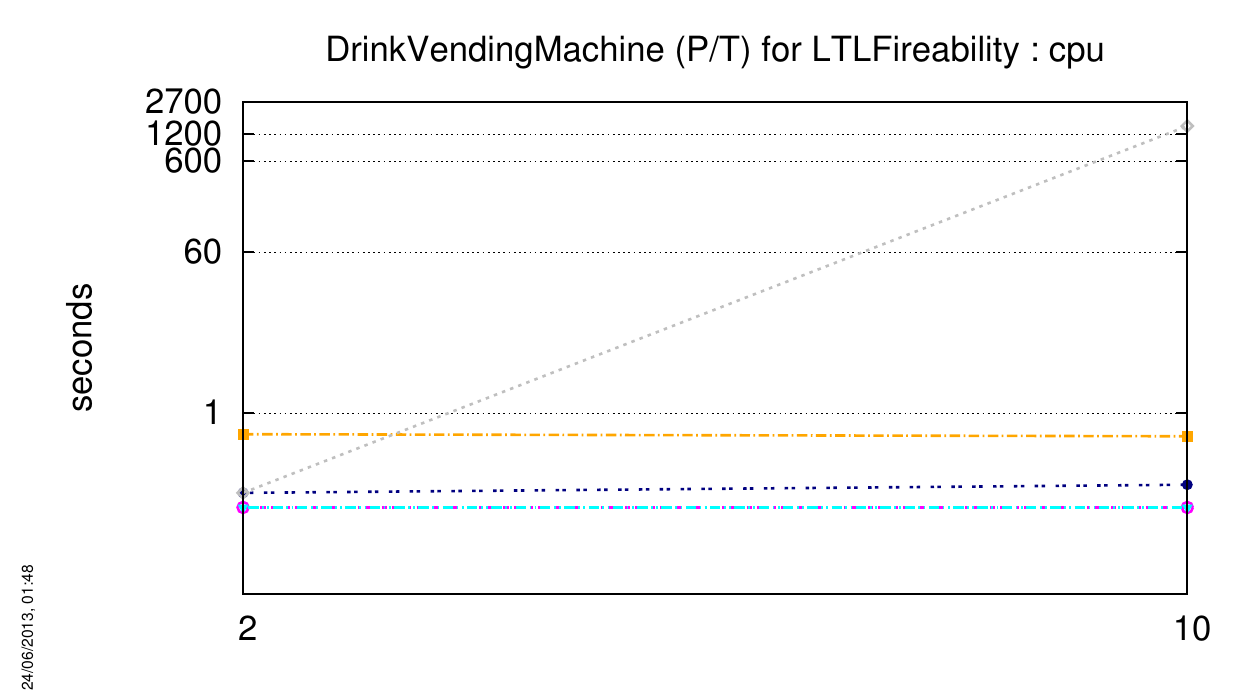}

   \includegraphics[height=1cm]{figures/tools-legend.pdf}
\end{center}

\subsubsection{\acs{Echo-PT}}
The charts below respectively show how tools compete with this ``Known'' model (memory and CPU).

\index{Performances!LTLFireability!Echo (P/T)}
\begin{center}
   \includegraphics[width=7.2cm]{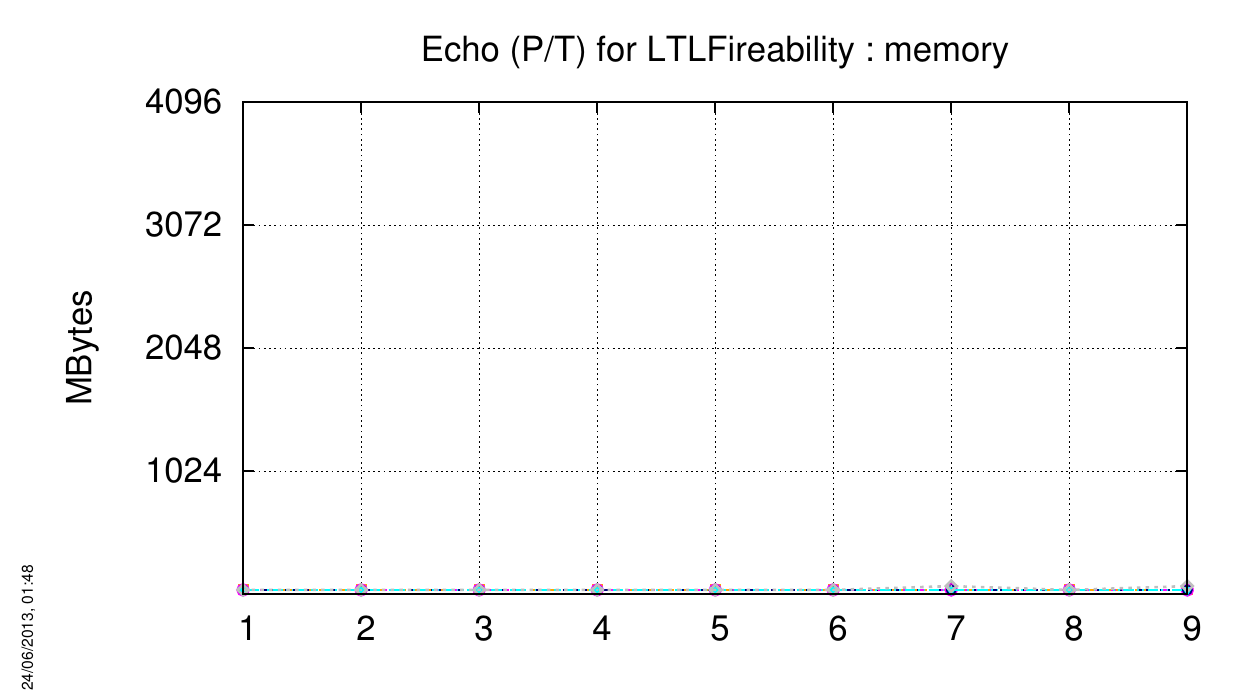}
   \includegraphics[width=7.2cm]{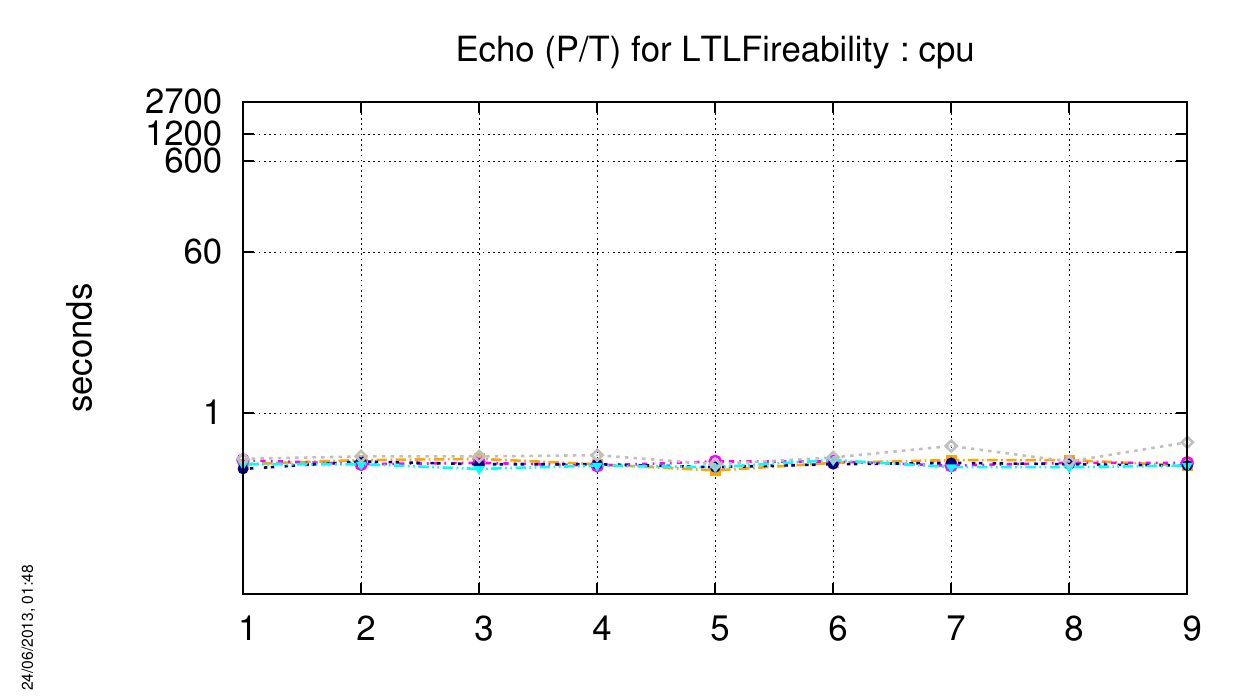}

   \includegraphics[height=1cm]{figures/tools-legend.pdf}
\end{center}

\subsubsection{\acs{Eratosthenes-PT}}
The charts below respectively show how tools compete with this ``Known'' model (memory and CPU).

\index{Performances!LTLFireability!Eratosthenes (P/T)}
\begin{center}
   \includegraphics[width=7.2cm]{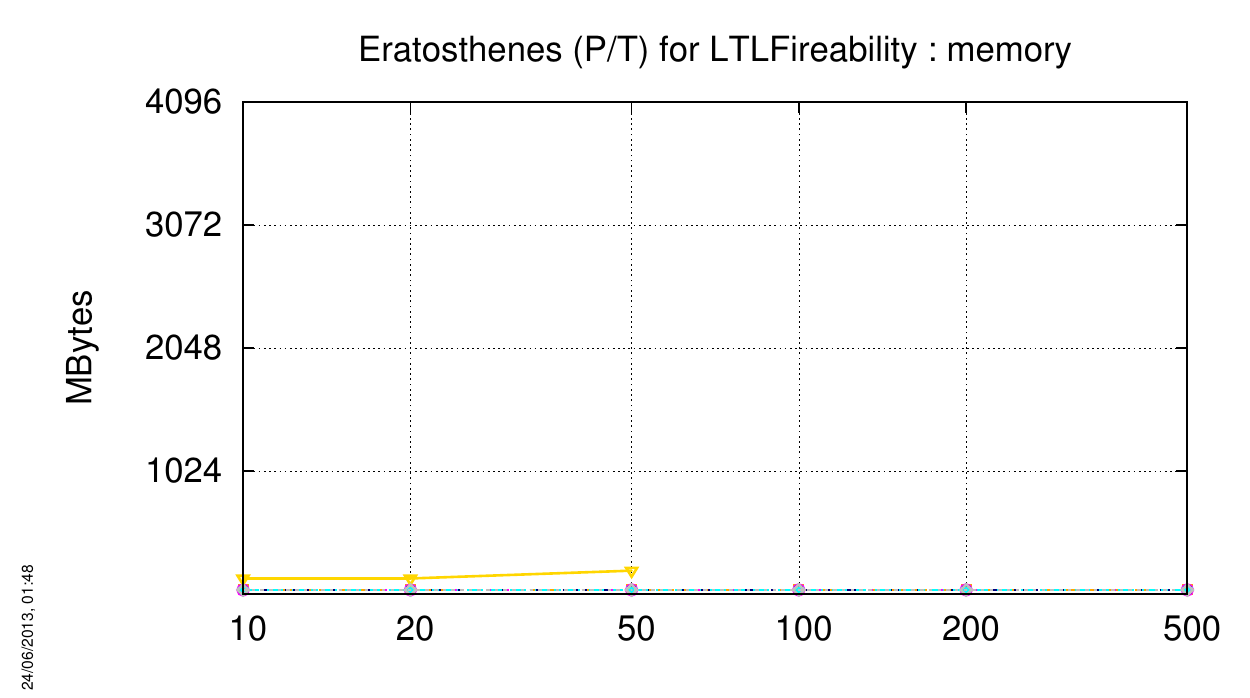}
   \includegraphics[width=7.2cm]{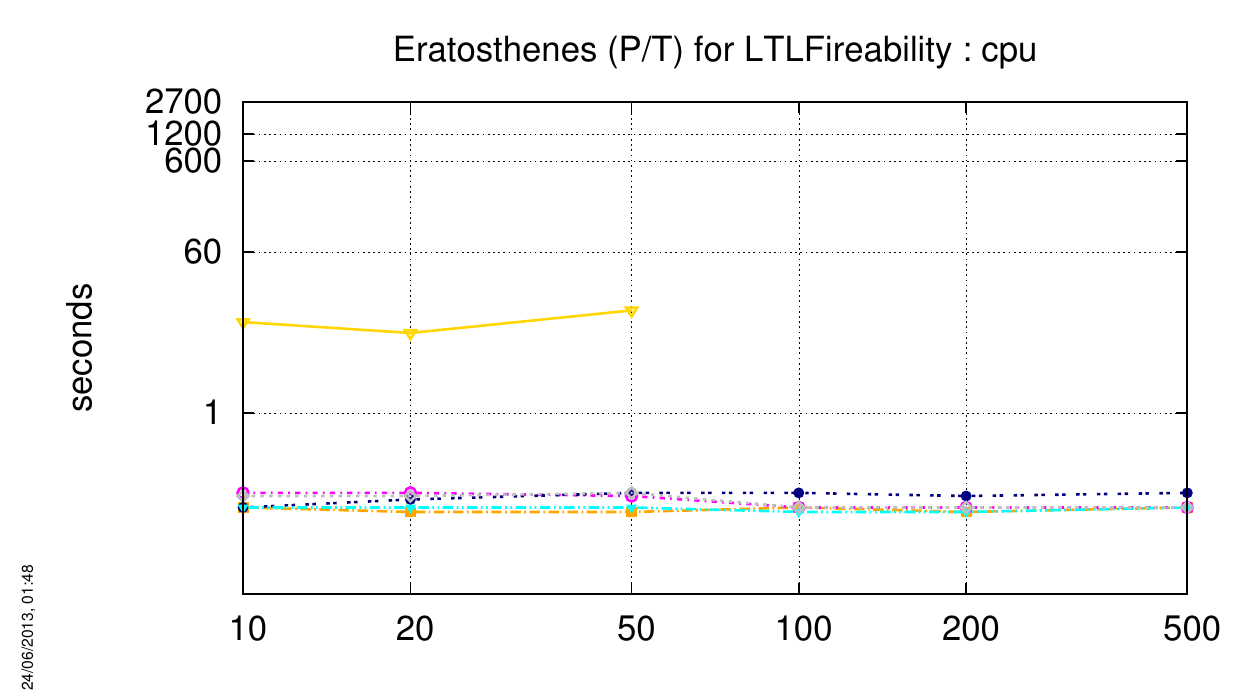}

   \includegraphics[height=1cm]{figures/tools-legend.pdf}
\end{center}

\subsubsection{\acs{FMS-PT}}
The charts below respectively show how tools compete with this ``Known'' model (memory and CPU).

\index{Performances!LTLFireability!FMS (P/T)}
\begin{center}
   \includegraphics[width=7.2cm]{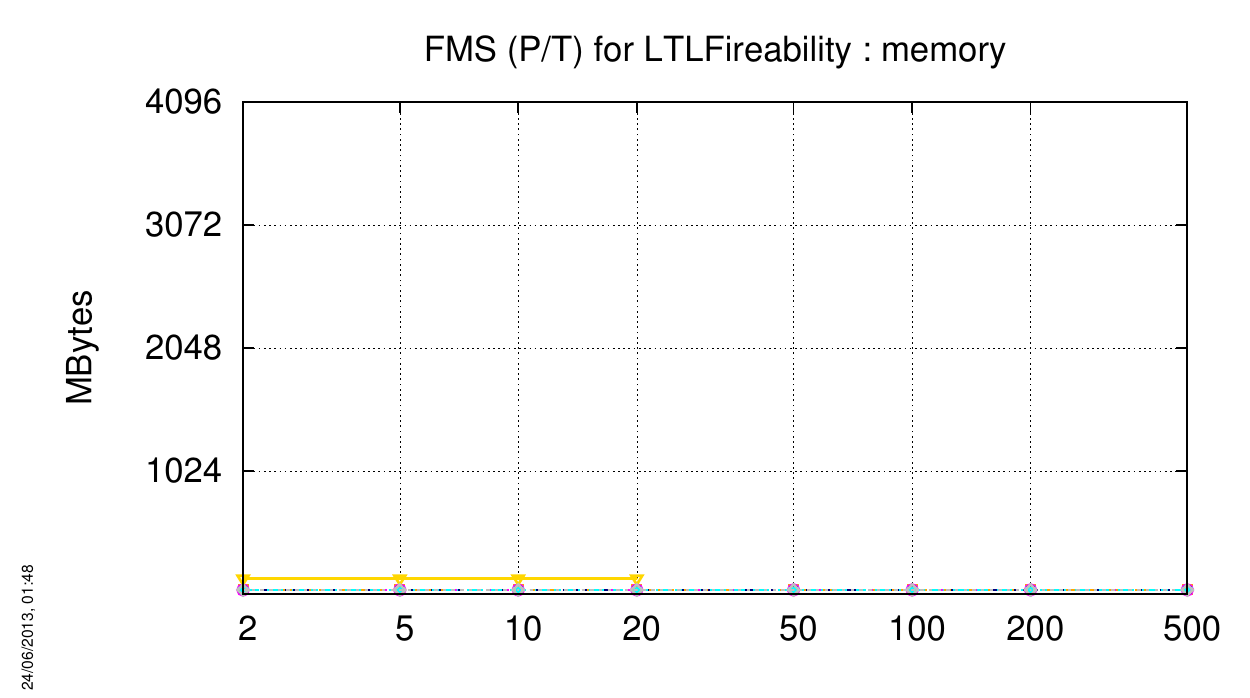}
   \includegraphics[width=7.2cm]{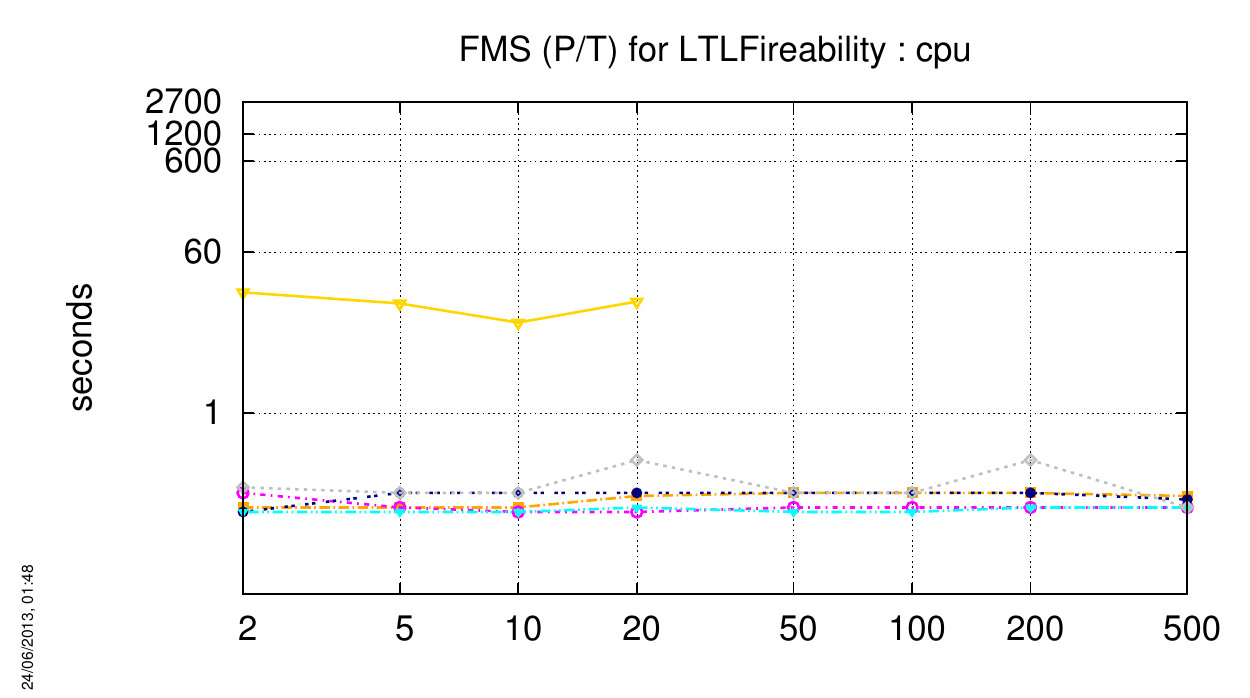}

   \includegraphics[height=1cm]{figures/tools-legend.pdf}
\end{center}

\subsubsection{\acs{GlobalRessAlloc-COL}}
No instance of this model could be computed for the \textbf{LTLFireability} examination.

\subsubsection{\acs{GlobalRessAlloc-PT}}
The charts below respectively show how tools compete with this ``Known'' model (memory and CPU).

\index{Performances!LTLFireability!GlobalRessAlloc (P/T)}
\begin{center}
   \includegraphics[width=7.2cm]{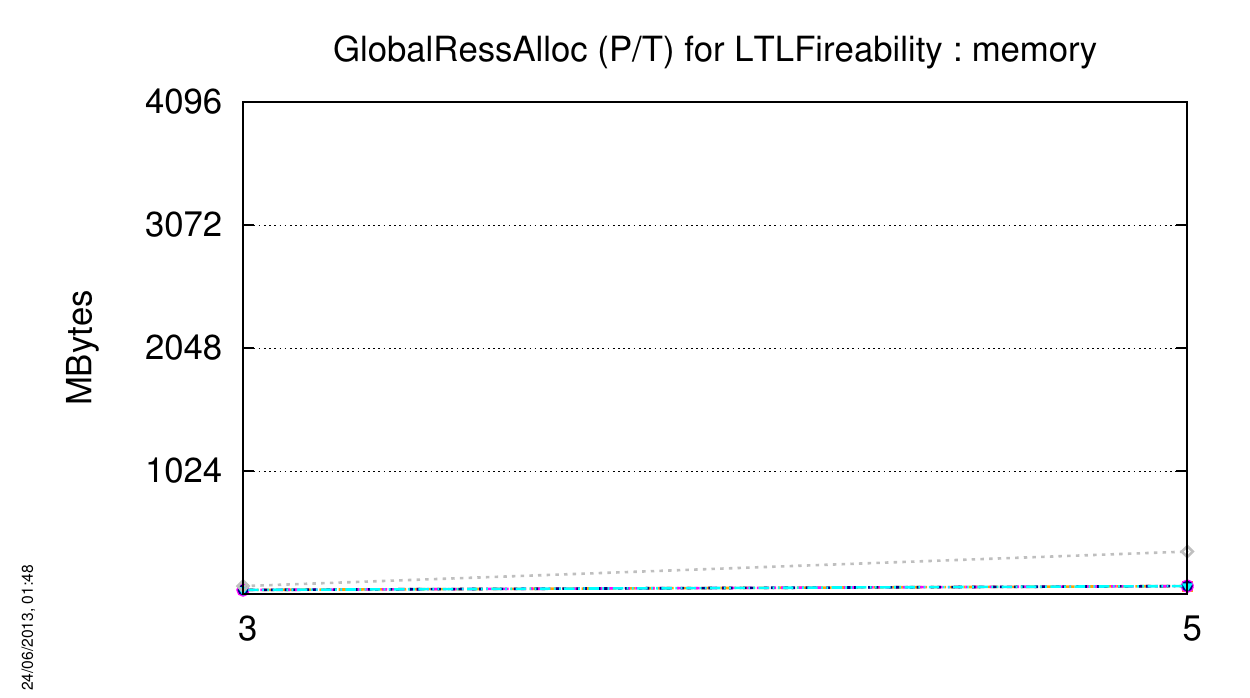}
   \includegraphics[width=7.2cm]{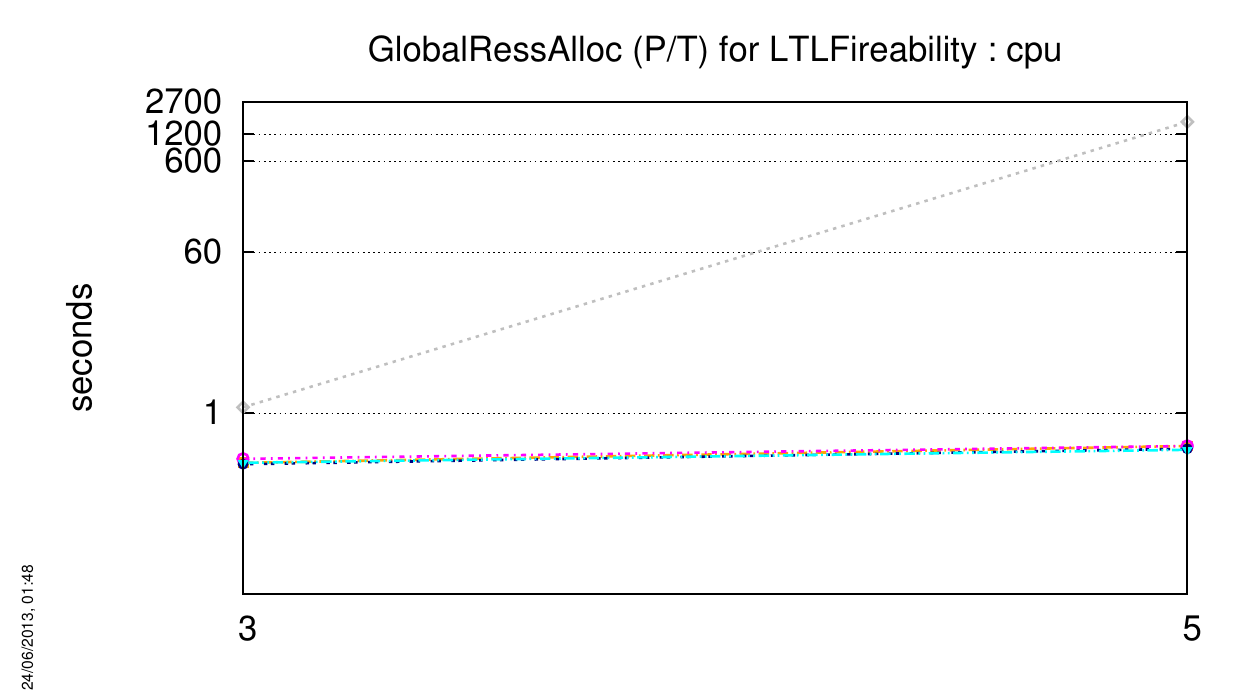}

   \includegraphics[height=1cm]{figures/tools-legend.pdf}
\end{center}

\subsubsection{\acs{Kanban-PT}}
The charts below respectively show how tools compete with this ``Known'' model (memory and CPU).

\index{Performances!LTLFireability!Kanban (P/T)}
\begin{center}
   \includegraphics[width=7.2cm]{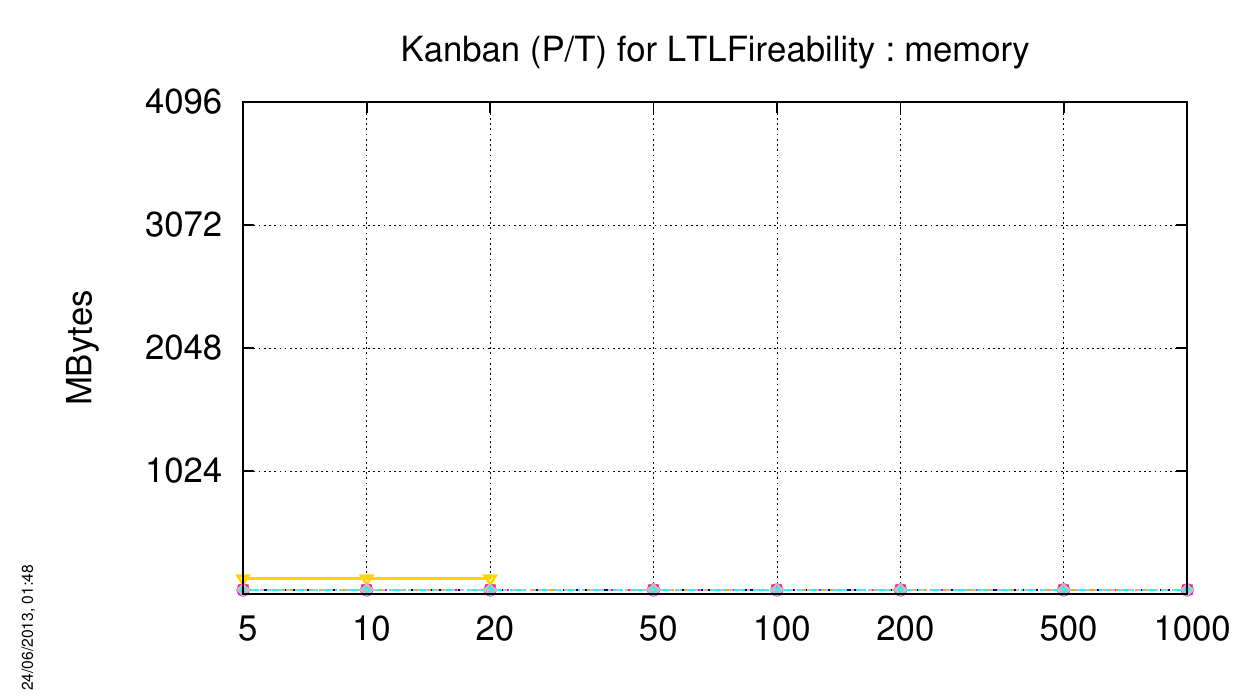}
   \includegraphics[width=7.2cm]{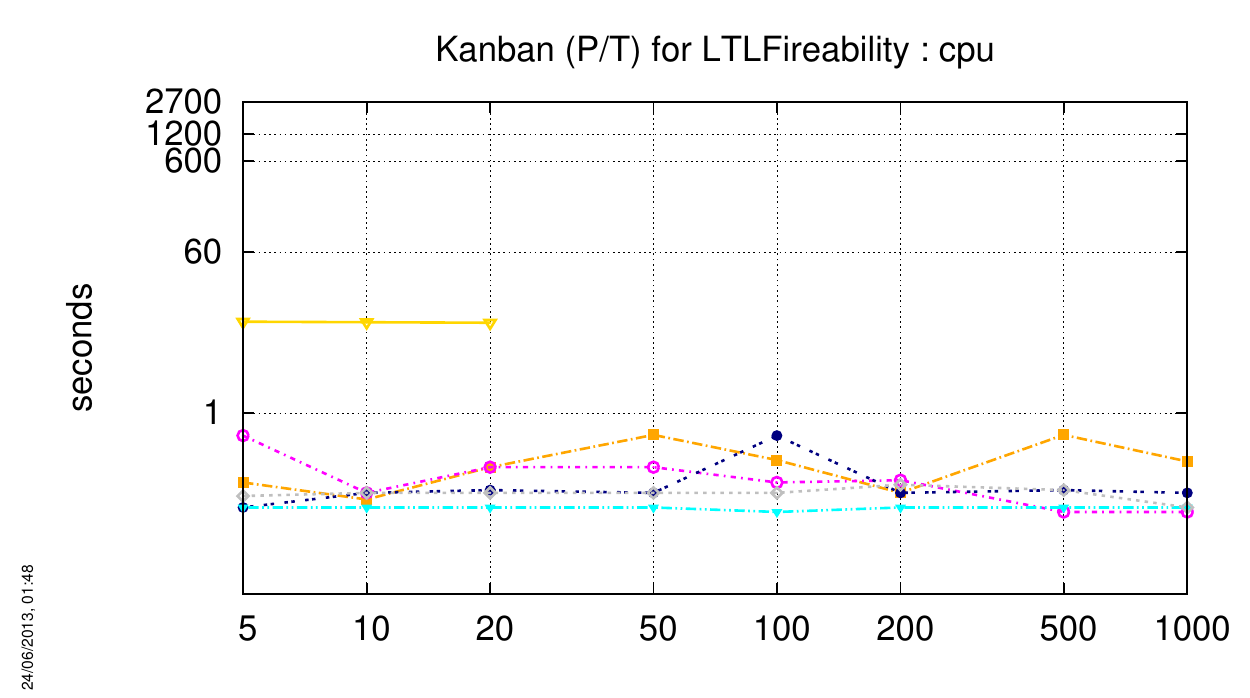}

   \includegraphics[height=1cm]{figures/tools-legend.pdf}
\end{center}

\subsubsection{\acs{LamportFastMutEx-COL}}
No instance of this model could be computed for the \textbf{LTLFireability} examination.

\subsubsection{\acs{LamportFastMutEx-PT}}
The charts below respectively show how tools compete with this ``Known'' model (memory and CPU).

\index{Performances!LTLFireability!LamportFastMutEx (P/T)}
\begin{center}
   \includegraphics[width=7.2cm]{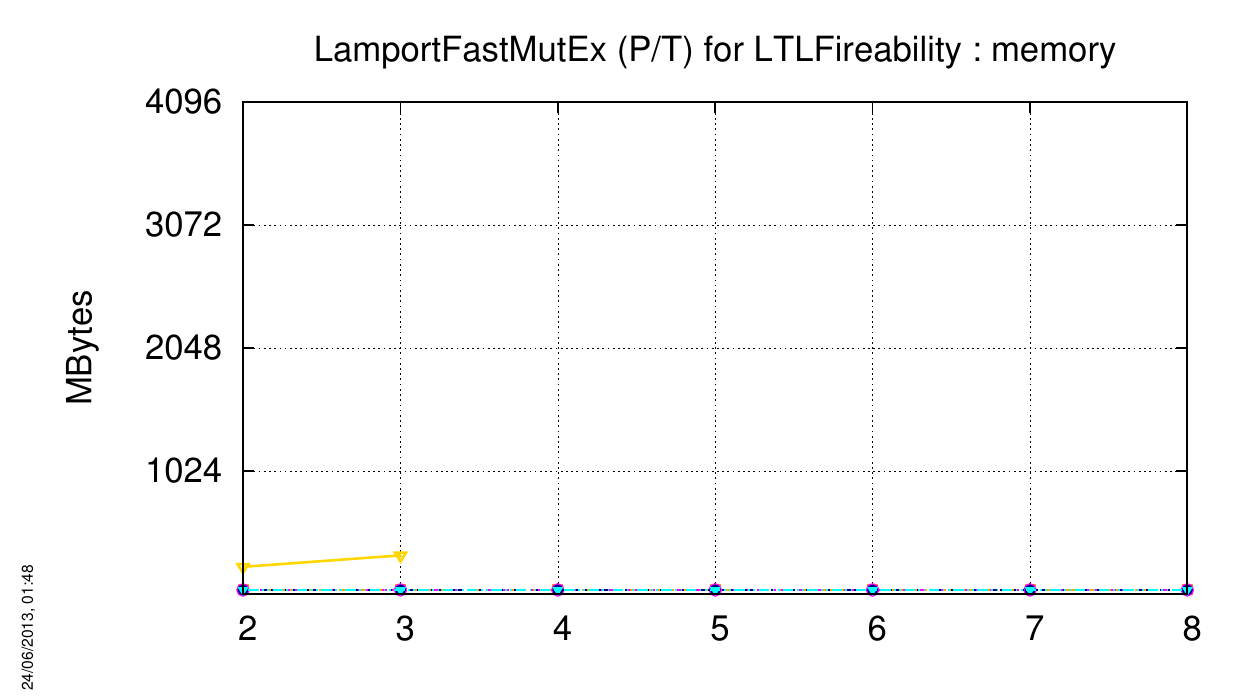}
   \includegraphics[width=7.2cm]{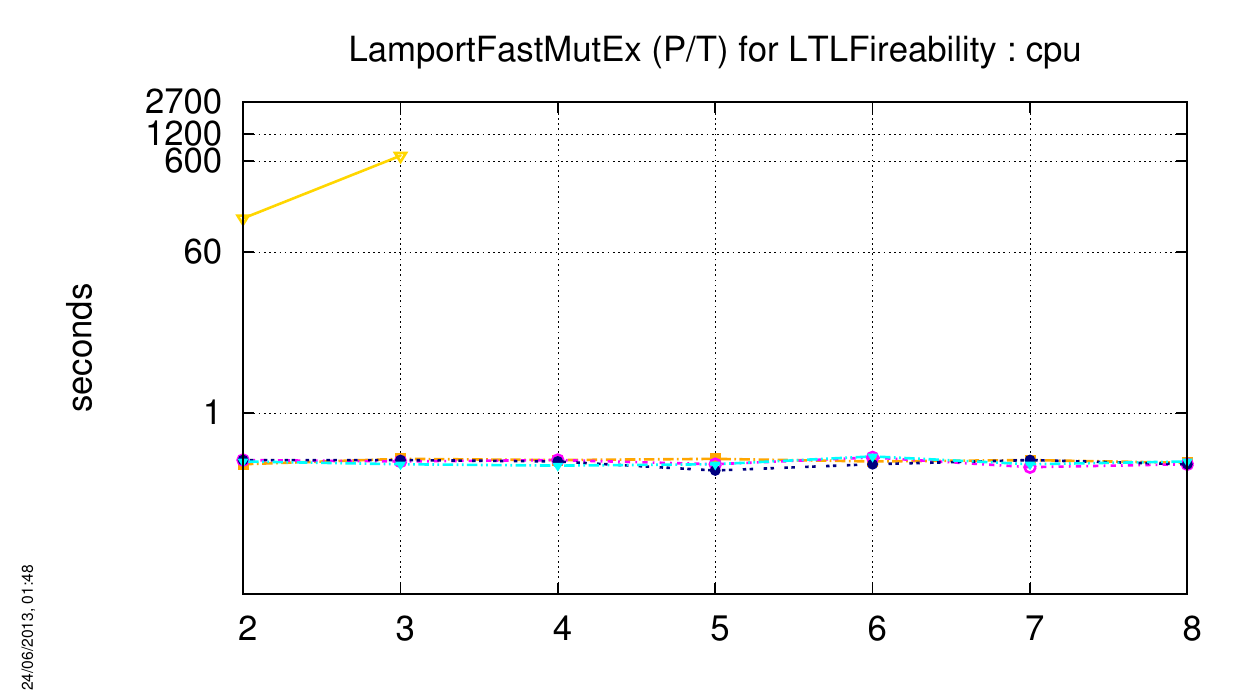}

   \includegraphics[height=1cm]{figures/tools-legend.pdf}
\end{center}

\subsubsection{\acs{MAPK-PT}}
The charts below respectively show how tools compete with this ``Known'' model (memory and CPU).

\index{Performances!LTLFireability!MAPK (P/T)}
\begin{center}
   \includegraphics[width=7.2cm]{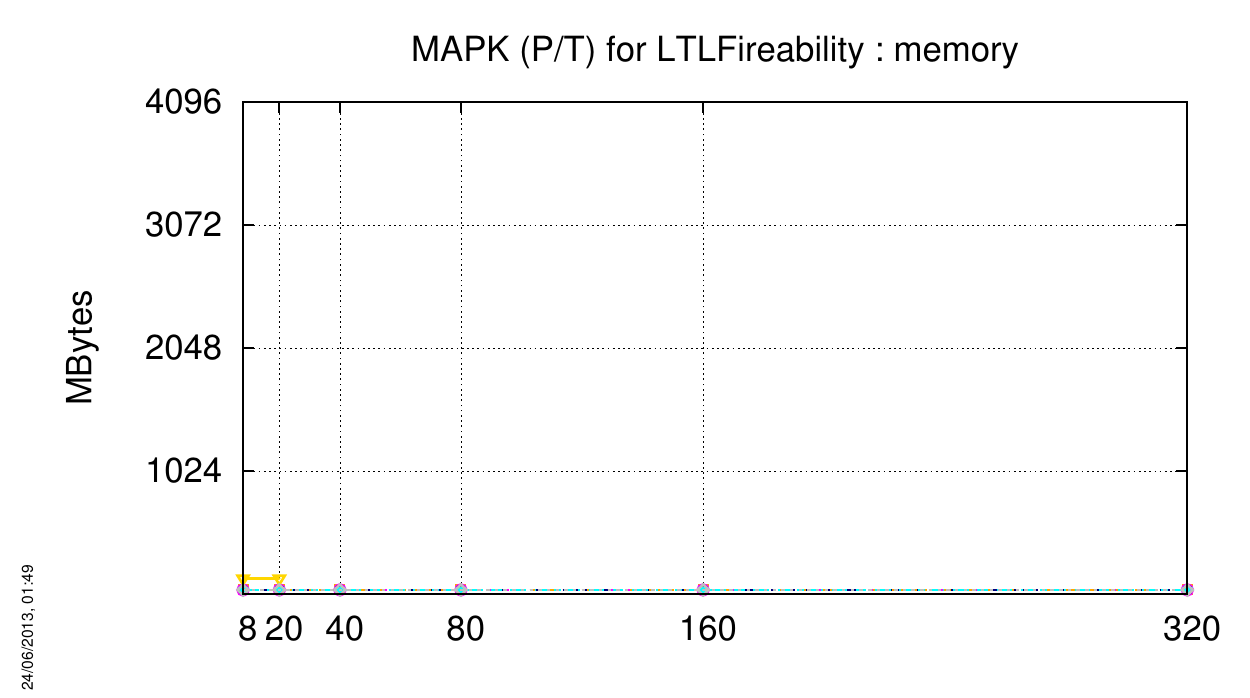}
   \includegraphics[width=7.2cm]{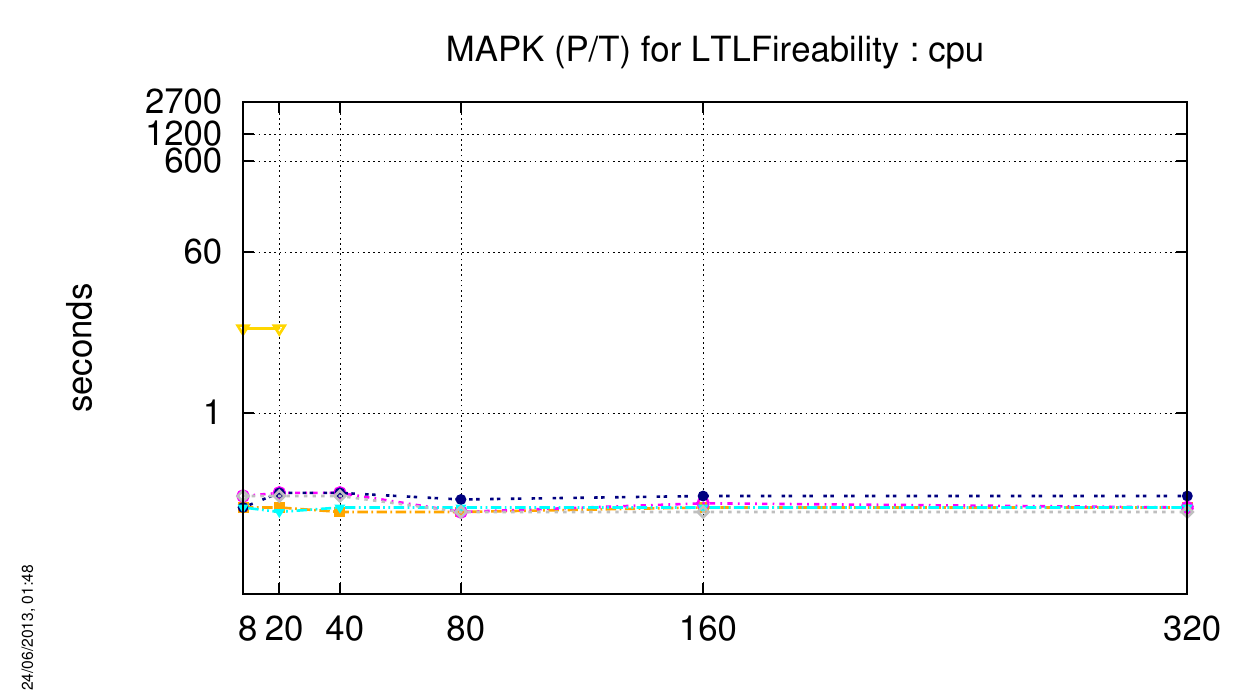}

   \includegraphics[height=1cm]{figures/tools-legend.pdf}
\end{center}

\subsubsection{\acs{NeoElection-COL}}
No instance of this model could be computed for the \textbf{LTLFireability} examination.

\subsubsection{\acs{NeoElection-PT}}
The charts below respectively show how tools compete with this ``Known'' model (memory and CPU).

\index{Performances!LTLFireability!NeoElection (P/T)}
\begin{center}
   \includegraphics[width=7.2cm]{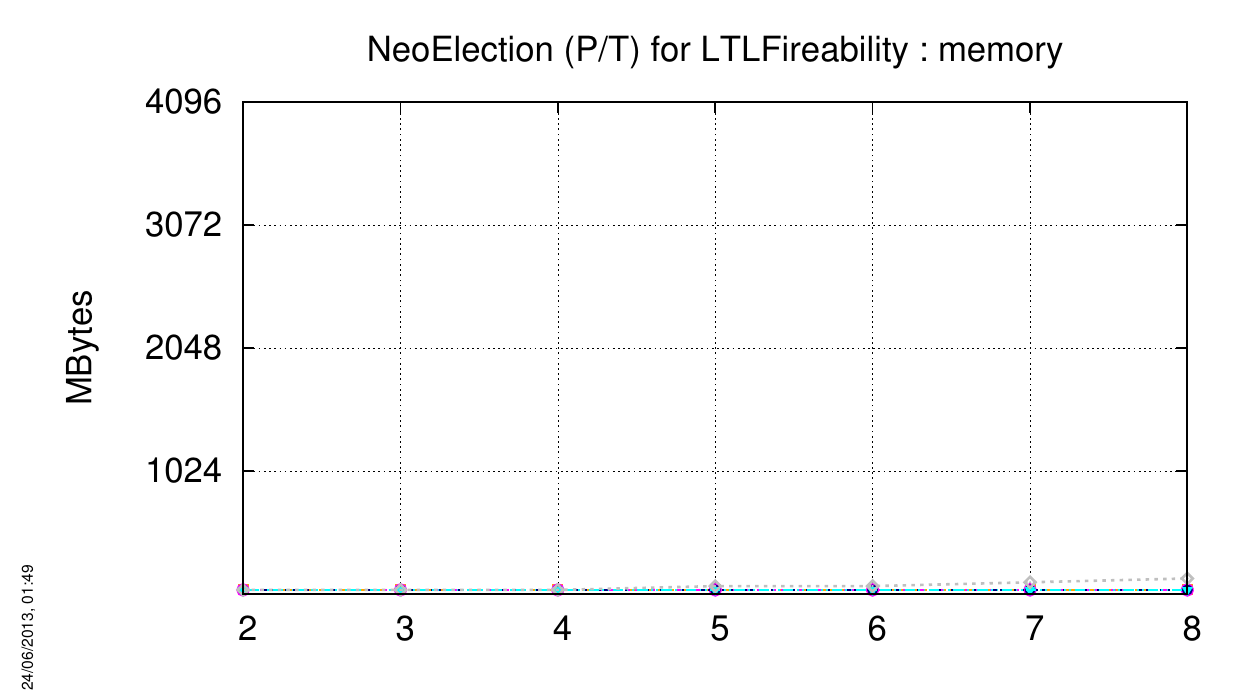}
   \includegraphics[width=7.2cm]{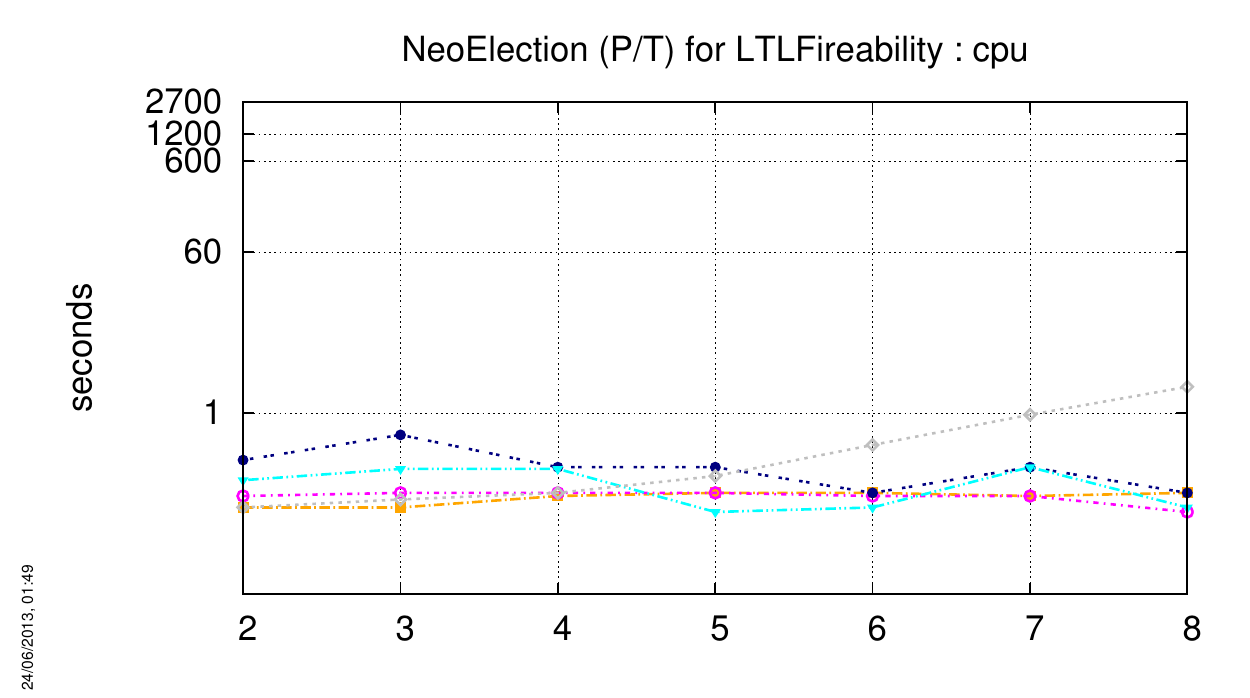}

   \includegraphics[height=1cm]{figures/tools-legend.pdf}
\end{center}

\subsubsection{\acs{PermAdmissibility-COL}}
No instance of this model could be computed for the \textbf{LTLFireability} examination.

\subsubsection{\acs{PermAdmissibility-PT}}
The charts below respectively show how tools compete with this ``Known'' model (memory and CPU).

\index{Performances!LTLFireability!PermAdmissibility (P/T)}
\begin{center}
   \includegraphics[width=7.2cm]{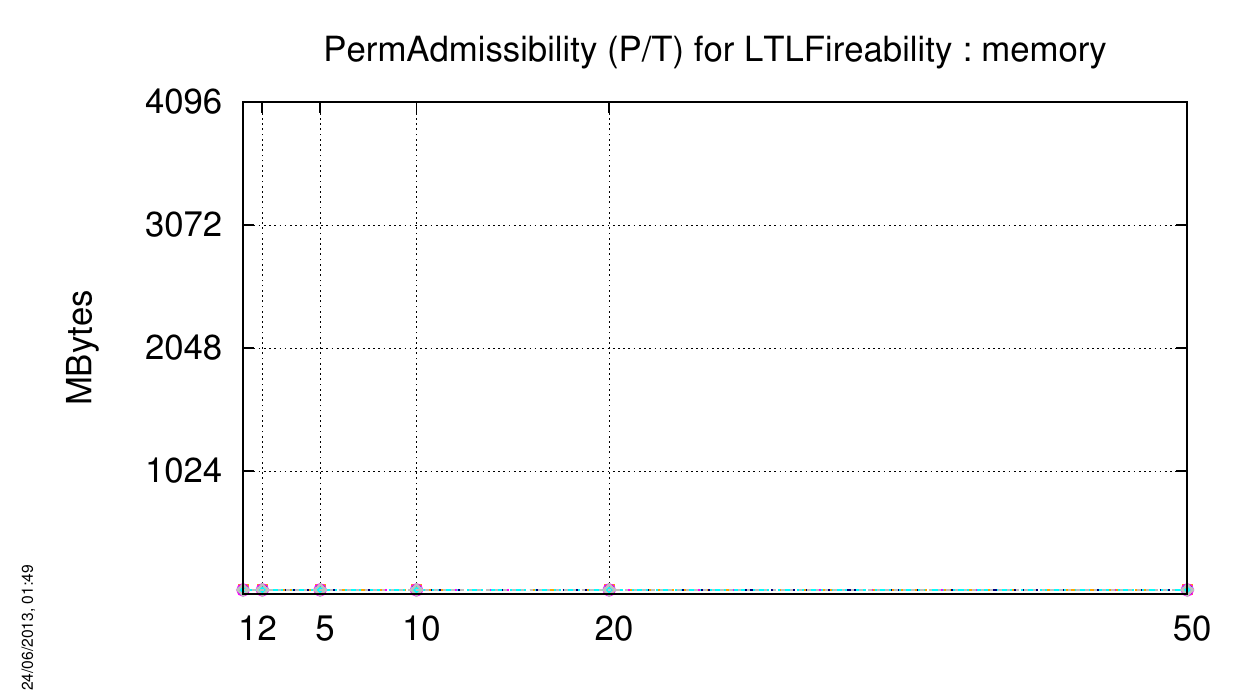}
   \includegraphics[width=7.2cm]{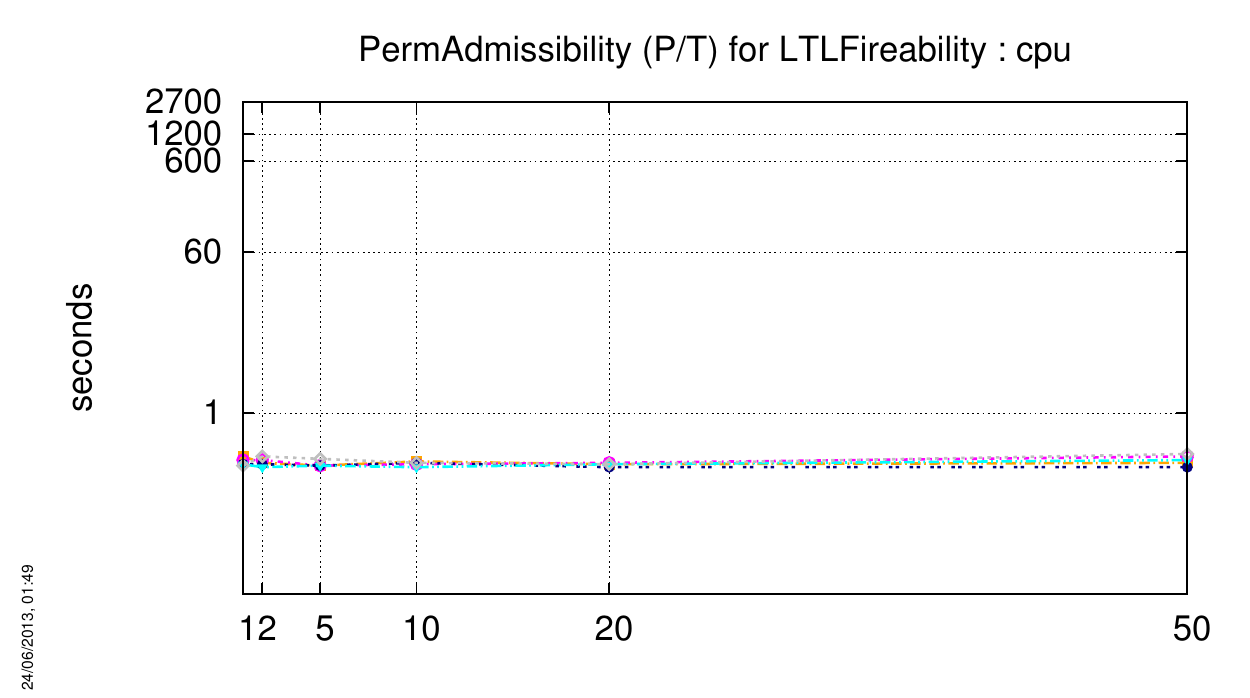}

   \includegraphics[height=1cm]{figures/tools-legend.pdf}
\end{center}

\subsubsection{\acs{Peterson-COL}}
No instance of this model could be computed for the \textbf{LTLFireability} examination.

\subsubsection{\acs{Peterson-PT}}
The charts below respectively show how tools compete with this ``Known'' model (memory and CPU).

\index{Performances!LTLFireability!Peterson (P/T)}
\begin{center}
   \includegraphics[width=7.2cm]{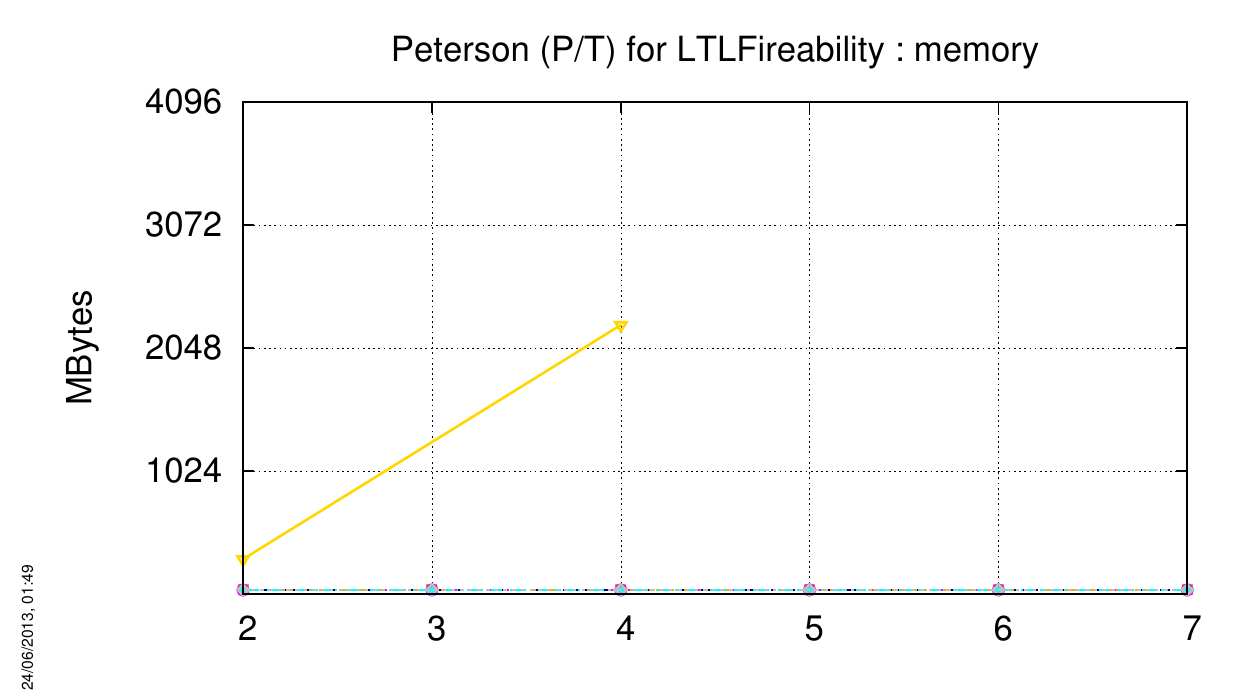}
   \includegraphics[width=7.2cm]{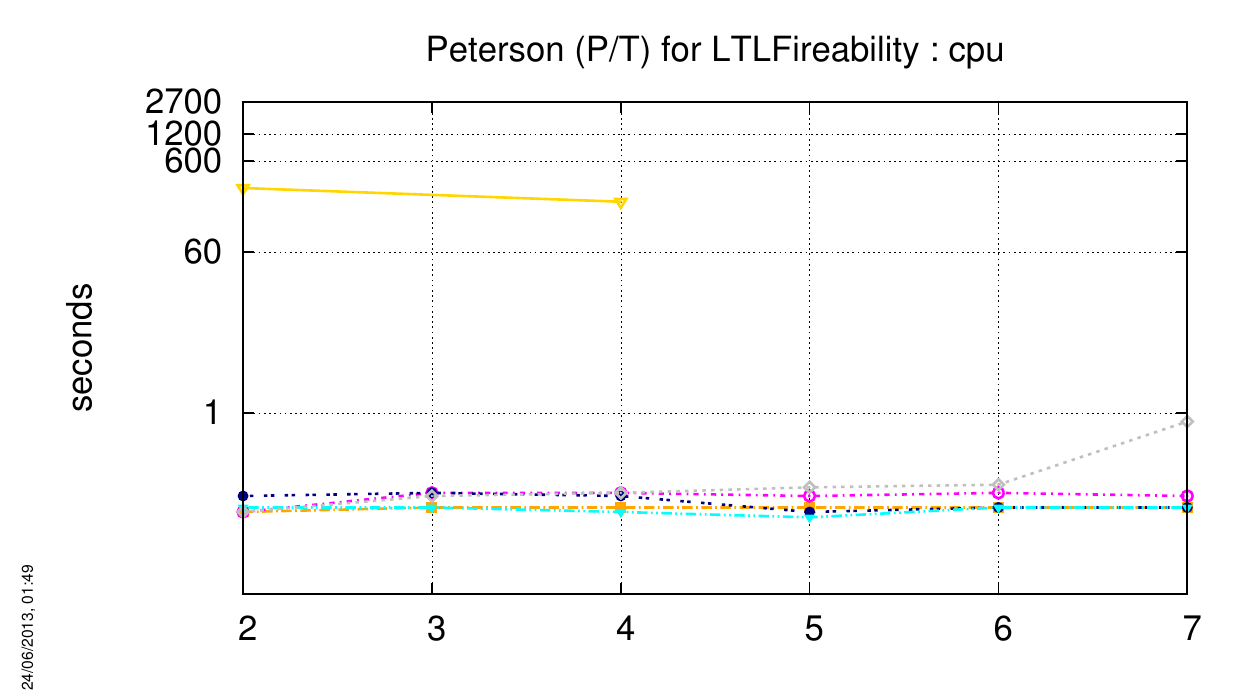}

   \includegraphics[height=1cm]{figures/tools-legend.pdf}
\end{center}

\subsubsection{\acs{Philosophers-COL}}
No instance of this model could be computed for the \textbf{LTLFireability} examination.

\subsubsection{\acs{Philosophers-PT}}
The charts below respectively show how tools compete with this ``Known'' model (memory and CPU).

\index{Performances!LTLFireability!Philosophers (P/T)}
\begin{center}
   \includegraphics[width=7.2cm]{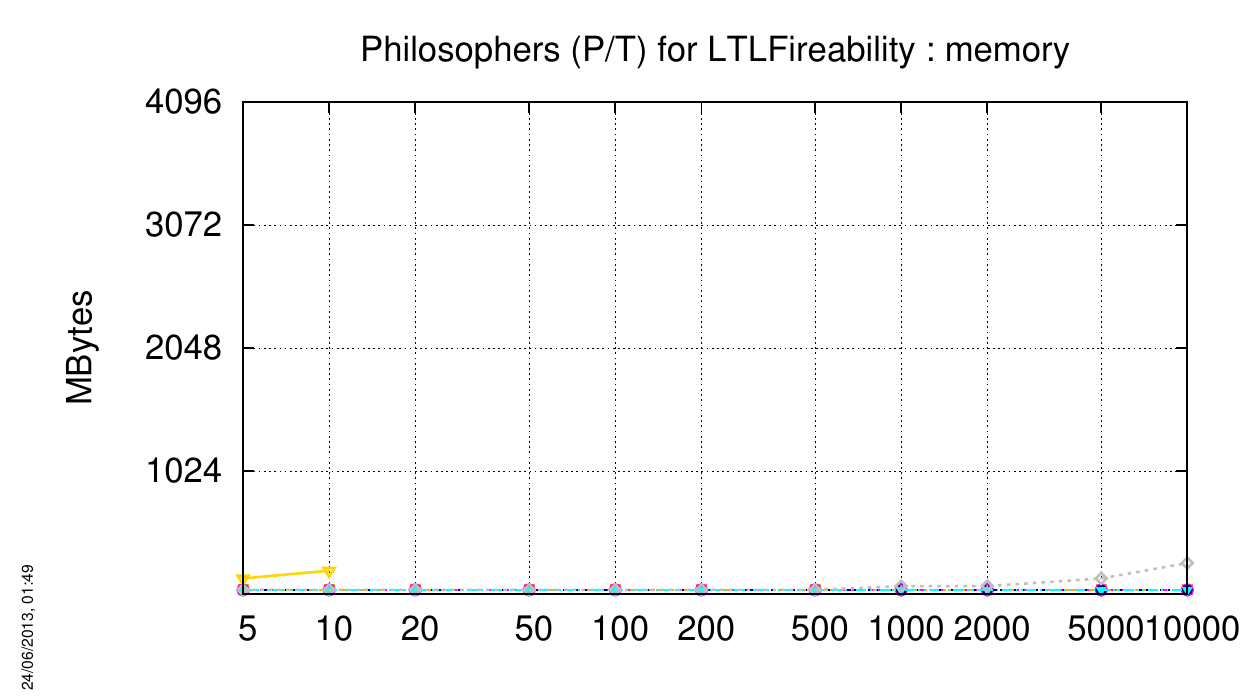}
   \includegraphics[width=7.2cm]{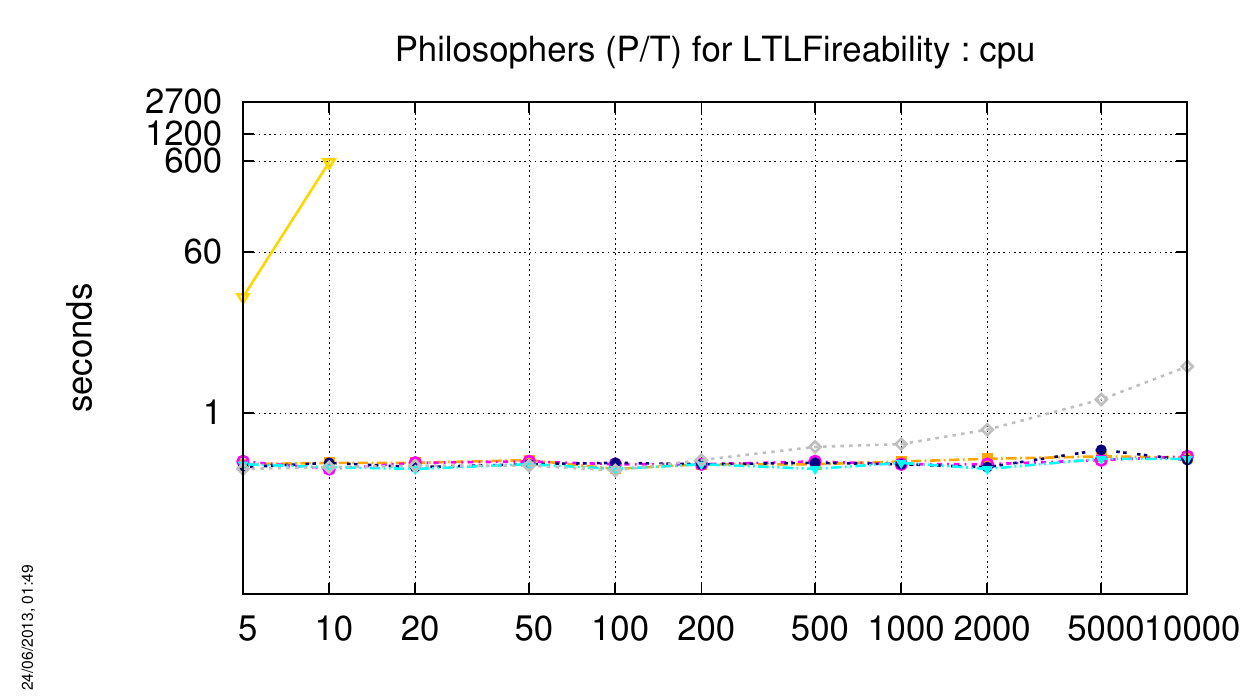}

   \includegraphics[height=1cm]{figures/tools-legend.pdf}
\end{center}

\subsubsection{\acs{PhilosophersDyn-COL}}
No instance of this model could be computed for the \textbf{LTLFireability} examination.

\subsubsection{\acs{PhilosophersDyn-PT}}
The charts below respectively show how tools compete with this ``Known'' model (memory and CPU).

\index{Performances!LTLFireability!PhilosophersDyn (P/T)}
\begin{center}
   \includegraphics[width=7.2cm]{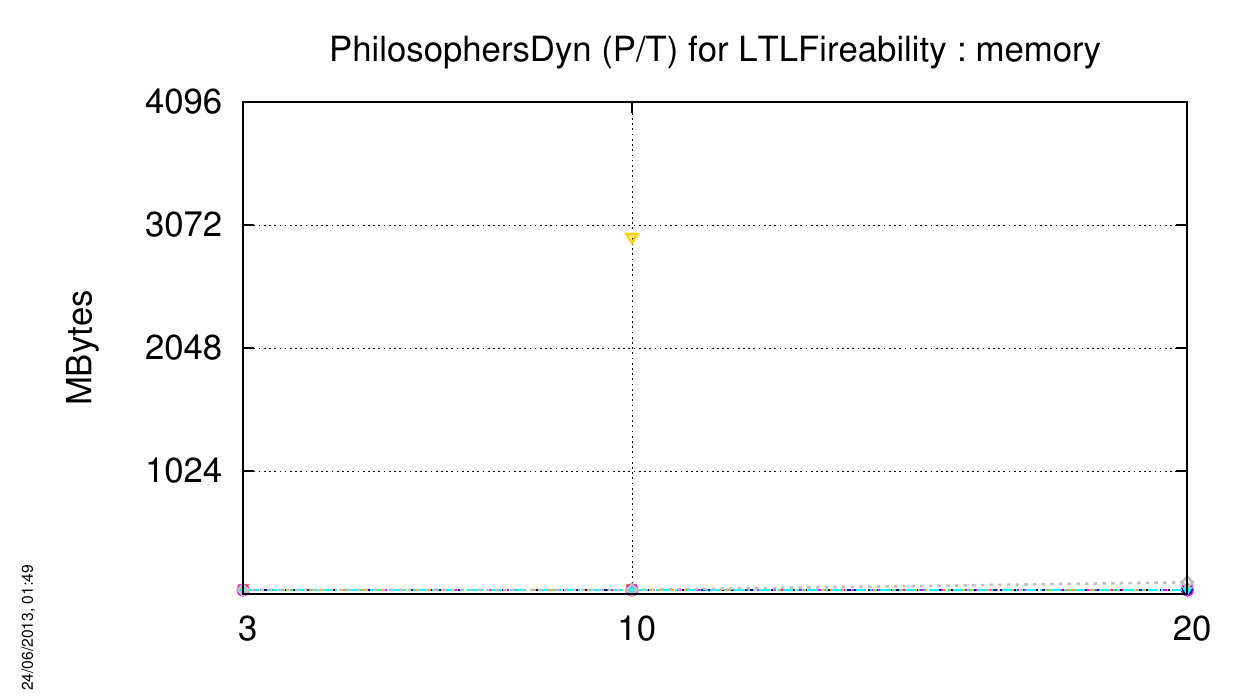}
   \includegraphics[width=7.2cm]{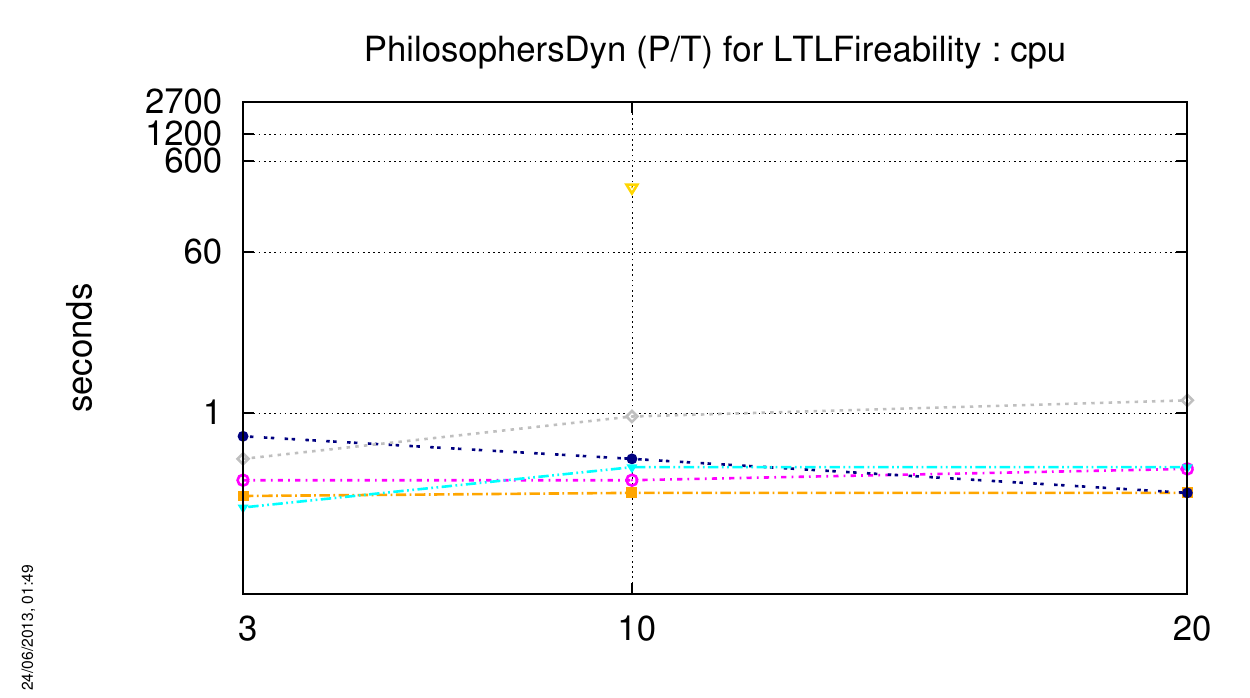}

   \includegraphics[height=1cm]{figures/tools-legend.pdf}
\end{center}

\subsubsection{\acs{Planning-PT}}
No instance of this model could be computed for the \textbf{LTLFireability} examination.

\subsubsection{\acs{Railroad-PT}}
The charts below respectively show how tools compete with this ``Known'' model (memory and CPU).

\index{Performances!LTLFireability!Railroad (P/T)}
\begin{center}
   \includegraphics[width=7.2cm]{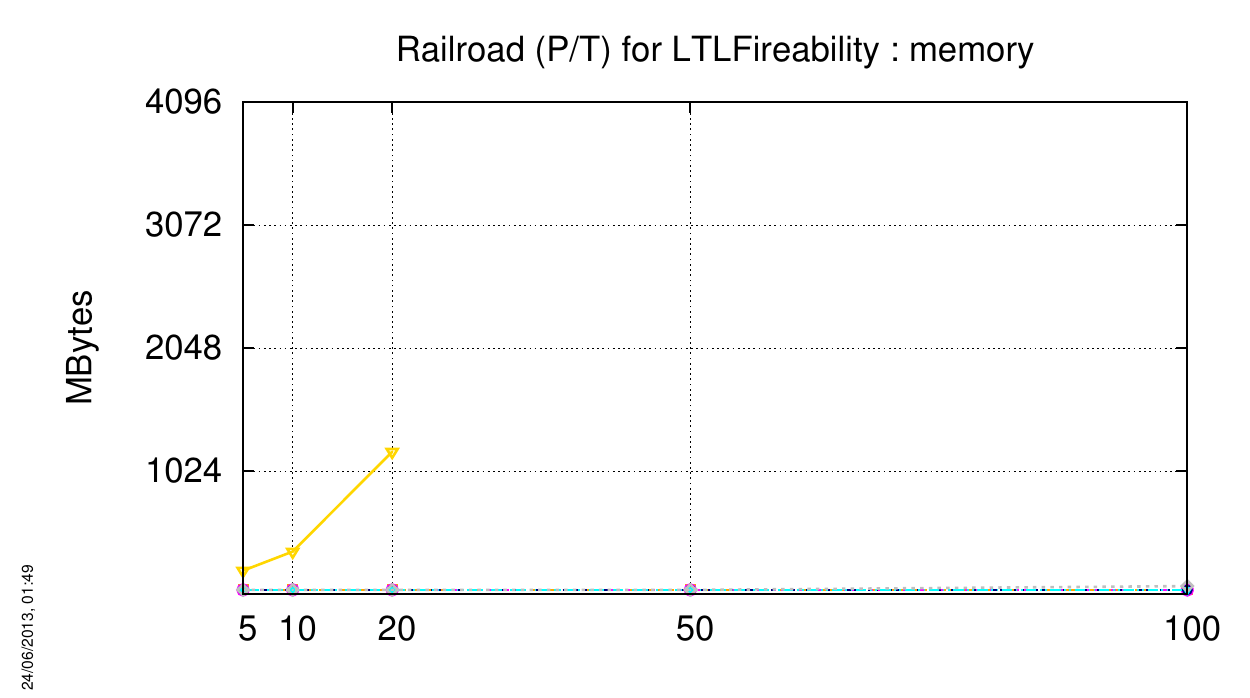}
   \includegraphics[width=7.2cm]{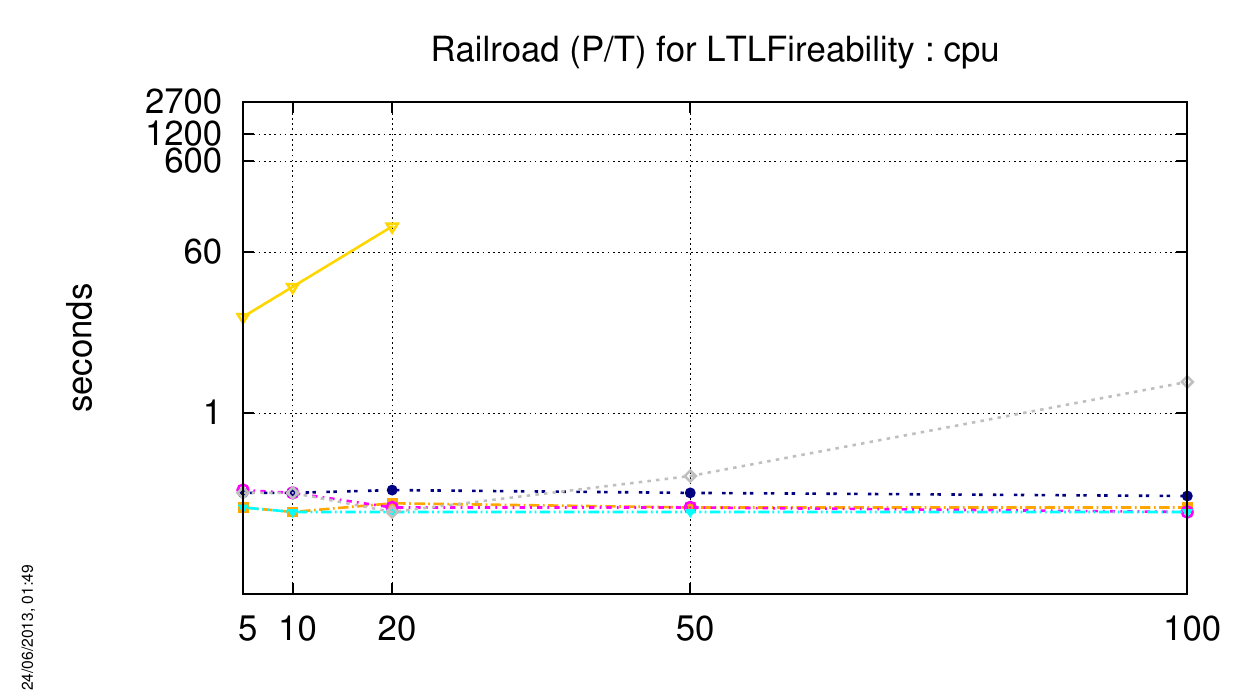}

   \includegraphics[height=1cm]{figures/tools-legend.pdf}
\end{center}

\subsubsection{\acs{RessAllocation-PT}}
The charts below respectively show how tools compete with this ``Known'' model (memory and CPU).

\index{Performances!LTLFireability!RessAllocation (P/T)}
\begin{center}
   \includegraphics[width=7.2cm]{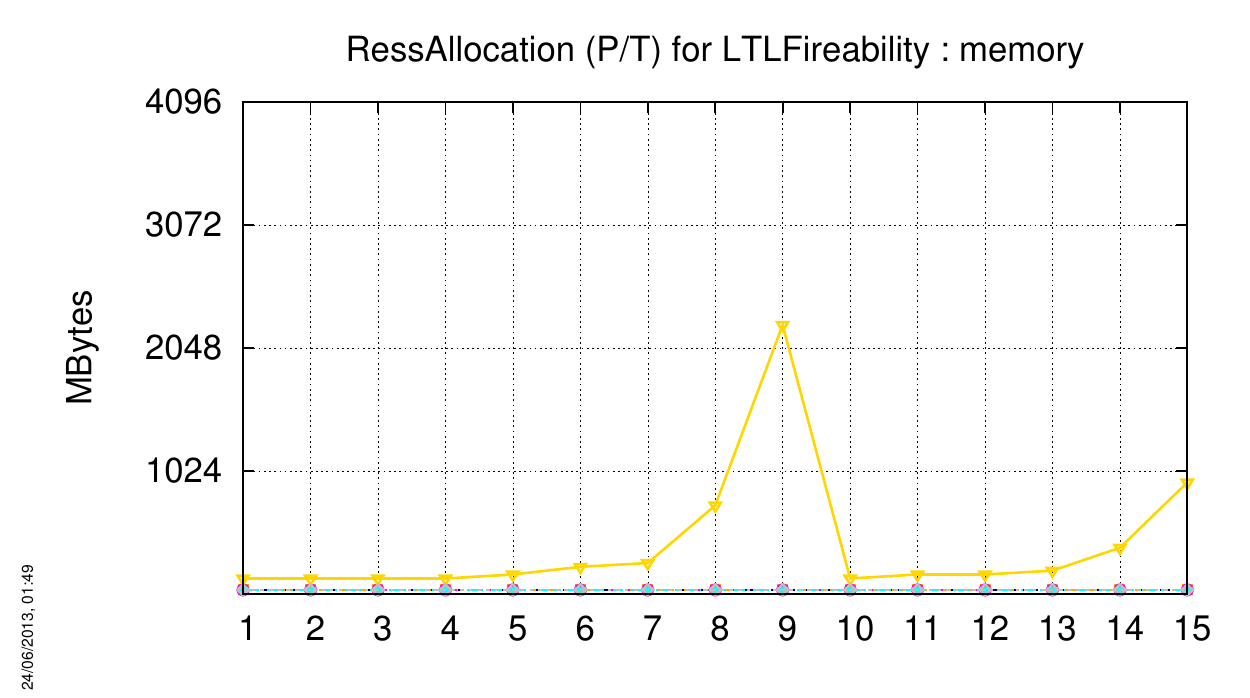}
   \includegraphics[width=7.2cm]{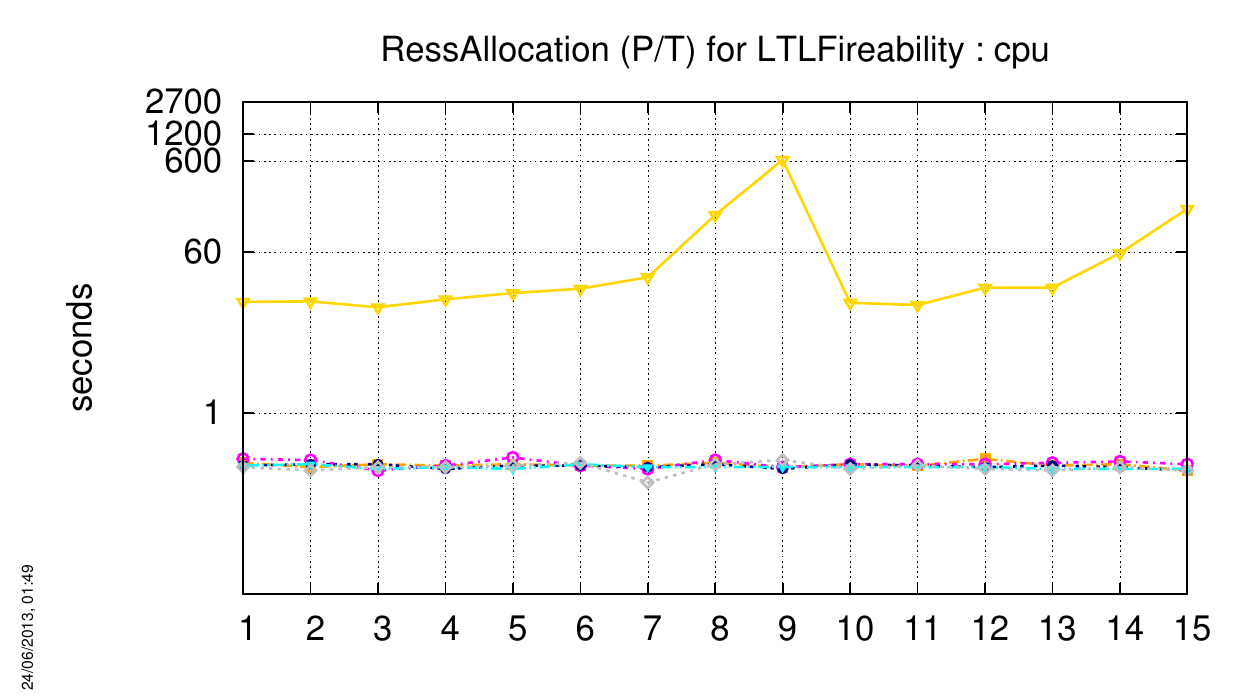}

   \includegraphics[height=1cm]{figures/tools-legend.pdf}
\end{center}

\subsubsection{\acs{Ring-PT}}
The charts below respectively show how tools compete with this ``Known'' model (memory and CPU).

\index{Performances!LTLFireability!Ring (P/T)}
\begin{center}
   \includegraphics[width=7.2cm]{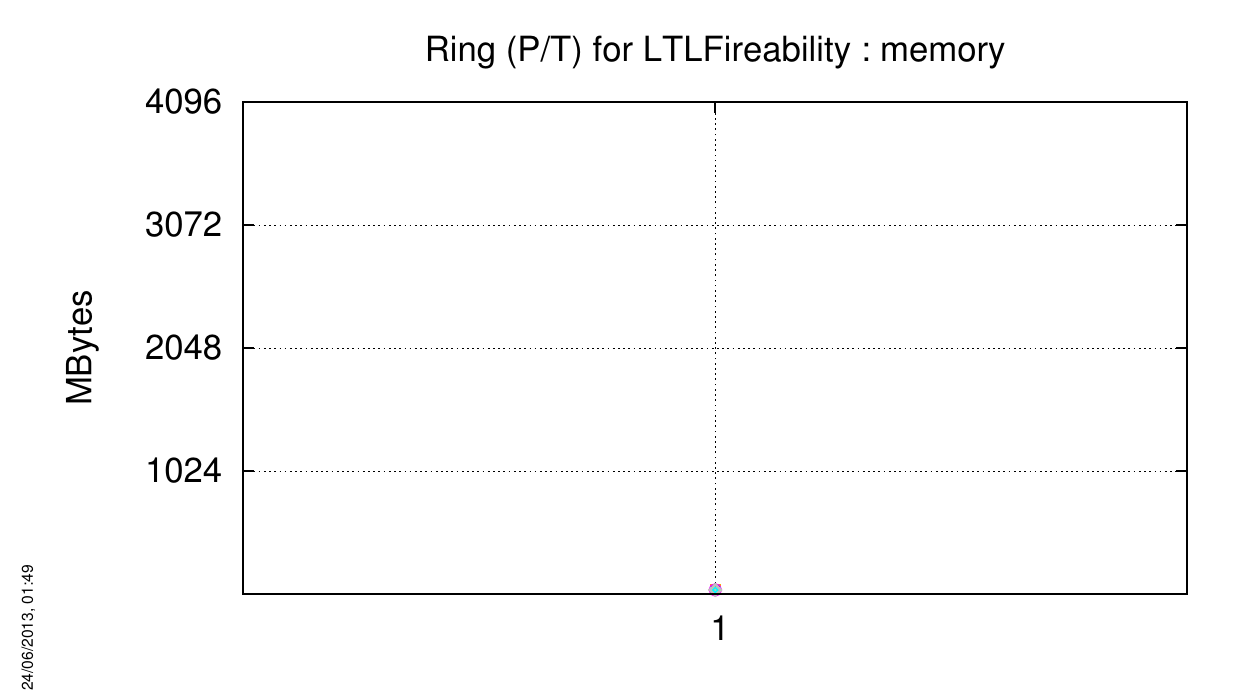}
   \includegraphics[width=7.2cm]{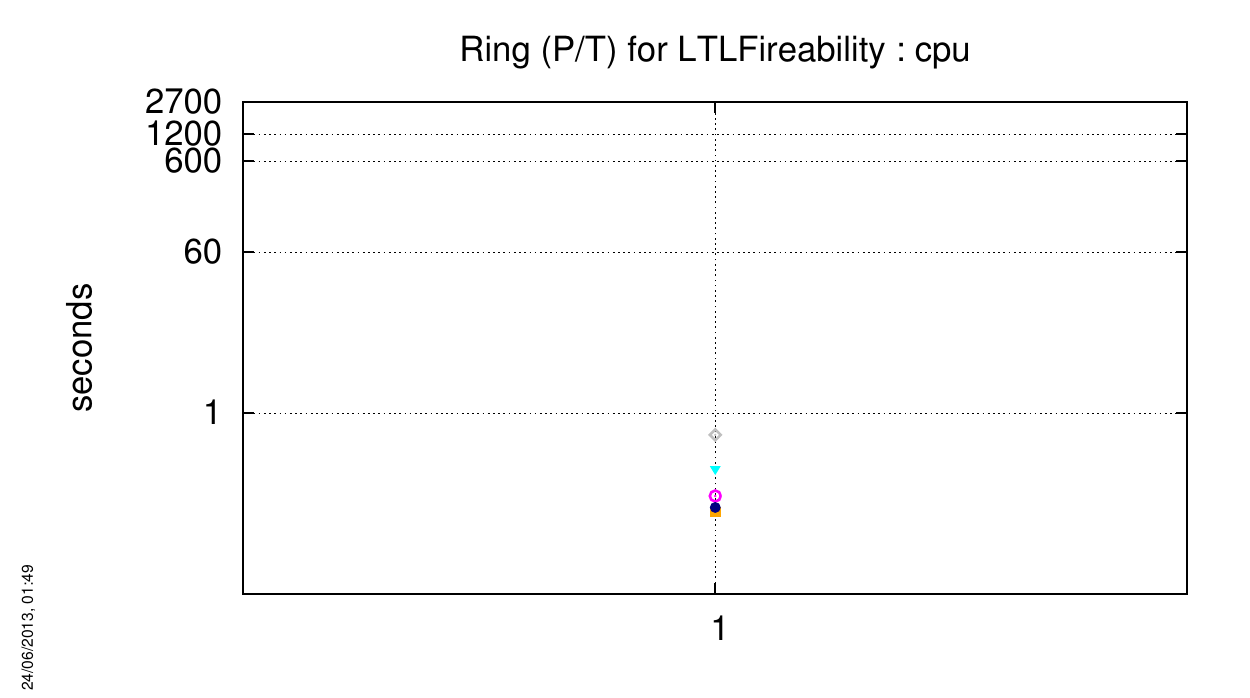}

   \includegraphics[height=1cm]{figures/tools-legend.pdf}
\end{center}

\subsubsection{\acs{RwMutex-PT}}
The charts below respectively show how tools compete with this ``Known'' model (memory and CPU).

\index{Performances!LTLFireability!RwMutex (P/T)}
\begin{center}
   \includegraphics[width=7.2cm]{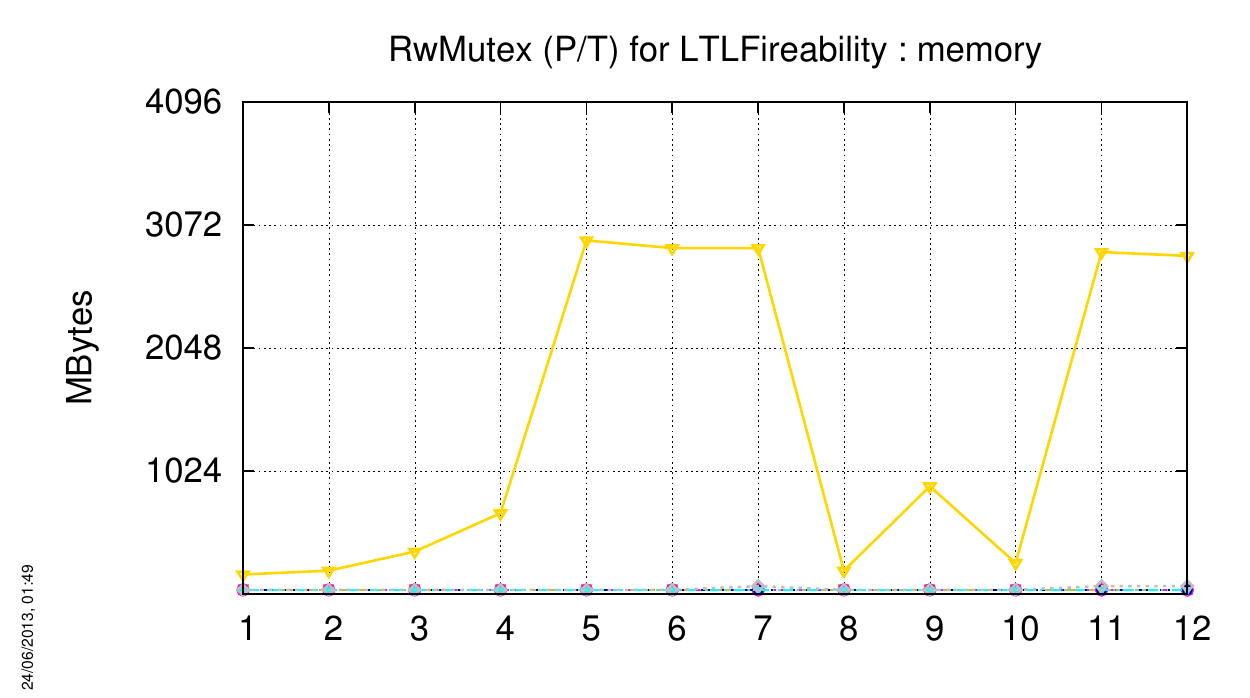}
   \includegraphics[width=7.2cm]{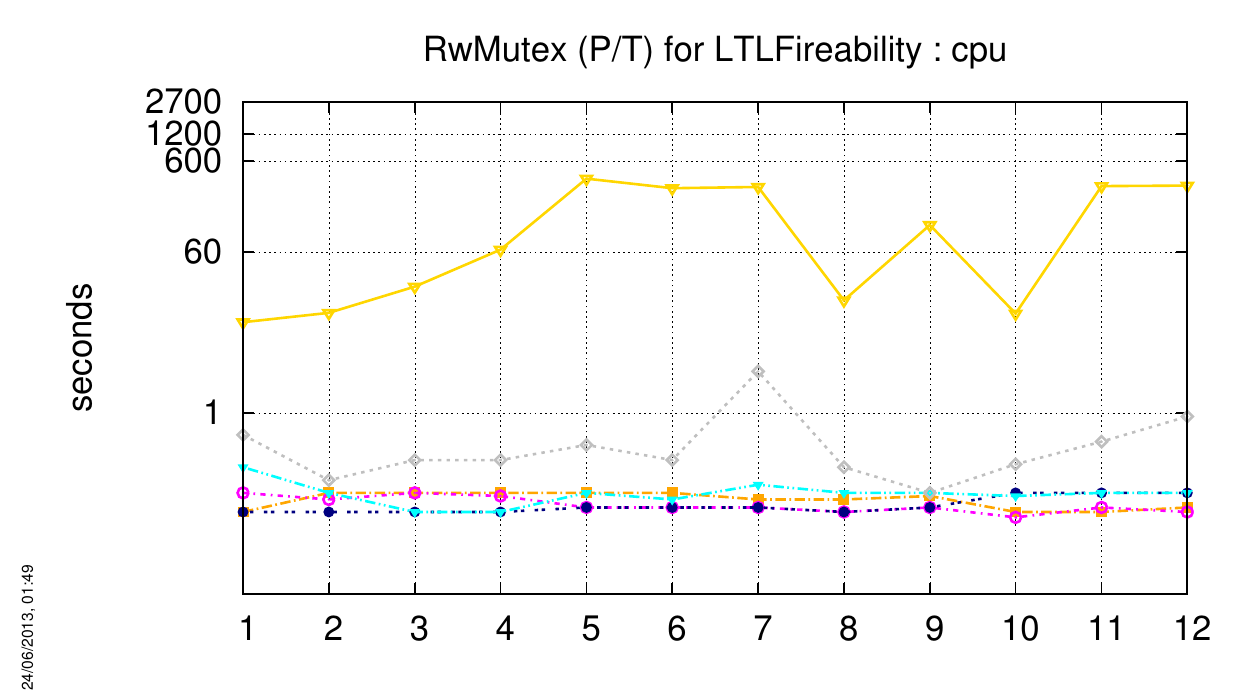}

   \includegraphics[height=1cm]{figures/tools-legend.pdf}
\end{center}

\subsubsection{\acs{SharedMemory-COL}}
No instance of this model could be computed for the \textbf{LTLFireability} examination.

\subsubsection{\acs{SharedMemory-PT}}
The charts below respectively show how tools compete with this ``Known'' model (memory and CPU).

\index{Performances!LTLFireability!SharedMemory (P/T)}
\begin{center}
   \includegraphics[width=7.2cm]{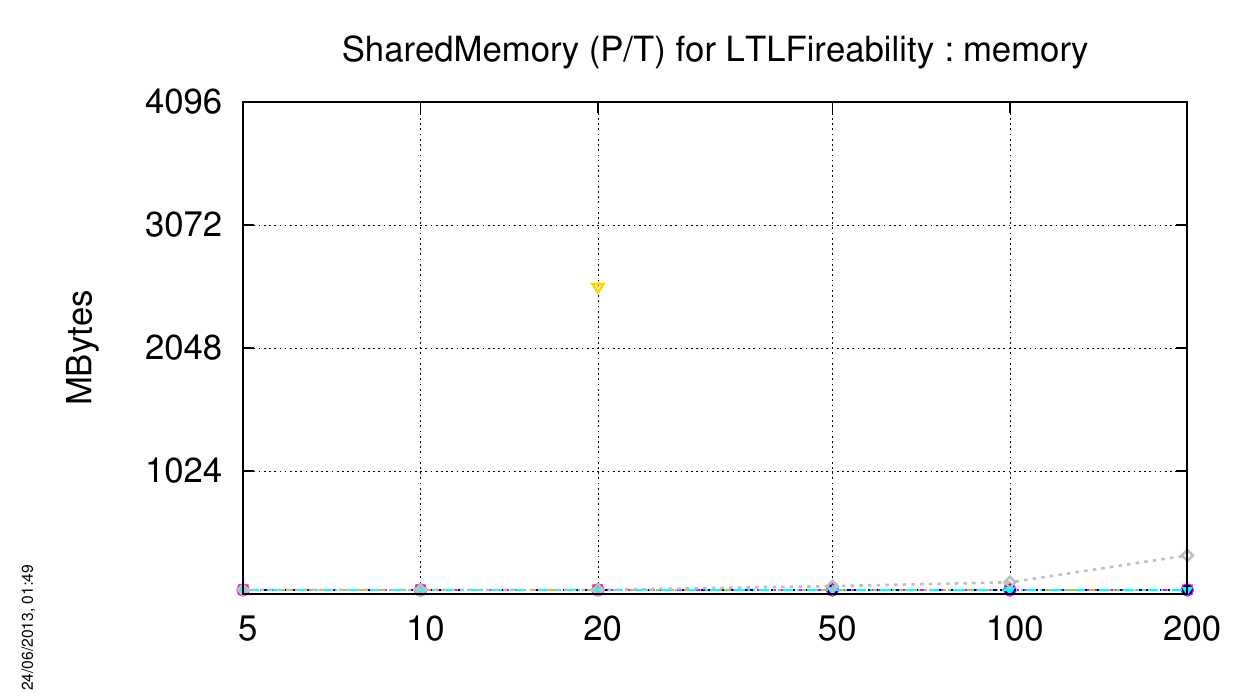}
   \includegraphics[width=7.2cm]{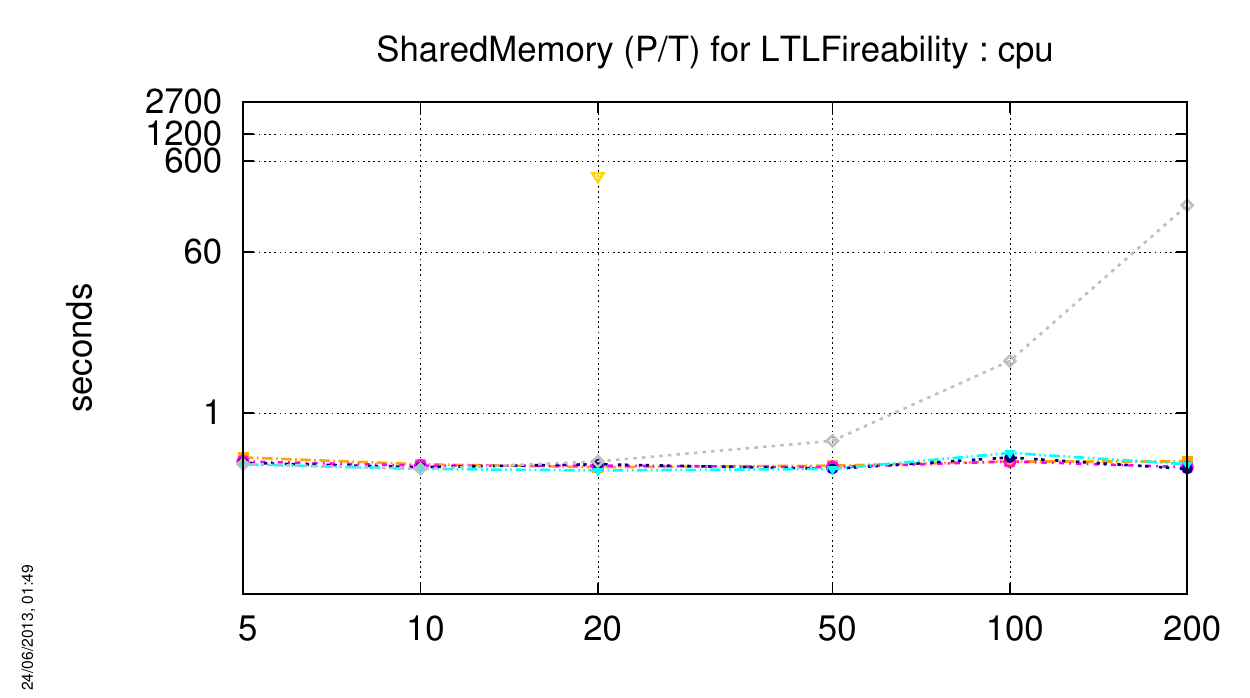}

   \includegraphics[height=1cm]{figures/tools-legend.pdf}
\end{center}

\subsubsection{\acs{SimpleLoadBal-COL}}
No instance of this model could be computed for the \textbf{LTLFireability} examination.

\subsubsection{\acs{SimpleLoadBal-PT}}
The charts below respectively show how tools compete with this ``Known'' model (memory and CPU).

\index{Performances!LTLFireability!SimpleLoadBal (P/T)}
\begin{center}
   \includegraphics[width=7.2cm]{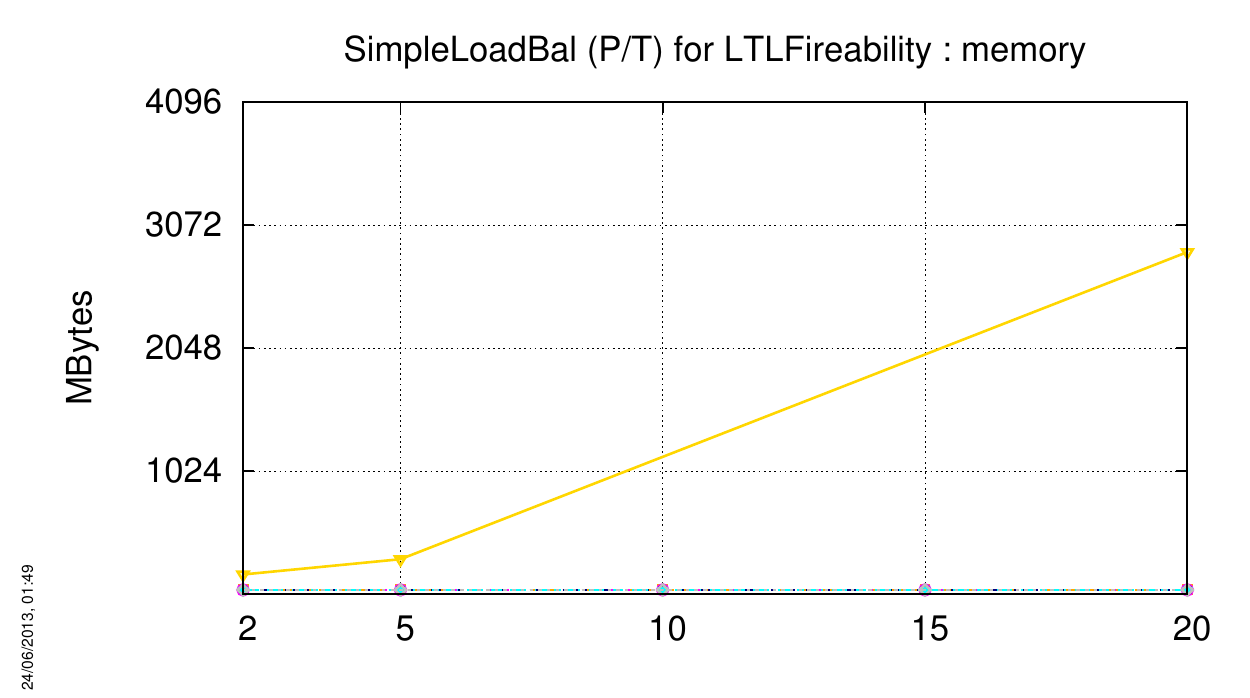}
   \includegraphics[width=7.2cm]{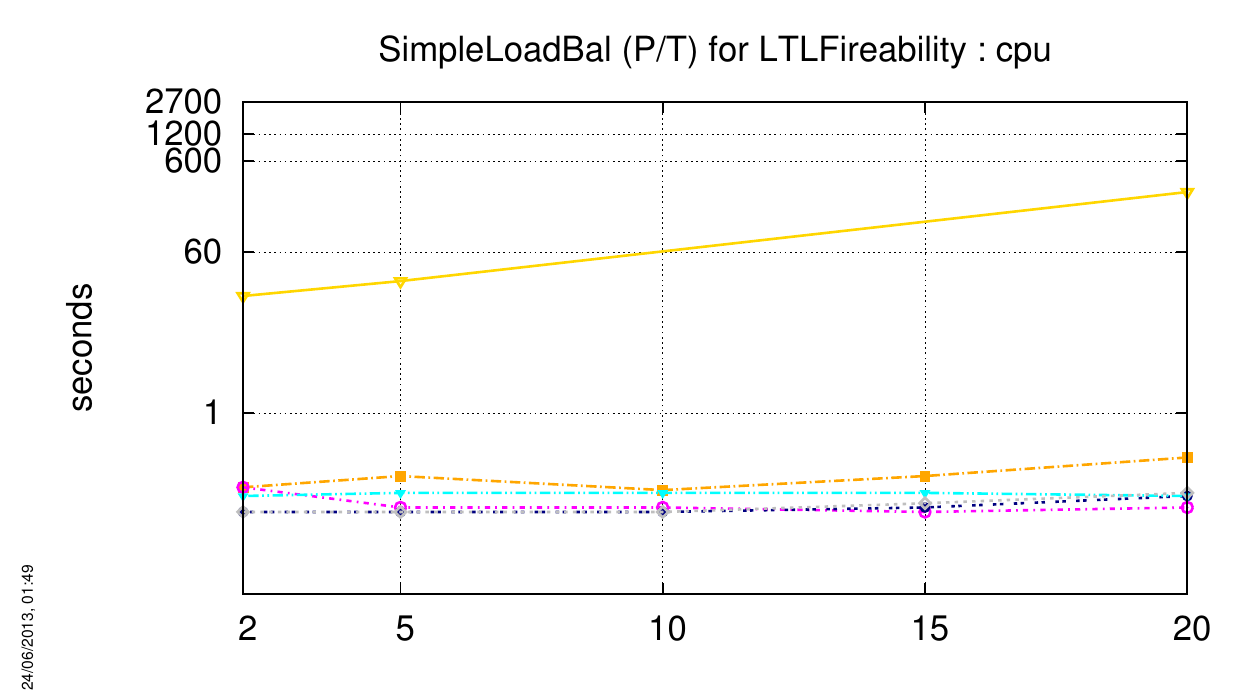}

   \includegraphics[height=1cm]{figures/tools-legend.pdf}
\end{center}

\subsubsection{\acs{TokenRing-COL}}
No instance of this model could be computed for the \textbf{LTLFireability} examination.

\subsubsection{\acs{TokenRing-PT}}
The charts below respectively show how tools compete with this ``Known'' model (memory and CPU).

\index{Performances!LTLFireability!TokenRing (P/T)}
\begin{center}
   \includegraphics[width=7.2cm]{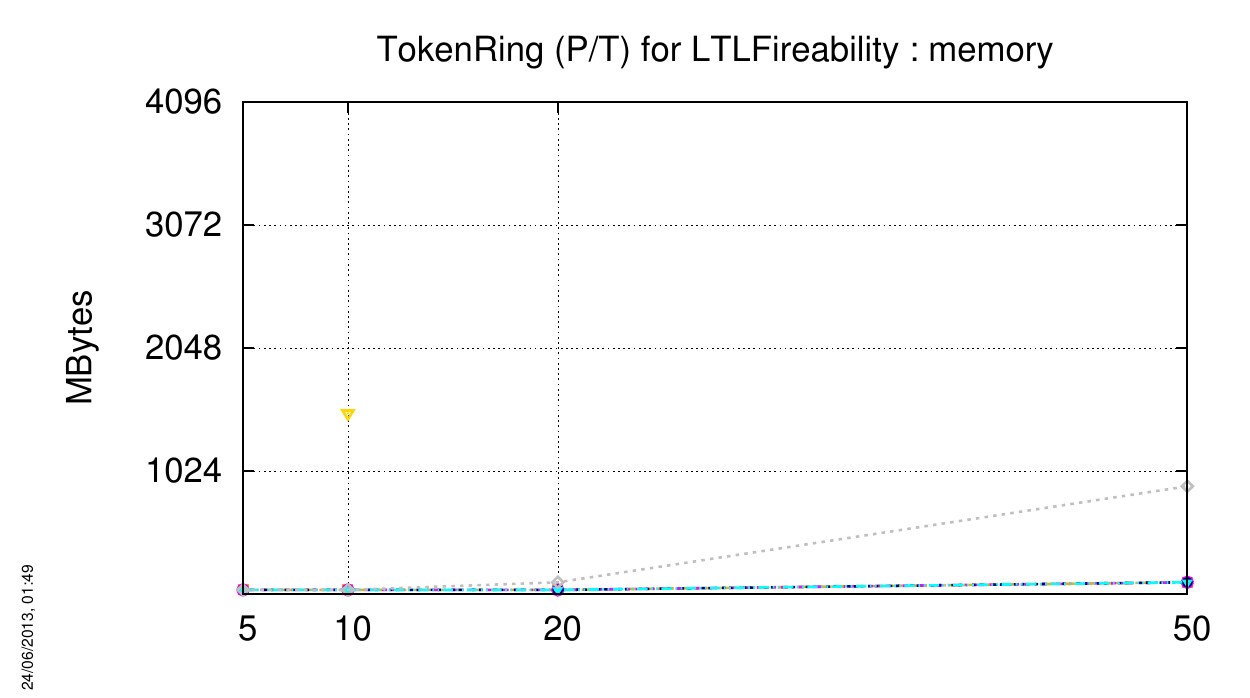}
   \includegraphics[width=7.2cm]{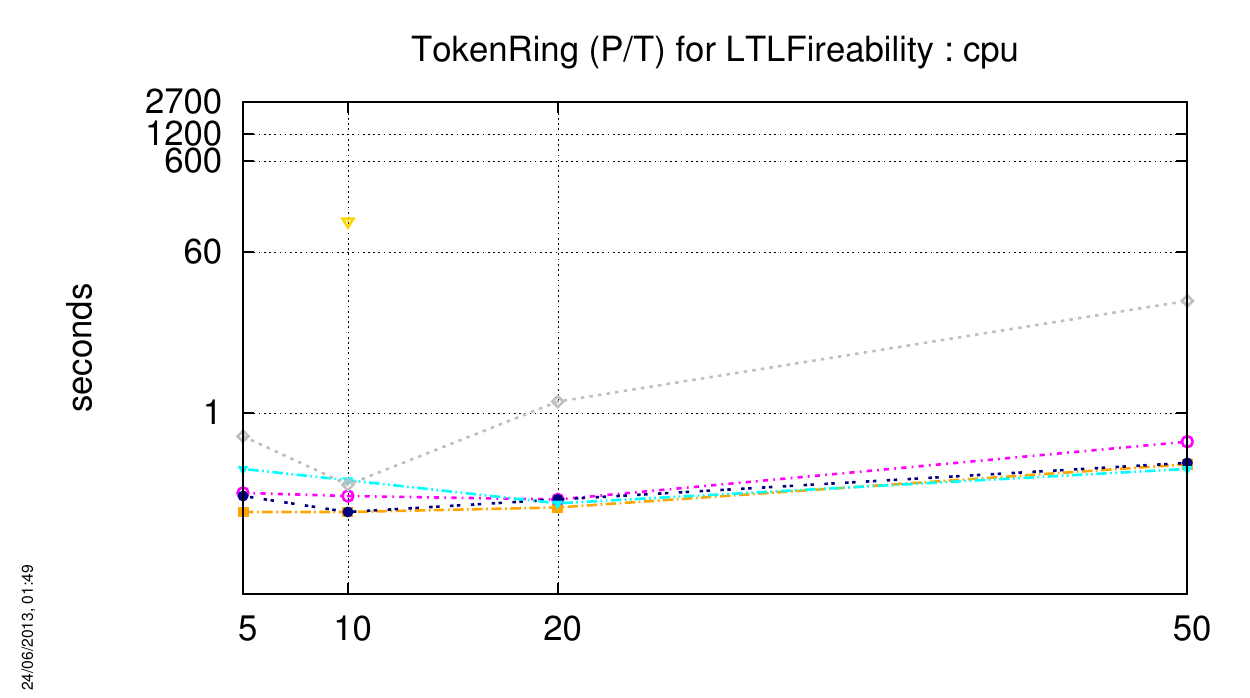}

   \includegraphics[height=1cm]{figures/tools-legend.pdf}
\end{center}

\subsubsection{\acs{HouseConstruction-PT}}
The charts below respectively show how tools compete with this ``Suprise'' model (memory and CPU).

\index{Performances!LTLFireability!HouseConstruction (P/T)}
\begin{center}
   \includegraphics[width=7.2cm]{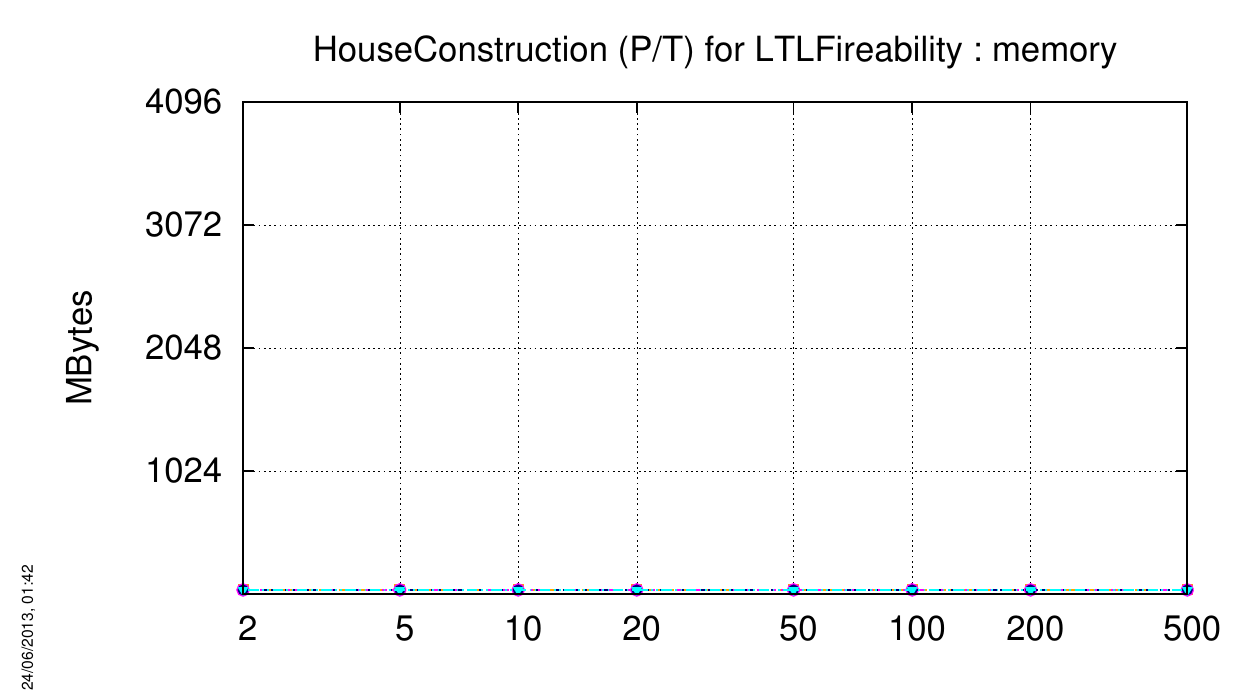}
   \includegraphics[width=7.2cm]{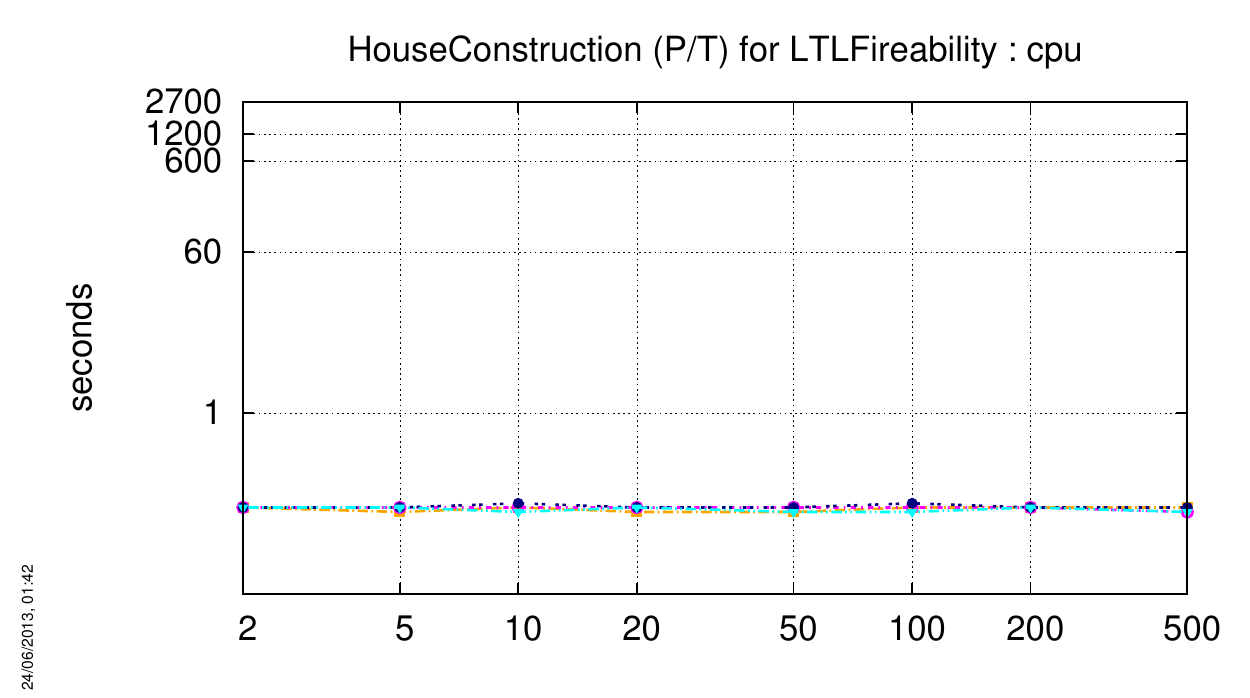}

   \includegraphics[height=1cm]{figures/tools-legend.pdf}
\end{center}

\subsubsection{\acs{IBMB2S565S3960-PT}}
The charts below respectively show how tools compete with this ``Suprise'' model (memory and CPU).

\index{Performances!LTLFireability!IBMB2S565S3960 (P/T)}
\begin{center}
   \includegraphics[width=7.2cm]{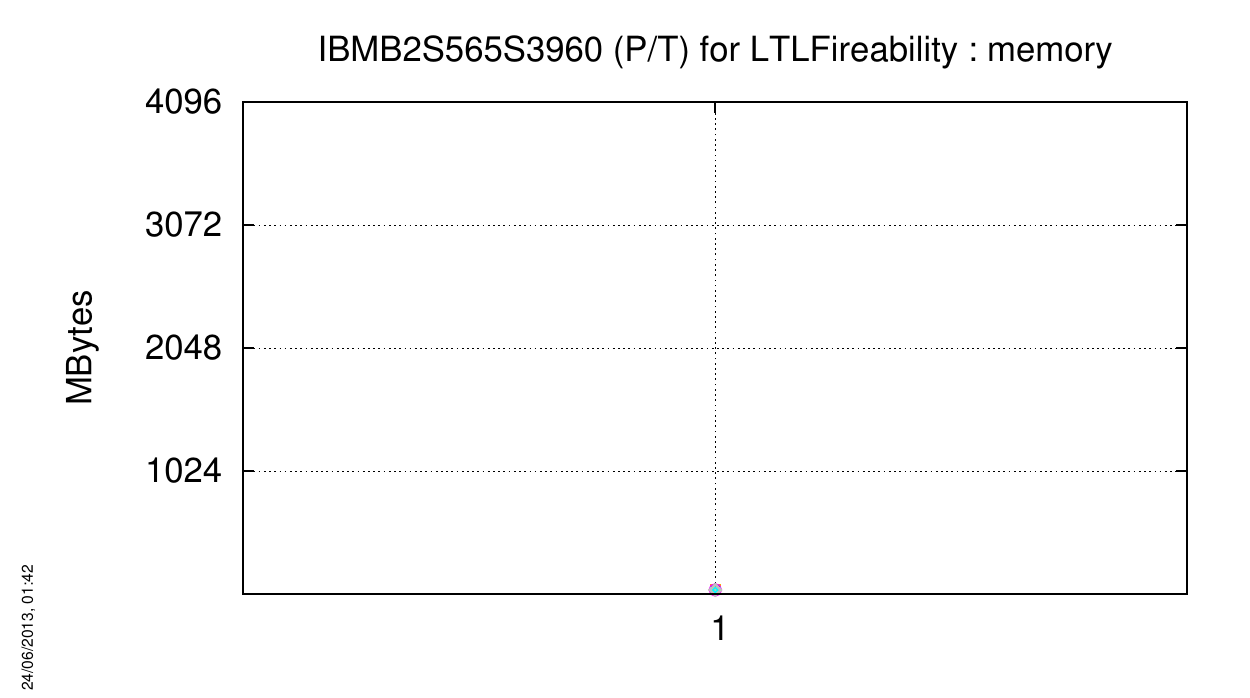}
   \includegraphics[width=7.2cm]{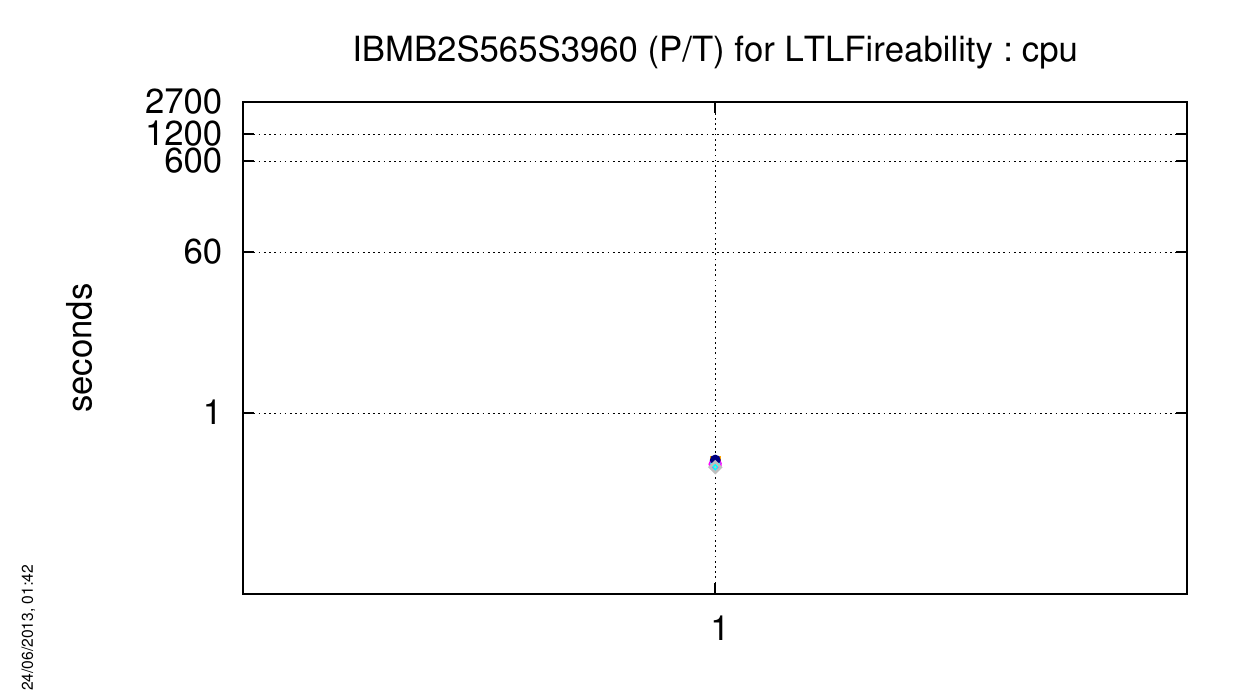}

   \includegraphics[height=1cm]{figures/tools-legend.pdf}
\end{center}

\subsubsection{\acs{QuasiCertifProtocol-COL}}
No instance of this model could be computed for the \textbf{LTLFireability} examination.

\subsubsection{\acs{QuasiCertifProtocol-PT}}
The charts below respectively show how tools compete with this ``Suprise'' model (memory and CPU).

\index{Performances!LTLFireability!QuasiCertifProtocol (P/T)}
\begin{center}
   \includegraphics[width=7.2cm]{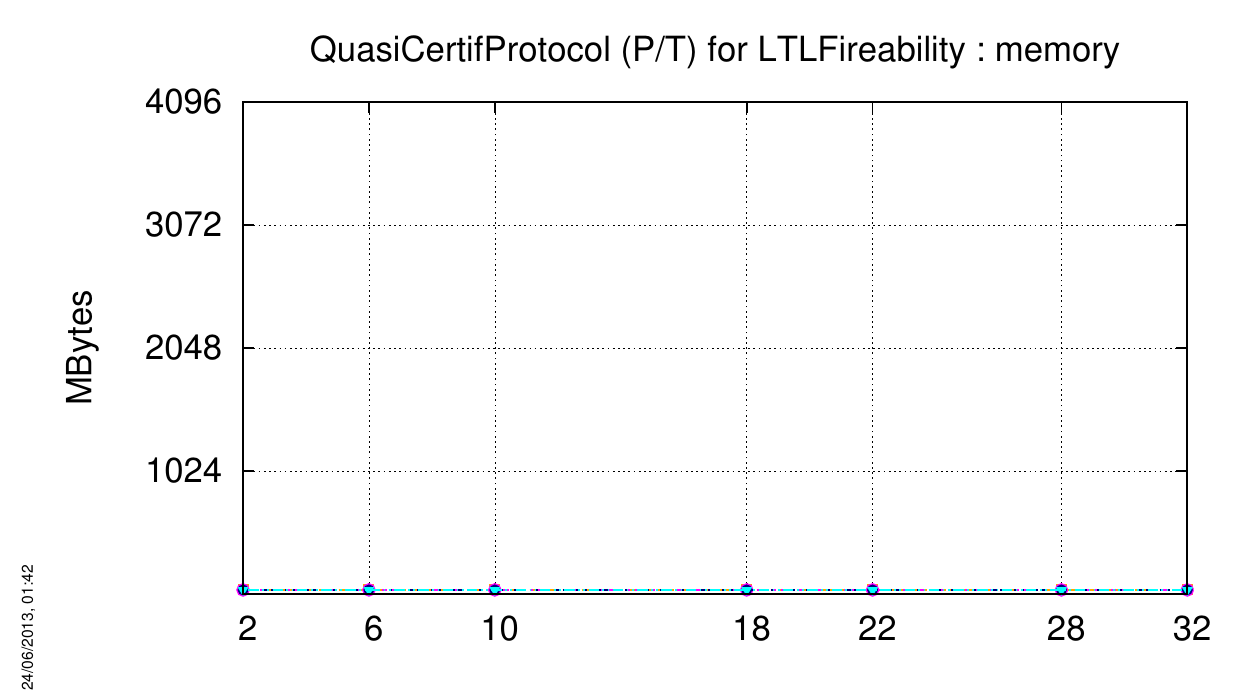}
   \includegraphics[width=7.2cm]{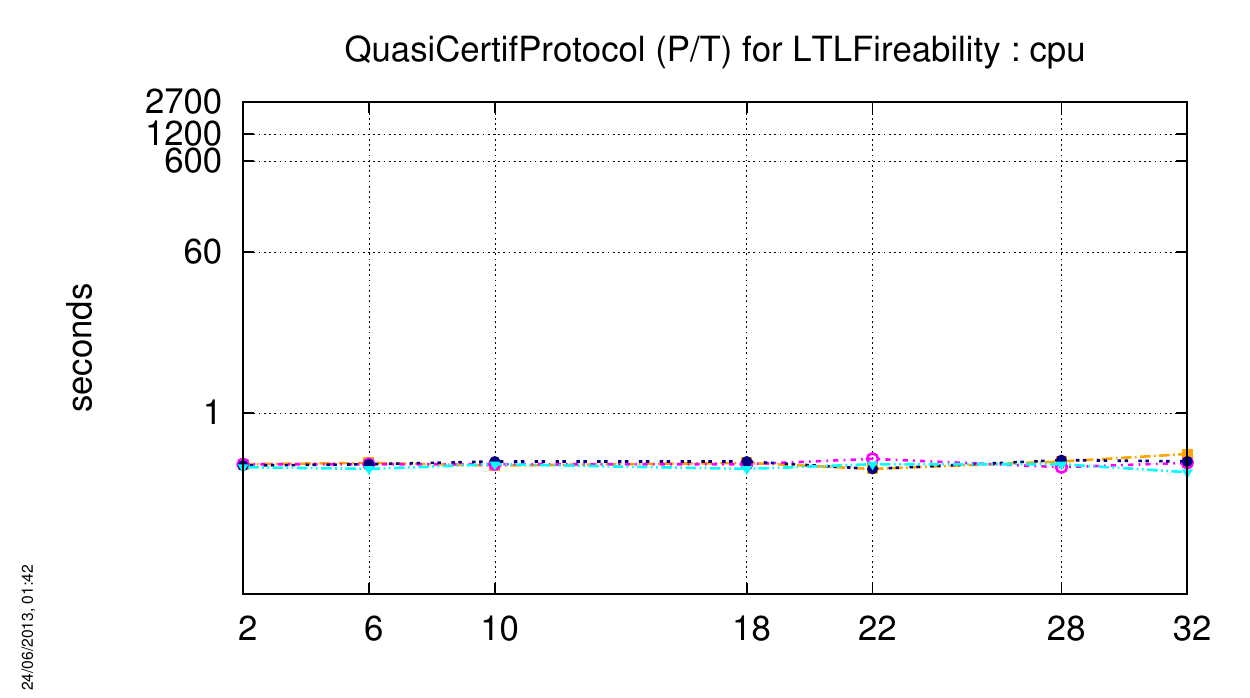}

   \includegraphics[height=1cm]{figures/tools-legend.pdf}
\end{center}

\subsubsection{\acs{Vasy2003-PT}}
The charts below respectively show how tools compete with this ``Suprise'' model (memory and CPU).

\index{Performances!LTLFireability!Vasy2003 (P/T)}
\begin{center}
   \includegraphics[width=7.2cm]{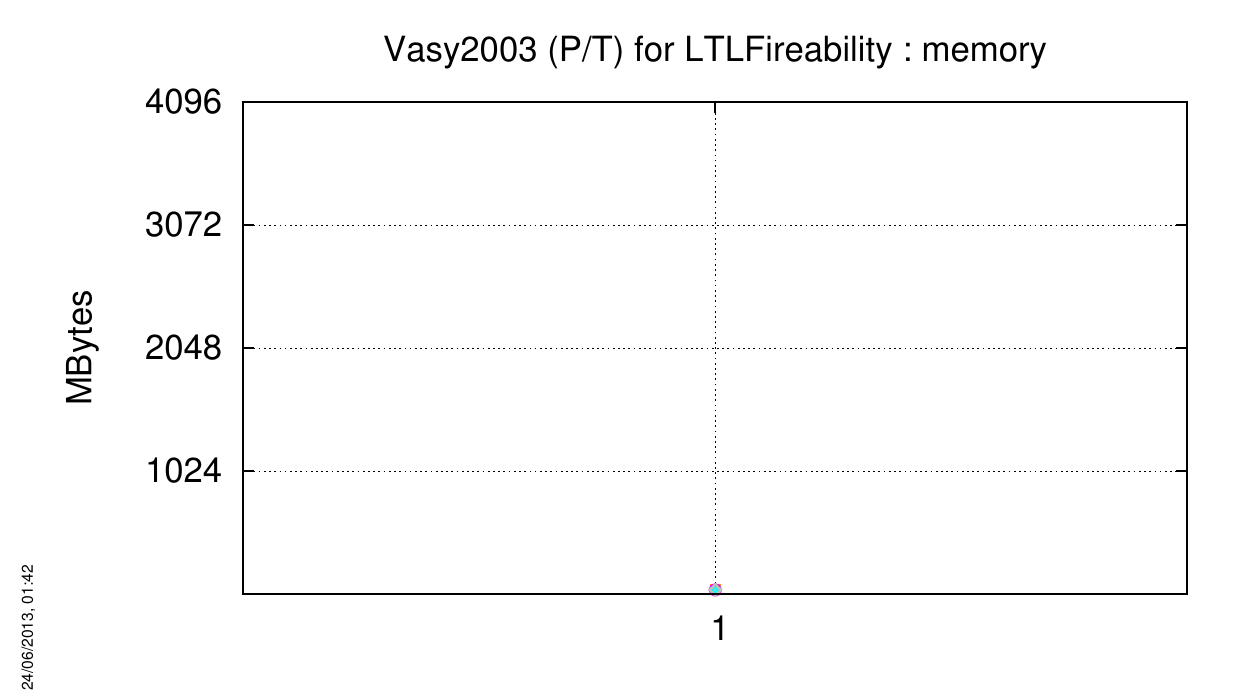}
   \includegraphics[width=7.2cm]{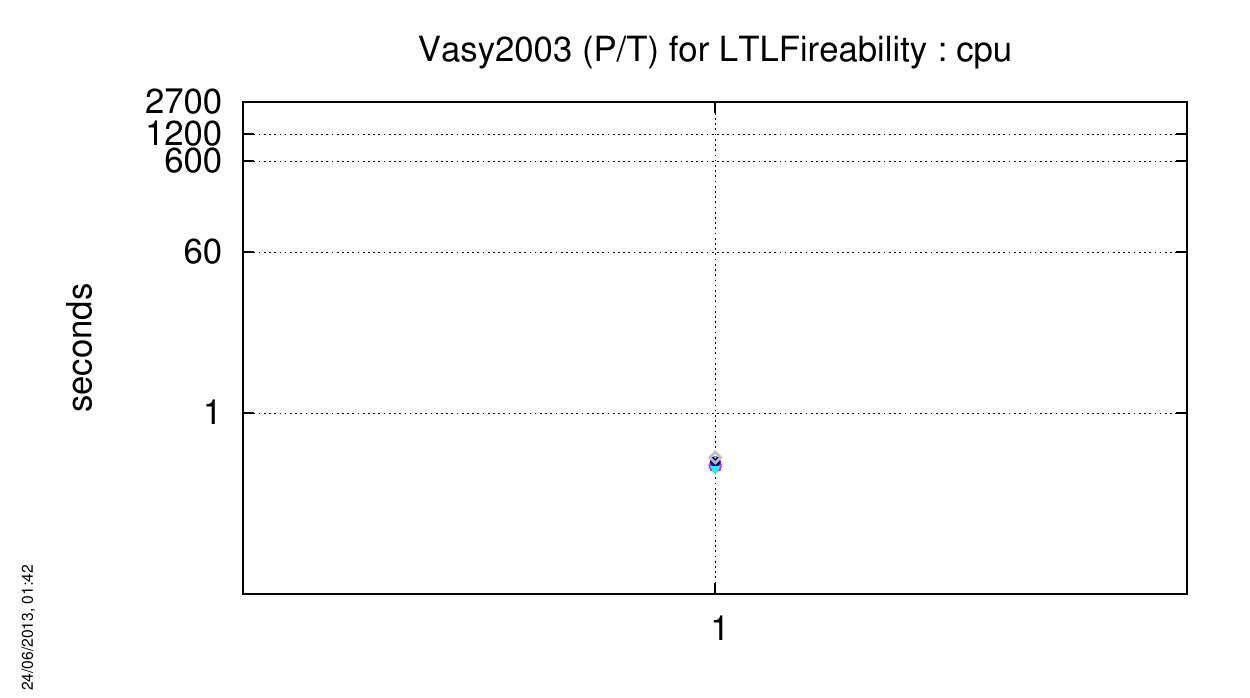}

   \includegraphics[height=1cm]{figures/tools-legend.pdf}
\end{center}

\subsection{Outputs for the LTLFireability Examination}
\index{Outputs!LTLFireability}

Please find enclosed the brute results for this examination (``Known'' and ``Surprise'' models).
We display only the score of tools that provide a results for at least one instance of one model.
The legend for the values is provided below:
\begin{itemize}
   \item\textbf{nc}: the tool does not compete this examination for this model/instance,
   \item\textbf{cc}: the tool cannot compute this examination for this model/instance,
   \item\textbf{to}: the tool cannot compute this examination for this model/instance within the maximum allowed time,
   \item\textbf{mp}: the tool encountered a memory problem (stack overflow or memory full),
   \item\textbf{nf}: there is no formula available for this type of examination (typically, this concerns P/T nets where
       comparing marking cardinality has no signification when there is no equivalent colored net).
\end{itemize}

\textbf{Note on the display of results for formulas:} each formula is considered as a flag (F if false, T if true, - or ?
when the value cannot be determined). These values are concatenated in the order they appear (we assume it is the order of formulas as they were provided).

\subsubsection{``Known'' Models}

\input{result_known_LTLFireability.tex}

\subsubsection{``Surprise'' Models}

\input{result_surprise_LTLFireability.tex}

\subsection{Score for the LTLFireability Examination}
\index{Scores!LTLFireability}

Please find enclosed the scores for this examination (``Known'' and ``Surprise'' models).
We display only the score of tools that provide a results for at least one instance of one model.
The total is first listed in the table below followed by a detail, for each proposed model.
Meaning of the line labels are:
\begin{itemize}
\item\textbf{1st instance}: the tool gets a bonus for having processed the first instance of this model (+1 point),
\item\textbf{instances}: the tool gets 1 point per instances treated 
(for that, we assume that at least one formula has been successfully computed),
\item\textbf{max reached}: the tool could process all the instances for the model (+2 points),
\item\textbf{best}: the tool is among the ones that processed a maximum of instances within the time and memory confinement (+2 points).
\end{itemize}

\subsubsection{``Known'' Models}

\input{score_known_LTLFireability.tex}

\subsubsection{``Surprise'' Models}

\input{score_surprise_LTLFireability.tex}

\subsection{Trophies for this Examination}
\index{Trophies!LTLFireability}

Trophies are divided in three categories: ``Known'' models,
``Surprise'' models, and the global trophies (formula is then
$score_{global} = score_{known} + 2 \times score_{surprise}$).

\subsubsection{For ``Known'' Models} \ \\

\begin{tabular}{c}
      1 \\
   \includegraphics[width=2cm]{figures/gold.jpg} \\
   \acs{neco} \\
   88 points \\
\end{tabular}

\subsubsection{For ``Surprise'' Models}\  \\

No tool could complete this examination.

\subsubsection{Global} \ \\

\begin{tabular}{c}
      1 \\
   \includegraphics[width=2cm]{figures/gold.jpg} \\
   \acs{neco} \\
   88 points \\
\end{tabular}

\newpage

\section{The LTLMarkingComparison Examination}
\label{sec:exam:LTLMarkingComparison}
\index{Results!LTLMarkingComparison}

This examination deals with LTL properties dealing with marking comparison only.
We first show a summary on the handling of models by the participating tools.
Then, we present the computed outputs and the associated scores for this
examination prior to a summary of relevant executions.

\subsection{Handling of Models by Tools}
\index{Performances!LTLMarkingComparison}

\subsubsection{\acs{CSRepetitions-COL}}
No instance of this model could be computed for the \textbf{LTLMarkingComparison} examination.

\subsubsection{\acs{CSRepetitions-PT}}
The charts below respectively show how tools compete with this ``Known'' model (memory and CPU).

\index{Performances!LTLMarkingComparison!CSRepetitions (P/T)}
\begin{center}
   \includegraphics[width=7.2cm]{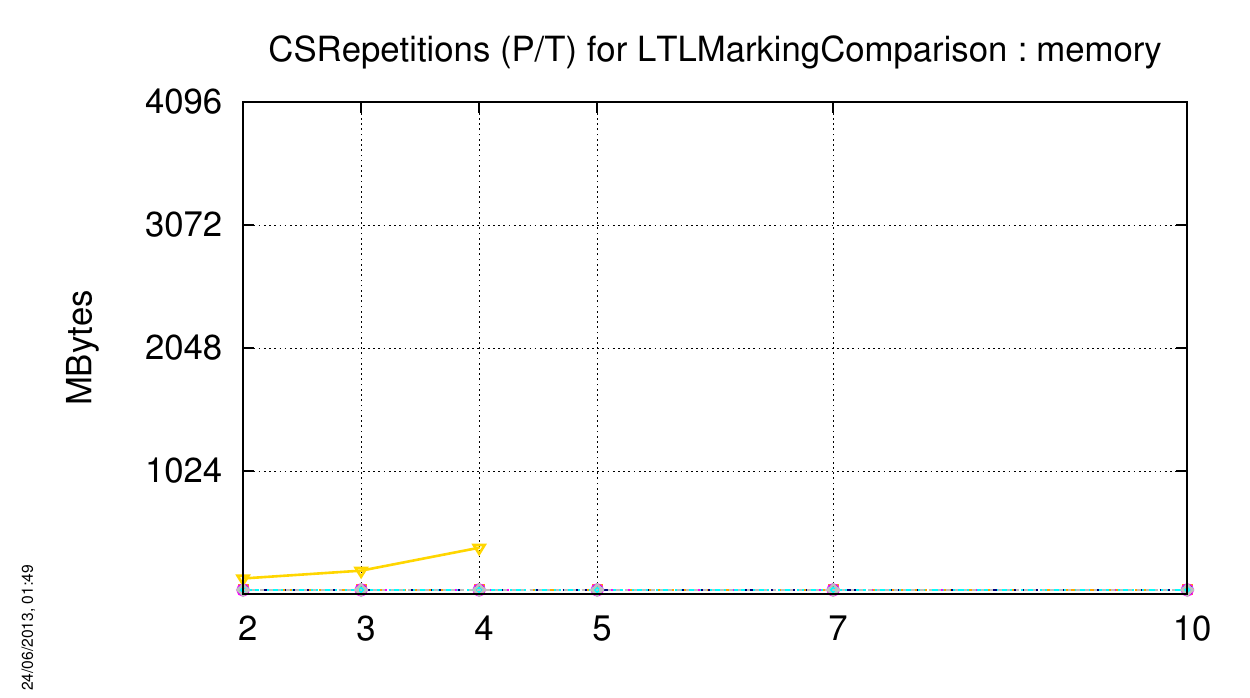}
   \includegraphics[width=7.2cm]{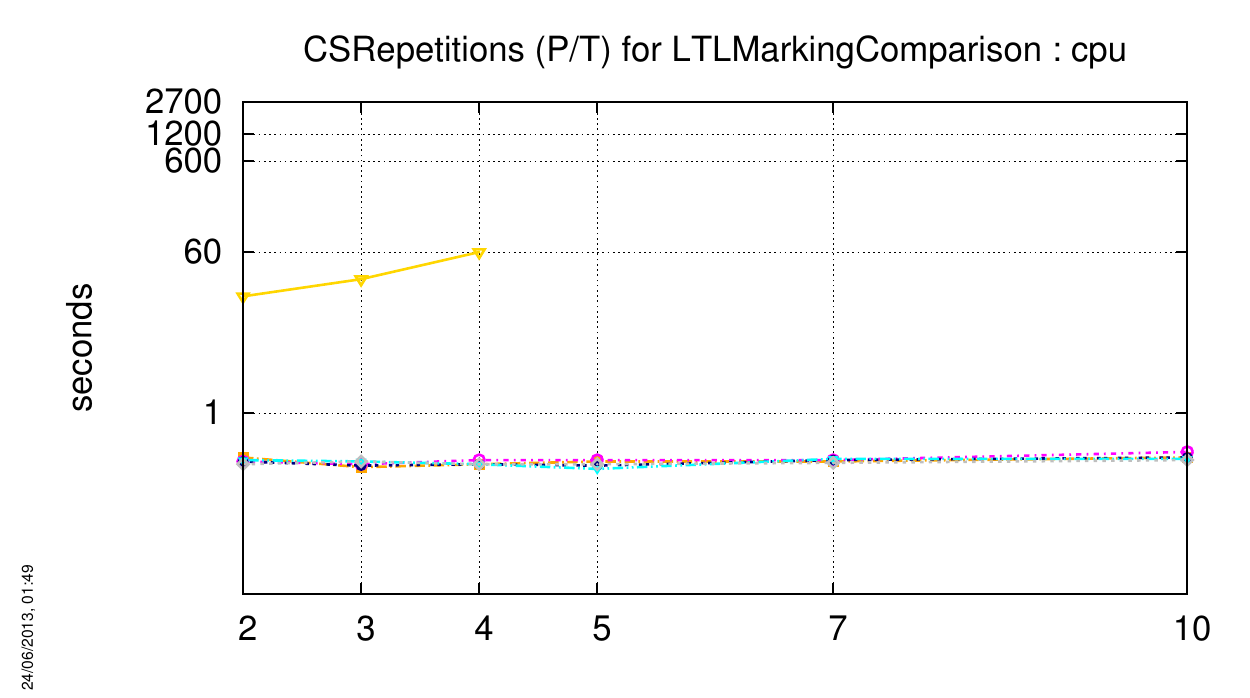}

   \includegraphics[height=1cm]{figures/tools-legend.pdf}
\end{center}

\subsubsection{\acs{Dekker-PT}}
No instance of this model could be computed for the \textbf{LTLMarkingComparison} examination.

\subsubsection{\acs{DotAndBoxes-COL}}
No instance of this model could be computed for the \textbf{LTLMarkingComparison} examination.

\subsubsection{\acs{DrinkVendingMachine-COL}}
No instance of this model could be computed for the \textbf{LTLMarkingComparison} examination.

\subsubsection{\acs{DrinkVendingMachine-PT}}
The charts below respectively show how tools compete with this ``Known'' model (memory and CPU).

\index{Performances!LTLMarkingComparison!DrinkVendingMachine (P/T)}
\begin{center}
   \includegraphics[width=7.2cm]{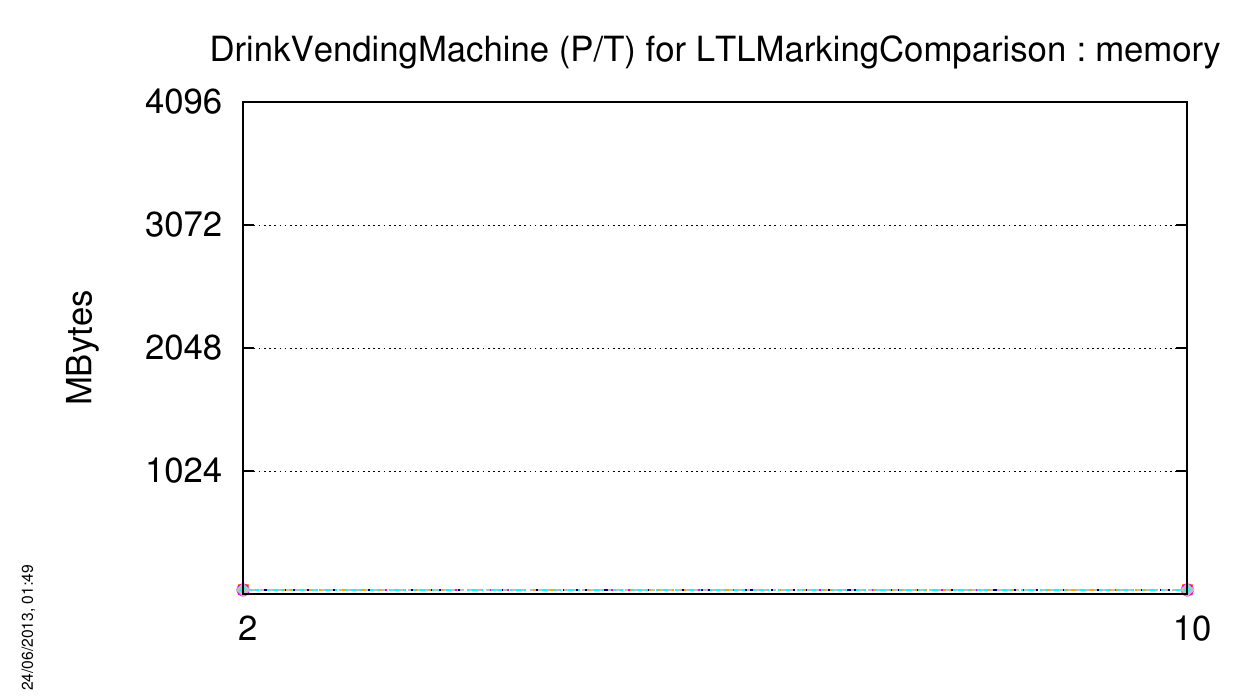}
   \includegraphics[width=7.2cm]{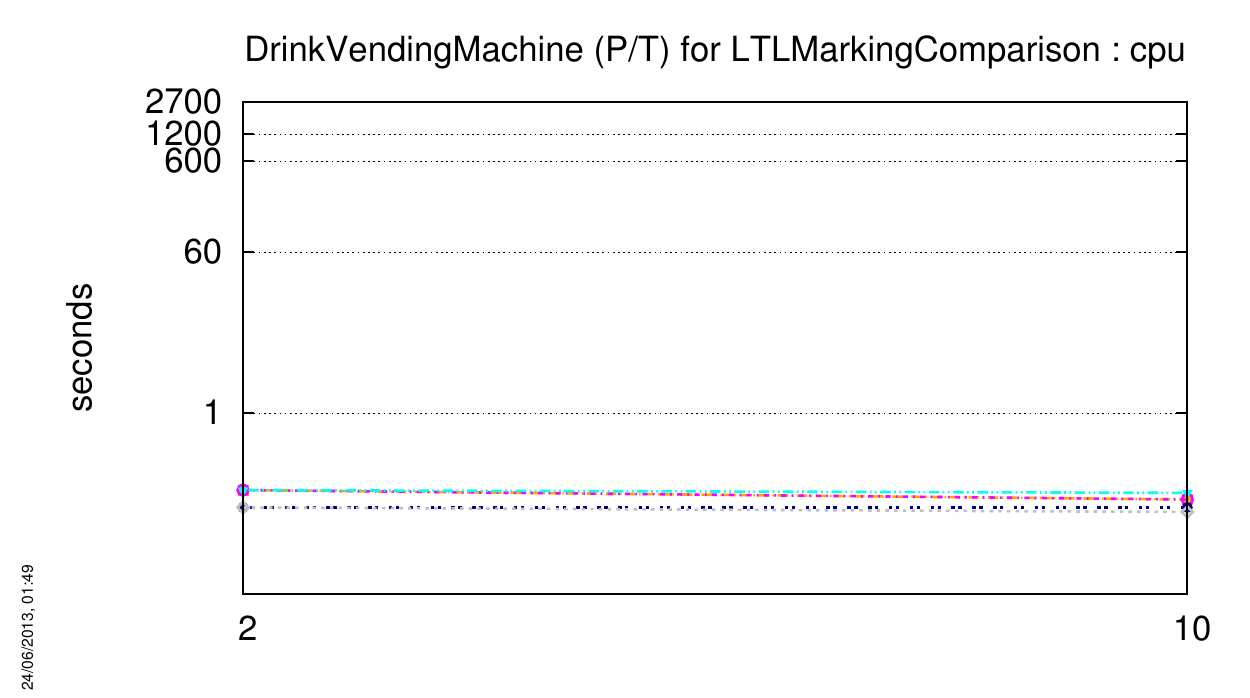}

   \includegraphics[height=1cm]{figures/tools-legend.pdf}
\end{center}

\subsubsection{\acs{Echo-PT}}
No instance of this model could be computed for the \textbf{LTLMarkingComparison} examination.

\subsubsection{\acs{Eratosthenes-PT}}
No instance of this model could be computed for the \textbf{LTLMarkingComparison} examination.

\subsubsection{\acs{FMS-PT}}
No instance of this model could be computed for the \textbf{LTLMarkingComparison} examination.

\subsubsection{\acs{GlobalRessAlloc-COL}}
No instance of this model could be computed for the \textbf{LTLMarkingComparison} examination.

\subsubsection{\acs{GlobalRessAlloc-PT}}
The charts below respectively show how tools compete with this ``Known'' model (memory and CPU).

\index{Performances!LTLMarkingComparison!GlobalRessAlloc (P/T)}
\begin{center}
   \includegraphics[width=7.2cm]{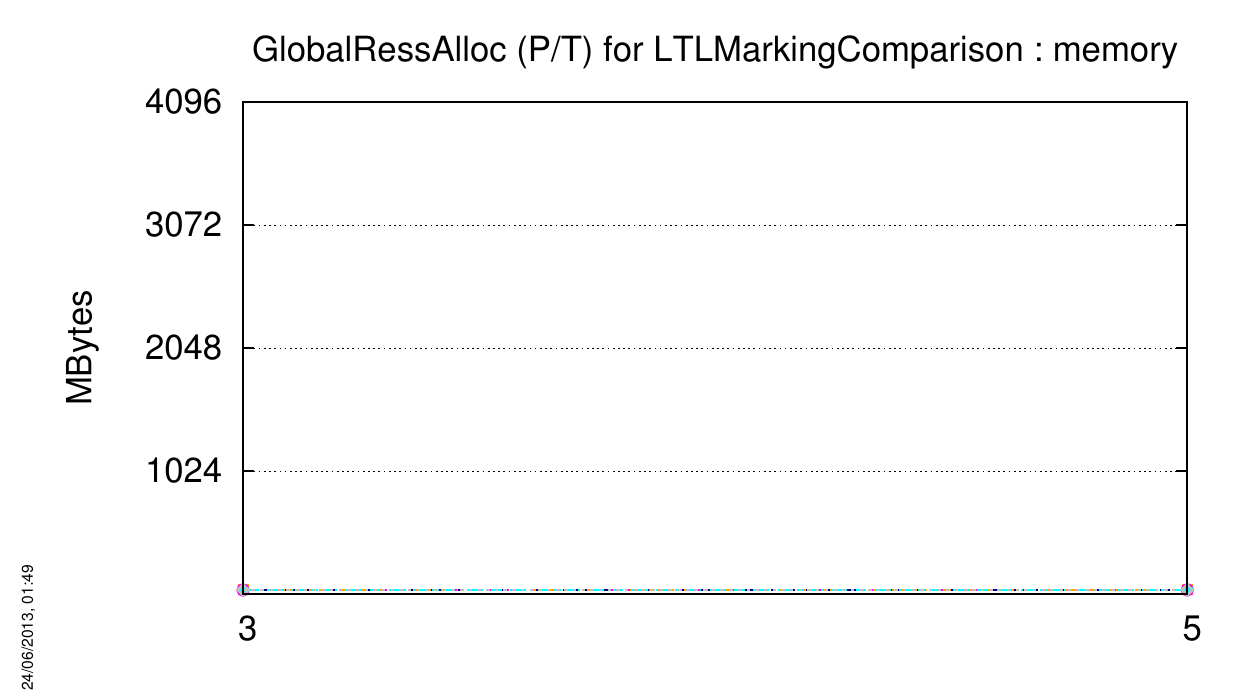}
   \includegraphics[width=7.2cm]{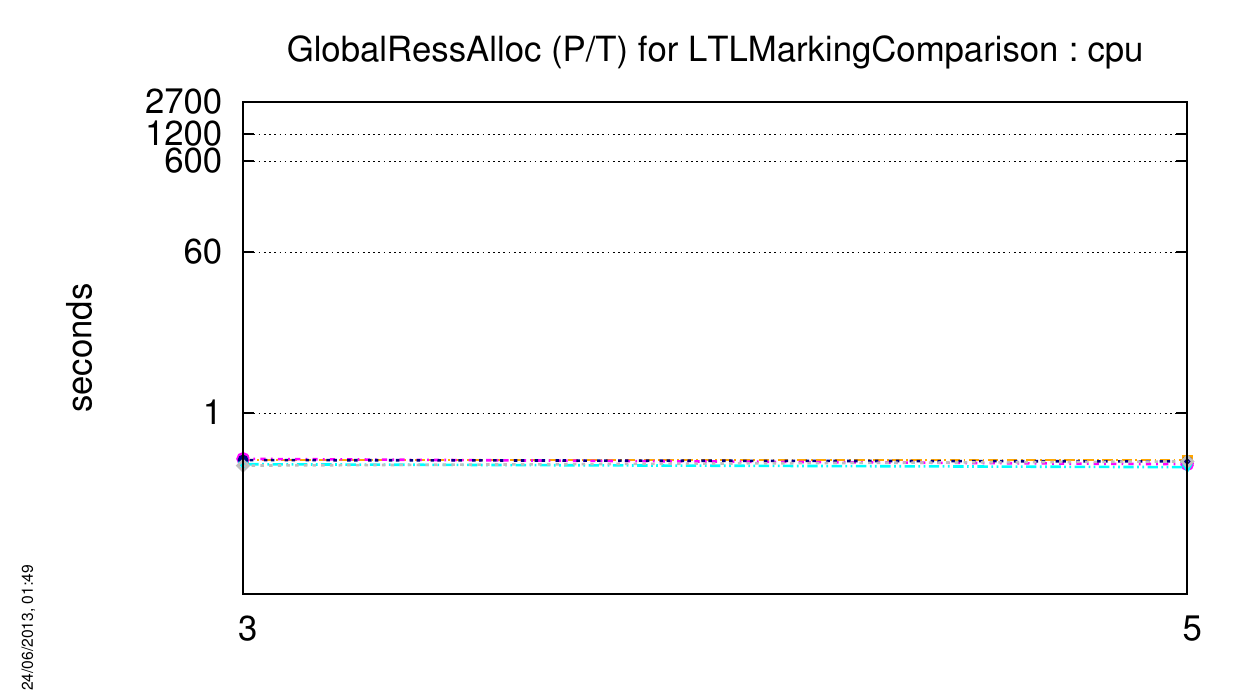}

   \includegraphics[height=1cm]{figures/tools-legend.pdf}
\end{center}

\subsubsection{\acs{Kanban-PT}}
No instance of this model could be computed for the \textbf{LTLMarkingComparison} examination.

\subsubsection{\acs{LamportFastMutEx-COL}}
No instance of this model could be computed for the \textbf{LTLMarkingComparison} examination.

\subsubsection{\acs{LamportFastMutEx-PT}}
The charts below respectively show how tools compete with this ``Known'' model (memory and CPU).

\index{Performances!LTLMarkingComparison!LamportFastMutEx (P/T)}
\begin{center}
   \includegraphics[width=7.2cm]{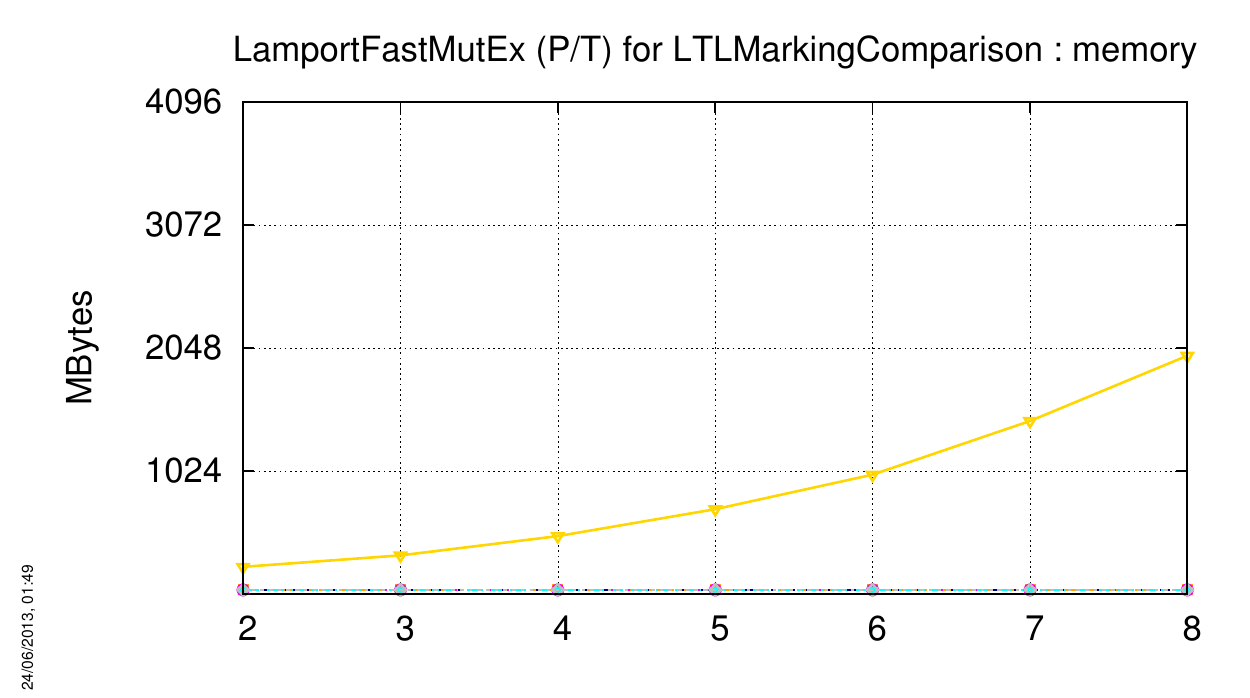}
   \includegraphics[width=7.2cm]{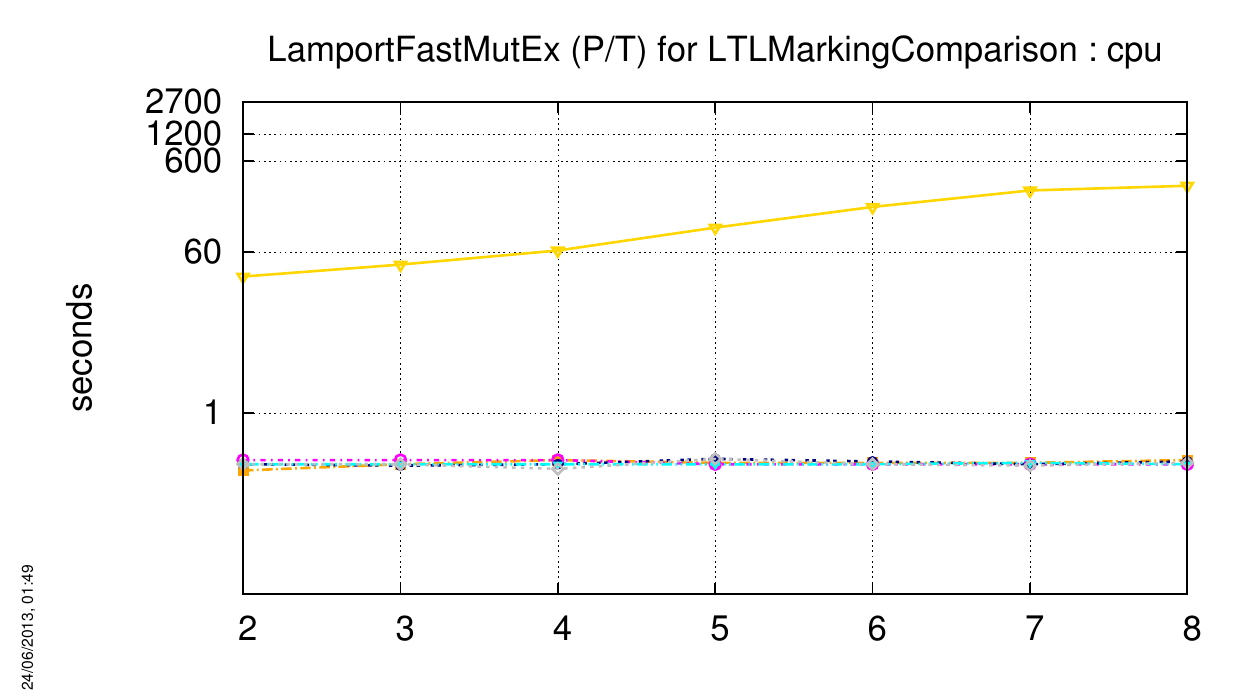}

   \includegraphics[height=1cm]{figures/tools-legend.pdf}
\end{center}

\subsubsection{\acs{MAPK-PT}}
No instance of this model could be computed for the \textbf{LTLMarkingComparison} examination.

\subsubsection{\acs{NeoElection-COL}}
No instance of this model could be computed for the \textbf{LTLMarkingComparison} examination.

\subsubsection{\acs{NeoElection-PT}}
The charts below respectively show how tools compete with this ``Known'' model (memory and CPU).

\index{Performances!LTLMarkingComparison!NeoElection (P/T)}
\begin{center}
   \includegraphics[width=7.2cm]{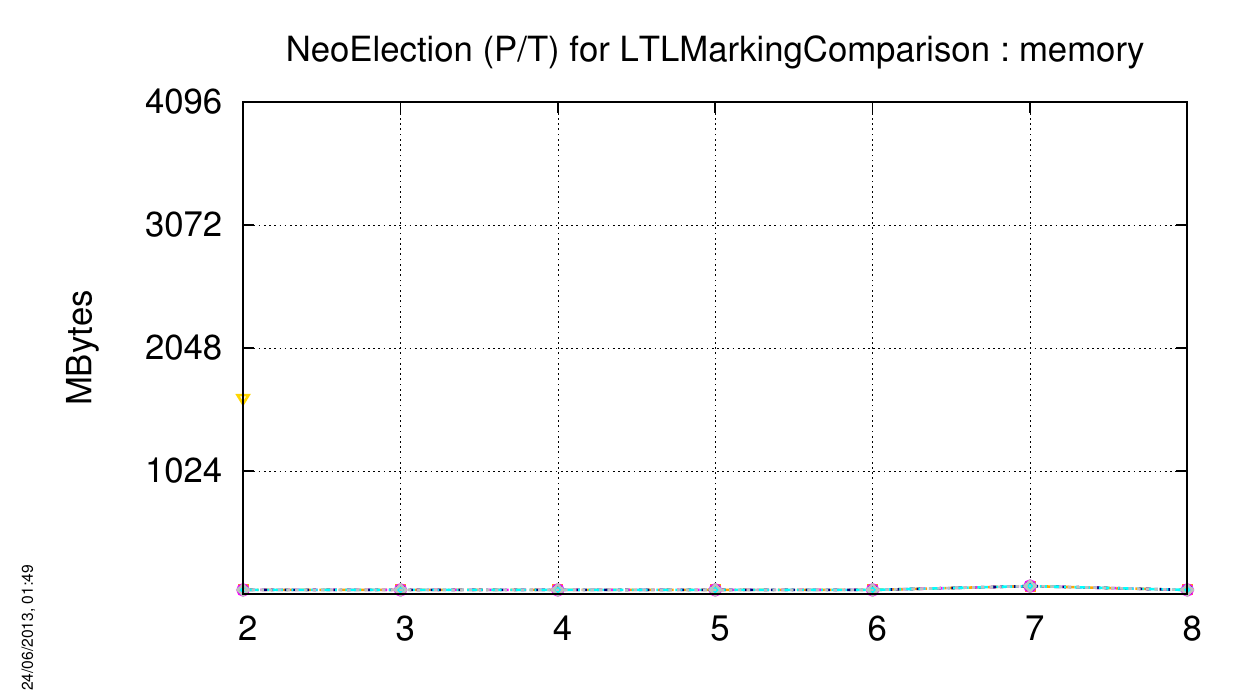}
   \includegraphics[width=7.2cm]{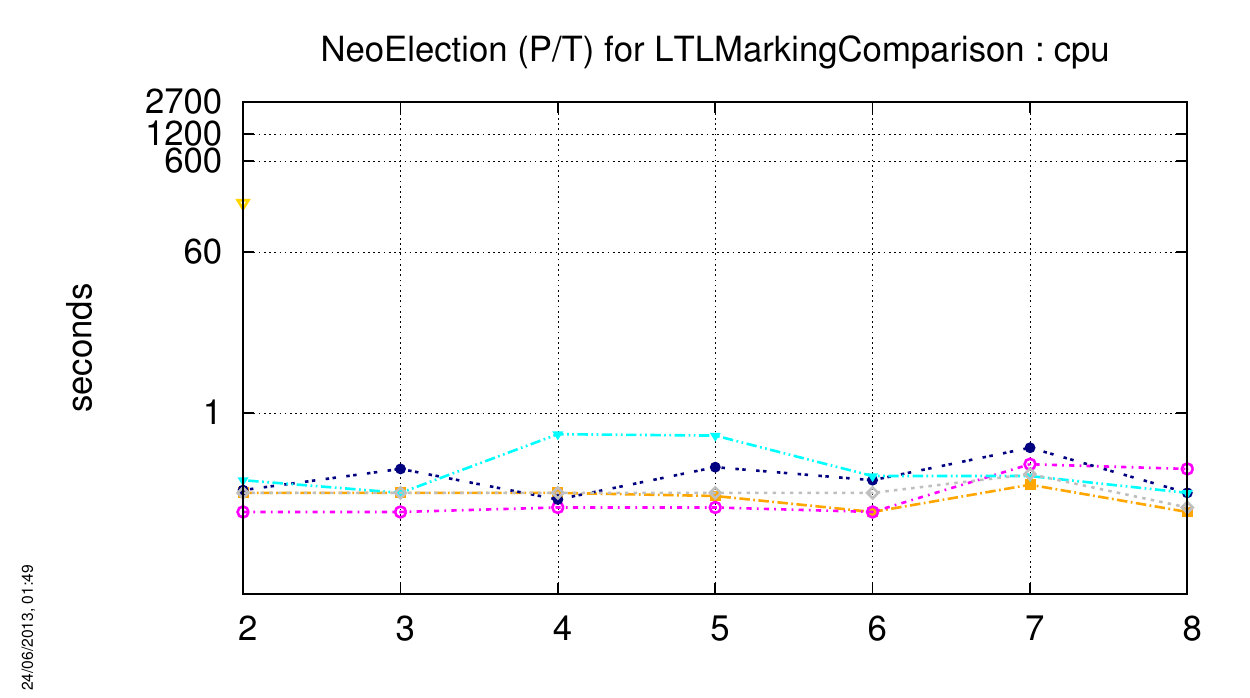}

   \includegraphics[height=1cm]{figures/tools-legend.pdf}
\end{center}

\subsubsection{\acs{PermAdmissibility-COL}}
No instance of this model could be computed for the \textbf{LTLMarkingComparison} examination.

\subsubsection{\acs{PermAdmissibility-PT}}
The charts below respectively show how tools compete with this ``Known'' model (memory and CPU).

\index{Performances!LTLMarkingComparison!PermAdmissibility (P/T)}
\begin{center}
   \includegraphics[width=7.2cm]{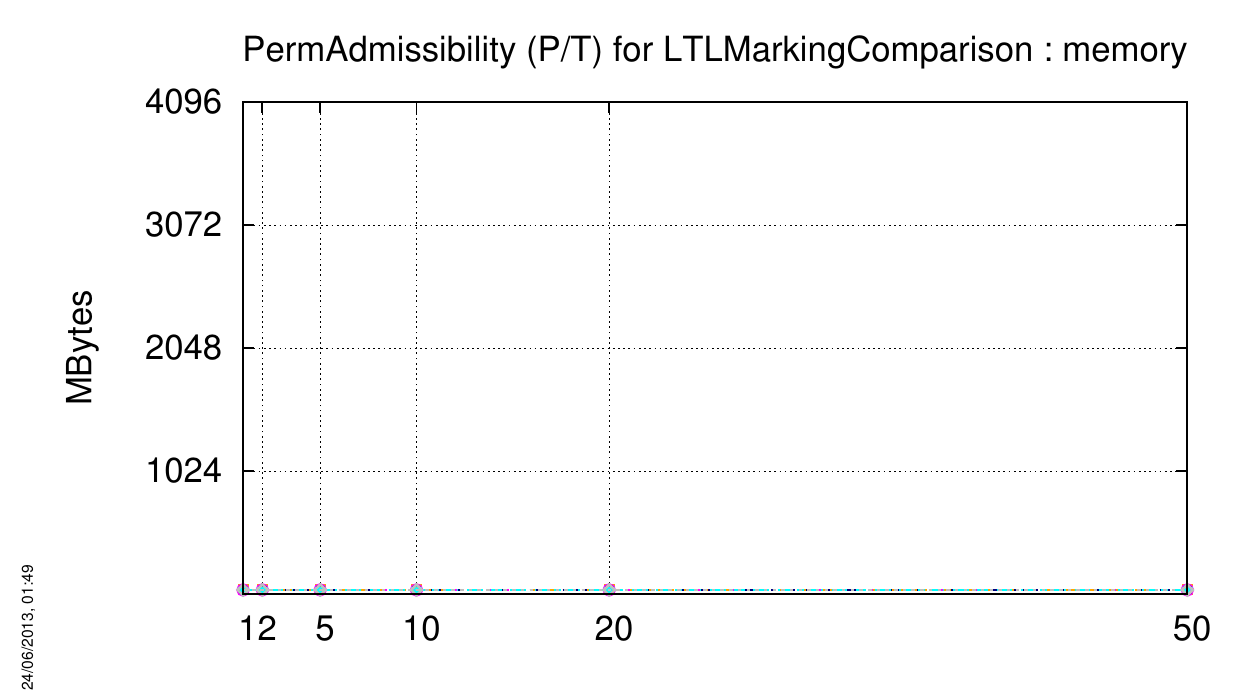}
   \includegraphics[width=7.2cm]{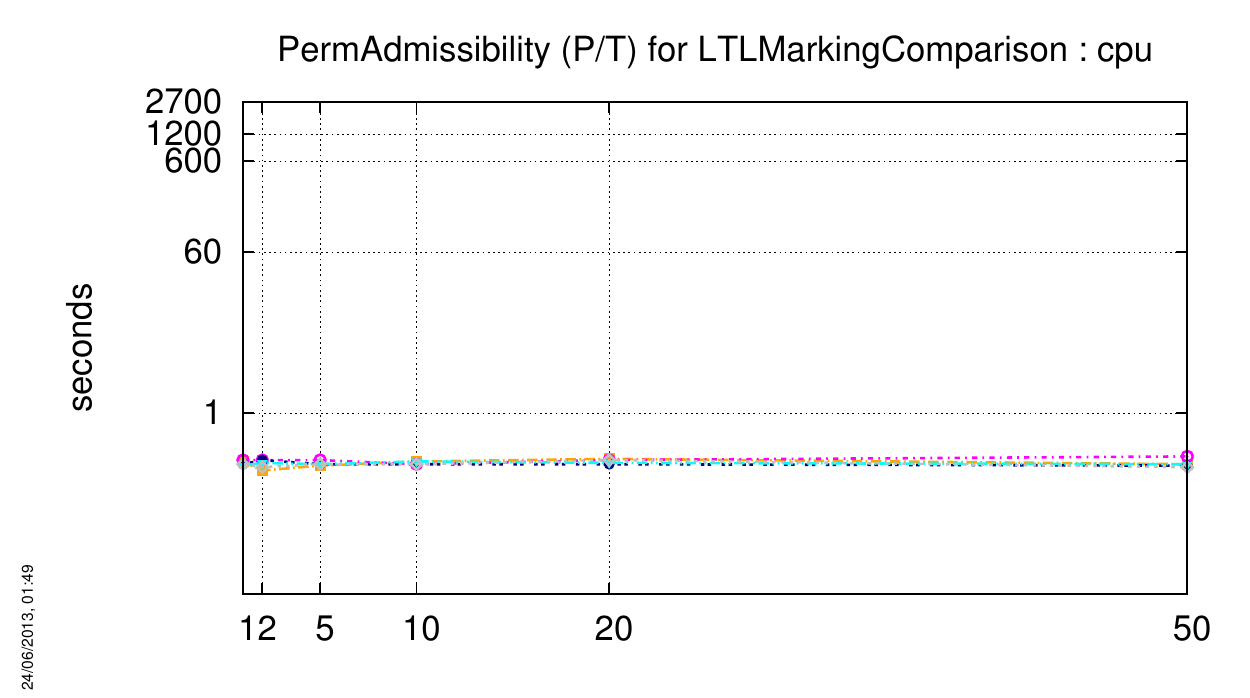}

   \includegraphics[height=1cm]{figures/tools-legend.pdf}
\end{center}

\subsubsection{\acs{Peterson-COL}}
No instance of this model could be computed for the \textbf{LTLMarkingComparison} examination.

\subsubsection{\acs{Peterson-PT}}
The charts below respectively show how tools compete with this ``Known'' model (memory and CPU).

\index{Performances!LTLMarkingComparison!Peterson (P/T)}
\begin{center}
   \includegraphics[width=7.2cm]{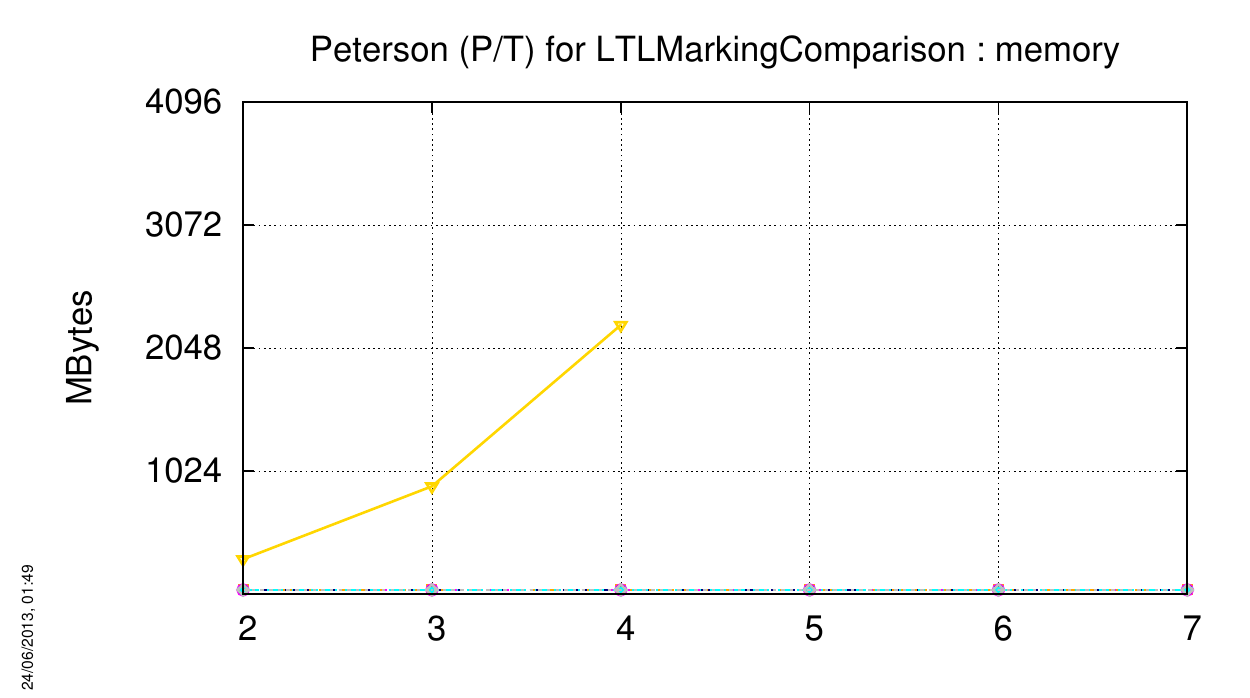}
   \includegraphics[width=7.2cm]{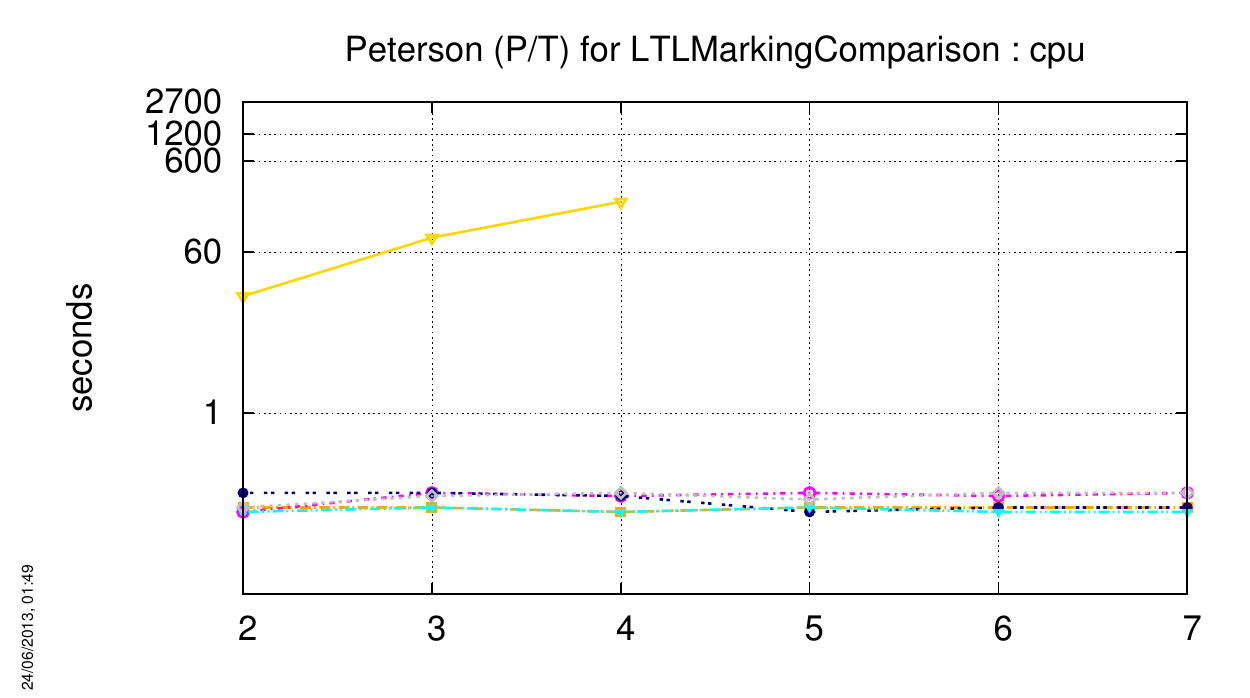}

   \includegraphics[height=1cm]{figures/tools-legend.pdf}
\end{center}

\subsubsection{\acs{Philosophers-COL}}
No instance of this model could be computed for the \textbf{LTLMarkingComparison} examination.

\subsubsection{\acs{Philosophers-PT}}
The charts below respectively show how tools compete with this ``Known'' model (memory and CPU).

\index{Performances!LTLMarkingComparison!Philosophers (P/T)}
\begin{center}
   \includegraphics[width=7.2cm]{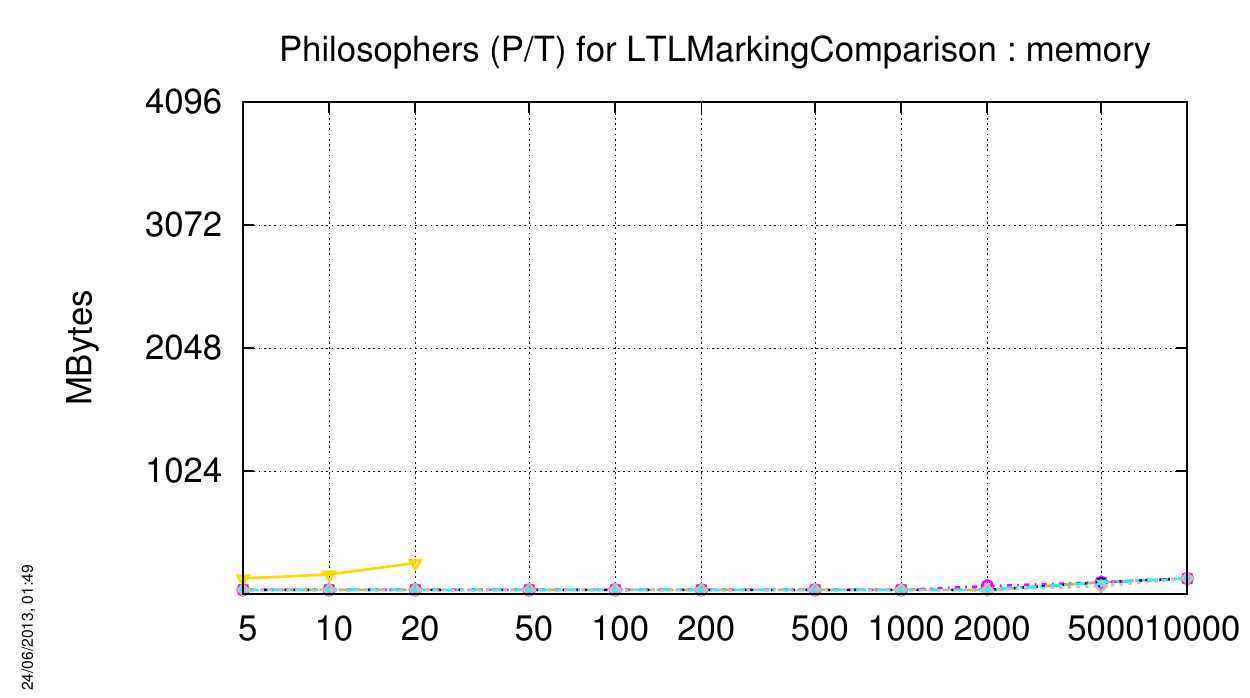}
   \includegraphics[width=7.2cm]{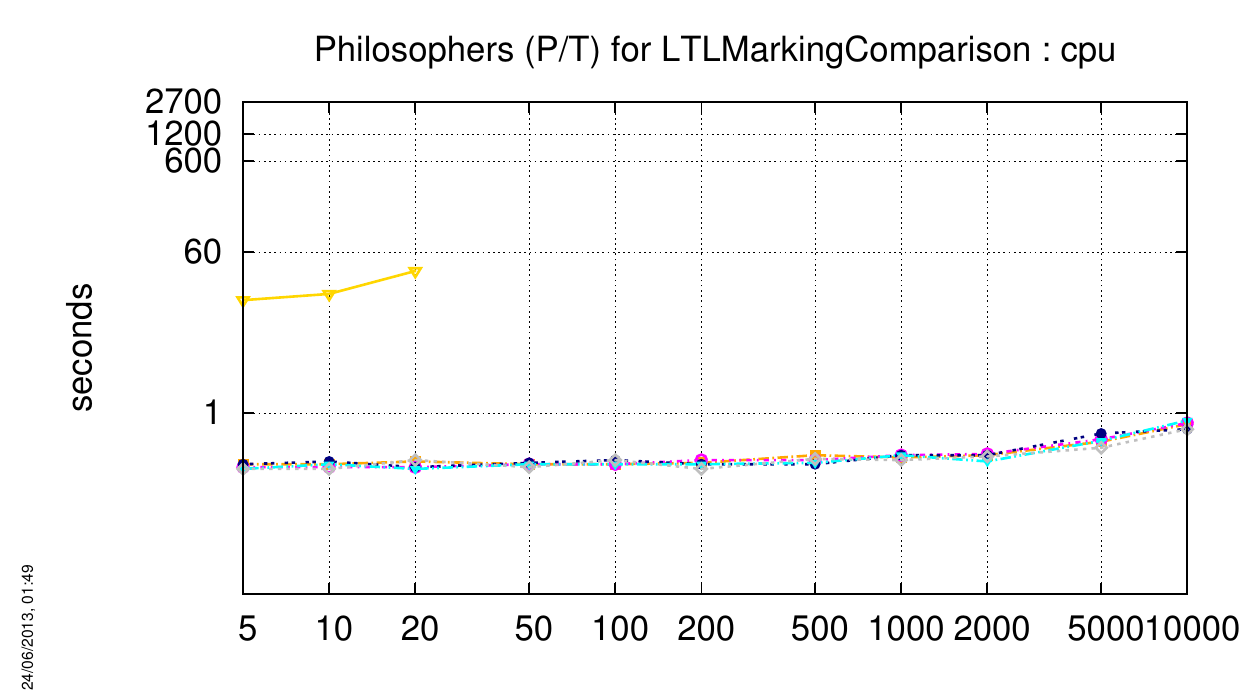}

   \includegraphics[height=1cm]{figures/tools-legend.pdf}
\end{center}

\subsubsection{\acs{PhilosophersDyn-COL}}
No instance of this model could be computed for the \textbf{LTLMarkingComparison} examination.

\subsubsection{\acs{PhilosophersDyn-PT}}
The charts below respectively show how tools compete with this ``Known'' model (memory and CPU).

\index{Performances!LTLMarkingComparison!PhilosophersDyn (P/T)}
\begin{center}
   \includegraphics[width=7.2cm]{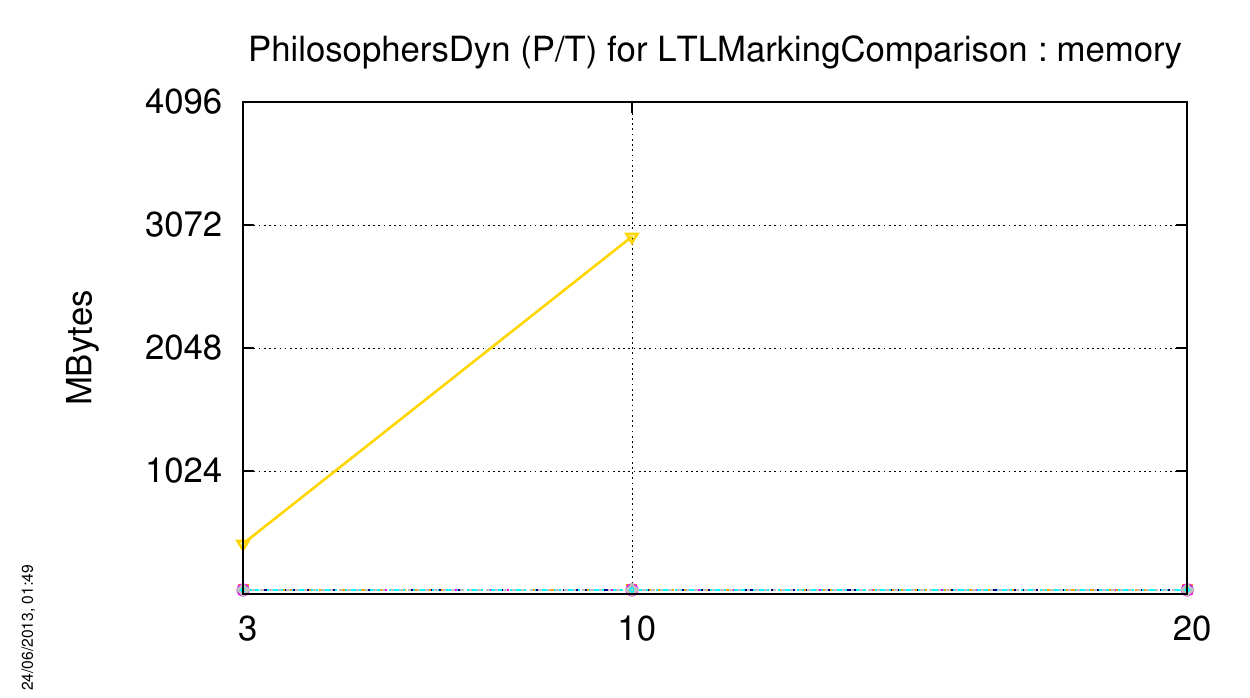}
   \includegraphics[width=7.2cm]{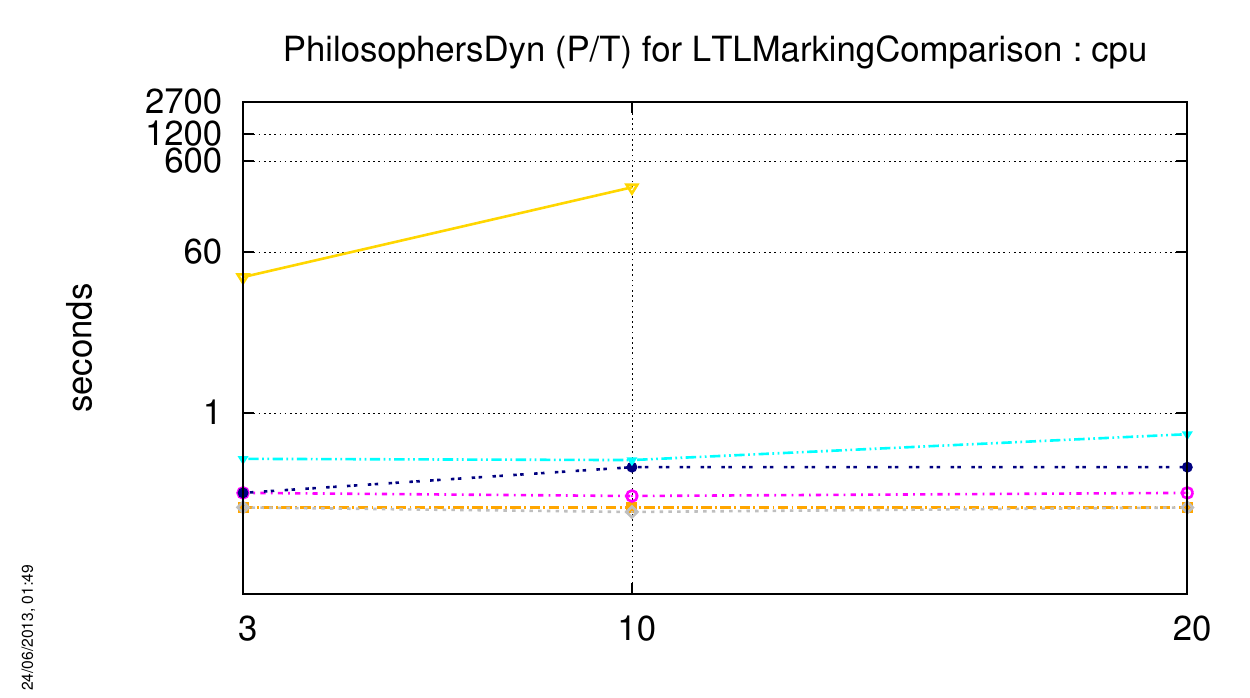}

   \includegraphics[height=1cm]{figures/tools-legend.pdf}
\end{center}

\subsubsection{\acs{Planning-PT}}
No instance of this model could be computed for the \textbf{LTLMarkingComparison} examination.

\subsubsection{\acs{Railroad-PT}}
No instance of this model could be computed for the \textbf{LTLMarkingComparison} examination.

\subsubsection{\acs{RessAllocation-PT}}
No instance of this model could be computed for the \textbf{LTLMarkingComparison} examination.

\subsubsection{\acs{Ring-PT}}
No instance of this model could be computed for the \textbf{LTLMarkingComparison} examination.

\subsubsection{\acs{RwMutex-PT}}
No instance of this model could be computed for the \textbf{LTLMarkingComparison} examination.

\subsubsection{\acs{SharedMemory-COL}}
No instance of this model could be computed for the \textbf{LTLMarkingComparison} examination.

\subsubsection{\acs{SharedMemory-PT}}
The charts below respectively show how tools compete with this ``Known'' model (memory and CPU).

\index{Performances!LTLMarkingComparison!SharedMemory (P/T)}
\begin{center}
   \includegraphics[width=7.2cm]{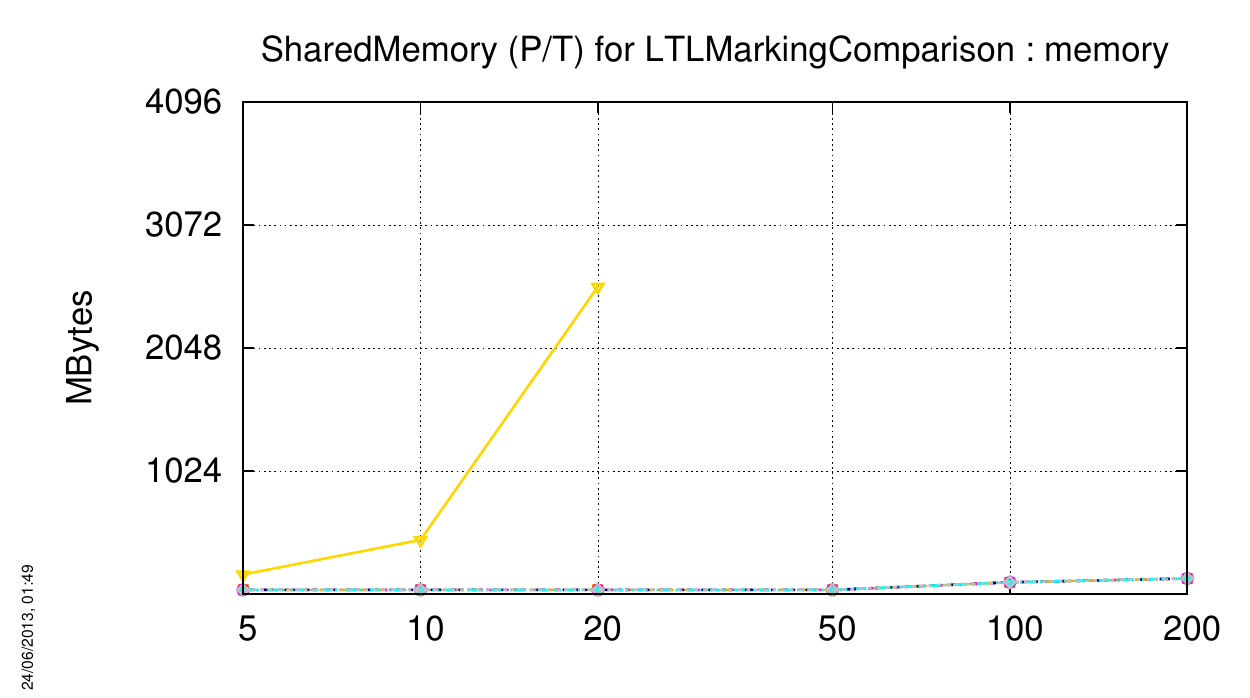}
   \includegraphics[width=7.2cm]{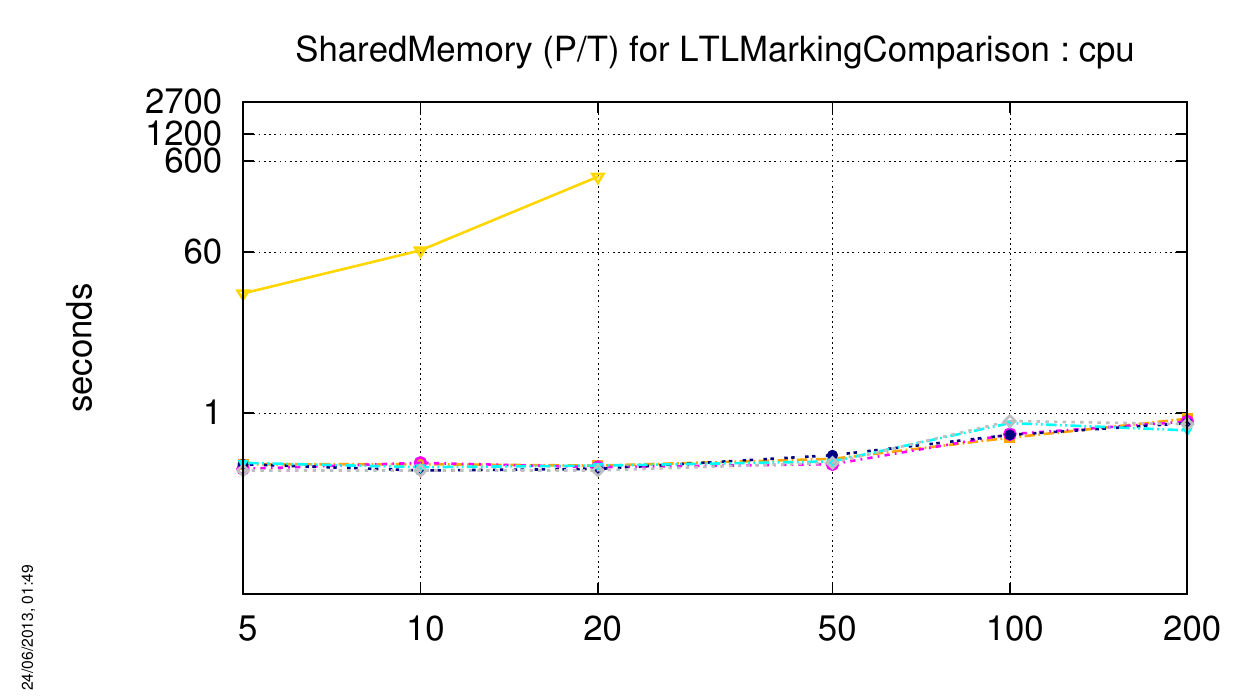}

   \includegraphics[height=1cm]{figures/tools-legend.pdf}
\end{center}

\subsubsection{\acs{SimpleLoadBal-COL}}
No instance of this model could be computed for the \textbf{LTLMarkingComparison} examination.

\subsubsection{\acs{SimpleLoadBal-PT}}
The charts below respectively show how tools compete with this ``Known'' model (memory and CPU).

\index{Performances!LTLMarkingComparison!SimpleLoadBal (P/T)}
\begin{center}
   \includegraphics[width=7.2cm]{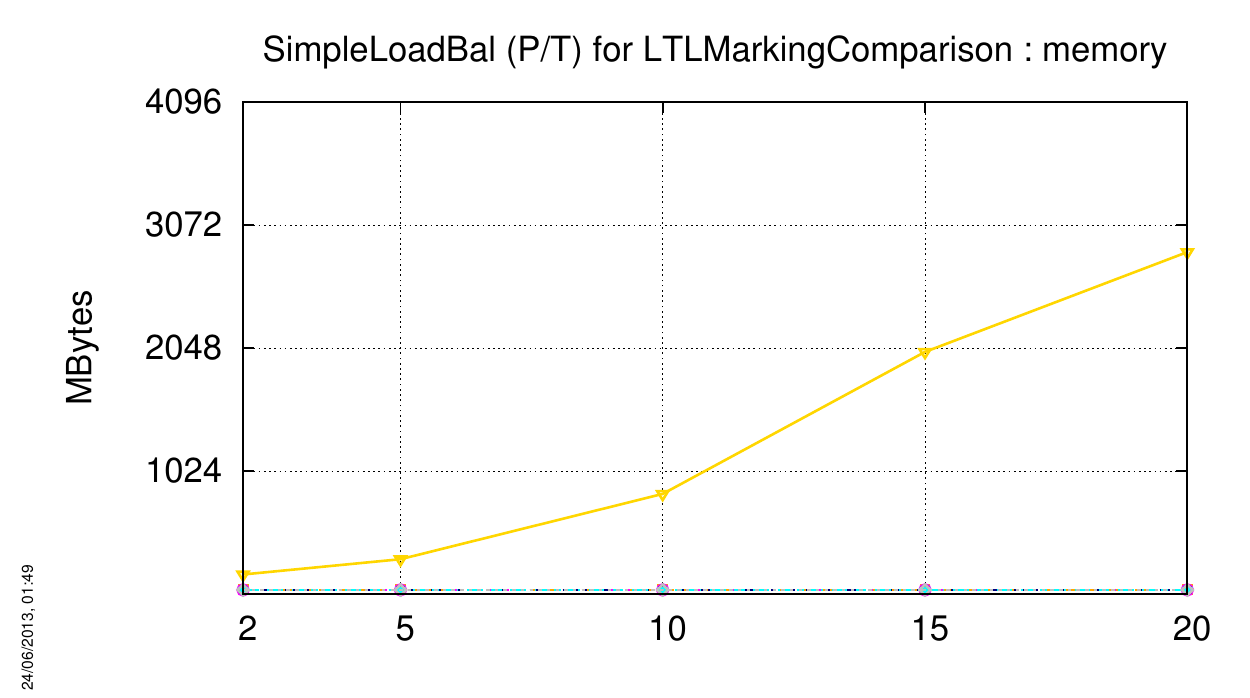}
   \includegraphics[width=7.2cm]{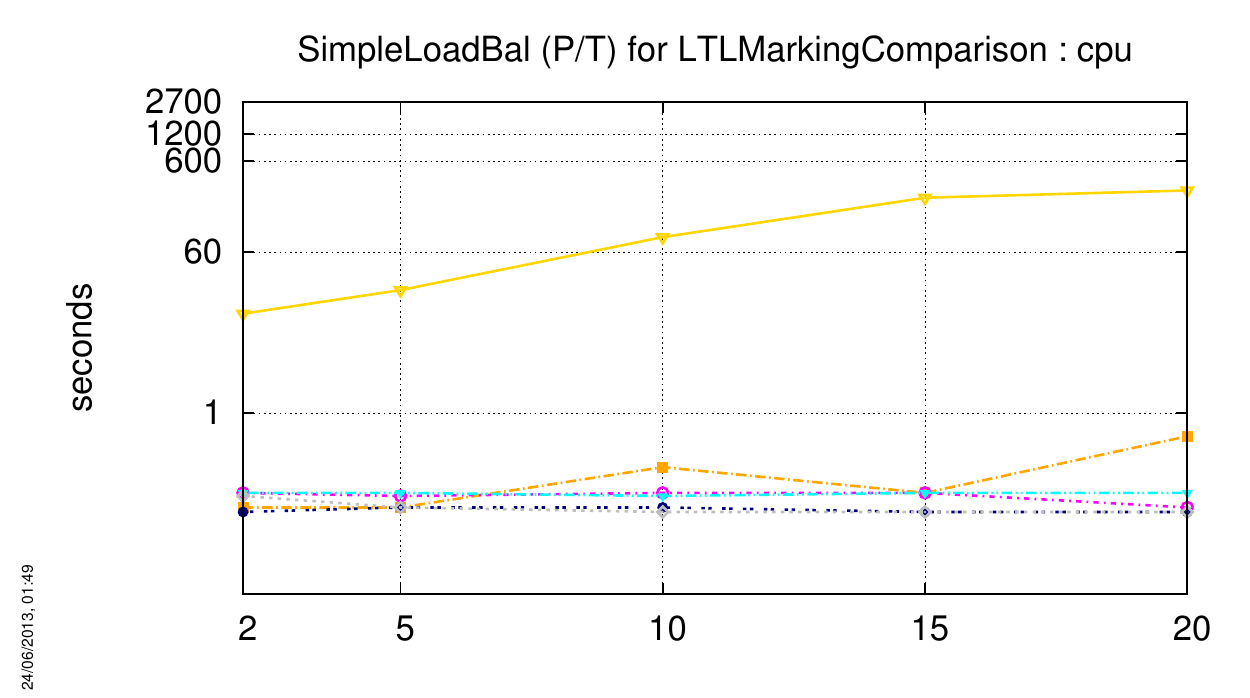}

   \includegraphics[height=1cm]{figures/tools-legend.pdf}
\end{center}

\subsubsection{\acs{TokenRing-COL}}
No instance of this model could be computed for the \textbf{LTLMarkingComparison} examination.

\subsubsection{\acs{TokenRing-PT}}
The charts below respectively show how tools compete with this ``Known'' model (memory and CPU).

\index{Performances!LTLMarkingComparison!TokenRing (P/T)}
\begin{center}
   \includegraphics[width=7.2cm]{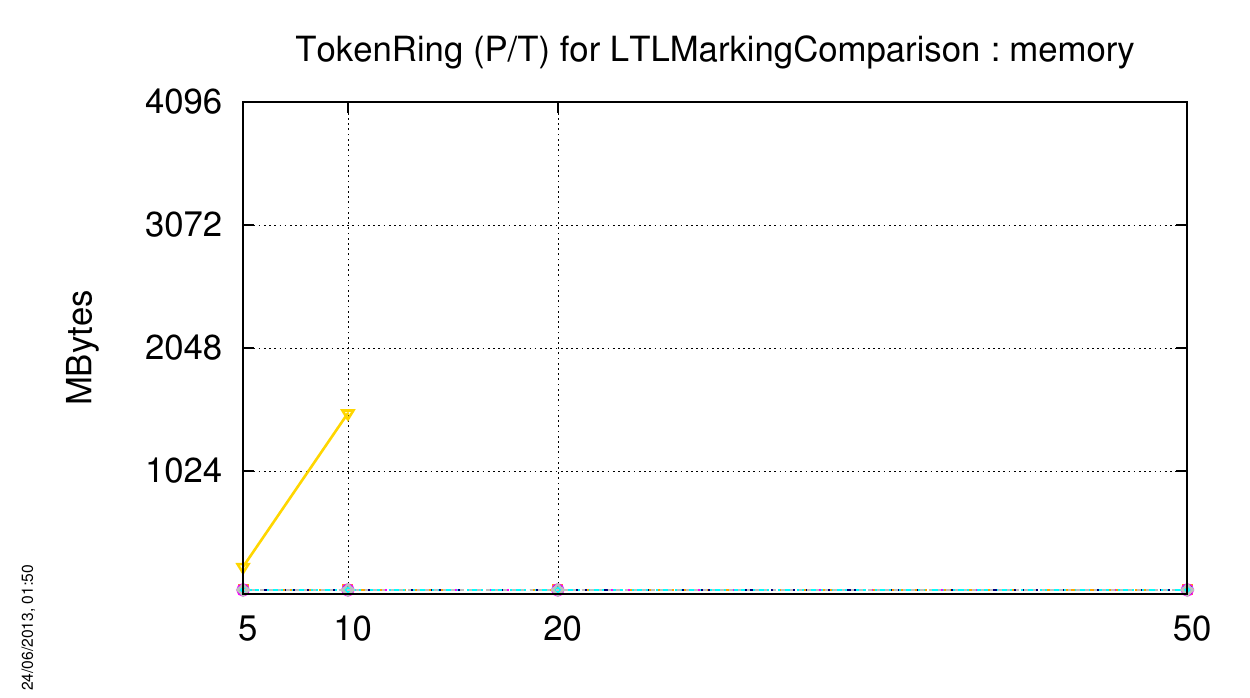}
   \includegraphics[width=7.2cm]{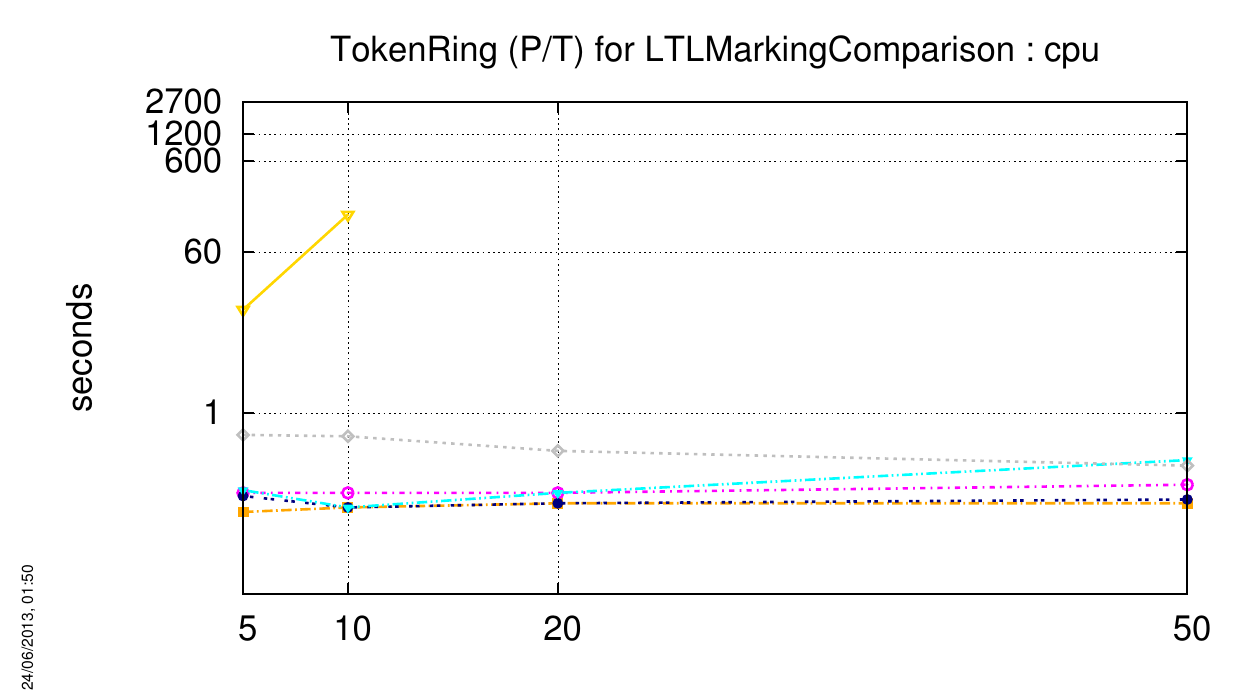}

   \includegraphics[height=1cm]{figures/tools-legend.pdf}
\end{center}

\subsubsection{\acs{HouseConstruction-PT}}
No instance of this model could be computed for the \textbf{LTLMarkingComparison} examination.

\subsubsection{\acs{IBMB2S565S3960-PT}}
No instance of this model could be computed for the \textbf{LTLMarkingComparison} examination.

\subsubsection{\acs{QuasiCertifProtocol-COL}}
No instance of this model could be computed for the \textbf{LTLMarkingComparison} examination.

\subsubsection{\acs{QuasiCertifProtocol-PT}}
The charts below respectively show how tools compete with this ``Suprise'' model (memory and CPU).

\index{Performances!LTLMarkingComparison!QuasiCertifProtocol (P/T)}
\begin{center}
   \includegraphics[width=7.2cm]{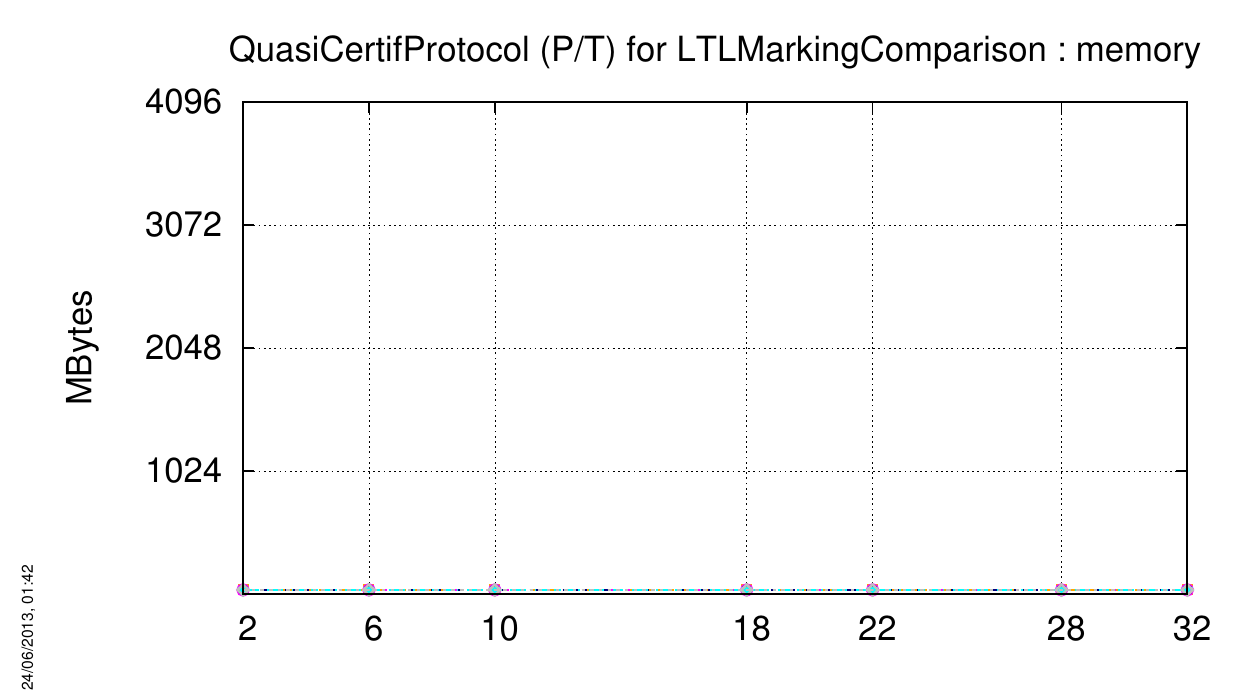}
   \includegraphics[width=7.2cm]{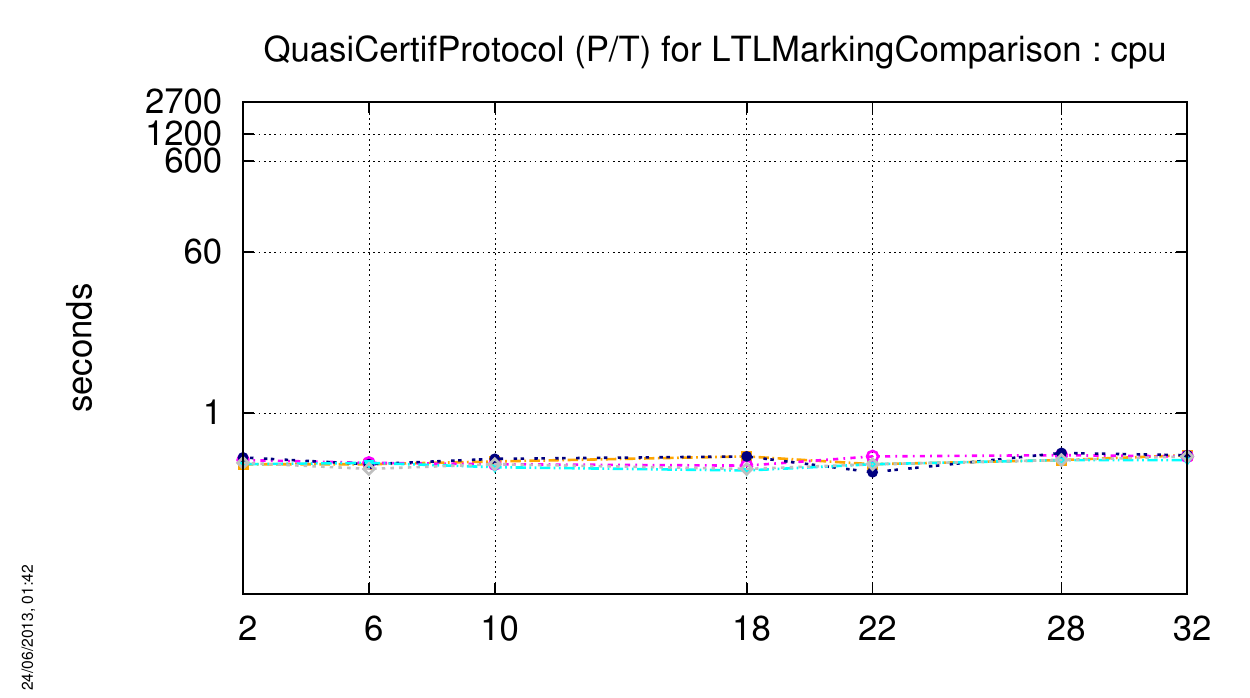}

   \includegraphics[height=1cm]{figures/tools-legend.pdf}
\end{center}

\subsubsection{\acs{Vasy2003-PT}}
No instance of this model could be computed for the \textbf{LTLMarkingComparison} examination.

\subsection{Outputs for the LTLMarkingComparison Examination}
\index{Outputs!LTLMarkingComparison}

Please find enclosed the brute results for this examination (``Known'' and ``Surprise'' models).
We display only the score of tools that provide a results for at least one instance of one model.
The legend for the values is provided below:
\begin{itemize}
   \item\textbf{nc}: the tool does not compete this examination for this model/instance,
   \item\textbf{cc}: the tool cannot compute this examination for this model/instance,
   \item\textbf{to}: the tool cannot compute this examination for this model/instance within the maximum allowed time,
   \item\textbf{mp}: the tool encountered a memory problem (stack overflow or memory full),
   \item\textbf{nf}: there is no formula available for this type of examination (typically, this concerns P/T nets where
       comparing marking cardinality has no signification when there is no equivalent colored net).
\end{itemize}

\textbf{Note on the display of results for formulas:} each formula is considered as a flag (F if false, T if true, - or ?
when the value cannot be determined). These values are concatenated in the order they appear (we assume it is the order of formulas as they were provided).

\subsubsection{``Known'' Models}

\input{result_known_LTLMarkingComparison.tex}

\subsubsection{``Surprise'' Models}

\input{result_surprise_LTLMarkingComparison.tex}

\subsection{Score for the LTLMarkingComparison Examination}
\index{Scores!LTLMarkingComparison}

Please find enclosed the scores for this examination (``Known'' and ``Surprise'' models).
We display only the score of tools that provide a results for at least one instance of one model.
The total is first listed in the table below followed by a detail, for each proposed model.
Meaning of the line labels are:
\begin{itemize}
\item\textbf{1st instance}: the tool gets a bonus for having processed the first instance of this model (+1 point),
\item\textbf{instances}: the tool gets 1 point per instances treated 
(for that, we assume that at least one formula has been successfully computed),
\item\textbf{max reached}: the tool could process all the instances for the model (+2 points),
\item\textbf{best}: the tool is among the ones that processed a maximum of instances within the time and memory confinement (+2 points).
\end{itemize}

\subsubsection{``Known'' Models}

\input{score_known_LTLMarkingComparison.tex}

\subsubsection{``Surprise'' Models}

\input{score_surprise_LTLMarkingComparison.tex}

\subsection{Trophies for this Examination}
\index{Trophies!LTLMarkingComparison}

Trophies are divided in three categories: ``Known'' models,
``Surprise'' models, and the global trophies (formula is then
$score_{global} = score_{known} + 2 \times score_{surprise}$).

\subsubsection{For ``Known'' Models} \ \\

\begin{tabular}{c}
      1 \\
   \includegraphics[width=2cm]{figures/gold.jpg} \\
   \acs{neco} \\
   54 points \\
\end{tabular}

\subsubsection{For ``Surprise'' Models}\  \\

No tool could complete this examination.

\subsubsection{Global} \ \\

\begin{tabular}{c}
      1 \\
   \includegraphics[width=2cm]{figures/gold.jpg} \\
   \acs{neco} \\
   54 points \\
\end{tabular}

\newpage

\section{The LTLPlaceComparison Examination}
\label{sec:exam:LTLPlaceComparison}
\index{Results!LTLPlaceComparison}

This examination deals with LTL properties dealing with the comparison of places marking only.
We first show a summary on the handling of models by the participating tools.
Then, we present the computed outputs and the associated scores for this
examination prior to a summary of relevant executions.

\subsection{Handling of Models by Tools}
\index{Performances!LTLPlaceComparison}

\subsubsection{\acs{CSRepetitions-COL}}
No instance of this model could be computed for the \textbf{LTLPlaceComparison} examination.

\subsubsection{\acs{CSRepetitions-PT}}
The charts below respectively show how tools compete with this ``Known'' model (memory and CPU).

\index{Performances!LTLPlaceComparison!CSRepetitions (P/T)}
\begin{center}
   \includegraphics[width=7.2cm]{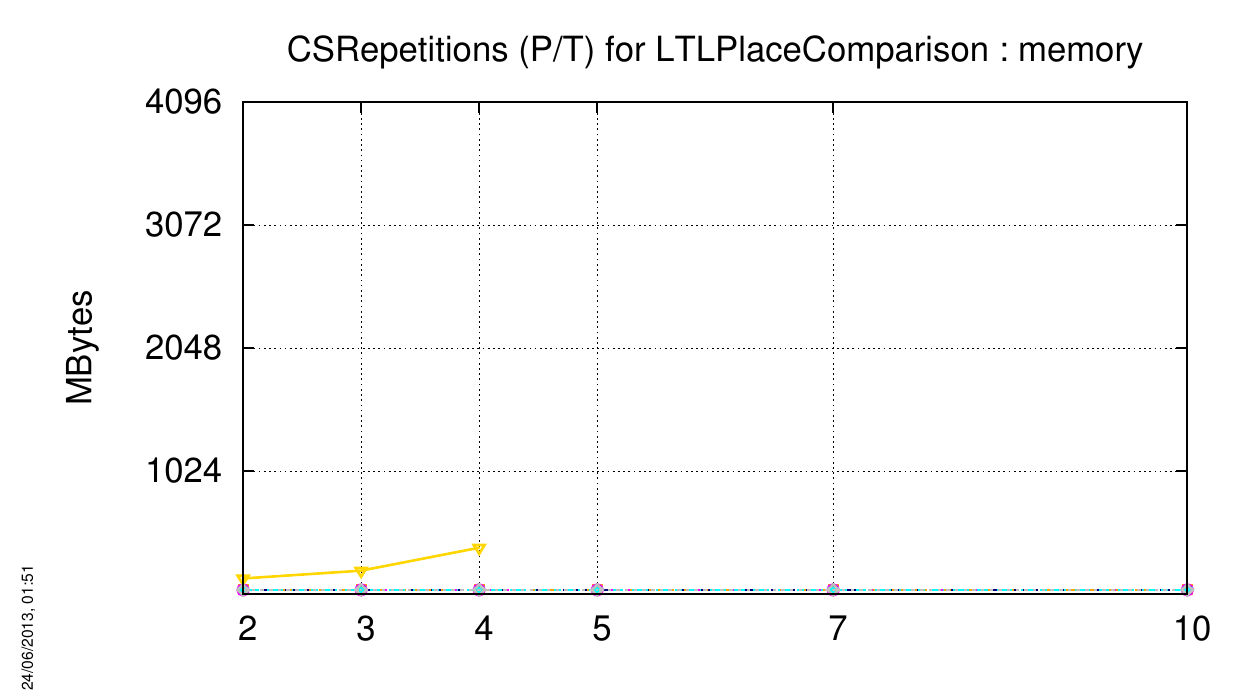}
   \includegraphics[width=7.2cm]{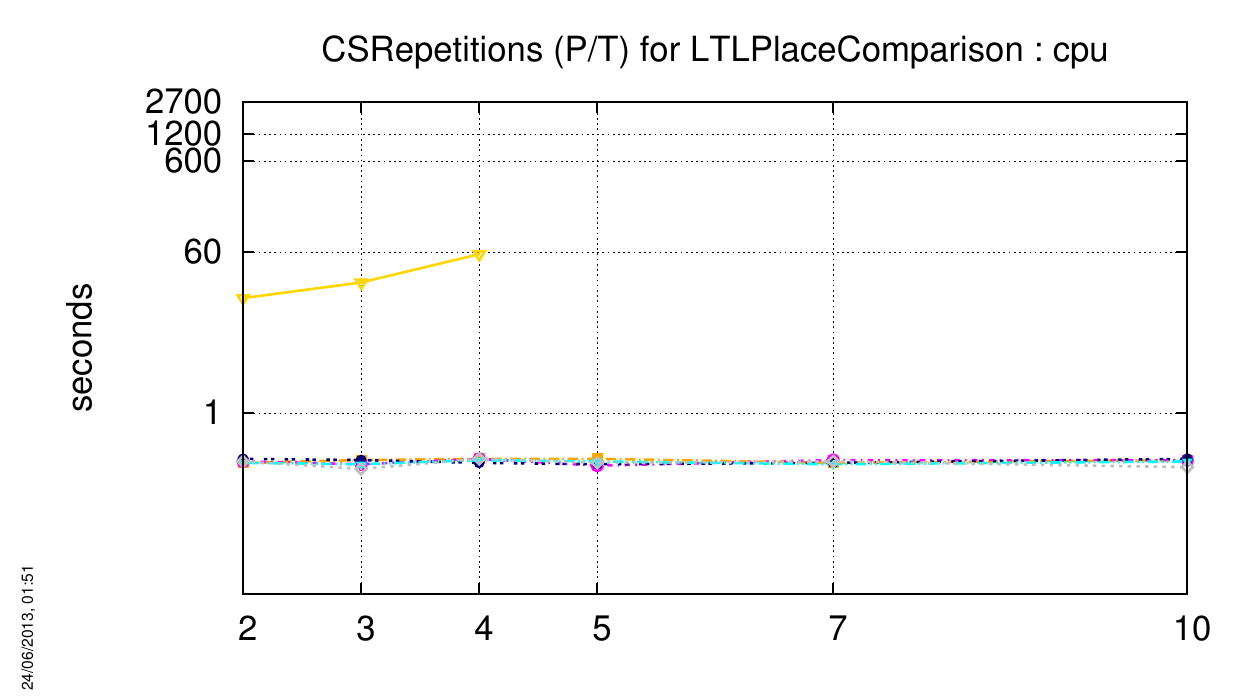}

   \includegraphics[height=1cm]{figures/tools-legend.pdf}
\end{center}

\subsubsection{\acs{Dekker-PT}}
The charts below respectively show how tools compete with this ``Known'' model (memory and CPU).

\index{Performances!LTLPlaceComparison!Dekker (P/T)}
\begin{center}
   \includegraphics[width=7.2cm]{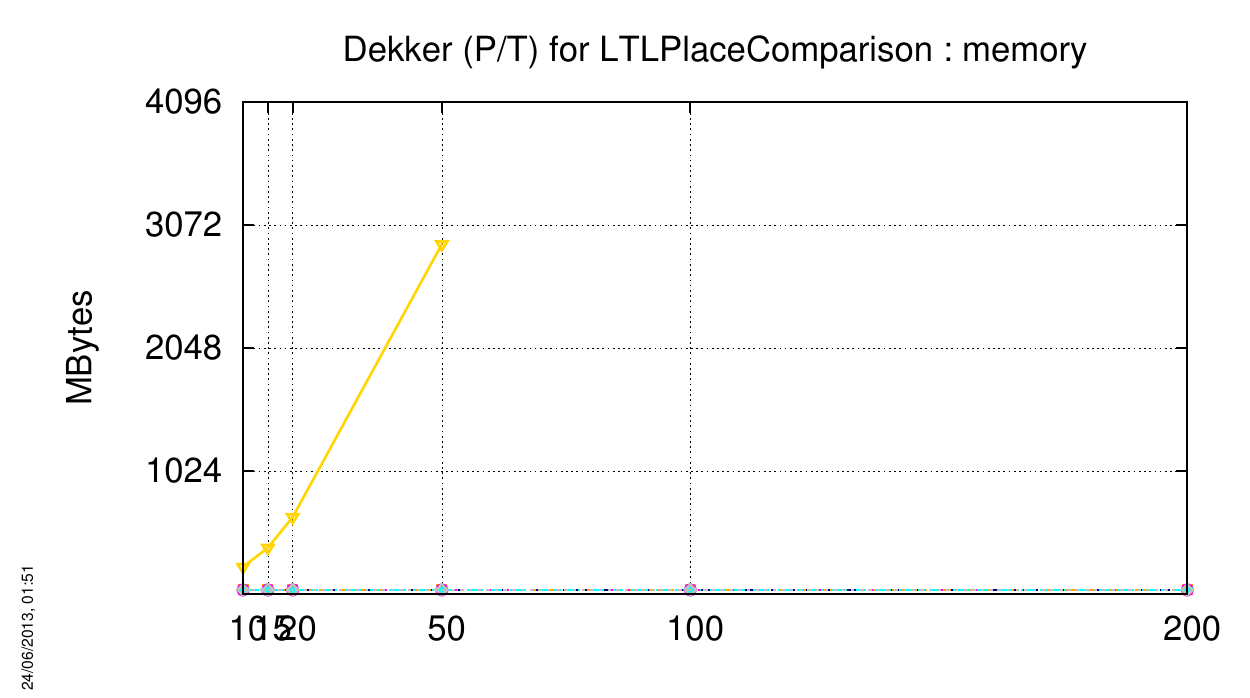}
   \includegraphics[width=7.2cm]{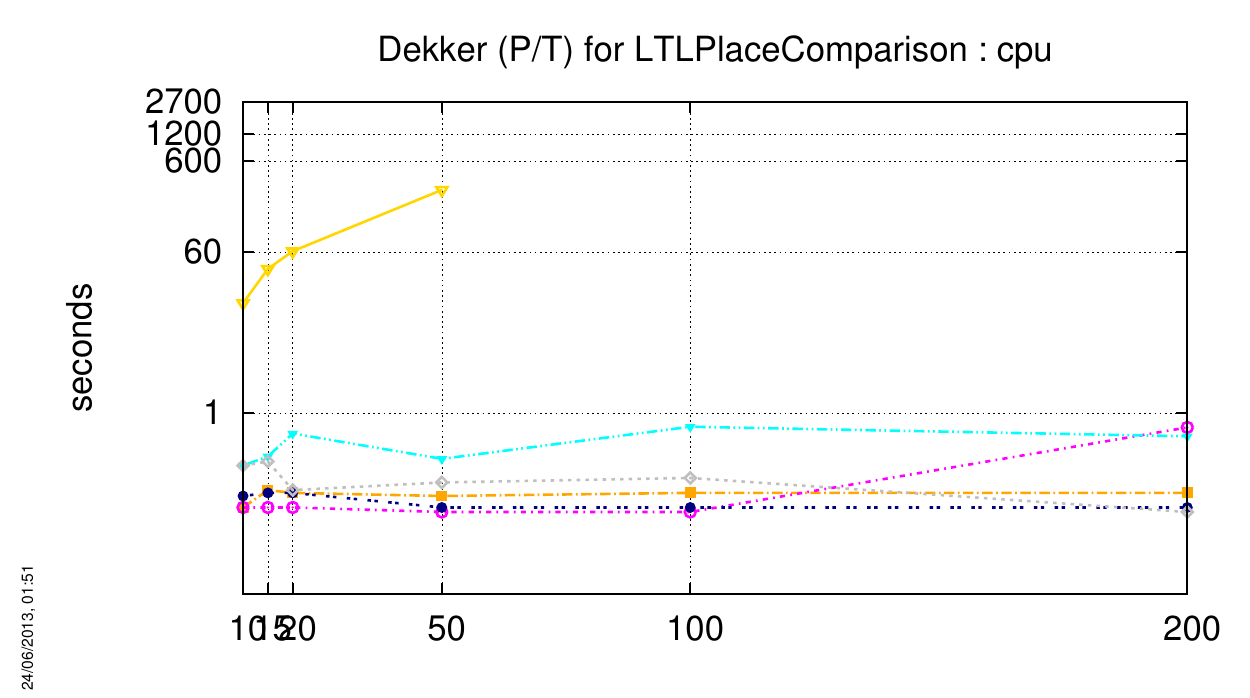}

   \includegraphics[height=1cm]{figures/tools-legend.pdf}
\end{center}

\subsubsection{\acs{DotAndBoxes-COL}}
No instance of this model could be computed for the \textbf{LTLPlaceComparison} examination.

\subsubsection{\acs{DrinkVendingMachine-COL}}
No instance of this model could be computed for the \textbf{LTLPlaceComparison} examination.

\subsubsection{\acs{DrinkVendingMachine-PT}}
The charts below respectively show how tools compete with this ``Known'' model (memory and CPU).

\index{Performances!LTLPlaceComparison!DrinkVendingMachine (P/T)}
\begin{center}
   \includegraphics[width=7.2cm]{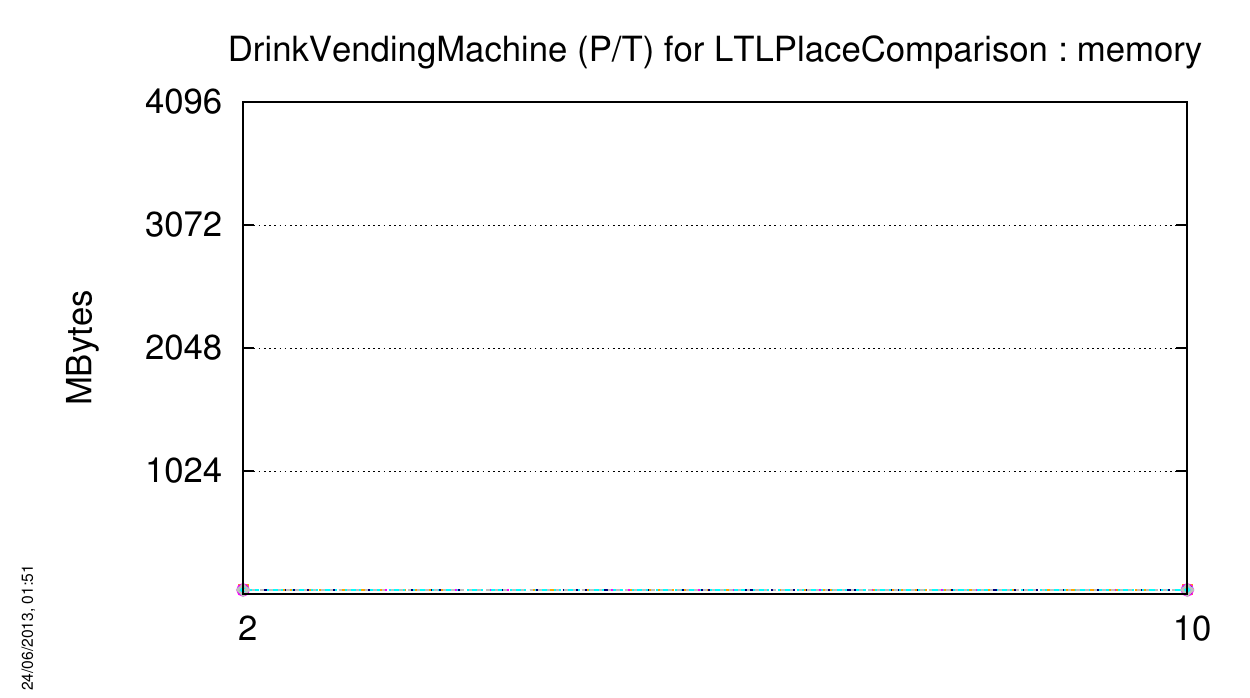}
   \includegraphics[width=7.2cm]{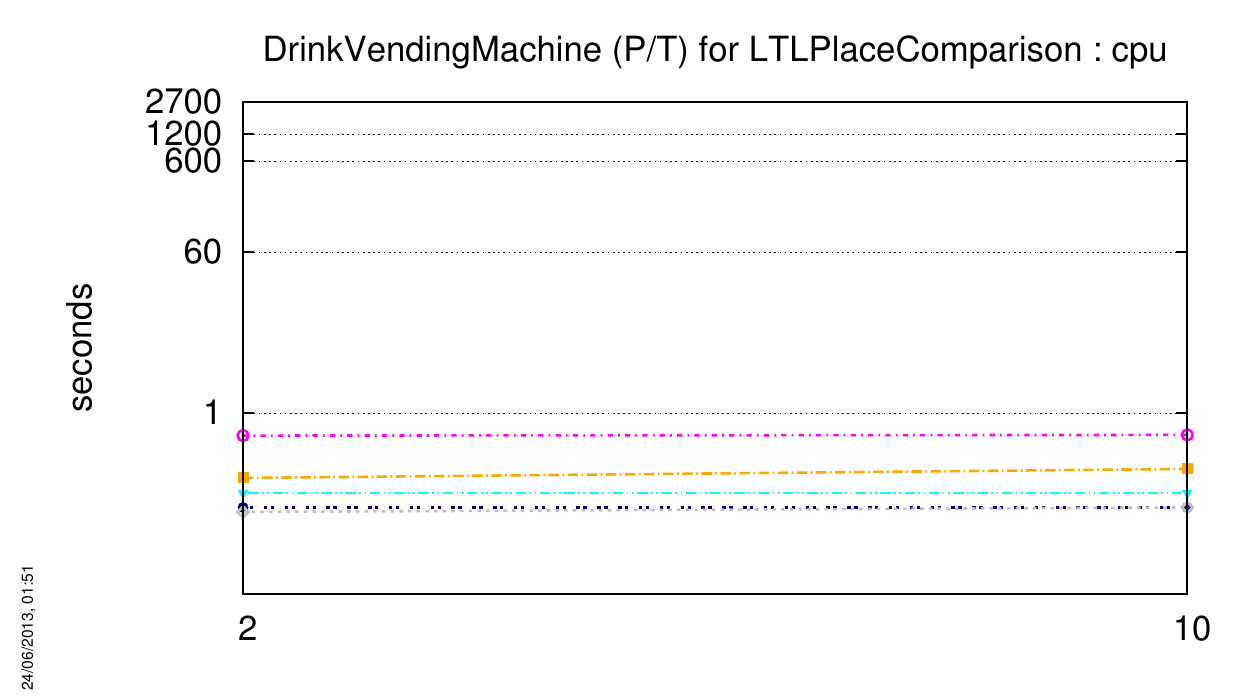}

   \includegraphics[height=1cm]{figures/tools-legend.pdf}
\end{center}

\subsubsection{\acs{Echo-PT}}
The charts below respectively show how tools compete with this ``Known'' model (memory and CPU).

\index{Performances!LTLPlaceComparison!Echo (P/T)}
\begin{center}
   \includegraphics[width=7.2cm]{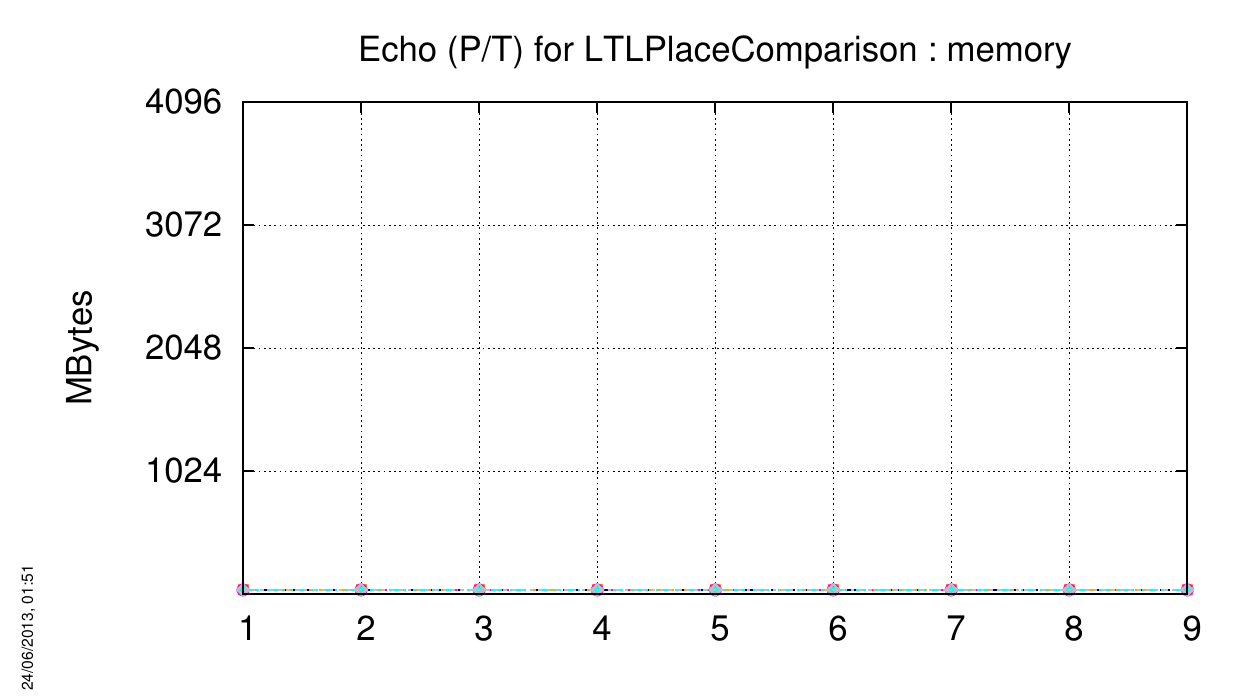}
   \includegraphics[width=7.2cm]{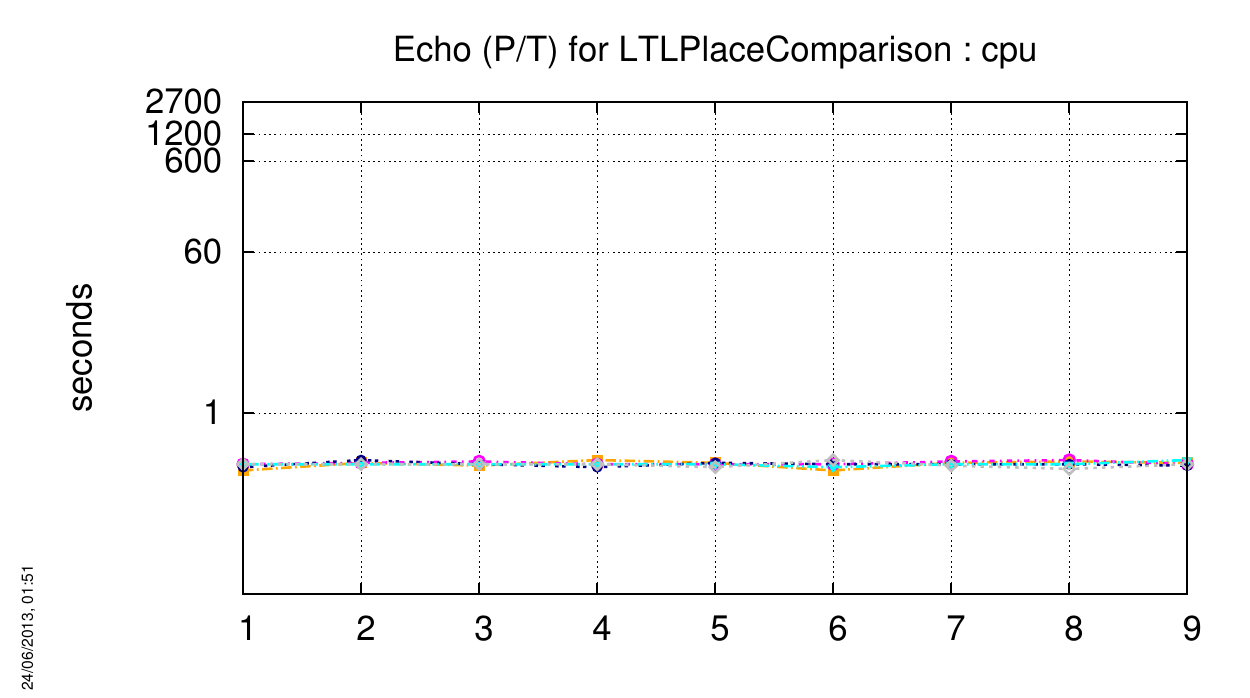}

   \includegraphics[height=1cm]{figures/tools-legend.pdf}
\end{center}

\subsubsection{\acs{Eratosthenes-PT}}
The charts below respectively show how tools compete with this ``Known'' model (memory and CPU).

\index{Performances!LTLPlaceComparison!Eratosthenes (P/T)}
\begin{center}
   \includegraphics[width=7.2cm]{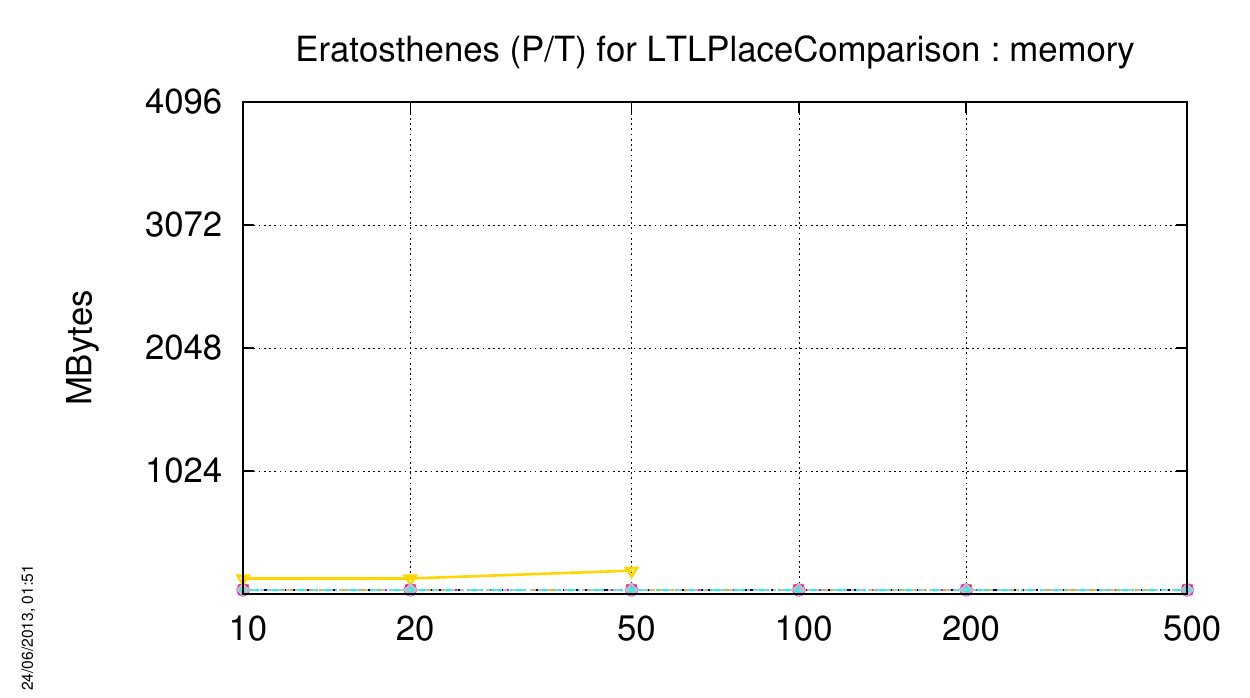}
   \includegraphics[width=7.2cm]{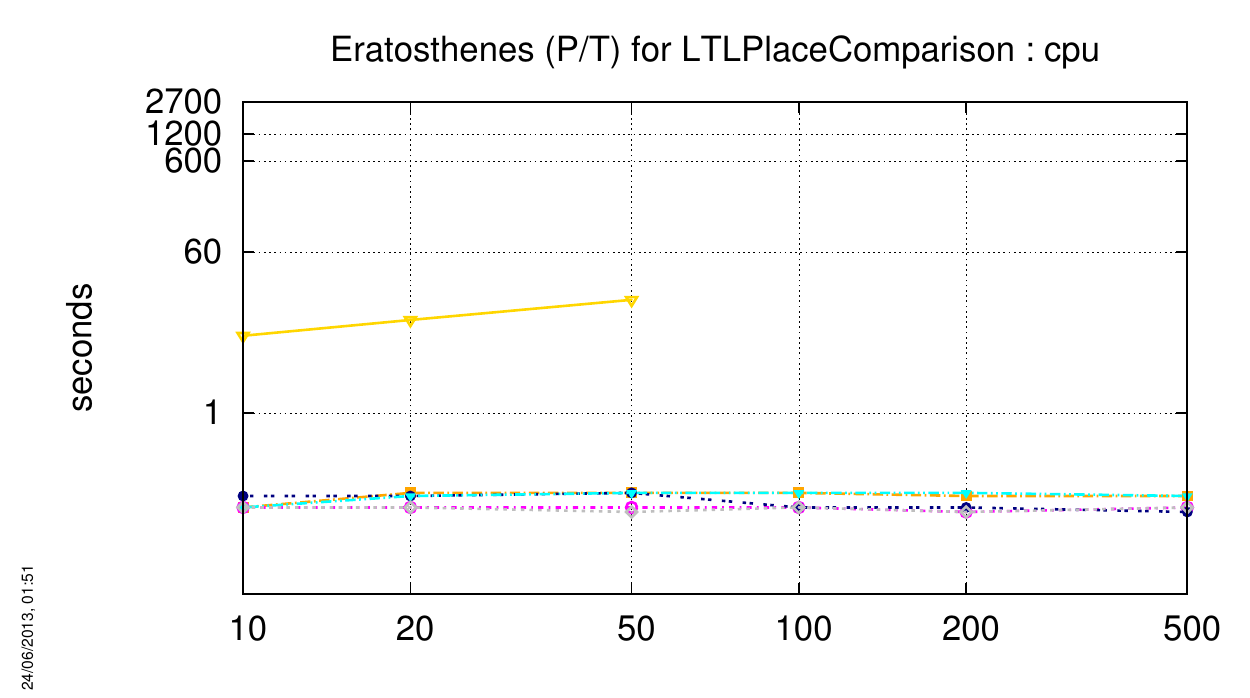}

   \includegraphics[height=1cm]{figures/tools-legend.pdf}
\end{center}

\subsubsection{\acs{FMS-PT}}
The charts below respectively show how tools compete with this ``Known'' model (memory and CPU).

\index{Performances!LTLPlaceComparison!FMS (P/T)}
\begin{center}
   \includegraphics[width=7.2cm]{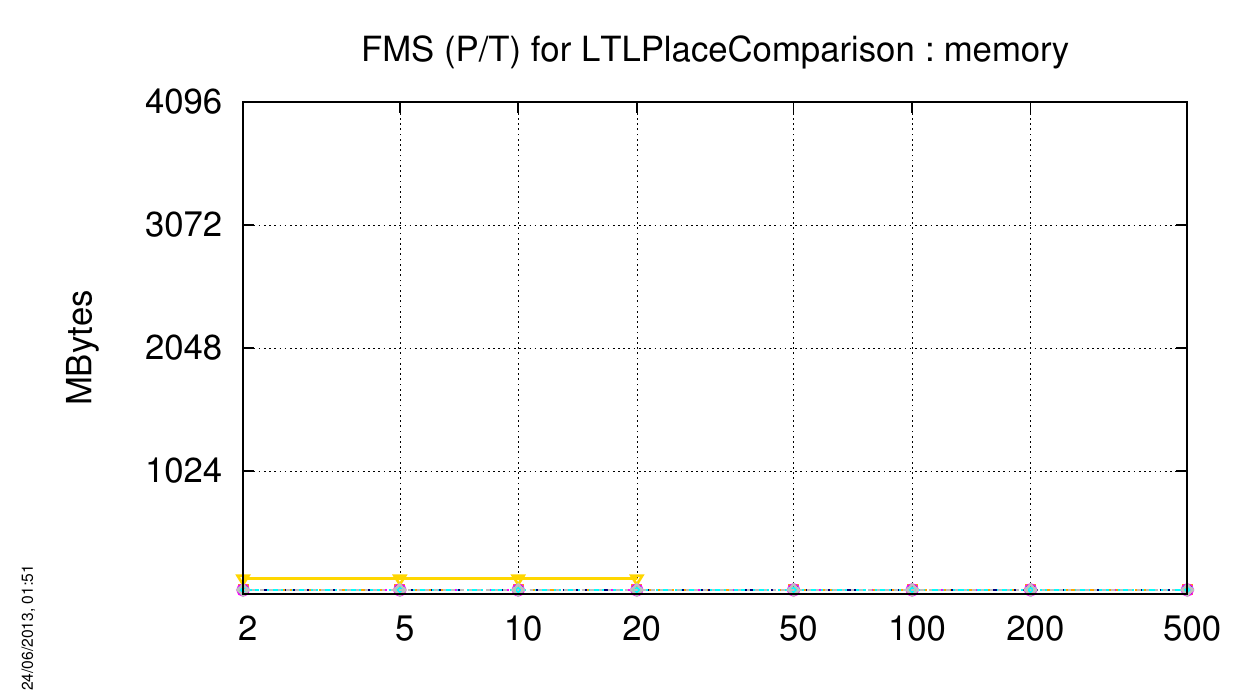}
   \includegraphics[width=7.2cm]{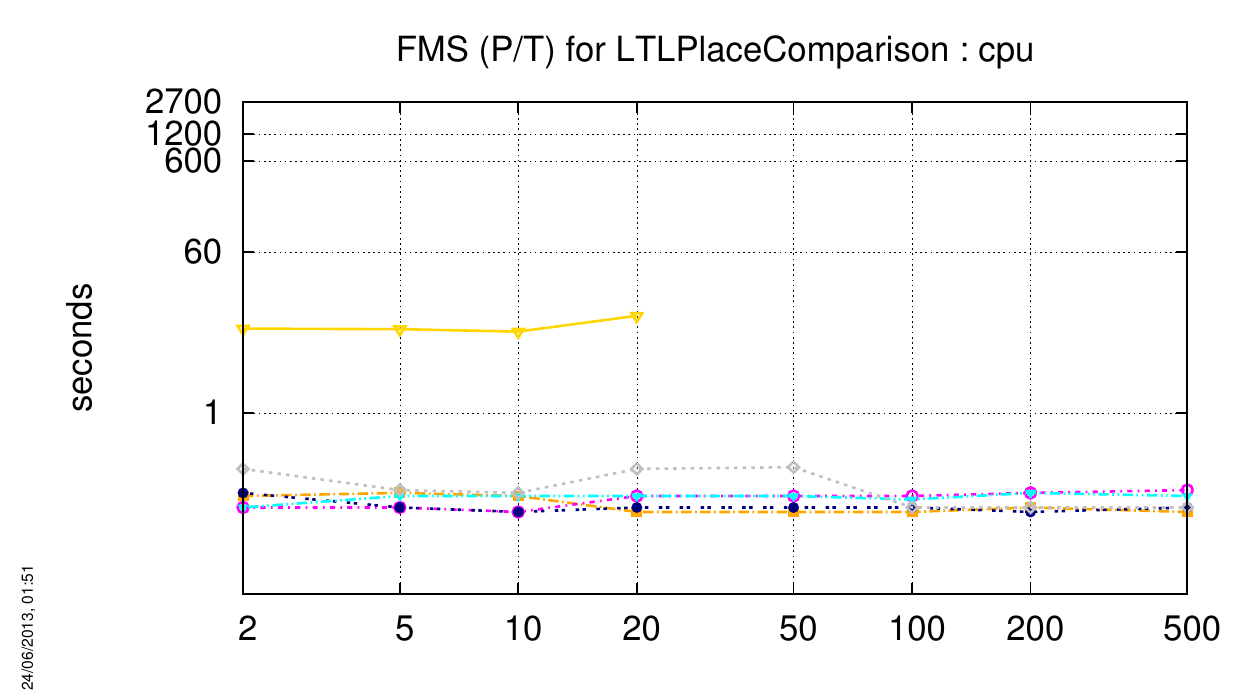}

   \includegraphics[height=1cm]{figures/tools-legend.pdf}
\end{center}

\subsubsection{\acs{GlobalRessAlloc-COL}}
No instance of this model could be computed for the \textbf{LTLPlaceComparison} examination.

\subsubsection{\acs{GlobalRessAlloc-PT}}
The charts below respectively show how tools compete with this ``Known'' model (memory and CPU).

\index{Performances!LTLPlaceComparison!GlobalRessAlloc (P/T)}
\begin{center}
   \includegraphics[width=7.2cm]{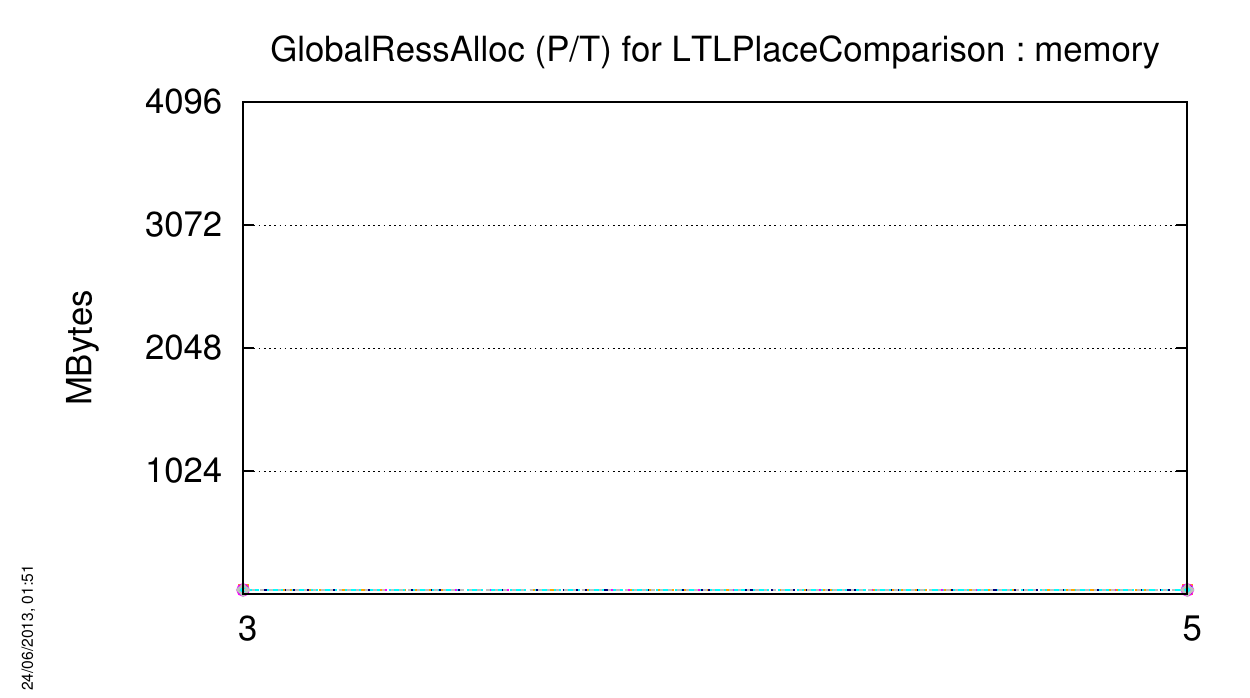}
   \includegraphics[width=7.2cm]{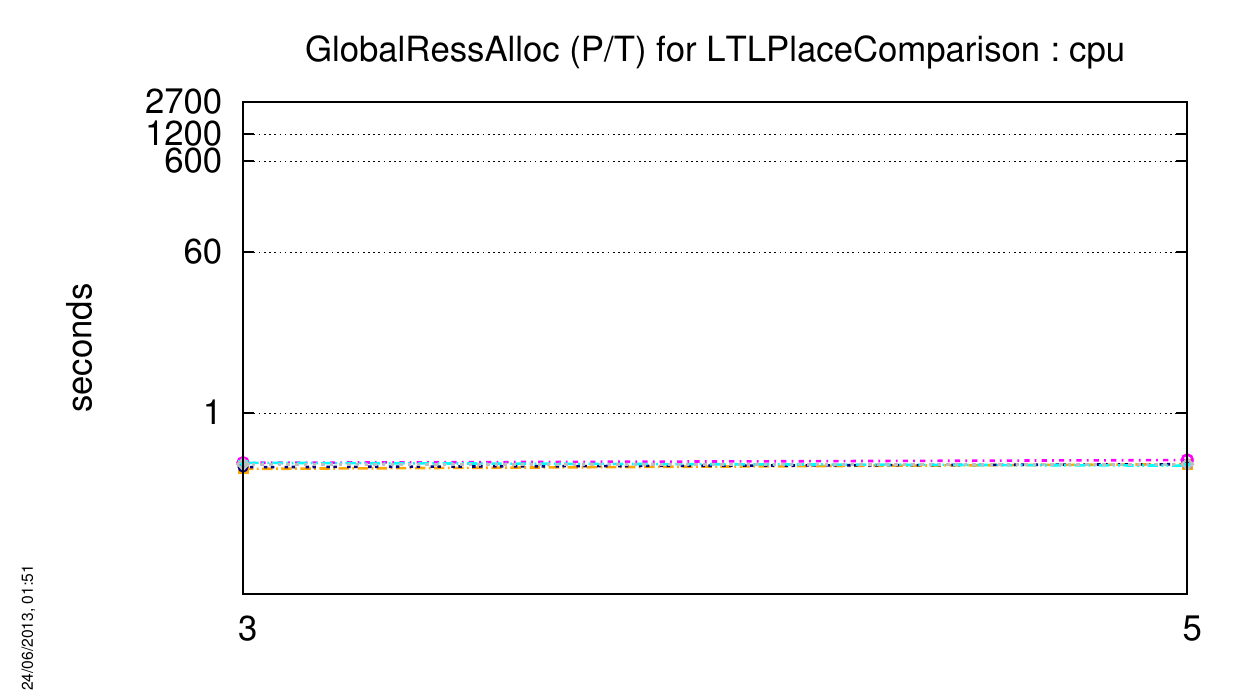}

   \includegraphics[height=1cm]{figures/tools-legend.pdf}
\end{center}

\subsubsection{\acs{Kanban-PT}}
The charts below respectively show how tools compete with this ``Known'' model (memory and CPU).

\index{Performances!LTLPlaceComparison!Kanban (P/T)}
\begin{center}
   \includegraphics[width=7.2cm]{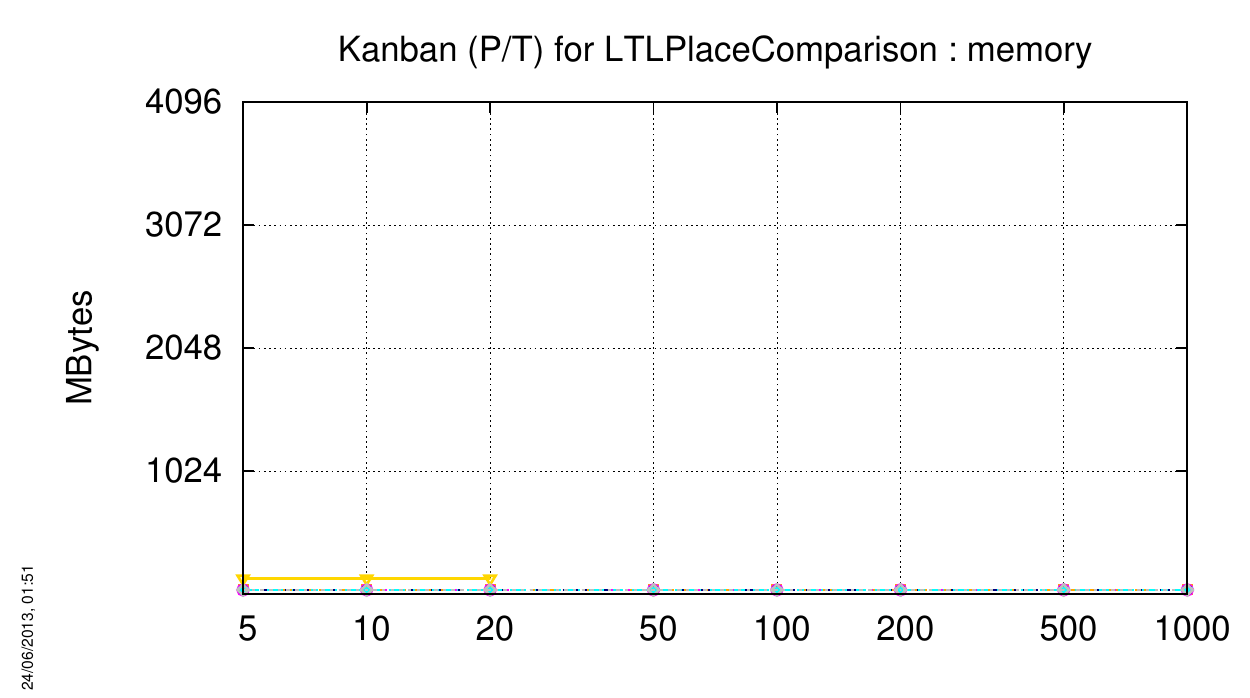}
   \includegraphics[width=7.2cm]{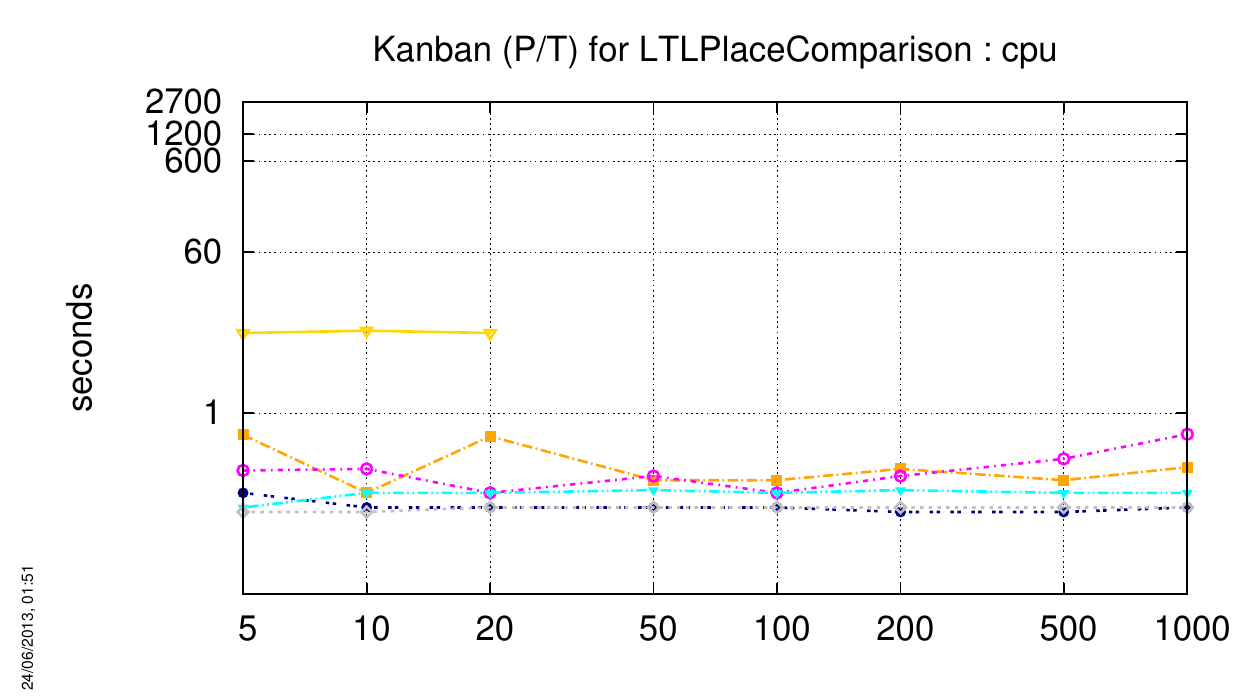}

   \includegraphics[height=1cm]{figures/tools-legend.pdf}
\end{center}

\subsubsection{\acs{LamportFastMutEx-COL}}
No instance of this model could be computed for the \textbf{LTLPlaceComparison} examination.

\subsubsection{\acs{LamportFastMutEx-PT}}
The charts below respectively show how tools compete with this ``Known'' model (memory and CPU).

\index{Performances!LTLPlaceComparison!LamportFastMutEx (P/T)}
\begin{center}
   \includegraphics[width=7.2cm]{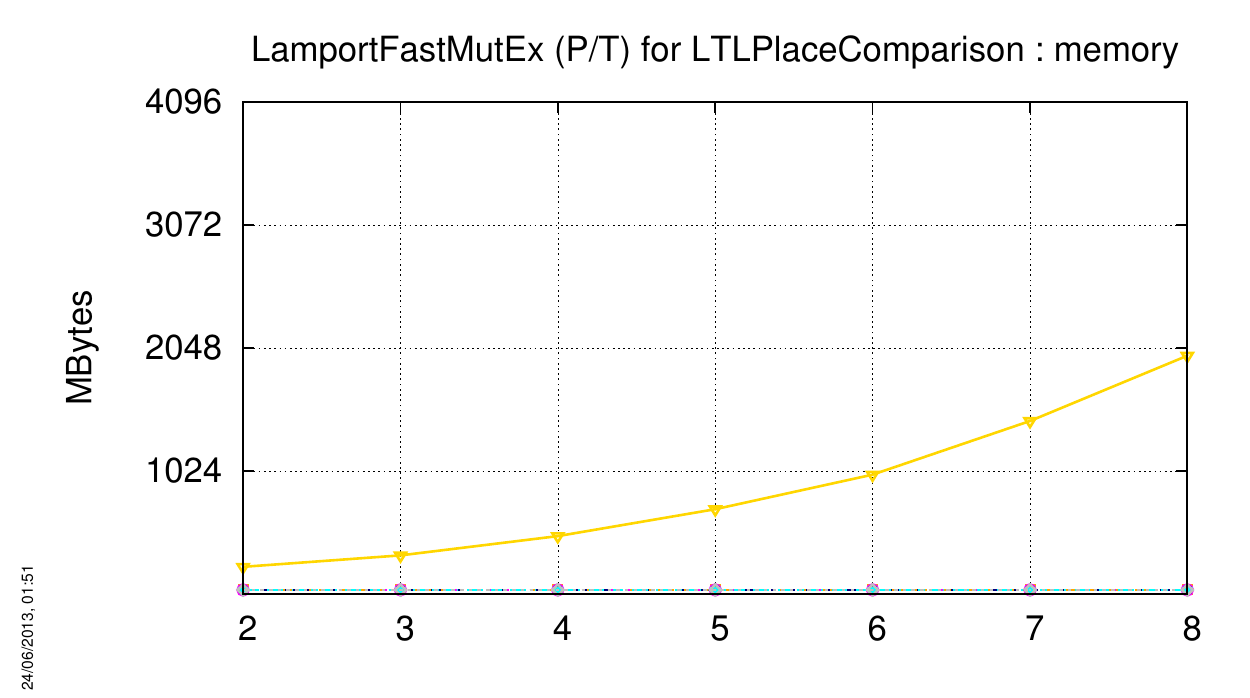}
   \includegraphics[width=7.2cm]{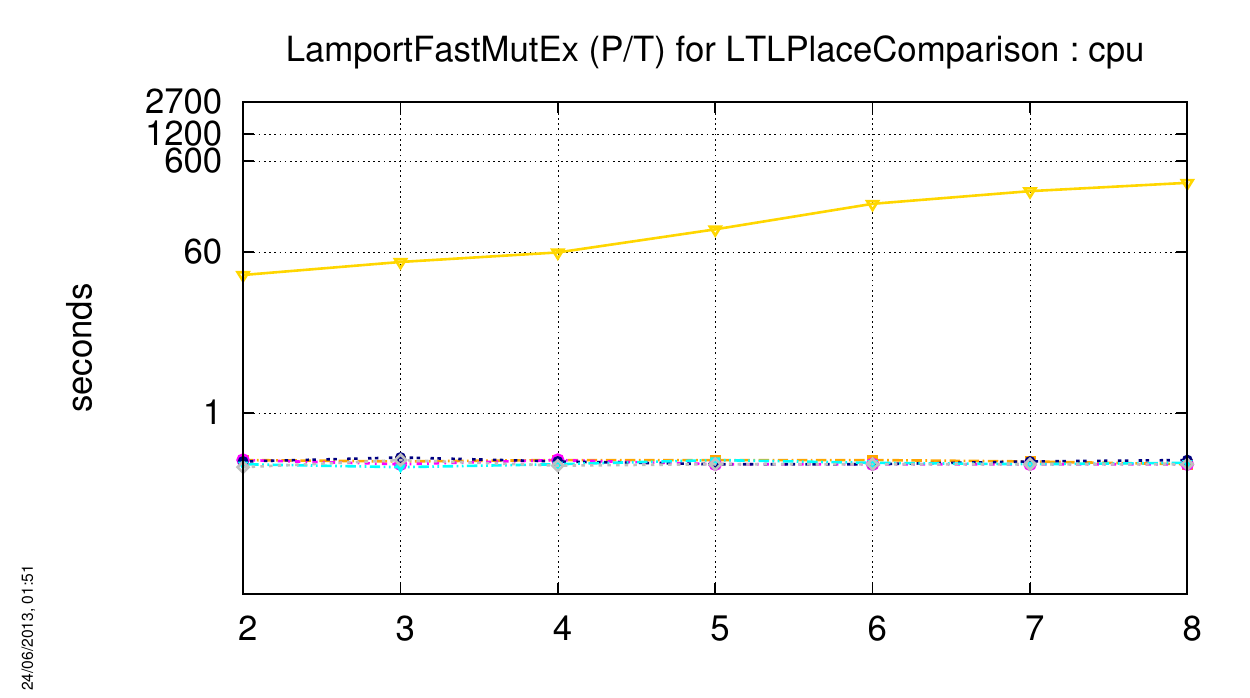}

   \includegraphics[height=1cm]{figures/tools-legend.pdf}
\end{center}

\subsubsection{\acs{MAPK-PT}}
The charts below respectively show how tools compete with this ``Known'' model (memory and CPU).

\index{Performances!LTLPlaceComparison!MAPK (P/T)}
\begin{center}
   \includegraphics[width=7.2cm]{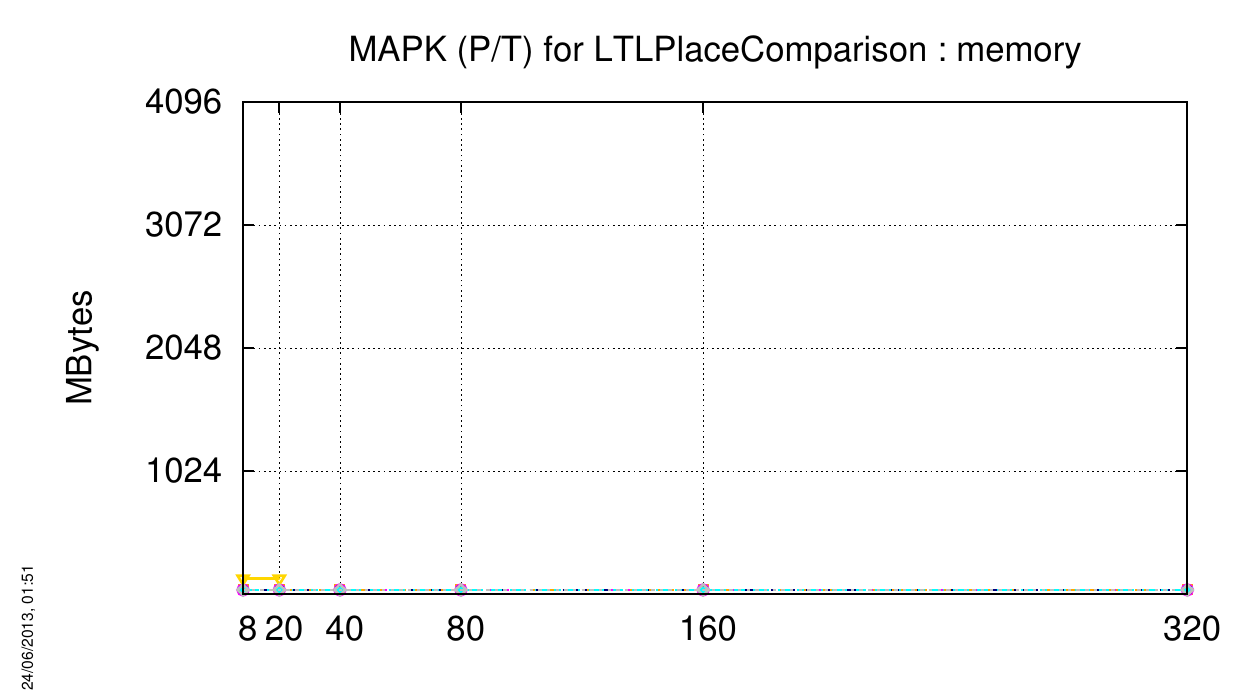}
   \includegraphics[width=7.2cm]{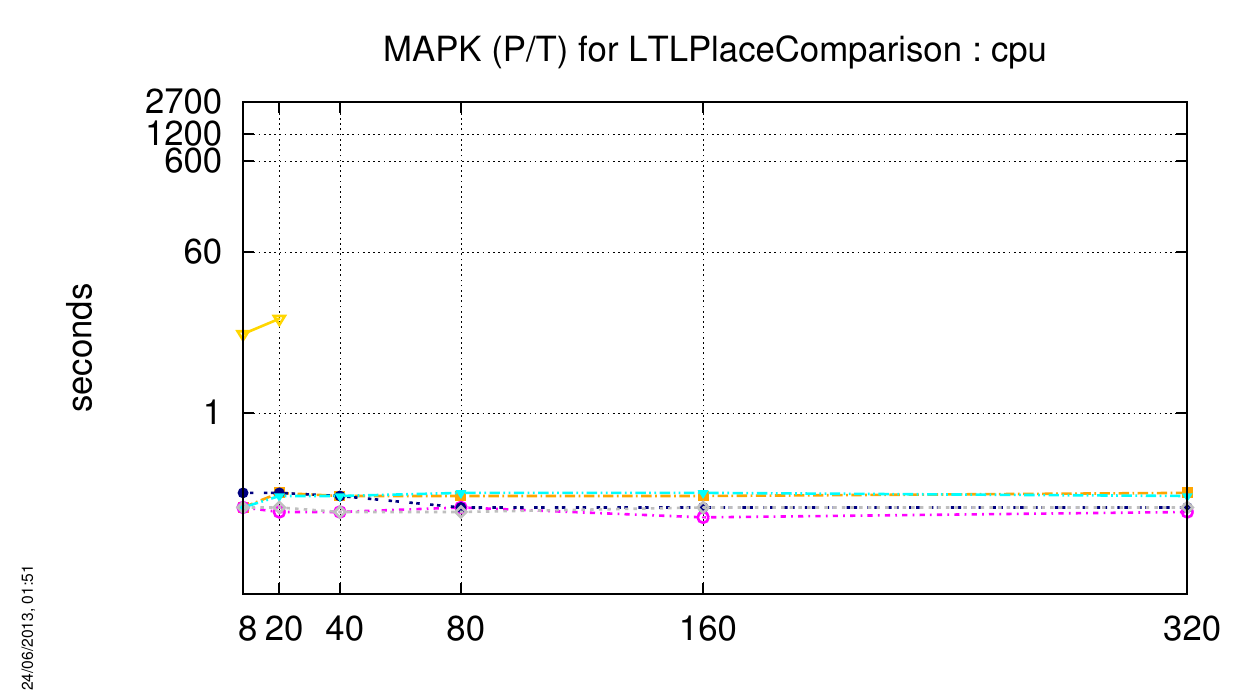}

   \includegraphics[height=1cm]{figures/tools-legend.pdf}
\end{center}

\subsubsection{\acs{NeoElection-COL}}
No instance of this model could be computed for the \textbf{LTLPlaceComparison} examination.

\subsubsection{\acs{NeoElection-PT}}
The charts below respectively show how tools compete with this ``Known'' model (memory and CPU).

\index{Performances!LTLPlaceComparison!NeoElection (P/T)}
\begin{center}
   \includegraphics[width=7.2cm]{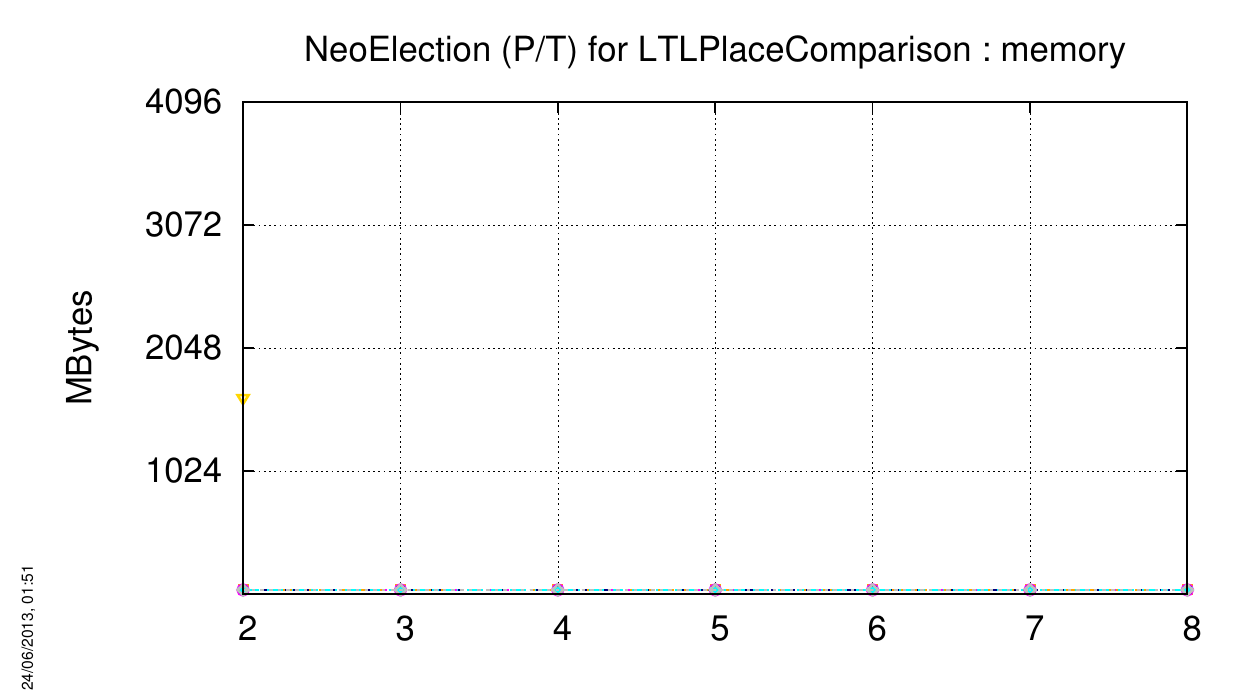}
   \includegraphics[width=7.2cm]{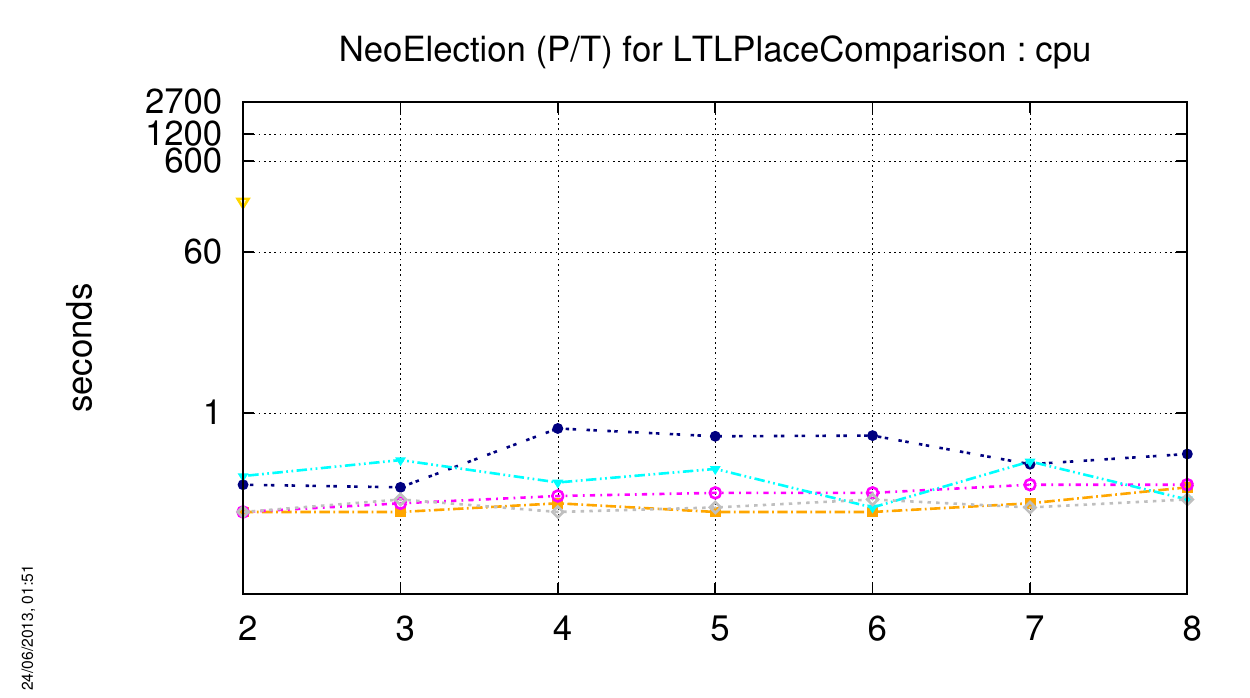}

   \includegraphics[height=1cm]{figures/tools-legend.pdf}
\end{center}

\subsubsection{\acs{PermAdmissibility-COL}}
No instance of this model could be computed for the \textbf{LTLPlaceComparison} examination.

\subsubsection{\acs{PermAdmissibility-PT}}
The charts below respectively show how tools compete with this ``Known'' model (memory and CPU).

\index{Performances!LTLPlaceComparison!PermAdmissibility (P/T)}
\begin{center}
   \includegraphics[width=7.2cm]{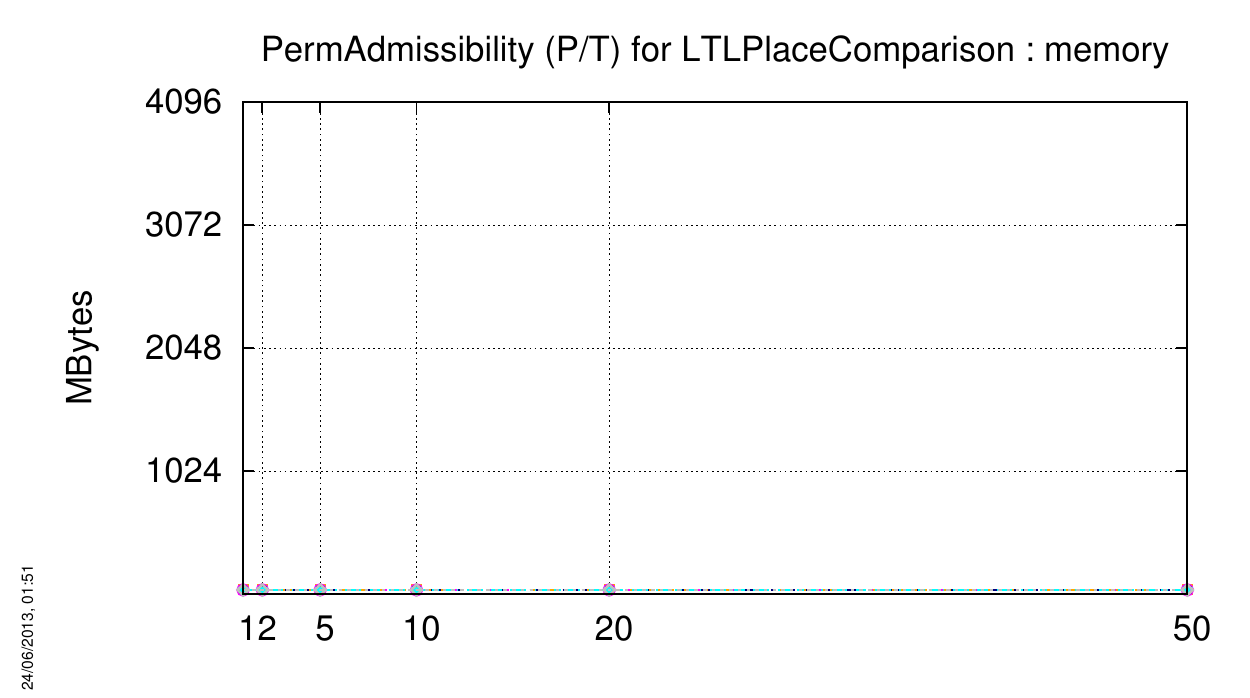}
   \includegraphics[width=7.2cm]{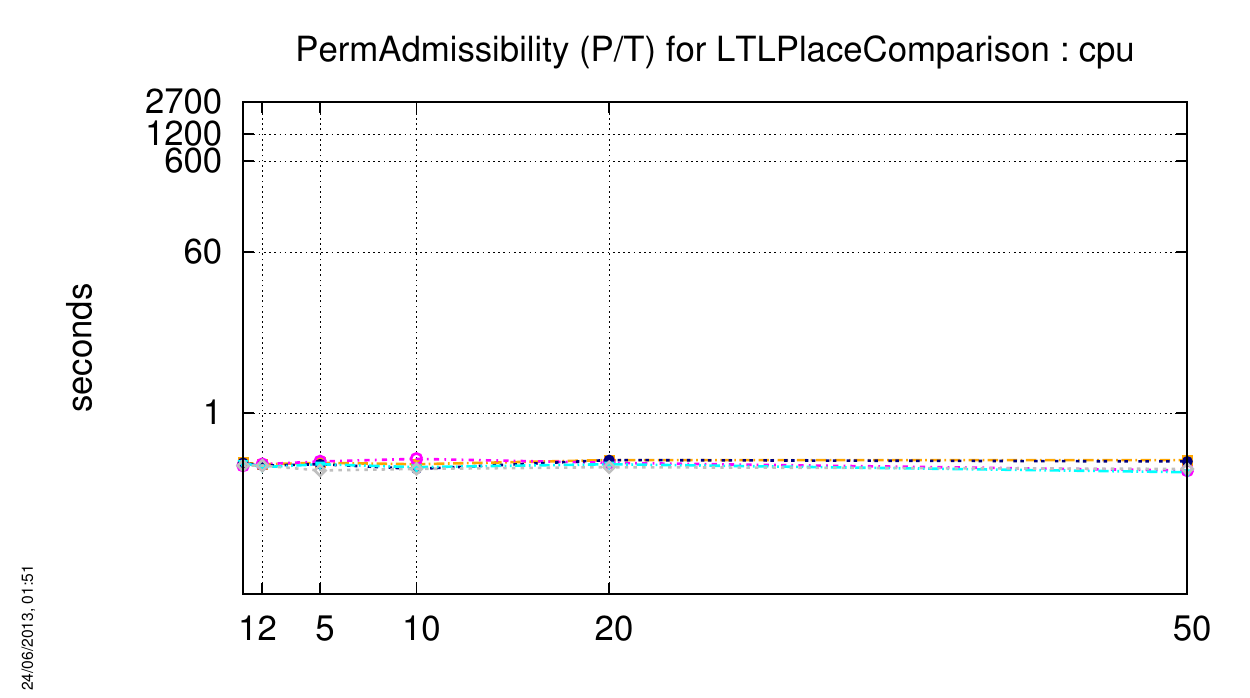}

   \includegraphics[height=1cm]{figures/tools-legend.pdf}
\end{center}

\subsubsection{\acs{Peterson-COL}}
No instance of this model could be computed for the \textbf{LTLPlaceComparison} examination.

\subsubsection{\acs{Peterson-PT}}
The charts below respectively show how tools compete with this ``Known'' model (memory and CPU).

\index{Performances!LTLPlaceComparison!Peterson (P/T)}
\begin{center}
   \includegraphics[width=7.2cm]{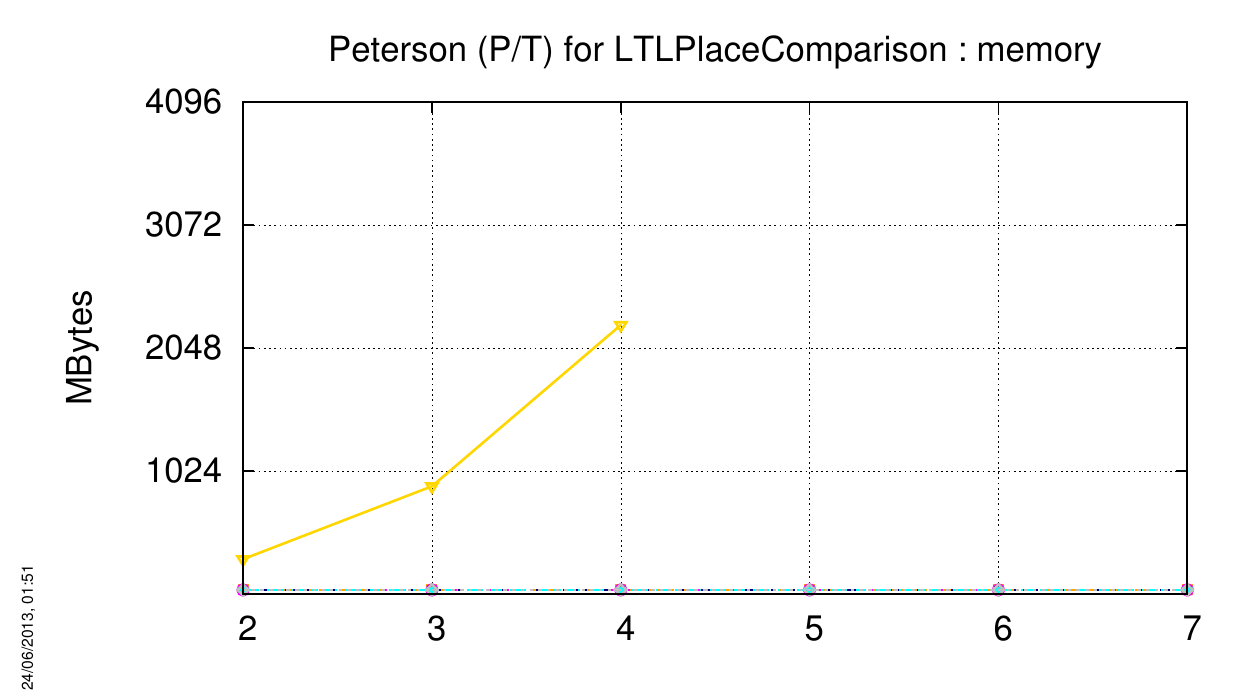}
   \includegraphics[width=7.2cm]{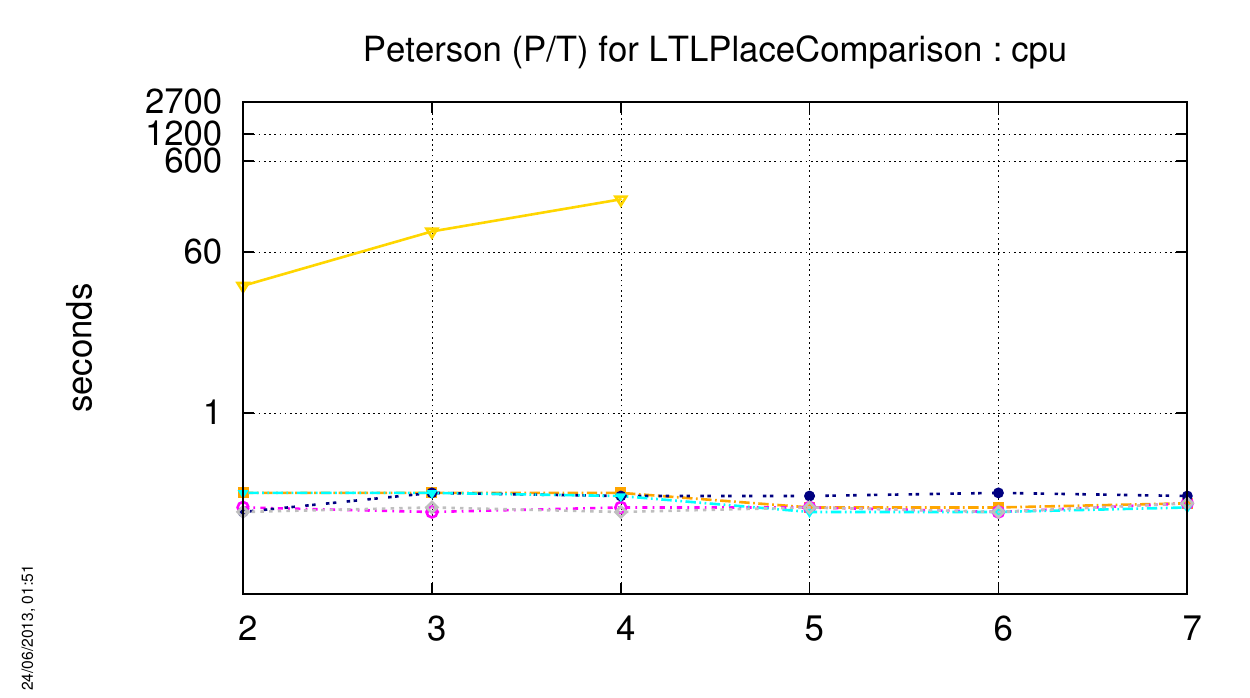}

   \includegraphics[height=1cm]{figures/tools-legend.pdf}
\end{center}

\subsubsection{\acs{Philosophers-COL}}
No instance of this model could be computed for the \textbf{LTLPlaceComparison} examination.

\subsubsection{\acs{Philosophers-PT}}
The charts below respectively show how tools compete with this ``Known'' model (memory and CPU).

\index{Performances!LTLPlaceComparison!Philosophers (P/T)}
\begin{center}
   \includegraphics[width=7.2cm]{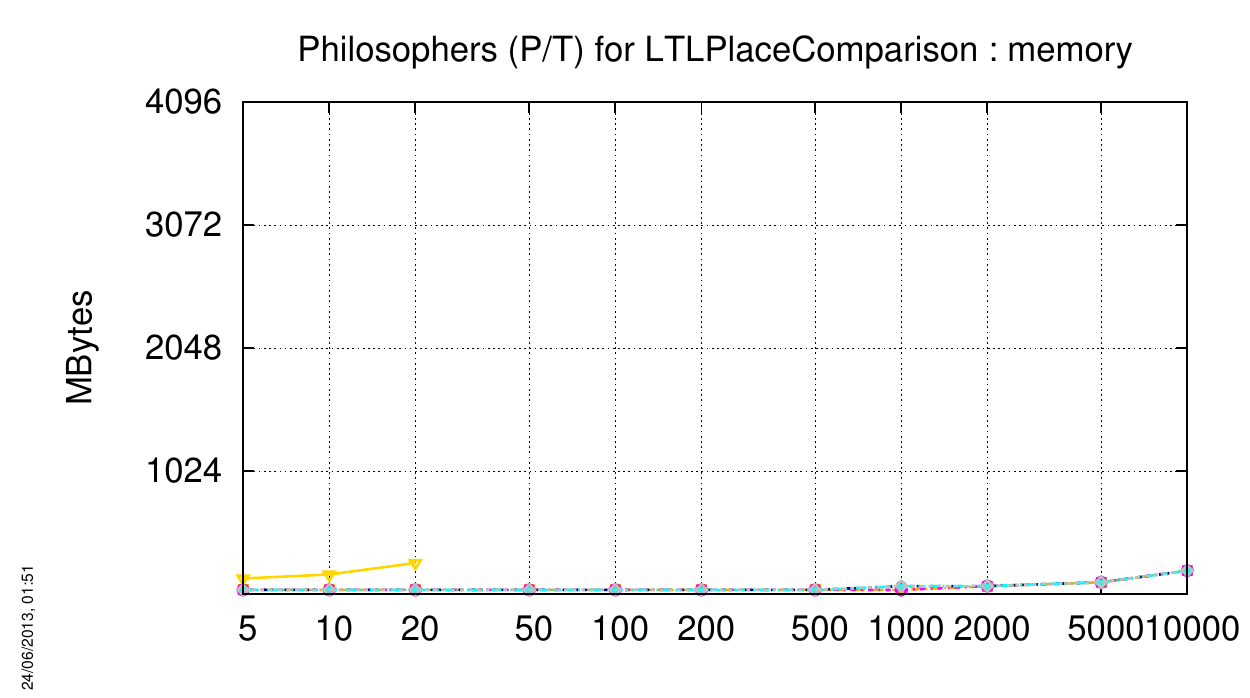}
   \includegraphics[width=7.2cm]{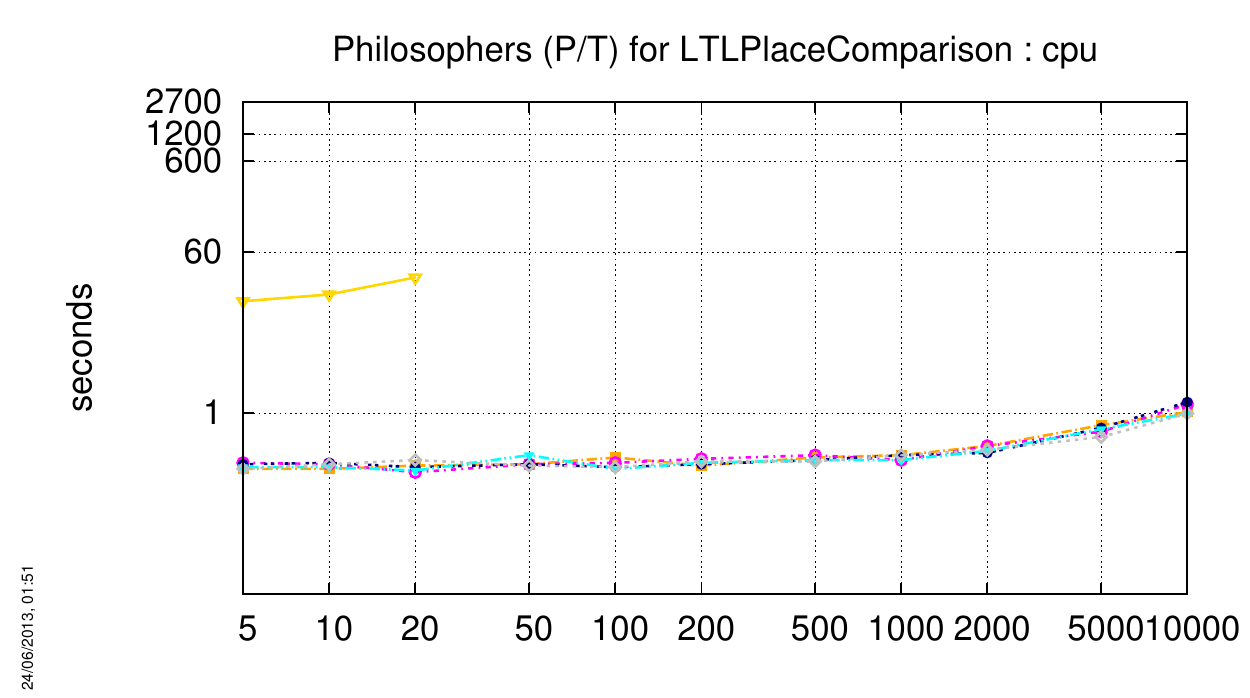}

   \includegraphics[height=1cm]{figures/tools-legend.pdf}
\end{center}

\subsubsection{\acs{PhilosophersDyn-COL}}
No instance of this model could be computed for the \textbf{LTLPlaceComparison} examination.

\subsubsection{\acs{PhilosophersDyn-PT}}
The charts below respectively show how tools compete with this ``Known'' model (memory and CPU).

\index{Performances!LTLPlaceComparison!PhilosophersDyn (P/T)}
\begin{center}
   \includegraphics[width=7.2cm]{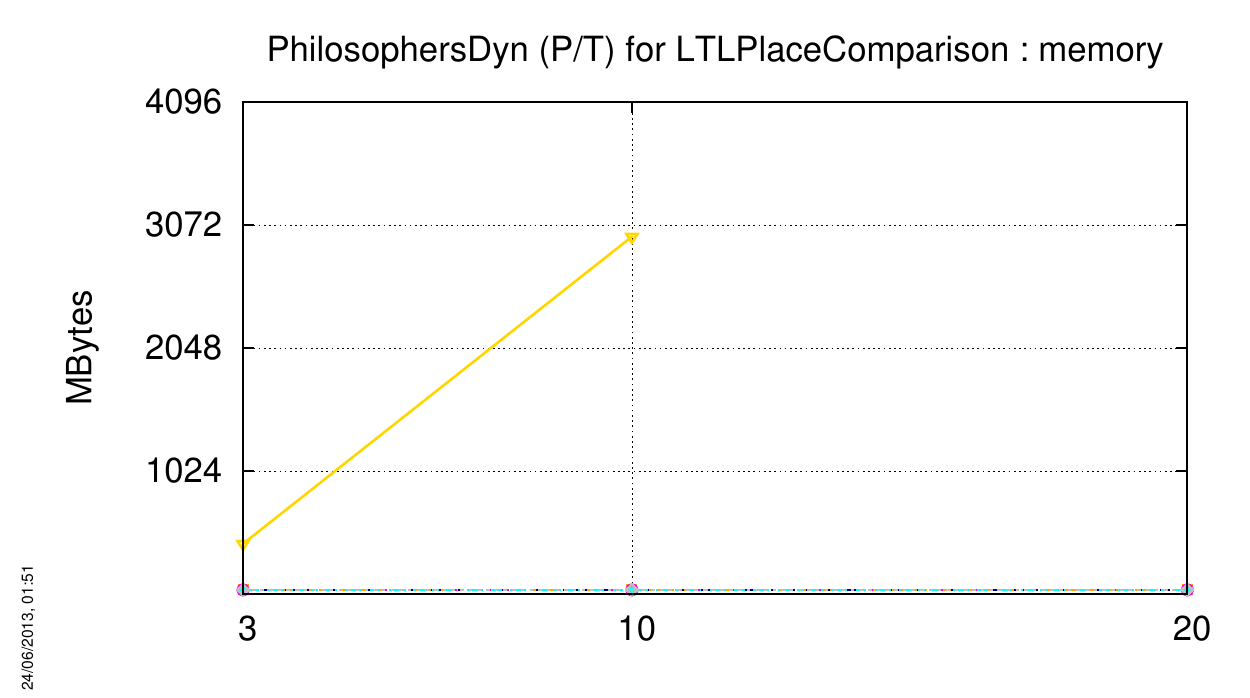}
   \includegraphics[width=7.2cm]{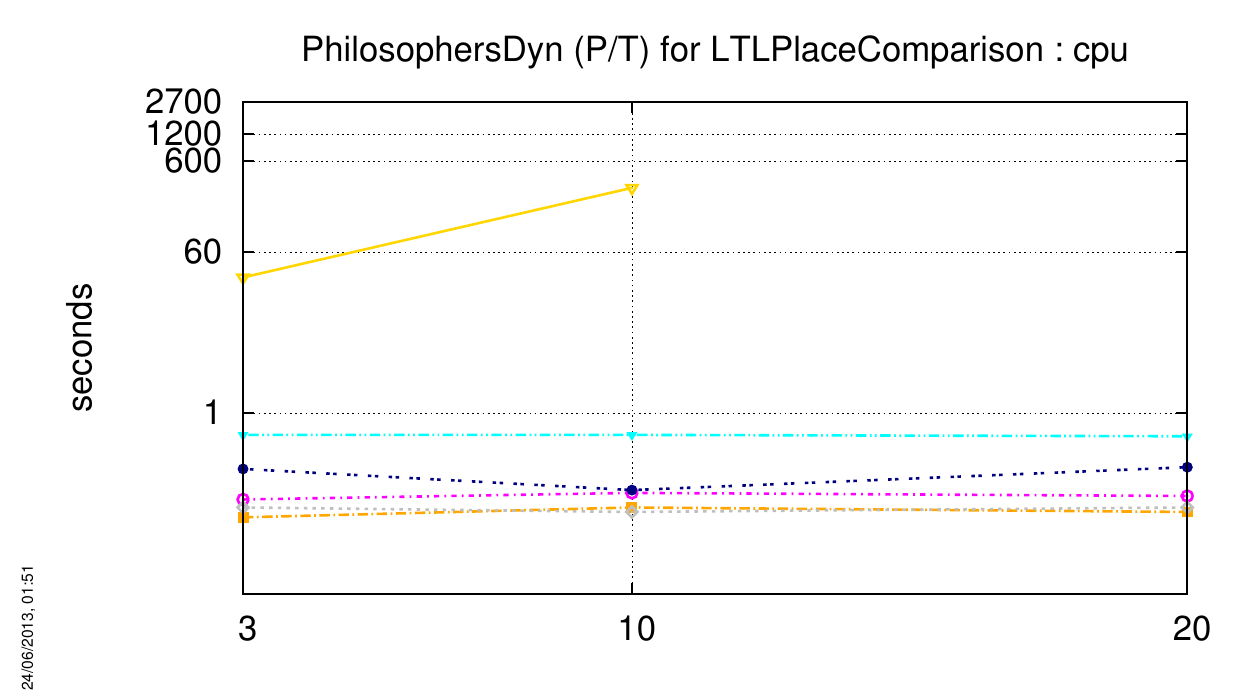}

   \includegraphics[height=1cm]{figures/tools-legend.pdf}
\end{center}

\subsubsection{\acs{Planning-PT}}
No instance of this model could be computed for the \textbf{LTLPlaceComparison} examination.

\subsubsection{\acs{Railroad-PT}}
The charts below respectively show how tools compete with this ``Known'' model (memory and CPU).

\index{Performances!LTLPlaceComparison!Railroad (P/T)}
\begin{center}
   \includegraphics[width=7.2cm]{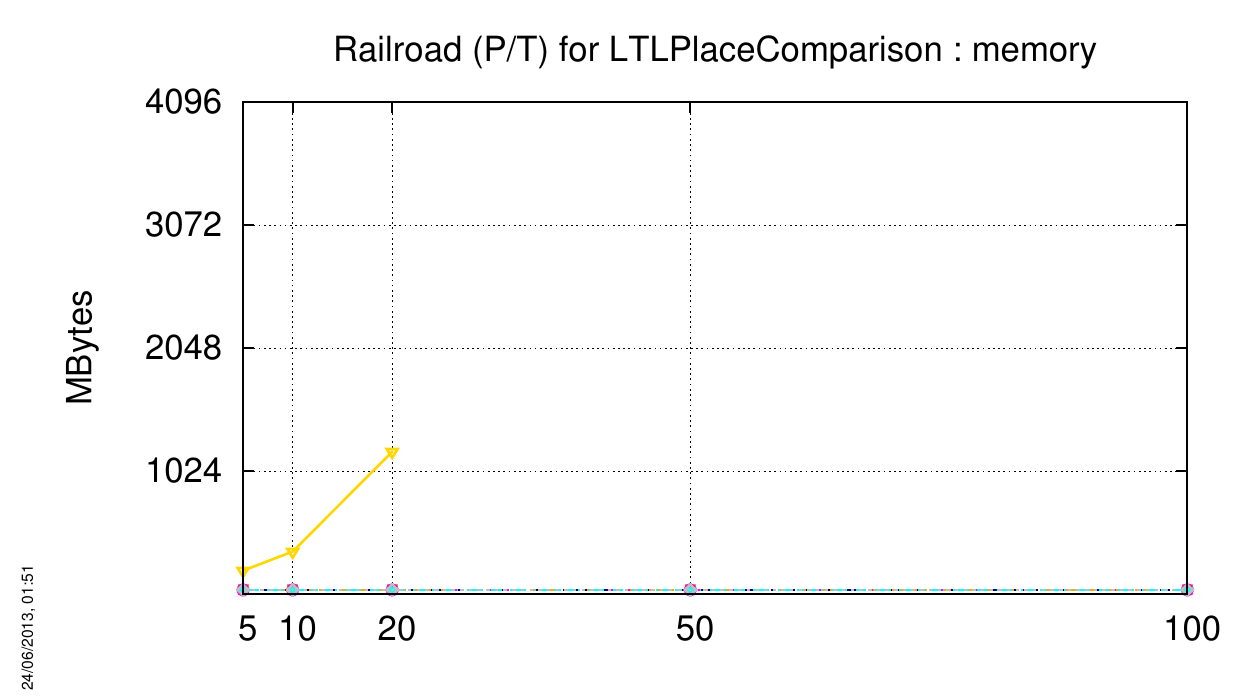}
   \includegraphics[width=7.2cm]{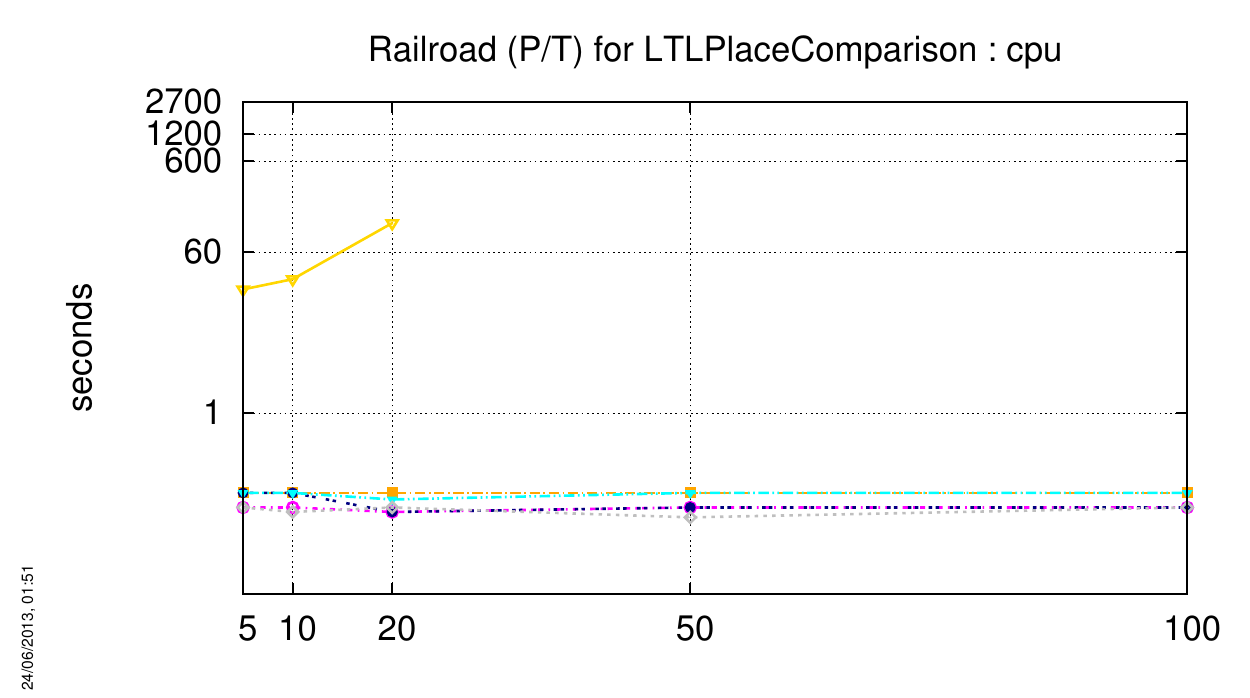}

   \includegraphics[height=1cm]{figures/tools-legend.pdf}
\end{center}

\subsubsection{\acs{RessAllocation-PT}}
The charts below respectively show how tools compete with this ``Known'' model (memory and CPU).

\index{Performances!LTLPlaceComparison!RessAllocation (P/T)}
\begin{center}
   \includegraphics[width=7.2cm]{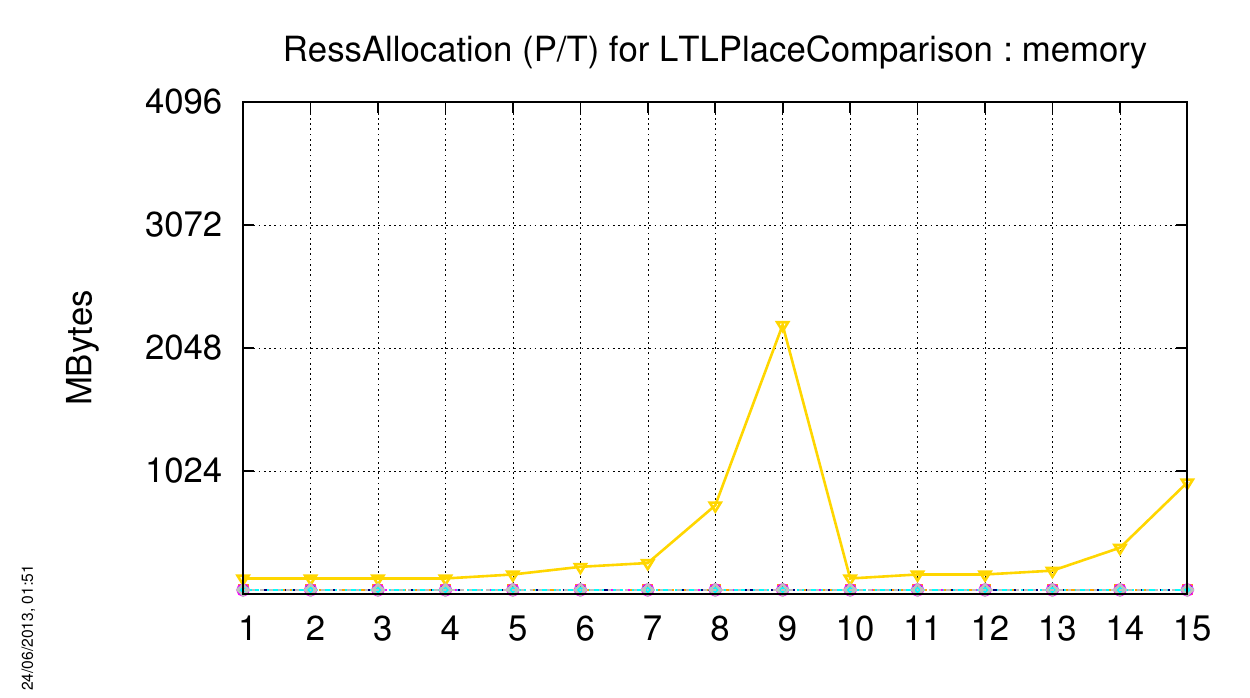}
   \includegraphics[width=7.2cm]{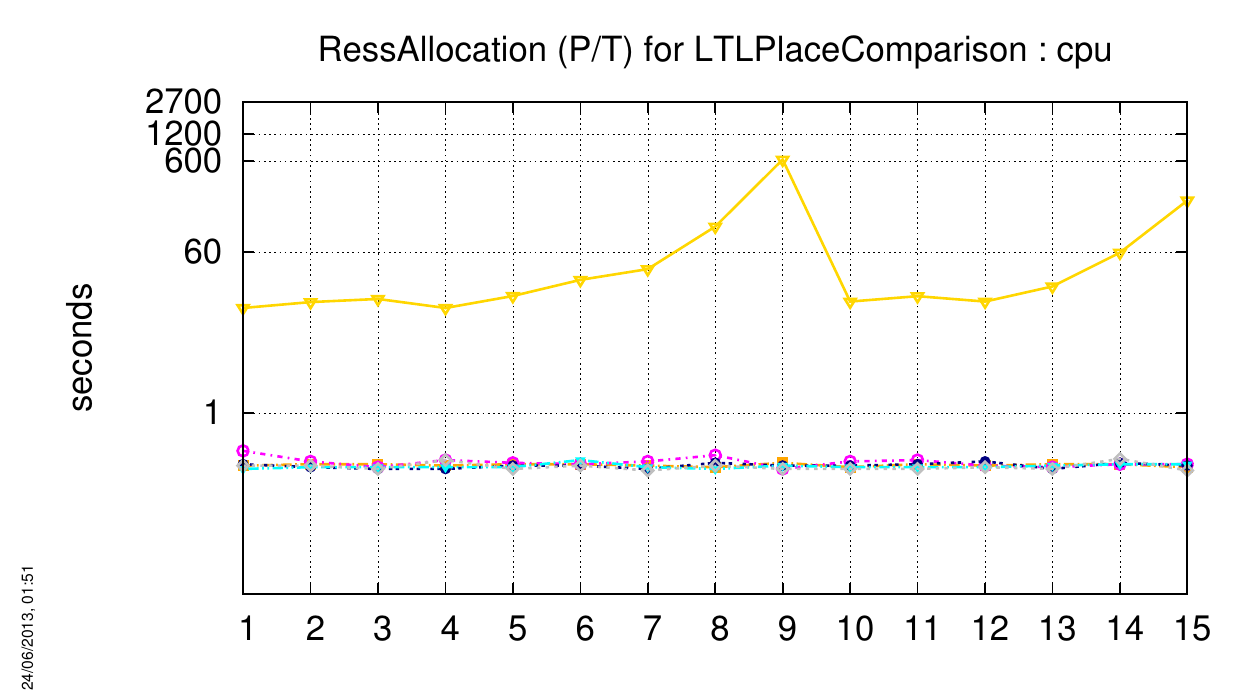}

   \includegraphics[height=1cm]{figures/tools-legend.pdf}
\end{center}

\subsubsection{\acs{Ring-PT}}
The charts below respectively show how tools compete with this ``Known'' model (memory and CPU).

\index{Performances!LTLPlaceComparison!Ring (P/T)}
\begin{center}
   \includegraphics[width=7.2cm]{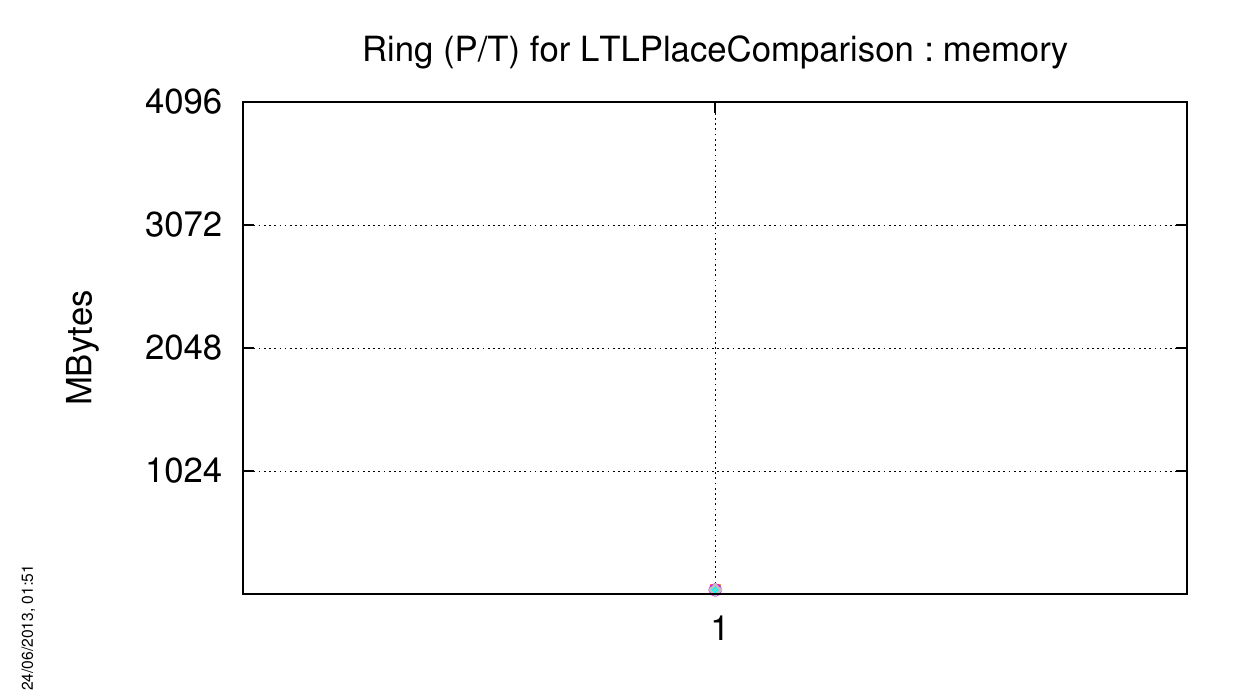}
   \includegraphics[width=7.2cm]{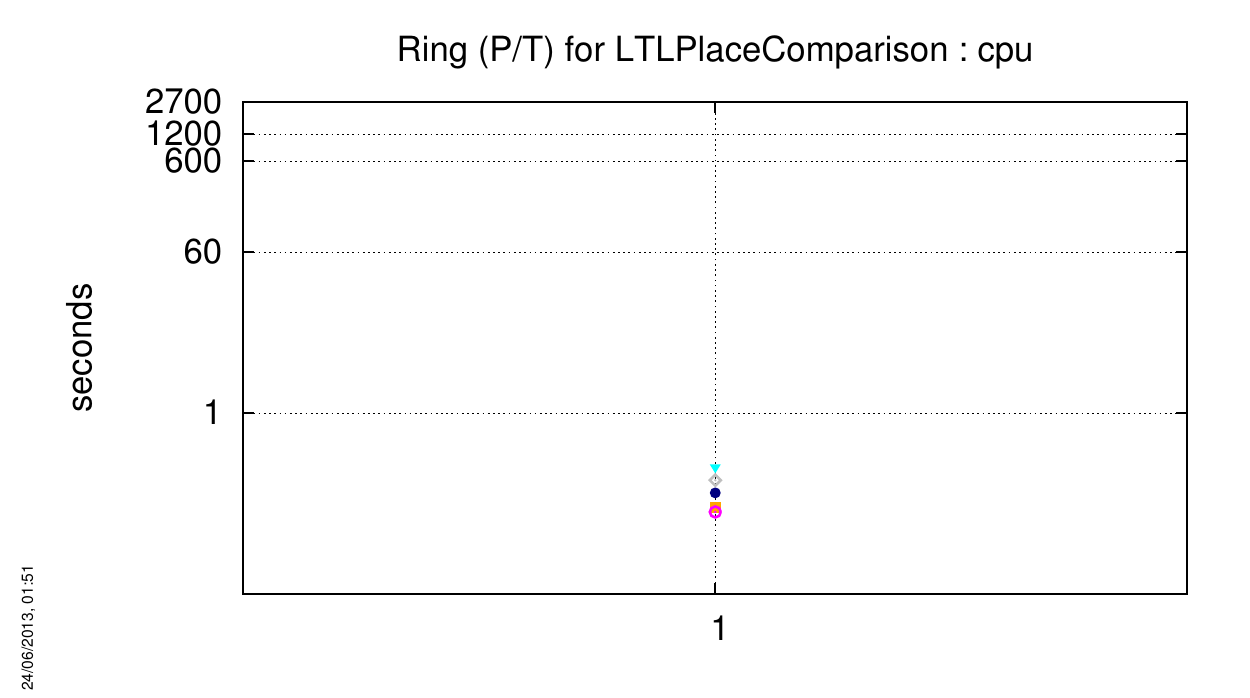}

   \includegraphics[height=1cm]{figures/tools-legend.pdf}
\end{center}

\subsubsection{\acs{RwMutex-PT}}
The charts below respectively show how tools compete with this ``Known'' model (memory and CPU).

\index{Performances!LTLPlaceComparison!RwMutex (P/T)}
\begin{center}
   \includegraphics[width=7.2cm]{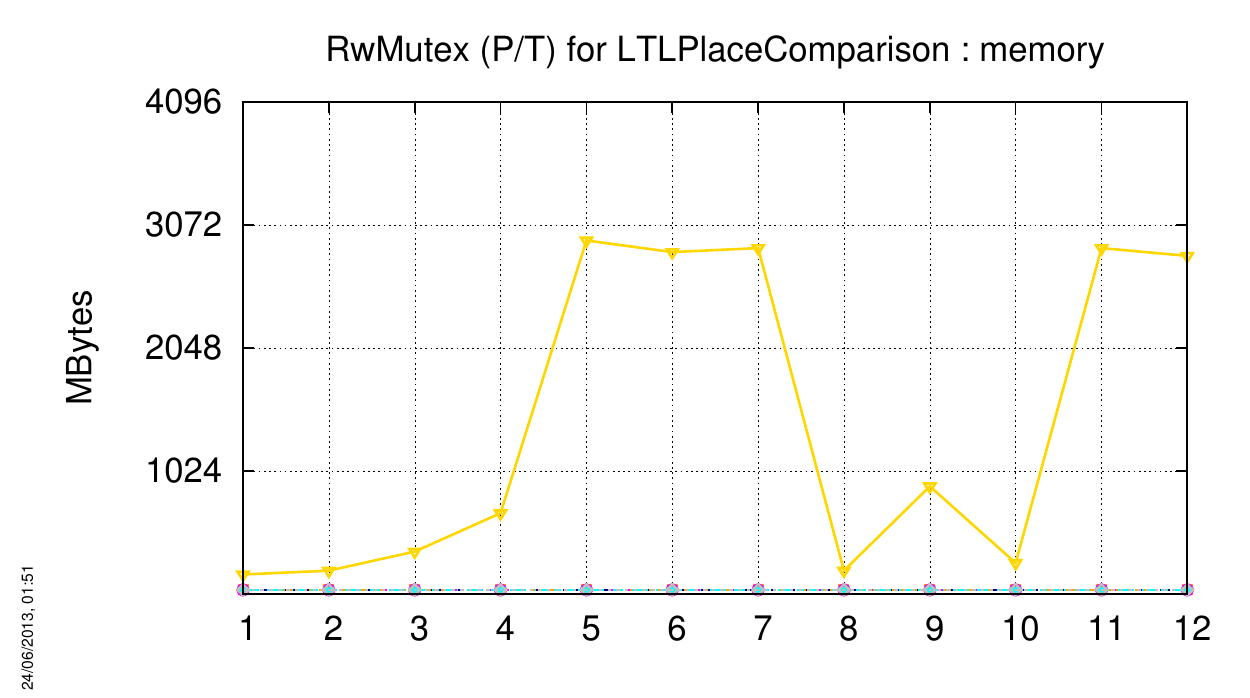}
   \includegraphics[width=7.2cm]{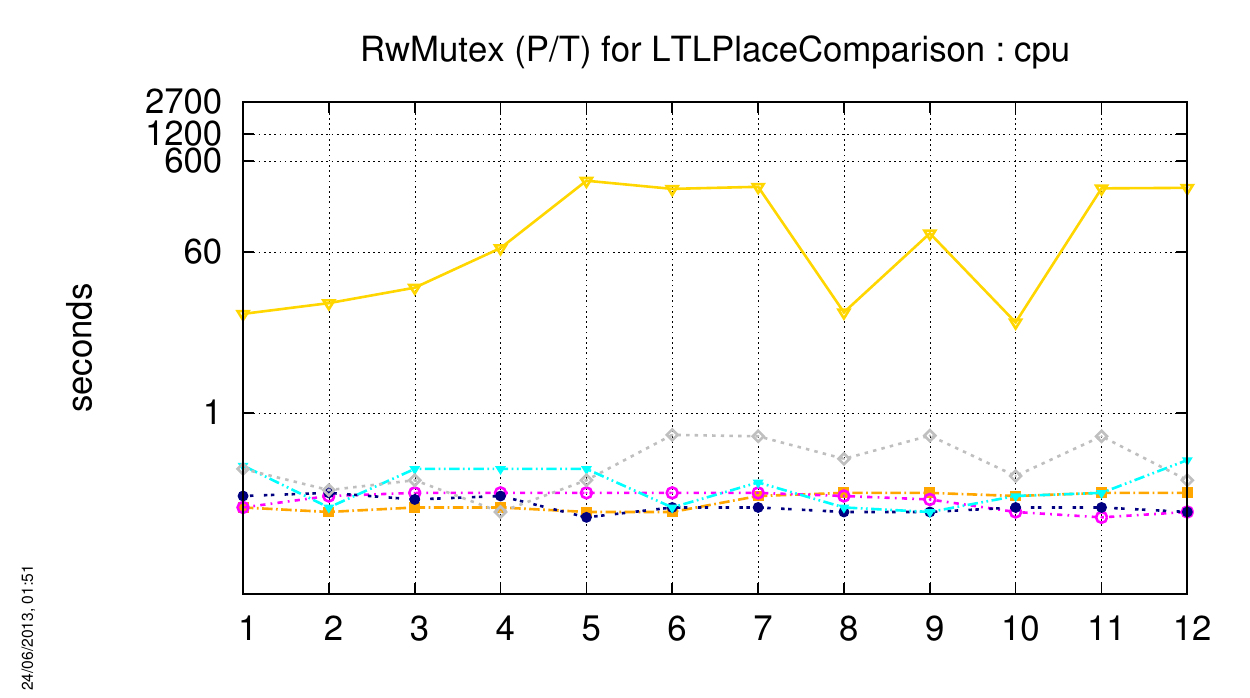}

   \includegraphics[height=1cm]{figures/tools-legend.pdf}
\end{center}

\subsubsection{\acs{SharedMemory-COL}}
No instance of this model could be computed for the \textbf{LTLPlaceComparison} examination.

\subsubsection{\acs{SharedMemory-PT}}
The charts below respectively show how tools compete with this ``Known'' model (memory and CPU).

\index{Performances!LTLPlaceComparison!SharedMemory (P/T)}
\begin{center}
   \includegraphics[width=7.2cm]{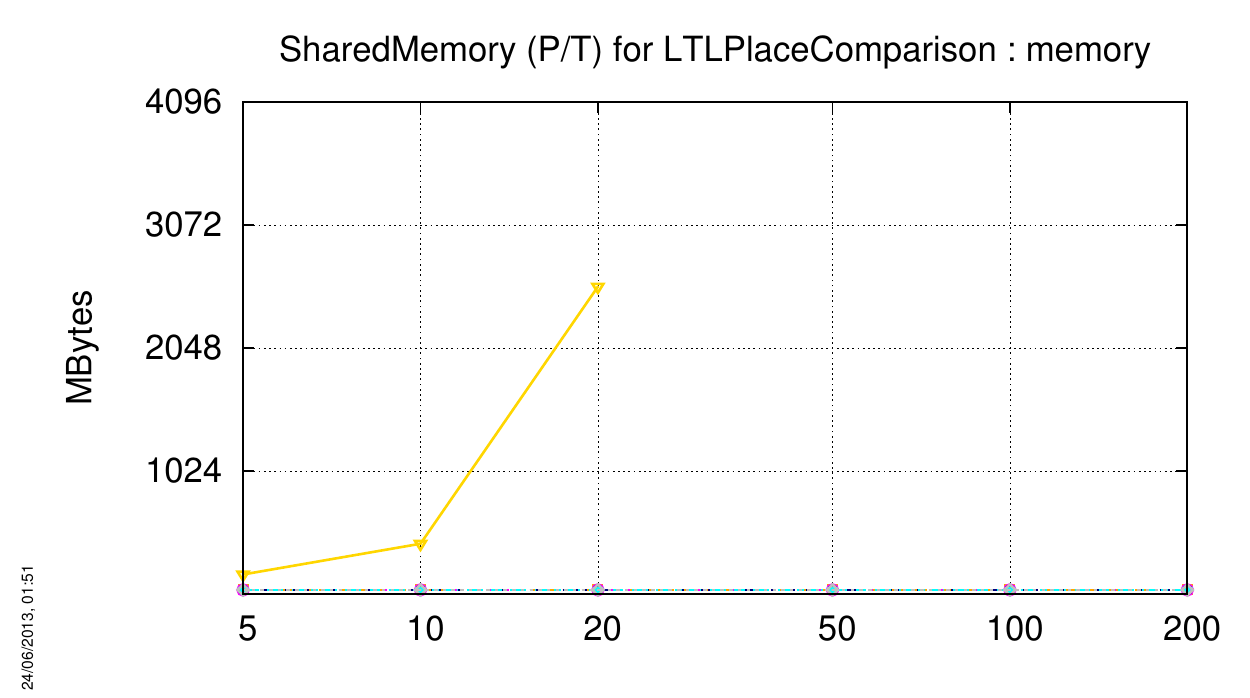}
   \includegraphics[width=7.2cm]{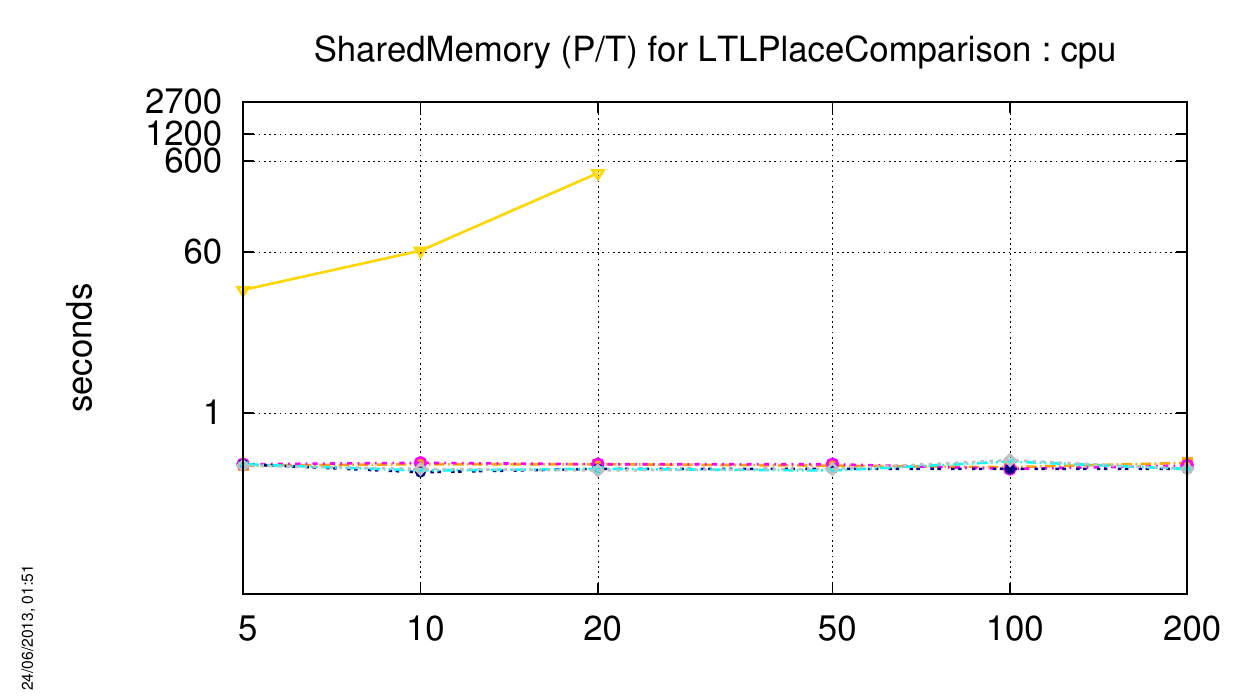}

   \includegraphics[height=1cm]{figures/tools-legend.pdf}
\end{center}

\subsubsection{\acs{SimpleLoadBal-COL}}
No instance of this model could be computed for the \textbf{LTLPlaceComparison} examination.

\subsubsection{\acs{SimpleLoadBal-PT}}
The charts below respectively show how tools compete with this ``Known'' model (memory and CPU).

\index{Performances!LTLPlaceComparison!SimpleLoadBal (P/T)}
\begin{center}
   \includegraphics[width=7.2cm]{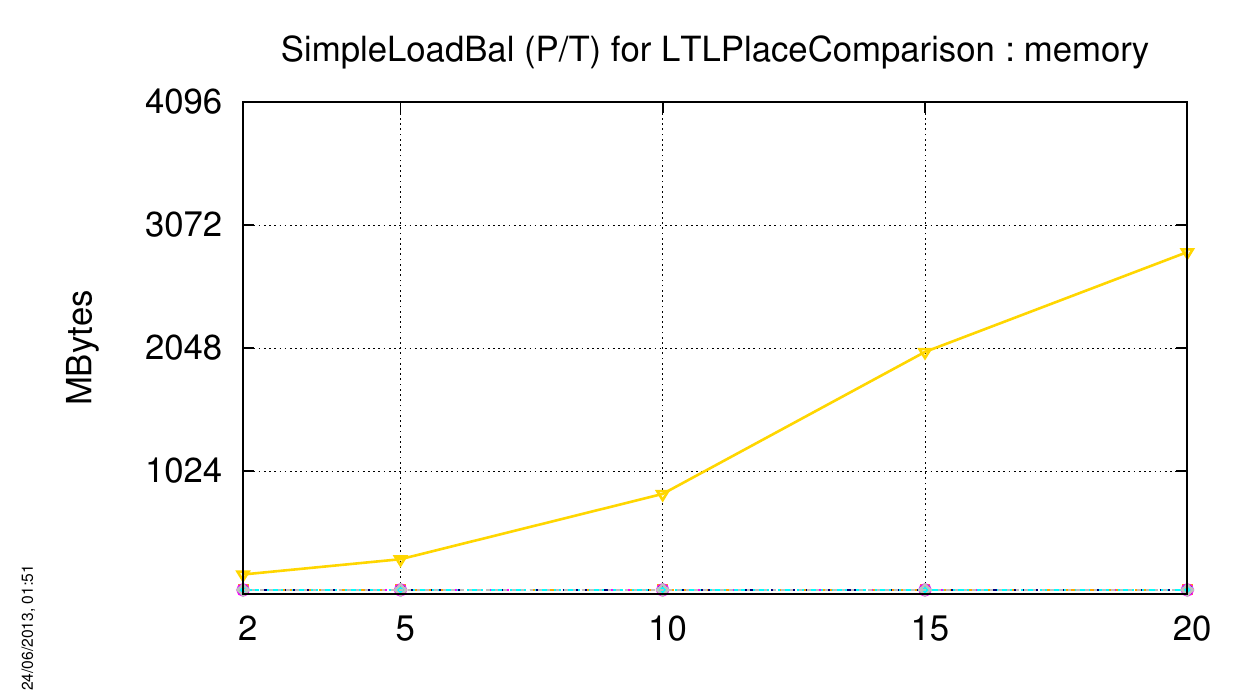}
   \includegraphics[width=7.2cm]{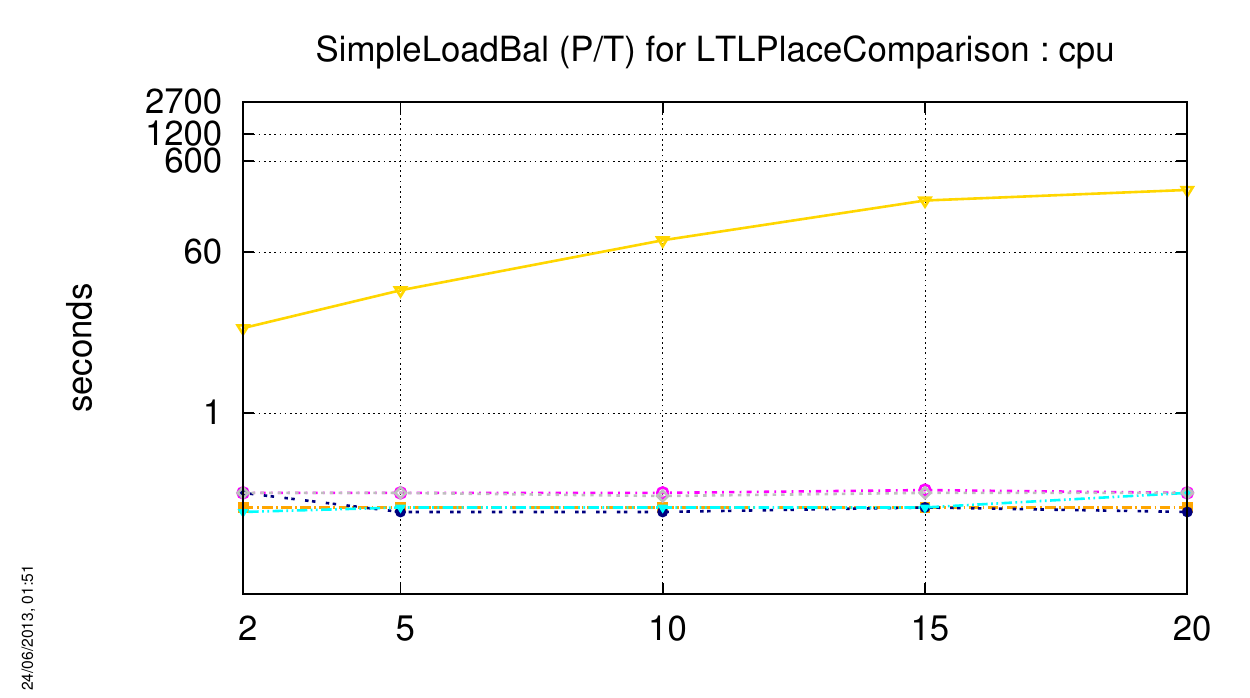}

   \includegraphics[height=1cm]{figures/tools-legend.pdf}
\end{center}

\subsubsection{\acs{TokenRing-COL}}
No instance of this model could be computed for the \textbf{LTLPlaceComparison} examination.

\subsubsection{\acs{TokenRing-PT}}
No instance of this model could be computed for the \textbf{LTLPlaceComparison} examination.

\subsubsection{\acs{HouseConstruction-PT}}
The charts below respectively show how tools compete with this ``Suprise'' model (memory and CPU).

\index{Performances!LTLPlaceComparison!HouseConstruction (P/T)}
\begin{center}
   \includegraphics[width=7.2cm]{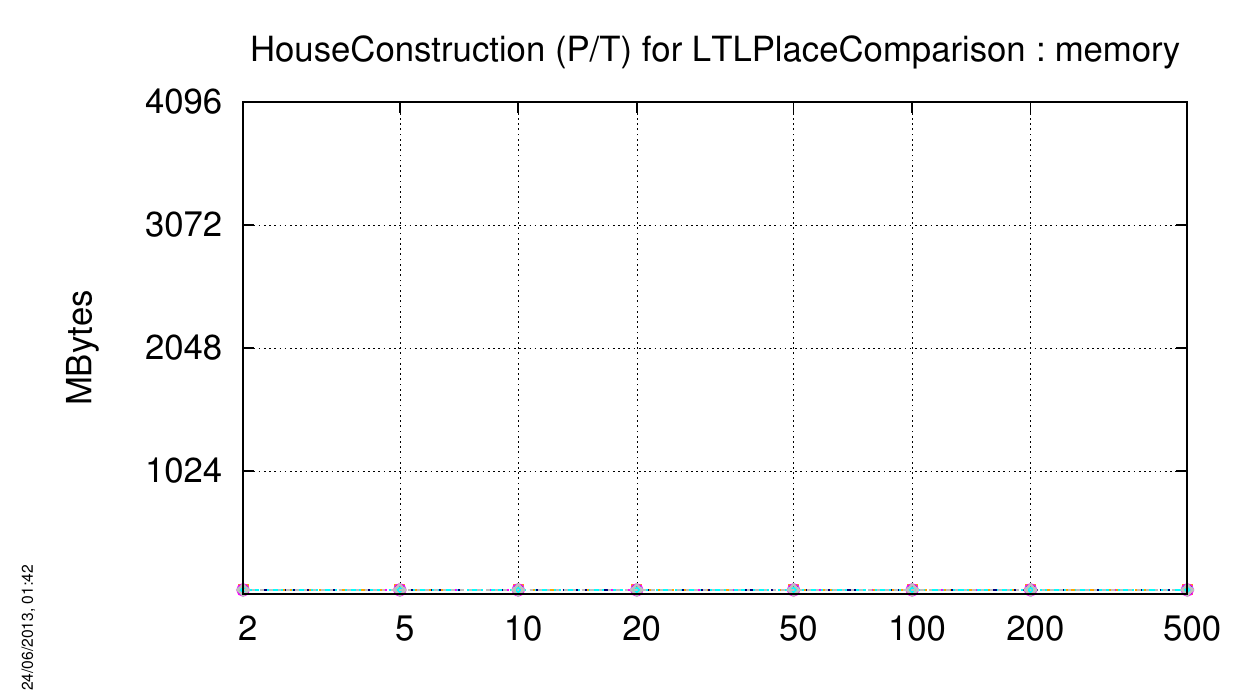}
   \includegraphics[width=7.2cm]{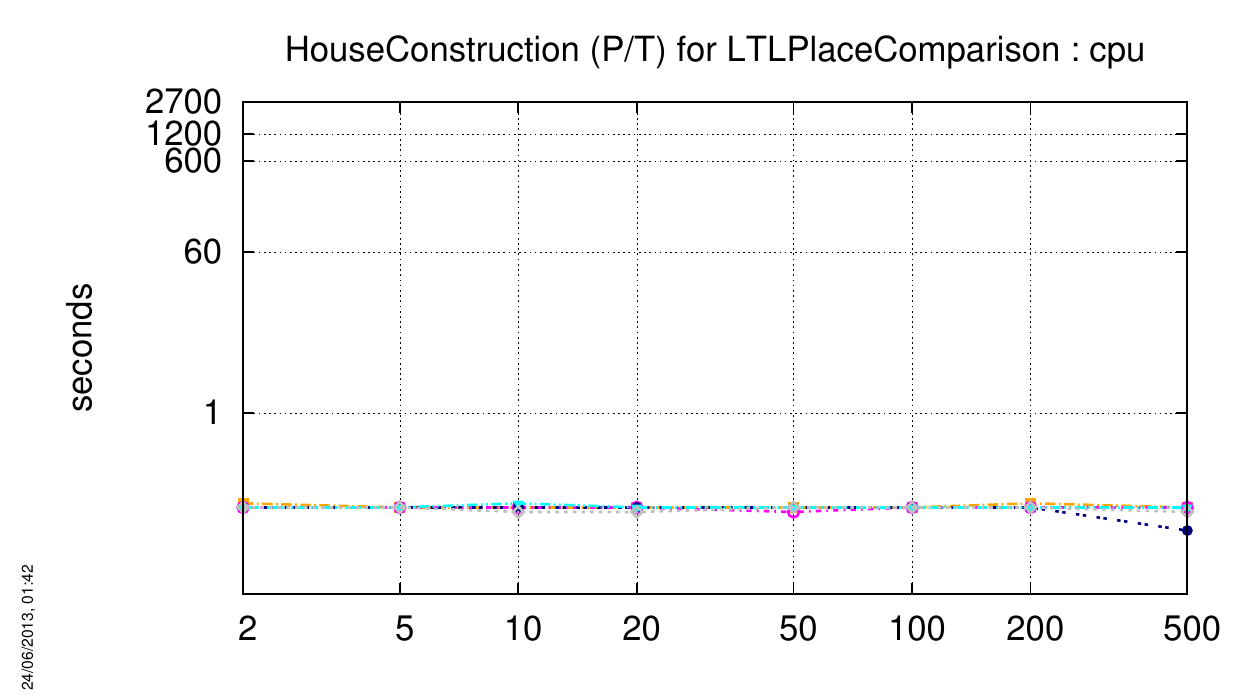}

   \includegraphics[height=1cm]{figures/tools-legend.pdf}
\end{center}

\subsubsection{\acs{IBMB2S565S3960-PT}}
The charts below respectively show how tools compete with this ``Suprise'' model (memory and CPU).

\index{Performances!LTLPlaceComparison!IBMB2S565S3960 (P/T)}
\begin{center}
   \includegraphics[width=7.2cm]{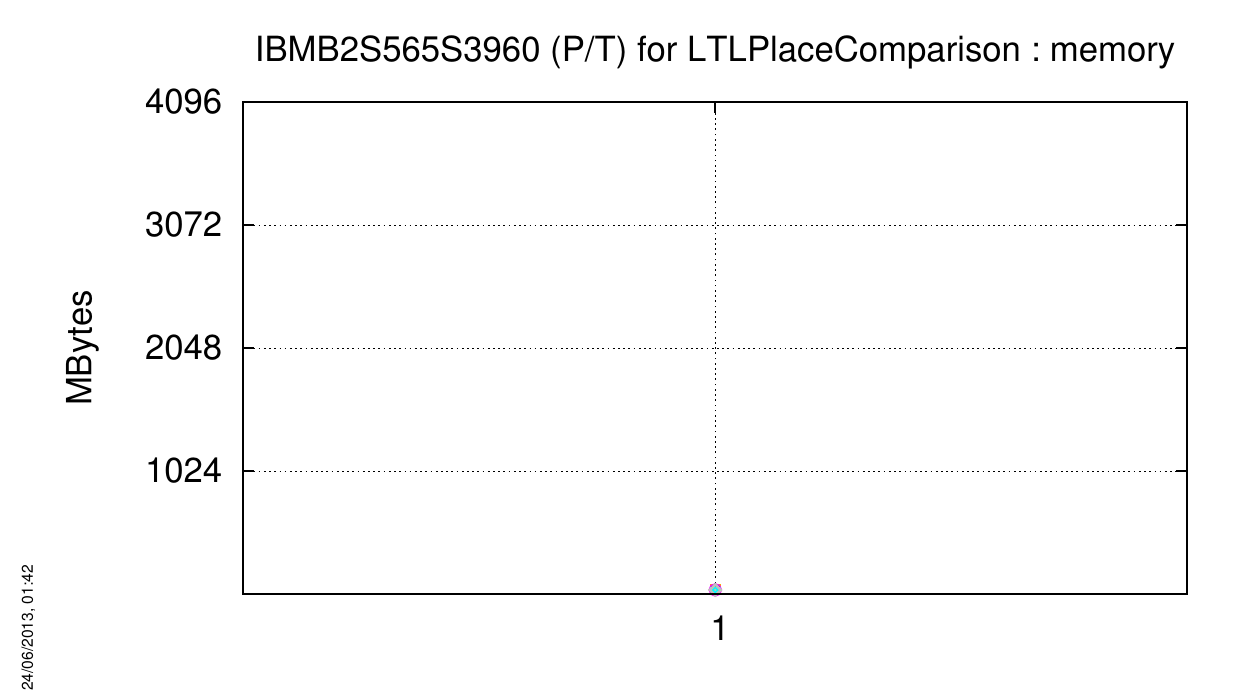}
   \includegraphics[width=7.2cm]{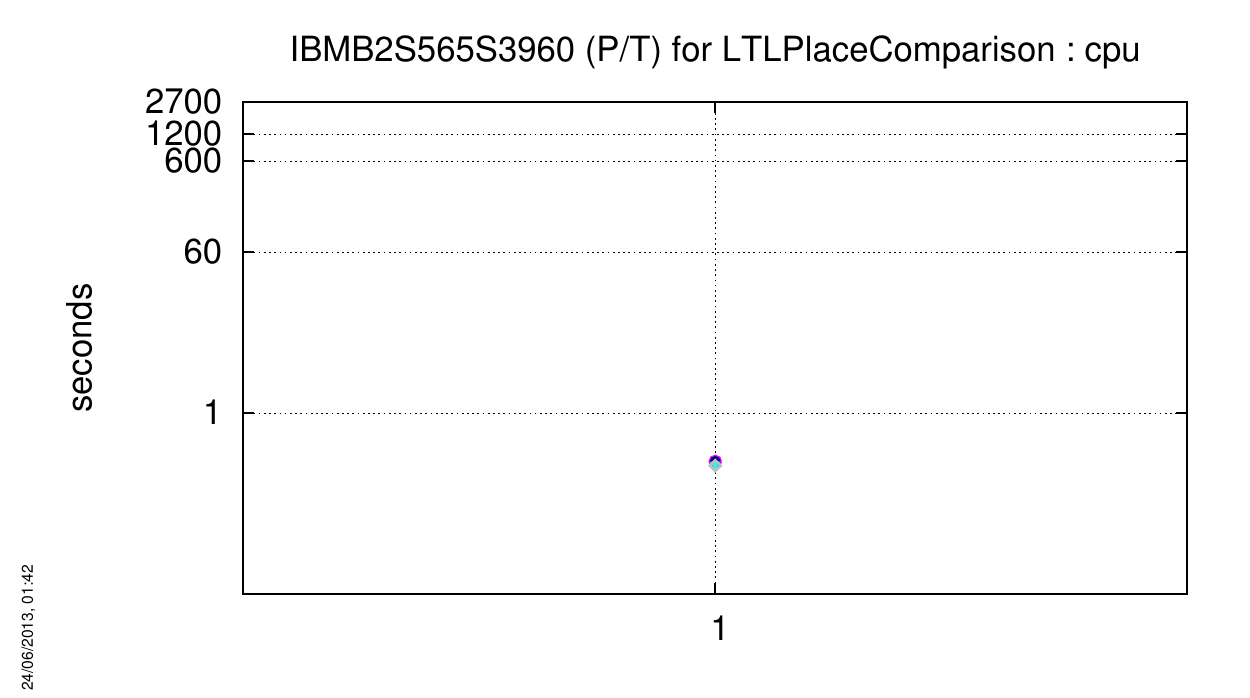}

   \includegraphics[height=1cm]{figures/tools-legend.pdf}
\end{center}

\subsubsection{\acs{QuasiCertifProtocol-COL}}
No instance of this model could be computed for the \textbf{LTLPlaceComparison} examination.

\subsubsection{\acs{QuasiCertifProtocol-PT}}
The charts below respectively show how tools compete with this ``Suprise'' model (memory and CPU).

\index{Performances!LTLPlaceComparison!QuasiCertifProtocol (P/T)}
\begin{center}
   \includegraphics[width=7.2cm]{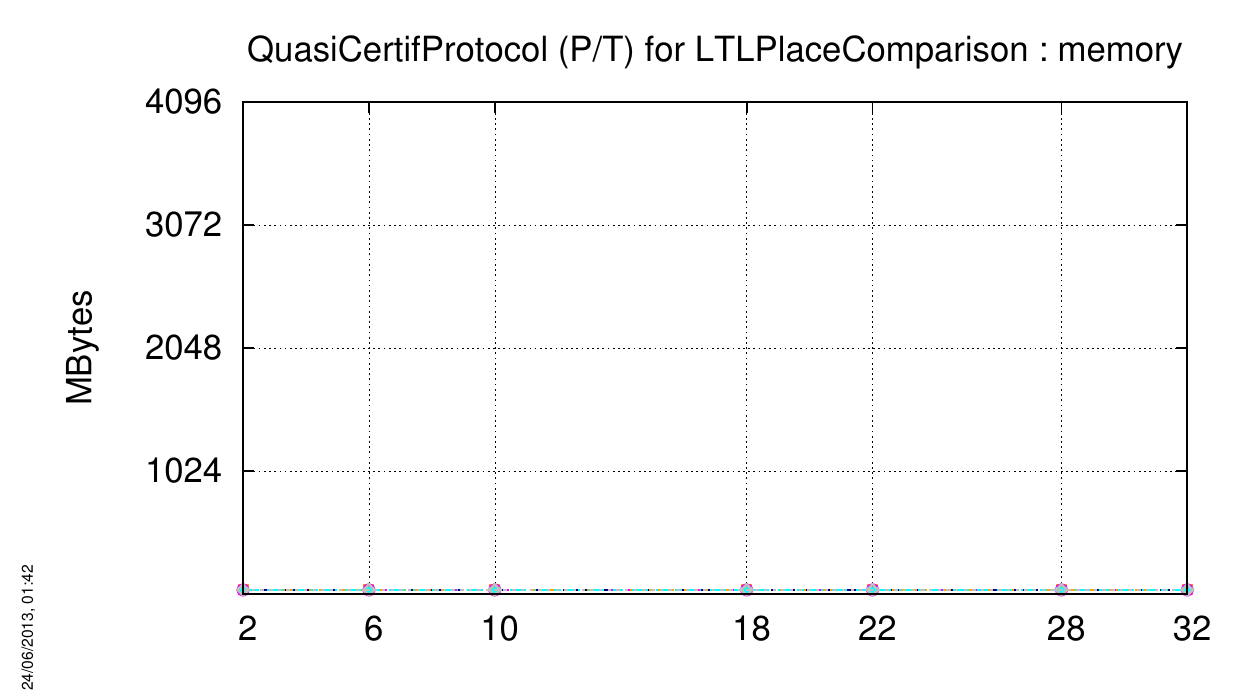}
   \includegraphics[width=7.2cm]{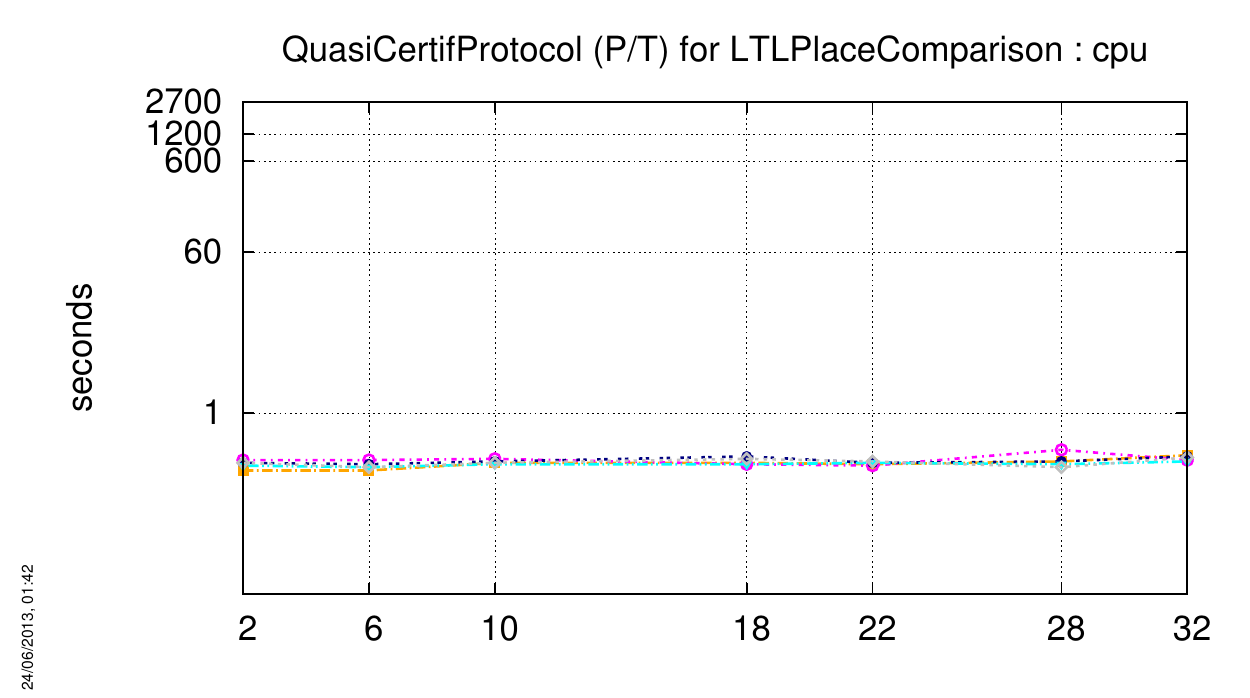}

   \includegraphics[height=1cm]{figures/tools-legend.pdf}
\end{center}

\subsubsection{\acs{Vasy2003-PT}}
The charts below respectively show how tools compete with this ``Suprise'' model (memory and CPU).

\index{Performances!LTLPlaceComparison!Vasy2003 (P/T)}
\begin{center}
   \includegraphics[width=7.2cm]{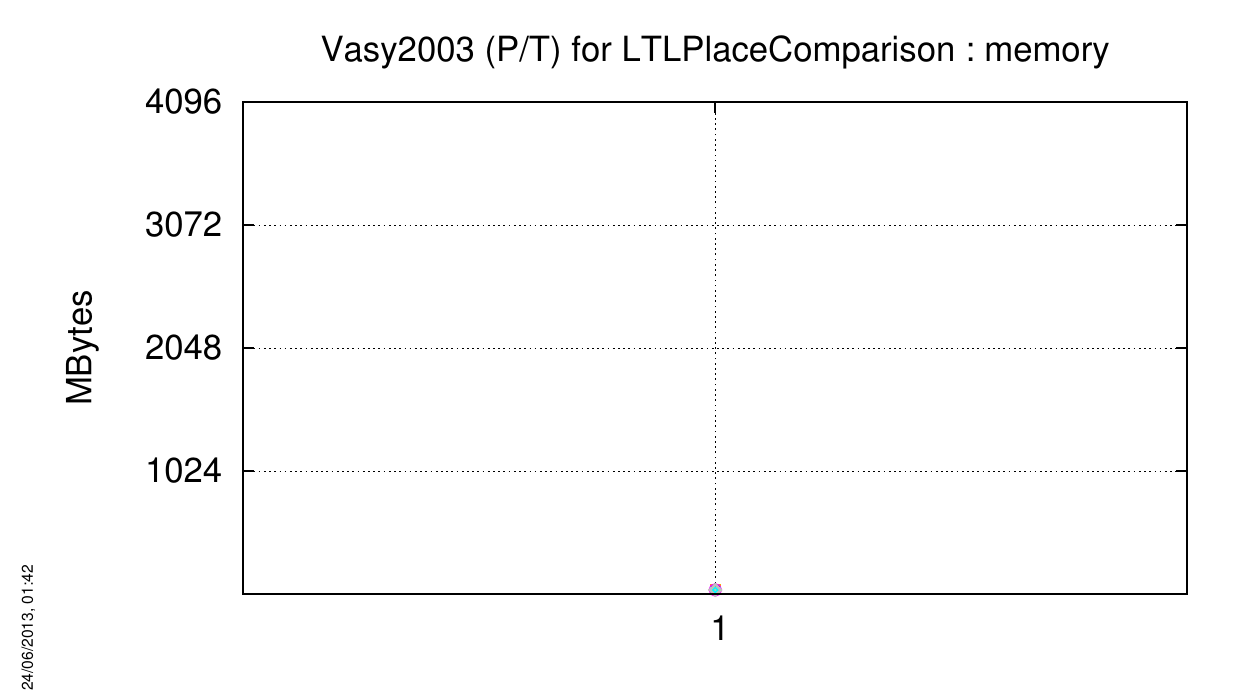}
   \includegraphics[width=7.2cm]{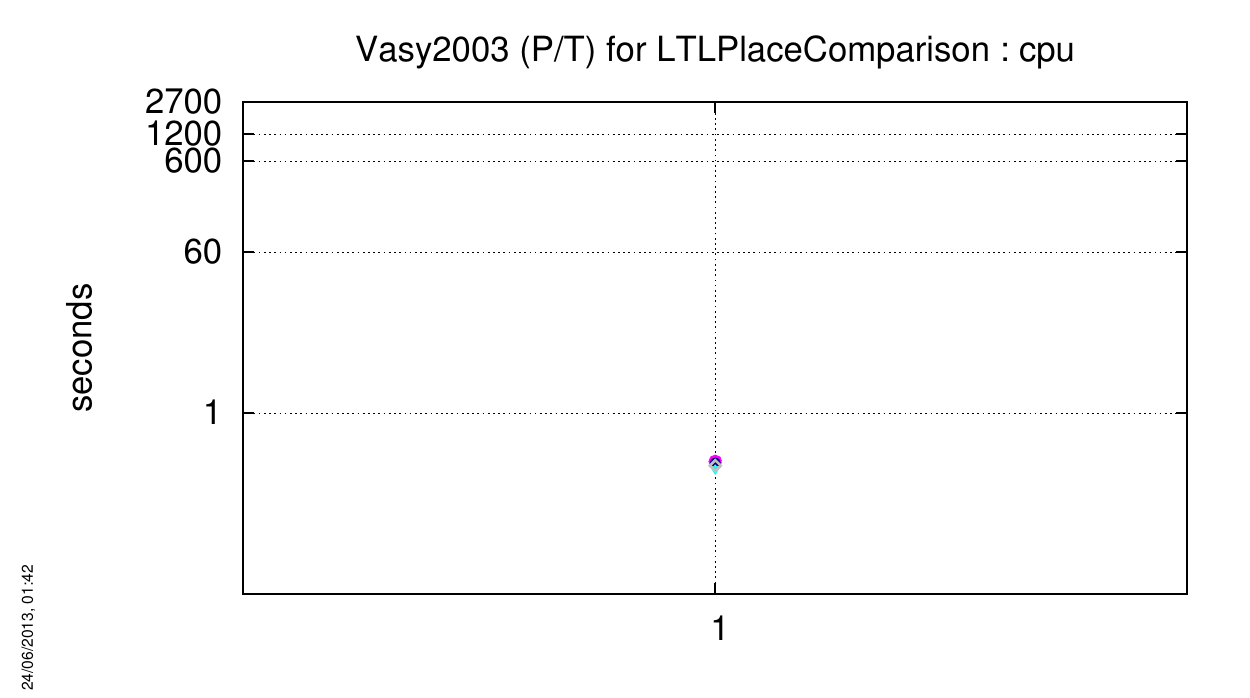}

   \includegraphics[height=1cm]{figures/tools-legend.pdf}
\end{center}

\subsection{Outputs for the LTLPlaceComparison Examination}
\index{Outputs!LTLPlaceComparison}

Please find enclosed the brute results for this examination (``Known'' and ``Surprise'' models).
We display only the score of tools that provide a results for at least one instance of one model.
The legend for the values is provided below:
\begin{itemize}
   \item\textbf{nc}: the tool does not compete this examination for this model/instance,
   \item\textbf{cc}: the tool cannot compute this examination for this model/instance,
   \item\textbf{to}: the tool cannot compute this examination for this model/instance within the maximum allowed time,
   \item\textbf{mp}: the tool encountered a memory problem (stack overflow or memory full),
   \item\textbf{nf}: there is no formula available for this type of examination (typically, this concerns P/T nets where
       comparing marking cardinality has no signification when there is no equivalent colored net).
\end{itemize}

\textbf{Note on the display of results for formulas:} each formula is considered as a flag (F if false, T if true, - or ?
when the value cannot be determined). These values are concatenated in the order they appear (we assume it is the order of formulas as they were provided).

\subsubsection{``Known'' Models}

\input{result_known_LTLPlaceComparison.tex}

\subsubsection{``Surprise'' Models}

\input{result_surprise_LTLPlaceComparison.tex}

\subsection{Score for the LTLPlaceComparison Examination}
\index{Scores!LTLPlaceComparison}

Please find enclosed the scores for this examination (``Known'' and ``Surprise'' models).
We display only the score of tools that provide a results for at least one instance of one model.
The total is first listed in the table below followed by a detail, for each proposed model.
Meaning of the line labels are:
\begin{itemize}
\item\textbf{1st instance}: the tool gets a bonus for having processed the first instance of this model (+1 point),
\item\textbf{instances}: the tool gets 1 point per instances treated 
(for that, we assume that at least one formula has been successfully computed),
\item\textbf{max reached}: the tool could process all the instances for the model (+2 points),
\item\textbf{best}: the tool is among the ones that processed a maximum of instances within the time and memory confinement (+2 points).
\end{itemize}

\subsubsection{``Known'' Models}

\input{score_known_LTLPlaceComparison.tex}

\subsubsection{``Surprise'' Models}

\input{score_surprise_LTLPlaceComparison.tex}

\subsection{Trophies for this Examination}
\index{Trophies!LTLPlaceComparison}

Trophies are divided in three categories: ``Known'' models,
``Surprise'' models, and the global trophies (formula is then
$score_{global} = score_{known} + 2 \times score_{surprise}$).

\subsubsection{For ``Known'' Models} \ \\

\begin{tabular}{c}
      1 \\
   \includegraphics[width=2cm]{figures/gold.jpg} \\
   \acs{neco} \\
   114 points \\
\end{tabular}

\subsubsection{For ``Surprise'' Models}\  \\

No tool could complete this examination.

\subsubsection{Global} \ \\

\begin{tabular}{c}
      1 \\
   \includegraphics[width=2cm]{figures/gold.jpg} \\
   \acs{neco} \\
   114 points \\
\end{tabular}

\newpage

\section{The LTLMix Examination}
\label{sec:exam:LTLMix}
\index{Results!LTLMix}

This examination deals with LTL properties dealing with all the previous type of atomic proposition.
We first show a summary on the handling of models by the participating tools.
Then, we present the computed outputs and the associated scores for this
examination prior to a summary of relevant executions.

\subsection{Handling of Models by Tools}
\index{Performances!LTLMix}

\subsubsection{\acs{CSRepetitions-COL}}
No instance of this model could be computed for the \textbf{LTLMix} examination.

\subsubsection{\acs{CSRepetitions-PT}}
The charts below respectively show how tools compete with this ``Known'' model (memory and CPU).

\index{Performances!LTLMix!CSRepetitions (P/T)}
\begin{center}
   \includegraphics[width=7.2cm]{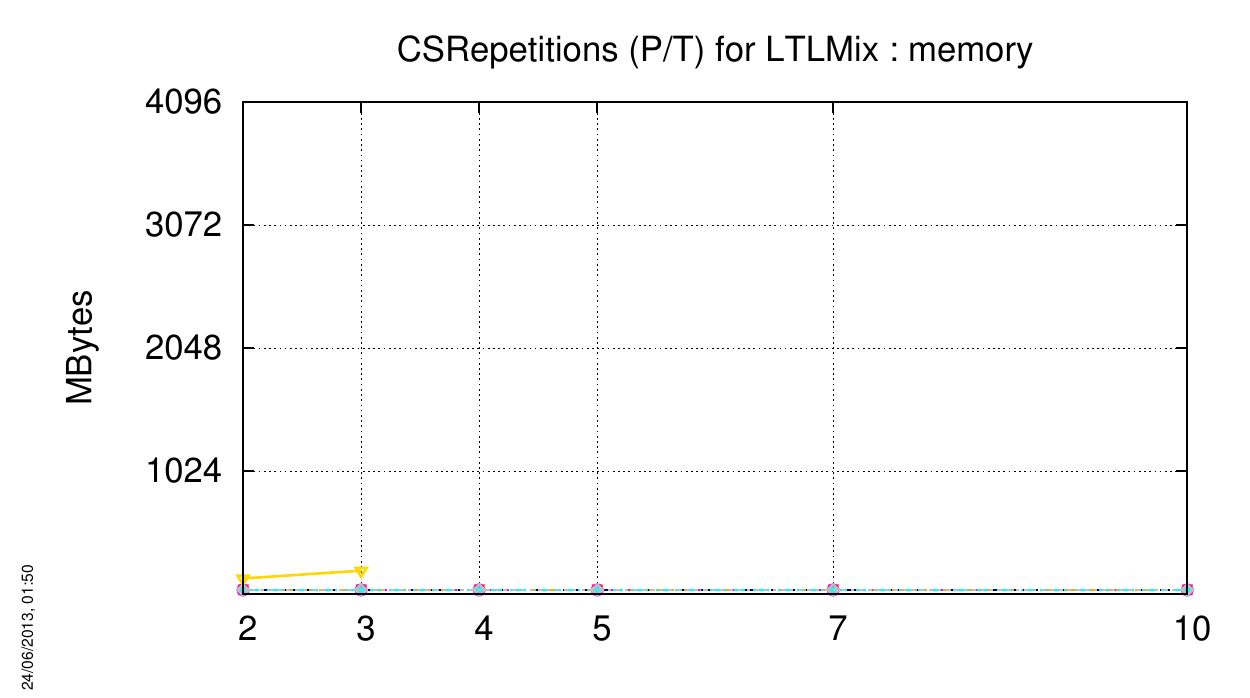}
   \includegraphics[width=7.2cm]{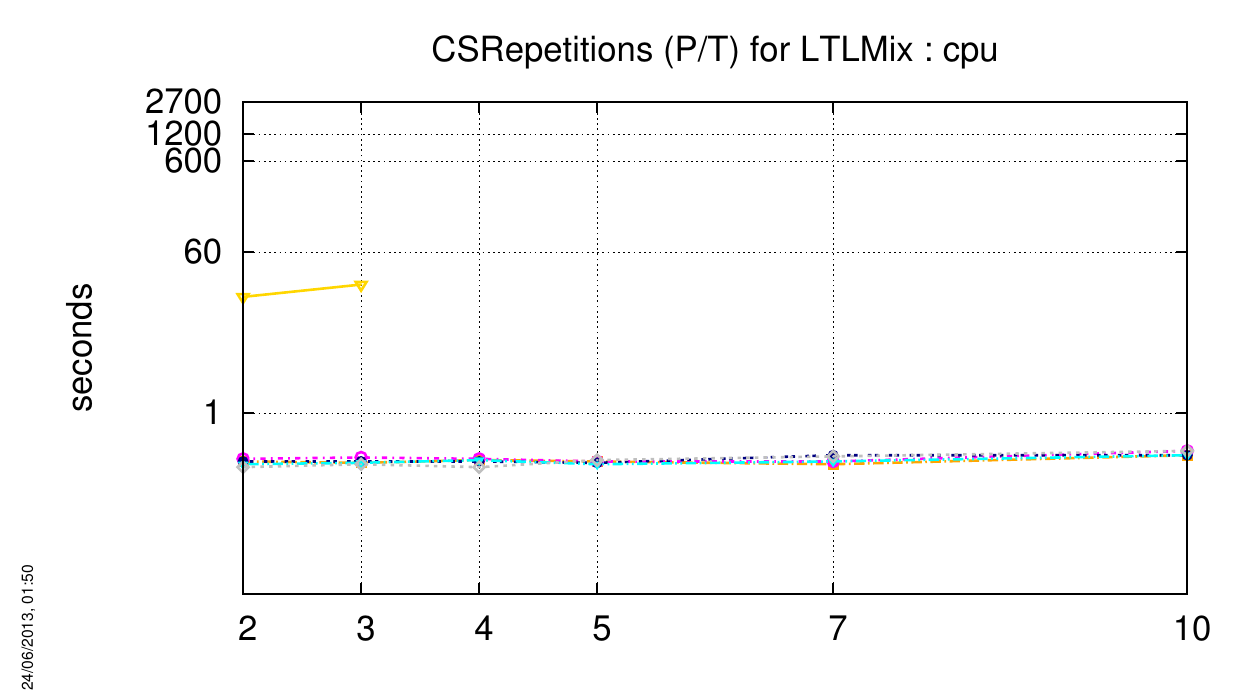}

   \includegraphics[height=1cm]{figures/tools-legend.pdf}
\end{center}

\subsubsection{\acs{Dekker-PT}}
The charts below respectively show how tools compete with this ``Known'' model (memory and CPU).

\index{Performances!LTLMix!Dekker (P/T)}
\begin{center}
   \includegraphics[width=7.2cm]{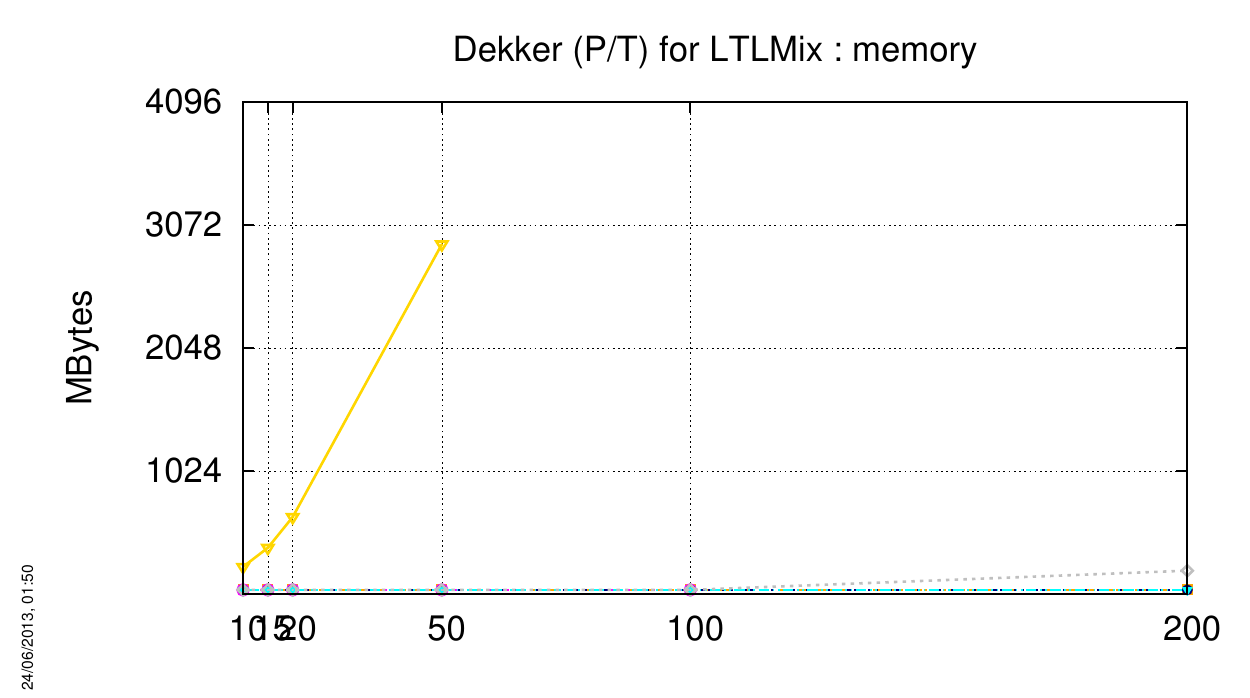}
   \includegraphics[width=7.2cm]{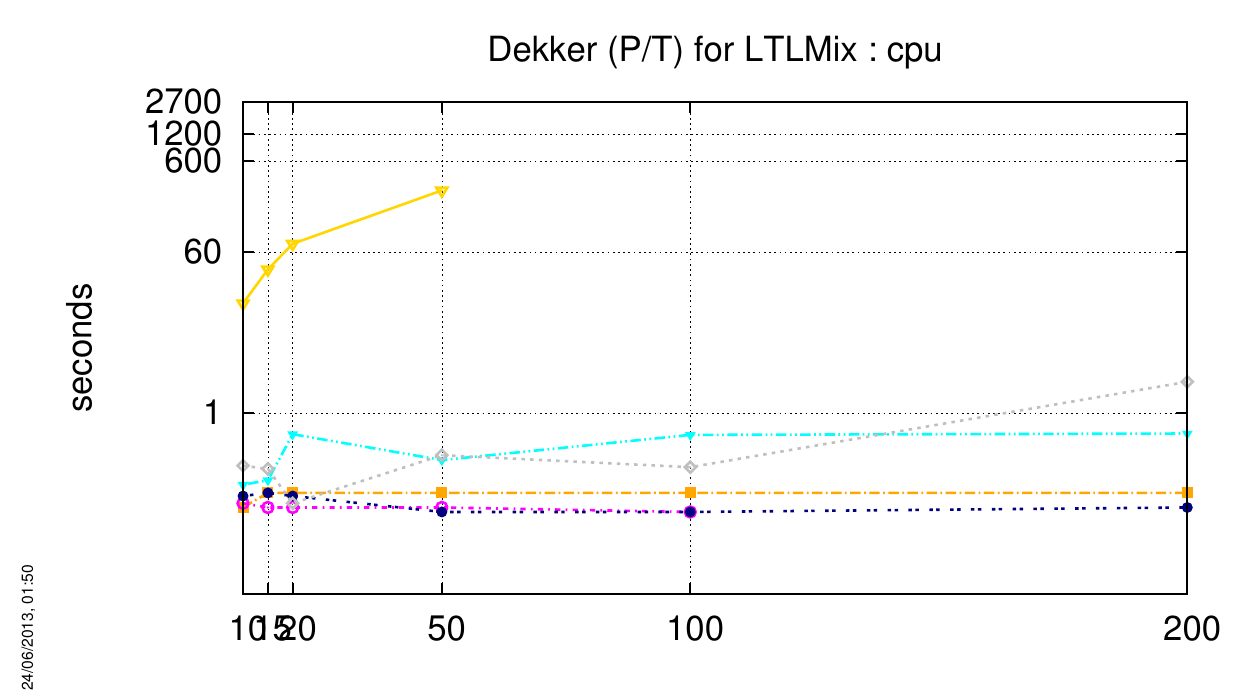}

   \includegraphics[height=1cm]{figures/tools-legend.pdf}
\end{center}

\subsubsection{\acs{DotAndBoxes-COL}}
No instance of this model could be computed for the \textbf{LTLMix} examination.

\subsubsection{\acs{DrinkVendingMachine-COL}}
No instance of this model could be computed for the \textbf{LTLMix} examination.

\subsubsection{\acs{DrinkVendingMachine-PT}}
The charts below respectively show how tools compete with this ``Known'' model (memory and CPU).

\index{Performances!LTLMix!DrinkVendingMachine (P/T)}
\begin{center}
   \includegraphics[width=7.2cm]{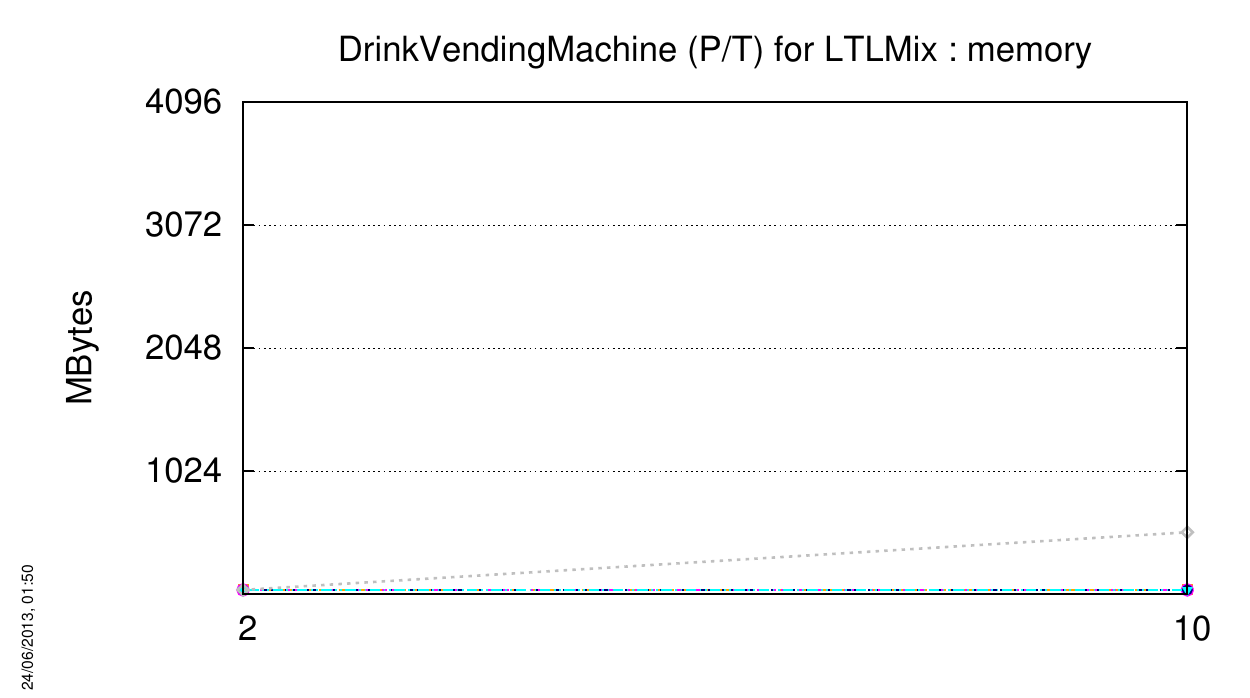}
   \includegraphics[width=7.2cm]{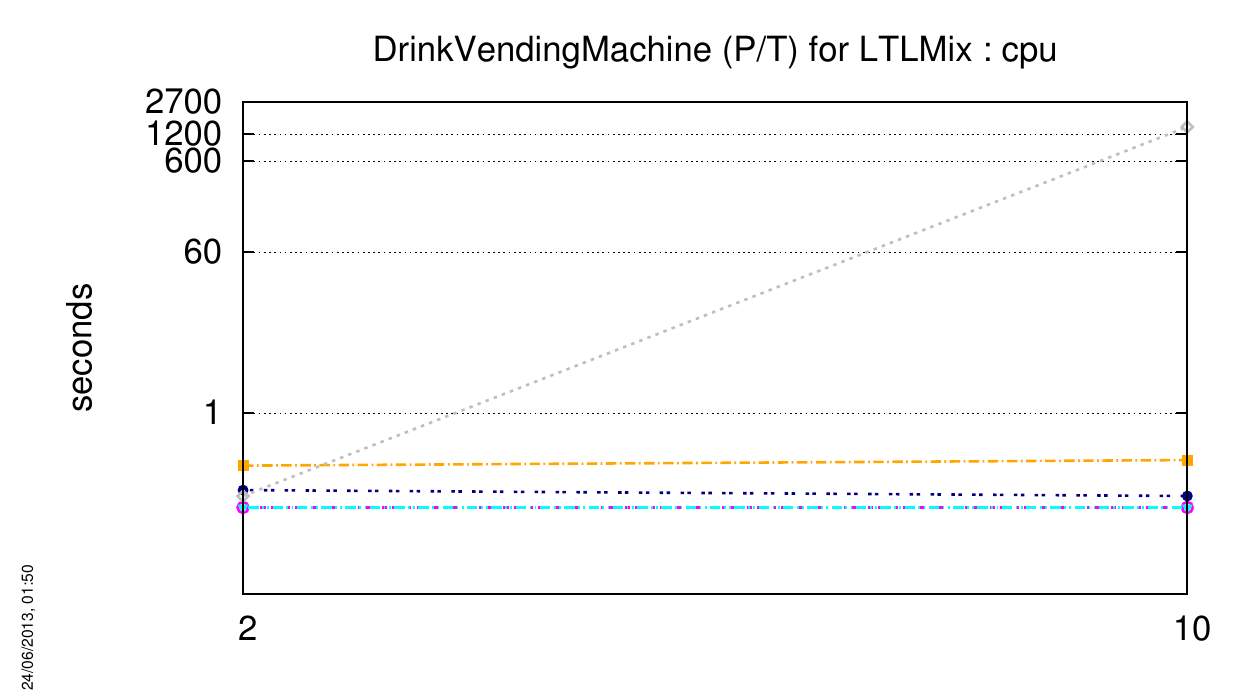}

   \includegraphics[height=1cm]{figures/tools-legend.pdf}
\end{center}

\subsubsection{\acs{Echo-PT}}
The charts below respectively show how tools compete with this ``Known'' model (memory and CPU).

\index{Performances!LTLMix!Echo (P/T)}
\begin{center}
   \includegraphics[width=7.2cm]{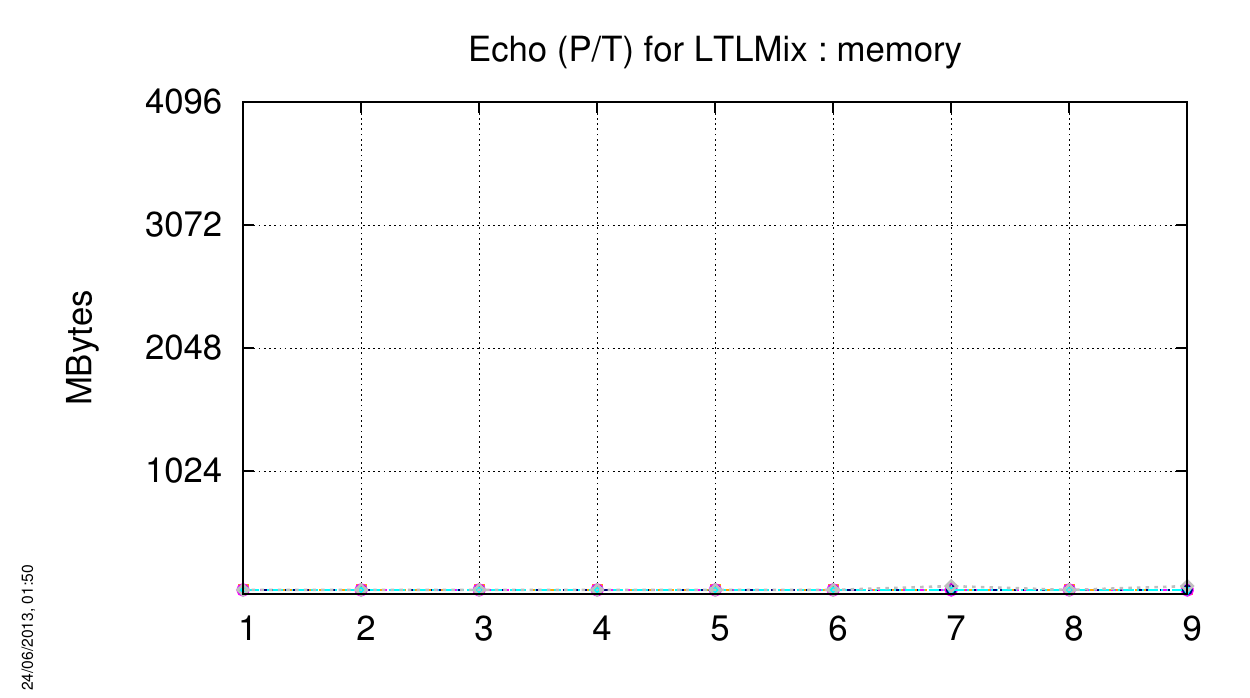}
   \includegraphics[width=7.2cm]{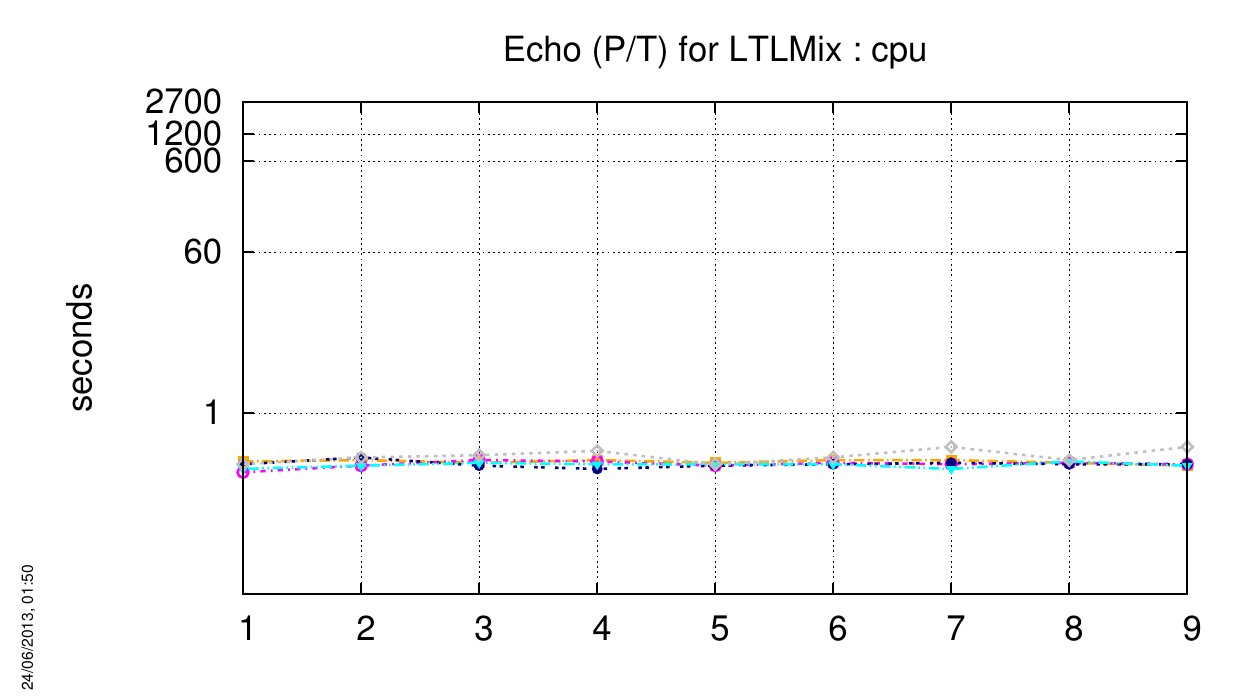}

   \includegraphics[height=1cm]{figures/tools-legend.pdf}
\end{center}

\subsubsection{\acs{Eratosthenes-PT}}
The charts below respectively show how tools compete with this ``Known'' model (memory and CPU).

\index{Performances!LTLMix!Eratosthenes (P/T)}
\begin{center}
   \includegraphics[width=7.2cm]{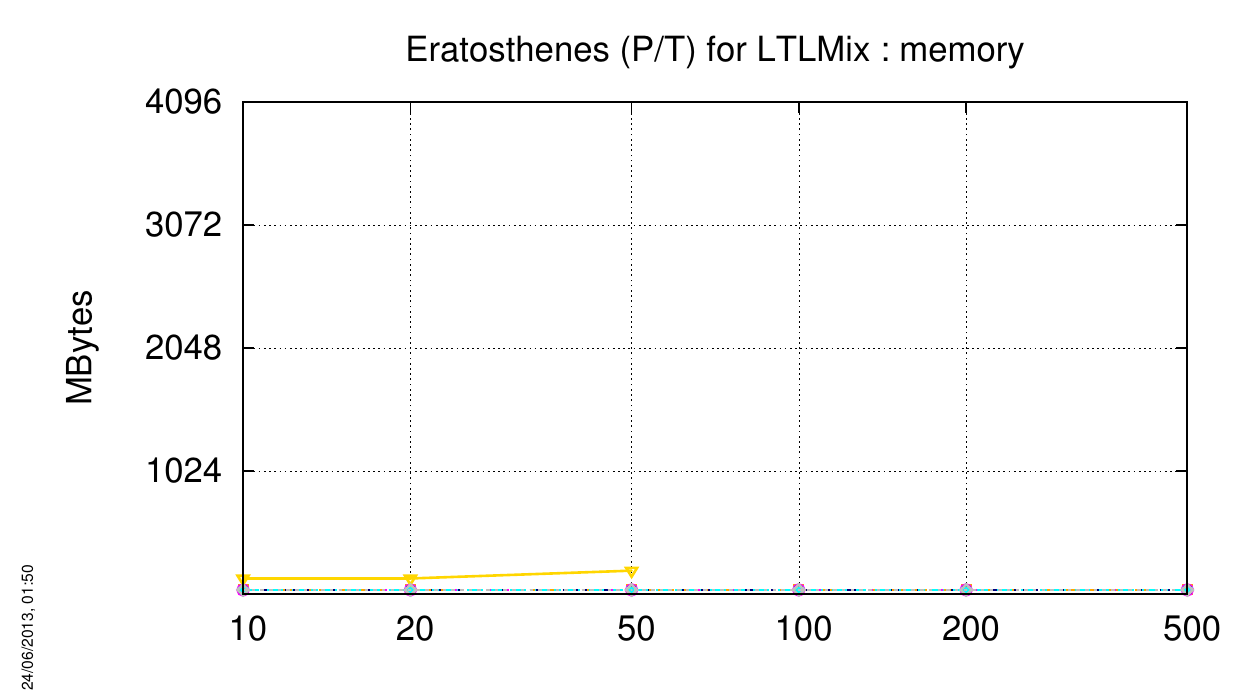}
   \includegraphics[width=7.2cm]{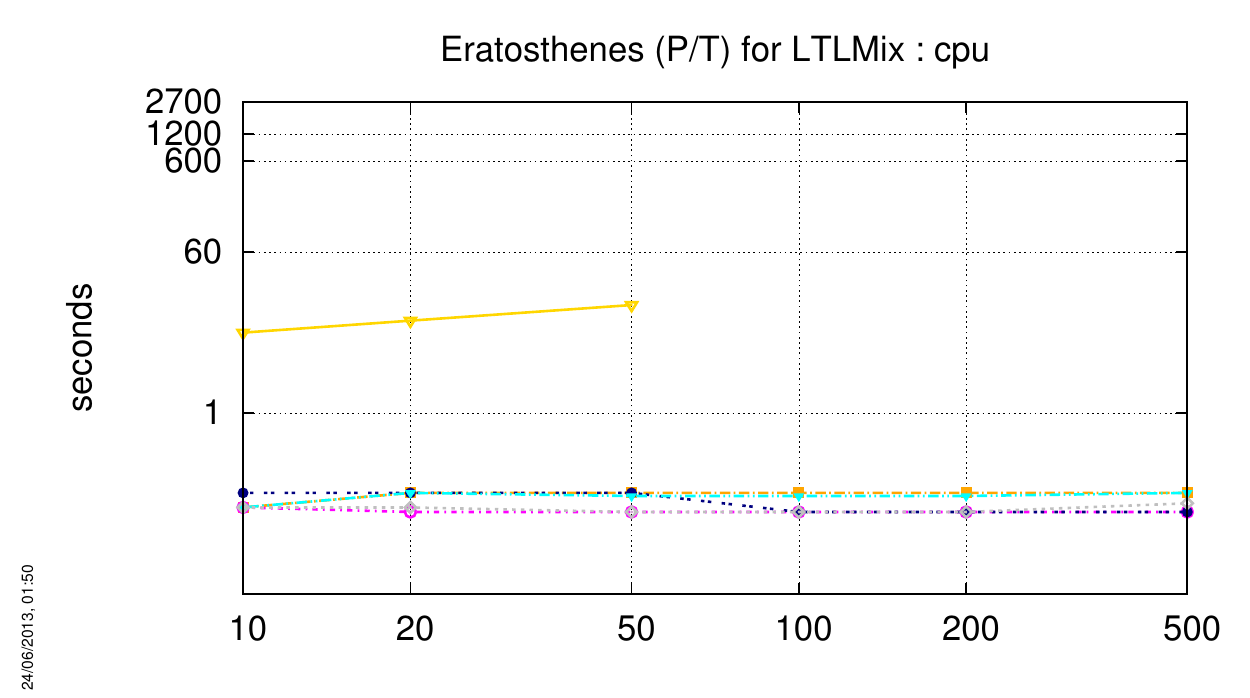}

   \includegraphics[height=1cm]{figures/tools-legend.pdf}
\end{center}

\subsubsection{\acs{FMS-PT}}
The charts below respectively show how tools compete with this ``Known'' model (memory and CPU).

\index{Performances!LTLMix!FMS (P/T)}
\begin{center}
   \includegraphics[width=7.2cm]{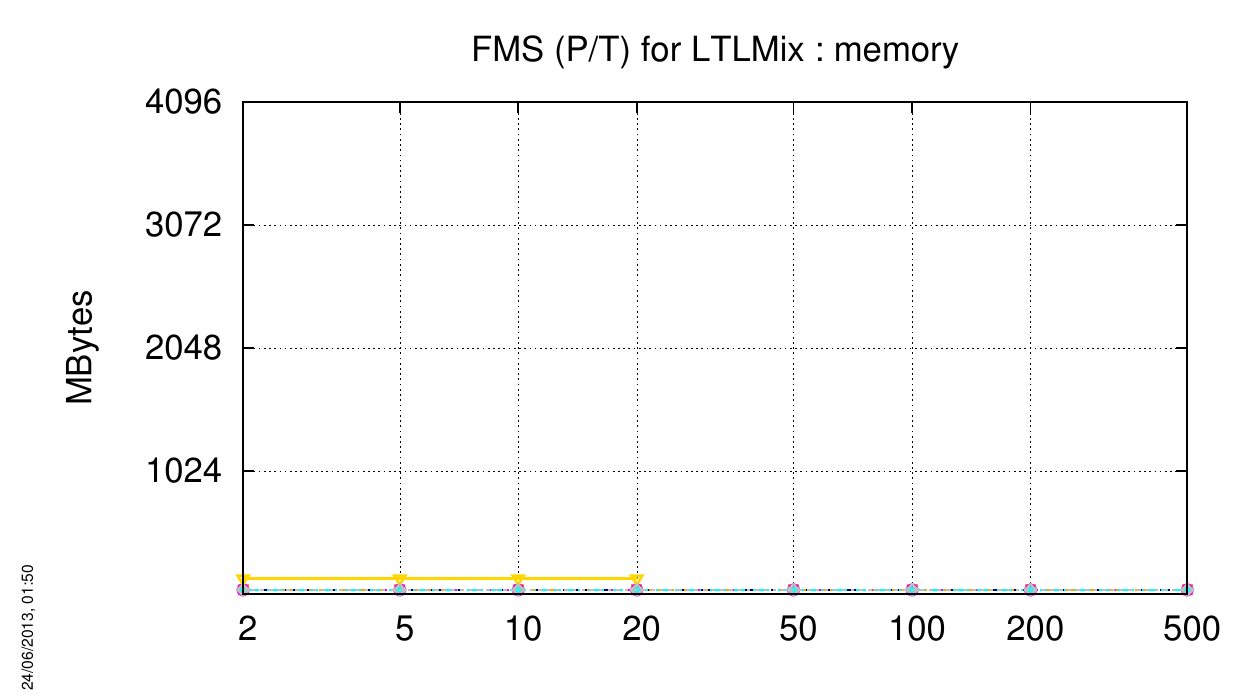}
   \includegraphics[width=7.2cm]{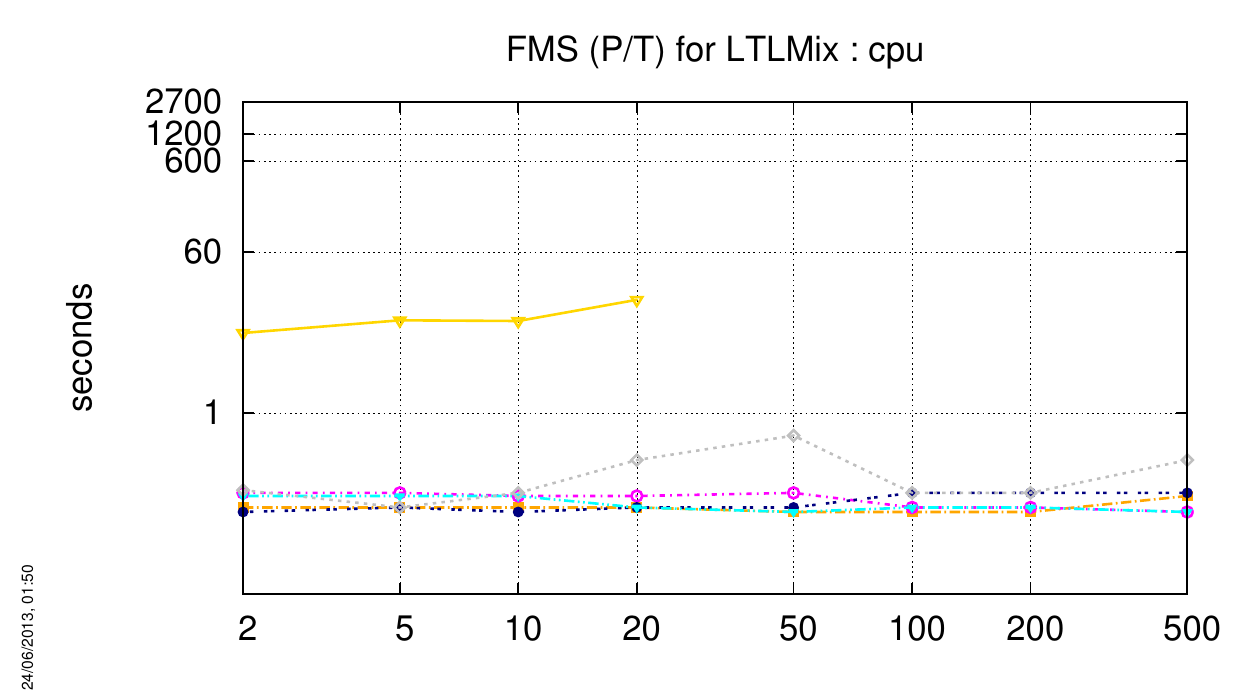}

   \includegraphics[height=1cm]{figures/tools-legend.pdf}
\end{center}

\subsubsection{\acs{GlobalRessAlloc-COL}}
No instance of this model could be computed for the \textbf{LTLMix} examination.

\subsubsection{\acs{GlobalRessAlloc-PT}}
The charts below respectively show how tools compete with this ``Known'' model (memory and CPU).

\index{Performances!LTLMix!GlobalRessAlloc (P/T)}
\begin{center}
   \includegraphics[width=7.2cm]{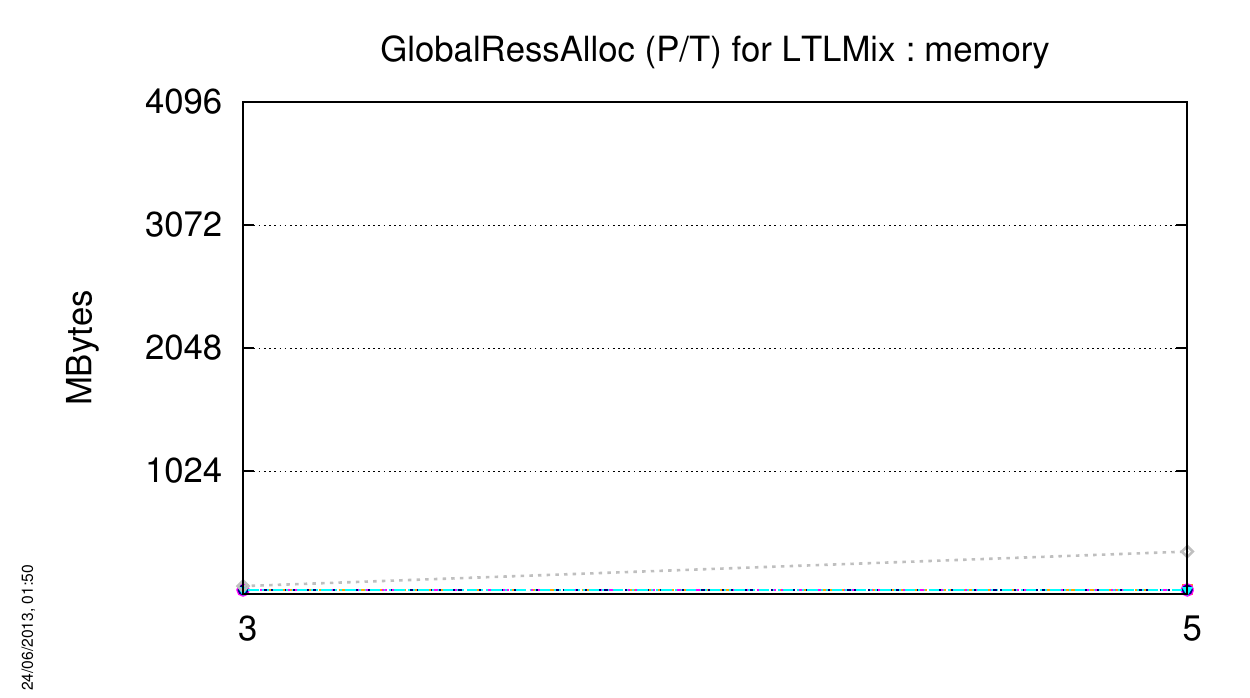}
   \includegraphics[width=7.2cm]{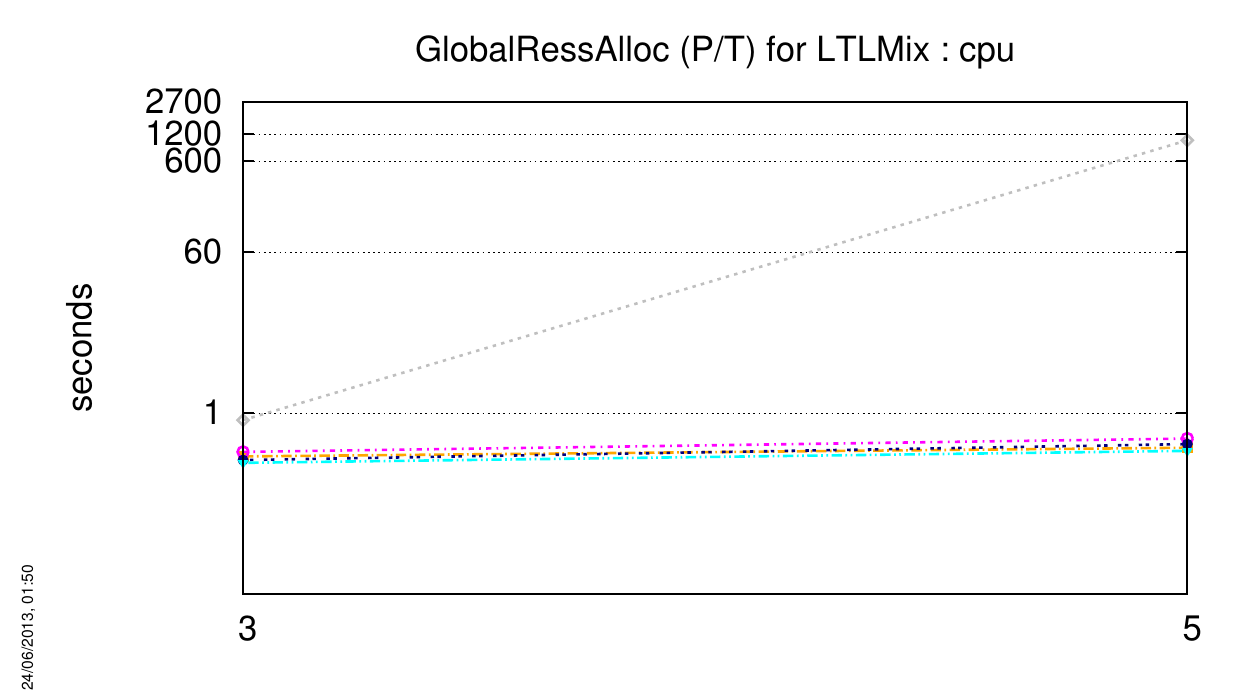}

   \includegraphics[height=1cm]{figures/tools-legend.pdf}
\end{center}

\subsubsection{\acs{Kanban-PT}}
The charts below respectively show how tools compete with this ``Known'' model (memory and CPU).

\index{Performances!LTLMix!Kanban (P/T)}
\begin{center}
   \includegraphics[width=7.2cm]{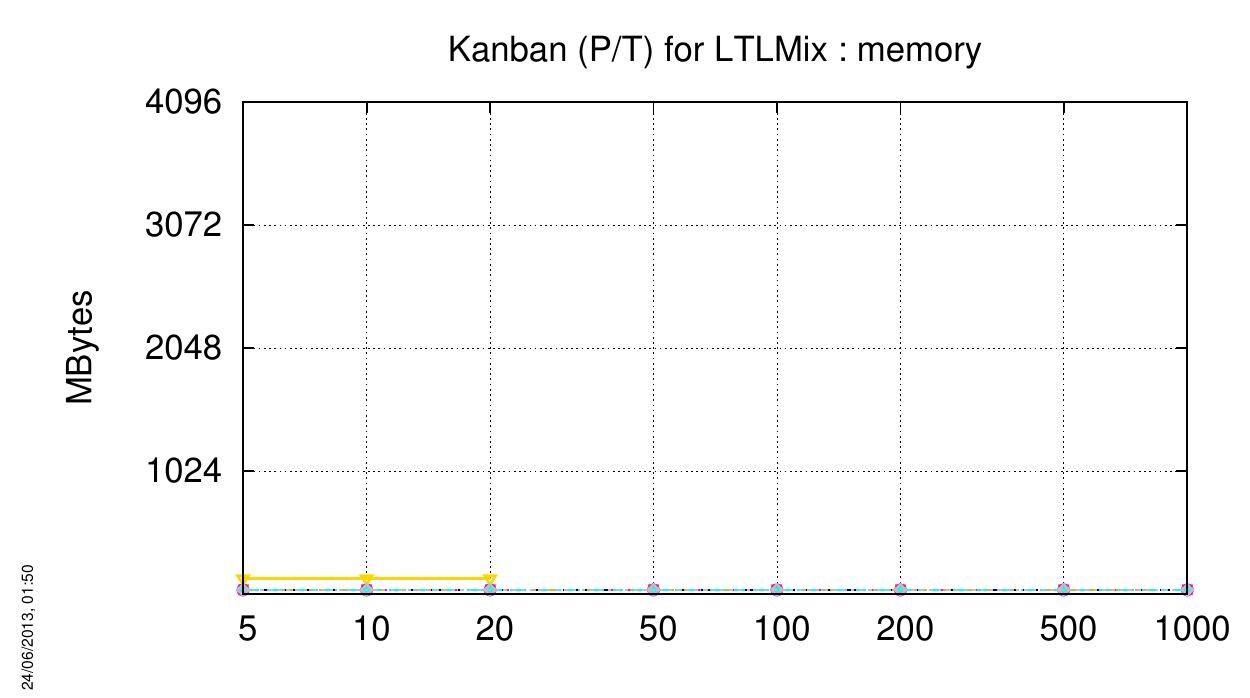}
   \includegraphics[width=7.2cm]{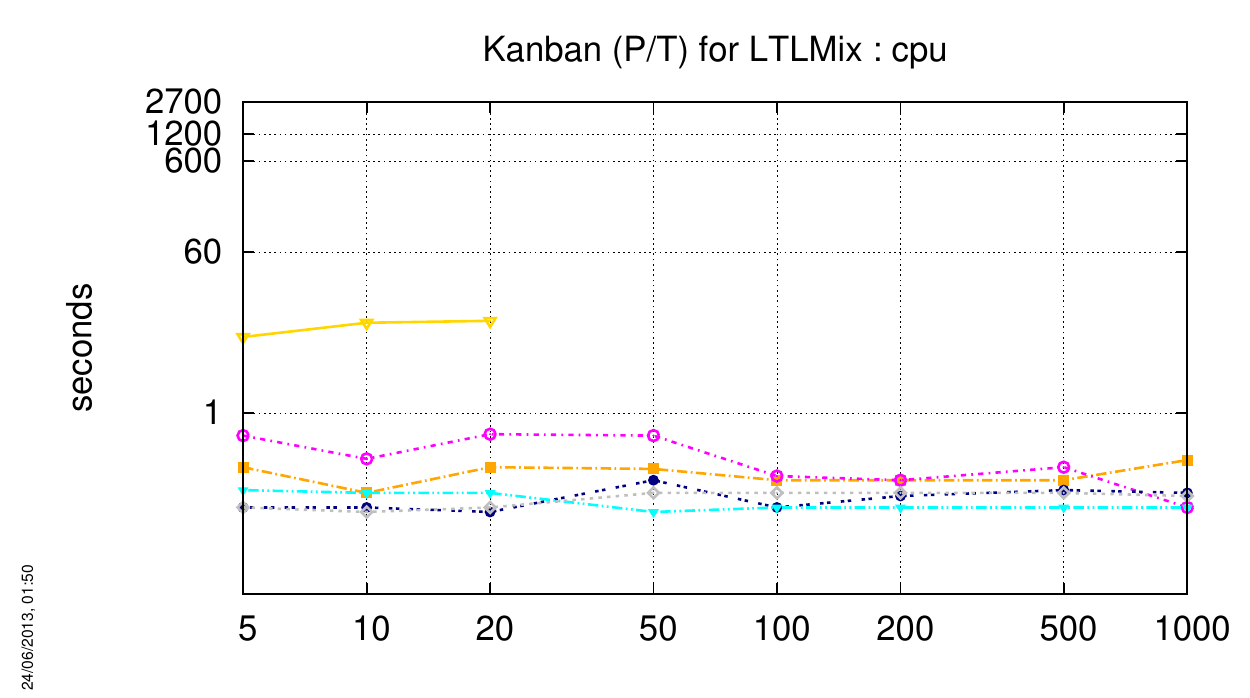}

   \includegraphics[height=1cm]{figures/tools-legend.pdf}
\end{center}

\subsubsection{\acs{LamportFastMutEx-COL}}
No instance of this model could be computed for the \textbf{LTLMix} examination.

\subsubsection{\acs{LamportFastMutEx-PT}}
The charts below respectively show how tools compete with this ``Known'' model (memory and CPU).

\index{Performances!LTLMix!LamportFastMutEx (P/T)}
\begin{center}
   \includegraphics[width=7.2cm]{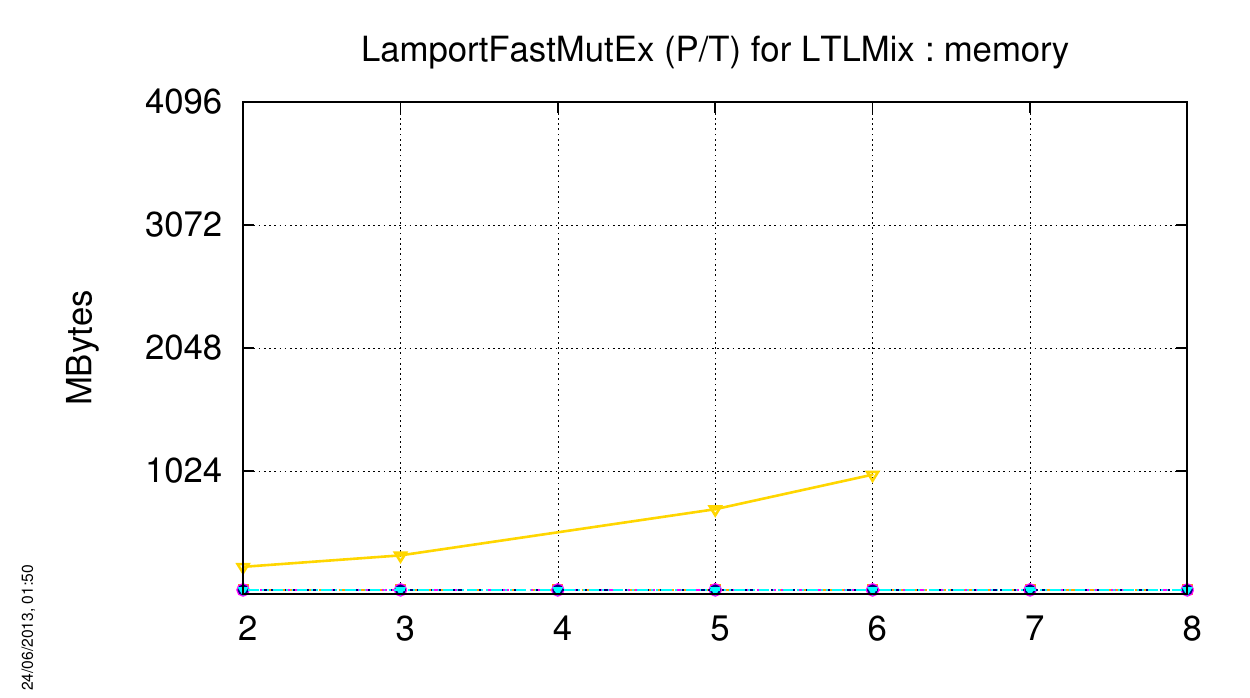}
   \includegraphics[width=7.2cm]{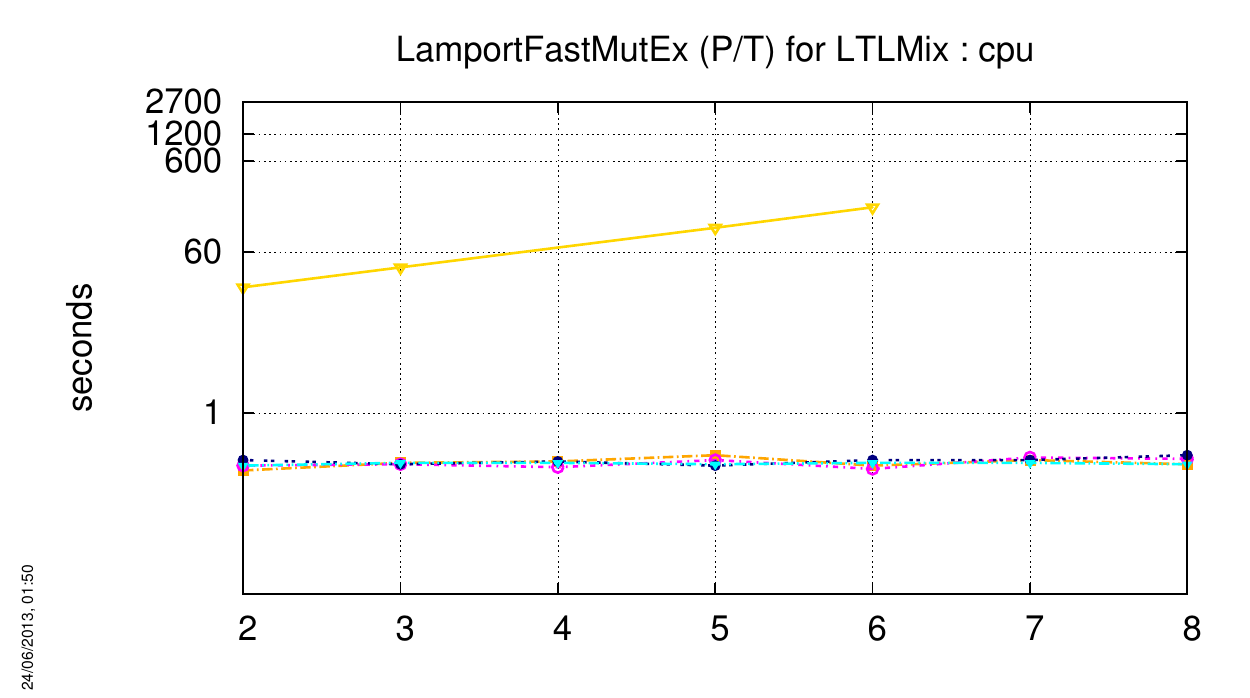}

   \includegraphics[height=1cm]{figures/tools-legend.pdf}
\end{center}

\subsubsection{\acs{MAPK-PT}}
The charts below respectively show how tools compete with this ``Known'' model (memory and CPU).

\index{Performances!LTLMix!MAPK (P/T)}
\begin{center}
   \includegraphics[width=7.2cm]{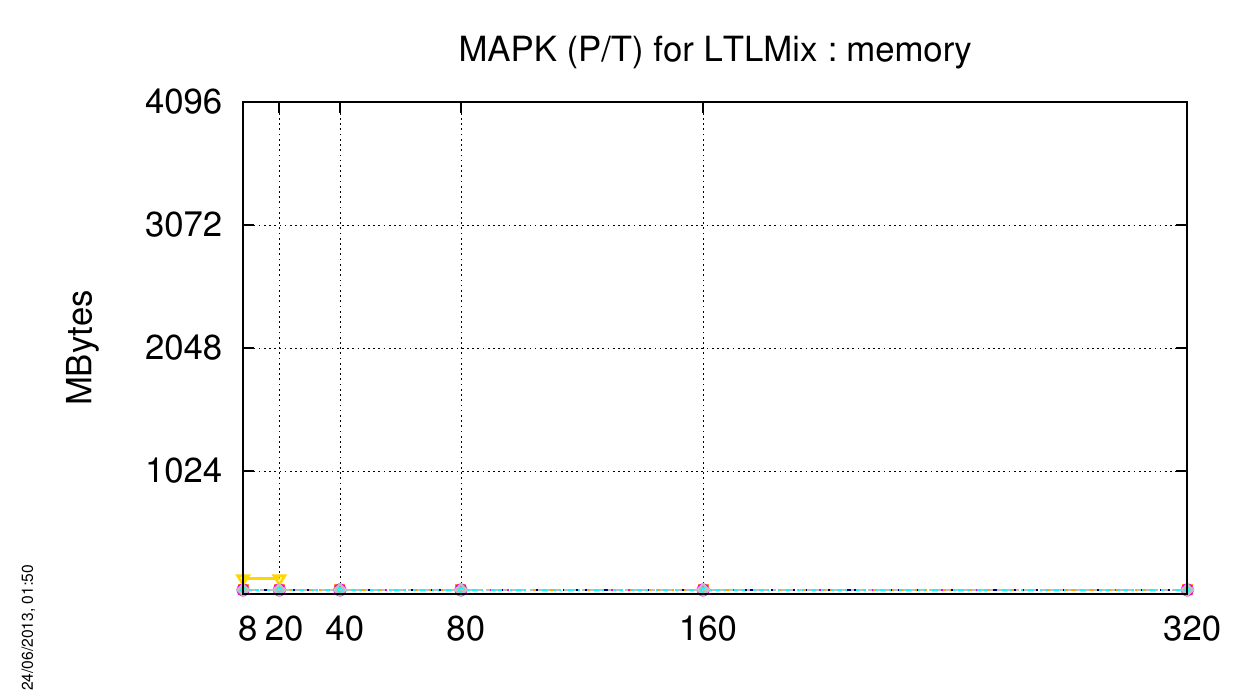}
   \includegraphics[width=7.2cm]{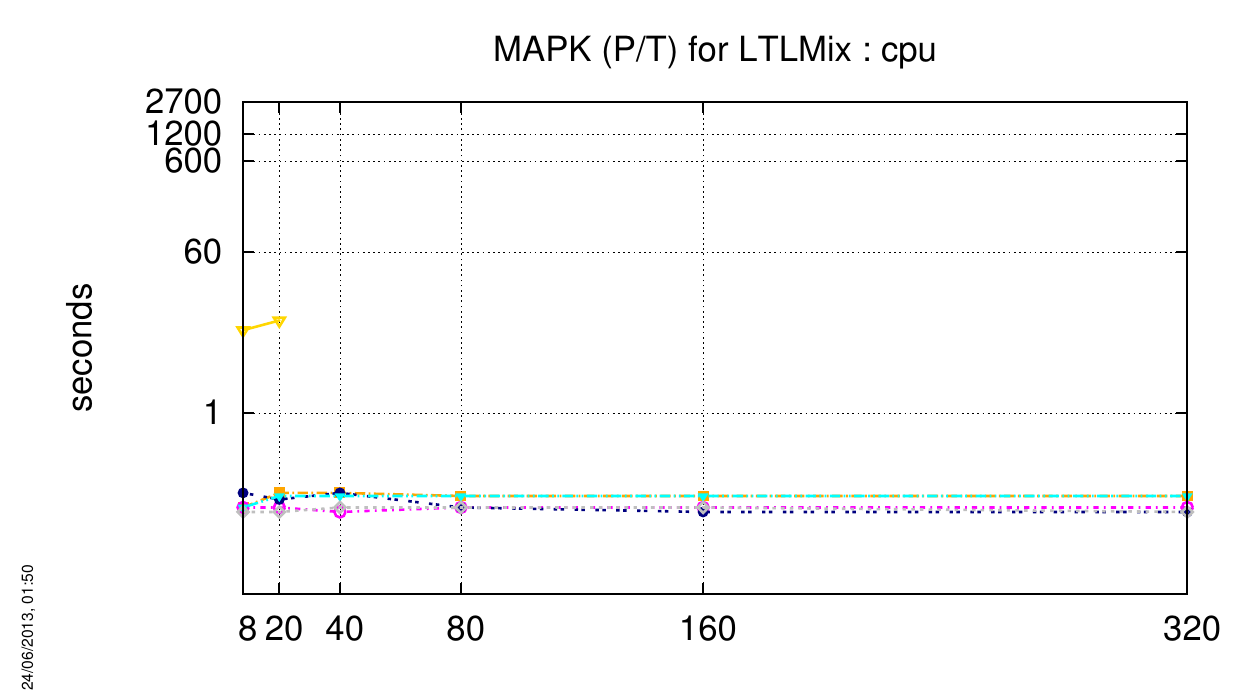}

   \includegraphics[height=1cm]{figures/tools-legend.pdf}
\end{center}

\subsubsection{\acs{NeoElection-COL}}
No instance of this model could be computed for the \textbf{LTLMix} examination.

\subsubsection{\acs{NeoElection-PT}}
The charts below respectively show how tools compete with this ``Known'' model (memory and CPU).

\index{Performances!LTLMix!NeoElection (P/T)}
\begin{center}
   \includegraphics[width=7.2cm]{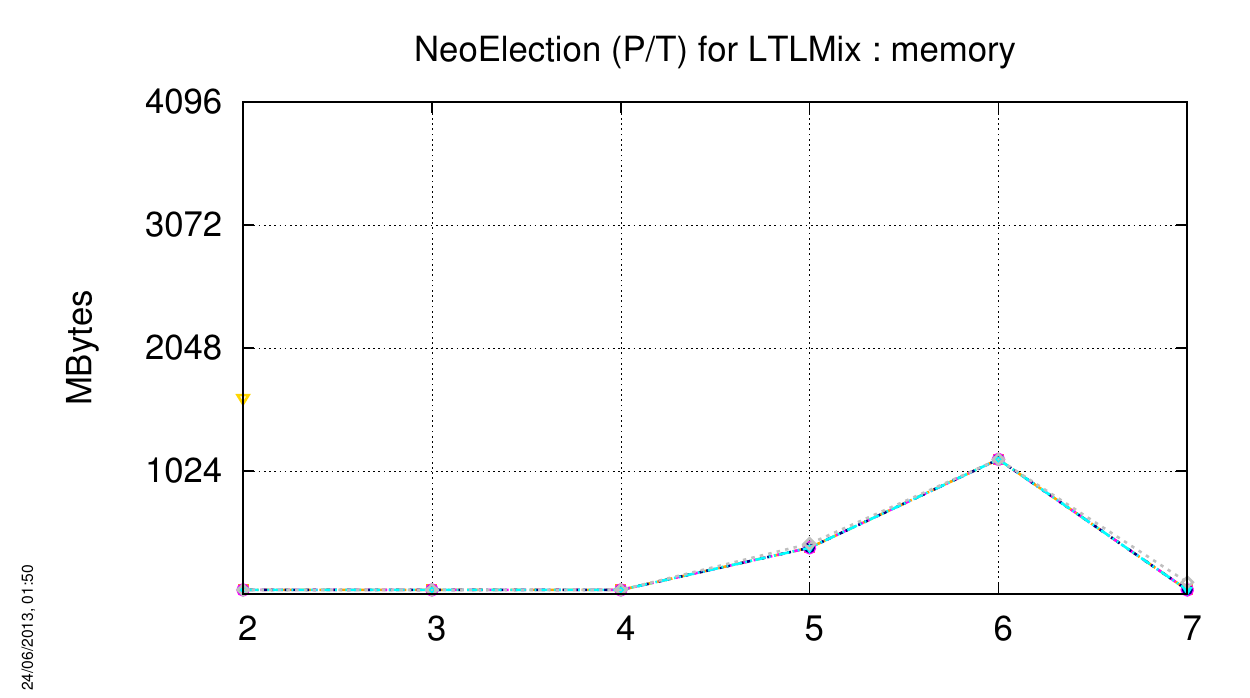}
   \includegraphics[width=7.2cm]{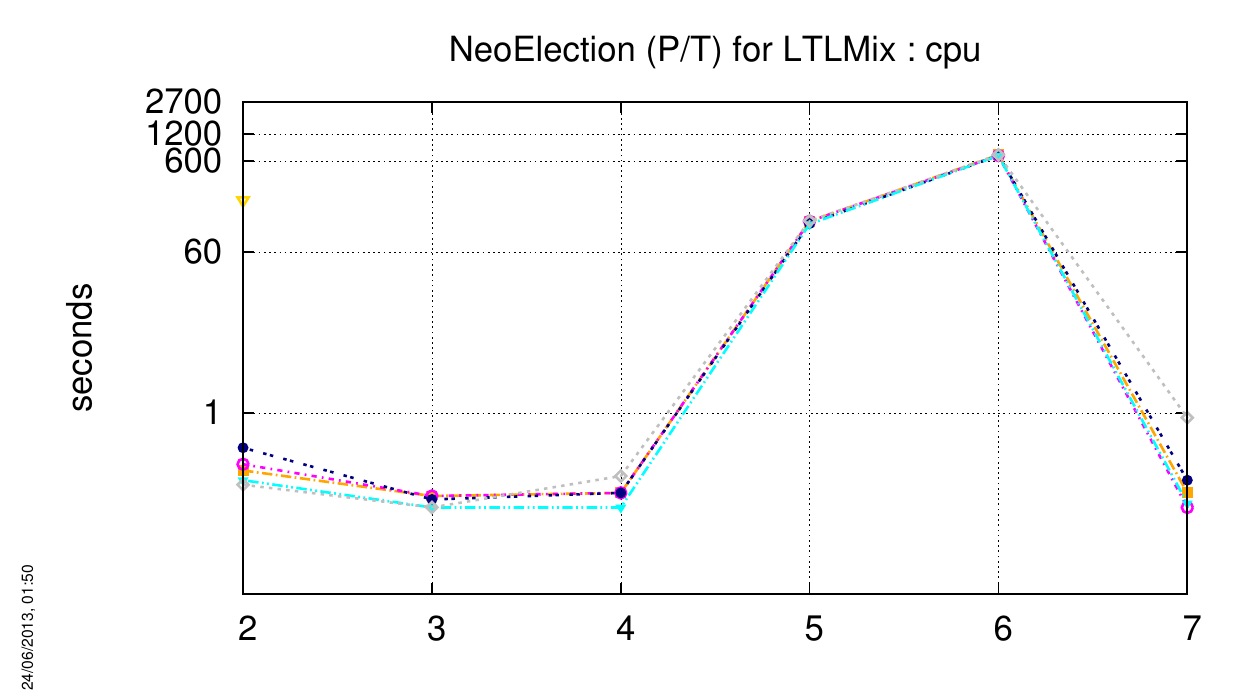}

   \includegraphics[height=1cm]{figures/tools-legend.pdf}
\end{center}

\subsubsection{\acs{PermAdmissibility-COL}}
No instance of this model could be computed for the \textbf{LTLMix} examination.

\subsubsection{\acs{PermAdmissibility-PT}}
The charts below respectively show how tools compete with this ``Known'' model (memory and CPU).

\index{Performances!LTLMix!PermAdmissibility (P/T)}
\begin{center}
   \includegraphics[width=7.2cm]{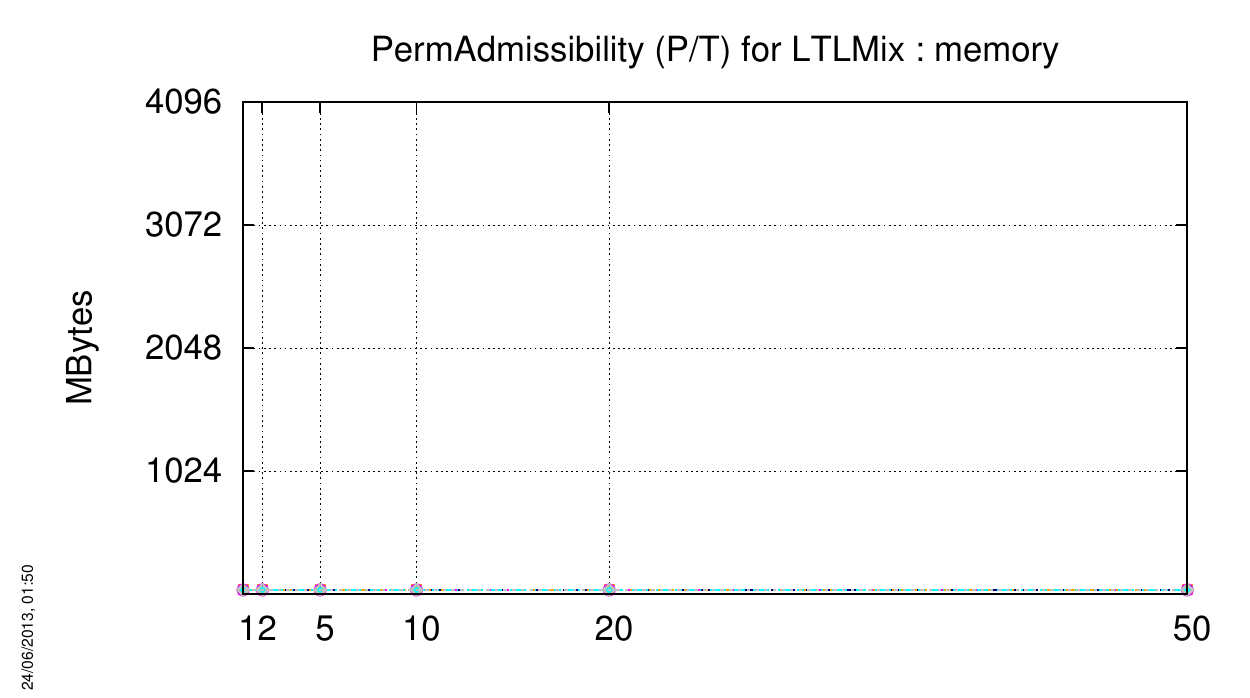}
   \includegraphics[width=7.2cm]{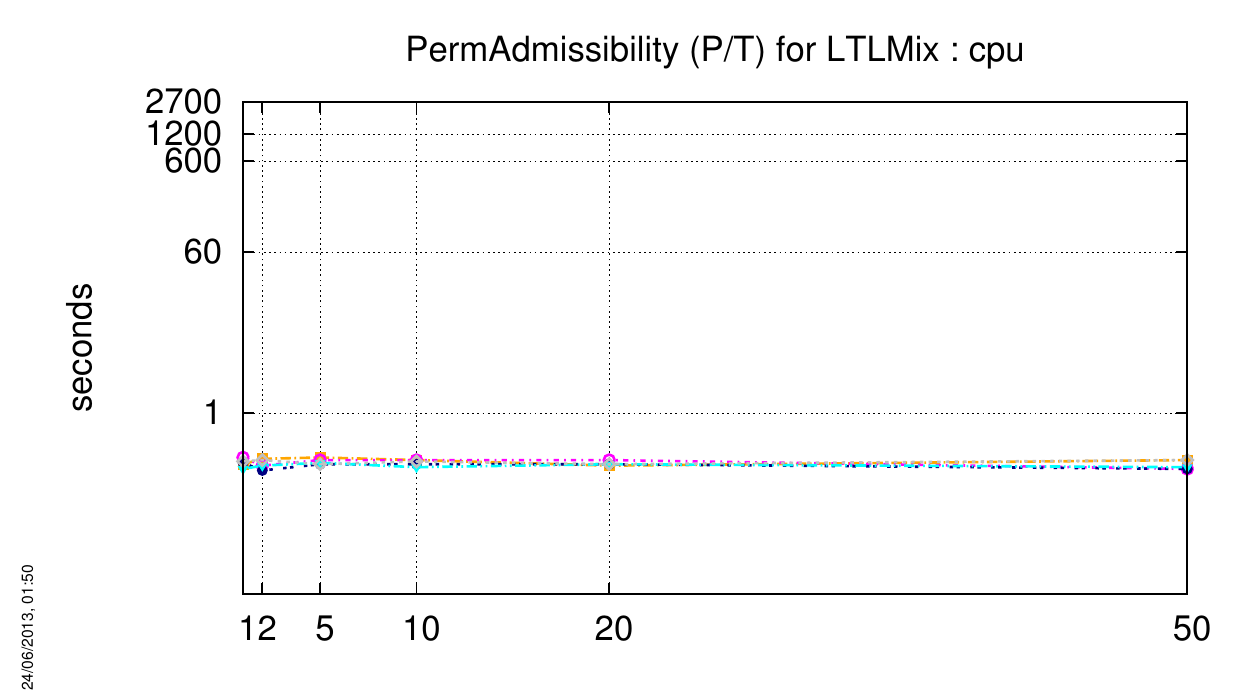}

   \includegraphics[height=1cm]{figures/tools-legend.pdf}
\end{center}

\subsubsection{\acs{Peterson-COL}}
No instance of this model could be computed for the \textbf{LTLMix} examination.

\subsubsection{\acs{Peterson-PT}}
The charts below respectively show how tools compete with this ``Known'' model (memory and CPU).

\index{Performances!LTLMix!Peterson (P/T)}
\begin{center}
   \includegraphics[width=7.2cm]{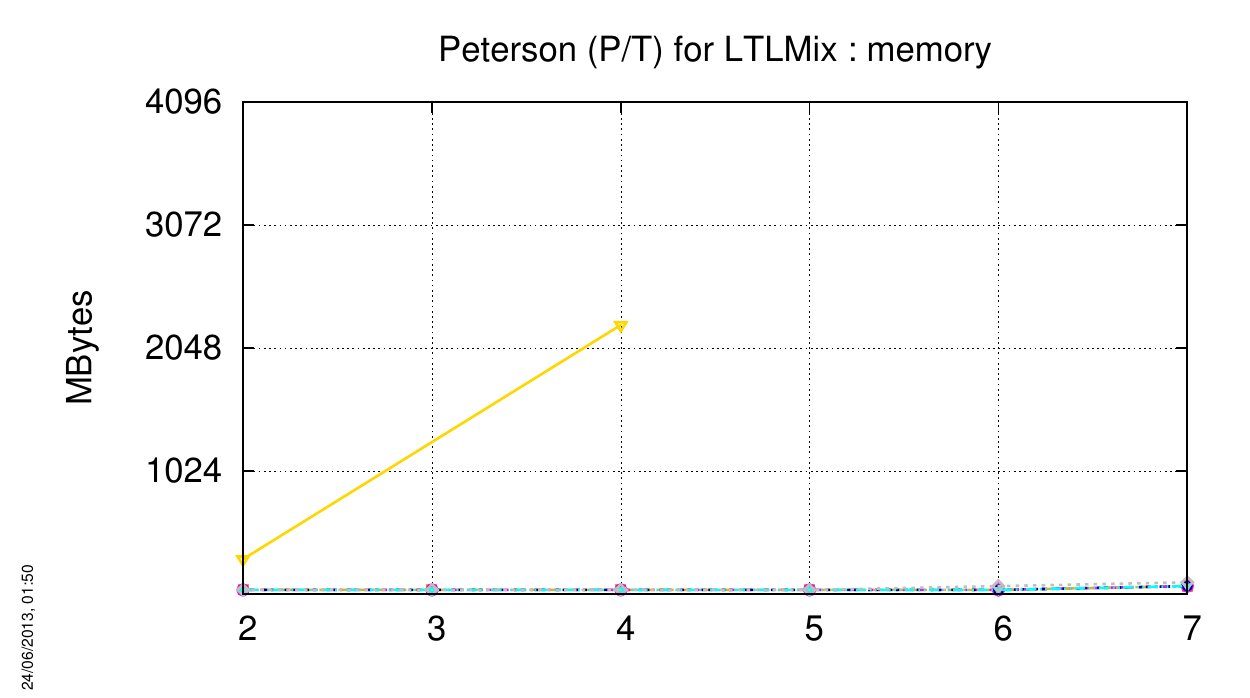}
   \includegraphics[width=7.2cm]{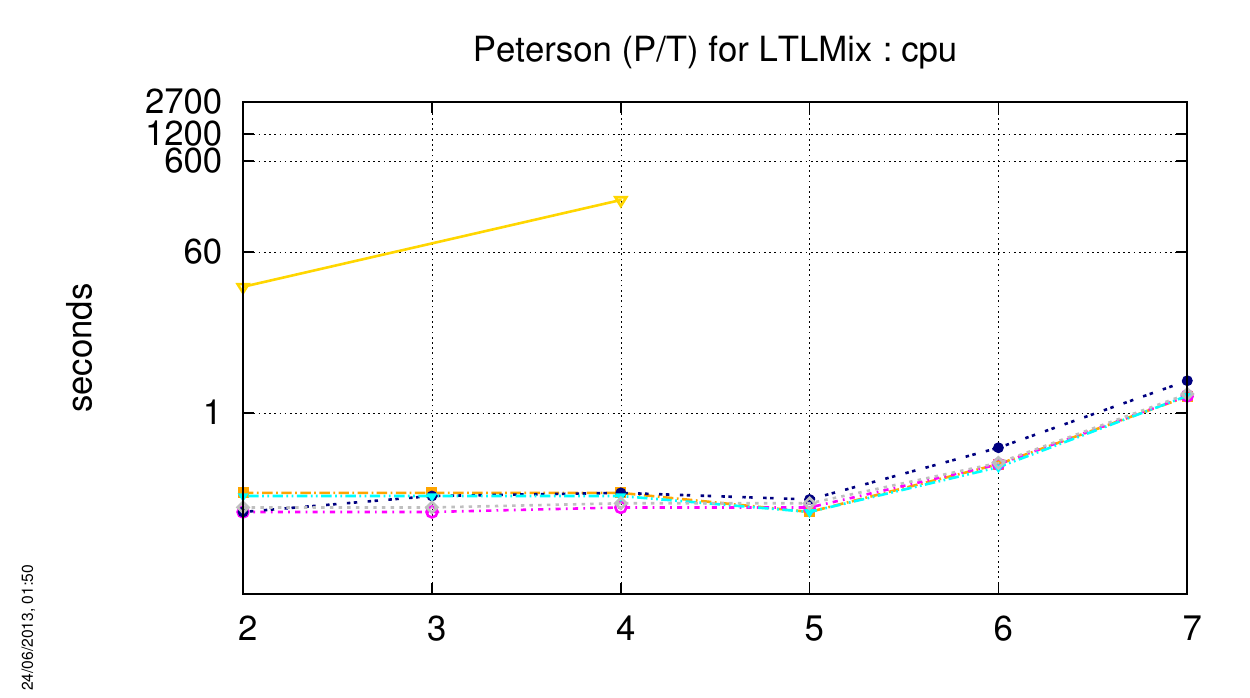}

   \includegraphics[height=1cm]{figures/tools-legend.pdf}
\end{center}

\subsubsection{\acs{Philosophers-COL}}
No instance of this model could be computed for the \textbf{LTLMix} examination.

\subsubsection{\acs{Philosophers-PT}}
The charts below respectively show how tools compete with this ``Known'' model (memory and CPU).

\index{Performances!LTLMix!Philosophers (P/T)}
\begin{center}
   \includegraphics[width=7.2cm]{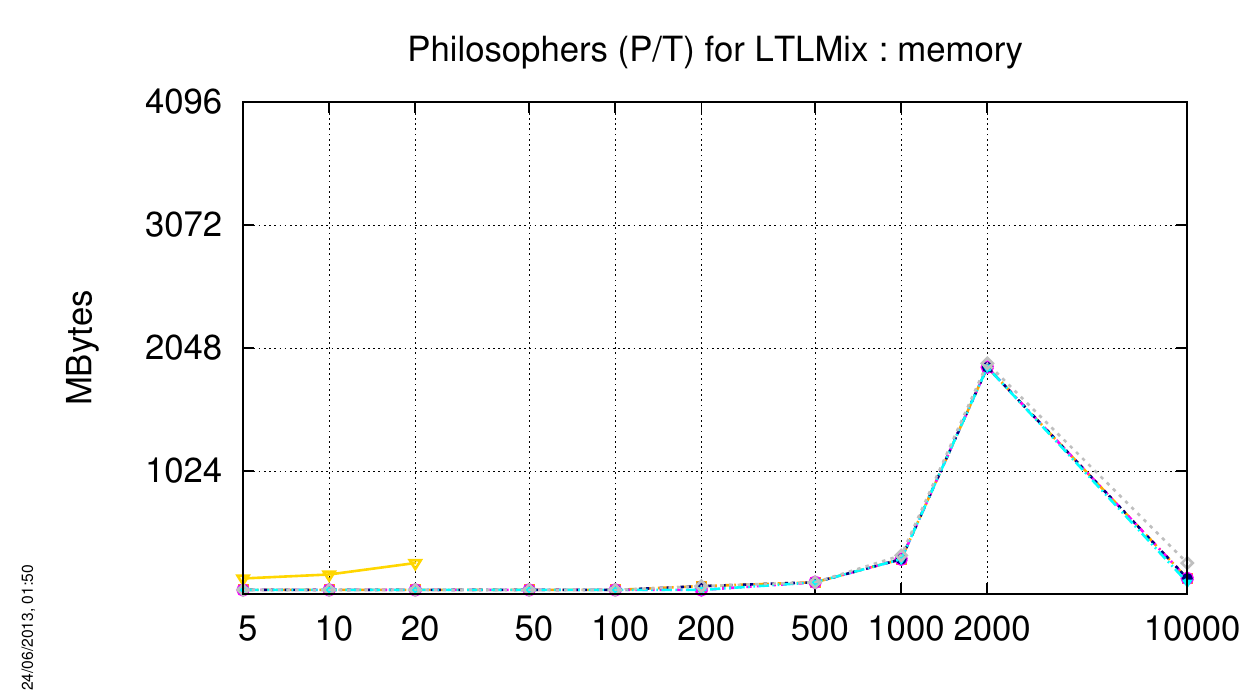}
   \includegraphics[width=7.2cm]{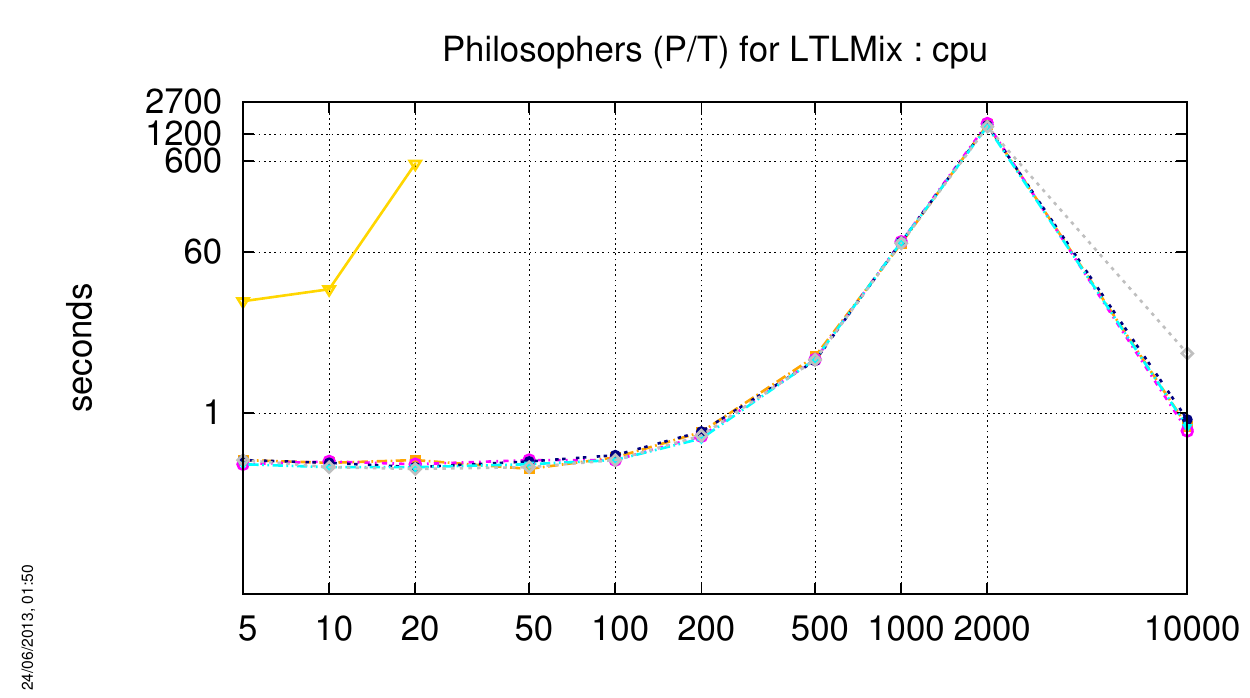}

   \includegraphics[height=1cm]{figures/tools-legend.pdf}
\end{center}

\subsubsection{\acs{PhilosophersDyn-COL}}
No instance of this model could be computed for the \textbf{LTLMix} examination.

\subsubsection{\acs{PhilosophersDyn-PT}}
The charts below respectively show how tools compete with this ``Known'' model (memory and CPU).

\index{Performances!LTLMix!PhilosophersDyn (P/T)}
\begin{center}
   \includegraphics[width=7.2cm]{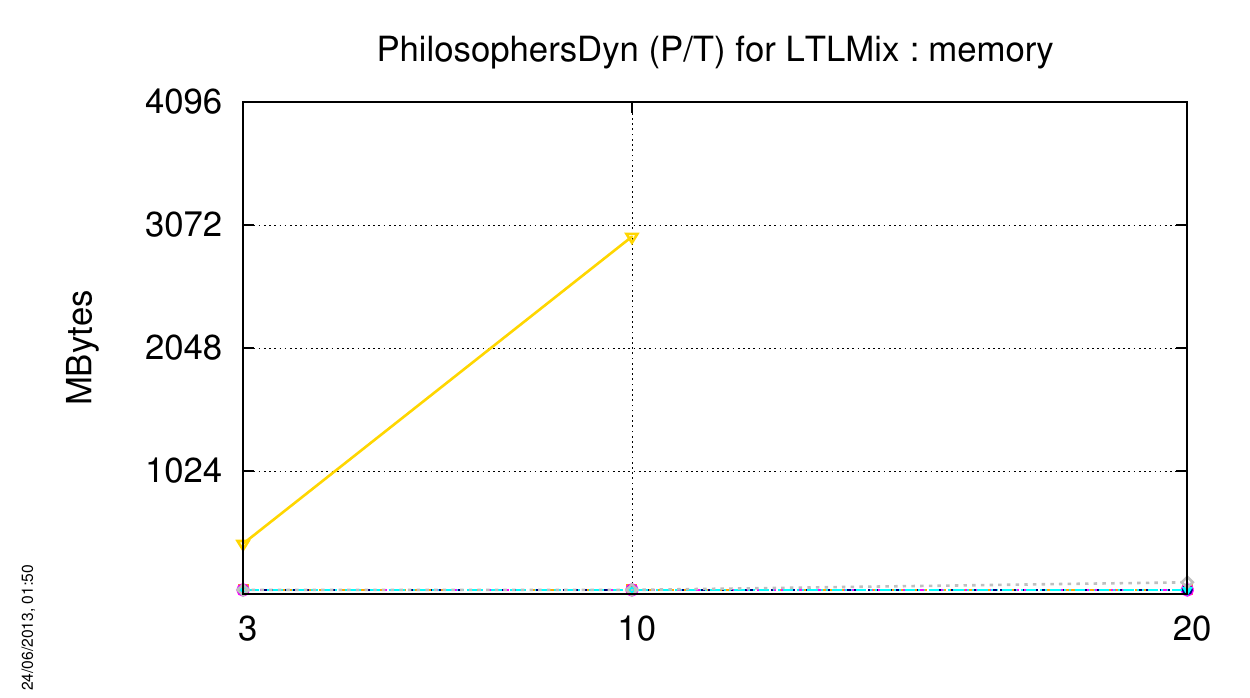}
   \includegraphics[width=7.2cm]{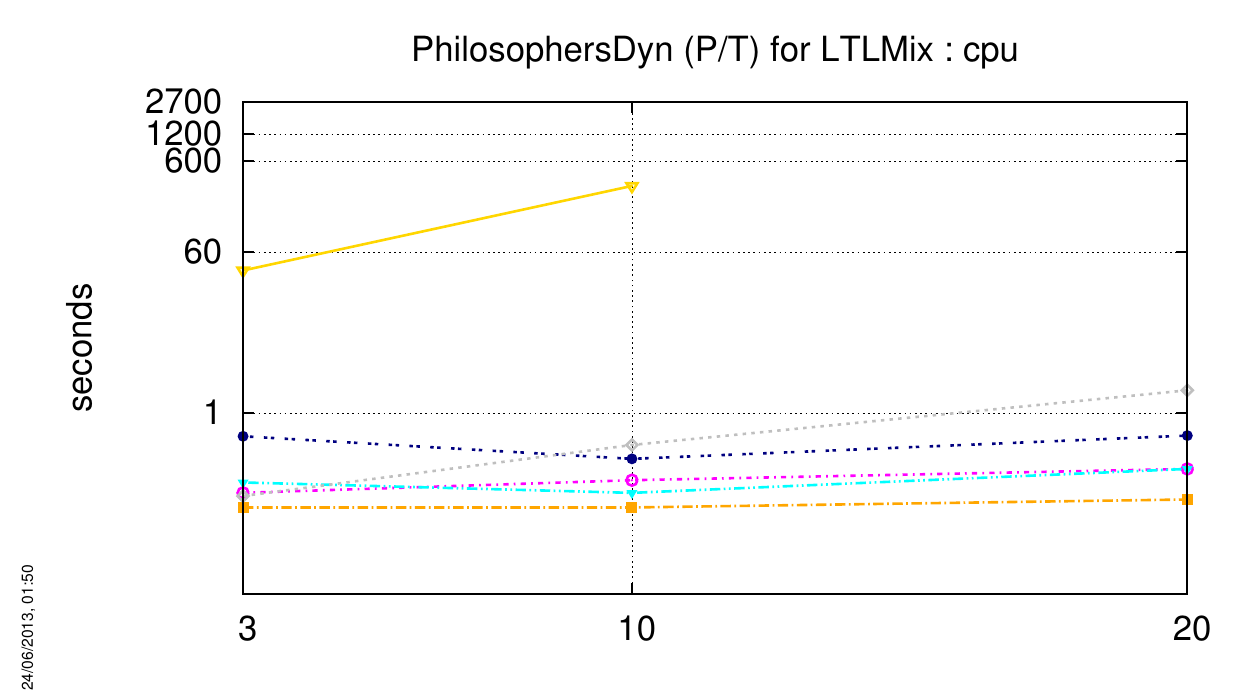}

   \includegraphics[height=1cm]{figures/tools-legend.pdf}
\end{center}

\subsubsection{\acs{Planning-PT}}
No instance of this model could be computed for the \textbf{LTLMix} examination.

\subsubsection{\acs{Railroad-PT}}
The charts below respectively show how tools compete with this ``Known'' model (memory and CPU).

\index{Performances!LTLMix!Railroad (P/T)}
\begin{center}
   \includegraphics[width=7.2cm]{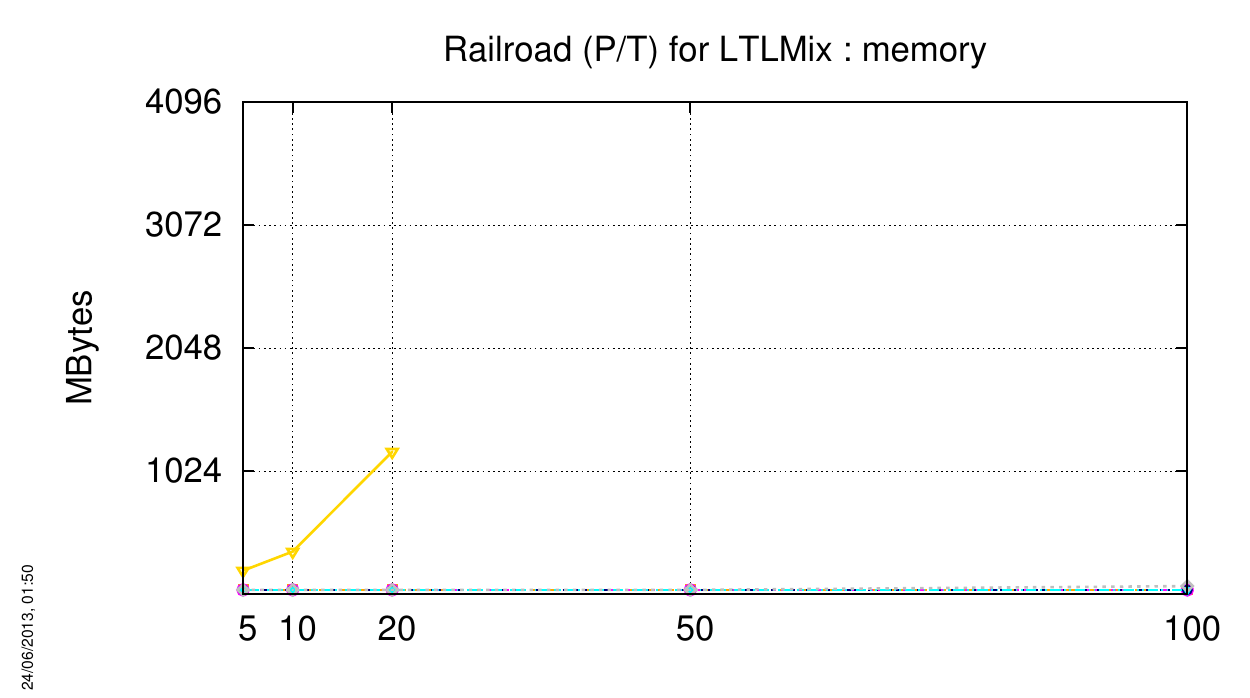}
   \includegraphics[width=7.2cm]{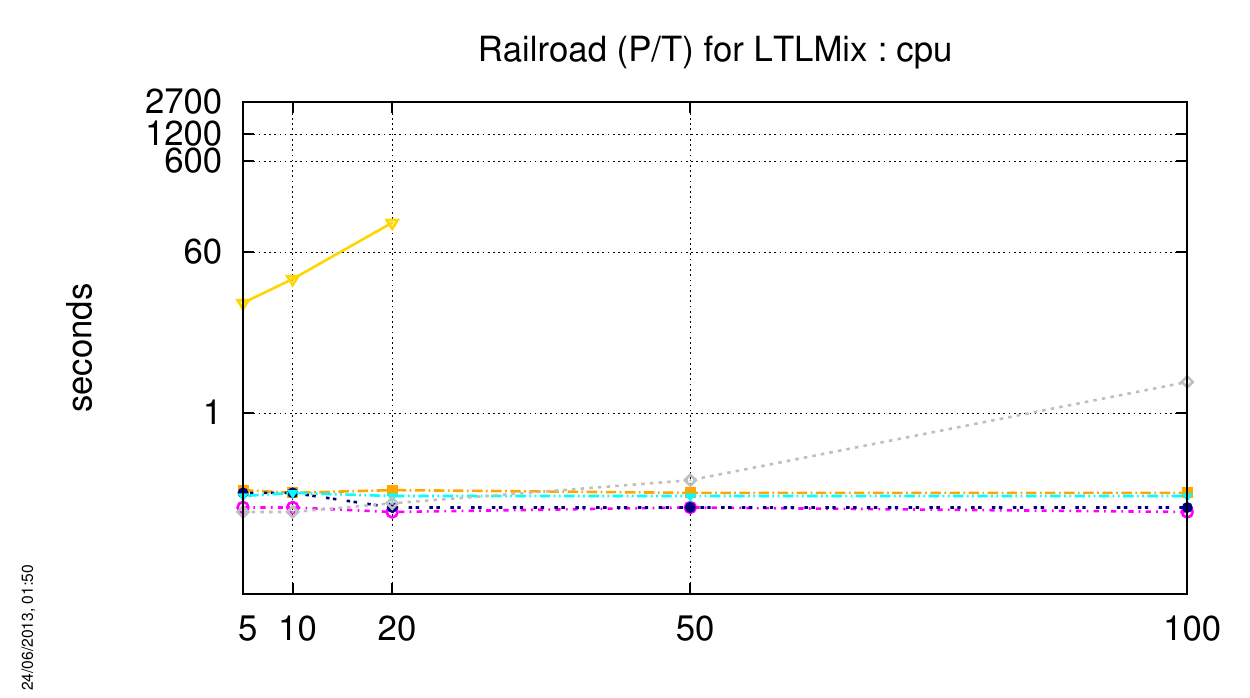}

   \includegraphics[height=1cm]{figures/tools-legend.pdf}
\end{center}

\subsubsection{\acs{RessAllocation-PT}}
The charts below respectively show how tools compete with this ``Known'' model (memory and CPU).

\index{Performances!LTLMix!RessAllocation (P/T)}
\begin{center}
   \includegraphics[width=7.2cm]{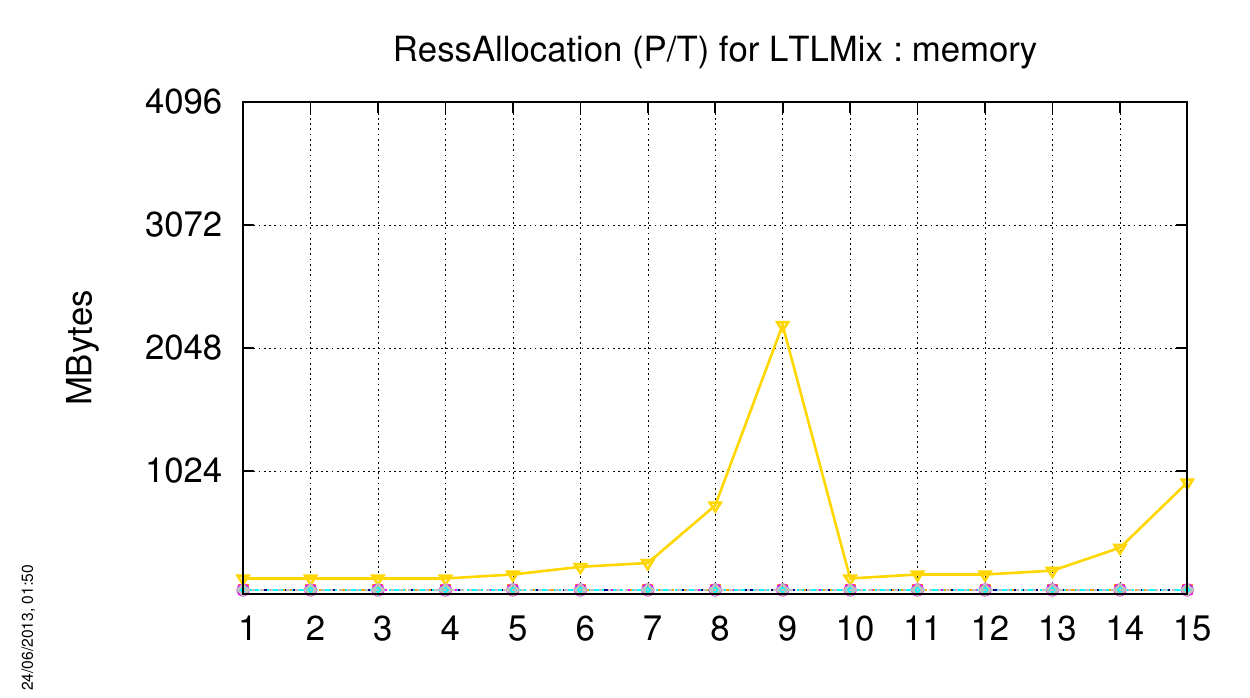}
   \includegraphics[width=7.2cm]{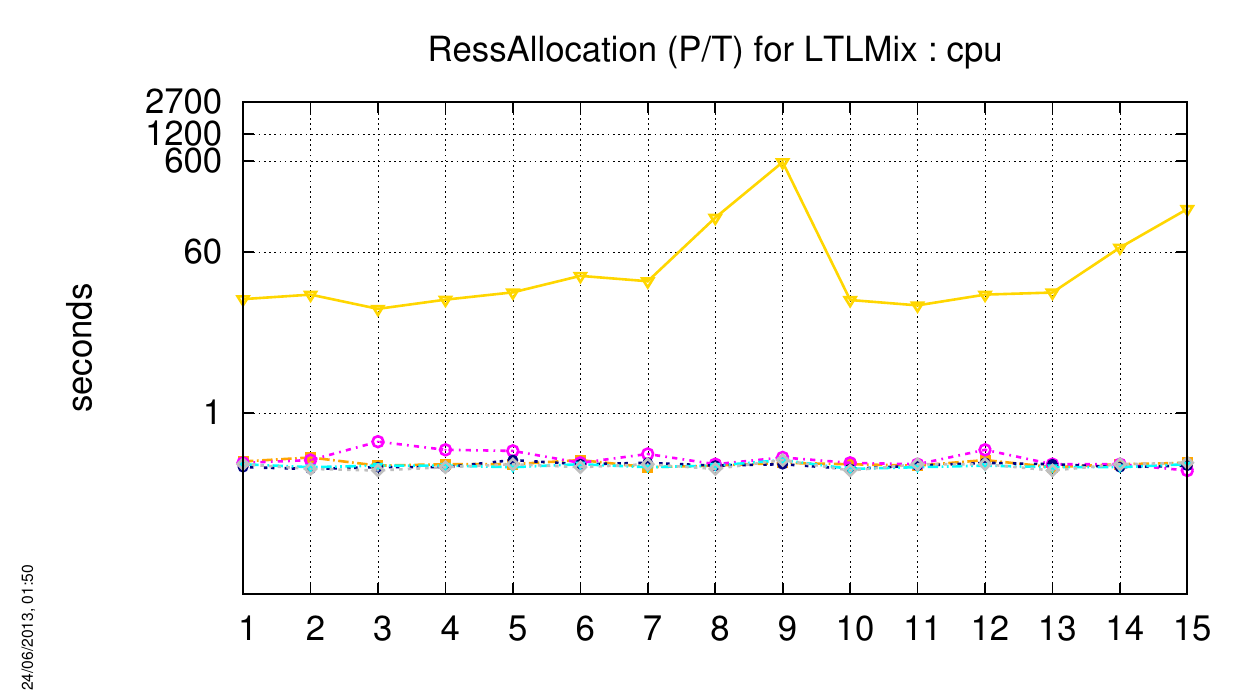}

   \includegraphics[height=1cm]{figures/tools-legend.pdf}
\end{center}

\subsubsection{\acs{Ring-PT}}
The charts below respectively show how tools compete with this ``Known'' model (memory and CPU).

\index{Performances!LTLMix!Ring (P/T)}
\begin{center}
   \includegraphics[width=7.2cm]{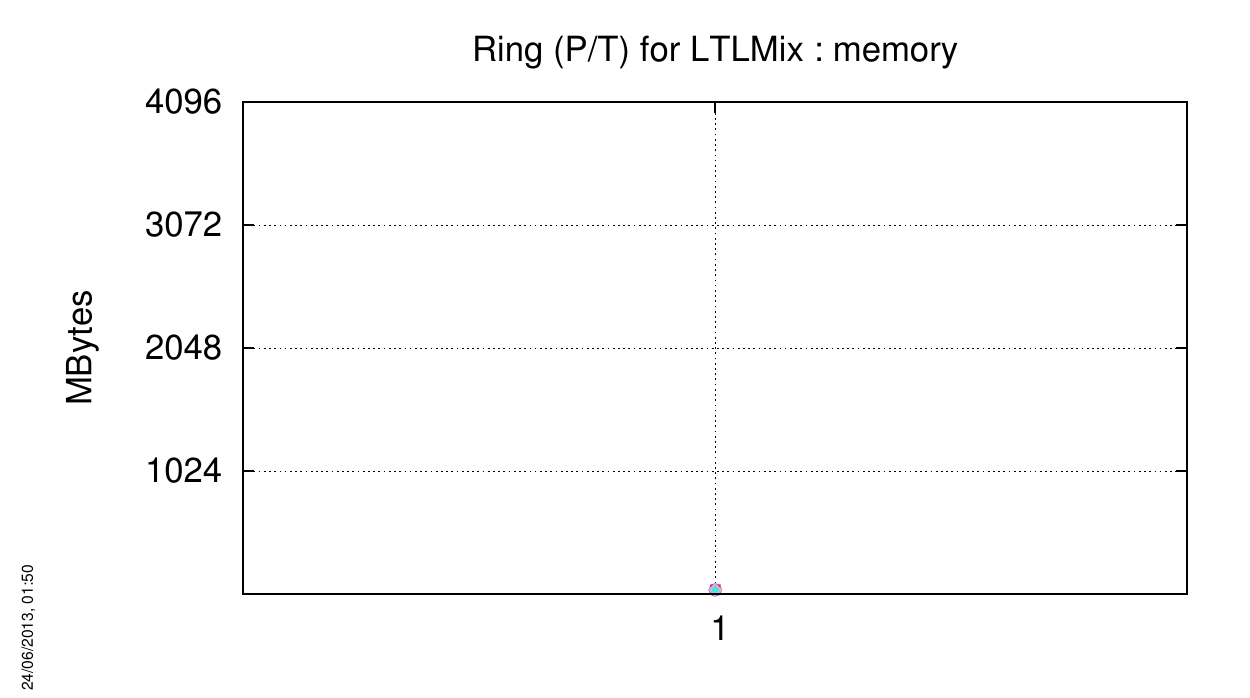}
   \includegraphics[width=7.2cm]{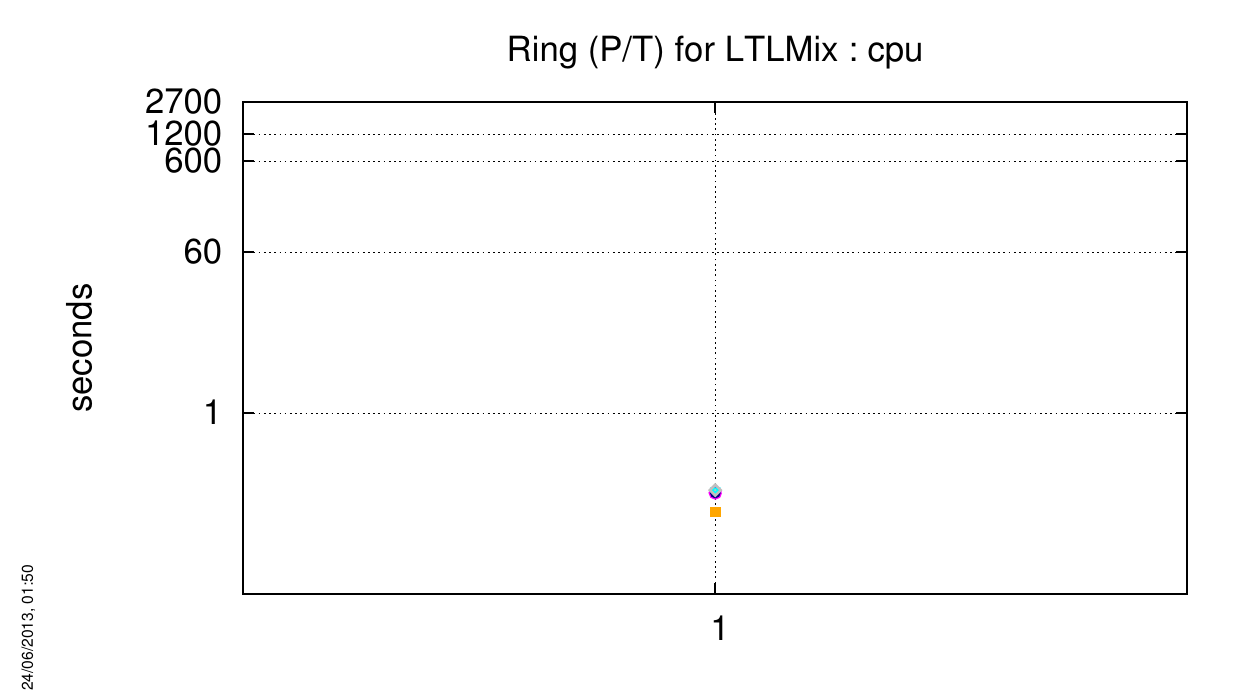}

   \includegraphics[height=1cm]{figures/tools-legend.pdf}
\end{center}

\subsubsection{\acs{RwMutex-PT}}
The charts below respectively show how tools compete with this ``Known'' model (memory and CPU).

\index{Performances!LTLMix!RwMutex (P/T)}
\begin{center}
   \includegraphics[width=7.2cm]{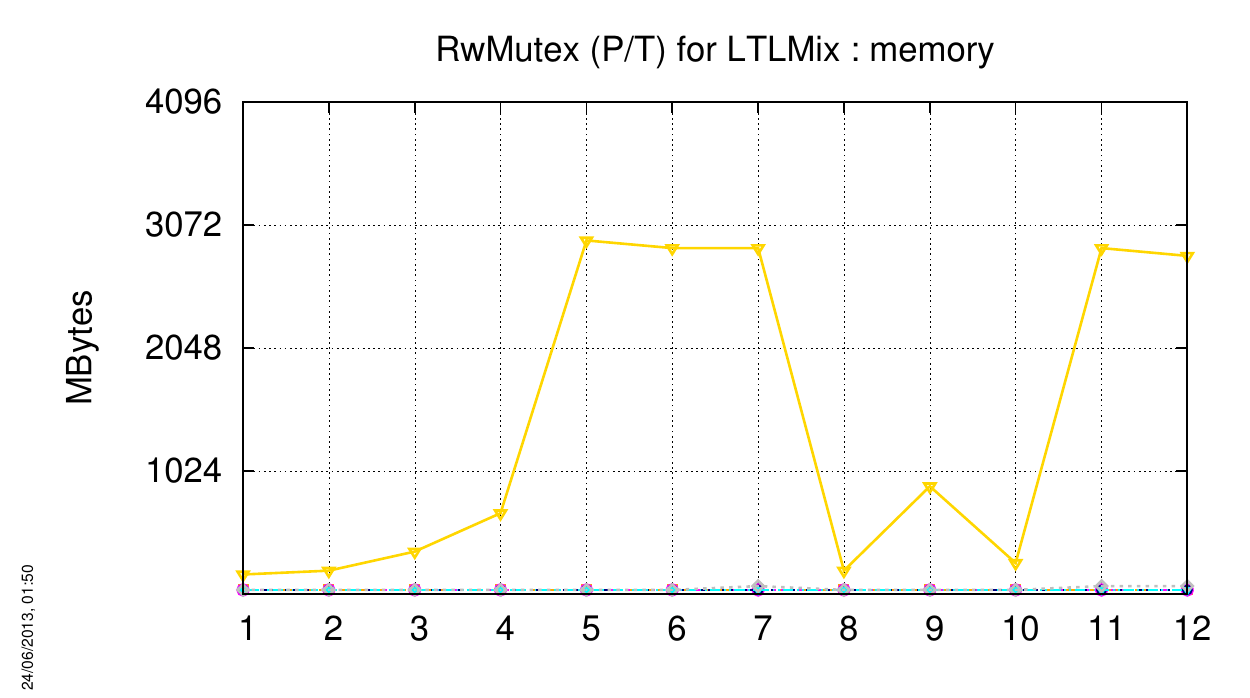}
   \includegraphics[width=7.2cm]{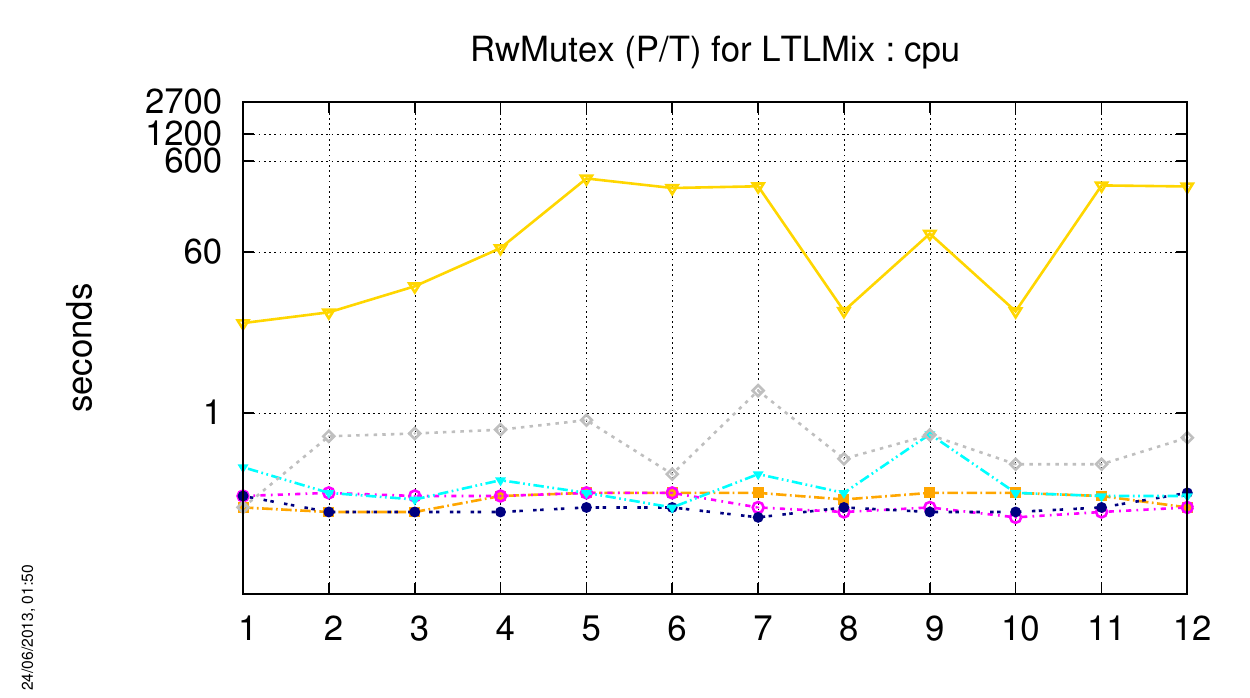}

   \includegraphics[height=1cm]{figures/tools-legend.pdf}
\end{center}

\subsubsection{\acs{SharedMemory-COL}}
No instance of this model could be computed for the \textbf{LTLMix} examination.

\subsubsection{\acs{SharedMemory-PT}}
The charts below respectively show how tools compete with this ``Known'' model (memory and CPU).

\index{Performances!LTLMix!SharedMemory (P/T)}
\begin{center}
   \includegraphics[width=7.2cm]{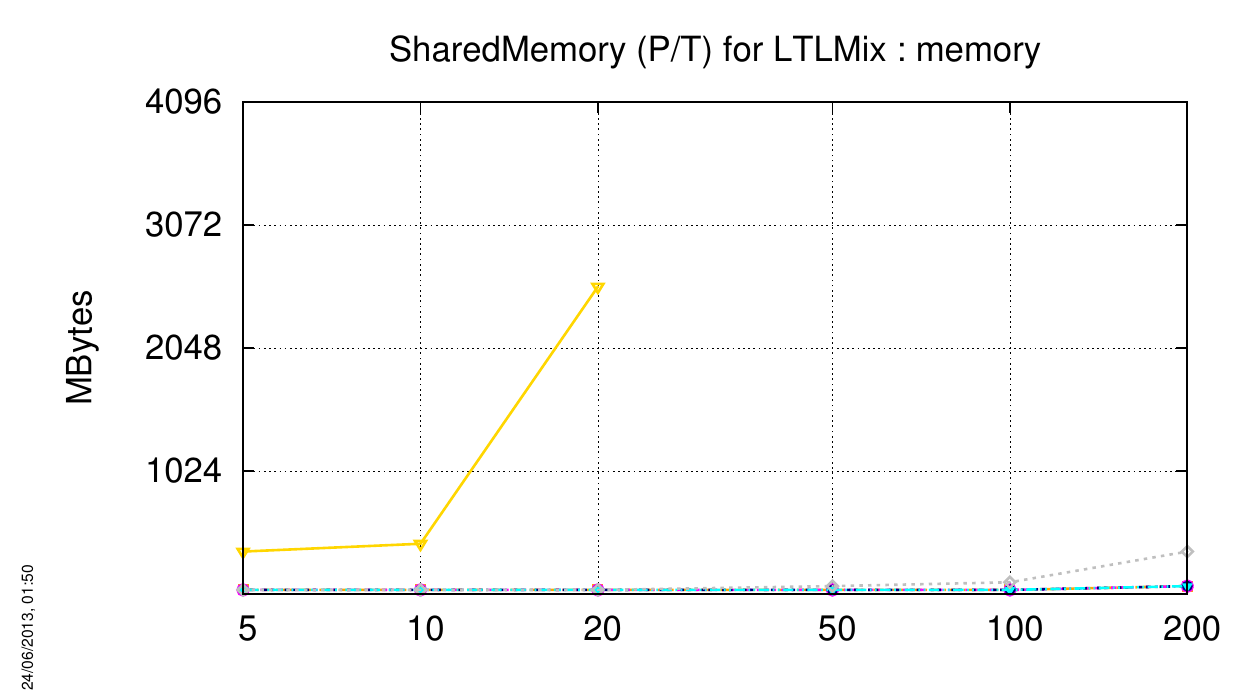}
   \includegraphics[width=7.2cm]{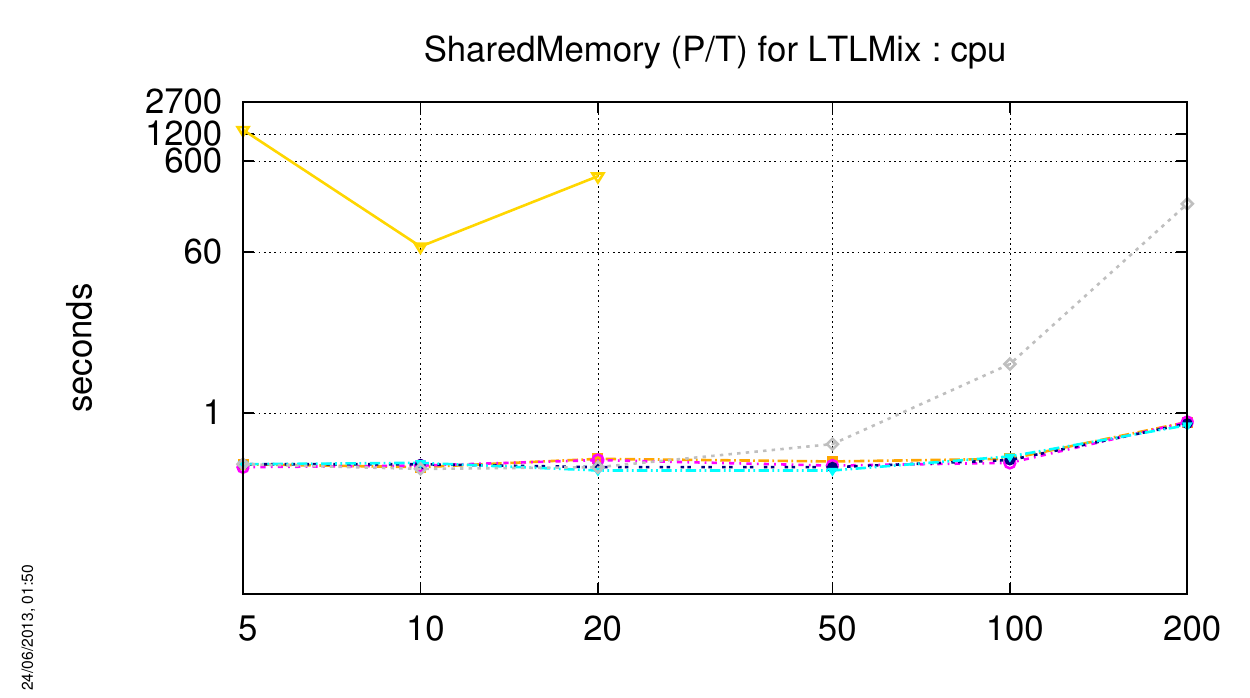}

   \includegraphics[height=1cm]{figures/tools-legend.pdf}
\end{center}

\subsubsection{\acs{SimpleLoadBal-COL}}
No instance of this model could be computed for the \textbf{LTLMix} examination.

\subsubsection{\acs{SimpleLoadBal-PT}}
The charts below respectively show how tools compete with this ``Known'' model (memory and CPU).

\index{Performances!LTLMix!SimpleLoadBal (P/T)}
\begin{center}
   \includegraphics[width=7.2cm]{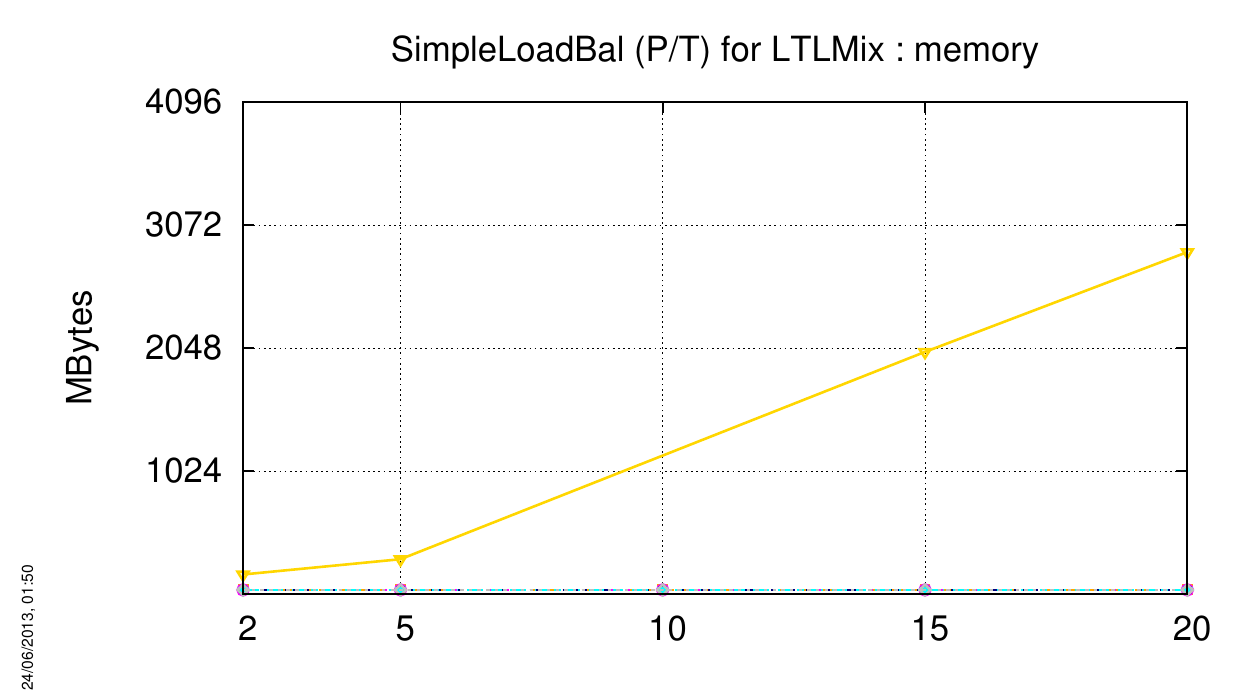}
   \includegraphics[width=7.2cm]{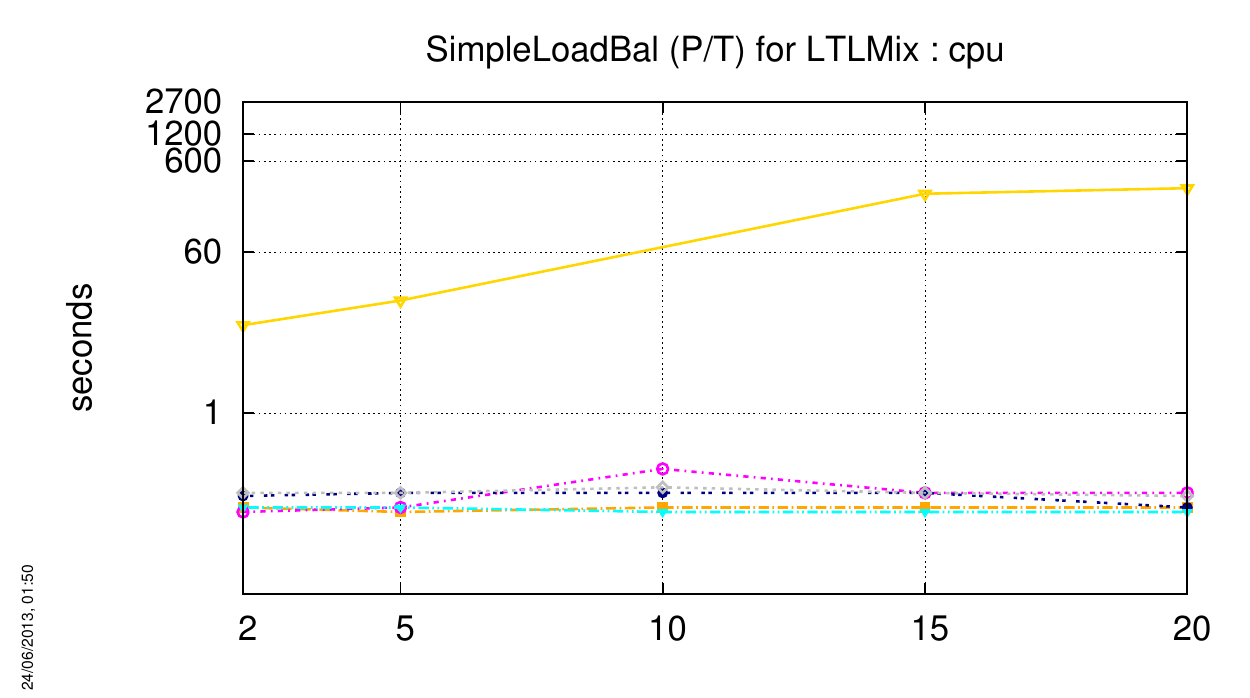}

   \includegraphics[height=1cm]{figures/tools-legend.pdf}
\end{center}

\subsubsection{\acs{TokenRing-COL}}
No instance of this model could be computed for the \textbf{LTLMix} examination.

\subsubsection{\acs{TokenRing-PT}}
The charts below respectively show how tools compete with this ``Known'' model (memory and CPU).

\index{Performances!LTLMix!TokenRing (P/T)}
\begin{center}
   \includegraphics[width=7.2cm]{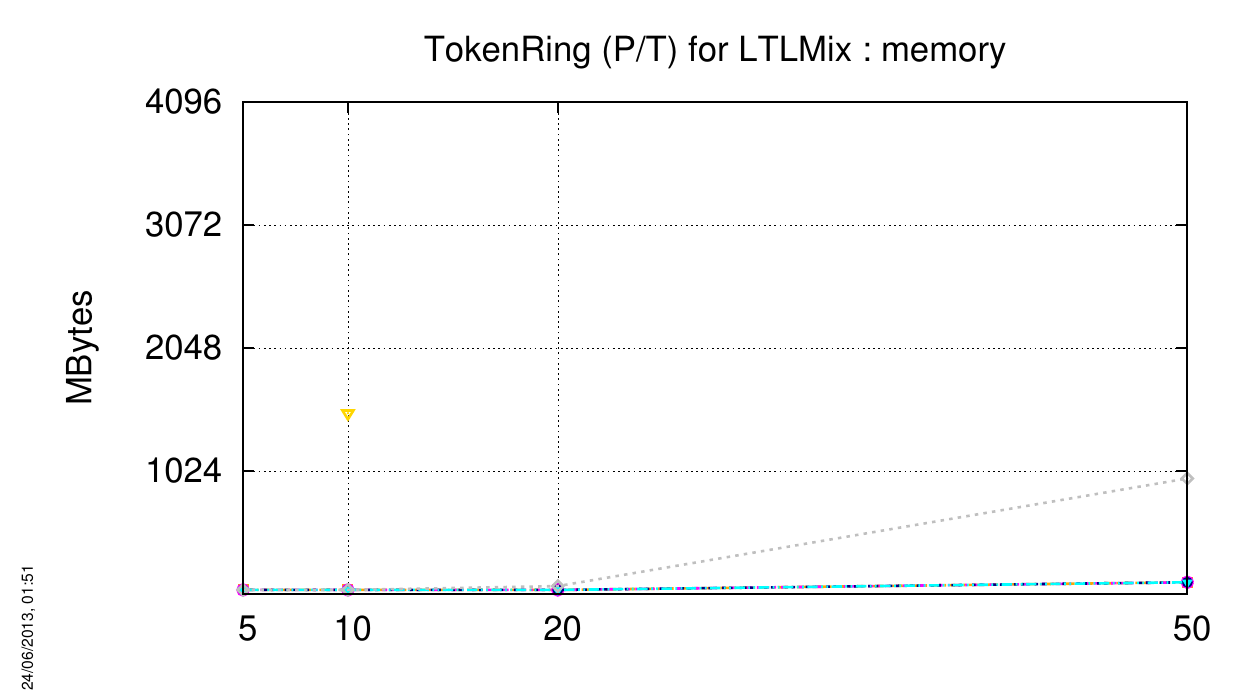}
   \includegraphics[width=7.2cm]{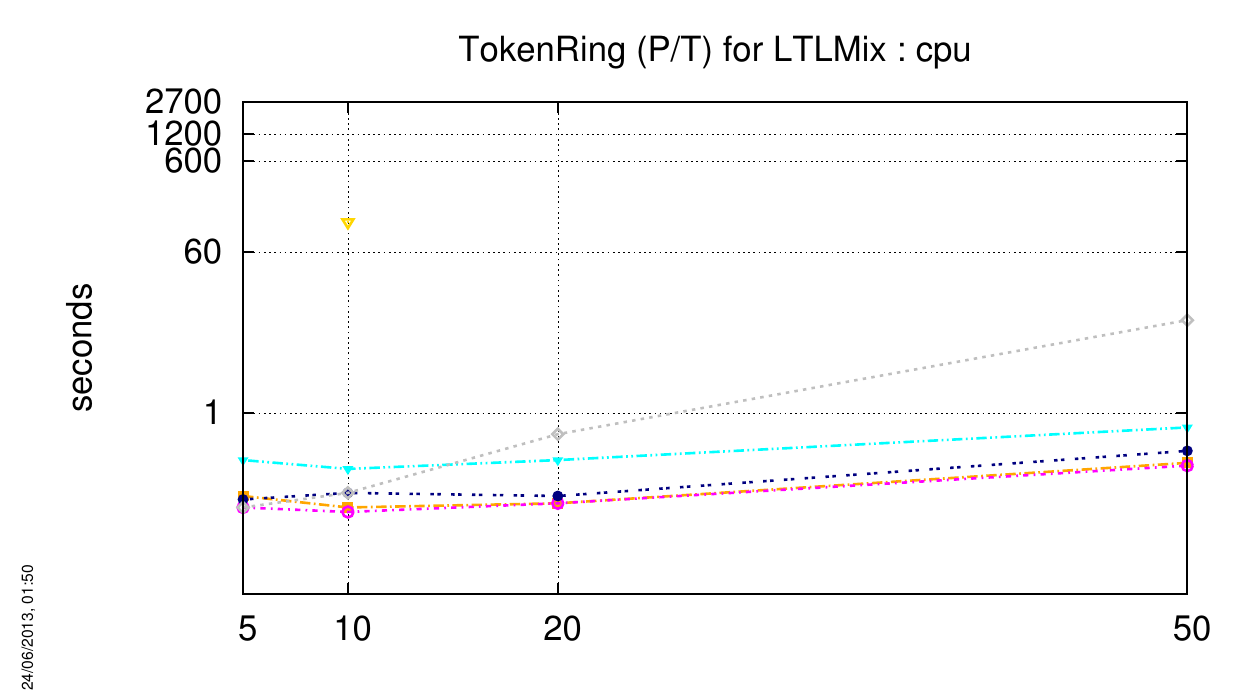}

   \includegraphics[height=1cm]{figures/tools-legend.pdf}
\end{center}

\subsubsection{\acs{HouseConstruction-PT}}
The charts below respectively show how tools compete with this ``Suprise'' model (memory and CPU).

\index{Performances!LTLMix!HouseConstruction (P/T)}
\begin{center}
   \includegraphics[width=7.2cm]{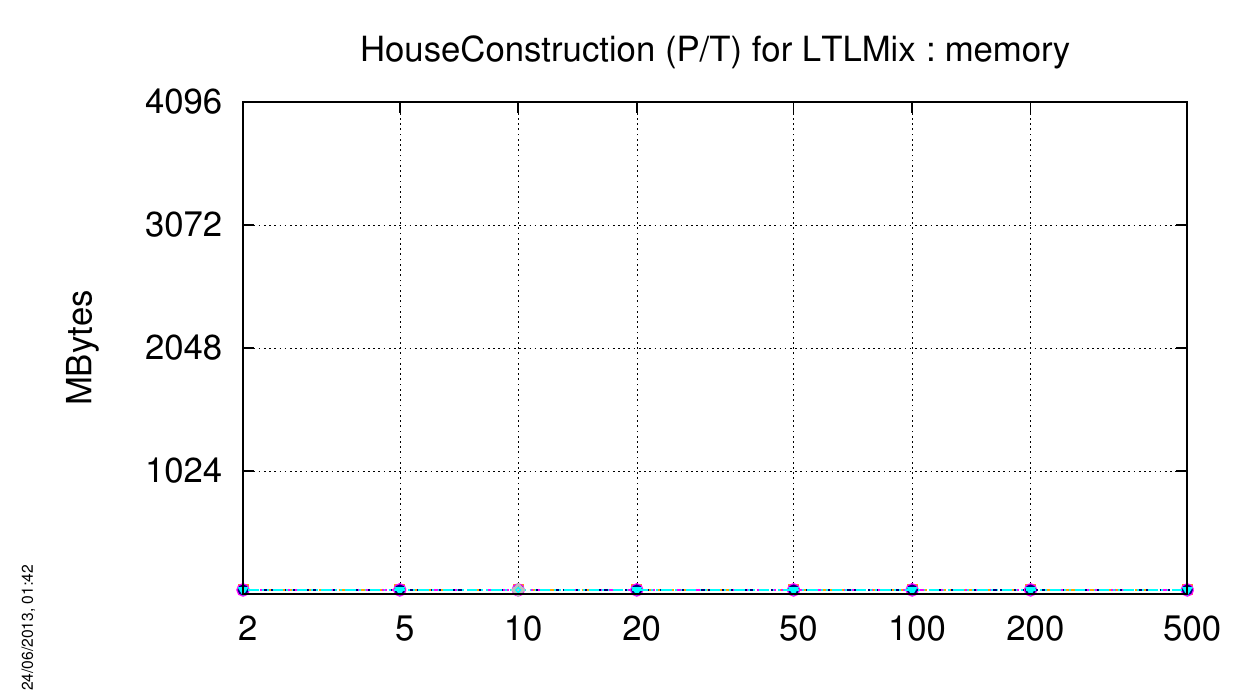}
   \includegraphics[width=7.2cm]{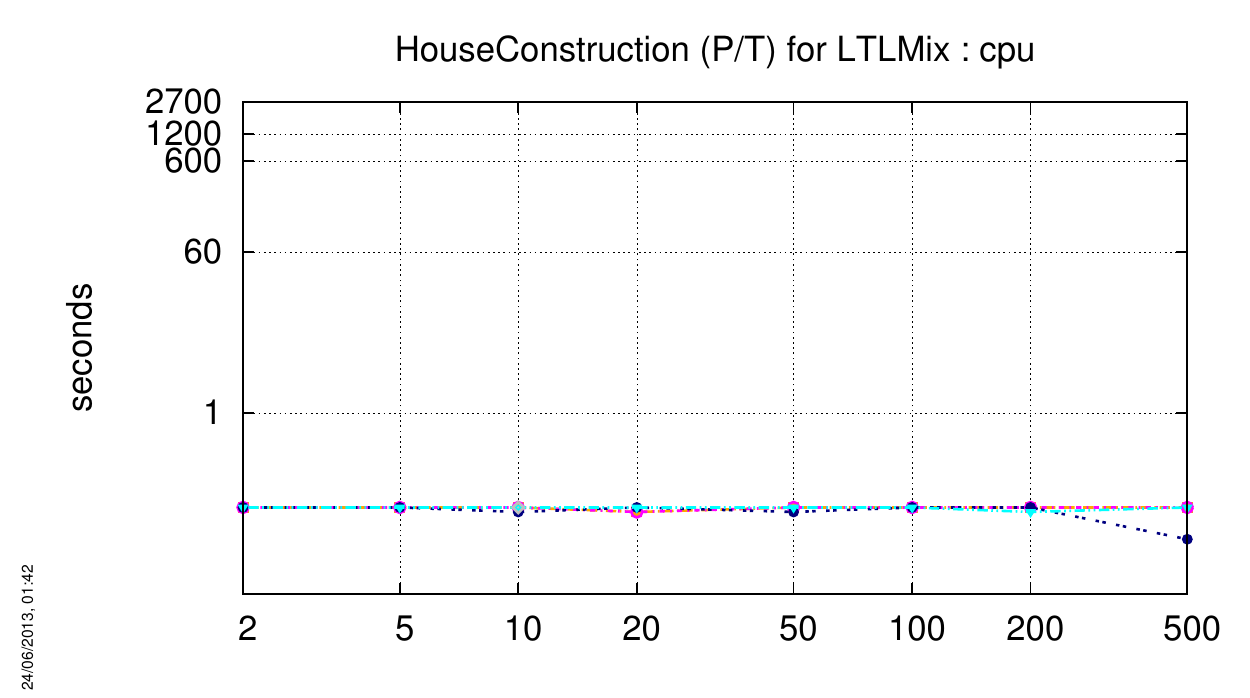}

   \includegraphics[height=1cm]{figures/tools-legend.pdf}
\end{center}

\subsubsection{\acs{IBMB2S565S3960-PT}}
The charts below respectively show how tools compete with this ``Suprise'' model (memory and CPU).

\index{Performances!LTLMix!IBMB2S565S3960 (P/T)}
\begin{center}
   \includegraphics[width=7.2cm]{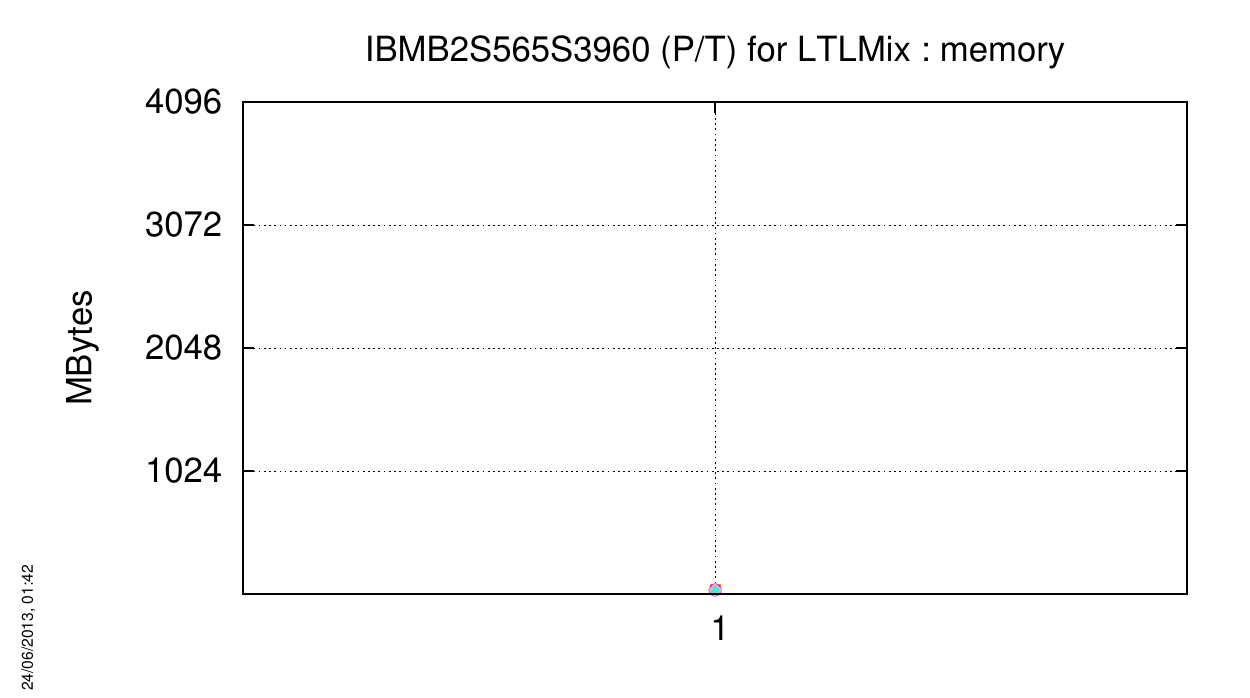}
   \includegraphics[width=7.2cm]{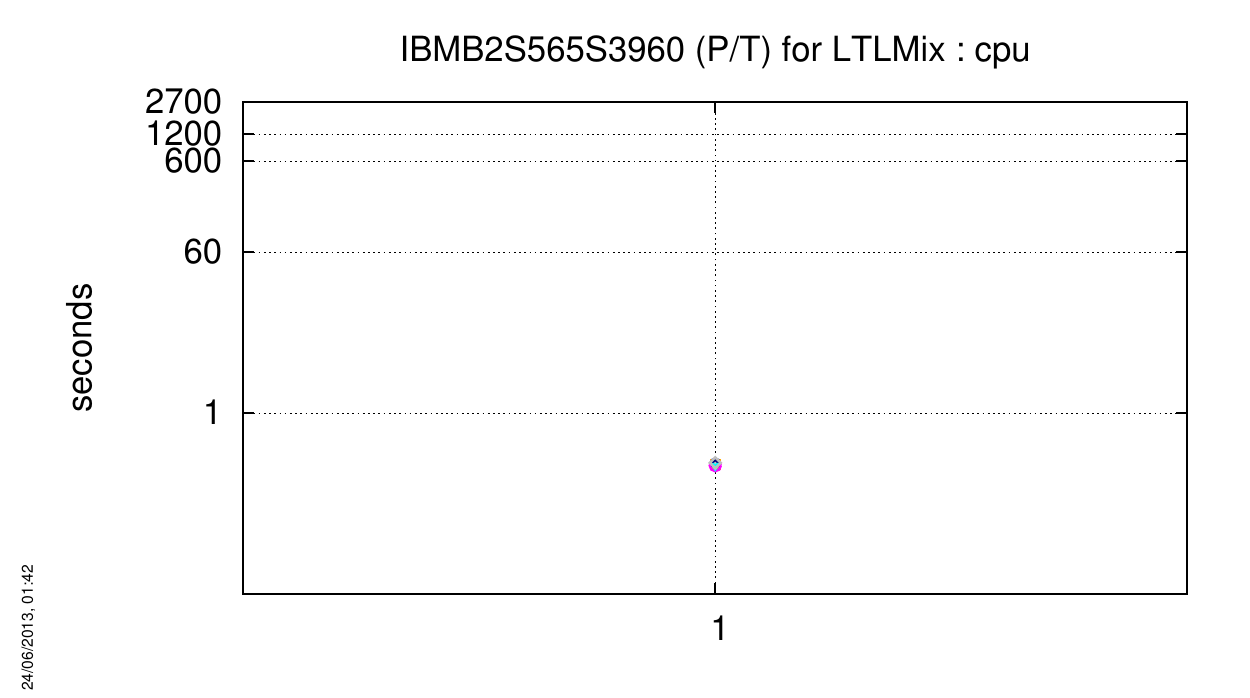}

   \includegraphics[height=1cm]{figures/tools-legend.pdf}
\end{center}

\subsubsection{\acs{QuasiCertifProtocol-COL}}
No instance of this model could be computed for the \textbf{LTLMix} examination.

\subsubsection{\acs{QuasiCertifProtocol-PT}}
The charts below respectively show how tools compete with this ``Suprise'' model (memory and CPU).

\index{Performances!LTLMix!QuasiCertifProtocol (P/T)}
\begin{center}
   \includegraphics[width=7.2cm]{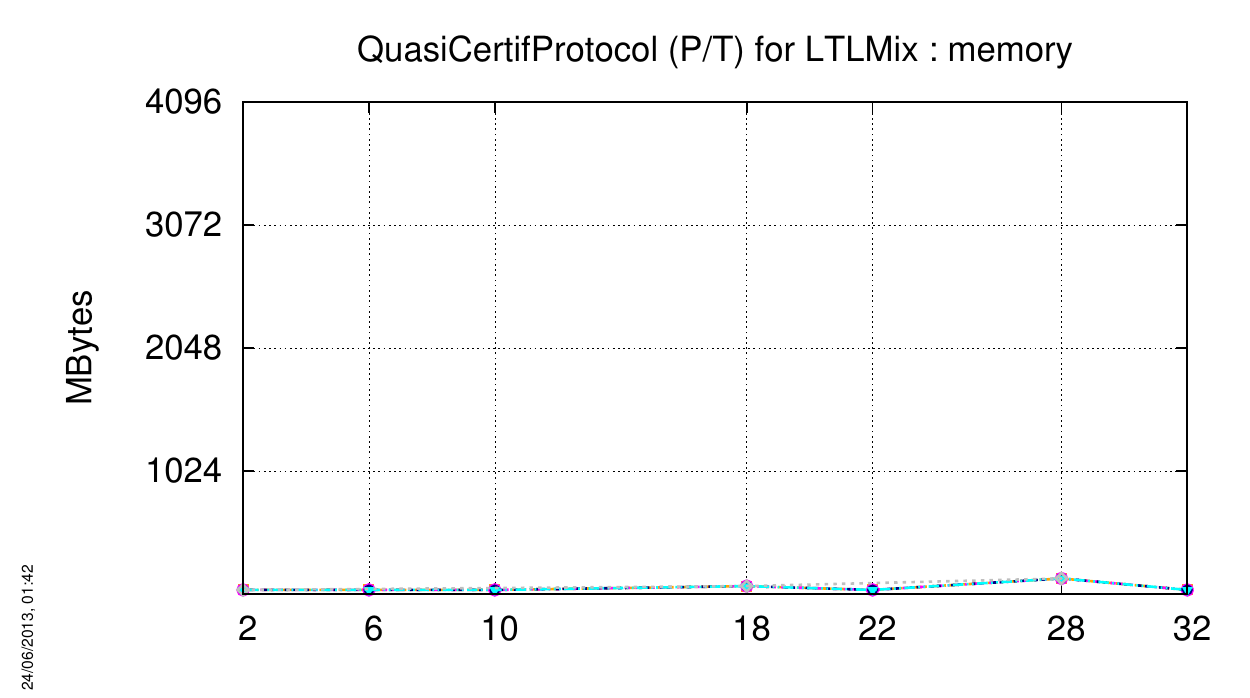}
   \includegraphics[width=7.2cm]{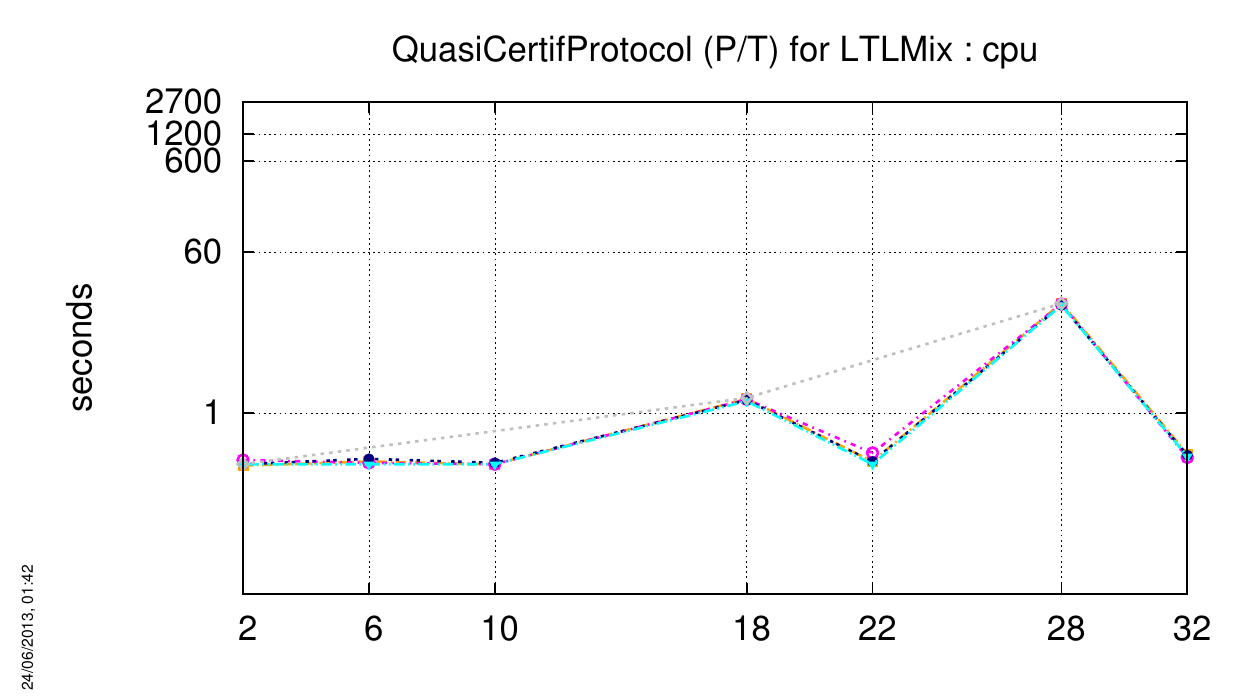}

   \includegraphics[height=1cm]{figures/tools-legend.pdf}
\end{center}

\subsubsection{\acs{Vasy2003-PT}}
The charts below respectively show how tools compete with this ``Suprise'' model (memory and CPU).

\index{Performances!LTLMix!Vasy2003 (P/T)}
\begin{center}
   \includegraphics[width=7.2cm]{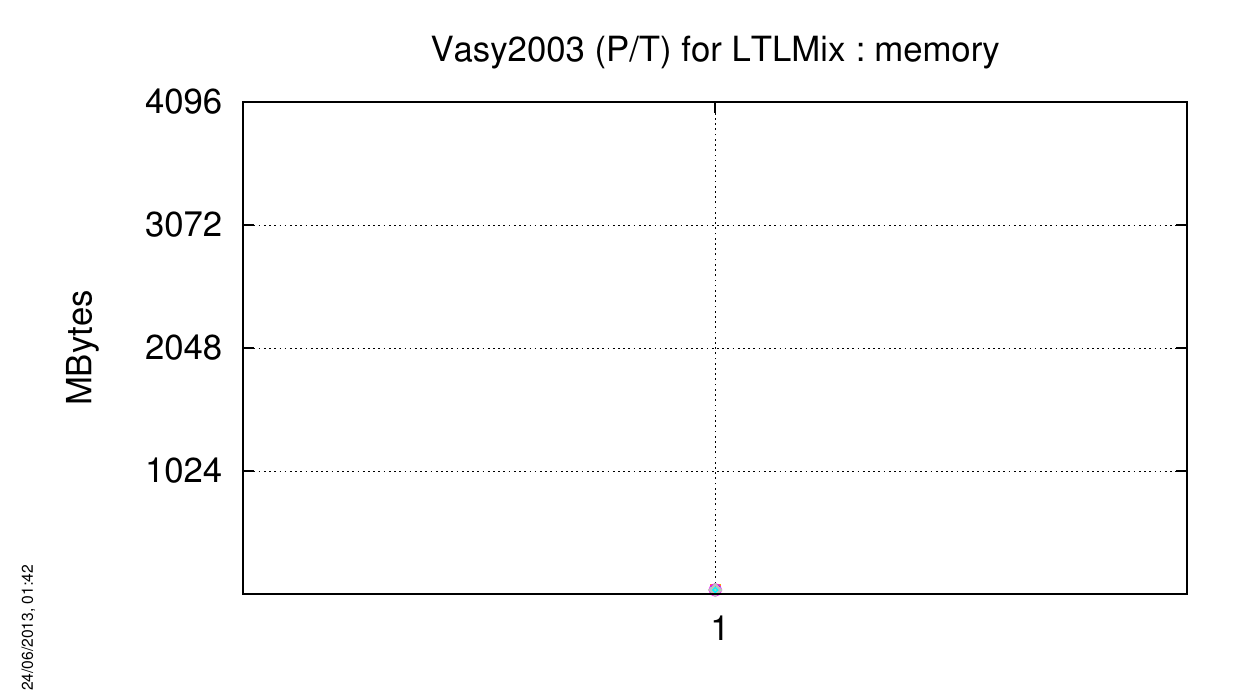}
   \includegraphics[width=7.2cm]{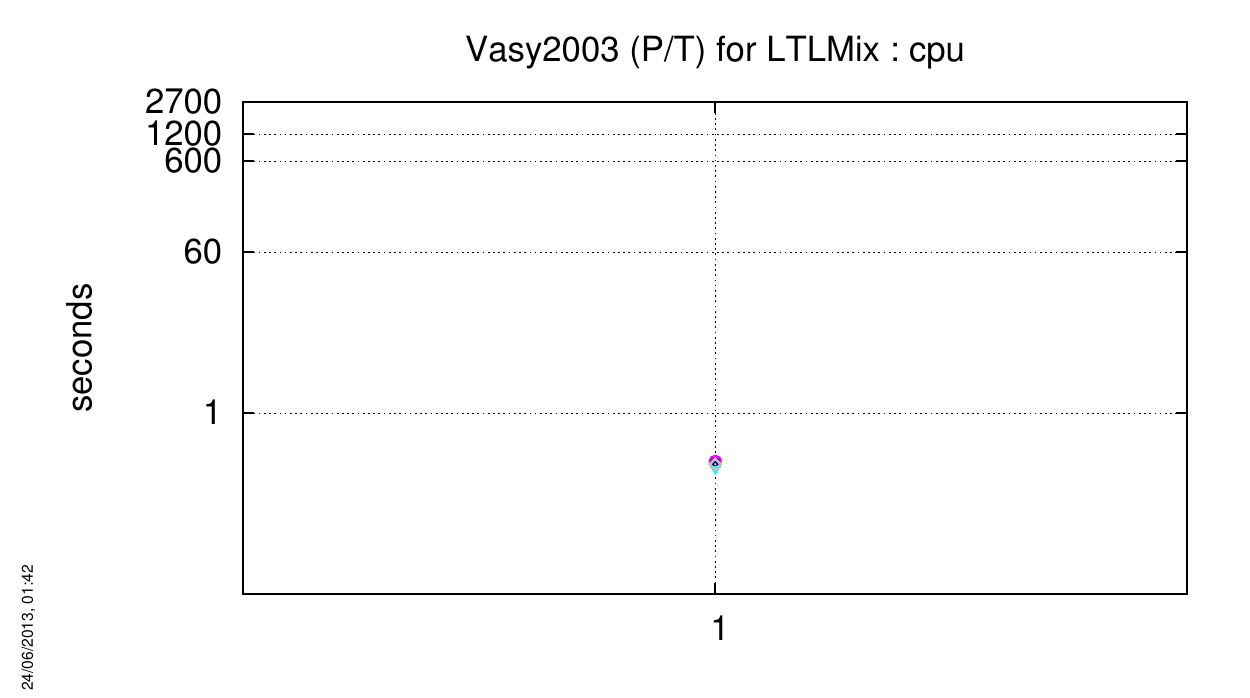}

   \includegraphics[height=1cm]{figures/tools-legend.pdf}
\end{center}

\subsection{Outputs for the LTLMix Examination}
\index{Outputs!LTLMix}

Please find enclosed the brute results for this examination (``Known'' and ``Surprise'' models).
We display only the score of tools that provide a results for at least one instance of one model.
The legend for the values is provided below:
\begin{itemize}
   \item\textbf{nc}: the tool does not compete this examination for this model/instance,
   \item\textbf{cc}: the tool cannot compute this examination for this model/instance,
   \item\textbf{to}: the tool cannot compute this examination for this model/instance within the maximum allowed time,
   \item\textbf{mp}: the tool encountered a memory problem (stack overflow or memory full),
   \item\textbf{nf}: there is no formula available for this type of examination (typically, this concerns P/T nets where
       comparing marking cardinality has no signification when there is no equivalent colored net).
\end{itemize}

\textbf{Note on the display of results for formulas:} each formula is considered as a flag (F if false, T if true, - or ?
when the value cannot be determined). These values are concatenated in the order they appear (we assume it is the order of formulas as they were provided).

\subsubsection{``Known'' Models}

\input{result_known_LTLMix.tex}

\subsubsection{``Surprise'' Models}

\input{result_surprise_LTLMix.tex}

\subsection{Score for the LTLMix Examination}
\index{Scores!LTLMix}

Please find enclosed the scores for this examination (``Known'' and ``Surprise'' models).
We display only the score of tools that provide a results for at least one instance of one model.
The total is first listed in the table below followed by a detail, for each proposed model.
Meaning of the line labels are:
\begin{itemize}
\item\textbf{1st instance}: the tool gets a bonus for having processed the first instance of this model (+1 point),
\item\textbf{instances}: the tool gets 1 point per instances treated 
(for that, we assume that at least one formula has been successfully computed),
\item\textbf{max reached}: the tool could process all the instances for the model (+2 points),
\item\textbf{best}: the tool is among the ones that processed a maximum of instances within the time and memory confinement (+2 points).
\end{itemize}

\subsubsection{``Known'' Models}

\input{score_known_LTLMix.tex}

\subsubsection{``Surprise'' Models}

\input{score_surprise_LTLMix.tex}

\subsection{Trophies for this Examination}
\index{Trophies!LTLMix}

Trophies are divided in three categories: ``Known'' models,
``Surprise'' models, and the global trophies (formula is then
$score_{global} = score_{known} + 2 \times score_{surprise}$).

\subsubsection{For ``Known'' Models} \ \\

\begin{tabular}{c}
      1 \\
   \includegraphics[width=2cm]{figures/gold.jpg} \\
   \acs{neco} \\
   110 points \\
\end{tabular}

\subsubsection{For ``Surprise'' Models}\  \\

No tool could complete this examination.

\subsubsection{Global} \ \\

\begin{tabular}{c}
      1 \\
   \includegraphics[width=2cm]{figures/gold.jpg} \\
   \acs{neco} \\
   110 points \\
\end{tabular}

\cleardoublepage


\cleardoublepage
\bibliographystyle{splncs03}
\addcontentsline{toc}{chapter}{References}
\bibliography{mcc-report}

{\footnotesize\printindex}

\cleardoublepage
\thispagestyle{empty}
\mbox{}
\clearpage
\thispagestyle{empty}
\renewcommand\CoverPicture{
\put(-5,-100){
\parbox[b][\paperheight]{\paperwidth}{%
\vfill
\centering
\includegraphics[width=23.5cm,keepaspectratio]{cover/cover-MCC.jpg}%
\vfill
}}}
\AddToShipoutPicture{\CoverPicture}
\mbox{}
\end{document}